\def\issubmit{1}
\def \issubmit{1}

\ifdefined \issubmit \else \def \issubmit{0} \fi

\newif \ifsubmit {} \if \issubmit 0 \submitfalse \else \submittrue \fi

\documentclass[acmsmall,screen]{acmart}
\setcopyright{acmcopyright}
\acmPrice{15.00}
\acmDOI{10.1145/000000}
\copyrightyear{2023}
\acmJournal{TOPLAS}
\acmVolume{1}
\acmNumber{1}
\acmMonth{1}
\acmYear{2023}

\usepackage{yhmath}
\usepackage[T5,T1]{fontenc}

\usepackage{longtable}
\usepackage{multirow}
\usepackage{xspace}

\usepackage{amsmath}

\usepackage{graphicx}
\graphicspath{ {../images/} }

\usepackage{wrapfig}
\usepackage{soul}

\makeatletter
\newcommand*\bigcdot{\mathpalette\bigcdot@{.5}}
\newcommand*\bigcdot@[2]{\mathbin{\vcenter{\hbox{\scalebox{#2}{$\m@th#1\bullet$}}}}}
\makeatother

\usepackage{microtype}                % Better typesetting and layout algorithms

\usepackage{float}                    % Float layout placement 
\usepackage{placeins}                 % and management 

\usepackage{hyperref}                 % Hyperlinked references
\hypersetup{
    colorlinks,
    linkcolor={red!50!black},
    citecolor={blue!50!black},
    urlcolor={blue!80!black}
}

\usepackage{amsmath}                  % AMS math tools
\usepackage{amsfonts}                 % ...
\usepackage{amsthm}                   % ...
\usepackage{mathtools}                % ...

\newtheorem*{fact*}{Fact}                       % ...
\newtheorem{definition}{Definition}[section]    % ...
\newtheorem*{definition*}{Definition}           % ...
\newtheorem{proposition}{Proposition}[section]  % ...
\newtheorem*{proposition*}{Proposition}         % ...
\newtheorem{theorem}{Theorem}[section]          % ...
\newtheorem*{theorem*}{Theorem}                 % ...
\newtheorem{lemma}[theorem]{Lemma}              % ...
\newtheorem*{lemma*}{Lemma}                     % ...
        % ...
\newtheorem*{sublemma*}{Sublemma}               % ...
\newtheorem{corrolary}{Corrolary}[theorem]      % ...
\newtheorem*{corrolary*}{Corrolary}             % ...

\usepackage{natbib}                   % Better bibs

\usepackage{stmaryrd}                 % More symbols
\usepackage{pifont}                   % ...
\usepackage{textcomp}                 % ...

\usepackage{centernot}                % slashing things

\usepackage{stackengine}              % Symbol Stacking
\stackMath

\usepackage{nicefrac}                 % Nice Fractions

\usepackage{array}                    % Array Layout
\usepackage{tabularx}                 % Improved Tabular Layout

\usepackage{framed}                   % Framed Environment

\usepackage{lipsum}                   % Easy placeholder text generation

\usepackage{enumitem}                 % Better control over itemize/enumerate

\usepackage{mathpartir}               % Flexible math display layout

\usepackage{galois}                   % Galois connections

\usepackage{geometry}                 % Margins etc. (define this locally)

\usepackage{thmtools}
\usepackage{tikz}

\usepackage{relsize}

\newcommand{\mtext}[1]{\ifmmode\operatorname{\mathrm{#1}}\else\textnormal{#1}\fi}
\newcommand{\mtexttt}[1]{\ifmmode\operatorname{\mathtt{#1}}\else\textnormal{\texttt{#1}}\fi}
\newcommand{\mtextit}[1]{\ifmmode\operatorname{\mathit{#1}}\else\textnormal{\textit{#1}}\fi}
\newcommand{\mtextbf}[1]{\ifmmode\operatorname{\mathbf{#1}}\else\textnormal{\textbf{#1}}\fi}
\newcommand{\mtextsc}[1]{\ifmmode\operatorname{\textsc{\smaller #1}}\else\textnormal{\textsc{\smaller #1}}\fi}

\definecolor{cbsafeABright}{RGB}{0,72,158} % tory blue
\definecolor{cbsafeADark}{RGB}{0,48,111}   % madison

\definecolor{cbasifABright}{RGB}{59,59,158} % **tory blue
\definecolor{cbasifADark}{RGB}{39,39,111}   % **madison

\definecolor{cbsafeBBright}{RGB}{0,149,3}  % india green
\definecolor{cbsafeBDark}{RGB}{0,90,1}     % san felix

\definecolor{cbasifBBright}{RGB}{124,124,24} % **india green
\definecolor{cbasifBDark}{RGB}{74,74,11}   % **san felix

\definecolor{cbsafeCBright}{RGB}{199,0,124} % magenta
\definecolor{cbsafeCDark}{RGB}{86,1,51}    % mulberry

\definecolor{cbasifCBright}{RGB}{120,120,119} % **magenta
\definecolor{cbasifCDark}{RGB}{49,49,49}   % **mulberry

\newcommand{\colorMATHA}{cbsafeABright}
\newcommand{\colorSYNTAXA}{cbsafeABright}
\newcommand{\colorMATHB}{cbsafeBBright}
\newcommand{\colorSYNTAXB}{cbsafeBBright}
\newcommand{\colorMATHC}{cbsafeCBright}
\newcommand{\colorSYNTAXC}{cbsafeCBright}

\newcommand{\colorTEXT}{black}

\newcommand{\colorMATH}{\colorMATHA}
\newcommand{\colorSYNTAX}{\colorSYNTAXA}

\usepackage{upgreek}
\usepackage[normalem]{ulem}

\ifsubmit
\newcommand{\mynote}[2]{}
\else
\newcommand{\mynote}[2]
    {{\color{red} \fbox{\bfseries\sffamily\scriptsize#1}
    {\small$\blacktriangleright$\textsf{\emph{#2}}$\blacktriangleleft$}}~}
\fi

\newcommand{\mt}[1]{\mynote{MT}{#1}}

\ifsubmit
\newcommand{\add}[1]{}
\newcommand{\remove}[1]{}
\newcommand{\replace}[2]{}
\else
\newcommand{\add}[1]{{\color{green!75!black}#1}}
\newcommand{\remove}[1]{{\color{red}\sout{#1}}}
\newcommand{\replace}[2]{\add{#2}\remove{#1}}
\fi

\newcommand{\fuzz}{\textsc{Fuzz}\xspace}
\newcommand{\dfuzz}{\textsc{DFuzz}\xspace}
\newcommand{\fuzzed}{\textsc{Fuzz}$^{\epsilon\delta}$\xspace}
\newcommand{\fuzzi}{\textsc{Fuzzi}\xspace}
\newcommand{\duet}{\textsc{Duet}\xspace}
\newcommand{\system}{\textsc{Jazz}\xspace}
\newcommand{\ssystem}{\textsc{Sax}\xspace}
\newcommand{\hoaresq}{\textsc{HOARe$^2$}\xspace}
\newcommand{\apRHL}{\textsc{apRHL}\xspace}
\newcommand{\apRHLplus}{\textsc{apRHL$^+$}\xspace}
\newcommand{\spanapRHL}{\textsc{span-apRHL}\xspace}

\newcommand{\oversetcustom}[2]{\overset{#1}{\vphantom{._1}\smash{{}#2{}}}}

\newcommand{\lang}{\lambda_{\text{\textsc{J}}}\xspace}

\newcommand{\instE}[2]{#1 \mathord{\cdotp } #2}

\newcommand{\addProduct}[2]{(#1,#2)}

\newcommand{\slambda}{{\color{\colorSYNTAXB}\lambda^{\!^{{}_{\mathrm{s}}}}}}
\newcommand{\sS}{\Sigma}
\newcommand{\se}{e}
\newcommand{\sv}{v}
\newcommand{\sss}{s}

\newcommand{\plambda}{{\color{\colorSYNTAXC}\lambda^{\!^{{}_{\mathrm{p}}}}}}
\newcommand{\pS}{\Upsigma}
\newcommand{\pe}{\mathsf{e}}

\newcommand{\xrightarrowdbl}[2][]{%
  \xrightarrow[#1]{#2}\mathrel{\mkern-14mu}\rightarrow
}

\newcommand{\xrightarrowS}[1]{\mathrel{{\begingroup\renewcommand\colorMATH{\colorMATHB}\renewcommand\colorSYNTAX{\colorSYNTAXB}{{\color{\colorMATH}\ensuremath{\xrightarrow{{\begingroup\renewcommand\colorMATH{\colorMATHA}\renewcommand\colorSYNTAX{\colorSYNTAXA}{{\color{\colorMATH}\ensuremath{#1}}}\endgroup }}}}}\endgroup }}}
\newcommand{\xrightarrowP}[1]{\mathrel{{\begingroup\renewcommand\colorMATH{\colorMATHC}\renewcommand\colorSYNTAX{\colorSYNTAXC}{{\color{\colorMATH}\ensuremath{\xrightarrowdbl{{\begingroup\renewcommand\colorMATH{\colorMATHA}\renewcommand\colorSYNTAX{\colorSYNTAXA}{{\color{\colorMATH}\ensuremath{#1}}}\endgroup }}}}}\endgroup }}}

\newcommand{\inr}{{\begingroup\renewcommand\colorMATH{\colorMATHB}\renewcommand\colorSYNTAX{\colorSYNTAXB}{{\color{\colorSYNTAX}\texttt{inr}}}\endgroup }}
\newcommand{\inl}{{\begingroup\renewcommand\colorMATH{\colorMATHB}\renewcommand\colorSYNTAX{\colorSYNTAXB}{{\color{\colorSYNTAX}\texttt{inl}}}\endgroup }}
\newcommand{\inlr}{{\begingroup\renewcommand\colorMATH{\colorMATHB}\renewcommand\colorSYNTAX{\colorSYNTAXB}{{\color{\colorSYNTAX}\texttt{in(l/r)}}}\endgroup }}
\newcommand{\ccase}{{\begingroup\renewcommand\colorMATH{\colorMATHB}\renewcommand\colorSYNTAX{\colorSYNTAXB}{{\color{\colorSYNTAX}\texttt{case}}}\endgroup }}
\newcommand{\of}{{\begingroup\renewcommand\colorMATH{\colorMATHB}\renewcommand\colorSYNTAX{\colorSYNTAXB}{{\color{\colorSYNTAX}\texttt{of}}}\endgroup }}
\newcommand{\fst}{{\begingroup\renewcommand\colorMATH{\colorMATHB}\renewcommand\colorSYNTAX{\colorSYNTAXB}{{\color{\colorSYNTAX}\texttt{fst}}}\endgroup }}
\newcommand{\snd}{{\begingroup\renewcommand\colorMATH{\colorMATHB}\renewcommand\colorSYNTAX{\colorSYNTAXB}{{\color{\colorSYNTAX}\texttt{snd}}}\endgroup }}
\newcommand{\tlet}{{\begingroup\renewcommand\colorMATH{\colorMATHB}\renewcommand\colorSYNTAX{\colorSYNTAXB}{{\color{\colorSYNTAX}\texttt{let}}}\endgroup }}
\newcommand{\tin}{{\begingroup\renewcommand\colorMATH{\colorMATHB}\renewcommand\colorSYNTAX{\colorSYNTAXB}{{\color{\colorSYNTAX}\texttt{in}}}\endgroup }}
\newcommand{\ttt}{{\begingroup\renewcommand\colorMATH{\colorMATHB}\renewcommand\colorSYNTAX{\colorSYNTAXB}{{\color{\colorSYNTAX}\texttt{tt}}}\endgroup }}
\newcommand{\sif}{{\begingroup\renewcommand\colorMATH{\colorMATHB}\renewcommand\colorSYNTAX{\colorSYNTAXB}{{\color{\colorSYNTAX}\texttt{if}}}\endgroup }}
\newcommand{\sthen}{{\begingroup\renewcommand\colorMATH{\colorMATHB}\renewcommand\colorSYNTAX{\colorSYNTAXB}{{\color{\colorSYNTAX}\texttt{then}}}\endgroup }}
\newcommand{\selse}{{\begingroup\renewcommand\colorMATH{\colorMATHB}\renewcommand\colorSYNTAX{\colorSYNTAXB}{{\color{\colorSYNTAX}\texttt{else}}}\endgroup }}
\newcommand{\distance}{d}
\newcommand{\Distance}{\Delta}
\newcommand{\distanceName}{relational distance\xspace}
\newcommand{\distanceBoundName}{relational distance\xspace}
\newcommand{\subst}[3][{\begingroup\renewcommand\colorMATH{\colorMATHB}\renewcommand\colorSYNTAX{\colorSYNTAXB}{{\color{\colorMATH}\ensuremath{\Distance}}}\endgroup }]{[#2/#3]}
\newcommand{\Jc}[3][]{{\underset{#1}{\bigsqcup}#2#3}}

\newcommand{\instPE}[2]{#1 {\begingroup\renewcommand\colorMATH{\colorMATHC}\renewcommand\colorSYNTAX{\colorSYNTAXC}{{\color{\colorMATH}\ensuremath{\bigcdot}}}\endgroup } #2}

\newcommand{\longnamep}{}

\newcommand{\toplas}[1]{{#1}}
\newcommand{\toplass}[1]{{#1}}
\newcommand{\toplasss}[1]{{#1}}
\newcommand{\toplassss}[1]{{ #1}}

\newcommand{\dom}{\mathit{dom}}
\newcommand{\multProd}[2]{#1 + #2}
\newcommand{\addProd}[2]{#1 \sqcup  #2}
\newcommand{\dist}[1][]{D_{#1}}
\newcommand{\distp}[1][]{D'_{#1}}
\newcommand{\distpp}[1][]{D''_{#1}}
\newcommand{\Sup}[1]{\mathtt{Sup}(#1)}



\begin{document}

% {-{ authors

\title{
  \ifsubmit
  \else
  !!\ COMMENTS ARE ON --- DO NOT SUBMIT\ !! \\
  \fi
  Contextual Linear Types for Differential Privacy
  }
  \titlenote{This work is partially funded by ANID FONDECYT Projects 11181208, 1190058, 3200583,
ANID Millennium Science Initiative Program code ICN17\_002, and NSF award CCF-2119939.}

%\author{Mat\'ias Toro, David Darais, Chike Abuah, Joe Near, Federico Olmedo, and \'Eric tanter}
% Authors:
%  Matías Toro
%  David Darais
%  Chike Abuah
%  Joe Near
%  Federico Olmedo
%  \'Eric Tanter

\author{Mat\'ias Toro}
\affiliation{%
  \institution{Computer Science Department (DCC), University of Chile}
  \city{Santiago}
  \country{Chile}
  % \email{mtoro@dcc.uchile.cl}
}

\author{David Darais}
\affiliation{%
  \institution{Galois, Inc.}
  \city{Portland}
  \country{USA}
  % \email{david.darais@uvm.edu}
}
\authornote{Work done in part while at University of Vermont.}

\author{Chike Abuah}
\affiliation{%
  \institution{Computer Science Department, University of Vermont}
  \city{Burlington}
  \country{USA}
  % \email{cabuah@uvm.edu}
}

\author{Joe Near}
\affiliation{%
  \institution{Computer Science Department, University of Vermont}
  \city{Burlington}
  \country{USA}
  % \email{jnear@uvm.edu}
}

\author{Dami\'an \'Arquez}
\affiliation{%
  \institution{Computer Science Department (DCC), University of Chile \& IMFD}
  \city{Santiago}
  \country{Chile}
  % \email{folmedo@dcc.uchile.cl}
}

\author{Federico Olmedo}
\affiliation{%
  \institution{Computer Science Department (DCC), University of Chile
    \& IMFD}
  \city{Santiago}
  \country{Chile}
  % \email{folmedo@dcc.uchile.cl}
}

\author{\'Eric Tanter}
\affiliation{%
  \institution{Computer Science Department (DCC), University of Chile \& IMFD}
  \city{Santiago}
  \country{Chile}
  % \email{etanter@dcc.uchile.cl}
}

% }-}

\begin{abstract} % {-{
Language support for differentially-private programming is both crucial and delicate. While elaborate program logics can be very expressive, type-system based approaches using linear types tend to be more lightweight and amenable to automatic checking and inference, and in particular in the presence of higher-order programming. Since the seminal design of \fuzz, which is restricted to $\epsilon $-differential privacy \toplass{in its original design}, \toplass{significant progress} has been made to support more advanced variants of differential privacy, like $(\epsilon ,\delta )$-differential privacy. 
However, supporting these advanced privacy variants while also supporting higher-order programming in full has proven to be challenging.
%However, no existing type system supports these advanced privacy variants while also supporting higher-order programming in full generality. 
We present \system, a language and type system which uses linear types and latent contextual effects to support both advanced variants of differential privacy and higher-order programming. \toplass{Latent contextual effects allow delaying the payment of effects for connectives such as products, sums and functions, yielding advantages in terms of precision of the analysis and annotation burden upon elimination, as well as modularity.}
 We formalize the core of \system, prove it sound for privacy via a logical relation for metric preservation, and
illustrate its expressive power through a number of case studies drawn from the recent differential privacy literature.
\end{abstract} % }-}

\maketitle

\section{Introduction} % {-{
\paragraph{Note} {\textit{This paper uses colorblind-friendly colors in
notation to convey information, and is best consumed using an
electronic device or color printer.}}

Over the past decade, differential privacy~\cite{dwork2014algorithmic} has become the de-facto gold standard in protecting the privacy of individuals when processing sensitive data. In contrast to traditional approaches like de-identification, differential privacy provides a formal, composable privacy guarantee. Differentially private algorithms typically protect privacy by selecting from a handful of basic \emph{mechanisms} to perturb their outputs. For example, the \emph{Laplace mechanism} can be used to add noise to the population count of a city to prevent an adversary from successfully guessing whether or not a particular individual lives in that city. Most programming-language-based approaches to differential privacy are applied to verifying either the {\textit{implementation}} of a mechanism, such as the Exponential mechanism, or the {\textit{composition}} of multiple uses of mechanisms, such as computing a histogram using the Laplace mechanism (multiple times) as a primitive.

There are two challenges when writing differentially private
programs. First, noise must be added to the right values in the
program in order to achieve {\textit{some}} guarantee of privacy; this
includes the final output of the program, as well as many
intermediate program values.  Second, the {\textit{correct amount of noise}}
must be added in those places to achieve the {\textit{desired amount}} of
privacy. In the differential privacy framework, privacy is a
\toplas{quantitative feature}---more noise gives more privacy. Adding too little noise is as
ineffective as adding no noise at all, and adding too much noise
renders the result of the computation useless.
Programmers must therefore ensure they have added {\textit{enough noise}}, {\textit{in the
right places}}, and that the noise is {\textit{minimal}}---a daunting task.

Since differential privacy is a probabilistic, multi-run (hyper~\cite{hyperproperties}) property, it is not
straightforward to develop test cases for differentially private
algorithms. Consequently, differentially private algorithms are
usually developed by experts in the field, and these experts produce
manual proofs of privacy for each new algorithm. This reliance on
experts is limiting. First, there is a practical need for developing
privacy-preserving applications without access to an expert in
differential privacy. Even still, experts aren't perfect: for
example, several incorrect versions of the Sparse Vector
Technique~\cite{svt,dwork2014algorithmic} have appeared in published
papers~\cite{DBLP:journals/pvldb/LyuSL17}, despite being authored and peer-reviewed by experts
in differential privacy.

Due to these challenges, verifying differential privacy in programs via type checking has received considerable attention. The first such approach, \fuzz~\cite{reed2010distance}, uses linear types to verify pure $\epsilon $-differential privacy. \fuzz and its successor \dfuzz~\cite{gaboardi2013linear} have a number of attractive properties, including support for automation and higher-order programming. \fuzz was the first to use linear types to bound \emph{function sensitivity}: how much a function's output changes given a change to its input. Sensitivity is then used to determine the (minimal) amount of noise required to achieve privacy. \fuzz uses the same sensitivity type system to also track privacy, which is advantageous due to its simplicity, but as a consequence is unable to support advanced variants of differential privacy, like $(\epsilon ,\delta )$. \toplass{A recent approach \cite{de2019probabilistic} extends the terminating fragment of \fuzz using \toplasss{graded comonadic liftings} to support advanced variants such as $(\epsilon ,\delta )$-differential privacy. In the following, we call this extended language \fuzzed.} Another approach, \hoaresq, uses relational refinement types to encode differential privacy~\cite{Barthe:POPL:15}, and improves on \fuzz-like systems in its ability to support advanced variants.
%, but has limited support for automation. 
In general, type-based approaches like \fuzz and \hoaresq are used to verify programs which compose mechanisms, and not the implementations of mechanisms.

An alternative set of approaches use program logics~\cite{Barthe:POPL12,Barthe:TOPLAS:13,Barthe:LICS16,Sato:LICS19}
to verify both the implementations of mechanisms and simple forms of composing mechanisms, while also supporting advanced
variants like $(\epsilon , \delta )$-differential
privacy~\cite{dwork2014algorithmic}, zero-concentrated differential
privacy~\cite{bun2016concentrated}, and R\'enyi differential
privacy~\cite{mironov2017renyi}. However, these benefits come at the expense of
automation and support for higher-order programming.

The \duet language~\cite{near2019duet} and type system strikes a new
balance in this space by building on the designs of \fuzz and \dfuzz.
Like \fuzz, \duet supports automation and higher-order programming, and like \toplass{\fuzzed,} \hoaresq and recently developed program logics, \duet supports advanced variants of differential privacy.
Like all type-based approaches, \duet cannot be used to verify implementations of
mechanisms, however even when verifying programs which compose
mechanisms there is still room to improve: \duet is not expressive
enough to support higher-order programming in full
generality---something \fuzz, \dfuzz, \toplass{\fuzzed} and \hoaresq are each able to achieve.

This paper presents \system, the successor to \duet which significantly improves upon its design. \system is a linear type system \toplass{with support for} {\textit{latent contextual effects}} for function sensitivity and differential privacy; this combination supports advanced privacy variants (like \duet, \toplass{\fuzzed} and \hoaresq), automation (like \duet and \dfuzz), and \toplass{fully general} higher-order programming (like \fuzz/\dfuzz, \toplass{\fuzzed} and \hoaresq).
% \system (like \fuzz) while also  (like \duet and approaches based on program logics).
Like \duet, the \system language is built from two mutually-embedded sublanguages---one for sensitivity, and one for privacy---which allows it to support advanced variants of differential privacy automatically through typechecking. Also like \duet (and \fuzz/\dfuzz, \toplass{\fuzzed and \hoaresq}), \system is designed for verifying the composition of mechanisms, and not their direct implementation.
% \et{this last bit doesn't seem a useful characterization:} but is otherwise a completely redesigned language.

The key insight of \system is the incorporation of {\textit{latent contextual effects}} into a linear type system. A {\textit{latent}} effect is one that is {\textit{deferred}} or {\textit{delayed}}; rather than accounting for the effect immediately, it is tracked and accounted for later. A {\textit{contextual}} effect is one that tracks effect information for each variable in the context, including closure variables used in higher-order function bodies. \toplas{Technically, this is similar to the {\textit{open closure types}} introduced by Scherer and Hoffmann~\cite{schererHoffmann:lpar2013}, specialized to the tracking of sensitivity and privacy, and generalized to positive type constructors such as sums and products.}
In addition to supporting higher-order programming in the presence of advanced privacy variants, these latent contextual effects also 
\toplass{can yield advantages in terms of precision of the analysis, annotation burden, and modularity.}

\paragraph{The challenge of higher-order programming.}
Consider the {{\color{\colorMATH}\ensuremath{n}}}-iteration loop combinator in \fuzz, {{\color{\colorMATH}\ensuremath{{\text{loop}}_{n}}}}, which has type
{{\color{\colorMATH}\ensuremath{{{\color{\colorSYNTAX}\texttt{{\ensuremath{{{\color{\colorMATH}\ensuremath{\tau }}} \rightarrow  {{\color{\colorMATH}\ensuremath{({{\color{\colorSYNTAX}\texttt{{\ensuremath{{{\color{\colorMATH}\ensuremath{\tau }}} \rightarrow  {\scriptstyle \bigcirc } {{\color{\colorMATH}\ensuremath{\tau }}}}}}}})}}} \multimap _{{{\color{\colorMATH}\ensuremath{n}}}} {\scriptstyle \bigcirc } {{\color{\colorMATH}\ensuremath{\tau }}}}}}}}}}}. This type describes a two
argument function that takes some value of type {{\color{\colorMATH}\ensuremath{\tau }}} as the first
argument, a function as second argument (which accepts and returns
values of type {{\color{\colorMATH}\ensuremath{\tau }}}), and returns a final value of type {{\color{\colorMATH}\ensuremath{\tau }}}. The
modality {{\color{\colorMATH}\ensuremath{{\scriptstyle \bigcirc }}}} in the return type for the function argument and final
return type indicates that the function is probabilistic (due to \toplas{the} use
of differential privacy mechanisms), and \toplas{when appearing in the codomain of a linear arrow {{\color{\colorMATH}\ensuremath{\multimap }}}} indicates that the function
satisfies differential privacy.

Both function sensitivity and differential privacy are two-run (hyper)properties of
a function output {w.r.t.} some particular input. For example, \toplas{a function of body} {{\color{\colorMATH}\ensuremath{2x +
3y}}} is {{\color{\colorMATH}\ensuremath{2}}}-sensitive in {{\color{\colorMATH}\ensuremath{x}}} and {{\color{\colorMATH}\ensuremath{3}}}-sensitive in {{\color{\colorMATH}\ensuremath{y}}}, meaning that if {e.g.}
input {{\color{\colorMATH}\ensuremath{x}}} varies by at most {{\color{\colorMATH}\ensuremath{{\begingroup\renewcommand\colorMATH{\colorMATHB}\renewcommand\colorSYNTAX{\colorSYNTAXB}{{\color{\colorMATH}\ensuremath{\distance}}}\endgroup }}}} \toplasss{and {{\color{\colorMATH}\ensuremath{y}}} is held constant}, then the function output varies at
most by {{\color{\colorMATH}\ensuremath{2{\begingroup\renewcommand\colorMATH{\colorMATHB}\renewcommand\colorSYNTAX{\colorSYNTAXB}{{\color{\colorMATH}\ensuremath{\distance}}}\endgroup }}}}. When a closure
is created, the closure captures sensitivities
as well as values, so the sensitivity of the closure {{\color{\colorSYNTAX}\texttt{{\ensuremath{\lambda {{\color{\colorMATH}\ensuremath{x}}}.\hspace*{0.33em} {{\color{\colorMATH}\ensuremath{2x + 3y}}}}}}}}
would be ``{{\color{\colorMATH}\ensuremath{3}}} in {{\color{\colorMATH}\ensuremath{y}}}''. The situation is analogous when tracking
privacy and creating closures which capture privacy costs.
Looking back to the type of {{\color{\colorMATH}\ensuremath{{\text{loop}}_{n}}}} in \fuzz, the second argument
will be a closure whose captured environment tracks
a privacy cost for each closure variable.
The interpretation of the linear function type
{{\color{\colorMATH}\ensuremath{\multimap _{n}}}} is to {\textit{scale}} the privacy effects in the closure
environment of the looping function of type {{\color{\colorSYNTAX}\texttt{{\ensuremath{ {{\color{\colorMATH}\ensuremath{\tau }}} \rightarrow  {\scriptstyle \bigcirc } {{\color{\colorMATH}\ensuremath{\tau }}} }}}}} by {{\color{\colorMATH}\ensuremath{n}}}.
\toplas{We call}
this scaling {\textit{implicit}} and {\textit{pervasive}} in \fuzz{}
because it occurs at every let-binding and function call. \toplass{In the original \fuzz language,} such scaling is
only sound and precise for pure $\epsilon $-differential privacy, and as a consequence of this
pervasive scaling, \fuzz~ \toplass{could not} be instantiated to advanced
differential privacy variants\toplass{, until recently, where \fuzzed now support advanced variants such as $(\epsilon ,\delta )$-differential privacy through a path metric construction.}

The \duet language prohibits this pervasive scaling in its type system in
order to support advanced differential privacy variants, but
% as a consequence \toplass{{{\color{\colorMATH}\ensuremath{{\text{loop}}_{n}}}} cannot be assigned a ``plain'' type in \duet, and instead it must be given a custom typing rule}.
as a consequence \toplasss{it cannot initially derive a type for {{\color{\colorMATH}\ensuremath{{\text{loop}}_{n}}}}, and instead it must define a custom typing rule for {{\color{\colorMATH}\ensuremath{{\text{loop}}_{n}}}}}.
The issue is that \duet prohibits {\textit{all}} scaling of privacy
quantities. However, scaling {\textit{is}} allowable
({i.e.}, sound) in special restricted instances when using advanced
variants. The challenge is then to disallow implicit pervasive scaling
while allowing explicit restricted scaling.
\toplass{Because no type can be written for {{\color{\colorMATH}\ensuremath{{\text{loop}}_{n}}}} in \duet, it (and many other
higher-order functions) must be given explicit typing rules. This poses a significant restriction on higher-order programming, for instance {{\color{\colorMATH}\ensuremath{{\text{loop}}_{n}}}} cannot be lambda-abstracted in \duet.}

\system directly solves the challenge of encoding the explicit,
restricted scaling that is required to support both advanced
privacy variants and higher-order programming.
In \system, the type of the {{\color{\colorMATH}\ensuremath{n}}}-iteration construct
is: {{\color{\colorMATH}\ensuremath{{\text{loop}}_{n} \mathrel{:} {{\color{\colorSYNTAX}\texttt{{\ensuremath{ {{\color{\colorMATH}\ensuremath{\tau }}} \mathrel{{\begingroup\renewcommand\colorMATH{\colorMATHB}\renewcommand\colorSYNTAX{\colorSYNTAXB}{{\color{\colorMATH}\ensuremath{\rightarrow }}}\endgroup }} {{\color{\colorMATH}\ensuremath{({{\color{\colorSYNTAX}\texttt{{\ensuremath{{{\color{\colorMATH}\ensuremath{\tau }}} \xrightarrowP {{\begingroup\renewcommand\colorMATH{\colorMATHC}\renewcommand\colorSYNTAX{\colorSYNTAXC}{{\color{\colorMATH}\ensuremath{\rceil \Sigma \lceil ^{{{\color{\colorSYNTAX}\texttt{{\ensuremath{{\begingroup\renewcommand\colorMATH{\colorMATHA}\renewcommand\colorSYNTAX{\colorSYNTAXA}{{\color{\colorMATH}\ensuremath{\epsilon }}}\endgroup },{\begingroup\renewcommand\colorMATH{\colorMATHA}\renewcommand\colorSYNTAX{\colorSYNTAXA}{{\color{\colorMATH}\ensuremath{\delta }}}\endgroup }}}}}}}}}}\endgroup }} {{\color{\colorMATH}\ensuremath{\tau }}}}}}}})}}}
\xrightarrowP {{\begingroup\renewcommand\colorMATH{\colorMATHC}\renewcommand\colorSYNTAX{\colorSYNTAXC}{{\color{\colorMATH}\ensuremath{\rceil \Sigma \lceil ^{{{\color{\colorSYNTAX}\texttt{{\ensuremath{{\begingroup\renewcommand\colorMATH{\colorMATHA}\renewcommand\colorSYNTAX{\colorSYNTAXA}{{\color{\colorMATH}\ensuremath{n\epsilon }}}\endgroup },{\begingroup\renewcommand\colorMATH{\colorMATHA}\renewcommand\colorSYNTAX{\colorSYNTAXA}{{\color{\colorMATH}\ensuremath{n\delta }}}\endgroup }}}}}}}}}}\endgroup }} {{\color{\colorMATH}\ensuremath{\tau }}}   }}}}}}}}. In this type, the privacy effect on
the closure is given an explicit representation notated
{\begingroup\renewcommand\colorMATH{\colorMATHC}\renewcommand\colorSYNTAX{\colorSYNTAXC}{{\color{\colorMATH}\ensuremath{\rceil \Sigma \lceil ^{{{\color{\colorSYNTAX}\texttt{{\ensuremath{{\begingroup\renewcommand\colorMATH{\colorMATHA}\renewcommand\colorSYNTAX{\colorSYNTAXA}{{\color{\colorMATH}\ensuremath{\epsilon }}}\endgroup },{\begingroup\renewcommand\colorMATH{\colorMATHA}\renewcommand\colorSYNTAX{\colorSYNTAXA}{{\color{\colorMATH}\ensuremath{\delta }}}\endgroup }}}}}}}}}}\endgroup }, which means ``({{\color{\colorMATH}\ensuremath{\epsilon }}},{{\color{\colorMATH}\ensuremath{\delta }}})-privacy for variables
in the closure environment {\begingroup\renewcommand\colorMATH{\colorMATHC}\renewcommand\colorSYNTAX{\colorSYNTAXC}{{\color{\colorMATH}\ensuremath{\Sigma }}}\endgroup }''. This effect is {\textit{latent}} because the effect isn't
``paid for'' until (and each time) the function is called, and it is
{\textit{contextual}} because it includes a privacy effect (which may be
``zero'') for each free \toplas{variable} in the context.
This effect is then explicitly
scaled by {{\color{\colorMATH}\ensuremath{n}}}, the number of loop iterations, in the final effect of
applying the function.\footnote{In addition to the color red, we notate the arrow in privacy
function types with a double head {{\color{\colorMATH}\ensuremath{\mathrel{{\begingroup\renewcommand\colorMATH{\colorMATHC}\renewcommand\colorSYNTAX{\colorSYNTAXC}{{\color{\colorMATH}\ensuremath{\twoheadrightarrow }}}\endgroup }}}}} to further visually
distinguish them from sensitivity function arrows {{\color{\colorMATH}\ensuremath{\mathrel{{\begingroup\renewcommand\colorMATH{\colorMATHB}\renewcommand\colorSYNTAX{\colorSYNTAXB}{{\color{\colorMATH}\ensuremath{\rightarrow }}}\endgroup }}}}}. We describe
details such as the definition of the
{\begingroup\renewcommand\colorMATH{\colorMATHC}\renewcommand\colorSYNTAX{\colorSYNTAXC}{{\color{\colorMATH}\ensuremath{\rceil {\begingroup\renewcommand\colorMATH{\colorMATHA}\renewcommand\colorSYNTAX{\colorSYNTAXA}{{\color{\colorMATH}\ensuremath{\underline{\hspace{0.66em}}}}}\endgroup }\lceil ^{{\begingroup\renewcommand\colorMATH{\colorMATHA}\renewcommand\colorSYNTAX{\colorSYNTAXA}{{\color{\colorMATH}\ensuremath{\underline{\hspace{0.66em}}}}}\endgroup }}}}}\endgroup } notation later.}
More powerful looping combinators such as advanced
composition can also be encoded with these latent contextual effects; such combinators cannot be described in any prior linear type system---including \fuzz and \dfuzz.

\paragraph{Contributions.} {\textit{\system supports writing {\textbf{higher-order
programs}}, and {\textbf{automatically}} verifying that such programs satisfy
{\textbf{advanced variants}} of differential privacy}}.
%---a significant improvement over prior work. % \dd{the whole
% paragraph before this note is new and replaced the one about laplace, which
% wasn't quite right.}
The novel features of \system---linear types with latent contextual effects---are crucial
for practical differentially-private programming. We illustrate this expressive
power by showing how to encode numerous mechanisms and tools for differential
privacy as \system primitives, including the Laplace, Gaussian,
and Exponential mechanisms, advanced composition, and privacy amplification by
subsampling. We also demonstrate the use of \system to verify larger algorithms
in two case studies: the MWEM algorithm~\cite{hardt2012simple} and a
recently-proposed differentially-private machine learning algorithm based on
gradient descent with adaptive gradient
clipping~\cite{thakkar2019differentially}. Note that these examples are expressible in \duet only by adding new core typing rules for each primitive used, which strictly speaking requires re-proving the metatheory of the extended language. In contrast, \system subsumes \duet and supports all these examples without having to add new typing rules, and with a much smaller core language.
Finally, \system is amenable to reasonably
efficient  automated typechecking; we have implemented a typechecker for the language that can verify privacy costs for our case studies in milliseconds.

We prove the type soundness of \system using a step-indexed logical relation over a mixed big-step/denotational semantics with embedded discrete probability
distributions as probability mass functions (PMFs).\footnote{\toplass{We restrict ourselves to discrete distributions because considering continuous distributions would complicate the language semantics (continuous distributions interact badly with higher-order functions) and our main focus here is on the type system.}}
In summary, the contributions of this paper are:

\begin{itemize}
\item \system, a practical, higher-order, general purpose programming language for writing differentially private programs, which supports advanced variants of differential privacy.
% \mt{Do we show other than {{\color{\colorMATH}\ensuremath{(\epsilon ,\delta }}}-DP?}
\item A novel linear type system for \system which includes latent contextual effects,
\toplass{allowing to delay the payment of effects of connectives such as product, sums and functions, until actually  eliminated; e.g. if the second element of a pair is never used, it does not contribute to the effect of the program.}
% a function that is never applied 
% \et{can it be worse?}\dd{no, we will never be less precise than \fuzz or \duet, but we don't make this argument in the paper.}
\item A formalization and proof of type soundness of $\lang$, the core language of \system, based on a proof technique with step-indexed logical relations.
\item A prototype implementation of the \system typechecker, together with a library of primitives for differential privacy, and case studies that demonstrate the expressive power and practicality of \system.
% 's type system (included with the executable typechecker).
\end{itemize}
We first briefly introduce some key concepts of differential privacy
(\S ~\ref{sec:background}) and then give an overview of key
design choices and benefits of contextual linear types in \system
(\S ~\ref{sec:overview}). \system is a two-language design, and
what follows is a presentation of each sub-language in two
multi-section arcs. First, we present the sensitivity-only language
design (\S ~\ref{sec:sensitivity-design}) and metatheory (\S ~\ref{sec:sensitivity-formalism}). This language does not
include differential privacy operations in the language, or privacy
quantities in types. Next, building on this sensitivity core,
we present the full privacy
language design and metatheory (\S ~\ref{sec:privacy-design} and
\ref{sec:privacy-formalism}). Finally, we discuss implementation
details including gaps between the actual implementation of \system and its formal
model (\S ~\ref{sec:system}), present a few case studies in
\system (\S ~\ref{sec:casestudies}), discuss related
work (\S ~\ref{sec:relatedwork}), and conclude (\S ~\ref{sec:conclusion}).

% Section~\ref{sec:privacy-formalism} develops $\lang$ and its metatheory,
% while Section~\ref{sec:system} briefly discusses the differences
% between $\lang$ and \system. Section~\ref{sec:casestudies} describes
% the applications of \system, Section~\ref{sec:relatedwork} discusses
% related work, and Section~\ref{sec:conclusion} concludes. The
% Appendix is provided as anonymous supplementary material.

% }-}

\section{A Differential Privacy Primer} % {-{
\label{sec:background}

Differential privacy is a mathematical definition of what it means for a computation over sensitive data to preserve privacy~\cite{dwork2014algorithmic}. It interprets privacy as a form of plausible deniability and relies on the use of randomization to achieve it. Informally, a randomized algorithm is differentially private if the probability that it outputs a particular value remains almost the same with or without a single individual's data used as part of the input. Formally, the definition is parameterized by two \emph{privacy parameters} {{\color{\colorMATH}\ensuremath{\epsilon }}} and {{\color{\colorMATH}\ensuremath{\delta }}} that specify to what extent two probabilities are ``almost the same'', and by a \emph{distance metric} over the algorithm's (sensitive) input whose role we discuss shortly.
\begin{definition}[Differential privacy]
  Given a randomized algorithm (or \emph{mechanism}) {{\color{\colorMATH}\ensuremath{{\mathcal{M}} \in  A \rightarrow  B }}} and a distance metric {{\color{\colorMATH}\ensuremath{{\mathfrak{D}} _{A} \in  A \times  A \rightarrow  {\mathbb{R}}}}}, the algorithm {{\color{\colorMATH}\ensuremath{{\mathcal{M}}}}} satisfies ({{\color{\colorMATH}\ensuremath{\epsilon , \delta }}})-differential privacy if for all {{\color{\colorMATH}\ensuremath{x, x^{\prime} \in  A }}} such that {{\color{\colorMATH}\ensuremath{{\mathfrak{D}} _{A}(x, x^{\prime}) \leq  1}}} and all possible sets {{\color{\colorMATH}\ensuremath{S \subseteq  B }}} of outcomes, {{\color{\colorMATH}\ensuremath{ {\text{Pr}}[{\mathcal{M}}(x) \in  S] \leq  e^{\epsilon } {\text{Pr}}[{\mathcal{M}}(x^{\prime}) \in  S] + \delta  }}}.
\end{definition}
\noindent
The paramenter {{\color{\colorMATH}\ensuremath{\epsilon }}} quantifies the adversary ability to distinguish two neighbouring inputs upon observing the corresponding algorithm outputs. It represents the privacy guarantee provided by the algorithm---the smaller, the less information is leaked about its input. On the other hand, the parameter {{\color{\colorMATH}\ensuremath{\delta }}} represents a \emph{failure} probability: with probability at most {{\color{\colorMATH}\ensuremath{\delta }}}, the algorithm is allowed to violate privacy altogether. In combination, {{\color{\colorMATH}\ensuremath{\epsilon }}} and {{\color{\colorMATH}\ensuremath{\delta }}} are typically understood as the ``privacy cost'' incurred by publicly releasing the algorithm output, associated to a given sensitive input. The case where {{\color{\colorMATH}\ensuremath{\delta  = 0}}} is called \emph{pure} (or \emph{pure} $\epsilon $-) differential privacy, and the case where {{\color{\colorMATH}\ensuremath{\delta  > 0}}} is called \emph{approximate} differential privacy. Several other recently-proposed variants of the definition build on the advantages of $(\epsilon ,\delta )$-differential privacy while eliminating the potential for failure; these include R\'enyi differential privacy (RDP)~\cite{mironov2017renyi}, zero-concentrated differential privacy (zCDP)~\cite{bun2016concentrated}, and truncated concentrated differential privacy (tCDP)~\cite{bun2018composable}.

Two algorithm inputs are said to be \emph{neighbors} if the distance between them is bounded by {{\color{\colorMATH}\ensuremath{1}}} (i.e. {{\color{\colorMATH}\ensuremath{{\mathfrak{D}} (x,x^{\prime}) \leq  1}}}). In order for the formal definition to match our informal statement, the distance metric {{\color{\colorMATH}\ensuremath{{\mathfrak{D}} _{A}}}} should ensure that neighboring inputs differ by at most one individual's data. Formalizing this notion depends heavily on the domain, so different definitions of {{\color{\colorMATH}\ensuremath{{\mathfrak{D}} }}} are used in different domains. When considering a relational database table represented as a bag of tuples, one commonly-used definition for {{\color{\colorMATH}\ensuremath{{\mathfrak{D}} _{{\text{DB}}}}}} is symmetric difference~\cite{DBLP:conf/sigmod/McSherry09}: %
{{\color{\colorMATH}\ensuremath{ {\mathfrak{D}} _{{\text{DB}}}(x, x^{\prime}) = |(x - x^{\prime}) \cup  (x^{\prime} - x)| }}}. %
Under this definition, {{\color{\colorMATH}\ensuremath{{\mathfrak{D}} _{{\text{DB}}}(x, x^{\prime}) = 1}}} for tables that differ in one \emph{row}; if the data contributed by each individual is bounded to a single row, then this is a good approximation of neighboring inputs.
% In other domains, designing a good approximation is more challenging.

The definition of differential privacy implies two key properties: post-processing and composition. Post-processing means that the output of a differentially-private mechanism \emph{stays} differentially private, no matter what additional processing is applied. Composition allows bounding the privacy cost of multiple computations over the same underlaying data: running an ({{\color{\colorMATH}\ensuremath{\epsilon _{1}}}}, {{\color{\colorMATH}\ensuremath{\delta _{1}}}})-differentially private mechanism followed by an ({{\color{\colorMATH}\ensuremath{\epsilon _{2}}}}, {{\color{\colorMATH}\ensuremath{\delta _{2}}}})-differentially private mechanism satisfies ({{\color{\colorMATH}\ensuremath{\epsilon _{1}+\epsilon _{2}}}}, {{\color{\colorMATH}\ensuremath{\delta _{1}+\delta _{2}}}})-differential privacy.
The privacy parameters {{\color{\colorMATH}\ensuremath{\epsilon }}} and {{\color{\colorMATH}\ensuremath{\delta }}} are often called the {\em privacy cost} because of the additive nature of composition.

%\subsubsection{The Laplace Mechanism and Function Sensitivity}

\paragraph{Basic Mechanisms, composition, and scaling.}
Differential privacy mechanisms typically add noise to the output of a deterministic function scaled to the function's \emph{sensitivity}~\cite{dwork2014algorithmic}. A function {{\color{\colorMATH}\ensuremath{f \in  A \rightarrow  B }}} with distance metrics {{\color{\colorMATH}\ensuremath{{\mathfrak{D}} _{A}}}} and {{\color{\colorMATH}\ensuremath{{\mathfrak{D}} _{B}}}} is called {{\color{\colorMATH}\ensuremath{s}}}\emph{-sensitive} if {{\color{\colorMATH}\ensuremath{{\mathfrak{D}} _{A}(x,y) \leq  d \implies  {\mathfrak{D}} _{B}(f(x),f(y)) \leq  sd}}} for every {{\color{\colorMATH}\ensuremath{d \in  {\mathbb{R}}}}} and every {{\color{\colorMATH}\ensuremath{ x,y \in  A }}}.
Two commonly-used mechanisms are the \emph{Laplace}~\cite{dwork2014algorithmic} and the \emph{Gaussian}~\cite{dwork2014algorithmic, balle2018improving} mechanisms.
Given an {{\color{\colorMATH}\ensuremath{s}}}-sensitive function {{\color{\colorMATH}\ensuremath{f \in  A \rightarrow  {\mathbb{R}}}}}, the Laplace mechanism releases {{\color{\colorMATH}\ensuremath{ f(x) + {\text{Lap}}\big(\frac{s}{\epsilon }\big) }}}, where {{\color{\colorMATH}\ensuremath{{\text{Lap}}(b)}}} denotes a random sample from the Laplace distribution centered at {{\color{\colorMATH}\ensuremath{0}}} with scale {{\color{\colorMATH}\ensuremath{b}}}; it satisfies {{\color{\colorMATH}\ensuremath{\epsilon }}}-differential privacy.
The Gaussian mechanism releases {{\color{\colorMATH}\ensuremath{ f(x) + {\mathcal{N}}\Big(\frac{2s ^{2}\ln (1.25/\delta )}{\epsilon ^{2}}\Big) }}}, where {{\color{\colorMATH}\ensuremath{{\mathcal{N}}(\sigma ^{2})}}} denotes a random sample from the Gaussian distribution centered at {{\color{\colorMATH}\ensuremath{0}}} with variance {{\color{\colorMATH}\ensuremath{\sigma ^{2}}}} and {{\color{\colorMATH}\ensuremath{\epsilon , \delta  \in  (0, 1)}}}; it satisfies ({{\color{\colorMATH}\ensuremath{\epsilon ,\delta }}})-differential privacy. While the original Gaussian mechanism requires {{\color{\colorMATH}\ensuremath{\epsilon  < 1}}}, \citet{balle2018improving} \toplass{introduce a variant---called \emph{analytic} Gaussian mechanism---that drops the requirement that {{\color{\colorMATH}\ensuremath{\epsilon  < 1}}}.} 

\toplass{For implementation purposes, naive floating-point truncations of the real-valued Laplace distribution lead to fatal privacy breaches~\cite{Mironov12}. \citet{DiscreteGauss} have shown that the Laplacian and Gaussian mechanisms can both be discretized, while still providing formal privacy guarantees.}

% \paragraph{Advanced Composition.}
Advanced composition~\cite{dwork2014algorithmic} yields tighter bounds on privacy cost for many iterative algorithms, but requires ($\epsilon $, $\delta $)-differential privacy. For {{\color{\colorMATH}\ensuremath{\epsilon ,\delta ,\delta ^{\prime} \geq  0}}}, the class of ({{\color{\colorMATH}\ensuremath{\epsilon }}}, {{\color{\colorMATH}\ensuremath{\delta }}})-differentially private mechanisms satisfies ({{\color{\colorMATH}\ensuremath{\epsilon ^{\prime}}}}, {{\color{\colorMATH}\ensuremath{k\delta  + \delta ^{\prime}}}})-differential privacy under \emph{{{\color{\colorMATH}\ensuremath{k}}}-fold adaptive composition} (e.g. a loop with {{\color{\colorMATH}\ensuremath{k}}} iterations) where {{\color{\colorMATH}\ensuremath{\epsilon ^{\prime} = k\epsilon (e^\epsilon -1) + \epsilon \sqrt {2 k \ln (1/\delta ^{\prime})}}}}.
Advanced composition is especially useful for iterative algorithms that perform many differentially private steps in sequence (e.g. iterative machine learning algorithms).

% \paragraph{Scaling Privacy with Distance.}
Differential privacy is stated in terms of neighboring inputs,
i.e.~inputs {{\color{\colorMATH}\ensuremath{x}}} and {{\color{\colorMATH}\ensuremath{x^{\prime}}}} such that {{\color{\colorMATH}\ensuremath{{\mathfrak{D}} _{A}(x, x^{\prime}) \leq  1}}}. When {{\color{\colorMATH}\ensuremath{{\mathfrak{D}} _{A}(x,
x^{\prime}) > 1}}}, an {{\color{\colorMATH}\ensuremath{\epsilon }}}-differentially private mechanism provides {{\color{\colorMATH}\ensuremath{{\mathfrak{D}} _{A}(x,
x^{\prime})\mathord{\cdotp }\epsilon }}}-differential privacy. Distances larger than one are typically
interpreted as groups of individuals, e.g., {{\color{\colorMATH}\ensuremath{{\mathfrak{D}} _{A}(x,x^{\prime}) = k}}}
represents a change to {{\color{\colorMATH}\ensuremath{k}}} individual's input data. Therefore, an
{{\color{\colorMATH}\ensuremath{\epsilon }}}-differentially private mechanism provides {{\color{\colorMATH}\ensuremath{k\epsilon }}}-differential
{\textit{group privacy}}~\cite{dwork2014algorithmic} for groups of size {{\color{\colorMATH}\ensuremath{k}}}.
A similar property holds for ($\epsilon $, $\delta $)-differential privacy and
the more recently developed advanced variants, but the scaling of privacy cost
is \emph{nonlinear} for all of these variants%
%, e.g., worse than ({{\color{\colorMATH}\ensuremath{k\epsilon }}},{{\color{\colorMATH}\ensuremath{k\delta }}}) for ($\epsilon ,\delta $)-differential privacy.
. \toplas{For example, on inputs at distance {{\color{\colorMATH}\ensuremath{k}}}, an algorithm satisfying ($\epsilon ,\delta $)-differential privacy yields outputs that are only ({{\color{\colorMATH}\ensuremath{k\epsilon }}},{{\color{\colorMATH}\ensuremath{k\delta e^{(k-1)\epsilon }}}})-close--instead of ({{\color{\colorMATH}\ensuremath{k\epsilon }}},{{\color{\colorMATH}\ensuremath{k\delta }}})-close.}
This nonlinearity
makes it difficult to apply techniques based on linear types (which
generally internalize linear scaling of
costs~\cite{reed2010distance,gaboardi2013linear}) for these variants
of differential privacy.

\paragraph{Verification Techniques for Differential Privacy.}
A number of techniques have been proposed for verifying that a program satisfies differential privacy, including approaches based on linear logic~\cite{reed2010distance, gaboardi2013linear, de2019probabilistic, zhang2019fuzzi, near2019duet}, couplings and program logics~\cite{DBLP:journals/lmcs/BartheEHSS19, Barthe:POPL12,Barthe:TOPLAS:13, Barthe:LICS16, Sato:LICS19, DBLP:journals/pacmpl/AlbarghouthiH18, Barthe:CSF14}, and randomness alignments~\cite{zhang2017lightdp, wang2019proving}. Our work is most closely related to \fuzz~\cite{reed2010distance} and its descendants \dfuzz~\cite{gaboardi2013linear}, \duet~\cite{near2019duet}, \toplass{and \fuzzed~\cite{de2019probabilistic},} which are based on linear type systems. In particular, these approaches focus heavily on fully automated verification of differential privacy properties through typechecking, and are typically less expressive than program logics, which by contrast support significantly less (or no) automation. We defer a more complete discussion of related work to Section~\ref{sec:relatedwork}.

% }-}

\section{Overview of \system} % {-{
\label{sec:overview}

\system builds on the linear type system of
\fuzz~\cite{reed2010distance} and the two-language design of
\duet~\cite{near2019duet} by
introducing {\textit{latent contextual effects}}.
This section introduces and motivates the design
of \system's
two languages---
one for sensitivity and one for privacy---using simple examples.
The design of each language is described in
Sections~\ref{sec:sensitivity-design} and \ref{sec:privacy-design}
respectively, and each metatheory is described in
Sections~\ref{sec:sensitivity-formalism} and
\ref{sec:privacy-formalism}.

\subsection{A Two-Language Design}
% \mt{R2:  the words "implicit" and "pervasive" do not make too much
%  sense until I gain a deeper understanding of Fuzz, Duet and Jazz.}

\system follows \duet in being structured as two mutually-embedded sublanguages, one for sensitivity and one for privacy. In a nutshell, this is because supporting {\em scaling} of both sensitivity and privacy \toplas{in a uniform and tight fashion} is sound only for {{\color{\colorMATH}\ensuremath{\epsilon }}}-differential privacy, and \emph{not} for ({{\color{\colorMATH}\ensuremath{\epsilon }}}, {{\color{\colorMATH}\ensuremath{\delta }}})-differential privacy, which has nonlinear group privacy, as discussed in Section~\ref{sec:background}. %Therefore, a single language for both sensitivity and privacy, as in \fuzz/\dfuzz, does not support ({{\color{\colorMATH}\ensuremath{\epsilon }}}, {{\color{\colorMATH}\ensuremath{\delta }}})-differential privacy, and cannot be easily extended to do so.

\toplas{Let us elaborate on this.
Recall that the framework of differential privacy builds on randomization to achieve privacy, and the randomization is typically calibrated according to the sensitivity of the function whose output one wants to protect. Therefore, in a language to describe differentially private computations, we can mostly distinguish two class of functions: On the \toplass{one} hand, randomized (effectful) functions that are annotated with privacy information, and on the other hand pure functions that are annotated with sensitivity information.
% To star with, observe that in a language to describe differentially private computations we can mostly distinguish two class of functions, which are combined---typically  following complex patterns---to achieve differential privacy. On the one hand, there are pure functions that are annotated with sensitivity information, and on the other hand there are randomized functions that are annotated with privacy information (recall that the framework of differential privacy builds on randomization to provide privacy guarantees).
However, when composing functions, the way their information is combined highly depends on the class of the composed functions.

For example, when composing two functions from the sensitivity fragment, their information is naturally combined via scaling. Consider, for instance, a {{\color{\colorMATH}\ensuremath{3}}}-sensitive function {{\color{\colorMATH}\ensuremath{f}}}; it is not hard to see that the expression
\begingroup\color{\colorMATH}\begin{gather*} f (x + x) \end{gather*}\endgroup
is {{\color{\colorMATH}\ensuremath{(2 \cdot 3)}}}-sensitive in {{\color{\colorMATH}\ensuremath{x}}}. The same scaling pattern remains valid when composing a function from the sensitivity fragment with a function from the privacy fragment, provided the latter satisfies pure differential privacy: If {{\color{\colorMATH}\ensuremath{g}}} is an {e.g.} {{\color{\colorMATH}\ensuremath{\epsilon }}}-differentially private function, then the computation
\begingroup\color{\colorMATH}\begin{gather*} g (x + x) \end{gather*}\endgroup
is {{\color{\colorMATH}\ensuremath{(2 \cdot \epsilon)}}}-differentially private in {{\color{\colorMATH}\ensuremath{x}}}. This uniform scaling behavior, which besides being sound is also tight, lies at the heart of \fuzz/\dfuzz design. In fact, in  \fuzz/\dfuzz both class of functions live within the same space, following the same typing rules. To enable this uniform treatment, the languages rely on the two fundamental ingredients: i) the presence of a monad/modality to encode randomization, and ii) the association of a metric space to each type, which in the case of monadic types is tailored to {\textit{encode}} differential privacy. Said otherwise, in \fuzz/\dfuzz differential privacy is encoded as a sensitivity claim about functions (with a monadic return type).

Unfortunately, this linear scaling---pervasive in \fuzz/\dfuzz---is no longer sound when composing a function from the sensitivity fragment with a function from the privacy fragment that satisfies \emph{approximate}---rather than \emph{pure}---differential privacy. Returning to the previous example, if {{\color{\colorMATH}\ensuremath{g}}} is instead ({{\color{\colorMATH}\ensuremath{\epsilon }}}, {{\color{\colorMATH}\ensuremath{\delta }}})-differentially private, then {{\color{\colorMATH}\ensuremath{g (x + x)}}} does \emph{not} satisfy ({{\color{\colorMATH}\ensuremath{2 \epsilon }}}, {{\color{\colorMATH}\ensuremath{2 \delta }}})-differential privacy in {{\color{\colorMATH}\ensuremath{x}}}, but only ({{\color{\colorMATH}\ensuremath{2 \epsilon }}}, {{\color{\colorMATH}\ensuremath{2 \delta  e^\epsilon }}})-differential privacy as dictated by the group privacy property for approximate differential privacy (see Section~\ref{sec:background} and \citet[\S2.3]{dwork2014algorithmic}). In effect, this is why  approximate differential privacy lies out of the scope of \fuzz/\dfuzz``uniform'' design---\toplass{although recently, \citet{de2019probabilistic} has shown that using different metrics allows approximate differential privacy functions to be linear, extending a terminating subset of \fuzz to support ({{\color{\colorMATH}\ensuremath{\epsilon }}}, {{\color{\colorMATH}\ensuremath{\delta }}})-differentially privacy via a path metric construction.}}

In view of this, \duet introduced a two-language design, separating a sensitivity sublanguage in which scaling remains {\textit{implicit}} \toplas{(internalized by the typing rules)} and {\textit{pervasive}} \toplas{(modeling function composition)}, and a
privacy sublanguage in which scaling is {\textit{explicit and restricted}} \toplas{(ad-hoc typing rules are needed {e.g.} to model some privacy combinators)}.
\system builds upon this approach and
% In other words, a function with the
% output type {{\color{\colorMATH}\ensuremath{{\scriptstyle \bigcirc }{\mathbb{R}}}}} which is {{\color{\colorMATH}\ensuremath{\epsilon }}}-sensitive in its argument will satisfy
significantly improves both sublanguages thanks to latent contextual effects, as we illustrate next.

\paragraph{Coloring convention.}
As noted in the introduction, this paper uses colorblind-friendly
colors in notation to convey information, and is best consumed using
an electronic device or color printer. \system consists of two
mutually embedded sublanguages, and each language is given its own
color. Furthermore, these two languages share the same language of
types, so we use a third color for the shared fragment.
Consequently, we use three color schemes throughout the paper: (1) {\begingroup\renewcommand\colorMATH{\colorMATHA}\renewcommand\colorSYNTAX{\colorSYNTAXA}{{\color{\colorMATH}\ensuremath{{\text{blue}}}}}\endgroup }
for general math notation and the type system shared between
languages; (2)
{\begingroup\renewcommand\colorMATH{\colorMATHB}\renewcommand\colorSYNTAX{\colorSYNTAXB}{{\color{\colorMATH}\ensuremath{{\text{green}}}}}\endgroup } for the sensitivity language; and (3) {\begingroup\renewcommand\colorMATH{\colorMATHC}\renewcommand\colorSYNTAX{\colorSYNTAXC}{{\color{\colorMATH}\ensuremath{{\text{red}}}}}\endgroup } for the
privacy language. We have
carefully chosen the schemes to be distinguishable
(as much as possible) for persons with various forms of
color blindness.\footnote{We chose colors following the 24-Color
Palette from \url{http://mkweb.bcgsc.ca/colorblind}. E.g., to persons
with deuteranopia (the most common form of color blindness),
colorschemes
{\color{\colorMATHA}$\blacksquare$}/%
{\color{\colorMATHB}$\blacksquare$}/%
{\color{\colorMATHC}$\blacksquare$}
appear as
{\color{cbasifABright}$\blacksquare$}/%
{\color{cbasifBBright}$\blacksquare$}/%
{\color{cbasifCBright}$\blacksquare$}
respectively.}

\subsection{Sensitivity}
\label{sec:sensitivity-intro}

In the sensitivity sublanguage of \system, the identity function is written {\begingroup\renewcommand\colorMATH{\colorMATHB}\renewcommand\colorSYNTAX{\colorSYNTAXB}{{\color{\colorSYNTAX}\texttt{{\ensuremath{{\begingroup\renewcommand\colorMATH{\colorMATHB}\renewcommand\colorSYNTAX{\colorSYNTAXB}{{\color{\colorMATH}\ensuremath{\slambda}}}\endgroup } {\begingroup\renewcommand\colorMATH{\colorMATHA}\renewcommand\colorSYNTAX{\colorSYNTAXA}{{\color{\colorMATH}\ensuremath{x}}}\endgroup }.\hspace*{0.33em}{\begingroup\renewcommand\colorMATH{\colorMATHA}\renewcommand\colorSYNTAX{\colorSYNTAXA}{{\color{\colorMATH}\ensuremath{x}}}\endgroup }}}}}}\endgroup }.
We write sensitivity lambdas as {\begingroup\renewcommand\colorMATH{\colorMATHB}\renewcommand\colorSYNTAX{\colorSYNTAXB}{{\color{\colorSYNTAX}\texttt{{\ensuremath{{\begingroup\renewcommand\colorMATH{\colorMATHB}\renewcommand\colorSYNTAX{\colorSYNTAXB}{{\color{\colorMATH}\ensuremath{\slambda}}}\endgroup } }}}}}\endgroup } to more easily
distinguish them from privacy lambdas, written {\begingroup\renewcommand\colorMATH{\colorMATHC}\renewcommand\colorSYNTAX{\colorSYNTAXC}{{\color{\colorSYNTAX}\texttt{{\ensuremath{{\begingroup\renewcommand\colorMATH{\colorMATHC}\renewcommand\colorSYNTAX{\colorSYNTAXC}{{\color{\colorMATH}\ensuremath{\plambda}}}\endgroup } }}}}}\endgroup } (described later).
The identity function is {{\color{\colorMATH}\ensuremath{1}}}-sensitive in its argument {\begingroup\renewcommand\colorMATH{\colorMATHA}\renewcommand\colorSYNTAX{\colorSYNTAXA}{{\color{\colorMATH}\ensuremath{x}}}\endgroup }: if
{\begingroup\renewcommand\colorMATH{\colorMATHA}\renewcommand\colorSYNTAX{\colorSYNTAXA}{{\color{\colorMATH}\ensuremath{x}}}\endgroup } changes by {\begingroup\renewcommand\colorMATH{\colorMATHA}\renewcommand\colorSYNTAX{\colorSYNTAXA}{{\color{\colorMATH}\ensuremath{{\begingroup\renewcommand\colorMATH{\colorMATHB}\renewcommand\colorSYNTAX{\colorSYNTAXB}{{\color{\colorMATH}\ensuremath{\distance}}}\endgroup }}}}\endgroup }, then the function's output also changes by
{\begingroup\renewcommand\colorMATH{\colorMATHA}\renewcommand\colorSYNTAX{\colorSYNTAXA}{{\color{\colorMATH}\ensuremath{{\begingroup\renewcommand\colorMATH{\colorMATHB}\renewcommand\colorSYNTAX{\colorSYNTAXB}{{\color{\colorMATH}\ensuremath{\distance}}}\endgroup }}}}\endgroup }.
Similarly to the identity function, the doubling function
{{\color{\colorMATH}\ensuremath{{\begingroup\renewcommand\colorMATH{\colorMATHB}\renewcommand\colorSYNTAX{\colorSYNTAXB}{{\color{\colorMATH}\ensuremath{\slambda}}}\endgroup }  x.\hspace*{0.33em}x+x}}} is {{\color{\colorMATH}\ensuremath{2}}}-sensitive in {\begingroup\renewcommand\colorMATH{\colorMATHA}\renewcommand\colorSYNTAX{\colorSYNTAXA}{{\color{\colorMATH}\ensuremath{x}}}\endgroup }: if {\begingroup\renewcommand\colorMATH{\colorMATHA}\renewcommand\colorSYNTAX{\colorSYNTAXA}{{\color{\colorMATH}\ensuremath{x}}}\endgroup } changes
by {\begingroup\renewcommand\colorMATH{\colorMATHA}\renewcommand\colorSYNTAX{\colorSYNTAXA}{{\color{\colorMATH}\ensuremath{{\begingroup\renewcommand\colorMATH{\colorMATHB}\renewcommand\colorSYNTAX{\colorSYNTAXB}{{\color{\colorMATH}\ensuremath{\distance}}}\endgroup }}}}\endgroup }, then the function's output changes by {\begingroup\renewcommand\colorMATH{\colorMATHA}\renewcommand\colorSYNTAX{\colorSYNTAXA}{{\color{\colorMATH}\ensuremath{2{\begingroup\renewcommand\colorMATH{\colorMATHB}\renewcommand\colorSYNTAX{\colorSYNTAXB}{{\color{\colorMATH}\ensuremath{\distance}}}\endgroup }}}}\endgroup }.

\toplas{\fuzz extends the notion of sensitivity to multi-argument
  functions by assigning a sensitivity to each argument.}
For example,
the curried function {{\color{\colorMATH}\ensuremath{{\begingroup\renewcommand\colorMATH{\colorMATHB}\renewcommand\colorSYNTAX{\colorSYNTAXB}{{\color{\colorMATH}\ensuremath{\slambda}}}\endgroup } x.\hspace*{0.33em}{\begingroup\renewcommand\colorMATH{\colorMATHB}\renewcommand\colorSYNTAX{\colorSYNTAXB}{{\color{\colorMATH}\ensuremath{\slambda}}}\endgroup } y.\hspace*{0.33em}x+x+y}}} is
{{\color{\colorMATH}\ensuremath{2}}}-sensitive in {{\color{\colorMATH}\ensuremath{x}}} and {{\color{\colorMATH}\ensuremath{1}}}-sensitive in {{\color{\colorMATH}\ensuremath{y}}}. If {{\color{\colorMATH}\ensuremath{x}}} changes by
{{\color{\colorMATH}\ensuremath{{\begingroup\renewcommand\colorMATH{\colorMATHB}\renewcommand\colorSYNTAX{\colorSYNTAXB}{{\color{\colorMATH}\ensuremath{\distance_{x}}}}\endgroup }}}} and {{\color{\colorMATH}\ensuremath{y}}} changes by {{\color{\colorMATH}\ensuremath{{\begingroup\renewcommand\colorMATH{\colorMATHB}\renewcommand\colorSYNTAX{\colorSYNTAXB}{{\color{\colorMATH}\ensuremath{\distance_{y}}}}\endgroup }}}}, then the function's output
changes by {{\color{\colorMATH}\ensuremath{2{\begingroup\renewcommand\colorMATH{\colorMATHB}\renewcommand\colorSYNTAX{\colorSYNTAXB}{{\color{\colorMATH}\ensuremath{\distance_{x}}}}\endgroup } +1{\begingroup\renewcommand\colorMATH{\colorMATHB}\renewcommand\colorSYNTAX{\colorSYNTAXB}{{\color{\colorMATH}\ensuremath{\distance_{y}}}}\endgroup }}}}.

In general, the sensitivity of a function can be written as a linear
combination of the changes in its inputs. In \system, we express
function sensitivities as linear formulas over the function's input
variables, using the variable name itself as a placeholder for the
change in that input. \system's type system gives the following types
for the three examples we have seen so far:
\begingroup
\begingroup\color{\colorMATH}\begin{gather*}
\begin{array}{rcl
} ({\begingroup\renewcommand\colorMATH{\colorMATHB}\renewcommand\colorSYNTAX{\colorSYNTAXB}{{\color{\colorMATH}\ensuremath{\slambda}}}\endgroup } x.\hspace*{0.33em}x)         &{}\mathrel{:}{}& (x \mathrel{:} {\mathbb{R}}) \xrightarrowS {x} {\mathbb{R}}
\cr  ({\begingroup\renewcommand\colorMATH{\colorMATHB}\renewcommand\colorSYNTAX{\colorSYNTAXB}{{\color{\colorMATH}\ensuremath{\slambda}}}\endgroup } x.\hspace*{0.33em}x+x)       &{}\mathrel{:}{}& (x \mathrel{:} {\mathbb{R}}) \xrightarrowS {2x} {\mathbb{R}}
\cr  ({\begingroup\renewcommand\colorMATH{\colorMATHB}\renewcommand\colorSYNTAX{\colorSYNTAXB}{{\color{\colorMATH}\ensuremath{\slambda}}}\endgroup } x.\hspace*{0.33em}{\begingroup\renewcommand\colorMATH{\colorMATHB}\renewcommand\colorSYNTAX{\colorSYNTAXB}{{\color{\colorMATH}\ensuremath{\slambda}}}\endgroup } y.\hspace*{0.33em}x+x+y) &{}\mathrel{:}{}& (x \mathrel{:} {\mathbb{R}}) \xrightarrowS {0x} (y \mathrel{:} {\mathbb{R}}) \xrightarrowS {2x + y} {\mathbb{R}}
\end{array}
\end{gather*}\endgroup
\endgroup
The linear formulas written above function arrows represent the
\emph{sensitivity effect} of the corresponding function.
The general form of sensitivity function types is {{\color{\colorSYNTAX}\texttt{{\ensuremath{({{\color{\colorMATH}\ensuremath{x}}} : {{\color{\colorMATH}\ensuremath{\tau _{1}}}})
\xrightarrowS {{\begingroup\renewcommand\colorMATH{\colorMATHB}\renewcommand\colorSYNTAX{\colorSYNTAXB}{{\color{\colorMATH}\ensuremath{\Sigma }}}\endgroup }} {{\color{\colorMATH}\ensuremath{\tau _{2}}}}}}}}}, where {\begingroup\renewcommand\colorMATH{\colorMATHB}\renewcommand\colorSYNTAX{\colorSYNTAXB}{{\color{\colorMATH}\ensuremath{\Sigma }}}\endgroup } is the {\em sensitivity effect} of the function,
expressed as a linear formula.
Note that {{\color{\colorMATH}\ensuremath{x}}} is in scope for {\begingroup\renewcommand\colorMATH{\colorMATHB}\renewcommand\colorSYNTAX{\colorSYNTAXB}{{\color{\colorMATH}\ensuremath{\Sigma }}}\endgroup } and {{\color{\colorMATH}\ensuremath{\tau _{2}}}}. Importantly, occurrences of
{{\color{\colorMATH}\ensuremath{x}}} in {\begingroup\renewcommand\colorMATH{\colorMATHB}\renewcommand\colorSYNTAX{\colorSYNTAXB}{{\color{\colorMATH}\ensuremath{\Sigma }}}\endgroup } and {{\color{\colorMATH}\ensuremath{\tau _{2}}}} represent the {\textit{sensitivity}} of the variable {{\color{\colorMATH}\ensuremath{x}}},
rather than its value, so \system supports {\em sensitivity-dependent} types.
% \footnote{Prior type systems for differential privacy
% like \dfuzz~\cite{gaboardi2013linear} use value-dependent types, which is different from the
%  supported by \system.}
Also, we usually drop ``null'' effects over function arrows such as {{\color{\colorSYNTAX}\texttt{{\ensuremath{\xrightarrowS {{\begingroup\renewcommand\colorMATH{\colorMATHA}\renewcommand\colorSYNTAX{\colorSYNTAXA}{{\color{\colorMATH}\ensuremath{0x}}}\endgroup }}}}}}} above, and instead just write {{\color{\colorSYNTAX}\texttt{{\ensuremath{\mathrel{{\begingroup\renewcommand\colorMATH{\colorMATHB}\renewcommand\colorSYNTAX{\colorSYNTAXB}{{\color{\colorMATH}\ensuremath{\rightarrow }}}\endgroup }}}}}}}.
%
% The general form of sensitivity function types is {{\color{\colorSYNTAX}\texttt{{\ensuremath{({{\color{\colorMATH}\ensuremath{x}}} : {{\color{\colorMATH}\ensuremath{\tau _{1}}}})
% \xrightarrowS {{\begingroup\renewcommand\colorMATH{\colorMATHB}\renewcommand\colorSYNTAX{\colorSYNTAXB}{{\color{\colorMATH}\ensuremath{\Sigma }}}\endgroup }} {{\color{\colorMATH}\ensuremath{\tau _{2}}}}}}}}}, where {\begingroup\renewcommand\colorMATH{\colorMATHB}\renewcommand\colorSYNTAX{\colorSYNTAXB}{{\color{\colorMATH}\ensuremath{\Sigma }}}\endgroup } is the sensitivity effect expressed as a
% linear formula,

As usual, \system accommodates higher-order functions by scaling
sensitivities. For example, applying a {{\color{\colorMATH}\ensuremath{2}}}-sensitive function twice yields a {{\color{\colorMATH}\ensuremath{4}}}-sensitive function:
\begingroup\color{\colorMATH}\begin{gather*} ({\begingroup\renewcommand\colorMATH{\colorMATHB}\renewcommand\colorSYNTAX{\colorSYNTAXB}{{\color{\colorMATH}\ensuremath{\slambda}}}\endgroup }  f.\hspace*{0.33em}{\begingroup\renewcommand\colorMATH{\colorMATHB}\renewcommand\colorSYNTAX{\colorSYNTAXB}{{\color{\colorMATH}\ensuremath{\slambda}}}\endgroup }  y.\hspace*{0.33em} f\hspace*{0.33em}y+f\hspace*{0.33em}y)\hspace*{0.33em}({\begingroup\renewcommand\colorMATH{\colorMATHB}\renewcommand\colorSYNTAX{\colorSYNTAXB}{{\color{\colorMATH}\ensuremath{\slambda}}}\endgroup }  x.\hspace*{0.33em} x+x) \mathrel{:} (y \mathrel{:} {\mathbb{R}}) \xrightarrowS {4y} {\mathbb{R}} \end{gather*}\endgroup
In addition to function types, other
type connectives in \system like sums and products also carry sensitivity
effects, such as {{\color{\colorSYNTAX}\texttt{{\ensuremath{{{\color{\colorMATH}\ensuremath{\tau _{1}}}} \mathrel{^{{\begingroup\renewcommand\colorMATH{\colorMATHB}\renewcommand\colorSYNTAX{\colorSYNTAXB}{{\color{\colorMATH}\ensuremath{\Sigma _{1}}}}\endgroup }}\oplus ^{{\begingroup\renewcommand\colorMATH{\colorMATHB}\renewcommand\colorSYNTAX{\colorSYNTAXB}{{\color{\colorMATH}\ensuremath{\Sigma _{2}}}}\endgroup }}} {{\color{\colorMATH}\ensuremath{\tau _{2}}}}}}}}} for sums, {{\color{\colorSYNTAX}\texttt{{\ensuremath{{{\color{\colorMATH}\ensuremath{\tau _{1}}}}
\mathrel{^{{\begingroup\renewcommand\colorMATH{\colorMATHB}\renewcommand\colorSYNTAX{\colorSYNTAXB}{{\color{\colorMATH}\ensuremath{\Sigma _{1}}}}\endgroup }}\otimes ^{{\begingroup\renewcommand\colorMATH{\colorMATHB}\renewcommand\colorSYNTAX{\colorSYNTAXB}{{\color{\colorMATH}\ensuremath{\Sigma _{2}}}}\endgroup }}} {{\color{\colorMATH}\ensuremath{\tau _{2}}}}}}}}} for multiplicative products, and {{\color{\colorSYNTAX}\texttt{{\ensuremath{{{\color{\colorMATH}\ensuremath{\tau _{1}}}}
\mathrel{^{{\begingroup\renewcommand\colorMATH{\colorMATHB}\renewcommand\colorSYNTAX{\colorSYNTAXB}{{\color{\colorMATH}\ensuremath{\Sigma _{1}}}}\endgroup }}\&^{{\begingroup\renewcommand\colorMATH{\colorMATHB}\renewcommand\colorSYNTAX{\colorSYNTAXB}{{\color{\colorMATH}\ensuremath{\Sigma _{2}}}}\endgroup }}} {{\color{\colorMATH}\ensuremath{\tau _{2}}}}}}}}} for additive products.
These connectives are extensions of the
linear type connectives {{\color{\colorSYNTAX}\texttt{{\ensuremath{{{\color{\colorMATH}\ensuremath{\tau _{1}}}} \oplus  {{\color{\colorMATH}\ensuremath{\tau _{2}}}}}}}}}, {{\color{\colorSYNTAX}\texttt{{\ensuremath{{{\color{\colorMATH}\ensuremath{\tau _{1}}}} \otimes  {{\color{\colorMATH}\ensuremath{\tau _{2}}}}}}}}} and {{\color{\colorSYNTAX}\texttt{{\ensuremath{{{\color{\colorMATH}\ensuremath{\tau _{1}}}} \mathrel{\&}
{{\color{\colorMATH}\ensuremath{\tau _{2}}}}}}}}} from \fuzz, augmented with latent sensitivity effects.

We say that sensitivity effects in \system are {\em latent} because they only contribute to the sensitivity of an expression when the type connective is actually eliminated.
For instance, in the third example above---a curried function---the sensitivity effect on
the first argument is delayed until the second argument is
received. If a second argument is \emph{never} received, then the
sensitivity effect on the first argument can be ignored.
Likewise, the annotations {\begingroup\renewcommand\colorMATH{\colorMATHB}\renewcommand\colorSYNTAX{\colorSYNTAXB}{{\color{\colorMATH}\ensuremath{\Sigma _{1}}}}\endgroup } and {\begingroup\renewcommand\colorMATH{\colorMATHB}\renewcommand\colorSYNTAX{\colorSYNTAXB}{{\color{\colorMATH}\ensuremath{\Sigma _{2}}}}\endgroup }
in the type {{\color{\colorSYNTAX}\texttt{{\ensuremath{{{\color{\colorMATH}\ensuremath{\tau _{1}}}} \mathrel{^{{\begingroup\renewcommand\colorMATH{\colorMATHB}\renewcommand\colorSYNTAX{\colorSYNTAXB}{{\color{\colorMATH}\ensuremath{\Sigma _{1}}}}\endgroup }}\otimes ^{{\begingroup\renewcommand\colorMATH{\colorMATHB}\renewcommand\colorSYNTAX{\colorSYNTAXB}{{\color{\colorMATH}\ensuremath{\Sigma _{2}}}}\endgroup }}} {{\color{\colorMATH}\ensuremath{\tau _{2}}}}}}}}} encode the latent
sensitivity cost for each component of the connective: for {{\color{\colorMATH}\ensuremath{\tau _{1}}}} (the left) and
{{\color{\colorMATH}\ensuremath{\tau _{2}}}} (the right) respectively. In contrast to \fuzz, creating a pair in \system~\toplass{ can have} no immediate sensitivity cost: only projecting out of a pair has a cost in sensitivity, depending on which component is projected.
Additionally, we say that sensitivity effects are {\em contextual} because {\begingroup\renewcommand\colorMATH{\colorMATHB}\renewcommand\colorSYNTAX{\colorSYNTAXB}{{\color{\colorMATH}\ensuremath{\Sigma }}}\endgroup } can refer to variables in scope.
% \system is the first linear type system for differential
% privacy with the ability to delay sensitivity (and privacy) effects
% this way.

% As we illustrate in Section~\ref{sec:sensitivity-design}, latent contextual effects
% are the key mechanism that enable strictly more precise analyses in \system than in prior systems.
% The same technique also applies to the privacy tracking overviewed below.

\subsection{Privacy}
\label{sec:privacy_jazz_intro}
To encode differential privacy, \system makes use of {\textit{privacy functions}} (notated {{\color{\colorMATH}\ensuremath{\mathrel{{\begingroup\renewcommand\colorMATH{\colorMATHC}\renewcommand\colorSYNTAX{\colorSYNTAXC}{{\color{\colorMATH}\ensuremath{\twoheadrightarrow }}}\endgroup }}}}}) rather than of {\textit{sensitivity functions}} (notated {{\color{\colorMATH}\ensuremath{\mathrel{{\begingroup\renewcommand\colorMATH{\colorMATHB}\renewcommand\colorSYNTAX{\colorSYNTAXB}{{\color{\colorMATH}\ensuremath{\rightarrow }}}\endgroup }}}}}) as in the previous examples. As such, privacy functions are annotated with {\textit{privacy}}---rather than {\textit{sensitivity}}---effects.
The type of a function from {{\color{\colorMATH}\ensuremath{\tau _{1}}}} to {{\color{\colorMATH}\ensuremath{\tau _{2}}}} which is ({{\color{\colorMATH}\ensuremath{\epsilon }}},{{\color{\colorMATH}\ensuremath{\delta }}})-differentially private in its argument is as follows:
\begingroup\color{\colorMATH}\begin{gather*}
%f \mathrel{:}
{{\color{\colorSYNTAX}\texttt{{\ensuremath{ ({{\color{\colorMATH}\ensuremath{x}}} \mathrel{:} \tau _{1}\mathord{\cdotp }{\begingroup\renewcommand\colorMATH{\colorMATHB}\renewcommand\colorSYNTAX{\colorSYNTAXB}{{\color{\colorMATH}\ensuremath{\distance}}}\endgroup }) \xrightarrowP {({{\color{\colorSYNTAX}\texttt{{\ensuremath{{\begingroup\renewcommand\colorMATH{\colorMATHA}\renewcommand\colorSYNTAX{\colorSYNTAXA}{{\color{\colorMATH}\ensuremath{\epsilon }}}\endgroup },{\begingroup\renewcommand\colorMATH{\colorMATHA}\renewcommand\colorSYNTAX{\colorSYNTAXA}{{\color{\colorMATH}\ensuremath{\delta }}}\endgroup }}}}}}){\begingroup\renewcommand\colorMATH{\colorMATHA}\renewcommand\colorSYNTAX{\colorSYNTAXA}{{\color{\colorMATH}\ensuremath{x}}}\endgroup }} \tau _{2}}}}}}
\end{gather*}\endgroup
Semantically, this type describes
functions {{\color{\colorMATH}\ensuremath{f}}}, where if {{\color{\colorMATH}\ensuremath{{\mathfrak{D}} _{\tau _{1}}(x,x^{\prime}) \leq  {\begingroup\renewcommand\colorMATH{\colorMATHB}\renewcommand\colorSYNTAX{\colorSYNTAXB}{{\color{\colorMATH}\ensuremath{\distance}}}\endgroup }}}} then {{\color{\colorMATH}\ensuremath{f(x)}}} and {{\color{\colorMATH}\ensuremath{f(x^{\prime})}}}
yield distributions which are ``({{\color{\colorMATH}\ensuremath{\epsilon }}},{{\color{\colorMATH}\ensuremath{\delta }}})-close'' according to the
definition of  ({{\color{\colorMATH}\ensuremath{\epsilon }}}, {{\color{\colorMATH}\ensuremath{\delta }}})-differential privacy.

The annotation {{\color{\colorMATH}\ensuremath{{\begingroup\renewcommand\colorMATH{\colorMATHB}\renewcommand\colorSYNTAX{\colorSYNTAXB}{{\color{\colorMATH}\ensuremath{\distance}}}\endgroup }}}} is necessary to support (and unique to) advanced
variants of differential privacy like  ({{\color{\colorMATH}\ensuremath{\epsilon }}}, {{\color{\colorMATH}\ensuremath{\delta }}})-differential privacy.
In the pure  {{\color{\colorMATH}\ensuremath{\epsilon }}}-differential privacy framework, it is common to first
establish the property for {{\color{\colorMATH}\ensuremath{{\begingroup\renewcommand\colorMATH{\colorMATHB}\renewcommand\colorSYNTAX{\colorSYNTAXB}{{\color{\colorMATH}\ensuremath{\distance}}}\endgroup } = 1}}}, that is, if {{\color{\colorMATH}\ensuremath{{\mathfrak{D}} _{\tau _{1}}(x,x^{\prime}) \leq  1}}}
then {{\color{\colorMATH}\ensuremath{f(x)}}} and {{\color{\colorMATH}\ensuremath{f(x^{\prime})}}} are {{\color{\colorMATH}\ensuremath{\epsilon }}}-close. Once established, this
property then implies that if {{\color{\colorMATH}\ensuremath{{\mathfrak{D}} _{\tau _{1}}(x,x^{\prime}) \leq  {\begingroup\renewcommand\colorMATH{\colorMATHB}\renewcommand\colorSYNTAX{\colorSYNTAXB}{{\color{\colorMATH}\ensuremath{\distance}}}\endgroup }}}} then {{\color{\colorMATH}\ensuremath{f(x)}}} and
{{\color{\colorMATH}\ensuremath{f(x^{\prime})}}} are {{\color{\colorMATH}\ensuremath{{\begingroup\renewcommand\colorMATH{\colorMATHB}\renewcommand\colorSYNTAX{\colorSYNTAXB}{{\color{\colorMATH}\ensuremath{\distance}}}\endgroup }\epsilon }}}-close, for any {{\color{\colorMATH}\ensuremath{{\begingroup\renewcommand\colorMATH{\colorMATHB}\renewcommand\colorSYNTAX{\colorSYNTAXB}{{\color{\colorMATH}\ensuremath{\distance}}}\endgroup }}}}. However, this linear scaling does {\textit{not}} carry over to advanced variants like  ({{\color{\colorMATH}\ensuremath{\epsilon }}}, {{\color{\colorMATH}\ensuremath{\delta }}})-differential
privacy. As a consequence, {{\color{\colorMATH}\ensuremath{{\begingroup\renewcommand\colorMATH{\colorMATHB}\renewcommand\colorSYNTAX{\colorSYNTAXB}{{\color{\colorMATH}\ensuremath{\distance}}}\endgroup }}}} must be specified directly as a
parameter and cannot be recovered by scaling the property instantiated to {{\color{\colorMATH}\ensuremath{{\begingroup\renewcommand\colorMATH{\colorMATHB}\renewcommand\colorSYNTAX{\colorSYNTAXB}{{\color{\colorMATH}\ensuremath{\distance}}}\endgroup } =
1}}}. We refer to this distance---{{\color{\colorMATH}\ensuremath{{\begingroup\renewcommand\colorMATH{\colorMATHB}\renewcommand\colorSYNTAX{\colorSYNTAXB}{{\color{\colorMATH}\ensuremath{\distance}}}\endgroup }}}}---as the {\textit{relational distance}} since
it pertains to the (two-run) relational property of differential
privacy, and specifically, the distance between inputs {{\color{\colorMATH}\ensuremath{x}}} and {{\color{\colorMATH}\ensuremath{x^{\prime}}}}
for each of the two executions {{\color{\colorMATH}\ensuremath{f(x)}}} and {{\color{\colorMATH}\ensuremath{f(x^{\prime})}}}. We also use this
terminology in the context of sensitivity, e.g., an
{{\color{\colorMATH}\ensuremath{s}}}-sensitive function is one which upon inputs within relational
distance {{\color{\colorMATH}\ensuremath{{\begingroup\renewcommand\colorMATH{\colorMATHB}\renewcommand\colorSYNTAX{\colorSYNTAXB}{{\color{\colorMATH}\ensuremath{\distance}}}\endgroup }}}} returns outputs within relational distance {{\color{\colorMATH}\ensuremath{s{\begingroup\renewcommand\colorMATH{\colorMATHB}\renewcommand\colorSYNTAX{\colorSYNTAXB}{{\color{\colorMATH}\ensuremath{\distance}}}\endgroup }}}}.

As explained in Section~\ref{sec:background}, differential privacy is usually
achieved by the use of mechanisms like the Laplace (for {{\color{\colorMATH}\ensuremath{\epsilon }}}-differential privacy) or the Gaussian mechanism (for  ({{\color{\colorMATH}\ensuremath{\epsilon }}}, {{\color{\colorMATH}\ensuremath{\delta }}})-differential privacy). %---in order to add noise depending on the sensitivity of
%the computation that produced the value to protect.
%
In \system, the primitive
function implementing the Laplace mechanism has the following type:
\begingroup\color{\colorMATH}\begin{gather*} {\begingroup\renewcommand\colorMATH{\colorMATHC}\renewcommand\colorSYNTAX{\colorSYNTAXC}{{\color{\colorMATH}\ensuremath{{\text{laplace}}}}}\endgroup }
   \mathrel{:}
   \forall  (\hat d \mathrel{:} {\mathbb{R}})\hspace*{0.33em}(\hat \epsilon  \mathrel{:} {\mathbb{R}}).\hspace*{0.33em}
     (d \mathrel{:} {\mathbb{R}}[\hat d])
     \mathrel{{\begingroup\renewcommand\colorMATH{\colorMATHB}\renewcommand\colorSYNTAX{\colorSYNTAXB}{{\color{\colorMATH}\ensuremath{\rightarrow }}}\endgroup }} (\epsilon  \mathrel{:} {\mathbb{R}}[\hat \epsilon ])
     \mathrel{{\begingroup\renewcommand\colorMATH{\colorMATHB}\renewcommand\colorSYNTAX{\colorSYNTAXB}{{\color{\colorMATH}\ensuremath{\rightarrow }}}\endgroup }} (x \mathrel{:} {\mathbb{R}}\mathord{\cdotp }\hat d)
     \xrightarrowP {\infty (d + \epsilon ) + (\hat \epsilon ,0)x}
     {\mathbb{R}}
\end{gather*}\endgroup
There are three logical parameters to the {\begingroup\renewcommand\colorMATH{\colorMATHC}\renewcommand\colorSYNTAX{\colorSYNTAXC}{{\color{\colorMATH}\ensuremath{{\text{laplace}}}}}\endgroup } function: {{\color{\colorMATH}\ensuremath{d}}}
is the relational distance (explained above) used in the statement of
privacy satisfied by the function,
%(e.g., often {{\color{\colorMATH}\ensuremath{1}}},
%per the definition of differential privacy)
{{\color{\colorMATH}\ensuremath{\epsilon }}} is the
privacy \toplas{level} we want to enforce, and {{\color{\colorMATH}\ensuremath{x}}} is the value we (want to protect and) are
adding noise to. When {\textit{executing}} {\begingroup\renewcommand\colorMATH{\colorMATHC}\renewcommand\colorSYNTAX{\colorSYNTAXC}{{\color{\colorMATH}\ensuremath{{\text{laplace}}}}}\endgroup }, the amount of noise
added is {{\color{\colorMATH}\ensuremath{{\text{Lap}}\left(\frac{d}{\epsilon }\right)}}}
which depends on both {{\color{\colorMATH}\ensuremath{d}}} and {{\color{\colorMATH}\ensuremath{\epsilon }}}, so they must be {\textit{runtime values}}.
Also, when {\textit{typechecking}} {\begingroup\renewcommand\colorMATH{\colorMATHC}\renewcommand\colorSYNTAX{\colorSYNTAXC}{{\color{\colorMATH}\ensuremath{{\text{laplace}}}}}\endgroup }, the amount of privacy
obtained depends on {{\color{\colorMATH}\ensuremath{\epsilon }}}, and the distance {{\color{\colorMATH}\ensuremath{d}}} must also be tracked
to enforce that the computation \toplas{feeding {\begingroup\renewcommand\colorMATH{\colorMATHC}\renewcommand\colorSYNTAX{\colorSYNTAXC}{{\color{\colorMATH}\ensuremath{{\text{laplace}}}}}\endgroup } with its argument {{\color{\colorMATH}\ensuremath{x}}}} produces values within relational distance no larger than {{\color{\colorMATH}\ensuremath{d}}}.
Because the values of {{\color{\colorMATH}\ensuremath{d}}} and {{\color{\colorMATH}\ensuremath{\epsilon }}} are required for both runtime execution and
type checking, we require a form of dependent types.
%  in some form to support the
% dependence of types on the values of function arguments.

To support dependent types, we use a {\textit{singletons}} approach---a
technique initially developed for dependently typed programming in
Haskell~\cite{hajashi91,singletons}, and which we borrow directly from \dfuzz
in the context of supporting parameterized differentially private
functions~\cite{gaboardi2013linear}. In this approach, each dependent argument has
two representations---one for the type and term level respectively. In
the type of {\begingroup\renewcommand\colorMATH{\colorMATHC}\renewcommand\colorSYNTAX{\colorSYNTAXC}{{\color{\colorMATH}\ensuremath{{\text{laplace}}}}}\endgroup }, {{\color{\colorMATH}\ensuremath{\hat d}}} is the type-level representation of
term-level variable {{\color{\colorMATH}\ensuremath{d}}}, and likewise for {{\color{\colorMATH}\ensuremath{\hat \epsilon }}} and {{\color{\colorMATH}\ensuremath{\epsilon }}}. (We further discuss
singletons and their implementation in Section~\ref{sec:system}.)

The final argument {{\color{\colorMATH}\ensuremath{x \mathrel{:} {\mathbb{R}}\mathord{\cdotp }\hat d}}} will
have Laplace noise added to it and then returned as the result of
the {\begingroup\renewcommand\colorMATH{\colorMATHC}\renewcommand\colorSYNTAX{\colorSYNTAXC}{{\color{\colorMATH}\ensuremath{{\text{laplace}}}}}\endgroup } function.
% (specifically,
% {{\color{\colorMATH}\ensuremath{{\text{Lap}}\left(\frac{{{\color{\colorMATH}\ensuremath{\hat s}}}}{{{\color{\colorMATH}\ensuremath{\hat \epsilon }}}}\right)}}}) \fo{Isn't {{\color{\colorMATH}\ensuremath{{\text{Lap}}\left(\frac{{{\color{\colorMATH}\ensuremath{s}}}}{{{\color{\colorMATH}\ensuremath{\epsilon }}}}\right)}}} more appropriate?}, and then returned as the output to
% the function.
The annotation ``{{\color{\colorMATH}\ensuremath{\mathord{\cdotp }\hat d}}}'' in the type of {{\color{\colorMATH}\ensuremath{x}}}
places a {\textit{precondition}} on the {\textit{computation}} used to supply the
value to protect: its output must have relational distance no larger than {{\color{\colorMATH}\ensuremath{\hat {\begingroup\renewcommand\colorMATH{\colorMATHB}\renewcommand\colorSYNTAX{\colorSYNTAXB}{{\color{\colorMATH}\ensuremath{\distance}}}\endgroup }}}}.
% \fo{We
% should clarify the sensitivity wrt which variable we are referring
% (the one we want to protect --the DP mechanism input)}\dd{I tried to
% clarify this. ok?};
After all, the noise added is only guaranteed to
give {{\color{\colorMATH}\ensuremath{\epsilon }}}-differential privacy for values that result from
computations with relational distance {{\color{\colorMATH}\ensuremath{\hat {\begingroup\renewcommand\colorMATH{\colorMATHB}\renewcommand\colorSYNTAX{\colorSYNTAXB}{{\color{\colorMATH}\ensuremath{\distance}}}\endgroup }}}}.

The final privacy effect for the function is {{\color{\colorMATH}\ensuremath{\infty (d + \epsilon ) + (\hat \epsilon ,0)x}}},
indicating that no privacy promises are made for the values {{\color{\colorMATH}\ensuremath{d}}} and
{{\color{\colorMATH}\ensuremath{\epsilon }}}, and that privacy is promised for input {{\color{\colorMATH}\ensuremath{x}}} with cost {{\color{\colorMATH}\ensuremath{(\hat \epsilon ,0)}}};
we write {{\color{\colorMATH}\ensuremath{\infty (d + \epsilon )}}} as shorthand for {{\color{\colorMATH}\ensuremath{\infty d + \infty \epsilon }}}.

\toplas{Like \fuzz, \dfuzz, and \duet, \system extends the notion of
  differential privacy from single-argument to multi-argument
  functions, assigning each argument a privacy cost (e.g. the privacy
  effect {{\color{\colorMATH}\ensuremath{\infty (d + \epsilon ) + (\hat \epsilon ,0)x}}} for the {\begingroup\renewcommand\colorMATH{\colorMATHC}\renewcommand\colorSYNTAX{\colorSYNTAXC}{{\color{\colorMATH}\ensuremath{{\text{laplace}}}}}\endgroup } function
  describes privacy costs for {{\color{\colorMATH}\ensuremath{d}}}, {{\color{\colorMATH}\ensuremath{\epsilon }}}, and {{\color{\colorMATH}\ensuremath{x}}}). This approach is
  formalized in Section~\ref{sec:soundness}. By convention, most
  differentially private programs expect a single input to contain the
  sensitive data, and the privacy cost assigned to this argument is
  most important in ensuring privacy. The costs associated with the
  other arguments are typically infinite, indicating that the program
  does not preserve privacy for these inputs.}

We can give a
similar type to the {{\color{\colorMATH}\ensuremath{{\text{gauss}}}}} function, which provides {{\color{\colorMATH}\ensuremath{(\epsilon ,
\delta )}}}-differential privacy by adding Gaussian noise drawn from
{{\color{\colorMATH}\ensuremath{{\mathcal{N}}\left(\frac{2{\begingroup\renewcommand\colorMATH{\colorMATHB}\renewcommand\colorSYNTAX{\colorSYNTAXB}{{\color{\colorMATH}\ensuremath{\distance}}}\endgroup } ^{2}\ln (1.25/\delta )}{\epsilon ^{2}}\right)}}}:
\begingroup\color{\colorMATH}\begin{gather*} {\begingroup\renewcommand\colorMATH{\colorMATHC}\renewcommand\colorSYNTAX{\colorSYNTAXC}{{\color{\colorMATH}\ensuremath{{\text{gauss}}}}}\endgroup }
   \mathrel{:}
   \forall  (\hat d \mathrel{:} {\mathbb{R}})\hspace*{0.33em}(\hat \epsilon  \mathrel{:} {\mathbb{R}})\hspace*{0.33em}(\hat \delta  \mathrel{:} {\mathbb{R}}).\hspace*{0.33em}
     (d \mathrel{:} {\mathbb{R}}[\hat d]) \mathrel{{\begingroup\renewcommand\colorMATH{\colorMATHB}\renewcommand\colorSYNTAX{\colorSYNTAXB}{{\color{\colorMATH}\ensuremath{\rightarrow }}}\endgroup }} (\epsilon  \mathrel{:} {\mathbb{R}}[\hat \epsilon ]) \mathrel{{\begingroup\renewcommand\colorMATH{\colorMATHB}\renewcommand\colorSYNTAX{\colorSYNTAXB}{{\color{\colorMATH}\ensuremath{\rightarrow }}}\endgroup }} (\delta  \mathrel{:} {\mathbb{R}}[\hat \delta ]) \mathrel{{\begingroup\renewcommand\colorMATH{\colorMATHB}\renewcommand\colorSYNTAX{\colorSYNTAXB}{{\color{\colorMATH}\ensuremath{\rightarrow }}}\endgroup }} (x \mathrel{:} {\mathbb{R}}\mathord{\cdotp }\hat d) \xrightarrowP {\infty (d + \epsilon  + \delta ) + (\hat \epsilon ,\hat \delta )x} {\mathbb{R}}
\end{gather*}\endgroup
In \system, privacy primitives are used in the privacy
  sublanguage. For example, the following privacy-sublanguage
expression partially applies the Gaussian mechanism to values for {{\color{\colorMATH}\ensuremath{{\begingroup\renewcommand\colorMATH{\colorMATHB}\renewcommand\colorSYNTAX{\colorSYNTAXB}{{\color{\colorMATH}\ensuremath{\distance}}}\endgroup }}}}, {{\color{\colorMATH}\ensuremath{\epsilon }}}
and {{\color{\colorMATH}\ensuremath{\delta }}}, resulting in a privacy function that satisfies {{\color{\colorMATH}\ensuremath{(1.5,
10^{-5})}}}-differential privacy \toplas{for any input at relational distance {{\color{\colorMATH}\ensuremath{4}}}}:
% (We use
% an artificially low value of {{\color{\colorMATH}\ensuremath{\delta }}} for readability.) \dd{isn't the
% value artificially high, and the privacy guarantee artificially
% low??}:
%
\begingroup\color{\colorMATH}\begin{gather*} {\text{gauss}}\hspace*{0.33em}4\hspace*{0.33em}1.5\hspace*{0.33em}10^{-5} \mathrel{:} {{\color{\colorSYNTAX}\texttt{{\ensuremath{({{\color{\colorMATH}\ensuremath{x}}} \mathrel{:} {\mathbb{R}}\mathord{\cdotp }{{\color{\colorMATH}\ensuremath{4}}}) \xrightarrowP {({{\color{\colorSYNTAX}\texttt{{\ensuremath{{\begingroup\renewcommand\colorMATH{\colorMATHA}\renewcommand\colorSYNTAX{\colorSYNTAXA}{{\color{\colorMATH}\ensuremath{1.5}}}\endgroup },{\begingroup\renewcommand\colorMATH{\colorMATHA}\renewcommand\colorSYNTAX{\colorSYNTAXA}{{\color{\colorMATH}\ensuremath{10^{-5}}}}\endgroup }}}}}}){\begingroup\renewcommand\colorMATH{\colorMATHA}\renewcommand\colorSYNTAX{\colorSYNTAXA}{{\color{\colorMATH}\ensuremath{x}}}\endgroup }} {\mathbb{R}}}}}}}
\end{gather*}\endgroup
Note that we omit the instantiation of forall-quantified type variables
{{\color{\colorMATH}\ensuremath{\hat {\begingroup\renewcommand\colorMATH{\colorMATHB}\renewcommand\colorSYNTAX{\colorSYNTAXB}{{\color{\colorMATH}\ensuremath{\distance}}}\endgroup }}}}, {{\color{\colorMATH}\ensuremath{\hat \epsilon }}} and {{\color{\colorMATH}\ensuremath{\hat \delta }}} to type-level constants {{\color{\colorMATH}\ensuremath{4}}}, {{\color{\colorMATH}\ensuremath{1.5}}} and
{{\color{\colorMATH}\ensuremath{10^{-5}}}}, as they can be inferred from the value-level arguments {{\color{\colorMATH}\ensuremath{{\begingroup\renewcommand\colorMATH{\colorMATHB}\renewcommand\colorSYNTAX{\colorSYNTAXB}{{\color{\colorMATH}\ensuremath{\distance}}}\endgroup }}}},
{{\color{\colorMATH}\ensuremath{\epsilon }}} and {{\color{\colorMATH}\ensuremath{\delta }}}.

The privacy sublanguage also contains monadic \emph{bind} (notated
{\begingroup\renewcommand\colorMATH{\colorMATHC}\renewcommand\colorSYNTAX{\colorSYNTAXC}{{\color{\colorSYNTAX}\texttt{{\ensuremath{\leftarrow }}}}}\endgroup }) and \emph{return} constructs for composing differentially
private computations. Privacy
functions {{\color{\colorMATH}\ensuremath{{{\color{\colorSYNTAX}\texttt{{\ensuremath{{\begingroup\renewcommand\colorMATH{\colorMATHC}\renewcommand\colorSYNTAX{\colorSYNTAXC}{{\color{\colorMATH}\ensuremath{\plambda}}}\endgroup } }}}}}(x\mathord{\cdotp }{\begingroup\renewcommand\colorMATH{\colorMATHB}\renewcommand\colorSYNTAX{\colorSYNTAXB}{{\color{\colorMATH}\ensuremath{\distance}}}\endgroup }).\hspace*{0.33em} {\begingroup\renewcommand\colorMATH{\colorMATHC}\renewcommand\colorSYNTAX{\colorSYNTAXC}{{\color{\colorMATH}\ensuremath{e}}}\endgroup }}}} are created in the sensitivity sublanguage
(because function creation is ``pure''), although the function body
{\begingroup\renewcommand\colorMATH{\colorMATHC}\renewcommand\colorSYNTAX{\colorSYNTAXC}{{\color{\colorMATH}\ensuremath{e}}}\endgroup } lives in the privacy sublanguage.
The annotation {{\color{\colorMATH}\ensuremath{{\begingroup\renewcommand\colorMATH{\colorMATHB}\renewcommand\colorSYNTAX{\colorSYNTAXB}{{\color{\colorMATH}\ensuremath{\distance}}}\endgroup }}}} is the relational distance explained previously
for privacy function types. For example, the
following function computes two differentially private results and
adds them together:
\begingroup\color{\colorMATH}\begin{gather*}
\begin{tabularx}{\linewidth}{>{\centering\arraybackslash\(}X<{\)}}
\hfill\hspace{0pt}
\hfill\hspace{0pt}
\begin{array}{l
} {\begingroup\renewcommand\colorMATH{\colorMATHC}\renewcommand\colorSYNTAX{\colorSYNTAXC}{{\color{\colorMATH}\ensuremath{\plambda}}}\endgroup } (x \mathord{\cdotp } 1).\hspace*{0.33em}
   \begin{array}[t]{l
   } r_{1} \mathrel{{\begingroup\renewcommand\colorMATH{\colorMATHC}\renewcommand\colorSYNTAX{\colorSYNTAXC}{{\color{\colorMATH}\ensuremath{\leftarrow }}}\endgroup }} {\text{gauss}}\hspace*{0.33em}1\hspace*{0.33em}1.5\hspace*{0.33em}0.001\hspace*{0.33em}x {\begingroup\renewcommand\colorMATH{\colorMATHC}\renewcommand\colorSYNTAX{\colorSYNTAXC}{{\color{\colorMATH}\ensuremath{;}}}\endgroup }
   \cr  r_{2} \mathrel{{\begingroup\renewcommand\colorMATH{\colorMATHC}\renewcommand\colorSYNTAX{\colorSYNTAXC}{{\color{\colorMATH}\ensuremath{\leftarrow }}}\endgroup }} {\text{gauss}}\hspace*{0.33em}2\hspace*{0.33em}0.5\hspace*{0.33em}0.001\hspace*{0.33em}(x + x) {\begingroup\renewcommand\colorMATH{\colorMATHC}\renewcommand\colorSYNTAX{\colorSYNTAXC}{{\color{\colorMATH}\ensuremath{;}}}\endgroup }
   \cr  {\begingroup\renewcommand\colorMATH{\colorMATHC}\renewcommand\colorSYNTAX{\colorSYNTAXC}{{\color{\colorSYNTAX}\texttt{return}}}\endgroup }\hspace*{0.33em}(r_{1} + r_{2})
   \end{array}
\end{array}
\hfill\hspace{0pt}
\mathrel{:}
\hfill\hspace{0pt}
(x \mathrel{:} {\mathbb{R}}\mathord{\cdotp }1) \xrightarrowP {(2.0,0.002)x} {\mathbb{R}}
\hfill\hspace{0pt}
\hfill\hspace{0pt}
\end{tabularx}
\end{gather*}\endgroup
The \emph{bind} operator encodes the sequential composition
property of differential privacy (Section~\ref{sec:background}), adding up the {{\color{\colorMATH}\ensuremath{\epsilon }}} and {{\color{\colorMATH}\ensuremath{\delta }}} values of
subcomputations. The \emph{return} operator encodes the
post-processing property of differential privacy.
The relational distance parameter of {{\color{\colorMATH}\ensuremath{1}}} is in general inferrable
during type checking; we include it as a
visible term-level parameter for presentation purposes.

\paragraph{Beyond \duet} The privacy sublanguage of \system briefly introduced here lifts a number of important limitations of the privacy sublanguage of \duet.
We sketch two of these here, and postpone further comparison to later sections.

First, to avoid scaling in the privacy sublanguage, \duet requires the
arguments to privacy functions to have a maximum sensitivity \toplas{or relational distance} of
{{\color{\colorMATH}\ensuremath{1}}}. This limitation makes it impossible to give general types to the
{\begingroup\renewcommand\colorMATH{\colorMATHC}\renewcommand\colorSYNTAX{\colorSYNTAXC}{{\color{\colorMATH}\ensuremath{{\text{gauss}}}}}\endgroup } and {\begingroup\renewcommand\colorMATH{\colorMATHC}\renewcommand\colorSYNTAX{\colorSYNTAXC}{{\color{\colorMATH}\ensuremath{{\text{laplace}}}}}\endgroup } functions as we just shown in \system.
% , as we do in \system above---the \duet version is
% limited to {{\color{\colorMATH}\ensuremath{1}}}-sensitive arguments.
As a result, \duet includes a
dedicated {\textit{type rule}} for each basic differential
privacy mechanism, where each rule is parametric in the sensitivity \toplas{or relational distance} of the
argument. \system's addition of \toplas{relational distance} annotation ``{{\color{\colorSYNTAX}\texttt{{\ensuremath{\mathord{\cdotp }{{\color{\colorMATH}\ensuremath{{\begingroup\renewcommand\colorMATH{\colorMATHB}\renewcommand\colorSYNTAX{\colorSYNTAXB}{{\color{\colorMATH}\ensuremath{\distance}}}\endgroup }}}}}}}}}'' in
the types of function arguments eliminates the need for special type rules, and mechanisms can instead be encoded as primitives with an axiomatized type.
The primary benefit of this is that the metatheory need not be extended each time a new mechanism is considered.

Second, while pervasive scaling is generally undesirable for privacy costs, some
constructs such as advanced composition rely on the ability to scale
privacy costs in controlled ways that are supported by \toplass{theorems
specific to that privacy model}.
Because \duet's privacy language disallows scaling entirely,
these constructs are impossible to encode as functions and must also
be given special typing rules. The latent privacy effects in \system
allow constructs like advanced composition to be given regular
function types.
% so they do not require special type rules in
% \system.\jn{can someone check that I used the right PL words
% here?}\dd{Looks good to me!}
Overall, the \system design makes differential privacy by typing in the
presence of higher-order programming possible for advanced differential privacy
variants. The following sections dive into these benefits, by focusing first on the sensitivity sublanguage (Section~\ref{sec:sensitivity-design}), and then the privacy sublanguage (Section~\ref{sec:privacy-design}). Sections~\ref{sec:sensitivity-formalism} and~\ref{sec:privacy-formalism} develop the metatheory of each respective sublanguage.

\section{Design of  \system's Sensitivity Type System}
\label{sec:sensitivity-design}

\system builds upon prior approaches to encoding differential privacy using linear types. In
this section, we first overview some limitations of these approaches related
to the \emph{tracking of sensitivities}, and then discuss how they \toplass{can be}
addressed by \system. In this section we color expressions and metavariables
{\begingroup\renewcommand\colorMATH{\colorMATHB}\renewcommand\colorSYNTAX{\colorSYNTAXB}{{\color{\colorMATH}\ensuremath{{\text{green}}}}}\endgroup } as they pertain to the sensitivity fragment of \system.

\subsection{Linear Products and Sums}
Existing approaches based on linear types
\cite{reed2010distance,gaboardi2013linear,near2019duet} provide elementary
datatype abstractions to programmers such as pairs (products) and tagged unions
(sums). However, \toplass{some of} the sensitivity analysis they implement for these
datatypes can lead to overly imprecise---or even unsound---approximations
in some circumstances.

We now briefly overview these datatype abstractions; a summary is provided in Table~\ref{tab:prodsum}.

% For the sake of self-containedness, we briefly overview these datatype
% abstractions; a summary is provided in Table~\ref{tab:prodsum}.

%
\begin{table}[t]
{\small
\begin{tabular}[t]{p{1.6cm} @{\hspace{1.5em}}l @{\hspace{1.5em}}l @{\hspace{1.5em}}l}
\toprule
\textbf{Datatype}
& \textbf{Intro.}
&  \textbf{Elimination}
& \textbf{Distance}
\\
\bottomrule
\end{tabular}}
\caption{Datatype abstractions provided by systems based on linear types~\cite{reed2010distance}.
\newline{\footnotesize For defining the distance associated to the
datatypes (table last column), we assume that {{\color{\colorMATH}\ensuremath{e_{1 1}, e_{2 1} : \tau _{1}}}} and
{{\color{\colorMATH}\ensuremath{e_{1 2}, e_{2 2} : \tau _{2}}}}.}}
\label{tab:prodsum}
\end{table}

\subsubsection*{Linear products}

Because existing systems are based on intuitionistic linear logic, two product
types emerge: \emph{multiplicative} products {{\color{\colorMATH}\ensuremath{\otimes }}} and \emph{additive} products {{\color{\colorMATH}\ensuremath{\mathrel{\&}}}}. Multiplicative pairs encode two resources, both of which can be used. Additive pairs encode two resources, but in contrast to multiplicative pairs, only one of them can be
used at a time---a computation may use either their left or right component, but not
both.
\toplas{This constraint is reflected on the management of type environments in their typing rules.
Consider, for instance, the multiplicative product {{\color{\colorMATH}\ensuremath{\langle x,x\rangle }}} and the additive product {{\color{\colorMATH}\ensuremath{\addProduct{x}{x}}}}. \fuzz generates the following type derivations for the pairs:
\begingroup\color{\colorMATH}\begin{mathpar}
   \inferrule*[lab={\textsc{ $\otimes $I}}
   ]{ x:_{1}{\mathbb{R}} \hspace*{0.33em}\vdash \hspace*{0.33em}x : {\mathbb{R}}
   \\ x:_{1}{\mathbb{R}} \hspace*{0.33em}\vdash \hspace*{0.33em}x: {\mathbb{R}}
      }{
      x:_{2}{\mathbb{R}} \hspace*{0.33em}\vdash \hspace*{0.33em} \langle x,x\rangle  : {\mathbb{R}}\otimes {\mathbb{R}}
   }
   \and \inferrule*[lab={\textsc{ $\mathrel{\&}$I}}
   ]{ x:_{1}{\mathbb{R}} \hspace*{0.33em}\vdash \hspace*{0.33em}x : {\mathbb{R}}
   \\ x:_{1}{\mathbb{R}} \hspace*{0.33em}\vdash \hspace*{0.33em}x: {\mathbb{R}}
      }{
      x:_{1}{\mathbb{R}} \hspace*{0.33em}\vdash \hspace*{0.33em} \addProduct{x}{x} : {\mathbb{R}}\mathrel{\&}{\mathbb{R}}
   }
\end{mathpar}\endgroup
where judgment {{\color{\colorMATH}\ensuremath{x :_{\begingroup\renewcommand\colorMATH{\colorMATHB}\renewcommand\colorSYNTAX{\colorSYNTAXB}{{\color{\colorMATH}\ensuremath{\sss}}}\endgroup } \tau  \vdash  e : \tau '}}} denotes that expression {{\color{\colorMATH}\ensuremath{e}}} is an {{\color{\colorMATH}\ensuremath{{\begingroup\renewcommand\colorMATH{\colorMATHB}\renewcommand\colorSYNTAX{\colorSYNTAXB}{{\color{\colorMATH}\ensuremath{\sss}}}\endgroup }}}}-sensitive computation on {{\color{\colorMATH}\ensuremath{x}}} (and has type {{\color{\colorMATH}\ensuremath{\tau '}}} assuming that {{\color{\colorMATH}\ensuremath{x}}} has type {{\color{\colorMATH}\ensuremath{\tau }}}). The type derivation on the left (for multiplicative products) adds (variablewise) the environments of both components ({{\color{\colorMATH}\ensuremath{x:_{2}{\mathbb{R}} = x:_{1}{\mathbb{R}} + x:_{1}{\mathbb{R}}}}}), reporting a sensitivity of {{\color{\colorMATH}\ensuremath{2}}} in {{\color{\colorMATH}\ensuremath{x}}}. On the other hand, the type derivation on the right (for additive products) calculates the maximum (variablewise) between the environments of both components ({{\color{\colorMATH}\ensuremath{x:_{1}{\mathbb{R}} = \mathit{max}(x:_{1}{\mathbb{R}}, x:_{1}{\mathbb{R}})}}}), reporting a sensitivity of {{\color{\colorMATH}\ensuremath{1}}} in {{\color{\colorMATH}\ensuremath{x}}}.
The elimination rules also follow these principles:} while
a multiplicative pair is destructed via pattern matching giving access to both
its components, an additive product is destructed via projection operators
that give access to a single component.

When applied to \toplas{sensitivity} analysis, these type connectives no longer encode
{\textit{accessibility}} of a pair of resources, rather they encode an abstraction of the
\toplas{{\textit{sensitivities}}} of each component of the pair. The \toplas{sensitivity} for the whole
pair is coarse and either tracks the sum of \toplas{sensitivities} of each component (in
the case of multiplicative products) or their maximum (in the case of additive
products), as reflected in the last column of Table~\ref{tab:prodsum}.
%\fo{In this paragraph, don't we mean \emph{sensitivity} rather than \emph{distance}? In the end, \fuzz tracks sensitivities.}\mt{A reviewer was asking to change this to distance.}

% As a final remark, we note that multiplicative products are adjoint to the
% function space i.e., ``currying'' in the linear function space uses multiplicative
% pairs.\fo{Not sure whether this is relevant to the point we are making here.}

\subsubsection*{Linear sums} \label{sec:prelimSums}
Rather than a simultaneous occurrence of resources, sums
encode an \emph{alternative} occurrence of resources. Sums are introduced via
{\begingroup\renewcommand\colorMATH{\colorMATHB}\renewcommand\colorSYNTAX{\colorSYNTAXB}{{\color{\colorSYNTAX}\texttt{inl}}}\endgroup } and {\begingroup\renewcommand\colorMATH{\colorMATHB}\renewcommand\colorSYNTAX{\colorSYNTAXB}{{\color{\colorSYNTAX}\texttt{inr}}}\endgroup } constructors, and destructed via a {{\color{\colorMATH}\ensuremath{\ccase}}} expression with one branch for each of the constructors.

In the context of sensitivity analysis, the
sensitivity of a sum \toplas{{\begingroup\renewcommand\colorMATH{\colorMATHB}\renewcommand\colorSYNTAX{\colorSYNTAXB}{{\color{\colorMATH}\ensuremath{\inlr\hspace*{0.33em}e}}}\endgroup }} encodes {\textit{both}} the sensitivities of the
contained expression {\begingroup\renewcommand\colorMATH{\colorMATHB}\renewcommand\colorSYNTAX{\colorSYNTAXB}{{\color{\colorMATH}\ensuremath{e}}}\endgroup }, as well as the sensitivities for the {\textit{direction}} of the injection (left or right).
For example, {{\color{\colorMATH}\ensuremath{\inl\hspace*{0.33em}(x + x)}}} is {{\color{\colorMATH}\ensuremath{2}}}-sensitive in {{\color{\colorMATH}\ensuremath{x}}},
however {{\color{\colorMATH}\ensuremath{{\begingroup\renewcommand\colorMATH{\colorMATHB}\renewcommand\colorSYNTAX{\colorSYNTAXB}{{\color{\colorSYNTAX}\texttt{if}}}\endgroup }\hspace*{0.33em}y\leq 10\hspace*{0.33em}{\begingroup\renewcommand\colorMATH{\colorMATHB}\renewcommand\colorSYNTAX{\colorSYNTAXB}{{\color{\colorSYNTAX}\texttt{then}}}\endgroup }\hspace*{0.33em}\inl\hspace*{0.33em}x\hspace*{0.33em}{\begingroup\renewcommand\colorMATH{\colorMATHB}\renewcommand\colorSYNTAX{\colorSYNTAXB}{{\color{\colorSYNTAX}\texttt{else}}}\endgroup }\hspace*{0.33em}\inr\hspace*{0.33em}x}}}
is {{\color{\colorMATH}\ensuremath{\infty }}}-sensitive in {{\color{\colorMATH}\ensuremath{y}}} because a change in {{\color{\colorMATH}\ensuremath{y}}} could change the
direction of the injection.

As usual, these systems leverage sums to encode boolean values, e.g., the boolean type is
encoded as {{\color{\colorMATH}\ensuremath{{\mathbb{B}} \triangleq  {{\color{\colorSYNTAX}\texttt{{\ensuremath{{{\color{\colorSYNTAX}\texttt{unit}}} \oplus  {{\color{\colorSYNTAX}\texttt{unit}}}}}}}}}}}, where {{\color{\colorSYNTAX}\texttt{unit}}} represents the
unit \emph{type}, inhabited by unit \emph{value} {\begingroup\renewcommand\colorMATH{\colorMATHB}\renewcommand\colorSYNTAX{\colorSYNTAXB}{{\color{\colorSYNTAX}\texttt{tt}}}\endgroup }. Under this
encoding, an {{\color{\colorMATH}\ensuremath{{\begingroup\renewcommand\colorMATH{\colorMATHB}\renewcommand\colorSYNTAX{\colorSYNTAXB}{{\color{\colorSYNTAX}\texttt{if}}}\endgroup }{-}{\begingroup\renewcommand\colorMATH{\colorMATHB}\renewcommand\colorSYNTAX{\colorSYNTAXB}{{\color{\colorSYNTAX}\texttt{then}}}\endgroup }{-}{\begingroup\renewcommand\colorMATH{\colorMATHB}\renewcommand\colorSYNTAX{\colorSYNTAXB}{{\color{\colorSYNTAX}\texttt{else}}}\endgroup }}}} expression becomes syntactic sugar
for a {\begingroup\renewcommand\colorMATH{\colorMATHB}\renewcommand\colorSYNTAX{\colorSYNTAXB}{{\color{\colorSYNTAX}\texttt{case}}}\endgroup } expression. Also note that boolean values {{\color{\colorMATH}\ensuremath{{\text{true}} \triangleq 
{\begingroup\renewcommand\colorMATH{\colorMATHB}\renewcommand\colorSYNTAX{\colorSYNTAXB}{{\color{\colorSYNTAX}\texttt{inl}}}\endgroup }\hspace*{0.33em}{\begingroup\renewcommand\colorMATH{\colorMATHB}\renewcommand\colorSYNTAX{\colorSYNTAXB}{{\color{\colorSYNTAX}\texttt{tt}}}\endgroup }}}} and {{\color{\colorMATH}\ensuremath{{\text{false}} \triangleq  {\begingroup\renewcommand\colorMATH{\colorMATHB}\renewcommand\colorSYNTAX{\colorSYNTAXB}{{\color{\colorSYNTAX}\texttt{inr}}}\endgroup }\hspace*{0.33em}{\begingroup\renewcommand\colorMATH{\colorMATHB}\renewcommand\colorSYNTAX{\colorSYNTAXB}{{\color{\colorSYNTAX}\texttt{tt}}}\endgroup }}}} are at distance {{\color{\colorMATH}\ensuremath{\infty }}} from each other
in this encoding. This observation will be particularly relevant in some of the
forthcoming examples.

% \remove{An important corollary of this is
% that programs are trivially {{\color{\colorMATH}\ensuremath{1}}}-sensitive with respect to their boolean
% variables (or more generally, {{\color{\colorMATH}\ensuremath{s}}}-sensitive for any {{\color{\colorMATH}\ensuremath{s > 0}}}). Both
% these observations will be particularly relevant in some of the
%forthcoming examples.}\dd{I think this is more confusing than it is
%helpful. E.g., there are also 0-sensitive functions in boolean
%arguments, it's just that any {{\color{\colorMATH}\ensuremath{n}}}-sensitive function in a boolean for
%{{\color{\colorMATH}\ensuremath{n>0}}} is also {{\color{\colorMATH}\ensuremath{m}}}-sensitive for {{\color{\colorMATH}\ensuremath{m>0}}}.}\fo{True}

\subsection{Limitations of Prior Sensitivity Linear Type Systems} % {-{
\fuzz~\cite{reed2010distance} is the first work to leverage linear (or affine) types for reasoning about program sensitivity. Since its introduction, other systems based on linear types were developed to address different limitations of \fuzz. These primarily \toplas{comprise} \dfuzz~\cite{gaboardi2013linear}, which allows value-dependent sensitivities and privacy costs, and \duet~\cite{near2019duet}, which allows advanced variants of differential privacy.

Being based on the same underlying sensitivity analysis, all these systems suffer from common limitations related to the sensitivity tracking for products and sums. Through a series of minimal---yet instructive---examples, we now discuss the limitations we have identified.  %now pinpoint these limitations.

 %We primarily compare \system to \fuzz~\cite{reed2010distance},
 %\dfuzz~\cite{gaboardi2013linear} and \duet~\cite{near2019duet}--e.g.~leaving out \hoaresq~\cite{Barthe:POPL:15} and
 %\lightdp~\cite{zhang2017lightdp}--because \system aims to achieve
 %\emph{lightweight} automated verification of compositional and higher-order
 %differentially-private programs, which is out of reach for systems not
 %built on linear typing\et{maybe this sounds a bit too "definite", I don't know}. \fo{Indeed, \hoareseq features an automated type checker and supports compositional and higher-order
 %differentially-private programs}

\subsubsection*{Limitations related to linear products} % {-{

In \fuzz-like systems, each product and sum type introduces an approximation for the sensitivity analysis they underpin.
When using pair types, this approximation forces the programmer to predict how
each pair will be used in later parts of the program, and select the right one
to achieve precision: if only one component of the pair is used, then the
additive product will give perfect precision; conversely, if both components of the pair are
used with the same sensitivity, then the multiplicative product will give
perfect precision. % In any other case, the use of either product can result in
% precision lost, as illustrated by the following examples.
This is limiting for
abstraction, e.g., a library author must commit to one product
type, and clients of the library may turn out to require the other.

Imprecision issues remain even if functions can be inlined: (1) the optimal
product choice may be influenced by the dynamic control flow of the program, which
 cannot be predicted statically in general; and (2) for multiplicative products in particular, if
both components of the pair are used with different sensitivities in the body of the
pattern match, the sensitivity estimation may give imprecise results. To illustrate these limitations, consider the following examples as seen by \fuzz.

\begin{example}[dynamic control]\label{ex:dyncontrol}
The program below contains a branch on a boolean variable, which determines the usage pattern of an additive pair:  while one branch uses one component of the pair, the other branch uses both.
\begingroup\color{\colorMATH}\begin{gather*}
\begin{array}{l
} \textit{\textcolor{gray}{\small // variant using an additive pair {{\color{\colorMATH}\ensuremath{(\cdot,\cdot)}}}}}
\cr  {\begingroup\renewcommand\colorMATH{\colorMATHB}\renewcommand\colorSYNTAX{\colorSYNTAXB}{{\color{\colorSYNTAX}\texttt{let}}}\endgroup }\hspace*{0.33em}p = (2*x, x)\hspace*{0.33em}{\begingroup\renewcommand\colorMATH{\colorMATHB}\renewcommand\colorSYNTAX{\colorSYNTAXB}{{\color{\colorSYNTAX}\texttt{in}}}\endgroup }
\cr  {\begingroup\renewcommand\colorMATH{\colorMATHB}\renewcommand\colorSYNTAX{\colorSYNTAXB}{{\color{\colorSYNTAX}\texttt{if}}}\endgroup }\hspace*{0.33em}b\hspace*{0.33em} \begin{array}[t]{l
             } {\begingroup\renewcommand\colorMATH{\colorMATHB}\renewcommand\colorSYNTAX{\colorSYNTAXB}{{\color{\colorSYNTAX}\texttt{then}}}\endgroup }\hspace*{0.33em}3 * {\begingroup\renewcommand\colorMATH{\colorMATHB}\renewcommand\colorSYNTAX{\colorSYNTAXB}{{\color{\colorSYNTAX}\texttt{fst}}}\endgroup }\hspace*{0.33em}p
             \cr  {\begingroup\renewcommand\colorMATH{\colorMATHB}\renewcommand\colorSYNTAX{\colorSYNTAXB}{{\color{\colorSYNTAX}\texttt{else}}}\endgroup }\hspace*{0.33em}2 * ({\begingroup\renewcommand\colorMATH{\colorMATHB}\renewcommand\colorSYNTAX{\colorSYNTAXB}{{\color{\colorSYNTAX}\texttt{fst}}}\endgroup }\hspace*{0.33em}p + {\begingroup\renewcommand\colorMATH{\colorMATHB}\renewcommand\colorSYNTAX{\colorSYNTAXB}{{\color{\colorSYNTAX}\texttt{snd}}}\endgroup }\hspace*{0.33em}p)
             \end{array}
\end{array}
\end{gather*}\endgroup
%\mt{Careful, we are using {{\color{\colorMATH}\ensuremath{*}}} symbol for multiplication and that does not match with our syntax. Better to use {{\color{\colorMATH}\ensuremath{\mathord{\cdotp }}}} as {{\color{\colorMATH}\ensuremath{*}}} is used for the special multiplication operator.}
%
%
% \noindent For concreteness, the example is
% written with additive pairs; it induces an imprecise analysis. If rewritten
% using multiplicative pairs, the analysis would also be imprecise.
%
%
First, observe that the program is semantically equivalent to {{\color{\colorMATH}\ensuremath{6*x}}}, which is {{\color{\colorMATH}\ensuremath{6}}}-sensitive in {{\color{\colorMATH}\ensuremath{x}}}. For the sensitivity analysis \`a la \fuzz,
the pair {{\color{\colorMATH}\ensuremath{p}}} is assigned {{\color{\colorMATH}\ensuremath{2}}}-sensitivity in {{\color{\colorMATH}\ensuremath{x}}} (the max of each side).
The \xspace {{\color{\colorMATH}\ensuremath{{{\color{\colorSYNTAX}\texttt{if}}}}}} rule pessimistically takes the maximum between the sensitivities
of each branch. This maximum sensitivity is attained by the {\begingroup\renewcommand\colorMATH{\colorMATHB}\renewcommand\colorSYNTAX{\colorSYNTAXB}{{\color{\colorSYNTAX}\texttt{else}}}\endgroup }-branch and gives {{\color{\colorMATH}\ensuremath{8=2 \cdot (\underline{2}+\underline{2})}}}, where the underlined {{\color{\colorMATH}\ensuremath{\underline{2}}}} corresponds to the sensitivity of pair {{\color{\colorMATH}\ensuremath{p}}} in variable {{\color{\colorMATH}\ensuremath{x}}}.

% To begin with, observe that the program is semantically equivalent to {{\color{\colorMATH}\ensuremath{6*x}}}, which is {{\color{\colorMATH}\ensuremath{6}}}-sensitive in {{\color{\colorMATH}\ensuremath{x}}}. For the sensitivity analysis, \fuzz \xspace {{\color{\colorMATH}\ensuremath{{{\color{\colorSYNTAX}\texttt{if}}}}}} rule pessimistically takes the maximum between the sensitivities
%of each branch. For the program above written using additive pairs, the maximum sensitivity is attained by the {{\color{\colorMATH}\ensuremath{{{\color{\colorSYNTAX}\texttt{else}}}}}} branch and gives {{\color{\colorMATH}\ensuremath{8=2 \cdot (\underline{2}+\underline{2})}}}, where the underlined {{\color{\colorMATH}\ensuremath{\underline{2}}}} corresponds to the sensitivity of pair {{\color{\colorMATH}\ensuremath{p}}} in variable {{\color{\colorMATH}\ensuremath{x}}}, which is computed as the \emph{maximum} of the sensitivities of each component, 2 and 1, respectively.

Now assume that we rewrite the program using a multiplicative---rather than additive---pair:
\begingroup\color{\colorMATH}\begin{gather*}
\begin{array}{l
} \textit{\textcolor{gray}{\small // variant using a multiplicative pair {{\color{\colorMATH}\ensuremath{\langle \cdot,\cdot\rangle }}}}}
\cr  {\begingroup\renewcommand\colorMATH{\colorMATHB}\renewcommand\colorSYNTAX{\colorSYNTAXB}{{\color{\colorSYNTAX}\texttt{let}}}\endgroup }\hspace*{0.33em} x_{1}, x_{2} = \langle 2*x, x\rangle \hspace*{0.33em}{\begingroup\renewcommand\colorMATH{\colorMATHB}\renewcommand\colorSYNTAX{\colorSYNTAXB}{{\color{\colorSYNTAX}\texttt{in}}}\endgroup }
\cr  {\begingroup\renewcommand\colorMATH{\colorMATHB}\renewcommand\colorSYNTAX{\colorSYNTAXB}{{\color{\colorSYNTAX}\texttt{if}}}\endgroup }\hspace*{0.33em}b\hspace*{0.33em}\begin{array}[t]{l
            } {\begingroup\renewcommand\colorMATH{\colorMATHB}\renewcommand\colorSYNTAX{\colorSYNTAXB}{{\color{\colorSYNTAX}\texttt{then}}}\endgroup }\hspace*{0.33em}3 * x_{1}
            \cr  {\begingroup\renewcommand\colorMATH{\colorMATHB}\renewcommand\colorSYNTAX{\colorSYNTAXB}{{\color{\colorSYNTAX}\texttt{else}}}\endgroup }\hspace*{0.33em}2 * (x_{1} + x_{2})
            \end{array}
\end{array}
\end{gather*}\endgroup
In this case, the pair is considered {{\color{\colorMATH}\ensuremath{3}}}-sensitive (rather than {{\color{\colorMATH}\ensuremath{2}}}-sensitive) in {{\color{\colorMATH}\ensuremath{x}}}, an estimate that is obtained by \emph{adding} the sensitivities of its two components, instead of taking their maximum. To obtain the overall program sensitivity, the pair sensitivity is scaled by the maximum sensitivity of the two branches in either component of the pair; this maximum is attained by the {\begingroup\renewcommand\colorMATH{\colorMATHB}\renewcommand\colorSYNTAX{\colorSYNTAXB}{{\color{\colorSYNTAX}\texttt{then}}}\endgroup }-branch and gives 3 (since the {\begingroup\renewcommand\colorMATH{\colorMATHB}\renewcommand\colorSYNTAX{\colorSYNTAXB}{{\color{\colorSYNTAX}\texttt{else}}}\endgroup }-branch has sensitivity {{\color{\colorMATH}\ensuremath{2}}} in both pair components). Overall, this gives an even worse sensitivity in variable {{\color{\colorMATH}\ensuremath{x}}} of {{\color{\colorMATH}\ensuremath{9 = 3 \cdot 3}}}.

% If on the other hand the program is rewritten using multiplicative pairs \mt{Maybe present the alternative program}, the analysis slightly varies. The pair is considered  {{\color{\colorMATH}\ensuremath{3}}}-sensitive (rather than {{\color{\colorMATH}\ensuremath{2}}}-sensitive) in {{\color{\colorMATH}\ensuremath{x}}}, estimate that is obtained by \emph{adding} (rather than "max-\emph{ing}") the sensitivities of its two components. To obtain the overall program sensitivity, the pair sensitivity is scaled by the maximum sensitivity of the two branches in either component of the pair; this maximum is attained by the {{\color{\colorMATH}\ensuremath{{\texttt{then}}}}} branch and gives 3 (since the {{\color{\colorMATH}\ensuremath{{\texttt{else}}}}} branch has sensitivity 2 in both pair components). Overall, this gives an even worse sensitivity in variable {{\color{\colorMATH}\ensuremath{x}}} of {{\color{\colorMATH}\ensuremath{9 = 3 \cdot 3}}}.

In summary, following \fuzz-like analysis, there is no choice of product connective that yields the precise sensitivity bound in {{\color{\colorMATH}\ensuremath{x}}} of {{\color{\colorMATH}\ensuremath{6}}}.
% \toplas{However, in \fuzz we can recover the precise sensitivity bound by adapting the program in terms of the the metric scaling operator {{\color{\colorMATH}\ensuremath{!}}} as
% {{\color{\colorMATH}\ensuremath{{\begingroup\renewcommand\colorMATH{\colorMATHB}\renewcommand\colorSYNTAX{\colorSYNTAXB}{{\color{\colorSYNTAX}\texttt{let}}}\endgroup }\hspace*{0.33em} x_{1}, x_{2} = \langle 2*x, x\rangle \hspace*{0.33em}{\begingroup\renewcommand\colorMATH{\colorMATHB}\renewcommand\colorSYNTAX{\colorSYNTAXB}{{\color{\colorSYNTAX}\texttt{in}}}\endgroup } \hspace*{0.33em} !({\begingroup\renewcommand\colorMATH{\colorMATHB}\renewcommand\colorSYNTAX{\colorSYNTAXB}{{\color{\colorSYNTAX}\texttt{if}}}\endgroup } ...)}}}. The downside of this approach is that it requires a non-deterministic type derivation (typing rules for the metric scaling operator {{\color{\colorMATH}\ensuremath{!}}} require ``guessing'' the appropriate scaling factor) and as a consequence of the scaling, the program is assigned type {{\color{\colorMATH}\ensuremath{!_{\frac{3}{4}}{\mathbb{R}}}}} instead of just {{\color{\colorMATH}\ensuremath{{\mathbb{R}}}}}.
% %(we could go even further and derive that the expression is {{\color{\colorMATH}\ensuremath{2}}}-sensitive in {{\color{\colorMATH}\ensuremath{x}}}, but under a return type {{\color{\colorMATH}\ensuremath{!_{\frac{1}{4}}{\mathbb{R}}}}}).
% }
\qed
\end{example}

\begin{example}[imprecise scaling]\label{ex:scaling}
This example shows how imprecision can arise when components of a
pair are scaled before introduction, and then in an asymmetric way
after elimination. We only show the multiplicative pair variant.
\begingroup\color{\colorMATH}\begin{gather*}
\begin{array}{l
} {\begingroup\renewcommand\colorMATH{\colorMATHB}\renewcommand\colorSYNTAX{\colorSYNTAXB}{{\color{\colorSYNTAX}\texttt{let}}}\endgroup }\hspace*{0.33em}x_{1}, x_{2} = \langle 2 * x,y\rangle  \hspace*{0.33em}{\begingroup\renewcommand\colorMATH{\colorMATHB}\renewcommand\colorSYNTAX{\colorSYNTAXB}{{\color{\colorSYNTAX}\texttt{in}}}\endgroup }
\cr  x_{1} + 2 * x_{2}
\end{array}
\end{gather*}\endgroup
The above program is semantically equivalent to {{\color{\colorMATH}\ensuremath{2*x + 2*y}}}, which is {{\color{\colorMATH}\ensuremath{2}}}-sensitive in {{\color{\colorMATH}\ensuremath{x}}} and {{\color{\colorMATH}\ensuremath{2}}}-sensitive in {{\color{\colorMATH}\ensuremath{y}}}. However, the type-based analysis yields a sensitivity bound of {{\color{\colorMATH}\ensuremath{4}}} in {{\color{\colorMATH}\ensuremath{x}}}, doubling its actual value. The analysis proceeds roughly as follows. The left component of the pair
is {{\color{\colorMATH}\ensuremath{2}}}-sensitive in {{\color{\colorMATH}\ensuremath{x}}}, and the right component is {{\color{\colorMATH}\ensuremath{1}}}-sensitive
in {{\color{\colorMATH}\ensuremath{y}}}. As hinted in the previous example, for multiplicative pairs \fuzz-like systems \emph{sum} the sensitivities of each component to yield the sensitivity of the whole, so the resulting pair is {{\color{\colorMATH}\ensuremath{2}}}-sensitive in {{\color{\colorMATH}\ensuremath{x}}} and
{{\color{\colorMATH}\ensuremath{1}}}-sensitive in {{\color{\colorMATH}\ensuremath{y}}}; note that {{\color{\colorMATH}\ensuremath{\langle 2*x + y, 0\rangle }}},
{{\color{\colorMATH}\ensuremath{\langle 0, 2*x + y\rangle }}} or even {{\color{\colorMATH}\ensuremath{\langle x, x + y\rangle }}} would
also result in the exact same sensitivity analysis. The effect of
eliminating the pair via pattern matching is to scale the pair
sensitivity by the maximum sensitivity of the body
({{\color{\colorMATH}\ensuremath{x_{1} + 2 * x_{2}}}})
in the pattern variables ({{\color{\colorMATH}\ensuremath{x_{1}}}} and {{\color{\colorMATH}\ensuremath{x_{2}}}}), {{\color{\colorMATH}\ensuremath{2}}} in this case. The result
is a final sensitivity of {{\color{\colorMATH}\ensuremath{4 = 2 \cdot 2}}} in {{\color{\colorMATH}\ensuremath{x}}} and {{\color{\colorMATH}\ensuremath{2= 2 \cdot 1}}}
in {{\color{\colorMATH}\ensuremath{y}}}, which is precise for {{\color{\colorMATH}\ensuremath{y}}}, but imprecise for {{\color{\colorMATH}\ensuremath{x}}}.

If the program is converted to instead use additive pairs, the
sensitivity of the pair construction is {{\color{\colorMATH}\ensuremath{2}}} in {{\color{\colorMATH}\ensuremath{x}}} and {{\color{\colorMATH}\ensuremath{1}}} in {{\color{\colorMATH}\ensuremath{y}}}
(the pointwise max from of each side), and the sensitivity of the
whole expression is {{\color{\colorMATH}\ensuremath{6}}} in {{\color{\colorMATH}\ensuremath{x}}} and {{\color{\colorMATH}\ensuremath{3}}} in {{\color{\colorMATH}\ensuremath{y}}}---strictly worse than
the analysis when using multiplicative pairs.

\toplas{We could fix this program in \fuzz, just like in the previous example, by rewriting the program to use the scaling operator: either {{\color{\colorMATH}\ensuremath{{\begingroup\renewcommand\colorMATH{\colorMATHB}\renewcommand\colorSYNTAX{\colorSYNTAXB}{{\color{\colorSYNTAX}\texttt{let}}}\endgroup }\hspace*{0.33em}x_{1}, x_{2} = \langle !2 * x,y\rangle  \hspace*{0.33em}{\begingroup\renewcommand\colorMATH{\colorMATHB}\renewcommand\colorSYNTAX{\colorSYNTAXB}{{\color{\colorSYNTAX}\texttt{in}}}\endgroup }\hspace*{0.33em} {\begingroup\renewcommand\colorMATH{\colorMATHB}\renewcommand\colorSYNTAX{\colorSYNTAXB}{{\color{\colorSYNTAX}\texttt{let}}}\endgroup }\hspace*{0.33em}x_{2}' = x_{2} \hspace*{0.33em}{\begingroup\renewcommand\colorMATH{\colorMATHB}\renewcommand\colorSYNTAX{\colorSYNTAXB}{{\color{\colorSYNTAX}\texttt{in}}}\endgroup }\hspace*{0.33em} x_{1} + 2 * x_{2}}}} or {{\color{\colorMATH}\ensuremath{{\begingroup\renewcommand\colorMATH{\colorMATHB}\renewcommand\colorSYNTAX{\colorSYNTAXB}{{\color{\colorSYNTAX}\texttt{let}}}\endgroup }\hspace*{0.33em}x_{1}, x_{2} = \langle 2 * x,!y\rangle  \hspace*{0.33em}{\begingroup\renewcommand\colorMATH{\colorMATHB}\renewcommand\colorSYNTAX{\colorSYNTAXB}{{\color{\colorSYNTAX}\texttt{in}}}\endgroup }\hspace*{0.33em}{\begingroup\renewcommand\colorMATH{\colorMATHB}\renewcommand\colorSYNTAX{\colorSYNTAXB}{{\color{\colorSYNTAX}\texttt{let}}}\endgroup }\hspace*{0.33em}x_{2}' = x_{2} \hspace*{0.33em}{\begingroup\renewcommand\colorMATH{\colorMATHB}\renewcommand\colorSYNTAX{\colorSYNTAXB}{{\color{\colorSYNTAX}\texttt{in}}}\endgroup }\hspace*{0.33em} x_{1} + 2 * x_{2}}}}.
%, but at the cost of typing {{\color{\colorMATH}\ensuremath{x}}} as {{\color{\colorMATH}\ensuremath{!_{\frac{1}{2}}{\mathbb{R}}}}}, or {{\color{\colorMATH}\ensuremath{y}}} as {{\color{\colorMATH}\ensuremath{!_{2}{\mathbb{R}}}}} respectively.
This may be considered as an annotation burden for programmers because (1) the programmer must know beforehand that the analysis is imprecise (which might be hard for long and complex programs), and (2) the programmer must manually know where to apply scaling to achieve better precision. Also, this process relies in an algorithmic version of the type system of \fuzz, which is not trivial to achieve~\cite{DFuzzTypeChecking}.
Finally, note that scaling in \fuzz is restricted to non-zero sensitivities. This means that a program such as {{\color{\colorMATH}\ensuremath{{\begingroup\renewcommand\colorMATH{\colorMATHB}\renewcommand\colorSYNTAX{\colorSYNTAXB}{{\color{\colorSYNTAX}\texttt{let}}}\endgroup }\hspace*{0.33em}x_{1}, x_{2} = \langle 2 * x,!y\rangle  \hspace*{0.33em}{\begingroup\renewcommand\colorMATH{\colorMATHB}\renewcommand\colorSYNTAX{\colorSYNTAXB}{{\color{\colorSYNTAX}\texttt{in}}}\endgroup }\hspace*{0.33em} x_{1}}}} would be pessimistically considered to be {{\color{\colorMATH}\ensuremath{2}}}-sensitive in {{\color{\colorMATH}\ensuremath{x}}} and {{\color{\colorMATH}\ensuremath{2}}}-sensitive in {{\color{\colorMATH}\ensuremath{y}}}, although the program is really {{\color{\colorMATH}\ensuremath{2}}}-sensitive in {{\color{\colorMATH}\ensuremath{x}}} and {{\color{\colorMATH}\ensuremath{0}}}-sensitive in {{\color{\colorMATH}\ensuremath{y}}}.
}
\qed
\end{example}

\subsubsection*{Limitations related to linear sums} \label{sec:senslimit} % {-{
\label{sec:linear-products-and-sums-lim}
In addition to imprecision with the product types,
\fuzz-like systems also exhibit imprecision with sum types. In these
systems, the sensitivity analysis for a sum introduction is
straightforward: the sensitivity of \toplas{{\begingroup\renewcommand\colorMATH{\colorMATHB}\renewcommand\colorSYNTAX{\colorSYNTAXB}{{\color{\colorMATH}\ensuremath{\inlr\hspace*{0.33em}e}}}\endgroup }} is simply the
sensitivity of {\begingroup\renewcommand\colorMATH{\colorMATHB}\renewcommand\colorSYNTAX{\colorSYNTAXB}{{\color{\colorMATH}\ensuremath{e}}}\endgroup }. The sensitivity analysis for a sum
elimination via expression {{\color{\colorMATH}\ensuremath{{\begingroup\renewcommand\colorMATH{\colorMATHB}\renewcommand\colorSYNTAX{\colorSYNTAXB}{{\color{\colorSYNTAX}\texttt{case}}}\endgroup }\hspace*{0.33em}{\begingroup\renewcommand\colorMATH{\colorMATHB}\renewcommand\colorSYNTAX{\colorSYNTAXB}{{\color{\colorMATH}\ensuremath{e}}}\endgroup }\hspace*{0.33em}{\begingroup\renewcommand\colorMATH{\colorMATHB}\renewcommand\colorSYNTAX{\colorSYNTAXB}{{\color{\colorSYNTAX}\texttt{of}}}\endgroup }\hspace*{0.33em}\{ x_{1} \Rightarrow  {\begingroup\renewcommand\colorMATH{\colorMATHB}\renewcommand\colorSYNTAX{\colorSYNTAXB}{{\color{\colorMATH}\ensuremath{e_{1}}}}\endgroup }\} \{ x_{2}
\Rightarrow  {\begingroup\renewcommand\colorMATH{\colorMATHB}\renewcommand\colorSYNTAX{\colorSYNTAXB}{{\color{\colorMATH}\ensuremath{e_{2}}}}\endgroup }\} }}} is, however, more involved. First, it computes the
sensitivity of {{\color{\colorMATH}\ensuremath{{\begingroup\renewcommand\colorMATH{\colorMATHB}\renewcommand\colorSYNTAX{\colorSYNTAXB}{{\color{\colorMATH}\ensuremath{e}}}\endgroup }_{i}}}} in binder {{\color{\colorMATH}\ensuremath{x_{i}}}} for {{\color{\colorMATH}\ensuremath{i={\begingroup\renewcommand\colorMATH{\colorMATHB}\renewcommand\colorSYNTAX{\colorSYNTAXB}{{\color{\colorMATH}\ensuremath{1}}}\endgroup },{\begingroup\renewcommand\colorMATH{\colorMATHB}\renewcommand\colorSYNTAX{\colorSYNTAXB}{{\color{\colorMATH}\ensuremath{2}}}\endgroup }}}} and retains the
greatest, say {{\color{\colorMATH}\ensuremath{r}}}. The sensitivity of the overall {\begingroup\renewcommand\colorMATH{\colorMATHB}\renewcommand\colorSYNTAX{\colorSYNTAXB}{{\color{\colorSYNTAX}\texttt{case}}}\endgroup } expression
in some variable, say {{\color{\colorMATH}\ensuremath{x}}}, is then computed as the sum between (1)
the max sensitivity of {{\color{\colorMATH}\ensuremath{{\begingroup\renewcommand\colorMATH{\colorMATHB}\renewcommand\colorSYNTAX{\colorSYNTAXB}{{\color{\colorMATH}\ensuremath{e}}}\endgroup }_{i}}}} in {{\color{\colorMATH}\ensuremath{x}}} for {{\color{\colorMATH}\ensuremath{i={\begingroup\renewcommand\colorMATH{\colorMATHB}\renewcommand\colorSYNTAX{\colorSYNTAXB}{{\color{\colorMATH}\ensuremath{1}}}\endgroup },{\begingroup\renewcommand\colorMATH{\colorMATHB}\renewcommand\colorSYNTAX{\colorSYNTAXB}{{\color{\colorMATH}\ensuremath{2}}}\endgroup }}}}, and (2) the
sensitivity of {\begingroup\renewcommand\colorMATH{\colorMATHB}\renewcommand\colorSYNTAX{\colorSYNTAXB}{{\color{\colorMATH}\ensuremath{e}}}\endgroup } in {{\color{\colorMATH}\ensuremath{x}}}, scaled by {{\color{\colorMATH}\ensuremath{r}}}. This brings both unsound
and imprecise estimations.
%\mt{is this sentence applies to all existing Fuzz-like approaches? Answered below}.

\begin{example}[discontinuous predicate]\label{ex:discont}
An unsound corner case of the above sensitivity analysis arises, for
example, for the program:
\begingroup\color{\colorMATH}\begin{gather*}
  {\begingroup\renewcommand\colorMATH{\colorMATHB}\renewcommand\colorSYNTAX{\colorSYNTAXB}{{\color{\colorSYNTAX}\texttt{if}}}\endgroup }\hspace*{0.33em}(x \leq  10)\hspace*{0.33em}{\begingroup\renewcommand\colorMATH{\colorMATHB}\renewcommand\colorSYNTAX{\colorSYNTAXB}{{\color{\colorSYNTAX}\texttt{then}}}\endgroup }\hspace*{0.33em}{\text{true}}\hspace*{0.33em}{\begingroup\renewcommand\colorMATH{\colorMATHB}\renewcommand\colorSYNTAX{\colorSYNTAXB}{{\color{\colorSYNTAX}\texttt{else}}}\endgroup }\hspace*{0.33em}{\text{false}}
\end{gather*}\endgroup
The program, which desugars to {{\color{\colorMATH}\ensuremath{\ccase\hspace*{0.33em}(x \leq  10)\hspace*{0.33em}\{ x_{1} \Rightarrow 
{\text{true}}\} \{ x_{2} \Rightarrow  {\text{false}}\} }}}, is {\textit{semantically}} {{\color{\colorMATH}\ensuremath{\infty }}}-sensitive in
{{\color{\colorMATH}\ensuremath{x}}} because changing {{\color{\colorMATH}\ensuremath{x}}} by, say {{\color{\colorMATH}\ensuremath{1}}}, could change the result from
{{\color{\colorMATH}\ensuremath{{\text{true}}}}} to {{\color{\colorMATH}\ensuremath{{\text{false}}}}}, which are infinitely far apart values.
Intuitively, we can attribute this to the discontinuity of the
program at {{\color{\colorMATH}\ensuremath{x=10}}}. \toplas{
As for \dfuzz and derivative systems like \duet (which support null  sensitivities), they derive a
sensitivity of {{\color{\colorMATH}\ensuremath{0}}} in {{\color{\colorMATH}\ensuremath{x}}}.
To illustrate this, let us consider the corresponding type derivation in \duet:
\begingroup\color{\colorMATH}\begin{mathpar}
   \inferrule*[lab={\textsc{ $\uplus$-E}}
   ]{ x:_{\infty }{\mathbb{R}} \hspace*{0.33em}\vdash \hspace*{0.33em}x \leq  10 : {\mathbb{B}}
   \\ x:_{0}{\mathbb{R}}, x_{1}:_{0}{\mathbb{R}} \hspace*{0.33em}\vdash \hspace*{0.33em}{\text{true}} : {\mathbb{B}}
   \\ x:_{0}{\mathbb{R}}, x_{2}:_{0}{\mathbb{R}} \hspace*{0.33em}\vdash \hspace*{0.33em}{\text{false}} : {\mathbb{B}}
      }{
      x:_{0}{\mathbb{R}} \hspace*{0.33em}\vdash \hspace*{0.33em} \ccase\hspace*{0.33em}(x \leq  10)\hspace*{0.33em}\{ x_{1} \Rightarrow  {\text{true}}\} \{ x_{2} \Rightarrow  {\text{false}}\}  : {\mathbb{B}}
   }
\end{mathpar}\endgroup
The reported sensitivity environment is {{\color{\colorMATH}\ensuremath{ 0 \mathord{\cdotp } x:_{\infty }{\mathbb{R}} + x:_{0}{\mathbb{R}} = x:_{(0 \mathord{\cdotp } \infty  + 0)}{\mathbb{R}} = x:_{0}{\mathbb{R}}}}}: the left summand {{\color{\colorMATH}\ensuremath{0 \mathord{\cdotp } \infty }}} originates from the fact that branches are {{\color{\colorMATH}\ensuremath{0}}}-sensitive in their binders, and expression {{\color{\colorMATH}\ensuremath{x \leq  10}}} is
{{\color{\colorMATH}\ensuremath{\infty }}}-sensitive in {{\color{\colorMATH}\ensuremath{x}}}, and the right summand {{\color{\colorMATH}\ensuremath{0}}} originates from
the fact that both branches are {{\color{\colorMATH}\ensuremath{0}}}-sensitive in {{\color{\colorMATH}\ensuremath{x}}}. Since the product operation (for sensitivities)
adopted by \toplas{\dfuzz} regards {{\color{\colorMATH}\ensuremath{0 \mathord{\cdotp } \infty  = 0}}}, the analysis wrongly infers an
overall sensitivity of {{\color{\colorMATH}\ensuremath{0}}} in {{\color{\colorMATH}\ensuremath{x}}}.\qed}
\end{example}

\toplas{Although \dfuzz and derivative systems do not account for this corner case and are, therefore,
unsound, this soundness problem is not present in \fuzz as its type system is constrained to non-null sensitivities (therefore, leaving the program out of its scope)}.
%thus deriving {{\color{\colorMATH}\ensuremath{\infty }}}-sensitive in {{\color{\colorMATH}\ensuremath{x}}}.
\toplasss{Follow-up work such as ~\cite{DBLP:journals/corr/AmorimAGH15} and \fuzzed}
introduces rules that recover the analysis soundness by interpreting
{{\color{\colorMATH}\ensuremath{\infty  \mathord{\cdotp } 0  = 0 \mathord{\cdotp } \infty  = \infty }}} rather than {{\color{\colorMATH}\ensuremath{\infty  \mathord{\cdotp } 0 = 0}}}, but this leads to imprecision
elsewhere in the system. \toplas{For example, with this fix the program
{{\color{\colorMATH}\ensuremath{{\begingroup\renewcommand\colorMATH{\colorMATHB}\renewcommand\colorSYNTAX{\colorSYNTAXB}{{\color{\colorSYNTAX}\texttt{let}}}\endgroup }\hspace*{0.33em}y = x \leq  10 \hspace*{0.33em}{\begingroup\renewcommand\colorMATH{\colorMATHB}\renewcommand\colorSYNTAX{\colorSYNTAXB}{{\color{\colorSYNTAX}\texttt{in}}}\endgroup }\hspace*{0.33em}1}}} reports
sensitivity {{\color{\colorMATH}\ensuremath{\infty }}} in {{\color{\colorMATH}\ensuremath{x}}} despite the term being equivalent to the
constant {{\color{\colorMATH}\ensuremath{1}}}:
\begingroup\color{\colorMATH}\begin{mathpar}
   \inferrule*[lab={\textsc{ }}
   ]{ x:_{\infty }{\mathbb{R}} \hspace*{0.33em}\vdash \hspace*{0.33em}x \leq  10 : {\mathbb{B}}
   \\ x:_{0}{\mathbb{R}}, y:_{0}{\mathbb{B}} \hspace*{0.33em}\vdash \hspace*{0.33em}1 : {\mathbb{R}}
      }{
      x:_{0 \mathord{\cdotp } \infty }{\mathbb{R}} \hspace*{0.33em}\vdash \hspace*{0.33em} {\begingroup\renewcommand\colorMATH{\colorMATHB}\renewcommand\colorSYNTAX{\colorSYNTAXB}{{\color{\colorSYNTAX}\texttt{let}}}\endgroup }\hspace*{0.33em}y = x \leq  10 \hspace*{0.33em}{\begingroup\renewcommand\colorMATH{\colorMATHB}\renewcommand\colorSYNTAX{\colorSYNTAXB}{{\color{\colorSYNTAX}\texttt{in}}}\endgroup }\hspace*{0.33em}1 : {\mathbb{R}}
   }
\end{mathpar}\endgroup
}

%\fo{We may want to give some intuition on the origin of
%this imprecision.}
A more recent
work~\cite{amorim:popl2017} defines a non-commutative multiplication
operator where {{\color{\colorMATH}\ensuremath{0\mathord{\cdotp }\infty  = \infty }}} but {{\color{\colorMATH}\ensuremath{\infty \mathord{\cdotp }0 = 0}}}.
In doing so, it addresses the soundness problem for
{{\color{\colorMATH}\ensuremath{\ccase}}} expressions, \toplas{and even though not supporting {{\color{\colorMATH}\ensuremath{\tlet}}}-like operations, it could be extended to do so in a precise manner (e.g. {{\color{\colorMATH}\ensuremath{x:_{\infty  \mathord{\cdotp } 0}{\mathbb{R}} \hspace*{0.33em}\vdash \hspace*{0.33em} {\begingroup\renewcommand\colorMATH{\colorMATHB}\renewcommand\colorSYNTAX{\colorSYNTAXB}{{\color{\colorSYNTAX}\texttt{let}}}\endgroup }\hspace*{0.33em}y = x \leq  10 \hspace*{0.33em}{\begingroup\renewcommand\colorMATH{\colorMATHB}\renewcommand\colorSYNTAX{\colorSYNTAXB}{{\color{\colorSYNTAX}\texttt{in}}}\endgroup }\hspace*{0.33em}1 : {\mathbb{R}}}}})}.
% In doing so, it addresses the soundness problem for {{\color{\colorMATH}\ensuremath{\ccase}}} expressions while also \toplas{could prevent} imprecisions for {{\color{\colorMATH}\ensuremath{\tlet}}}-like operations \toplas{(e.g. {{\color{\colorMATH}\ensuremath{x:_{\infty  \mathord{\cdotp } 0}{\mathbb{R}} \hspace*{0.33em}\vdash \hspace*{0.33em} {\begingroup\renewcommand\colorMATH{\colorMATHB}\renewcommand\colorSYNTAX{\colorSYNTAXB}{{\color{\colorSYNTAX}\texttt{let}}}\endgroup }\hspace*{0.33em}y = x \leq  10 \hspace*{0.33em}{\begingroup\renewcommand\colorMATH{\colorMATHB}\renewcommand\colorSYNTAX{\colorSYNTAXB}{{\color{\colorSYNTAX}\texttt{in}}}\endgroup }\hspace*{0.33em}1 : {\mathbb{R}}}}})}.
This multiplication operator is, however, awkward to manipulate and not amenable to automation due to lack of
support for non-commutative ring theories in SMT solvers. Even still, imprecisions continue to arise in this design, as we will see in the forthcoming Example \ref{ex:confbranches}.  %\mt{not presented yet, maybe say ``as shown next' Done!}.

Besides the corner case described above leading to unsound estimates, \emph{imprecise} estimates can also arise when eliminating sums. Imprecision arises because, loosely speaking, the analysis approximates the sensitivity of a sum elimination via a {\begingroup\renewcommand\colorMATH{\colorMATHB}\renewcommand\colorSYNTAX{\colorSYNTAXB}{{\color{\colorSYNTAX}\texttt{case}}}\endgroup } expression as the maximum sensitivity of its branches. As illustrated by the following example, this analysis can dismiss significant information.

% \begin{example}[dead branch]\label{ex:deadbranch}
% Consider program
% \[
% \ccase\hspace*{0.33em}\inl\hspace*{0.33em}(2*x+y)\hspace*{0.33em}\{ x_{1} \Rightarrow  10*x\} \{ x_{2} \Rightarrow  100 * x\} 
% \]
% The left injection of {{\color{\colorMATH}\ensuremath{x+y}}} creates a sum that is eliminated with a left branch {{\color{\colorMATH}\ensuremath{10}}}-sensitive in {{\color{\colorMATH}\ensuremath{x}}} and a right branch {{\color{\colorMATH}\ensuremath{100}}}-sensitive in {{\color{\colorMATH}\ensuremath{x}}}. Even though the right branch ---of greater sensitivity--- is never taken, \fuzz analysis conservatively takes the maximum sensitivity of the two branches leading to an overall program sensitivity of {{\color{\colorMATH}\ensuremath{100}}} in {{\color{\colorMATH}\ensuremath{x}}}, whereas the program is in fact {{\color{\colorMATH}\ensuremath{10}}}-sensitive in {{\color{\colorMATH}\ensuremath{x}}}.\qed
% \end{example}

% The left injection of {{\color{\colorMATH}\ensuremath{x}}} creates a sum that is {{\color{\colorMATH}\ensuremath{1}}}-sensitive in {{\color{\colorMATH}\ensuremath{x}}}; the sum is then eliminated with a left branch

\begin{example}[conflated branches]\label{ex:confbranches}
%\dd{maybe {{\color{\colorMATH}\ensuremath{s}}} is a bad variable name for this, as we use it for sensitivity so often?} \fo{Right! Addressed.}
Consider the following program:
\begingroup\color{\colorMATH}\begin{gather*}
\begin{array}[t]{l
} {\begingroup\renewcommand\colorMATH{\colorMATHB}\renewcommand\colorSYNTAX{\colorSYNTAXB}{{\color{\colorSYNTAX}\texttt{let}}}\endgroup }\hspace*{0.33em}a = {\begingroup\renewcommand\colorMATH{\colorMATHB}\renewcommand\colorSYNTAX{\colorSYNTAXB}{{\color{\colorSYNTAX}\texttt{if}}}\endgroup }\hspace*{0.33em}b\hspace*{0.33em}{\begingroup\renewcommand\colorMATH{\colorMATHB}\renewcommand\colorSYNTAX{\colorSYNTAXB}{{\color{\colorSYNTAX}\texttt{then}}}\endgroup }\hspace*{0.33em}\inl\hspace*{0.33em}(x * x)\hspace*{0.33em}{\begingroup\renewcommand\colorMATH{\colorMATHB}\renewcommand\colorSYNTAX{\colorSYNTAXB}{{\color{\colorSYNTAX}\texttt{else}}}\endgroup }\hspace*{0.33em}\inr\hspace*{0.33em}x\hspace*{0.33em}{{\color{\colorSYNTAX}\texttt{in}}}
\cr  \ccase\hspace*{0.33em}a\hspace*{0.33em}\of\hspace*{0.33em}\{ x_{1} \Rightarrow  0\} \{ x_{2} \Rightarrow  x_{2}\} 
\end{array}
\end{gather*}\endgroup
A sum is created as either the left injection of an expression
that is {{\color{\colorMATH}\ensuremath{\infty }}}-sensitive in {{\color{\colorMATH}\ensuremath{x}}} (since {{\color{\colorMATH}\ensuremath{x * x}}} is so), or the right injection of an expression {{\color{\colorMATH}\ensuremath{1}}}-sensitive
in {{\color{\colorMATH}\ensuremath{x}}}. In \fuzz-like systems, such a sum is conservatively deemed {{\color{\colorMATH}\ensuremath{\infty }}}-sensitive in {{\color{\colorMATH}\ensuremath{x}}}. The sum is then eliminated with a constant left branch,
and a right branch that is {{\color{\colorMATH}\ensuremath{1}}}-sensitive in its binder. The ground truth for the program is that it is {{\color{\colorMATH}\ensuremath{1}}}-sensitive in {{\color{\colorMATH}\ensuremath{x}}}, as the left injection {{\color{\colorMATH}\ensuremath{\infty }}}-sensitive in {{\color{\colorMATH}\ensuremath{x}}} is eliminated to a constant. However, the usual linear typing discipline does not match the sensitivities of each injection with the {\begingroup\renewcommand\colorMATH{\colorMATHB}\renewcommand\colorSYNTAX{\colorSYNTAXB}{{\color{\colorSYNTAX}\texttt{case}}}\endgroup }-branch that each injection  would see, reporting an imprecise final sensitivity of {{\color{\colorMATH}\ensuremath{\infty }}} in {{\color{\colorMATH}\ensuremath{x}}}.\qed
\end{example}

\subsection{Latent Contextual Effects for Precise Sensitivity Tracking in \system} % {-{
\label{sec:advantages-of-sensitivity-ts}

\toplass{\system adopts a novel approach to sensitivity tracking for product and sum types, which can address the previous limitations without the need to rely on scaling of types.}
%\toplass{\system improves on prior \fuzz-like systems in reducing annotation burden, and in its ability to achieve better precise sensitivity tracking without the need of scaling types} for product and sum types
The key insight is to {\textit{delay}} the tracking of
sensitivities whenever possible, and to {\textit{split}} it into two separate
analyses: one for each side of the product or sum. Technically, the main idea is to encode
latent sensitivity effects at the type-connective level. For
instance, for multiplicative pairs \system has type
{{\color{\colorSYNTAX}\texttt{{\ensuremath{{{\color{\colorMATH}\ensuremath{\tau _{1}}}} \mathrel{^{{\begingroup\renewcommand\colorMATH{\colorMATHB}\renewcommand\colorSYNTAX{\colorSYNTAXB}{{\color{\colorMATH}\ensuremath{\Sigma _{1}}}}\endgroup }}\otimes ^{{\begingroup\renewcommand\colorMATH{\colorMATHB}\renewcommand\colorSYNTAX{\colorSYNTAXB}{{\color{\colorMATH}\ensuremath{\Sigma _{2}}}}\endgroup }}} {{\color{\colorMATH}\ensuremath{\tau _{2}}}}}}}}},
where {{\color{\colorMATH}\ensuremath{{\begingroup\renewcommand\colorMATH{\colorMATHB}\renewcommand\colorSYNTAX{\colorSYNTAXB}{{\color{\colorMATH}\ensuremath{\Sigma _{1}}}}\endgroup }}}} and {{\color{\colorMATH}\ensuremath{{\begingroup\renewcommand\colorMATH{\colorMATHB}\renewcommand\colorSYNTAX{\colorSYNTAXB}{{\color{\colorMATH}\ensuremath{\Sigma _{2}}}}\endgroup }}}} denote the
latent sensitivity effects of each of the pair components. This is in contrast to
the \fuzz type {{\color{\colorSYNTAX}\texttt{{\ensuremath{{{\color{\colorMATH}\ensuremath{\tau _{1}}}} \otimes  {{\color{\colorMATH}\ensuremath{\tau _{2}}}}}}}}}, which pays for all of its
sensitivity effects upfront, when the pair is created.

\subsubsection*{Precise products}
Consider the three related multiplicative pair constructions:
\begingroup\color{\colorMATH}\begin{gather*}
{\begingroup\renewcommand\colorMATH{\colorMATHB}\renewcommand\colorSYNTAX{\colorSYNTAXB}{{\color{\colorMATH}\ensuremath{e_{1}}}}\endgroup } \triangleq  \langle 2*x + y, 0\rangle 
\qquad
{\begingroup\renewcommand\colorMATH{\colorMATHB}\renewcommand\colorSYNTAX{\colorSYNTAXB}{{\color{\colorMATH}\ensuremath{e_{2}}}}\endgroup } \triangleq  \langle 0, 2*x + y\rangle 
\qquad
{\begingroup\renewcommand\colorMATH{\colorMATHB}\renewcommand\colorSYNTAX{\colorSYNTAXB}{{\color{\colorMATH}\ensuremath{e_{3}}}}\endgroup } \triangleq  \langle x , x + y\rangle 
\end{gather*}\endgroup
For the purpose of sensitivity analysis, \fuzz is unable to distinguish them, as it derives the very same type judgment for all three, namely
\begingroup\color{\colorMATH}\begin{gather*}
x  \mathrel{:}_{2} {\mathbb{R}}, y \mathrel{:}_{1} {\mathbb{R}} \vdash   {\begingroup\renewcommand\colorMATH{\colorMATHB}\renewcommand\colorSYNTAX{\colorSYNTAXB}{{\color{\colorMATH}\ensuremath{e}}}\endgroup } \mathrel{:} {\mathbb{R}} \otimes  {\mathbb{R}}
\qquad
{{\color{\colorTEXT}\textnormal{for {{\color{\colorMATH}\ensuremath{{\begingroup\renewcommand\colorMATH{\colorMATHB}\renewcommand\colorSYNTAX{\colorSYNTAXB}{{\color{\colorMATH}\ensuremath{e}}}\endgroup } \in  \{ {\begingroup\renewcommand\colorMATH{\colorMATHB}\renewcommand\colorSYNTAX{\colorSYNTAXB}{{\color{\colorMATH}\ensuremath{e_{1}}}}\endgroup }, {\begingroup\renewcommand\colorMATH{\colorMATHB}\renewcommand\colorSYNTAX{\colorSYNTAXB}{{\color{\colorMATH}\ensuremath{e_{2}}}}\endgroup } , {\begingroup\renewcommand\colorMATH{\colorMATHB}\renewcommand\colorSYNTAX{\colorSYNTAXB}{{\color{\colorMATH}\ensuremath{e_{3}}}}\endgroup }\} }}}}}}
\end{gather*}\endgroup
The type judgment says that the pairs are {{\color{\colorMATH}\ensuremath{2}}}-sensitive in {{\color{\colorMATH}\ensuremath{x}}} and
{{\color{\colorMATH}\ensuremath{1}}}-sensitive in {{\color{\colorMATH}\ensuremath{y}}} (the subscript annotations in the type environment), but does not say how this sensitivity effect is
distributed between the pair components. In other words, \fuzz treats
pairs as a whole. In contrast, \system can derive three different type
judgments, precisely capturing the sensitivity of each pair
component:
\begingroup\color{\colorMATH}\begin{gather*}
x \mathrel{:} {\mathbb{R}}, y \mathrel{:} {\mathbb{R}} \hspace*{0.33em} {\begingroup\renewcommand\colorMATH{\colorMATHB}\renewcommand\colorSYNTAX{\colorSYNTAXB}{{\color{\colorMATH}\ensuremath{\vdash }}}\endgroup }\hspace*{0.33em} {\begingroup\renewcommand\colorMATH{\colorMATHB}\renewcommand\colorSYNTAX{\colorSYNTAXB}{{\color{\colorMATH}\ensuremath{e_{1}}}}\endgroup } \mathrel{:} {\mathbb{R}} \mathrel{^{2x+y}\otimes } {\mathbb{R}}
\qquad
x \mathrel{:} {\mathbb{R}}, y \mathrel{:} {\mathbb{R}} \hspace*{0.33em} {\begingroup\renewcommand\colorMATH{\colorMATHB}\renewcommand\colorSYNTAX{\colorSYNTAXB}{{\color{\colorMATH}\ensuremath{\vdash }}}\endgroup }\hspace*{0.33em} {\begingroup\renewcommand\colorMATH{\colorMATHB}\renewcommand\colorSYNTAX{\colorSYNTAXB}{{\color{\colorMATH}\ensuremath{e_{2}}}}\endgroup } \mathrel{:} {\mathbb{R}} \mathrel{\otimes ^{2x+y}} {\mathbb{R}}
\qquad
x \mathrel{:} {\mathbb{R}}, y \mathrel{:} {\mathbb{R}} \hspace*{0.33em} {\begingroup\renewcommand\colorMATH{\colorMATHB}\renewcommand\colorSYNTAX{\colorSYNTAXB}{{\color{\colorMATH}\ensuremath{\vdash }}}\endgroup }\hspace*{0.33em} {\begingroup\renewcommand\colorMATH{\colorMATHB}\renewcommand\colorSYNTAX{\colorSYNTAXB}{{\color{\colorMATH}\ensuremath{e_{3}}}}\endgroup } \mathrel{:} {\mathbb{R}} \mathrel{^{x}\otimes ^{x + y}} {\mathbb{R}}
\end{gather*}\endgroup
Recall from Section~\ref{sec:sensitivity-intro} that in \system we use linear formulas to denote sensitivity effects and therefore, in {e.g.} the first type judgment above, {{\color{\colorMATH}\ensuremath{2x + y}}} refers to the sensitivity effect {{\color{\colorMATH}\ensuremath{{\begingroup\renewcommand\colorMATH{\colorMATHB}\renewcommand\colorSYNTAX{\colorSYNTAXB}{{\color{\colorMATH}\ensuremath{\Sigma }}}\endgroup } \triangleq  \{ x \mapsto  2,y \mapsto  1\} }}}, meaning {{\color{\colorMATH}\ensuremath{2}}}-sensitive in {{\color{\colorMATH}\ensuremath{x}}}, and {{\color{\colorMATH}\ensuremath{1}}}-sensitive in {{\color{\colorMATH}\ensuremath{y}}}. Moreover, we elide null sensitivity effects likes {{\color{\colorMATH}\ensuremath{0x + 0y}}}. This fine-grained tracking of the sensitivity of each pair component allows, in turn, deferring the payment of the pair sensitivity effect to the precise point where the pair is used, {i.e.} eliminated, and therefore paying only for what (and how it) is used. For example, if pair {\begingroup\renewcommand\colorMATH{\colorMATHB}\renewcommand\colorSYNTAX{\colorSYNTAXB}{{\color{\colorMATH}\ensuremath{e_{3}}}}\endgroup } is used in a context where only its first component is referred, we pay for sensitivity effect {{\color{\colorMATH}\ensuremath{x}}}. \fuzz, in contrast, would always pay {{\color{\colorMATH}\ensuremath{2x+y}}}.

Let us discuss the benefits that this fine-grained tracking brings to Examples~\ref{ex:dyncontrol} and \ref{ex:scaling}.  % \mt{recall examples so we don't have to look back? The reason why you enclose programs in numbered environment is precisely to avoid that.}
Consider first the program from Example~\ref{ex:dyncontrol}, more concretely, the variant with additive pairs. The sensitivity of the {\begingroup\renewcommand\colorMATH{\colorMATHB}\renewcommand\colorSYNTAX{\colorSYNTAXB}{{\color{\colorSYNTAX}\texttt{then}}}\endgroup }-branch is calculated as {{\color{\colorMATH}\ensuremath{6x}}} from scaling
by {{\color{\colorMATH}\ensuremath{3}}} the (latent) sensitivity effect {{\color{\colorMATH}\ensuremath{2x}}} of the left component of pair {{\color{\colorMATH}\ensuremath{p}}}. Likewise, the sensitivity of the
{\begingroup\renewcommand\colorMATH{\colorMATHB}\renewcommand\colorSYNTAX{\colorSYNTAXB}{{\color{\colorSYNTAX}\texttt{else}}}\endgroup }-branch is calculated also as {{\color{\colorMATH}\ensuremath{6x}}} from scaling by {{\color{\colorMATH}\ensuremath{2}}} the sum of (latent)
sensitivity effects {{\color{\colorMATH}\ensuremath{2x}}} and {{\color{\colorMATH}\ensuremath{x}}} of the respective left and right component of the pair. %\mt{I would first re-present the example, explain what is the sensitivity of {{\color{\colorMATH}\ensuremath{p}}}, then the sensitivity of each branch, and finally the final sensitivity of the program. At this point it is preferable a high-level explanation rather than one drived by "low-level" type-derivation, which is addressed in the following section.}
As a result, \system reports the precise sensitivity of {{\color{\colorMATH}\ensuremath{6x}}} for the whole program. An analogous
fine-grained tracking for the program from Example~\ref{ex:scaling} gives also precise sensitivity {{\color{\colorMATH}\ensuremath{2x + 2y}}}.
% \mt{Maybe present Table 2 beforehand. The reaoson to place it at the end is to be more punching, when closing the section.}

\subsubsection*{Precise sums}
The use of latent sensitivity effects yields tighter sensitivity bounds also for sums. However, the handling of sums impose an additional technical challenge related to the impossibility of delaying sensitivity effects. To illustrate this phenomenon, consider expressions:
\begingroup\color{\colorMATH}\begin{gather*}
{\begingroup\renewcommand\colorMATH{\colorMATHB}\renewcommand\colorSYNTAX{\colorSYNTAXB}{{\color{\colorMATH}\ensuremath{e_{4}}}}\endgroup } \triangleq  {\begingroup\renewcommand\colorMATH{\colorMATHB}\renewcommand\colorSYNTAX{\colorSYNTAXB}{{\color{\colorSYNTAX}\texttt{inl}}}\endgroup }\hspace*{0.33em}(x * x)
\qquad
{\begingroup\renewcommand\colorMATH{\colorMATHB}\renewcommand\colorSYNTAX{\colorSYNTAXB}{{\color{\colorMATH}\ensuremath{e_{5}}}}\endgroup } \triangleq  {\begingroup\renewcommand\colorMATH{\colorMATHB}\renewcommand\colorSYNTAX{\colorSYNTAXB}{{\color{\colorSYNTAX}\texttt{inr}}}\endgroup }\hspace*{0.33em}(x * x)
\qquad
{\begingroup\renewcommand\colorMATH{\colorMATHB}\renewcommand\colorSYNTAX{\colorSYNTAXB}{{\color{\colorMATH}\ensuremath{e_{6}}}}\endgroup } \triangleq  {\begingroup\renewcommand\colorMATH{\colorMATHB}\renewcommand\colorSYNTAX{\colorSYNTAXB}{{\color{\colorSYNTAX}\texttt{if}}}\endgroup }\hspace*{0.33em}(x \leq  10)\hspace*{0.33em}{\begingroup\renewcommand\colorMATH{\colorMATHB}\renewcommand\colorSYNTAX{\colorSYNTAXB}{{\color{\colorSYNTAX}\texttt{then}}}\endgroup }\hspace*{0.33em}{\begingroup\renewcommand\colorMATH{\colorMATHB}\renewcommand\colorSYNTAX{\colorSYNTAXB}{{\color{\colorSYNTAX}\texttt{inl}}}\endgroup }\hspace*{0.33em}1\hspace*{0.33em}{\begingroup\renewcommand\colorMATH{\colorMATHB}\renewcommand\colorSYNTAX{\colorSYNTAXB}{{\color{\colorSYNTAX}\texttt{else}}}\endgroup }\hspace*{0.33em}{\begingroup\renewcommand\colorMATH{\colorMATHB}\renewcommand\colorSYNTAX{\colorSYNTAXB}{{\color{\colorSYNTAX}\texttt{inr}}}\endgroup }\hspace*{0.33em}1
\end{gather*}\endgroup
All three expressions are {{\color{\colorMATH}\ensuremath{\infty }}}-sensitive in {{\color{\colorMATH}\ensuremath{x}}}. \fuzz sensitivity
analysis conflates \toplass{the three expressions to the same type, and some \toplas{\fuzz} derivative systems with support for {{\color{\colorMATH}\ensuremath{0}}}-sensitivities such as \dfuzz, derive an
unsound type ({w.r.t.}~the embodied sensitivity analysis) for {\begingroup\renewcommand\colorMATH{\colorMATHB}\renewcommand\colorSYNTAX{\colorSYNTAXB}{{\color{\colorMATH}\ensuremath{e_{6}}}}\endgroup }: {{\color{\colorMATH}\ensuremath{x  \mathrel{:}_{0} {\mathbb{R}} \vdash   {\begingroup\renewcommand\colorMATH{\colorMATHB}\renewcommand\colorSYNTAX{\colorSYNTAXB}{{\color{\colorMATH}\ensuremath{e_{6}}}}\endgroup } \mathrel{:} {\mathbb{R}} \oplus  {\mathbb{R}}}}}}.
\begingroup\color{\colorMATH}\begin{gather*}
x  \mathrel{:}_{\infty } {\mathbb{R}} \vdash   {\begingroup\renewcommand\colorMATH{\colorMATHB}\renewcommand\colorSYNTAX{\colorSYNTAXB}{{\color{\colorMATH}\ensuremath{e_{4}}}}\endgroup } \mathrel{:} {\mathbb{R}} \oplus  {\mathbb{R}}
\qquad
x  \mathrel{:}_{\infty } {\mathbb{R}} \vdash   {\begingroup\renewcommand\colorMATH{\colorMATHB}\renewcommand\colorSYNTAX{\colorSYNTAXB}{{\color{\colorMATH}\ensuremath{e_{5}}}}\endgroup } \mathrel{:} {\mathbb{R}} \oplus  {\mathbb{R}}
\qquad
x  \mathrel{:}_{\infty } {\mathbb{R}} \vdash   {\begingroup\renewcommand\colorMATH{\colorMATHB}\renewcommand\colorSYNTAX{\colorSYNTAXB}{{\color{\colorMATH}\ensuremath{e_{6}}}}\endgroup } \mathrel{:} {\mathbb{R}} \oplus  {\mathbb{R}}
\end{gather*}\endgroup
\system derives instead:
\begingroup\color{\colorMATH}\begin{gather*}
x  \mathrel{:} {\mathbb{R}} \hspace*{0.33em}{\begingroup\renewcommand\colorMATH{\colorMATHB}\renewcommand\colorSYNTAX{\colorSYNTAXB}{{\color{\colorMATH}\ensuremath{\vdash }}}\endgroup }\hspace*{0.33em} {\begingroup\renewcommand\colorMATH{\colorMATHB}\renewcommand\colorSYNTAX{\colorSYNTAXB}{{\color{\colorMATH}\ensuremath{e_{4}}}}\endgroup } \mathrel{:} {\mathbb{R}} \mathrel{^{\infty x}\oplus } {\mathbb{R}}
\qquad
x  \mathrel{:} {\mathbb{R}} \hspace*{0.33em}{\begingroup\renewcommand\colorMATH{\colorMATHB}\renewcommand\colorSYNTAX{\colorSYNTAXB}{{\color{\colorMATH}\ensuremath{\vdash }}}\endgroup }\hspace*{0.33em} {\begingroup\renewcommand\colorMATH{\colorMATHB}\renewcommand\colorSYNTAX{\colorSYNTAXB}{{\color{\colorMATH}\ensuremath{e_{5}}}}\endgroup } \mathrel{:} {\mathbb{R}} \mathrel{\oplus ^{\infty x}} {\mathbb{R}}
\qquad
x  \mathrel{:} {\mathbb{R}} \hspace*{0.33em}{\begingroup\renewcommand\colorMATH{\colorMATHB}\renewcommand\colorSYNTAX{\colorSYNTAXB}{{\color{\colorMATH}\ensuremath{\vdash }}}\endgroup }\hspace*{0.33em} {\begingroup\renewcommand\colorMATH{\colorMATHB}\renewcommand\colorSYNTAX{\colorSYNTAXB}{{\color{\colorMATH}\ensuremath{e_{6}}}}\endgroup } \mathrel{:} {\mathbb{R}} \oplus  {\mathbb{R}}
\end{gather*}\endgroup
The types of {\begingroup\renewcommand\colorMATH{\colorMATHB}\renewcommand\colorSYNTAX{\colorSYNTAXB}{{\color{\colorMATH}\ensuremath{e_{4}}}}\endgroup } and {\begingroup\renewcommand\colorMATH{\colorMATHB}\renewcommand\colorSYNTAX{\colorSYNTAXB}{{\color{\colorMATH}\ensuremath{e_{5}}}}\endgroup } encode a latent sensitivity effect for each side of the
sum. In contrast, the type of {\begingroup\renewcommand\colorMATH{\colorMATHB}\renewcommand\colorSYNTAX{\colorSYNTAXB}{{\color{\colorMATH}\ensuremath{e_{6}}}}\endgroup } is not able to represent its {{\color{\colorMATH}\ensuremath{\infty }}}-sensitivity in {{\color{\colorMATH}\ensuremath{x}}} as a latent effect because {{\color{\colorMATH}\ensuremath{x}}} influences which injection is used to create
the sum itself, not the value inside the injection. Instead, the
effect must be paid for {\textit{eagerly}} in the so-called \emph{ambient}
sensitivity effect (which was \toplass{elided} in previous examples).
Therefore type judgments in \system have shape {\begingroup\renewcommand\colorMATH{\colorMATHB}\renewcommand\colorSYNTAX{\colorSYNTAXB}{{\color{\colorSYNTAX}\texttt{{\ensuremath{{\begingroup\renewcommand\colorMATH{\colorMATHA}\renewcommand\colorSYNTAX{\colorSYNTAXA}{{\color{\colorMATH}\ensuremath{\Gamma }}}\endgroup } \vdash  {{\color{\colorMATH}\ensuremath{e}}} \mathrel{:} {\begingroup\renewcommand\colorMATH{\colorMATHA}\renewcommand\colorSYNTAX{\colorSYNTAXA}{{\color{\colorMATH}\ensuremath{\tau }}}\endgroup } \mathrel{;} {{\color{\colorMATH}\ensuremath{\Sigma }}}}}}}}\endgroup },
where {\begingroup\renewcommand\colorMATH{\colorMATHB}\renewcommand\colorSYNTAX{\colorSYNTAXB}{{\color{\colorMATH}\ensuremath{\Sigma }}}\endgroup } represents the ambient sensitivity effect and {{\color{\colorMATH}\ensuremath{\Gamma }}} is a
``traditional'' environment, mapping variables to types. Thus,
expression {\begingroup\renewcommand\colorMATH{\colorMATHB}\renewcommand\colorSYNTAX{\colorSYNTAXB}{{\color{\colorMATH}\ensuremath{e_{6}}}}\endgroup } is formally typed as:
\begingroup\color{\colorMATH}\begin{gather*}
{\begingroup\renewcommand\colorMATH{\colorMATHB}\renewcommand\colorSYNTAX{\colorSYNTAXB}{{\color{\colorSYNTAX}\texttt{{\ensuremath{ \toplas{x  \mathrel{:} {\mathbb{R}}} \vdash  {{\color{\colorMATH}\ensuremath{e_{6}}}} \mathrel{:} {\begingroup\renewcommand\colorMATH{\colorMATHA}\renewcommand\colorSYNTAX{\colorSYNTAXA}{{\color{\colorSYNTAX}\texttt{{\ensuremath{ {\mathbb{R}} \oplus  {\mathbb{R}} }}}}}\endgroup } \mathrel{;} {\begingroup\renewcommand\colorMATH{\colorMATHA}\renewcommand\colorSYNTAX{\colorSYNTAXA}{{\color{\colorMATH}\ensuremath{\infty x}}}\endgroup } }}}}}\endgroup }
\end{gather*}\endgroup
%\mt{Either change {{\color{\colorMATH}\ensuremath{e_{6}}}} to be {{\color{\colorMATH}\ensuremath{{{\color{\colorSYNTAX}\texttt{if}}}\hspace*{0.33em}x\leq 10\hspace*{0.33em}{{\color{\colorSYNTAX}\texttt{then}}}\hspace*{0.33em}{{\color{\colorSYNTAX}\texttt{inl}}}\hspace*{0.33em}0\hspace*{0.33em}{{\color{\colorSYNTAX}\texttt{else}}}\hspace*{0.33em}{{\color{\colorSYNTAX}\texttt{inr}}}\hspace*{0.33em}1}}}, or {{\color{\colorMATH}\ensuremath{\vdash   e_{6} \mathrel{:} {\mathbb{R}} \oplus  {\mathbb{R}} \mathrel{;} x}}} (see below explanation for similar example 4.4). Done!}
%
with ambient sensitivity effect {{\color{\colorMATH}\ensuremath{\infty x}}}. {\begingroup\renewcommand\colorMATH{\colorMATHB}\renewcommand\colorSYNTAX{\colorSYNTAXB}{{\color{\colorMATH}\ensuremath{e_{4}}}}\endgroup } is typed as
{\begingroup\renewcommand\colorMATH{\colorMATHB}\renewcommand\colorSYNTAX{\colorSYNTAXB}{{\color{\colorSYNTAX}\texttt{{\ensuremath{\toplas{x  \mathrel{:} {\mathbb{R}}} \vdash  {{\color{\colorMATH}\ensuremath{e_{4}}}} \mathrel{:} {\begingroup\renewcommand\colorMATH{\colorMATHA}\renewcommand\colorSYNTAX{\colorSYNTAXA}{{\color{\colorSYNTAX}\texttt{{\ensuremath{ {\mathbb{R}} \mathrel{^{{{\color{\colorMATH}\ensuremath{\infty x}}}}\oplus } {\mathbb{R}} }}}}}\endgroup } \mathrel{;} {\begingroup\renewcommand\colorMATH{\colorMATHA}\renewcommand\colorSYNTAX{\colorSYNTAXA}{{\color{\colorMATH}\ensuremath{\varnothing }}}\endgroup }}}}}}\endgroup }, i.e., with an empty ambient
sensitivity effect, and analogously for {\begingroup\renewcommand\colorMATH{\colorMATHB}\renewcommand\colorSYNTAX{\colorSYNTAXB}{{\color{\colorMATH}\ensuremath{e_{5}}}}\endgroup }.

To showcase the benefits of this design, let us re-examine Example~\ref{ex:confbranches}. In \system, the type for {{\color{\colorMATH}\ensuremath{a}}} is {\begingroup\renewcommand\colorMATH{\colorMATHA}\renewcommand\colorSYNTAX{\colorSYNTAXA}{{\color{\colorSYNTAX}\texttt{{\ensuremath{ {\mathbb{R}} \mathrel{^{{{\color{\colorMATH}\ensuremath{\infty x}}}}\oplus ^{{{\color{\colorMATH}\ensuremath{x}}}}} {\mathbb{R}} }}}}}\endgroup } with ambient effect {{\color{\colorMATH}\ensuremath{b}}}.\footnote{At first sight, one might think that {{\color{\colorMATH}\ensuremath{a}}} is {{\color{\colorMATH}\ensuremath{\infty}}}-sensitive in {{\color{\colorMATH}\ensuremath{b}}} because a change in {{\color{\colorMATH}\ensuremath{b}}} may flip the direction of the returned injection. However, any change on the value of {{\color{\colorMATH}\ensuremath{b}}} necessarily results in an infinite variation since {{\color{\colorMATH}\ensuremath{{\text{true}}}}} and {{\color{\colorMATH}\ensuremath{{\text{false}}}}} are {{\color{\colorMATH}\ensuremath{\infty}}} far apart. Therefore, the induced variation on the value of {{\color{\colorMATH}\ensuremath{a}}} is trivially bounded by {{\color{\colorMATH}\ensuremath{\infty}}}, scaled by {{\color{\colorMATH}\ensuremath{1}}}, turning {{\color{\colorMATH}\ensuremath{a}}} {{\color{\colorMATH}\ensuremath{1}}}-sensitive in {{\color{\colorMATH}\ensuremath{b}}}.}
% \fo{Clarify why {{\color{\colorMATH}\ensuremath{b}}} and not {{\color{\colorMATH}\ensuremath{\infty b}}}.} \mt{Intuitively, given two executions, if {{\color{\colorMATH}\ensuremath{b}}} are the same, then the output will not depend on {{\color{\colorMATH}\ensuremath{b}}}, if both {{\color{\colorMATH}\ensuremath{b}}} are different, then the sensitivity will be the distance between the two different {{\color{\colorMATH}\ensuremath{b}}}, which is in this case infinite. This is crisply captured in our metric preservation property} %\mt{This is just {{\color{\colorMATH}\ensuremath{b}}}, not infinite. Either two different {{\color{\colorMATH}\ensuremath{b}}} are infinitely apart, or branching on a {{\color{\colorMATH}\ensuremath{b}}} scales {{\color{\colorMATH}\ensuremath{b}}} to infinite. We choose the former. Corrected!}
 To compute the sensitivity of the {{\color{\colorMATH}\ensuremath{\ccase}}}-expression over {{\color{\colorMATH}\ensuremath{a}}}, we join---by taking the variable-wise maximum---the ambient effect of {{\color{\colorMATH}\ensuremath{a}}}, namely {{\color{\colorMATH}\ensuremath{b}}}, with the ``global'' sensitivity effect of the second branch, namely {{\color{\colorMATH}\ensuremath{[b+x/x_{2}]x_{2}}}}---the first branch is dismissed because it has no ambient effect. To compute the purported sensitivity effect of the second branch, we take its ambient effect {{\color{\colorMATH}\ensuremath{x_{2}}}} and replace every occurrence of the branch binder, also {{\color{\colorMATH}\ensuremath{x_{2}}}}, with the effect {{\color{\colorMATH}\ensuremath{b+x}}} of the right component of {{\color{\colorMATH}\ensuremath{a}}}, computed as the sum between its ambient effect {{\color{\colorMATH}\ensuremath{b}}} and its latent effect {{\color{\colorMATH}\ensuremath{x}}}. This yields an overall sensitivity of {{\color{\colorMATH}\ensuremath{b + x = b \sqcup  [b+x/x_{2}]x_{2}}}} for the {{\color{\colorMATH}\ensuremath{\ccase}}}-expression.

% Roughly speaking, the {{\color{\colorMATH}\ensuremath{\ccase}}}-expression over {{\color{\colorMATH}\ensuremath{s}}} acts by scaling the delayed sensitivity effects from the type of {{\color{\colorMATH}\ensuremath{s}}} ({{\color{\colorMATH}\ensuremath{\infty x}}} and {{\color{\colorMATH}\ensuremath{x}}}) by the sensitivity of the corresponding branch in its binder ({{\color{\colorMATH}\ensuremath{0}}} and {{\color{\colorMATH}\ensuremath{1}}}). Accounting also for the ambient effect {{\color{\colorMATH}\ensuremath{b}}} yields an overall sensitivity of {{\color{\colorMATH}\ensuremath{x + b}}}. Two technical points are worth noting here. First, this scaling is done using an ``ordinary'' product operation, noted ``{{\color{\colorMATH}\ensuremath{\cdot}}}'', where {{\color{\colorMATH}\ensuremath{0 \mathord{\cdotp } \infty  = 0}}}. Second, formally, the ambient effect {{\color{\colorMATH}\ensuremath{b}}} is also scaled (using, however, the ``special'' product operation ``{{\color{\colorMATH}\ensuremath{*}}}'' that we introduce below to avoid corner cases), but in this example the scaling has no net effect (as we scale by {{\color{\colorMATH}\ensuremath{0}}} and {{\color{\colorMATH}\ensuremath{1}}}, and keep the maximum of the resulting sensitivities).

Consider now Example~\ref{ex:discont}. The guard {{\color{\colorMATH}\ensuremath{x \leq 10}}} of the conditional expression has type {{\color{\colorMATH}\ensuremath{{{\color{\colorSYNTAX}\texttt{unit}}} \oplus  {{\color{\colorSYNTAX}\texttt{unit}}}}}} with ambient effect {{\color{\colorMATH}\ensuremath{\infty x}}}. Since the branches are constant and have no ambient effect, they do not contribute to the sensitivity of the conditional. \system analysis then concludes that the sensitivity of the conditional reduces to the ambient sensitivity of the guard, namely {{\color{\colorMATH}\ensuremath{\infty x}}}, recovering soundness (and precision).

\begin{table}[t]
{\small
\toplas{
\begin{tabular}{ b{1.6cm} @{\hspace{1.2em}}b{1.4cm} b{0.2em} b{1.6cm}  @{\hspace{1.2em}}b{1.7cm}  b{0.2em} b{1.6cm}   @{\hspace{1.2em}}b{1.8cm} }
\toprule
\multicolumn{2}{c}{\toplas{\textbf{\fuzz}(*)}} && \multicolumn{2}{c}{\toplas{\textbf{\dfuzz-like TS(**)}}}     && \multicolumn{2}{c}{\textbf{\system}}    \\ \cmidrule[\lightrulewidth]{1-2} \cmidrule[\lightrulewidth]{4-5} \cmidrule[\lightrulewidth]{7-8}
{\footnotesize\textbf{Reported Sensitivity}} & {\footnotesize\textbf{Bound}\newline \textbf{Quality}}
&& {\footnotesize\textbf{Reported Sensitivity}} & {\footnotesize\textbf{Bound}\newline \textbf{Quality}} && {\footnotesize\textbf{Reported Sensitivity}} & {\footnotesize\textbf{Bound\quad Quality}} \\ \midrule
\multicolumn{8}{c}{{Example~\ref{ex:dyncontrol} {\scriptsize (additive)}}}\\\midrule
{{\color{\colorMATH}\ensuremath{8x}}} & loose     &&  {{\color{\colorMATH}\ensuremath{8x}}}         &  loose        && {{\color{\colorMATH}\ensuremath{6x}}}        &  tight \\
% & typed {{\color{\colorMATH}\ensuremath{!_{\frac{3}{4}}{\mathbb{R}}}}}&&&&&&\\
\midrule
\multicolumn{8}{c}{{Example~\ref{ex:scaling} {\scriptsize (multiplicative)}}}\\\midrule
{{\color{\colorMATH}\ensuremath{2x + 2y}}} &  tight   &&  {{\color{\colorMATH}\ensuremath{4x + 2y}}}    &  loose (in {{\color{\colorMATH}\ensuremath{x}}})       && {{\color{\colorMATH}\ensuremath{2x + 2y}}}   &  tight \\
%& {{\color{\colorMATH}\ensuremath{y: !_{2}{\mathbb{R}}}}}&&&&&&\\
\midrule
\multicolumn{8}{c}{{Example~\ref{ex:discont}}}\\\midrule
{{\color{\colorMATH}\ensuremath{\infty x}}} & tight  &&  {{\color{\colorMATH}\ensuremath{0x}}}         &  unsound      && {{\color{\colorMATH}\ensuremath{\infty x}}}        &  tight \\
\bottomrule
\end{tabular}
}}
\\
{\toplas{\footnotesize(*): sensitivities strictly greater than 0, programs transformed using scaling.}}\\
{\toplas{\footnotesize(**): sensitivities can be greater or equal to 0, no scaling allowed.}}
\caption{Comparison of sensitivity type-system: \toplas{\fuzz and \dfuzz-like type systems} vs \system.}
\label{table:fuzzvsjazz}
\end{table}

\mbox{}
\system recovers soundness and precision for all four examples discussed in Section~\ref{sec:senslimit}, as summarized in Table~\ref{table:fuzzvsjazz}. With this observation, we conclude our motivation for the design of the \system sensitivity type system, based on latent contextual effects.

\begin{example}[Prepayment of effects]\label{ex:prepayment}
\toplass{
We remark that the use of latent contextual effects does not always yield better precision than eager (\fuzz-like) systems.
Consider the following program:
\begingroup\color{\colorMATH}\begin{gather*} 
\cr  {\begingroup\renewcommand\colorMATH{\colorMATHB}\renewcommand\colorSYNTAX{\colorSYNTAXB}{{\color{\colorSYNTAX}\texttt{let}}}\endgroup }\hspace*{0.33em}y_{1},y_{2} = ({\begingroup\renewcommand\colorMATH{\colorMATHB}\renewcommand\colorSYNTAX{\colorSYNTAXB}{{\color{\colorSYNTAX}\texttt{let}}}\endgroup }\hspace*{0.33em}x_{1},x_{2}\hspace*{0.33em} = p\hspace*{0.33em}{\begingroup\renewcommand\colorMATH{\colorMATHB}\renewcommand\colorSYNTAX{\colorSYNTAXB}{{\color{\colorSYNTAX}\texttt{in}}}\endgroup }\hspace*{0.33em}\langle x_{1},x_{2}\rangle ) \hspace*{0.33em}{\begingroup\renewcommand\colorMATH{\colorMATHB}\renewcommand\colorSYNTAX{\colorSYNTAXB}{{\color{\colorSYNTAX}\texttt{in}}}\endgroup } \hspace*{0.33em}y_{1} + y_{2}
\end{gather*}\endgroup
Using latent effects, the subexpression {{\color{\colorMATH}\ensuremath{\langle x_{1},x_{2}\rangle }}} has type {{\color{\colorMATH}\ensuremath{{\mathbb{R}} \mathrel{^{x_{1}}\otimes ^{x_{2}}} {\mathbb{R}};\varnothing }}}. Thus the subexpression {{\color{\colorMATH}\ensuremath{{\begingroup\renewcommand\colorMATH{\colorMATHB}\renewcommand\colorSYNTAX{\colorSYNTAXB}{{\color{\colorSYNTAX}\texttt{let}}}\endgroup }\hspace*{0.33em}x_{1},x_{2}\hspace*{0.33em} = p\hspace*{0.33em}{\begingroup\renewcommand\colorMATH{\colorMATHB}\renewcommand\colorSYNTAX{\colorSYNTAXB}{{\color{\colorSYNTAX}\texttt{in}}}\endgroup }\hspace*{0.33em}\langle x_{1},x_{2}\rangle }}} has type {{\color{\colorMATH}\ensuremath{{\mathbb{R}} \mathrel{^{p}\otimes ^{p}} {\mathbb{R}};\varnothing }}}, i.e. it represents a pure expression where the cost of accessing either of its component is {{\color{\colorMATH}\ensuremath{p}}}. The ambient effect of the whole expression is the sum of the cost of accessing the pair ({{\color{\colorMATH}\ensuremath{\varnothing }}}), plus the cost of accessing {{\color{\colorMATH}\ensuremath{y_{1}}}} ({{\color{\colorMATH}\ensuremath{p}}}), plus the cost of accessing {{\color{\colorMATH}\ensuremath{y_{2}}}} ({{\color{\colorMATH}\ensuremath{p}}}), yielding effect {{\color{\colorMATH}\ensuremath{2p}}}. In \fuzz, the same program reports sensitivity {{\color{\colorMATH}\ensuremath{p}}}, yielding better precision than \system.
}
% which allow for very
% precise sensitivity bounds, the distinction between multiplicative and additive products is
% particularly important to maintain in presence of type-level polymorphism. \mt{I don't undertand why we wrote this.}
% Indeed, effects may need to be paid for ``upfront'' if a latent effect would
% otherwise be substituted into an incompatible scoping environment. 
% When paying
% for these sensitivities upfront, the usual advantages of having two product
% type connectives apply---one pays for the maximum sensitivity of arguments
% while the other pays for the sum.
\toplas{To recover \fuzz's precision, \system allows effects of products, sums, and functions to be paid for eagerly, by combining contextual and linear effects: parts of the sensitivity effect of each component of the product can contribute to the ambient effect of the product. For instance, consider environment {{\color{\colorMATH}\ensuremath{\Gamma  = x \mathrel{:} {\mathbb{R}}, y \mathrel{:} {\mathbb{R}}}}}. \system can produce the following type derivations for expression {{\color{\colorMATH}\ensuremath{{\begingroup\renewcommand\colorMATH{\colorMATHB}\renewcommand\colorSYNTAX{\colorSYNTAXB}{{\color{\colorMATH}\ensuremath{\se}}}\endgroup } = \langle x, y\rangle }}}:
\begingroup\color{\colorMATH}\begin{gather*}
\Gamma  \hspace*{0.33em} {\begingroup\renewcommand\colorMATH{\colorMATHB}\renewcommand\colorSYNTAX{\colorSYNTAXB}{{\color{\colorMATH}\ensuremath{\vdash }}}\endgroup }\hspace*{0.33em} {\begingroup\renewcommand\colorMATH{\colorMATHB}\renewcommand\colorSYNTAX{\colorSYNTAXB}{{\color{\colorMATH}\ensuremath{e}}}\endgroup } \mathrel{:} {\mathbb{R}} \mathrel{^{x}\otimes ^{y}} {\mathbb{R}}; \varnothing 
\qquad
\Gamma  \hspace*{0.33em} {\begingroup\renewcommand\colorMATH{\colorMATHB}\renewcommand\colorSYNTAX{\colorSYNTAXB}{{\color{\colorMATH}\ensuremath{\vdash }}}\endgroup }\hspace*{0.33em} {\begingroup\renewcommand\colorMATH{\colorMATHB}\renewcommand\colorSYNTAX{\colorSYNTAXB}{{\color{\colorMATH}\ensuremath{e_{2}}}}\endgroup } \mathrel{:} {\mathbb{R}} \mathrel{\otimes ^{y}} {\mathbb{R}};x
\qquad
\Gamma  \hspace*{0.33em} {\begingroup\renewcommand\colorMATH{\colorMATHB}\renewcommand\colorSYNTAX{\colorSYNTAXB}{{\color{\colorMATH}\ensuremath{\vdash }}}\endgroup }\hspace*{0.33em} {\begingroup\renewcommand\colorMATH{\colorMATHB}\renewcommand\colorSYNTAX{\colorSYNTAXB}{{\color{\colorMATH}\ensuremath{e_{3}}}}\endgroup } \mathrel{:} {\mathbb{R}} \mathrel{^{x}\otimes } {\mathbb{R}}; y
\qquad
\Gamma  \hspace*{0.33em} {\begingroup\renewcommand\colorMATH{\colorMATHB}\renewcommand\colorSYNTAX{\colorSYNTAXB}{{\color{\colorMATH}\ensuremath{\vdash }}}\endgroup }\hspace*{0.33em} {\begingroup\renewcommand\colorMATH{\colorMATHB}\renewcommand\colorSYNTAX{\colorSYNTAXB}{{\color{\colorMATH}\ensuremath{e_{3}}}}\endgroup } \mathrel{:} {\mathbb{R}} \mathrel{\otimes } {\mathbb{R}}; x+y
\end{gather*}\endgroup
In the first type derivation, the effect of both components are latent, and thus the ambient effect is empty.
In the second ({resp.} third) type derivation, the latent effect of the type is the ambient effect of the right ({resp.} left) component, and the ambient effect of the product is the ambient effect of the left ({resp.}right) component.
In the last type derivation, the latent effect of the type is empty, everything is paid upfront, coinciding with \fuzz-like type systems.
}
\toplass{
  Going back to the example, if we prepay the effects of the subexpression
  {{\color{\colorMATH}\ensuremath{\langle x_{1},x_{2}\rangle }}} then the product has type {{\color{\colorMATH}\ensuremath{{\mathbb{R}} \mathrel{^{\varnothing }\otimes ^{\varnothing }} {\mathbb{R}};x_{1}+x_{2}}}}. 
  Now the subexpression {{\color{\colorMATH}\ensuremath{{\begingroup\renewcommand\colorMATH{\colorMATHB}\renewcommand\colorSYNTAX{\colorSYNTAXB}{{\color{\colorSYNTAX}\texttt{let}}}\endgroup }\hspace*{0.33em}x_{1},x_{2}\hspace*{0.33em} = p\hspace*{0.33em}{\begingroup\renewcommand\colorMATH{\colorMATHB}\renewcommand\colorSYNTAX{\colorSYNTAXB}{{\color{\colorSYNTAX}\texttt{in}}}\endgroup }\hspace*{0.33em}\langle x_{1},x_{2}\rangle }}} has type {{\color{\colorMATH}\ensuremath{{\mathbb{R}} \mathrel{^{\varnothing }\otimes ^{\varnothing }} {\mathbb{R}};p}}}, because using multiplicative products we only pay for {{\color{\colorMATH}\ensuremath{p}}} proportional to the maximum sensitivity between {{\color{\colorMATH}\ensuremath{x_{1}}}} and {{\color{\colorMATH}\ensuremath{x_{2}}}}, i.e. {{\color{\colorMATH}\ensuremath{(1\sqcup 1)p}}}. The ambient effect of the whole expression is the sum of the cost of accessing the pair ({{\color{\colorMATH}\ensuremath{p}}}), plus the cost of accessing {{\color{\colorMATH}\ensuremath{y_{1}}}} and {{\color{\colorMATH}\ensuremath{y_{2}}}} ({{\color{\colorMATH}\ensuremath{\varnothing }}}), yielding the tight ambient effect {{\color{\colorMATH}\ensuremath{p}}}.

}
\end{example}

\toplass{As a final remark, note that contrary to \fuzz, \system does not currently support recursive types\toplasss{; such functions} are required to be primitives, as illustrated in \S~\ref{sec:system}.}

The following section presents the formal development of latent contextual effects for sensitivity typing, and includes a step-by-step type derivation for all four examples.

\section{\ssystem: \system's Sensitivity Type System, Formally} % {-{
\label{sec:sensitivity-formalism}

In this section, we present a core sensitivity sublanguage of \system, called $\ssystem$, for which we develop the sensitivity metatheory.
In particular, we prove the type soundness property known as {\em sensitivity metric preservation}~\cite{reed2010distance}. 
The core subset of \system that extends \ssystem with privacy is presented in later sections.

\subsection{Syntax and Type System}
\label{sec:sensitivity-static-semantics}

The $\ssystem$ type system is technically a
type-and-effect system~\cite{GiffordLucassen}. It supports 
real numbers, functions, sums and products.
As $\ssystem$ only deals with ambient effects, all metavariables and keywords are typeset in {\begingroup\renewcommand\colorMATH{\colorMATHB}\renewcommand\colorSYNTAX{\colorSYNTAXB}{{\color{\colorMATH}\ensuremath{{\text{green}}}}}\endgroup }.
\paragraph{Syntax} Figure~\ref{fig:syntax} presents the syntax of $\ssystem$. %\fo{Presenting the language syntax within the (static) semantic section was a bit unexpected to me.}
\begin{figure}[t]
  \begin{small}
  \begin{framed}
\begingroup\color{\colorMATH}\begin{gather*}
\end{gather*}\endgroup
\end{framed}
  \end{small}
  \caption{$\ssystem$: Syntax}
  \label{fig:sensitivity-syntax} 
\end{figure}
Expressions {{\color{\colorMATH}\ensuremath{{\begingroup\renewcommand\colorMATH{\colorMATHB}\renewcommand\colorSYNTAX{\colorSYNTAXB}{{\color{\colorMATH}\ensuremath{\se}}}\endgroup }}}} are mostly standard and include: real number {{\color{\colorMATH}\ensuremath{{\begingroup\renewcommand\colorMATH{\colorMATHB}\renewcommand\colorSYNTAX{\colorSYNTAXB}{{\color{\colorMATH}\ensuremath{r}}}\endgroup }}}}, 
addition {{\color{\colorMATH}\ensuremath{{\begingroup\renewcommand\colorMATH{\colorMATHB}\renewcommand\colorSYNTAX{\colorSYNTAXB}{{\color{\colorMATH}\ensuremath{\se}}}\endgroup } + {\begingroup\renewcommand\colorMATH{\colorMATHB}\renewcommand\colorSYNTAX{\colorSYNTAXB}{{\color{\colorMATH}\ensuremath{\se}}}\endgroup }}}}, multiplication {{\color{\colorMATH}\ensuremath{{\begingroup\renewcommand\colorMATH{\colorMATHB}\renewcommand\colorSYNTAX{\colorSYNTAXB}{{\color{\colorMATH}\ensuremath{\se}}}\endgroup } * {\begingroup\renewcommand\colorMATH{\colorMATHB}\renewcommand\colorSYNTAX{\colorSYNTAXB}{{\color{\colorMATH}\ensuremath{\se}}}\endgroup }}}}, comparison {{\color{\colorMATH}\ensuremath{{\begingroup\renewcommand\colorMATH{\colorMATHB}\renewcommand\colorSYNTAX{\colorSYNTAXB}{{\color{\colorMATH}\ensuremath{\se}}}\endgroup } \leq  {\begingroup\renewcommand\colorMATH{\colorMATHB}\renewcommand\colorSYNTAX{\colorSYNTAXB}{{\color{\colorMATH}\ensuremath{\se}}}\endgroup }}}}, 
variable {{\color{\colorMATH}\ensuremath{x}}}, sensitivity lambda {{\color{\colorMATH}\ensuremath{{\begingroup\renewcommand\colorMATH{\colorMATHB}\renewcommand\colorSYNTAX{\colorSYNTAXB}{{\color{\colorMATH}\ensuremath{\slambda}}}\endgroup } (x\mathrel{:}\tau ).\hspace*{0.33em}{\begingroup\renewcommand\colorMATH{\colorMATHB}\renewcommand\colorSYNTAX{\colorSYNTAXB}{{\color{\colorMATH}\ensuremath{\se}}}\endgroup }}}}, application {{\color{\colorMATH}\ensuremath{{\begingroup\renewcommand\colorMATH{\colorMATHB}\renewcommand\colorSYNTAX{\colorSYNTAXB}{{\color{\colorMATH}\ensuremath{\se}}}\endgroup }\hspace*{0.33em}{\begingroup\renewcommand\colorMATH{\colorMATHB}\renewcommand\colorSYNTAX{\colorSYNTAXB}{{\color{\colorMATH}\ensuremath{\se}}}\endgroup }}}}, unit value
{{\color{\colorMATH}\ensuremath{\ttt}}}, sum constructors {{\color{\colorMATH}\ensuremath{\inl^{\tau _{2}}\hspace*{0.33em}{\begingroup\renewcommand\colorMATH{\colorMATHB}\renewcommand\colorSYNTAX{\colorSYNTAXB}{{\color{\colorMATH}\ensuremath{\se}}}\endgroup }}}} and {{\color{\colorMATH}\ensuremath{\inr^{\tau _{1}}\hspace*{0.33em}{\begingroup\renewcommand\colorMATH{\colorMATHB}\renewcommand\colorSYNTAX{\colorSYNTAXB}{{\color{\colorMATH}\ensuremath{\se}}}\endgroup }}}}, and the sum destructor
{{\color{\colorMATH}\ensuremath{\ccase\hspace*{0.33em}{\begingroup\renewcommand\colorMATH{\colorMATHB}\renewcommand\colorSYNTAX{\colorSYNTAXB}{{\color{\colorMATH}\ensuremath{\se}}}\endgroup }\hspace*{0.33em}\of\hspace*{0.33em}\{ x\Rightarrow {\begingroup\renewcommand\colorMATH{\colorMATHB}\renewcommand\colorSYNTAX{\colorSYNTAXB}{{\color{\colorMATH}\ensuremath{\se}}}\endgroup }\} \hspace*{0.33em}\{ x\Rightarrow {\begingroup\renewcommand\colorMATH{\colorMATHB}\renewcommand\colorSYNTAX{\colorSYNTAXB}{{\color{\colorMATH}\ensuremath{\se}}}\endgroup }\} }}}.

$\ssystem$ also supports two linear products types: additive and multiplicative. With
additive products, the sensitivity of a pair may be approximated as the {\em max} of the
sensitivities of each side; this sensitivity is paid for every
projection. With multiplicative products, the sensitivity of a pair may be
approximated as the {\em sum} of the sensitivities of each side; this sensitivity
is paid for every tuple pattern match, scaled by the sensitivities of pattern
variables in the body. We write additive product constructions {{\color{\colorMATH}\ensuremath{{{\color{\colorSYNTAX}\texttt{{\ensuremath{\addProduct{{\begingroup\renewcommand\colorMATH{\colorMATHB}\renewcommand\colorSYNTAX{\colorSYNTAXB}{{\color{\colorMATH}\ensuremath{\se}}}\endgroup }}{{\begingroup\renewcommand\colorMATH{\colorMATHB}\renewcommand\colorSYNTAX{\colorSYNTAXB}{{\color{\colorMATH}\ensuremath{\se}}}\endgroup }}}}}}}}}} and
destructions {{\color{\colorMATH}\ensuremath{\fst\hspace*{0.33em}{\begingroup\renewcommand\colorMATH{\colorMATHB}\renewcommand\colorSYNTAX{\colorSYNTAXB}{{\color{\colorMATH}\ensuremath{\se}}}\endgroup }}}} and {{\color{\colorMATH}\ensuremath{\snd\hspace*{0.33em}{\begingroup\renewcommand\colorMATH{\colorMATHB}\renewcommand\colorSYNTAX{\colorSYNTAXB}{{\color{\colorMATH}\ensuremath{\se}}}\endgroup }}}}, and multiplicative product constructions
{{\color{\colorMATH}\ensuremath{{{\color{\colorSYNTAX}\texttt{{\ensuremath{\langle {\begingroup\renewcommand\colorMATH{\colorMATHB}\renewcommand\colorSYNTAX{\colorSYNTAXB}{{\color{\colorMATH}\ensuremath{\se}}}\endgroup },{\begingroup\renewcommand\colorMATH{\colorMATHB}\renewcommand\colorSYNTAX{\colorSYNTAXB}{{\color{\colorMATH}\ensuremath{\se}}}\endgroup }\rangle }}}}}}}} and destructions {{\color{\colorMATH}\ensuremath{\tlet\hspace*{0.33em}x,x={\begingroup\renewcommand\colorMATH{\colorMATHB}\renewcommand\colorSYNTAX{\colorSYNTAXB}{{\color{\colorMATH}\ensuremath{\se}}}\endgroup }\hspace*{0.33em}\tin\hspace*{0.33em}{\begingroup\renewcommand\colorMATH{\colorMATHB}\renewcommand\colorSYNTAX{\colorSYNTAXB}{{\color{\colorMATH}\ensuremath{\se}}}\endgroup }}}}. 

Finally, an expression {{\color{\colorMATH}\ensuremath{{\begingroup\renewcommand\colorMATH{\colorMATHB}\renewcommand\colorSYNTAX{\colorSYNTAXB}{{\color{\colorMATH}\ensuremath{\se}}}\endgroup }}}}
can be an ascription {{\color{\colorMATH}\ensuremath{{\begingroup\renewcommand\colorMATH{\colorMATHB}\renewcommand\colorSYNTAX{\colorSYNTAXB}{{\color{\colorMATH}\ensuremath{\se}}}\endgroup } \mathrel{:: } {\begingroup\renewcommand\colorMATH{\colorMATHB}\renewcommand\colorSYNTAX{\colorSYNTAXB}{{\color{\colorMATH}\ensuremath{\tau }}}\endgroup }}}}, or a derived expressions such as a boolean
{{\color{\colorMATH}\ensuremath{b}}}, a conditional {{\color{\colorMATH}\ensuremath{\sif\hspace*{0.33em}{\begingroup\renewcommand\colorMATH{\colorMATHB}\renewcommand\colorSYNTAX{\colorSYNTAXB}{{\color{\colorMATH}\ensuremath{\se}}}\endgroup }\hspace*{0.33em}\sthen\hspace*{0.33em}{\begingroup\renewcommand\colorMATH{\colorMATHB}\renewcommand\colorSYNTAX{\colorSYNTAXB}{{\color{\colorMATH}\ensuremath{\se}}}\endgroup }\hspace*{0.33em}\selse\hspace*{0.33em}{\begingroup\renewcommand\colorMATH{\colorMATHB}\renewcommand\colorSYNTAX{\colorSYNTAXB}{{\color{\colorMATH}\ensuremath{\se}}}\endgroup }}}}, or a let expression
{{\color{\colorMATH}\ensuremath{\tlet\hspace*{0.33em}x={\begingroup\renewcommand\colorMATH{\colorMATHB}\renewcommand\colorSYNTAX{\colorSYNTAXB}{{\color{\colorMATH}\ensuremath{\se}}}\endgroup }\hspace*{0.33em}\tin\hspace*{0.33em}{\begingroup\renewcommand\colorMATH{\colorMATHB}\renewcommand\colorSYNTAX{\colorSYNTAXB}{{\color{\colorMATH}\ensuremath{\se}}}\endgroup }}}}. Booleans are encoded as {{\color{\colorMATH}\ensuremath{{\text{true}} \triangleq  \inl\hspace*{0.33em}\ttt}}}, {{\color{\colorMATH}\ensuremath{{\text{false}} \triangleq 
\inr\hspace*{0.33em}\ttt}}}, \toplas{and {{\color{\colorMATH}\ensuremath{{\mathbb{B}}}}} as {{\color{\colorMATH}\ensuremath{{{\color{\colorSYNTAX}\texttt{unit}}}^{\varnothing }\oplus ^{\varnothing }{{\color{\colorSYNTAX}\texttt{unit}}}}}}}, conditionals as {{\color{\colorMATH}\ensuremath{\sif\hspace*{0.33em}{\begingroup\renewcommand\colorMATH{\colorMATHB}\renewcommand\colorSYNTAX{\colorSYNTAXB}{{\color{\colorMATH}\ensuremath{\se_{1}}}}\endgroup }\hspace*{0.33em}\sthen\hspace*{0.33em}{\begingroup\renewcommand\colorMATH{\colorMATHB}\renewcommand\colorSYNTAX{\colorSYNTAXB}{{\color{\colorMATH}\ensuremath{\se_{2}}}}\endgroup }\hspace*{0.33em}\selse\hspace*{0.33em}{\begingroup\renewcommand\colorMATH{\colorMATHB}\renewcommand\colorSYNTAX{\colorSYNTAXB}{{\color{\colorMATH}\ensuremath{\se_{3}}}}\endgroup } \triangleq  \ccase\hspace*{0.33em}{\begingroup\renewcommand\colorMATH{\colorMATHB}\renewcommand\colorSYNTAX{\colorSYNTAXB}{{\color{\colorMATH}\ensuremath{\se_{1}}}}\endgroup }\hspace*{0.33em}\of\hspace*{0.33em}\{ x\Rightarrow {\begingroup\renewcommand\colorMATH{\colorMATHB}\renewcommand\colorSYNTAX{\colorSYNTAXB}{{\color{\colorMATH}\ensuremath{\se_{2}}}}\endgroup }\} \hspace*{0.33em}\{ y\Rightarrow {\begingroup\renewcommand\colorMATH{\colorMATHB}\renewcommand\colorSYNTAX{\colorSYNTAXB}{{\color{\colorMATH}\ensuremath{\se_{3}}}}\endgroup }\} }}}, and
let expressions as {{\color{\colorMATH}\ensuremath{\tlet\hspace*{0.33em}x={\begingroup\renewcommand\colorMATH{\colorMATHB}\renewcommand\colorSYNTAX{\colorSYNTAXB}{{\color{\colorMATH}\ensuremath{\se_{1}}}}\endgroup }\hspace*{0.33em}\tin\hspace*{0.33em}{\begingroup\renewcommand\colorMATH{\colorMATHB}\renewcommand\colorSYNTAX{\colorSYNTAXB}{{\color{\colorMATH}\ensuremath{\se_{2}}}}\endgroup } \triangleq  ({\begingroup\renewcommand\colorMATH{\colorMATHB}\renewcommand\colorSYNTAX{\colorSYNTAXB}{{\color{\colorMATH}\ensuremath{\slambda}}}\endgroup } (x:\tau _{1}). {\begingroup\renewcommand\colorMATH{\colorMATHB}\renewcommand\colorSYNTAX{\colorSYNTAXB}{{\color{\colorMATH}\ensuremath{\se_{2}}}}\endgroup })\hspace*{0.33em}{\begingroup\renewcommand\colorMATH{\colorMATHB}\renewcommand\colorSYNTAX{\colorSYNTAXB}{{\color{\colorMATH}\ensuremath{\se_{1}}}}\endgroup }}}}.

A
sensitivity {{\color{\colorMATH}\ensuremath{{\begingroup\renewcommand\colorMATH{\colorMATHB}\renewcommand\colorSYNTAX{\colorSYNTAXB}{{\color{\colorMATH}\ensuremath{\sss}}}\endgroup }}}} is either a non-negative real number or the symbol {{\color{\colorMATH}\ensuremath{\infty }}}, which represents an
unbounded sensitivity; we notate this set {{\color{\colorMATH}\ensuremath{{\begingroup\renewcommand\colorMATH{\colorMATHA}\renewcommand\colorSYNTAX{\colorSYNTAXA}{{\color{\colorSYNTAX}\texttt{{\ensuremath{{\mathbb{R}}}}}}}\endgroup }^{\infty }_{\geq 0} \triangleq  {\begingroup\renewcommand\colorMATH{\colorMATHA}\renewcommand\colorSYNTAX{\colorSYNTAXA}{{\color{\colorSYNTAX}\texttt{{\ensuremath{{\mathbb{R}}}}}}}\endgroup }_{\geq 0} \uplus  \{ \infty \} }}}.
A sensitivity environment {{\color{\colorMATH}\ensuremath{{\begingroup\renewcommand\colorMATH{\colorMATHB}\renewcommand\colorSYNTAX{\colorSYNTAXB}{{\color{\colorMATH}\ensuremath{\sS}}}\endgroup }}}} is a mapping from variables to their
sensitivities. For convenience, we write sensitivity environments as
first-order polynomials, e.g. {{\color{\colorMATH}\ensuremath{{\begingroup\renewcommand\colorMATH{\colorMATHB}\renewcommand\colorSYNTAX{\colorSYNTAXB}{{\color{\colorMATH}\ensuremath{\sS}}}\endgroup } = 1x + 2y}}} corresponds to an environment {{\color{\colorMATH}\ensuremath{{\begingroup\renewcommand\colorMATH{\colorMATHB}\renewcommand\colorSYNTAX{\colorSYNTAXB}{{\color{\colorMATH}\ensuremath{\sS}}}\endgroup }}}}
such that {{\color{\colorMATH}\ensuremath{{\begingroup\renewcommand\colorMATH{\colorMATHB}\renewcommand\colorSYNTAX{\colorSYNTAXB}{{\color{\colorMATH}\ensuremath{\sS}}}\endgroup }(x) = 1}}} and {{\color{\colorMATH}\ensuremath{{\begingroup\renewcommand\colorMATH{\colorMATHB}\renewcommand\colorSYNTAX{\colorSYNTAXB}{{\color{\colorMATH}\ensuremath{\sS}}}\endgroup }(y) = 2}}}.
%, e.g. {{\color{\colorMATH}\ensuremath{{\begingroup\renewcommand\colorMATH{\colorMATHC}\renewcommand\colorSYNTAX{\colorSYNTAXC}{{\color{\colorMATH}\ensuremath{\pS}}}\endgroup } = {\begingroup\renewcommand\colorMATH{\colorMATHC}\renewcommand\colorSYNTAX{\colorSYNTAXC}{{\color{\colorMATH}\ensuremath{p_{1}}}}\endgroup }x + {\begingroup\renewcommand\colorMATH{\colorMATHC}\renewcommand\colorSYNTAX{\colorSYNTAXC}{{\color{\colorMATH}\ensuremath{p_{2}}}}\endgroup }y}}} correspond to environment such that {{\color{\colorMATH}\ensuremath{{\begingroup\renewcommand\colorMATH{\colorMATHC}\renewcommand\colorSYNTAX{\colorSYNTAXC}{{\color{\colorMATH}\ensuremath{\pS}}}\endgroup }(x) = {\begingroup\renewcommand\colorMATH{\colorMATHC}\renewcommand\colorSYNTAX{\colorSYNTAXC}{{\color{\colorMATH}\ensuremath{p_{1}}}}\endgroup }}}} and
%{{\color{\colorMATH}\ensuremath{{\begingroup\renewcommand\colorMATH{\colorMATHC}\renewcommand\colorSYNTAX{\colorSYNTAXC}{{\color{\colorMATH}\ensuremath{\pS}}}\endgroup }(y) = {\begingroup\renewcommand\colorMATH{\colorMATHC}\renewcommand\colorSYNTAX{\colorSYNTAXC}{{\color{\colorMATH}\ensuremath{p_{2}}}}\endgroup }}}}.
A type {{\color{\colorMATH}\ensuremath{\tau }}} is either the real number type {{\color{\colorMATH}\ensuremath{{\begingroup\renewcommand\colorMATH{\colorMATHA}\renewcommand\colorSYNTAX{\colorSYNTAXA}{{\color{\colorSYNTAX}\texttt{{\ensuremath{{\mathbb{R}}}}}}}\endgroup }}}}, the boolean type {{\color{\colorMATH}\ensuremath{{\begingroup\renewcommand\colorMATH{\colorMATHA}\renewcommand\colorSYNTAX{\colorSYNTAXA}{{\color{\colorSYNTAX}\texttt{{\ensuremath{{\mathbb{B}}}}}}}\endgroup }}}}, the unit
type {{\color{\colorMATH}\ensuremath{{{\color{\colorSYNTAX}\texttt{unit}}}}}}, a function type {{\color{\colorMATH}\ensuremath{(x\mathrel{:}\tau ) \xrightarrowS {{\begingroup\renewcommand\colorMATH{\colorMATHB}\renewcommand\colorSYNTAX{\colorSYNTAXB}{{\color{\colorMATH}\ensuremath{\sS}}}\endgroup }} \tau }}}, a sum type {{\color{\colorMATH}\ensuremath{\tau  \mathrel{^{{\begingroup\renewcommand\colorMATH{\colorMATHB}\renewcommand\colorSYNTAX{\colorSYNTAXB}{{\color{\colorMATH}\ensuremath{\sS}}}\endgroup }}\oplus ^{{\begingroup\renewcommand\colorMATH{\colorMATHB}\renewcommand\colorSYNTAX{\colorSYNTAXB}{{\color{\colorMATH}\ensuremath{\sS}}}\endgroup }}} \tau }}}, an additive
product type {{\color{\colorMATH}\ensuremath{\tau  \mathrel{^{{\begingroup\renewcommand\colorMATH{\colorMATHB}\renewcommand\colorSYNTAX{\colorSYNTAXB}{{\color{\colorMATH}\ensuremath{\sS}}}\endgroup }}\&^{{\begingroup\renewcommand\colorMATH{\colorMATHB}\renewcommand\colorSYNTAX{\colorSYNTAXB}{{\color{\colorMATH}\ensuremath{\sS}}}\endgroup }}} \tau }}}, or a multiplicative product type {{\color{\colorMATH}\ensuremath{\tau  \mathrel{^{{\begingroup\renewcommand\colorMATH{\colorMATHB}\renewcommand\colorSYNTAX{\colorSYNTAXB}{{\color{\colorMATH}\ensuremath{\sS}}}\endgroup }}\otimes ^{{\begingroup\renewcommand\colorMATH{\colorMATHB}\renewcommand\colorSYNTAX{\colorSYNTAXB}{{\color{\colorMATH}\ensuremath{\sS}}}\endgroup }}} \tau }}}.
The sensitivity environment annotation {{\color{\colorMATH}\ensuremath{{\begingroup\renewcommand\colorMATH{\colorMATHB}\renewcommand\colorSYNTAX{\colorSYNTAXB}{{\color{\colorMATH}\ensuremath{\sS}}}\endgroup }}}} is called the
{\em latent contextual sensitivity effect} (also called latent effect when clear from the context) and represents a \emph{delayed} effect that emerges when a term of said type is eliminated. The latent effect {{\color{\colorMATH}\ensuremath{{\begingroup\renewcommand\colorMATH{\colorMATHB}\renewcommand\colorSYNTAX{\colorSYNTAXB}{{\color{\colorMATH}\ensuremath{\sS}}}\endgroup }}}} 
of a function of type {{\color{\colorMATH}\ensuremath{(x\mathrel{:}\tau ) \xrightarrowS {{\begingroup\renewcommand\colorMATH{\colorMATHB}\renewcommand\colorSYNTAX{\colorSYNTAXB}{{\color{\colorMATH}\ensuremath{\sS}}}\endgroup }} \tau }}} corresponds to the effects of applying the function,
{i.e.}, a static approximation of the sensitivity of each variable used in its body. The sensitivity environment {{\color{\colorMATH}\ensuremath{{\begingroup\renewcommand\colorMATH{\colorMATHB}\renewcommand\colorSYNTAX{\colorSYNTAXB}{{\color{\colorMATH}\ensuremath{\sS_{1}}}}\endgroup }}}} (resp. {{\color{\colorMATH}\ensuremath{{\begingroup\renewcommand\colorMATH{\colorMATHB}\renewcommand\colorSYNTAX{\colorSYNTAXB}{{\color{\colorMATH}\ensuremath{\sS_{2}}}}\endgroup }}}}) in {{\color{\colorMATH}\ensuremath{\tau _{1} \mathrel{^{{\begingroup\renewcommand\colorMATH{\colorMATHB}\renewcommand\colorSYNTAX{\colorSYNTAXB}{{\color{\colorMATH}\ensuremath{\sS_{1}}}}\endgroup }}\oplus ^{{\begingroup\renewcommand\colorMATH{\colorMATHB}\renewcommand\colorSYNTAX{\colorSYNTAXB}{{\color{\colorMATH}\ensuremath{\sS_{2}}}}\endgroup }}} \tau _{2}}}} corresponds to the latent effect of the injected value using \inl\ (resp. \inr). And similarly, {{\color{\colorMATH}\ensuremath{{\begingroup\renewcommand\colorMATH{\colorMATHB}\renewcommand\colorSYNTAX{\colorSYNTAXB}{{\color{\colorMATH}\ensuremath{\sS_{1}}}}\endgroup }}}} and
{{\color{\colorMATH}\ensuremath{{\begingroup\renewcommand\colorMATH{\colorMATHB}\renewcommand\colorSYNTAX{\colorSYNTAXB}{{\color{\colorMATH}\ensuremath{\sS_{2}}}}\endgroup }}}} in {{\color{\colorMATH}\ensuremath{\tau _{1} \mathrel{^{{\begingroup\renewcommand\colorMATH{\colorMATHB}\renewcommand\colorSYNTAX{\colorSYNTAXB}{{\color{\colorMATH}\ensuremath{\sS_{1}}}}\endgroup }}\&^{{\begingroup\renewcommand\colorMATH{\colorMATHB}\renewcommand\colorSYNTAX{\colorSYNTAXB}{{\color{\colorMATH}\ensuremath{\sS_{2}}}}\endgroup }}} \tau _{2}}}} or {{\color{\colorMATH}\ensuremath{\tau _{1} \mathrel{^{{\begingroup\renewcommand\colorMATH{\colorMATHB}\renewcommand\colorSYNTAX{\colorSYNTAXB}{{\color{\colorMATH}\ensuremath{\sS_{1}}}}\endgroup }}\otimes ^{{\begingroup\renewcommand\colorMATH{\colorMATHB}\renewcommand\colorSYNTAX{\colorSYNTAXB}{{\color{\colorMATH}\ensuremath{\sS_{2}}}}\endgroup }}} \tau _{2}}}} correspond to the latent effect of accessing the first and second components of the pair, respectively. Finally, a type
environment {{\color{\colorMATH}\ensuremath{\Gamma }}} is, as usual, a mapping from variables to types.

\begin{figure}[t]
  \begin{small}
  \begin{framed}
  \input{sensitivity-simple-type-system1}
  \end{framed}
  \end{small}
  \caption{$\ssystem$: Type system}
  \label{fig:sensitivity-simple-type-system1}
\end{figure}
% \begin{figure}[t]
%   \begin{small}
%   \input{sensitivity-type-system2}
%   \end{small}
%   \caption{$\ssystem$: Type system of the sensitivity sublanguage (part 2)}
%   \label{fig:sensitivity-type-system2}
% \end{figure}
\paragraph{Type system}
The \ssystem type system is presented in
Figure~\ref{fig:sensitivity-simple-type-system1}.
% and~\ref{fig:sensitivity-type-system2}.
The judgment {{\color{\colorMATH}\ensuremath{{{\begingroup\renewcommand\colorMATH{\colorMATHA}\renewcommand\colorSYNTAX{\colorSYNTAXA}{{\color{\colorMATH}\ensuremath{\Gamma }}}\endgroup }\hspace*{0.33em}{\begingroup\renewcommand\colorMATH{\colorMATHB}\renewcommand\colorSYNTAX{\colorSYNTAXB}{{\color{\colorMATH}\ensuremath{\vdash }}}\endgroup }\hspace*{0.33em}{\begingroup\renewcommand\colorMATH{\colorMATHB}\renewcommand\colorSYNTAX{\colorSYNTAXB}{{\color{\colorMATH}\ensuremath{\se}}}\endgroup } \mathrel{:} {\begingroup\renewcommand\colorMATH{\colorMATHA}\renewcommand\colorSYNTAX{\colorSYNTAXA}{{\color{\colorMATH}\ensuremath{\tau }}}\endgroup } \mathrel{;} {\begingroup\renewcommand\colorMATH{\colorMATHB}\renewcommand\colorSYNTAX{\colorSYNTAXB}{{\color{\colorMATH}\ensuremath{\sS}}}\endgroup }}}}} says that the term  {{\color{\colorMATH}\ensuremath{{\begingroup\renewcommand\colorMATH{\colorMATHB}\renewcommand\colorSYNTAX{\colorSYNTAXB}{{\color{\colorMATH}\ensuremath{\se}}}\endgroup }}}} has
type {{\color{\colorMATH}\ensuremath{\tau }}} and ambient sensitivity effect {{\color{\colorMATH}\ensuremath{{\begingroup\renewcommand\colorMATH{\colorMATHB}\renewcommand\colorSYNTAX{\colorSYNTAXB}{{\color{\colorMATH}\ensuremath{\sS}}}\endgroup }}}} (or ambient effect when clear from the context) under type environment {{\color{\colorMATH}\ensuremath{\Gamma }}}. The ambient effect {{\color{\colorMATH}\ensuremath{{\begingroup\renewcommand\colorMATH{\colorMATHB}\renewcommand\colorSYNTAX{\colorSYNTAXB}{{\color{\colorMATH}\ensuremath{\sS}}}\endgroup }}}} represents an upper bound (conservative approximation)  of the real sensitivity of {{\color{\colorMATH}\ensuremath{{\begingroup\renewcommand\colorMATH{\colorMATHB}\renewcommand\colorSYNTAX{\colorSYNTAXB}{{\color{\colorMATH}\ensuremath{\se}}}\endgroup }}}} after executing the program. \toplas{The use of a sensitivity environment {{\color{\colorMATH}\ensuremath{{\begingroup\renewcommand\colorMATH{\colorMATHB}\renewcommand\colorSYNTAX{\colorSYNTAXB}{{\color{\colorMATH}\ensuremath{\sS}}}\endgroup }}}} is different from \duet, where sensitivities are tracked in {{\color{\colorMATH}\ensuremath{\Gamma }}} and presented as a necessary condition to type check the expression. In other words, in \ssystem {{\color{\colorMATH}\ensuremath{{\begingroup\renewcommand\colorMATH{\colorMATHB}\renewcommand\colorSYNTAX{\colorSYNTAXB}{{\color{\colorMATH}\ensuremath{\sS}}}\endgroup }}}} is used to infer sensitivities, whereas in \duet~{{\color{\colorMATH}\ensuremath{\Gamma }}} is used to check sensitivities.} 

\noindent 
\begin{itemize}[label=\textbf{-},leftmargin=*]\item  Rules{\textsc{ rlit}} and{\textsc{ unit}} are standard and report no effect, as no variable is accessed. \toplas{These two rules present no novelty with respect to \duet.}
\item  Rule{\textsc{ var}} is mostly standard; it reports an ambient effect {{\color{\colorMATH}\ensuremath{1x}}}.
   
   For example,
   \begingroup\color{\colorMATH}\begin{gather*} 
     \inferrule*[lab={\textsc{ var}}
     ]{ (x\mathrel{:}{\begingroup\renewcommand\colorMATH{\colorMATHA}\renewcommand\colorSYNTAX{\colorSYNTAXA}{{\color{\colorSYNTAX}\texttt{{\ensuremath{{\mathbb{R}}}}}}}\endgroup })(x) = {\begingroup\renewcommand\colorMATH{\colorMATHA}\renewcommand\colorSYNTAX{\colorSYNTAXA}{{\color{\colorSYNTAX}\texttt{{\ensuremath{{\mathbb{R}}}}}}}\endgroup }
        }{
        x\mathrel{:}{\begingroup\renewcommand\colorMATH{\colorMATHA}\renewcommand\colorSYNTAX{\colorSYNTAXA}{{\color{\colorSYNTAX}\texttt{{\ensuremath{{\mathbb{R}}}}}}}\endgroup }\hspace*{0.33em} {\begingroup\renewcommand\colorMATH{\colorMATHB}\renewcommand\colorSYNTAX{\colorSYNTAXB}{{\color{\colorMATH}\ensuremath{\vdash }}}\endgroup }\hspace*{0.33em}x \mathrel{:} {\begingroup\renewcommand\colorMATH{\colorMATHA}\renewcommand\colorSYNTAX{\colorSYNTAXA}{{\color{\colorSYNTAX}\texttt{{\ensuremath{{\mathbb{R}}}}}}}\endgroup } \mathrel{;} x
     }
   \end{gather*}\endgroup

\item  Rule{\textsc{ plus}} computes the resulting ambient effect as the addition of the
   ambient effects of both subterms. To add sensitivity environments we use
   the {{\color{\colorMATH}\ensuremath{+}}} operator, which is simply defined as the addition of
   polynomials, e.g. {{\color{\colorMATH}\ensuremath{(1x + 2y) + (3x) = 4x + 2y}}}.

   For example, in the following type derivation
   \begingroup\color{\colorMATH}\begin{gather*} 
     \inferrule*[lab={\textsc{ plus}}
     ]{ x\mathrel{:}{\begingroup\renewcommand\colorMATH{\colorMATHA}\renewcommand\colorSYNTAX{\colorSYNTAXA}{{\color{\colorSYNTAX}\texttt{{\ensuremath{{\mathbb{R}}}}}}}\endgroup } \hspace*{0.33em}{\begingroup\renewcommand\colorMATH{\colorMATHB}\renewcommand\colorSYNTAX{\colorSYNTAXB}{{\color{\colorMATH}\ensuremath{\vdash }}}\endgroup }\hspace*{0.33em}x \mathrel{:} {\begingroup\renewcommand\colorMATH{\colorMATHA}\renewcommand\colorSYNTAX{\colorSYNTAXA}{{\color{\colorSYNTAX}\texttt{{\ensuremath{{\mathbb{R}}}}}}}\endgroup } \mathrel{;} x
     \\ x\mathrel{:}{\begingroup\renewcommand\colorMATH{\colorMATHA}\renewcommand\colorSYNTAX{\colorSYNTAXA}{{\color{\colorSYNTAX}\texttt{{\ensuremath{{\mathbb{R}}}}}}}\endgroup } \hspace*{0.33em}{\begingroup\renewcommand\colorMATH{\colorMATHB}\renewcommand\colorSYNTAX{\colorSYNTAXB}{{\color{\colorMATH}\ensuremath{\vdash }}}\endgroup }\hspace*{0.33em}x \mathrel{:} {\begingroup\renewcommand\colorMATH{\colorMATHA}\renewcommand\colorSYNTAX{\colorSYNTAXA}{{\color{\colorSYNTAX}\texttt{{\ensuremath{{\mathbb{R}}}}}}}\endgroup } \mathrel{;} x
        }{
        x\mathrel{:}{\begingroup\renewcommand\colorMATH{\colorMATHA}\renewcommand\colorSYNTAX{\colorSYNTAXA}{{\color{\colorSYNTAX}\texttt{{\ensuremath{{\mathbb{R}}}}}}}\endgroup }\hspace*{0.33em} {\begingroup\renewcommand\colorMATH{\colorMATHB}\renewcommand\colorSYNTAX{\colorSYNTAXB}{{\color{\colorMATH}\ensuremath{\vdash }}}\endgroup }\hspace*{0.33em}x+x \mathrel{:} {\begingroup\renewcommand\colorMATH{\colorMATHA}\renewcommand\colorSYNTAX{\colorSYNTAXA}{{\color{\colorSYNTAX}\texttt{{\ensuremath{{\mathbb{R}}}}}}}\endgroup } \mathrel{;} 2x
     }
   \end{gather*}\endgroup
   we write {{\color{\colorMATH}\ensuremath{2x}}} instead of {{\color{\colorMATH}\ensuremath{x+x}}}.

\item  Rules{\textsc{ times}} and{\textsc{ leq}} are similar to{\textsc{ plus}}, but the resulting sensitivity
   effect is scaled by infinity because (1) the sensitivity of a multiplication
   when neither side is a constant is unbounded, and (2) the distance between
   distinct boolean values is deemed infinite, as explained in Section~\ref{sec:linear-products-and-sums-lim}. 
   Scaling a sensitivity
   environment {{\color{\colorMATH}\ensuremath{{\begingroup\renewcommand\colorMATH{\colorMATHB}\renewcommand\colorSYNTAX{\colorSYNTAXB}{{\color{\colorMATH}\ensuremath{\sS}}}\endgroup }}}} by sensitivity {{\color{\colorMATH}\ensuremath{{\begingroup\renewcommand\colorMATH{\colorMATHB}\renewcommand\colorSYNTAX{\colorSYNTAXB}{{\color{\colorMATH}\ensuremath{\sss}}}\endgroup }}}}, written {{\color{\colorMATH}\ensuremath{{\begingroup\renewcommand\colorMATH{\colorMATHB}\renewcommand\colorSYNTAX{\colorSYNTAXB}{{\color{\colorMATH}\ensuremath{\sss}}}\endgroup }{\begingroup\renewcommand\colorMATH{\colorMATHB}\renewcommand\colorSYNTAX{\colorSYNTAXB}{{\color{\colorMATH}\ensuremath{\sS}}}\endgroup }}}}, produces a new
   sensitivity environment in which each sensitivity in {{\color{\colorMATH}\ensuremath{{\begingroup\renewcommand\colorMATH{\colorMATHB}\renewcommand\colorSYNTAX{\colorSYNTAXB}{{\color{\colorMATH}\ensuremath{\sS}}}\endgroup }}}} is multiplied by {{\color{\colorMATH}\ensuremath{{\begingroup\renewcommand\colorMATH{\colorMATHB}\renewcommand\colorSYNTAX{\colorSYNTAXB}{{\color{\colorMATH}\ensuremath{\sss}}}\endgroup }}}}.
   For multiplication we assume that {{\color{\colorMATH}\ensuremath{0s = s0 = 0}}} for all {{\color{\colorMATH}\ensuremath{s \in {\begingroup\renewcommand\colorMATH{\colorMATHA}\renewcommand\colorSYNTAX{\colorSYNTAXA}{{\color{\colorSYNTAX}\texttt{{\ensuremath{{\mathbb{R}}}}}}}\endgroup }^{\infty }_{\geq 0}}}} and we deem {{\color{\colorMATH}\ensuremath{\infty s = s\infty  = \infty }}} for
   {{\color{\colorMATH}\ensuremath{s\neq 0}}}.

  Rules{\textsc{ l-scale}} and{\textsc{ r-scale}} address the overapproximation yielded by rule{\textsc{ times}} when one of the factors is a real number. For instance, for program {{\color{\colorMATH}\ensuremath{0.5*x}}} rule{\textsc{ l-scale}} reports a (precise) sensitivity of {{\color{\colorMATH}\ensuremath{0.5x}}}, whereas rules{\textsc{ times}} would report {{\color{\colorMATH}\ensuremath{\infty x}}}.

\item  Rule{\textsc{ lam}} typechecks sensitivity functions \toplas{and is novel with respect to \duet}.  The type
   of the function is annotated with a latent effect {{\color{\colorMATH}\ensuremath{{\begingroup\renewcommand\colorMATH{\colorMATHB}\renewcommand\colorSYNTAX{\colorSYNTAXB}{{\color{\colorMATH}\ensuremath{\sS}}}\endgroup }}}}, computed as \toplasss{a subset of} the effect of its body.
   \toplasss{
   On a fully-latent discipline, the whole body effect is left as latent and the ambient effect {{\color{\colorMATH}\ensuremath{{\begingroup\renewcommand\colorMATH{\colorMATHB}\renewcommand\colorSYNTAX{\colorSYNTAXB}{{\color{\colorMATH}\ensuremath{\sS'}}}\endgroup }}}} of the function is empty.
   On the other hand, full eagerness of effects, as in \duet, is achieved when the latent effect {{\color{\colorMATH}\ensuremath{{\begingroup\renewcommand\colorMATH{\colorMATHB}\renewcommand\colorSYNTAX{\colorSYNTAXB}{{\color{\colorMATH}\ensuremath{\sS}}}\endgroup }}}} is empty and the full effect of the body is paid upon construction.

   Since the splitting of {{\color{\colorMATH}\ensuremath{{\begingroup\renewcommand\colorMATH{\colorMATHB}\renewcommand\colorSYNTAX{\colorSYNTAXB}{{\color{\colorMATH}\ensuremath{\sS}}}\endgroup } + {\begingroup\renewcommand\colorMATH{\colorMATHB}\renewcommand\colorSYNTAX{\colorSYNTAXB}{{\color{\colorMATH}\ensuremath{\sS'}}}\endgroup }}}} is non-deterministic, a lambda expression can be given many types, ranging from fully-latent to fully-eager disciplines.
   We show this behavior later when explaining rules{\textsc{ pair}} and{\textsc{ tup}}.  
   The implementation addresses this issue through the use of additional type annotations.
   Without loss of generality, in this paper, we assume the fully-latent derivation for all lambdas unless stated otherwise.
   The same applies to other language constructs that exhibit this kind of non-deterministic prepayment of latent effects.
   }

   For example, consider program {{\color{\colorMATH}\ensuremath{{\begingroup\renewcommand\colorMATH{\colorMATHB}\renewcommand\colorSYNTAX{\colorSYNTAXB}{{\color{\colorMATH}\ensuremath{\slambda}}}\endgroup } (x\mathrel{:}{\begingroup\renewcommand\colorMATH{\colorMATHA}\renewcommand\colorSYNTAX{\colorSYNTAXA}{{\color{\colorSYNTAX}\texttt{{\ensuremath{{\mathbb{R}}}}}}}\endgroup }).\hspace*{0.33em}x+x}}} and its type derivation:
   \begingroup\color{\colorMATH}\begin{gather*} 
     \inferrule*[lab={\textsc{ lam}}
     ]{ x\mathrel{:}{\begingroup\renewcommand\colorMATH{\colorMATHA}\renewcommand\colorSYNTAX{\colorSYNTAXA}{{\color{\colorSYNTAX}\texttt{{\ensuremath{{\mathbb{R}}}}}}}\endgroup } \hspace*{0.33em}{\begingroup\renewcommand\colorMATH{\colorMATHB}\renewcommand\colorSYNTAX{\colorSYNTAXB}{{\color{\colorMATH}\ensuremath{\vdash }}}\endgroup }\hspace*{0.33em}x+x \mathrel{:} {\begingroup\renewcommand\colorMATH{\colorMATHA}\renewcommand\colorSYNTAX{\colorSYNTAXA}{{\color{\colorSYNTAX}\texttt{{\ensuremath{{\mathbb{R}}}}}}}\endgroup } \mathrel{;} 2x
        }{
        \varnothing \hspace*{0.33em} {\begingroup\renewcommand\colorMATH{\colorMATHB}\renewcommand\colorSYNTAX{\colorSYNTAXB}{{\color{\colorMATH}\ensuremath{\vdash }}}\endgroup }\hspace*{0.33em}{\begingroup\renewcommand\colorMATH{\colorMATHB}\renewcommand\colorSYNTAX{\colorSYNTAXB}{{\color{\colorMATH}\ensuremath{\slambda}}}\endgroup } (x\mathrel{:}{\begingroup\renewcommand\colorMATH{\colorMATHA}\renewcommand\colorSYNTAX{\colorSYNTAXA}{{\color{\colorSYNTAX}\texttt{{\ensuremath{{\mathbb{R}}}}}}}\endgroup }).\hspace*{0.33em}x+x \mathrel{:} (x\mathrel{:}{\begingroup\renewcommand\colorMATH{\colorMATHA}\renewcommand\colorSYNTAX{\colorSYNTAXA}{{\color{\colorSYNTAX}\texttt{{\ensuremath{{\mathbb{R}}}}}}}\endgroup }) \xrightarrowS {2x} {\begingroup\renewcommand\colorMATH{\colorMATHA}\renewcommand\colorSYNTAX{\colorSYNTAXA}{{\color{\colorSYNTAX}\texttt{{\ensuremath{{\mathbb{R}}}}}}}\endgroup } \mathrel{;} {\begingroup\renewcommand\colorMATH{\colorMATHB}\renewcommand\colorSYNTAX{\colorSYNTAXB}{{\color{\colorMATH}\ensuremath{\varnothing }}}\endgroup }
     }
   \end{gather*}\endgroup
   The ambient effect of the program is empty (values are pure) but its latent effect is {{\color{\colorMATH}\ensuremath{2x}}}, the ambient effect of its body.

\item  Rule{\textsc{ app}} deals with function application. \toplas{Unlike \duet, as} variable {{\color{\colorMATH}\ensuremath{x}}} may be free
   in {{\color{\colorMATH}\ensuremath{\tau _{2}}}} (e.g. {{\color{\colorMATH}\ensuremath{\tau _{2}}}} can be a function type whose latent effect includes
   {{\color{\colorMATH}\ensuremath{x}}}), the resulting type replaces {{\color{\colorMATH}\ensuremath{x}}} with the ambient effect {{\color{\colorMATH}\ensuremath{{\begingroup\renewcommand\colorMATH{\colorMATHB}\renewcommand\colorSYNTAX{\colorSYNTAXB}{{\color{\colorMATH}\ensuremath{\sS_{2}}}}\endgroup }}}} of its
   argument using the sensitivity environment substitution operator
   defined in Figure~\ref{fig:sensitivity-simple-statics-auxiliary-definitions}. For instance, consider type {{\color{\colorMATH}\ensuremath{(x:{\begingroup\renewcommand\colorMATH{\colorMATHA}\renewcommand\colorSYNTAX{\colorSYNTAXA}{{\color{\colorSYNTAX}\texttt{{\ensuremath{{\mathbb{R}}}}}}}\endgroup }) \xrightarrowS {\varnothing } (z:{\begingroup\renewcommand\colorMATH{\colorMATHA}\renewcommand\colorSYNTAX{\colorSYNTAXA}{{\color{\colorSYNTAX}\texttt{{\ensuremath{{\mathbb{R}}}}}}}\endgroup }) \xrightarrowS {2x+y+z} {\begingroup\renewcommand\colorMATH{\colorMATHA}\renewcommand\colorSYNTAX{\colorSYNTAXA}{{\color{\colorSYNTAX}\texttt{{\ensuremath{{\mathbb{R}}}}}}}\endgroup }}}}. After application, if {{\color{\colorMATH}\ensuremath{{\begingroup\renewcommand\colorMATH{\colorMATHB}\renewcommand\colorSYNTAX{\colorSYNTAXB}{{\color{\colorMATH}\ensuremath{\sS_{2}}}}\endgroup } = 3y}}}, the resulting type  
   would be {{\color{\colorMATH}\ensuremath{[3y/x]((z:{\begingroup\renewcommand\colorMATH{\colorMATHA}\renewcommand\colorSYNTAX{\colorSYNTAXA}{{\color{\colorSYNTAX}\texttt{{\ensuremath{{\mathbb{R}}}}}}}\endgroup }) \xrightarrowS {2x+y+z} {\begingroup\renewcommand\colorMATH{\colorMATHA}\renewcommand\colorSYNTAX{\colorSYNTAXA}{{\color{\colorSYNTAX}\texttt{{\ensuremath{{\mathbb{R}}}}}}}\endgroup }) = ((z:{\begingroup\renewcommand\colorMATH{\colorMATHA}\renewcommand\colorSYNTAX{\colorSYNTAXA}{{\color{\colorSYNTAX}\texttt{{\ensuremath{{\mathbb{R}}}}}}}\endgroup }) \xrightarrowS {6y+y+z} {\begingroup\renewcommand\colorMATH{\colorMATHA}\renewcommand\colorSYNTAX{\colorSYNTAXA}{{\color{\colorSYNTAX}\texttt{{\ensuremath{{\mathbb{R}}}}}}}\endgroup }) = ((z:{\begingroup\renewcommand\colorMATH{\colorMATHA}\renewcommand\colorSYNTAX{\colorSYNTAXA}{{\color{\colorSYNTAX}\texttt{{\ensuremath{{\mathbb{R}}}}}}}\endgroup }) \xrightarrowS {7y+y} {\begingroup\renewcommand\colorMATH{\colorMATHA}\renewcommand\colorSYNTAX{\colorSYNTAXA}{{\color{\colorSYNTAX}\texttt{{\ensuremath{{\mathbb{R}}}}}}}\endgroup })}}}.
   \begin{figure}[t]
     \begin{small}
     \begin{framed}
\begingroup\color{\colorMATH}\begin{gather*}
\begin{tabularx}{\linewidth}{>{\centering\arraybackslash\(}X<{\)}}\hfill\hspace{0pt}\begingroup\color{\colorTEXT}\boxed{\begingroup\color{\colorMATH} [{\begingroup\renewcommand\colorMATH{\colorMATHB}\renewcommand\colorSYNTAX{\colorSYNTAXB}{{\color{\colorMATH}\ensuremath{\sS}}}\endgroup }/x]\tau  \endgroup}\endgroup %{{\color{\colorTEXT}\textnormal{{\textit{Sensitivity environment substitution}}}}}
  \cr
\end{tabularx}
\end{gather*}\endgroup
\end{framed}

     \end{small}
     \caption{$\ssystem$: Auxiliary definitions of the static semantics (selected rules)}
     \label{fig:sensitivity-simple-statics-auxiliary-definitions}
   \end{figure}
   The ambient effect of an application is computed as the ambient effect of
   the function {{\color{\colorMATH}\ensuremath{{\begingroup\renewcommand\colorMATH{\colorMATHB}\renewcommand\colorSYNTAX{\colorSYNTAXB}{{\color{\colorMATH}\ensuremath{\sS_{1}}}}\endgroup }}}}, plus its latent effect; but as {{\color{\colorMATH}\ensuremath{x}}} is free we substitute it
   by {{\color{\colorMATH}\ensuremath{{\begingroup\renewcommand\colorMATH{\colorMATHB}\renewcommand\colorSYNTAX{\colorSYNTAXB}{{\color{\colorMATH}\ensuremath{\sS_{2}}}}\endgroup }}}}, e.g. if {{\color{\colorMATH}\ensuremath{{\begingroup\renewcommand\colorMATH{\colorMATHB}\renewcommand\colorSYNTAX{\colorSYNTAXB}{{\color{\colorMATH}\ensuremath{\sS}}}\endgroup } = {\begingroup\renewcommand\colorMATH{\colorMATHB}\renewcommand\colorSYNTAX{\colorSYNTAXB}{{\color{\colorMATH}\ensuremath{\sS'}}}\endgroup }+{\begingroup\renewcommand\colorMATH{\colorMATHB}\renewcommand\colorSYNTAX{\colorSYNTAXB}{{\color{\colorMATH}\ensuremath{\sss}}}\endgroup }x}}}, then  {{\color{\colorMATH}\ensuremath{[{\begingroup\renewcommand\colorMATH{\colorMATHB}\renewcommand\colorSYNTAX{\colorSYNTAXB}{{\color{\colorMATH}\ensuremath{\sS_{2}}}}\endgroup }/x]({\begingroup\renewcommand\colorMATH{\colorMATHB}\renewcommand\colorSYNTAX{\colorSYNTAXB}{{\color{\colorMATH}\ensuremath{\sS_{1}}}}\endgroup } + {\begingroup\renewcommand\colorMATH{\colorMATHB}\renewcommand\colorSYNTAX{\colorSYNTAXB}{{\color{\colorMATH}\ensuremath{\sS'}}}\endgroup } + {\begingroup\renewcommand\colorMATH{\colorMATHB}\renewcommand\colorSYNTAX{\colorSYNTAXB}{{\color{\colorMATH}\ensuremath{\sss}}}\endgroup }x) = {\begingroup\renewcommand\colorMATH{\colorMATHB}\renewcommand\colorSYNTAX{\colorSYNTAXB}{{\color{\colorMATH}\ensuremath{\sS_{1}}}}\endgroup } + ({\begingroup\renewcommand\colorMATH{\colorMATHB}\renewcommand\colorSYNTAX{\colorSYNTAXB}{{\color{\colorMATH}\ensuremath{\sS'}}}\endgroup } + {\begingroup\renewcommand\colorMATH{\colorMATHB}\renewcommand\colorSYNTAX{\colorSYNTAXB}{{\color{\colorMATH}\ensuremath{\sss}}}\endgroup }{\begingroup\renewcommand\colorMATH{\colorMATHB}\renewcommand\colorSYNTAX{\colorSYNTAXB}{{\color{\colorMATH}\ensuremath{\sS_{2}}}}\endgroup })}}}. \toplas{This is different from \duet as, where the latent effect of the function {{\color{\colorMATH}\ensuremath{{\begingroup\renewcommand\colorMATH{\colorMATHB}\renewcommand\colorSYNTAX{\colorSYNTAXB}{{\color{\colorMATH}\ensuremath{\sS}}}\endgroup }}}} is paid when the function is created.}

   For instance, consider the open program {{\color{\colorMATH}\ensuremath{({\begingroup\renewcommand\colorMATH{\colorMATHB}\renewcommand\colorSYNTAX{\colorSYNTAXB}{{\color{\colorMATH}\ensuremath{\slambda}}}\endgroup } (x\mathrel{:}{\begingroup\renewcommand\colorMATH{\colorMATHA}\renewcommand\colorSYNTAX{\colorSYNTAXA}{{\color{\colorSYNTAX}\texttt{{\ensuremath{{\mathbb{R}}}}}}}\endgroup }).\hspace*{0.33em}2*x+y)\ (3*y)}}} and the following type derivation:
  \begingroup\color{\colorMATH}\begin{gather*} 
     \inferrule*[lab={\textsc{ app}}
     ]{ y:{\begingroup\renewcommand\colorMATH{\colorMATHA}\renewcommand\colorSYNTAX{\colorSYNTAXA}{{\color{\colorSYNTAX}\texttt{{\ensuremath{{\mathbb{R}}}}}}}\endgroup }\hspace*{0.33em} {\begingroup\renewcommand\colorMATH{\colorMATHB}\renewcommand\colorSYNTAX{\colorSYNTAXB}{{\color{\colorMATH}\ensuremath{\vdash }}}\endgroup }\hspace*{0.33em}{\begingroup\renewcommand\colorMATH{\colorMATHB}\renewcommand\colorSYNTAX{\colorSYNTAXB}{{\color{\colorMATH}\ensuremath{\slambda}}}\endgroup } (x\mathrel{:}{\begingroup\renewcommand\colorMATH{\colorMATHA}\renewcommand\colorSYNTAX{\colorSYNTAXA}{{\color{\colorSYNTAX}\texttt{{\ensuremath{{\mathbb{R}}}}}}}\endgroup }).\hspace*{0.33em}2*x+y \mathrel{:} (x\mathrel{:}{\begingroup\renewcommand\colorMATH{\colorMATHA}\renewcommand\colorSYNTAX{\colorSYNTAXA}{{\color{\colorSYNTAX}\texttt{{\ensuremath{{\mathbb{R}}}}}}}\endgroup }) \xrightarrowS {2x+y} {\begingroup\renewcommand\colorMATH{\colorMATHA}\renewcommand\colorSYNTAX{\colorSYNTAXA}{{\color{\colorSYNTAX}\texttt{{\ensuremath{{\mathbb{R}}}}}}}\endgroup } \mathrel{;} {\begingroup\renewcommand\colorMATH{\colorMATHB}\renewcommand\colorSYNTAX{\colorSYNTAXB}{{\color{\colorMATH}\ensuremath{\varnothing }}}\endgroup }
     \\ y:{\begingroup\renewcommand\colorMATH{\colorMATHA}\renewcommand\colorSYNTAX{\colorSYNTAXA}{{\color{\colorSYNTAX}\texttt{{\ensuremath{{\mathbb{R}}}}}}}\endgroup }\hspace*{0.33em} {\begingroup\renewcommand\colorMATH{\colorMATHB}\renewcommand\colorSYNTAX{\colorSYNTAXB}{{\color{\colorMATH}\ensuremath{\vdash }}}\endgroup }\hspace*{0.33em}3*y : {\begingroup\renewcommand\colorMATH{\colorMATHA}\renewcommand\colorSYNTAX{\colorSYNTAXA}{{\color{\colorSYNTAX}\texttt{{\ensuremath{{\mathbb{R}}}}}}}\endgroup } ; 3y
        }{
        y:{\begingroup\renewcommand\colorMATH{\colorMATHA}\renewcommand\colorSYNTAX{\colorSYNTAXA}{{\color{\colorSYNTAX}\texttt{{\ensuremath{{\mathbb{R}}}}}}}\endgroup }\hspace*{0.33em} {\begingroup\renewcommand\colorMATH{\colorMATHB}\renewcommand\colorSYNTAX{\colorSYNTAXB}{{\color{\colorMATH}\ensuremath{\vdash }}}\endgroup }\hspace*{0.33em}({\begingroup\renewcommand\colorMATH{\colorMATHB}\renewcommand\colorSYNTAX{\colorSYNTAXB}{{\color{\colorMATH}\ensuremath{\slambda}}}\endgroup } (x\mathrel{:}{\begingroup\renewcommand\colorMATH{\colorMATHA}\renewcommand\colorSYNTAX{\colorSYNTAXA}{{\color{\colorSYNTAX}\texttt{{\ensuremath{{\mathbb{R}}}}}}}\endgroup }).\hspace*{0.33em}2*x+y)\ (3*y) \mathrel{:} {\begingroup\renewcommand\colorMATH{\colorMATHA}\renewcommand\colorSYNTAX{\colorSYNTAXA}{{\color{\colorSYNTAX}\texttt{{\ensuremath{{\mathbb{R}}}}}}}\endgroup } \mathrel{;} 7y
     }
   \end{gather*}\endgroup
   The resulting ambient effect cannot depend on {{\color{\colorMATH}\ensuremath{x}}} (otherwise it would be free), therefore it is computed as 
   the substitution of {{\color{\colorMATH}\ensuremath{x}}} by the \toplas{ambient} effect of the argument {{\color{\colorMATH}\ensuremath{[3y/x](2x+y) = 7y}}}.

\item  Contrary to previous work~\cite{reed2010distance,gaboardi2013linear} \toplas{, and in particular \duet}, Rule{\textsc{ inl}} 
   \toplasss{
      does not necessarily report the effect of its body.
      The payment of effects for the subexpression (or a subset of it) can be {\em delayed}, and eventually payed only if the sum is accessed or used. 
   }
   The term is tagged with type {{\color{\colorMATH}\ensuremath{\tau _{2}}}}  to aid type
   inference. The resulting type is just a sum type where the latent effect of
   the left type is {{\color{\colorMATH}\ensuremath{{\begingroup\renewcommand\colorMATH{\colorMATHB}\renewcommand\colorSYNTAX{\colorSYNTAXB}{{\color{\colorMATH}\ensuremath{\sS}}}\endgroup }}}}, \toplasss{a subset of} the ambient effect of its subterm, and the latent
   effect of the right type is empty (as it will never be used/accessed so we
   choose the tighter ambient effect). 
   \toplasss{
      Non-determinism is addressed similarly to rule{\textsc{ lam}}. 
   }
   Rule{\textsc{ inr}} is defined similarly.

For instance, consider the type derivations of expressions {{\color{\colorMATH}\ensuremath{e_{4} = \inl\hspace*{0.33em}(x*x)}}} and {{\color{\colorMATH}\ensuremath{e_{5} = \inr\hspace*{0.33em}(x*x)}}} of Section~\ref{sec:advantages-of-sensitivity-ts}:
\begingroup\color{\colorMATH}\begin{mathpar} 
     \inferrule*[lab={\textsc{ inl}}
     ]{ x\mathrel{:}{\begingroup\renewcommand\colorMATH{\colorMATHA}\renewcommand\colorSYNTAX{\colorSYNTAXA}{{\color{\colorSYNTAX}\texttt{{\ensuremath{{\mathbb{R}}}}}}}\endgroup } \hspace*{0.33em}{\begingroup\renewcommand\colorMATH{\colorMATHB}\renewcommand\colorSYNTAX{\colorSYNTAXB}{{\color{\colorMATH}\ensuremath{\vdash }}}\endgroup }\hspace*{0.33em}(x*x) \mathrel{:} {\begingroup\renewcommand\colorMATH{\colorMATHA}\renewcommand\colorSYNTAX{\colorSYNTAXA}{{\color{\colorSYNTAX}\texttt{{\ensuremath{{\mathbb{R}}}}}}}\endgroup } \mathrel{;} \infty x
        }{
        \varnothing \hspace*{0.33em} {\begingroup\renewcommand\colorMATH{\colorMATHB}\renewcommand\colorSYNTAX{\colorSYNTAXB}{{\color{\colorMATH}\ensuremath{\vdash }}}\endgroup }\hspace*{0.33em}\inl^{{\begingroup\renewcommand\colorMATH{\colorMATHA}\renewcommand\colorSYNTAX{\colorSYNTAXA}{{\color{\colorSYNTAX}\texttt{{\ensuremath{{\mathbb{R}}}}}}}\endgroup }}\hspace*{0.33em}(x*x) \mathrel{:} {\begingroup\renewcommand\colorMATH{\colorMATHA}\renewcommand\colorSYNTAX{\colorSYNTAXA}{{\color{\colorSYNTAX}\texttt{{\ensuremath{{\mathbb{R}}}}}}}\endgroup } \mathrel{^{\infty x}\oplus ^{{\begingroup\renewcommand\colorMATH{\colorMATHB}\renewcommand\colorSYNTAX{\colorSYNTAXB}{{\color{\colorMATH}\ensuremath{\varnothing }}}\endgroup }}} {\begingroup\renewcommand\colorMATH{\colorMATHA}\renewcommand\colorSYNTAX{\colorSYNTAXA}{{\color{\colorSYNTAX}\texttt{{\ensuremath{{\mathbb{R}}}}}}}\endgroup } \mathrel{;} {\begingroup\renewcommand\colorMATH{\colorMATHB}\renewcommand\colorSYNTAX{\colorSYNTAXB}{{\color{\colorMATH}\ensuremath{\varnothing }}}\endgroup }
     }
  \and \inferrule*[lab={\textsc{ inr}}
     ]{ x\mathrel{:}{\begingroup\renewcommand\colorMATH{\colorMATHA}\renewcommand\colorSYNTAX{\colorSYNTAXA}{{\color{\colorSYNTAX}\texttt{{\ensuremath{{\mathbb{R}}}}}}}\endgroup } \hspace*{0.33em}{\begingroup\renewcommand\colorMATH{\colorMATHB}\renewcommand\colorSYNTAX{\colorSYNTAXB}{{\color{\colorMATH}\ensuremath{\vdash }}}\endgroup }\hspace*{0.33em}x*x \mathrel{:} {\begingroup\renewcommand\colorMATH{\colorMATHA}\renewcommand\colorSYNTAX{\colorSYNTAXA}{{\color{\colorSYNTAX}\texttt{{\ensuremath{{\mathbb{R}}}}}}}\endgroup } \mathrel{;} \infty x
        }{
        \varnothing \hspace*{0.33em} {\begingroup\renewcommand\colorMATH{\colorMATHB}\renewcommand\colorSYNTAX{\colorSYNTAXB}{{\color{\colorMATH}\ensuremath{\vdash }}}\endgroup }\hspace*{0.33em}\inr^{{\begingroup\renewcommand\colorMATH{\colorMATHA}\renewcommand\colorSYNTAX{\colorSYNTAXA}{{\color{\colorSYNTAX}\texttt{{\ensuremath{{\mathbb{R}}}}}}}\endgroup }}\hspace*{0.33em}(x*x) \mathrel{:} {\begingroup\renewcommand\colorMATH{\colorMATHA}\renewcommand\colorSYNTAX{\colorSYNTAXA}{{\color{\colorSYNTAX}\texttt{{\ensuremath{{\mathbb{R}}}}}}}\endgroup } \mathrel{^{{\begingroup\renewcommand\colorMATH{\colorMATHB}\renewcommand\colorSYNTAX{\colorSYNTAXB}{{\color{\colorMATH}\ensuremath{\varnothing }}}\endgroup }}\oplus ^{\infty x}} {\begingroup\renewcommand\colorMATH{\colorMATHA}\renewcommand\colorSYNTAX{\colorSYNTAXA}{{\color{\colorSYNTAX}\texttt{{\ensuremath{{\mathbb{R}}}}}}}\endgroup } \mathrel{;} {\begingroup\renewcommand\colorMATH{\colorMATHB}\renewcommand\colorSYNTAX{\colorSYNTAXB}{{\color{\colorMATH}\ensuremath{\varnothing }}}\endgroup }
     }
\end{mathpar}\endgroup
For expression {{\color{\colorMATH}\ensuremath{e_{4}}}}, the latent effect of the left type is {{\color{\colorMATH}\ensuremath{\infty x}}}, and of the right type is empty (it is the tighter upper bound as the right component cannot be accessed). An analogous argument is used for {{\color{\colorMATH}\ensuremath{e_{5}}}}.

\item  Rule{\textsc{ case}} is more involved. The
   resulting type of the case is just the least upper bound (join) of the branch types
   {{\color{\colorMATH}\ensuremath{\tau _{2}}}} and {{\color{\colorMATH}\ensuremath{\tau _{3}}}}. The join operator is defined in
   Figure~\ref{fig:sensitivity-simple-statics-join}.
  \begin{figure}[t]
     \begin{small}
     \begin{framed}
	 \begingroup\color{\colorMATH}\begin{gather*}
\begin{tabularx}{\linewidth}{>{\centering\arraybackslash\(}X<{\)}}\hfill\hspace{0pt}\begingroup\color{\colorTEXT}\boxed{\begingroup\color{\colorMATH} \tau  \sqcup  \tau  \endgroup}\endgroup
	  \cr
\end{tabularx}
	\end{gather*}\endgroup
\end{framed}
     \end{small}
     \caption{$\ssystem$: Join and Meet of types and sensitivity environments}
     \label{fig:sensitivity-simple-statics-join}
   \end{figure}
   Note that similarly to rule {\textsc{ app}}, {{\color{\colorMATH}\ensuremath{\tau _{2}}}} and {{\color{\colorMATH}\ensuremath{\tau _{3}}}} 
   may have {{\color{\colorMATH}\ensuremath{x}}} and {{\color{\colorMATH}\ensuremath{y}}} as free variables respectively, thus we replace 
   those variables with the ambient effects of using the sum term {{\color{\colorMATH}\ensuremath{{\begingroup\renewcommand\colorMATH{\colorMATHB}\renewcommand\colorSYNTAX{\colorSYNTAXB}{{\color{\colorMATH}\ensuremath{\se_{1}}}}\endgroup }}}}:  {{\color{\colorMATH}\ensuremath{{\begingroup\renewcommand\colorMATH{\colorMATHB}\renewcommand\colorSYNTAX{\colorSYNTAXB}{{\color{\colorMATH}\ensuremath{\sS_{1}}}}\endgroup } + {\begingroup\renewcommand\colorMATH{\colorMATHB}\renewcommand\colorSYNTAX{\colorSYNTAXB}{{\color{\colorMATH}\ensuremath{\sS_{1 1}}}}\endgroup }}}} and {{\color{\colorMATH}\ensuremath{{\begingroup\renewcommand\colorMATH{\colorMATHB}\renewcommand\colorSYNTAX{\colorSYNTAXB}{{\color{\colorMATH}\ensuremath{\sS_{1}}}}\endgroup } + {\begingroup\renewcommand\colorMATH{\colorMATHB}\renewcommand\colorSYNTAX{\colorSYNTAXB}{{\color{\colorMATH}\ensuremath{\sS_{1 2}}}}\endgroup }}}} respectively. 
   The
   resulting ambient effect is computed as follows: we join the cost of reducing {{\color{\colorMATH}\ensuremath{{\begingroup\renewcommand\colorMATH{\colorMATHB}\renewcommand\colorSYNTAX{\colorSYNTAXB}{{\color{\colorMATH}\ensuremath{\se_{1}}}}\endgroup }}}}: {{\color{\colorMATH}\ensuremath{{\begingroup\renewcommand\colorMATH{\colorMATHB}\renewcommand\colorSYNTAX{\colorSYNTAXB}{{\color{\colorMATH}\ensuremath{\sS_{1}}}}\endgroup }}}}, with the join of the cost of
   taking each branch. \toplas{This is different from \duet, where {{\color{\colorMATH}\ensuremath{{\begingroup\renewcommand\colorMATH{\colorMATHB}\renewcommand\colorSYNTAX{\colorSYNTAXB}{{\color{\colorMATH}\ensuremath{\sS_{1}}}}\endgroup }}}} is {\bf added} to the cost of taking each branch, leading to a looser bound.}
   Similarly to types {{\color{\colorMATH}\ensuremath{\tau _{2}}}} and {{\color{\colorMATH}\ensuremath{\tau _{3}}}}, ambient effects {{\color{\colorMATH}\ensuremath{{\begingroup\renewcommand\colorMATH{\colorMATHB}\renewcommand\colorSYNTAX{\colorSYNTAXB}{{\color{\colorMATH}\ensuremath{\sS_{2}}}}\endgroup }}}} and {{\color{\colorMATH}\ensuremath{{\begingroup\renewcommand\colorMATH{\colorMATHB}\renewcommand\colorSYNTAX{\colorSYNTAXB}{{\color{\colorMATH}\ensuremath{\sS_{3}}}}\endgroup }}}} may have {{\color{\colorMATH}\ensuremath{x}}} and {{\color{\colorMATH}\ensuremath{y}}} free, so we substitute them away from the effects.
   Note that we use the join between {\begingroup\renewcommand\colorMATH{\colorMATHB}\renewcommand\colorSYNTAX{\colorSYNTAXB}{{\color{\colorMATH}\ensuremath{\sS_{1}}}}\endgroup } and the cost of the branches (instead of the addition for instance), otherwise the result would be less precise when the branches use {{\color{\colorMATH}\ensuremath{x}}} or {{\color{\colorMATH}\ensuremath{y}}}. 

   For instance, 
  the type derivation of Example~\ref{ex:confbranches} is described below:
  \begingroup\color{\colorMATH}\begin{gather*}
     \inferrule*[lab={\textsc{ case}}
     ]{ \Gamma  \hspace*{0.33em}{\begingroup\renewcommand\colorMATH{\colorMATHB}\renewcommand\colorSYNTAX{\colorSYNTAXB}{{\color{\colorMATH}\ensuremath{\vdash }}}\endgroup }\hspace*{0.33em}{\begingroup\renewcommand\colorMATH{\colorMATHB}\renewcommand\colorSYNTAX{\colorSYNTAXB}{{\color{\colorMATH}\ensuremath{\se}}}\endgroup } : {\begingroup\renewcommand\colorMATH{\colorMATHA}\renewcommand\colorSYNTAX{\colorSYNTAXA}{{\color{\colorSYNTAX}\texttt{{\ensuremath{{\mathbb{R}}}}}}}\endgroup } \mathrel{^{\infty x}\oplus ^{x}} {\begingroup\renewcommand\colorMATH{\colorMATHA}\renewcommand\colorSYNTAX{\colorSYNTAXA}{{\color{\colorSYNTAX}\texttt{{\ensuremath{{\mathbb{R}}}}}}}\endgroup }; b
     \\ \Gamma , x_{1}:{\begingroup\renewcommand\colorMATH{\colorMATHA}\renewcommand\colorSYNTAX{\colorSYNTAXA}{{\color{\colorSYNTAX}\texttt{{\ensuremath{{\mathbb{R}}}}}}}\endgroup } \hspace*{0.33em}{\begingroup\renewcommand\colorMATH{\colorMATHB}\renewcommand\colorSYNTAX{\colorSYNTAXB}{{\color{\colorMATH}\ensuremath{\vdash }}}\endgroup }\hspace*{0.33em} 0: {\begingroup\renewcommand\colorMATH{\colorMATHA}\renewcommand\colorSYNTAX{\colorSYNTAXA}{{\color{\colorSYNTAX}\texttt{{\ensuremath{{\mathbb{R}}}}}}}\endgroup }; \varnothing 
     \\ \Gamma , x_{2}:{\begingroup\renewcommand\colorMATH{\colorMATHA}\renewcommand\colorSYNTAX{\colorSYNTAXA}{{\color{\colorSYNTAX}\texttt{{\ensuremath{{\mathbb{R}}}}}}}\endgroup } \hspace*{0.33em}{\begingroup\renewcommand\colorMATH{\colorMATHB}\renewcommand\colorSYNTAX{\colorSYNTAXB}{{\color{\colorMATH}\ensuremath{\vdash }}}\endgroup }\hspace*{0.33em} x_{2}: {\begingroup\renewcommand\colorMATH{\colorMATHA}\renewcommand\colorSYNTAX{\colorSYNTAXA}{{\color{\colorSYNTAX}\texttt{{\ensuremath{{\mathbb{R}}}}}}}\endgroup }; x_{2}
        }{
        \Gamma  \hspace*{0.33em}{\begingroup\renewcommand\colorMATH{\colorMATHB}\renewcommand\colorSYNTAX{\colorSYNTAXB}{{\color{\colorMATH}\ensuremath{\vdash }}}\endgroup }\hspace*{0.33em} \ccase\hspace*{0.33em}{\begingroup\renewcommand\colorMATH{\colorMATHB}\renewcommand\colorSYNTAX{\colorSYNTAXB}{{\color{\colorMATH}\ensuremath{\se}}}\endgroup }\hspace*{0.33em}\of\hspace*{0.33em}\{ x_{1} \Rightarrow  0\} \{ x_{2} \Rightarrow  x_{2}\}  : {\begingroup\renewcommand\colorMATH{\colorMATHA}\renewcommand\colorSYNTAX{\colorSYNTAXA}{{\color{\colorSYNTAX}\texttt{{\ensuremath{{\mathbb{R}}}}}}}\endgroup } ; b \sqcup  (0b + 0(\infty x)) \sqcup  (1b + 1(x))
     }
   \end{gather*}\endgroup
   where {{\color{\colorMATH}\ensuremath{\Gamma  = b: {\begingroup\renewcommand\colorMATH{\colorMATHA}\renewcommand\colorSYNTAX{\colorSYNTAXA}{{\color{\colorSYNTAX}\texttt{{\ensuremath{{\mathbb{B}}}}}}}\endgroup }, x:{\begingroup\renewcommand\colorMATH{\colorMATHA}\renewcommand\colorSYNTAX{\colorSYNTAXA}{{\color{\colorSYNTAX}\texttt{{\ensuremath{{\mathbb{R}}}}}}}\endgroup }}}}, and {{\color{\colorMATH}\ensuremath{{\begingroup\renewcommand\colorMATH{\colorMATHB}\renewcommand\colorSYNTAX{\colorSYNTAXB}{{\color{\colorMATH}\ensuremath{\se}}}\endgroup } = {\begingroup\renewcommand\colorMATH{\colorMATHB}\renewcommand\colorSYNTAX{\colorSYNTAXB}{{\color{\colorSYNTAX}\texttt{{\ensuremath{{\texttt{if}}}}}}}\endgroup }\hspace*{0.33em}b\hspace*{0.33em}{\begingroup\renewcommand\colorMATH{\colorMATHB}\renewcommand\colorSYNTAX{\colorSYNTAXB}{{\color{\colorSYNTAX}\texttt{{\ensuremath{{\texttt{then}}}}}}}\endgroup }\hspace*{0.33em}\{\inl\hspace*{0.33em}(x * x)\}\hspace*{0.33em}{\begingroup\renewcommand\colorMATH{\colorMATHB}\renewcommand\colorSYNTAX{\colorSYNTAXB}{{\color{\colorSYNTAX}\texttt{{\ensuremath{{\texttt{else}}}}}}}\endgroup }\hspace*{0.33em}\{\inr\hspace*{0.33em}x\}}}}.
   \toplas{As {{\color{\colorMATH}\ensuremath{0\infty  = 0}}}, the} resulting ambient effect is {{\color{\colorMATH}\ensuremath{b \sqcup  (0b + 0(\infty x)) \sqcup  (1b + 1(x)) = b \sqcup  (0b + 0x) \sqcup  (b + x) = b + x}}}, where previous work reported {{\color{\colorMATH}\ensuremath{\infty }}} on {{\color{\colorMATH}\ensuremath{x}}}.

   Notice that if we change the program to {{\color{\colorMATH}\ensuremath{\ccase\hspace*{0.33em}{\begingroup\renewcommand\colorMATH{\colorMATHB}\renewcommand\colorSYNTAX{\colorSYNTAXB}{{\color{\colorMATH}\ensuremath{\se}}}\endgroup }\hspace*{0.33em}\of\hspace*{0.33em}\{ x_{1} \Rightarrow  0\} \{ x_{2} \Rightarrow  1\} }}}, then the resulting ambient effect is {{\color{\colorMATH}\ensuremath{b \sqcup  (0b + 0(\infty x)) \sqcup  (0b + 0(x)) = b}}}, i.e. the payment is not zero but {{\color{\colorMATH}\ensuremath{b}}}, the cost of reducing expression {{\color{\colorMATH}\ensuremath{{\begingroup\renewcommand\colorMATH{\colorMATHB}\renewcommand\colorSYNTAX{\colorSYNTAXB}{{\color{\colorMATH}\ensuremath{\se}}}\endgroup }}}} to a value.

   Example~\ref{ex:discont} is desugared and type checked as follows:
   \begingroup\color{\colorMATH}\begin{gather*}
     \inferrule*[lab={\textsc{ case}}
     ]{ \Gamma  \hspace*{0.33em}{\begingroup\renewcommand\colorMATH{\colorMATHB}\renewcommand\colorSYNTAX{\colorSYNTAXB}{{\color{\colorMATH}\ensuremath{\vdash }}}\endgroup }\hspace*{0.33em}x \leq  10 : {\begingroup\renewcommand\colorMATH{\colorMATHA}\renewcommand\colorSYNTAX{\colorSYNTAXA}{{\color{\colorSYNTAX}\texttt{{\ensuremath{{\mathbb{B}}}}}}}\endgroup }; \infty x
     \\ \Gamma , x_{1}:{\begingroup\renewcommand\colorMATH{\colorMATHA}\renewcommand\colorSYNTAX{\colorSYNTAXA}{{\color{\colorSYNTAX}\texttt{{\ensuremath{{\mathbb{R}}}}}}}\endgroup } \hspace*{0.33em}{\begingroup\renewcommand\colorMATH{\colorMATHB}\renewcommand\colorSYNTAX{\colorSYNTAXB}{{\color{\colorMATH}\ensuremath{\vdash }}}\endgroup }\hspace*{0.33em} {\text{true}}: {\begingroup\renewcommand\colorMATH{\colorMATHA}\renewcommand\colorSYNTAX{\colorSYNTAXA}{{\color{\colorSYNTAX}\texttt{{\ensuremath{{\mathbb{R}}}}}}}\endgroup }; \varnothing 
     \\ \Gamma , x_{2}:{\begingroup\renewcommand\colorMATH{\colorMATHA}\renewcommand\colorSYNTAX{\colorSYNTAXA}{{\color{\colorSYNTAX}\texttt{{\ensuremath{{\mathbb{R}}}}}}}\endgroup } \hspace*{0.33em}{\begingroup\renewcommand\colorMATH{\colorMATHB}\renewcommand\colorSYNTAX{\colorSYNTAXB}{{\color{\colorMATH}\ensuremath{\vdash }}}\endgroup }\hspace*{0.33em} {\text{false}}: {\begingroup\renewcommand\colorMATH{\colorMATHA}\renewcommand\colorSYNTAX{\colorSYNTAXA}{{\color{\colorSYNTAX}\texttt{{\ensuremath{{\mathbb{R}}}}}}}\endgroup }; \varnothing 
        }{
        \Gamma  \hspace*{0.33em}{\begingroup\renewcommand\colorMATH{\colorMATHB}\renewcommand\colorSYNTAX{\colorSYNTAXB}{{\color{\colorMATH}\ensuremath{\vdash }}}\endgroup }\hspace*{0.33em} \ccase\hspace*{0.33em}x\leq 10\hspace*{0.33em}\of\hspace*{0.33em}\{ x_{1} \Rightarrow  {\text{true}}\} \{ x_{2} \Rightarrow  {\text{false}}\}  : {\begingroup\renewcommand\colorMATH{\colorMATHA}\renewcommand\colorSYNTAX{\colorSYNTAXA}{{\color{\colorSYNTAX}\texttt{{\ensuremath{{\mathbb{B}}}}}}}\endgroup } ; \infty x \sqcup  (0(\infty x) + 0(\varnothing )) \sqcup  (0(\infty x) + 0(\varnothing ))
     }
   \end{gather*}\endgroup
   where {{\color{\colorMATH}\ensuremath{{\begingroup\renewcommand\colorMATH{\colorMATHA}\renewcommand\colorSYNTAX{\colorSYNTAXA}{{\color{\colorSYNTAX}\texttt{{\ensuremath{{\mathbb{B}}}}}}}\endgroup } = {{\color{\colorSYNTAX}\texttt{unit}}} \mathrel{^{\varnothing }\oplus ^{\varnothing }} {{\color{\colorSYNTAX}\texttt{unit}}}}}}.
   As {{\color{\colorMATH}\ensuremath{\infty x \sqcup  (0(\infty x) + 0(\varnothing )) \sqcup  (0(\infty x) + 0(\varnothing )) = \infty x \sqcup  0x = \infty x}}}, the expression is {{\color{\colorMATH}\ensuremath{\infty }}}-sensitive in {{\color{\colorMATH}\ensuremath{x}}}.

\item  %Similarly to Rules{\textsc{ inl}} and{\textsc{ inr}}, Rules{\textsc{ pair}} and{\textsc{ tup}} report no
   \toplas{Rules{\textsc{ pair}} and{\textsc{ tup}} are novel and non-deterministic: the ambient effects are computed using subsets of the ambient effect of each component.
   If the ambient effects of the left component is {{\color{\colorMATH}\ensuremath{{\begingroup\renewcommand\colorMATH{\colorMATHB}\renewcommand\colorSYNTAX{\colorSYNTAXB}{{\color{\colorMATH}\ensuremath{\sS_{1}}}}\endgroup }+{\begingroup\renewcommand\colorMATH{\colorMATHB}\renewcommand\colorSYNTAX{\colorSYNTAXB}{{\color{\colorMATH}\ensuremath{\sS'_{1}}}}\endgroup }}}} and of the right component is {{\color{\colorMATH}\ensuremath{{\begingroup\renewcommand\colorMATH{\colorMATHB}\renewcommand\colorSYNTAX{\colorSYNTAXB}{{\color{\colorMATH}\ensuremath{\sS_{2}}}}\endgroup }+{\begingroup\renewcommand\colorMATH{\colorMATHB}\renewcommand\colorSYNTAX{\colorSYNTAXB}{{\color{\colorMATH}\ensuremath{\sS'_{2}}}}\endgroup }}}} (for some {{\color{\colorMATH}\ensuremath{{\begingroup\renewcommand\colorMATH{\colorMATHB}\renewcommand\colorSYNTAX{\colorSYNTAXB}{{\color{\colorMATH}\ensuremath{\sS_{1}}}}\endgroup }, {\begingroup\renewcommand\colorMATH{\colorMATHB}\renewcommand\colorSYNTAX{\colorSYNTAXB}{{\color{\colorMATH}\ensuremath{\sS'_{1}}}}\endgroup }, {\begingroup\renewcommand\colorMATH{\colorMATHB}\renewcommand\colorSYNTAX{\colorSYNTAXB}{{\color{\colorMATH}\ensuremath{\sS_{2}}}}\endgroup }, {\begingroup\renewcommand\colorMATH{\colorMATHB}\renewcommand\colorSYNTAX{\colorSYNTAXB}{{\color{\colorMATH}\ensuremath{\sS'_{2}}}}\endgroup }}}}), then the latent effects of using the left component is {{\color{\colorMATH}\ensuremath{{\begingroup\renewcommand\colorMATH{\colorMATHB}\renewcommand\colorSYNTAX{\colorSYNTAXB}{{\color{\colorMATH}\ensuremath{\sS_{1}}}}\endgroup }}}}, and for the right component is {{\color{\colorMATH}\ensuremath{{\begingroup\renewcommand\colorMATH{\colorMATHB}\renewcommand\colorSYNTAX{\colorSYNTAXB}{{\color{\colorMATH}\ensuremath{\sS_{2}}}}\endgroup }}}}. For additive products, the ambient effect is the maximum between {{\color{\colorMATH}\ensuremath{{\begingroup\renewcommand\colorMATH{\colorMATHB}\renewcommand\colorSYNTAX{\colorSYNTAXB}{{\color{\colorMATH}\ensuremath{\sS'_{1}}}}\endgroup }}}} and {{\color{\colorMATH}\ensuremath{{\begingroup\renewcommand\colorMATH{\colorMATHB}\renewcommand\colorSYNTAX{\colorSYNTAXB}{{\color{\colorMATH}\ensuremath{\sS'_{2}}}}\endgroup }}}}, and for multiplicative products, the sum between {{\color{\colorMATH}\ensuremath{{\begingroup\renewcommand\colorMATH{\colorMATHB}\renewcommand\colorSYNTAX{\colorSYNTAXB}{{\color{\colorMATH}\ensuremath{\sS'_{1}}}}\endgroup }}}} and {{\color{\colorMATH}\ensuremath{{\begingroup\renewcommand\colorMATH{\colorMATHB}\renewcommand\colorSYNTAX{\colorSYNTAXB}{{\color{\colorMATH}\ensuremath{\sS'_{2}}}}\endgroup }}}}. }

   For instance, let us consider examples {{\color{\colorMATH}\ensuremath{e_{1} = \addProduct{2*x+y}{0}}}}, {{\color{\colorMATH}\ensuremath{e_{2} = \addProduct{0}{2*x+y}}}}, and {{\color{\colorMATH}\ensuremath{e_{3} = \addProduct{x}{x+y}}}} from Section~\ref{sec:advantages-of-sensitivity-ts}. \toplas{We present next ``lazy'' type derivations for each of the examples}:
\begingroup\color{\colorMATH}\begin{mathpar} 
   \inferrule*[lab={\textsc{ pair}}
   ]{ \Gamma \hspace*{0.33em} {\begingroup\renewcommand\colorMATH{\colorMATHB}\renewcommand\colorSYNTAX{\colorSYNTAXB}{{\color{\colorMATH}\ensuremath{ \vdash  }}}\endgroup }\hspace*{0.33em} 2*x+y \mathrel{:} {\begingroup\renewcommand\colorMATH{\colorMATHA}\renewcommand\colorSYNTAX{\colorSYNTAXA}{{\color{\colorSYNTAX}\texttt{{\ensuremath{{\mathbb{R}}}}}}}\endgroup } \mathrel{;} 2x+y
   \\ \Gamma \hspace*{0.33em} {\begingroup\renewcommand\colorMATH{\colorMATHB}\renewcommand\colorSYNTAX{\colorSYNTAXB}{{\color{\colorMATH}\ensuremath{ \vdash  }}}\endgroup }\hspace*{0.33em} 0 \mathrel{:} {\begingroup\renewcommand\colorMATH{\colorMATHA}\renewcommand\colorSYNTAX{\colorSYNTAXA}{{\color{\colorSYNTAX}\texttt{{\ensuremath{{\mathbb{R}}}}}}}\endgroup } \mathrel{;} \varnothing 
      }{
      \Gamma \hspace*{0.33em} {\begingroup\renewcommand\colorMATH{\colorMATHB}\renewcommand\colorSYNTAX{\colorSYNTAXB}{{\color{\colorMATH}\ensuremath{ \vdash  }}}\endgroup }\hspace*{0.33em}\addProduct{2*x+y}{0} \mathrel{:} {\begingroup\renewcommand\colorMATH{\colorMATHA}\renewcommand\colorSYNTAX{\colorSYNTAXA}{{\color{\colorSYNTAX}\texttt{{\ensuremath{{\mathbb{R}}}}}}}\endgroup } \mathrel{^{2x+y}\&^{\varnothing }} {\begingroup\renewcommand\colorMATH{\colorMATHA}\renewcommand\colorSYNTAX{\colorSYNTAXA}{{\color{\colorSYNTAX}\texttt{{\ensuremath{{\mathbb{R}}}}}}}\endgroup } \mathrel{;} {\begingroup\renewcommand\colorMATH{\colorMATHB}\renewcommand\colorSYNTAX{\colorSYNTAXB}{{\color{\colorMATH}\ensuremath{\varnothing }}}\endgroup }
   }
  \and \inferrule*[lab={\textsc{ pair}}
   ]{ \Gamma \hspace*{0.33em} {\begingroup\renewcommand\colorMATH{\colorMATHB}\renewcommand\colorSYNTAX{\colorSYNTAXB}{{\color{\colorMATH}\ensuremath{ \vdash  }}}\endgroup }\hspace*{0.33em} 0 \mathrel{:} {\begingroup\renewcommand\colorMATH{\colorMATHA}\renewcommand\colorSYNTAX{\colorSYNTAXA}{{\color{\colorSYNTAX}\texttt{{\ensuremath{{\mathbb{R}}}}}}}\endgroup } \mathrel{;} \varnothing 
   \\ \Gamma \hspace*{0.33em} {\begingroup\renewcommand\colorMATH{\colorMATHB}\renewcommand\colorSYNTAX{\colorSYNTAXB}{{\color{\colorMATH}\ensuremath{ \vdash  }}}\endgroup }\hspace*{0.33em} 2*x+y \mathrel{:} {\begingroup\renewcommand\colorMATH{\colorMATHA}\renewcommand\colorSYNTAX{\colorSYNTAXA}{{\color{\colorSYNTAX}\texttt{{\ensuremath{{\mathbb{R}}}}}}}\endgroup } \mathrel{;} 2x+y
      }{
      \Gamma \hspace*{0.33em} {\begingroup\renewcommand\colorMATH{\colorMATHB}\renewcommand\colorSYNTAX{\colorSYNTAXB}{{\color{\colorMATH}\ensuremath{ \vdash  }}}\endgroup }\hspace*{0.33em}\addProduct{0}{2*x+y} \mathrel{:} {\begingroup\renewcommand\colorMATH{\colorMATHA}\renewcommand\colorSYNTAX{\colorSYNTAXA}{{\color{\colorSYNTAX}\texttt{{\ensuremath{{\mathbb{R}}}}}}}\endgroup } \mathrel{^{\varnothing }\&^{2x+y}} {\begingroup\renewcommand\colorMATH{\colorMATHA}\renewcommand\colorSYNTAX{\colorSYNTAXA}{{\color{\colorSYNTAX}\texttt{{\ensuremath{{\mathbb{R}}}}}}}\endgroup } \mathrel{;} {\begingroup\renewcommand\colorMATH{\colorMATHB}\renewcommand\colorSYNTAX{\colorSYNTAXB}{{\color{\colorMATH}\ensuremath{\varnothing }}}\endgroup }
   }
  \and \inferrule*[lab={\textsc{ pair}}
   ]{ \Gamma \hspace*{0.33em} {\begingroup\renewcommand\colorMATH{\colorMATHB}\renewcommand\colorSYNTAX{\colorSYNTAXB}{{\color{\colorMATH}\ensuremath{ \vdash  }}}\endgroup }\hspace*{0.33em} x \mathrel{:} {\begingroup\renewcommand\colorMATH{\colorMATHA}\renewcommand\colorSYNTAX{\colorSYNTAXA}{{\color{\colorSYNTAX}\texttt{{\ensuremath{{\mathbb{R}}}}}}}\endgroup } \mathrel{;} x
   \\ \Gamma \hspace*{0.33em} {\begingroup\renewcommand\colorMATH{\colorMATHB}\renewcommand\colorSYNTAX{\colorSYNTAXB}{{\color{\colorMATH}\ensuremath{ \vdash  }}}\endgroup }\hspace*{0.33em} x+y \mathrel{:} {\begingroup\renewcommand\colorMATH{\colorMATHA}\renewcommand\colorSYNTAX{\colorSYNTAXA}{{\color{\colorSYNTAX}\texttt{{\ensuremath{{\mathbb{R}}}}}}}\endgroup } \mathrel{;} x+y
      }{
      \Gamma \hspace*{0.33em} {\begingroup\renewcommand\colorMATH{\colorMATHB}\renewcommand\colorSYNTAX{\colorSYNTAXB}{{\color{\colorMATH}\ensuremath{ \vdash  }}}\endgroup }\hspace*{0.33em}\addProduct{x}{2*x+y} \mathrel{:} {\begingroup\renewcommand\colorMATH{\colorMATHA}\renewcommand\colorSYNTAX{\colorSYNTAXA}{{\color{\colorSYNTAX}\texttt{{\ensuremath{{\mathbb{R}}}}}}}\endgroup } \mathrel{^{x}\&^{x+y}} {\begingroup\renewcommand\colorMATH{\colorMATHA}\renewcommand\colorSYNTAX{\colorSYNTAXA}{{\color{\colorSYNTAX}\texttt{{\ensuremath{{\mathbb{R}}}}}}}\endgroup } \mathrel{;} {\begingroup\renewcommand\colorMATH{\colorMATHB}\renewcommand\colorSYNTAX{\colorSYNTAXB}{{\color{\colorMATH}\ensuremath{\varnothing }}}\endgroup }
   }
\end{mathpar}\endgroup
where {{\color{\colorMATH}\ensuremath{\Gamma  = x:{\begingroup\renewcommand\colorMATH{\colorMATHA}\renewcommand\colorSYNTAX{\colorSYNTAXA}{{\color{\colorSYNTAX}\texttt{{\ensuremath{{\mathbb{R}}}}}}}\endgroup }, y:{\begingroup\renewcommand\colorMATH{\colorMATHA}\renewcommand\colorSYNTAX{\colorSYNTAXA}{{\color{\colorSYNTAX}\texttt{{\ensuremath{{\mathbb{R}}}}}}}\endgroup }}}}.

\toplas{
Now, let us consider examples {{\color{\colorMATH}\ensuremath{\addProduct{2x}{x}}}} and {{\color{\colorMATH}\ensuremath{\langle 2x,x\rangle }}}. Here are six possible type derivations of paying eagerly for effects:
\begingroup\color{\colorMATH}\begin{mathpar} 
  \and \inferrule*[lab={\textsc{ pair}}
   ]{ \Gamma \hspace*{0.33em} {\begingroup\renewcommand\colorMATH{\colorMATHB}\renewcommand\colorSYNTAX{\colorSYNTAXB}{{\color{\colorMATH}\ensuremath{ \vdash  }}}\endgroup }\hspace*{0.33em} 2x \mathrel{:} {\begingroup\renewcommand\colorMATH{\colorMATHA}\renewcommand\colorSYNTAX{\colorSYNTAXA}{{\color{\colorSYNTAX}\texttt{{\ensuremath{{\mathbb{R}}}}}}}\endgroup } \mathrel{;} \varnothing  + 2x
   \\ \Gamma \hspace*{0.33em} {\begingroup\renewcommand\colorMATH{\colorMATHB}\renewcommand\colorSYNTAX{\colorSYNTAXB}{{\color{\colorMATH}\ensuremath{ \vdash  }}}\endgroup }\hspace*{0.33em} x \mathrel{:} {\begingroup\renewcommand\colorMATH{\colorMATHA}\renewcommand\colorSYNTAX{\colorSYNTAXA}{{\color{\colorSYNTAX}\texttt{{\ensuremath{{\mathbb{R}}}}}}}\endgroup } \mathrel{;} x + \varnothing 
      }{
      \Gamma \hspace*{0.33em} {\begingroup\renewcommand\colorMATH{\colorMATHB}\renewcommand\colorSYNTAX{\colorSYNTAXB}{{\color{\colorMATH}\ensuremath{ \vdash  }}}\endgroup }\hspace*{0.33em}\addProduct{2x}{x} \mathrel{:} {\begingroup\renewcommand\colorMATH{\colorMATHA}\renewcommand\colorSYNTAX{\colorSYNTAXA}{{\color{\colorSYNTAX}\texttt{{\ensuremath{{\mathbb{R}}}}}}}\endgroup } \mathrel{^{}\&^{x}} {\begingroup\renewcommand\colorMATH{\colorMATHA}\renewcommand\colorSYNTAX{\colorSYNTAXA}{{\color{\colorSYNTAX}\texttt{{\ensuremath{{\mathbb{R}}}}}}}\endgroup } \mathrel{;} 2x
   }
  \and \inferrule*[lab={\textsc{ tup}}
   ]{ \Gamma \hspace*{0.33em} {\begingroup\renewcommand\colorMATH{\colorMATHB}\renewcommand\colorSYNTAX{\colorSYNTAXB}{{\color{\colorMATH}\ensuremath{ \vdash  }}}\endgroup }\hspace*{0.33em} 2x \mathrel{:} {\begingroup\renewcommand\colorMATH{\colorMATHA}\renewcommand\colorSYNTAX{\colorSYNTAXA}{{\color{\colorSYNTAX}\texttt{{\ensuremath{{\mathbb{R}}}}}}}\endgroup } \mathrel{;} \varnothing  + 2x
   \\ \Gamma \hspace*{0.33em} {\begingroup\renewcommand\colorMATH{\colorMATHB}\renewcommand\colorSYNTAX{\colorSYNTAXB}{{\color{\colorMATH}\ensuremath{ \vdash  }}}\endgroup }\hspace*{0.33em} x \mathrel{:} {\begingroup\renewcommand\colorMATH{\colorMATHA}\renewcommand\colorSYNTAX{\colorSYNTAXA}{{\color{\colorSYNTAX}\texttt{{\ensuremath{{\mathbb{R}}}}}}}\endgroup } \mathrel{;} x + \varnothing 
      }{
      \Gamma \hspace*{0.33em} {\begingroup\renewcommand\colorMATH{\colorMATHB}\renewcommand\colorSYNTAX{\colorSYNTAXB}{{\color{\colorMATH}\ensuremath{ \vdash  }}}\endgroup }\hspace*{0.33em}\langle 2x,x\rangle  \mathrel{:} {\begingroup\renewcommand\colorMATH{\colorMATHA}\renewcommand\colorSYNTAX{\colorSYNTAXA}{{\color{\colorSYNTAX}\texttt{{\ensuremath{{\mathbb{R}}}}}}}\endgroup } \mathrel{^{}\otimes ^{x}} {\begingroup\renewcommand\colorMATH{\colorMATHA}\renewcommand\colorSYNTAX{\colorSYNTAXA}{{\color{\colorSYNTAX}\texttt{{\ensuremath{{\mathbb{R}}}}}}}\endgroup } \mathrel{;} 2x
   }
   \and \inferrule*[lab={\textsc{ pair}}
   ]{ \Gamma \hspace*{0.33em} {\begingroup\renewcommand\colorMATH{\colorMATHB}\renewcommand\colorSYNTAX{\colorSYNTAXB}{{\color{\colorMATH}\ensuremath{ \vdash  }}}\endgroup }\hspace*{0.33em} 2x \mathrel{:} {\begingroup\renewcommand\colorMATH{\colorMATHA}\renewcommand\colorSYNTAX{\colorSYNTAXA}{{\color{\colorSYNTAX}\texttt{{\ensuremath{{\mathbb{R}}}}}}}\endgroup } \mathrel{;} x + x
   \\ \Gamma \hspace*{0.33em} {\begingroup\renewcommand\colorMATH{\colorMATHB}\renewcommand\colorSYNTAX{\colorSYNTAXB}{{\color{\colorMATH}\ensuremath{ \vdash  }}}\endgroup }\hspace*{0.33em} x \mathrel{:} {\begingroup\renewcommand\colorMATH{\colorMATHA}\renewcommand\colorSYNTAX{\colorSYNTAXA}{{\color{\colorSYNTAX}\texttt{{\ensuremath{{\mathbb{R}}}}}}}\endgroup } \mathrel{;} x + \varnothing 
      }{
      \Gamma \hspace*{0.33em} {\begingroup\renewcommand\colorMATH{\colorMATHB}\renewcommand\colorSYNTAX{\colorSYNTAXB}{{\color{\colorMATH}\ensuremath{ \vdash  }}}\endgroup }\hspace*{0.33em}\addProduct{2x}{x} \mathrel{:} {\begingroup\renewcommand\colorMATH{\colorMATHA}\renewcommand\colorSYNTAX{\colorSYNTAXA}{{\color{\colorSYNTAX}\texttt{{\ensuremath{{\mathbb{R}}}}}}}\endgroup } \mathrel{^{x}\&^{x}} {\begingroup\renewcommand\colorMATH{\colorMATHA}\renewcommand\colorSYNTAX{\colorSYNTAXA}{{\color{\colorSYNTAX}\texttt{{\ensuremath{{\mathbb{R}}}}}}}\endgroup } \mathrel{;} x
   } 
   \and \inferrule*[lab={\textsc{ tup}}
   ]{ \Gamma \hspace*{0.33em} {\begingroup\renewcommand\colorMATH{\colorMATHB}\renewcommand\colorSYNTAX{\colorSYNTAXB}{{\color{\colorMATH}\ensuremath{ \vdash  }}}\endgroup }\hspace*{0.33em} 2x \mathrel{:} {\begingroup\renewcommand\colorMATH{\colorMATHA}\renewcommand\colorSYNTAX{\colorSYNTAXA}{{\color{\colorSYNTAX}\texttt{{\ensuremath{{\mathbb{R}}}}}}}\endgroup } \mathrel{;} x + x
   \\ \Gamma \hspace*{0.33em} {\begingroup\renewcommand\colorMATH{\colorMATHB}\renewcommand\colorSYNTAX{\colorSYNTAXB}{{\color{\colorMATH}\ensuremath{ \vdash  }}}\endgroup }\hspace*{0.33em} x \mathrel{:} {\begingroup\renewcommand\colorMATH{\colorMATHA}\renewcommand\colorSYNTAX{\colorSYNTAXA}{{\color{\colorSYNTAX}\texttt{{\ensuremath{{\mathbb{R}}}}}}}\endgroup } \mathrel{;} x + \varnothing 
      }{
      \Gamma \hspace*{0.33em} {\begingroup\renewcommand\colorMATH{\colorMATHB}\renewcommand\colorSYNTAX{\colorSYNTAXB}{{\color{\colorMATH}\ensuremath{ \vdash  }}}\endgroup }\hspace*{0.33em}\langle 2x,x\rangle  \mathrel{:} {\begingroup\renewcommand\colorMATH{\colorMATHA}\renewcommand\colorSYNTAX{\colorSYNTAXA}{{\color{\colorSYNTAX}\texttt{{\ensuremath{{\mathbb{R}}}}}}}\endgroup } \mathrel{^{x}\otimes ^{x}} {\begingroup\renewcommand\colorMATH{\colorMATHA}\renewcommand\colorSYNTAX{\colorSYNTAXA}{{\color{\colorSYNTAX}\texttt{{\ensuremath{{\mathbb{R}}}}}}}\endgroup } \mathrel{;} x
   } 
   \and \inferrule*[lab={\textsc{ pair}}
   ]{ \Gamma \hspace*{0.33em} {\begingroup\renewcommand\colorMATH{\colorMATHB}\renewcommand\colorSYNTAX{\colorSYNTAXB}{{\color{\colorMATH}\ensuremath{ \vdash  }}}\endgroup }\hspace*{0.33em} 2x \mathrel{:} {\begingroup\renewcommand\colorMATH{\colorMATHA}\renewcommand\colorSYNTAX{\colorSYNTAXA}{{\color{\colorSYNTAX}\texttt{{\ensuremath{{\mathbb{R}}}}}}}\endgroup } \mathrel{;} \varnothing  + 2x
   \\ \Gamma \hspace*{0.33em} {\begingroup\renewcommand\colorMATH{\colorMATHB}\renewcommand\colorSYNTAX{\colorSYNTAXB}{{\color{\colorMATH}\ensuremath{ \vdash  }}}\endgroup }\hspace*{0.33em} x \mathrel{:} {\begingroup\renewcommand\colorMATH{\colorMATHA}\renewcommand\colorSYNTAX{\colorSYNTAXA}{{\color{\colorSYNTAX}\texttt{{\ensuremath{{\mathbb{R}}}}}}}\endgroup } \mathrel{;} \varnothing  + x
      }{
      \Gamma \hspace*{0.33em} {\begingroup\renewcommand\colorMATH{\colorMATHB}\renewcommand\colorSYNTAX{\colorSYNTAXB}{{\color{\colorMATH}\ensuremath{ \vdash  }}}\endgroup }\hspace*{0.33em}\addProduct{2x}{x} \mathrel{:} {\begingroup\renewcommand\colorMATH{\colorMATHA}\renewcommand\colorSYNTAX{\colorSYNTAXA}{{\color{\colorSYNTAX}\texttt{{\ensuremath{{\mathbb{R}}}}}}}\endgroup } \mathrel{^{}\&^{}} {\begingroup\renewcommand\colorMATH{\colorMATHA}\renewcommand\colorSYNTAX{\colorSYNTAXA}{{\color{\colorSYNTAX}\texttt{{\ensuremath{{\mathbb{R}}}}}}}\endgroup } \mathrel{;} 2x
   }
   \and \inferrule*[lab={\textsc{ tup}}
   ]{ \Gamma \hspace*{0.33em} {\begingroup\renewcommand\colorMATH{\colorMATHB}\renewcommand\colorSYNTAX{\colorSYNTAXB}{{\color{\colorMATH}\ensuremath{ \vdash  }}}\endgroup }\hspace*{0.33em} 2x \mathrel{:} {\begingroup\renewcommand\colorMATH{\colorMATHA}\renewcommand\colorSYNTAX{\colorSYNTAXA}{{\color{\colorSYNTAX}\texttt{{\ensuremath{{\mathbb{R}}}}}}}\endgroup } \mathrel{;} \varnothing  + 2x
   \\ \Gamma \hspace*{0.33em} {\begingroup\renewcommand\colorMATH{\colorMATHB}\renewcommand\colorSYNTAX{\colorSYNTAXB}{{\color{\colorMATH}\ensuremath{ \vdash  }}}\endgroup }\hspace*{0.33em} x \mathrel{:} {\begingroup\renewcommand\colorMATH{\colorMATHA}\renewcommand\colorSYNTAX{\colorSYNTAXA}{{\color{\colorSYNTAX}\texttt{{\ensuremath{{\mathbb{R}}}}}}}\endgroup } \mathrel{;} \varnothing  + x
      }{
      \Gamma \hspace*{0.33em} {\begingroup\renewcommand\colorMATH{\colorMATHB}\renewcommand\colorSYNTAX{\colorSYNTAXB}{{\color{\colorMATH}\ensuremath{ \vdash  }}}\endgroup }\hspace*{0.33em}\langle 2x,x\rangle  \mathrel{:} {\begingroup\renewcommand\colorMATH{\colorMATHA}\renewcommand\colorSYNTAX{\colorSYNTAXA}{{\color{\colorSYNTAX}\texttt{{\ensuremath{{\mathbb{R}}}}}}}\endgroup } \mathrel{^{}\otimes ^{}} {\begingroup\renewcommand\colorMATH{\colorMATHA}\renewcommand\colorSYNTAX{\colorSYNTAXA}{{\color{\colorSYNTAX}\texttt{{\ensuremath{{\mathbb{R}}}}}}}\endgroup } \mathrel{;} 3x
   }
\end{mathpar}\endgroup
Note that the difference between the two form of products is only present when effects are paid eagerly for both components.
}
% 2x+y \varnothing 
% \varnothing  2x+y
% x x+y

\item  Rules{\textsc{ proj1}} and{\textsc{ proj2}} type check the deconstruction of an additive
   product. The ambient effect is computed as the cost of reducing the
   product ({{\color{\colorMATH}\ensuremath{{\begingroup\renewcommand\colorMATH{\colorMATHB}\renewcommand\colorSYNTAX{\colorSYNTAXB}{{\color{\colorMATH}\ensuremath{\sS}}}\endgroup }}}}), plus the cost of accessing either the first or the second
   component correspondingly ({{\color{\colorMATH}\ensuremath{{\begingroup\renewcommand\colorMATH{\colorMATHB}\renewcommand\colorSYNTAX{\colorSYNTAXB}{{\color{\colorMATH}\ensuremath{\sS_{1}}}}\endgroup }}}} or {{\color{\colorMATH}\ensuremath{{\begingroup\renewcommand\colorMATH{\colorMATHB}\renewcommand\colorSYNTAX{\colorSYNTAXB}{{\color{\colorMATH}\ensuremath{\sS_{2}}}}\endgroup }}}}). \toplas{This differs from \duet where, and conservatively, the cost of accessing both components are paid when the pair is created. In \ssystem we only paid for the component we are accessing.}

   For instance, let us consider the first projections of last examples:
\begingroup\color{\colorMATH}\begin{mathpar} 
   \inferrule*[lab={\textsc{ proj1}}
   ]{ \Gamma \hspace*{0.33em} {\begingroup\renewcommand\colorMATH{\colorMATHB}\renewcommand\colorSYNTAX{\colorSYNTAXB}{{\color{\colorMATH}\ensuremath{ \vdash  }}}\endgroup }\hspace*{0.33em} \addProduct{2*x+y}{0} \mathrel{:} {\begingroup\renewcommand\colorMATH{\colorMATHA}\renewcommand\colorSYNTAX{\colorSYNTAXA}{{\color{\colorSYNTAX}\texttt{{\ensuremath{{\mathbb{R}}}}}}}\endgroup } \mathrel{^{2x+y}\&^{\varnothing }} {\begingroup\renewcommand\colorMATH{\colorMATHA}\renewcommand\colorSYNTAX{\colorSYNTAXA}{{\color{\colorSYNTAX}\texttt{{\ensuremath{{\mathbb{R}}}}}}}\endgroup } \mathrel{;} \varnothing 
      }{
      \Gamma \hspace*{0.33em} {\begingroup\renewcommand\colorMATH{\colorMATHB}\renewcommand\colorSYNTAX{\colorSYNTAXB}{{\color{\colorMATH}\ensuremath{ \vdash  }}}\endgroup }\hspace*{0.33em} \fst\hspace*{0.33em}\addProduct{2*x+y}{0} \mathrel{:} {\begingroup\renewcommand\colorMATH{\colorMATHA}\renewcommand\colorSYNTAX{\colorSYNTAXA}{{\color{\colorSYNTAX}\texttt{{\ensuremath{{\mathbb{R}}}}}}}\endgroup } \mathrel{;} 2x+y
   }
  \and \inferrule*[lab={\textsc{ proj1}}
   ]{ \Gamma \hspace*{0.33em} {\begingroup\renewcommand\colorMATH{\colorMATHB}\renewcommand\colorSYNTAX{\colorSYNTAXB}{{\color{\colorMATH}\ensuremath{ \vdash  }}}\endgroup }\hspace*{0.33em} \addProduct{0}{2*x+y} \mathrel{:} {\begingroup\renewcommand\colorMATH{\colorMATHA}\renewcommand\colorSYNTAX{\colorSYNTAXA}{{\color{\colorSYNTAX}\texttt{{\ensuremath{{\mathbb{R}}}}}}}\endgroup } \mathrel{^{\varnothing }\&^{2x+y}} {\begingroup\renewcommand\colorMATH{\colorMATHA}\renewcommand\colorSYNTAX{\colorSYNTAXA}{{\color{\colorSYNTAX}\texttt{{\ensuremath{{\mathbb{R}}}}}}}\endgroup } \mathrel{;} \varnothing 
      }{
      \Gamma \hspace*{0.33em} {\begingroup\renewcommand\colorMATH{\colorMATHB}\renewcommand\colorSYNTAX{\colorSYNTAXB}{{\color{\colorMATH}\ensuremath{ \vdash  }}}\endgroup }\hspace*{0.33em} \fst\hspace*{0.33em}\addProduct{0}{2*x+y} \mathrel{:} {\begingroup\renewcommand\colorMATH{\colorMATHA}\renewcommand\colorSYNTAX{\colorSYNTAXA}{{\color{\colorSYNTAX}\texttt{{\ensuremath{{\mathbb{R}}}}}}}\endgroup } \mathrel{;} \varnothing 
   }
  \and \inferrule*[lab={\textsc{ proj1}}
   ]{ \Gamma \hspace*{0.33em} {\begingroup\renewcommand\colorMATH{\colorMATHB}\renewcommand\colorSYNTAX{\colorSYNTAXB}{{\color{\colorMATH}\ensuremath{ \vdash  }}}\endgroup }\hspace*{0.33em} \addProduct{x}{2*x+y} \mathrel{:} {\begingroup\renewcommand\colorMATH{\colorMATHA}\renewcommand\colorSYNTAX{\colorSYNTAXA}{{\color{\colorSYNTAX}\texttt{{\ensuremath{{\mathbb{R}}}}}}}\endgroup } \mathrel{^{x}\&^{x+y}} {\begingroup\renewcommand\colorMATH{\colorMATHA}\renewcommand\colorSYNTAX{\colorSYNTAXA}{{\color{\colorSYNTAX}\texttt{{\ensuremath{{\mathbb{R}}}}}}}\endgroup } \mathrel{;} \varnothing 
      }{
      \Gamma \hspace*{0.33em} {\begingroup\renewcommand\colorMATH{\colorMATHB}\renewcommand\colorSYNTAX{\colorSYNTAXB}{{\color{\colorMATH}\ensuremath{ \vdash  }}}\endgroup }\hspace*{0.33em} \fst\hspace*{0.33em}\addProduct{x}{2*x+y} \mathrel{:} {\begingroup\renewcommand\colorMATH{\colorMATHA}\renewcommand\colorSYNTAX{\colorSYNTAXA}{{\color{\colorSYNTAX}\texttt{{\ensuremath{{\mathbb{R}}}}}}}\endgroup } \mathrel{;} x
   } 
\end{mathpar}\endgroup
Contrary to previous work, the ambient effects of all three projections are different, as they capture precisely the variables accessed on the corresponding component.

\item  Rule{\textsc{ untup}} typechecks the deconstruction of a multiplicative product, and is a little more involved.
\toplass{To compute the ambient effect we start by paying for {{\color{\colorMATH}\ensuremath{{\begingroup\renewcommand\colorMATH{\colorMATHB}\renewcommand\colorSYNTAX{\colorSYNTAXB}{{\color{\colorMATH}\ensuremath{\sS_{2}}}}\endgroup }}}}, the ambient effect of subexpression {{\color{\colorMATH}\ensuremath{{\begingroup\renewcommand\colorMATH{\colorMATHB}\renewcommand\colorSYNTAX{\colorSYNTAXB}{{\color{\colorMATH}\ensuremath{\se_{2}}}}\endgroup }}}}. We also want to pay {{\color{\colorMATH}\ensuremath{{\begingroup\renewcommand\colorMATH{\colorMATHB}\renewcommand\colorSYNTAX{\colorSYNTAXB}{{\color{\colorMATH}\ensuremath{\sS_{1 1}}}}\endgroup }}}} and {{\color{\colorMATH}\ensuremath{{\begingroup\renewcommand\colorMATH{\colorMATHB}\renewcommand\colorSYNTAX{\colorSYNTAXB}{{\color{\colorMATH}\ensuremath{\sS_{1 2}}}}\endgroup }}}}, the cost of accessing the left and the right components respectively, proportionally to the sensitivity of the left and right variables {{\color{\colorMATH}\ensuremath{x_{1}}}} and {{\color{\colorMATH}\ensuremath{x_{2}}}} in {{\color{\colorMATH}\ensuremath{{\begingroup\renewcommand\colorMATH{\colorMATHB}\renewcommand\colorSYNTAX{\colorSYNTAXB}{{\color{\colorMATH}\ensuremath{\se_{2}}}}\endgroup }}}}, i.e. {{\color{\colorMATH}\ensuremath{{\begingroup\renewcommand\colorMATH{\colorMATHB}\renewcommand\colorSYNTAX{\colorSYNTAXB}{{\color{\colorMATH}\ensuremath{\sss_{1}}}}\endgroup }{\begingroup\renewcommand\colorMATH{\colorMATHB}\renewcommand\colorSYNTAX{\colorSYNTAXB}{{\color{\colorMATH}\ensuremath{\sS_{1 1}}}}\endgroup } + {\begingroup\renewcommand\colorMATH{\colorMATHB}\renewcommand\colorSYNTAX{\colorSYNTAXB}{{\color{\colorMATH}\ensuremath{\sss_{2}}}}\endgroup }{\begingroup\renewcommand\colorMATH{\colorMATHB}\renewcommand\colorSYNTAX{\colorSYNTAXB}{{\color{\colorMATH}\ensuremath{\sS_{1 2}}}}\endgroup }}}}. But we also have to pay for {{\color{\colorMATH}\ensuremath{{\begingroup\renewcommand\colorMATH{\colorMATHB}\renewcommand\colorSYNTAX{\colorSYNTAXB}{{\color{\colorMATH}\ensuremath{\sS_{1}}}}\endgroup }}}}, the ambient effect of {{\color{\colorMATH}\ensuremath{{\begingroup\renewcommand\colorMATH{\colorMATHB}\renewcommand\colorSYNTAX{\colorSYNTAXB}{{\color{\colorMATH}\ensuremath{\se_{1}}}}\endgroup }}}}. We could pay {{\color{\colorMATH}\ensuremath{({\begingroup\renewcommand\colorMATH{\colorMATHB}\renewcommand\colorSYNTAX{\colorSYNTAXB}{{\color{\colorMATH}\ensuremath{\sss_{1}}}}\endgroup } + {\begingroup\renewcommand\colorMATH{\colorMATHB}\renewcommand\colorSYNTAX{\colorSYNTAXB}{{\color{\colorMATH}\ensuremath{\sss_{2}}}}\endgroup }){\begingroup\renewcommand\colorMATH{\colorMATHB}\renewcommand\colorSYNTAX{\colorSYNTAXB}{{\color{\colorMATH}\ensuremath{\sS_{1}}}}\endgroup }}}}, but that would be an unnecessary over-approximation. For instance, program {{\color{\colorMATH}\ensuremath{\tlet\hspace*{0.33em}x_{1}, x_{2} = p \hspace*{0.33em} \tin \hspace*{0.33em} x_{1} + x_{2}}}} would pay twice for {{\color{\colorMATH}\ensuremath{p}}} (the ambient effect of {{\color{\colorMATH}\ensuremath{{\begingroup\renewcommand\colorMATH{\colorMATHB}\renewcommand\colorSYNTAX{\colorSYNTAXB}{{\color{\colorMATH}\ensuremath{\se_{1}}}}\endgroup }}}}), even though the whole pair is used only once. Instead we want to pay proportional to the the maximum sensitivity between {{\color{\colorMATH}\ensuremath{x_{1}}}} and {{\color{\colorMATH}\ensuremath{x_{2}}}}, i.e. {{\color{\colorMATH}\ensuremath{({\begingroup\renewcommand\colorMATH{\colorMATHB}\renewcommand\colorSYNTAX{\colorSYNTAXB}{{\color{\colorMATH}\ensuremath{\sss_{1}}}}\endgroup } \sqcup  {\begingroup\renewcommand\colorMATH{\colorMATHB}\renewcommand\colorSYNTAX{\colorSYNTAXB}{{\color{\colorMATH}\ensuremath{\sss_{2}}}}\endgroup }) {\begingroup\renewcommand\colorMATH{\colorMATHB}\renewcommand\colorSYNTAX{\colorSYNTAXB}{{\color{\colorMATH}\ensuremath{\sS_{1}}}}\endgroup }}}}. Finally, the ambient effect of the let expression  is {{\color{\colorMATH}\ensuremath{({\begingroup\renewcommand\colorMATH{\colorMATHB}\renewcommand\colorSYNTAX{\colorSYNTAXB}{{\color{\colorMATH}\ensuremath{\sss_{1}}}}\endgroup } \sqcup  {\begingroup\renewcommand\colorMATH{\colorMATHB}\renewcommand\colorSYNTAX{\colorSYNTAXB}{{\color{\colorMATH}\ensuremath{\sss_{2}}}}\endgroup }) {\begingroup\renewcommand\colorMATH{\colorMATHB}\renewcommand\colorSYNTAX{\colorSYNTAXB}{{\color{\colorMATH}\ensuremath{\sS_{1}}}}\endgroup } + {\begingroup\renewcommand\colorMATH{\colorMATHB}\renewcommand\colorSYNTAX{\colorSYNTAXB}{{\color{\colorMATH}\ensuremath{\sss_{1}}}}\endgroup }{\begingroup\renewcommand\colorMATH{\colorMATHB}\renewcommand\colorSYNTAX{\colorSYNTAXB}{{\color{\colorMATH}\ensuremath{\sS_{1 1}}}}\endgroup } + {\begingroup\renewcommand\colorMATH{\colorMATHB}\renewcommand\colorSYNTAX{\colorSYNTAXB}{{\color{\colorMATH}\ensuremath{\sss_{2}}}}\endgroup }{\begingroup\renewcommand\colorMATH{\colorMATHB}\renewcommand\colorSYNTAX{\colorSYNTAXB}{{\color{\colorMATH}\ensuremath{\sS_{1 2}}}}\endgroup } + {\begingroup\renewcommand\colorMATH{\colorMATHB}\renewcommand\colorSYNTAX{\colorSYNTAXB}{{\color{\colorMATH}\ensuremath{\sS_{2}}}}\endgroup }}}}. 
%Because the resulting type {{\color{\colorMATH}\ensuremath{\tau _{2}}}} may have {{\color{\colorMATH}\ensuremath{x_{1}}}} and {{\color{\colorMATH}\ensuremath{x_{2}}}} free in latent annotations, we substitute them away using the same reasoning to compute the ambient effect. For this reason, we define a substitution function XXX over types and effects, defined as follows: 
}

   \toplass{For instance, let us consider the typing derivation of Example~\ref{ex:scaling}:}
   \begingroup\color{\colorMATH}\begin{gather*} 
     \inferrule*[lab={\textsc{ untup}}
     ]{ \Gamma \hspace*{0.33em} {\begingroup\renewcommand\colorMATH{\colorMATHB}\renewcommand\colorSYNTAX{\colorSYNTAXB}{{\color{\colorMATH}\ensuremath{ \vdash  }}}\endgroup }\hspace*{0.33em} \addProduct{2*x}{y} \mathrel{:} {\begingroup\renewcommand\colorMATH{\colorMATHA}\renewcommand\colorSYNTAX{\colorSYNTAXA}{{\color{\colorSYNTAX}\texttt{{\ensuremath{{\mathbb{R}}}}}}}\endgroup } \mathrel{^{2x}\otimes ^{y}} {\begingroup\renewcommand\colorMATH{\colorMATHA}\renewcommand\colorSYNTAX{\colorSYNTAXA}{{\color{\colorSYNTAX}\texttt{{\ensuremath{{\mathbb{R}}}}}}}\endgroup } \mathrel{;} \varnothing 
     \\ \Gamma ,x_{1}\mathrel{:}{\begingroup\renewcommand\colorMATH{\colorMATHA}\renewcommand\colorSYNTAX{\colorSYNTAXA}{{\color{\colorSYNTAX}\texttt{{\ensuremath{{\mathbb{R}}}}}}}\endgroup },x_{2}\mathrel{:}{\begingroup\renewcommand\colorMATH{\colorMATHA}\renewcommand\colorSYNTAX{\colorSYNTAXA}{{\color{\colorSYNTAX}\texttt{{\ensuremath{{\mathbb{R}}}}}}}\endgroup }\hspace*{0.33em} {\begingroup\renewcommand\colorMATH{\colorMATHB}\renewcommand\colorSYNTAX{\colorSYNTAXB}{{\color{\colorMATH}\ensuremath{ \vdash  }}}\endgroup } \hspace*{0.33em} x_{1}+2*x_{2} \mathrel{:} {\begingroup\renewcommand\colorMATH{\colorMATHA}\renewcommand\colorSYNTAX{\colorSYNTAXA}{{\color{\colorSYNTAX}\texttt{{\ensuremath{{\mathbb{R}}}}}}}\endgroup } \mathrel{;} x_{1}+2x_{2}
        }{
        \Gamma \hspace*{0.33em} {\begingroup\renewcommand\colorMATH{\colorMATHB}\renewcommand\colorSYNTAX{\colorSYNTAXB}{{\color{\colorMATH}\ensuremath{ \vdash  }}}\endgroup }\hspace*{0.33em}\tlet\hspace*{0.33em}x_{1},x_{2}=\addProduct{2*x}{y}\hspace*{0.33em}\tin\hspace*{0.33em}x_{1}+2*x_{2} \mathrel{:} {\begingroup\renewcommand\colorMATH{\colorMATHA}\renewcommand\colorSYNTAX{\colorSYNTAXA}{{\color{\colorSYNTAX}\texttt{{\ensuremath{{\mathbb{R}}}}}}}\endgroup } \mathrel{;} 1(2x)+ 2(y)
     }
   \end{gather*}\endgroup
   where {{\color{\colorMATH}\ensuremath{\Gamma  = x:{\begingroup\renewcommand\colorMATH{\colorMATHA}\renewcommand\colorSYNTAX{\colorSYNTAXA}{{\color{\colorSYNTAX}\texttt{{\ensuremath{{\mathbb{R}}}}}}}\endgroup }, y:{\begingroup\renewcommand\colorMATH{\colorMATHA}\renewcommand\colorSYNTAX{\colorSYNTAXA}{{\color{\colorSYNTAX}\texttt{{\ensuremath{{\mathbb{R}}}}}}}\endgroup }}}}. The resulting ambient effect is {{\color{\colorMATH}\ensuremath{1(2x)+ 2(y) = 2x + 2y}}}.

   \toplass{
   Now consider {{\color{\colorMATH}\ensuremath{\Gamma  = p \mathrel{:} {\begingroup\renewcommand\colorMATH{\colorMATHA}\renewcommand\colorSYNTAX{\colorSYNTAXA}{{\color{\colorSYNTAX}\texttt{{\ensuremath{{\mathbb{R}}}}}}}\endgroup } \mathrel{^{\varnothing }\otimes ^{\varnothing }} {\begingroup\renewcommand\colorMATH{\colorMATHA}\renewcommand\colorSYNTAX{\colorSYNTAXA}{{\color{\colorSYNTAX}\texttt{{\ensuremath{{\mathbb{R}}}}}}}\endgroup }}}}, and the following typing derivation:
   \begingroup\color{\colorMATH}\begin{gather*} 
     \inferrule*[lab={\textsc{ untup}}
     ]{ 
        \inferrule*[lab=
        ]{ \Gamma \hspace*{0.33em} {\begingroup\renewcommand\colorMATH{\colorMATHB}\renewcommand\colorSYNTAX{\colorSYNTAXB}{{\color{\colorMATH}\ensuremath{ \vdash  }}}\endgroup }\hspace*{0.33em} p \mathrel{:} {\begingroup\renewcommand\colorMATH{\colorMATHA}\renewcommand\colorSYNTAX{\colorSYNTAXA}{{\color{\colorSYNTAX}\texttt{{\ensuremath{{\mathbb{R}}}}}}}\endgroup } \mathrel{^{\varnothing }\otimes ^{\varnothing }} {\begingroup\renewcommand\colorMATH{\colorMATHA}\renewcommand\colorSYNTAX{\colorSYNTAXA}{{\color{\colorSYNTAX}\texttt{{\ensuremath{{\mathbb{R}}}}}}}\endgroup } \mathrel{;} p
        \\ \Gamma ,x_{1}\mathrel{:}{\begingroup\renewcommand\colorMATH{\colorMATHA}\renewcommand\colorSYNTAX{\colorSYNTAXA}{{\color{\colorSYNTAX}\texttt{{\ensuremath{{\mathbb{R}}}}}}}\endgroup },x_{2}\mathrel{:}{\begingroup\renewcommand\colorMATH{\colorMATHA}\renewcommand\colorSYNTAX{\colorSYNTAXA}{{\color{\colorSYNTAX}\texttt{{\ensuremath{{\mathbb{R}}}}}}}\endgroup }\hspace*{0.33em} {\begingroup\renewcommand\colorMATH{\colorMATHB}\renewcommand\colorSYNTAX{\colorSYNTAXB}{{\color{\colorMATH}\ensuremath{ \vdash  }}}\endgroup } \hspace*{0.33em} \langle x_{1},x_{2}\rangle  \mathrel{:} {\begingroup\renewcommand\colorMATH{\colorMATHA}\renewcommand\colorSYNTAX{\colorSYNTAXA}{{\color{\colorSYNTAX}\texttt{{\ensuremath{{\mathbb{R}}}}}}}\endgroup } \mathrel{^{{\begingroup\renewcommand\colorMATH{\colorMATHB}\renewcommand\colorSYNTAX{\colorSYNTAXB}{{\color{\colorMATH}\ensuremath{\sS_{1}}}}\endgroup }}\otimes ^{{\begingroup\renewcommand\colorMATH{\colorMATHB}\renewcommand\colorSYNTAX{\colorSYNTAXB}{{\color{\colorMATH}\ensuremath{\sS_{2}}}}\endgroup }}} {\begingroup\renewcommand\colorMATH{\colorMATHA}\renewcommand\colorSYNTAX{\colorSYNTAXA}{{\color{\colorSYNTAX}\texttt{{\ensuremath{{\mathbb{R}}}}}}}\endgroup } \mathrel{;} {\begingroup\renewcommand\colorMATH{\colorMATHB}\renewcommand\colorSYNTAX{\colorSYNTAXB}{{\color{\colorMATH}\ensuremath{\sS_{3}}}}\endgroup }
           }{
           \Gamma \hspace*{0.33em} {\begingroup\renewcommand\colorMATH{\colorMATHB}\renewcommand\colorSYNTAX{\colorSYNTAXB}{{\color{\colorMATH}\ensuremath{ \vdash  }}}\endgroup }\hspace*{0.33em}\tlet\hspace*{0.33em}x_{1},x_{2}=p\hspace*{0.33em}\tin\hspace*{0.33em}\langle x_{1},x_{2}\rangle  \mathrel{:} {\begingroup\renewcommand\colorMATH{\colorMATHA}\renewcommand\colorSYNTAX{\colorSYNTAXA}{{\color{\colorSYNTAX}\texttt{{\ensuremath{{\mathbb{R}}}}}}}\endgroup } \mathrel{^{{\begingroup\renewcommand\colorMATH{\colorMATHB}\renewcommand\colorSYNTAX{\colorSYNTAXB}{{\color{\colorMATH}\ensuremath{\sS'_{1}}}}\endgroup }}\otimes ^{{\begingroup\renewcommand\colorMATH{\colorMATHB}\renewcommand\colorSYNTAX{\colorSYNTAXB}{{\color{\colorMATH}\ensuremath{\sS'_{2}}}}\endgroup }}} {\begingroup\renewcommand\colorMATH{\colorMATHA}\renewcommand\colorSYNTAX{\colorSYNTAXA}{{\color{\colorSYNTAX}\texttt{{\ensuremath{{\mathbb{R}}}}}}}\endgroup } \mathrel{;} {\begingroup\renewcommand\colorMATH{\colorMATHB}\renewcommand\colorSYNTAX{\colorSYNTAXB}{{\color{\colorMATH}\ensuremath{\sS'_{3}}}}\endgroup }
        }
     \\ \Gamma ,y_{1}\mathrel{:}{\begingroup\renewcommand\colorMATH{\colorMATHA}\renewcommand\colorSYNTAX{\colorSYNTAXA}{{\color{\colorSYNTAX}\texttt{{\ensuremath{{\mathbb{R}}}}}}}\endgroup },y_{2}\mathrel{:}{\begingroup\renewcommand\colorMATH{\colorMATHA}\renewcommand\colorSYNTAX{\colorSYNTAXA}{{\color{\colorSYNTAX}\texttt{{\ensuremath{{\mathbb{R}}}}}}}\endgroup }\hspace*{0.33em} {\begingroup\renewcommand\colorMATH{\colorMATHB}\renewcommand\colorSYNTAX{\colorSYNTAXB}{{\color{\colorMATH}\ensuremath{ \vdash  }}}\endgroup } \hspace*{0.33em} y_{1}+y_{2} \mathrel{:} {\begingroup\renewcommand\colorMATH{\colorMATHA}\renewcommand\colorSYNTAX{\colorSYNTAXA}{{\color{\colorSYNTAX}\texttt{{\ensuremath{{\mathbb{R}}}}}}}\endgroup } \mathrel{;} y_{1}+y_{2}
        }{
        \Gamma \hspace*{0.33em} {\begingroup\renewcommand\colorMATH{\colorMATHB}\renewcommand\colorSYNTAX{\colorSYNTAXB}{{\color{\colorMATH}\ensuremath{ \vdash  }}}\endgroup }\hspace*{0.33em}\tlet\hspace*{0.33em} y_{1},y_{2} = (\tlet\hspace*{0.33em}x_{1},x_{2}=p\hspace*{0.33em}\tin\hspace*{0.33em}\langle x_{1},x_{2}\rangle ) \hspace*{0.33em}  \tin \hspace*{0.33em} y_{1}+y_{2} \mathrel{:} {\begingroup\renewcommand\colorMATH{\colorMATHA}\renewcommand\colorSYNTAX{\colorSYNTAXA}{{\color{\colorSYNTAX}\texttt{{\ensuremath{{\mathbb{R}}}}}}}\endgroup } \mathrel{;} {\begingroup\renewcommand\colorMATH{\colorMATHB}\renewcommand\colorSYNTAX{\colorSYNTAXB}{{\color{\colorMATH}\ensuremath{\sS'_{3}}}}\endgroup } + {\begingroup\renewcommand\colorMATH{\colorMATHB}\renewcommand\colorSYNTAX{\colorSYNTAXB}{{\color{\colorMATH}\ensuremath{\sS'_{1}}}}\endgroup } + {\begingroup\renewcommand\colorMATH{\colorMATHB}\renewcommand\colorSYNTAX{\colorSYNTAXB}{{\color{\colorMATH}\ensuremath{\sS'_{2}}}}\endgroup }
     }
   \end{gather*}\endgroup
   We can typecheck subexpression {{\color{\colorMATH}\ensuremath{\tlet\hspace*{0.33em}x_{1},x_{2}=p\hspace*{0.33em}\tin\hspace*{0.33em}(x_{1},x_{2})}}} in different ways. If we do not prepay effects, then {{\color{\colorMATH}\ensuremath{{\begingroup\renewcommand\colorMATH{\colorMATHB}\renewcommand\colorSYNTAX{\colorSYNTAXB}{{\color{\colorMATH}\ensuremath{\sS_{1}}}}\endgroup } = x_{1}, {\begingroup\renewcommand\colorMATH{\colorMATHB}\renewcommand\colorSYNTAX{\colorSYNTAXB}{{\color{\colorMATH}\ensuremath{\sS_{2}}}}\endgroup } = x_{2}}}}, and {{\color{\colorMATH}\ensuremath{{\begingroup\renewcommand\colorMATH{\colorMATHB}\renewcommand\colorSYNTAX{\colorSYNTAXB}{{\color{\colorMATH}\ensuremath{\sS_{3}}}}\endgroup } = \varnothing }}}. Thus {{\color{\colorMATH}\ensuremath{{\begingroup\renewcommand\colorMATH{\colorMATHB}\renewcommand\colorSYNTAX{\colorSYNTAXB}{{\color{\colorMATH}\ensuremath{\sS'_{1}}}}\endgroup } = {\begingroup\renewcommand\colorMATH{\colorMATHB}\renewcommand\colorSYNTAX{\colorSYNTAXB}{{\color{\colorMATH}\ensuremath{\sS'_{2}}}}\endgroup } = p}}}, and {{\color{\colorMATH}\ensuremath{{\begingroup\renewcommand\colorMATH{\colorMATHB}\renewcommand\colorSYNTAX{\colorSYNTAXB}{{\color{\colorMATH}\ensuremath{\sS'_{3}}}}\endgroup } = \varnothing }}}. Finally the ambient effect of the program is {{\color{\colorMATH}\ensuremath{2p}}}.

   If we prepay the accesses of {{\color{\colorMATH}\ensuremath{x_{1}}}} and {{\color{\colorMATH}\ensuremath{x_{2}}}}, then 
   {{\color{\colorMATH}\ensuremath{{\begingroup\renewcommand\colorMATH{\colorMATHB}\renewcommand\colorSYNTAX{\colorSYNTAXB}{{\color{\colorMATH}\ensuremath{\sS_{1}}}}\endgroup } = \varnothing , {\begingroup\renewcommand\colorMATH{\colorMATHB}\renewcommand\colorSYNTAX{\colorSYNTAXB}{{\color{\colorMATH}\ensuremath{\sS_{2}}}}\endgroup } = \varnothing }}}, and {{\color{\colorMATH}\ensuremath{{\begingroup\renewcommand\colorMATH{\colorMATHB}\renewcommand\colorSYNTAX{\colorSYNTAXB}{{\color{\colorMATH}\ensuremath{\sS_{2}}}}\endgroup } = x_{1} + x_{2}}}}. Thus {{\color{\colorMATH}\ensuremath{{\begingroup\renewcommand\colorMATH{\colorMATHB}\renewcommand\colorSYNTAX{\colorSYNTAXB}{{\color{\colorMATH}\ensuremath{\sS'_{1}}}}\endgroup } = {\begingroup\renewcommand\colorMATH{\colorMATHB}\renewcommand\colorSYNTAX{\colorSYNTAXB}{{\color{\colorMATH}\ensuremath{\sS'_{2}}}}\endgroup } = \varnothing }}}, and {{\color{\colorMATH}\ensuremath{{\begingroup\renewcommand\colorMATH{\colorMATHB}\renewcommand\colorSYNTAX{\colorSYNTAXB}{{\color{\colorMATH}\ensuremath{\sS'_{3}}}}\endgroup } = (1 \sqcup  1)p = p}}}. Finally the ambient effect of the program is {{\color{\colorMATH}\ensuremath{p}}}.
    
}

\item  Rules for booleans, conditionals and let expressions are derived rules from sums, case, and application rules respectively, and can be found in Figure~\ref{fig:sensitivity-simple-derived-rules}.
\begin{figure}[t]
   \begin{small}
   \input{sensitivity-simple-derived-rules}
   \end{small}
   \caption{$\ssystem$: Derived type rules}
   \label{fig:sensitivity-simple-derived-rules}
 \end{figure}

\item  Finally, Rule{\textsc{ ascr}} is the only rule that supports the use of subtyping, \toplas{and takes the role of \emph{checking} whether the subexpression is subtype of a given type.} 
\end{itemize}

\begin{figure}[t]
   \begin{small}
   \begin{framed}
 \begingroup\color{\colorMATH}\begin{gather*}
\begin{tabularx}{\linewidth}{>{\centering\arraybackslash\(}X<{\)}}\hfill\hspace{0pt}\begingroup\color{\colorTEXT}\boxed{\begingroup\color{\colorMATH} {\begingroup\renewcommand\colorMATH{\colorMATHB}\renewcommand\colorSYNTAX{\colorSYNTAXB}{{\color{\colorMATH}\ensuremath{\sS}}}\endgroup } <: {\begingroup\renewcommand\colorMATH{\colorMATHB}\renewcommand\colorSYNTAX{\colorSYNTAXB}{{\color{\colorMATH}\ensuremath{\sS}}}\endgroup } \endgroup}\endgroup
	\cr \hfill\hspace{0pt}{\begingroup\renewcommand\colorMATH{\colorMATHB}\renewcommand\colorSYNTAX{\colorSYNTAXB}{{\color{\colorMATH}\ensuremath{\sS}}}\endgroup } <: {\begingroup\renewcommand\colorMATH{\colorMATHB}\renewcommand\colorSYNTAX{\colorSYNTAXB}{{\color{\colorMATH}\ensuremath{\sS^{\prime}}}}\endgroup } \overset \vartriangle \iff     \forall x.\hspace*{0.33em} {\begingroup\renewcommand\colorMATH{\colorMATHB}\renewcommand\colorSYNTAX{\colorSYNTAXB}{{\color{\colorMATH}\ensuremath{\sS}}}\endgroup }(x) \leq  {\begingroup\renewcommand\colorMATH{\colorMATHB}\renewcommand\colorSYNTAX{\colorSYNTAXB}{{\color{\colorMATH}\ensuremath{\sS^{\prime}}}}\endgroup }(x)
	  \hfill\hspace{0pt}
	\cr \hfill\hspace{0pt}\begingroup\color{\colorTEXT}\boxed{\begingroup\color{\colorMATH} \tau  <: \tau  \endgroup}\endgroup
	\cr
\end{tabularx}
	\end{gather*}\endgroup
\end{framed}
   \end{small}
   \caption{$\ssystem$: Subtyping}
   \label{fig:sensitivity-simple-subtyping}
 \end{figure}
Subtyping for
types and sensitivity environments is presented in
Figure~\ref{fig:sensitivity-simple-subtyping}, and is mostly standard. 
We only allow subtyping for the sensitivity
parts of types. A sensitivity environment is subtype of another if their
sensitivities are less than or equal than the other for each variable.
For instance,
{{\color{\colorMATH}\ensuremath{(x:{\begingroup\renewcommand\colorMATH{\colorMATHA}\renewcommand\colorSYNTAX{\colorSYNTAXA}{{\color{\colorSYNTAX}\texttt{{\ensuremath{{\mathbb{R}}}}}}}\endgroup }) \xrightarrowS {x+y} {\begingroup\renewcommand\colorMATH{\colorMATHA}\renewcommand\colorSYNTAX{\colorSYNTAXA}{{\color{\colorSYNTAX}\texttt{{\ensuremath{{\mathbb{R}}}}}}}\endgroup } <: (x:{\begingroup\renewcommand\colorMATH{\colorMATHA}\renewcommand\colorSYNTAX{\colorSYNTAXA}{{\color{\colorSYNTAX}\texttt{{\ensuremath{{\mathbb{R}}}}}}}\endgroup }) \xrightarrowS {x+2y+3z} {\begingroup\renewcommand\colorMATH{\colorMATHA}\renewcommand\colorSYNTAX{\colorSYNTAXA}{{\color{\colorSYNTAX}\texttt{{\ensuremath{{\mathbb{R}}}}}}}\endgroup }}}} because {{\color{\colorMATH}\ensuremath{x+y <: x+2y+3z}}} ({{\color{\colorMATH}\ensuremath{x \leq  x}}}, {{\color{\colorMATH}\ensuremath{y \leq  2y}}}, and {{\color{\colorMATH}\ensuremath{0z \leq  3z}}}).

\subsection{Type Safety}
\label{sec:sensitivity-simple-type-safety}

Type safety is established relative to the runtime semantics of $\ssystem$. We adopt a big-step semantics
with explicit substitutions. Concretely, we use {{\color{\colorMATH}\ensuremath{\gamma  \vdash  {\begingroup\renewcommand\colorMATH{\colorMATHB}\renewcommand\colorSYNTAX{\colorSYNTAXB}{{\color{\colorMATH}\ensuremath{\se}}}\endgroup } \Downarrow  {\begingroup\renewcommand\colorMATH{\colorMATHB}\renewcommand\colorSYNTAX{\colorSYNTAXB}{{\color{\colorMATH}\ensuremath{\sv}}}\endgroup }}}} to represent that \emph{configuration} {{\color{\colorMATH}\ensuremath{\gamma  \vdash  {\begingroup\renewcommand\colorMATH{\colorMATHB}\renewcommand\colorSYNTAX{\colorSYNTAXB}{{\color{\colorMATH}\ensuremath{\se}}}\endgroup }}}} ---formed by expression {{\color{\colorMATH}\ensuremath{{\begingroup\renewcommand\colorMATH{\colorMATHB}\renewcommand\colorSYNTAX{\colorSYNTAXB}{{\color{\colorMATH}\ensuremath{\se}}}\endgroup }}}} and value environment
{{\color{\colorMATH}\ensuremath{\gamma }}} mapping variables to values--- reduces to value {{\color{\colorMATH}\ensuremath{{\begingroup\renewcommand\colorMATH{\colorMATHB}\renewcommand\colorSYNTAX{\colorSYNTAXB}{{\color{\colorMATH}\ensuremath{\sv}}}\endgroup }}}} after some number of steps. Reducion rules are rather standard and can be found in Appendix~\ref{asec:dynamic-semantics}, Figure~\ref{afig:sensitivity-dynamic-semantics}.

To establish $\ssystem$ type safety, we employ simple unary logical relations, called the \emph{type safety logical relations}, that characterize well-typed, non-stuck execution. 
\toplasss{This relation is well defined because it is defined by induction over the structure of types}.
The type safety result itself is derived as a corollary of the fundamental property of the type safety logical relations.

\begin{figure}[t]
\begin{small}
\begin{framed}
\begingroup\color{\colorMATH}\begin{mathpar}
   \inferrule*[lab={\textsc{ }}
   ]{ \varnothing  \vdash  {\begingroup\renewcommand\colorMATH{\colorMATHB}\renewcommand\colorSYNTAX{\colorSYNTAXB}{{\color{\colorMATH}\ensuremath{\sv}}}\endgroup } : \tau '; \varnothing 
   \\  \tau ' <: \tau 
      }{
      {\begingroup\renewcommand\colorMATH{\colorMATHB}\renewcommand\colorSYNTAX{\colorSYNTAXB}{{\color{\colorMATH}\ensuremath{\sv}}}\endgroup } \in  Atom\llbracket \tau \rrbracket 
   }
\and \inferrule*[lab={\textsc{ }}
   ]{ {\begingroup\renewcommand\colorMATH{\colorMATHB}\renewcommand\colorSYNTAX{\colorSYNTAXB}{{\color{\colorMATH}\ensuremath{r}}}\endgroup } \in  Atom\llbracket {\begingroup\renewcommand\colorMATH{\colorMATHA}\renewcommand\colorSYNTAX{\colorSYNTAXA}{{\color{\colorSYNTAX}\texttt{{\ensuremath{{\mathbb{R}}}}}}}\endgroup }\rrbracket 
      }{
      {\begingroup\renewcommand\colorMATH{\colorMATHB}\renewcommand\colorSYNTAX{\colorSYNTAXB}{{\color{\colorMATH}\ensuremath{r}}}\endgroup } \in  {\mathcal{V}}\llbracket {\begingroup\renewcommand\colorMATH{\colorMATHA}\renewcommand\colorSYNTAX{\colorSYNTAXA}{{\color{\colorSYNTAX}\texttt{{\ensuremath{{\mathbb{R}}}}}}}\endgroup }\rrbracket 
   } 
\and \inferrule*[lab={\textsc{ }}
   ]{ \ttt \in  Atom\llbracket {{\color{\colorSYNTAX}\texttt{unit}}}\rrbracket 
      }{
      \ttt \in  {\mathcal{V}}\llbracket {{\color{\colorSYNTAX}\texttt{unit}}}\rrbracket 
   }
\and \inferrule*[lab={\textsc{ }} 
   ]{ \inl^{\tau '_{2}}\hspace*{0.33em}{\begingroup\renewcommand\colorMATH{\colorMATHB}\renewcommand\colorSYNTAX{\colorSYNTAXB}{{\color{\colorMATH}\ensuremath{\sv}}}\endgroup } \in  Atom\llbracket \tau _{1} \mathrel{^{{\begingroup\renewcommand\colorMATH{\colorMATHB}\renewcommand\colorSYNTAX{\colorSYNTAXB}{{\color{\colorMATH}\ensuremath{\varnothing }}}\endgroup }}\oplus ^{{\begingroup\renewcommand\colorMATH{\colorMATHB}\renewcommand\colorSYNTAX{\colorSYNTAXB}{{\color{\colorMATH}\ensuremath{\varnothing }}}\endgroup }}} \tau _{2}\rrbracket  
   \\ {\begingroup\renewcommand\colorMATH{\colorMATHB}\renewcommand\colorSYNTAX{\colorSYNTAXB}{{\color{\colorMATH}\ensuremath{\sv}}}\endgroup } \in  {\mathcal{V}}\llbracket \tau _{1}\rrbracket 
      }{
      \inl^{\tau '_{2}}\hspace*{0.33em}{\begingroup\renewcommand\colorMATH{\colorMATHB}\renewcommand\colorSYNTAX{\colorSYNTAXB}{{\color{\colorMATH}\ensuremath{\sv}}}\endgroup } \in  {\mathcal{V}}\llbracket \tau _{1} \mathrel{^{{\begingroup\renewcommand\colorMATH{\colorMATHB}\renewcommand\colorSYNTAX{\colorSYNTAXB}{{\color{\colorMATH}\ensuremath{\varnothing }}}\endgroup }}\oplus ^{{\begingroup\renewcommand\colorMATH{\colorMATHB}\renewcommand\colorSYNTAX{\colorSYNTAXB}{{\color{\colorMATH}\ensuremath{\varnothing }}}\endgroup }}} \tau _{2}\rrbracket 
   }
\and \inferrule*[lab={\textsc{ }}
   ]{ \inr^{\tau '_{1}}\hspace*{0.33em}{\begingroup\renewcommand\colorMATH{\colorMATHB}\renewcommand\colorSYNTAX{\colorSYNTAXB}{{\color{\colorMATH}\ensuremath{\sv}}}\endgroup } \in  Atom\llbracket \tau _{1} \mathrel{^{{\begingroup\renewcommand\colorMATH{\colorMATHB}\renewcommand\colorSYNTAX{\colorSYNTAXB}{{\color{\colorMATH}\ensuremath{\varnothing }}}\endgroup }}\oplus ^{{\begingroup\renewcommand\colorMATH{\colorMATHB}\renewcommand\colorSYNTAX{\colorSYNTAXB}{{\color{\colorMATH}\ensuremath{\varnothing }}}\endgroup }}} \tau _{2}\rrbracket  
   \\ {\begingroup\renewcommand\colorMATH{\colorMATHB}\renewcommand\colorSYNTAX{\colorSYNTAXB}{{\color{\colorMATH}\ensuremath{\sv}}}\endgroup } \in  {\mathcal{V}}\llbracket \tau _{2}\rrbracket 
      }{
      \inr^{\tau '_{1}}\hspace*{0.33em}{\begingroup\renewcommand\colorMATH{\colorMATHB}\renewcommand\colorSYNTAX{\colorSYNTAXB}{{\color{\colorMATH}\ensuremath{\sv}}}\endgroup } \in  {\mathcal{V}}\llbracket \tau _{1} \mathrel{^{{\begingroup\renewcommand\colorMATH{\colorMATHB}\renewcommand\colorSYNTAX{\colorSYNTAXB}{{\color{\colorMATH}\ensuremath{\varnothing }}}\endgroup }}\oplus ^{{\begingroup\renewcommand\colorMATH{\colorMATHB}\renewcommand\colorSYNTAX{\colorSYNTAXB}{{\color{\colorMATH}\ensuremath{\varnothing }}}\endgroup }}} \tau _{2}\rrbracket 
   }
\and \inferrule*[lab={\textsc{ }}
   ]{ \addProduct{{\begingroup\renewcommand\colorMATH{\colorMATHB}\renewcommand\colorSYNTAX{\colorSYNTAXB}{{\color{\colorMATH}\ensuremath{\sv_{1}}}}\endgroup }}{{\begingroup\renewcommand\colorMATH{\colorMATHB}\renewcommand\colorSYNTAX{\colorSYNTAXB}{{\color{\colorMATH}\ensuremath{\sv_{2}}}}\endgroup }} \in  Atom\llbracket \tau _{1} \mathrel{^{{\begingroup\renewcommand\colorMATH{\colorMATHB}\renewcommand\colorSYNTAX{\colorSYNTAXB}{{\color{\colorMATH}\ensuremath{\varnothing }}}\endgroup }}\&^{{\begingroup\renewcommand\colorMATH{\colorMATHB}\renewcommand\colorSYNTAX{\colorSYNTAXB}{{\color{\colorMATH}\ensuremath{\varnothing }}}\endgroup }}} \tau _{2}\rrbracket  
   \\ {\begingroup\renewcommand\colorMATH{\colorMATHB}\renewcommand\colorSYNTAX{\colorSYNTAXB}{{\color{\colorMATH}\ensuremath{\sv_{1}}}}\endgroup } \in  {\mathcal{V}}\llbracket \tau _{1}\rrbracket 
   \\ {\begingroup\renewcommand\colorMATH{\colorMATHB}\renewcommand\colorSYNTAX{\colorSYNTAXB}{{\color{\colorMATH}\ensuremath{\sv_{2}}}}\endgroup } \in  {\mathcal{V}}\llbracket \tau _{2}\rrbracket 
      }{
      \addProduct{{\begingroup\renewcommand\colorMATH{\colorMATHB}\renewcommand\colorSYNTAX{\colorSYNTAXB}{{\color{\colorMATH}\ensuremath{\sv_{1}}}}\endgroup }}{{\begingroup\renewcommand\colorMATH{\colorMATHB}\renewcommand\colorSYNTAX{\colorSYNTAXB}{{\color{\colorMATH}\ensuremath{\sv_{2}}}}\endgroup }} \in  {\mathcal{V}}\llbracket \tau _{1} \mathrel{^{{\begingroup\renewcommand\colorMATH{\colorMATHB}\renewcommand\colorSYNTAX{\colorSYNTAXB}{{\color{\colorMATH}\ensuremath{\varnothing }}}\endgroup }}\&^{{\begingroup\renewcommand\colorMATH{\colorMATHB}\renewcommand\colorSYNTAX{\colorSYNTAXB}{{\color{\colorMATH}\ensuremath{\varnothing }}}\endgroup }}} \tau _{2}\rrbracket 
   }
\and \inferrule*[lab={\textsc{ }}
   ]{ \langle {\begingroup\renewcommand\colorMATH{\colorMATHB}\renewcommand\colorSYNTAX{\colorSYNTAXB}{{\color{\colorMATH}\ensuremath{\sv_{1}}}}\endgroup },{\begingroup\renewcommand\colorMATH{\colorMATHB}\renewcommand\colorSYNTAX{\colorSYNTAXB}{{\color{\colorMATH}\ensuremath{\sv_{2}}}}\endgroup }\rangle  \in  Atom\llbracket \tau _{1} \mathrel{^{{\begingroup\renewcommand\colorMATH{\colorMATHB}\renewcommand\colorSYNTAX{\colorSYNTAXB}{{\color{\colorMATH}\ensuremath{\varnothing }}}\endgroup }}\otimes ^{{\begingroup\renewcommand\colorMATH{\colorMATHB}\renewcommand\colorSYNTAX{\colorSYNTAXB}{{\color{\colorMATH}\ensuremath{\varnothing }}}\endgroup }}} \tau _{2}\rrbracket  
   \\ {\begingroup\renewcommand\colorMATH{\colorMATHB}\renewcommand\colorSYNTAX{\colorSYNTAXB}{{\color{\colorMATH}\ensuremath{\sv_{1}}}}\endgroup } \in  {\mathcal{V}}\llbracket \tau _{1}\rrbracket 
   \\ {\begingroup\renewcommand\colorMATH{\colorMATHB}\renewcommand\colorSYNTAX{\colorSYNTAXB}{{\color{\colorMATH}\ensuremath{\sv_{2}}}}\endgroup } \in  {\mathcal{V}}\llbracket \tau _{2}\rrbracket 
      }{
      \langle {\begingroup\renewcommand\colorMATH{\colorMATHB}\renewcommand\colorSYNTAX{\colorSYNTAXB}{{\color{\colorMATH}\ensuremath{\sv_{1}}}}\endgroup },{\begingroup\renewcommand\colorMATH{\colorMATHB}\renewcommand\colorSYNTAX{\colorSYNTAXB}{{\color{\colorMATH}\ensuremath{\sv_{2}}}}\endgroup }\rangle  \in  {\mathcal{V}}\llbracket \tau _{1} \mathrel{^{{\begingroup\renewcommand\colorMATH{\colorMATHB}\renewcommand\colorSYNTAX{\colorSYNTAXB}{{\color{\colorMATH}\ensuremath{\varnothing }}}\endgroup }}\otimes ^{{\begingroup\renewcommand\colorMATH{\colorMATHB}\renewcommand\colorSYNTAX{\colorSYNTAXB}{{\color{\colorMATH}\ensuremath{\varnothing }}}\endgroup }}} \tau _{2}\rrbracket 
   }
\and \inferrule*[lab={\textsc{ }}
   ]{ \langle {\begingroup\renewcommand\colorMATH{\colorMATHB}\renewcommand\colorSYNTAX{\colorSYNTAXB}{{\color{\colorMATH}\ensuremath{\slambda}}}\endgroup } x:\tau . {\begingroup\renewcommand\colorMATH{\colorMATHB}\renewcommand\colorSYNTAX{\colorSYNTAXB}{{\color{\colorMATH}\ensuremath{\se}}}\endgroup }, \gamma \rangle  \in  Atom\llbracket (x\mathrel{:}\tau _{1}) \xrightarrowS {{\begingroup\renewcommand\colorMATH{\colorMATHB}\renewcommand\colorSYNTAX{\colorSYNTAXB}{{\color{\colorMATH}\ensuremath{\sss}}}\endgroup }x} \tau _{2}\rrbracket 
   \\ \forall   {\begingroup\renewcommand\colorMATH{\colorMATHB}\renewcommand\colorSYNTAX{\colorSYNTAXB}{{\color{\colorMATH}\ensuremath{\sv}}}\endgroup } \in  {\mathcal{V}}\llbracket \tau _{1}\rrbracket . \gamma [x \mapsto  {\begingroup\renewcommand\colorMATH{\colorMATHB}\renewcommand\colorSYNTAX{\colorSYNTAXB}{{\color{\colorMATH}\ensuremath{\sv}}}\endgroup }] \vdash  {\begingroup\renewcommand\colorMATH{\colorMATHB}\renewcommand\colorSYNTAX{\colorSYNTAXB}{{\color{\colorMATH}\ensuremath{\se}}}\endgroup } \in  {\mathcal{E}}\llbracket \tau _{2}/(x:\tau _{1})\rrbracket 
      }{
      \langle {\begingroup\renewcommand\colorMATH{\colorMATHB}\renewcommand\colorSYNTAX{\colorSYNTAXB}{{\color{\colorMATH}\ensuremath{\slambda}}}\endgroup } x:\tau _{1}. {\begingroup\renewcommand\colorMATH{\colorMATHB}\renewcommand\colorSYNTAX{\colorSYNTAXB}{{\color{\colorMATH}\ensuremath{\se}}}\endgroup }, \gamma \rangle  \in  {\mathcal{V}}\llbracket (x\mathrel{:}\tau _{1}) \xrightarrowS {{\begingroup\renewcommand\colorMATH{\colorMATHB}\renewcommand\colorSYNTAX{\colorSYNTAXB}{{\color{\colorMATH}\ensuremath{\sss}}}\endgroup }x} \tau _{2}\rrbracket 
   }
\and \inferrule*[lab={\textsc{ }}
   ]{ 
   \gamma  \vdash  {\begingroup\renewcommand\colorMATH{\colorMATHB}\renewcommand\colorSYNTAX{\colorSYNTAXB}{{\color{\colorMATH}\ensuremath{\se}}}\endgroup } \Downarrow  {\begingroup\renewcommand\colorMATH{\colorMATHB}\renewcommand\colorSYNTAX{\colorSYNTAXB}{{\color{\colorMATH}\ensuremath{\sv}}}\endgroup }
   \\  {\begingroup\renewcommand\colorMATH{\colorMATHB}\renewcommand\colorSYNTAX{\colorSYNTAXB}{{\color{\colorMATH}\ensuremath{\sv}}}\endgroup } \in  {\mathcal{V}}\llbracket \tau \rrbracket 
      }{
      \gamma  \vdash  {\begingroup\renewcommand\colorMATH{\colorMATHB}\renewcommand\colorSYNTAX{\colorSYNTAXB}{{\color{\colorMATH}\ensuremath{\se}}}\endgroup } \in  {\mathcal{E}}\llbracket \tau \rrbracket 
   }
\and \inferrule*[lab={\textsc{ }}
   ]{ dom(\Gamma ) = dom(\gamma ) 
   \\ \forall  x \in  dom(\gamma ). 
    \gamma (x) \in  {\mathcal{V}}\llbracket \Gamma (x)/\Gamma \rrbracket  
      }{
      \gamma  \in  {\mathcal{G}}\llbracket \Gamma \rrbracket 
   }
\end{mathpar}\endgroup
\end{framed}
\end{small}
\caption{$\ssystem$: Type Safety Logical Relations}
\label{fig:type-safety-lr-1}
\end{figure}

The type safety logical relations is defined in Figure~\ref{fig:type-safety-lr-1}. For simplicity, we only present the cases for real numbers, variables, functions, and sums. The other cases are similar and straightforward. The unary logical relations are split into mutually recursive value relations {{\color{\colorMATH}\ensuremath{{\mathcal{V}}}}}, computation relation {{\color{\colorMATH}\ensuremath{{\mathcal{E}}}}}, and environment relation {{\color{\colorMATH}\ensuremath{{\mathcal{G}}}}}, and defined as follows:

\begin{itemize}[label=\textbf{-},leftmargin=*]\item  
Any value is in {{\color{\colorMATH}\ensuremath{Atom\llbracket \tau \rrbracket }}}  if the value type checks to some {{\color{\colorMATH}\ensuremath{\tau ' <: \tau }}} under an empty type environment.
\item  A real number is in the value relation at type {{\color{\colorMATH}\ensuremath{{\begingroup\renewcommand\colorMATH{\colorMATHA}\renewcommand\colorSYNTAX{\colorSYNTAXA}{{\color{\colorSYNTAX}\texttt{{\ensuremath{{\mathbb{R}}}}}}}\endgroup }}}} 
if the number is in {{\color{\colorMATH}\ensuremath{Atom\llbracket {\begingroup\renewcommand\colorMATH{\colorMATHA}\renewcommand\colorSYNTAX{\colorSYNTAXA}{{\color{\colorSYNTAX}\texttt{{\ensuremath{{\mathbb{R}}}}}}}\endgroup }\rrbracket }}}.

\item  Similarly, a unit value {{\color{\colorMATH}\ensuremath{\ttt}}} is always related at type {{\color{\colorMATH}\ensuremath{{{\color{\colorSYNTAX}\texttt{unit}}}}}}.

\item  An \inl (resp. \inr) value is in the value relation
 at {{\color{\colorMATH}\ensuremath{\tau _{1} \mathrel{^{\varnothing }\oplus ^{\varnothing }} \tau _{2}}}} if the value is in {{\color{\colorMATH}\ensuremath{Atom\llbracket \tau _{1} \mathrel{^{\varnothing }\oplus ^{\varnothing }} \tau _{2}\rrbracket }}} and the underlying value {{\color{\colorMATH}\ensuremath{v}}} is in the value relation at {{\color{\colorMATH}\ensuremath{\tau _{1}}}} (resp. {{\color{\colorMATH}\ensuremath{\tau _{2}}}}).

\item  A closure is in the value relation at type
{{\color{\colorMATH}\ensuremath{(x\mathrel{:}\tau _{1}) \xrightarrowS {{\begingroup\renewcommand\colorMATH{\colorMATHB}\renewcommand\colorSYNTAX{\colorSYNTAXB}{{\color{\colorMATH}\ensuremath{\sss}}}\endgroup }x} \tau _{2}}}} if it satisfies {{\color{\colorMATH}\ensuremath{Atom\llbracket (x\mathrel{:}\tau _{1}) \xrightarrowS {{\begingroup\renewcommand\colorMATH{\colorMATHB}\renewcommand\colorSYNTAX{\colorSYNTAXB}{{\color{\colorMATH}\ensuremath{\sss}}}\endgroup }x} \tau _{2}\rrbracket }}}, and given any value {{\color{\colorMATH}\ensuremath{{\begingroup\renewcommand\colorMATH{\colorMATHB}\renewcommand\colorSYNTAX{\colorSYNTAXB}{{\color{\colorMATH}\ensuremath{\sv}}}\endgroup }}}} in the value relation at argument type {{\color{\colorMATH}\ensuremath{\tau _{1}}}}, the extended configuration {{\color{\colorMATH}\ensuremath{\gamma [x \mapsto  {\begingroup\renewcommand\colorMATH{\colorMATHB}\renewcommand\colorSYNTAX{\colorSYNTAXB}{{\color{\colorMATH}\ensuremath{\sv}}}\endgroup }] \vdash  {\begingroup\renewcommand\colorMATH{\colorMATHB}\renewcommand\colorSYNTAX{\colorSYNTAXB}{{\color{\colorMATH}\ensuremath{\se}}}\endgroup }}}} is in the computation relation at type {{\color{\colorMATH}\ensuremath{\tau _{2}/(x:\tau _{1})}}}.
We use the {{\color{\colorMATH}\ensuremath{./\Gamma }}} operator to remove variables from a type and is defined as follows:
\begingroup\color{\colorMATH}\begin{gather*} \tau /\Gamma  = [\varnothing /x_{1},...,\varnothing /x_{n}]\tau , \forall  x_{i} \in  dom(\Gamma )
\end{gather*}\endgroup
\item  A configuration is in the computation relation at type {{\color{\colorMATH}\ensuremath{\tau }}}, if  the configuration reduces to some value {{\color{\colorMATH}\ensuremath{v}}}, which is itself in the value relation at type {{\color{\colorMATH}\ensuremath{\tau }}}.

\item  Finally, a value environment {{\color{\colorMATH}\ensuremath{\gamma }}} is in the environment relation at {{\color{\colorMATH}\ensuremath{\Gamma }}} if the domains of {{\color{\colorMATH}\ensuremath{\gamma }}} and {{\color{\colorMATH}\ensuremath{\Gamma }}} are the same, and for each variable in the domain of {{\color{\colorMATH}\ensuremath{\gamma }}} the underlying value {{\color{\colorMATH}\ensuremath{\gamma (x)}}} is in the value relation at type {{\color{\colorMATH}\ensuremath{\Gamma (x)/\Gamma }}} (we use the {{\color{\colorMATH}\ensuremath{./\Gamma }}} operator to \toplas{emphasize} that the type is closed).

\end{itemize}

As usual, the fundamental property of the type safety logical relation states that well-typed open terms are in the relation closed by an adequate environment {{\color{\colorMATH}\ensuremath{\gamma }}}:

\begin{proposition}[Fundamental Property of the Type Safety Logical Relation]
  \label{lm:sensitivity-simple-type-safety-FP}\ \\
  Let {{\color{\colorMATH}\ensuremath{\Gamma  \vdash  {\begingroup\renewcommand\colorMATH{\colorMATHB}\renewcommand\colorSYNTAX{\colorSYNTAXB}{{\color{\colorMATH}\ensuremath{\se}}}\endgroup } : \tau  ; {\begingroup\renewcommand\colorMATH{\colorMATHB}\renewcommand\colorSYNTAX{\colorSYNTAXB}{{\color{\colorMATH}\ensuremath{\sS}}}\endgroup }}}}, and {{\color{\colorMATH}\ensuremath{\gamma  \in  {\mathcal{G}}\llbracket \Gamma \rrbracket }}}. Then
    {{\color{\colorMATH}\ensuremath{\gamma \vdash  {\begingroup\renewcommand\colorMATH{\colorMATHB}\renewcommand\colorSYNTAX{\colorSYNTAXB}{{\color{\colorMATH}\ensuremath{\se}}}\endgroup } \in  {\mathcal{E}}\llbracket \tau /\Gamma \rrbracket }}}.
\end{proposition}

% \begin{lemma}[Weakening]\;
%   \label{lm:weakening-type-safety}
%   If {{\color{\colorMATH}\ensuremath{{\begingroup\renewcommand\colorMATH{\colorMATHB}\renewcommand\colorSYNTAX{\colorSYNTAXB}{{\color{\colorMATH}\ensuremath{\sv}}}\endgroup } \in  {\mathcal{V}}\llbracket \tau \rrbracket }}} and {{\color{\colorMATH}\ensuremath{\tau  <: \tau '}}}, {{\color{\colorMATH}\ensuremath{FV(\tau ') = \varnothing }}}, then {{\color{\colorMATH}\ensuremath{{\begingroup\renewcommand\colorMATH{\colorMATHB}\renewcommand\colorSYNTAX{\colorSYNTAXB}{{\color{\colorMATH}\ensuremath{\sv}}}\endgroup } \in  {\mathcal{V}}\llbracket \tau '\rrbracket }}}.
% \end{lemma}

Type safety for closed terms follows immediately as a corollary:

\begin{corollary}[Type Safety and Normalization of $\ssystem$]\ \\
  \label{lm:sensitivity-simple-type-preservation}
  Let {{\color{\colorMATH}\ensuremath{\vdash  {\begingroup\renewcommand\colorMATH{\colorMATHB}\renewcommand\colorSYNTAX{\colorSYNTAXB}{{\color{\colorMATH}\ensuremath{\se}}}\endgroup } : \tau  ; \varnothing }}}, then 
      {{\color{\colorMATH}\ensuremath{\vdash  {\begingroup\renewcommand\colorMATH{\colorMATHB}\renewcommand\colorSYNTAX{\colorSYNTAXB}{{\color{\colorMATH}\ensuremath{\se}}}\endgroup } \Downarrow  {\begingroup\renewcommand\colorMATH{\colorMATHB}\renewcommand\colorSYNTAX{\colorSYNTAXB}{{\color{\colorMATH}\ensuremath{\sv}}}\endgroup }}}} \toplas{for some {{\color{\colorMATH}\ensuremath{{\begingroup\renewcommand\colorMATH{\colorMATHB}\renewcommand\colorSYNTAX{\colorSYNTAXB}{{\color{\colorMATH}\ensuremath{\sv}}}\endgroup }}}} and {{\color{\colorMATH}\ensuremath{\tau '}}}, such that} {{\color{\colorMATH}\ensuremath{\vdash  {\begingroup\renewcommand\colorMATH{\colorMATHB}\renewcommand\colorSYNTAX{\colorSYNTAXB}{{\color{\colorMATH}\ensuremath{\sv}}}\endgroup }: \tau ';\varnothing }}} and
      {{\color{\colorMATH}\ensuremath{\tau ' <: \tau }}}.
\end{corollary}

\subsection{Type Soundness}
\label{sec:sensitivity-simple-soundness}

% \mt{TODO: Explain that for simplicity this section does not use {{\color{\colorMATH}\ensuremath{\Phi }}} if required.}

This section establishes the \emph{type soundness} of $\ssystem$, stated in terms of a \emph{metric
preservation} result. Loosely speaking, metric preservation captures the maximum variation of an open term
when it is closed under two different (but related) environments.

\paragraph{Logical relations} To establish this soundness result, we make use of logical
relations~\cite{appelMcAllester:toplas2001,ahmed:esop2006}. In particular, we define (mutually recursive) logical relations for sensitivity values, computations and environments; see Figure~\ref{fig:sensitivity-simple-logical-relation}.
\begin{figure}[t!]
\begin{small}
\begin{framed}
  \begingroup\color{\colorMATH}\begin{gather*}
\end{gather*}\endgroup
\end{framed}
\end{small}
\caption{$\ssystem$: logical relations for metric preservation}
\label{fig:sensitivity-simple-logical-relation}
\end{figure}
The logical relations for values ({{\color{\colorMATH}\ensuremath{{\mathcal{V}}_{{\begingroup\renewcommand\colorMATH{\colorMATHB}\renewcommand\colorSYNTAX{\colorSYNTAXB}{{\color{\colorMATH}\ensuremath{\distance}}}\endgroup }}\llbracket \sigma \rrbracket }}}) and computations ({{\color{\colorMATH}\ensuremath{{\mathcal{E}}_{{\begingroup\renewcommand\colorMATH{\colorMATHB}\renewcommand\colorSYNTAX{\colorSYNTAXB}{{\color{\colorMATH}\ensuremath{\distance}}}\endgroup }}\llbracket \sigma \rrbracket }}}) are indexed by a relational distance {{\color{\colorMATH}\ensuremath{{\begingroup\renewcommand\colorMATH{\colorMATHB}\renewcommand\colorSYNTAX{\colorSYNTAXB}{{\color{\colorMATH}\ensuremath{\distance}}}\endgroup } \in {\begingroup\renewcommand\colorMATH{\colorMATHA}\renewcommand\colorSYNTAX{\colorSYNTAXA}{{\color{\colorSYNTAX}\texttt{{\ensuremath{{\mathbb{R}}}}}}}\endgroup }^{\infty }_{\geq 0}}}} and a 
 so called \emph{\distanceName type} {{\color{\colorMATH}\ensuremath{\sigma }}}, which is a regular type where sensitivity environments are enriched with a constant {{\color{\colorMATH}\ensuremath{{\begingroup\renewcommand\colorMATH{\colorMATHB}\renewcommand\colorSYNTAX{\colorSYNTAXB}{{\color{\colorMATH}\ensuremath{\distance}}}\endgroup } \in {\begingroup\renewcommand\colorMATH{\colorMATHA}\renewcommand\colorSYNTAX{\colorSYNTAXA}{{\color{\colorSYNTAX}\texttt{{\ensuremath{{\mathbb{R}}}}}}}\endgroup }^{\infty }_{\geq 0}}}} denoting the distance induced by pair of substitutions. Formally, the syntax of \distanceName types is
defined as follows:
\begingroup\color{\colorMATH}\begin{gather*}

\end{gather*}\endgroup
\toplass{Notice that the logical relations do not mention sensitivity environments {{\color{\colorMATH}\ensuremath{{\begingroup\renewcommand\colorMATH{\colorMATHB}\renewcommand\colorSYNTAX{\colorSYNTAXB}{{\color{\colorMATH}\ensuremath{\sS}}}\endgroup }}}} because they are defined over closed terms and values. Nevertheless, \distanceName types {{\color{\colorMATH}\ensuremath{\sigma }}} do mention sensitivity environments {{\color{\colorMATH}\ensuremath{{\begingroup\renewcommand\colorMATH{\colorMATHB}\renewcommand\colorSYNTAX{\colorSYNTAXB}{{\color{\colorMATH}\ensuremath{\sS}}}\endgroup }}}}.}
\toplas{We use a combination of sensitivity environments and relational distances ({{\color{\colorMATH}\ensuremath{{\begingroup\renewcommand\colorMATH{\colorMATHB}\renewcommand\colorSYNTAX{\colorSYNTAXB}{{\color{\colorMATH}\ensuremath{\sS}}}\endgroup } + {\begingroup\renewcommand\colorMATH{\colorMATHB}\renewcommand\colorSYNTAX{\colorSYNTAXB}{{\color{\colorMATH}\ensuremath{\distance}}}\endgroup }}}}), because functions types introduce binders that cannot be substituted until application. For instance, consider type {{\color{\colorMATH}\ensuremath{(x\mathrel{:}{\mathbb{R}}) \xrightarrowS {x + y} ({\mathbb{R}} \mathrel{^{x + 3z}\oplus ^{2x + 2y}} {\mathbb{R}})}}}, and two pair of substitutions for {{\color{\colorMATH}\ensuremath{y}}} and {{\color{\colorMATH}\ensuremath{z}}}, at distance {{\color{\colorMATH}\ensuremath{2}}} and {{\color{\colorMATH}\ensuremath{1}}} respectively\toplass{, e.g. $\gamma_1 = y \mapsto 1, z \mapsto 1$ and $\gamma_2 = y \mapsto 3, z \mapsto 2$, where $|\gamma_1(y) - \gamma_2(y)| \leq 2$ and $|\gamma_1(z) - \gamma_2(z)| \leq 1$}. The corresponding relational distance type after substitution is {{\color{\colorMATH}\ensuremath{(x\mathrel{:}{\mathbb{R}}) \xrightarrowS {x + 2} ({\mathbb{R}} \mathrel{^{x + 3}\oplus ^{2x + 4}} {\mathbb{R}})}}}.}
For notation simplicity, in the rest of the section we name \distanceName types as types when the acompanying relational distances can be inferred from the context. Also we omit the environment notations when they are empty. On the other hand, the logical relation for environments ({{\color{\colorMATH}\ensuremath{{\mathcal{G}}_{{\begingroup\renewcommand\colorMATH{\colorMATHB}\renewcommand\colorSYNTAX{\colorSYNTAXB}{{\color{\colorMATH}\ensuremath{\Distance}}}\endgroup }}\llbracket \Gamma \rrbracket }}}) is indexed by a \emph{relational distance environment} {{\color{\colorMATH}\ensuremath{{\begingroup\renewcommand\colorMATH{\colorMATHB}\renewcommand\colorSYNTAX{\colorSYNTAXB}{{\color{\colorMATH}\ensuremath{\Distance}}}\endgroup }}}}, mapping variables to relational distances in {{\color{\colorMATH}\ensuremath{{\begingroup\renewcommand\colorMATH{\colorMATHA}\renewcommand\colorSYNTAX{\colorSYNTAXA}{{\color{\colorSYNTAX}\texttt{{\ensuremath{{\mathbb{R}}}}}}}\endgroup }^{\infty }_{\geq 0}}}}  and a type environment {{\color{\colorMATH}\ensuremath{\Gamma }}}. We use {{\color{\colorMATH}\ensuremath{({\begingroup\renewcommand\colorMATH{\colorMATHB}\renewcommand\colorSYNTAX{\colorSYNTAXB}{{\color{\colorMATH}\ensuremath{\sv_{1}}}}\endgroup },{\begingroup\renewcommand\colorMATH{\colorMATHB}\renewcommand\colorSYNTAX{\colorSYNTAXB}{{\color{\colorMATH}\ensuremath{\sv_{2}}}}\endgroup }) \in  {\mathcal{V}}_{{\begingroup\renewcommand\colorMATH{\colorMATHB}\renewcommand\colorSYNTAX{\colorSYNTAXB}{{\color{\colorMATH}\ensuremath{\distance}}}\endgroup }}\llbracket \sigma \rrbracket }}} to denote that value {{\color{\colorMATH}\ensuremath{{\begingroup\renewcommand\colorMATH{\colorMATHB}\renewcommand\colorSYNTAX{\colorSYNTAXB}{{\color{\colorMATH}\ensuremath{\sv_{1}}}}\endgroup }}}} is related to value {{\color{\colorMATH}\ensuremath{{\begingroup\renewcommand\colorMATH{\colorMATHB}\renewcommand\colorSYNTAX{\colorSYNTAXB}{{\color{\colorMATH}\ensuremath{\sv_{2}}}}\endgroup }}}} at type {{\color{\colorMATH}\ensuremath{\sigma }}} and relational distance {{\color{\colorMATH}\ensuremath{{\begingroup\renewcommand\colorMATH{\colorMATHB}\renewcommand\colorSYNTAX{\colorSYNTAXB}{{\color{\colorMATH}\ensuremath{\distance}}}\endgroup }}}}, and likewise for expressions (i.e.~computations) and environments.

%Note that instead of distances {{\color{\colorMATH}\ensuremath{{\begingroup\renewcommand\colorMATH{\colorMATHB}\renewcommand\colorSYNTAX{\colorSYNTAXB}{{\color{\colorMATH}\ensuremath{\distance}}}\endgroup }}}} and distance environments {{\color{\colorMATH}\ensuremath{{\begingroup\renewcommand\colorMATH{\colorMATHB}\renewcommand\colorSYNTAX{\colorSYNTAXB}{{\color{\colorMATH}\ensuremath{\Distance}}}\endgroup }}}} we could have used sensitivities {{\color{\colorMATH}\ensuremath{{\begingroup\renewcommand\colorMATH{\colorMATHB}\renewcommand\colorSYNTAX{\colorSYNTAXB}{{\color{\colorMATH}\ensuremath{\sss}}}\endgroup }}}} and sensitivity environments {{\color{\colorMATH}\ensuremath{{\begingroup\renewcommand\colorMATH{\colorMATHB}\renewcommand\colorSYNTAX{\colorSYNTAXB}{{\color{\colorMATH}\ensuremath{\sS}}}\endgroup }}}} respectively. We use different symbols for better readability and explanations.
% , where {{\color{\colorMATH}\ensuremath{{\begingroup\renewcommand\colorMATH{\colorMATHB}\renewcommand\colorSYNTAX{\colorSYNTAXB}{{\color{\colorMATH}\ensuremath{\sss}}}\endgroup }}}} intuitively represents the \emph{distance} between values {{\color{\colorMATH}\ensuremath{{\begingroup\renewcommand\colorMATH{\colorMATHB}\renewcommand\colorSYNTAX{\colorSYNTAXB}{{\color{\colorMATH}\ensuremath{\sv_{1}}}}\endgroup }}}}
% and {{\color{\colorMATH}\ensuremath{{\begingroup\renewcommand\colorMATH{\colorMATHB}\renewcommand\colorSYNTAX{\colorSYNTAXB}{{\color{\colorMATH}\ensuremath{\sv_{2}}}}\endgroup }}}}.

To define logical relations we also make use of {\textit{relational distance instantiations}}, which have shape {{\color{\colorMATH}\ensuremath{{\begingroup\renewcommand\colorMATH{\colorMATHB}\renewcommand\colorSYNTAX{\colorSYNTAXB}{{\color{\colorMATH}\ensuremath{\Distance}}}\endgroup }\mathord{\cdotp }({\begingroup\renewcommand\colorMATH{\colorMATHB}\renewcommand\colorSYNTAX{\colorSYNTAXB}{{\color{\colorMATH}\ensuremath{\sS}}}\endgroup } + {\begingroup\renewcommand\colorMATH{\colorMATHB}\renewcommand\colorSYNTAX{\colorSYNTAXB}{{\color{\colorMATH}\ensuremath{\distance}}}\endgroup })}}} and act by replacing free variables in sensitivity environment {{\color{\colorMATH}\ensuremath{{\begingroup\renewcommand\colorMATH{\colorMATHB}\renewcommand\colorSYNTAX{\colorSYNTAXB}{{\color{\colorMATH}\ensuremath{\sS}}}\endgroup }}}} with the distances provided by distance environment {{\color{\colorMATH}\ensuremath{{\begingroup\renewcommand\colorMATH{\colorMATHB}\renewcommand\colorSYNTAX{\colorSYNTAXB}{{\color{\colorMATH}\ensuremath{\Distance}}}\endgroup }}}}. Relational distance instantiations only close variables defined in {{\color{\colorMATH}\ensuremath{{\begingroup\renewcommand\colorMATH{\colorMATHB}\renewcommand\colorSYNTAX{\colorSYNTAXB}{{\color{\colorMATH}\ensuremath{\Distance}}}\endgroup }}}} and are formally defined as:  
\begingroup\color{\colorMATH}\begin{gather*}

\end{gather*}\endgroup
Furthermore, to close a type under a sensitivity environment we use the
\distanceName type instantiation operator {{\color{\colorMATH}\ensuremath{{\begingroup\renewcommand\colorMATH{\colorMATHB}\renewcommand\colorSYNTAX{\colorSYNTAXB}{{\color{\colorMATH}\ensuremath{\Distance}}}\endgroup }(\sigma )}}} (note that a {{\color{\colorMATH}\ensuremath{\tau }}} is also an {{\color{\colorMATH}\ensuremath{\sigma }}} assuming that the ``default'' relational distance {{\color{\colorMATH}\ensuremath{{\begingroup\renewcommand\colorMATH{\colorMATHB}\renewcommand\colorSYNTAX{\colorSYNTAXB}{{\color{\colorMATH}\ensuremath{\distance}}}\endgroup }}}} is {{\color{\colorMATH}\ensuremath{0}}})
defined below.
\begingroup\color{\colorMATH}\begin{gather*}

\end{gather*}\endgroup

Now that we have all the prerequisite, we briefly go through the definition of the logical relations (in Figure~\ref{fig:sensitivity-simple-logical-relation})\footnote{for simplicity we use ``distance'' instead of ''relational distance''}:

\begin{itemize}[label=\textbf{-},leftmargin=*]\item  Two real numbers are related at type {{\color{\colorMATH}\ensuremath{{\begingroup\renewcommand\colorMATH{\colorMATHA}\renewcommand\colorSYNTAX{\colorSYNTAXA}{{\color{\colorSYNTAX}\texttt{{\ensuremath{{\mathbb{R}}}}}}}\endgroup }}}} and distance {{\color{\colorMATH}\ensuremath{{\begingroup\renewcommand\colorMATH{\colorMATHB}\renewcommand\colorSYNTAX{\colorSYNTAXB}{{\color{\colorMATH}\ensuremath{\distance}}}\endgroup }}}}, if and only if the absolute difference between both numbers is at
   most {{\color{\colorMATH}\ensuremath{{\begingroup\renewcommand\colorMATH{\colorMATHB}\renewcommand\colorSYNTAX{\colorSYNTAXB}{{\color{\colorMATH}\ensuremath{\distance}}}\endgroup }}}}. For instance, {{\color{\colorMATH}\ensuremath{(1,3) \in  {\mathcal{V}}_{2}\llbracket {\begingroup\renewcommand\colorMATH{\colorMATHA}\renewcommand\colorSYNTAX{\colorSYNTAXA}{{\color{\colorSYNTAX}\texttt{{\ensuremath{{\mathbb{R}}}}}}}\endgroup }\rrbracket }}} and {{\color{\colorMATH}\ensuremath{(3, 1) \in  {\mathcal{V}}_{2}\llbracket {\begingroup\renewcommand\colorMATH{\colorMATHA}\renewcommand\colorSYNTAX{\colorSYNTAXA}{{\color{\colorSYNTAX}\texttt{{\ensuremath{{\mathbb{R}}}}}}}\endgroup }\rrbracket }}}, as
   the logical relations are reflexive.

\item  Unit value {{\color{\colorMATH}\ensuremath{\ttt}}} is always related to itself at type {{\color{\colorMATH}\ensuremath{{{\color{\colorSYNTAX}\texttt{unit}}}}}} under any
   distance.

\item  Two \inl \ (resp. \inr) values are related at {{\color{\colorMATH}\ensuremath{\sigma _{1} \mathrel{^{{\begingroup\renewcommand\colorMATH{\colorMATHB}\renewcommand\colorSYNTAX{\colorSYNTAXB}{{\color{\colorMATH}\ensuremath{\distance_{1}}}}\endgroup }}\oplus ^{{\begingroup\renewcommand\colorMATH{\colorMATHB}\renewcommand\colorSYNTAX{\colorSYNTAXB}{{\color{\colorMATH}\ensuremath{\distance_{2}}}}\endgroup }}} \sigma _{2}}}} and
   distance {{\color{\colorMATH}\ensuremath{{\begingroup\renewcommand\colorMATH{\colorMATHB}\renewcommand\colorSYNTAX{\colorSYNTAXB}{{\color{\colorMATH}\ensuremath{\distance}}}\endgroup }}}} if the underlying values are related at type {{\color{\colorMATH}\ensuremath{\sigma _{1}}}}
   (resp. {{\color{\colorMATH}\ensuremath{\sigma _{2}}}}) and distance {{\color{\colorMATH}\ensuremath{ {\begingroup\renewcommand\colorMATH{\colorMATHB}\renewcommand\colorSYNTAX{\colorSYNTAXB}{{\color{\colorMATH}\ensuremath{\distance}}}\endgroup } + {\begingroup\renewcommand\colorMATH{\colorMATHB}\renewcommand\colorSYNTAX{\colorSYNTAXB}{{\color{\colorMATH}\ensuremath{\distance_{1}}}}\endgroup } }}} (resp. {{\color{\colorMATH}\ensuremath{ {\begingroup\renewcommand\colorMATH{\colorMATHB}\renewcommand\colorSYNTAX{\colorSYNTAXB}{{\color{\colorMATH}\ensuremath{\distance}}}\endgroup } + {\begingroup\renewcommand\colorMATH{\colorMATHB}\renewcommand\colorSYNTAX{\colorSYNTAXB}{{\color{\colorMATH}\ensuremath{\distance_{2}}}}\endgroup } }}}). The intuition is
   that {{\color{\colorMATH}\ensuremath{{\begingroup\renewcommand\colorMATH{\colorMATHB}\renewcommand\colorSYNTAX{\colorSYNTAXB}{{\color{\colorMATH}\ensuremath{\distance}}}\endgroup }}}} can be treated as the distance between two computations that reduce
   to the given sums, and {{\color{\colorMATH}\ensuremath{ {\begingroup\renewcommand\colorMATH{\colorMATHB}\renewcommand\colorSYNTAX{\colorSYNTAXB}{{\color{\colorMATH}\ensuremath{\distance_{1}}}}\endgroup } }}} can be treated as the distance between the
   underlying values; thus the total cost is the addition of both
   distances. For instance, for any {{\color{\colorMATH}\ensuremath{{\begingroup\renewcommand\colorMATH{\colorMATHB}\renewcommand\colorSYNTAX{\colorSYNTAXB}{{\color{\colorMATH}\ensuremath{\distance}}}\endgroup }}}} and {{\color{\colorMATH}\ensuremath{\sigma }}}, we have {{\color{\colorMATH}\ensuremath{(\inl\hspace*{0.33em}1,
   \inl\hspace*{0.33em}3) \in  {\mathcal{V}}_{0}\llbracket {\begingroup\renewcommand\colorMATH{\colorMATHA}\renewcommand\colorSYNTAX{\colorSYNTAXA}{{\color{\colorSYNTAX}\texttt{{\ensuremath{{\mathbb{R}}}}}}}\endgroup } \mathrel{^{2}\oplus ^{{\begingroup\renewcommand\colorMATH{\colorMATHB}\renewcommand\colorSYNTAX{\colorSYNTAXB}{{\color{\colorMATH}\ensuremath{\distance}}}\endgroup }}} \sigma \rrbracket }}} because they are at immediate distance
   zero (both are \inl) and latent distance 2; instead of delaying the
   distance, one also has {{\color{\colorMATH}\ensuremath{(\inl\hspace*{0.33em}1, \inl\hspace*{0.33em}3) \in  {\mathcal{V}}_{2}\llbracket {\begingroup\renewcommand\colorMATH{\colorMATHA}\renewcommand\colorSYNTAX{\colorSYNTAXA}{{\color{\colorSYNTAX}\texttt{{\ensuremath{{\mathbb{R}}}}}}}\endgroup } \mathrel{^{0}\oplus ^{{\begingroup\renewcommand\colorMATH{\colorMATHB}\renewcommand\colorSYNTAX{\colorSYNTAXB}{{\color{\colorMATH}\ensuremath{\distance}}}\endgroup }}} \sigma \rrbracket }}}, i.e.
   both values are at distance 2 with zero latent distance between their
   content.
   %\et{not sure about the "immediate/latent distance" terminology I'm using here}

% \item  Any \inl \ value is considered to be at infinite distance from any \inr \ 
%    value, although they may have the same type. For instance, {{\color{\colorMATH}\ensuremath{{{\color{\colorMATH}\ensuremath{(\inl\hspace*{0.33em}1,
%    \inr\hspace*{0.33em}\ttt) \in  {\mathcal{V}}_{\infty }\llbracket {\begingroup\renewcommand\colorMATH{\colorMATHA}\renewcommand\colorSYNTAX{\colorSYNTAXA}{{\color{\colorSYNTAX}\texttt{{\ensuremath{{\mathbb{R}}}}}}}\endgroup } \mathrel{^{{\begingroup\renewcommand\colorMATH{\colorMATHB}\renewcommand\colorSYNTAX{\colorSYNTAXB}{{\color{\colorMATH}\ensuremath{\distance_{1}}}}\endgroup }}\oplus ^{{\begingroup\renewcommand\colorMATH{\colorMATHB}\renewcommand\colorSYNTAX{\colorSYNTAXB}{{\color{\colorMATH}\ensuremath{\distance_{2}}}}\endgroup }}} {{\color{\colorSYNTAX}\texttt{unit}}}\rrbracket }}}}}}, for any {{\color{\colorMATH}\ensuremath{ {\begingroup\renewcommand\colorMATH{\colorMATHB}\renewcommand\colorSYNTAX{\colorSYNTAXB}{{\color{\colorMATH}\ensuremath{\distance_{1}}}}\endgroup }, {\begingroup\renewcommand\colorMATH{\colorMATHB}\renewcommand\colorSYNTAX{\colorSYNTAXB}{{\color{\colorMATH}\ensuremath{\distance_{2}}}}\endgroup } }}}.

\item  \toplas{Two additive (resp. multiplicative) products are
   related at type {{\color{\colorMATH}\ensuremath{\sigma _{1} \mathrel{^{{\begingroup\renewcommand\colorMATH{\colorMATHB}\renewcommand\colorSYNTAX{\colorSYNTAXB}{{\color{\colorMATH}\ensuremath{\distance_{1}}}}\endgroup }}\&^{{\begingroup\renewcommand\colorMATH{\colorMATHB}\renewcommand\colorSYNTAX{\colorSYNTAXB}{{\color{\colorMATH}\ensuremath{\distance_{2}}}}\endgroup }}} \sigma _{2}}}} (resp. {{\color{\colorMATH}\ensuremath{\sigma _{1} \mathrel{^{{\begingroup\renewcommand\colorMATH{\colorMATHB}\renewcommand\colorSYNTAX{\colorSYNTAXB}{{\color{\colorMATH}\ensuremath{\distance_{1}}}}\endgroup }}\otimes ^{{\begingroup\renewcommand\colorMATH{\colorMATHB}\renewcommand\colorSYNTAX{\colorSYNTAXB}{{\color{\colorMATH}\ensuremath{\distance_{2}}}}\endgroup }}} \sigma _{2}}}}) and
   distance {{\color{\colorMATH}\ensuremath{{\begingroup\renewcommand\colorMATH{\colorMATHB}\renewcommand\colorSYNTAX{\colorSYNTAXB}{{\color{\colorMATH}\ensuremath{\distance}}}\endgroup }+\addProd{{\begingroup\renewcommand\colorMATH{\colorMATHB}\renewcommand\colorSYNTAX{\colorSYNTAXB}{{\color{\colorMATH}\ensuremath{\distance'_{1}}}}\endgroup }}{{\begingroup\renewcommand\colorMATH{\colorMATHB}\renewcommand\colorSYNTAX{\colorSYNTAXB}{{\color{\colorMATH}\ensuremath{\distance'_{2}}}}\endgroup }}}}} (resp. {{\color{\colorMATH}\ensuremath{{\begingroup\renewcommand\colorMATH{\colorMATHB}\renewcommand\colorSYNTAX{\colorSYNTAXB}{{\color{\colorMATH}\ensuremath{\distance}}}\endgroup }+\multProd{{\begingroup\renewcommand\colorMATH{\colorMATHB}\renewcommand\colorSYNTAX{\colorSYNTAXB}{{\color{\colorMATH}\ensuremath{\distance'_{1}}}}\endgroup }}{{\begingroup\renewcommand\colorMATH{\colorMATHB}\renewcommand\colorSYNTAX{\colorSYNTAXB}{{\color{\colorMATH}\ensuremath{\distance'_{2}}}}\endgroup }}}}}), if both first components are related at type {{\color{\colorMATH}\ensuremath{\sigma _{1}}}} and
   distance {{\color{\colorMATH}\ensuremath{{\begingroup\renewcommand\colorMATH{\colorMATHB}\renewcommand\colorSYNTAX{\colorSYNTAXB}{{\color{\colorMATH}\ensuremath{\distance}}}\endgroup } + {\begingroup\renewcommand\colorMATH{\colorMATHB}\renewcommand\colorSYNTAX{\colorSYNTAXB}{{\color{\colorMATH}\ensuremath{\distance_{1}}}}\endgroup }+{\begingroup\renewcommand\colorMATH{\colorMATHB}\renewcommand\colorSYNTAX{\colorSYNTAXB}{{\color{\colorMATH}\ensuremath{\distance'_{1}}}}\endgroup }}}}, and both second components are related at type {{\color{\colorMATH}\ensuremath{\sigma _{2}}}} and
   distance {{\color{\colorMATH}\ensuremath{{\begingroup\renewcommand\colorMATH{\colorMATHB}\renewcommand\colorSYNTAX{\colorSYNTAXB}{{\color{\colorMATH}\ensuremath{\distance}}}\endgroup } + {\begingroup\renewcommand\colorMATH{\colorMATHB}\renewcommand\colorSYNTAX{\colorSYNTAXB}{{\color{\colorMATH}\ensuremath{\distance_{2}}}}\endgroup }+{\begingroup\renewcommand\colorMATH{\colorMATHB}\renewcommand\colorSYNTAX{\colorSYNTAXB}{{\color{\colorMATH}\ensuremath{\distance'_{2}}}}\endgroup }}}}. For instance, {{\color{\colorMATH}\ensuremath{(\addProduct{\inl\hspace*{0.33em}1}{ 4}, \addProduct{\inl\hspace*{0.33em}3}{ 5}) \in 
   {\mathcal{V}}_{0}\llbracket ({\begingroup\renewcommand\colorMATH{\colorMATHA}\renewcommand\colorSYNTAX{\colorSYNTAXA}{{\color{\colorSYNTAX}\texttt{{\ensuremath{{\mathbb{R}}}}}}}\endgroup } \mathrel{^{2}\oplus ^{0}} \sigma ) \mathrel{^{0}\&^{1}} {\begingroup\renewcommand\colorMATH{\colorMATHA}\renewcommand\colorSYNTAX{\colorSYNTAXA}{{\color{\colorSYNTAX}\texttt{{\ensuremath{{\mathbb{R}}}}}}}\endgroup }\rrbracket }}} are at distance {{\color{\colorMATH}\ensuremath{0}}} and {{\color{\colorMATH}\ensuremath{(\inl\hspace*{0.33em}1,
   \inl\hspace*{0.33em}3) \in  {\mathcal{V}}_{0}\llbracket {\begingroup\renewcommand\colorMATH{\colorMATHA}\renewcommand\colorSYNTAX{\colorSYNTAXA}{{\color{\colorSYNTAX}\texttt{{\ensuremath{{\mathbb{R}}}}}}}\endgroup } \mathrel{^{2}\oplus ^{0}} \sigma \rrbracket }}} and {{\color{\colorMATH}\ensuremath{(4,5) \in  {\mathcal{V}}_{1}\llbracket {\begingroup\renewcommand\colorMATH{\colorMATHA}\renewcommand\colorSYNTAX{\colorSYNTAXA}{{\color{\colorSYNTAX}\texttt{{\ensuremath{{\mathbb{R}}}}}}}\endgroup }\rrbracket }}}.}

\item  Two sensitivity closures are related if, given related inputs, they produce
   related computations. In more detail, first the environments has to be related at some {{\color{\colorMATH}\ensuremath{\Gamma }}}
   and distance environment {{\color{\colorMATH}\ensuremath{{\begingroup\renewcommand\colorMATH{\colorMATHB}\renewcommand\colorSYNTAX{\colorSYNTAXB}{{\color{\colorMATH}\ensuremath{\Distance}}}\endgroup }}}}. Note that {{\color{\colorMATH}\ensuremath{{\begingroup\renewcommand\colorMATH{\colorMATHB}\renewcommand\colorSYNTAX{\colorSYNTAXB}{{\color{\colorMATH}\ensuremath{\Distance}}}\endgroup }}}} has to
   be the same environment that closes the latent effect of the function {{\color{\colorMATH}\ensuremath{{\begingroup\renewcommand\colorMATH{\colorMATHB}\renewcommand\colorSYNTAX{\colorSYNTAXB}{{\color{\colorMATH}\ensuremath{\Distance}}}\endgroup }\mathord{\cdotp }{\begingroup\renewcommand\colorMATH{\colorMATHB}\renewcommand\colorSYNTAX{\colorSYNTAXB}{{\color{\colorMATH}\ensuremath{\sS}}}\endgroup }
   + {\begingroup\renewcommand\colorMATH{\colorMATHB}\renewcommand\colorSYNTAX{\colorSYNTAXB}{{\color{\colorMATH}\ensuremath{\sss'}}}\endgroup }x}}}, and the one that closes the input type ({{\color{\colorMATH}\ensuremath{\sigma _{1} = {\begingroup\renewcommand\colorMATH{\colorMATHB}\renewcommand\colorSYNTAX{\colorSYNTAXB}{{\color{\colorMATH}\ensuremath{\Distance}}}\endgroup }(\tau _{1})}}}). Second, inputs {{\color{\colorMATH}\ensuremath{{\begingroup\renewcommand\colorMATH{\colorMATHB}\renewcommand\colorSYNTAX{\colorSYNTAXB}{{\color{\colorMATH}\ensuremath{\sv'_{1}}}}\endgroup }}}} and {{\color{\colorMATH}\ensuremath{{\begingroup\renewcommand\colorMATH{\colorMATHB}\renewcommand\colorSYNTAX{\colorSYNTAXB}{{\color{\colorMATH}\ensuremath{\sv'_{2}}}}\endgroup }}}} have to be related at argument type {{\color{\colorMATH}\ensuremath{\sigma _{1}}}} and any distance {{\color{\colorMATH}\ensuremath{{\begingroup\renewcommand\colorMATH{\colorMATHB}\renewcommand\colorSYNTAX{\colorSYNTAXB}{{\color{\colorMATH}\ensuremath{\distance'}}}\endgroup }}}}. Finally, the body of the functions in environments extended with inputs {{\color{\colorMATH}\ensuremath{{\begingroup\renewcommand\colorMATH{\colorMATHB}\renewcommand\colorSYNTAX{\colorSYNTAXB}{{\color{\colorMATH}\ensuremath{\sv'_{1}}}}\endgroup }}}} and {{\color{\colorMATH}\ensuremath{{\begingroup\renewcommand\colorMATH{\colorMATHB}\renewcommand\colorSYNTAX{\colorSYNTAXB}{{\color{\colorMATH}\ensuremath{\sv'_{2}}}}\endgroup }}}} have to be related computations at type {{\color{\colorMATH}\ensuremath{{\begingroup\renewcommand\colorMATH{\colorMATHB}\renewcommand\colorSYNTAX{\colorSYNTAXB}{{\color{\colorMATH}\ensuremath{\distance'}}}\endgroup }x(\sigma _{2})}}} and distance {{\color{\colorMATH}\ensuremath{{\begingroup\renewcommand\colorMATH{\colorMATHB}\renewcommand\colorSYNTAX{\colorSYNTAXB}{{\color{\colorMATH}\ensuremath{\distance}}}\endgroup }+{\begingroup\renewcommand\colorMATH{\colorMATHB}\renewcommand\colorSYNTAX{\colorSYNTAXB}{{\color{\colorMATH}\ensuremath{\Distance}}}\endgroup }\mathord{\cdotp }{\begingroup\renewcommand\colorMATH{\colorMATHB}\renewcommand\colorSYNTAX{\colorSYNTAXB}{{\color{\colorMATH}\ensuremath{\sS}}}\endgroup }+{\begingroup\renewcommand\colorMATH{\colorMATHB}\renewcommand\colorSYNTAX{\colorSYNTAXB}{{\color{\colorMATH}\ensuremath{\sss}}}\endgroup }{\begingroup\renewcommand\colorMATH{\colorMATHB}\renewcommand\colorSYNTAX{\colorSYNTAXB}{{\color{\colorMATH}\ensuremath{\distance'}}}\endgroup }}}}. Note that, as the variable {{\color{\colorMATH}\ensuremath{x}}} is out of scope after the application, we replace any instance of {{\color{\colorMATH}\ensuremath{x}}} with the distance of the inputs {{\color{\colorMATH}\ensuremath{ {\begingroup\renewcommand\colorMATH{\colorMATHB}\renewcommand\colorSYNTAX{\colorSYNTAXB}{{\color{\colorMATH}\ensuremath{\distance'}}}\endgroup }}}}, using the distance type instantiation operator. The new distance at which both computations are now related is computed as the addition of the distance of the values {{\color{\colorMATH}\ensuremath{{\begingroup\renewcommand\colorMATH{\colorMATHB}\renewcommand\colorSYNTAX{\colorSYNTAXB}{{\color{\colorMATH}\ensuremath{\distance}}}\endgroup }}}}, and the closed latent effect {{\color{\colorMATH}\ensuremath{{\begingroup\renewcommand\colorMATH{\colorMATHB}\renewcommand\colorSYNTAX{\colorSYNTAXB}{{\color{\colorMATH}\ensuremath{\distance'}}}\endgroup }x({\begingroup\renewcommand\colorMATH{\colorMATHB}\renewcommand\colorSYNTAX{\colorSYNTAXB}{{\color{\colorMATH}\ensuremath{\Distance}}}\endgroup }\mathord{\cdotp }{\begingroup\renewcommand\colorMATH{\colorMATHB}\renewcommand\colorSYNTAX{\colorSYNTAXB}{{\color{\colorMATH}\ensuremath{\sS}}}\endgroup }+{\begingroup\renewcommand\colorMATH{\colorMATHB}\renewcommand\colorSYNTAX{\colorSYNTAXB}{{\color{\colorMATH}\ensuremath{\sss}}}\endgroup }x) = {\begingroup\renewcommand\colorMATH{\colorMATHB}\renewcommand\colorSYNTAX{\colorSYNTAXB}{{\color{\colorMATH}\ensuremath{\Distance}}}\endgroup }\mathord{\cdotp }{\begingroup\renewcommand\colorMATH{\colorMATHB}\renewcommand\colorSYNTAX{\colorSYNTAXB}{{\color{\colorMATH}\ensuremath{\sS}}}\endgroup }+ {\begingroup\renewcommand\colorMATH{\colorMATHB}\renewcommand\colorSYNTAX{\colorSYNTAXB}{{\color{\colorMATH}\ensuremath{\sss}}}\endgroup }{\begingroup\renewcommand\colorMATH{\colorMATHB}\renewcommand\colorSYNTAX{\colorSYNTAXB}{{\color{\colorMATH}\ensuremath{\distance'}}}\endgroup }}}}. For instance, {{\color{\colorMATH}\ensuremath{(\langle {\begingroup\renewcommand\colorMATH{\colorMATHB}\renewcommand\colorSYNTAX{\colorSYNTAXB}{{\color{\colorMATH}\ensuremath{\slambda}}}\endgroup } x: {\begingroup\renewcommand\colorMATH{\colorMATHA}\renewcommand\colorSYNTAX{\colorSYNTAXA}{{\color{\colorSYNTAX}\texttt{{\ensuremath{{\mathbb{R}}}}}}}\endgroup }. x+y), y \mapsto  1\rangle , \langle {\begingroup\renewcommand\colorMATH{\colorMATHB}\renewcommand\colorSYNTAX{\colorSYNTAXB}{{\color{\colorMATH}\ensuremath{\slambda}}}\endgroup } x: {\begingroup\renewcommand\colorMATH{\colorMATHA}\renewcommand\colorSYNTAX{\colorSYNTAXA}{{\color{\colorSYNTAX}\texttt{{\ensuremath{{\mathbb{R}}}}}}}\endgroup }. x+y), y \mapsto  3\rangle  \in  {\mathcal{V}}_{0}\llbracket (x: {\begingroup\renewcommand\colorMATH{\colorMATHA}\renewcommand\colorSYNTAX{\colorSYNTAXA}{{\color{\colorSYNTAX}\texttt{{\ensuremath{{\mathbb{R}}}}}}}\endgroup }) \xrightarrowS {2+1x} {\begingroup\renewcommand\colorMATH{\colorMATHA}\renewcommand\colorSYNTAX{\colorSYNTAXA}{{\color{\colorSYNTAX}\texttt{{\ensuremath{{\mathbb{R}}}}}}}\endgroup }\rrbracket }}}, as in this case {{\color{\colorMATH}\ensuremath{{\begingroup\renewcommand\colorMATH{\colorMATHB}\renewcommand\colorSYNTAX{\colorSYNTAXB}{{\color{\colorMATH}\ensuremath{\Distance}}}\endgroup } = 2y}}}, {{\color{\colorMATH}\ensuremath{{\begingroup\renewcommand\colorMATH{\colorMATHB}\renewcommand\colorSYNTAX{\colorSYNTAXB}{{\color{\colorMATH}\ensuremath{\sS}}}\endgroup } = 1y}}}, and {{\color{\colorMATH}\ensuremath{{\begingroup\renewcommand\colorMATH{\colorMATHB}\renewcommand\colorSYNTAX{\colorSYNTAXB}{{\color{\colorMATH}\ensuremath{\Distance}}}\endgroup } \mathord{\cdotp } {\begingroup\renewcommand\colorMATH{\colorMATHB}\renewcommand\colorSYNTAX{\colorSYNTAXB}{{\color{\colorMATH}\ensuremath{\sS}}}\endgroup }  = 2y \mathord{\cdotp } 1y = 2}}}.

\item  Two sensitivity configurations are related computations at type {{\color{\colorMATH}\ensuremath{\sigma }}} and distance {{\color{\colorMATH}\ensuremath{{\begingroup\renewcommand\colorMATH{\colorMATHB}\renewcommand\colorSYNTAX{\colorSYNTAXB}{{\color{\colorMATH}\ensuremath{\distance}}}\endgroup }}}}, noted {{\color{\colorMATH}\ensuremath{(\gamma _{1} \vdash  {\begingroup\renewcommand\colorMATH{\colorMATHB}\renewcommand\colorSYNTAX{\colorSYNTAXB}{{\color{\colorMATH}\ensuremath{\se_{1}}}}\endgroup }, \gamma _{2} \vdash  {\begingroup\renewcommand\colorMATH{\colorMATHB}\renewcommand\colorSYNTAX{\colorSYNTAXB}{{\color{\colorMATH}\ensuremath{\se_{2}}}}\endgroup }) \in  {\mathcal{E}}_{{\begingroup\renewcommand\colorMATH{\colorMATHB}\renewcommand\colorSYNTAX{\colorSYNTAXB}{{\color{\colorMATH}\ensuremath{\distance}}}\endgroup }}\llbracket \sigma \rrbracket }}}, when \toplass{the distance is infinite, or if the first configuration reduces to a value, then the second configuration also reduces to a value, and} these values are related at type {{\color{\colorMATH}\ensuremath{\sigma }}} and distance {{\color{\colorMATH}\ensuremath{{\begingroup\renewcommand\colorMATH{\colorMATHB}\renewcommand\colorSYNTAX{\colorSYNTAXB}{{\color{\colorMATH}\ensuremath{\distance}}}\endgroup }}}}. 

\item  Finally, value environment {{\color{\colorMATH}\ensuremath{\gamma _{1}}}} is related to value environment {{\color{\colorMATH}\ensuremath{\gamma _{2}}}} at type environment {{\color{\colorMATH}\ensuremath{\Gamma }}} and distance environment {{\color{\colorMATH}\ensuremath{{\begingroup\renewcommand\colorMATH{\colorMATHB}\renewcommand\colorSYNTAX{\colorSYNTAXB}{{\color{\colorMATH}\ensuremath{\Distance}}}\endgroup }}}}, written {{\color{\colorMATH}\ensuremath{(\gamma _{1},\gamma _{2}) \in  {\mathcal{G}}_{{\begingroup\renewcommand\colorMATH{\colorMATHB}\renewcommand\colorSYNTAX{\colorSYNTAXB}{{\color{\colorMATH}\ensuremath{\Distance}}}\endgroup }}\llbracket \Gamma \rrbracket }}}, if they both map each variable {{\color{\colorMATH}\ensuremath{x}}} in the type environment to values related at their corresponding type (closed with {{\color{\colorMATH}\ensuremath{{\begingroup\renewcommand\colorMATH{\colorMATHB}\renewcommand\colorSYNTAX{\colorSYNTAXB}{{\color{\colorMATH}\ensuremath{\Distance}}}\endgroup }}}}) and at distance {{\color{\colorMATH}\ensuremath{{\begingroup\renewcommand\colorMATH{\colorMATHB}\renewcommand\colorSYNTAX{\colorSYNTAXB}{{\color{\colorMATH}\ensuremath{\Distance}}}\endgroup }(x)}}}.
\end{itemize}

\paragraph{Sensitivity Metric Preservation} Armed with these logical relations, we can establish the notion of type soundness, and prove the fundamental property---well-typed terms are related with themselves---which corresponds to metric preservation~\cite{reed2010distance}. As usual, we state this property appealing to open terms, where free variables indicate input parameters, which are then closed by related value environments.

\begin{restatable}[Sensitivity Metric Preservation]{theorem}{FundamentalPropertySensitivity}
  \label{lm:sensitivity-simple-fp}
If \ {{\color{\colorMATH}\ensuremath{\Gamma  \vdash  {\begingroup\renewcommand\colorMATH{\colorMATHB}\renewcommand\colorSYNTAX{\colorSYNTAXB}{{\color{\colorMATH}\ensuremath{\se}}}\endgroup } \mathrel{:} \tau  \mathrel{;} {\begingroup\renewcommand\colorMATH{\colorMATHB}\renewcommand\colorSYNTAX{\colorSYNTAXB}{{\color{\colorMATH}\ensuremath{\sS}}}\endgroup }}}}, then for any distance environment {{\color{\colorMATH}\ensuremath{{\begingroup\renewcommand\colorMATH{\colorMATHB}\renewcommand\colorSYNTAX{\colorSYNTAXB}{{\color{\colorMATH}\ensuremath{\Distance}}}\endgroup }}}} with {{\color{\colorMATH}\ensuremath{dom(\Gamma ) \subseteq  dom({\begingroup\renewcommand\colorMATH{\colorMATHB}\renewcommand\colorSYNTAX{\colorSYNTAXB}{{\color{\colorMATH}\ensuremath{\Distance}}}\endgroup })}}} and any pair of value environments {{\color{\colorMATH}\ensuremath{(\gamma _{1},\gamma _{2}) \in  {\mathcal{G}}_{{\begingroup\renewcommand\colorMATH{\colorMATHB}\renewcommand\colorSYNTAX{\colorSYNTAXB}{{\color{\colorMATH}\ensuremath{\Distance}}}\endgroup }}\llbracket \Gamma \rrbracket }}}, it holds that {{\color{\colorMATH}\ensuremath{(\gamma _{1}\vdash {\begingroup\renewcommand\colorMATH{\colorMATHB}\renewcommand\colorSYNTAX{\colorSYNTAXB}{{\color{\colorMATH}\ensuremath{\se}}}\endgroup },\gamma _{2}\vdash {\begingroup\renewcommand\colorMATH{\colorMATHB}\renewcommand\colorSYNTAX{\colorSYNTAXB}{{\color{\colorMATH}\ensuremath{\se}}}\endgroup }) \in  {\mathcal{E}}_{{\begingroup\renewcommand\colorMATH{\colorMATHB}\renewcommand\colorSYNTAX{\colorSYNTAXB}{{\color{\colorMATH}\ensuremath{\Distance}}}\endgroup }\mathord{\cdotp }{\begingroup\renewcommand\colorMATH{\colorMATHB}\renewcommand\colorSYNTAX{\colorSYNTAXB}{{\color{\colorMATH}\ensuremath{\sS}}}\endgroup }}\llbracket {\begingroup\renewcommand\colorMATH{\colorMATHB}\renewcommand\colorSYNTAX{\colorSYNTAXB}{{\color{\colorMATH}\ensuremath{\Distance}}}\endgroup }(\tau )\rrbracket }}}.
\end{restatable}

\noindent In other words, if a sensitivity term is well-typed, then for any valid distance environment {{\color{\colorMATH}\ensuremath{{\begingroup\renewcommand\colorMATH{\colorMATHB}\renewcommand\colorSYNTAX{\colorSYNTAXB}{{\color{\colorMATH}\ensuremath{\Distance}}}\endgroup }}}} (that ``fits'' {{\color{\colorMATH}\ensuremath{\Gamma }}}) and any two value environments {{\color{\colorMATH}\ensuremath{\gamma _{1},\gamma _{2}}}} related at {{\color{\colorMATH}\ensuremath{\Gamma }}} and {{\color{\colorMATH}\ensuremath{{\begingroup\renewcommand\colorMATH{\colorMATHB}\renewcommand\colorSYNTAX{\colorSYNTAXB}{{\color{\colorMATH}\ensuremath{\Distance}}}\endgroup }}}}, configurations {{\color{\colorMATH}\ensuremath{\gamma _{1}\vdash {\begingroup\renewcommand\colorMATH{\colorMATHB}\renewcommand\colorSYNTAX{\colorSYNTAXB}{{\color{\colorMATH}\ensuremath{\se}}}\endgroup },\gamma _{2}\vdash {\begingroup\renewcommand\colorMATH{\colorMATHB}\renewcommand\colorSYNTAX{\colorSYNTAXB}{{\color{\colorMATH}\ensuremath{\se}}}\endgroup }}}} represent related computations at type {{\color{\colorMATH}\ensuremath{{\begingroup\renewcommand\colorMATH{\colorMATHB}\renewcommand\colorSYNTAX{\colorSYNTAXB}{{\color{\colorMATH}\ensuremath{\Distance}}}\endgroup }(\tau )}}} (closing all free variables) and distance {{\color{\colorMATH}\ensuremath{{\begingroup\renewcommand\colorMATH{\colorMATHB}\renewcommand\colorSYNTAX{\colorSYNTAXB}{{\color{\colorMATH}\ensuremath{\Distance}}}\endgroup }\mathord{\cdotp }{\begingroup\renewcommand\colorMATH{\colorMATHB}\renewcommand\colorSYNTAX{\colorSYNTAXB}{{\color{\colorMATH}\ensuremath{\sS}}}\endgroup }}}}. Note that since {{\color{\colorMATH}\ensuremath{dom({\begingroup\renewcommand\colorMATH{\colorMATHB}\renewcommand\colorSYNTAX{\colorSYNTAXB}{{\color{\colorMATH}\ensuremath{\sS}}}\endgroup }) \subseteq  dom(\Gamma ) \subseteq  dom({\begingroup\renewcommand\colorMATH{\colorMATHB}\renewcommand\colorSYNTAX{\colorSYNTAXB}{{\color{\colorMATH}\ensuremath{\Distance}}}\endgroup })}}}, we have {{\color{\colorMATH}\ensuremath{{\begingroup\renewcommand\colorMATH{\colorMATHB}\renewcommand\colorSYNTAX{\colorSYNTAXB}{{\color{\colorMATH}\ensuremath{\Distance}}}\endgroup }\mathord{\cdotp }{\begingroup\renewcommand\colorMATH{\colorMATHB}\renewcommand\colorSYNTAX{\colorSYNTAXB}{{\color{\colorMATH}\ensuremath{\sS}}}\endgroup } \in  {\begingroup\renewcommand\colorMATH{\colorMATHA}\renewcommand\colorSYNTAX{\colorSYNTAXA}{{\color{\colorSYNTAX}\texttt{{\ensuremath{{\mathbb{R}}}}}}}\endgroup }^{\infty }_{\geq 0}}}}.

From the above theorem it is easy to derive a corollary that only characterizes closed terms:

\begin{corrolary}[FP for closed sensitivity terms]\
  If {{\color{\colorMATH}\ensuremath{\varnothing  \vdash  {\begingroup\renewcommand\colorMATH{\colorMATHB}\renewcommand\colorSYNTAX{\colorSYNTAXB}{{\color{\colorMATH}\ensuremath{\se}}}\endgroup } \mathrel{:} \tau  \mathrel{;} \varnothing }}}, then {{\color{\colorMATH}\ensuremath{(\varnothing  \vdash  {\begingroup\renewcommand\colorMATH{\colorMATHB}\renewcommand\colorSYNTAX{\colorSYNTAXB}{{\color{\colorMATH}\ensuremath{\se}}}\endgroup }, \varnothing  \vdash {\begingroup\renewcommand\colorMATH{\colorMATHB}\renewcommand\colorSYNTAX{\colorSYNTAXB}{{\color{\colorMATH}\ensuremath{\se}}}\endgroup }) \in  {\mathcal{E}}_{{\begingroup\renewcommand\colorMATH{\colorMATHB}\renewcommand\colorSYNTAX{\colorSYNTAXB}{{\color{\colorMATH}\ensuremath{_{0}}}}\endgroup }}\llbracket \tau \rrbracket }}}. 
\end{corrolary}

As a direct consequence of Theorem~\ref{lm:sensitivity-simple-fp} we can also establish the sensitivity type soundness at base types:
\begin{restatable}[Sensitivity Type Soundness at Base Types]{theorem}{SensitivityTypeSoundnessBaseTypesSimple}
  \label{thm:sensitivity-simple-SensitivityTypeSoundnessBaseTypes}
  If {{\color{\colorMATH}\ensuremath{\varnothing  \vdash  {\begingroup\renewcommand\colorMATH{\colorMATHB}\renewcommand\colorSYNTAX{\colorSYNTAXB}{{\color{\colorMATH}\ensuremath{\se}}}\endgroup } \mathrel{:} (x\mathrel{:} {\begingroup\renewcommand\colorMATH{\colorMATHA}\renewcommand\colorSYNTAX{\colorSYNTAXA}{{\color{\colorSYNTAX}\texttt{{\ensuremath{{\mathbb{R}}}}}}}\endgroup }) \xrightarrowS {{\begingroup\renewcommand\colorMATH{\colorMATHB}\renewcommand\colorSYNTAX{\colorSYNTAXB}{{\color{\colorMATH}\ensuremath{\sss}}}\endgroup }x} {\begingroup\renewcommand\colorMATH{\colorMATHA}\renewcommand\colorSYNTAX{\colorSYNTAXA}{{\color{\colorSYNTAX}\texttt{{\ensuremath{{\mathbb{R}}}}}}}\endgroup } \mathrel{;} \varnothing }}},\\
  {{\color{\colorMATH}\ensuremath{|{\begingroup\renewcommand\colorMATH{\colorMATHB}\renewcommand\colorSYNTAX{\colorSYNTAXB}{{\color{\colorMATH}\ensuremath{r_{1}}}}\endgroup }-{\begingroup\renewcommand\colorMATH{\colorMATHB}\renewcommand\colorSYNTAX{\colorSYNTAXB}{{\color{\colorMATH}\ensuremath{r_{2}}}}\endgroup }| \leq  {\begingroup\renewcommand\colorMATH{\colorMATHB}\renewcommand\colorSYNTAX{\colorSYNTAXB}{{\color{\colorMATH}\ensuremath{\distance}}}\endgroup }}}}, {{\color{\colorMATH}\ensuremath{\varnothing  \vdash  {\begingroup\renewcommand\colorMATH{\colorMATHB}\renewcommand\colorSYNTAX{\colorSYNTAXB}{{\color{\colorMATH}\ensuremath{\se}}}\endgroup }\hspace*{0.33em}{\begingroup\renewcommand\colorMATH{\colorMATHB}\renewcommand\colorSYNTAX{\colorSYNTAXB}{{\color{\colorMATH}\ensuremath{r_{1}}}}\endgroup } \Downarrow  {\begingroup\renewcommand\colorMATH{\colorMATHB}\renewcommand\colorSYNTAX{\colorSYNTAXB}{{\color{\colorMATH}\ensuremath{r_{1}^{\prime}}}}\endgroup }}}}, {{\color{\colorMATH}\ensuremath{\varnothing  \vdash  {\begingroup\renewcommand\colorMATH{\colorMATHB}\renewcommand\colorSYNTAX{\colorSYNTAXB}{{\color{\colorMATH}\ensuremath{\se}}}\endgroup }\hspace*{0.33em}{\begingroup\renewcommand\colorMATH{\colorMATHB}\renewcommand\colorSYNTAX{\colorSYNTAXB}{{\color{\colorMATH}\ensuremath{r_{2}}}}\endgroup } \Downarrow  {\begingroup\renewcommand\colorMATH{\colorMATHB}\renewcommand\colorSYNTAX{\colorSYNTAXB}{{\color{\colorMATH}\ensuremath{r_{2}^{\prime}}}}\endgroup } }}}, then {{\color{\colorMATH}\ensuremath{|{\begingroup\renewcommand\colorMATH{\colorMATHB}\renewcommand\colorSYNTAX{\colorSYNTAXB}{{\color{\colorMATH}\ensuremath{r_{1}^{\prime}}}}\endgroup }-{\begingroup\renewcommand\colorMATH{\colorMATHB}\renewcommand\colorSYNTAX{\colorSYNTAXB}{{\color{\colorMATH}\ensuremath{r_{2}^{\prime}}}}\endgroup }| \leq  {\begingroup\renewcommand\colorMATH{\colorMATHB}\renewcommand\colorSYNTAX{\colorSYNTAXB}{{\color{\colorMATH}\ensuremath{\sss}}}\endgroup }{\begingroup\renewcommand\colorMATH{\colorMATHB}\renewcommand\colorSYNTAX{\colorSYNTAXB}{{\color{\colorMATH}\ensuremath{\distance}}}\endgroup }}}}.
\end{restatable}

\toplas{
Let us illustrate metric preservation by revisiting some examples. Consider example \ref{ex:scaling}:
\begingroup\color{\colorMATH}\begin{gather*}
x: {\mathbb{R}}, y: {\mathbb{R}} \vdash  {\begingroup\renewcommand\colorMATH{\colorMATHB}\renewcommand\colorSYNTAX{\colorSYNTAXB}{{\color{\colorSYNTAX}\texttt{let}}}\endgroup }\hspace*{0.33em}x_{1}, x_{2} = \langle 2 * x,y\rangle  \hspace*{0.33em}{\begingroup\renewcommand\colorMATH{\colorMATHB}\renewcommand\colorSYNTAX{\colorSYNTAXB}{{\color{\colorSYNTAX}\texttt{in}}}\endgroup } \hspace*{0.33em} x_{1} + 2 * x_{2}: {\mathbb{R}}; 2x+2y
\end{gather*}\endgroup
If we know that in two different executions {{\color{\colorMATH}\ensuremath{x}}} may differ in at most {{\color{\colorMATH}\ensuremath{1}}}, and {{\color{\colorMATH}\ensuremath{y}}} in at most {{\color{\colorMATH}\ensuremath{3}}}, i.e. {{\color{\colorMATH}\ensuremath{\Delta  = 1x+3y}}}, then the result will differ in at most {{\color{\colorMATH}\ensuremath{\Delta  \mathord{\cdotp } (2x + 2y) = 1\mathord{\cdotp }2 + 3\mathord{\cdotp }2 = 8}}}. For instance, if in one execution {{\color{\colorMATH}\ensuremath{x}}} is bound to {{\color{\colorMATH}\ensuremath{0}}} and {{\color{\colorMATH}\ensuremath{y}}} to {{\color{\colorMATH}\ensuremath{4}}} then the result will be {{\color{\colorMATH}\ensuremath{8}}}. In a second execution, if {{\color{\colorMATH}\ensuremath{x}}} is bound to {{\color{\colorMATH}\ensuremath{1}}} and {{\color{\colorMATH}\ensuremath{y}}} to {{\color{\colorMATH}\ensuremath{6}}} then the result will be {{\color{\colorMATH}\ensuremath{14}}}. Comparing both results we get {{\color{\colorMATH}\ensuremath{|8-14| = 6 \leq  8}}}. Finally, in a third execution, if {{\color{\colorMATH}\ensuremath{x}}} is bound to {{\color{\colorMATH}\ensuremath{1}}} and {{\color{\colorMATH}\ensuremath{y}}} to {{\color{\colorMATH}\ensuremath{7}}} then the result will be {{\color{\colorMATH}\ensuremath{16}}}. Comparing with the first execution we have {{\color{\colorMATH}\ensuremath{|8-16| = 8 \leq  8}}}, and with the second {{\color{\colorMATH}\ensuremath{|14-16| = 2 \leq  8}}}.
}

\toplas{
Now consider example \ref{ex:discont}:
\begingroup\color{\colorMATH}\begin{gather*}
x:{\mathbb{R}} \hspace*{0.33em}\vdash \hspace*{0.33em} \ccase\hspace*{0.33em}(x \leq  10)\hspace*{0.33em}\{ x_{1} \Rightarrow  {\text{true}}\} \{ x_{2} \Rightarrow  {\text{false}}\}  : {\mathbb{B}} ; \infty x
\end{gather*}\endgroup
In this case if {{\color{\colorMATH}\ensuremath{x}}} varies in two different executions {{\color{\colorMATH}\ensuremath{(\Delta (x) > 0)}}} then the outcome will differ in at most {{\color{\colorMATH}\ensuremath{\Delta (x)\mathord{\cdotp }\infty =\infty }}}. For instance, if in one execution {{\color{\colorMATH}\ensuremath{x}}} is bound to {{\color{\colorMATH}\ensuremath{0}}} the result will be {{\color{\colorMATH}\ensuremath{{\text{true}}}}}, and if in a second execution {{\color{\colorMATH}\ensuremath{x}}} is bound to {{\color{\colorMATH}\ensuremath{1}}}, then the result is also going to be {{\color{\colorMATH}\ensuremath{{\text{true}}}}}, and {{\color{\colorMATH}\ensuremath{{\text{true}}}}} is at distance zero with respect to itself, and {{\color{\colorMATH}\ensuremath{0 \leq  \infty }}}. If in a third execution {{\color{\colorMATH}\ensuremath{x}}} is bound to {{\color{\colorMATH}\ensuremath{11}}}, then the result will be {{\color{\colorMATH}\ensuremath{{\text{false}}}}}, and {{\color{\colorMATH}\ensuremath{{\text{false}}}}} is at distance infinity with respect to {{\color{\colorMATH}\ensuremath{{\text{true}}}}}.
Now, if we now that {{\color{\colorMATH}\ensuremath{x}}} is constant across multiple executions {{\color{\colorMATH}\ensuremath{(\Delta (x) = 0)}}}, then we know from metric preservation that the result will differ in {{\color{\colorMATH}\ensuremath{0\mathord{\cdotp }\infty  = 0}}}, i.e. the result will be constant.
}

% }-}

% }-}

\section{Design of \system's Privacy Type System} % {-{
\label{sec:privacy-design}

In this section, we review the limitations of prior approaches related to the tracking of privacy, and then discuss how they are addressed by Jazz. In this section, we color expressions and metavariables {\begingroup\renewcommand\colorMATH{\colorMATHC}\renewcommand\colorSYNTAX{\colorSYNTAXC}{{\color{\colorMATH}\ensuremath{red}}}\endgroup } as they pertain to the privacy fragment of Jazz.

\label{sec:advant-syst-diff}

% y\mapsto 1 \vdash 
% let f: (x:{\mathbb{R}} y^\otimes ^2y {\mathbb{R}}) ->^y {\mathbb{R}} = (\lambda x:{\mathbb{R}} y^\otimes ^2y {\mathbb{R}}. \pi _{1}(x)) in
% f (y, 2y)

% y\mapsto 1 \vdash 

% (\lambda y:{\mathbb{R}}.
%   3y \vdash  let p = (y, 2y) : {\mathbb{R}}\otimes {\mathbb{R}}
%   . \vdash  let f: (x:{\mathbb{R}}\otimes {\mathbb{R}}) \multimap  {\mathbb{R}} = (\lambda x:{\mathbb{R}}\otimes {\mathbb{R}}. \pi _{1}(x)) in
%   3y \vdash  let r = f(p) : {\mathbb{R}}
%   y \vdash  !r : ![1/3]{\mathbb{R}}
% ) : {\mathbb{R}} \multimap  {\mathbb{R}}

% 100x,100y \vdash 

%   !(x,y \vdash  <!2x,y> : (![1/2] {\mathbb{R}}) \otimes  {\mathbb{R}}

%    let (x1, x2) = \langle !(100x), y\rangle  :(![1/100] {\mathbb{R}}) \otimes  {\mathbb{R}} in
%  100(x1) \vdash   let !x1' = x1  in
%    x1 \vdash  x1' + x2 : {\mathbb{R}}) :

% x1 : ![1/100]{\mathbb{R}}

% r*1/100 = 1
% r = 100

% 2x,2y
%  \vdash 
% let (x1, x2) = \langle 2x, !y\rangle  in
% let x2' = x2 in
% 1x1, 2r*x2 \vdash  x1 + 2x2

% r*2 = 1
% r = 1

% 4x,2y \vdash 
% let (x1, x2) = \langle 2x, y\rangle  in
% x1 + 2x2

% programmer has to manually add annotations where necessary:
% 2x,2y \vdash 
% let (x1, x2) = \langle !(2x), y\rangle  in
% x1 + 2x2

% if you don't use a variable, you still pay for it:
% 2x,2y \vdash 
% let (x1, x2) = \langle !(2x), y\rangle  in
% x1

% * in unsound
% Add columns for Fuzz and DFuzz separately

\subsection{Privacy Closures} % {-{
\label{sec:priv_closures}

% \system improves on prior systems by supporting both advanced variants of
% differential privacy such as {{\color{\colorMATH}\ensuremath{(\epsilon ,\delta )}}}, as well as bounded privacy for
% higher-order programming. \fuzz and \dfuzz support higher-order programming
% but not advanced privacy variants. \duet supports both advanced privacy variants and higher-order programs, but fails to bound the privacy effects for many uses of higher-order functions.
Consider a family of looping combinators parameterized by the number of loop
iterations {{\color{\colorMATH}\ensuremath{n}}}, \toplas{e.g., where {{\color{\colorMATH}\ensuremath{{\text{loop}}_{3}\hspace*{0.33em}x\hspace*{0.33em}f = f\hspace*{0.33em}(f\hspace*{0.33em}(f\hspace*{0.33em}x))}}}.
In \fuzz, {{\color{\colorMATH}\ensuremath{{\text{loop}}_{n}}}} would have the type} {{\color{\colorMATH}\ensuremath{{{\color{\colorSYNTAX}\texttt{{\ensuremath{{{\color{\colorMATH}\ensuremath{\tau }}} \rightarrow 
{{\color{\colorMATH}\ensuremath{({{\color{\colorSYNTAX}\texttt{{\ensuremath{{{\color{\colorMATH}\ensuremath{\tau }}} \rightarrow  {\scriptstyle \bigcirc } {{\color{\colorMATH}\ensuremath{\tau }}}}}}}})}}} \multimap _{{{\color{\colorMATH}\ensuremath{n}}}} {\scriptstyle \bigcirc } {{\color{\colorMATH}\ensuremath{\tau }}}}}}}}}}}. In this type, regular arrows {{\color{\colorSYNTAX}\texttt{{\ensuremath{\rightarrow }}}}}
mean no sensitivity is tracked for the
argument. The linear arrow {{\color{\colorSYNTAX}\texttt{{\ensuremath{\multimap _{{{\color{\colorMATH}\ensuremath{n}}}}}}}}} means the result is {{\color{\colorMATH}\ensuremath{n}}}-sensitive (where {{\color{\colorMATH}\ensuremath{n}}} is
the number of loop iterations) in the {\textit{closure variables}} of the supplied
function of type {{\color{\colorMATH}\ensuremath{({{\color{\colorSYNTAX}\texttt{{\ensuremath{{{\color{\colorMATH}\ensuremath{\tau }}} \rightarrow  {\scriptstyle \bigcirc } {{\color{\colorMATH}\ensuremath{\tau }}}}}}}})}}}. This allows for instantiating
{{\color{\colorMATH}\ensuremath{{\text{loop}}}}} with a closure
capturing a sensitive variable, like {{\color{\colorMATH}\ensuremath{{\text{db}}}}}. So
{{\color{\colorMATH}\ensuremath{{\text{loop}}_{n}\hspace*{0.33em}0\hspace*{0.33em}({{\color{\colorSYNTAX}\texttt{{\ensuremath{\lambda {{\color{\colorMATH}\ensuremath{x}}}.\hspace*{0.33em}{{\color{\colorMATH}\ensuremath{x+{\text{laplace}}_{\epsilon }\hspace*{0.33em}db}}}}}}}})}}} will give {{\color{\colorMATH}\ensuremath{n\epsilon }}} differential privacy for
{{\color{\colorMATH}\ensuremath{db}}} by scaling {{\color{\colorMATH}\ensuremath{\epsilon }}}---the privacy cost of closure variable {{\color{\colorMATH}\ensuremath{db}}}---by the loop
iteration {{\color{\colorMATH}\ensuremath{n}}}. \toplass{When supporting advanced variants of
differential privacy like $(\epsilon ,\delta )$, a different metric must be chosen to recover this kind of scaling; otherwise this  argument only holds for pure $\epsilon $-differential privacy.} %\mt{check}

In \duet, in order to support $(\epsilon ,\delta )$-differential privacy (and disallow
problematic scaling), privacy closures immediately report unbounded
privacy ({\begingroup\renewcommand\colorMATH{\colorMATHC}\renewcommand\colorSYNTAX{\colorSYNTAXC}{{\color{\colorSYNTAX}\texttt{{\ensuremath{\infty }}}}}\endgroup })
for any captured variables in privacy lambdas. The {\textit{principle}} of {{\color{\colorMATH}\ensuremath{{\text{loop}}}}}'s
type above is justified in \duet, but not via
a scaling argument, and instead via a primitive type {\textit{rule}}---it cannot be expressed
as a {\textit{type}}. This is problematic for two reasons: first, it is not possible to
extend \duet's implementation with new looping primitives by adding terms with
axiomatically justified types, leading to a bloated set of core typing rules, and second,
it is not possible to lambda abstract looping combinators, e.g., to chain or
compose them in helper functions.

To see the root cause for the limitation in \duet, we show the type rules for
looping (advanced composition) and function introduction (from~\cite{near2019duet}):
\begingroup\color{\colorMATH}\begin{gather*} % {-{

\end{gather*}\endgroup % }-}
In the rule for advanced composition shown above \toplas{(left)}, {\begingroup\renewcommand\colorMATH{\colorMATHB}\renewcommand\colorSYNTAX{\colorSYNTAXB}{{\color{\colorMATH}\ensuremath{e_{1}}}}\endgroup } is the initial value
for the looping state of type {{\color{\colorMATH}\ensuremath{\tau }}}, and {\begingroup\renewcommand\colorMATH{\colorMATHC}\renewcommand\colorSYNTAX{\colorSYNTAXC}{{\color{\colorMATH}\ensuremath{e_{2}}}}\endgroup } is the loop body which updates
the looping state {{\color{\colorSYNTAX}\texttt{{\ensuremath{{{\color{\colorMATH}\ensuremath{\tau }}} \rightarrow  {{\color{\colorMATH}\ensuremath{\tau }}}}}}}} and may mention closure variables in {\begingroup\renewcommand\colorMATH{\colorMATHC}\renewcommand\colorSYNTAX{\colorSYNTAXC}{{\color{\colorMATH}\ensuremath{\Gamma _{2}}}}\endgroup }.
Parameter {{\color{\colorMATH}\ensuremath{\delta ^{\prime}}}} is a meta-parameter for the advanced composition formula---this
parameter is unique to looping in {{\color{\colorMATH}\ensuremath{(\epsilon ,\delta )}}}-differential privacy. The notation
{\begingroup\renewcommand\colorMATH{\colorMATHC}\renewcommand\colorSYNTAX{\colorSYNTAXC}{{\color{\colorMATH}\ensuremath{\rceil \Gamma _{2}\lceil ^{{{\color{\colorSYNTAX}\texttt{{\ensuremath{{\begingroup\renewcommand\colorMATH{\colorMATHA}\renewcommand\colorSYNTAX{\colorSYNTAXA}{{\color{\colorMATH}\ensuremath{\epsilon }}}\endgroup },{\begingroup\renewcommand\colorMATH{\colorMATHA}\renewcommand\colorSYNTAX{\colorSYNTAXA}{{\color{\colorMATH}\ensuremath{\delta }}}\endgroup }}}}}}}}}}\endgroup } means there must exist some privacy cost {{\color{\colorMATH}\ensuremath{\epsilon }}} and {{\color{\colorMATH}\ensuremath{\delta }}}
which upper-bounds any individual cost for each of these closure variables. The
privacy cost of the whole loop is calculated based on this upper bound for
closure variables with the formula {\begingroup\renewcommand\colorMATH{\colorMATHC}\renewcommand\colorSYNTAX{\colorSYNTAXC}{{\color{\colorSYNTAX}\texttt{{\ensuremath{{\begingroup\renewcommand\colorMATH{\colorMATHA}\renewcommand\colorSYNTAX{\colorSYNTAXA}{{\color{\colorMATH}\ensuremath{2\epsilon \sqrt {2n\ln (1/\delta ^{\prime})}}}}\endgroup },{\begingroup\renewcommand\colorMATH{\colorMATHA}\renewcommand\colorSYNTAX{\colorSYNTAXA}{{\color{\colorMATH}\ensuremath{\delta ^{\prime}+n\delta }}}\endgroup }}}}}}\endgroup }. An
attempt to turn {\begingroup\renewcommand\colorMATH{\colorMATHC}\renewcommand\colorSYNTAX{\colorSYNTAXC}{{\color{\colorSYNTAX}\texttt{loop}}}\endgroup } into a primitive (or abstract over {\begingroup\renewcommand\colorMATH{\colorMATHC}\renewcommand\colorSYNTAX{\colorSYNTAXC}{{\color{\colorSYNTAX}\texttt{loop}}}\endgroup }, e.g.,
eta-expand via lambda abstraction) fails because privacy types in \duet do not
track privacy effects for closure variables; instead, they are just thrown away.
In the rule for function introduction \toplas{shown above (right), the function type
{{\color{\colorMATH}\ensuremath{\tau _{1}@(\epsilon ,\delta ) \multimap ^{*} \tau _{2}}}} is a probabilistic function from elements in {{\color{\colorMATH}\ensuremath{\tau _{1}}}} to elements
in {{\color{\colorMATH}\ensuremath{\tau _{2}}}} which satisfies {{\color{\colorMATH}\ensuremath{(\epsilon ,\delta )}}}-differential privacy in its argument.}
Notice the closure
environment {\begingroup\renewcommand\colorMATH{\colorMATHC}\renewcommand\colorSYNTAX{\colorSYNTAXC}{{\color{\colorMATH}\ensuremath{\Gamma }}}\endgroup } above the line which is bumped to {\begingroup\renewcommand\colorMATH{\colorMATHB}\renewcommand\colorSYNTAX{\colorSYNTAXB}{{\color{\colorSYNTAX}\texttt{{\ensuremath{\infty }}}}}\endgroup } in
{\begingroup\renewcommand\colorMATH{\colorMATHB}\renewcommand\colorSYNTAX{\colorSYNTAXB}{{\color{\colorMATH}\ensuremath{\rceil {\begingroup\renewcommand\colorMATH{\colorMATHC}\renewcommand\colorSYNTAX{\colorSYNTAXC}{{\color{\colorMATH}\ensuremath{\Gamma }}}\endgroup }\lceil ^{{{\color{\colorSYNTAX}\texttt{{\ensuremath{\infty }}}}}}}}}\endgroup } below the line. This has the effect of tossing out privacy
bounds for anything with non-zero privacy in {\begingroup\renewcommand\colorMATH{\colorMATHC}\renewcommand\colorSYNTAX{\colorSYNTAXC}{{\color{\colorMATH}\ensuremath{\Gamma }}}\endgroup }, \toplas{i.e., any closure variables that are used in the function's definition}. Privacy is only tracked for
the function parameter {{\color{\colorMATH}\ensuremath{x}}} (or possibly multiple parameters; privacy functions
in \duet are n-ary).

A deeper limitation in \duet is that the iterated 1-ary function space does not
generalize to support encoding n-ary functions (i.e., currification is not supported).
For this reason, n-ary
functions are primitive in \duet. Implementing n-ary from 1-ary functions is
computationally possible in \duet, but results in discarding bounds on privacy
effects. For example, the \duet term
\mbox{{{\color{\colorMATH}\ensuremath{{\begingroup\renewcommand\colorMATH{\colorMATHC}\renewcommand\colorSYNTAX{\colorSYNTAXC}{{\color{\colorMATH}\ensuremath{\plambda}}}\endgroup } (x \mathrel{:} \tau _{1}).\hspace*{0.33em} {\begingroup\renewcommand\colorMATH{\colorMATHC}\renewcommand\colorSYNTAX{\colorSYNTAXC}{{\color{\colorSYNTAX}\texttt{return}}}\endgroup }\hspace*{0.33em}({\begingroup\renewcommand\colorMATH{\colorMATHC}\renewcommand\colorSYNTAX{\colorSYNTAXC}{{\color{\colorMATH}\ensuremath{\plambda}}}\endgroup } (y \mathrel{:} \tau _{2}).\hspace*{0.33em} f(x, y))}}}}
in a
context where {{\color{\colorMATH}\ensuremath{f \mathrel{:} (\tau _{1}@(\epsilon _{1},\delta _{1}), \tau _{2}@(\epsilon _{2},\delta _{2})) \multimap ^{*} \tau _{3}}}} has type
{{\color{\colorMATH}\ensuremath{(\tau _{1}@\infty ) \multimap ^{*} (\tau _{2}@(\epsilon _{2},\delta _{2})) \multimap ^{*} \tau _{3}}}}, i.e., the privacy bounds
{{\color{\colorMATH}\ensuremath{(\epsilon _{1},\delta _{1})}}} for the first argument {{\color{\colorMATH}\ensuremath{\tau _{1}}}} get discarded due to the {\textsc{ \duet:
Privacy-Fun-I}} rule.

\paragraph{Privacy Closures in \system} % {-{

In \system, both privacy and sensitivity effects are delayed and attached to
type-level connectives, including for privacy functions. Whereas in \duet
privacy functions are written {{\color{\colorSYNTAX}\texttt{{\ensuremath{({{\color{\colorMATH}\ensuremath{\tau _{1}}}}@{\begingroup\renewcommand\colorMATH{\colorMATHC}\renewcommand\colorSYNTAX{\colorSYNTAXC}{{\color{\colorMATH}\ensuremath{p_{1}}}}\endgroup },{{\color{\colorMATH}\ensuremath{\ldots }}},{{\color{\colorMATH}\ensuremath{\tau _{n}}}}@{\begingroup\renewcommand\colorMATH{\colorMATHC}\renewcommand\colorSYNTAX{\colorSYNTAXC}{{\color{\colorMATH}\ensuremath{p_{n}}}}\endgroup }) \multimap ^{*} {{\color{\colorMATH}\ensuremath{\tau }}}}}}}}, privacy
functions in \system are written simply {{\color{\colorSYNTAX}\texttt{{\ensuremath{({{\color{\colorMATH}\ensuremath{x}}} \mathrel{:} {{\color{\colorMATH}\ensuremath{\tau _{1}}}}) \overset {{\begingroup\renewcommand\colorMATH{\colorMATHC}\renewcommand\colorSYNTAX{\colorSYNTAXC}{{\color{\colorMATH}\ensuremath{\Sigma }}}\endgroup }}{\mathrel{{\begingroup\renewcommand\colorMATH{\colorMATHC}\renewcommand\colorSYNTAX{\colorSYNTAXC}{{\color{\colorMATH}\ensuremath{\twoheadrightarrow }}}\endgroup }}} {{\color{\colorMATH}\ensuremath{\tau _{2}}}}}}}}} where
{\begingroup\renewcommand\colorMATH{\colorMATHC}\renewcommand\colorSYNTAX{\colorSYNTAXC}{{\color{\colorMATH}\ensuremath{\Sigma }}}\endgroup } is a latent contextual \longnamep effect that can mention {{\color{\colorMATH}\ensuremath{x}}}. A type can now be
given to {{\color{\colorMATH}\ensuremath{{\text{loop}}}}} (a named constant, analogous to the {\begingroup\renewcommand\colorMATH{\colorMATHC}\renewcommand\colorSYNTAX{\colorSYNTAXC}{{\color{\colorSYNTAX}\texttt{loop}}}\endgroup } {\textit{primitive}} from \duet) in \system, and abstracting over {{\color{\colorMATH}\ensuremath{{\text{loop}}}}} is
possible due to the function introduction rule, also shown below.
\begingroup\color{\colorMATH}\begin{gather*} % {-{

\end{gather*}\endgroup % }-}
N-ary functions are now recoverable from 1-ary ones using latent contextual \longnamep
effects in closures. The relational distance {{\color{\colorMATH}\ensuremath{{\begingroup\renewcommand\colorMATH{\colorMATHB}\renewcommand\colorSYNTAX{\colorSYNTAXB}{{\color{\colorMATH}\ensuremath{{\begingroup\renewcommand\colorMATH{\colorMATHB}\renewcommand\colorSYNTAX{\colorSYNTAXB}{{\color{\colorMATH}\ensuremath{\distance}}}\endgroup }}}}\endgroup }}}} defaults to {{\color{\colorMATH}\ensuremath{1}}} when omitted. The encoding
of lambda-abstracted {{\color{\colorMATH}\ensuremath{{\text{gauss}}}}} then follows the usual approach of
nested lambda abstractions, but with
sensitivity lambdas on the outside with a single privacy lambda on
the inside. A 3-ary abstraction of the Gaussian mechanism applied to
the sum of three arguments is as follows:

{{\color{\colorMATH}\ensuremath{
\begin{array}{l
} ({\begingroup\renewcommand\colorMATH{\colorMATHB}\renewcommand\colorSYNTAX{\colorSYNTAXB}{{\color{\colorMATH}\ensuremath{\slambda}}}\endgroup } (x\mathrel{:}{\mathbb{R}} \mathord{\cdotp } 1).\hspace*{0.33em} {\begingroup\renewcommand\colorMATH{\colorMATHB}\renewcommand\colorSYNTAX{\colorSYNTAXB}{{\color{\colorMATH}\ensuremath{\slambda}}}\endgroup } (y\mathrel{:}{\mathbb{R}} \mathord{\cdotp } 1).\hspace*{0.33em} {\begingroup\renewcommand\colorMATH{\colorMATHC}\renewcommand\colorSYNTAX{\colorSYNTAXC}{{\color{\colorMATH}\ensuremath{\plambda}}}\endgroup } (z\mathrel{:}{\mathbb{R}} \mathord{\cdotp } 1).\hspace*{0.33em}
\cr  \hspace*{1.00em}\hspace*{1.00em}\hspace*{1.00em}\hspace*{1.00em}\hspace*{1.00em}\hspace*{1.00em} {\text{gauss}}\hspace*{0.33em}\underbracket{3}_{\mathclap{{{\color{\colorTEXT}\textnormal{sensitivity sum of {{\color{\colorMATH}\ensuremath{(x + y + z)}}}}}}}}\hspace*{0.33em}
\overbracket{\epsilon \hspace*{0.33em}\delta }^{\mathclap{{{\color{\colorTEXT}\textnormal{desired privacy}}}}}\hspace*{0.33em}(x + y + z))
\mathrel{:}
(x\mathrel{:}{\mathbb{R}} \mathord{\cdotp } 1) \mathrel{{\begingroup\renewcommand\colorMATH{\colorMATHB}\renewcommand\colorSYNTAX{\colorSYNTAXB}{{\color{\colorMATH}\ensuremath{\rightarrow }}}\endgroup }} (y\mathrel{:}{\mathbb{R}} \mathord{\cdotp } 1) \mathrel{{\begingroup\renewcommand\colorMATH{\colorMATHB}\renewcommand\colorSYNTAX{\colorSYNTAXB}{{\color{\colorMATH}\ensuremath{\rightarrow }}}\endgroup }} (z\mathrel{:}{\mathbb{R}} \mathord{\cdotp } 1) \xrightarrowP {(\epsilon ,\delta )x \sqcup  (\epsilon ,\delta )y \sqcup  (\epsilon ,\delta )z} {\mathbb{R}}
\end{array}
}}}

Notice here that the latent contextual \longnamep effect is computed using a syntactic {\emph{join}} operator {{\color{\colorMATH}\ensuremath{(\epsilon ,\delta )x \sqcup  (\epsilon ,\delta )y \sqcup  (\epsilon ,\delta )z}}}, which computes the pointwise maximum, instead of the sum ({{\color{\colorMATH}\ensuremath{(\epsilon ,\delta )x + (\epsilon ,\delta )y + (\epsilon ,\delta )z}}}).
% Otherwise, if only one of the three variables wiggle, then the resulting privacy would be {{\color{\colorMATH}\ensuremath{(\epsilon ,\delta )}}} (and it would coincide with the current literature).
One of the novelties of \system is that we can reason about two executions where more than one input is at relational distance greater than 0. In particular, if {{\color{\colorMATH}\ensuremath{x}}}, {{\color{\colorMATH}\ensuremath{y}}} and {{\color{\colorMATH}\ensuremath{z}}} are at relational distance {{\color{\colorMATH}\ensuremath{1}}}, i.e. the argument of {{\color{\colorMATH}\ensuremath{{\text{gauss}}\hspace*{0.33em}3\hspace*{0.33em}\epsilon \hspace*{0.33em}\delta }}} is at relational distance {{\color{\colorMATH}\ensuremath{3}}}, then using addition would yield an over-approximated latent privacy of {{\color{\colorMATH}\ensuremath{(3\epsilon ,3\delta )}}}, while using the join, we obtain a latent privacy of {{\color{\colorMATH}\ensuremath{(\epsilon ,\delta )}}} as desired.

% }-}

\paragraph{Abstracting Privacy Mechanisms}  % {-{

Even with support for privacy closures, there are still challenges in
supporting lambda abstraction around privacy mechanisms in full generality.
In \fuzz, the type assigned to the family of Laplace differential privacy
mechanisms parameterized by {{\color{\colorMATH}\ensuremath{\epsilon }}} is {{\color{\colorMATH}\ensuremath{{\text{laplace}}_{{{\color{\colorMATH}\ensuremath{\epsilon }}}} \mathrel{:} {\mathbb{R}} \multimap _{\epsilon } {\scriptstyle \bigcirc } {\mathbb{R}}}}} for achieved
privacy {{\color{\colorMATH}\ensuremath{\epsilon }}}. This mechanism does not need a dedicated type rule in the core
calculus---it can be axiomatized as a primitive with the right type---and
lambda-abstracting this primitive is natural via eta-expansion {{\color{\colorSYNTAX}\texttt{{\ensuremath{ \lambda ({{\color{\colorMATH}\ensuremath{x}}}\mathrel{:}{\mathbb{R}}).\hspace*{0.33em}
{\text{laplace}}_{{{\color{\colorMATH}\ensuremath{\epsilon }}}}\hspace*{0.33em}x }}}}} resulting in the same type and guarantee for privacy. 
However this approach does not support $(\epsilon ,\delta )$-differential privacy 
\toplass{directly. \fuzzed shows how to extend \fuzz to recover $(\epsilon ,\delta )$-differential privacy, by using \toplasss{graded comonadic liftings}, and path construction. In particular, the type assigned to the family of Gaussian differential privacy mechanisms parametrized by {{\color{\colorMATH}\ensuremath{\varepsilon }}} and {{\color{\colorMATH}\ensuremath{\delta }}} is 
{{\color{\colorMATH}\ensuremath{{\text{gauss}}_{{{\color{\colorMATH}\ensuremath{(\epsilon ,\delta )}}}} \mathrel{:} \lceil{\mathbb{R}}\rceil \multimap  {\scriptstyle \bigcirc }_{(\epsilon ,\delta )} {\mathbb{R}}}}}, where {{\color{\colorMATH}\ensuremath{\lceil{\mathbb{R}}\rceil}}} is {{\color{\colorMATH}\ensuremath{{\mathbb{R}}}}} but with the metric rounded up to the nearest integer. 
%Now, the eta-expansion {{\color{\colorSYNTAX}\texttt{{\ensuremath{ \lambda ({{\color{\colorMATH}\ensuremath{x}}}\mathrel{:}{\mathbb{R}}).\hspace*{0.33em}
%{\text{gauss}}_{{{\color{\colorMATH}\ensuremath{(\epsilon ,\delta )}}}}\hspace*{0.33em}x }}}}} results in the same type and guarantee for privacy.
} %\mt{check}
In
\duet, in order to support $(\epsilon ,\delta )$-differential privacy, the Gaussian
mechanism requires its own typing rule, shown below.
Furthermore, a use of the mechanism looks like {\begingroup\renewcommand\colorMATH{\colorMATHC}\renewcommand\colorSYNTAX{\colorSYNTAXC}{{\color{\colorSYNTAX}\texttt{{\ensuremath{ {{\color{\colorSYNTAX}\texttt{gauss}}}_{{\begingroup\renewcommand\colorMATH{\colorMATHA}\renewcommand\colorSYNTAX{\colorSYNTAXA}{{\color{\colorMATH}\ensuremath{\epsilon ,\delta }}}\endgroup }}^{{\begingroup\renewcommand\colorMATH{\colorMATHA}\renewcommand\colorSYNTAX{\colorSYNTAXA}{{\color{\colorMATH}\ensuremath{s}}}\endgroup }}\hspace*{0.33em}{\begingroup\renewcommand\colorMATH{\colorMATHB}\renewcommand\colorSYNTAX{\colorSYNTAXB}{{\color{\colorMATH}\ensuremath{e}}}\endgroup } }}}}}\endgroup }
where the argument {\begingroup\renewcommand\colorMATH{\colorMATHB}\renewcommand\colorSYNTAX{\colorSYNTAXB}{{\color{\colorMATH}\ensuremath{e}}}\endgroup } is a term in the sensitivity language with sensitivity
bounded by {\begingroup\renewcommand\colorMATH{\colorMATHA}\renewcommand\colorSYNTAX{\colorSYNTAXA}{{\color{\colorMATH}\ensuremath{s}}}\endgroup }. Using privacy closures as described above, we can write
{{\color{\colorMATH}\ensuremath{{\begingroup\renewcommand\colorMATH{\colorMATHC}\renewcommand\colorSYNTAX{\colorSYNTAXC}{{\color{\colorMATH}\ensuremath{\plambda}}}\endgroup } (x\mathrel{:}{\mathbb{R}}).\hspace*{0.33em} {\text{gauss}}\hspace*{0.33em}1\hspace*{0.33em}\epsilon \hspace*{0.33em}\delta \hspace*{0.33em}x}}}, however note that
we have lost the ability to be parametric in {\begingroup\renewcommand\colorMATH{\colorMATHA}\renewcommand\colorSYNTAX{\colorSYNTAXA}{{\color{\colorMATH}\ensuremath{s}}}\endgroup }---it must be fixed to {\begingroup\renewcommand\colorMATH{\colorMATHA}\renewcommand\colorSYNTAX{\colorSYNTAXA}{{\color{\colorMATH}\ensuremath{1}}}\endgroup }.
This assumption that {{\color{\colorMATH}\ensuremath{{\text{gauss}}}}} will be called only with a {{\color{\colorMATH}\ensuremath{1}}}-sensitive
argument is enforced in the function application rule in \duet, also shown
below.
\begingroup\color{\colorMATH}\begin{gather*} % {-{

\end{gather*}\endgroup % }-}
In the rule for {\begingroup\renewcommand\colorMATH{\colorMATHC}\renewcommand\colorSYNTAX{\colorSYNTAXC}{{\color{\colorSYNTAX}\texttt{gauss}}}\endgroup } (left) it allows an argument of any sensitivity
{{\color{\colorMATH}\ensuremath{s}}}, however the privacy function application rule (right) restricts that
arguments must have sensitivity equal to {{\color{\colorMATH}\ensuremath{1}}}. Restricting {\begingroup\renewcommand\colorMATH{\colorMATHC}\renewcommand\colorSYNTAX{\colorSYNTAXC}{{\color{\colorSYNTAX}\texttt{gauss}}}\endgroup } to only
{{\color{\colorMATH}\ensuremath{1}}}-distance arguments can be overly restrictive (e.g.,
{{\color{\colorMATH}\ensuremath{{\begingroup\renewcommand\colorMATH{\colorMATHC}\renewcommand\colorSYNTAX{\colorSYNTAXC}{{\color{\colorSYNTAX}\texttt{gauss}}}\endgroup }\hspace*{0.33em}2\hspace*{0.33em}\epsilon \hspace*{0.33em}\delta \hspace*{0.33em}(x + x)}}}), and relaxing the restriction on function
application to an arbitrary {{\color{\colorMATH}\ensuremath{s \neq  1}}} in \duet would be unsound.

In \system, we extend function introduction to include an explicit bound on
the sensitivity of the parameter, and enforce this restriction in the application
rule. Function introduction syntax introduces the bound, and allows us to
eta-expand the Gaussian mechanism with relational distance {{\color{\colorMATH}\ensuremath{{\begingroup\renewcommand\colorMATH{\colorMATHB}\renewcommand\colorSYNTAX{\colorSYNTAXB}{{\color{\colorMATH}\ensuremath{\distance}}}\endgroup }}}} as a parameter, as shown below. The
bound {{\color{\colorMATH}\ensuremath{{\begingroup\renewcommand\colorMATH{\colorMATHB}\renewcommand\colorSYNTAX{\colorSYNTAXB}{{\color{\colorMATH}\ensuremath{\distance}}}\endgroup }}}} for the lambda argument is then enforced in function application as
the upper bound of argument relational distance, instead of being fixed to {{\color{\colorMATH}\ensuremath{1}}} as in
\duet. Now the use of a variable---like {{\color{\colorMATH}\ensuremath{x}}} in the body of eta-expanded
{\begingroup\renewcommand\colorMATH{\colorMATHC}\renewcommand\colorSYNTAX{\colorSYNTAXC}{{\color{\colorSYNTAX}\texttt{gauss}}}\endgroup }
below---is not always considered {{\color{\colorMATH}\ensuremath{1}}}-sensitive. To communicate non-zero
sensitivities to variables in the type system, an environment of relational distances on
lambda arguments must be threaded through the type system, which we notate
{{\color{\colorMATH}\ensuremath{{\begingroup\renewcommand\colorMATH{\colorMATHB}\renewcommand\colorSYNTAX{\colorSYNTAXB}{{\color{\colorMATH}\ensuremath{\Distance}}}\endgroup }}}}. After extending this {{\color{\colorMATH}\ensuremath{{\begingroup\renewcommand\colorMATH{\colorMATHB}\renewcommand\colorSYNTAX{\colorSYNTAXB}{{\color{\colorMATH}\ensuremath{\Distance}}}\endgroup }}}} to remember that {{\color{\colorMATH}\ensuremath{x}}} has relational distance {{\color{\colorMATH}\ensuremath{{\begingroup\renewcommand\colorMATH{\colorMATHB}\renewcommand\colorSYNTAX{\colorSYNTAXB}{{\color{\colorMATH}\ensuremath{\distance}}}\endgroup }}}} in
\system lambda abstraction, {{\color{\colorMATH}\ensuremath{{\text{gauss}}\hspace*{0.33em}s\hspace*{0.33em}\epsilon \hspace*{0.33em}\delta \hspace*{0.33em}x}}} will see {{\color{\colorMATH}\ensuremath{x}}} as {{\color{\colorMATH}\ensuremath{{\begingroup\renewcommand\colorMATH{\colorMATHB}\renewcommand\colorSYNTAX{\colorSYNTAXB}{{\color{\colorMATH}\ensuremath{\distance}}}\endgroup }}}}
distant inside the lambda body. To do this, we allow
lambda-abstracting gauss (including the distance parameter {{\color{\colorMATH}\ensuremath{{\begingroup\renewcommand\colorMATH{\colorMATHB}\renewcommand\colorSYNTAX{\colorSYNTAXB}{{\color{\colorMATH}\ensuremath{\distance}}}\endgroup }}}}, via
singleton types), and
extend the structure of typing for sensitivity and privacy terms
respectively as follows:
\begingroup\color{\colorMATH}\begin{gather*} % {-{

\end{gather*}\endgroup % }-}

% }-}

% }-}

\subsection{Sensitivity Binding in Privacy Contexts} % {-{
\label{sec:delay-sens}

\system improves on prior systems by supporting let-binding intermediate
sensitivity computations within the privacy language, while also supporting
{{\color{\colorMATH}\ensuremath{(\epsilon ,\delta )}}}-differential privacy. \fuzz{} and \dfuzz{} \toplas{encode let-binding through function application, which scales the sensitivity of the right-hand-side of the let with the}
sensitivity of the let-variable in the body. So {{\color{\colorMATH}\ensuremath{{{\color{\colorSYNTAX}\texttt{let}}}\hspace*{0.33em}y = 2 * x \hspace*{0.33em}{{\color{\colorSYNTAX}\texttt{in}}}\hspace*{0.33em} 3 * y}}}
is {{\color{\colorMATH}\ensuremath{6}}}-sensitive in {{\color{\colorMATH}\ensuremath{x}}} because the right-hand-side is {{\color{\colorMATH}\ensuremath{2}}}-sensitive, and this
is scaled by {{\color{\colorMATH}\ensuremath{3}}}, the sensitivity of {{\color{\colorMATH}\ensuremath{y}}} in the body. \toplas{However, monadic {{\color{\colorSYNTAX}\texttt{return}}} and bind in \fuzz can also be used to encode let-binding, e.g., {{\color{\colorMATH}\ensuremath{x \leftarrow  {\text{return}}\hspace*{0.33em}e_{1} \mathrel{;} e_{2}}}} instead of {{\color{\colorMATH}\ensuremath{{{\color{\colorSYNTAX}\texttt{let}}}\hspace*{0.33em}x=e_{1}\hspace*{0.33em}{{\color{\colorSYNTAX}\texttt{in}}}\hspace*{0.33em}e_{2}}}}.
Unfortunately, this encoding of {{\color{\colorSYNTAX}\texttt{let}}} using}
{{\color{\colorSYNTAX}\texttt{return}}} and monadic bind does not preserve typeability in \fuzz; instead it destroys
the sensitivity/privacy analysis of the right-hand-side, bumping its privacy
cost unnecessarily to {{\color{\colorMATH}\ensuremath{\infty }}}.  \toplas{For this reason, let statements are encoded exclusively through function application in \fuzz, and not through monadic {{\color{\colorSYNTAX}\texttt{return}}}/bind.}

In \fuzz, let-binding a sensitivity computation (the pure fragment) inside a
privacy computation (the monadic fragment)\toplas{---via encoding through function application---}is supported seamlessly without the
addition of extra rules\toplass{. This flexibility can be extended to advanced privacy variants as shown by \citet{de2019probabilistic}}. In \duet, however, the privacy/monadic fragment of
\fuzz is pulled out into its own language with explicit typing rules; the
primary reason to do this is to place restrictions on function application \toplas{in order to support advanced privacy variants}, as
described in the previous subsection. This leaves the need for either an
explicit typing rule for let-binding inside the privacy language, or an
escape hatch so that privacy analysis is not destroyed for let-binding in
privacy contexts a la \fuzz. \duet solves this issue by introducing a {\textit{boxed}}
type which delays the payment of a sensitivity term at the point it is ``boxed'',
and pays for it later when it is ``unboxed''. This avoids the issue but is
unfriendly to program with: every let-binding requires an explicit box, and
every use of a let-bound variable requires an explicit unbox. So instead of writing the program below on the left, \duet programmers are forced to write the
program on the right.
\begingroup\color{\colorMATH}\begin{gather*} % {-{

\end{gather*}\endgroup % }-}
In this program it is essential to let-bind the expensive result, since
inlining it would unnecessarily duplicate the computation, and many real
programs in differential privacy require support for this
pattern~\cite{near2019duet}.

In \system, we \toplas{recover the expressiveness that box types provide, while
eliminating the need for the programmer to explicitly introduce and eliminate them.
In this way, our design can also be seen as a powerful box-inference capability,
although we do not demonstrate explicit embeddings between a core language with
box types.
To recover the expressiveness of boxes without requiring the programmer to write
them down,} we add new information to
typing judgments that has the effect of automatically boxing let-bound
variables in privacy contexts, and unboxing them at their use. The added
information extends typing judgments with a new component {\begingroup\renewcommand\colorMATH{\colorMATHB}\renewcommand\colorSYNTAX{\colorSYNTAXB}{{\color{\colorMATH}\ensuremath{\Phi }}}\endgroup } that tracks
the sensitivities of all let-bound variables {w.r.t.} the sensitivities of
all lambda-bound variables. All sensitivity contexts that mention both
let-bound and lambda-bound variables are then reduced using {\begingroup\renewcommand\colorMATH{\colorMATHB}\renewcommand\colorSYNTAX{\colorSYNTAXB}{{\color{\colorMATH}\ensuremath{\Phi }}}\endgroup } as needed to
contexts that only mention lambda-bound variables. {\begingroup\renewcommand\colorMATH{\colorMATHB}\renewcommand\colorSYNTAX{\colorSYNTAXB}{{\color{\colorMATH}\ensuremath{\Phi }}}\endgroup } can be seen as a
matrix, and the reduction of contexts to only lambda-bound variables is then
just matrix multiplication---a beautiful coincidence for a linear type system.
The final form of type judgments for the sensitivity and privacy type
systems are then:
\begingroup\color{\colorMATH}\begin{gather*}
\begin{tabularx}{\linewidth}{>{\centering\arraybackslash\(}X<{\)}}
\hfill\hspace{0pt}
\Gamma \mathrel{;}{\begingroup\renewcommand\colorMATH{\colorMATHB}\renewcommand\colorSYNTAX{\colorSYNTAXB}{{\color{\colorMATH}\ensuremath{\Distance}}}\endgroup }\mathrel{;}{\begingroup\renewcommand\colorMATH{\colorMATHB}\renewcommand\colorSYNTAX{\colorSYNTAXB}{{\color{\colorMATH}\ensuremath{\Phi }}}\endgroup } \vdash  {\begingroup\renewcommand\colorMATH{\colorMATHB}\renewcommand\colorSYNTAX{\colorSYNTAXB}{{\color{\colorMATH}\ensuremath{e}}}\endgroup } \mathrel{:} \tau  \mathrel{;} {\begingroup\renewcommand\colorMATH{\colorMATHB}\renewcommand\colorSYNTAX{\colorSYNTAXB}{{\color{\colorMATH}\ensuremath{\Sigma }}}\endgroup }
\hfill\hspace{0pt}
\Gamma \mathrel{;}{\begingroup\renewcommand\colorMATH{\colorMATHB}\renewcommand\colorSYNTAX{\colorSYNTAXB}{{\color{\colorMATH}\ensuremath{\Distance}}}\endgroup }\mathrel{;}{\begingroup\renewcommand\colorMATH{\colorMATHB}\renewcommand\colorSYNTAX{\colorSYNTAXB}{{\color{\colorMATH}\ensuremath{\Phi }}}\endgroup } \vdash  {\begingroup\renewcommand\colorMATH{\colorMATHC}\renewcommand\colorSYNTAX{\colorSYNTAXC}{{\color{\colorMATH}\ensuremath{e}}}\endgroup } \mathrel{:} \tau  \mathrel{;} {\begingroup\renewcommand\colorMATH{\colorMATHC}\renewcommand\colorSYNTAX{\colorSYNTAXC}{{\color{\colorMATH}\ensuremath{\Sigma }}}\endgroup }
\hfill\hspace{0pt}
\end{tabularx}
\end{gather*}\endgroup
Although the prototype implementation adopts the typing rules
with {\begingroup\renewcommand\colorMATH{\colorMATHB}\renewcommand\colorSYNTAX{\colorSYNTAXB}{{\color{\colorMATH}\ensuremath{\Phi }}}\endgroup },  and because the manipulation of {\begingroup\renewcommand\colorMATH{\colorMATHB}\renewcommand\colorSYNTAX{\colorSYNTAXB}{{\color{\colorMATH}\ensuremath{\Phi }}}\endgroup } is \toplas{more tedious than insightful}, we omit it in
the following technical presentation.

\section{\system's Differential Privacy Type System, Formally} % {-{
\label{sec:privacy-formalism}

In this section, we present a core subset of \system, dubbed $\lang$. $\lang$ is an extension of \ssystem with support for reasoning about differential privacy.
Similarly to \ssystem, we prove the type safety and type soundness property of $\lang$. We discuss how to bridge the gap between $\lang$ and \system in Section~\ref{sec:system}.
Note that our formalism is fixed to {{\color{\colorMATH}\ensuremath{(\epsilon ,\delta )}}}-differential privacy, but our design can be instantiated to other forms of  advanced differential privacy disciplines as illustrated in Section~\ref{sec:system}.

\subsection{Syntax and Type System}
\label{sec:static-semantics}

$\lang$ is divided in two mutually embedded sublanguages: the \emph{sensitivity}
sublanguage ---an extension of \ssystem--- used to reason about the sensitivity of computations, and the
\emph{privacy} sublanguage used to reason about differential privacy.
Thus, the type system
of $\lang$ contains two mutually embedded type systems, one for each of the
sublanguages. Expressions of the sensitivity sublanguage remain typeset in {\begingroup\renewcommand\colorMATH{\colorMATHB}\renewcommand\colorSYNTAX{\colorSYNTAXB}{{\color{\colorMATH}\ensuremath{{\text{green}}}}}\endgroup } and expressions of the 
privacy sublanguage are typeset in {\begingroup\renewcommand\colorMATH{\colorMATHC}\renewcommand\colorSYNTAX{\colorSYNTAXC}{{\color{\colorMATH}\ensuremath{{\text{red}}}}}\endgroup }.
\paragraph{Syntax} Figure~\ref{fig:syntax} presents the syntax of $\lang$.
\begin{figure}[t]
  \begin{small}
  \begin{framed}
\begingroup\color{\colorMATH}\begin{gather*}
\end{gather*}\endgroup
\end{framed}
  \end{small}
  \caption{$\lang$: Syntax}
  \label{fig:syntax}
\end{figure}
Expressions of the language are divided into two mutually embedded expressions:
sensitivity expressions {{\color{\colorMATH}\ensuremath{{\begingroup\renewcommand\colorMATH{\colorMATHB}\renewcommand\colorSYNTAX{\colorSYNTAXB}{{\color{\colorMATH}\ensuremath{\se}}}\endgroup }}}} and privacy expressions {{\color{\colorMATH}\ensuremath{{\begingroup\renewcommand\colorMATH{\colorMATHC}\renewcommand\colorSYNTAX{\colorSYNTAXC}{{\color{\colorMATH}\ensuremath{\pe}}}\endgroup }}}}.
Sensitivity expressions are defined the same way as in \ssystem, except that functions are split into sensitivity lambdas {{\color{\colorMATH}\ensuremath{{\begingroup\renewcommand\colorMATH{\colorMATHB}\renewcommand\colorSYNTAX{\colorSYNTAXB}{{\color{\colorMATH}\ensuremath{\slambda}}}\endgroup } (x\mathrel{:}\tau  \mathord{\cdotp } {\begingroup\renewcommand\colorMATH{\colorMATHB}\renewcommand\colorSYNTAX{\colorSYNTAXB}{{\color{\colorMATH}\ensuremath{\distance}}}\endgroup }).\hspace*{0.33em}{\begingroup\renewcommand\colorMATH{\colorMATHB}\renewcommand\colorSYNTAX{\colorSYNTAXB}{{\color{\colorMATH}\ensuremath{\se}}}\endgroup }}}} and privacy lambdas {{\color{\colorMATH}\ensuremath{{\begingroup\renewcommand\colorMATH{\colorMATHC}\renewcommand\colorSYNTAX{\colorSYNTAXC}{{\color{\colorMATH}\ensuremath{\plambda}}}\endgroup } (x\mathrel{:}\tau  \mathord{\cdotp } {\begingroup\renewcommand\colorMATH{\colorMATHB}\renewcommand\colorSYNTAX{\colorSYNTAXB}{{\color{\colorMATH}\ensuremath{\distance}}}\endgroup }).\hspace*{0.33em}{\begingroup\renewcommand\colorMATH{\colorMATHC}\renewcommand\colorSYNTAX{\colorSYNTAXC}{{\color{\colorMATH}\ensuremath{\pe}}}\endgroup }}}}.
Note that the only difference between a sensitivity lambda and a privacy lambda
is that the body of a privacy lambda is a privacy expression {{\color{\colorMATH}\ensuremath{{\begingroup\renewcommand\colorMATH{\colorMATHC}\renewcommand\colorSYNTAX{\colorSYNTAXC}{{\color{\colorMATH}\ensuremath{\pe}}}\endgroup }}}}.
Also, both
sensitivity lambdas and privacy lambdas are parametrized by a relational distance {{\color{\colorMATH}\ensuremath{{\begingroup\renewcommand\colorMATH{\colorMATHB}\renewcommand\colorSYNTAX{\colorSYNTAXB}{{\color{\colorMATH}\ensuremath{\distance}}}\endgroup }}}}
which represents an upper bound on distance between inputs 
pertained to the binary relational property of differential privacy: the maximum argument variation for each of two executions. 
% an upper bound on the allowed argument variation when functions are invoked with two different arguments. 

A privacy
expression {{\color{\colorMATH}\ensuremath{{\begingroup\renewcommand\colorMATH{\colorMATHC}\renewcommand\colorSYNTAX{\colorSYNTAXC}{{\color{\colorMATH}\ensuremath{\pe}}}\endgroup }}}} can be a point distribution {{\color{\colorMATH}\ensuremath{{\begingroup\renewcommand\colorMATH{\colorMATHC}\renewcommand\colorSYNTAX{\colorSYNTAXC}{{\color{\colorSYNTAX}\texttt{return}}}\endgroup }\hspace*{0.33em}{\begingroup\renewcommand\colorMATH{\colorMATHB}\renewcommand\colorSYNTAX{\colorSYNTAXB}{{\color{\colorMATH}\ensuremath{\se}}}\endgroup }}}}, a sequential
composition {{\color{\colorMATH}\ensuremath{x:\tau \hspace*{0.33em}\leftarrow \hspace*{0.33em}{\begingroup\renewcommand\colorMATH{\colorMATHC}\renewcommand\colorSYNTAX{\colorSYNTAXC}{{\color{\colorMATH}\ensuremath{\pe}}}\endgroup }\mathrel{;}{\begingroup\renewcommand\colorMATH{\colorMATHC}\renewcommand\colorSYNTAX{\colorSYNTAXC}{{\color{\colorMATH}\ensuremath{\pe}}}\endgroup }}}}, an application {{\color{\colorMATH}\ensuremath{{\begingroup\renewcommand\colorMATH{\colorMATHC}\renewcommand\colorSYNTAX{\colorSYNTAXC}{{\color{\colorMATH}\ensuremath{\pe}}}\endgroup }\hspace*{0.33em}{\begingroup\renewcommand\colorMATH{\colorMATHC}\renewcommand\colorSYNTAX{\colorSYNTAXC}{{\color{\colorMATH}\ensuremath{\pe}}}\endgroup }}}}, a conditional {{\color{\colorMATH}\ensuremath{{\begingroup\renewcommand\colorMATH{\colorMATHC}\renewcommand\colorSYNTAX{\colorSYNTAXC}{{\color{\colorSYNTAX}\texttt{if}}}\endgroup }\hspace*{0.33em}{\begingroup\renewcommand\colorMATH{\colorMATHB}\renewcommand\colorSYNTAX{\colorSYNTAXB}{{\color{\colorMATH}\ensuremath{\se}}}\endgroup }\hspace*{0.33em}{\begingroup\renewcommand\colorMATH{\colorMATHC}\renewcommand\colorSYNTAX{\colorSYNTAXC}{{\color{\colorSYNTAX}\texttt{then}}}\endgroup }\hspace*{0.33em}{\begingroup\renewcommand\colorMATH{\colorMATHC}\renewcommand\colorSYNTAX{\colorSYNTAXC}{{\color{\colorMATH}\ensuremath{\pe}}}\endgroup }\hspace*{0.33em}{\begingroup\renewcommand\colorMATH{\colorMATHC}\renewcommand\colorSYNTAX{\colorSYNTAXC}{{\color{\colorSYNTAX}\texttt{else}}}\endgroup }\hspace*{0.33em}{\begingroup\renewcommand\colorMATH{\colorMATHC}\renewcommand\colorSYNTAX{\colorSYNTAXC}{{\color{\colorMATH}\ensuremath{\pe}}}\endgroup }}}}, a
case expression {{\color{\colorMATH}\ensuremath{{\begingroup\renewcommand\colorMATH{\colorMATHC}\renewcommand\colorSYNTAX{\colorSYNTAXC}{{\color{\colorSYNTAX}\texttt{case}}}\endgroup }\hspace*{0.33em}{\begingroup\renewcommand\colorMATH{\colorMATHB}\renewcommand\colorSYNTAX{\colorSYNTAXB}{{\color{\colorMATH}\ensuremath{\se}}}\endgroup }\hspace*{0.33em}{\begingroup\renewcommand\colorMATH{\colorMATHC}\renewcommand\colorSYNTAX{\colorSYNTAXC}{{\color{\colorSYNTAX}\texttt{of}}}\endgroup }\hspace*{0.33em}\{ x\Rightarrow {\begingroup\renewcommand\colorMATH{\colorMATHC}\renewcommand\colorSYNTAX{\colorSYNTAXC}{{\color{\colorMATH}\ensuremath{\pe}}}\endgroup }\} \hspace*{0.33em}\{ y\Rightarrow  {\begingroup\renewcommand\colorMATH{\colorMATHC}\renewcommand\colorSYNTAX{\colorSYNTAXC}{{\color{\colorMATH}\ensuremath{\pe}}}\endgroup }\} }}}, or a let {{\color{\colorMATH}\ensuremath{{\begingroup\renewcommand\colorMATH{\colorMATHC}\renewcommand\colorSYNTAX{\colorSYNTAXC}{{\color{\colorSYNTAX}\texttt{let}}}\endgroup }\hspace*{0.33em}x={\begingroup\renewcommand\colorMATH{\colorMATHB}\renewcommand\colorSYNTAX{\colorSYNTAXB}{{\color{\colorMATH}\ensuremath{\se}}}\endgroup }\hspace*{0.33em}{\begingroup\renewcommand\colorMATH{\colorMATHC}\renewcommand\colorSYNTAX{\colorSYNTAXC}{{\color{\colorSYNTAX}\texttt{in}}}\endgroup }\hspace*{0.33em}{\begingroup\renewcommand\colorMATH{\colorMATHC}\renewcommand\colorSYNTAX{\colorSYNTAXC}{{\color{\colorMATH}\ensuremath{\pe}}}\endgroup }}}}.

A privacy cost {{\color{\colorMATH}\ensuremath{{\begingroup\renewcommand\colorMATH{\colorMATHC}\renewcommand\colorSYNTAX{\colorSYNTAXC}{{\color{\colorMATH}\ensuremath{p}}}\endgroup } = ({\begingroup\renewcommand\colorMATH{\colorMATHC}\renewcommand\colorSYNTAX{\colorSYNTAXC}{{\color{\colorMATH}\ensuremath{\epsilon }}}\endgroup }, {\begingroup\renewcommand\colorMATH{\colorMATHC}\renewcommand\colorSYNTAX{\colorSYNTAXC}{{\color{\colorMATH}\ensuremath{\delta }}}\endgroup })}}} is a pair of
two (possibly-infinite) real numbers, where the first component corresponds to
the epsilon, and the second to the delta in {{\color{\colorMATH}\ensuremath{(\epsilon ,\delta )}}}-differential privacy. We
use notation {{\color{\colorMATH}\ensuremath{{\begingroup\renewcommand\colorMATH{\colorMATHC}\renewcommand\colorSYNTAX{\colorSYNTAXC}{{\color{\colorMATH}\ensuremath{p}}}\endgroup }.{\begingroup\renewcommand\colorMATH{\colorMATHC}\renewcommand\colorSYNTAX{\colorSYNTAXC}{{\color{\colorMATH}\ensuremath{\epsilon }}}\endgroup }}}} and {{\color{\colorMATH}\ensuremath{{\begingroup\renewcommand\colorMATH{\colorMATHC}\renewcommand\colorSYNTAX{\colorSYNTAXC}{{\color{\colorMATH}\ensuremath{p}}}\endgroup }.{\begingroup\renewcommand\colorMATH{\colorMATHC}\renewcommand\colorSYNTAX{\colorSYNTAXC}{{\color{\colorMATH}\ensuremath{\delta }}}\endgroup }}}} to extract {{\color{\colorMATH}\ensuremath{{\begingroup\renewcommand\colorMATH{\colorMATHC}\renewcommand\colorSYNTAX{\colorSYNTAXC}{{\color{\colorMATH}\ensuremath{\epsilon }}}\endgroup }}}} and {{\color{\colorMATH}\ensuremath{{\begingroup\renewcommand\colorMATH{\colorMATHC}\renewcommand\colorSYNTAX{\colorSYNTAXC}{{\color{\colorMATH}\ensuremath{\delta }}}\endgroup }}}} respectively. A privacy
environment {{\color{\colorMATH}\ensuremath{{\begingroup\renewcommand\colorMATH{\colorMATHC}\renewcommand\colorSYNTAX{\colorSYNTAXC}{{\color{\colorMATH}\ensuremath{\pS}}}\endgroup }}}} is either an empty environment {{\color{\colorMATH}\ensuremath{\varnothing }}}, a pair {{\color{\colorMATH}\ensuremath{{\begingroup\renewcommand\colorMATH{\colorMATHC}\renewcommand\colorSYNTAX{\colorSYNTAXC}{{\color{\colorMATH}\ensuremath{p}}}\endgroup }x}}} representing that variable {{\color{\colorMATH}\ensuremath{x}}} has privacy cost {{\color{\colorMATH}\ensuremath{{\begingroup\renewcommand\colorMATH{\colorMATHC}\renewcommand\colorSYNTAX{\colorSYNTAXC}{{\color{\colorMATH}\ensuremath{p}}}\endgroup }}}}, the addition {{\color{\colorMATH}\ensuremath{{\begingroup\renewcommand\colorMATH{\colorMATHC}\renewcommand\colorSYNTAX{\colorSYNTAXC}{{\color{\colorMATH}\ensuremath{\pS}}}\endgroup } + {\begingroup\renewcommand\colorMATH{\colorMATHC}\renewcommand\colorSYNTAX{\colorSYNTAXC}{{\color{\colorMATH}\ensuremath{\pS}}}\endgroup }}}} of two privacy \toplas{environments}, the join {{\color{\colorMATH}\ensuremath{{\begingroup\renewcommand\colorMATH{\colorMATHC}\renewcommand\colorSYNTAX{\colorSYNTAXC}{{\color{\colorMATH}\ensuremath{\pS}}}\endgroup } \sqcup  {\begingroup\renewcommand\colorMATH{\colorMATHC}\renewcommand\colorSYNTAX{\colorSYNTAXC}{{\color{\colorMATH}\ensuremath{\pS}}}\endgroup }}}} of two privacy \toplas{environments}, and the meet {{\color{\colorMATH}\ensuremath{{\begingroup\renewcommand\colorMATH{\colorMATHC}\renewcommand\colorSYNTAX{\colorSYNTAXC}{{\color{\colorMATH}\ensuremath{\pS}}}\endgroup } \sqcap  {\begingroup\renewcommand\colorMATH{\colorMATHC}\renewcommand\colorSYNTAX{\colorSYNTAXC}{{\color{\colorMATH}\ensuremath{\pS}}}\endgroup }}}} of two privacy \toplas{environments}. 
Similarly to
sensitivity environments, we also write privacy environments as first-order
polynomials when possible. For instance {{\color{\colorMATH}\ensuremath{{\begingroup\renewcommand\colorMATH{\colorMATHC}\renewcommand\colorSYNTAX{\colorSYNTAXC}{{\color{\colorMATH}\ensuremath{p_{1}}}}\endgroup }x + {\begingroup\renewcommand\colorMATH{\colorMATHC}\renewcommand\colorSYNTAX{\colorSYNTAXC}{{\color{\colorMATH}\ensuremath{p_{2}}}}\endgroup }x}}} can be written as {{\color{\colorMATH}\ensuremath{({\begingroup\renewcommand\colorMATH{\colorMATHC}\renewcommand\colorSYNTAX{\colorSYNTAXC}{{\color{\colorMATH}\ensuremath{p_{1}}}}\endgroup }+{\begingroup\renewcommand\colorMATH{\colorMATHC}\renewcommand\colorSYNTAX{\colorSYNTAXC}{{\color{\colorMATH}\ensuremath{p_{2}}}}\endgroup })x}}}, but 
{{\color{\colorMATH}\ensuremath{{\begingroup\renewcommand\colorMATH{\colorMATHC}\renewcommand\colorSYNTAX{\colorSYNTAXC}{{\color{\colorMATH}\ensuremath{p_{1}}}}\endgroup }x + ({\begingroup\renewcommand\colorMATH{\colorMATHC}\renewcommand\colorSYNTAX{\colorSYNTAXC}{{\color{\colorMATH}\ensuremath{p_{2}}}}\endgroup }x\sqcup {\begingroup\renewcommand\colorMATH{\colorMATHC}\renewcommand\colorSYNTAX{\colorSYNTAXC}{{\color{\colorMATH}\ensuremath{p_{3}}}}\endgroup }y)}}} cannot be rewritten as a polynomial without \toplass{losing} precision.
%, e.g. {{\color{\colorMATH}\ensuremath{{\begingroup\renewcommand\colorMATH{\colorMATHC}\renewcommand\colorSYNTAX{\colorSYNTAXC}{{\color{\colorMATH}\ensuremath{\pS}}}\endgroup } = {\begingroup\renewcommand\colorMATH{\colorMATHC}\renewcommand\colorSYNTAX{\colorSYNTAXC}{{\color{\colorMATH}\ensuremath{p_{1}}}}\endgroup }x + {\begingroup\renewcommand\colorMATH{\colorMATHC}\renewcommand\colorSYNTAX{\colorSYNTAXC}{{\color{\colorMATH}\ensuremath{p_{2}}}}\endgroup }y}}} correspond to environment such that {{\color{\colorMATH}\ensuremath{{\begingroup\renewcommand\colorMATH{\colorMATHC}\renewcommand\colorSYNTAX{\colorSYNTAXC}{{\color{\colorMATH}\ensuremath{\pS}}}\endgroup }(x) = {\begingroup\renewcommand\colorMATH{\colorMATHC}\renewcommand\colorSYNTAX{\colorSYNTAXC}{{\color{\colorMATH}\ensuremath{p_{1}}}}\endgroup }}}} and
%{{\color{\colorMATH}\ensuremath{{\begingroup\renewcommand\colorMATH{\colorMATHC}\renewcommand\colorSYNTAX{\colorSYNTAXC}{{\color{\colorMATH}\ensuremath{\pS}}}\endgroup }(y) = {\begingroup\renewcommand\colorMATH{\colorMATHC}\renewcommand\colorSYNTAX{\colorSYNTAXC}{{\color{\colorMATH}\ensuremath{p_{2}}}}\endgroup }}}}.
Function types are now divided into sensitivity function types {{\color{\colorMATH}\ensuremath{(x\mathrel{:}\tau \mathord{\cdotp }{\begingroup\renewcommand\colorMATH{\colorMATHB}\renewcommand\colorSYNTAX{\colorSYNTAXB}{{\color{\colorMATH}\ensuremath{\distance}}}\endgroup }) \xrightarrowS {{\begingroup\renewcommand\colorMATH{\colorMATHB}\renewcommand\colorSYNTAX{\colorSYNTAXB}{{\color{\colorMATH}\ensuremath{\sS}}}\endgroup }} \tau }}}, and privacy function types {{\color{\colorMATH}\ensuremath{(x\mathrel{:}\tau \mathord{\cdotp }{\begingroup\renewcommand\colorMATH{\colorMATHB}\renewcommand\colorSYNTAX{\colorSYNTAXB}{{\color{\colorMATH}\ensuremath{\distance}}}\endgroup }) \xrightarrowP {{\begingroup\renewcommand\colorMATH{\colorMATHC}\renewcommand\colorSYNTAX{\colorSYNTAXC}{{\color{\colorMATH}\ensuremath{\pS}}}\endgroup }} \tau }}}.

\begin{figure}[t]
  \begin{small}
  \begin{framed}
  \input{extension-sensitivity-type-system1}
  \end{framed}
  \end{small}
  \caption{$\lang$: Type system of the sensitivity sublanguage (extract)}
  \label{fig:sensitivity-type-system1}
\end{figure}
% \begin{figure}[t]
%   \begin{small}
%   \input{sensitivity-type-system2}
%   \end{small}
%   \caption{$\lang$: Type system of the sensitivity sublanguage (part 2)}
%   \label{fig:sensitivity-type-system2}
% \end{figure}
\paragraph{Sensitivity type system}
The type system for the sensitivity sublanguage is presented in
Figure~\ref{fig:sensitivity-type-system1}.
% and~\ref{fig:sensitivity-type-system2}.
The judgment {{\color{\colorMATH}\ensuremath{{{\begingroup\renewcommand\colorMATH{\colorMATHA}\renewcommand\colorSYNTAX{\colorSYNTAXA}{{\color{\colorMATH}\ensuremath{\Gamma }}}\endgroup } \mathrel{;} {\begingroup\renewcommand\colorMATH{\colorMATHB}\renewcommand\colorSYNTAX{\colorSYNTAXB}{{\color{\colorMATH}\ensuremath{\Distance}}}\endgroup }\hspace*{0.33em}{\begingroup\renewcommand\colorMATH{\colorMATHB}\renewcommand\colorSYNTAX{\colorSYNTAXB}{{\color{\colorMATH}\ensuremath{\vdash }}}\endgroup }\hspace*{0.33em}{\begingroup\renewcommand\colorMATH{\colorMATHB}\renewcommand\colorSYNTAX{\colorSYNTAXB}{{\color{\colorMATH}\ensuremath{\se}}}\endgroup } \mathrel{:} {\begingroup\renewcommand\colorMATH{\colorMATHA}\renewcommand\colorSYNTAX{\colorSYNTAXA}{{\color{\colorMATH}\ensuremath{\tau }}}\endgroup } \mathrel{;} {\begingroup\renewcommand\colorMATH{\colorMATHB}\renewcommand\colorSYNTAX{\colorSYNTAXB}{{\color{\colorMATH}\ensuremath{\sS}}}\endgroup }}}}} now includes a \toplas{novel} \distanceBoundName
environment {{\color{\colorMATH}\ensuremath{{\begingroup\renewcommand\colorMATH{\colorMATHB}\renewcommand\colorSYNTAX{\colorSYNTAXB}{{\color{\colorMATH}\ensuremath{\Distance}}}\endgroup }}}}. The \distanceBoundName environment {{\color{\colorMATH}\ensuremath{{\begingroup\renewcommand\colorMATH{\colorMATHB}\renewcommand\colorSYNTAX{\colorSYNTAXB}{{\color{\colorMATH}\ensuremath{\Distance}}}\endgroup }}}} stores how much each variable in {{\color{\colorMATH}\ensuremath{\Gamma }}} can vary in every two executions of a program.
Most of the rules are straightforward extensions of the type system of \ssystem to include \distanceBoundName environments. We only present interesting cases.

Some of the rules use the sensitivity environment substitution operator {{\color{\colorMATH}\ensuremath{[{\begingroup\renewcommand\colorMATH{\colorMATHB}\renewcommand\colorSYNTAX{\colorSYNTAXB}{{\color{\colorMATH}\ensuremath{\sS}}}\endgroup }/x]\tau }}}. We extend the definition of \ssystem to support privacy functions as shown in Figure~\ref{fig:privacy-statics-auxiliary-definitions}.
\begin{figure}[t]
 \begin{small}
 \begin{framed}
\begingroup\color{\colorMATH}\begin{gather*}
\begin{tabularx}{\linewidth}{>{\centering\arraybackslash\(}X<{\)}}\hfill\hspace{0pt}\begingroup\color{\colorTEXT}\boxed{\begingroup\color{\colorMATH} \subst{{\begingroup\renewcommand\colorMATH{\colorMATHB}\renewcommand\colorSYNTAX{\colorSYNTAXB}{{\color{\colorMATH}\ensuremath{\sS}}}\endgroup }}{x}\tau  \endgroup}\endgroup
  \cr \begin{array}{rcl
    } ... &{}={}& ...
	    \cr  \subst{{\begingroup\renewcommand\colorMATH{\colorMATHB}\renewcommand\colorSYNTAX{\colorSYNTAXB}{{\color{\colorMATH}\ensuremath{\sS}}}\endgroup }}{x}((y\mathrel{:}\tau _{1}\mathord{\cdotp }{\begingroup\renewcommand\colorMATH{\colorMATHB}\renewcommand\colorSYNTAX{\colorSYNTAXB}{{\color{\colorMATH}\ensuremath{\distance}}}\endgroup }) \xrightarrowP {{\begingroup\renewcommand\colorMATH{\colorMATHC}\renewcommand\colorSYNTAX{\colorSYNTAXC}{{\color{\colorMATH}\ensuremath{\pS'}}}\endgroup }} \tau _{2}) &{}={}& (y\mathrel{:}\subst{{\begingroup\renewcommand\colorMATH{\colorMATHB}\renewcommand\colorSYNTAX{\colorSYNTAXB}{{\color{\colorMATH}\ensuremath{\sS}}}\endgroup }}{x}\tau _{1}\mathord{\cdotp }{\begingroup\renewcommand\colorMATH{\colorMATHB}\renewcommand\colorSYNTAX{\colorSYNTAXB}{{\color{\colorMATH}\ensuremath{\distance}}}\endgroup }) \xrightarrowP {\subst{{\begingroup\renewcommand\colorMATH{\colorMATHB}\renewcommand\colorSYNTAX{\colorSYNTAXB}{{\color{\colorMATH}\ensuremath{\sS}}}\endgroup }}{x}{\begingroup\renewcommand\colorMATH{\colorMATHC}\renewcommand\colorSYNTAX{\colorSYNTAXC}{{\color{\colorMATH}\ensuremath{\pS'}}}\endgroup }} \subst{{\begingroup\renewcommand\colorMATH{\colorMATHB}\renewcommand\colorSYNTAX{\colorSYNTAXB}{{\color{\colorMATH}\ensuremath{\sS}}}\endgroup }}{x}\tau _{2}
	    \end{array}
  \cr
\end{tabularx}
\end{gather*}\endgroup
\end{framed}

 \end{small}
 \caption{$\lang$: Auxiliary definitions of the static semantics}
 \label{fig:privacy-statics-auxiliary-definitions}
\end{figure}
Substitution on privacy function types depends on the definition of sensitivity environment substitution on privacy environments {{\color{\colorMATH}\ensuremath{[{\begingroup\renewcommand\colorMATH{\colorMATHB}\renewcommand\colorSYNTAX{\colorSYNTAXB}{{\color{\colorMATH}\ensuremath{\sS}}}\endgroup }/x]{\begingroup\renewcommand\colorMATH{\colorMATHC}\renewcommand\colorSYNTAX{\colorSYNTAXC}{{\color{\colorMATH}\ensuremath{\pS}}}\endgroup }}}}.
{{\color{\colorMATH}\ensuremath{[{\begingroup\renewcommand\colorMATH{\colorMATHB}\renewcommand\colorSYNTAX{\colorSYNTAXB}{{\color{\colorMATH}\ensuremath{\sS}}}\endgroup }/x]{\begingroup\renewcommand\colorMATH{\colorMATHC}\renewcommand\colorSYNTAX{\colorSYNTAXC}{{\color{\colorMATH}\ensuremath{\pS}}}\endgroup }}}} is defined inductively on the structure of {{\color{\colorMATH}\ensuremath{{\begingroup\renewcommand\colorMATH{\colorMATHC}\renewcommand\colorSYNTAX{\colorSYNTAXC}{{\color{\colorMATH}\ensuremath{\pS}}}\endgroup }}}}, where the only interesting case is when {{\color{\colorMATH}\ensuremath{{\begingroup\renewcommand\colorMATH{\colorMATHC}\renewcommand\colorSYNTAX{\colorSYNTAXC}{{\color{\colorMATH}\ensuremath{\pS}}}\endgroup } = {\begingroup\renewcommand\colorMATH{\colorMATHC}\renewcommand\colorSYNTAX{\colorSYNTAXC}{{\color{\colorMATH}\ensuremath{p}}}\endgroup }x}}}. Substitution {{\color{\colorMATH}\ensuremath{[{\begingroup\renewcommand\colorMATH{\colorMATHB}\renewcommand\colorSYNTAX{\colorSYNTAXB}{{\color{\colorMATH}\ensuremath{\sS}}}\endgroup }/x]{\begingroup\renewcommand\colorMATH{\colorMATHC}\renewcommand\colorSYNTAX{\colorSYNTAXC}{{\color{\colorMATH}\ensuremath{p}}}\endgroup }x}}} is defined using the lift operator: {{\color{\colorMATH}\ensuremath{{\begingroup\renewcommand\colorMATH{\colorMATHC}\renewcommand\colorSYNTAX{\colorSYNTAXC}{{\color{\colorMATH}\ensuremath{\rceil {\begingroup\renewcommand\colorMATH{\colorMATHA}\renewcommand\colorSYNTAX{\colorSYNTAXA}{{\color{\colorMATH}\ensuremath{{\begingroup\renewcommand\colorMATH{\colorMATHB}\renewcommand\colorSYNTAX{\colorSYNTAXB}{{\color{\colorMATH}\ensuremath{\sS}}}\endgroup }}}}\endgroup }\lceil ^{p}}}}\endgroup }}}}.
Intuitively, if we wiggle~\footnote{We show how to wiggle variables on privacy environment with the relational distance instantiation operator, later on Section~\ref{sec:soundness}.} {{\color{\colorMATH}\ensuremath{x}}} on {{\color{\colorMATH}\ensuremath{{\begingroup\renewcommand\colorMATH{\colorMATHC}\renewcommand\colorSYNTAX{\colorSYNTAXC}{{\color{\colorMATH}\ensuremath{p}}}\endgroup }x}}}, then the privacy obtained is at most {{\color{\colorMATH}\ensuremath{{\begingroup\renewcommand\colorMATH{\colorMATHC}\renewcommand\colorSYNTAX{\colorSYNTAXC}{{\color{\colorMATH}\ensuremath{p}}}\endgroup }}}} (no scaling, and zero if {{\color{\colorMATH}\ensuremath{x}}} does not change). After substitution, as {{\color{\colorMATH}\ensuremath{x}}} depends on all variables on {{\color{\colorMATH}\ensuremath{{\begingroup\renewcommand\colorMATH{\colorMATHB}\renewcommand\colorSYNTAX{\colorSYNTAXB}{{\color{\colorMATH}\ensuremath{\sS}}}\endgroup }}}}, if we wiggle all \toplas{variables} in {{\color{\colorMATH}\ensuremath{{\begingroup\renewcommand\colorMATH{\colorMATHB}\renewcommand\colorSYNTAX{\colorSYNTAXB}{{\color{\colorMATH}\ensuremath{\sS}}}\endgroup }}}} at the same time, then the privacy obtained should still be {{\color{\colorMATH}\ensuremath{{\begingroup\renewcommand\colorMATH{\colorMATHC}\renewcommand\colorSYNTAX{\colorSYNTAXC}{{\color{\colorMATH}\ensuremath{p}}}\endgroup }}}} (scaling {{\color{\colorMATH}\ensuremath{{\begingroup\renewcommand\colorMATH{\colorMATHC}\renewcommand\colorSYNTAX{\colorSYNTAXC}{{\color{\colorMATH}\ensuremath{p}}}\endgroup }}}} would be an over approximation). 
Because of this, {{\color{\colorMATH}\ensuremath{{\begingroup\renewcommand\colorMATH{\colorMATHC}\renewcommand\colorSYNTAX{\colorSYNTAXC}{{\color{\colorMATH}\ensuremath{\rceil {\begingroup\renewcommand\colorMATH{\colorMATHA}\renewcommand\colorSYNTAX{\colorSYNTAXA}{{\color{\colorMATH}\ensuremath{{\begingroup\renewcommand\colorMATH{\colorMATHB}\renewcommand\colorSYNTAX{\colorSYNTAXB}{{\color{\colorMATH}\ensuremath{\sS}}}\endgroup }}}}\endgroup }\lceil ^{p}}}}\endgroup }}}} is defined as the join {{\color{\colorMATH}\ensuremath{{\begingroup\renewcommand\colorMATH{\colorMATHC}\renewcommand\colorSYNTAX{\colorSYNTAXC}{{\color{\colorMATH}\ensuremath{p}}}\endgroup }x_{1} \sqcup  ... \sqcup  {\begingroup\renewcommand\colorMATH{\colorMATHC}\renewcommand\colorSYNTAX{\colorSYNTAXC}{{\color{\colorMATH}\ensuremath{p}}}\endgroup }x_{n}}}}, where {{\color{\colorMATH}\ensuremath{x_{i} \in  dom({\begingroup\renewcommand\colorMATH{\colorMATHB}\renewcommand\colorSYNTAX{\colorSYNTAXB}{{\color{\colorMATH}\ensuremath{\sS}}}\endgroup })}}}. 
If all {{\color{\colorMATH}\ensuremath{x_{i}}}} wiggle, then the privacy obtained would be at most {{\color{\colorMATH}\ensuremath{{\begingroup\renewcommand\colorMATH{\colorMATHC}\renewcommand\colorSYNTAX{\colorSYNTAXC}{{\color{\colorMATH}\ensuremath{p}}}\endgroup }}}}.
But as any {{\color{\colorMATH}\ensuremath{{\begingroup\renewcommand\colorMATH{\colorMATHB}\renewcommand\colorSYNTAX{\colorSYNTAXB}{{\color{\colorMATH}\ensuremath{\sS}}}\endgroup }(x_{i})}}} can be zero (it means that variable {{\color{\colorMATH}\ensuremath{x_{i}}}} is not used), we multiply each {{\color{\colorMATH}\ensuremath{{\begingroup\renewcommand\colorMATH{\colorMATHC}\renewcommand\colorSYNTAX{\colorSYNTAXC}{{\color{\colorMATH}\ensuremath{p}}}\endgroup }}}} in {{\color{\colorMATH}\ensuremath{{\begingroup\renewcommand\colorMATH{\colorMATHC}\renewcommand\colorSYNTAX{\colorSYNTAXC}{{\color{\colorMATH}\ensuremath{p}}}\endgroup }x_{i}}}}, by {{\color{\colorMATH}\ensuremath{{\begingroup\renewcommand\colorMATH{\colorMATHB}\renewcommand\colorSYNTAX{\colorSYNTAXB}{{\color{\colorMATH}\ensuremath{\rceil {\begingroup\renewcommand\colorMATH{\colorMATHA}\renewcommand\colorSYNTAX{\colorSYNTAXA}{{\color{\colorMATH}\ensuremath{{\begingroup\renewcommand\colorMATH{\colorMATHB}\renewcommand\colorSYNTAX{\colorSYNTAXB}{{\color{\colorMATH}\ensuremath{\sS}}}\endgroup }(x_{i})}}}\endgroup }\lceil ^{1}}}}\endgroup }}}}, to remove those variables from the resulting privacy environment.
{{\color{\colorMATH}\ensuremath{{\begingroup\renewcommand\colorMATH{\colorMATHB}\renewcommand\colorSYNTAX{\colorSYNTAXB}{{\color{\colorMATH}\ensuremath{\rceil {\begingroup\renewcommand\colorMATH{\colorMATHA}\renewcommand\colorSYNTAX{\colorSYNTAXA}{{\color{\colorMATH}\ensuremath{s}}}\endgroup }\lceil ^{s'}}}}\endgroup }}}} along other lift operators used by~\citet{near2019duet} are defined in Figure~\ref{fig:privacy-statics-auxiliary-definitions}.
For instance, suppose that {{\color{\colorMATH}\ensuremath{x}}} depends on {{\color{\colorMATH}\ensuremath{2y+0z}}}, then {{\color{\colorMATH}\ensuremath{[(2y+0z)/x]{\begingroup\renewcommand\colorMATH{\colorMATHC}\renewcommand\colorSYNTAX{\colorSYNTAXC}{{\color{\colorMATH}\ensuremath{p}}}\endgroup }x}}} is computed as {{\color{\colorMATH}\ensuremath{{\begingroup\renewcommand\colorMATH{\colorMATHC}\renewcommand\colorSYNTAX{\colorSYNTAXC}{{\color{\colorMATH}\ensuremath{\rceil {\begingroup\renewcommand\colorMATH{\colorMATHA}\renewcommand\colorSYNTAX{\colorSYNTAXA}{{\color{\colorMATH}\ensuremath{2y+0z}}}\endgroup }\lceil ^{p}}}}\endgroup } = {\begingroup\renewcommand\colorMATH{\colorMATHC}\renewcommand\colorSYNTAX{\colorSYNTAXC}{{\color{\colorMATH}\ensuremath{p}}}\endgroup }y \sqcup  0z = {\begingroup\renewcommand\colorMATH{\colorMATHC}\renewcommand\colorSYNTAX{\colorSYNTAXC}{{\color{\colorMATH}\ensuremath{p}}}\endgroup }y}}}.

We now turn to describe the main changes of each type rule with respect to \ssystem.
\begin{itemize}[label=\textbf{-},leftmargin=*]\item  Rule{\textsc{ var}} now requires variable {{\color{\colorMATH}\ensuremath{x}}} to be present in the \distanceBoundName
   environment {{\color{\colorMATH}\ensuremath{{\begingroup\renewcommand\colorMATH{\colorMATHB}\renewcommand\colorSYNTAX{\colorSYNTAXB}{{\color{\colorMATH}\ensuremath{\Distance}}}\endgroup }}}}. This way, if {{\color{\colorMATH}\ensuremath{\Gamma ; {\begingroup\renewcommand\colorMATH{\colorMATHB}\renewcommand\colorSYNTAX{\colorSYNTAXB}{{\color{\colorMATH}\ensuremath{\Distance}}}\endgroup } \vdash  {\begingroup\renewcommand\colorMATH{\colorMATHB}\renewcommand\colorSYNTAX{\colorSYNTAXB}{{\color{\colorMATH}\ensuremath{\se}}}\endgroup }: \tau ; {\begingroup\renewcommand\colorMATH{\colorMATHB}\renewcommand\colorSYNTAX{\colorSYNTAXB}{{\color{\colorMATH}\ensuremath{\sS}}}\endgroup }}}}, we can compute how much
   the result of evaluating {{\color{\colorMATH}\ensuremath{{\begingroup\renewcommand\colorMATH{\colorMATHB}\renewcommand\colorSYNTAX{\colorSYNTAXB}{{\color{\colorMATH}\ensuremath{\se}}}\endgroup }}}} can change if we wiggle input {{\color{\colorMATH}\ensuremath{x}}}, by
   multiplying {{\color{\colorMATH}\ensuremath{ {\begingroup\renewcommand\colorMATH{\colorMATHB}\renewcommand\colorSYNTAX{\colorSYNTAXB}{{\color{\colorMATH}\ensuremath{\Distance}}}\endgroup }(x)}}} by {{\color{\colorMATH}\ensuremath{{\begingroup\renewcommand\colorMATH{\colorMATHB}\renewcommand\colorSYNTAX{\colorSYNTAXB}{{\color{\colorMATH}\ensuremath{\sS}}}\endgroup }(x)}}}.
   For instance, consider program {{\color{\colorMATH}\ensuremath{x+x}}} and the following type derivations
   \begingroup\color{\colorMATH}\begin{mathpar} {\mathcal{D}} = \infer[var]{(x\mathrel{:}{\begingroup\renewcommand\colorMATH{\colorMATHA}\renewcommand\colorSYNTAX{\colorSYNTAXA}{{\color{\colorSYNTAX}\texttt{{\ensuremath{{\mathbb{R}}}}}}}\endgroup })(x) = {\begingroup\renewcommand\colorMATH{\colorMATHA}\renewcommand\colorSYNTAX{\colorSYNTAXA}{{\color{\colorSYNTAX}\texttt{{\ensuremath{{\mathbb{R}}}}}}}\endgroup }}{x\mathrel{:}{\begingroup\renewcommand\colorMATH{\colorMATHA}\renewcommand\colorSYNTAX{\colorSYNTAXA}{{\color{\colorSYNTAX}\texttt{{\ensuremath{{\mathbb{R}}}}}}}\endgroup }; 3x\hspace*{0.33em} {\begingroup\renewcommand\colorMATH{\colorMATHB}\renewcommand\colorSYNTAX{\colorSYNTAXB}{{\color{\colorMATH}\ensuremath{\vdash }}}\endgroup }\hspace*{0.33em}x \mathrel{:} {\begingroup\renewcommand\colorMATH{\colorMATHA}\renewcommand\colorSYNTAX{\colorSYNTAXA}{{\color{\colorSYNTAX}\texttt{{\ensuremath{{\mathbb{R}}}}}}}\endgroup } \mathrel{;} x}
   \and \infer[plus]{{\mathcal{D}} \and {\mathcal{D}}}{x\mathrel{:}{\begingroup\renewcommand\colorMATH{\colorMATHA}\renewcommand\colorSYNTAX{\colorSYNTAXA}{{\color{\colorSYNTAX}\texttt{{\ensuremath{{\mathbb{R}}}}}}}\endgroup }; 3x\hspace*{0.33em} {\begingroup\renewcommand\colorMATH{\colorMATHB}\renewcommand\colorSYNTAX{\colorSYNTAXB}{{\color{\colorMATH}\ensuremath{\vdash }}}\endgroup }\hspace*{0.33em}x+x \mathrel{:} {\begingroup\renewcommand\colorMATH{\colorMATHA}\renewcommand\colorSYNTAX{\colorSYNTAXA}{{\color{\colorSYNTAX}\texttt{{\ensuremath{{\mathbb{R}}}}}}}\endgroup } \mathrel{;} 2x}
   \end{mathpar}\endgroup
   Then we know that (1) {{\color{\colorMATH}\ensuremath{x}}} can change at most by
   {{\color{\colorMATH}\ensuremath{3}}}, and (2) the expression is {{\color{\colorMATH}\ensuremath{2}}}-sensitive in {{\color{\colorMATH}\ensuremath{x}}}, therefore the result can change at most
   by {{\color{\colorMATH}\ensuremath{6}}}.
\item  Rule{\textsc{ s-lam}} type checks sensitivity functions. As the body of the lambda has
   {{\color{\colorMATH}\ensuremath{x}}} as a free variable, the \distanceBoundName environment {{\color{\colorMATH}\ensuremath{{\begingroup\renewcommand\colorMATH{\colorMATHB}\renewcommand\colorSYNTAX{\colorSYNTAXB}{{\color{\colorMATH}\ensuremath{\Distance}}}\endgroup }}}} is extended with
   distance {{\color{\colorMATH}\ensuremath{{\begingroup\renewcommand\colorMATH{\colorMATHB}\renewcommand\colorSYNTAX{\colorSYNTAXB}{{\color{\colorMATH}\ensuremath{\distance}}}\endgroup }}}} obtained from the type annotation on the argument.
   For example, consider program {{\color{\colorMATH}\ensuremath{{\begingroup\renewcommand\colorMATH{\colorMATHB}\renewcommand\colorSYNTAX{\colorSYNTAXB}{{\color{\colorMATH}\ensuremath{\slambda}}}\endgroup } (x\mathrel{:}{\begingroup\renewcommand\colorMATH{\colorMATHA}\renewcommand\colorSYNTAX{\colorSYNTAXA}{{\color{\colorSYNTAX}\texttt{{\ensuremath{{\mathbb{R}}}}}}}\endgroup }\mathord{\cdotp }2).\hspace*{0.33em}x+x}}} and its type derivation:
   \begingroup\color{\colorMATH}\begin{gather*} 
     \inferrule*[lab={\textsc{ s-lam}}
     ]{ x\mathrel{:}{\begingroup\renewcommand\colorMATH{\colorMATHA}\renewcommand\colorSYNTAX{\colorSYNTAXA}{{\color{\colorSYNTAX}\texttt{{\ensuremath{{\mathbb{R}}}}}}}\endgroup }; 2x \hspace*{0.33em}{\begingroup\renewcommand\colorMATH{\colorMATHB}\renewcommand\colorSYNTAX{\colorSYNTAXB}{{\color{\colorMATH}\ensuremath{\vdash }}}\endgroup }\hspace*{0.33em}x+x \mathrel{:} {\begingroup\renewcommand\colorMATH{\colorMATHA}\renewcommand\colorSYNTAX{\colorSYNTAXA}{{\color{\colorSYNTAX}\texttt{{\ensuremath{{\mathbb{R}}}}}}}\endgroup } \mathrel{;} 2x
        }{
        \varnothing \hspace*{0.33em}; \varnothing  {\begingroup\renewcommand\colorMATH{\colorMATHB}\renewcommand\colorSYNTAX{\colorSYNTAXB}{{\color{\colorMATH}\ensuremath{\vdash }}}\endgroup }\hspace*{0.33em}{\begingroup\renewcommand\colorMATH{\colorMATHB}\renewcommand\colorSYNTAX{\colorSYNTAXB}{{\color{\colorMATH}\ensuremath{\slambda}}}\endgroup } (x\mathrel{:}{\begingroup\renewcommand\colorMATH{\colorMATHA}\renewcommand\colorSYNTAX{\colorSYNTAXA}{{\color{\colorSYNTAX}\texttt{{\ensuremath{{\mathbb{R}}}}}}}\endgroup }\mathord{\cdotp }2).\hspace*{0.33em}x+x \mathrel{:} (x\mathrel{:}{\begingroup\renewcommand\colorMATH{\colorMATHA}\renewcommand\colorSYNTAX{\colorSYNTAXA}{{\color{\colorSYNTAX}\texttt{{\ensuremath{{\mathbb{R}}}}}}}\endgroup }\mathord{\cdotp }2) \xrightarrowS {2x} {\begingroup\renewcommand\colorMATH{\colorMATHA}\renewcommand\colorSYNTAX{\colorSYNTAXA}{{\color{\colorSYNTAX}\texttt{{\ensuremath{{\mathbb{R}}}}}}}\endgroup } \mathrel{;} {\begingroup\renewcommand\colorMATH{\colorMATHB}\renewcommand\colorSYNTAX{\colorSYNTAXB}{{\color{\colorMATH}\ensuremath{\varnothing }}}\endgroup }
     }
   \end{gather*}\endgroup
   The program is a sensitivity lambda that takes as argument a real with an allowed variation of at most {{\color{\colorMATH}\ensuremath{2}}}, and has a latent contextual effect of {{\color{\colorMATH}\ensuremath{2x}}}.
\item  Rule{\textsc{ p-lam}} is defined analogously to {\textsc{ s-lam}}, except that its body is a
   privacy term, therefore it is type checked using the privacy type system,
   explained later.
\item  Rule{\textsc{ s-app}} deals with sensitivity applications. Note that from the type of the function we
   know that {{\color{\colorMATH}\ensuremath{{\begingroup\renewcommand\colorMATH{\colorMATHB}\renewcommand\colorSYNTAX{\colorSYNTAXB}{{\color{\colorMATH}\ensuremath{\distance}}}\endgroup }}}} is an upper bound on the allowed argument variation, therefore we require
   that the dot product between the \distanceBoundName environment and the sensitivity
   effect of the argument be less or equal than {{\color{\colorMATH}\ensuremath{{\begingroup\renewcommand\colorMATH{\colorMATHB}\renewcommand\colorSYNTAX{\colorSYNTAXB}{{\color{\colorMATH}\ensuremath{\distance}}}\endgroup }}}}. Intuitively, as {{\color{\colorMATH}\ensuremath{{\begingroup\renewcommand\colorMATH{\colorMATHB}\renewcommand\colorSYNTAX{\colorSYNTAXB}{{\color{\colorMATH}\ensuremath{\Distance}}}\endgroup }}}}
   represents how much the input can change, and {{\color{\colorMATH}\ensuremath{{\begingroup\renewcommand\colorMATH{\colorMATHB}\renewcommand\colorSYNTAX{\colorSYNTAXB}{{\color{\colorMATH}\ensuremath{\sS_{2}}}}\endgroup }}}} represent the sensitivity of
   variables used in the argument, {{\color{\colorMATH}\ensuremath{{\begingroup\renewcommand\colorMATH{\colorMATHB}\renewcommand\colorSYNTAX{\colorSYNTAXB}{{\color{\colorMATH}\ensuremath{\Distance}}}\endgroup } \mathord{\cdotp } {\begingroup\renewcommand\colorMATH{\colorMATHB}\renewcommand\colorSYNTAX{\colorSYNTAXB}{{\color{\colorMATH}\ensuremath{\sS_{2}}}}\endgroup }}}} represents how much the argument can change.
   For example, consider program {{\color{\colorMATH}\ensuremath{{\begingroup\renewcommand\colorMATH{\colorMATHB}\renewcommand\colorSYNTAX{\colorSYNTAXB}{{\color{\colorMATH}\ensuremath{\slambda}}}\endgroup } (y:{\begingroup\renewcommand\colorMATH{\colorMATHA}\renewcommand\colorSYNTAX{\colorSYNTAXA}{{\color{\colorSYNTAX}\texttt{{\ensuremath{{\mathbb{R}}}}}}}\endgroup }\mathord{\cdotp }1).\hspace*{0.33em}({\begingroup\renewcommand\colorMATH{\colorMATHB}\renewcommand\colorSYNTAX{\colorSYNTAXB}{{\color{\colorMATH}\ensuremath{\slambda}}}\endgroup } (x\mathrel{:}{\begingroup\renewcommand\colorMATH{\colorMATHA}\renewcommand\colorSYNTAX{\colorSYNTAXA}{{\color{\colorSYNTAX}\texttt{{\ensuremath{{\mathbb{R}}}}}}}\endgroup }\mathord{\cdotp }2).\hspace*{0.33em}x+x) (y+y)}}}  and its type derivation:

   \begingroup\color{\colorMATH}\begin{gather*}
     \inferrule*[lab={\textsc{ s-lam}}
     ]{
        \inferrule*[lab={\textsc{ s-app}}
        ]{ y:{\begingroup\renewcommand\colorMATH{\colorMATHA}\renewcommand\colorSYNTAX{\colorSYNTAXA}{{\color{\colorSYNTAX}\texttt{{\ensuremath{{\mathbb{R}}}}}}}\endgroup }\hspace*{0.33em}; 1y \hspace*{0.33em} {\begingroup\renewcommand\colorMATH{\colorMATHB}\renewcommand\colorSYNTAX{\colorSYNTAXB}{{\color{\colorMATH}\ensuremath{\vdash }}}\endgroup }\hspace*{0.33em}({\begingroup\renewcommand\colorMATH{\colorMATHB}\renewcommand\colorSYNTAX{\colorSYNTAXB}{{\color{\colorMATH}\ensuremath{\slambda}}}\endgroup } (x\mathrel{:}{\begingroup\renewcommand\colorMATH{\colorMATHA}\renewcommand\colorSYNTAX{\colorSYNTAXA}{{\color{\colorSYNTAX}\texttt{{\ensuremath{{\mathbb{R}}}}}}}\endgroup }\mathord{\cdotp }2).\hspace*{0.33em}x+x) : (x\mathrel{:}{\begingroup\renewcommand\colorMATH{\colorMATHA}\renewcommand\colorSYNTAX{\colorSYNTAXA}{{\color{\colorSYNTAX}\texttt{{\ensuremath{{\mathbb{R}}}}}}}\endgroup }\mathord{\cdotp }2) \xrightarrowS {2x} {\begingroup\renewcommand\colorMATH{\colorMATHA}\renewcommand\colorSYNTAX{\colorSYNTAXA}{{\color{\colorSYNTAX}\texttt{{\ensuremath{{\mathbb{R}}}}}}}\endgroup }
        \\ y:{\begingroup\renewcommand\colorMATH{\colorMATHA}\renewcommand\colorSYNTAX{\colorSYNTAXA}{{\color{\colorSYNTAX}\texttt{{\ensuremath{{\mathbb{R}}}}}}}\endgroup }\hspace*{0.33em}; 1y \hspace*{0.33em} {\begingroup\renewcommand\colorMATH{\colorMATHB}\renewcommand\colorSYNTAX{\colorSYNTAXB}{{\color{\colorMATH}\ensuremath{\vdash }}}\endgroup } (2*y) : {\begingroup\renewcommand\colorMATH{\colorMATHA}\renewcommand\colorSYNTAX{\colorSYNTAXA}{{\color{\colorSYNTAX}\texttt{{\ensuremath{{\mathbb{R}}}}}}}\endgroup } ; 2y
        \\ \begingroup\color{\colorTEXT}\boxed{\begingroup\color{\colorMATH} (1y \mathord{\cdotp } 2y) \leq  2 \endgroup}\endgroup
          }{
          y:{\begingroup\renewcommand\colorMATH{\colorMATHA}\renewcommand\colorSYNTAX{\colorSYNTAXA}{{\color{\colorSYNTAX}\texttt{{\ensuremath{{\mathbb{R}}}}}}}\endgroup }\hspace*{0.33em}; 1y \hspace*{0.33em} {\begingroup\renewcommand\colorMATH{\colorMATHB}\renewcommand\colorSYNTAX{\colorSYNTAXB}{{\color{\colorMATH}\ensuremath{\vdash }}}\endgroup }\hspace*{0.33em}({\begingroup\renewcommand\colorMATH{\colorMATHB}\renewcommand\colorSYNTAX{\colorSYNTAXB}{{\color{\colorMATH}\ensuremath{\slambda}}}\endgroup } (x\mathrel{:}{\begingroup\renewcommand\colorMATH{\colorMATHA}\renewcommand\colorSYNTAX{\colorSYNTAXA}{{\color{\colorSYNTAX}\texttt{{\ensuremath{{\mathbb{R}}}}}}}\endgroup }\mathord{\cdotp }2).\hspace*{0.33em}x+x) (2*y) \mathrel{:} {\begingroup\renewcommand\colorMATH{\colorMATHA}\renewcommand\colorSYNTAX{\colorSYNTAXA}{{\color{\colorSYNTAX}\texttt{{\ensuremath{{\mathbb{R}}}}}}}\endgroup } \mathrel{;} 4y
        }
        }{
        \varnothing \hspace*{0.33em}; \varnothing  {\begingroup\renewcommand\colorMATH{\colorMATHB}\renewcommand\colorSYNTAX{\colorSYNTAXB}{{\color{\colorMATH}\ensuremath{\vdash }}}\endgroup }\hspace*{0.33em}{\begingroup\renewcommand\colorMATH{\colorMATHB}\renewcommand\colorSYNTAX{\colorSYNTAXB}{{\color{\colorMATH}\ensuremath{\slambda}}}\endgroup } (y:{\begingroup\renewcommand\colorMATH{\colorMATHA}\renewcommand\colorSYNTAX{\colorSYNTAXA}{{\color{\colorSYNTAX}\texttt{{\ensuremath{{\mathbb{R}}}}}}}\endgroup }\mathord{\cdotp }1).\hspace*{0.33em} ({\begingroup\renewcommand\colorMATH{\colorMATHB}\renewcommand\colorSYNTAX{\colorSYNTAXB}{{\color{\colorMATH}\ensuremath{\slambda}}}\endgroup } (x\mathrel{:}{\begingroup\renewcommand\colorMATH{\colorMATHA}\renewcommand\colorSYNTAX{\colorSYNTAXA}{{\color{\colorSYNTAX}\texttt{{\ensuremath{{\mathbb{R}}}}}}}\endgroup }\mathord{\cdotp }2).\hspace*{0.33em}x+x) (2*y) \mathrel{:} (y\mathrel{:}{\begingroup\renewcommand\colorMATH{\colorMATHA}\renewcommand\colorSYNTAX{\colorSYNTAXA}{{\color{\colorSYNTAX}\texttt{{\ensuremath{{\mathbb{R}}}}}}}\endgroup }) \xrightarrowS {4y} {\begingroup\renewcommand\colorMATH{\colorMATHA}\renewcommand\colorSYNTAX{\colorSYNTAXA}{{\color{\colorSYNTAX}\texttt{{\ensuremath{{\mathbb{R}}}}}}}\endgroup } \mathrel{;} {\begingroup\renewcommand\colorMATH{\colorMATHB}\renewcommand\colorSYNTAX{\colorSYNTAXB}{{\color{\colorMATH}\ensuremath{\varnothing }}}\endgroup }
     }
   \end{gather*}\endgroup
   The outermost lambda allows a maximum variation of {{\color{\colorMATH}\ensuremath{1}}} in its argument {{\color{\colorMATH}\ensuremath{y}}}. The inner lambda allows a maximum variation of {{\color{\colorMATH}\ensuremath{2}}} on its argument, and its being applied to {{\color{\colorMATH}\ensuremath{2*y}}}. As {{\color{\colorMATH}\ensuremath{2*y}}} is {{\color{\colorMATH}\ensuremath{2}}}-sensitive on {{\color{\colorMATH}\ensuremath{y}}} and {{\color{\colorMATH}\ensuremath{y}}} can wiggle at most by {{\color{\colorMATH}\ensuremath{1}}}, then we know that the argument is going to wiggle at most by {{\color{\colorMATH}\ensuremath{2}}}, which matches the maximum permitted argument variation. If the argument were {{\color{\colorMATH}\ensuremath{3*y}}}, then the program would not type check as the argument of the application could wiggle at most by {{\color{\colorMATH}\ensuremath{1*3=3}}}.

\item  Rule{\textsc{ s-case}} type checks subterms {{\color{\colorMATH}\ensuremath{{\begingroup\renewcommand\colorMATH{\colorMATHB}\renewcommand\colorSYNTAX{\colorSYNTAXB}{{\color{\colorMATH}\ensuremath{\se_{2}}}}\endgroup }}}} and {{\color{\colorMATH}\ensuremath{{\begingroup\renewcommand\colorMATH{\colorMATHB}\renewcommand\colorSYNTAX{\colorSYNTAXB}{{\color{\colorMATH}\ensuremath{\se_{3}}}}\endgroup }}}} by
   extending the \distanceBoundName environment with a sound bound for {{\color{\colorMATH}\ensuremath{x}}} and {{\color{\colorMATH}\ensuremath{y}}}
   respectively. For {{\color{\colorMATH}\ensuremath{x}}} (resp.~{{\color{\colorMATH}\ensuremath{y}}}) we use the dot product between the relational distance on all
   variables in scope {{\color{\colorMATH}\ensuremath{{\begingroup\renewcommand\colorMATH{\colorMATHB}\renewcommand\colorSYNTAX{\colorSYNTAXB}{{\color{\colorMATH}\ensuremath{\Distance}}}\endgroup }}}}, and the cost of using {{\color{\colorMATH}\ensuremath{{\begingroup\renewcommand\colorMATH{\colorMATHB}\renewcommand\colorSYNTAX{\colorSYNTAXB}{{\color{\colorMATH}\ensuremath{\se_{1}}}}\endgroup }}}}: the cost {{\color{\colorMATH}\ensuremath{{\begingroup\renewcommand\colorMATH{\colorMATHB}\renewcommand\colorSYNTAX{\colorSYNTAXB}{{\color{\colorMATH}\ensuremath{\sS_{1}}}}\endgroup }}}} of reducing
   the expression, plus the latent cost of using its subterm {{\color{\colorMATH}\ensuremath{{\begingroup\renewcommand\colorMATH{\colorMATHB}\renewcommand\colorSYNTAX{\colorSYNTAXB}{{\color{\colorMATH}\ensuremath{\sS_{1 1}}}}\endgroup }}}} (resp.~{{\color{\colorMATH}\ensuremath{{\begingroup\renewcommand\colorMATH{\colorMATHB}\renewcommand\colorSYNTAX{\colorSYNTAXB}{{\color{\colorMATH}\ensuremath{\sS_{1 2}}}}\endgroup }}}}).
   For example, consider the type derivation of Example~\ref{ex:confbranches} given e.g.~\distanceBoundName environment {{\color{\colorMATH}\ensuremath{x+2b}}}:

   \begingroup\color{\colorMATH}\begin{gather*}
     \inferrule*[lab={\textsc{ case}}
     ]{ \Gamma ;x+2b \hspace*{0.33em}{\begingroup\renewcommand\colorMATH{\colorMATHB}\renewcommand\colorSYNTAX{\colorSYNTAXB}{{\color{\colorMATH}\ensuremath{\vdash }}}\endgroup }\hspace*{0.33em}{\begingroup\renewcommand\colorMATH{\colorMATHB}\renewcommand\colorSYNTAX{\colorSYNTAXB}{{\color{\colorMATH}\ensuremath{\se}}}\endgroup } : {\begingroup\renewcommand\colorMATH{\colorMATHA}\renewcommand\colorSYNTAX{\colorSYNTAXA}{{\color{\colorSYNTAX}\texttt{{\ensuremath{{\mathbb{R}}}}}}}\endgroup } \mathrel{^{\infty x}\oplus ^{x}} {\begingroup\renewcommand\colorMATH{\colorMATHA}\renewcommand\colorSYNTAX{\colorSYNTAXA}{{\color{\colorSYNTAX}\texttt{{\ensuremath{{\mathbb{R}}}}}}}\endgroup }; b
     \\ \Gamma , x_{1}:{\begingroup\renewcommand\colorMATH{\colorMATHA}\renewcommand\colorSYNTAX{\colorSYNTAXA}{{\color{\colorSYNTAX}\texttt{{\ensuremath{{\mathbb{R}}}}}}}\endgroup };x+2b+ \begingroup\color{\colorTEXT}\boxed{\begingroup\color{\colorMATH}\infty x_{1}\endgroup}\endgroup \hspace*{0.33em}{\begingroup\renewcommand\colorMATH{\colorMATHB}\renewcommand\colorSYNTAX{\colorSYNTAXB}{{\color{\colorMATH}\ensuremath{\vdash }}}\endgroup }\hspace*{0.33em} 0: {\begingroup\renewcommand\colorMATH{\colorMATHA}\renewcommand\colorSYNTAX{\colorSYNTAXA}{{\color{\colorSYNTAX}\texttt{{\ensuremath{{\mathbb{R}}}}}}}\endgroup }; \varnothing 
     \\ \Gamma , x_{2}:{\begingroup\renewcommand\colorMATH{\colorMATHA}\renewcommand\colorSYNTAX{\colorSYNTAXA}{{\color{\colorSYNTAX}\texttt{{\ensuremath{{\mathbb{R}}}}}}}\endgroup };x+2b+ \begingroup\color{\colorTEXT}\boxed{\begingroup\color{\colorMATH}3x_{2}\endgroup}\endgroup\hspace*{0.33em}{\begingroup\renewcommand\colorMATH{\colorMATHB}\renewcommand\colorSYNTAX{\colorSYNTAXB}{{\color{\colorMATH}\ensuremath{\vdash }}}\endgroup }\hspace*{0.33em} x_{2}: {\begingroup\renewcommand\colorMATH{\colorMATHA}\renewcommand\colorSYNTAX{\colorSYNTAXA}{{\color{\colorSYNTAX}\texttt{{\ensuremath{{\mathbb{R}}}}}}}\endgroup }; x_{2}
        }{
        \Gamma ;x+2b\hspace*{0.33em}{\begingroup\renewcommand\colorMATH{\colorMATHB}\renewcommand\colorSYNTAX{\colorSYNTAXB}{{\color{\colorMATH}\ensuremath{\vdash }}}\endgroup }\hspace*{0.33em} \ccase\hspace*{0.33em}{\begingroup\renewcommand\colorMATH{\colorMATHB}\renewcommand\colorSYNTAX{\colorSYNTAXB}{{\color{\colorMATH}\ensuremath{\se}}}\endgroup }\hspace*{0.33em}\of\hspace*{0.33em}\{ x_{1} \Rightarrow  0\} \{ x_{2} \Rightarrow  x_{2}\}  : {\begingroup\renewcommand\colorMATH{\colorMATHA}\renewcommand\colorSYNTAX{\colorSYNTAXA}{{\color{\colorSYNTAX}\texttt{{\ensuremath{{\mathbb{R}}}}}}}\endgroup } ; b+x
     }
   \end{gather*}\endgroup
   The relational distance for {{\color{\colorMATH}\ensuremath{x_{1}}}} on the first branch is computed as the dot product between the maximum distance of all variables in scope, {{\color{\colorMATH}\ensuremath{x+2b}}}, and the cost of using {{\color{\colorMATH}\ensuremath{{\begingroup\renewcommand\colorMATH{\colorMATHB}\renewcommand\colorSYNTAX{\colorSYNTAXB}{{\color{\colorMATH}\ensuremath{\se}}}\endgroup }}}} if it were an \inl\ expression, i.e. {{\color{\colorMATH}\ensuremath{((x+2b)\mathord{\cdotp }(\infty x+b)) = \infty }}}.
   Analogously, the bound for {{\color{\colorMATH}\ensuremath{x_{2}}}} on the second branch is computed as {{\color{\colorMATH}\ensuremath{((x+2b)\mathord{\cdotp }(x+b)) = 3}}}.

\item  In Rule{\textsc{ untup}}, as
   expression {{\color{\colorMATH}\ensuremath{{\begingroup\renewcommand\colorMATH{\colorMATHB}\renewcommand\colorSYNTAX{\colorSYNTAXB}{{\color{\colorMATH}\ensuremath{\se_{2}}}}\endgroup }}}} has in scope new variables {{\color{\colorMATH}\ensuremath{x_{1}}}} and {{\color{\colorMATH}\ensuremath{x_{2}}}},
   the \distanceBoundName environment {{\color{\colorMATH}\ensuremath{{\begingroup\renewcommand\colorMATH{\colorMATHB}\renewcommand\colorSYNTAX{\colorSYNTAXB}{{\color{\colorMATH}\ensuremath{\Distance}}}\endgroup }}}} is extended accordingly. The relational distance for {{\color{\colorMATH}\ensuremath{x_{1}}}} is
   computed as the dot product between the \distanceBoundName environment {{\color{\colorMATH}\ensuremath{{\begingroup\renewcommand\colorMATH{\colorMATHB}\renewcommand\colorSYNTAX{\colorSYNTAXB}{{\color{\colorMATH}\ensuremath{\Distance}}}\endgroup }}}} and the cost {{\color{\colorMATH}\ensuremath{{\begingroup\renewcommand\colorMATH{\colorMATHB}\renewcommand\colorSYNTAX{\colorSYNTAXB}{{\color{\colorMATH}\ensuremath{\sS_{1}}}}\endgroup } + {\begingroup\renewcommand\colorMATH{\colorMATHB}\renewcommand\colorSYNTAX{\colorSYNTAXB}{{\color{\colorMATH}\ensuremath{\sS_{1 1}}}}\endgroup }}}}
   of accessing the first component (we proceed similarly with {{\color{\colorMATH}\ensuremath{x_{2}}}}).
   For instance, consider the type derivation of Example~\ref{ex:scaling}  given some arbitrary \distanceBoundName environment {{\color{\colorMATH}\ensuremath{2x+3y}}}
   \begingroup\color{\colorMATH}\begin{gather*}
     \inferrule*[lab={\textsc{ untup}}
     ]{ \Gamma ;2x+3y\hspace*{0.33em} {\begingroup\renewcommand\colorMATH{\colorMATHB}\renewcommand\colorSYNTAX{\colorSYNTAXB}{{\color{\colorMATH}\ensuremath{ \vdash  }}}\endgroup }\hspace*{0.33em} \addProduct{2*x}{y} \mathrel{:} {\begingroup\renewcommand\colorMATH{\colorMATHA}\renewcommand\colorSYNTAX{\colorSYNTAXA}{{\color{\colorSYNTAX}\texttt{{\ensuremath{{\mathbb{R}}}}}}}\endgroup } \mathrel{^{2x}\otimes ^{y}} {\begingroup\renewcommand\colorMATH{\colorMATHA}\renewcommand\colorSYNTAX{\colorSYNTAXA}{{\color{\colorSYNTAX}\texttt{{\ensuremath{{\mathbb{R}}}}}}}\endgroup } \mathrel{;} \varnothing 
     \\ \Gamma ,x_{1}\mathrel{:}{\begingroup\renewcommand\colorMATH{\colorMATHA}\renewcommand\colorSYNTAX{\colorSYNTAXA}{{\color{\colorSYNTAX}\texttt{{\ensuremath{{\mathbb{R}}}}}}}\endgroup },x_{2}\mathrel{:}{\begingroup\renewcommand\colorMATH{\colorMATHA}\renewcommand\colorSYNTAX{\colorSYNTAXA}{{\color{\colorSYNTAX}\texttt{{\ensuremath{{\mathbb{R}}}}}}}\endgroup };2x+3y+\begingroup\color{\colorTEXT}\boxed{\begingroup\color{\colorMATH}4x_{1}+3x_{2}\endgroup}\endgroup\hspace*{0.33em} {\begingroup\renewcommand\colorMATH{\colorMATHB}\renewcommand\colorSYNTAX{\colorSYNTAXB}{{\color{\colorMATH}\ensuremath{ \vdash  }}}\endgroup }\hspace*{0.33em} x_{1}+2*x_{2} \mathrel{:} {\begingroup\renewcommand\colorMATH{\colorMATHA}\renewcommand\colorSYNTAX{\colorSYNTAXA}{{\color{\colorSYNTAX}\texttt{{\ensuremath{{\mathbb{R}}}}}}}\endgroup } \mathrel{;} x_{1}+2x_{2}
        }{
        \Gamma ;2x+3y\hspace*{0.33em} {\begingroup\renewcommand\colorMATH{\colorMATHB}\renewcommand\colorSYNTAX{\colorSYNTAXB}{{\color{\colorMATH}\ensuremath{ \vdash  }}}\endgroup }\hspace*{0.33em}\tlet\hspace*{0.33em}x_{1},x_{2}=\addProduct{2*x}{y}\hspace*{0.33em}\tin\hspace*{0.33em}x_{1}+2*x_{2} \mathrel{:} {\begingroup\renewcommand\colorMATH{\colorMATHA}\renewcommand\colorSYNTAX{\colorSYNTAXA}{{\color{\colorSYNTAX}\texttt{{\ensuremath{{\mathbb{R}}}}}}}\endgroup } \mathrel{;} 2x+ 2y
     }
   \end{gather*}\endgroup
   The relational distance for {{\color{\colorMATH}\ensuremath{x_{1}}}} is computed as the dot product between the relational distance of all variables in scope {{\color{\colorMATH}\ensuremath{2x+3y}}} and the effect of using the left component of the pair {{\color{\colorMATH}\ensuremath{2x+0y}}}, i.e. {{\color{\colorMATH}\ensuremath{(2x+3y) \mathord{\cdotp } (2x+0y) = 4}}}. Similarly, the bound of {{\color{\colorMATH}\ensuremath{x_{2}}}} is computed as {{\color{\colorMATH}\ensuremath{(2x+3y) \mathord{\cdotp } (y+0x) = 3}}}
\end{itemize}

\begin{figure}[t]
  \begin{small}
  \begin{framed}
  \hfill\hspace{0pt}\begingroup\color{\colorTEXT}\boxed{\begingroup\color{\colorMATH} \tau  <: \tau  \endgroup}\endgroup
  \begingroup\color{\colorMATH}\begin{gather*}

\end{gather*}\endgroup
\hfill\hspace{0pt}\begingroup\color{\colorTEXT}\boxed{\begingroup\color{\colorMATH} {\begingroup\renewcommand\colorMATH{\colorMATHC}\renewcommand\colorSYNTAX{\colorSYNTAXC}{{\color{\colorMATH}\ensuremath{\pS}}}\endgroup } <: {\begingroup\renewcommand\colorMATH{\colorMATHC}\renewcommand\colorSYNTAX{\colorSYNTAXC}{{\color{\colorMATH}\ensuremath{\pS}}}\endgroup } \endgroup}\endgroup
\begingroup\color{\colorMATH}\begin{gather*}
  \begin{tabularx}{\linewidth}{>{\centering\arraybackslash\(}X<{\)}}\parbox{\linewidth}{
  \begingroup\color{\colorMATH}\begin{mathpar}\inferrule*[lab=
    ]{ \forall  {\begingroup\renewcommand\colorMATH{\colorMATHB}\renewcommand\colorSYNTAX{\colorSYNTAXB}{{\color{\colorMATH}\ensuremath{\Distance}}}\endgroup }, dom({\begingroup\renewcommand\colorMATH{\colorMATHC}\renewcommand\colorSYNTAX{\colorSYNTAXC}{{\color{\colorMATH}\ensuremath{\pS}}}\endgroup }) \subseteq  dom({\begingroup\renewcommand\colorMATH{\colorMATHB}\renewcommand\colorSYNTAX{\colorSYNTAXB}{{\color{\colorMATH}\ensuremath{\Distance}}}\endgroup }). {\begingroup\renewcommand\colorMATH{\colorMATHB}\renewcommand\colorSYNTAX{\colorSYNTAXB}{{\color{\colorMATH}\ensuremath{\Distance}}}\endgroup } {\begingroup\renewcommand\colorMATH{\colorMATHC}\renewcommand\colorSYNTAX{\colorSYNTAXC}{{\color{\colorMATH}\ensuremath{\bigcdot}}}\endgroup } {\begingroup\renewcommand\colorMATH{\colorMATHC}\renewcommand\colorSYNTAX{\colorSYNTAXC}{{\color{\colorMATH}\ensuremath{\pS}}}\endgroup } \leq  {\begingroup\renewcommand\colorMATH{\colorMATHB}\renewcommand\colorSYNTAX{\colorSYNTAXB}{{\color{\colorMATH}\ensuremath{\Distance}}}\endgroup } {\begingroup\renewcommand\colorMATH{\colorMATHC}\renewcommand\colorSYNTAX{\colorSYNTAXC}{{\color{\colorMATH}\ensuremath{\bigcdot}}}\endgroup } {\begingroup\renewcommand\colorMATH{\colorMATHC}\renewcommand\colorSYNTAX{\colorSYNTAXC}{{\color{\colorMATH}\ensuremath{\pS'}}}\endgroup }
      }{
       {\begingroup\renewcommand\colorMATH{\colorMATHC}\renewcommand\colorSYNTAX{\colorSYNTAXC}{{\color{\colorMATH}\ensuremath{\pS}}}\endgroup } <: {\begingroup\renewcommand\colorMATH{\colorMATHC}\renewcommand\colorSYNTAX{\colorSYNTAXC}{{\color{\colorMATH}\ensuremath{\pS'}}}\endgroup }
    }
  \end{mathpar}\endgroup
  }
  \end{tabularx}
\end{gather*}\endgroup
\end{framed}
  \end{small}
  \caption{$\lang$: Subtyping}
  \label{fig:subtyping-extension}
\end{figure}
Subtyping is extended accordingly and presented in Figure~\ref{fig:subtyping-extension}. 
Parameterized relational distances on function types are \toplass{contravariant}, and subtyping for privacy function types relies on the definition of subtyping for privacy environment also defined in Figure~\ref{fig:subtyping-extension}, where 
%\toplas{{{\color{\colorMATH}\ensuremath{\mathrel{\| }\mathord{\cdotp }\mathrel{\| }_{1}}}} stands for the {{\color{\colorMATH}\ensuremath{\ell^{1}}}} norm.}
{{\color{\colorMATH}\ensuremath{{\begingroup\renewcommand\colorMATH{\colorMATHC}\renewcommand\colorSYNTAX{\colorSYNTAXC}{{\color{\colorMATH}\ensuremath{\bigcdot}}}\endgroup }}}} is an operator to close privacy environments defined below:
\begingroup\color{\colorMATH}\begin{gather*}

\end{gather*}\endgroup

%parameterized bound on argument sensitivity

\begin{figure}[t]
  \begin{small}
  \input{privacy-type-system}
  \end{small}
  \caption{$\lang$: Type system of the privacy sublanguage}
  \label{fig:privacy-type-system}
\end{figure}
\paragraph{Privacy type system}
The type system of the privacy part of the language is presented in
Figure~\ref{fig:privacy-type-system}. The judgment {{\color{\colorMATH}\ensuremath{\Gamma  \mathrel{;} {\begingroup\renewcommand\colorMATH{\colorMATHB}\renewcommand\colorSYNTAX{\colorSYNTAXB}{{\color{\colorMATH}\ensuremath{\Distance}}}\endgroup } \hspace*{0.33em}{\begingroup\renewcommand\colorMATH{\colorMATHC}\renewcommand\colorSYNTAX{\colorSYNTAXC}{{\color{\colorMATH}\ensuremath{\vdash }}}\endgroup }\hspace*{0.33em}{\begingroup\renewcommand\colorMATH{\colorMATHC}\renewcommand\colorSYNTAX{\colorSYNTAXC}{{\color{\colorMATH}\ensuremath{\pe}}}\endgroup } \mathrel{:} \tau  \mathrel{;} {\begingroup\renewcommand\colorMATH{\colorMATHC}\renewcommand\colorSYNTAX{\colorSYNTAXC}{{\color{\colorMATH}\ensuremath{\pS}}}\endgroup }}}}
says that privacy term  {{\color{\colorMATH}\ensuremath{{\begingroup\renewcommand\colorMATH{\colorMATHC}\renewcommand\colorSYNTAX{\colorSYNTAXC}{{\color{\colorMATH}\ensuremath{\pe}}}\endgroup }}}} has type {{\color{\colorMATH}\ensuremath{\tau }}} and ambient privacy effect {{\color{\colorMATH}\ensuremath{{\begingroup\renewcommand\colorMATH{\colorMATHC}\renewcommand\colorSYNTAX{\colorSYNTAXC}{{\color{\colorMATH}\ensuremath{\pS}}}\endgroup }}}} under type
environment {{\color{\colorMATH}\ensuremath{\Gamma }}}, and \distanceBoundName environment {{\color{\colorMATH}\ensuremath{{\begingroup\renewcommand\colorMATH{\colorMATHB}\renewcommand\colorSYNTAX{\colorSYNTAXB}{{\color{\colorMATH}\ensuremath{\Distance}}}\endgroup }}}}.
% to justify these rules intuitively.}
\begin{itemize}[label=\textbf{-},leftmargin=*]\item  Rule{\textsc{ return}} uses the type system of the sensitivity language to type check
   its subexpression {{\color{\colorMATH}\ensuremath{{\begingroup\renewcommand\colorMATH{\colorMATHB}\renewcommand\colorSYNTAX{\colorSYNTAXB}{{\color{\colorMATH}\ensuremath{\se}}}\endgroup }}}}. Operationally, {\begingroup\renewcommand\colorMATH{\colorMATHC}\renewcommand\colorSYNTAX{\colorSYNTAXC}{{\color{\colorMATH}\ensuremath{{{\color{\colorSYNTAX}\texttt{return}}}}}}\endgroup } constructs a
   point-distribution, and any sensitive variables in the subexpression {{\color{\colorMATH}\ensuremath{{\begingroup\renewcommand\colorMATH{\colorMATHB}\renewcommand\colorSYNTAX{\colorSYNTAXB}{{\color{\colorMATH}\ensuremath{\se}}}\endgroup }}}}
   will have their privacy violated, i.e., privacy cost {\begingroup\renewcommand\colorMATH{\colorMATHC}\renewcommand\colorSYNTAX{\colorSYNTAXC}{{\color{\colorMATH}\ensuremath{\infty }}}\endgroup }. 
   Notice that {{\color{\colorMATH}\ensuremath{{\begingroup\renewcommand\colorMATH{\colorMATHC}\renewcommand\colorSYNTAX{\colorSYNTAXC}{{\color{\colorMATH}\ensuremath{\infty }}}\endgroup }}}} \toplas{corresponds} to the pair {{\color{\colorMATH}\ensuremath{({\begingroup\renewcommand\colorMATH{\colorMATHC}\renewcommand\colorSYNTAX{\colorSYNTAXC}{{\color{\colorMATH}\ensuremath{\epsilon }}}\endgroup }, {\begingroup\renewcommand\colorMATH{\colorMATHC}\renewcommand\colorSYNTAX{\colorSYNTAXC}{{\color{\colorMATH}\ensuremath{\delta }}}\endgroup }) = ({\begingroup\renewcommand\colorMATH{\colorMATHC}\renewcommand\colorSYNTAX{\colorSYNTAXC}{{\color{\colorMATH}\ensuremath{\infty }}}\endgroup }, {\begingroup\renewcommand\colorMATH{\colorMATHC}\renewcommand\colorSYNTAX{\colorSYNTAXC}{{\color{\colorMATH}\ensuremath{\infty }}}\endgroup })}}}.
   The resulting
   ambient privacy effect is computed by lifting \toplas{to infinity the ambient effect of the subexpression as well as all free variable in {{\color{\colorMATH}\ensuremath{{\begingroup\renewcommand\colorMATH{\colorMATHB}\renewcommand\colorSYNTAX{\colorSYNTAXB}{{\color{\colorMATH}\ensuremath{\se}}}\endgroup }}}}. } 
   %variables  sensitivity of every
   %variable with non-zero sensitivity in {{\color{\colorMATH}\ensuremath{{\begingroup\renewcommand\colorMATH{\colorMATHB}\renewcommand\colorSYNTAX{\colorSYNTAXB}{{\color{\colorMATH}\ensuremath{\se}}}\endgroup }}}} \toplas{to infinity}
   %, as well as variables with
   %non-zero sensitivity or privacy in free variables of {{\color{\colorMATH}\ensuremath{\tau }}}, the type of {{\color{\colorMATH}\ensuremath{{\begingroup\renewcommand\colorMATH{\colorMATHB}\renewcommand\colorSYNTAX{\colorSYNTAXB}{{\color{\colorMATH}\ensuremath{\se}}}\endgroup }}}}
   The operator
   that lift to {{\color{\colorMATH}\ensuremath{{\begingroup\renewcommand\colorMATH{\colorMATHC}\renewcommand\colorSYNTAX{\colorSYNTAXC}{{\color{\colorMATH}\ensuremath{\infty }}}\endgroup }}}} free variables is written {{\color{\colorMATH}\ensuremath{{\text{FP}}^\infty }}} and defined in
   Figure~\ref{fig:privacy-type-system}. 
   %The free variables, sensitivities and \toplas{privacy costs}
   %in {{\color{\colorMATH}\ensuremath{\tau }}} represent the possible latent contextual effect of the subexpression. 
   As we pay infinity for every free variable in {{\color{\colorMATH}\ensuremath{{\begingroup\renewcommand\colorMATH{\colorMATHB}\renewcommand\colorSYNTAX{\colorSYNTAXB}{{\color{\colorMATH}\ensuremath{\se}}}\endgroup }}}}, 
   we remove those variables from the reported type {{\color{\colorMATH}\ensuremath{\tau }}} using the
   sensitivity environment substitution operator defined in
   Figure~\ref{fig:privacy-statics-auxiliary-definitions}.
   For instance, consider the following type derivation,
   \begingroup\color{\colorMATH}\begin{gather*}
     \inferrule*[lab={\textsc{ return}}
     ]{ \Gamma ;y+z\hspace*{0.33em} {\begingroup\renewcommand\colorMATH{\colorMATHB}\renewcommand\colorSYNTAX{\colorSYNTAXB}{{\color{\colorMATH}\ensuremath{ \vdash  }}}\endgroup }\hspace*{0.33em} {\begingroup\renewcommand\colorMATH{\colorMATHB}\renewcommand\colorSYNTAX{\colorSYNTAXB}{{\color{\colorMATH}\ensuremath{\slambda}}}\endgroup } (x\mathrel{:}{\begingroup\renewcommand\colorMATH{\colorMATHA}\renewcommand\colorSYNTAX{\colorSYNTAXA}{{\color{\colorSYNTAX}\texttt{{\ensuremath{{\mathbb{R}}}}}}}\endgroup }\mathord{\cdotp }1).\hspace*{0.33em}2x+y \mathrel{:} (x:{\begingroup\renewcommand\colorMATH{\colorMATHA}\renewcommand\colorSYNTAX{\colorSYNTAXA}{{\color{\colorSYNTAX}\texttt{{\ensuremath{{\mathbb{R}}}}}}}\endgroup }\mathord{\cdotp }1) \xrightarrowS {2x+y} {\begingroup\renewcommand\colorMATH{\colorMATHA}\renewcommand\colorSYNTAX{\colorSYNTAXA}{{\color{\colorSYNTAX}\texttt{{\ensuremath{{\mathbb{R}}}}}}}\endgroup }; \varnothing 
        }{
        \Gamma ;y+z\hspace*{0.33em} {\begingroup\renewcommand\colorMATH{\colorMATHC}\renewcommand\colorSYNTAX{\colorSYNTAXC}{{\color{\colorMATH}\ensuremath{ \vdash  }}}\endgroup }\hspace*{0.33em} {\begingroup\renewcommand\colorMATH{\colorMATHC}\renewcommand\colorSYNTAX{\colorSYNTAXC}{{\color{\colorSYNTAX}\texttt{return}}}\endgroup }\hspace*{0.33em} {\begingroup\renewcommand\colorMATH{\colorMATHB}\renewcommand\colorSYNTAX{\colorSYNTAXB}{{\color{\colorMATH}\ensuremath{\slambda}}}\endgroup } (x\mathrel{:}{\begingroup\renewcommand\colorMATH{\colorMATHA}\renewcommand\colorSYNTAX{\colorSYNTAXA}{{\color{\colorSYNTAX}\texttt{{\ensuremath{{\mathbb{R}}}}}}}\endgroup }\mathord{\cdotp }1).\hspace*{0.33em}2x+y \mathrel{:} (x:{\begingroup\renewcommand\colorMATH{\colorMATHA}\renewcommand\colorSYNTAX{\colorSYNTAXA}{{\color{\colorSYNTAX}\texttt{{\ensuremath{{\mathbb{R}}}}}}}\endgroup }\mathord{\cdotp }1) \xrightarrowS {2x} {\begingroup\renewcommand\colorMATH{\colorMATHA}\renewcommand\colorSYNTAX{\colorSYNTAXA}{{\color{\colorSYNTAX}\texttt{{\ensuremath{{\mathbb{R}}}}}}}\endgroup }; {\begingroup\renewcommand\colorMATH{\colorMATHC}\renewcommand\colorSYNTAX{\colorSYNTAXC}{{\color{\colorMATH}\ensuremath{\infty }}}\endgroup }y
     }
   \end{gather*}\endgroup
   The resulting type and effect environment is computed by paying in advance for the free variables in scope:
   the type {{\color{\colorMATH}\ensuremath{[\varnothing /y](2x+y) = 2x}}} is computed by erasing the free variables, and the effect environment {{\color{\colorMATH}\ensuremath{{\begingroup\renewcommand\colorMATH{\colorMATHC}\renewcommand\colorSYNTAX{\colorSYNTAXC}{{\color{\colorMATH}\ensuremath{\infty }}}\endgroup }y}}} is computed by lifting \toplas{the free variables to infinity}. 

\item  Rule{\textsc{ bind}} type checks both subexpressions using the type system for the
   privacy language as they are privacy expressions. To type check {{\color{\colorMATH}\ensuremath{{\begingroup\renewcommand\colorMATH{\colorMATHC}\renewcommand\colorSYNTAX{\colorSYNTAXC}{{\color{\colorMATH}\ensuremath{\pe_{2}}}}\endgroup }}}} we
   extend type environment with variable {{\color{\colorMATH}\ensuremath{x}}}, therefore the \distanceBoundName environment
   {{\color{\colorMATH}\ensuremath{{\begingroup\renewcommand\colorMATH{\colorMATHB}\renewcommand\colorSYNTAX{\colorSYNTAXB}{{\color{\colorMATH}\ensuremath{\Distance}}}\endgroup }}}} is also extended. We extend {{\color{\colorMATH}\ensuremath{{\begingroup\renewcommand\colorMATH{\colorMATHB}\renewcommand\colorSYNTAX{\colorSYNTAXB}{{\color{\colorMATH}\ensuremath{\Distance}}}\endgroup }}}} with {{\color{\colorMATH}\ensuremath{{\begingroup\renewcommand\colorMATH{\colorMATHB}\renewcommand\colorSYNTAX{\colorSYNTAXB}{{\color{\colorMATH}\ensuremath{0}}}\endgroup }x}}} as the value bound to {{\color{\colorMATH}\ensuremath{x}}} is
   no longer considered sensitive---it has been declassified and can be used
   without restriction. 
   Finally, as {{\color{\colorMATH}\ensuremath{x}}} is out of scope we remove it from {{\color{\colorMATH}\ensuremath{\tau _{2}}}}
   and from the resulting ambient privacy effect.
   For instance, consider the type derivation of program {{\color{\colorMATH}\ensuremath{y: {\begingroup\renewcommand\colorMATH{\colorMATHA}\renewcommand\colorSYNTAX{\colorSYNTAXA}{{\color{\colorSYNTAX}\texttt{{\ensuremath{{\mathbb{R}}}}}}}\endgroup } \leftarrow  {\text{laplace}}_{\epsilon _{1}} x\mathrel{;} z: {\begingroup\renewcommand\colorMATH{\colorMATHA}\renewcommand\colorSYNTAX{\colorSYNTAXA}{{\color{\colorSYNTAX}\texttt{{\ensuremath{{\mathbb{R}}}}}}}\endgroup } \leftarrow  {\text{laplace}}_{\epsilon _{2}} x; {\begingroup\renewcommand\colorMATH{\colorMATHC}\renewcommand\colorSYNTAX{\colorSYNTAXC}{{\color{\colorSYNTAX}\texttt{return}}}\endgroup } \hspace*{0.33em} y+z}}}, similar to the example presented in Section~\ref{sec:delay-sens}, given an arbitrary \distanceBoundName environment {{\color{\colorMATH}\ensuremath{x}}}. 
   \begingroup\color{\colorMATH}\begin{gather*} 
     \inferrule*[lab={\textsc{ bind}}
    ]{ \Gamma  \mathrel{;} x\hspace*{0.33em}{\begingroup\renewcommand\colorMATH{\colorMATHC}\renewcommand\colorSYNTAX{\colorSYNTAXC}{{\color{\colorMATH}\ensuremath{\vdash }}}\endgroup }\hspace*{0.33em} {\text{laplace}}_{\epsilon _{1}} x \mathrel{:} {\begingroup\renewcommand\colorMATH{\colorMATHA}\renewcommand\colorSYNTAX{\colorSYNTAXA}{{\color{\colorSYNTAX}\texttt{{\ensuremath{{\mathbb{R}}}}}}}\endgroup } \mathrel{;} ({\begingroup\renewcommand\colorMATH{\colorMATHC}\renewcommand\colorSYNTAX{\colorSYNTAXC}{{\color{\colorMATH}\ensuremath{\epsilon _{1}}}}\endgroup },0)x
    \\ 
      \inferrule*[lab={\textsc{ bind}}
      ]{ \Gamma ,y\mathrel{:}{\begingroup\renewcommand\colorMATH{\colorMATHA}\renewcommand\colorSYNTAX{\colorSYNTAXA}{{\color{\colorSYNTAX}\texttt{{\ensuremath{{\mathbb{R}}}}}}}\endgroup } \mathrel{;} x + {\begingroup\renewcommand\colorMATH{\colorMATHB}\renewcommand\colorSYNTAX{\colorSYNTAXB}{{\color{\colorMATH}\ensuremath{0}}}\endgroup }y   \hspace*{0.33em}{\begingroup\renewcommand\colorMATH{\colorMATHC}\renewcommand\colorSYNTAX{\colorSYNTAXC}{{\color{\colorMATH}\ensuremath{\vdash }}}\endgroup }\hspace*{0.33em} {\text{laplace}}_{\epsilon _{2}} x \mathrel{:} {\begingroup\renewcommand\colorMATH{\colorMATHA}\renewcommand\colorSYNTAX{\colorSYNTAXA}{{\color{\colorSYNTAX}\texttt{{\ensuremath{{\mathbb{R}}}}}}}\endgroup } \mathrel{;} ({\begingroup\renewcommand\colorMATH{\colorMATHC}\renewcommand\colorSYNTAX{\colorSYNTAXC}{{\color{\colorMATH}\ensuremath{\epsilon _{2}}}}\endgroup },0)x
      \\ 
      \Gamma ,y\mathrel{:}{\begingroup\renewcommand\colorMATH{\colorMATHA}\renewcommand\colorSYNTAX{\colorSYNTAXA}{{\color{\colorSYNTAX}\texttt{{\ensuremath{{\mathbb{R}}}}}}}\endgroup } \mathrel{;} x + {\begingroup\renewcommand\colorMATH{\colorMATHB}\renewcommand\colorSYNTAX{\colorSYNTAXB}{{\color{\colorMATH}\ensuremath{0}}}\endgroup }y + {\begingroup\renewcommand\colorMATH{\colorMATHB}\renewcommand\colorSYNTAX{\colorSYNTAXB}{{\color{\colorMATH}\ensuremath{0}}}\endgroup }z \hspace*{0.33em}{\begingroup\renewcommand\colorMATH{\colorMATHC}\renewcommand\colorSYNTAX{\colorSYNTAXC}{{\color{\colorMATH}\ensuremath{\vdash }}}\endgroup }\hspace*{0.33em} {\begingroup\renewcommand\colorMATH{\colorMATHC}\renewcommand\colorSYNTAX{\colorSYNTAXC}{{\color{\colorSYNTAX}\texttt{return}}}\endgroup }\hspace*{0.33em}y+z \mathrel{:} {\begingroup\renewcommand\colorMATH{\colorMATHA}\renewcommand\colorSYNTAX{\colorSYNTAXA}{{\color{\colorSYNTAX}\texttt{{\ensuremath{{\mathbb{R}}}}}}}\endgroup } \mathrel{;} ({\begingroup\renewcommand\colorMATH{\colorMATHC}\renewcommand\colorSYNTAX{\colorSYNTAXC}{{\color{\colorMATH}\ensuremath{\infty }}}\endgroup },0)y + ({\begingroup\renewcommand\colorMATH{\colorMATHC}\renewcommand\colorSYNTAX{\colorSYNTAXC}{{\color{\colorMATH}\ensuremath{\infty }}}\endgroup },0)z
        }{
         \Gamma ,y\mathrel{:}{\begingroup\renewcommand\colorMATH{\colorMATHA}\renewcommand\colorSYNTAX{\colorSYNTAXA}{{\color{\colorSYNTAX}\texttt{{\ensuremath{{\mathbb{R}}}}}}}\endgroup } \mathrel{;} x + {\begingroup\renewcommand\colorMATH{\colorMATHB}\renewcommand\colorSYNTAX{\colorSYNTAXB}{{\color{\colorMATH}\ensuremath{0}}}\endgroup }y   \hspace*{0.33em}{\begingroup\renewcommand\colorMATH{\colorMATHC}\renewcommand\colorSYNTAX{\colorSYNTAXC}{{\color{\colorMATH}\ensuremath{\vdash }}}\endgroup }\hspace*{0.33em} z: {\begingroup\renewcommand\colorMATH{\colorMATHA}\renewcommand\colorSYNTAX{\colorSYNTAXA}{{\color{\colorSYNTAX}\texttt{{\ensuremath{{\mathbb{R}}}}}}}\endgroup } \leftarrow  {\text{laplace}}_{\epsilon _{2}} x; {\begingroup\renewcommand\colorMATH{\colorMATHC}\renewcommand\colorSYNTAX{\colorSYNTAXC}{{\color{\colorSYNTAX}\texttt{return}}}\endgroup }\hspace*{0.33em}y+z \mathrel{:} {\begingroup\renewcommand\colorMATH{\colorMATHA}\renewcommand\colorSYNTAX{\colorSYNTAXA}{{\color{\colorSYNTAX}\texttt{{\ensuremath{{\mathbb{R}}}}}}}\endgroup } \mathrel{;} ({\begingroup\renewcommand\colorMATH{\colorMATHC}\renewcommand\colorSYNTAX{\colorSYNTAXC}{{\color{\colorMATH}\ensuremath{\epsilon _{2}}}}\endgroup },0)x + ({\begingroup\renewcommand\colorMATH{\colorMATHC}\renewcommand\colorSYNTAX{\colorSYNTAXC}{{\color{\colorMATH}\ensuremath{\infty }}}\endgroup },0)y
      }    
      }{
       \Gamma  \mathrel{;} x\hspace*{0.33em}{\begingroup\renewcommand\colorMATH{\colorMATHC}\renewcommand\colorSYNTAX{\colorSYNTAXC}{{\color{\colorMATH}\ensuremath{\vdash }}}\endgroup }\hspace*{0.33em} y: {\begingroup\renewcommand\colorMATH{\colorMATHA}\renewcommand\colorSYNTAX{\colorSYNTAXA}{{\color{\colorSYNTAX}\texttt{{\ensuremath{{\mathbb{R}}}}}}}\endgroup } \leftarrow  {\text{laplace}}_{\epsilon _{1}} x\mathrel{;} z: {\begingroup\renewcommand\colorMATH{\colorMATHA}\renewcommand\colorSYNTAX{\colorSYNTAXA}{{\color{\colorSYNTAX}\texttt{{\ensuremath{{\mathbb{R}}}}}}}\endgroup } \leftarrow  {\text{laplace}}_{\epsilon _{2}} x; {\begingroup\renewcommand\colorMATH{\colorMATHC}\renewcommand\colorSYNTAX{\colorSYNTAXC}{{\color{\colorSYNTAX}\texttt{return}}}\endgroup } \hspace*{0.33em} y+z \mathrel{:} {\begingroup\renewcommand\colorMATH{\colorMATHA}\renewcommand\colorSYNTAX{\colorSYNTAXA}{{\color{\colorSYNTAX}\texttt{{\ensuremath{{\mathbb{R}}}}}}}\endgroup } \mathrel{;} ({\begingroup\renewcommand\colorMATH{\colorMATHC}\renewcommand\colorSYNTAX{\colorSYNTAXC}{{\color{\colorMATH}\ensuremath{\epsilon _{1}}}}\endgroup } + {\begingroup\renewcommand\colorMATH{\colorMATHC}\renewcommand\colorSYNTAX{\colorSYNTAXC}{{\color{\colorMATH}\ensuremath{\epsilon _{2}}}}\endgroup },0)x
    }
   \end{gather*}\endgroup
    Each {{\color{\colorMATH}\ensuremath{{\text{laplace}}}}} call has an effect environment of {{\color{\colorMATH}\ensuremath{({\begingroup\renewcommand\colorMATH{\colorMATHC}\renewcommand\colorSYNTAX{\colorSYNTAXC}{{\color{\colorMATH}\ensuremath{\epsilon _{1}}}}\endgroup },0)x}}} and {{\color{\colorMATH}\ensuremath{({\begingroup\renewcommand\colorMATH{\colorMATHC}\renewcommand\colorSYNTAX{\colorSYNTAXC}{{\color{\colorMATH}\ensuremath{\epsilon _{2}}}}\endgroup },0)x}}} respectively.
    The {{\color{\colorMATH}\ensuremath{{\begingroup\renewcommand\colorMATH{\colorMATHC}\renewcommand\colorSYNTAX{\colorSYNTAXC}{{\color{\colorSYNTAX}\texttt{return}}}\endgroup }}}} subexpression lifts to infinite the privacy of variables {{\color{\colorMATH}\ensuremath{y}}} and {{\color{\colorMATH}\ensuremath{z}}}, but to typecheck the innermost bind expression the privacy on  {{\color{\colorMATH}\ensuremath{z}}} is dropped: {{\color{\colorMATH}\ensuremath{({\begingroup\renewcommand\colorMATH{\colorMATHC}\renewcommand\colorSYNTAX{\colorSYNTAXC}{{\color{\colorMATH}\ensuremath{\epsilon _{2}}}}\endgroup },0)x + ({\begingroup\renewcommand\colorMATH{\colorMATHC}\renewcommand\colorSYNTAX{\colorSYNTAXC}{{\color{\colorMATH}\ensuremath{\infty }}}\endgroup },0)y}}}. Then to typecheck the outermost bind expression, now the privacy on {{\color{\colorMATH}\ensuremath{y}}} is dropped getting a final effect environment of {{\color{\colorMATH}\ensuremath{({\begingroup\renewcommand\colorMATH{\colorMATHC}\renewcommand\colorSYNTAX{\colorSYNTAXC}{{\color{\colorMATH}\ensuremath{\epsilon _{1}}}}\endgroup },0)x + ({\begingroup\renewcommand\colorMATH{\colorMATHC}\renewcommand\colorSYNTAX{\colorSYNTAXC}{{\color{\colorMATH}\ensuremath{\epsilon _{2}}}}\endgroup },0)x = ({\begingroup\renewcommand\colorMATH{\colorMATHC}\renewcommand\colorSYNTAX{\colorSYNTAXC}{{\color{\colorMATH}\ensuremath{\epsilon _{1}}}}\endgroup } + {\begingroup\renewcommand\colorMATH{\colorMATHC}\renewcommand\colorSYNTAX{\colorSYNTAXC}{{\color{\colorMATH}\ensuremath{\epsilon _{2}}}}\endgroup },0)x}}}.
\item  Rule{\textsc{ p-case}} is similar to rule{\textsc{ s-case}}. Here we lift the ambient sensitivity effect of
   the sum expression to infinity, {i.e.}, we pay infinity for all
   non-zero-sensitive variables used in {{\color{\colorMATH}\ensuremath{{\begingroup\renewcommand\colorMATH{\colorMATHB}\renewcommand\colorSYNTAX{\colorSYNTAXB}{{\color{\colorMATH}\ensuremath{\se_{1}}}}\endgroup }}}}.
   % We also scale {{\color{\colorMATH}\ensuremath{{\begingroup\renewcommand\colorMATH{\colorMATHB}\renewcommand\colorSYNTAX{\colorSYNTAXB}{{\color{\colorMATH}\ensuremath{\sS_{1}}}}\endgroup }}}} with \distanceBoundName environment {{\color{\colorMATH}\ensuremath{{\begingroup\renewcommand\colorMATH{\colorMATHB}\renewcommand\colorSYNTAX{\colorSYNTAXB}{{\color{\colorMATH}\ensuremath{\Distance}}}\endgroup }}}} using the operator {{\color{\colorMATH}\ensuremath{{\begingroup\renewcommand\colorMATH{\colorMATHB}\renewcommand\colorSYNTAX{\colorSYNTAXB}{{\color{\colorMATH}\ensuremath{\Distance}}}\endgroup } \times  {\begingroup\renewcommand\colorMATH{\colorMATHB}\renewcommand\colorSYNTAX{\colorSYNTAXB}{{\color{\colorMATH}\ensuremath{\sS_{1}}}}\endgroup }}}}
   % defined as {{\color{\colorMATH}\ensuremath{\forall  x, {\begingroup\renewcommand\colorMATH{\colorMATHB}\renewcommand\colorSYNTAX{\colorSYNTAXB}{{\color{\colorMATH}\ensuremath{\Distance}}}\endgroup } \times  {\begingroup\renewcommand\colorMATH{\colorMATHB}\renewcommand\colorSYNTAX{\colorSYNTAXB}{{\color{\colorMATH}\ensuremath{\sS_{1}}}}\endgroup }(x) = {\begingroup\renewcommand\colorMATH{\colorMATHB}\renewcommand\colorSYNTAX{\colorSYNTAXB}{{\color{\colorMATH}\ensuremath{\Distance}}}\endgroup }(x){\begingroup\renewcommand\colorMATH{\colorMATHB}\renewcommand\colorSYNTAX{\colorSYNTAXB}{{\color{\colorMATH}\ensuremath{\sS_{1}}}}\endgroup }(x)}}}. This is to not pay infinity when
   % we know that variables have zero sensitivity, {i.e.}, when they are
   % ``constant''. 
   For the additional cost of each branch, we compute the join
   between the cost of each branch by substituting each binder by their appropriated cost: {{\color{\colorMATH}\ensuremath{[{\begingroup\renewcommand\colorMATH{\colorMATHB}\renewcommand\colorSYNTAX{\colorSYNTAXB}{{\color{\colorMATH}\ensuremath{\sS_{1 1}}}}\endgroup }/x]{\begingroup\renewcommand\colorMATH{\colorMATHC}\renewcommand\colorSYNTAX{\colorSYNTAXC}{{\color{\colorMATH}\ensuremath{\pS_{2}}}}\endgroup }}}} for the first branch, and {{\color{\colorMATH}\ensuremath{[{\begingroup\renewcommand\colorMATH{\colorMATHB}\renewcommand\colorSYNTAX{\colorSYNTAXB}{{\color{\colorMATH}\ensuremath{\sS_{1 2}}}}\endgroup }/y]{\begingroup\renewcommand\colorMATH{\colorMATHC}\renewcommand\colorSYNTAX{\colorSYNTAXC}{{\color{\colorMATH}\ensuremath{\pS_{3}}}}\endgroup }}}} for the second. Note that we do not use {{\color{\colorMATH}\ensuremath{[{\begingroup\renewcommand\colorMATH{\colorMATHB}\renewcommand\colorSYNTAX{\colorSYNTAXB}{{\color{\colorMATH}\ensuremath{\sS_{1}}}}\endgroup }+{\begingroup\renewcommand\colorMATH{\colorMATHB}\renewcommand\colorSYNTAX{\colorSYNTAXB}{{\color{\colorMATH}\ensuremath{\sS_{1 1}}}}\endgroup }/x]{\begingroup\renewcommand\colorMATH{\colorMATHC}\renewcommand\colorSYNTAX{\colorSYNTAXC}{{\color{\colorMATH}\ensuremath{\pS_{2}}}}\endgroup }}}} and {{\color{\colorMATH}\ensuremath{[{\begingroup\renewcommand\colorMATH{\colorMATHB}\renewcommand\colorSYNTAX{\colorSYNTAXB}{{\color{\colorMATH}\ensuremath{\sS_{1}}}}\endgroup }+{\begingroup\renewcommand\colorMATH{\colorMATHB}\renewcommand\colorSYNTAX{\colorSYNTAXB}{{\color{\colorMATH}\ensuremath{\sS_{1 2}}}}\endgroup }/x]{\begingroup\renewcommand\colorMATH{\colorMATHC}\renewcommand\colorSYNTAX{\colorSYNTAXC}{{\color{\colorMATH}\ensuremath{\pS_{3}}}}\endgroup }}}} as we do in rule{\textsc{ s-case}}, because we are already lifting to infinity (or paying for) every cost associated with {{\color{\colorMATH}\ensuremath{{\begingroup\renewcommand\colorMATH{\colorMATHB}\renewcommand\colorSYNTAX{\colorSYNTAXB}{{\color{\colorMATH}\ensuremath{\sS_{1}}}}\endgroup }}}}.
   For example, consider the following type derivation

   \begingroup\color{\colorMATH}\begin{gather*}
     \inferrule*[lab={\textsc{ p-case}}
     ]{ \Gamma  \mathrel{;} x+2y\hspace*{0.33em}{\begingroup\renewcommand\colorMATH{\colorMATHB}\renewcommand\colorSYNTAX{\colorSYNTAXB}{{\color{\colorMATH}\ensuremath{\vdash }}}\endgroup }\hspace*{0.33em}x \mathrel{:} {\begingroup\renewcommand\colorMATH{\colorMATHA}\renewcommand\colorSYNTAX{\colorSYNTAXA}{{\color{\colorSYNTAX}\texttt{{\ensuremath{{\mathbb{R}}}}}}}\endgroup } \mathrel{^{y}\oplus ^{\varnothing }} {\begingroup\renewcommand\colorMATH{\colorMATHA}\renewcommand\colorSYNTAX{\colorSYNTAXA}{{\color{\colorSYNTAX}\texttt{{\ensuremath{{\mathbb{R}}}}}}}\endgroup } \mathrel{;} x
     \\ \Gamma ,x_{1}\mathrel{:}{\begingroup\renewcommand\colorMATH{\colorMATHA}\renewcommand\colorSYNTAX{\colorSYNTAXA}{{\color{\colorSYNTAX}\texttt{{\ensuremath{{\mathbb{R}}}}}}}\endgroup } \mathrel{;} x+2y + 3x_{1}   \hspace*{0.33em}{\begingroup\renewcommand\colorMATH{\colorMATHC}\renewcommand\colorSYNTAX{\colorSYNTAXC}{{\color{\colorMATH}\ensuremath{\vdash }}}\endgroup }\hspace*{0.33em} {\begingroup\renewcommand\colorMATH{\colorMATHC}\renewcommand\colorSYNTAX{\colorSYNTAXC}{{\color{\colorMATH}\ensuremath{\pe_{2}}}}\endgroup } \mathrel{:} {\begingroup\renewcommand\colorMATH{\colorMATHA}\renewcommand\colorSYNTAX{\colorSYNTAXA}{{\color{\colorSYNTAX}\texttt{{\ensuremath{{\mathbb{R}}}}}}}\endgroup } \mathrel{;} {\begingroup\renewcommand\colorMATH{\colorMATHC}\renewcommand\colorSYNTAX{\colorSYNTAXC}{{\color{\colorMATH}\ensuremath{p_{x}}}}\endgroup }x+{\begingroup\renewcommand\colorMATH{\colorMATHC}\renewcommand\colorSYNTAX{\colorSYNTAXC}{{\color{\colorMATH}\ensuremath{p_{y}}}}\endgroup } y+{\begingroup\renewcommand\colorMATH{\colorMATHC}\renewcommand\colorSYNTAX{\colorSYNTAXC}{{\color{\colorMATH}\ensuremath{p_{2}}}}\endgroup }x_{1}
     \\ \Gamma ,x_{2}\mathrel{:}{\begingroup\renewcommand\colorMATH{\colorMATHA}\renewcommand\colorSYNTAX{\colorSYNTAXA}{{\color{\colorSYNTAX}\texttt{{\ensuremath{{\mathbb{R}}}}}}}\endgroup } \mathrel{;} x+2y + x_{2}   \hspace*{0.33em}{\begingroup\renewcommand\colorMATH{\colorMATHC}\renewcommand\colorSYNTAX{\colorSYNTAXC}{{\color{\colorMATH}\ensuremath{\vdash }}}\endgroup }\hspace*{0.33em} {\begingroup\renewcommand\colorMATH{\colorMATHC}\renewcommand\colorSYNTAX{\colorSYNTAXC}{{\color{\colorMATH}\ensuremath{\pe_{3}}}}\endgroup } \mathrel{:} {\begingroup\renewcommand\colorMATH{\colorMATHA}\renewcommand\colorSYNTAX{\colorSYNTAXA}{{\color{\colorSYNTAX}\texttt{{\ensuremath{{\mathbb{R}}}}}}}\endgroup } \mathrel{;} {\begingroup\renewcommand\colorMATH{\colorMATHC}\renewcommand\colorSYNTAX{\colorSYNTAXC}{{\color{\colorMATH}\ensuremath{p_{x}'}}}\endgroup }x+{\begingroup\renewcommand\colorMATH{\colorMATHC}\renewcommand\colorSYNTAX{\colorSYNTAXC}{{\color{\colorMATH}\ensuremath{p_{y}'}}}\endgroup }y+{\begingroup\renewcommand\colorMATH{\colorMATHC}\renewcommand\colorSYNTAX{\colorSYNTAXC}{{\color{\colorMATH}\ensuremath{p_{3}}}}\endgroup }x_{2}
        }{
        \Gamma  \mathrel{;} x+2y\hspace*{0.33em}{\begingroup\renewcommand\colorMATH{\colorMATHC}\renewcommand\colorSYNTAX{\colorSYNTAXC}{{\color{\colorMATH}\ensuremath{\vdash }}}\endgroup }\hspace*{0.33em} {\begingroup\renewcommand\colorMATH{\colorMATHC}\renewcommand\colorSYNTAX{\colorSYNTAXC}{{\color{\colorSYNTAX}\texttt{case}}}\endgroup }\hspace*{0.33em}x\hspace*{0.33em}{\begingroup\renewcommand\colorMATH{\colorMATHC}\renewcommand\colorSYNTAX{\colorSYNTAXC}{{\color{\colorSYNTAX}\texttt{of}}}\endgroup }\hspace*{0.33em}\{ x_{1}\Rightarrow {\begingroup\renewcommand\colorMATH{\colorMATHC}\renewcommand\colorSYNTAX{\colorSYNTAXC}{{\color{\colorMATH}\ensuremath{\pe_{2}}}}\endgroup }\} \hspace*{0.33em}\{ x_{2}\Rightarrow {\begingroup\renewcommand\colorMATH{\colorMATHC}\renewcommand\colorSYNTAX{\colorSYNTAXC}{{\color{\colorMATH}\ensuremath{\pe_{3}}}}\endgroup }\}  \mathrel{:} \qquad\qquad\qquad\qquad\qquad\qquad
      \\ \qquad\qquad\qquad {\begingroup\renewcommand\colorMATH{\colorMATHA}\renewcommand\colorSYNTAX{\colorSYNTAXA}{{\color{\colorSYNTAX}\texttt{{\ensuremath{{\mathbb{R}}}}}}}\endgroup } \mathrel{;}  {\begingroup\renewcommand\colorMATH{\colorMATHC}\renewcommand\colorSYNTAX{\colorSYNTAXC}{{\color{\colorMATH}\ensuremath{\rceil {\begingroup\renewcommand\colorMATH{\colorMATHA}\renewcommand\colorSYNTAX{\colorSYNTAXA}{{\color{\colorMATH}\ensuremath{x}}}\endgroup }\lceil ^{\infty }}}}\endgroup } \sqcup  ({\begingroup\renewcommand\colorMATH{\colorMATHC}\renewcommand\colorSYNTAX{\colorSYNTAXC}{{\color{\colorMATH}\ensuremath{p_{x}}}}\endgroup }x+{\begingroup\renewcommand\colorMATH{\colorMATHC}\renewcommand\colorSYNTAX{\colorSYNTAXC}{{\color{\colorMATH}\ensuremath{p_{y}}}}\endgroup } y+ {\begingroup\renewcommand\colorMATH{\colorMATHC}\renewcommand\colorSYNTAX{\colorSYNTAXC}{{\color{\colorMATH}\ensuremath{\rceil {\begingroup\renewcommand\colorMATH{\colorMATHA}\renewcommand\colorSYNTAX{\colorSYNTAXA}{{\color{\colorMATH}\ensuremath{y}}}\endgroup }\lceil ^{p_{2}}}}}\endgroup }) \sqcup  ({\begingroup\renewcommand\colorMATH{\colorMATHC}\renewcommand\colorSYNTAX{\colorSYNTAXC}{{\color{\colorMATH}\ensuremath{p_{x}'}}}\endgroup }x+{\begingroup\renewcommand\colorMATH{\colorMATHC}\renewcommand\colorSYNTAX{\colorSYNTAXC}{{\color{\colorMATH}\ensuremath{p_{y}'}}}\endgroup }y+{\begingroup\renewcommand\colorMATH{\colorMATHC}\renewcommand\colorSYNTAX{\colorSYNTAXC}{{\color{\colorMATH}\ensuremath{\rceil {\begingroup\renewcommand\colorMATH{\colorMATHA}\renewcommand\colorSYNTAX{\colorSYNTAXA}{{\color{\colorMATH}\ensuremath{\varnothing }}}\endgroup }\lceil ^{p_{3}}}}}\endgroup })
     }
   \end{gather*}\endgroup
   Note that the variation bound of {{\color{\colorMATH}\ensuremath{x_{1}}}} is computed as {{\color{\colorMATH}\ensuremath{(x+2y)\mathord{\cdotp }(x+y) = 3}}}, and that of {{\color{\colorMATH}\ensuremath{x_{2}}}} as {{\color{\colorMATH}\ensuremath{(x+2y)\mathord{\cdotp }(x) = 1}}}.
   As {{\color{\colorMATH}\ensuremath{{\begingroup\renewcommand\colorMATH{\colorMATHC}\renewcommand\colorSYNTAX{\colorSYNTAXC}{{\color{\colorMATH}\ensuremath{\rceil {\begingroup\renewcommand\colorMATH{\colorMATHA}\renewcommand\colorSYNTAX{\colorSYNTAXA}{{\color{\colorMATH}\ensuremath{x}}}\endgroup }\lceil ^{\infty }}}}\endgroup } = {\begingroup\renewcommand\colorMATH{\colorMATHC}\renewcommand\colorSYNTAX{\colorSYNTAXC}{{\color{\colorMATH}\ensuremath{\infty }}}\endgroup }x}}}, {{\color{\colorMATH}\ensuremath{{\begingroup\renewcommand\colorMATH{\colorMATHC}\renewcommand\colorSYNTAX{\colorSYNTAXC}{{\color{\colorMATH}\ensuremath{\rceil {\begingroup\renewcommand\colorMATH{\colorMATHA}\renewcommand\colorSYNTAX{\colorSYNTAXA}{{\color{\colorMATH}\ensuremath{y}}}\endgroup }\lceil ^{p_{2}}}}}\endgroup } = {\begingroup\renewcommand\colorMATH{\colorMATHC}\renewcommand\colorSYNTAX{\colorSYNTAXC}{{\color{\colorMATH}\ensuremath{p_{2}}}}\endgroup }y}}}, and {{\color{\colorMATH}\ensuremath{{\begingroup\renewcommand\colorMATH{\colorMATHC}\renewcommand\colorSYNTAX{\colorSYNTAXC}{{\color{\colorMATH}\ensuremath{\rceil {\begingroup\renewcommand\colorMATH{\colorMATHA}\renewcommand\colorSYNTAX{\colorSYNTAXA}{{\color{\colorMATH}\ensuremath{\varnothing }}}\endgroup }\lceil ^{p_{3}}}}}\endgroup } = \varnothing }}}, then
   the resulting effect environment is {{\color{\colorMATH}\ensuremath{{\begingroup\renewcommand\colorMATH{\colorMATHC}\renewcommand\colorSYNTAX{\colorSYNTAXC}{{\color{\colorMATH}\ensuremath{\infty }}}\endgroup }x \sqcup  ({\begingroup\renewcommand\colorMATH{\colorMATHC}\renewcommand\colorSYNTAX{\colorSYNTAXC}{{\color{\colorMATH}\ensuremath{p_{x}}}}\endgroup }x+({\begingroup\renewcommand\colorMATH{\colorMATHC}\renewcommand\colorSYNTAX{\colorSYNTAXC}{{\color{\colorMATH}\ensuremath{p_{y}}}}\endgroup } + {\begingroup\renewcommand\colorMATH{\colorMATHC}\renewcommand\colorSYNTAX{\colorSYNTAXC}{{\color{\colorMATH}\ensuremath{p_{2}}}}\endgroup })y) \sqcup  ({\begingroup\renewcommand\colorMATH{\colorMATHC}\renewcommand\colorSYNTAX{\colorSYNTAXC}{{\color{\colorMATH}\ensuremath{p_{x}'}}}\endgroup }x+{\begingroup\renewcommand\colorMATH{\colorMATHC}\renewcommand\colorSYNTAX{\colorSYNTAXC}{{\color{\colorMATH}\ensuremath{p_{y}'}}}\endgroup }y) = {\begingroup\renewcommand\colorMATH{\colorMATHC}\renewcommand\colorSYNTAX{\colorSYNTAXC}{{\color{\colorMATH}\ensuremath{\infty }}}\endgroup }x \sqcup  (({\begingroup\renewcommand\colorMATH{\colorMATHC}\renewcommand\colorSYNTAX{\colorSYNTAXC}{{\color{\colorMATH}\ensuremath{p_{y}}}}\endgroup }+{\begingroup\renewcommand\colorMATH{\colorMATHC}\renewcommand\colorSYNTAX{\colorSYNTAXC}{{\color{\colorMATH}\ensuremath{p_{2}}}}\endgroup }) \sqcup  {\begingroup\renewcommand\colorMATH{\colorMATHC}\renewcommand\colorSYNTAX{\colorSYNTAXC}{{\color{\colorMATH}\ensuremath{p_{y}'}}}\endgroup })y}}}.

\item  Rule{\textsc{ p-app}} uses the sensitivity type system to typecheck both subterms.
   The first subterm has to be typed as a privacy function. Just as {\textsc{ s-app}}, it checks that the sensitivity cost of the argument is bounded by {{\color{\colorMATH}\ensuremath{{\begingroup\renewcommand\colorMATH{\colorMATHB}\renewcommand\colorSYNTAX{\colorSYNTAXB}{{\color{\colorMATH}\ensuremath{\distance}}}\endgroup }}}} by
   computing the dot operation {{\color{\colorMATH}\ensuremath{{\begingroup\renewcommand\colorMATH{\colorMATHB}\renewcommand\colorSYNTAX{\colorSYNTAXB}{{\color{\colorMATH}\ensuremath{\Distance}}}\endgroup } \mathord{\cdotp } {\begingroup\renewcommand\colorMATH{\colorMATHB}\renewcommand\colorSYNTAX{\colorSYNTAXB}{{\color{\colorMATH}\ensuremath{\sS_{2}}}}\endgroup }}}} between \distanceBoundName environment {{\color{\colorMATH}\ensuremath{{\begingroup\renewcommand\colorMATH{\colorMATHB}\renewcommand\colorSYNTAX{\colorSYNTAXB}{{\color{\colorMATH}\ensuremath{\Distance}}}\endgroup }}}} and sensitivity environment {{\color{\colorMATH}\ensuremath{{\begingroup\renewcommand\colorMATH{\colorMATHB}\renewcommand\colorSYNTAX{\colorSYNTAXB}{{\color{\colorMATH}\ensuremath{\sS_{2}}}}\endgroup }}}}. The
   resulting ambient privacy effect is computed as the lift to infinite of the ambient sensitivity effect of {{\color{\colorMATH}\ensuremath{{\begingroup\renewcommand\colorMATH{\colorMATHB}\renewcommand\colorSYNTAX{\colorSYNTAXB}{{\color{\colorMATH}\ensuremath{\se_{1}}}}\endgroup }}}}, 
   plus the latent contextual effect of the privacy function, where we substitute {{\color{\colorMATH}\ensuremath{{\begingroup\renewcommand\colorMATH{\colorMATHB}\renewcommand\colorSYNTAX{\colorSYNTAXB}{{\color{\colorMATH}\ensuremath{\sS_{2}}}}\endgroup }}}} by {{\color{\colorMATH}\ensuremath{x}}}. 
   Similarly to rule{\textsc{ s-app}}, rule{\textsc{ p-app}} also 
   enforces that the relational distance of the argument is bounded by {{\color{\colorMATH}\ensuremath{{\begingroup\renewcommand\colorMATH{\colorMATHB}\renewcommand\colorSYNTAX{\colorSYNTAXB}{{\color{\colorMATH}\ensuremath{\distance}}}\endgroup }}}}, i.e. {{\color{\colorMATH}\ensuremath{{\begingroup\renewcommand\colorMATH{\colorMATHB}\renewcommand\colorSYNTAX{\colorSYNTAXB}{{\color{\colorMATH}\ensuremath{\Distance}}}\endgroup } \mathord{\cdotp } {\begingroup\renewcommand\colorMATH{\colorMATHB}\renewcommand\colorSYNTAX{\colorSYNTAXB}{{\color{\colorMATH}\ensuremath{\sS_{2}}}}\endgroup } \leq  {\begingroup\renewcommand\colorMATH{\colorMATHB}\renewcommand\colorSYNTAX{\colorSYNTAXB}{{\color{\colorMATH}\ensuremath{\distance}}}\endgroup }}}}.
   For instance, consider the following type derivation:

   \begingroup\color{\colorMATH}\begin{gather*}
     \inferrule*[lab={\textsc{ p-app}}
      ]{ \Gamma  \mathrel{;} y\hspace*{0.33em}{\begingroup\renewcommand\colorMATH{\colorMATHB}\renewcommand\colorSYNTAX{\colorSYNTAXB}{{\color{\colorMATH}\ensuremath{\vdash }}}\endgroup }\hspace*{0.33em} {\begingroup\renewcommand\colorMATH{\colorMATHC}\renewcommand\colorSYNTAX{\colorSYNTAXC}{{\color{\colorSYNTAX}\texttt{if}}}\endgroup }\hspace*{0.33em}y\hspace*{0.33em}{\begingroup\renewcommand\colorMATH{\colorMATHC}\renewcommand\colorSYNTAX{\colorSYNTAXC}{{\color{\colorSYNTAX}\texttt{then}}}\endgroup }\hspace*{0.33em}{\begingroup\renewcommand\colorMATH{\colorMATHB}\renewcommand\colorSYNTAX{\colorSYNTAXB}{{\color{\colorMATH}\ensuremath{\se_{2}}}}\endgroup }\hspace*{0.33em}{\begingroup\renewcommand\colorMATH{\colorMATHC}\renewcommand\colorSYNTAX{\colorSYNTAXC}{{\color{\colorSYNTAX}\texttt{else}}}\endgroup }\hspace*{0.33em}{\begingroup\renewcommand\colorMATH{\colorMATHB}\renewcommand\colorSYNTAX{\colorSYNTAXB}{{\color{\colorMATH}\ensuremath{\se_{3}}}}\endgroup } \mathrel{:} (x\mathrel{:}{\begingroup\renewcommand\colorMATH{\colorMATHA}\renewcommand\colorSYNTAX{\colorSYNTAXA}{{\color{\colorSYNTAX}\texttt{{\ensuremath{{\mathbb{R}}}}}}}\endgroup }\mathord{\cdotp }4) \xrightarrowP {{\begingroup\renewcommand\colorMATH{\colorMATHC}\renewcommand\colorSYNTAX{\colorSYNTAXC}{{\color{\colorMATH}\ensuremath{p_{1}}}}\endgroup }y\sqcup {\begingroup\renewcommand\colorMATH{\colorMATHC}\renewcommand\colorSYNTAX{\colorSYNTAXC}{{\color{\colorMATH}\ensuremath{p_{2}}}}\endgroup }x} {\begingroup\renewcommand\colorMATH{\colorMATHA}\renewcommand\colorSYNTAX{\colorSYNTAXA}{{\color{\colorSYNTAX}\texttt{{\ensuremath{{\mathbb{R}}}}}}}\endgroup } \mathrel{;} y
      \\ \Gamma  \mathrel{;} y\hspace*{0.33em}{\begingroup\renewcommand\colorMATH{\colorMATHB}\renewcommand\colorSYNTAX{\colorSYNTAXB}{{\color{\colorMATH}\ensuremath{\vdash }}}\endgroup }\hspace*{0.33em}2*y \mathrel{:} {\begingroup\renewcommand\colorMATH{\colorMATHA}\renewcommand\colorSYNTAX{\colorSYNTAXA}{{\color{\colorSYNTAX}\texttt{{\ensuremath{{\mathbb{R}}}}}}}\endgroup } \mathrel{;} 2y
      \\ 2 \leq  4
         }{
         \Gamma  \mathrel{;} y\hspace*{0.33em}{\begingroup\renewcommand\colorMATH{\colorMATHC}\renewcommand\colorSYNTAX{\colorSYNTAXC}{{\color{\colorMATH}\ensuremath{\vdash }}}\endgroup }\hspace*{0.33em} ({\begingroup\renewcommand\colorMATH{\colorMATHC}\renewcommand\colorSYNTAX{\colorSYNTAXC}{{\color{\colorSYNTAX}\texttt{if}}}\endgroup }\hspace*{0.33em}y\hspace*{0.33em}{\begingroup\renewcommand\colorMATH{\colorMATHC}\renewcommand\colorSYNTAX{\colorSYNTAXC}{{\color{\colorSYNTAX}\texttt{then}}}\endgroup }\hspace*{0.33em}{\begingroup\renewcommand\colorMATH{\colorMATHB}\renewcommand\colorSYNTAX{\colorSYNTAXB}{{\color{\colorMATH}\ensuremath{\se_{2}}}}\endgroup }\hspace*{0.33em}{\begingroup\renewcommand\colorMATH{\colorMATHC}\renewcommand\colorSYNTAX{\colorSYNTAXC}{{\color{\colorSYNTAX}\texttt{else}}}\endgroup }\hspace*{0.33em}{\begingroup\renewcommand\colorMATH{\colorMATHB}\renewcommand\colorSYNTAX{\colorSYNTAXB}{{\color{\colorMATH}\ensuremath{\se_{3}}}}\endgroup }\} )\hspace*{0.33em}(2*y) \mathrel{:} {\begingroup\renewcommand\colorMATH{\colorMATHA}\renewcommand\colorSYNTAX{\colorSYNTAXA}{{\color{\colorSYNTAX}\texttt{{\ensuremath{{\mathbb{R}}}}}}}\endgroup } \mathrel{;}  {\begingroup\renewcommand\colorMATH{\colorMATHC}\renewcommand\colorSYNTAX{\colorSYNTAXC}{{\color{\colorMATH}\ensuremath{\rceil {\begingroup\renewcommand\colorMATH{\colorMATHA}\renewcommand\colorSYNTAX{\colorSYNTAXA}{{\color{\colorMATH}\ensuremath{y}}}\endgroup }\lceil ^{\infty }}}}\endgroup } + ({\begingroup\renewcommand\colorMATH{\colorMATHC}\renewcommand\colorSYNTAX{\colorSYNTAXC}{{\color{\colorMATH}\ensuremath{p_{1}}}}\endgroup }y \sqcup  {\begingroup\renewcommand\colorMATH{\colorMATHC}\renewcommand\colorSYNTAX{\colorSYNTAXC}{{\color{\colorMATH}\ensuremath{\rceil {\begingroup\renewcommand\colorMATH{\colorMATHA}\renewcommand\colorSYNTAX{\colorSYNTAXA}{{\color{\colorMATH}\ensuremath{2y}}}\endgroup }\lceil ^{p_{2}}}}}\endgroup }) 
      }
   \end{gather*}\endgroup
   % Note that the \distanceBoundName environment is {{\color{\colorMATH}\ensuremath{0y}}}, this means that {{\color{\colorMATH}\ensuremath{y}}} cannot wiggle, i.e. {{\color{\colorMATH}\ensuremath{y}}} will behave as a constant value. 
   The resulting effect environment is computed as {{\color{\colorMATH}\ensuremath{{\begingroup\renewcommand\colorMATH{\colorMATHC}\renewcommand\colorSYNTAX{\colorSYNTAXC}{{\color{\colorMATH}\ensuremath{\infty }}}\endgroup }y + [2y/x]({\begingroup\renewcommand\colorMATH{\colorMATHC}\renewcommand\colorSYNTAX{\colorSYNTAXC}{{\color{\colorMATH}\ensuremath{p_{1}}}}\endgroup }y\sqcup {\begingroup\renewcommand\colorMATH{\colorMATHC}\renewcommand\colorSYNTAX{\colorSYNTAXC}{{\color{\colorMATH}\ensuremath{p_{2}}}}\endgroup }x) = {\begingroup\renewcommand\colorMATH{\colorMATHC}\renewcommand\colorSYNTAX{\colorSYNTAXC}{{\color{\colorMATH}\ensuremath{\infty }}}\endgroup }y + {\begingroup\renewcommand\colorMATH{\colorMATHC}\renewcommand\colorSYNTAX{\colorSYNTAXC}{{\color{\colorMATH}\ensuremath{p_{1}}}}\endgroup }y\sqcup {\begingroup\renewcommand\colorMATH{\colorMATHC}\renewcommand\colorSYNTAX{\colorSYNTAXC}{{\color{\colorMATH}\ensuremath{\rceil {\begingroup\renewcommand\colorMATH{\colorMATHA}\renewcommand\colorSYNTAX{\colorSYNTAXA}{{\color{\colorMATH}\ensuremath{2y}}}\endgroup }\lceil ^{p_{2}}}}}\endgroup } = {\begingroup\renewcommand\colorMATH{\colorMATHC}\renewcommand\colorSYNTAX{\colorSYNTAXC}{{\color{\colorMATH}\ensuremath{\infty }}}\endgroup }y + ({\begingroup\renewcommand\colorMATH{\colorMATHC}\renewcommand\colorSYNTAX{\colorSYNTAXC}{{\color{\colorMATH}\ensuremath{p_{1}}}}\endgroup } \sqcup  {\begingroup\renewcommand\colorMATH{\colorMATHC}\renewcommand\colorSYNTAX{\colorSYNTAXC}{{\color{\colorMATH}\ensuremath{p_{2}}}}\endgroup })y}}}, which is equivalent to {{\color{\colorMATH}\ensuremath{{\begingroup\renewcommand\colorMATH{\colorMATHC}\renewcommand\colorSYNTAX{\colorSYNTAXC}{{\color{\colorMATH}\ensuremath{\infty }}}\endgroup }y}}}.
   If {{\color{\colorMATH}\ensuremath{y}}} does not wiggle, then the ambient privacy effect will be zero.
   If the \distanceBoundName environment for {{\color{\colorMATH}\ensuremath{y}}} were {{\color{\colorMATH}\ensuremath{3}}}, then this program would be ill-typed since {{\color{\colorMATH}\ensuremath{3y\mathord{\cdotp }2y \not\leq 4}}}.
\end{itemize}

\subsection{$\lang$: Type Safety}
\label{sec:type-safety}
Type safety is defined in the same line of Section~\ref{sec:sensitivity-simple-type-safety}.
To establish type safety of the privacy language, we define a non-deterministic sampling big-step semantics of privacy expressions; see Figure~\ref{fig:interpreter-prob-sem}.
\begin{figure}[t]
\begin{small}
\begin{framed}
\begingroup\color{\colorMATH}\begin{mathpar}\inferrule*[lab={\textsc{ return}}
  ]{ \gamma  \vdash  {\begingroup\renewcommand\colorMATH{\colorMATHB}\renewcommand\colorSYNTAX{\colorSYNTAXB}{{\color{\colorMATH}\ensuremath{\se}}}\endgroup } \Downarrow  {\begingroup\renewcommand\colorMATH{\colorMATHB}\renewcommand\colorSYNTAX{\colorSYNTAXB}{{\color{\colorMATH}\ensuremath{\sv}}}\endgroup }
    }{
     \gamma  \vdash  {\begingroup\renewcommand\colorMATH{\colorMATHC}\renewcommand\colorSYNTAX{\colorSYNTAXC}{{\color{\colorSYNTAX}\texttt{return}}}\endgroup }\hspace*{0.33em}{\begingroup\renewcommand\colorMATH{\colorMATHB}\renewcommand\colorSYNTAX{\colorSYNTAXB}{{\color{\colorMATH}\ensuremath{\se}}}\endgroup } \Downarrow  {\begingroup\renewcommand\colorMATH{\colorMATHB}\renewcommand\colorSYNTAX{\colorSYNTAXB}{{\color{\colorMATH}\ensuremath{\sv}}}\endgroup }
  }
\and\inferrule*[lab={\textsc{ bind}}
  ]{  \gamma  \vdash  {\begingroup\renewcommand\colorMATH{\colorMATHC}\renewcommand\colorSYNTAX{\colorSYNTAXC}{{\color{\colorMATH}\ensuremath{\pe_{1}}}}\endgroup } \Downarrow  {\begingroup\renewcommand\colorMATH{\colorMATHB}\renewcommand\colorSYNTAX{\colorSYNTAXB}{{\color{\colorMATH}\ensuremath{\sv_{1}}}}\endgroup }
  \\  \gamma [x \mapsto  {\begingroup\renewcommand\colorMATH{\colorMATHB}\renewcommand\colorSYNTAX{\colorSYNTAXB}{{\color{\colorMATH}\ensuremath{\sv_{1}}}}\endgroup }] \vdash  {\begingroup\renewcommand\colorMATH{\colorMATHC}\renewcommand\colorSYNTAX{\colorSYNTAXC}{{\color{\colorMATH}\ensuremath{\pe_{2}}}}\endgroup } \Downarrow  {\begingroup\renewcommand\colorMATH{\colorMATHB}\renewcommand\colorSYNTAX{\colorSYNTAXB}{{\color{\colorMATH}\ensuremath{\sv_{2}}}}\endgroup }
    }{
      \gamma  \vdash  x: \tau _{1} \leftarrow  {\begingroup\renewcommand\colorMATH{\colorMATHC}\renewcommand\colorSYNTAX{\colorSYNTAXC}{{\color{\colorMATH}\ensuremath{\pe_{1}}}}\endgroup }\mathrel{;}{\begingroup\renewcommand\colorMATH{\colorMATHC}\renewcommand\colorSYNTAX{\colorSYNTAXC}{{\color{\colorMATH}\ensuremath{\pe_{2}}}}\endgroup } \Downarrow  {\begingroup\renewcommand\colorMATH{\colorMATHB}\renewcommand\colorSYNTAX{\colorSYNTAXB}{{\color{\colorMATH}\ensuremath{\sv_{2}}}}\endgroup }
  }
\and\inferrule*[lab={\textsc{ gauss}}
  ]{  {\begingroup\renewcommand\colorMATH{\colorMATHB}\renewcommand\colorSYNTAX{\colorSYNTAXB}{{\color{\colorMATH}\ensuremath{r}}}\endgroup } \in  {\begingroup\renewcommand\colorMATH{\colorMATHA}\renewcommand\colorSYNTAX{\colorSYNTAXA}{{\color{\colorSYNTAX}\texttt{{\ensuremath{{\mathbb{R}}}}}}}\endgroup }
    }{
      \gamma  \vdash  {\begingroup\renewcommand\colorMATH{\colorMATHC}\renewcommand\colorSYNTAX{\colorSYNTAXC}{{\color{\colorSYNTAX}\texttt{gauss}}}\endgroup } \hspace*{0.33em} \mu  \hspace*{0.33em} \sigma ^{2} \Downarrow  {\begingroup\renewcommand\colorMATH{\colorMATHB}\renewcommand\colorSYNTAX{\colorSYNTAXB}{{\color{\colorMATH}\ensuremath{r}}}\endgroup }
  }
\and\inferrule*[lab={\textsc{ if-true}}
  ]{ \gamma  \vdash  {\begingroup\renewcommand\colorMATH{\colorMATHB}\renewcommand\colorSYNTAX{\colorSYNTAXB}{{\color{\colorMATH}\ensuremath{\se_{1}}}}\endgroup } \Downarrow  {\text{true}}
  \\ \gamma  \vdash  {\begingroup\renewcommand\colorMATH{\colorMATHC}\renewcommand\colorSYNTAX{\colorSYNTAXC}{{\color{\colorMATH}\ensuremath{\pe_{2}}}}\endgroup } \Downarrow  {\begingroup\renewcommand\colorMATH{\colorMATHB}\renewcommand\colorSYNTAX{\colorSYNTAXB}{{\color{\colorMATH}\ensuremath{\sv_{2}}}}\endgroup }
     }{
      \gamma  \vdash  {\begingroup\renewcommand\colorMATH{\colorMATHC}\renewcommand\colorSYNTAX{\colorSYNTAXC}{{\color{\colorSYNTAX}\texttt{if}}}\endgroup }\hspace*{0.33em}{\begingroup\renewcommand\colorMATH{\colorMATHB}\renewcommand\colorSYNTAX{\colorSYNTAXB}{{\color{\colorMATH}\ensuremath{\se_{1}}}}\endgroup }\hspace*{0.33em}{\begingroup\renewcommand\colorMATH{\colorMATHC}\renewcommand\colorSYNTAX{\colorSYNTAXC}{{\color{\colorSYNTAX}\texttt{then}}}\endgroup }\hspace*{0.33em}{\begingroup\renewcommand\colorMATH{\colorMATHC}\renewcommand\colorSYNTAX{\colorSYNTAXC}{{\color{\colorMATH}\ensuremath{\pe_{2}}}}\endgroup }\hspace*{0.33em}{\begingroup\renewcommand\colorMATH{\colorMATHC}\renewcommand\colorSYNTAX{\colorSYNTAXC}{{\color{\colorSYNTAX}\texttt{else}}}\endgroup }\hspace*{0.33em}{\begingroup\renewcommand\colorMATH{\colorMATHC}\renewcommand\colorSYNTAX{\colorSYNTAXC}{{\color{\colorMATH}\ensuremath{\pe_{3}}}}\endgroup } \Downarrow  {\begingroup\renewcommand\colorMATH{\colorMATHB}\renewcommand\colorSYNTAX{\colorSYNTAXB}{{\color{\colorMATH}\ensuremath{\sv_{2}}}}\endgroup }
  }
\and\inferrule*[lab={\textsc{ if-false}}
  ]{ \gamma  \vdash  {\begingroup\renewcommand\colorMATH{\colorMATHB}\renewcommand\colorSYNTAX{\colorSYNTAXB}{{\color{\colorMATH}\ensuremath{\se_{1}}}}\endgroup } \Downarrow  {\text{false}}
  \\ \gamma  \vdash  {\begingroup\renewcommand\colorMATH{\colorMATHC}\renewcommand\colorSYNTAX{\colorSYNTAXC}{{\color{\colorMATH}\ensuremath{\pe_{3}}}}\endgroup } \Downarrow  {\begingroup\renewcommand\colorMATH{\colorMATHB}\renewcommand\colorSYNTAX{\colorSYNTAXB}{{\color{\colorMATH}\ensuremath{\sv_{3}}}}\endgroup }
     }{
      \gamma  \vdash  {\begingroup\renewcommand\colorMATH{\colorMATHC}\renewcommand\colorSYNTAX{\colorSYNTAXC}{{\color{\colorSYNTAX}\texttt{if}}}\endgroup }\hspace*{0.33em}{\begingroup\renewcommand\colorMATH{\colorMATHB}\renewcommand\colorSYNTAX{\colorSYNTAXB}{{\color{\colorMATH}\ensuremath{\se_{1}}}}\endgroup }\hspace*{0.33em}{\begingroup\renewcommand\colorMATH{\colorMATHC}\renewcommand\colorSYNTAX{\colorSYNTAXC}{{\color{\colorSYNTAX}\texttt{then}}}\endgroup }\hspace*{0.33em}{\begingroup\renewcommand\colorMATH{\colorMATHC}\renewcommand\colorSYNTAX{\colorSYNTAXC}{{\color{\colorMATH}\ensuremath{\pe_{2}}}}\endgroup }\hspace*{0.33em}{\begingroup\renewcommand\colorMATH{\colorMATHC}\renewcommand\colorSYNTAX{\colorSYNTAXC}{{\color{\colorSYNTAX}\texttt{else}}}\endgroup }\hspace*{0.33em}{\begingroup\renewcommand\colorMATH{\colorMATHC}\renewcommand\colorSYNTAX{\colorSYNTAXC}{{\color{\colorMATH}\ensuremath{\pe_{3}}}}\endgroup } \Downarrow  {\begingroup\renewcommand\colorMATH{\colorMATHB}\renewcommand\colorSYNTAX{\colorSYNTAXB}{{\color{\colorMATH}\ensuremath{\sv_{3}}}}\endgroup }
  }
\and \inferrule*[lab={\textsc{ case-left}}
   ]{ \gamma \vdash  {\begingroup\renewcommand\colorMATH{\colorMATHC}\renewcommand\colorSYNTAX{\colorSYNTAXC}{{\color{\colorMATH}\ensuremath{\pe}}}\endgroup }  \Downarrow  \inl\hspace*{0.33em}{\begingroup\renewcommand\colorMATH{\colorMATHB}\renewcommand\colorSYNTAX{\colorSYNTAXB}{{\color{\colorMATH}\ensuremath{\sv}}}\endgroup }
   \\ \gamma [x\mapsto {\begingroup\renewcommand\colorMATH{\colorMATHB}\renewcommand\colorSYNTAX{\colorSYNTAXB}{{\color{\colorMATH}\ensuremath{\sv}}}\endgroup } ]\vdash {\begingroup\renewcommand\colorMATH{\colorMATHC}\renewcommand\colorSYNTAX{\colorSYNTAXC}{{\color{\colorMATH}\ensuremath{\pe_{2}}}}\endgroup }  \Downarrow  {\begingroup\renewcommand\colorMATH{\colorMATHB}\renewcommand\colorSYNTAX{\colorSYNTAXB}{{\color{\colorMATH}\ensuremath{\sv_{2}}}}\endgroup }
      }{
      \gamma \vdash {\begingroup\renewcommand\colorMATH{\colorMATHC}\renewcommand\colorSYNTAX{\colorSYNTAXC}{{\color{\colorSYNTAX}\texttt{case}}}\endgroup }\hspace*{0.33em}{\begingroup\renewcommand\colorMATH{\colorMATHC}\renewcommand\colorSYNTAX{\colorSYNTAXC}{{\color{\colorMATH}\ensuremath{\pe}}}\endgroup }\hspace*{0.33em}{\begingroup\renewcommand\colorMATH{\colorMATHC}\renewcommand\colorSYNTAX{\colorSYNTAXC}{{\color{\colorSYNTAX}\texttt{of}}}\endgroup }\hspace*{0.33em}\{ x\Rightarrow  {\begingroup\renewcommand\colorMATH{\colorMATHC}\renewcommand\colorSYNTAX{\colorSYNTAXC}{{\color{\colorMATH}\ensuremath{\pe_{2}}}}\endgroup } \} \hspace*{0.33em}\{ x\Rightarrow  {\begingroup\renewcommand\colorMATH{\colorMATHC}\renewcommand\colorSYNTAX{\colorSYNTAXC}{{\color{\colorMATH}\ensuremath{\pe_{3}}}}\endgroup }\}  \Downarrow  {\begingroup\renewcommand\colorMATH{\colorMATHB}\renewcommand\colorSYNTAX{\colorSYNTAXB}{{\color{\colorMATH}\ensuremath{\sv_{2}}}}\endgroup }
   }
\and \inferrule*[lab={\textsc{ case-right}}
   ]{ \gamma \vdash  {\begingroup\renewcommand\colorMATH{\colorMATHC}\renewcommand\colorSYNTAX{\colorSYNTAXC}{{\color{\colorMATH}\ensuremath{\pe}}}\endgroup }  \Downarrow  \inr\hspace*{0.33em}{\begingroup\renewcommand\colorMATH{\colorMATHB}\renewcommand\colorSYNTAX{\colorSYNTAXB}{{\color{\colorMATH}\ensuremath{\sv}}}\endgroup }
   \\ \gamma [x\mapsto {\begingroup\renewcommand\colorMATH{\colorMATHB}\renewcommand\colorSYNTAX{\colorSYNTAXB}{{\color{\colorMATH}\ensuremath{\sv}}}\endgroup } ]\vdash {\begingroup\renewcommand\colorMATH{\colorMATHC}\renewcommand\colorSYNTAX{\colorSYNTAXC}{{\color{\colorMATH}\ensuremath{\pe_{3}}}}\endgroup }  \Downarrow  {\begingroup\renewcommand\colorMATH{\colorMATHB}\renewcommand\colorSYNTAX{\colorSYNTAXB}{{\color{\colorMATH}\ensuremath{\sv_{3}}}}\endgroup }
      }{
      \gamma \vdash {\begingroup\renewcommand\colorMATH{\colorMATHC}\renewcommand\colorSYNTAX{\colorSYNTAXC}{{\color{\colorSYNTAX}\texttt{case}}}\endgroup }\hspace*{0.33em}{\begingroup\renewcommand\colorMATH{\colorMATHC}\renewcommand\colorSYNTAX{\colorSYNTAXC}{{\color{\colorMATH}\ensuremath{\pe}}}\endgroup }\hspace*{0.33em}{\begingroup\renewcommand\colorMATH{\colorMATHC}\renewcommand\colorSYNTAX{\colorSYNTAXC}{{\color{\colorSYNTAX}\texttt{of}}}\endgroup }\hspace*{0.33em}\{ x\Rightarrow  {\begingroup\renewcommand\colorMATH{\colorMATHC}\renewcommand\colorSYNTAX{\colorSYNTAXC}{{\color{\colorMATH}\ensuremath{\pe_{2}}}}\endgroup } \} \hspace*{0.33em}\{ x\Rightarrow  {\begingroup\renewcommand\colorMATH{\colorMATHC}\renewcommand\colorSYNTAX{\colorSYNTAXC}{{\color{\colorMATH}\ensuremath{\pe_{3}}}}\endgroup }\}  \Downarrow  {\begingroup\renewcommand\colorMATH{\colorMATHB}\renewcommand\colorSYNTAX{\colorSYNTAXB}{{\color{\colorMATH}\ensuremath{\sv_{3}}}}\endgroup }
   }
\and \inferrule*[lab={\textsc{ app}}
   ]{ \gamma \vdash  {\begingroup\renewcommand\colorMATH{\colorMATHC}\renewcommand\colorSYNTAX{\colorSYNTAXC}{{\color{\colorMATH}\ensuremath{\pe_{1}}}}\endgroup }  \Downarrow  \langle {\begingroup\renewcommand\colorMATH{\colorMATHC}\renewcommand\colorSYNTAX{\colorSYNTAXC}{{\color{\colorMATH}\ensuremath{\plambda}}}\endgroup } x:\tau \mathord{\cdotp } {\begingroup\renewcommand\colorMATH{\colorMATHB}\renewcommand\colorSYNTAX{\colorSYNTAXB}{{\color{\colorMATH}\ensuremath{\sss}}}\endgroup }.\hspace*{0.33em}{\begingroup\renewcommand\colorMATH{\colorMATHC}\renewcommand\colorSYNTAX{\colorSYNTAXC}{{\color{\colorMATH}\ensuremath{\pe^{\prime}}}}\endgroup },\gamma ^{\prime}\rangle 
   \\ \gamma \vdash  {\begingroup\renewcommand\colorMATH{\colorMATHC}\renewcommand\colorSYNTAX{\colorSYNTAXC}{{\color{\colorMATH}\ensuremath{\pe_{2}}}}\endgroup }  \Downarrow  {\begingroup\renewcommand\colorMATH{\colorMATHB}\renewcommand\colorSYNTAX{\colorSYNTAXB}{{\color{\colorMATH}\ensuremath{\sv}}}\endgroup }
   \\ \gamma ^{\prime}[x\mapsto {\begingroup\renewcommand\colorMATH{\colorMATHB}\renewcommand\colorSYNTAX{\colorSYNTAXB}{{\color{\colorMATH}\ensuremath{\sv}}}\endgroup } ]\vdash  {\begingroup\renewcommand\colorMATH{\colorMATHC}\renewcommand\colorSYNTAX{\colorSYNTAXC}{{\color{\colorMATH}\ensuremath{\pe^{\prime}}}}\endgroup } \Downarrow  {\begingroup\renewcommand\colorMATH{\colorMATHB}\renewcommand\colorSYNTAX{\colorSYNTAXB}{{\color{\colorMATH}\ensuremath{\sv'}}}\endgroup }
      }{
      \gamma \vdash  {\begingroup\renewcommand\colorMATH{\colorMATHC}\renewcommand\colorSYNTAX{\colorSYNTAXC}{{\color{\colorMATH}\ensuremath{\pe_{1}}}}\endgroup }\hspace*{0.33em}{\begingroup\renewcommand\colorMATH{\colorMATHC}\renewcommand\colorSYNTAX{\colorSYNTAXC}{{\color{\colorMATH}\ensuremath{\pe_{2}}}}\endgroup }  \Downarrow  {\begingroup\renewcommand\colorMATH{\colorMATHB}\renewcommand\colorSYNTAX{\colorSYNTAXB}{{\color{\colorMATH}\ensuremath{\sv'}}}\endgroup }
   }
\end{mathpar}\endgroup
\end{framed}
\end{small}
\caption{Non-deterministic sampling semantics for privacy expressions}
\label{fig:interpreter-prob-sem}
\end{figure}
We naturally extend the type safety logical relations of Figure~\ref{fig:sensitivity-simple-logical-relation} to support for both sensitivity and privacy lambdas, and privacy expressions as shown in Figure~\ref{fig:type-safety-lr}.
\begin{figure}[t]
\begin{small}
\begin{framed}
\begingroup\color{\colorMATH}\begin{mathpar} \inferrule*[lab={\textsc{ }}
   ]{ \langle {\begingroup\renewcommand\colorMATH{\colorMATHB}\renewcommand\colorSYNTAX{\colorSYNTAXB}{{\color{\colorMATH}\ensuremath{\slambda}}}\endgroup } x:\tau \mathord{\cdotp }{\begingroup\renewcommand\colorMATH{\colorMATHB}\renewcommand\colorSYNTAX{\colorSYNTAXB}{{\color{\colorMATH}\ensuremath{\distance'}}}\endgroup }. {\begingroup\renewcommand\colorMATH{\colorMATHB}\renewcommand\colorSYNTAX{\colorSYNTAXB}{{\color{\colorMATH}\ensuremath{\se}}}\endgroup }, \gamma \rangle  \in  Atom\llbracket (x\mathrel{:}\tau _{1}\mathord{\cdotp }{\begingroup\renewcommand\colorMATH{\colorMATHB}\renewcommand\colorSYNTAX{\colorSYNTAXB}{{\color{\colorMATH}\ensuremath{\distance'}}}\endgroup }) \xrightarrowS {{\begingroup\renewcommand\colorMATH{\colorMATHB}\renewcommand\colorSYNTAX{\colorSYNTAXB}{{\color{\colorMATH}\ensuremath{\sss}}}\endgroup }x} \tau _{2}\rrbracket 
   \\ \forall   {\begingroup\renewcommand\colorMATH{\colorMATHB}\renewcommand\colorSYNTAX{\colorSYNTAXB}{{\color{\colorMATH}\ensuremath{\sv}}}\endgroup } \in  {\mathcal{V}}\llbracket \tau _{1}\rrbracket . \gamma [x \mapsto  {\begingroup\renewcommand\colorMATH{\colorMATHB}\renewcommand\colorSYNTAX{\colorSYNTAXB}{{\color{\colorMATH}\ensuremath{\sv}}}\endgroup }] \vdash  {\begingroup\renewcommand\colorMATH{\colorMATHB}\renewcommand\colorSYNTAX{\colorSYNTAXB}{{\color{\colorMATH}\ensuremath{\se}}}\endgroup } \in  {\mathcal{E}}\llbracket \tau _{2}/(x:\tau _{1})\rrbracket 
      }{
      \langle {\begingroup\renewcommand\colorMATH{\colorMATHB}\renewcommand\colorSYNTAX{\colorSYNTAXB}{{\color{\colorMATH}\ensuremath{\slambda}}}\endgroup } x:\tau _{1}\mathord{\cdotp }{\begingroup\renewcommand\colorMATH{\colorMATHB}\renewcommand\colorSYNTAX{\colorSYNTAXB}{{\color{\colorMATH}\ensuremath{\distance'}}}\endgroup }. {\begingroup\renewcommand\colorMATH{\colorMATHB}\renewcommand\colorSYNTAX{\colorSYNTAXB}{{\color{\colorMATH}\ensuremath{\se}}}\endgroup }, \gamma \rangle  \in  {\mathcal{V}}\llbracket (x\mathrel{:}\tau _{1}\mathord{\cdotp }{\begingroup\renewcommand\colorMATH{\colorMATHB}\renewcommand\colorSYNTAX{\colorSYNTAXB}{{\color{\colorMATH}\ensuremath{\distance'}}}\endgroup }) \xrightarrowS {{\begingroup\renewcommand\colorMATH{\colorMATHB}\renewcommand\colorSYNTAX{\colorSYNTAXB}{{\color{\colorMATH}\ensuremath{\sss}}}\endgroup }x} \tau _{2}\rrbracket 
   }
\and \inferrule*[lab={\textsc{ }}
   ]{ \langle {\begingroup\renewcommand\colorMATH{\colorMATHC}\renewcommand\colorSYNTAX{\colorSYNTAXC}{{\color{\colorMATH}\ensuremath{\plambda}}}\endgroup } x:\tau \mathord{\cdotp }{\begingroup\renewcommand\colorMATH{\colorMATHB}\renewcommand\colorSYNTAX{\colorSYNTAXB}{{\color{\colorMATH}\ensuremath{\distance}}}\endgroup }. {\begingroup\renewcommand\colorMATH{\colorMATHC}\renewcommand\colorSYNTAX{\colorSYNTAXC}{{\color{\colorMATH}\ensuremath{\pe}}}\endgroup }, \gamma \rangle  \in  Atom\llbracket (x\mathrel{:}\tau _{1}\mathord{\cdotp }{\begingroup\renewcommand\colorMATH{\colorMATHB}\renewcommand\colorSYNTAX{\colorSYNTAXB}{{\color{\colorMATH}\ensuremath{\distance}}}\endgroup }) \xrightarrowP {{\begingroup\renewcommand\colorMATH{\colorMATHC}\renewcommand\colorSYNTAX{\colorSYNTAXC}{{\color{\colorMATH}\ensuremath{\pS}}}\endgroup }} \tau _{2}\rrbracket 
   \\ \forall   {\begingroup\renewcommand\colorMATH{\colorMATHB}\renewcommand\colorSYNTAX{\colorSYNTAXB}{{\color{\colorMATH}\ensuremath{\sv}}}\endgroup } \in  {\mathcal{V}}\llbracket \tau _{1}\rrbracket . \gamma [x \mapsto  {\begingroup\renewcommand\colorMATH{\colorMATHB}\renewcommand\colorSYNTAX{\colorSYNTAXB}{{\color{\colorMATH}\ensuremath{\sv}}}\endgroup }] \vdash  {\begingroup\renewcommand\colorMATH{\colorMATHC}\renewcommand\colorSYNTAX{\colorSYNTAXC}{{\color{\colorMATH}\ensuremath{\pe}}}\endgroup } \in  {\mathcal{E}}\llbracket \tau _{2}/(x:\tau _{1})\rrbracket 
      }{
      \langle {\begingroup\renewcommand\colorMATH{\colorMATHC}\renewcommand\colorSYNTAX{\colorSYNTAXC}{{\color{\colorMATH}\ensuremath{\plambda}}}\endgroup } x:\tau _{1}\mathord{\cdotp }{\begingroup\renewcommand\colorMATH{\colorMATHB}\renewcommand\colorSYNTAX{\colorSYNTAXB}{{\color{\colorMATH}\ensuremath{\distance}}}\endgroup }. {\begingroup\renewcommand\colorMATH{\colorMATHC}\renewcommand\colorSYNTAX{\colorSYNTAXC}{{\color{\colorMATH}\ensuremath{\pe}}}\endgroup }, \gamma \rangle  \in  {\mathcal{V}}\llbracket (x\mathrel{:}\tau _{1}\mathord{\cdotp }{\begingroup\renewcommand\colorMATH{\colorMATHB}\renewcommand\colorSYNTAX{\colorSYNTAXB}{{\color{\colorMATH}\ensuremath{\distance}}}\endgroup }) \xrightarrowP {{\begingroup\renewcommand\colorMATH{\colorMATHC}\renewcommand\colorSYNTAX{\colorSYNTAXC}{{\color{\colorMATH}\ensuremath{\pS}}}\endgroup }} \tau _{2}\rrbracket 
   }
\and \inferrule*[lab={\textsc{ }}
   ]{
   \forall {\begingroup\renewcommand\colorMATH{\colorMATHB}\renewcommand\colorSYNTAX{\colorSYNTAXB}{{\color{\colorMATH}\ensuremath{\sv}}}\endgroup }, \gamma  \vdash  {\begingroup\renewcommand\colorMATH{\colorMATHC}\renewcommand\colorSYNTAX{\colorSYNTAXC}{{\color{\colorMATH}\ensuremath{\pe}}}\endgroup } \Downarrow  {\begingroup\renewcommand\colorMATH{\colorMATHB}\renewcommand\colorSYNTAX{\colorSYNTAXB}{{\color{\colorMATH}\ensuremath{\sv}}}\endgroup } \implies {\begingroup\renewcommand\colorMATH{\colorMATHB}\renewcommand\colorSYNTAX{\colorSYNTAXB}{{\color{\colorMATH}\ensuremath{\sv}}}\endgroup } \in  {\mathcal{V}}\llbracket \tau \rrbracket 
      }{
      \gamma  \vdash  {\begingroup\renewcommand\colorMATH{\colorMATHC}\renewcommand\colorSYNTAX{\colorSYNTAXC}{{\color{\colorMATH}\ensuremath{\pe}}}\endgroup } \in  {\mathcal{E}}\llbracket \tau \rrbracket 
   }
\end{mathpar}\endgroup
\end{framed}
\end{small}
\caption{$\lang$: Type Safety Logical Relation (selected rules)}
\label{fig:type-safety-lr}
\end{figure}
The fundamental property of the type safety logical relation is defined similarly to Proposition~\ref{lm:sensitivity-simple-fp}, but now accounting for \distanceBoundName environments and expressions:
\begin{proposition}[Fundamental Property of the Type Safety Logical Relation]
  \label{lm:type-safety-FP-paper}\;
  \begin{enumerate}[label=(\alph*)]
  \item Let {{\color{\colorMATH}\ensuremath{\Gamma ;{\begingroup\renewcommand\colorMATH{\colorMATHB}\renewcommand\colorSYNTAX{\colorSYNTAXB}{{\color{\colorMATH}\ensuremath{\Distance}}}\endgroup } \vdash  {\begingroup\renewcommand\colorMATH{\colorMATHB}\renewcommand\colorSYNTAX{\colorSYNTAXB}{{\color{\colorMATH}\ensuremath{\se}}}\endgroup } : \tau  ; {\begingroup\renewcommand\colorMATH{\colorMATHB}\renewcommand\colorSYNTAX{\colorSYNTAXB}{{\color{\colorMATH}\ensuremath{\sS}}}\endgroup }}}}, and {{\color{\colorMATH}\ensuremath{\gamma  \in  {\mathcal{G}}\llbracket \Gamma \rrbracket }}}. Then
    {{\color{\colorMATH}\ensuremath{\gamma \vdash  {\begingroup\renewcommand\colorMATH{\colorMATHB}\renewcommand\colorSYNTAX{\colorSYNTAXB}{{\color{\colorMATH}\ensuremath{\se}}}\endgroup } \in  {\mathcal{E}}\llbracket \tau /\Gamma \rrbracket }}}.
  \item Let {{\color{\colorMATH}\ensuremath{\Gamma ;{\begingroup\renewcommand\colorMATH{\colorMATHB}\renewcommand\colorSYNTAX{\colorSYNTAXB}{{\color{\colorMATH}\ensuremath{\Distance}}}\endgroup } \vdash  {\begingroup\renewcommand\colorMATH{\colorMATHC}\renewcommand\colorSYNTAX{\colorSYNTAXC}{{\color{\colorMATH}\ensuremath{\pe}}}\endgroup } : \tau  ; {\begingroup\renewcommand\colorMATH{\colorMATHC}\renewcommand\colorSYNTAX{\colorSYNTAXC}{{\color{\colorMATH}\ensuremath{\pS}}}\endgroup }}}}, and {{\color{\colorMATH}\ensuremath{\gamma  \in  {\mathcal{G}}\llbracket \Gamma \rrbracket }}}.  Then
    {{\color{\colorMATH}\ensuremath{\gamma \vdash  {\begingroup\renewcommand\colorMATH{\colorMATHC}\renewcommand\colorSYNTAX{\colorSYNTAXC}{{\color{\colorMATH}\ensuremath{\pe}}}\endgroup } \in  {\mathcal{E}}\llbracket \tau /\Gamma \rrbracket }}}.
  \end{enumerate}
\end{proposition}

Finally type safety for closed terms is just a corollary of the fundamental property above:
\begin{corollary}[Type Safety and Normalization of $\lang$]\;
  \label{lm:type-preservation-sensitivity-paper}
  \begin{enumerate}[label=(\alph*)]
    \item Let {{\color{\colorMATH}\ensuremath{\vdash  {\begingroup\renewcommand\colorMATH{\colorMATHB}\renewcommand\colorSYNTAX{\colorSYNTAXB}{{\color{\colorMATH}\ensuremath{\se}}}\endgroup } : \tau  ; \varnothing }}}, then
      {{\color{\colorMATH}\ensuremath{\vdash  {\begingroup\renewcommand\colorMATH{\colorMATHB}\renewcommand\colorSYNTAX{\colorSYNTAXB}{{\color{\colorMATH}\ensuremath{\se}}}\endgroup } \Downarrow  {\begingroup\renewcommand\colorMATH{\colorMATHB}\renewcommand\colorSYNTAX{\colorSYNTAXB}{{\color{\colorMATH}\ensuremath{\sv}}}\endgroup }}}}, and {{\color{\colorMATH}\ensuremath{\vdash  {\begingroup\renewcommand\colorMATH{\colorMATHB}\renewcommand\colorSYNTAX{\colorSYNTAXB}{{\color{\colorMATH}\ensuremath{\sv}}}\endgroup }: \tau ';\varnothing }}}, where
      {{\color{\colorMATH}\ensuremath{\tau ' <: \tau }}}.
    \item Let {{\color{\colorMATH}\ensuremath{\vdash  {\begingroup\renewcommand\colorMATH{\colorMATHC}\renewcommand\colorSYNTAX{\colorSYNTAXC}{{\color{\colorMATH}\ensuremath{\pe}}}\endgroup } : \tau  ; \varnothing }}}, then
      {{\color{\colorMATH}\ensuremath{\vdash  {\begingroup\renewcommand\colorMATH{\colorMATHC}\renewcommand\colorSYNTAX{\colorSYNTAXC}{{\color{\colorMATH}\ensuremath{\pe}}}\endgroup } \Downarrow  {\begingroup\renewcommand\colorMATH{\colorMATHB}\renewcommand\colorSYNTAX{\colorSYNTAXB}{{\color{\colorMATH}\ensuremath{\sv}}}\endgroup }}}}, and {{\color{\colorMATH}\ensuremath{\vdash  {\begingroup\renewcommand\colorMATH{\colorMATHB}\renewcommand\colorSYNTAX{\colorSYNTAXB}{{\color{\colorMATH}\ensuremath{\sv}}}\endgroup }: \tau ';\varnothing }}}, where
      {{\color{\colorMATH}\ensuremath{\tau ' <: \tau }}}.
  \end{enumerate}
\end{corollary}

\subsection{Soundness of $\lang$: Metric Preservation}
\label{sec:soundness}

This section establishes the \emph{soundness} of $\lang$, named \emph{metric
preservation}. Metric preservation for $\lang$ extends the notion of metric preservation of \ssystem. In addition to \toplas{reasoning} about sensitivity terms, given a privacy term with free variables, we can reason about the achieved privacy level when closing the privacy term under different (but related) environments.

Contrary to \ssystem, we establish soundness for $\lang$ using a \emph{step-indexed} logical
relation~\cite{ahmed}. Although $\lang$ is a strongly-normalizing language,
step indexing is still required to prove the bind case of the fundamental
property of the logical relation.
% Also, primitives which introduce
% nontermination can easily be added to out metatheory without modification.

\subsubsection*{Probabilistic Semantics}

\toplas{A first step to define the soundness property of $\lang$ is to endow privacy expressions with a probabilistic semantics. An important observation here is that even though to match their traditional (theoretical) presentation, we have introduced the Laplacian and Gaussian mechanisms as sampling from the (uncountable) set of real numbers, for the formal account of the language we consider \emph{discrete} versions thereof over the set of integers~\cite{DiscreteGauss}. This discretization is not only a natural but also a necessary requirement for any \emph{implementation} of the language (on ``finite'' computers), since it is well-known that the na\"ive use of finite-precision approximations may result in fatal privacy breaches~\cite{Mironov12}. (However, for the sake of uniformity, in the rest of the presentation we refer to these mechanisms---at the type level---as operating over the set of real numbers.)
Therefore, privacy expressions in $\lang$ sample values only from discrete distributions and can be interpreted as discrete distributions over values.}

\toplass{The probabilistic semantics of a privacy expression {{\color{\colorMATH}\ensuremath{{\begingroup\renewcommand\colorMATH{\colorMATHC}\renewcommand\colorSYNTAX{\colorSYNTAXC}{{\color{\colorMATH}\ensuremath{\pe}}}\endgroup }}}} is formally defined in Figure~\ref{fig:probabilistic-semantics}.
Judgment {{\color{\colorMATH}\ensuremath{\gamma  \vdash  {\begingroup\renewcommand\colorMATH{\colorMATHC}\renewcommand\colorSYNTAX{\colorSYNTAXC}{{\color{\colorMATH}\ensuremath{\pe}}}\endgroup } \Downarrow ^{k} \dist}}} denotes that privacy expression {{\color{\colorMATH}\ensuremath{{\begingroup\renewcommand\colorMATH{\colorMATHC}\renewcommand\colorSYNTAX{\colorSYNTAXC}{{\color{\colorMATH}\ensuremath{\pe}}}\endgroup }}}}
reduces to distribution {{\color{\colorMATH}\ensuremath{\dist}}} \toplassss{within} {{\color{\colorMATH}\ensuremath{k}}} steps; the probability that the privacy configuration {{\color{\colorMATH}\ensuremath{\gamma  \vdash  {\begingroup\renewcommand\colorMATH{\colorMATHC}\renewcommand\colorSYNTAX{\colorSYNTAXC}{{\color{\colorMATH}\ensuremath{\pe}}}\endgroup }}}} reduces to value {{\color{\colorMATH}\ensuremath{{\begingroup\renewcommand\colorMATH{\colorMATHB}\renewcommand\colorSYNTAX{\colorSYNTAXB}{{\color{\colorMATH}\ensuremath{\sv}}}\endgroup }}}} is then computed as {{\color{\colorMATH}\ensuremath{\dist({\begingroup\renewcommand\colorMATH{\colorMATHB}\renewcommand\colorSYNTAX{\colorSYNTAXB}{{\color{\colorMATH}\ensuremath{\sv}}}\endgroup })}}}.
} 
%We encode \toplas{subdistribution} {{\color{\colorMATH}\ensuremath{\llbracket {\begingroup\renewcommand\colorMATH{\colorMATHC}\renewcommand\colorSYNTAX{\colorSYNTAXC}{{\color{\colorMATH}\ensuremath{\pe}}}\endgroup }\rrbracket _{\gamma }^{k}}}} as a \emph{\toplas{(sub)}probability mass function} (PMF) over values, {i.e.} as an element of the set {{\color{\colorMATH}\ensuremath{{\mathcal{D}}({\textit{val}}) = \{ f \colon {\textit{val}} \rightarrow  [0,1] \mid \sum_{v \in {\textit{val}}} f(v) \,\toplas{\leq}\, 1\}}}}.
\toplas{We encode discrete distributions {{\color{\colorMATH}\ensuremath{\dist}}} as \emph{probability mass functions} (PMF), {i.e.} a discrete distribution over $A$ is modeled as an element of the set {{\color{\colorMATH}\ensuremath{{\mathcal{D}}(A) = \{ f \colon A \rightarrow  [0,1] \mid \sum_{a \in A} f(a) \,=\, 1\}}}}.}

\begin{figure}[t]
\begin{small}
\toplass{
\begin{framed}
\begingroup\color{\colorMATH}\begin{mathpar}\inferrule*[lab={\textsc{ return}}
  ]{ \gamma  \vdash  {\begingroup\renewcommand\colorMATH{\colorMATHB}\renewcommand\colorSYNTAX{\colorSYNTAXB}{{\color{\colorMATH}\ensuremath{\se}}}\endgroup } \Downarrow ^{k} {\begingroup\renewcommand\colorMATH{\colorMATHB}\renewcommand\colorSYNTAX{\colorSYNTAXB}{{\color{\colorMATH}\ensuremath{\sv}}}\endgroup } 
    }{
     \gamma  \vdash  {\begingroup\renewcommand\colorMATH{\colorMATHC}\renewcommand\colorSYNTAX{\colorSYNTAXC}{{\color{\colorSYNTAX}\texttt{return}}}\endgroup }\hspace*{0.33em}{\begingroup\renewcommand\colorMATH{\colorMATHB}\renewcommand\colorSYNTAX{\colorSYNTAXB}{{\color{\colorMATH}\ensuremath{\se}}}\endgroup } \Downarrow ^{k} \begin{array}[t]{l
                   } \lambda  x .\hspace*{0.33em}\left\{ \begin{array}{l@{\hspace*{1.00em}}c@{\hspace*{1.00em}}l
                     } 1 &{}{\textit{when}}{}& x = {\begingroup\renewcommand\colorMATH{\colorMATHB}\renewcommand\colorSYNTAX{\colorSYNTAXB}{{\color{\colorMATH}\ensuremath{\sv}}}\endgroup }
                     \cr  0 &{}{\textit{otherwise}}{}&
                     \end{array}\right.
                   \end{array}
  }
\and\inferrule*[lab={\textsc{ bind}} 
  ]{  \gamma  \vdash  {\begingroup\renewcommand\colorMATH{\colorMATHC}\renewcommand\colorSYNTAX{\colorSYNTAXC}{{\color{\colorMATH}\ensuremath{\pe_{1}}}}\endgroup } \Downarrow ^{k} \dist[1]
  \\  \forall  {\begingroup\renewcommand\colorMATH{\colorMATHB}\renewcommand\colorSYNTAX{\colorSYNTAXB}{{\color{\colorMATH}\ensuremath{\sv_{i}}}}\endgroup } \in  \Sup{\dist[1]}, \gamma [x \mapsto  {\begingroup\renewcommand\colorMATH{\colorMATHB}\renewcommand\colorSYNTAX{\colorSYNTAXB}{{\color{\colorMATH}\ensuremath{\sv_{i}}}}\endgroup }] \vdash  {\begingroup\renewcommand\colorMATH{\colorMATHC}\renewcommand\colorSYNTAX{\colorSYNTAXC}{{\color{\colorMATH}\ensuremath{\pe_{2}}}}\endgroup } \Downarrow ^{k_i} \dist[2i]
    }{
      \gamma  \vdash  x: \tau _{1} \leftarrow  {\begingroup\renewcommand\colorMATH{\colorMATHC}\renewcommand\colorSYNTAX{\colorSYNTAXC}{{\color{\colorMATH}\ensuremath{\pe_{1}}}}\endgroup }\mathrel{;}{\begingroup\renewcommand\colorMATH{\colorMATHC}\renewcommand\colorSYNTAX{\colorSYNTAXC}{{\color{\colorMATH}\ensuremath{\pe_{2}}}}\endgroup } \Downarrow ^{k+\mathsf{max}_{i} k_i} \lambda x. \sum_{{\begingroup\renewcommand\colorMATH{\colorMATHB}\renewcommand\colorSYNTAX{\colorSYNTAXB}{{\color{\colorMATH}\ensuremath{\sv_{i}}}}\endgroup } \in  \Sup{\dist[1]}} \dist[1]({\begingroup\renewcommand\colorMATH{\colorMATHB}\renewcommand\colorSYNTAX{\colorSYNTAXB}{{\color{\colorMATH}\ensuremath{\sv_{i}}}}\endgroup })\mathord{\cdotp }\dist[2i](x)
  }
% \and\inferrule*[lab={\textsc{ gauss}}
%   ]{  
%     }{
%       \gamma  \vdash  {\begingroup\renewcommand\colorMATH{\colorMATHC}\renewcommand\colorSYNTAX{\colorSYNTAXC}{{\color{\colorSYNTAX}\texttt{gauss}}}\endgroup } \hspace*{0.33em} \mu  \hspace*{0.33em} \sigma ^{2} \Downarrow ^{1} \lambda  x .\hspace*{0.33em} \frac{e^{\nicefrac{-(x - \mu )^{2}}{2\sigma ^{2}}}}{\sum_{y \in \mathbb{Z}} e^{\nicefrac{-(y - \mu )^{2}}{2\sigma ^{2}}}}
%   }
  \and \inferrule*[lab={\textsc{ app}}
   ]{ \gamma \vdash  {\begingroup\renewcommand\colorMATH{\colorMATHB}\renewcommand\colorSYNTAX{\colorSYNTAXB}{{\color{\colorMATH}\ensuremath{\se_{1}}}}\endgroup }  \Downarrow ^{k_{1}} \langle {\begingroup\renewcommand\colorMATH{\colorMATHC}\renewcommand\colorSYNTAX{\colorSYNTAXC}{{\color{\colorMATH}\ensuremath{\plambda}}}\endgroup } x:\tau \mathord{\cdotp } {\begingroup\renewcommand\colorMATH{\colorMATHB}\renewcommand\colorSYNTAX{\colorSYNTAXB}{{\color{\colorMATH}\ensuremath{\sss}}}\endgroup }.\hspace*{0.33em}{\begingroup\renewcommand\colorMATH{\colorMATHC}\renewcommand\colorSYNTAX{\colorSYNTAXC}{{\color{\colorMATH}\ensuremath{\pe^{\prime}}}}\endgroup },\gamma ^{\prime}\rangle 
   \\ \gamma \vdash  {\begingroup\renewcommand\colorMATH{\colorMATHB}\renewcommand\colorSYNTAX{\colorSYNTAXB}{{\color{\colorMATH}\ensuremath{\se_{2}}}}\endgroup }  \Downarrow ^{k_{2}} {\begingroup\renewcommand\colorMATH{\colorMATHB}\renewcommand\colorSYNTAX{\colorSYNTAXB}{{\color{\colorMATH}\ensuremath{\sv}}}\endgroup }
   \\ \gamma ^{\prime}[x\mapsto {\begingroup\renewcommand\colorMATH{\colorMATHB}\renewcommand\colorSYNTAX{\colorSYNTAXB}{{\color{\colorMATH}\ensuremath{\sv}}}\endgroup } ]\vdash  {\begingroup\renewcommand\colorMATH{\colorMATHC}\renewcommand\colorSYNTAX{\colorSYNTAXC}{{\color{\colorMATH}\ensuremath{\pe^{\prime}}}}\endgroup } \Downarrow ^{k_{3}} \dist
      }{
      \gamma \vdash  {\begingroup\renewcommand\colorMATH{\colorMATHC}\renewcommand\colorSYNTAX{\colorSYNTAXC}{{\color{\colorMATH}\ensuremath{\pe_{1}}}}\endgroup }\hspace*{0.33em}{\begingroup\renewcommand\colorMATH{\colorMATHC}\renewcommand\colorSYNTAX{\colorSYNTAXC}{{\color{\colorMATH}\ensuremath{\pe_{2}}}}\endgroup }  \Downarrow ^{k_{1}+k_{2}+k_{3}} \dist
   }
\and \inferrule*[lab={\textsc{ case-left}}
   ]{ \gamma \vdash  {\begingroup\renewcommand\colorMATH{\colorMATHB}\renewcommand\colorSYNTAX{\colorSYNTAXB}{{\color{\colorMATH}\ensuremath{\se}}}\endgroup }  \Downarrow ^{k_{1}} \inl\hspace*{0.33em}{\begingroup\renewcommand\colorMATH{\colorMATHB}\renewcommand\colorSYNTAX{\colorSYNTAXB}{{\color{\colorMATH}\ensuremath{\sv}}}\endgroup }
   \\ \gamma [x\mapsto {\begingroup\renewcommand\colorMATH{\colorMATHB}\renewcommand\colorSYNTAX{\colorSYNTAXB}{{\color{\colorMATH}\ensuremath{\sv}}}\endgroup } ]\vdash {\begingroup\renewcommand\colorMATH{\colorMATHC}\renewcommand\colorSYNTAX{\colorSYNTAXC}{{\color{\colorMATH}\ensuremath{\pe_{2}}}}\endgroup }  \Downarrow ^{k_{2}} \dist
      }{
      \gamma \vdash {\begingroup\renewcommand\colorMATH{\colorMATHC}\renewcommand\colorSYNTAX{\colorSYNTAXC}{{\color{\colorSYNTAX}\texttt{case}}}\endgroup }\hspace*{0.33em}{\begingroup\renewcommand\colorMATH{\colorMATHC}\renewcommand\colorSYNTAX{\colorSYNTAXC}{{\color{\colorMATH}\ensuremath{\pe}}}\endgroup }\hspace*{0.33em}{\begingroup\renewcommand\colorMATH{\colorMATHC}\renewcommand\colorSYNTAX{\colorSYNTAXC}{{\color{\colorSYNTAX}\texttt{of}}}\endgroup }\hspace*{0.33em}\{ x\Rightarrow  {\begingroup\renewcommand\colorMATH{\colorMATHC}\renewcommand\colorSYNTAX{\colorSYNTAXC}{{\color{\colorMATH}\ensuremath{\pe_{2}}}}\endgroup } \} \hspace*{0.33em}\{ x\Rightarrow  {\begingroup\renewcommand\colorMATH{\colorMATHC}\renewcommand\colorSYNTAX{\colorSYNTAXC}{{\color{\colorMATH}\ensuremath{\pe_{3}}}}\endgroup }\}  \Downarrow ^{k_{1}+k_{2}} \dist
   }
% \and \inferrule*[lab={\textsc{ case-right}}
%    ]{ \gamma \vdash  {\begingroup\renewcommand\colorMATH{\colorMATHC}\renewcommand\colorSYNTAX{\colorSYNTAXC}{{\color{\colorMATH}\ensuremath{\pe}}}\endgroup }  \Downarrow ^{k_{1}} \inr\hspace*{0.33em}{\begingroup\renewcommand\colorMATH{\colorMATHB}\renewcommand\colorSYNTAX{\colorSYNTAXB}{{\color{\colorMATH}\ensuremath{\sv}}}\endgroup }
%    \\ \gamma [x\mapsto {\begingroup\renewcommand\colorMATH{\colorMATHB}\renewcommand\colorSYNTAX{\colorSYNTAXB}{{\color{\colorMATH}\ensuremath{\sv}}}\endgroup } ]\vdash {\begingroup\renewcommand\colorMATH{\colorMATHC}\renewcommand\colorSYNTAX{\colorSYNTAXC}{{\color{\colorMATH}\ensuremath{\pe_{3}}}}\endgroup }  \Downarrow ^{k_{2}} \dist
%       }{
%       \gamma \vdash {\begingroup\renewcommand\colorMATH{\colorMATHC}\renewcommand\colorSYNTAX{\colorSYNTAXC}{{\color{\colorSYNTAX}\texttt{case}}}\endgroup }\hspace*{0.33em}{\begingroup\renewcommand\colorMATH{\colorMATHC}\renewcommand\colorSYNTAX{\colorSYNTAXC}{{\color{\colorMATH}\ensuremath{\pe}}}\endgroup }\hspace*{0.33em}{\begingroup\renewcommand\colorMATH{\colorMATHC}\renewcommand\colorSYNTAX{\colorSYNTAXC}{{\color{\colorSYNTAX}\texttt{of}}}\endgroup }\hspace*{0.33em}\{ x\Rightarrow  {\begingroup\renewcommand\colorMATH{\colorMATHC}\renewcommand\colorSYNTAX{\colorSYNTAXC}{{\color{\colorMATH}\ensuremath{\pe_{2}}}}\endgroup } \} \hspace*{0.33em}\{ x\Rightarrow  {\begingroup\renewcommand\colorMATH{\colorMATHC}\renewcommand\colorSYNTAX{\colorSYNTAXC}{{\color{\colorMATH}\ensuremath{\pe_{3}}}}\endgroup }\}  \Downarrow ^{k_{1}+k_{2}} \dist
%    }
\end{mathpar}\endgroup
\end{framed}
}
\end{small}
\caption{Probabilistic semantics of privacy expressions (selected rules)}
\label{fig:probabilistic-semantics}
\end{figure}

Let us briefly explain the set of rules in Figure~\ref{fig:probabilistic-semantics}. \toplas{For simplicity, we omit the underlying step indices.} The probabilistic semantics of {{\color{\colorMATH}\ensuremath{{\begingroup\renewcommand\colorMATH{\colorMATHC}\renewcommand\colorSYNTAX{\colorSYNTAXC}{{\color{\colorSYNTAX}\texttt{return}}}\endgroup }\hspace*{0.33em}{\begingroup\renewcommand\colorMATH{\colorMATHB}\renewcommand\colorSYNTAX{\colorSYNTAXB}{{\color{\colorMATH}\ensuremath{\se}}}\endgroup }}}} assigns probability 1 to the (necessarily unique) value to which expression {{\color{\colorMATH}\ensuremath{{\begingroup\renewcommand\colorMATH{\colorMATHB}\renewcommand\colorSYNTAX{\colorSYNTAXB}{{\color{\colorMATH}\ensuremath{\se}}}\endgroup }}}} reduces.
The probabilistic semantics of a bind {{\color{\colorMATH}\ensuremath{y: \tau _{1} \leftarrow  {\begingroup\renewcommand\colorMATH{\colorMATHC}\renewcommand\colorSYNTAX{\colorSYNTAXC}{{\color{\colorMATH}\ensuremath{\pe_{1}}}}\endgroup }\mathrel{;} {\begingroup\renewcommand\colorMATH{\colorMATHC}\renewcommand\colorSYNTAX{\colorSYNTAXC}{{\color{\colorMATH}\ensuremath{\pe_{2}}}}\endgroup }}}} operates as follows: To compute the probability that it assigns to {{\color{\colorMATH}\ensuremath{x}}}, it ranges over the set of values \toplass{in the support of {{\color{\colorMATH}\ensuremath{\dist[1]}}} denoted as {{\color{\colorMATH}\ensuremath{\Sup{\dist[1]}}}}, i.e. the set of values {{\color{\colorMATH}\ensuremath{{\begingroup\renewcommand\colorMATH{\colorMATHB}\renewcommand\colorSYNTAX{\colorSYNTAXB}{{\color{\colorMATH}\ensuremath{\sv}}}\endgroup }}}} such that {{\color{\colorMATH}\ensuremath{\dist[1]({\begingroup\renewcommand\colorMATH{\colorMATHB}\renewcommand\colorSYNTAX{\colorSYNTAXB}{{\color{\colorMATH}\ensuremath{\sv}}}\endgroup })>0}}}}, 
%and which by type safety of type {{\color{\colorMATH}\ensuremath{\tau /\gamma }}} ({{\color{\colorMATH}\ensuremath{\tau /\gamma }}} removes free variables from type {{\color{\colorMATH}\ensuremath{\tau }}} and is formally defined as {{\color{\colorMATH}\ensuremath{[\varnothing /x_{1},...,\varnothing /x_{n}]\tau }}} for {{\color{\colorMATH}\ensuremath{x_{i} \in  dom(\gamma )}}}), 
and for each {{\color{\colorMATH}\ensuremath{{\begingroup\renewcommand\colorMATH{\colorMATHB}\renewcommand\colorSYNTAX{\colorSYNTAXB}{{\color{\colorMATH}\ensuremath{\sv_{i}}}}\endgroup } \in \Sup{\dist[1]}}}} it sums the product between the probability that {{\color{\colorMATH}\ensuremath{{\begingroup\renewcommand\colorMATH{\colorMATHC}\renewcommand\colorSYNTAX{\colorSYNTAXC}{{\color{\colorMATH}\ensuremath{\pe_{1}}}}\endgroup }}}} reduces to {{\color{\colorMATH}\ensuremath{{\begingroup\renewcommand\colorMATH{\colorMATHB}\renewcommand\colorSYNTAX{\colorSYNTAXB}{{\color{\colorMATH}\ensuremath{\sv_{i}}}}\endgroup }}}} with the probability that {{\color{\colorMATH}\ensuremath{{\begingroup\renewcommand\colorMATH{\colorMATHC}\renewcommand\colorSYNTAX{\colorSYNTAXC}{{\color{\colorMATH}\ensuremath{\pe_{2}}}}\endgroup }}}} reduces to {{\color{\colorMATH}\ensuremath{x}}} in an extended environment where {{\color{\colorMATH}\ensuremath{y}}} is bound to {{\color{\colorMATH}\ensuremath{{\begingroup\renewcommand\colorMATH{\colorMATHB}\renewcommand\colorSYNTAX{\colorSYNTAXB}{{\color{\colorMATH}\ensuremath{\sv_{i}}}}\endgroup }}}}. \toplas{The discrete Gauss distribution with mean {{\color{\colorMATH}\ensuremath{\mu }}} and scale {{\color{\colorMATH}\ensuremath{\sigma ^{2}}}} assigns probability proportional to {{\color{\colorMATH}\ensuremath{e^{\nicefrac{-(x - \mu )^{2}}{2\sigma ^{2}}}}}} to each integer {{\color{\colorMATH}\ensuremath{x}}}.} The probabilistic semantics of a privacy application is defined as the probabilistic semantics of the body of the resulting privacy closure, in an extended environment where the closure formal argument is bound to the value of the real argument. Finally, the probabilistic semantics of a {{\color{\colorMATH}\ensuremath{\ccase}}} term is simply the probabilistic semantics of the corresponding branch \toplas{in an extended environment with the corresponding association for the branch binder variable}.

\toplassss{We consider a step-indexed semantics to establish the language metatheory. In particular, the metric preservation theorem is proved by induction on the step index of the logical relation. However, step indices might not interact very well with the bind rule: when reducing {{\color{\colorMATH}\ensuremath{y: \tau _{1} \leftarrow  {\begingroup\renewcommand\colorMATH{\colorMATHC}\renewcommand\colorSYNTAX{\colorSYNTAXC}{{\color{\colorMATH}\ensuremath{\pe_{1}}}}\endgroup }\mathrel{;} {\begingroup\renewcommand\colorMATH{\colorMATHC}\renewcommand\colorSYNTAX{\colorSYNTAXC}{{\color{\colorMATH}\ensuremath{\pe_{2}}}}\endgroup }}}} , the reduction of {{\color{\colorMATH}\ensuremath{{\begingroup\renewcommand\colorMATH{\colorMATHC}\renewcommand\colorSYNTAX{\colorSYNTAXC}{{\color{\colorMATH}\ensuremath{\pe_{2}}}}\endgroup }}}} requires a possibly different number of steps ({{\color{\colorMATH}\ensuremath{k_i}}}) for each value (${\begingroup\renewcommand\colorMATH{\colorMATHB}\renewcommand\colorSYNTAX{\colorSYNTAXB}{{\color{\colorMATH}\ensuremath{\sv_{i}}}}\endgroup }$) to which {{\color{\colorMATH}\ensuremath{{\begingroup\renewcommand\colorMATH{\colorMATHC}\renewcommand\colorSYNTAX{\colorSYNTAXC}{{\color{\colorMATH}\ensuremath{\pe_{1}}}}\endgroup }}}} reduces. This set of
steps could in principle be unbounded, making $\mathsf{max}_{i}~k_i$ undefined and thus rendering the semantics partial. However, this is not an issue for our technical development because all formal results are concerned with programs that reach the distribution of final
values within a {\em finite} number of steps, only. % Said otherwise, all our
%results remain vacuously valid for programs that require an unbounded number
%of steps to reach all of their final values.
}

\toplass{For convenience throughout this section, we also introduce  {{\color{\colorMATH}\ensuremath{\dist(S)}}} to denote the probability of observing {{\color{\colorMATH}\ensuremath{S}}} in {{\color{\colorMATH}\ensuremath{\dist}}}, computed as {{\color{\colorMATH}\ensuremath{\sum_{{\begingroup\renewcommand\colorMATH{\colorMATHB}\renewcommand\colorSYNTAX{\colorSYNTAXB}{{\color{\colorMATH}\ensuremath{\sv}}}\endgroup } \in  S} \dist({\begingroup\renewcommand\colorMATH{\colorMATHB}\renewcommand\colorSYNTAX{\colorSYNTAXB}{{\color{\colorMATH}\ensuremath{\sv}}}\endgroup })}}}. Also we define {{\color{\colorMATH}\ensuremath{{\text{Pr}}[\varnothing  \vdash  {\begingroup\renewcommand\colorMATH{\colorMATHC}\renewcommand\colorSYNTAX{\colorSYNTAXC}{{\color{\colorMATH}\ensuremath{\pe}}}\endgroup } \Downarrow ^{k} {\begingroup\renewcommand\colorMATH{\colorMATHB}\renewcommand\colorSYNTAX{\colorSYNTAXB}{{\color{\colorMATH}\ensuremath{\sv}}}\endgroup }]}}} as {{\color{\colorMATH}\ensuremath{\dist({\begingroup\renewcommand\colorMATH{\colorMATHB}\renewcommand\colorSYNTAX{\colorSYNTAXB}{{\color{\colorMATH}\ensuremath{\sv}}}\endgroup })}}} if {{\color{\colorMATH}\ensuremath{\varnothing  \vdash  {\begingroup\renewcommand\colorMATH{\colorMATHC}\renewcommand\colorSYNTAX{\colorSYNTAXC}{{\color{\colorMATH}\ensuremath{\pe}}}\endgroup } \Downarrow ^{k'} \dist}}} for some {{\color{\colorMATH}\ensuremath{k' \leq  k}}} (undefined otherwise).}

\subsubsection*{Logical relation} The logical relations for sensitivity computations, privacy computations, values, and environments are mutually recursive and presented in Figure~\ref{fig:logical-relation}.
\begin{figure}[t!]
\begin{small}
\begin{framed}
  \begingroup\color{\colorMATH}\begin{gather*}
\end{gather*}\endgroup
\end{framed}

% gauss: (z:{\mathbb{R}}) -->{({\begingroup\renewcommand\colorMATH{\colorMATHC}\renewcommand\colorSYNTAX{\colorSYNTAXC}{{\color{\colorMATH}\ensuremath{\pe}}}\endgroup },\delta )z} {\mathbb{R}}

% (\lambda x:{\mathbb{R}}\mathord{\cdotp }1. (\lambda y:{\mathbb{R}}\mathord{\cdotp }1. gauss x+y))

% gauss x+y : {\mathbb{R}}; ({\begingroup\renewcommand\colorMATH{\colorMATHC}\renewcommand\colorSYNTAX{\colorSYNTAXC}{{\color{\colorMATH}\ensuremath{\pe}}}\endgroup },\delta )x \sqcup  ({\begingroup\renewcommand\colorMATH{\colorMATHC}\renewcommand\colorSYNTAX{\colorSYNTAXC}{{\color{\colorMATH}\ensuremath{\pe}}}\endgroup },\delta )y

% (\lambda x:{\mathbb{R}}. x :: z) z

\end{small}
\caption{$\lang$: logical relations for metric preservation}
\label{fig:logical-relation}
\end{figure}

Note that each logical relation is also indexed by a \distanceName type that now accounts for sensitivity and privacy lambdas:
\begingroup\color{\colorMATH}\begin{gather*}
  \begin{array}{rcl
  } \sigma  &{}={}& ... \mathrel{|} (x\mathrel{:}\sigma \mathord{\cdotp }{\begingroup\renewcommand\colorMATH{\colorMATHB}\renewcommand\colorSYNTAX{\colorSYNTAXB}{{\color{\colorMATH}\ensuremath{\distance}}}\endgroup }) \xrightarrowS {{\begingroup\renewcommand\colorMATH{\colorMATHB}\renewcommand\colorSYNTAX{\colorSYNTAXB}{{\color{\colorMATH}\ensuremath{\sS}}}\endgroup } + {\begingroup\renewcommand\colorMATH{\colorMATHB}\renewcommand\colorSYNTAX{\colorSYNTAXB}{{\color{\colorMATH}\ensuremath{\distance}}}\endgroup }} \sigma  \mathrel{|} (x\mathrel{:}\sigma \mathord{\cdotp }{\begingroup\renewcommand\colorMATH{\colorMATHB}\renewcommand\colorSYNTAX{\colorSYNTAXB}{{\color{\colorSYNTAX}\texttt{{{\color{\colorMATH}\ensuremath{s}}}}}}\endgroup }) \xrightarrowP {{\begingroup\renewcommand\colorMATH{\colorMATHC}\renewcommand\colorSYNTAX{\colorSYNTAXC}{{\color{\colorMATH}\ensuremath{\pS}}}\endgroup }} \sigma 
  \cr  {\begingroup\renewcommand\colorMATH{\colorMATHC}\renewcommand\colorSYNTAX{\colorSYNTAXC}{{\color{\colorMATH}\ensuremath{\pS}}}\endgroup } &{}={}& ... \mathrel{|} {\begingroup\renewcommand\colorMATH{\colorMATHC}\renewcommand\colorSYNTAX{\colorSYNTAXC}{{\color{\colorMATH}\ensuremath{p}}}\endgroup }
  \end{array}
\end{gather*}\endgroup
Note that similarly to sensitivity environments, privacy \toplas{environments} {{\color{\colorMATH}\ensuremath{{\begingroup\renewcommand\colorMATH{\colorMATHC}\renewcommand\colorSYNTAX{\colorSYNTAXC}{{\color{\colorMATH}\ensuremath{\pS}}}\endgroup }}}} are also extended to include partially instantiated data, for instance {{\color{\colorMATH}\ensuremath{{\begingroup\renewcommand\colorMATH{\colorMATHC}\renewcommand\colorSYNTAX{\colorSYNTAXC}{{\color{\colorMATH}\ensuremath{p}}}\endgroup }x + {\begingroup\renewcommand\colorMATH{\colorMATHC}\renewcommand\colorSYNTAX{\colorSYNTAXC}{{\color{\colorMATH}\ensuremath{p'}}}\endgroup }}}}.
Notation {{\color{\colorMATH}\ensuremath{({\begingroup\renewcommand\colorMATH{\colorMATHB}\renewcommand\colorSYNTAX{\colorSYNTAXB}{{\color{\colorMATH}\ensuremath{\sv_{1}}}}\endgroup },{\begingroup\renewcommand\colorMATH{\colorMATHB}\renewcommand\colorSYNTAX{\colorSYNTAXB}{{\color{\colorMATH}\ensuremath{\sv_{2}}}}\endgroup }) \in  {\mathcal{V}}_{{\begingroup\renewcommand\colorMATH{\colorMATHB}\renewcommand\colorSYNTAX{\colorSYNTAXB}{{\color{\colorMATH}\ensuremath{\distance}}}\endgroup }}^{k}\llbracket \sigma \rrbracket }}} indicates
that value {{\color{\colorMATH}\ensuremath{{\begingroup\renewcommand\colorMATH{\colorMATHB}\renewcommand\colorSYNTAX{\colorSYNTAXB}{{\color{\colorMATH}\ensuremath{\sv_{1}}}}\endgroup }}}} is related to {{\color{\colorMATH}\ensuremath{{\begingroup\renewcommand\colorMATH{\colorMATHB}\renewcommand\colorSYNTAX{\colorSYNTAXB}{{\color{\colorMATH}\ensuremath{\sv_{2}}}}\endgroup }}}} at type {{\color{\colorMATH}\ensuremath{\sigma }}} and distance {{\color{\colorMATH}\ensuremath{{\begingroup\renewcommand\colorMATH{\colorMATHB}\renewcommand\colorSYNTAX{\colorSYNTAXB}{{\color{\colorMATH}\ensuremath{\distance}}}\endgroup }}}} for {{\color{\colorMATH}\ensuremath{k}}}
steps.

The sensitivity parts of the logical relations are defined analogously to Figure~\ref{fig:sensitivity-simple-logical-relation} with the addition of a step index {{\color{\colorMATH}\ensuremath{k}}}. We only present relevant changes:

\begin{itemize}[label=\textbf{-},leftmargin=*]\item Two sensitivity closures are also related if, given related inputs, they produce
   related computations. Specifically, first the environments \toplas{have} to be related for any step {{\color{\colorMATH}\ensuremath{j < k}}}.
Second, inputs {{\color{\colorMATH}\ensuremath{{\begingroup\renewcommand\colorMATH{\colorMATHB}\renewcommand\colorSYNTAX{\colorSYNTAXB}{{\color{\colorMATH}\ensuremath{\sv'_{1}}}}\endgroup }}}} and {{\color{\colorMATH}\ensuremath{{\begingroup\renewcommand\colorMATH{\colorMATHB}\renewcommand\colorSYNTAX{\colorSYNTAXB}{{\color{\colorMATH}\ensuremath{\sv'_{2}}}}\endgroup }}}} have to be related at distance {{\color{\colorMATH}\ensuremath{{\begingroup\renewcommand\colorMATH{\colorMATHB}\renewcommand\colorSYNTAX{\colorSYNTAXB}{{\color{\colorMATH}\ensuremath{\distance^{\prime \prime}}}}\endgroup }}}} not greater than {{\color{\colorMATH}\ensuremath{{\begingroup\renewcommand\colorMATH{\colorMATHB}\renewcommand\colorSYNTAX{\colorSYNTAXB}{{\color{\colorMATH}\ensuremath{\distance}}}\endgroup }´}}}, and for {{\color{\colorMATH}\ensuremath{j}}} steps.
Finally, the \toplas{bodies} of the functions in extended environments have to be related computations for {{\color{\colorMATH}\ensuremath{j}}} steps. 

\item  Similarly to sensitivity closures, two privacy closures are related if they produce related computations when applied to related inputs. The computations are related at privacy {{\color{\colorMATH}\ensuremath{{\begingroup\renewcommand\colorMATH{\colorMATHC}\renewcommand\colorSYNTAX{\colorSYNTAXC}{{\color{\colorMATH}\ensuremath{\rceil {\begingroup\renewcommand\colorMATH{\colorMATHB}\renewcommand\colorSYNTAX{\colorSYNTAXB}{{\color{\colorMATH}\ensuremath{\distance}}}\endgroup }\lceil ^{\infty }}}}\endgroup } + ({\begingroup\renewcommand\colorMATH{\colorMATHB}\renewcommand\colorSYNTAX{\colorSYNTAXB}{{\color{\colorMATH}\ensuremath{\Distance}}}\endgroup } + {\begingroup\renewcommand\colorMATH{\colorMATHB}\renewcommand\colorSYNTAX{\colorSYNTAXB}{{\color{\colorMATH}\ensuremath{\distance^{\prime \prime}}}}\endgroup }x) {\begingroup\renewcommand\colorMATH{\colorMATHC}\renewcommand\colorSYNTAX{\colorSYNTAXC}{{\color{\colorMATH}\ensuremath{\bigcdot}}}\endgroup } {\begingroup\renewcommand\colorMATH{\colorMATHC}\renewcommand\colorSYNTAX{\colorSYNTAXC}{{\color{\colorMATH}\ensuremath{\pS}}}\endgroup }}}}.
%as {{\color{\colorMATH}\ensuremath{{\begingroup\renewcommand\colorMATH{\colorMATHB}\renewcommand\colorSYNTAX{\colorSYNTAXB}{{\color{\colorMATH}\ensuremath{\sS}}}\endgroup } {\begingroup\renewcommand\colorMATH{\colorMATHC}\renewcommand\colorSYNTAX{\colorSYNTAXC}{{\color{\colorMATH}\ensuremath{\bigcdot}}}\endgroup } {\begingroup\renewcommand\colorMATH{\colorMATHC}\renewcommand\colorSYNTAX{\colorSYNTAXC}{{\color{\colorMATH}\ensuremath{\pS}}}\endgroup } = \rceil {\begingroup\renewcommand\colorMATH{\colorMATHB}\renewcommand\colorSYNTAX{\colorSYNTAXB}{{\color{\colorMATH}\ensuremath{\sS}}}\endgroup }\lceil ^{1}\mathord{\cdotp }{\begingroup\renewcommand\colorMATH{\colorMATHC}\renewcommand\colorSYNTAX{\colorSYNTAXC}{{\color{\colorMATH}\ensuremath{\pS}}}\endgroup }}}}.
Note that we lift {{\color{\colorMATH}\ensuremath{{\begingroup\renewcommand\colorMATH{\colorMATHB}\renewcommand\colorSYNTAX{\colorSYNTAXB}{{\color{\colorMATH}\ensuremath{\distance}}}\endgroup }}}} to infinite because we cannot record relational distances as a privacy result.
% we have to pay infinite cost for the variables that are used when the values {{\color{\colorMATH}\ensuremath{{\begingroup\renewcommand\colorMATH{\colorMATHB}\renewcommand\colorSYNTAX{\colorSYNTAXB}{{\color{\colorMATH}\ensuremath{\sv_{1}}}}\endgroup }}}} and {{\color{\colorMATH}\ensuremath{{\begingroup\renewcommand\colorMATH{\colorMATHB}\renewcommand\colorSYNTAX{\colorSYNTAXB}{{\color{\colorMATH}\ensuremath{\sv_{2}}}}\endgroup }}}} are reduced\et{"the values are reduced" sounds weird, values do not reduce}.
In addition to that, we also pay the latent contextual effect of the function instantiated to {{\color{\colorMATH}\ensuremath{{\begingroup\renewcommand\colorMATH{\colorMATHB}\renewcommand\colorSYNTAX{\colorSYNTAXB}{{\color{\colorMATH}\ensuremath{\distance^{\prime \prime}}}}\endgroup }x}}}, i.e. {{\color{\colorMATH}\ensuremath{({\begingroup\renewcommand\colorMATH{\colorMATHB}\renewcommand\colorSYNTAX{\colorSYNTAXB}{{\color{\colorMATH}\ensuremath{\Distance}}}\endgroup } + {\begingroup\renewcommand\colorMATH{\colorMATHB}\renewcommand\colorSYNTAX{\colorSYNTAXB}{{\color{\colorMATH}\ensuremath{\distance^{\prime \prime}}}}\endgroup }x) {\begingroup\renewcommand\colorMATH{\colorMATHC}\renewcommand\colorSYNTAX{\colorSYNTAXC}{{\color{\colorMATH}\ensuremath{\bigcdot}}}\endgroup } {\begingroup\renewcommand\colorMATH{\colorMATHC}\renewcommand\colorSYNTAX{\colorSYNTAXC}{{\color{\colorMATH}\ensuremath{\pS}}}\endgroup }}}}.
\end{itemize}

Two sensitivity configurations are related computations at type {{\color{\colorMATH}\ensuremath{\sigma }}} and distance {{\color{\colorMATH}\ensuremath{{\begingroup\renewcommand\colorMATH{\colorMATHB}\renewcommand\colorSYNTAX{\colorSYNTAXB}{{\color{\colorMATH}\ensuremath{\distance}}}\endgroup }}}} for {{\color{\colorMATH}\ensuremath{k}}} steps, noted {{\color{\colorMATH}\ensuremath{(\gamma _{1} \vdash  {\begingroup\renewcommand\colorMATH{\colorMATHB}\renewcommand\colorSYNTAX{\colorSYNTAXB}{{\color{\colorMATH}\ensuremath{\se_{1}}}}\endgroup }, \gamma _{2} \vdash  {\begingroup\renewcommand\colorMATH{\colorMATHB}\renewcommand\colorSYNTAX{\colorSYNTAXB}{{\color{\colorMATH}\ensuremath{\se_{2}}}}\endgroup }) \in  {\mathcal{E}}_{{\begingroup\renewcommand\colorMATH{\colorMATHB}\renewcommand\colorSYNTAX{\colorSYNTAXB}{{\color{\colorMATH}\ensuremath{\distance}}}\endgroup }}^{k}\llbracket \sigma \rrbracket }}}, when for any {{\color{\colorMATH}\ensuremath{j < k}}}, if \toplass{the first configuration reduces in {{\color{\colorMATH}\ensuremath{j}}} steps to a value, then the second configuration also reduces to a value in any number of steps, and} these values are related for the remaining {{\color{\colorMATH}\ensuremath{k-j}}} steps at type {{\color{\colorMATH}\ensuremath{\sigma }}} and distance {{\color{\colorMATH}\ensuremath{{\begingroup\renewcommand\colorMATH{\colorMATHB}\renewcommand\colorSYNTAX{\colorSYNTAXB}{{\color{\colorMATH}\ensuremath{\distance}}}\endgroup }}}}. We write {{\color{\colorMATH}\ensuremath{\gamma  \vdash  {\begingroup\renewcommand\colorMATH{\colorMATHB}\renewcommand\colorSYNTAX{\colorSYNTAXB}{{\color{\colorMATH}\ensuremath{\se}}}\endgroup } \Downarrow ^{k} {\begingroup\renewcommand\colorMATH{\colorMATHB}\renewcommand\colorSYNTAX{\colorSYNTAXB}{{\color{\colorMATH}\ensuremath{\sv}}}\endgroup }}}} to say that the configuration {{\color{\colorMATH}\ensuremath{\gamma  \vdash  {\begingroup\renewcommand\colorMATH{\colorMATHB}\renewcommand\colorSYNTAX{\colorSYNTAXB}{{\color{\colorMATH}\ensuremath{\se}}}\endgroup }}}} reduces to value {{\color{\colorMATH}\ensuremath{{\begingroup\renewcommand\colorMATH{\colorMATHB}\renewcommand\colorSYNTAX{\colorSYNTAXB}{{\color{\colorMATH}\ensuremath{\sv}}}\endgroup }}}} in {{\color{\colorMATH}\ensuremath{k}}} steps.

\toplas{We now turn to the definition of related privacy computations.} 
Notation {{\color{\colorMATH}\ensuremath{(\gamma _{1} \vdash  {\begingroup\renewcommand\colorMATH{\colorMATHC}\renewcommand\colorSYNTAX{\colorSYNTAXC}{{\color{\colorMATH}\ensuremath{\pe_{1}}}}\endgroup }, \gamma _{2} \vdash  {\begingroup\renewcommand\colorMATH{\colorMATHC}\renewcommand\colorSYNTAX{\colorSYNTAXC}{{\color{\colorMATH}\ensuremath{\pe_{2}}}}\endgroup }) \in  {\mathcal{E}}_{{\begingroup\renewcommand\colorMATH{\colorMATHC}\renewcommand\colorSYNTAX{\colorSYNTAXC}{{\color{\colorMATH}\ensuremath{p}}}\endgroup }}^{k}\llbracket \sigma \rrbracket }}} indicates that two privacy configurations are related computations at type {{\color{\colorMATH}\ensuremath{\sigma }}} and privacy {{\color{\colorMATH}\ensuremath{{\begingroup\renewcommand\colorMATH{\colorMATHC}\renewcommand\colorSYNTAX{\colorSYNTAXC}{{\color{\colorMATH}\ensuremath{p}}}\endgroup } = ({\begingroup\renewcommand\colorMATH{\colorMATHC}\renewcommand\colorSYNTAX{\colorSYNTAXC}{{\color{\colorMATH}\ensuremath{\epsilon }}}\endgroup }, {\begingroup\renewcommand\colorMATH{\colorMATHC}\renewcommand\colorSYNTAX{\colorSYNTAXC}{{\color{\colorMATH}\ensuremath{\delta }}}\endgroup })}}} for {{\color{\colorMATH}\ensuremath{k}}} steps. 
\toplass{Two privacy configurations are related when, for any {{\color{\colorMATH}\ensuremath{j < k}}}, if the first privacy configuration reduces in {{\color{\colorMATH}\ensuremath{j}}} steps to a distribution, then the second configuration also reduces to a distribution in any number of steps, 
and the probability of observing {{\color{\colorMATH}\ensuremath{S}}} in the fist distribution is no greater than {{\color{\colorMATH}\ensuremath{e^{{\begingroup\renewcommand\colorMATH{\colorMATHC}\renewcommand\colorSYNTAX{\colorSYNTAXC}{{\color{\colorMATH}\ensuremath{\epsilon }}}\endgroup }}}}} times the probability of observing {{\color{\colorMATH}\ensuremath{S}}} in the second distribution, plus {{\color{\colorMATH}\ensuremath{{\begingroup\renewcommand\colorMATH{\colorMATHC}\renewcommand\colorSYNTAX{\colorSYNTAXC}{{\color{\colorMATH}\ensuremath{\delta }}}\endgroup }}}}.}

% This means that for any subset {{\color{\colorMATH}\ensuremath{S}}} of {{\color{\colorMATH}\ensuremath{val(*)}}}\toplass{, where {{\color{\colorMATH}\ensuremath{val(*)}}} is the set of all values of any type},
%   if the probabilities of reducing the configurations to {{\color{\colorMATH}\ensuremath{S}}} are defined for {{\color{\colorMATH}\ensuremath{k}}} steps (noted {{\color{\colorMATH}\ensuremath{({\text{Pr}}[\gamma _{1} \vdash  {\begingroup\renewcommand\colorMATH{\colorMATHC}\renewcommand\colorSYNTAX{\colorSYNTAXC}{{\color{\colorMATH}\ensuremath{\pe_{1}}}}\endgroup } \Downarrow ^{k} S] \wedge  {\text{Pr}}[\gamma _{2} \vdash  {\begingroup\renewcommand\colorMATH{\colorMATHC}\renewcommand\colorSYNTAX{\colorSYNTAXC}{{\color{\colorMATH}\ensuremath{\pe_{2}}}}\endgroup } \Downarrow ^{k} S])}}}), then the probability of reducing the first configuration to an element of set {{\color{\colorMATH}\ensuremath{S}}} is no greater than {{\color{\colorMATH}\ensuremath{e^{{\begingroup\renewcommand\colorMATH{\colorMATHC}\renewcommand\colorSYNTAX{\colorSYNTAXC}{{\color{\colorMATH}\ensuremath{\epsilon }}}\endgroup }}}}} times the probability of reducing the second configuration to an element of the same set {{\color{\colorMATH}\ensuremath{S}}}, plus {{\color{\colorMATH}\ensuremath{{\begingroup\renewcommand\colorMATH{\colorMATHC}\renewcommand\colorSYNTAX{\colorSYNTAXC}{{\color{\colorMATH}\ensuremath{\delta }}}\endgroup }}}} (and vice versa).

\paragraph{Metric Preservation} Armed with these logical relations, we can establish the notion of type soundness for $\lang$, and prove the fundamental property.

\begin{restatable}[Metric Preservation]{theorem}{FundamentalProperty}\
  \label{lm:fp}
  \begin{enumerate}\item  {{\color{\colorMATH}\ensuremath{ \Gamma ,{\begingroup\renewcommand\colorMATH{\colorMATHB}\renewcommand\colorSYNTAX{\colorSYNTAXB}{{\color{\colorMATH}\ensuremath{\Distance}}}\endgroup } \vdash  {\begingroup\renewcommand\colorMATH{\colorMATHB}\renewcommand\colorSYNTAX{\colorSYNTAXB}{{\color{\colorMATH}\ensuremath{\se}}}\endgroup } \mathrel{:} \tau  \mathrel{;} {\begingroup\renewcommand\colorMATH{\colorMATHB}\renewcommand\colorSYNTAX{\colorSYNTAXB}{{\color{\colorMATH}\ensuremath{\sS}}}\endgroup } \Rightarrow  \forall  k \geq  0, \forall {\begingroup\renewcommand\colorMATH{\colorMATHB}\renewcommand\colorSYNTAX{\colorSYNTAXB}{{\color{\colorMATH}\ensuremath{\Distance^{\prime}}}}\endgroup } \sqsubseteq  {\begingroup\renewcommand\colorMATH{\colorMATHB}\renewcommand\colorSYNTAX{\colorSYNTAXB}{{\color{\colorMATH}\ensuremath{\Distance}}}\endgroup }, \forall (\gamma _{1},\gamma _{2}) \in  {\mathcal{G}}_{{\begingroup\renewcommand\colorMATH{\colorMATHB}\renewcommand\colorSYNTAX{\colorSYNTAXB}{{\color{\colorMATH}\ensuremath{\Distance^{\prime}}}}\endgroup }}^{k}\llbracket \Gamma \rrbracket . (\gamma _{1}\vdash {\begingroup\renewcommand\colorMATH{\colorMATHB}\renewcommand\colorSYNTAX{\colorSYNTAXB}{{\color{\colorMATH}\ensuremath{\se}}}\endgroup },\gamma _{2}\vdash {\begingroup\renewcommand\colorMATH{\colorMATHB}\renewcommand\colorSYNTAX{\colorSYNTAXB}{{\color{\colorMATH}\ensuremath{\se}}}\endgroup }) \in  {\mathcal{E}}_{{\begingroup\renewcommand\colorMATH{\colorMATHB}\renewcommand\colorSYNTAX{\colorSYNTAXB}{{\color{\colorMATH}\ensuremath{\Distance^{\prime}}}}\endgroup }\mathord{\cdotp }{\begingroup\renewcommand\colorMATH{\colorMATHB}\renewcommand\colorSYNTAX{\colorSYNTAXB}{{\color{\colorMATH}\ensuremath{\sS}}}\endgroup }}^{k}\llbracket {\begingroup\renewcommand\colorMATH{\colorMATHB}\renewcommand\colorSYNTAX{\colorSYNTAXB}{{\color{\colorMATH}\ensuremath{\Distance^{\prime}}}}\endgroup }(\tau )\rrbracket }}}
  \item  {{\color{\colorMATH}\ensuremath{ \Gamma ,{\begingroup\renewcommand\colorMATH{\colorMATHB}\renewcommand\colorSYNTAX{\colorSYNTAXB}{{\color{\colorMATH}\ensuremath{\Distance}}}\endgroup } \vdash  {\begingroup\renewcommand\colorMATH{\colorMATHC}\renewcommand\colorSYNTAX{\colorSYNTAXC}{{\color{\colorMATH}\ensuremath{\pe}}}\endgroup } \mathrel{:} \tau  \mathrel{;} {\begingroup\renewcommand\colorMATH{\colorMATHC}\renewcommand\colorSYNTAX{\colorSYNTAXC}{{\color{\colorMATH}\ensuremath{\pS}}}\endgroup } \Rightarrow  \forall  k \geq  0, \forall {\begingroup\renewcommand\colorMATH{\colorMATHB}\renewcommand\colorSYNTAX{\colorSYNTAXB}{{\color{\colorMATH}\ensuremath{\Distance^{\prime}}}}\endgroup } \sqsubseteq  {\begingroup\renewcommand\colorMATH{\colorMATHB}\renewcommand\colorSYNTAX{\colorSYNTAXB}{{\color{\colorMATH}\ensuremath{\Distance}}}\endgroup }, \forall (\gamma _{1},\gamma _{2}) \in  {\mathcal{G}}_{{\begingroup\renewcommand\colorMATH{\colorMATHB}\renewcommand\colorSYNTAX{\colorSYNTAXB}{{\color{\colorMATH}\ensuremath{\Distance^{\prime}}}}\endgroup }}^{k}\llbracket \Gamma \rrbracket . (\gamma _{1}\vdash {\begingroup\renewcommand\colorMATH{\colorMATHC}\renewcommand\colorSYNTAX{\colorSYNTAXC}{{\color{\colorMATH}\ensuremath{\pe}}}\endgroup },\gamma _{2}\vdash {\begingroup\renewcommand\colorMATH{\colorMATHC}\renewcommand\colorSYNTAX{\colorSYNTAXC}{{\color{\colorMATH}\ensuremath{\pe}}}\endgroup }) \in  {\mathcal{E}}_{{\begingroup\renewcommand\colorMATH{\colorMATHB}\renewcommand\colorSYNTAX{\colorSYNTAXB}{{\color{\colorMATH}\ensuremath{\Distance^{\prime}}}}\endgroup }{\begingroup\renewcommand\colorMATH{\colorMATHC}\renewcommand\colorSYNTAX{\colorSYNTAXC}{{\color{\colorMATH}\ensuremath{\bigcdot}}}\endgroup }{\begingroup\renewcommand\colorMATH{\colorMATHC}\renewcommand\colorSYNTAX{\colorSYNTAXC}{{\color{\colorMATH}\ensuremath{\pS}}}\endgroup }}^{k}\llbracket {\begingroup\renewcommand\colorMATH{\colorMATHB}\renewcommand\colorSYNTAX{\colorSYNTAXB}{{\color{\colorMATH}\ensuremath{\Distance^{\prime}}}}\endgroup }(\tau )\rrbracket }}}
  \end{enumerate}
\end{restatable}
where {{\color{\colorMATH}\ensuremath{{\begingroup\renewcommand\colorMATH{\colorMATHB}\renewcommand\colorSYNTAX{\colorSYNTAXB}{{\color{\colorMATH}\ensuremath{\Distance^{\prime}}}}\endgroup } \sqsubseteq  {\begingroup\renewcommand\colorMATH{\colorMATHB}\renewcommand\colorSYNTAX{\colorSYNTAXB}{{\color{\colorMATH}\ensuremath{\Distance}}}\endgroup } \iff  dom({\begingroup\renewcommand\colorMATH{\colorMATHB}\renewcommand\colorSYNTAX{\colorSYNTAXB}{{\color{\colorMATH}\ensuremath{\Distance^{\prime}}}}\endgroup }) = dom({\begingroup\renewcommand\colorMATH{\colorMATHB}\renewcommand\colorSYNTAX{\colorSYNTAXB}{{\color{\colorMATH}\ensuremath{\Distance}}}\endgroup }) \wedge  \forall  x \in  dom({\begingroup\renewcommand\colorMATH{\colorMATHB}\renewcommand\colorSYNTAX{\colorSYNTAXB}{{\color{\colorMATH}\ensuremath{\Distance^{\prime}}}}\endgroup }), {\begingroup\renewcommand\colorMATH{\colorMATHB}\renewcommand\colorSYNTAX{\colorSYNTAXB}{{\color{\colorMATH}\ensuremath{\Distance^{\prime}}}}\endgroup }(x) \leq  {\begingroup\renewcommand\colorMATH{\colorMATHB}\renewcommand\colorSYNTAX{\colorSYNTAXB}{{\color{\colorMATH}\ensuremath{\Distance}}}\endgroup }(x)}}}.
The theorem says that if a sensitivity term (resp. privacy term) is well-typed, then for any number of steps {{\color{\colorMATH}\ensuremath{k}}}, valid relational distance environment {{\color{\colorMATH}\ensuremath{{\begingroup\renewcommand\colorMATH{\colorMATHB}\renewcommand\colorSYNTAX{\colorSYNTAXB}{{\color{\colorMATH}\ensuremath{\Distance^{\prime}}}}\endgroup }}}} (not greater than {{\color{\colorMATH}\ensuremath{{\begingroup\renewcommand\colorMATH{\colorMATHB}\renewcommand\colorSYNTAX{\colorSYNTAXB}{{\color{\colorMATH}\ensuremath{\Distance}}}\endgroup }}}}) and value environments {{\color{\colorMATH}\ensuremath{(\gamma _{1},\gamma _{2})}}}, the two configurations are related computations at distance type {{\color{\colorMATH}\ensuremath{{\begingroup\renewcommand\colorMATH{\colorMATHB}\renewcommand\colorSYNTAX{\colorSYNTAXB}{{\color{\colorMATH}\ensuremath{\Distance^{\prime}}}}\endgroup }(\tau )}}} (closing all free sensitivity or privacy variables), and at relational distance {{\color{\colorMATH}\ensuremath{{\begingroup\renewcommand\colorMATH{\colorMATHB}\renewcommand\colorSYNTAX{\colorSYNTAXB}{{\color{\colorMATH}\ensuremath{\Distance^{\prime}}}}\endgroup }\mathord{\cdotp }{\begingroup\renewcommand\colorMATH{\colorMATHB}\renewcommand\colorSYNTAX{\colorSYNTAXB}{{\color{\colorMATH}\ensuremath{\sS}}}\endgroup }}}} (resp. {\begingroup\renewcommand\colorMATH{\colorMATHB}\renewcommand\colorSYNTAX{\colorSYNTAXB}{{\color{\colorMATH}\ensuremath{\Distance^{\prime}}}}\endgroup }{\begingroup\renewcommand\colorMATH{\colorMATHC}\renewcommand\colorSYNTAX{\colorSYNTAXC}{{\color{\colorMATH}\ensuremath{\bigcdot}}}\endgroup }{\begingroup\renewcommand\colorMATH{\colorMATHC}\renewcommand\colorSYNTAX{\colorSYNTAXC}{{\color{\colorMATH}\ensuremath{\pS}}}\endgroup }). Note that as {{\color{\colorMATH}\ensuremath{dom({\begingroup\renewcommand\colorMATH{\colorMATHB}\renewcommand\colorSYNTAX{\colorSYNTAXB}{{\color{\colorMATH}\ensuremath{\sS}}}\endgroup }) \subseteq  dom({\begingroup\renewcommand\colorMATH{\colorMATHB}\renewcommand\colorSYNTAX{\colorSYNTAXB}{{\color{\colorMATH}\ensuremath{\Distance^{\prime}}}}\endgroup }) = dom({\begingroup\renewcommand\colorMATH{\colorMATHB}\renewcommand\colorSYNTAX{\colorSYNTAXB}{{\color{\colorMATH}\ensuremath{\Distance}}}\endgroup })}}} and {{\color{\colorMATH}\ensuremath{dom({\begingroup\renewcommand\colorMATH{\colorMATHC}\renewcommand\colorSYNTAX{\colorSYNTAXC}{{\color{\colorMATH}\ensuremath{\pS}}}\endgroup }) \subseteq  dom({\begingroup\renewcommand\colorMATH{\colorMATHB}\renewcommand\colorSYNTAX{\colorSYNTAXB}{{\color{\colorMATH}\ensuremath{\Distance^{\prime}}}}\endgroup }) = dom({\begingroup\renewcommand\colorMATH{\colorMATHB}\renewcommand\colorSYNTAX{\colorSYNTAXB}{{\color{\colorMATH}\ensuremath{\Distance}}}\endgroup })}}}, then {{\color{\colorMATH}\ensuremath{{\begingroup\renewcommand\colorMATH{\colorMATHB}\renewcommand\colorSYNTAX{\colorSYNTAXB}{{\color{\colorMATH}\ensuremath{\Distance^{\prime}}}}\endgroup }\mathord{\cdotp }{\begingroup\renewcommand\colorMATH{\colorMATHB}\renewcommand\colorSYNTAX{\colorSYNTAXB}{{\color{\colorMATH}\ensuremath{\sS}}}\endgroup } \in  {\mathbb{R}}^{\infty }}}} and {{\color{\colorMATH}\ensuremath{{\begingroup\renewcommand\colorMATH{\colorMATHB}\renewcommand\colorSYNTAX{\colorSYNTAXB}{{\color{\colorMATH}\ensuremath{\Distance^{\prime}}}}\endgroup }{\begingroup\renewcommand\colorMATH{\colorMATHC}\renewcommand\colorSYNTAX{\colorSYNTAXC}{{\color{\colorMATH}\ensuremath{\bigcdot}}}\endgroup }{\begingroup\renewcommand\colorMATH{\colorMATHC}\renewcommand\colorSYNTAX{\colorSYNTAXC}{{\color{\colorMATH}\ensuremath{\pS}}}\endgroup } \in  {\text{{\begingroup\renewcommand\colorMATH{\colorMATHC}\renewcommand\colorSYNTAX{\colorSYNTAXC}{{\color{\colorSYNTAX}\texttt{{{\color{\colorMATH}\ensuremath{priv}}}}}}\endgroup }}}}}}.

To prove the fundamental property,
we rely on the following three lemmas which connect types, \toplas{sensitivity and privacy environments} from the type system, with distances and \toplas{privacy costs} from the logical relations:

\begin{restatable}{lemma}{lmdistrdot}
\label{lm:distrdot}
If {{\color{\colorMATH}\ensuremath{\instE{{\begingroup\renewcommand\colorMATH{\colorMATHB}\renewcommand\colorSYNTAX{\colorSYNTAXB}{{\color{\colorMATH}\ensuremath{\Distance}}}\endgroup }}{{\begingroup\renewcommand\colorMATH{\colorMATHB}\renewcommand\colorSYNTAX{\colorSYNTAXB}{{\color{\colorMATH}\ensuremath{\sS}}}\endgroup }} = {\begingroup\renewcommand\colorMATH{\colorMATHB}\renewcommand\colorSYNTAX{\colorSYNTAXB}{{\color{\colorMATH}\ensuremath{\distance}}}\endgroup }}}} and {{\color{\colorMATH}\ensuremath{x \notin  dom({\begingroup\renewcommand\colorMATH{\colorMATHB}\renewcommand\colorSYNTAX{\colorSYNTAXB}{{\color{\colorMATH}\ensuremath{\sS}}}\endgroup })\cup  dom({\begingroup\renewcommand\colorMATH{\colorMATHB}\renewcommand\colorSYNTAX{\colorSYNTAXB}{{\color{\colorMATH}\ensuremath{\Distance}}}\endgroup })}}}, then
  {{\color{\colorMATH}\ensuremath{\instE{{\begingroup\renewcommand\colorMATH{\colorMATHB}\renewcommand\colorSYNTAX{\colorSYNTAXB}{{\color{\colorMATH}\ensuremath{\Distance}}}\endgroup }}{([{\begingroup\renewcommand\colorMATH{\colorMATHB}\renewcommand\colorSYNTAX{\colorSYNTAXB}{{\color{\colorMATH}\ensuremath{\sS}}}\endgroup }/x]{\begingroup\renewcommand\colorMATH{\colorMATHB}\renewcommand\colorSYNTAX{\colorSYNTAXB}{{\color{\colorMATH}\ensuremath{\sS'}}}\endgroup })} = \instE{({\begingroup\renewcommand\colorMATH{\colorMATHB}\renewcommand\colorSYNTAX{\colorSYNTAXB}{{\color{\colorMATH}\ensuremath{\Distance}}}\endgroup } + {\begingroup\renewcommand\colorMATH{\colorMATHB}\renewcommand\colorSYNTAX{\colorSYNTAXB}{{\color{\colorMATH}\ensuremath{\distance}}}\endgroup }x)}{{\begingroup\renewcommand\colorMATH{\colorMATHB}\renewcommand\colorSYNTAX{\colorSYNTAXB}{{\color{\colorMATH}\ensuremath{\sS'}}}\endgroup }}}}}
\end{restatable}

\begin{restatable}{lemma}{lmdistrdotpp}
\label{lm:distrdotpp}
If {{\color{\colorMATH}\ensuremath{\instE{{\begingroup\renewcommand\colorMATH{\colorMATHB}\renewcommand\colorSYNTAX{\colorSYNTAXB}{{\color{\colorMATH}\ensuremath{\Distance}}}\endgroup }}{{\begingroup\renewcommand\colorMATH{\colorMATHB}\renewcommand\colorSYNTAX{\colorSYNTAXB}{{\color{\colorMATH}\ensuremath{\sS}}}\endgroup }} = {\begingroup\renewcommand\colorMATH{\colorMATHB}\renewcommand\colorSYNTAX{\colorSYNTAXB}{{\color{\colorMATH}\ensuremath{\distance}}}\endgroup }}}} and {{\color{\colorMATH}\ensuremath{x \notin  dom({\begingroup\renewcommand\colorMATH{\colorMATHB}\renewcommand\colorSYNTAX{\colorSYNTAXB}{{\color{\colorMATH}\ensuremath{\sS}}}\endgroup })\cup  dom({\begingroup\renewcommand\colorMATH{\colorMATHB}\renewcommand\colorSYNTAX{\colorSYNTAXB}{{\color{\colorMATH}\ensuremath{\Distance}}}\endgroup })}}}, then
  {{\color{\colorMATH}\ensuremath{\instPE{{\begingroup\renewcommand\colorMATH{\colorMATHB}\renewcommand\colorSYNTAX{\colorSYNTAXB}{{\color{\colorMATH}\ensuremath{\Distance}}}\endgroup }}{(\subst[{\begingroup\renewcommand\colorMATH{\colorMATHB}\renewcommand\colorSYNTAX{\colorSYNTAXB}{{\color{\colorMATH}\ensuremath{\Distance}}}\endgroup }]{{\begingroup\renewcommand\colorMATH{\colorMATHB}\renewcommand\colorSYNTAX{\colorSYNTAXB}{{\color{\colorMATH}\ensuremath{\sS}}}\endgroup }}{x}{\begingroup\renewcommand\colorMATH{\colorMATHC}\renewcommand\colorSYNTAX{\colorSYNTAXC}{{\color{\colorMATH}\ensuremath{\pS'}}}\endgroup })} = \instPE{({\begingroup\renewcommand\colorMATH{\colorMATHB}\renewcommand\colorSYNTAX{\colorSYNTAXB}{{\color{\colorMATH}\ensuremath{\Distance}}}\endgroup } + {\begingroup\renewcommand\colorMATH{\colorMATHB}\renewcommand\colorSYNTAX{\colorSYNTAXB}{{\color{\colorMATH}\ensuremath{\distance}}}\endgroup }x)}{{\begingroup\renewcommand\colorMATH{\colorMATHC}\renewcommand\colorSYNTAX{\colorSYNTAXC}{{\color{\colorMATH}\ensuremath{\pS'}}}\endgroup }}}}}
\end{restatable}

\begin{restatable}{lemma}{lmdistrinst}
\label{lm:distrinst}
Let {{\color{\colorMATH}\ensuremath{{\begingroup\renewcommand\colorMATH{\colorMATHB}\renewcommand\colorSYNTAX{\colorSYNTAXB}{{\color{\colorMATH}\ensuremath{\Distance}}}\endgroup }\mathord{\cdotp }{\begingroup\renewcommand\colorMATH{\colorMATHB}\renewcommand\colorSYNTAX{\colorSYNTAXB}{{\color{\colorMATH}\ensuremath{\sS}}}\endgroup } = {\begingroup\renewcommand\colorMATH{\colorMATHB}\renewcommand\colorSYNTAX{\colorSYNTAXB}{{\color{\colorMATH}\ensuremath{\distance}}}\endgroup }}}} and {{\color{\colorMATH}\ensuremath{x \notin  dom({\begingroup\renewcommand\colorMATH{\colorMATHB}\renewcommand\colorSYNTAX{\colorSYNTAXB}{{\color{\colorMATH}\ensuremath{\sS}}}\endgroup })\cup  dom({\begingroup\renewcommand\colorMATH{\colorMATHB}\renewcommand\colorSYNTAX{\colorSYNTAXB}{{\color{\colorMATH}\ensuremath{\Distance}}}\endgroup })}}}, then
  {{\color{\colorMATH}\ensuremath{{\begingroup\renewcommand\colorMATH{\colorMATHB}\renewcommand\colorSYNTAX{\colorSYNTAXB}{{\color{\colorMATH}\ensuremath{\Distance}}}\endgroup }([{\begingroup\renewcommand\colorMATH{\colorMATHB}\renewcommand\colorSYNTAX{\colorSYNTAXB}{{\color{\colorMATH}\ensuremath{\sS}}}\endgroup }/x]\tau ) = {\begingroup\renewcommand\colorMATH{\colorMATHB}\renewcommand\colorSYNTAX{\colorSYNTAXB}{{\color{\colorMATH}\ensuremath{\distance}}}\endgroup }x({\begingroup\renewcommand\colorMATH{\colorMATHB}\renewcommand\colorSYNTAX{\colorSYNTAXB}{{\color{\colorMATH}\ensuremath{\Distance}}}\endgroup }(\tau ))}}} 
\end{restatable}

We can also derive from the fundamental property some corollaries about closed terms.
\begin{corollary}[FP for closed sensitivity terms]\
  If {{\color{\colorMATH}\ensuremath{\varnothing ; \varnothing  \vdash  {\begingroup\renewcommand\colorMATH{\colorMATHB}\renewcommand\colorSYNTAX{\colorSYNTAXB}{{\color{\colorMATH}\ensuremath{\se}}}\endgroup } \mathrel{:} \tau  \mathrel{;} \varnothing }}}, then\\  {{\color{\colorMATH}\ensuremath{\forall k\geq 0,(\varnothing  \vdash  {\begingroup\renewcommand\colorMATH{\colorMATHB}\renewcommand\colorSYNTAX{\colorSYNTAXB}{{\color{\colorMATH}\ensuremath{\se}}}\endgroup }, \varnothing  \vdash {\begingroup\renewcommand\colorMATH{\colorMATHB}\renewcommand\colorSYNTAX{\colorSYNTAXB}{{\color{\colorMATH}\ensuremath{\se}}}\endgroup }) \in  {\mathcal{E}}^k_{{\begingroup\renewcommand\colorMATH{\colorMATHB}\renewcommand\colorSYNTAX{\colorSYNTAXB}{{\color{\colorMATH}\ensuremath{_{0}}}}\endgroup }}\llbracket \tau \rrbracket }}}
\end{corollary}

\begin{corollary}[FP for closed privacy terms]\
  If {{\color{\colorMATH}\ensuremath{\varnothing ; \varnothing  \vdash  {\begingroup\renewcommand\colorMATH{\colorMATHC}\renewcommand\colorSYNTAX{\colorSYNTAXC}{{\color{\colorMATH}\ensuremath{\pe}}}\endgroup } \mathrel{:} \tau  \mathrel{;} \varnothing }}}, then\\ {{\color{\colorMATH}\ensuremath{\forall k\geq 0,(\varnothing  \vdash  {\begingroup\renewcommand\colorMATH{\colorMATHC}\renewcommand\colorSYNTAX{\colorSYNTAXC}{{\color{\colorMATH}\ensuremath{\pe}}}\endgroup }, \varnothing  \vdash {\begingroup\renewcommand\colorMATH{\colorMATHC}\renewcommand\colorSYNTAX{\colorSYNTAXC}{{\color{\colorMATH}\ensuremath{\pe}}}\endgroup }) \in  {\mathcal{E}}^k_{{\begingroup\renewcommand\colorMATH{\colorMATHC}\renewcommand\colorSYNTAX{\colorSYNTAXC}{{\color{\colorMATH}\ensuremath{(_{0},_{0})}}}\endgroup }}\llbracket \tau \rrbracket }}}
\end{corollary}

In addition to sensitivity type soundness at base types (Prop~\ref{thm:sensitivity-simple-SensitivityTypeSoundnessBaseTypes}), from the fundamental property we can now establish privacy type soundness at base types:
%\et{is there a similar result at the privacy level?}
\begin{restatable}[Privacy Type Soundness at Base Types]{theorem}{PrivacyTypeSoundnessBaseTypes}
  \label{thm:PrivacyTypeSoundnessBaseTypes}
  If {{\color{\colorMATH}\ensuremath{\varnothing  \vdash  {\begingroup\renewcommand\colorMATH{\colorMATHB}\renewcommand\colorSYNTAX{\colorSYNTAXB}{{\color{\colorMATH}\ensuremath{\se}}}\endgroup } \mathrel{:} (x\mathrel{:} {\begingroup\renewcommand\colorMATH{\colorMATHA}\renewcommand\colorSYNTAX{\colorSYNTAXA}{{\color{\colorSYNTAX}\texttt{{\ensuremath{{\mathbb{R}}}}}}}\endgroup } \mathord{\cdotp } 1) \xrightarrowP {({\begingroup\renewcommand\colorMATH{\colorMATHC}\renewcommand\colorSYNTAX{\colorSYNTAXC}{{\color{\colorMATH}\ensuremath{\epsilon }}}\endgroup }, {\begingroup\renewcommand\colorMATH{\colorMATHC}\renewcommand\colorSYNTAX{\colorSYNTAXC}{{\color{\colorMATH}\ensuremath{\delta }}}\endgroup }) x} {\begingroup\renewcommand\colorMATH{\colorMATHA}\renewcommand\colorSYNTAX{\colorSYNTAXA}{{\color{\colorSYNTAX}\texttt{{\ensuremath{{\mathbb{R}}}}}}}\endgroup } \mathrel{;} \varnothing }}},\\
  {{\color{\colorMATH}\ensuremath{|{\begingroup\renewcommand\colorMATH{\colorMATHB}\renewcommand\colorSYNTAX{\colorSYNTAXB}{{\color{\colorMATH}\ensuremath{r_{1}}}}\endgroup }-{\begingroup\renewcommand\colorMATH{\colorMATHB}\renewcommand\colorSYNTAX{\colorSYNTAXB}{{\color{\colorMATH}\ensuremath{r_{2}}}}\endgroup }| \leq  1}}}, {{\color{\colorMATH}\ensuremath{\forall {\begingroup\renewcommand\colorMATH{\colorMATHB}\renewcommand\colorSYNTAX{\colorSYNTAXB}{{\color{\colorMATH}\ensuremath{r}}}\endgroup }}}},
  {{\color{\colorMATH}\ensuremath{{\text{Pr}}[\varnothing  \vdash  {\begingroup\renewcommand\colorMATH{\colorMATHC}\renewcommand\colorSYNTAX{\colorSYNTAXC}{{\color{\colorMATH}\ensuremath{\pe}}}\endgroup }\hspace*{0.33em}{\begingroup\renewcommand\colorMATH{\colorMATHC}\renewcommand\colorSYNTAX{\colorSYNTAXC}{{\color{\colorMATH}\ensuremath{r_{1}}}}\endgroup } \Downarrow ^{\infty } {\begingroup\renewcommand\colorMATH{\colorMATHB}\renewcommand\colorSYNTAX{\colorSYNTAXB}{{\color{\colorMATH}\ensuremath{r}}}\endgroup }] \leq  e^{{\begingroup\renewcommand\colorMATH{\colorMATHC}\renewcommand\colorSYNTAX{\colorSYNTAXC}{{\color{\colorMATH}\ensuremath{\epsilon }}}\endgroup }} {\text{Pr}}[\varnothing  \vdash  {\begingroup\renewcommand\colorMATH{\colorMATHC}\renewcommand\colorSYNTAX{\colorSYNTAXC}{{\color{\colorMATH}\ensuremath{\pe}}}\endgroup }\hspace*{0.33em}{\begingroup\renewcommand\colorMATH{\colorMATHC}\renewcommand\colorSYNTAX{\colorSYNTAXC}{{\color{\colorMATH}\ensuremath{r_{2}}}}\endgroup } \Downarrow ^{\infty } {\begingroup\renewcommand\colorMATH{\colorMATHB}\renewcommand\colorSYNTAX{\colorSYNTAXB}{{\color{\colorMATH}\ensuremath{r}}}\endgroup }] + {\begingroup\renewcommand\colorMATH{\colorMATHC}\renewcommand\colorSYNTAX{\colorSYNTAXC}{{\color{\colorMATH}\ensuremath{\delta }}}\endgroup }}}}
\end{restatable}

\toplasss{
  Finally, we observe that our technical development relies on the specific variant of differential privacy considered in restricted places:
  the bind case of the soundness theorem (Theorem~\ref{alm:fp});
  the definition of subtyping for privacy costs, specially the base case (Figure~\ref{afig:subtyping});
  operations over privacies $p$, such as dot product, addition, meet, join, lifting (Figures \ref{afig:lift-operators} and \ref{afig:types-join-meet});
  monotonicity of meet and join of privacies w.r.t subtyping (Lemmas \ref{lm:joinmeetgreen} and \ref{lm:joinmeetred});
  monotonicity of dot product w.r.t. privacy ordering (Lemma~\ref{lm:dot-subtp});
  distributivity of dot product w.r.t. substitution (Lemma~\ref{lm:distrdotpp});
  weakening of related private computations (Lemma~\ref{lm:lrweakening-sensitivity}).
  The remaining definitions and lemmas are independent and could be reused as such in order to adapt this work to deal with other variants of differential privacy.
}

%\mt{In this case it should not be a problem as real numbers are semantically equivalent for {{\color{\colorMATH}\ensuremath{\infty }}} steps.}

% If non termination then several changes must be done: change in the probabilistic semantics, change in the logical relation to be termination sensitive. \et{keep that for discussion/future work somewhere}

% }-}

\section{From $\lang$ to \system} % {-{
\label{sec:system}

The full prototype implementation of \system includes several extensions to the core language $\lang$\toplas{, and address the non-determinism of multiplicative and additive products by using type annotations}.

\paragraph{Type Polymorphism.}
\system implements System F (universal \toplas{quantification} over well-kinded types) and
parametric polymorphism over all compound types, including vector/matrix schemas, allowing all data objects and functions in the language to be fully generic.
This feature requires the use of type-level quantifiers and application.

\paragraph{Value Dependency.} \system supports type-level dependency
on values through singleton types---an approach we borrow directly
from \dfuzz~\cite{gaboardi2013linear}. This allows differentially private
algorithms to be verified with respect to privacy parameters which
are not fixed, and instead are function arguments.

More specifically, singleton types~\cite{singletons} are a technique
for supporting limited forms of value dependency which builds on
standard (System-F-style) polymorphic type system features and an
enriched kind system. In a type system with native support for
dependent types, a dependent function with a real-valued argument is
written {{\color{\colorSYNTAX}\texttt{{\ensuremath{({{\color{\colorMATH}\ensuremath{x}}} \mathrel{:} {\mathbb{R}}) \rightarrow  {{\color{\colorMATH}\ensuremath{\tau }}}}}}}} where the return type {{\color{\colorMATH}\ensuremath{\tau }}} can use {{\color{\colorMATH}\ensuremath{x}}} to
refer symbolically to the eventual runtime value of {{\color{\colorMATH}\ensuremath{x}}}. In a
singleton type encoding of dependent types, the same function is
written {{\color{\colorSYNTAX}\texttt{{\ensuremath{{{\color{\colorMATH}\ensuremath{x}}} \mathrel{:} {\mathbb{R}}[{{\color{\colorMATH}\ensuremath{\hat x}}}] \rightarrow  {{\color{\colorMATH}\ensuremath{\tau }}}}}}}}, where the return type {{\color{\colorMATH}\ensuremath{\tau }}} can use
{{\color{\colorMATH}\ensuremath{\hat x}}} to refer symbolically to the eventual runtime value of {{\color{\colorMATH}\ensuremath{x}}}. In
essence, there is still a syntactic split between term-level
variables ({{\color{\colorMATH}\ensuremath{x}}}) and type-level variables ({{\color{\colorMATH}\ensuremath{\hat x}}}), and the type
declaration {{\color{\colorSYNTAX}\texttt{{\ensuremath{({{\color{\colorMATH}\ensuremath{x}}} \mathrel{:} {\mathbb{R}}[{{\color{\colorMATH}\ensuremath{\hat x}}}])}}}}} links them, so {{\color{\colorMATH}\ensuremath{\hat x}}} is the
type-level {\textit{proxy}} for the term-level variable {{\color{\colorMATH}\ensuremath{x}}}.

\paragraph{Let-binding Sensitivity Terms in Privacy Contexts.}
As described in Section~\ref{sec:delay-sens} we implement latent sensitivity via local bindings in
the privacy language.
We implement this feature using an environment {{\color{\colorMATH}\ensuremath{{\begingroup\renewcommand\colorMATH{\colorMATHB}\renewcommand\colorSYNTAX{\colorSYNTAXB}{{\color{\colorMATH}\ensuremath{\Phi }}}\endgroup }}}} to delay the ``payments'' of a value's sensitivity, which fulfills the same role as the boxed type introduced by~\citet{near2019duet}. Unlike boxed types, this feature requires no additional annotations---sensitivity is inferred automatically.
The complete type systems that includes {{\color{\colorMATH}\ensuremath{{\begingroup\renewcommand\colorMATH{\colorMATHB}\renewcommand\colorSYNTAX{\colorSYNTAXB}{{\color{\colorMATH}\ensuremath{\Phi }}}\endgroup }}}} are presented in Appendix~\ref{asec:static-semantics}, Figures~\ref{afig:sensitivity-type-system-1} and~\ref{afig:sensitivity-type-system-2}.

\paragraph{Context Polymorphism.}
\toplas{When implementing flexible primitives in \system, it becomes convenient
to abstract over latent contextual effects. We label this form of
abstraction as context polymorphism. This form of polymorphism in our language
is what enables us to give a single generalized type to the \toplass{primitives} {\begingroup\renewcommand\colorMATH{\colorMATHC}\renewcommand\colorSYNTAX{\colorSYNTAXC}{{\color{\colorMATH}\ensuremath{{\text{gauss}}}}}\endgroup } and {\begingroup\renewcommand\colorMATH{\colorMATHC}\renewcommand\colorSYNTAX{\colorSYNTAXC}{{\color{\colorMATH}\ensuremath{{\text{seqloop}}}}}\endgroup } \toplass{(a looping combinator that uses sequential composition, further discussed in Section~\ref{sec:mwem})}.
Because \system implements quantification over latent contextual effects, it is possible to afford privacy in type signatures to closed-over variables involved in the differentially private computation. This feature requires the use of type-level quantifiers, application, substitution and annotations for context schemas. A context schema is an angle bracket enclosed list of variables. Angle bracket context schemas in \system denote the set of variables that we care about preserving privacy for, and are used in the introduction forms for sums, pairs, and functions, as well as in type-level application. For example, {\begingroup\renewcommand\colorMATH{\colorMATHC}\renewcommand\colorSYNTAX{\colorSYNTAXC}{{\color{\colorMATH}\ensuremath{{\text{gauss}}}}}\endgroup } <{{\color{\colorMATH}\ensuremath{x}}}> ({{\color{\colorMATH}\ensuremath{x}}} + {{\color{\colorMATH}\ensuremath{y}}} * {{\color{\colorMATH}\ensuremath{y}}}) is 1-sensitive in {{\color{\colorMATH}\ensuremath{x}}}, and bumps {{\color{\colorMATH}\ensuremath{y}}} to infinity privacy. This demonstrates the use of context polymorphism to indicate which variables we care about preserving privacy for. The use of {\begingroup\renewcommand\colorMATH{\colorMATHC}\renewcommand\colorSYNTAX{\colorSYNTAXC}{{\color{\colorMATH}\ensuremath{{\text{seqloop}}}}}\endgroup } in Section~\ref{sec:casestudies} provides another example of context polymorphism in action. Note that context-polymorphic functions in \system are required to be primitives.}

\paragraph{Variants of Differential Privacy.}
In addition to {{\color{\colorMATH}\ensuremath{(\epsilon , \delta )}}}-differential privacy, \system supports
zero-concentrated differential privacy~\cite{bun2016concentrated} and
R\'enyi differential privacy~\cite{mironov2017renyi}, and has built-in
constructs for mixing the variants. Each variant has different privacy
parameters and rules for composition, but all of them follow the same
basic pattern as {{\color{\colorMATH}\ensuremath{(\epsilon , \delta )}}}-differential privacy. For example, we can
give the Gaussian mechanism the following types for R\'enyi differential
privacy ({RDP}; privacy parameters {{\color{\colorMATH}\ensuremath{\alpha }}} and {{\color{\colorMATH}\ensuremath{\epsilon }}}) and zero-concentrated
differential privacy ({zCDP}; privacy parameter {{\color{\colorMATH}\ensuremath{\rho }}}):
\begingroup\color{\colorMATH}\begin{gather*}
\begin{array}{rcl
   } {\begingroup\renewcommand\colorMATH{\colorMATHC}\renewcommand\colorSYNTAX{\colorSYNTAXC}{{\color{\colorMATH}\ensuremath{\overset{{\text{RDP}}}{{\text{gauss}}}}}}\endgroup } &{}\mathrel{:}{}&
      \forall \hspace*{0.33em}(\hat d \mathrel{:} {\mathbb{R}})\hspace*{0.33em}(\hat \alpha  \mathrel{:} {\mathbb{R}})\hspace*{0.33em}(\hat \epsilon  \mathrel{:} {\mathbb{R}}).\hspace*{0.33em}
         (d \mathrel{:} {\mathbb{R}}[\hat d])
      \mathrel{{\begingroup\renewcommand\colorMATH{\colorMATHB}\renewcommand\colorSYNTAX{\colorSYNTAXB}{{\color{\colorMATH}\ensuremath{\rightarrow }}}\endgroup }} (\alpha  \mathrel{:} {\mathbb{R}}[\hat \alpha ])
      \mathrel{{\begingroup\renewcommand\colorMATH{\colorMATHB}\renewcommand\colorSYNTAX{\colorSYNTAXB}{{\color{\colorMATH}\ensuremath{\rightarrow }}}\endgroup }} (\epsilon  \mathrel{:} {\mathbb{R}}[\hat \epsilon ])
      \mathrel{{\begingroup\renewcommand\colorMATH{\colorMATHB}\renewcommand\colorSYNTAX{\colorSYNTAXB}{{\color{\colorMATH}\ensuremath{\rightarrow }}}\endgroup }} (x \mathrel{:} {\mathbb{R}}\mathord{\cdotp }\hat d)
      \xrightarrowP {\infty (d + \alpha  + \epsilon ) + (\hat \alpha ,\hat \epsilon )x}
      {\mathbb{R}}
   \cr
\end{array}
\end{gather*}\endgroup

Since RDP and zCDP guarantees can be converted to {{\color{\colorMATH}\ensuremath{(\epsilon , \delta )}}} guarantees,
\system provides constructs for converting between variants. For
example, the following code uses the Gaussian mechanism twice, each
time satisfying {{\color{\colorMATH}\ensuremath{(20, 0.25)}}}-RDP. By sequential composition, the total
cost is {{\color{\colorMATH}\ensuremath{(20, 0.5)}}}-RDP. The program then converts this guarantee to
{{\color{\colorMATH}\ensuremath{(\epsilon , \delta )}}}-differential privacy, using {{\color{\colorMATH}\ensuremath{\delta  = 10^{-5}}}}.
\begingroup\color{\colorMATH}\begin{gather*}
\begin{tabularx}{\linewidth}{>{\centering\arraybackslash\(}X<{\)}}
\hfill\hspace{0pt}
\hfill\hspace{0pt}
\begin{array}{l
} {\begingroup\renewcommand\colorMATH{\colorMATHC}\renewcommand\colorSYNTAX{\colorSYNTAXC}{{\color{\colorMATH}\ensuremath{\plambda}}}\endgroup } (x \mathord{\cdotp } 1).\hspace*{0.33em} {\begingroup\renewcommand\colorMATH{\colorMATHC}\renewcommand\colorSYNTAX{\colorSYNTAXC}{{\color{\colorSYNTAX}\texttt{RENYI}}}\endgroup } [\delta  = 10^{-5}] \{ \hspace*{0.33em}
   \begin{array}[t]{l
   } r_{1} \mathrel{{\begingroup\renewcommand\colorMATH{\colorMATHC}\renewcommand\colorSYNTAX{\colorSYNTAXC}{{\color{\colorMATH}\ensuremath{\leftarrow }}}\endgroup }} {\begingroup\renewcommand\colorMATH{\colorMATHC}\renewcommand\colorSYNTAX{\colorSYNTAXC}{{\color{\colorMATH}\ensuremath{{\text{gauss}}}}}\endgroup }\hspace*{0.33em}1\hspace*{0.33em}20\hspace*{0.33em}0.25\hspace*{0.33em}x {\begingroup\renewcommand\colorMATH{\colorMATHC}\renewcommand\colorSYNTAX{\colorSYNTAXC}{{\color{\colorMATH}\ensuremath{;}}}\endgroup }
   \cr  r_{2} \mathrel{{\begingroup\renewcommand\colorMATH{\colorMATHC}\renewcommand\colorSYNTAX{\colorSYNTAXC}{{\color{\colorMATH}\ensuremath{\leftarrow }}}\endgroup }} {\begingroup\renewcommand\colorMATH{\colorMATHC}\renewcommand\colorSYNTAX{\colorSYNTAXC}{{\color{\colorMATH}\ensuremath{{\text{gauss}}}}}\endgroup }\hspace*{0.33em}2\hspace*{0.33em}20\hspace*{0.33em}0.25\hspace*{0.33em}(x + x) {\begingroup\renewcommand\colorMATH{\colorMATHC}\renewcommand\colorSYNTAX{\colorSYNTAXC}{{\color{\colorMATH}\ensuremath{;}}}\endgroup }
   \cr  {\begingroup\renewcommand\colorMATH{\colorMATHC}\renewcommand\colorSYNTAX{\colorSYNTAXC}{{\color{\colorSYNTAX}\texttt{return}}}\endgroup }\hspace*{0.33em}(r_{1} + r_{2}) \hspace*{0.33em} \} 
   \end{array}
\end{array}
\hfill\hspace{0pt}
\mathrel{:}
\hfill\hspace{0pt}
(x \mathrel{:} {\mathbb{R}}\mathord{\cdotp }1) \xrightarrowP {(1.08,10^{-5})x} {\mathbb{R}}
\hfill\hspace{0pt}
\hfill\hspace{0pt}
\end{tabularx}
\end{gather*}\endgroup
\noindent \system automatically finds the privacy cost of this
function, in {{\color{\colorMATH}\ensuremath{(\epsilon , \delta )}}}-differential privacy, by performing the
appropriate conversion. The ability to mix privacy variants in \system
makes it easy to frame the privacy guarantee of any program in terms
of {{\color{\colorMATH}\ensuremath{(\epsilon , \delta )}}} privacy cost, allowing privacy costs to be directly
compared. In addition, it enables embedding iterative RDP and zCDP
algorithms inside of {{\color{\colorMATH}\ensuremath{(\epsilon , \delta )}}} programs, allowing these programs to
take advantage of the improved composition properties RDP and zCDP
provide. This approach---leveraging recent variants for composition, but
reporting privacy costs in terms of {{\color{\colorMATH}\ensuremath{\epsilon }}} and {{\color{\colorMATH}\ensuremath{\delta }}}---is extremely common in
recent work on differentially private machine learning~\cite{abadi2016deep}. We
make extensive use of variant-mixing in our case studies, described
next.

% }-}

\section{Implementation \& Case Studies} % {-{
\label{sec:casestudies}

\toplas{
\system enables programmers to implement and verify largely the same set of
{\textit{applications}} as \duet, but \system empowers the programmer to construct these
applications in {\textit{simpler ways}}, e.g., via composition of reusable library
functions. This is possible because \system gives types to many privacy
functions and looping combinators that required custom typing rules in \duet.
Our case studies demonstrate that instead of
encoding these applications as a single monolithic function, \system enables
their implementation through composition of multiple helper functions, and their
verification without the use of custom typing rules.
}

\toplass{ In particular, we highlight two important features of
  \system that enable refactoring programs to reuse library
  functions:
  \begin{itemize}
  \item \system gives types to privacy primitives and looping
    combinators that are not typeable in \duet, enabling privacy
    functions to be parameterized by these components (see
    Section~\ref{sec:mwem}).
  \item \system's privacy functions allow sensitivity arguments with
    arbitrary sensitivity bounds, enabling code reuse in more places
    than \duet's sensitivity-1 privacy functions (see
    Section~\ref{sec:adaptive-clipping}).
  \end{itemize}
  In addition, we select realistic algorithms previously verified
  using other systems, to demonstrate that \system maintains the
  capabilities of previous work.}

A summary of our case study programs appears in Table~\ref{tbl:case_studies}. 
We present two new representative case study algorithms we have implemented and verified using \system: the MWEM algorithm~\cite{hardt2012simple} for a workload of linear queries, and a recently proposed algorithm for differentially private deep learning with adaptive clipping~\cite{thakkar2019differentially}. In both case studies, privacy mechanisms (e.g. {\begingroup\renewcommand\colorMATH{\colorMATHC}\renewcommand\colorSYNTAX{\colorSYNTAXC}{{\color{\colorMATH}\ensuremath{{\text{laplace}}}}}\endgroup } and {\begingroup\renewcommand\colorMATH{\colorMATHC}\renewcommand\colorSYNTAX{\colorSYNTAXC}{{\color{\colorMATH}\ensuremath{{\text{exponential}}}}}\endgroup }) and looping constructs (e.g. {{\color{\colorSYNTAX}\texttt{aloop}}}---advanced composition) can be expressed with regular functions, provided in a library of primitives. We mark these two case studies with a $^*$ in Table~\ref{tbl:case_studies}, and describe them in detail later in this section. The other case study programs are available in our source code repository.

\toplas{
\subsection{Implementation}

We have implemented a prototype of the \system typechecker in Haskell, and used it to verify the case studies from Table~\ref{tbl:case_studies}.
The prototype implementation is available on GitHub\footnote{\url{https://github.com/uvm-plaid/contextual-duet/}}. Table~\ref{tbl:case_studies} lists the time needed to typecheck each of the case studies; our typechecker takes just a few milliseconds for each one.

\paragraph{Type inference \& annotations.}
Our prototype implements type inference for both \ssystem and \system. Type annotations for sensitivity and privacy are required for \emph{inputs} at top-level functions and lambda-expressions, but no additional annotations are required for function outputs or elsewhere in the program. The case studies described later in this section have been typeset for readability, but are otherwise identical to the input for our prototype; in particular, the actual examples typechecked by our prototype have the same annotations as the examples in this section, except for the input sensitivity annotations on inputs to top-level functions. The types given for primitives in this section are also drawn directly from our implementation.

\paragraph{Constraint solving.}
As described earlier, and detailed in Section~\ref{sec:relatedwork}, prior work has made extensive use of SMT solvers for the equations over real expressions which arise in type inference for sensitivity. Because SMT solvers are incomplete for non-linear operations (like the logarithms and square roots used in advanced composition), our implementation does not follow the same path.

Instead, we implement a custom solver for inequalities over symbolic real expressions, based on the solver from \duet~\cite{near2019duet}. Our custom solver is based on a simple decidable
(but incomplete) theory; it supports logarithms, square roots, and polynomial formulas over real numbers. The solver is transparent to the programmer, and produces readable output expressions for the privacy costs in our case studies.
}

\begin{table}
\centering
{\small
\begin{tabular}{l l l l}
  & & & \textbf{Typecheck}\\
  \textbf{Technique} & \textbf{Ref.} & \textbf{Privacy Concept} & \textbf{Time}\\
  \hline
  \textit{Machine Learning Algorithms} & & \\
  \hline
  Noisy Gradient Descent & \cite{BST} & Composition & \toplas{4.1 ms} \\
  Gradient Descent w/ Output Perturbation & \cite{PSGD} & Parallel {comp.} ({sens.}) & \toplas{4.2 ms} \\
  Noisy Frank-Wolfe & \cite{ttz16} & Exponential mechanism & \toplas{5.9 ms} \\[2mm]

  \hline
  \textit{Variations on Gradient Descent} & & \\
  \hline
  Minibatching & \cite{BST} &  Privacy amplification & \toplas{5.5 ms} \\
  Parallel-composition minibatching & --- &  Parallel composition & \toplas{5.9 ms} \\
  Gradient clipping & \cite{abadi2016deep} &  Sensitivity bounds & \toplas{4.5 ms} \\[1mm]
  Adaptive gradient clipping$^*$ \toplas{(\S\ref{sec:adaptive-clipping})} & \cite{thakkar2019differentially} & Advanced variants & \toplas{5.6 ms} \\[2mm]

  \hline
  \textit{Preprocessing \& Deployment} & & \\
  \hline
  Hyperparameter tuning & \cite{chaudhuri2013stability} & Exponential mechanism & \toplas{6.9 ms} \\
  Adaptive clipping & --- & Sparse Vector Technique & \toplas{7.7 ms} \\
  Z-Score normalization & \cite{sklearn_normalization} & Composition & \toplas{6.9 ms} \\[2mm]

  \hline
  \textit{Algorithms for Linear Queries} & & \\
  \hline
  Multiplicative Weights (MWEM)$^*$ \toplas{(\S\ref{sec:mwem})} & \cite{hardt2012simple} & Exponential mechanism & \toplas{5.2 ms} \\
%  \textbf{Combining All of the Above} &  & \ref{sec:example_sys} & Composition \\
\end{tabular}
}
\bigskip

\caption{List of case studies included with the \system
  implementation. Case studies marked with a $^*$ are described in
  detail in this section.}
\label{tbl:case_studies}
\end{table}

\subsection{MWEM}
\label{sec:mwem}
The MWEM algorithm~\cite{hardt2012simple} generates differentially private synthetic data approximating the target sensitive data by iteratively optimizing the accuracy of a set of workload queries on the synthetic data. In each iteration, the algorithm uses the exponential mechanism to pick a query from the workload for which the synthetic data produces an \emph{inaccurate} result, uses the Laplace mechanism to run that query on the real data, and uses the result to update the synthetic data via the \emph{multiplicative weights update rule}.
The \system program shown below implements the MWEM algorithm. Its inputs are a sensitive dataset {{\color{\colorMATH}\ensuremath{X}}} over a domain {{\color{\colorMATH}\ensuremath{D}}}, a workload {{\color{\colorMATH}\ensuremath{Q}}} of linear queries, the number of iterations to be performed {{\color{\colorMATH}\ensuremath{k}}}, the privacy parameter {{\color{\colorMATH}\ensuremath{\epsilon }}}, initial synthetic data {{\color{\colorMATH}\ensuremath{ Y_{0} }}} ({{\color{\colorMATH}\ensuremath{n}}} times the uniform distribution over {{\color{\colorMATH}\ensuremath{D}}}), and the dimensions of the input data set (matrix) {{\color{\colorMATH}\ensuremath{m}}} rows by {{\color{\colorMATH}\ensuremath{n}}} columns. The algorithm performs {{\color{\colorMATH}\ensuremath{k}}} iterations, invoking {\begingroup\renewcommand\colorMATH{\colorMATHC}\renewcommand\colorSYNTAX{\colorSYNTAXC}{{\color{\colorMATH}\ensuremath{{\text{laplace}}}}}\endgroup } and {\begingroup\renewcommand\colorMATH{\colorMATHC}\renewcommand\colorSYNTAX{\colorSYNTAXC}{{\color{\colorMATH}\ensuremath{{\text{exponential}}}}}\endgroup } in each iteration. The privacy parameter for each invocation is ${\begingroup\renewcommand\colorMATH{\colorMATHC}\renewcommand\colorSYNTAX{\colorSYNTAXC}{{\color{\colorMATH}\ensuremath{\epsilon }}}\endgroup }/{{\color{\colorMATH}\ensuremath{2k}}}$, yielding a total privacy cost of {\begingroup\renewcommand\colorMATH{\colorMATHC}\renewcommand\colorSYNTAX{\colorSYNTAXC}{{\color{\colorMATH}\ensuremath{\epsilon }}}\endgroup }.
\toplas{We omit the sensitivity annotations on the function's inputs for readability.}
% The flexibility of the \system language will allow us to obtain even
% tighter privacy bounds when working with vector-valued queries by substituting {{\color{\colorSYNTAX}\texttt{laplace}}} for {{\color{\colorSYNTAX}\texttt{gauss}}} and sequential composition for advanced composition, both of which can also be expressed as
% primitive library functions in \system.

\begingroup\color{\colorMATH}\begin{gather*}
\begin{array}{l
}{\begingroup\renewcommand\colorMATH{\colorMATHC}\renewcommand\colorSYNTAX{\colorSYNTAXC}{{\color{\colorMATH}\ensuremath{{\text{MWEM}}}}}\endgroup } \triangleq  {\begingroup\renewcommand\colorMATH{\colorMATHB}\renewcommand\colorSYNTAX{\colorSYNTAXB}{{\color{\colorMATH}\ensuremath{\slambda}}}\endgroup }  n.\hspace*{0.33em} {\begingroup\renewcommand\colorMATH{\colorMATHB}\renewcommand\colorSYNTAX{\colorSYNTAXB}{{\color{\colorMATH}\ensuremath{\slambda}}}\endgroup }  k.\hspace*{0.33em} {\begingroup\renewcommand\colorMATH{\colorMATHB}\renewcommand\colorSYNTAX{\colorSYNTAXB}{{\color{\colorMATH}\ensuremath{\slambda}}}\endgroup }  \epsilon .\hspace*{0.33em} {\begingroup\renewcommand\colorMATH{\colorMATHB}\renewcommand\colorSYNTAX{\colorSYNTAXB}{{\color{\colorMATH}\ensuremath{\slambda}}}\endgroup }  X.\hspace*{0.33em} {\begingroup\renewcommand\colorMATH{\colorMATHB}\renewcommand\colorSYNTAX{\colorSYNTAXB}{{\color{\colorMATH}\ensuremath{\slambda}}}\endgroup }  Q.\hspace*{0.33em} {\begingroup\renewcommand\colorMATH{\colorMATHB}\renewcommand\colorSYNTAX{\colorSYNTAXB}{{\color{\colorMATH}\ensuremath{\slambda}}}\endgroup }  {\begingroup\renewcommand\colorMATH{\colorMATHC}\renewcommand\colorSYNTAX{\colorSYNTAXC}{{\color{\colorMATH}\ensuremath{{{\color{\colorSYNTAX}\texttt{loop}}}}}}\endgroup }.\hspace*{0.33em} {\begingroup\renewcommand\colorMATH{\colorMATHC}\renewcommand\colorSYNTAX{\colorSYNTAXC}{{\color{\colorMATH}\ensuremath{\plambda}}}\endgroup }  Y_{0}.\hspace*{0.33em}
\cr  \hspace*{1.00em} {\begingroup\renewcommand\colorMATH{\colorMATHC}\renewcommand\colorSYNTAX{\colorSYNTAXC}{{\color{\colorMATH}\ensuremath{{\text{loop}}}}}\endgroup }\hspace*{0.33em}k\hspace*{0.33em}Y_{0}\hspace*{0.33em}{<}X{>}\hspace*{0.33em}({\begingroup\renewcommand\colorMATH{\colorMATHC}\renewcommand\colorSYNTAX{\colorSYNTAXC}{{\color{\colorMATH}\ensuremath{\plambda}}}\endgroup } \hspace*{0.33em}A_{i - 1}.\hspace*{0.33em}
\cr
\end{array}
\end{gather*}\endgroup
The \system typechecker produces the following type for this implementation, indicating that the algorithm satisfies {{\color{\colorMATH}\ensuremath{\epsilon }}}-differential privacy. Note the homogenous matrix type notation used here is {{\color{\colorMATH}\ensuremath{{{\color{\colorSYNTAX}\texttt{{\ensuremath{{\mathbb{M}}}}}}} [m,n]\hspace*{0.33em} \tau }}} where {{\color{\colorMATH}\ensuremath{m}}} denotes the number of rows, {{\color{\colorMATH}\ensuremath{n}}} the number of columns, and {{\color{\colorMATH}\ensuremath{\tau }}} the type of each \toplas{entry}. \toplas{{{\color{\colorMATH}\ensuremath{{<}X{>}}}} is the context schema argument for type-application of {\begingroup\renewcommand\colorMATH{\colorMATHC}\renewcommand\colorSYNTAX{\colorSYNTAXC}{{\color{\colorMATH}\ensuremath{{\text{loop}}}}}\endgroup }, and indicates the program variable we want to preserve privacy for in this expression.}
\begingroup\color{\colorMATH}\begin{gather*}
\begin{array}{l
   } {\begingroup\renewcommand\colorMATH{\colorMATHC}\renewcommand\colorSYNTAX{\colorSYNTAXC}{{\color{\colorMATH}\ensuremath{{\text{MWEM}}}}}\endgroup } \mathrel{:}
      \forall \hspace*{0.33em}(\hat m\mathrel{:}{\mathbb{N}})\hspace*{0.33em}(\hat n\mathrel{:}{\mathbb{N}})\hspace*{0.33em}(\hat k\mathrel{:}{\mathbb{N}})\hspace*{0.33em}(\hat d\mathrel{:}{\mathbb{R}}^{+})\hspace*{0.33em}(\hat \epsilon \mathrel{:}{\mathbb{R}}^{+}).\hspace*{0.33em}
         \overbracket{(n\mathrel{:}{\mathbb{N}}[\hat n])}^{\clap{\smaller{{\color{\colorTEXT}\textnormal{columns}}}}}
      \mathrel{{\begingroup\renewcommand\colorMATH{\colorMATHB}\renewcommand\colorSYNTAX{\colorSYNTAXB}{{\color{\colorMATH}\ensuremath{\rightarrow }}}\endgroup }} \overbracket{(k\mathrel{:}{\mathbb{N}}[\hat k])}^{\clap{\smaller{{\color{\colorTEXT}\textnormal{iterations}}}}}
      \mathrel{{\begingroup\renewcommand\colorMATH{\colorMATHB}\renewcommand\colorSYNTAX{\colorSYNTAXB}{{\color{\colorMATH}\ensuremath{\rightarrow }}}\endgroup }} \overbracket{(d\mathrel{:}{\mathbb{R}}^{+}[\hat d])}^{\clap{\smaller{{\color{\colorTEXT}\textnormal{relational distance}}}}}
   \cr
\end{array}
\end{gather*}\endgroup
On an average of 10 runs, it takes the \system typechecker 5.2ms to produce this type for the MWEM algorithm.

% Intuitively {{\color{\colorSYNTAX}\texttt{MWEM}}} works by iteratively correcting an approximating distribution {{\color{\colorMATH}\ensuremath{Y_{0}}}} of the true dataset
% {{\color{\colorMATH}\ensuremath{ B }}} by selecting queries on which the approximate and true datasets differ most significantly, and then performing a
% Multiplicative Weights update on the approximate database using a differentially private result {{\color{\colorMATH}\ensuremath{m_{i}}}} of the
% chosen query on the true database. The query {{\color{\colorMATH}\ensuremath{q_{i}}}} selected is also chosen privately using the {{\color{\colorSYNTAX}\texttt{exponential}}}
% mechanism.
% {{\color{\colorSYNTAX}\texttt{MWEM}}} typechecks in \system in 16ms on average.
%
% \begin{center}
% \includegraphics[scale=0.3]{mwem}
% \end{center}

\toplass{
\paragraph{Beyond \duet.}
This example demonstrates \system's ability to define algorithms in
terms of library functions, and to parameterize algorithms by the
choice of component pieces. In this case, we define {\begingroup\renewcommand\colorMATH{\colorMATHC}\renewcommand\colorSYNTAX{\colorSYNTAXC}{{\color{\colorSYNTAX}\texttt{MWEM}}}\endgroup } in terms
of a generic looping privacy combinator {\begingroup\renewcommand\colorMATH{\colorMATHC}\renewcommand\colorSYNTAX{\colorSYNTAXC}{{\color{\colorSYNTAX}\texttt{loop}}}\endgroup }; the caller of
{\begingroup\renewcommand\colorMATH{\colorMATHC}\renewcommand\colorSYNTAX{\colorSYNTAXC}{{\color{\colorSYNTAX}\texttt{MWEM}}}\endgroup } can specify looping combinators based on sequential
composition, advanced composition, or even a custom combinator.
\system similarly allows functions like {\begingroup\renewcommand\colorMATH{\colorMATHC}\renewcommand\colorSYNTAX{\colorSYNTAXC}{{\color{\colorSYNTAX}\texttt{MWEM}}}\endgroup } to be parameterized
by the choice of basic mechanism (e.g. {\begingroup\renewcommand\colorMATH{\colorMATHC}\renewcommand\colorSYNTAX{\colorSYNTAXC}{{\color{\colorSYNTAX}\texttt{laplace}}}\endgroup } vs. {\begingroup\renewcommand\colorMATH{\colorMATHC}\renewcommand\colorSYNTAX{\colorSYNTAXC}{{\color{\colorSYNTAX}\texttt{gauss}}}\endgroup }) with
the appropriate privacy function type. In both cases, the relevant
functions can be pulled from libraries or defined by the programmer.

This kind of modularity is impossible in \duet. Functions cannot be
parameterized by basic privacy mechanisms or looping combinators,
because it is not possible to write their types in \duet. }

\paragraph{Primitives used.}
This case study demonstrates the composition of a complex iterative algorithm from basic privacy mechanisms encoded as \system primitives (e.g. {\begingroup\renewcommand\colorMATH{\colorMATHC}\renewcommand\colorSYNTAX{\colorSYNTAXC}{{\color{\colorMATH}\ensuremath{{\text{laplace}}}}}\endgroup } and {\begingroup\renewcommand\colorMATH{\colorMATHC}\renewcommand\colorSYNTAX{\colorSYNTAXC}{{\color{\colorMATH}\ensuremath{{\text{exponential}}}}}\endgroup }) and privacy combinators (e.g. {\begingroup\renewcommand\colorMATH{\colorMATHC}\renewcommand\colorSYNTAX{\colorSYNTAXC}{{\color{\colorMATH}\ensuremath{{\text{seqloop}}}}}\endgroup }, which implements looping with sequential composition for privacy). These primitives with types shown above would require explicit typing rules in the core \duet language. In \system, they can be given regular types, as shown below:
\begingroup\color{\colorMATH}\begin{gather*}
\begin{array}{l
  } {\begingroup\renewcommand\colorMATH{\colorMATHC}\renewcommand\colorSYNTAX{\colorSYNTAXC}{{\color{\colorMATH}\ensuremath{{\text{exponential}}}}}\endgroup } \mathrel{:}
     \forall \hspace*{0.33em}(\hat m\mathrel{:}{\mathbb{N}})\hspace*{0.33em}(\hat n\mathrel{:}{\mathbb{N}})\hspace*{0.33em}(\hat d\mathrel{:}{\mathbb{R}}^{+})\hspace*{0.33em}(\hat \epsilon \mathrel{:}{\mathbb{R}}^{+})\hspace*{0.33em}(\Gamma \mathrel{:}{{\color{\colorSYNTAX}\texttt{cxt}}})\hspace*{0.33em}(\tau \mathrel{:}\bigstar ).\hspace*{0.33em}
        \overbracket{(d \mathrel{:} {\mathbb{R}}^{+}[\hat d])}^{\clap{\smaller{{\color{\colorTEXT}\textnormal{relational distance}}}}}
     \mathrel{{\begingroup\renewcommand\colorMATH{\colorMATHB}\renewcommand\colorSYNTAX{\colorSYNTAXB}{{\color{\colorMATH}\ensuremath{\rightarrow }}}\endgroup }} \overbracket{(\epsilon \mathrel{:}{\mathbb{R}}^{+}[\hat \epsilon ])}^{\clap{\smaller{{\color{\colorTEXT}\textnormal{desired privacy}}}}}
  \cr
\end{array}
\end{gather*}\endgroup
To typecheck {\begingroup\renewcommand\colorMATH{\colorMATHC}\renewcommand\colorSYNTAX{\colorSYNTAXC}{{\color{\colorSYNTAX}\texttt{MWEM}}}\endgroup }, the privacy closure rule in \system creates a function
type for the {{\color{\colorSYNTAX}\texttt{{\ensuremath{{\begingroup\renewcommand\colorMATH{\colorMATHC}\renewcommand\colorSYNTAX{\colorSYNTAXC}{{\color{\colorMATH}\ensuremath{\plambda}}}\endgroup } }}}}} which has a privacy effect for the body of {{\color{\colorMATH}\ensuremath{\frac{\epsilon }{k}}}}
because of the two uses of mechanisms which give {{\color{\colorMATH}\ensuremath{\frac{\epsilon }{2k}}}} differential
privacy. If we pass {\begingroup\renewcommand\colorMATH{\colorMATHC}\renewcommand\colorSYNTAX{\colorSYNTAXC}{{\color{\colorSYNTAX}\texttt{seqloop}}}\endgroup } as the looping combinator {\begingroup\renewcommand\colorMATH{\colorMATHC}\renewcommand\colorSYNTAX{\colorSYNTAXC}{{\color{\colorSYNTAX}\texttt{loop}}}\endgroup }, then this
privacy effect is multiplied by the loop iteration number {{\color{\colorMATH}\ensuremath{k}}} as
a result of the type of {{\color{\colorMATH}\ensuremath{{\begingroup\renewcommand\colorMATH{\colorMATHC}\renewcommand\colorSYNTAX{\colorSYNTAXC}{{\color{\colorMATH}\ensuremath{{\text{seqloop}}}}}\endgroup } \mathrel{:} {\mathbb{N}}[k] \mathrel{{\begingroup\renewcommand\colorMATH{\colorMATHB}\renewcommand\colorSYNTAX{\colorSYNTAXB}{{\color{\colorMATH}\ensuremath{\rightarrow }}}\endgroup }} \tau  \mathrel{{\begingroup\renewcommand\colorMATH{\colorMATHB}\renewcommand\colorSYNTAX{\colorSYNTAXB}{{\color{\colorMATH}\ensuremath{\rightarrow }}}\endgroup }} (\tau  \xrightarrowP {\rceil {\begingroup\renewcommand\colorMATH{\colorMATHC}\renewcommand\colorSYNTAX{\colorSYNTAXC}{{\color{\colorMATH}\ensuremath{\Sigma }}}\endgroup }\lceil ^{\epsilon }}
\tau ) \xrightarrowP {\rceil {\begingroup\renewcommand\colorMATH{\colorMATHC}\renewcommand\colorSYNTAX{\colorSYNTAXC}{{\color{\colorMATH}\ensuremath{\Sigma }}}\endgroup }\lceil ^{k\epsilon }} \tau }}} and the type rule for privacy function
application. Finally, \toplas{the new let rule for the \system privacy fragment which}
tracks latent contextual sensitivities is used in the let-binding for {{\color{\colorMATH}\ensuremath{q_{i}}}} to
precompute an intermediate value which is used multiple times, without the
need for explicit boxing.

\subsection{Differentially Private Deep Learning with Adaptive Clipping}
\label{sec:adaptive-clipping}
The current state-of-the-art in differentially private machine learning is \emph{noisy gradient descent}~\cite{abadi2016deep}: at each iteration of training, compute the gradient, \emph{clip} the gradient to have bounded {{\color{\colorMATH}\ensuremath{L2}}} norm, and add noise in proportion to the clipping parameter.
 % (our gradient descent example from Section~\ref{sec:overview} followed this basic structure).
 The clipping parameter is typically treated as a hyperparameter, set by the analyst before training.
\begingroup\color{\colorMATH}\begin{gather*}

\end{gather*}\endgroup
Recent work by \citet{thakkar2019differentially} proposed an algorithm for \emph{adaptively} determining the clipping parameter during training, by adaptively improving the clipping parameter based on a differentially private estimate of the percentage of gradients clipped in each iteration.

% In this case study we demonstrate the process of differentially private adaptive learning {{\color{\colorSYNTAX}\texttt{DPAL}}}: using adaptive clipping to train learning models with differential privacy. We provide a simplified version of the
% algorithm as seen in
% Thakkar et al 2019, implemented in \system. Note that as in the previous example: looping constructs
% such as advanced composition ({{\color{\colorSYNTAX}\texttt{aloop}}}) can be expressed as generic looping combinators, provided as primitive library functions.

In each iteration, the implementation computes the gradients for a batch of examples {{\color{\colorMATH}\ensuremath{gs}}}, clips each gradient using the current parameter {{\color{\colorMATH}\ensuremath{C^{t}}}}, and uses the Gaussian mechanism to compute a differentially private average gradient {{\color{\colorMATH}\ensuremath{g_{p}}}}. Then, the algorithm updates the clipping parameter for the next iteration {{\color{\colorMATH}\ensuremath{C^{t \prime}}}} using {{\color{\colorMATH}\ensuremath{{\begingroup\renewcommand\colorMATH{\colorMATHC}\renewcommand\colorSYNTAX{\colorSYNTAXC}{{\color{\colorMATH}\ensuremath{{\text{clipUpdate}}}}}\endgroup }}}}, which computes a noisy count {{\color{\colorMATH}\ensuremath{\beta }}} of the number of gradients in {{\color{\colorMATH}\ensuremath{gs}}} that are clipped under the clipping parameter {{\color{\colorMATH}\ensuremath{C^{t}}}} and uses the count to update the parameter. The inputs to the algorithm are the training data {{\color{\colorMATH}\ensuremath{X}}}, the training labels {{\color{\colorMATH}\ensuremath{y}}}, the number of iterations {{\color{\colorMATH}\ensuremath{k}}}, and the target percentage of gradients remaining un-clipped {{\color{\colorMATH}\ensuremath{\gamma }}}.
{{\color{\colorMATH}\ensuremath{\epsilon }}}, {{\color{\colorMATH}\ensuremath{\delta }}}, and {{\color{\colorMATH}\ensuremath{\delta ^{\prime}}}} are the privacy cost parameters.
\toplas{We omit the sensitivity annotations on top-level function inputs for readability.}
The \system typechecker typechecks {{\color{\colorSYNTAX}\texttt{DPAL}}} in 5.6ms (averaged over 10 runs).

\toplass{
  \paragraph{Beyond \duet.}
  In this case study, we implement a library function for the
  differentially private average ({{\color{\colorMATH}\ensuremath{{\begingroup\renewcommand\colorMATH{\colorMATHC}\renewcommand\colorSYNTAX{\colorSYNTAXC}{{\color{\colorMATH}\ensuremath{{{\color{\colorSYNTAX}\texttt{DPMean}}}}}}\endgroup }}}}) and use it in two
  places. This refactoring is not possible in \duet, because \duet's
  privacy functions require all sensitive arguments to have a
  sensitivity of 1. In this algorithm, one of the uses of
  {{\color{\colorMATH}\ensuremath{{\begingroup\renewcommand\colorMATH{\colorMATHC}\renewcommand\colorSYNTAX{\colorSYNTAXC}{{\color{\colorMATH}\ensuremath{{{\color{\colorSYNTAX}\texttt{DPMean}}}}}}\endgroup }}}} in fact has \emph{data-dependent} sensitivity (the
  sensitivity of {{\color{\colorMATH}\ensuremath{{\begingroup\renewcommand\colorMATH{\colorMATHB}\renewcommand\colorSYNTAX{\colorSYNTAXB}{{\color{\colorMATH}\ensuremath{{\text{mclip}}}}}\endgroup }\hspace*{0.33em} {\begingroup\renewcommand\colorMATH{\colorMATHA}\renewcommand\colorSYNTAX{\colorSYNTAXA}{{\color{\colorMATH}\ensuremath{C^{t}}}}\endgroup }\hspace*{0.33em}{\begingroup\renewcommand\colorMATH{\colorMATHA}\renewcommand\colorSYNTAX{\colorSYNTAXA}{{\color{\colorMATH}\ensuremath{gs}}}\endgroup }}}} is {{\color{\colorMATH}\ensuremath{{\begingroup\renewcommand\colorMATH{\colorMATHA}\renewcommand\colorSYNTAX{\colorSYNTAXA}{{\color{\colorMATH}\ensuremath{C^{t}}}}\endgroup }}}}).

  Practical implementations of algorithms like this one often rely on
  libraries of differentially private functions like {{\color{\colorMATH}\ensuremath{{\begingroup\renewcommand\colorMATH{\colorMATHC}\renewcommand\colorSYNTAX{\colorSYNTAXC}{{\color{\colorMATH}\ensuremath{{{\color{\colorSYNTAX}\texttt{DPMean}}}}}}\endgroup }}}}
  (e.g. the Opacus library for differentially private deep
  learning~\cite{yousefpour2021opacus}, or the OpenDP library for
  differentially private analytics~\cite{gaboardi2020programming}). By
  lifting the limitations of \duet's privacy functions, \system makes
  it possible to implement and use such libraries.}

\paragraph{Primitives used.}
This algorithm demonstrates the use of privacy combinators (e.g. {\begingroup\renewcommand\colorMATH{\colorMATHC}\renewcommand\colorSYNTAX{\colorSYNTAXC}{{\color{\colorMATH}\ensuremath{ {{\color{\colorSYNTAX}\texttt{aloop}}} }}}\endgroup }) that can be specified as primitives in \system but require special typing rules in \duet:

\begingroup\color{\colorMATH}\begin{gather*}
\begin{array}{l
    } {\begingroup\renewcommand\colorMATH{\colorMATHC}\renewcommand\colorSYNTAX{\colorSYNTAXC}{{\color{\colorMATH}\ensuremath{{{\color{\colorSYNTAX}\texttt{aloop}}}}}}\endgroup } \mathrel{:}  \forall \hspace*{0.33em} \epsilon :{{\color{\colorSYNTAX}\texttt{{\ensuremath{{\mathbb{R}}^{+}}}}}}, \delta :{{\color{\colorSYNTAX}\texttt{{\ensuremath{{\mathbb{R}}^{+}}}}}}, \delta ^{\prime}:{{\color{\colorSYNTAX}\texttt{{\ensuremath{{\mathbb{R}}^{+}}}}}},s:{{\color{\colorSYNTAX}\texttt{{\ensuremath{{\mathbb{R}}^{+}}}}}},\Gamma \mathrel{:}{{\color{\colorSYNTAX}\texttt{cxt}}}.
    \overbracket{(\delta ^{\prime}:{{\color{\colorSYNTAX}\texttt{{\ensuremath{{\mathbb{R}}^{+}}}}}}[\delta ^{\prime}])}^{\clap{\smaller{{\color{\colorTEXT}\textnormal{{{\color{\colorMATH}\ensuremath{\delta }}} relaxation}}}}}
    \mathrel{{\begingroup\renewcommand\colorMATH{\colorMATHB}\renewcommand\colorSYNTAX{\colorSYNTAXB}{{\color{\colorMATH}\ensuremath{\rightarrow }}}\endgroup }} \overbracket{(k \mathrel{:}{{\color{\colorSYNTAX}\texttt{{\ensuremath{{\mathbb{N}}}}}}})}^{\clap{\smaller{{\color{\colorTEXT}\textnormal{iterations}}}}} 
    \mathrel{{\begingroup\renewcommand\colorMATH{\colorMATHB}\renewcommand\colorSYNTAX{\colorSYNTAXB}{{\color{\colorMATH}\ensuremath{\rightarrow }}}\endgroup }} \overbracket{(init\mathrel{:} \tau )}^{\clap{\smaller{{\color{\colorTEXT}\textnormal{initial value}}}}} 
    %\cr  \hspace*{1.00em}\hspace*{1.00em}\hspace*{1.00em} \rightarrow  (\Gamma \mathrel{:}{{\color{\colorSYNTAX}\texttt{cxt}}})S
    \cr
\end{array}
\end{gather*}\endgroup
\toplas{This type for {{\color{\colorMATH}\ensuremath{{\begingroup\renewcommand\colorMATH{\colorMATHC}\renewcommand\colorSYNTAX{\colorSYNTAXC}{{\color{\colorSYNTAX}\texttt{aloop}}}\endgroup }}}} encodes the advanced composition
  theorem (introduced in Section~\ref{sec:background}). Our advanced
  composition combinator runs an {{\color{\colorMATH}\ensuremath{(\epsilon , \delta )}}}-differentially private
  function ({{\color{\colorMATH}\ensuremath{{\mathcal{M}}}}}) representing the body of the loop {{\color{\colorMATH}\ensuremath{k}}} times, for a
  total privacy cost of {{\color{\colorMATH}\ensuremath{\Big( {\begingroup\renewcommand\colorMATH{\colorMATHC}\renewcommand\colorSYNTAX{\colorSYNTAXC}{{\color{\colorMATH}\ensuremath{2\epsilon (\sqrt {2k(\log  (1/\delta ^{\prime}))}), k\delta  + \delta ^{\prime} }}}\endgroup }
  \Big)}}} (a significant improvement over the sequential composition
  cost of {{\color{\colorMATH}\ensuremath{({\begingroup\renewcommand\colorMATH{\colorMATHC}\renewcommand\colorSYNTAX{\colorSYNTAXC}{{\color{\colorMATH}\ensuremath{k\epsilon , k\delta  }}}\endgroup })}}}).}

Our version of the adaptive clipping gradient descent algorithm uses
advanced composition and {{\color{\colorMATH}\ensuremath{(\epsilon , \delta )}}}-differential privacy, to demonstrate
the encoding of {{\color{\colorMATH}\ensuremath{{\begingroup\renewcommand\colorMATH{\colorMATHC}\renewcommand\colorSYNTAX{\colorSYNTAXC}{{\color{\colorSYNTAX}\texttt{aloop}}}\endgroup }}}} as a regular function in \system. Our
source code repository contains an alternative implementation that
uses zero-concentrated differential privacy for improved composition,
and converts the privacy guarantee to {{\color{\colorMATH}\ensuremath{(\epsilon , \delta )}}}-differential privacy at
the end of the algorithm.

\section{Related Work} % {-{
\label{sec:relatedwork}

\subsubsection*{Verification techniques based on type systems.}
There are two threads of prior work in type-system-based verification of
differential privacy for high-level programs: those based in linear types, and
those based on relational refinement types. Reed and
Pierce~\cite{reed2010distance} proposed \fuzz, the first type system for
differential privacy based on linear typing; its fundamental components are a
linear type system with an indexed ``scaling'' modality {{\color{\colorMATH}\ensuremath{!_{s}}}} for tracking the
sensitivity of programs and a monadic connective {{\color{\colorMATH}\ensuremath{{\scriptstyle \bigcirc }}}} to model randomized
computations. An {{\color{\colorMATH}\ensuremath{s}}}-sensitive function is encoded in \fuzz as a linear
function with scaled domain {{\color{\colorMATH}\ensuremath{ !_{s}A \multimap  B }}} and often notated {{\color{\colorMATH}\ensuremath{ A \multimap _{s} B }}}. An
{{\color{\colorMATH}\ensuremath{\epsilon }}}-differential privacy mechanism is represented as an {{\color{\colorMATH}\ensuremath{\epsilon }}}-sensitive function
with monadic return type as in {{\color{\colorMATH}\ensuremath{ A \multimap _{\epsilon } {\scriptstyle \bigcirc }B }}}. \dfuzz~\cite{gaboardi2013linear}
extends \fuzz with dependent types to encode sensitivity and privacy bounds
that depend on the values of function arguments. This allows e.g.~reasoning
about the privacy of iterative algorithms whose privacy cost depend on the
number of iterations. \fuzz and \dfuzz~ can be characterized by strong support
for higher-order programming and potential for automation via type inference.
They support pure differential privacy but approximate and any other recent
variants of differential privacy fall out of their scope due to nonlinear
scaling. Several recently-proposed approaches allow a \fuzz-like analysis for
({{\color{\colorMATH}\ensuremath{\epsilon ,\delta }}})-differentially private programs: Azevedo de Amorim et
al.~\cite{de2019probabilistic} leverage a \emph{path construction} and a
\fuzz-like type system. \fuzzi~\cite{zhang2019fuzzi} integrates a \fuzz-like
type system with an expressive program logic.
\toplas{\fuzzi directly connects (automated) type-based proofs of
  composition for sensitivity and privacy properties with (manual)
  \apRHL proofs for basic constructs like sequential composition and
  the Laplace mechanism. In our approach, on the other hand,
  properties of basic mechanisms must be axiomatized. The \fuzzi
  system targets imperative programs, and does not provide support for
  higher-order programming with privacy functions.}
%
%TODO: \fo{@David, anything else to comment about \cite{de2019probabilistic} and \fuzzi~\cite{zhang2019fuzzi}?}
Finally  \duet~\cite{near2019duet}
%, the predecesor of \system,
proposes a two-language design with linear types for tracking sensitivity and privacy; crucial restrictions in the typing rules of the privacy language allow encoding advanced
differential privacy variants such as approximate, R\'enyi, zero-concentrated and
truncated-concentrated differential privacy. In \duet privacy functions are {{\color{\colorMATH}\ensuremath{n}}}-ary and written
{{\color{\colorSYNTAX}\texttt{{\ensuremath{({{\color{\colorMATH}\ensuremath{\tau _{1}}}}@{\begingroup\renewcommand\colorMATH{\colorMATHC}\renewcommand\colorSYNTAX{\colorSYNTAXC}{{\color{\colorMATH}\ensuremath{p_{1}}}}\endgroup },\ldots ,{{\color{\colorMATH}\ensuremath{\tau _{n}}}}@{\begingroup\renewcommand\colorMATH{\colorMATHC}\renewcommand\colorSYNTAX{\colorSYNTAXC}{{\color{\colorMATH}\ensuremath{p_{n}}}}\endgroup }) \multimap ^{*} {{\color{\colorMATH}\ensuremath{\tau }}}}}}}} for privacy quantities {\begingroup\renewcommand\colorMATH{\colorMATHC}\renewcommand\colorSYNTAX{\colorSYNTAXC}{{\color{\colorMATH}\ensuremath{p_{i}}}}\endgroup } such as
{{\color{\colorMATH}\ensuremath{(\epsilon _{i},\delta _{i})}}} in the case of approximate differential privacy. % \duet also has built-in type rules for privacy combinators such as advanced composition and random s\cite{de2019probabilistic} and \fuzzi~\cite{zhang2019fuzzi}ubsampling with privacy amplification.

 Unlike previous works based on linear type systems, \hoaresq~\cite{Barthe:POPL:15} uses relational refinement types to encode arbitrary relational properties of programs, including differential privacy. In \hoaresq, an
{{\color{\colorMATH}\ensuremath{{\begingroup\renewcommand\colorMATH{\colorMATHB}\renewcommand\colorSYNTAX{\colorSYNTAXB}{{\color{\colorMATH}\ensuremath{\sss}}}\endgroup }}}}-sensitive function type is written {{\color{\colorMATH}\ensuremath{\Pi s^{\prime}.\hspace*{0.33em} \{ x \mathrel{:: } \tau _{1} \mathrel{|}
{\mathfrak{D}} _{\tau _{1}}(x_{\vartriangleleft },x_{\vartriangleright }) \leq  s^{\prime}\}  \rightarrow  \{ y \mathrel{:: } \tau _{2} \mathrel{|} {\mathcal{D}}_{\tau _{2}}(y_{\vartriangleleft },y_{\vartriangleright }) \leq  {\begingroup\renewcommand\colorMATH{\colorMATHB}\renewcommand\colorSYNTAX{\colorSYNTAXB}{{\color{\colorMATH}\ensuremath{\sss}}}\endgroup } s^{\prime}\} }}} where {{\color{\colorMATH}\ensuremath{{\mathfrak{D}} _{\tau }}}} is a
type-indexed distance metric, and {{\color{\colorMATH}\ensuremath{x_{\vartriangleleft }}}} and {{\color{\colorMATH}\ensuremath{x_{\vartriangleright }}}} are explicit symbolic
representation of the ``left'' and ``right'' execution of the program in support of
encoding relational properties. To account for probabilistic private computations, \hoaresq uses an indexed monad {{\color{\colorMATH}\ensuremath{{\mathfrak{M}} _{\epsilon ,\delta }[\tau ]}}}:
% which denotes a probabilistic
% computation that satisfies {{\color{\colorMATH}\ensuremath{(\epsilon ,\delta )}}}-differential privacy for the output up-to
% the relational invariants encoded in {{\color{\colorMATH}\ensuremath{\tau }}}, which must be instantiated to
% equality in the case of differential privacy. A differentially private function
% type doesn't quantify over distances between related input because the
% definition of differential privacy is fixed to {{\color{\colorMATH}\ensuremath{1}}} in the assumption of
% distance for related inputs. Putting these together,
the type of an {{\color{\colorMATH}\ensuremath{(\epsilon ,\delta )}}}-differentially private function is
written {{\color{\colorMATH}\ensuremath{\{ x \mathrel{:: } \tau _{1} \mathrel{|} {\mathfrak{D}} _{\tau _{1}}(x_{\vartriangleleft },x_{\vartriangleright }) \leq  1\}  \rightarrow  {\mathfrak{M}} _{\epsilon ,\delta }[\{ y \mathrel{:: } \tau _{2} \mathrel{|} y_{\vartriangleleft }
= y_{\vartriangleright }\} ]}}}. A limitation of this encoding is that a function of two arguments
which provides different privacy bounds for each argument \toplas{(as described in Section~\ref{sec:privacy_jazz_intro})}
will report a summed,
global privacy bound, because the tracking of privacy occurs in a single global
index to the privacy monad {{\color{\colorMATH}\ensuremath{{\mathfrak{M}} _{\epsilon ,\delta }}}}. \toplas{Because privacy is proved  as a relational property, rather than as a sensitivity/Lipschitz continuity property, \hoaresq is also capable of placing relational distance bounds on arguments to functions. Regarding the implementation, both \hoaresq and \system use dependent types to capture sizes.} Regarding typechecking, even though some automation has been achieved~\cite{DBLP:conf/pldi/CicekQBG019}, the automation relies heavily on what is
achievable with SMT solvers, and has limited application to programs which make
generous use of compositional or higher-order programming techniques, or
metric-distance relationships between values at non-base types.
\toplas{SMT solvers are not complete for non-linear operations, which are common in complex differential privacy mechanisms. Consider, for example, the expression for privacy cost under advanced composition: {{\color{\colorMATH}\ensuremath{\epsilon ^{\prime} = k\epsilon (e^\epsilon -1) + \epsilon \sqrt {2 k \ln (1/\delta ^{\prime})}}}}; SMT solvers are not capable of proving universally quantified qualities between equations like these, which limits their ability to automate reasoning about privacy cost.}

\toplas{It is important to note that all of these type systems support some form of recursion. Adding any form of recursion in the formalism of \system would make the technical development even more complicated. We just focused on a small core that could illustrate the main novelties of the latent or contextual approach.}

To conclude the overview about type-system-based verification techniques, we refer the reader to Table~\ref{tbl:typing-comparison}, comparing different aspects of the reviewed type systems.

\begin{table} % {-{ tbl:typing-comparison
  \begin{framed}
  \hspace*{-11pt}
  \centering
\begin{tabular}{r l l l l}
      \smaller{\textbf{System}} &
      \smaller{\textbf{Sens. Function}} &
      \smaller{\textbf{Sens. Typing}} &
      \smaller{\textbf{Priv. Function}} &
      \smaller{\textbf{Priv. Typing}} \\
    \hline
      {\textit{(D)}}\fuzz
      & {{\color{\colorMATH}\ensuremath{\tau _{1}\multimap _{s} \tau _{2}}}}
      & {{\color{\colorMATH}\ensuremath{\Gamma  \vdash  e \mathrel{:} \tau }}}
      & {{\color{\colorMATH}\ensuremath{\tau _{1}\multimap _{\epsilon } {\scriptstyle \bigcirc }\tau _{2}}}}
      & {{\color{\colorMATH}\ensuremath{\Gamma  \vdash  e \mathrel{:} {\scriptstyle \bigcirc }\tau }}}
      \\
    \hoaresq
      & {\smaller\smaller {{\color{\colorMATH}\ensuremath{\begin{array}{rcll
                           } &{}\Pi s^{\prime}.{}& \{ x \mathrel{:: } \tau _{1} \mathrel{|} {\mathfrak{D}} _{\tau _{1}}(x_{\vartriangleleft },x_{\vartriangleright }) \leq  s^{\prime} \} 
                           \cr     &{}\rightarrow {}& \{ y \mathrel{:: } \tau _{2} \mathrel{|} {\mathfrak{D}} _{\tau _{2}}(y_{\vartriangleleft },y_{\vartriangleright }) \leq  {\begingroup\renewcommand\colorMATH{\colorMATHB}\renewcommand\colorSYNTAX{\colorSYNTAXB}{{\color{\colorMATH}\ensuremath{\sss}}}\endgroup } s^{\prime}\} 
                           \end{array}}}}}
      & {{\color{\colorMATH}\ensuremath{{\mathfrak{G}}  \vdash  e \mathrel{:: } \tau }}}
      & {\smaller\smaller {{\color{\colorMATH}\ensuremath{\begin{array}{l
                           } \{ x \mathrel{:: } \tau _{1} \mathrel{|} {\mathfrak{D}} _{\tau _{1}}(x_{\vartriangleleft },x_{\vartriangleright }) \leq  1 \} 
                           \cr  \rightarrow  {\mathfrak{M}} _{\epsilon ,\delta }[\{ y \mathrel{:: } \tau _{2} \mathrel{|} y_{\vartriangleleft } = y_{\vartriangleright } \} ]
                           \end{array}}}}}
      & {{\color{\colorMATH}\ensuremath{{\mathfrak{G}}  \vdash  e \mathrel{:: } {\mathfrak{M}} _{\epsilon ,\delta }[\tau ]}}}
      \\
\end{tabular}
  \end{framed}
  \caption{How each system ---{\textit{(D)}}\fuzz,
  \hoaresq, \duet and \system (this paper)---(1)
  encodes function sensitivity in types, (2) structures typing judgments for
  function sensitivity, (3) encodes differential privacy in types, and (4)
  structures typing judgments for differential privacy.}
  \label{tbl:typing-comparison}
  \vspace{-2em}
\end{table}

\subsubsection*{Techniques based on couplings and program logics.}
Approximate couplings~\cite{DBLP:journals/lmcs/BartheEHSS19} are a probabilistic abstraction that witnesses differential privacy properties of programs and have been successfully exploited for verification purposes. The relational Hoare logic \apRHL~\cite{Barthe:POPL12} and its successors \apRHLplus~\cite{Barthe:LICS16} and \spanapRHL~\cite{Sato:LICS19} internalize the compositional construction of such couplings and capture from pure and approximate differential \toplas{privacy} to more recent variants such as R\'enyi, zero-concentrated and
truncated-concentrated differential privacy. While compared to other methods these program logics are rather expressive going beyond the composition of (a set of predefined) basic mechanisms, derivations in the logics involve complex quantitative reasoning, not always amenable to automation. Even though there has been a successful report on partial automation~\cite{DBLP:conf/csfw/BartheDGKB13}, e.g.~the synthesis of (quantitative relational) loop invariants for iterative algorithms remains challenging. To synthesise couplings, Albarghouthi and Hsu~\cite{DBLP:journals/pacmpl/AlbarghouthiH18} use an alternative approach based on constraint solving which is highly amenable to automation; the approach is however confined to {{\color{\colorMATH}\ensuremath{\epsilon }}}-DP. Finally, to verify programs that achieve {{\color{\colorMATH}\ensuremath{(\epsilon ,\delta )}}}-DP composing basic mechanisms, \citet{Barthe:CSF14} use a customised program product construction and traditional (non-relational and non-probabilistic) Hoare logic augmented with mechanism-specific rules. \toplas{In recent work, \citet{barthe2020deciding} show that checking differential privacy for imperative programs is undecidable in general, but present a reduction to a decidable fragment of first-order logic for a restricted class of programs. \citet{barthe2021deciding} also shows that checking accuracy bounds for differentially private programs is also undecidable (in general).} A common limitation of all these approaches is that they are restricted to first order imperative programs.

\subsubsection*{Techniques based on randomness alignment.}

LightDP~\cite{zhang2017lightdp} and ShadowDP~\cite{wang2019proving} take a third approach to verifying differential privacy based on \emph{randomness alignments}. A randomness alignment is an injective function relating the randomness from one execution of a differentially private mechanism to a second execution of the same mechanism (i.e. {{\color{\colorMATH}\ensuremath{{\mathcal{M}}(x)}}} outputs the same result with noise {{\color{\colorMATH}\ensuremath{H}}} as {{\color{\colorMATH}\ensuremath{{\mathcal{M}}(x^{\prime})}}} outputs with noise {{\color{\colorMATH}\ensuremath{f(H)}}}, where {{\color{\colorMATH}\ensuremath{f}}} is the randomness alignment). Both LightDP and ShadowDP are capable of verifying complex low-level mechanisms like the sparse vector technique in just a few seconds. However, both tools target a first-order imperative programming language, and have limited support for higher-order programming.

\subsubsection*{Techniques based on testing.}

Since differential privacy mechanisms are randomized, traditional methods of software testing do not apply. Two recent works by \citet{bichsel2018dp} and \citet{ding2018detecting} address this challenge by automatically generating neighboring inputs for the mechanism being tested, and sampling from their outputs many times to approximate their output distributions. For privacy mechanisms with major bugs, these tools are able to show that the approximated distributions do not satisfy the claimed differential privacy guarantee. \citet{wilson2020differentially} have implemented a testing tool based on this approach in their open-source library. \toplas{DPCheck~\cite{zhang2020testing} combines static analysis with instrumented concrete execution of the target program to detect bugs in even more complex algorithms.}

%TODO: \fo{Shall we mention the work of McIver and Morgan, reducing the problem of DP verification to a quantitative information flow problem.}

\toplas{
\subsubsection*{Type systems and contextual information.} The technical device of contextual latent effects used in Sax and Jazz is related to prior approaches to expose information about captured variables in function types. \citet{leroy:rr1992} and \citet{hannanAl:rr1997} use function types augmented with the set of captured variables, the former for tracking dangerous type variables for polymorphic generalization, and the latter for lifetime analysis.
\citet{schererHoffmann:lpar2013} introduce {\em open closure types} in order to track additional information about closed-over variables in first-class functions. An open closure type augments the traditional arrow type with a lexical environment of closed over variables, further decorated by a mapping characterizing the use of each variable. For instance, they formalize a system where the mapping marks each variable with a Boolean indicating whether the evaluation of the body of the function depends on the variable or not. An open closure type also includes the name of the function argument similarly decorated. For the considered system, they prove a non-interference property, which states that closing an open term with two valuations that coincide on used variables yield the same result, up to used variables.
Function types in Sax and Jazz can be seen as specific cases of open closure types, in which the information attached to captured variables (and arguments) is not a Boolean value, but sensitivity and privacy information. Consequently, the type soundness result we establish is more general than noninterference; for instance, in Sax, for two valuations that are at a given distance apart, the distance between the results are bounded by the sensitivities of each variables.
The technical development of contextual linear types shares many concerns with that of open closure types, notably regarding the proper handling of the typing environment---in which order matters---and of scoping. However, our soundness results rely on logical relations, while they adopt a more restricted technique, sufficient for the purpose of the simple type system tracking Boolean information.

As observed by Scherer and Hoffmann, using open closure types allows delaying the accounting of information flow into closures from abstraction time to application time. Likewise, contextual sensitivity in Sax delays sensitivity accounting to application time. We extend this principle to positive type constructors such as products and sums, generally deferring accounting to elimination forms; we believe this would likewise apply to the simpler information-flow control setting they study.

Recently, \citet{baoAl:oopsla21} use a similar context-annotation technique to track reachability information in types. For instance, they augment the type of reference cell with the variables in scope that alias the cell. This information is tracked on function types as well, in order to track the reachability information from closures.
In all these approaches, type information depends on variable names; this is quite different from dependent types, however, where types depend on arbitrary terms. The telescope nature of the typing environment is similar, but many of the deep challenges of dependent types do not manifest in this restricted setting.

Hoare Type Theory~\cite{nanevski:jfp2008} supports specifying effectful, heap-manipulating computations by introducing Hoare-style pre/postconditions in types. Computation types include contexts of variables and heap locations in order to track footprints of logical assertions. In a similar vein as \citet{leroy:rr1992}, the context of variables is not decorated with any information, so it is unclear whether one could handle contextual sensitivity and privacy in this approach.

Finally, it would be interesting to study if generic approaches such as {\em coeffects}~\cite{petricek:icfp14} and {\em graded modal types}~\cite{orchard:pacmpl2019} can express delayed sensitivity and privacy tracking as developed here. These generic approaches use resource algebras (such as the semiring of natural numbers) for capturing modalities in types. To the best of our knowledge, these are not contextual: for instance, the arrow type is annotated with a scalar, label, etc., not with a decorated environment as in open closure types and this work.
Extending these generic approaches to support contextual information on all type constructors could bring the benefits of lazy accounting to a wide range of type-based quantitative program reasoning.
}

\section{Conclusion}
\label{sec:conclusion}

We have presented \system, a language and type system for
differentially private programming with strong support for both
higher-order programming and advanced variants of differential
privacy. The key insight of our approach is latent contextual
tracking of both privacy and sensitivity, which enables sum, product and function types
to describe their privacy effects---even for closure variables---\toplass{making it possible to delay the payment of the effects until actual elimination, sometimes yielding advantages on the precision of the analysis and the annotation burden.}

We have formalized a core subset of \system and proved its soundness
using a step-indexed logical relation, following a novel strategy.
Case studies demonstrate the ability to encode basic privacy
mechanisms and privacy combinators as primitives in \system, and to
compose them to develop more complicated iterative differentially
private algorithms.

\system extends the expressive power of systems like
\fuzz~\cite{reed2010distance} to advanced variants of differential
privacy. Like \fuzz, it remains incapable of proving the correctness
of basic privacy mechanisms like the Laplace mechanism. One
interesting avenue for future work lies in combining \system with an
expressive program logic like
\apRHL~\cite{Barthe:POPL12,Barthe:TOPLAS:13} in the style of
\fuzzi~\cite{zhang2019fuzzi}. Such a combined system would provide a
complete programming framework supporting higher-order programming and
end-to-end privacy proofs.
\toplas{Another line of future work is to incorporate some form of recursion to \system such as recursive types.}

\bibliographystyle{../bst/ACM-Reference-Format}
\bibliography{../bib/acmart.bib}

\pagebreak

\clearpage
%\addtocontents{toc}{\protect\setcounter{tocdepth}{2}}
\appendix
\section*{Appendix}
%\setcounter{secnumdepth}{3}
%\setcounter{tocdepth}{2}
%\tableofcontents

Throughout the appendix we use symbol {{\color{\colorMATH}\ensuremath{{\begingroup\renewcommand\colorMATH{\colorMATHB}\renewcommand\colorSYNTAX{\colorSYNTAXB}{{\color{\colorMATH}\ensuremath{\sss}}}\endgroup }}}} instead of {{\color{\colorMATH}\ensuremath{{\begingroup\renewcommand\colorMATH{\colorMATHB}\renewcommand\colorSYNTAX{\colorSYNTAXB}{{\color{\colorMATH}\ensuremath{\distance}}}\endgroup }}}}, and {{\color{\colorMATH}\ensuremath{{\begingroup\renewcommand\colorMATH{\colorMATHB}\renewcommand\colorSYNTAX{\colorSYNTAXB}{{\color{\colorMATH}\ensuremath{\sS}}}\endgroup }}}} instead of {{\color{\colorMATH}\ensuremath{{\begingroup\renewcommand\colorMATH{\colorMATHB}\renewcommand\colorSYNTAX{\colorSYNTAXB}{{\color{\colorMATH}\ensuremath{\Distance}}}\endgroup }}}} as they are interchangeable.
\section{$\lang$: Static semantics}
\label{asec:static-semantics}
In this section we present some definitions of the static semantics of $\lang$ not presented in the main document.
Figure~\ref{afig:syntax-full} presents the syntax of $\lang$. 
Figures~\ref{afig:sensitivity-type-system-1} and~\ref{afig:sensitivity-type-system-1} present the complete sensitivity type system of $\lang$. 
Figure~\ref{afig:privacy-type-system} presents the complete type system of $\lang$.
They include the usage of {{\color{\colorMATH}\ensuremath{\Phi }}} and the type rules for the derived expressions: boolean, conditional and let expressions.
Figure~\ref{afig:sensitivity-instantiation} presents the sensitivity instantiation or dot product operator and the sensitivity type instantiation operator.
Figure~\ref{afig:subtyping} presents the subtyping rules for $\lang$.
Figure~\ref{afig:sensitivity-environment-substitution} presents the sensitivity environment substitution environment.
Figure~\ref{afig:lift-operators}, presents the different lifts operators.
Finally, Figure~\ref{afig:types-join-meet} presents the join and meet operator between types.

\begin{figure}[t]
\begin{framed}
 \begingroup\color{\colorMATH}\begin{gather*}
\end{gather*}\endgroup
\end{framed}
\caption{$\lang$: Syntax}
\label{afig:syntax-full} 
\end{figure}

\begin{figure}[t]
\begin{framed}
 \input{sensitivity-type-system-phi1}
\end{framed}
\caption{$\lang$: Complete sensitivity type system (part 1)}
\label{afig:sensitivity-type-system-1} 
\end{figure}

\begin{figure}[t]
\begin{framed}
 \input{sensitivity-type-system-phi2}
\end{framed}
\caption{$\lang$: Complete sensitivity type system (part 2)}
\label{afig:sensitivity-type-system-2} 
\end{figure}

\begin{figure}[t]
\begin{framed}
 \input{privacy-type-system-appendix}
\end{framed}
\caption{$\lang$: Complete privacy type system}
\label{afig:privacy-type-system} 
\end{figure}
\begin{figure}[t]
\begin{framed}
	\begingroup\color{\colorMATH}\begin{gather*}
\begin{tabularx}{\linewidth}{>{\centering\arraybackslash\(}X<{\)}}
	  %\cr  \_\mathord{\cdotp }\_: {\text{sensv}} \times  ({\text{sensv}} + {\text{sens}}) \rightarrow  {\text{sensv}} + {\text{sens}}
	  \cr
\end{tabularx}
	\end{gather*}\endgroup 

	\begingroup\color{\colorMATH}\begin{gather*}
	  \begin{tabularx}{\linewidth}{>{\centering\arraybackslash\(}X<{\)}}
	  %\cr  \_(\_): {\text{sensv}} \times  {\text{type^{*}}} \rightarrow  {\text{type^{*}}}
	  \cr
\end{tabularx}
	\end{gather*}\endgroup 
\end{framed}
\caption{Sensitivity instantiation / ``dot product'' and sensitivity type instantiation} 
\label{afig:sensitivity-instantiation}
\end{figure}

\begin{figure}[t]
\begin{framed}
 \begingroup\color{\colorMATH}\begin{gather*}
	\begin{tabularx}{\linewidth}{>{\centering\arraybackslash\(}X<{\)}}\hfill\hspace{0pt}\begingroup\color{\colorTEXT}\boxed{\begingroup\color{\colorMATH} {\begingroup\renewcommand\colorMATH{\colorMATHB}\renewcommand\colorSYNTAX{\colorSYNTAXB}{{\color{\colorMATH}\ensuremath{\sS}}}\endgroup } <: {\begingroup\renewcommand\colorMATH{\colorMATHB}\renewcommand\colorSYNTAX{\colorSYNTAXB}{{\color{\colorMATH}\ensuremath{\sS}}}\endgroup } \endgroup}\endgroup
	\cr \hfill\hspace{0pt}{\begingroup\renewcommand\colorMATH{\colorMATHB}\renewcommand\colorSYNTAX{\colorSYNTAXB}{{\color{\colorMATH}\ensuremath{\sS}}}\endgroup } <: {\begingroup\renewcommand\colorMATH{\colorMATHB}\renewcommand\colorSYNTAX{\colorSYNTAXB}{{\color{\colorMATH}\ensuremath{\sS^{\prime}}}}\endgroup } \overset \vartriangle \iff     \forall x.\hspace*{0.33em} {\begingroup\renewcommand\colorMATH{\colorMATHB}\renewcommand\colorSYNTAX{\colorSYNTAXB}{{\color{\colorMATH}\ensuremath{\sS}}}\endgroup }(x) \leq  {\begingroup\renewcommand\colorMATH{\colorMATHB}\renewcommand\colorSYNTAX{\colorSYNTAXB}{{\color{\colorMATH}\ensuremath{\sS^{\prime}}}}\endgroup }(x)
	  \hfill\hspace{0pt}
	\cr \hfill\hspace{0pt}\begingroup\color{\colorTEXT}\boxed{\begingroup\color{\colorMATH} {\begingroup\renewcommand\colorMATH{\colorMATHC}\renewcommand\colorSYNTAX{\colorSYNTAXC}{{\color{\colorMATH}\ensuremath{\pS}}}\endgroup } <: {\begingroup\renewcommand\colorMATH{\colorMATHC}\renewcommand\colorSYNTAX{\colorSYNTAXC}{{\color{\colorMATH}\ensuremath{\pS}}}\endgroup } \endgroup}\endgroup
	\cr
\end{tabularx}
	\end{gather*}\endgroup
\end{framed}
\caption{Subtyping}
\label{afig:subtyping} 
\end{figure}

\begin{figure}[t]
\begin{framed}
	\begingroup\color{\colorMATH}\begin{gather*}
	  \begin{tabularx}{\linewidth}{>{\centering\arraybackslash\(}X<{\)}}
	  \cr  [\_/x]\_: {\text{sensv}} \times  {\text{type}} \rightarrow  {\text{type}}
	  \cr
\end{tabularx}
	\end{gather*}\endgroup
\end{framed}
\caption{Sensitivity environment substitution}
\label{afig:sensitivity-environment-substitution} 
\end{figure}

\begin{figure}[t]
\begin{framed}
	\begingroup\color{\colorMATH}\begin{gather*}
	  \begin{tabularx}{\linewidth}{>{\centering\arraybackslash\(}X<{\)}}
	  \cr
\end{tabularx}
	\end{gather*}\endgroup
\end{framed}
\caption{Lift operators}
\label{afig:lift-operators} 
\end{figure}

\begin{figure}[t]
\begin{framed}
	 \begingroup\color{\colorMATH}\begin{gather*}
	  \begin{tabularx}{\linewidth}{>{\centering\arraybackslash\(}X<{\)}}
	  \cr  \_ \sqcup  \_: {\text{type}} \times  {\text{type}} \rightarrow  {\text{type}}
	  \cr
\end{tabularx}
	\end{gather*}\endgroup
\end{framed}
\caption{Join and Meet of types and sensitivity environments}
\label{afig:types-join-meet} 
\end{figure}

% 1.- {\begingroup\renewcommand\colorMATH{\colorMATHC}\renewcommand\colorSYNTAX{\colorSYNTAXC}{{\color{\colorMATH}\ensuremath{\pS}}}\endgroup } <: {\begingroup\renewcommand\colorMATH{\colorMATHC}\renewcommand\colorSYNTAX{\colorSYNTAXC}{{\color{\colorMATH}\ensuremath{\pS}}}\endgroup } \sqcup  {\begingroup\renewcommand\colorMATH{\colorMATHC}\renewcommand\colorSYNTAX{\colorSYNTAXC}{{\color{\colorMATH}\ensuremath{\pS'}}}\endgroup }
% 2.- {\begingroup\renewcommand\colorMATH{\colorMATHC}\renewcommand\colorSYNTAX{\colorSYNTAXC}{{\color{\colorMATH}\ensuremath{\pS}}}\endgroup } \sqcap  {\begingroup\renewcommand\colorMATH{\colorMATHC}\renewcommand\colorSYNTAX{\colorSYNTAXC}{{\color{\colorMATH}\ensuremath{\pS'}}}\endgroup } <: {\begingroup\renewcommand\colorMATH{\colorMATHC}\renewcommand\colorSYNTAX{\colorSYNTAXC}{{\color{\colorMATH}\ensuremath{\pS}}}\endgroup }

% (3x + 1x\sqcup 1y) \sqcup  2x =? 3x + 2x\sqcup 2y

% (1x + 2y) \sqcup  (3x + 1y) =? 3x + 3x\sqcup 1y + 3x\sqcup 2y + 1y + 1x\sqcup 1y + 1y\sqcup 2x	
% 3x + 2y

% 1x\sqcup y + 3x\sqcup y

% (1x + 2y) \sqcap  (3x + 1y) = 1x + 2y\sqcap 3x + 1y\sqcap 1x + 1y
% 1x + 1y

% (1x + 2y) \sqcap  (3x + 1y) = 1x + 1y

% (1x + 2y + 3x\sqcup y) \sqcap  (3x + 1y+ 2x\sqcup y) = 1x + 2y\sqcap 3x + 1y\sqcap 1x + 1y

% 1(x\sqcup y) + 3y + 1x + 3(x\sqcup y)
% (1(1x\sqcup 1y) + 3(0x\sqcup 1y)) \sqcap  (1(1x\sqcup 0y) + 3(1x\sqcup 1y)) = 1(1x\sqcup 0y) + 0 + 1(1x\sqcup 1y) + 3(0x\sqcup 1y)

% 1 + 5 \sqcap  2 + 7

\section{$\lang$: Dynamic semantics}
\label{asec:dynamic-semantics}
Figure~\ref{afig:sensitivity-dynamic-semantics}, presents the dynamic semantics of the sensitivity language.
Figure~\ref{afig:probabilistic-semantics}, presents the dynamic semantics of the privacy language.

\begin{figure}[t]
\begin{framed}
 
\begingroup\color{\colorMATH}\begin{gather*}

\end{gather*}\endgroup
\end{framed}
\caption{$\lang$: Sensitivity dynamic semantics}
\label{afig:sensitivity-dynamic-semantics} 
\end{figure}

\begin{figure}[t]
\begin{framed}
 \input{probabilistic-semantics}
\end{framed}
\caption{$\lang$: Probabilistic semantics}
\label{afig:probabilistic-semantics} 
\end{figure}

\newtheorem*{case}{Case}
\SetEnumitemKey{ncases}{itemindent=!,before=\let\makelabel\ncasesmakelabel,leftmargin=0cm}
\newcommand*\ncasesmakelabel[1]{Case #1}

\newenvironment{subproof}
  {\def\proofname{Subproof}%
   \def\qedsymbol{$\triangleleft$}%
   \proof}
{\endproof}

\section{$\lang$: Soundness}
\label{asec:soundness-step-indexing}
In this section we present auxiliary definitions used in Section~\ref{sec:soundness}, and the proof of the fundamental property.
Figure~\ref{afig:join-meet-sensitivity-environments-plus-sensitivities} presents the join and meet operators for sensitivity environments and sensitivities {{\color{\colorMATH}\ensuremath{{\begingroup\renewcommand\colorMATH{\colorMATHB}\renewcommand\colorSYNTAX{\colorSYNTAXB}{{\color{\colorMATH}\ensuremath{\sS}}}\endgroup } + {\begingroup\renewcommand\colorMATH{\colorMATHB}\renewcommand\colorSYNTAX{\colorSYNTAXB}{{\color{\colorMATH}\ensuremath{\sss}}}\endgroup }}}}.
Figure~\ref{afig:subtyping-sensible-types} presents the subtyping relation between sensible types.

\begin{figure}[t]
\begin{framed}
	 \begingroup\color{\colorMATH}\begin{gather*}
\begin{tabularx}{\linewidth}{>{\centering\arraybackslash\(}X<{\)}}
	  \cr
\end{tabularx}
	\end{gather*}\endgroup
\end{framed}
\caption{Join and Meet of sensitivity environment and sensitivities}
\label{afig:join-meet-sensitivity-environments-plus-sensitivities} 
\end{figure}

\begin{figure}[t]
\begin{framed}
	
 	\begingroup\color{\colorMATH}\begin{gather*}
	  \begin{tabularx}{\linewidth}{>{\centering\arraybackslash\(}X<{\)}}\hfill\hspace{0pt}\begingroup\color{\colorTEXT}\boxed{\begingroup\color{\colorMATH} \sigma  <: \sigma  \endgroup}\endgroup
	  \cr
\end{tabularx}
	\end{gather*}\endgroup
\end{framed}
\caption{Subtyping of sensible types}
\label{afig:subtyping-sensible-types} 
\end{figure}

\begin{lemma}[Associativity of the instantiation operator]
  \label{lm:associativity-inst}
  {{\color{\colorMATH}\ensuremath{\instE{{\begingroup\renewcommand\colorMATH{\colorMATHB}\renewcommand\colorSYNTAX{\colorSYNTAXB}{{\color{\colorMATH}\ensuremath{\sS}}}\endgroup }}{({\begingroup\renewcommand\colorMATH{\colorMATHB}\renewcommand\colorSYNTAX{\colorSYNTAXB}{{\color{\colorMATH}\ensuremath{\sS_{1}}}}\endgroup } + {\begingroup\renewcommand\colorMATH{\colorMATHB}\renewcommand\colorSYNTAX{\colorSYNTAXB}{{\color{\colorMATH}\ensuremath{\sS_{2}}}}\endgroup })} = \instE{{\begingroup\renewcommand\colorMATH{\colorMATHB}\renewcommand\colorSYNTAX{\colorSYNTAXB}{{\color{\colorMATH}\ensuremath{\sS}}}\endgroup }}{{\begingroup\renewcommand\colorMATH{\colorMATHB}\renewcommand\colorSYNTAX{\colorSYNTAXB}{{\color{\colorMATH}\ensuremath{\sS_{1}}}}\endgroup }} + \instE{{\begingroup\renewcommand\colorMATH{\colorMATHB}\renewcommand\colorSYNTAX{\colorSYNTAXB}{{\color{\colorMATH}\ensuremath{\sS}}}\endgroup }}{{\begingroup\renewcommand\colorMATH{\colorMATHB}\renewcommand\colorSYNTAX{\colorSYNTAXB}{{\color{\colorMATH}\ensuremath{\sS_{2}}}}\endgroup }}}}}
\end{lemma}
\begin{proof}
  By induction on {{\color{\colorMATH}\ensuremath{{\begingroup\renewcommand\colorMATH{\colorMATHB}\renewcommand\colorSYNTAX{\colorSYNTAXB}{{\color{\colorMATH}\ensuremath{\sS_{1}}}}\endgroup }}}}:
  \begin{enumerate}[ncases]\item  {{\color{\colorMATH}\ensuremath{{\begingroup\renewcommand\colorMATH{\colorMATHB}\renewcommand\colorSYNTAX{\colorSYNTAXB}{{\color{\colorMATH}\ensuremath{\sS_{1}}}}\endgroup } = \varnothing }}}
    \begin{subproof} 
      %Trivial as {{\color{\colorMATH}\ensuremath{\varnothing  \mathord{\cdotp } ({\begingroup\renewcommand\colorMATH{\colorMATHB}\renewcommand\colorSYNTAX{\colorSYNTAXB}{{\color{\colorMATH}\ensuremath{\sS_{1}}}}\endgroup } + {\begingroup\renewcommand\colorMATH{\colorMATHB}\renewcommand\colorSYNTAX{\colorSYNTAXB}{{\color{\colorMATH}\ensuremath{\sS_{2}}}}\endgroup }) = {\begingroup\renewcommand\colorMATH{\colorMATHB}\renewcommand\colorSYNTAX{\colorSYNTAXB}{{\color{\colorMATH}\ensuremath{\sS_{1}}}}\endgroup } + {\begingroup\renewcommand\colorMATH{\colorMATHB}\renewcommand\colorSYNTAX{\colorSYNTAXB}{{\color{\colorMATH}\ensuremath{\sS_{2}}}}\endgroup } = \varnothing  \mathord{\cdotp } {\begingroup\renewcommand\colorMATH{\colorMATHB}\renewcommand\colorSYNTAX{\colorSYNTAXB}{{\color{\colorMATH}\ensuremath{\sS_{1}}}}\endgroup } + \varnothing  \mathord{\cdotp } {\begingroup\renewcommand\colorMATH{\colorMATHB}\renewcommand\colorSYNTAX{\colorSYNTAXB}{{\color{\colorMATH}\ensuremath{\sS_{2}}}}\endgroup }}}}
      %Let us consider {{\color{\colorMATH}\ensuremath{{\begingroup\renewcommand\colorMATH{\colorMATHB}\renewcommand\colorSYNTAX{\colorSYNTAXB}{{\color{\colorMATH}\ensuremath{\sS_{1}}}}\endgroup } = \varnothing }}} (the other case is analogous).
      Trivial as {{\color{\colorMATH}\ensuremath{\instE{{\begingroup\renewcommand\colorMATH{\colorMATHB}\renewcommand\colorSYNTAX{\colorSYNTAXB}{{\color{\colorMATH}\ensuremath{\sS}}}\endgroup }}{(\varnothing  + {\begingroup\renewcommand\colorMATH{\colorMATHB}\renewcommand\colorSYNTAX{\colorSYNTAXB}{{\color{\colorMATH}\ensuremath{\sS_{2}}}}\endgroup })} = \instE{{\begingroup\renewcommand\colorMATH{\colorMATHB}\renewcommand\colorSYNTAX{\colorSYNTAXB}{{\color{\colorMATH}\ensuremath{\sS}}}\endgroup }}{{\begingroup\renewcommand\colorMATH{\colorMATHB}\renewcommand\colorSYNTAX{\colorSYNTAXB}{{\color{\colorMATH}\ensuremath{\sS_{2}}}}\endgroup }} = \instE{{\begingroup\renewcommand\colorMATH{\colorMATHB}\renewcommand\colorSYNTAX{\colorSYNTAXB}{{\color{\colorMATH}\ensuremath{\sS}}}\endgroup }}{\varnothing } + \instE{{\begingroup\renewcommand\colorMATH{\colorMATHB}\renewcommand\colorSYNTAX{\colorSYNTAXB}{{\color{\colorMATH}\ensuremath{\sS}}}\endgroup }}{{\begingroup\renewcommand\colorMATH{\colorMATHB}\renewcommand\colorSYNTAX{\colorSYNTAXB}{{\color{\colorMATH}\ensuremath{\sS_{2}}}}\endgroup }}}}}.
    \end{subproof}
  \item  {{\color{\colorMATH}\ensuremath{{\begingroup\renewcommand\colorMATH{\colorMATHB}\renewcommand\colorSYNTAX{\colorSYNTAXB}{{\color{\colorMATH}\ensuremath{\sS_{1}}}}\endgroup } = {\begingroup\renewcommand\colorMATH{\colorMATHB}\renewcommand\colorSYNTAX{\colorSYNTAXB}{{\color{\colorMATH}\ensuremath{\sS'_{1}}}}\endgroup } + s_{1}x}}}
    \begin{subproof} Let us assume {{\color{\colorMATH}\ensuremath{x \notin  {\begingroup\renewcommand\colorMATH{\colorMATHB}\renewcommand\colorSYNTAX{\colorSYNTAXB}{{\color{\colorMATH}\ensuremath{\sS_{2}}}}\endgroup }}}}, and that {{\color{\colorMATH}\ensuremath{r = {\begingroup\renewcommand\colorMATH{\colorMATHB}\renewcommand\colorSYNTAX{\colorSYNTAXB}{{\color{\colorMATH}\ensuremath{\sS}}}\endgroup }(x)}}} if {{\color{\colorMATH}\ensuremath{{\begingroup\renewcommand\colorMATH{\colorMATHB}\renewcommand\colorSYNTAX{\colorSYNTAXB}{{\color{\colorMATH}\ensuremath{\sS}}}\endgroup }(x)}}} is defined, otherwise {{\color{\colorMATH}\ensuremath{r = x}}}:
      \begingroup\color{\colorMATH}\begin{gather*}
\begin{array}{rclr
        } \instE{{\begingroup\renewcommand\colorMATH{\colorMATHB}\renewcommand\colorSYNTAX{\colorSYNTAXB}{{\color{\colorMATH}\ensuremath{\sS}}}\endgroup }}{(({\begingroup\renewcommand\colorMATH{\colorMATHB}\renewcommand\colorSYNTAX{\colorSYNTAXB}{{\color{\colorMATH}\ensuremath{\sS'_{1}}}}\endgroup } + s_{1}x)+{\begingroup\renewcommand\colorMATH{\colorMATHB}\renewcommand\colorSYNTAX{\colorSYNTAXB}{{\color{\colorMATH}\ensuremath{\sS_{2}}}}\endgroup })} &{}={}& \instE{{\begingroup\renewcommand\colorMATH{\colorMATHB}\renewcommand\colorSYNTAX{\colorSYNTAXB}{{\color{\colorMATH}\ensuremath{\sS}}}\endgroup }}{(({\begingroup\renewcommand\colorMATH{\colorMATHB}\renewcommand\colorSYNTAX{\colorSYNTAXB}{{\color{\colorMATH}\ensuremath{\sS'_{1}}}}\endgroup }+{\begingroup\renewcommand\colorMATH{\colorMATHB}\renewcommand\colorSYNTAX{\colorSYNTAXB}{{\color{\colorMATH}\ensuremath{\sS_{2}}}}\endgroup })+s_{1}x)}
        \cr   &{}={}& \instE{{\begingroup\renewcommand\colorMATH{\colorMATHB}\renewcommand\colorSYNTAX{\colorSYNTAXB}{{\color{\colorMATH}\ensuremath{\sS}}}\endgroup }}{({\begingroup\renewcommand\colorMATH{\colorMATHB}\renewcommand\colorSYNTAX{\colorSYNTAXB}{{\color{\colorMATH}\ensuremath{\sS'_{1}}}}\endgroup }+{\begingroup\renewcommand\colorMATH{\colorMATHB}\renewcommand\colorSYNTAX{\colorSYNTAXB}{{\color{\colorMATH}\ensuremath{\sS_{2}}}}\endgroup })} + s_{1}r & {{\color{\colorTEXT}\textnormal{ (by definition)}}}
        \cr
\end{array}
      \end{gather*}\endgroup
      and the result holds. Let us assume now that {{\color{\colorMATH}\ensuremath{{\begingroup\renewcommand\colorMATH{\colorMATHB}\renewcommand\colorSYNTAX{\colorSYNTAXB}{{\color{\colorMATH}\ensuremath{\sS_{2}}}}\endgroup } = {\begingroup\renewcommand\colorMATH{\colorMATHB}\renewcommand\colorSYNTAX{\colorSYNTAXB}{{\color{\colorMATH}\ensuremath{\sS'_{2}}}}\endgroup }+s_{2}x}}}, and that {{\color{\colorMATH}\ensuremath{r = {\begingroup\renewcommand\colorMATH{\colorMATHB}\renewcommand\colorSYNTAX{\colorSYNTAXB}{{\color{\colorMATH}\ensuremath{\sS}}}\endgroup }(x)}}} if {{\color{\colorMATH}\ensuremath{{\begingroup\renewcommand\colorMATH{\colorMATHB}\renewcommand\colorSYNTAX{\colorSYNTAXB}{{\color{\colorMATH}\ensuremath{\sS}}}\endgroup }(x)}}} is defined, otherwise {{\color{\colorMATH}\ensuremath{r = x}}}:
      \begingroup\color{\colorMATH}\begin{gather*}
\begin{array}{rclr
        } \instE{{\begingroup\renewcommand\colorMATH{\colorMATHB}\renewcommand\colorSYNTAX{\colorSYNTAXB}{{\color{\colorMATH}\ensuremath{\sS}}}\endgroup }}{(({\begingroup\renewcommand\colorMATH{\colorMATHB}\renewcommand\colorSYNTAX{\colorSYNTAXB}{{\color{\colorMATH}\ensuremath{\sS'_{1}}}}\endgroup } + s_{1}x)+({\begingroup\renewcommand\colorMATH{\colorMATHB}\renewcommand\colorSYNTAX{\colorSYNTAXB}{{\color{\colorMATH}\ensuremath{\sS'_{2}}}}\endgroup }+s_{2}x))} &{}={}& \instE{{\begingroup\renewcommand\colorMATH{\colorMATHB}\renewcommand\colorSYNTAX{\colorSYNTAXB}{{\color{\colorMATH}\ensuremath{\sS}}}\endgroup }}{(({\begingroup\renewcommand\colorMATH{\colorMATHB}\renewcommand\colorSYNTAX{\colorSYNTAXB}{{\color{\colorMATH}\ensuremath{\sS'_{1}}}}\endgroup }+{\begingroup\renewcommand\colorMATH{\colorMATHB}\renewcommand\colorSYNTAX{\colorSYNTAXB}{{\color{\colorMATH}\ensuremath{\sS'_{2}}}}\endgroup })+s_{1}x+s_{2}x)}
        \cr   &{}={}& \instE{{\begingroup\renewcommand\colorMATH{\colorMATHB}\renewcommand\colorSYNTAX{\colorSYNTAXB}{{\color{\colorMATH}\ensuremath{\sS}}}\endgroup }}{({\begingroup\renewcommand\colorMATH{\colorMATHB}\renewcommand\colorSYNTAX{\colorSYNTAXB}{{\color{\colorMATH}\ensuremath{\sS'_{1}}}}\endgroup }+{\begingroup\renewcommand\colorMATH{\colorMATHB}\renewcommand\colorSYNTAX{\colorSYNTAXB}{{\color{\colorMATH}\ensuremath{\sS'_{2}}}}\endgroup })} + (s_{1}+s_{2})r &
        \cr   &{}={}& \instE{{\begingroup\renewcommand\colorMATH{\colorMATHB}\renewcommand\colorSYNTAX{\colorSYNTAXB}{{\color{\colorMATH}\ensuremath{\sS}}}\endgroup }}{({\begingroup\renewcommand\colorMATH{\colorMATHB}\renewcommand\colorSYNTAX{\colorSYNTAXB}{{\color{\colorMATH}\ensuremath{\sS'_{1}}}}\endgroup }+{\begingroup\renewcommand\colorMATH{\colorMATHB}\renewcommand\colorSYNTAX{\colorSYNTAXB}{{\color{\colorMATH}\ensuremath{\sS'_{2}}}}\endgroup })} + (s_{1}+s_{2})r & {{\color{\colorTEXT}\textnormal{ (by definition)}}}
        \cr
\end{array}
      \end{gather*}\endgroup
      and the result holds.
    \end{subproof}
  \item  {{\color{\colorMATH}\ensuremath{{\begingroup\renewcommand\colorMATH{\colorMATHB}\renewcommand\colorSYNTAX{\colorSYNTAXB}{{\color{\colorMATH}\ensuremath{\sS_{1}}}}\endgroup } = {\begingroup\renewcommand\colorMATH{\colorMATHB}\renewcommand\colorSYNTAX{\colorSYNTAXB}{{\color{\colorMATH}\ensuremath{\sS'_{1}}}}\endgroup } + s_{1}}}}
    \begin{subproof} 
      Similar to previous case.
    \end{subproof}
  \end{enumerate}
\end{proof}

\begin{lemma}
\label{lm:equivsimplsubst}
If {{\color{\colorMATH}\ensuremath{x \notin  dom({\begingroup\renewcommand\colorMATH{\colorMATHB}\renewcommand\colorSYNTAX{\colorSYNTAXB}{{\color{\colorMATH}\ensuremath{\sS}}}\endgroup })}}}, then
{{\color{\colorMATH}\ensuremath{sx({\begingroup\renewcommand\colorMATH{\colorMATHB}\renewcommand\colorSYNTAX{\colorSYNTAXB}{{\color{\colorMATH}\ensuremath{\sS}}}\endgroup }(\tau )) = ({\begingroup\renewcommand\colorMATH{\colorMATHB}\renewcommand\colorSYNTAX{\colorSYNTAXB}{{\color{\colorMATH}\ensuremath{\sS}}}\endgroup } + sx)(\tau )}}}
\end{lemma}
\begin{proof}
  By induction on {{\color{\colorMATH}\ensuremath{\tau }}}.
\end{proof}

\lmdistrdot*
\begin{proof}
We prove: if {{\color{\colorMATH}\ensuremath{\instE{{\begingroup\renewcommand\colorMATH{\colorMATHB}\renewcommand\colorSYNTAX{\colorSYNTAXB}{{\color{\colorMATH}\ensuremath{\Distance}}}\endgroup }}{{\begingroup\renewcommand\colorMATH{\colorMATHB}\renewcommand\colorSYNTAX{\colorSYNTAXB}{{\color{\colorMATH}\ensuremath{\sS_{2}}}}\endgroup }} = {\begingroup\renewcommand\colorMATH{\colorMATHB}\renewcommand\colorSYNTAX{\colorSYNTAXB}{{\color{\colorMATH}\ensuremath{\distance}}}\endgroup } \in  {\text{sens}}}}} and {{\color{\colorMATH}\ensuremath{x \notin  dom({\begingroup\renewcommand\colorMATH{\colorMATHB}\renewcommand\colorSYNTAX{\colorSYNTAXB}{{\color{\colorMATH}\ensuremath{\sS_{2}}}}\endgroup })\cup  dom({\begingroup\renewcommand\colorMATH{\colorMATHB}\renewcommand\colorSYNTAX{\colorSYNTAXB}{{\color{\colorMATH}\ensuremath{\Distance}}}\endgroup })}}}, then
   {{\color{\colorMATH}\ensuremath{\instE{{\begingroup\renewcommand\colorMATH{\colorMATHB}\renewcommand\colorSYNTAX{\colorSYNTAXB}{{\color{\colorMATH}\ensuremath{\Distance}}}\endgroup }}{[{\begingroup\renewcommand\colorMATH{\colorMATHB}\renewcommand\colorSYNTAX{\colorSYNTAXB}{{\color{\colorMATH}\ensuremath{\sS_{2}}}}\endgroup }/x]{\begingroup\renewcommand\colorMATH{\colorMATHB}\renewcommand\colorSYNTAX{\colorSYNTAXB}{{\color{\colorMATH}\ensuremath{\sS_{3}}}}\endgroup }} = {\begingroup\renewcommand\colorMATH{\colorMATHB}\renewcommand\colorSYNTAX{\colorSYNTAXB}{{\color{\colorMATH}\ensuremath{\distance}}}\endgroup }x(\instE{{\begingroup\renewcommand\colorMATH{\colorMATHB}\renewcommand\colorSYNTAX{\colorSYNTAXB}{{\color{\colorMATH}\ensuremath{\Distance}}}\endgroup }}{{\begingroup\renewcommand\colorMATH{\colorMATHB}\renewcommand\colorSYNTAX{\colorSYNTAXB}{{\color{\colorMATH}\ensuremath{\sS_{3}}}}\endgroup }})}}}.
  We proceed by induction on the structure of {{\color{\colorMATH}\ensuremath{{\begingroup\renewcommand\colorMATH{\colorMATHB}\renewcommand\colorSYNTAX{\colorSYNTAXB}{{\color{\colorMATH}\ensuremath{\sS_{3}}}}\endgroup }}}}.
  \begin{enumerate}[ncases]\item  {{\color{\colorMATH}\ensuremath{{\begingroup\renewcommand\colorMATH{\colorMATHB}\renewcommand\colorSYNTAX{\colorSYNTAXB}{{\color{\colorMATH}\ensuremath{\sS_{3}}}}\endgroup }=\varnothing }}}
    \begin{subproof} 
      Trivial as {{\color{\colorMATH}\ensuremath{[{\begingroup\renewcommand\colorMATH{\colorMATHB}\renewcommand\colorSYNTAX{\colorSYNTAXB}{{\color{\colorMATH}\ensuremath{\sS_{2}}}}\endgroup }/x]\varnothing  = {\begingroup\renewcommand\colorMATH{\colorMATHB}\renewcommand\colorSYNTAX{\colorSYNTAXB}{{\color{\colorMATH}\ensuremath{\distance}}}\endgroup }x(\varnothing ) = \varnothing }}} and {{\color{\colorMATH}\ensuremath{\instE{{\begingroup\renewcommand\colorMATH{\colorMATHB}\renewcommand\colorSYNTAX{\colorSYNTAXB}{{\color{\colorMATH}\ensuremath{\Distance}}}\endgroup }}{\varnothing } = 0}}}.
    \end{subproof}
  \item  {{\color{\colorMATH}\ensuremath{x \notin  dom({\begingroup\renewcommand\colorMATH{\colorMATHB}\renewcommand\colorSYNTAX{\colorSYNTAXB}{{\color{\colorMATH}\ensuremath{\sS_{3}}}}\endgroup })}}}
    \begin{subproof} 
      Then {{\color{\colorMATH}\ensuremath{[{\begingroup\renewcommand\colorMATH{\colorMATHB}\renewcommand\colorSYNTAX{\colorSYNTAXB}{{\color{\colorMATH}\ensuremath{\sS_{2}}}}\endgroup }/x]{\begingroup\renewcommand\colorMATH{\colorMATHB}\renewcommand\colorSYNTAX{\colorSYNTAXB}{{\color{\colorMATH}\ensuremath{\sS_{3}}}}\endgroup } = {\begingroup\renewcommand\colorMATH{\colorMATHB}\renewcommand\colorSYNTAX{\colorSYNTAXB}{{\color{\colorMATH}\ensuremath{\sS_{3}}}}\endgroup }}}}, {{\color{\colorMATH}\ensuremath{\instE{{\begingroup\renewcommand\colorMATH{\colorMATHB}\renewcommand\colorSYNTAX{\colorSYNTAXB}{{\color{\colorMATH}\ensuremath{\Distance}}}\endgroup }}{[{\begingroup\renewcommand\colorMATH{\colorMATHB}\renewcommand\colorSYNTAX{\colorSYNTAXB}{{\color{\colorMATH}\ensuremath{\sS_{2}}}}\endgroup }/x]{\begingroup\renewcommand\colorMATH{\colorMATHB}\renewcommand\colorSYNTAX{\colorSYNTAXB}{{\color{\colorMATH}\ensuremath{\sS_{3}}}}\endgroup }} = \instE{{\begingroup\renewcommand\colorMATH{\colorMATHB}\renewcommand\colorSYNTAX{\colorSYNTAXB}{{\color{\colorMATH}\ensuremath{\Distance}}}\endgroup }}{{\begingroup\renewcommand\colorMATH{\colorMATHB}\renewcommand\colorSYNTAX{\colorSYNTAXB}{{\color{\colorMATH}\ensuremath{\sS_{3}}}}\endgroup }}}}}, and {{\color{\colorMATH}\ensuremath{[s/x](\instE{{\begingroup\renewcommand\colorMATH{\colorMATHB}\renewcommand\colorSYNTAX{\colorSYNTAXB}{{\color{\colorMATH}\ensuremath{\Distance}}}\endgroup }}{{\begingroup\renewcommand\colorMATH{\colorMATHB}\renewcommand\colorSYNTAX{\colorSYNTAXB}{{\color{\colorMATH}\ensuremath{\sS_{3}}}}\endgroup }}) = \instE{{\begingroup\renewcommand\colorMATH{\colorMATHB}\renewcommand\colorSYNTAX{\colorSYNTAXB}{{\color{\colorMATH}\ensuremath{\Distance}}}\endgroup }}{{\begingroup\renewcommand\colorMATH{\colorMATHB}\renewcommand\colorSYNTAX{\colorSYNTAXB}{{\color{\colorMATH}\ensuremath{\sS_{3}}}}\endgroup }}}}}, and the result holds.
    \end{subproof}
    \clearpage
  \item  {{\color{\colorMATH}\ensuremath{{\begingroup\renewcommand\colorMATH{\colorMATHB}\renewcommand\colorSYNTAX{\colorSYNTAXB}{{\color{\colorMATH}\ensuremath{\sS_{3}}}}\endgroup }={\begingroup\renewcommand\colorMATH{\colorMATHB}\renewcommand\colorSYNTAX{\colorSYNTAXB}{{\color{\colorMATH}\ensuremath{\sS_{4}}}}\endgroup }+{\begingroup\renewcommand\colorMATH{\colorMATHB}\renewcommand\colorSYNTAX{\colorSYNTAXB}{{\color{\colorMATH}\ensuremath{\distance'}}}\endgroup }x}}}
    \begin{subproof}
      By induction hypothesis on {{\color{\colorMATH}\ensuremath{{\begingroup\renewcommand\colorMATH{\colorMATHB}\renewcommand\colorSYNTAX{\colorSYNTAXB}{{\color{\colorMATH}\ensuremath{\sS_{4}}}}\endgroup }}}}, we know that {{\color{\colorMATH}\ensuremath{{\begingroup\renewcommand\colorMATH{\colorMATHB}\renewcommand\colorSYNTAX{\colorSYNTAXB}{{\color{\colorMATH}\ensuremath{\Distance}}}\endgroup }\mathord{\cdotp }([{\begingroup\renewcommand\colorMATH{\colorMATHB}\renewcommand\colorSYNTAX{\colorSYNTAXB}{{\color{\colorMATH}\ensuremath{\sS_{2}}}}\endgroup }/x]{\begingroup\renewcommand\colorMATH{\colorMATHB}\renewcommand\colorSYNTAX{\colorSYNTAXB}{{\color{\colorMATH}\ensuremath{\sS_{4}}}}\endgroup }) = {\begingroup\renewcommand\colorMATH{\colorMATHB}\renewcommand\colorSYNTAX{\colorSYNTAXB}{{\color{\colorMATH}\ensuremath{\distance}}}\endgroup }x(({\begingroup\renewcommand\colorMATH{\colorMATHB}\renewcommand\colorSYNTAX{\colorSYNTAXB}{{\color{\colorMATH}\ensuremath{\Distance}}}\endgroup }\mathord{\cdotp }{\begingroup\renewcommand\colorMATH{\colorMATHB}\renewcommand\colorSYNTAX{\colorSYNTAXB}{{\color{\colorMATH}\ensuremath{\sS_{4}}}}\endgroup }))}}}. Therefore
      \begingroup\color{\colorMATH}\begin{gather*}
\begin{array}{rclr
        } {\begingroup\renewcommand\colorMATH{\colorMATHB}\renewcommand\colorSYNTAX{\colorSYNTAXB}{{\color{\colorMATH}\ensuremath{\Distance}}}\endgroup }\mathord{\cdotp }([{\begingroup\renewcommand\colorMATH{\colorMATHB}\renewcommand\colorSYNTAX{\colorSYNTAXB}{{\color{\colorMATH}\ensuremath{\sS_{2}}}}\endgroup }/x]{\begingroup\renewcommand\colorMATH{\colorMATHB}\renewcommand\colorSYNTAX{\colorSYNTAXB}{{\color{\colorMATH}\ensuremath{\sS_{3}}}}\endgroup }) &{}={}& {\begingroup\renewcommand\colorMATH{\colorMATHB}\renewcommand\colorSYNTAX{\colorSYNTAXB}{{\color{\colorMATH}\ensuremath{\Distance}}}\endgroup }\mathord{\cdotp }([{\begingroup\renewcommand\colorMATH{\colorMATHB}\renewcommand\colorSYNTAX{\colorSYNTAXB}{{\color{\colorMATH}\ensuremath{\sS_{2}}}}\endgroup }/x]({\begingroup\renewcommand\colorMATH{\colorMATHB}\renewcommand\colorSYNTAX{\colorSYNTAXB}{{\color{\colorMATH}\ensuremath{\sS_{4}}}}\endgroup }+{\begingroup\renewcommand\colorMATH{\colorMATHB}\renewcommand\colorSYNTAX{\colorSYNTAXB}{{\color{\colorMATH}\ensuremath{\distance'}}}\endgroup }x)) & 
        \cr                &{}={}& {\begingroup\renewcommand\colorMATH{\colorMATHB}\renewcommand\colorSYNTAX{\colorSYNTAXB}{{\color{\colorMATH}\ensuremath{\Distance}}}\endgroup }\mathord{\cdotp }([{\begingroup\renewcommand\colorMATH{\colorMATHB}\renewcommand\colorSYNTAX{\colorSYNTAXB}{{\color{\colorMATH}\ensuremath{\sS_{2}}}}\endgroup }/x]{\begingroup\renewcommand\colorMATH{\colorMATHB}\renewcommand\colorSYNTAX{\colorSYNTAXB}{{\color{\colorMATH}\ensuremath{\sS_{4}}}}\endgroup }+{\begingroup\renewcommand\colorMATH{\colorMATHB}\renewcommand\colorSYNTAX{\colorSYNTAXB}{{\color{\colorMATH}\ensuremath{\distance'}}}\endgroup }{\begingroup\renewcommand\colorMATH{\colorMATHB}\renewcommand\colorSYNTAX{\colorSYNTAXB}{{\color{\colorMATH}\ensuremath{\sS_{2}}}}\endgroup })
        \cr
\end{array}
      \end{gather*}\endgroup
      And the result holds.
    \end{subproof} 
    \item  {{\color{\colorMATH}\ensuremath{{\begingroup\renewcommand\colorMATH{\colorMATHB}\renewcommand\colorSYNTAX{\colorSYNTAXB}{{\color{\colorMATH}\ensuremath{\sS_{3}}}}\endgroup }={\begingroup\renewcommand\colorMATH{\colorMATHB}\renewcommand\colorSYNTAX{\colorSYNTAXB}{{\color{\colorMATH}\ensuremath{\sS_{4}}}}\endgroup }+{\begingroup\renewcommand\colorMATH{\colorMATHB}\renewcommand\colorSYNTAX{\colorSYNTAXB}{{\color{\colorMATH}\ensuremath{\distance'}}}\endgroup }y}}}
    \begin{subproof}
      By induction hypothesis on {{\color{\colorMATH}\ensuremath{{\begingroup\renewcommand\colorMATH{\colorMATHB}\renewcommand\colorSYNTAX{\colorSYNTAXB}{{\color{\colorMATH}\ensuremath{\sS_{4}}}}\endgroup }}}}, we know that {{\color{\colorMATH}\ensuremath{{\begingroup\renewcommand\colorMATH{\colorMATHB}\renewcommand\colorSYNTAX{\colorSYNTAXB}{{\color{\colorMATH}\ensuremath{\Distance}}}\endgroup }\mathord{\cdotp }([{\begingroup\renewcommand\colorMATH{\colorMATHB}\renewcommand\colorSYNTAX{\colorSYNTAXB}{{\color{\colorMATH}\ensuremath{\sS_{2}}}}\endgroup }/x]{\begingroup\renewcommand\colorMATH{\colorMATHB}\renewcommand\colorSYNTAX{\colorSYNTAXB}{{\color{\colorMATH}\ensuremath{\sS_{4}}}}\endgroup }) = {\begingroup\renewcommand\colorMATH{\colorMATHB}\renewcommand\colorSYNTAX{\colorSYNTAXB}{{\color{\colorMATH}\ensuremath{\distance}}}\endgroup }x({\begingroup\renewcommand\colorMATH{\colorMATHB}\renewcommand\colorSYNTAX{\colorSYNTAXB}{{\color{\colorMATH}\ensuremath{\Distance}}}\endgroup }\mathord{\cdotp }{\begingroup\renewcommand\colorMATH{\colorMATHB}\renewcommand\colorSYNTAX{\colorSYNTAXB}{{\color{\colorMATH}\ensuremath{\sS_{4}}}}\endgroup })}}}. Therefore
      \begingroup\color{\colorMATH}\begin{gather*}
\begin{array}{rclr
        } {\begingroup\renewcommand\colorMATH{\colorMATHB}\renewcommand\colorSYNTAX{\colorSYNTAXB}{{\color{\colorMATH}\ensuremath{\Distance}}}\endgroup }\mathord{\cdotp }([{\begingroup\renewcommand\colorMATH{\colorMATHB}\renewcommand\colorSYNTAX{\colorSYNTAXB}{{\color{\colorMATH}\ensuremath{\sS_{2}}}}\endgroup }/x]{\begingroup\renewcommand\colorMATH{\colorMATHB}\renewcommand\colorSYNTAX{\colorSYNTAXB}{{\color{\colorMATH}\ensuremath{\sS_{3}}}}\endgroup }) &{}={}& {\begingroup\renewcommand\colorMATH{\colorMATHB}\renewcommand\colorSYNTAX{\colorSYNTAXB}{{\color{\colorMATH}\ensuremath{\Distance}}}\endgroup }\mathord{\cdotp }([{\begingroup\renewcommand\colorMATH{\colorMATHB}\renewcommand\colorSYNTAX{\colorSYNTAXB}{{\color{\colorMATH}\ensuremath{\sS_{2}}}}\endgroup }/x]({\begingroup\renewcommand\colorMATH{\colorMATHB}\renewcommand\colorSYNTAX{\colorSYNTAXB}{{\color{\colorMATH}\ensuremath{\sS_{4}}}}\endgroup }+{\begingroup\renewcommand\colorMATH{\colorMATHB}\renewcommand\colorSYNTAX{\colorSYNTAXB}{{\color{\colorMATH}\ensuremath{\distance'}}}\endgroup }y)) & 
        \cr                &{}={}& {\begingroup\renewcommand\colorMATH{\colorMATHB}\renewcommand\colorSYNTAX{\colorSYNTAXB}{{\color{\colorMATH}\ensuremath{\Distance}}}\endgroup }\mathord{\cdotp }([{\begingroup\renewcommand\colorMATH{\colorMATHB}\renewcommand\colorSYNTAX{\colorSYNTAXB}{{\color{\colorMATH}\ensuremath{\sS_{2}}}}\endgroup }/x]{\begingroup\renewcommand\colorMATH{\colorMATHB}\renewcommand\colorSYNTAX{\colorSYNTAXB}{{\color{\colorMATH}\ensuremath{\sS_{4}}}}\endgroup }+{\begingroup\renewcommand\colorMATH{\colorMATHB}\renewcommand\colorSYNTAX{\colorSYNTAXB}{{\color{\colorMATH}\ensuremath{\distance'}}}\endgroup }y)
        \cr
\end{array}
      \end{gather*}\endgroup
      And the result holds.  
    \end{subproof} 
    \item  {{\color{\colorMATH}\ensuremath{{\begingroup\renewcommand\colorMATH{\colorMATHB}\renewcommand\colorSYNTAX{\colorSYNTAXB}{{\color{\colorMATH}\ensuremath{\sS_{3}}}}\endgroup }={\begingroup\renewcommand\colorMATH{\colorMATHB}\renewcommand\colorSYNTAX{\colorSYNTAXB}{{\color{\colorMATH}\ensuremath{\sS_{4}}}}\endgroup }+{\begingroup\renewcommand\colorMATH{\colorMATHB}\renewcommand\colorSYNTAX{\colorSYNTAXB}{{\color{\colorMATH}\ensuremath{\distance'}}}\endgroup }}}}
    \begin{subproof}
      By induction hypothesis on {{\color{\colorMATH}\ensuremath{{\begingroup\renewcommand\colorMATH{\colorMATHB}\renewcommand\colorSYNTAX{\colorSYNTAXB}{{\color{\colorMATH}\ensuremath{\sS_{4}}}}\endgroup }}}}, we know that {{\color{\colorMATH}\ensuremath{{\begingroup\renewcommand\colorMATH{\colorMATHB}\renewcommand\colorSYNTAX{\colorSYNTAXB}{{\color{\colorMATH}\ensuremath{\Distance}}}\endgroup }\mathord{\cdotp }([{\begingroup\renewcommand\colorMATH{\colorMATHB}\renewcommand\colorSYNTAX{\colorSYNTAXB}{{\color{\colorMATH}\ensuremath{\sS_{2}}}}\endgroup }/x]{\begingroup\renewcommand\colorMATH{\colorMATHB}\renewcommand\colorSYNTAX{\colorSYNTAXB}{{\color{\colorMATH}\ensuremath{\sS_{4}}}}\endgroup }) = {\begingroup\renewcommand\colorMATH{\colorMATHB}\renewcommand\colorSYNTAX{\colorSYNTAXB}{{\color{\colorMATH}\ensuremath{\distance}}}\endgroup }x({\begingroup\renewcommand\colorMATH{\colorMATHB}\renewcommand\colorSYNTAX{\colorSYNTAXB}{{\color{\colorMATH}\ensuremath{\Distance}}}\endgroup }\mathord{\cdotp }{\begingroup\renewcommand\colorMATH{\colorMATHB}\renewcommand\colorSYNTAX{\colorSYNTAXB}{{\color{\colorMATH}\ensuremath{\sS_{4}}}}\endgroup })}}}. Therefore
      \begingroup\color{\colorMATH}\begin{gather*}
\begin{array}{rclr
        } {\begingroup\renewcommand\colorMATH{\colorMATHB}\renewcommand\colorSYNTAX{\colorSYNTAXB}{{\color{\colorMATH}\ensuremath{\Distance}}}\endgroup }\mathord{\cdotp }([{\begingroup\renewcommand\colorMATH{\colorMATHB}\renewcommand\colorSYNTAX{\colorSYNTAXB}{{\color{\colorMATH}\ensuremath{\sS_{2}}}}\endgroup }/x]{\begingroup\renewcommand\colorMATH{\colorMATHB}\renewcommand\colorSYNTAX{\colorSYNTAXB}{{\color{\colorMATH}\ensuremath{\sS_{3}}}}\endgroup }) &{}={}& {\begingroup\renewcommand\colorMATH{\colorMATHB}\renewcommand\colorSYNTAX{\colorSYNTAXB}{{\color{\colorMATH}\ensuremath{\Distance}}}\endgroup }\mathord{\cdotp }([{\begingroup\renewcommand\colorMATH{\colorMATHB}\renewcommand\colorSYNTAX{\colorSYNTAXB}{{\color{\colorMATH}\ensuremath{\sS_{2}}}}\endgroup }/x]({\begingroup\renewcommand\colorMATH{\colorMATHB}\renewcommand\colorSYNTAX{\colorSYNTAXB}{{\color{\colorMATH}\ensuremath{\sS_{4}}}}\endgroup }+{\begingroup\renewcommand\colorMATH{\colorMATHB}\renewcommand\colorSYNTAX{\colorSYNTAXB}{{\color{\colorMATH}\ensuremath{\distance'}}}\endgroup })) & 
        \cr                &{}={}& {\begingroup\renewcommand\colorMATH{\colorMATHB}\renewcommand\colorSYNTAX{\colorSYNTAXB}{{\color{\colorMATH}\ensuremath{\Distance}}}\endgroup }\mathord{\cdotp }([{\begingroup\renewcommand\colorMATH{\colorMATHB}\renewcommand\colorSYNTAX{\colorSYNTAXB}{{\color{\colorMATH}\ensuremath{\sS_{2}}}}\endgroup }/x]{\begingroup\renewcommand\colorMATH{\colorMATHB}\renewcommand\colorSYNTAX{\colorSYNTAXB}{{\color{\colorMATH}\ensuremath{\sS_{4}}}}\endgroup }+{\begingroup\renewcommand\colorMATH{\colorMATHB}\renewcommand\colorSYNTAX{\colorSYNTAXB}{{\color{\colorMATH}\ensuremath{\distance'}}}\endgroup })
        \cr
\end{array}
      \end{gather*}\endgroup
      And the result holds.  
    \end{subproof} 
  \end{enumerate}
\end{proof}

\lmdistrdotpp*
\begin{proof}
  We prove: if {{\color{\colorMATH}\ensuremath{\instE{{\begingroup\renewcommand\colorMATH{\colorMATHB}\renewcommand\colorSYNTAX{\colorSYNTAXB}{{\color{\colorMATH}\ensuremath{\Distance}}}\endgroup }}{{\begingroup\renewcommand\colorMATH{\colorMATHB}\renewcommand\colorSYNTAX{\colorSYNTAXB}{{\color{\colorMATH}\ensuremath{\sS_{2}}}}\endgroup }} = {\begingroup\renewcommand\colorMATH{\colorMATHB}\renewcommand\colorSYNTAX{\colorSYNTAXB}{{\color{\colorMATH}\ensuremath{\distance}}}\endgroup } \in  {\text{sens}}}}} and {{\color{\colorMATH}\ensuremath{x \notin  dom({\begingroup\renewcommand\colorMATH{\colorMATHB}\renewcommand\colorSYNTAX{\colorSYNTAXB}{{\color{\colorMATH}\ensuremath{\sS_{2}}}}\endgroup })\cup  dom({\begingroup\renewcommand\colorMATH{\colorMATHB}\renewcommand\colorSYNTAX{\colorSYNTAXB}{{\color{\colorMATH}\ensuremath{\Distance}}}\endgroup })}}}, then
   {{\color{\colorMATH}\ensuremath{\instPE{{\begingroup\renewcommand\colorMATH{\colorMATHB}\renewcommand\colorSYNTAX{\colorSYNTAXB}{{\color{\colorMATH}\ensuremath{\Distance}}}\endgroup }}{\subst[{\begingroup\renewcommand\colorMATH{\colorMATHB}\renewcommand\colorSYNTAX{\colorSYNTAXB}{{\color{\colorMATH}\ensuremath{\Distance}}}\endgroup }]{{\begingroup\renewcommand\colorMATH{\colorMATHB}\renewcommand\colorSYNTAX{\colorSYNTAXB}{{\color{\colorMATH}\ensuremath{\sS_{2}}}}\endgroup }}{x}{\begingroup\renewcommand\colorMATH{\colorMATHC}\renewcommand\colorSYNTAX{\colorSYNTAXC}{{\color{\colorMATH}\ensuremath{\pS_{3}}}}\endgroup }} = \instPE{{\begingroup\renewcommand\colorMATH{\colorMATHB}\renewcommand\colorSYNTAX{\colorSYNTAXB}{{\color{\colorMATH}\ensuremath{\distance}}}\endgroup }x}{(\instPE{{\begingroup\renewcommand\colorMATH{\colorMATHB}\renewcommand\colorSYNTAX{\colorSYNTAXB}{{\color{\colorMATH}\ensuremath{\Distance}}}\endgroup }}{{\begingroup\renewcommand\colorMATH{\colorMATHC}\renewcommand\colorSYNTAX{\colorSYNTAXC}{{\color{\colorMATH}\ensuremath{\pS_{3}}}}\endgroup }})}}}}.
  We proceed by induction on the structure of {{\color{\colorMATH}\ensuremath{{\begingroup\renewcommand\colorMATH{\colorMATHB}\renewcommand\colorSYNTAX{\colorSYNTAXB}{{\color{\colorMATH}\ensuremath{\sS_{3}}}}\endgroup }}}}.
  \begin{enumerate}[ncases]\item  {{\color{\colorMATH}\ensuremath{{\begingroup\renewcommand\colorMATH{\colorMATHC}\renewcommand\colorSYNTAX{\colorSYNTAXC}{{\color{\colorMATH}\ensuremath{\pS_{3}}}}\endgroup }=\varnothing }}}
    \begin{subproof} 
      Trivial as {{\color{\colorMATH}\ensuremath{\subst{{\begingroup\renewcommand\colorMATH{\colorMATHB}\renewcommand\colorSYNTAX{\colorSYNTAXB}{{\color{\colorMATH}\ensuremath{\sS_{2}}}}\endgroup }}{x}\varnothing  = {\begingroup\renewcommand\colorMATH{\colorMATHB}\renewcommand\colorSYNTAX{\colorSYNTAXB}{{\color{\colorMATH}\ensuremath{\distance}}}\endgroup }x(\varnothing ) = \varnothing }}} and {{\color{\colorMATH}\ensuremath{\instPE{{\begingroup\renewcommand\colorMATH{\colorMATHB}\renewcommand\colorSYNTAX{\colorSYNTAXB}{{\color{\colorMATH}\ensuremath{\Distance}}}\endgroup }}{\varnothing } = 0}}}.
    \end{subproof}
  \item  {{\color{\colorMATH}\ensuremath{{\begingroup\renewcommand\colorMATH{\colorMATHC}\renewcommand\colorSYNTAX{\colorSYNTAXC}{{\color{\colorMATH}\ensuremath{\pS_{3}}}}\endgroup }={\begingroup\renewcommand\colorMATH{\colorMATHC}\renewcommand\colorSYNTAX{\colorSYNTAXC}{{\color{\colorMATH}\ensuremath{p}}}\endgroup }y}}}
    \begin{subproof}
      Then {{\color{\colorMATH}\ensuremath{\subst{{\begingroup\renewcommand\colorMATH{\colorMATHB}\renewcommand\colorSYNTAX{\colorSYNTAXB}{{\color{\colorMATH}\ensuremath{\sS_{2}}}}\endgroup }}{x}{\begingroup\renewcommand\colorMATH{\colorMATHC}\renewcommand\colorSYNTAX{\colorSYNTAXC}{{\color{\colorMATH}\ensuremath{\pS_{3}}}}\endgroup } = {\begingroup\renewcommand\colorMATH{\colorMATHC}\renewcommand\colorSYNTAX{\colorSYNTAXC}{{\color{\colorMATH}\ensuremath{\pS_{3}}}}\endgroup }}}}, {{\color{\colorMATH}\ensuremath{\instPE{{\begingroup\renewcommand\colorMATH{\colorMATHB}\renewcommand\colorSYNTAX{\colorSYNTAXB}{{\color{\colorMATH}\ensuremath{\Distance}}}\endgroup }}{\subst{{\begingroup\renewcommand\colorMATH{\colorMATHB}\renewcommand\colorSYNTAX{\colorSYNTAXB}{{\color{\colorMATH}\ensuremath{\sS_{2}}}}\endgroup }}{x}{\begingroup\renewcommand\colorMATH{\colorMATHC}\renewcommand\colorSYNTAX{\colorSYNTAXC}{{\color{\colorMATH}\ensuremath{\pS_{3}}}}\endgroup }} = \instPE{{\begingroup\renewcommand\colorMATH{\colorMATHB}\renewcommand\colorSYNTAX{\colorSYNTAXB}{{\color{\colorMATH}\ensuremath{\Distance}}}\endgroup }}{{\begingroup\renewcommand\colorMATH{\colorMATHC}\renewcommand\colorSYNTAX{\colorSYNTAXC}{{\color{\colorMATH}\ensuremath{\pS_{3}}}}\endgroup }}}}}, and {{\color{\colorMATH}\ensuremath{\subst{{\begingroup\renewcommand\colorMATH{\colorMATHB}\renewcommand\colorSYNTAX{\colorSYNTAXB}{{\color{\colorMATH}\ensuremath{\distance}}}\endgroup }}{x}(\instPE{{\begingroup\renewcommand\colorMATH{\colorMATHB}\renewcommand\colorSYNTAX{\colorSYNTAXB}{{\color{\colorMATH}\ensuremath{\Distance}}}\endgroup }}{{\begingroup\renewcommand\colorMATH{\colorMATHC}\renewcommand\colorSYNTAX{\colorSYNTAXC}{{\color{\colorMATH}\ensuremath{\pS_{3}}}}\endgroup }}) = \instPE{{\begingroup\renewcommand\colorMATH{\colorMATHB}\renewcommand\colorSYNTAX{\colorSYNTAXB}{{\color{\colorMATH}\ensuremath{\Distance}}}\endgroup }}{{\begingroup\renewcommand\colorMATH{\colorMATHC}\renewcommand\colorSYNTAX{\colorSYNTAXC}{{\color{\colorMATH}\ensuremath{\pS_{3}}}}\endgroup }}}}}, and the result holds.
    \end{subproof}
  \item  {{\color{\colorMATH}\ensuremath{{\begingroup\renewcommand\colorMATH{\colorMATHC}\renewcommand\colorSYNTAX{\colorSYNTAXC}{{\color{\colorMATH}\ensuremath{\pS_{3}}}}\endgroup }={\begingroup\renewcommand\colorMATH{\colorMATHC}\renewcommand\colorSYNTAX{\colorSYNTAXC}{{\color{\colorMATH}\ensuremath{p}}}\endgroup }x}}}
    \begin{subproof}
      %By induction hypothesis on {{\color{\colorMATH}\ensuremath{{\begingroup\renewcommand\colorMATH{\colorMATHC}\renewcommand\colorSYNTAX{\colorSYNTAXC}{{\color{\colorMATH}\ensuremath{\pS_{4}}}}\endgroup }}}}, we know that {{\color{\colorMATH}\ensuremath{{\begingroup\renewcommand\colorMATH{\colorMATHB}\renewcommand\colorSYNTAX{\colorSYNTAXB}{{\color{\colorMATH}\ensuremath{\Distance}}}\endgroup }{\begingroup\renewcommand\colorMATH{\colorMATHC}\renewcommand\colorSYNTAX{\colorSYNTAXC}{{\color{\colorMATH}\ensuremath{\bigcdot}}}\endgroup }(\subst{{\begingroup\renewcommand\colorMATH{\colorMATHB}\renewcommand\colorSYNTAX{\colorSYNTAXB}{{\color{\colorMATH}\ensuremath{\sS_{2}}}}\endgroup }}{x}{\begingroup\renewcommand\colorMATH{\colorMATHC}\renewcommand\colorSYNTAX{\colorSYNTAXC}{{\color{\colorMATH}\ensuremath{\pS_{4}}}}\endgroup }) = {\begingroup\renewcommand\colorMATH{\colorMATHB}\renewcommand\colorSYNTAX{\colorSYNTAXB}{{\color{\colorMATH}\ensuremath{\distance}}}\endgroup }x({\begingroup\renewcommand\colorMATH{\colorMATHB}\renewcommand\colorSYNTAX{\colorSYNTAXB}{{\color{\colorMATH}\ensuremath{\Distance}}}\endgroup }\mathord{\cdotp }{\begingroup\renewcommand\colorMATH{\colorMATHC}\renewcommand\colorSYNTAX{\colorSYNTAXC}{{\color{\colorMATH}\ensuremath{\pS_{4}}}}\endgroup })}}}. 
      Then
      \begingroup\color{\colorMATH}\begin{gather*}
\begin{array}{rclr
        } \instPE{{\begingroup\renewcommand\colorMATH{\colorMATHB}\renewcommand\colorSYNTAX{\colorSYNTAXB}{{\color{\colorMATH}\ensuremath{\Distance}}}\endgroup }}{(\subst{{\begingroup\renewcommand\colorMATH{\colorMATHB}\renewcommand\colorSYNTAX{\colorSYNTAXB}{{\color{\colorMATH}\ensuremath{\sS_{2}}}}\endgroup }}{x}{\begingroup\renewcommand\colorMATH{\colorMATHC}\renewcommand\colorSYNTAX{\colorSYNTAXC}{{\color{\colorMATH}\ensuremath{\pS_{3}}}}\endgroup })} &{}={}& \instPE{{\begingroup\renewcommand\colorMATH{\colorMATHB}\renewcommand\colorSYNTAX{\colorSYNTAXB}{{\color{\colorMATH}\ensuremath{\Distance}}}\endgroup }}{(\subst{{\begingroup\renewcommand\colorMATH{\colorMATHB}\renewcommand\colorSYNTAX{\colorSYNTAXB}{{\color{\colorMATH}\ensuremath{\sS_{2}}}}\endgroup }}{x}({\begingroup\renewcommand\colorMATH{\colorMATHC}\renewcommand\colorSYNTAX{\colorSYNTAXC}{{\color{\colorMATH}\ensuremath{p}}}\endgroup }x))} & 
        \cr                &{}={}& \instPE{{\begingroup\renewcommand\colorMATH{\colorMATHB}\renewcommand\colorSYNTAX{\colorSYNTAXB}{{\color{\colorMATH}\ensuremath{\Distance}}}\endgroup }}{({\begingroup\renewcommand\colorMATH{\colorMATHC}\renewcommand\colorSYNTAX{\colorSYNTAXC}{{\color{\colorMATH}\ensuremath{\rceil {\begingroup\renewcommand\colorMATH{\colorMATHA}\renewcommand\colorSYNTAX{\colorSYNTAXA}{{\color{\colorMATH}\ensuremath{{\begingroup\renewcommand\colorMATH{\colorMATHB}\renewcommand\colorSYNTAX{\colorSYNTAXB}{{\color{\colorMATH}\ensuremath{\sS_{2}}}}\endgroup }}}}\endgroup }\lceil ^{{\begingroup\renewcommand\colorMATH{\colorMATHC}\renewcommand\colorSYNTAX{\colorSYNTAXC}{{\color{\colorMATH}\ensuremath{p}}}\endgroup }}}}}\endgroup })}
        \cr
\end{array}
      \end{gather*}\endgroup
      But {{\color{\colorMATH}\ensuremath{\instPE{({\begingroup\renewcommand\colorMATH{\colorMATHB}\renewcommand\colorSYNTAX{\colorSYNTAXB}{{\color{\colorMATH}\ensuremath{\Distance}}}\endgroup }+{\begingroup\renewcommand\colorMATH{\colorMATHB}\renewcommand\colorSYNTAX{\colorSYNTAXB}{{\color{\colorMATH}\ensuremath{\distance}}}\endgroup }x)}{{\begingroup\renewcommand\colorMATH{\colorMATHC}\renewcommand\colorSYNTAX{\colorSYNTAXC}{{\color{\colorMATH}\ensuremath{\pS_{3}}}}\endgroup }} = \instPE{({\begingroup\renewcommand\colorMATH{\colorMATHB}\renewcommand\colorSYNTAX{\colorSYNTAXB}{{\color{\colorMATH}\ensuremath{\Distance}}}\endgroup }+{\begingroup\renewcommand\colorMATH{\colorMATHB}\renewcommand\colorSYNTAX{\colorSYNTAXB}{{\color{\colorMATH}\ensuremath{\distance}}}\endgroup }x)}{{\begingroup\renewcommand\colorMATH{\colorMATHC}\renewcommand\colorSYNTAX{\colorSYNTAXC}{{\color{\colorMATH}\ensuremath{p}}}\endgroup }x} = {\begingroup\renewcommand\colorMATH{\colorMATHC}\renewcommand\colorSYNTAX{\colorSYNTAXC}{{\color{\colorMATH}\ensuremath{p}}}\endgroup }}}} and the result holds.
    \end{subproof} 
    \item  {{\color{\colorMATH}\ensuremath{{\begingroup\renewcommand\colorMATH{\colorMATHC}\renewcommand\colorSYNTAX{\colorSYNTAXC}{{\color{\colorMATH}\ensuremath{\pS_{3}}}}\endgroup }={\begingroup\renewcommand\colorMATH{\colorMATHC}\renewcommand\colorSYNTAX{\colorSYNTAXC}{{\color{\colorMATH}\ensuremath{\pS_{4}}}}\endgroup }+{\begingroup\renewcommand\colorMATH{\colorMATHC}\renewcommand\colorSYNTAX{\colorSYNTAXC}{{\color{\colorMATH}\ensuremath{\pS_{5}}}}\endgroup }}}}
    \begin{subproof}
      By induction hypothesis on {{\color{\colorMATH}\ensuremath{{\begingroup\renewcommand\colorMATH{\colorMATHC}\renewcommand\colorSYNTAX{\colorSYNTAXC}{{\color{\colorMATH}\ensuremath{\pS_{4}}}}\endgroup }}}} and {{\color{\colorMATH}\ensuremath{{\begingroup\renewcommand\colorMATH{\colorMATHC}\renewcommand\colorSYNTAX{\colorSYNTAXC}{{\color{\colorMATH}\ensuremath{\pS_{5}}}}\endgroup }}}}, we know that {{\color{\colorMATH}\ensuremath{\instPE{{\begingroup\renewcommand\colorMATH{\colorMATHB}\renewcommand\colorSYNTAX{\colorSYNTAXB}{{\color{\colorMATH}\ensuremath{\Distance}}}\endgroup }}{(\subst{{\begingroup\renewcommand\colorMATH{\colorMATHB}\renewcommand\colorSYNTAX{\colorSYNTAXB}{{\color{\colorMATH}\ensuremath{\sS_{2}}}}\endgroup }}{x}{\begingroup\renewcommand\colorMATH{\colorMATHC}\renewcommand\colorSYNTAX{\colorSYNTAXC}{{\color{\colorMATH}\ensuremath{\pS_{4}}}}\endgroup })} = \instPE{{\begingroup\renewcommand\colorMATH{\colorMATHB}\renewcommand\colorSYNTAX{\colorSYNTAXB}{{\color{\colorMATH}\ensuremath{\distance}}}\endgroup }x}{({\begingroup\renewcommand\colorMATH{\colorMATHB}\renewcommand\colorSYNTAX{\colorSYNTAXB}{{\color{\colorMATH}\ensuremath{\Distance}}}\endgroup }\mathord{\cdotp }{\begingroup\renewcommand\colorMATH{\colorMATHC}\renewcommand\colorSYNTAX{\colorSYNTAXC}{{\color{\colorMATH}\ensuremath{\pS_{4}}}}\endgroup })}}}}, and {{\color{\colorMATH}\ensuremath{\instPE{{\begingroup\renewcommand\colorMATH{\colorMATHB}\renewcommand\colorSYNTAX{\colorSYNTAXB}{{\color{\colorMATH}\ensuremath{\Distance}}}\endgroup }}{(\subst{{\begingroup\renewcommand\colorMATH{\colorMATHB}\renewcommand\colorSYNTAX{\colorSYNTAXB}{{\color{\colorMATH}\ensuremath{\sS_{2}}}}\endgroup }}{x}{\begingroup\renewcommand\colorMATH{\colorMATHC}\renewcommand\colorSYNTAX{\colorSYNTAXC}{{\color{\colorMATH}\ensuremath{\pS_{5}}}}\endgroup })} = \instPE{{\begingroup\renewcommand\colorMATH{\colorMATHB}\renewcommand\colorSYNTAX{\colorSYNTAXB}{{\color{\colorMATH}\ensuremath{\distance}}}\endgroup }x}{({\begingroup\renewcommand\colorMATH{\colorMATHB}\renewcommand\colorSYNTAX{\colorSYNTAXB}{{\color{\colorMATH}\ensuremath{\Distance}}}\endgroup }\mathord{\cdotp }{\begingroup\renewcommand\colorMATH{\colorMATHC}\renewcommand\colorSYNTAX{\colorSYNTAXC}{{\color{\colorMATH}\ensuremath{\pS_{5}}}}\endgroup })}}}}. Therefore
      \begingroup\color{\colorMATH}\begin{gather*}
\begin{array}{rclr
        } \instPE{{\begingroup\renewcommand\colorMATH{\colorMATHB}\renewcommand\colorSYNTAX{\colorSYNTAXB}{{\color{\colorMATH}\ensuremath{\Distance}}}\endgroup }}{(\subst{{\begingroup\renewcommand\colorMATH{\colorMATHB}\renewcommand\colorSYNTAX{\colorSYNTAXB}{{\color{\colorMATH}\ensuremath{\sS_{2}}}}\endgroup }}{x}{\begingroup\renewcommand\colorMATH{\colorMATHB}\renewcommand\colorSYNTAX{\colorSYNTAXB}{{\color{\colorMATH}\ensuremath{\sS_{3}}}}\endgroup })} &{}={}& \instPE{{\begingroup\renewcommand\colorMATH{\colorMATHB}\renewcommand\colorSYNTAX{\colorSYNTAXB}{{\color{\colorMATH}\ensuremath{\Distance}}}\endgroup }}{(\subst{{\begingroup\renewcommand\colorMATH{\colorMATHB}\renewcommand\colorSYNTAX{\colorSYNTAXB}{{\color{\colorMATH}\ensuremath{\sS_{2}}}}\endgroup }}{x}({\begingroup\renewcommand\colorMATH{\colorMATHC}\renewcommand\colorSYNTAX{\colorSYNTAXC}{{\color{\colorMATH}\ensuremath{\pS_{4}}}}\endgroup }+{\begingroup\renewcommand\colorMATH{\colorMATHC}\renewcommand\colorSYNTAX{\colorSYNTAXC}{{\color{\colorMATH}\ensuremath{\pS_{5}}}}\endgroup }))} & 
        \cr                &{}={}& \instPE{{\begingroup\renewcommand\colorMATH{\colorMATHB}\renewcommand\colorSYNTAX{\colorSYNTAXB}{{\color{\colorMATH}\ensuremath{\Distance}}}\endgroup }}{(\subst{{\begingroup\renewcommand\colorMATH{\colorMATHB}\renewcommand\colorSYNTAX{\colorSYNTAXB}{{\color{\colorMATH}\ensuremath{\sS_{2}}}}\endgroup }}{x}{\begingroup\renewcommand\colorMATH{\colorMATHC}\renewcommand\colorSYNTAX{\colorSYNTAXC}{{\color{\colorMATH}\ensuremath{\pS_{4}}}}\endgroup }+\subst{{\begingroup\renewcommand\colorMATH{\colorMATHB}\renewcommand\colorSYNTAX{\colorSYNTAXB}{{\color{\colorMATH}\ensuremath{\sS_{2}}}}\endgroup }}{x}{\begingroup\renewcommand\colorMATH{\colorMATHC}\renewcommand\colorSYNTAX{\colorSYNTAXC}{{\color{\colorMATH}\ensuremath{\pS_{5}}}}\endgroup })}
        \cr
\end{array}
      \end{gather*}\endgroup
      And the result holds.  
    \end{subproof}
    \item  {{\color{\colorMATH}\ensuremath{{\begingroup\renewcommand\colorMATH{\colorMATHC}\renewcommand\colorSYNTAX{\colorSYNTAXC}{{\color{\colorMATH}\ensuremath{\pS_{3}}}}\endgroup }={\begingroup\renewcommand\colorMATH{\colorMATHC}\renewcommand\colorSYNTAX{\colorSYNTAXC}{{\color{\colorMATH}\ensuremath{\pS_{4}}}}\endgroup }\sqcup {\begingroup\renewcommand\colorMATH{\colorMATHC}\renewcommand\colorSYNTAX{\colorSYNTAXC}{{\color{\colorMATH}\ensuremath{\pS_{5}}}}\endgroup }}}}
    \begin{subproof}
      By induction hypothesis on {{\color{\colorMATH}\ensuremath{{\begingroup\renewcommand\colorMATH{\colorMATHC}\renewcommand\colorSYNTAX{\colorSYNTAXC}{{\color{\colorMATH}\ensuremath{\pS_{4}}}}\endgroup }}}} and {{\color{\colorMATH}\ensuremath{{\begingroup\renewcommand\colorMATH{\colorMATHC}\renewcommand\colorSYNTAX{\colorSYNTAXC}{{\color{\colorMATH}\ensuremath{\pS_{5}}}}\endgroup }}}}, we know that {{\color{\colorMATH}\ensuremath{{\begingroup\renewcommand\colorMATH{\colorMATHB}\renewcommand\colorSYNTAX{\colorSYNTAXB}{{\color{\colorMATH}\ensuremath{\Distance}}}\endgroup }{\begingroup\renewcommand\colorMATH{\colorMATHC}\renewcommand\colorSYNTAX{\colorSYNTAXC}{{\color{\colorMATH}\ensuremath{\bigcdot}}}\endgroup }(\subst{{\begingroup\renewcommand\colorMATH{\colorMATHB}\renewcommand\colorSYNTAX{\colorSYNTAXB}{{\color{\colorMATH}\ensuremath{\sS_{2}}}}\endgroup }}{x}{\begingroup\renewcommand\colorMATH{\colorMATHC}\renewcommand\colorSYNTAX{\colorSYNTAXC}{{\color{\colorMATH}\ensuremath{\pS_{4}}}}\endgroup }) = {\begingroup\renewcommand\colorMATH{\colorMATHB}\renewcommand\colorSYNTAX{\colorSYNTAXB}{{\color{\colorMATH}\ensuremath{\distance}}}\endgroup }x{\begingroup\renewcommand\colorMATH{\colorMATHC}\renewcommand\colorSYNTAX{\colorSYNTAXC}{{\color{\colorMATH}\ensuremath{\bigcdot}}}\endgroup }({\begingroup\renewcommand\colorMATH{\colorMATHB}\renewcommand\colorSYNTAX{\colorSYNTAXB}{{\color{\colorMATH}\ensuremath{\Distance}}}\endgroup }\mathord{\cdotp }{\begingroup\renewcommand\colorMATH{\colorMATHC}\renewcommand\colorSYNTAX{\colorSYNTAXC}{{\color{\colorMATH}\ensuremath{\pS_{4}}}}\endgroup })}}}, and {{\color{\colorMATH}\ensuremath{{\begingroup\renewcommand\colorMATH{\colorMATHB}\renewcommand\colorSYNTAX{\colorSYNTAXB}{{\color{\colorMATH}\ensuremath{\Distance}}}\endgroup }{\begingroup\renewcommand\colorMATH{\colorMATHC}\renewcommand\colorSYNTAX{\colorSYNTAXC}{{\color{\colorMATH}\ensuremath{\bigcdot}}}\endgroup }(\subst{{\begingroup\renewcommand\colorMATH{\colorMATHB}\renewcommand\colorSYNTAX{\colorSYNTAXB}{{\color{\colorMATH}\ensuremath{\sS_{2}}}}\endgroup }}{x}{\begingroup\renewcommand\colorMATH{\colorMATHC}\renewcommand\colorSYNTAX{\colorSYNTAXC}{{\color{\colorMATH}\ensuremath{\pS_{5}}}}\endgroup }) = {\begingroup\renewcommand\colorMATH{\colorMATHB}\renewcommand\colorSYNTAX{\colorSYNTAXB}{{\color{\colorMATH}\ensuremath{\distance}}}\endgroup }x{\begingroup\renewcommand\colorMATH{\colorMATHC}\renewcommand\colorSYNTAX{\colorSYNTAXC}{{\color{\colorMATH}\ensuremath{\bigcdot}}}\endgroup }({\begingroup\renewcommand\colorMATH{\colorMATHB}\renewcommand\colorSYNTAX{\colorSYNTAXB}{{\color{\colorMATH}\ensuremath{\Distance}}}\endgroup }\mathord{\cdotp }{\begingroup\renewcommand\colorMATH{\colorMATHC}\renewcommand\colorSYNTAX{\colorSYNTAXC}{{\color{\colorMATH}\ensuremath{\pS_{5}}}}\endgroup })}}}. Therefore
      \begingroup\color{\colorMATH}\begin{gather*}
\begin{array}{rclr
        } {\begingroup\renewcommand\colorMATH{\colorMATHB}\renewcommand\colorSYNTAX{\colorSYNTAXB}{{\color{\colorMATH}\ensuremath{\Distance}}}\endgroup }{\begingroup\renewcommand\colorMATH{\colorMATHC}\renewcommand\colorSYNTAX{\colorSYNTAXC}{{\color{\colorMATH}\ensuremath{\bigcdot}}}\endgroup }(\subst{{\begingroup\renewcommand\colorMATH{\colorMATHB}\renewcommand\colorSYNTAX{\colorSYNTAXB}{{\color{\colorMATH}\ensuremath{\sS_{2}}}}\endgroup }}{x}{\begingroup\renewcommand\colorMATH{\colorMATHB}\renewcommand\colorSYNTAX{\colorSYNTAXB}{{\color{\colorMATH}\ensuremath{\sS_{3}}}}\endgroup }) &{}={}& {\begingroup\renewcommand\colorMATH{\colorMATHB}\renewcommand\colorSYNTAX{\colorSYNTAXB}{{\color{\colorMATH}\ensuremath{\Distance}}}\endgroup }{\begingroup\renewcommand\colorMATH{\colorMATHC}\renewcommand\colorSYNTAX{\colorSYNTAXC}{{\color{\colorMATH}\ensuremath{\bigcdot}}}\endgroup }(\subst{{\begingroup\renewcommand\colorMATH{\colorMATHB}\renewcommand\colorSYNTAX{\colorSYNTAXB}{{\color{\colorMATH}\ensuremath{\sS_{2}}}}\endgroup }}{x}({\begingroup\renewcommand\colorMATH{\colorMATHC}\renewcommand\colorSYNTAX{\colorSYNTAXC}{{\color{\colorMATH}\ensuremath{\pS_{4}}}}\endgroup }\sqcup {\begingroup\renewcommand\colorMATH{\colorMATHC}\renewcommand\colorSYNTAX{\colorSYNTAXC}{{\color{\colorMATH}\ensuremath{\pS_{5}}}}\endgroup })) & 
        \cr
\end{array}
      \end{gather*}\endgroup
      And the result holds.  
    \end{subproof}
    \item  {{\color{\colorMATH}\ensuremath{{\begingroup\renewcommand\colorMATH{\colorMATHC}\renewcommand\colorSYNTAX{\colorSYNTAXC}{{\color{\colorMATH}\ensuremath{\pS_{3}}}}\endgroup }={\begingroup\renewcommand\colorMATH{\colorMATHC}\renewcommand\colorSYNTAX{\colorSYNTAXC}{{\color{\colorMATH}\ensuremath{\pS_{4}}}}\endgroup }\sqcap {\begingroup\renewcommand\colorMATH{\colorMATHC}\renewcommand\colorSYNTAX{\colorSYNTAXC}{{\color{\colorMATH}\ensuremath{\pS_{5}}}}\endgroup }}}}
    \begin{subproof}
      By induction hypothesis on {{\color{\colorMATH}\ensuremath{{\begingroup\renewcommand\colorMATH{\colorMATHC}\renewcommand\colorSYNTAX{\colorSYNTAXC}{{\color{\colorMATH}\ensuremath{\pS_{4}}}}\endgroup }}}} and {{\color{\colorMATH}\ensuremath{{\begingroup\renewcommand\colorMATH{\colorMATHC}\renewcommand\colorSYNTAX{\colorSYNTAXC}{{\color{\colorMATH}\ensuremath{\pS_{5}}}}\endgroup }}}}, we know that {{\color{\colorMATH}\ensuremath{{\begingroup\renewcommand\colorMATH{\colorMATHB}\renewcommand\colorSYNTAX{\colorSYNTAXB}{{\color{\colorMATH}\ensuremath{\Distance}}}\endgroup }{\begingroup\renewcommand\colorMATH{\colorMATHC}\renewcommand\colorSYNTAX{\colorSYNTAXC}{{\color{\colorMATH}\ensuremath{\bigcdot}}}\endgroup }(\subst{{\begingroup\renewcommand\colorMATH{\colorMATHB}\renewcommand\colorSYNTAX{\colorSYNTAXB}{{\color{\colorMATH}\ensuremath{\sS_{2}}}}\endgroup }}{x}{\begingroup\renewcommand\colorMATH{\colorMATHC}\renewcommand\colorSYNTAX{\colorSYNTAXC}{{\color{\colorMATH}\ensuremath{\pS_{4}}}}\endgroup }) = {\begingroup\renewcommand\colorMATH{\colorMATHB}\renewcommand\colorSYNTAX{\colorSYNTAXB}{{\color{\colorMATH}\ensuremath{\distance}}}\endgroup }x{\begingroup\renewcommand\colorMATH{\colorMATHC}\renewcommand\colorSYNTAX{\colorSYNTAXC}{{\color{\colorMATH}\ensuremath{\bigcdot}}}\endgroup }({\begingroup\renewcommand\colorMATH{\colorMATHB}\renewcommand\colorSYNTAX{\colorSYNTAXB}{{\color{\colorMATH}\ensuremath{\Distance}}}\endgroup }\mathord{\cdotp }{\begingroup\renewcommand\colorMATH{\colorMATHC}\renewcommand\colorSYNTAX{\colorSYNTAXC}{{\color{\colorMATH}\ensuremath{\pS_{4}}}}\endgroup })}}}, and {{\color{\colorMATH}\ensuremath{{\begingroup\renewcommand\colorMATH{\colorMATHB}\renewcommand\colorSYNTAX{\colorSYNTAXB}{{\color{\colorMATH}\ensuremath{\Distance}}}\endgroup }{\begingroup\renewcommand\colorMATH{\colorMATHC}\renewcommand\colorSYNTAX{\colorSYNTAXC}{{\color{\colorMATH}\ensuremath{\bigcdot}}}\endgroup }(\subst{{\begingroup\renewcommand\colorMATH{\colorMATHB}\renewcommand\colorSYNTAX{\colorSYNTAXB}{{\color{\colorMATH}\ensuremath{\sS_{2}}}}\endgroup }}{x}{\begingroup\renewcommand\colorMATH{\colorMATHC}\renewcommand\colorSYNTAX{\colorSYNTAXC}{{\color{\colorMATH}\ensuremath{\pS_{5}}}}\endgroup }) = {\begingroup\renewcommand\colorMATH{\colorMATHB}\renewcommand\colorSYNTAX{\colorSYNTAXB}{{\color{\colorMATH}\ensuremath{\distance}}}\endgroup }x{\begingroup\renewcommand\colorMATH{\colorMATHC}\renewcommand\colorSYNTAX{\colorSYNTAXC}{{\color{\colorMATH}\ensuremath{\bigcdot}}}\endgroup }({\begingroup\renewcommand\colorMATH{\colorMATHB}\renewcommand\colorSYNTAX{\colorSYNTAXB}{{\color{\colorMATH}\ensuremath{\Distance}}}\endgroup }\mathord{\cdotp }{\begingroup\renewcommand\colorMATH{\colorMATHC}\renewcommand\colorSYNTAX{\colorSYNTAXC}{{\color{\colorMATH}\ensuremath{\pS_{5}}}}\endgroup })}}}. Therefore
      \begingroup\color{\colorMATH}\begin{gather*}
\begin{array}{rclr
        } {\begingroup\renewcommand\colorMATH{\colorMATHB}\renewcommand\colorSYNTAX{\colorSYNTAXB}{{\color{\colorMATH}\ensuremath{\Distance}}}\endgroup }{\begingroup\renewcommand\colorMATH{\colorMATHC}\renewcommand\colorSYNTAX{\colorSYNTAXC}{{\color{\colorMATH}\ensuremath{\bigcdot}}}\endgroup }(\subst{{\begingroup\renewcommand\colorMATH{\colorMATHB}\renewcommand\colorSYNTAX{\colorSYNTAXB}{{\color{\colorMATH}\ensuremath{\sS_{2}}}}\endgroup }}{x}{\begingroup\renewcommand\colorMATH{\colorMATHB}\renewcommand\colorSYNTAX{\colorSYNTAXB}{{\color{\colorMATH}\ensuremath{\sS_{3}}}}\endgroup }) &{}={}& {\begingroup\renewcommand\colorMATH{\colorMATHB}\renewcommand\colorSYNTAX{\colorSYNTAXB}{{\color{\colorMATH}\ensuremath{\Distance}}}\endgroup }{\begingroup\renewcommand\colorMATH{\colorMATHC}\renewcommand\colorSYNTAX{\colorSYNTAXC}{{\color{\colorMATH}\ensuremath{\bigcdot}}}\endgroup }(\subst{{\begingroup\renewcommand\colorMATH{\colorMATHB}\renewcommand\colorSYNTAX{\colorSYNTAXB}{{\color{\colorMATH}\ensuremath{\sS_{2}}}}\endgroup }}{x}({\begingroup\renewcommand\colorMATH{\colorMATHC}\renewcommand\colorSYNTAX{\colorSYNTAXC}{{\color{\colorMATH}\ensuremath{\pS_{4}}}}\endgroup }\sqcap {\begingroup\renewcommand\colorMATH{\colorMATHC}\renewcommand\colorSYNTAX{\colorSYNTAXC}{{\color{\colorMATH}\ensuremath{\pS_{5}}}}\endgroup })) & 
        \cr
\end{array}
      \end{gather*}\endgroup
      And the result holds.  
    \end{subproof}  
  \end{enumerate}
\end{proof}

\lmdistrinst*
\begin{proof}
  We prove: let {{\color{\colorMATH}\ensuremath{{\begingroup\renewcommand\colorMATH{\colorMATHB}\renewcommand\colorSYNTAX{\colorSYNTAXB}{{\color{\colorMATH}\ensuremath{\Distance}}}\endgroup }\mathord{\cdotp }{\begingroup\renewcommand\colorMATH{\colorMATHB}\renewcommand\colorSYNTAX{\colorSYNTAXB}{{\color{\colorMATH}\ensuremath{\sS'}}}\endgroup } = {\begingroup\renewcommand\colorMATH{\colorMATHB}\renewcommand\colorSYNTAX{\colorSYNTAXB}{{\color{\colorMATH}\ensuremath{\distance}}}\endgroup } \in  {\text{sens}}}}} and {{\color{\colorMATH}\ensuremath{x \notin  dom({\begingroup\renewcommand\colorMATH{\colorMATHB}\renewcommand\colorSYNTAX{\colorSYNTAXB}{{\color{\colorMATH}\ensuremath{\sS'}}}\endgroup })\cup  dom({\begingroup\renewcommand\colorMATH{\colorMATHB}\renewcommand\colorSYNTAX{\colorSYNTAXB}{{\color{\colorMATH}\ensuremath{\Distance}}}\endgroup })}}}, then
  {{\color{\colorMATH}\ensuremath{{\begingroup\renewcommand\colorMATH{\colorMATHB}\renewcommand\colorSYNTAX{\colorSYNTAXB}{{\color{\colorMATH}\ensuremath{\Distance}}}\endgroup }([{\begingroup\renewcommand\colorMATH{\colorMATHB}\renewcommand\colorSYNTAX{\colorSYNTAXB}{{\color{\colorMATH}\ensuremath{\sS'}}}\endgroup }/x]\tau ) = {\begingroup\renewcommand\colorMATH{\colorMATHB}\renewcommand\colorSYNTAX{\colorSYNTAXB}{{\color{\colorMATH}\ensuremath{\distance}}}\endgroup }x({\begingroup\renewcommand\colorMATH{\colorMATHB}\renewcommand\colorSYNTAX{\colorSYNTAXB}{{\color{\colorMATH}\ensuremath{\Distance}}}\endgroup }(\tau ))}}}.
  We proceed by induction on {{\color{\colorMATH}\ensuremath{\tau }}}.  
  \begin{enumerate}[ncases]\item  {{\color{\colorMATH}\ensuremath{\tau  \in  \{ {\begingroup\renewcommand\colorMATH{\colorMATHA}\renewcommand\colorSYNTAX{\colorSYNTAXA}{{\color{\colorSYNTAX}\texttt{{\ensuremath{{\mathbb{R}}}}}}}\endgroup }, {\mathbb{B}}, {{\color{\colorSYNTAX}\texttt{unit}}}\} }}} 
    \begin{subproof}
      Trivial as {{\color{\colorMATH}\ensuremath{{\begingroup\renewcommand\colorMATH{\colorMATHB}\renewcommand\colorSYNTAX{\colorSYNTAXB}{{\color{\colorMATH}\ensuremath{\Distance}}}\endgroup }(\tau ) = \tau }}} and {{\color{\colorMATH}\ensuremath{{\begingroup\renewcommand\colorMATH{\colorMATHB}\renewcommand\colorSYNTAX{\colorSYNTAXB}{{\color{\colorMATH}\ensuremath{\sss}}}\endgroup }x(\tau ) = \tau }}}. 
    \end{subproof}
  \item  {{\color{\colorMATH}\ensuremath{\tau =(x\mathrel{:}\tau _{1}) \xrightarrowS {{\begingroup\renewcommand\colorMATH{\colorMATHB}\renewcommand\colorSYNTAX{\colorSYNTAXB}{{\color{\colorMATH}\ensuremath{\sS''}}}\endgroup }} \tau _{2}}}}
    \begin{subproof} 
      \begingroup\color{\colorMATH}\begin{gather*} 
\begin{array}{rclr
        } {\begingroup\renewcommand\colorMATH{\colorMATHB}\renewcommand\colorSYNTAX{\colorSYNTAXB}{{\color{\colorMATH}\ensuremath{\Distance}}}\endgroup }([{\begingroup\renewcommand\colorMATH{\colorMATHB}\renewcommand\colorSYNTAX{\colorSYNTAXB}{{\color{\colorMATH}\ensuremath{\sS'}}}\endgroup }/x]((x\mathrel{:}\tau _{1}\mathord{\cdotp }{\begingroup\renewcommand\colorMATH{\colorMATHB}\renewcommand\colorSYNTAX{\colorSYNTAXB}{{\color{\colorMATH}\ensuremath{\distance'}}}\endgroup }) \xrightarrowS {{\begingroup\renewcommand\colorMATH{\colorMATHB}\renewcommand\colorSYNTAX{\colorSYNTAXB}{{\color{\colorMATH}\ensuremath{\sS''}}}\endgroup }} \tau _{2})) &{}={}& {\begingroup\renewcommand\colorMATH{\colorMATHB}\renewcommand\colorSYNTAX{\colorSYNTAXB}{{\color{\colorMATH}\ensuremath{\Distance}}}\endgroup }((x\mathrel{:}[{\begingroup\renewcommand\colorMATH{\colorMATHB}\renewcommand\colorSYNTAX{\colorSYNTAXB}{{\color{\colorMATH}\ensuremath{\sS'}}}\endgroup }/x]\tau _{1}\mathord{\cdotp }{\begingroup\renewcommand\colorMATH{\colorMATHB}\renewcommand\colorSYNTAX{\colorSYNTAXB}{{\color{\colorMATH}\ensuremath{\distance'}}}\endgroup }) \xrightarrowS {[{\begingroup\renewcommand\colorMATH{\colorMATHB}\renewcommand\colorSYNTAX{\colorSYNTAXB}{{\color{\colorMATH}\ensuremath{\sS'}}}\endgroup }/x]{\begingroup\renewcommand\colorMATH{\colorMATHB}\renewcommand\colorSYNTAX{\colorSYNTAXB}{{\color{\colorMATH}\ensuremath{\sS''}}}\endgroup }} [{\begingroup\renewcommand\colorMATH{\colorMATHB}\renewcommand\colorSYNTAX{\colorSYNTAXB}{{\color{\colorMATH}\ensuremath{\sS'}}}\endgroup }/x]\tau _{2}) &  {{\color{\colorTEXT}\textnormal{ (by def subst.)}}}
        \cr
\end{array}
      \end{gather*}\endgroup
      and the result holds.
    \end{subproof}
  \item  {{\color{\colorMATH}\ensuremath{\tau =(x\mathrel{:}\tau _{1}\mathord{\cdotp }{\begingroup\renewcommand\colorMATH{\colorMATHB}\renewcommand\colorSYNTAX{\colorSYNTAXB}{{\color{\colorMATH}\ensuremath{\distance'}}}\endgroup }) \xrightarrowS {{\begingroup\renewcommand\colorMATH{\colorMATHC}\renewcommand\colorSYNTAX{\colorSYNTAXC}{{\color{\colorMATH}\ensuremath{\pS''}}}\endgroup }} \tau _{2}}}}
    \begin{subproof} 
      \begingroup\color{\colorMATH}\begin{gather*} 
\begin{array}{rclr
        } {\begingroup\renewcommand\colorMATH{\colorMATHB}\renewcommand\colorSYNTAX{\colorSYNTAXB}{{\color{\colorMATH}\ensuremath{\Distance}}}\endgroup }([{\begingroup\renewcommand\colorMATH{\colorMATHB}\renewcommand\colorSYNTAX{\colorSYNTAXB}{{\color{\colorMATH}\ensuremath{\sS'}}}\endgroup }/x]((x\mathrel{:}\tau _{1}\mathord{\cdotp }{\begingroup\renewcommand\colorMATH{\colorMATHB}\renewcommand\colorSYNTAX{\colorSYNTAXB}{{\color{\colorMATH}\ensuremath{\distance'}}}\endgroup }) \xrightarrowS {{\begingroup\renewcommand\colorMATH{\colorMATHC}\renewcommand\colorSYNTAX{\colorSYNTAXC}{{\color{\colorMATH}\ensuremath{\pS''}}}\endgroup }} \tau _{2})) &{}={}& {\begingroup\renewcommand\colorMATH{\colorMATHB}\renewcommand\colorSYNTAX{\colorSYNTAXB}{{\color{\colorMATH}\ensuremath{\Distance}}}\endgroup }((x\mathrel{:}[{\begingroup\renewcommand\colorMATH{\colorMATHB}\renewcommand\colorSYNTAX{\colorSYNTAXB}{{\color{\colorMATH}\ensuremath{\sS'}}}\endgroup }/x]\tau _{1}\mathord{\cdotp }{\begingroup\renewcommand\colorMATH{\colorMATHB}\renewcommand\colorSYNTAX{\colorSYNTAXB}{{\color{\colorMATH}\ensuremath{\distance'}}}\endgroup }) \xrightarrowS {[{\begingroup\renewcommand\colorMATH{\colorMATHB}\renewcommand\colorSYNTAX{\colorSYNTAXB}{{\color{\colorMATH}\ensuremath{\sS'}}}\endgroup }/x]{\begingroup\renewcommand\colorMATH{\colorMATHC}\renewcommand\colorSYNTAX{\colorSYNTAXC}{{\color{\colorMATH}\ensuremath{\pS''}}}\endgroup }} [{\begingroup\renewcommand\colorMATH{\colorMATHB}\renewcommand\colorSYNTAX{\colorSYNTAXB}{{\color{\colorMATH}\ensuremath{\sS'}}}\endgroup }/x]\tau _{2}) &  {{\color{\colorTEXT}\textnormal{ (by def subst.)}}}
        \cr
\end{array}
      \end{gather*}\endgroup
      and the result holds.
    \end{subproof}
  \item  {{\color{\colorMATH}\ensuremath{\tau =\tau _{1} \mathrel{^{{\begingroup\renewcommand\colorMATH{\colorMATHB}\renewcommand\colorSYNTAX{\colorSYNTAXB}{{\color{\colorMATH}\ensuremath{\sS_{1}}}}\endgroup }}\oplus ^{{\begingroup\renewcommand\colorMATH{\colorMATHB}\renewcommand\colorSYNTAX{\colorSYNTAXB}{{\color{\colorMATH}\ensuremath{\sS_{2}}}}\endgroup }}} \tau _{2}}}}
    \begin{subproof} 
      \begingroup\color{\colorMATH}\begin{gather*} 
        \begin{array}{rclr
        } {\begingroup\renewcommand\colorMATH{\colorMATHB}\renewcommand\colorSYNTAX{\colorSYNTAXB}{{\color{\colorMATH}\ensuremath{\Distance}}}\endgroup }([{\begingroup\renewcommand\colorMATH{\colorMATHB}\renewcommand\colorSYNTAX{\colorSYNTAXB}{{\color{\colorMATH}\ensuremath{\sS'}}}\endgroup }/x](\tau _{1} \mathrel{^{{\begingroup\renewcommand\colorMATH{\colorMATHB}\renewcommand\colorSYNTAX{\colorSYNTAXB}{{\color{\colorMATH}\ensuremath{\sS_{1}}}}\endgroup }}\oplus ^{{\begingroup\renewcommand\colorMATH{\colorMATHB}\renewcommand\colorSYNTAX{\colorSYNTAXB}{{\color{\colorMATH}\ensuremath{\sS_{2}}}}\endgroup }}} \tau _{2})) &{}={}& {\begingroup\renewcommand\colorMATH{\colorMATHB}\renewcommand\colorSYNTAX{\colorSYNTAXB}{{\color{\colorMATH}\ensuremath{\Distance}}}\endgroup }([{\begingroup\renewcommand\colorMATH{\colorMATHB}\renewcommand\colorSYNTAX{\colorSYNTAXB}{{\color{\colorMATH}\ensuremath{\sS'}}}\endgroup }/x]\tau _{1} \mathrel{^{[{\begingroup\renewcommand\colorMATH{\colorMATHB}\renewcommand\colorSYNTAX{\colorSYNTAXB}{{\color{\colorMATH}\ensuremath{\sS'}}}\endgroup }/x]{\begingroup\renewcommand\colorMATH{\colorMATHB}\renewcommand\colorSYNTAX{\colorSYNTAXB}{{\color{\colorMATH}\ensuremath{\sS_{1}}}}\endgroup }}\oplus ^{[{\begingroup\renewcommand\colorMATH{\colorMATHB}\renewcommand\colorSYNTAX{\colorSYNTAXB}{{\color{\colorMATH}\ensuremath{\sS'}}}\endgroup }/x]{\begingroup\renewcommand\colorMATH{\colorMATHB}\renewcommand\colorSYNTAX{\colorSYNTAXB}{{\color{\colorMATH}\ensuremath{\sS_{2}}}}\endgroup }}} [{\begingroup\renewcommand\colorMATH{\colorMATHB}\renewcommand\colorSYNTAX{\colorSYNTAXB}{{\color{\colorMATH}\ensuremath{\sS'}}}\endgroup }/x]\tau _{2}) & {{\color{\colorTEXT}\textnormal{ (by def subst.)}}}
        \cr
\end{array}
      \end{gather*}\endgroup
      and the result holds.
    \end{subproof}
    \item  {{\color{\colorMATH}\ensuremath{\tau  = \tau _{1} \mathrel{^{{\begingroup\renewcommand\colorMATH{\colorMATHB}\renewcommand\colorSYNTAX{\colorSYNTAXB}{{\color{\colorMATH}\ensuremath{\sS_{1}}}}\endgroup }}\&^{{\begingroup\renewcommand\colorMATH{\colorMATHB}\renewcommand\colorSYNTAX{\colorSYNTAXB}{{\color{\colorMATH}\ensuremath{\sS_{2}}}}\endgroup }}} \tau _{2}}}}
      \begin{subproof} 
        Analogous to {{\color{\colorMATH}\ensuremath{\tau =\tau _{1} \mathrel{^{{\begingroup\renewcommand\colorMATH{\colorMATHB}\renewcommand\colorSYNTAX{\colorSYNTAXB}{{\color{\colorMATH}\ensuremath{\sS_{1}}}}\endgroup }}\oplus ^{{\begingroup\renewcommand\colorMATH{\colorMATHB}\renewcommand\colorSYNTAX{\colorSYNTAXB}{{\color{\colorMATH}\ensuremath{\sS_{2}}}}\endgroup }}} \tau _{2}}}} case.
      \end{subproof}
    \end{enumerate}
\end{proof}

\begin{lemma}
  \label{lm:subtypinginst}
  If {{\color{\colorMATH}\ensuremath{\tau  <: \tau '}}} then {{\color{\colorMATH}\ensuremath{{\begingroup\renewcommand\colorMATH{\colorMATHB}\renewcommand\colorSYNTAX{\colorSYNTAXB}{{\color{\colorMATH}\ensuremath{\sS}}}\endgroup }(\tau ) <: {\begingroup\renewcommand\colorMATH{\colorMATHB}\renewcommand\colorSYNTAX{\colorSYNTAXB}{{\color{\colorMATH}\ensuremath{\sS}}}\endgroup }(\tau ')}}}
\end{lemma}
\begin{proof}
  By induction on {{\color{\colorMATH}\ensuremath{\tau }}}:
  \begin{enumerate}[ncases]\item  {{\color{\colorMATH}\ensuremath{\tau  = {\begingroup\renewcommand\colorMATH{\colorMATHA}\renewcommand\colorSYNTAX{\colorSYNTAXA}{{\color{\colorSYNTAX}\texttt{{\ensuremath{{\mathbb{R}}}}}}}\endgroup }}}}
    \begin{subproof} 
      Then {{\color{\colorMATH}\ensuremath{\tau ' = {\begingroup\renewcommand\colorMATH{\colorMATHA}\renewcommand\colorSYNTAX{\colorSYNTAXA}{{\color{\colorSYNTAX}\texttt{{\ensuremath{{\mathbb{R}}}}}}}\endgroup }}}} so the result is trivial.
    \end{subproof}
  \item  {{\color{\colorMATH}\ensuremath{\tau  = {\mathbb{B}}}}}
    \begin{subproof} 
      Then {{\color{\colorMATH}\ensuremath{\tau ' = {\mathbb{B}}}}} so the result is trivial.
    \end{subproof}
  \item  {{\color{\colorMATH}\ensuremath{\tau  = {{\color{\colorSYNTAX}\texttt{unit}}}}}} 
    \begin{subproof} 
      Then {{\color{\colorMATH}\ensuremath{\tau ' = {{\color{\colorSYNTAX}\texttt{unit}}}}}} so the result is trivial.
    \end{subproof}
  \item  {{\color{\colorMATH}\ensuremath{\tau  = (x\mathrel{:}\tau _{1}\mathord{\cdotp }{\begingroup\renewcommand\colorMATH{\colorMATHB}\renewcommand\colorSYNTAX{\colorSYNTAXB}{{\color{\colorMATH}\ensuremath{\distance}}}\endgroup }) \xrightarrowS {{\begingroup\renewcommand\colorMATH{\colorMATHB}\renewcommand\colorSYNTAX{\colorSYNTAXB}{{\color{\colorMATH}\ensuremath{\sS'}}}\endgroup }} \tau _{2}}}} 
    \begin{subproof} 
      Then {{\color{\colorMATH}\ensuremath{\tau ' = (x\mathrel{:}\tau '_{1}\mathord{\cdotp }{\begingroup\renewcommand\colorMATH{\colorMATHB}\renewcommand\colorSYNTAX{\colorSYNTAXB}{{\color{\colorMATH}\ensuremath{\distance'}}}\endgroup }) \xrightarrowS {{\begingroup\renewcommand\colorMATH{\colorMATHB}\renewcommand\colorSYNTAX{\colorSYNTAXB}{{\color{\colorMATH}\ensuremath{\sS''}}}\endgroup }} \tau '_{2}}}} such that {{\color{\colorMATH}\ensuremath{{\begingroup\renewcommand\colorMATH{\colorMATHB}\renewcommand\colorSYNTAX{\colorSYNTAXB}{{\color{\colorMATH}\ensuremath{\distance'}}}\endgroup } <: {\begingroup\renewcommand\colorMATH{\colorMATHB}\renewcommand\colorSYNTAX{\colorSYNTAXB}{{\color{\colorMATH}\ensuremath{\distance}}}\endgroup }, \tau '_{1} <: \tau _{1}, {\begingroup\renewcommand\colorMATH{\colorMATHB}\renewcommand\colorSYNTAX{\colorSYNTAXB}{{\color{\colorMATH}\ensuremath{\sS'}}}\endgroup } <: {\begingroup\renewcommand\colorMATH{\colorMATHB}\renewcommand\colorSYNTAX{\colorSYNTAXB}{{\color{\colorMATH}\ensuremath{\sS''}}}\endgroup }}}}, and {{\color{\colorMATH}\ensuremath{\tau _{2} <: \tau '_{2}}}}.
      But  {{\color{\colorMATH}\ensuremath{{\begingroup\renewcommand\colorMATH{\colorMATHB}\renewcommand\colorSYNTAX{\colorSYNTAXB}{{\color{\colorMATH}\ensuremath{\sS}}}\endgroup }((x\mathrel{:}\tau _{1}\mathord{\cdotp }{\begingroup\renewcommand\colorMATH{\colorMATHB}\renewcommand\colorSYNTAX{\colorSYNTAXB}{{\color{\colorMATH}\ensuremath{\distance}}}\endgroup }) \xrightarrowS {{\begingroup\renewcommand\colorMATH{\colorMATHB}\renewcommand\colorSYNTAX{\colorSYNTAXB}{{\color{\colorMATH}\ensuremath{\sS'}}}\endgroup }} \tau _{2}) = (x\mathrel{:}{\begingroup\renewcommand\colorMATH{\colorMATHB}\renewcommand\colorSYNTAX{\colorSYNTAXB}{{\color{\colorMATH}\ensuremath{\sS}}}\endgroup }(\tau _{1})\mathord{\cdotp }{\begingroup\renewcommand\colorMATH{\colorMATHB}\renewcommand\colorSYNTAX{\colorSYNTAXB}{{\color{\colorMATH}\ensuremath{\distance}}}\endgroup }) \xrightarrowS {{\begingroup\renewcommand\colorMATH{\colorMATHB}\renewcommand\colorSYNTAX{\colorSYNTAXB}{{\color{\colorMATH}\ensuremath{\sS}}}\endgroup } \mathord{\cdotp } {\begingroup\renewcommand\colorMATH{\colorMATHB}\renewcommand\colorSYNTAX{\colorSYNTAXB}{{\color{\colorMATH}\ensuremath{\sS'}}}\endgroup }} {\begingroup\renewcommand\colorMATH{\colorMATHB}\renewcommand\colorSYNTAX{\colorSYNTAXB}{{\color{\colorMATH}\ensuremath{\sS}}}\endgroup }(\tau _{2})}}}, then by
      induction hypotheses {{\color{\colorMATH}\ensuremath{{\begingroup\renewcommand\colorMATH{\colorMATHB}\renewcommand\colorSYNTAX{\colorSYNTAXB}{{\color{\colorMATH}\ensuremath{\sS}}}\endgroup }(\tau '_{1}) <: {\begingroup\renewcommand\colorMATH{\colorMATHB}\renewcommand\colorSYNTAX{\colorSYNTAXB}{{\color{\colorMATH}\ensuremath{\sS}}}\endgroup }(\tau _{1})}}}, and {{\color{\colorMATH}\ensuremath{{\begingroup\renewcommand\colorMATH{\colorMATHB}\renewcommand\colorSYNTAX{\colorSYNTAXB}{{\color{\colorMATH}\ensuremath{\sS}}}\endgroup }(\tau _{2}) <: {\begingroup\renewcommand\colorMATH{\colorMATHB}\renewcommand\colorSYNTAX{\colorSYNTAXB}{{\color{\colorMATH}\ensuremath{\sS}}}\endgroup }(\tau '_{2})}}}. Also by Lemma~\ref{lm:dot-subt},
      {{\color{\colorMATH}\ensuremath{{\begingroup\renewcommand\colorMATH{\colorMATHB}\renewcommand\colorSYNTAX{\colorSYNTAXB}{{\color{\colorMATH}\ensuremath{\sS}}}\endgroup }\mathord{\cdotp }{\begingroup\renewcommand\colorMATH{\colorMATHB}\renewcommand\colorSYNTAX{\colorSYNTAXB}{{\color{\colorMATH}\ensuremath{\sS'}}}\endgroup } <: {\begingroup\renewcommand\colorMATH{\colorMATHB}\renewcommand\colorSYNTAX{\colorSYNTAXB}{{\color{\colorMATH}\ensuremath{\sS}}}\endgroup }\mathord{\cdotp }{\begingroup\renewcommand\colorMATH{\colorMATHB}\renewcommand\colorSYNTAX{\colorSYNTAXB}{{\color{\colorMATH}\ensuremath{\sS''}}}\endgroup }}}}, therefore\\ 
      {{\color{\colorMATH}\ensuremath{{\begingroup\renewcommand\colorMATH{\colorMATHB}\renewcommand\colorSYNTAX{\colorSYNTAXB}{{\color{\colorMATH}\ensuremath{\sS}}}\endgroup }((x\mathrel{:}\tau _{1}\mathord{\cdotp }{\begingroup\renewcommand\colorMATH{\colorMATHB}\renewcommand\colorSYNTAX{\colorSYNTAXB}{{\color{\colorMATH}\ensuremath{\distance}}}\endgroup }) \xrightarrowS {{\begingroup\renewcommand\colorMATH{\colorMATHB}\renewcommand\colorSYNTAX{\colorSYNTAXB}{{\color{\colorMATH}\ensuremath{\sS'}}}\endgroup }} \tau _{2}) = (x\mathrel{:}{\begingroup\renewcommand\colorMATH{\colorMATHB}\renewcommand\colorSYNTAX{\colorSYNTAXB}{{\color{\colorMATH}\ensuremath{\sS}}}\endgroup }(\tau _{1})\mathord{\cdotp }{\begingroup\renewcommand\colorMATH{\colorMATHB}\renewcommand\colorSYNTAX{\colorSYNTAXB}{{\color{\colorMATH}\ensuremath{\distance}}}\endgroup }) \xrightarrowS {{\begingroup\renewcommand\colorMATH{\colorMATHB}\renewcommand\colorSYNTAX{\colorSYNTAXB}{{\color{\colorMATH}\ensuremath{\sS}}}\endgroup } \mathord{\cdotp } {\begingroup\renewcommand\colorMATH{\colorMATHB}\renewcommand\colorSYNTAX{\colorSYNTAXB}{{\color{\colorMATH}\ensuremath{\sS'}}}\endgroup }} {\begingroup\renewcommand\colorMATH{\colorMATHB}\renewcommand\colorSYNTAX{\colorSYNTAXB}{{\color{\colorMATH}\ensuremath{\sS}}}\endgroup }(\tau _{2}) <: (x\mathrel{:}{\begingroup\renewcommand\colorMATH{\colorMATHB}\renewcommand\colorSYNTAX{\colorSYNTAXB}{{\color{\colorMATH}\ensuremath{\sS}}}\endgroup }(\tau '_{1})\mathord{\cdotp }{\begingroup\renewcommand\colorMATH{\colorMATHB}\renewcommand\colorSYNTAX{\colorSYNTAXB}{{\color{\colorMATH}\ensuremath{\distance'}}}\endgroup }) \xrightarrowS {{\begingroup\renewcommand\colorMATH{\colorMATHB}\renewcommand\colorSYNTAX{\colorSYNTAXB}{{\color{\colorMATH}\ensuremath{\sS}}}\endgroup } \mathord{\cdotp } {\begingroup\renewcommand\colorMATH{\colorMATHB}\renewcommand\colorSYNTAX{\colorSYNTAXB}{{\color{\colorMATH}\ensuremath{\sS''}}}\endgroup }} {\begingroup\renewcommand\colorMATH{\colorMATHB}\renewcommand\colorSYNTAX{\colorSYNTAXB}{{\color{\colorMATH}\ensuremath{\sS}}}\endgroup }(\tau '_{2}) = {\begingroup\renewcommand\colorMATH{\colorMATHB}\renewcommand\colorSYNTAX{\colorSYNTAXB}{{\color{\colorMATH}\ensuremath{\sS}}}\endgroup }((x\mathrel{:}\tau '_{1}\mathord{\cdotp }{\begingroup\renewcommand\colorMATH{\colorMATHB}\renewcommand\colorSYNTAX{\colorSYNTAXB}{{\color{\colorMATH}\ensuremath{\distance'}}}\endgroup }) \xrightarrowS {{\begingroup\renewcommand\colorMATH{\colorMATHB}\renewcommand\colorSYNTAX{\colorSYNTAXB}{{\color{\colorMATH}\ensuremath{\sS''}}}\endgroup }} \tau '_{2})}}} and the result holds.
    \end{subproof}
  \item  {{\color{\colorMATH}\ensuremath{\tau  = (x\mathrel{:}\tau _{1}\mathord{\cdotp }{\begingroup\renewcommand\colorMATH{\colorMATHB}\renewcommand\colorSYNTAX{\colorSYNTAXB}{{\color{\colorMATH}\ensuremath{\distance}}}\endgroup }) \xrightarrowP {{\begingroup\renewcommand\colorMATH{\colorMATHC}\renewcommand\colorSYNTAX{\colorSYNTAXC}{{\color{\colorMATH}\ensuremath{\pS'}}}\endgroup }} \tau _{2}}}} 
    \begin{subproof} 
      Then {{\color{\colorMATH}\ensuremath{\tau ' = (x\mathrel{:}\tau '_{1}\mathord{\cdotp }{\begingroup\renewcommand\colorMATH{\colorMATHB}\renewcommand\colorSYNTAX{\colorSYNTAXB}{{\color{\colorMATH}\ensuremath{\distance'}}}\endgroup }) \xrightarrowP {{\begingroup\renewcommand\colorMATH{\colorMATHC}\renewcommand\colorSYNTAX{\colorSYNTAXC}{{\color{\colorMATH}\ensuremath{\pS''}}}\endgroup }} \tau '_{2}}}} such that {{\color{\colorMATH}\ensuremath{{\begingroup\renewcommand\colorMATH{\colorMATHB}\renewcommand\colorSYNTAX{\colorSYNTAXB}{{\color{\colorMATH}\ensuremath{\distance'}}}\endgroup } <: {\begingroup\renewcommand\colorMATH{\colorMATHB}\renewcommand\colorSYNTAX{\colorSYNTAXB}{{\color{\colorMATH}\ensuremath{\distance}}}\endgroup }, \tau '_{1} <: \tau _{1}, {\begingroup\renewcommand\colorMATH{\colorMATHC}\renewcommand\colorSYNTAX{\colorSYNTAXC}{{\color{\colorMATH}\ensuremath{\pS'}}}\endgroup } <: {\begingroup\renewcommand\colorMATH{\colorMATHC}\renewcommand\colorSYNTAX{\colorSYNTAXC}{{\color{\colorMATH}\ensuremath{\pS''}}}\endgroup }}}}, and {{\color{\colorMATH}\ensuremath{\tau _{2} <: \tau '_{2}}}}.
      But  {{\color{\colorMATH}\ensuremath{{\begingroup\renewcommand\colorMATH{\colorMATHB}\renewcommand\colorSYNTAX{\colorSYNTAXB}{{\color{\colorMATH}\ensuremath{\sS}}}\endgroup }((x\mathrel{:}\tau _{1}\mathord{\cdotp }{\begingroup\renewcommand\colorMATH{\colorMATHB}\renewcommand\colorSYNTAX{\colorSYNTAXB}{{\color{\colorMATH}\ensuremath{\distance}}}\endgroup }) \xrightarrowP {{\begingroup\renewcommand\colorMATH{\colorMATHC}\renewcommand\colorSYNTAX{\colorSYNTAXC}{{\color{\colorMATH}\ensuremath{\pS'}}}\endgroup }} \tau _{2}) = (x\mathrel{:}{\begingroup\renewcommand\colorMATH{\colorMATHB}\renewcommand\colorSYNTAX{\colorSYNTAXB}{{\color{\colorMATH}\ensuremath{\sS}}}\endgroup }(\tau _{1})\mathord{\cdotp }{\begingroup\renewcommand\colorMATH{\colorMATHB}\renewcommand\colorSYNTAX{\colorSYNTAXB}{{\color{\colorMATH}\ensuremath{\distance}}}\endgroup }) \xrightarrowP {{\begingroup\renewcommand\colorMATH{\colorMATHB}\renewcommand\colorSYNTAX{\colorSYNTAXB}{{\color{\colorMATH}\ensuremath{\sS}}}\endgroup } \mathord{\cdotp } {\begingroup\renewcommand\colorMATH{\colorMATHC}\renewcommand\colorSYNTAX{\colorSYNTAXC}{{\color{\colorMATH}\ensuremath{\pS'}}}\endgroup }} {\begingroup\renewcommand\colorMATH{\colorMATHB}\renewcommand\colorSYNTAX{\colorSYNTAXB}{{\color{\colorMATH}\ensuremath{\sS}}}\endgroup }(\tau _{2})}}}, then by
      induction hypotheses {{\color{\colorMATH}\ensuremath{{\begingroup\renewcommand\colorMATH{\colorMATHB}\renewcommand\colorSYNTAX{\colorSYNTAXB}{{\color{\colorMATH}\ensuremath{\sS}}}\endgroup }(\tau '_{1}) <: {\begingroup\renewcommand\colorMATH{\colorMATHB}\renewcommand\colorSYNTAX{\colorSYNTAXB}{{\color{\colorMATH}\ensuremath{\sS}}}\endgroup }(\tau _{1})}}}, and {{\color{\colorMATH}\ensuremath{{\begingroup\renewcommand\colorMATH{\colorMATHB}\renewcommand\colorSYNTAX{\colorSYNTAXB}{{\color{\colorMATH}\ensuremath{\sS}}}\endgroup }(\tau _{2}) <: {\begingroup\renewcommand\colorMATH{\colorMATHB}\renewcommand\colorSYNTAX{\colorSYNTAXB}{{\color{\colorMATH}\ensuremath{\sS}}}\endgroup }(\tau '_{2})}}}. Also by Lemma~\ref{lm:dot-subtp},
      {{\color{\colorMATH}\ensuremath{{\begingroup\renewcommand\colorMATH{\colorMATHB}\renewcommand\colorSYNTAX{\colorSYNTAXB}{{\color{\colorMATH}\ensuremath{\sS}}}\endgroup }\mathord{\cdotp }{\begingroup\renewcommand\colorMATH{\colorMATHC}\renewcommand\colorSYNTAX{\colorSYNTAXC}{{\color{\colorMATH}\ensuremath{\pS'}}}\endgroup } <: {\begingroup\renewcommand\colorMATH{\colorMATHB}\renewcommand\colorSYNTAX{\colorSYNTAXB}{{\color{\colorMATH}\ensuremath{\sS}}}\endgroup }\mathord{\cdotp }{\begingroup\renewcommand\colorMATH{\colorMATHC}\renewcommand\colorSYNTAX{\colorSYNTAXC}{{\color{\colorMATH}\ensuremath{\pS''}}}\endgroup }}}}, therefore\\ 
      {{\color{\colorMATH}\ensuremath{{\begingroup\renewcommand\colorMATH{\colorMATHB}\renewcommand\colorSYNTAX{\colorSYNTAXB}{{\color{\colorMATH}\ensuremath{\sS}}}\endgroup }((x\mathrel{:}\tau _{1}\mathord{\cdotp }{\begingroup\renewcommand\colorMATH{\colorMATHB}\renewcommand\colorSYNTAX{\colorSYNTAXB}{{\color{\colorMATH}\ensuremath{\distance}}}\endgroup }) \xrightarrowP {{\begingroup\renewcommand\colorMATH{\colorMATHC}\renewcommand\colorSYNTAX{\colorSYNTAXC}{{\color{\colorMATH}\ensuremath{\pS'}}}\endgroup }} \tau _{2}) = (x\mathrel{:}{\begingroup\renewcommand\colorMATH{\colorMATHB}\renewcommand\colorSYNTAX{\colorSYNTAXB}{{\color{\colorMATH}\ensuremath{\sS}}}\endgroup }(\tau _{1})\mathord{\cdotp }{\begingroup\renewcommand\colorMATH{\colorMATHB}\renewcommand\colorSYNTAX{\colorSYNTAXB}{{\color{\colorMATH}\ensuremath{\distance}}}\endgroup }) \xrightarrowP {{\begingroup\renewcommand\colorMATH{\colorMATHB}\renewcommand\colorSYNTAX{\colorSYNTAXB}{{\color{\colorMATH}\ensuremath{\sS}}}\endgroup } \mathord{\cdotp } {\begingroup\renewcommand\colorMATH{\colorMATHC}\renewcommand\colorSYNTAX{\colorSYNTAXC}{{\color{\colorMATH}\ensuremath{\pS'}}}\endgroup }} {\begingroup\renewcommand\colorMATH{\colorMATHB}\renewcommand\colorSYNTAX{\colorSYNTAXB}{{\color{\colorMATH}\ensuremath{\sS}}}\endgroup }(\tau _{2}) <: (x\mathrel{:}{\begingroup\renewcommand\colorMATH{\colorMATHB}\renewcommand\colorSYNTAX{\colorSYNTAXB}{{\color{\colorMATH}\ensuremath{\sS}}}\endgroup }(\tau '_{1})\mathord{\cdotp }{\begingroup\renewcommand\colorMATH{\colorMATHB}\renewcommand\colorSYNTAX{\colorSYNTAXB}{{\color{\colorMATH}\ensuremath{\distance'}}}\endgroup }) \xrightarrowP {{\begingroup\renewcommand\colorMATH{\colorMATHB}\renewcommand\colorSYNTAX{\colorSYNTAXB}{{\color{\colorMATH}\ensuremath{\sS}}}\endgroup } \mathord{\cdotp } {\begingroup\renewcommand\colorMATH{\colorMATHC}\renewcommand\colorSYNTAX{\colorSYNTAXC}{{\color{\colorMATH}\ensuremath{\pS''}}}\endgroup }} {\begingroup\renewcommand\colorMATH{\colorMATHB}\renewcommand\colorSYNTAX{\colorSYNTAXB}{{\color{\colorMATH}\ensuremath{\sS}}}\endgroup }(\tau '_{2}) = {\begingroup\renewcommand\colorMATH{\colorMATHB}\renewcommand\colorSYNTAX{\colorSYNTAXB}{{\color{\colorMATH}\ensuremath{\sS}}}\endgroup }((x\mathrel{:}\tau '_{1}\mathord{\cdotp }{\begingroup\renewcommand\colorMATH{\colorMATHB}\renewcommand\colorSYNTAX{\colorSYNTAXB}{{\color{\colorMATH}\ensuremath{\distance'}}}\endgroup }) \xrightarrowP {{\begingroup\renewcommand\colorMATH{\colorMATHC}\renewcommand\colorSYNTAX{\colorSYNTAXC}{{\color{\colorMATH}\ensuremath{\pS''}}}\endgroup }} \tau '_{2})}}} and the result holds.
    \end{subproof}
  \item  {{\color{\colorMATH}\ensuremath{\tau  = \tau _{1} \mathrel{^{{\begingroup\renewcommand\colorMATH{\colorMATHB}\renewcommand\colorSYNTAX{\colorSYNTAXB}{{\color{\colorMATH}\ensuremath{\sS_{1}}}}\endgroup }}\oplus ^{{\begingroup\renewcommand\colorMATH{\colorMATHB}\renewcommand\colorSYNTAX{\colorSYNTAXB}{{\color{\colorMATH}\ensuremath{\sS_{2}}}}\endgroup }}} \tau _{2}}}} 
    \begin{subproof} 
      Then {{\color{\colorMATH}\ensuremath{\tau ' = \tau '_{1} \mathrel{^{{\begingroup\renewcommand\colorMATH{\colorMATHB}\renewcommand\colorSYNTAX{\colorSYNTAXB}{{\color{\colorMATH}\ensuremath{\sS'_{1}}}}\endgroup }}\oplus ^{{\begingroup\renewcommand\colorMATH{\colorMATHB}\renewcommand\colorSYNTAX{\colorSYNTAXB}{{\color{\colorMATH}\ensuremath{\sS'_{2}}}}\endgroup }}} \tau '_{2}}}} such that {{\color{\colorMATH}\ensuremath{\tau _{1} <: \tau '_{1}, {\begingroup\renewcommand\colorMATH{\colorMATHB}\renewcommand\colorSYNTAX{\colorSYNTAXB}{{\color{\colorMATH}\ensuremath{\sS_{1}}}}\endgroup } <: {\begingroup\renewcommand\colorMATH{\colorMATHB}\renewcommand\colorSYNTAX{\colorSYNTAXB}{{\color{\colorMATH}\ensuremath{\sS'_{1}}}}\endgroup }, {\begingroup\renewcommand\colorMATH{\colorMATHB}\renewcommand\colorSYNTAX{\colorSYNTAXB}{{\color{\colorMATH}\ensuremath{\sS'_{2}}}}\endgroup } <: {\begingroup\renewcommand\colorMATH{\colorMATHB}\renewcommand\colorSYNTAX{\colorSYNTAXB}{{\color{\colorMATH}\ensuremath{\sS'_{2}}}}\endgroup }}}}, and {{\color{\colorMATH}\ensuremath{\tau _{2} <: \tau '_{2}}}}.
      But  {{\color{\colorMATH}\ensuremath{{\begingroup\renewcommand\colorMATH{\colorMATHB}\renewcommand\colorSYNTAX{\colorSYNTAXB}{{\color{\colorMATH}\ensuremath{\sS}}}\endgroup }(\tau _{1} \mathrel{^{{\begingroup\renewcommand\colorMATH{\colorMATHB}\renewcommand\colorSYNTAX{\colorSYNTAXB}{{\color{\colorMATH}\ensuremath{\sS_{1}}}}\endgroup }}\oplus ^{{\begingroup\renewcommand\colorMATH{\colorMATHB}\renewcommand\colorSYNTAX{\colorSYNTAXB}{{\color{\colorMATH}\ensuremath{\sS_{2}}}}\endgroup }}} \tau _{2}) = {\begingroup\renewcommand\colorMATH{\colorMATHB}\renewcommand\colorSYNTAX{\colorSYNTAXB}{{\color{\colorMATH}\ensuremath{\sS}}}\endgroup }(\tau _{1}) \mathrel{^{{\begingroup\renewcommand\colorMATH{\colorMATHB}\renewcommand\colorSYNTAX{\colorSYNTAXB}{{\color{\colorMATH}\ensuremath{\sS}}}\endgroup } \mathord{\cdotp } {\begingroup\renewcommand\colorMATH{\colorMATHB}\renewcommand\colorSYNTAX{\colorSYNTAXB}{{\color{\colorMATH}\ensuremath{\sS_{1}}}}\endgroup }}\oplus ^{{\begingroup\renewcommand\colorMATH{\colorMATHB}\renewcommand\colorSYNTAX{\colorSYNTAXB}{{\color{\colorMATH}\ensuremath{\sS}}}\endgroup } \mathord{\cdotp } {\begingroup\renewcommand\colorMATH{\colorMATHB}\renewcommand\colorSYNTAX{\colorSYNTAXB}{{\color{\colorMATH}\ensuremath{\sS_{2}}}}\endgroup }}} {\begingroup\renewcommand\colorMATH{\colorMATHB}\renewcommand\colorSYNTAX{\colorSYNTAXB}{{\color{\colorMATH}\ensuremath{\sS}}}\endgroup }(\tau _{2})}}}, then by
      induction hypotheses {{\color{\colorMATH}\ensuremath{{\begingroup\renewcommand\colorMATH{\colorMATHB}\renewcommand\colorSYNTAX{\colorSYNTAXB}{{\color{\colorMATH}\ensuremath{\sS}}}\endgroup }(\tau _{1}) <: {\begingroup\renewcommand\colorMATH{\colorMATHB}\renewcommand\colorSYNTAX{\colorSYNTAXB}{{\color{\colorMATH}\ensuremath{\sS}}}\endgroup }(\tau '_{1})}}}, and {{\color{\colorMATH}\ensuremath{{\begingroup\renewcommand\colorMATH{\colorMATHB}\renewcommand\colorSYNTAX{\colorSYNTAXB}{{\color{\colorMATH}\ensuremath{\sS}}}\endgroup }(\tau _{2}) <: {\begingroup\renewcommand\colorMATH{\colorMATHB}\renewcommand\colorSYNTAX{\colorSYNTAXB}{{\color{\colorMATH}\ensuremath{\sS}}}\endgroup }(\tau '_{2})}}}. Also by Lemma~\ref{lm:dot-subt},
      {{\color{\colorMATH}\ensuremath{{\begingroup\renewcommand\colorMATH{\colorMATHB}\renewcommand\colorSYNTAX{\colorSYNTAXB}{{\color{\colorMATH}\ensuremath{\sS}}}\endgroup }\mathord{\cdotp }{\begingroup\renewcommand\colorMATH{\colorMATHB}\renewcommand\colorSYNTAX{\colorSYNTAXB}{{\color{\colorMATH}\ensuremath{\sS_{1}}}}\endgroup } <: {\begingroup\renewcommand\colorMATH{\colorMATHB}\renewcommand\colorSYNTAX{\colorSYNTAXB}{{\color{\colorMATH}\ensuremath{\sS}}}\endgroup }\mathord{\cdotp }{\begingroup\renewcommand\colorMATH{\colorMATHB}\renewcommand\colorSYNTAX{\colorSYNTAXB}{{\color{\colorMATH}\ensuremath{\sS'_{1}}}}\endgroup }}}} and {{\color{\colorMATH}\ensuremath{{\begingroup\renewcommand\colorMATH{\colorMATHB}\renewcommand\colorSYNTAX{\colorSYNTAXB}{{\color{\colorMATH}\ensuremath{\sS}}}\endgroup }\mathord{\cdotp }{\begingroup\renewcommand\colorMATH{\colorMATHB}\renewcommand\colorSYNTAX{\colorSYNTAXB}{{\color{\colorMATH}\ensuremath{\sS_{2}}}}\endgroup } <: {\begingroup\renewcommand\colorMATH{\colorMATHB}\renewcommand\colorSYNTAX{\colorSYNTAXB}{{\color{\colorMATH}\ensuremath{\sS}}}\endgroup }\mathord{\cdotp }{\begingroup\renewcommand\colorMATH{\colorMATHB}\renewcommand\colorSYNTAX{\colorSYNTAXB}{{\color{\colorMATH}\ensuremath{\sS'_{2}}}}\endgroup }}}}, therefore\\ 
      {{\color{\colorMATH}\ensuremath{{\begingroup\renewcommand\colorMATH{\colorMATHB}\renewcommand\colorSYNTAX{\colorSYNTAXB}{{\color{\colorMATH}\ensuremath{\sS}}}\endgroup }(\tau _{1} \mathrel{^{{\begingroup\renewcommand\colorMATH{\colorMATHB}\renewcommand\colorSYNTAX{\colorSYNTAXB}{{\color{\colorMATH}\ensuremath{\sS_{1}}}}\endgroup }}\oplus ^{{\begingroup\renewcommand\colorMATH{\colorMATHB}\renewcommand\colorSYNTAX{\colorSYNTAXB}{{\color{\colorMATH}\ensuremath{\sS_{2}}}}\endgroup }}} \tau _{2}) = {\begingroup\renewcommand\colorMATH{\colorMATHB}\renewcommand\colorSYNTAX{\colorSYNTAXB}{{\color{\colorMATH}\ensuremath{\sS}}}\endgroup }(\tau _{1}) \mathrel{^{{\begingroup\renewcommand\colorMATH{\colorMATHB}\renewcommand\colorSYNTAX{\colorSYNTAXB}{{\color{\colorMATH}\ensuremath{\sS}}}\endgroup } \mathord{\cdotp } {\begingroup\renewcommand\colorMATH{\colorMATHB}\renewcommand\colorSYNTAX{\colorSYNTAXB}{{\color{\colorMATH}\ensuremath{\sS_{1}}}}\endgroup }}\oplus ^{{\begingroup\renewcommand\colorMATH{\colorMATHB}\renewcommand\colorSYNTAX{\colorSYNTAXB}{{\color{\colorMATH}\ensuremath{\sS}}}\endgroup } \mathord{\cdotp } {\begingroup\renewcommand\colorMATH{\colorMATHB}\renewcommand\colorSYNTAX{\colorSYNTAXB}{{\color{\colorMATH}\ensuremath{\sS_{2}}}}\endgroup }}} {\begingroup\renewcommand\colorMATH{\colorMATHB}\renewcommand\colorSYNTAX{\colorSYNTAXB}{{\color{\colorMATH}\ensuremath{\sS}}}\endgroup }(\tau _{2}) <: {\begingroup\renewcommand\colorMATH{\colorMATHB}\renewcommand\colorSYNTAX{\colorSYNTAXB}{{\color{\colorMATH}\ensuremath{\sS}}}\endgroup }(\tau '_{1}) \mathrel{^{{\begingroup\renewcommand\colorMATH{\colorMATHB}\renewcommand\colorSYNTAX{\colorSYNTAXB}{{\color{\colorMATH}\ensuremath{\sS}}}\endgroup } \mathord{\cdotp } {\begingroup\renewcommand\colorMATH{\colorMATHB}\renewcommand\colorSYNTAX{\colorSYNTAXB}{{\color{\colorMATH}\ensuremath{\sS'_{1}}}}\endgroup }}\oplus ^{{\begingroup\renewcommand\colorMATH{\colorMATHB}\renewcommand\colorSYNTAX{\colorSYNTAXB}{{\color{\colorMATH}\ensuremath{\sS}}}\endgroup } \mathord{\cdotp } {\begingroup\renewcommand\colorMATH{\colorMATHB}\renewcommand\colorSYNTAX{\colorSYNTAXB}{{\color{\colorMATH}\ensuremath{\sS'_{2}}}}\endgroup }}} {\begingroup\renewcommand\colorMATH{\colorMATHB}\renewcommand\colorSYNTAX{\colorSYNTAXB}{{\color{\colorMATH}\ensuremath{\sS}}}\endgroup }(\tau '_{2}) = {\begingroup\renewcommand\colorMATH{\colorMATHB}\renewcommand\colorSYNTAX{\colorSYNTAXB}{{\color{\colorMATH}\ensuremath{\sS}}}\endgroup }(\tau '_{1} \mathrel{^{{\begingroup\renewcommand\colorMATH{\colorMATHB}\renewcommand\colorSYNTAX{\colorSYNTAXB}{{\color{\colorMATH}\ensuremath{\sS'_{1}}}}\endgroup }}\oplus ^{{\begingroup\renewcommand\colorMATH{\colorMATHB}\renewcommand\colorSYNTAX{\colorSYNTAXB}{{\color{\colorMATH}\ensuremath{\sS'_{2}}}}\endgroup }}} \tau '_{2})}}} and the result holds.
    \end{subproof}
  \item  {{\color{\colorMATH}\ensuremath{\tau  = \tau _{1} \mathrel{^{{\begingroup\renewcommand\colorMATH{\colorMATHB}\renewcommand\colorSYNTAX{\colorSYNTAXB}{{\color{\colorMATH}\ensuremath{\sS_{1}}}}\endgroup }}\&^{{\begingroup\renewcommand\colorMATH{\colorMATHB}\renewcommand\colorSYNTAX{\colorSYNTAXB}{{\color{\colorMATH}\ensuremath{\sS_{2}}}}\endgroup }}} \tau _{2}}}} 
    \begin{subproof} 
      Then {{\color{\colorMATH}\ensuremath{\tau ' = \tau '_{1} \mathrel{^{{\begingroup\renewcommand\colorMATH{\colorMATHB}\renewcommand\colorSYNTAX{\colorSYNTAXB}{{\color{\colorMATH}\ensuremath{\sS'_{1}}}}\endgroup }}\&^{{\begingroup\renewcommand\colorMATH{\colorMATHB}\renewcommand\colorSYNTAX{\colorSYNTAXB}{{\color{\colorMATH}\ensuremath{\sS'_{2}}}}\endgroup }}} \tau '_{2}}}} such that {{\color{\colorMATH}\ensuremath{\tau _{1} <: \tau '_{1}, {\begingroup\renewcommand\colorMATH{\colorMATHB}\renewcommand\colorSYNTAX{\colorSYNTAXB}{{\color{\colorMATH}\ensuremath{\sS_{1}}}}\endgroup } <: {\begingroup\renewcommand\colorMATH{\colorMATHB}\renewcommand\colorSYNTAX{\colorSYNTAXB}{{\color{\colorMATH}\ensuremath{\sS'_{1}}}}\endgroup }, {\begingroup\renewcommand\colorMATH{\colorMATHB}\renewcommand\colorSYNTAX{\colorSYNTAXB}{{\color{\colorMATH}\ensuremath{\sS'_{2}}}}\endgroup } <: {\begingroup\renewcommand\colorMATH{\colorMATHB}\renewcommand\colorSYNTAX{\colorSYNTAXB}{{\color{\colorMATH}\ensuremath{\sS'_{2}}}}\endgroup }}}}, and {{\color{\colorMATH}\ensuremath{\tau _{2} <: \tau '_{2}}}}.
      But  {{\color{\colorMATH}\ensuremath{{\begingroup\renewcommand\colorMATH{\colorMATHB}\renewcommand\colorSYNTAX{\colorSYNTAXB}{{\color{\colorMATH}\ensuremath{\sS}}}\endgroup }(\tau _{1} \mathrel{^{{\begingroup\renewcommand\colorMATH{\colorMATHB}\renewcommand\colorSYNTAX{\colorSYNTAXB}{{\color{\colorMATH}\ensuremath{\sS_{1}}}}\endgroup }}\&^{{\begingroup\renewcommand\colorMATH{\colorMATHB}\renewcommand\colorSYNTAX{\colorSYNTAXB}{{\color{\colorMATH}\ensuremath{\sS_{2}}}}\endgroup }}} \tau _{2}) = {\begingroup\renewcommand\colorMATH{\colorMATHB}\renewcommand\colorSYNTAX{\colorSYNTAXB}{{\color{\colorMATH}\ensuremath{\sS}}}\endgroup }(\tau _{1}) \mathrel{^{{\begingroup\renewcommand\colorMATH{\colorMATHB}\renewcommand\colorSYNTAX{\colorSYNTAXB}{{\color{\colorMATH}\ensuremath{\sS}}}\endgroup } \mathord{\cdotp } {\begingroup\renewcommand\colorMATH{\colorMATHB}\renewcommand\colorSYNTAX{\colorSYNTAXB}{{\color{\colorMATH}\ensuremath{\sS_{1}}}}\endgroup }}\&^{{\begingroup\renewcommand\colorMATH{\colorMATHB}\renewcommand\colorSYNTAX{\colorSYNTAXB}{{\color{\colorMATH}\ensuremath{\sS}}}\endgroup } \mathord{\cdotp } {\begingroup\renewcommand\colorMATH{\colorMATHB}\renewcommand\colorSYNTAX{\colorSYNTAXB}{{\color{\colorMATH}\ensuremath{\sS_{2}}}}\endgroup }}} {\begingroup\renewcommand\colorMATH{\colorMATHB}\renewcommand\colorSYNTAX{\colorSYNTAXB}{{\color{\colorMATH}\ensuremath{\sS}}}\endgroup }(\tau _{2})}}}, then by
      induction hypotheses {{\color{\colorMATH}\ensuremath{{\begingroup\renewcommand\colorMATH{\colorMATHB}\renewcommand\colorSYNTAX{\colorSYNTAXB}{{\color{\colorMATH}\ensuremath{\sS}}}\endgroup }(\tau _{1}) <: {\begingroup\renewcommand\colorMATH{\colorMATHB}\renewcommand\colorSYNTAX{\colorSYNTAXB}{{\color{\colorMATH}\ensuremath{\sS}}}\endgroup }(\tau '_{1})}}}, and {{\color{\colorMATH}\ensuremath{{\begingroup\renewcommand\colorMATH{\colorMATHB}\renewcommand\colorSYNTAX{\colorSYNTAXB}{{\color{\colorMATH}\ensuremath{\sS}}}\endgroup }(\tau _{2}) <: {\begingroup\renewcommand\colorMATH{\colorMATHB}\renewcommand\colorSYNTAX{\colorSYNTAXB}{{\color{\colorMATH}\ensuremath{\sS}}}\endgroup }(\tau '_{2})}}}. Also by Lemma~\ref{lm:dot-subt},
      {{\color{\colorMATH}\ensuremath{{\begingroup\renewcommand\colorMATH{\colorMATHB}\renewcommand\colorSYNTAX{\colorSYNTAXB}{{\color{\colorMATH}\ensuremath{\sS}}}\endgroup }\mathord{\cdotp }{\begingroup\renewcommand\colorMATH{\colorMATHB}\renewcommand\colorSYNTAX{\colorSYNTAXB}{{\color{\colorMATH}\ensuremath{\sS_{1}}}}\endgroup } <: {\begingroup\renewcommand\colorMATH{\colorMATHB}\renewcommand\colorSYNTAX{\colorSYNTAXB}{{\color{\colorMATH}\ensuremath{\sS}}}\endgroup }\mathord{\cdotp }{\begingroup\renewcommand\colorMATH{\colorMATHB}\renewcommand\colorSYNTAX{\colorSYNTAXB}{{\color{\colorMATH}\ensuremath{\sS'_{1}}}}\endgroup }}}} and {{\color{\colorMATH}\ensuremath{{\begingroup\renewcommand\colorMATH{\colorMATHB}\renewcommand\colorSYNTAX{\colorSYNTAXB}{{\color{\colorMATH}\ensuremath{\sS}}}\endgroup }\mathord{\cdotp }{\begingroup\renewcommand\colorMATH{\colorMATHB}\renewcommand\colorSYNTAX{\colorSYNTAXB}{{\color{\colorMATH}\ensuremath{\sS_{2}}}}\endgroup } <: {\begingroup\renewcommand\colorMATH{\colorMATHB}\renewcommand\colorSYNTAX{\colorSYNTAXB}{{\color{\colorMATH}\ensuremath{\sS}}}\endgroup }\mathord{\cdotp }{\begingroup\renewcommand\colorMATH{\colorMATHB}\renewcommand\colorSYNTAX{\colorSYNTAXB}{{\color{\colorMATH}\ensuremath{\sS'_{2}}}}\endgroup }}}}, therefore\\ 
      {{\color{\colorMATH}\ensuremath{{\begingroup\renewcommand\colorMATH{\colorMATHB}\renewcommand\colorSYNTAX{\colorSYNTAXB}{{\color{\colorMATH}\ensuremath{\sS}}}\endgroup }(\tau _{1} \mathrel{^{{\begingroup\renewcommand\colorMATH{\colorMATHB}\renewcommand\colorSYNTAX{\colorSYNTAXB}{{\color{\colorMATH}\ensuremath{\sS_{1}}}}\endgroup }}\&^{{\begingroup\renewcommand\colorMATH{\colorMATHB}\renewcommand\colorSYNTAX{\colorSYNTAXB}{{\color{\colorMATH}\ensuremath{\sS_{2}}}}\endgroup }}} \tau _{2}) = {\begingroup\renewcommand\colorMATH{\colorMATHB}\renewcommand\colorSYNTAX{\colorSYNTAXB}{{\color{\colorMATH}\ensuremath{\sS}}}\endgroup }(\tau _{1}) \mathrel{^{{\begingroup\renewcommand\colorMATH{\colorMATHB}\renewcommand\colorSYNTAX{\colorSYNTAXB}{{\color{\colorMATH}\ensuremath{\sS}}}\endgroup } \mathord{\cdotp } {\begingroup\renewcommand\colorMATH{\colorMATHB}\renewcommand\colorSYNTAX{\colorSYNTAXB}{{\color{\colorMATH}\ensuremath{\sS_{1}}}}\endgroup }}\&^{{\begingroup\renewcommand\colorMATH{\colorMATHB}\renewcommand\colorSYNTAX{\colorSYNTAXB}{{\color{\colorMATH}\ensuremath{\sS}}}\endgroup } \mathord{\cdotp } {\begingroup\renewcommand\colorMATH{\colorMATHB}\renewcommand\colorSYNTAX{\colorSYNTAXB}{{\color{\colorMATH}\ensuremath{\sS_{2}}}}\endgroup }}} {\begingroup\renewcommand\colorMATH{\colorMATHB}\renewcommand\colorSYNTAX{\colorSYNTAXB}{{\color{\colorMATH}\ensuremath{\sS}}}\endgroup }(\tau _{2}) <: {\begingroup\renewcommand\colorMATH{\colorMATHB}\renewcommand\colorSYNTAX{\colorSYNTAXB}{{\color{\colorMATH}\ensuremath{\sS}}}\endgroup }(\tau '_{1}) \mathrel{^{{\begingroup\renewcommand\colorMATH{\colorMATHB}\renewcommand\colorSYNTAX{\colorSYNTAXB}{{\color{\colorMATH}\ensuremath{\sS}}}\endgroup } \mathord{\cdotp } {\begingroup\renewcommand\colorMATH{\colorMATHB}\renewcommand\colorSYNTAX{\colorSYNTAXB}{{\color{\colorMATH}\ensuremath{\sS'_{1}}}}\endgroup }}\&^{{\begingroup\renewcommand\colorMATH{\colorMATHB}\renewcommand\colorSYNTAX{\colorSYNTAXB}{{\color{\colorMATH}\ensuremath{\sS}}}\endgroup } \mathord{\cdotp } {\begingroup\renewcommand\colorMATH{\colorMATHB}\renewcommand\colorSYNTAX{\colorSYNTAXB}{{\color{\colorMATH}\ensuremath{\sS'_{2}}}}\endgroup }}} {\begingroup\renewcommand\colorMATH{\colorMATHB}\renewcommand\colorSYNTAX{\colorSYNTAXB}{{\color{\colorMATH}\ensuremath{\sS}}}\endgroup }(\tau '_{2}) = {\begingroup\renewcommand\colorMATH{\colorMATHB}\renewcommand\colorSYNTAX{\colorSYNTAXB}{{\color{\colorMATH}\ensuremath{\sS}}}\endgroup }(\tau '_{1} \mathrel{^{{\begingroup\renewcommand\colorMATH{\colorMATHB}\renewcommand\colorSYNTAX{\colorSYNTAXB}{{\color{\colorMATH}\ensuremath{\sS'_{1}}}}\endgroup }}\&^{{\begingroup\renewcommand\colorMATH{\colorMATHB}\renewcommand\colorSYNTAX{\colorSYNTAXB}{{\color{\colorMATH}\ensuremath{\sS'_{2}}}}\endgroup }}} \tau '_{2})}}} and the result holds.
    \end{subproof}
  \end{enumerate}
\end{proof}

\begin{lemma}
  \label{lm:joinmeetgreen}
  Let {{\color{\colorMATH}\ensuremath{{\begingroup\renewcommand\colorMATH{\colorMATHB}\renewcommand\colorSYNTAX{\colorSYNTAXB}{{\color{\colorMATH}\ensuremath{\sS}}}\endgroup }}}} and {{\color{\colorMATH}\ensuremath{{\begingroup\renewcommand\colorMATH{\colorMATHB}\renewcommand\colorSYNTAX{\colorSYNTAXB}{{\color{\colorMATH}\ensuremath{\sS'}}}\endgroup }}}}
  Then 
  \begin{enumerate}\item  {{\color{\colorMATH}\ensuremath{{\begingroup\renewcommand\colorMATH{\colorMATHB}\renewcommand\colorSYNTAX{\colorSYNTAXB}{{\color{\colorMATH}\ensuremath{\sS}}}\endgroup } <: {\begingroup\renewcommand\colorMATH{\colorMATHB}\renewcommand\colorSYNTAX{\colorSYNTAXB}{{\color{\colorMATH}\ensuremath{\sS}}}\endgroup } \sqcup  {\begingroup\renewcommand\colorMATH{\colorMATHB}\renewcommand\colorSYNTAX{\colorSYNTAXB}{{\color{\colorMATH}\ensuremath{\sS'}}}\endgroup }}}}
  \item  {{\color{\colorMATH}\ensuremath{{\begingroup\renewcommand\colorMATH{\colorMATHB}\renewcommand\colorSYNTAX{\colorSYNTAXB}{{\color{\colorMATH}\ensuremath{\sS}}}\endgroup } \sqcap  {\begingroup\renewcommand\colorMATH{\colorMATHB}\renewcommand\colorSYNTAX{\colorSYNTAXB}{{\color{\colorMATH}\ensuremath{\sS'}}}\endgroup } <: {\begingroup\renewcommand\colorMATH{\colorMATHB}\renewcommand\colorSYNTAX{\colorSYNTAXB}{{\color{\colorMATH}\ensuremath{\sS}}}\endgroup }}}}
  \end{enumerate}
\end{lemma}
\begin{proof}
  By induction of {{\color{\colorMATH}\ensuremath{{\begingroup\renewcommand\colorMATH{\colorMATHB}\renewcommand\colorSYNTAX{\colorSYNTAXB}{{\color{\colorMATH}\ensuremath{\sS}}}\endgroup }}}}, and noticing that {{\color{\colorMATH}\ensuremath{{\begingroup\renewcommand\colorMATH{\colorMATHB}\renewcommand\colorSYNTAX{\colorSYNTAXB}{{\color{\colorMATH}\ensuremath{\sss}}}\endgroup } \leq  {\begingroup\renewcommand\colorMATH{\colorMATHB}\renewcommand\colorSYNTAX{\colorSYNTAXB}{{\color{\colorMATH}\ensuremath{\sss}}}\endgroup } \sqcup  {\begingroup\renewcommand\colorMATH{\colorMATHB}\renewcommand\colorSYNTAX{\colorSYNTAXB}{{\color{\colorMATH}\ensuremath{\sss'}}}\endgroup }}}}, and {{\color{\colorMATH}\ensuremath{{\begingroup\renewcommand\colorMATH{\colorMATHB}\renewcommand\colorSYNTAX{\colorSYNTAXB}{{\color{\colorMATH}\ensuremath{\sss}}}\endgroup } \sqcap  {\begingroup\renewcommand\colorMATH{\colorMATHB}\renewcommand\colorSYNTAX{\colorSYNTAXB}{{\color{\colorMATH}\ensuremath{\sss'}}}\endgroup } \leq  {\begingroup\renewcommand\colorMATH{\colorMATHB}\renewcommand\colorSYNTAX{\colorSYNTAXB}{{\color{\colorMATH}\ensuremath{\sss}}}\endgroup }}}}.
\end{proof}

\begin{lemma}
  \label{lm:joinmeetred}
  Let {{\color{\colorMATH}\ensuremath{{\begingroup\renewcommand\colorMATH{\colorMATHC}\renewcommand\colorSYNTAX{\colorSYNTAXC}{{\color{\colorMATH}\ensuremath{\pS}}}\endgroup }}}} and {{\color{\colorMATH}\ensuremath{{\begingroup\renewcommand\colorMATH{\colorMATHC}\renewcommand\colorSYNTAX{\colorSYNTAXC}{{\color{\colorMATH}\ensuremath{\pS'}}}\endgroup }}}}
  Then 
  \begin{enumerate}\item  {{\color{\colorMATH}\ensuremath{{\begingroup\renewcommand\colorMATH{\colorMATHC}\renewcommand\colorSYNTAX{\colorSYNTAXC}{{\color{\colorMATH}\ensuremath{\pS}}}\endgroup } <: {\begingroup\renewcommand\colorMATH{\colorMATHC}\renewcommand\colorSYNTAX{\colorSYNTAXC}{{\color{\colorMATH}\ensuremath{\pS}}}\endgroup } \sqcup  {\begingroup\renewcommand\colorMATH{\colorMATHC}\renewcommand\colorSYNTAX{\colorSYNTAXC}{{\color{\colorMATH}\ensuremath{\pS'}}}\endgroup }}}}
  \item  {{\color{\colorMATH}\ensuremath{{\begingroup\renewcommand\colorMATH{\colorMATHC}\renewcommand\colorSYNTAX{\colorSYNTAXC}{{\color{\colorMATH}\ensuremath{\pS}}}\endgroup } \sqcap  {\begingroup\renewcommand\colorMATH{\colorMATHC}\renewcommand\colorSYNTAX{\colorSYNTAXC}{{\color{\colorMATH}\ensuremath{\pS'}}}\endgroup } <: {\begingroup\renewcommand\colorMATH{\colorMATHC}\renewcommand\colorSYNTAX{\colorSYNTAXC}{{\color{\colorMATH}\ensuremath{\pS}}}\endgroup }}}}
  \end{enumerate}
\end{lemma}
\begin{proof}
  By definition of the {{\color{\colorMATH}\ensuremath{<:}}} operator for privacy environments.
\end{proof}

\begin{lemma}
  \label{lm:joinmeetprops}
  Let {{\color{\colorMATH}\ensuremath{\tau }}} and {{\color{\colorMATH}\ensuremath{\tau '}}}, such that {{\color{\colorMATH}\ensuremath{\tau  \sqcup  \tau '}}} and {{\color{\colorMATH}\ensuremath{\tau  \sqcap  \tau '}}} are defined. 
  Then 
  \begin{enumerate}\item  {{\color{\colorMATH}\ensuremath{\tau  <: \tau  \sqcup  \tau '}}}
  \item  {{\color{\colorMATH}\ensuremath{\tau  \sqcap  \tau ' <: \tau }}}
  \end{enumerate}
\end{lemma}
\begin{proof}
  Let us prove (1) first by induction by induction on {{\color{\colorMATH}\ensuremath{\tau }}}:
  \begin{enumerate}[ncases]\item  {{\color{\colorMATH}\ensuremath{\tau  = {\begingroup\renewcommand\colorMATH{\colorMATHA}\renewcommand\colorSYNTAX{\colorSYNTAXA}{{\color{\colorSYNTAX}\texttt{{\ensuremath{{\mathbb{R}}}}}}}\endgroup }}}}
    \begin{subproof} 
      Then {{\color{\colorMATH}\ensuremath{\tau ' = {\begingroup\renewcommand\colorMATH{\colorMATHA}\renewcommand\colorSYNTAX{\colorSYNTAXA}{{\color{\colorSYNTAX}\texttt{{\ensuremath{{\mathbb{R}}}}}}}\endgroup }}}} so the result is trivial.
    \end{subproof}
  \item  {{\color{\colorMATH}\ensuremath{\tau  = {\mathbb{B}}}}}
    \begin{subproof} 
      Then {{\color{\colorMATH}\ensuremath{\tau ' = {\mathbb{B}}}}} so the result is trivial.
    \end{subproof}
  \item  {{\color{\colorMATH}\ensuremath{\tau  = {{\color{\colorSYNTAX}\texttt{unit}}}}}} 
    \begin{subproof} 
      Then {{\color{\colorMATH}\ensuremath{\tau ' = {{\color{\colorSYNTAX}\texttt{unit}}}}}} so the result is trivial.
    \end{subproof}
  \item  {{\color{\colorMATH}\ensuremath{\tau  = (x\mathrel{:}\tau _{1}\mathord{\cdotp }{\begingroup\renewcommand\colorMATH{\colorMATHB}\renewcommand\colorSYNTAX{\colorSYNTAXB}{{\color{\colorMATH}\ensuremath{\distance}}}\endgroup }) \xrightarrowS {{\begingroup\renewcommand\colorMATH{\colorMATHB}\renewcommand\colorSYNTAX{\colorSYNTAXB}{{\color{\colorMATH}\ensuremath{\sS}}}\endgroup }} \tau _{2}}}} 
    \begin{subproof} 
      Then {{\color{\colorMATH}\ensuremath{\tau ' = (x\mathrel{:}\tau '_{1}\mathord{\cdotp }{\begingroup\renewcommand\colorMATH{\colorMATHB}\renewcommand\colorSYNTAX{\colorSYNTAXB}{{\color{\colorMATH}\ensuremath{\distance'}}}\endgroup }) \xrightarrowS {{\begingroup\renewcommand\colorMATH{\colorMATHB}\renewcommand\colorSYNTAX{\colorSYNTAXB}{{\color{\colorMATH}\ensuremath{\sS'}}}\endgroup }} \tau '_{2}}}}, and
      {{\color{\colorMATH}\ensuremath{\tau  \sqcup  \tau ' = (x\mathrel{:}(\tau _{1} \sqcap  \tau '_{1})\mathord{\cdotp }({\begingroup\renewcommand\colorMATH{\colorMATHB}\renewcommand\colorSYNTAX{\colorSYNTAXB}{{\color{\colorMATH}\ensuremath{\distance}}}\endgroup } \sqcap  {\begingroup\renewcommand\colorMATH{\colorMATHB}\renewcommand\colorSYNTAX{\colorSYNTAXB}{{\color{\colorMATH}\ensuremath{\distance'}}}\endgroup })) \xrightarrowS {{\begingroup\renewcommand\colorMATH{\colorMATHB}\renewcommand\colorSYNTAX{\colorSYNTAXB}{{\color{\colorMATH}\ensuremath{\sS}}}\endgroup } \sqcup  {\begingroup\renewcommand\colorMATH{\colorMATHB}\renewcommand\colorSYNTAX{\colorSYNTAXB}{{\color{\colorMATH}\ensuremath{\sS'}}}\endgroup }} (\tau _{2} \sqcup  \tau '_{2})}}}

      By induction hypothesis {{\color{\colorMATH}\ensuremath{(\tau _{1} \sqcap  \tau '_{1}) <: \tau _{1}}}}, we know that {{\color{\colorMATH}\ensuremath{{\begingroup\renewcommand\colorMATH{\colorMATHB}\renewcommand\colorSYNTAX{\colorSYNTAXB}{{\color{\colorMATH}\ensuremath{\distance}}}\endgroup } \sqcap  {\begingroup\renewcommand\colorMATH{\colorMATHB}\renewcommand\colorSYNTAX{\colorSYNTAXB}{{\color{\colorMATH}\ensuremath{\distance'}}}\endgroup } \leq  {\begingroup\renewcommand\colorMATH{\colorMATHB}\renewcommand\colorSYNTAX{\colorSYNTAXB}{{\color{\colorMATH}\ensuremath{\distance}}}\endgroup }}}}, 
      by Lemma~\ref{lm:joinmeetgreen}, {{\color{\colorMATH}\ensuremath{{\begingroup\renewcommand\colorMATH{\colorMATHB}\renewcommand\colorSYNTAX{\colorSYNTAXB}{{\color{\colorMATH}\ensuremath{\sS}}}\endgroup } <: {\begingroup\renewcommand\colorMATH{\colorMATHB}\renewcommand\colorSYNTAX{\colorSYNTAXB}{{\color{\colorMATH}\ensuremath{\sS}}}\endgroup } \sqcup  {\begingroup\renewcommand\colorMATH{\colorMATHB}\renewcommand\colorSYNTAX{\colorSYNTAXB}{{\color{\colorMATH}\ensuremath{\sS'}}}\endgroup }}}}, and by induction hypothesis {{\color{\colorMATH}\ensuremath{\tau _{2} <: \tau _{2} \sqcup  \tau '_{2}}}}.
      Then {{\color{\colorMATH}\ensuremath{(x\mathrel{:}\tau _{1}\mathord{\cdotp }{\begingroup\renewcommand\colorMATH{\colorMATHB}\renewcommand\colorSYNTAX{\colorSYNTAXB}{{\color{\colorMATH}\ensuremath{\distance}}}\endgroup }) \xrightarrowS {{\begingroup\renewcommand\colorMATH{\colorMATHB}\renewcommand\colorSYNTAX{\colorSYNTAXB}{{\color{\colorMATH}\ensuremath{\sS}}}\endgroup }} \tau _{2} <: (x\mathrel{:}(\tau _{1} \sqcap  \tau '_{1})\mathord{\cdotp }({\begingroup\renewcommand\colorMATH{\colorMATHB}\renewcommand\colorSYNTAX{\colorSYNTAXB}{{\color{\colorMATH}\ensuremath{\distance}}}\endgroup } \sqcap  {\begingroup\renewcommand\colorMATH{\colorMATHB}\renewcommand\colorSYNTAX{\colorSYNTAXB}{{\color{\colorMATH}\ensuremath{\distance'}}}\endgroup })) \xrightarrowS {{\begingroup\renewcommand\colorMATH{\colorMATHB}\renewcommand\colorSYNTAX{\colorSYNTAXB}{{\color{\colorMATH}\ensuremath{\sS}}}\endgroup } \sqcup  {\begingroup\renewcommand\colorMATH{\colorMATHB}\renewcommand\colorSYNTAX{\colorSYNTAXB}{{\color{\colorMATH}\ensuremath{\sS'}}}\endgroup }} (\tau _{2} \sqcup  \tau '_{2})}}} and the result holds.
    \end{subproof}
  \item  {{\color{\colorMATH}\ensuremath{\tau  = (x\mathrel{:}\tau _{1}\mathord{\cdotp }{\begingroup\renewcommand\colorMATH{\colorMATHB}\renewcommand\colorSYNTAX{\colorSYNTAXB}{{\color{\colorMATH}\ensuremath{\distance}}}\endgroup }) \xrightarrowP {{\begingroup\renewcommand\colorMATH{\colorMATHC}\renewcommand\colorSYNTAX{\colorSYNTAXC}{{\color{\colorMATH}\ensuremath{\pS'}}}\endgroup }} \tau _{2}}}} 
    \begin{subproof} 
      Then {{\color{\colorMATH}\ensuremath{\tau ' = (x\mathrel{:}\tau '_{1}\mathord{\cdotp }{\begingroup\renewcommand\colorMATH{\colorMATHB}\renewcommand\colorSYNTAX{\colorSYNTAXB}{{\color{\colorMATH}\ensuremath{\distance'}}}\endgroup }) \xrightarrowP {{\begingroup\renewcommand\colorMATH{\colorMATHC}\renewcommand\colorSYNTAX{\colorSYNTAXC}{{\color{\colorMATH}\ensuremath{\pS'}}}\endgroup }} \tau '_{2}}}}, and
      {{\color{\colorMATH}\ensuremath{\tau  \sqcup  \tau ' = (x\mathrel{:}(\tau _{1} \sqcap  \tau '_{1})\mathord{\cdotp }({\begingroup\renewcommand\colorMATH{\colorMATHB}\renewcommand\colorSYNTAX{\colorSYNTAXB}{{\color{\colorMATH}\ensuremath{\distance}}}\endgroup } \sqcap  {\begingroup\renewcommand\colorMATH{\colorMATHB}\renewcommand\colorSYNTAX{\colorSYNTAXB}{{\color{\colorMATH}\ensuremath{\distance'}}}\endgroup })) \xrightarrowP {{\begingroup\renewcommand\colorMATH{\colorMATHC}\renewcommand\colorSYNTAX{\colorSYNTAXC}{{\color{\colorMATH}\ensuremath{\pS}}}\endgroup } \sqcup  {\begingroup\renewcommand\colorMATH{\colorMATHC}\renewcommand\colorSYNTAX{\colorSYNTAXC}{{\color{\colorMATH}\ensuremath{\pS'}}}\endgroup }} (\tau _{2} \sqcup  \tau '_{2})}}}

      By induction hypothesis {{\color{\colorMATH}\ensuremath{(\tau _{1} \sqcap  \tau '_{1}) <: \tau _{1}}}}, we know that {{\color{\colorMATH}\ensuremath{{\begingroup\renewcommand\colorMATH{\colorMATHB}\renewcommand\colorSYNTAX{\colorSYNTAXB}{{\color{\colorMATH}\ensuremath{\distance}}}\endgroup } \sqcap  {\begingroup\renewcommand\colorMATH{\colorMATHB}\renewcommand\colorSYNTAX{\colorSYNTAXB}{{\color{\colorMATH}\ensuremath{\distance'}}}\endgroup } \leq  {\begingroup\renewcommand\colorMATH{\colorMATHB}\renewcommand\colorSYNTAX{\colorSYNTAXB}{{\color{\colorMATH}\ensuremath{\distance}}}\endgroup }}}}, 
      by Lemma~\ref{lm:joinmeetred}, {{\color{\colorMATH}\ensuremath{{\begingroup\renewcommand\colorMATH{\colorMATHC}\renewcommand\colorSYNTAX{\colorSYNTAXC}{{\color{\colorMATH}\ensuremath{\pS}}}\endgroup } <: {\begingroup\renewcommand\colorMATH{\colorMATHC}\renewcommand\colorSYNTAX{\colorSYNTAXC}{{\color{\colorMATH}\ensuremath{\pS}}}\endgroup } \sqcup  {\begingroup\renewcommand\colorMATH{\colorMATHC}\renewcommand\colorSYNTAX{\colorSYNTAXC}{{\color{\colorMATH}\ensuremath{\pS'}}}\endgroup }}}}, and by induction hypothesis {{\color{\colorMATH}\ensuremath{\tau _{2} <: \tau _{2} \sqcup  \tau '_{2}}}}.
      Then {{\color{\colorMATH}\ensuremath{(x\mathrel{:}\tau _{1}\mathord{\cdotp }{\begingroup\renewcommand\colorMATH{\colorMATHB}\renewcommand\colorSYNTAX{\colorSYNTAXB}{{\color{\colorMATH}\ensuremath{\distance}}}\endgroup }) \xrightarrowP {{\begingroup\renewcommand\colorMATH{\colorMATHC}\renewcommand\colorSYNTAX{\colorSYNTAXC}{{\color{\colorMATH}\ensuremath{\pS}}}\endgroup }} \tau _{2} <: (x\mathrel{:}(\tau _{1} \sqcap  \tau '_{1})\mathord{\cdotp }({\begingroup\renewcommand\colorMATH{\colorMATHB}\renewcommand\colorSYNTAX{\colorSYNTAXB}{{\color{\colorMATH}\ensuremath{\distance}}}\endgroup } \sqcap  {\begingroup\renewcommand\colorMATH{\colorMATHB}\renewcommand\colorSYNTAX{\colorSYNTAXB}{{\color{\colorMATH}\ensuremath{\distance'}}}\endgroup })) \xrightarrowP {{\begingroup\renewcommand\colorMATH{\colorMATHC}\renewcommand\colorSYNTAX{\colorSYNTAXC}{{\color{\colorMATH}\ensuremath{\pS}}}\endgroup } \sqcup  {\begingroup\renewcommand\colorMATH{\colorMATHC}\renewcommand\colorSYNTAX{\colorSYNTAXC}{{\color{\colorMATH}\ensuremath{\pS'}}}\endgroup }} (\tau _{2} \sqcup  \tau '_{2})}}} and the result holds.
    \end{subproof}
  \item  {{\color{\colorMATH}\ensuremath{\tau  = \tau _{1} \mathrel{^{{\begingroup\renewcommand\colorMATH{\colorMATHB}\renewcommand\colorSYNTAX{\colorSYNTAXB}{{\color{\colorMATH}\ensuremath{\sS_{1}}}}\endgroup }}\oplus ^{{\begingroup\renewcommand\colorMATH{\colorMATHB}\renewcommand\colorSYNTAX{\colorSYNTAXB}{{\color{\colorMATH}\ensuremath{\sS_{2}}}}\endgroup }}} \tau _{2}}}} 
    \begin{subproof} 
      Then {{\color{\colorMATH}\ensuremath{\tau ' = \tau '_{1} \mathrel{^{{\begingroup\renewcommand\colorMATH{\colorMATHB}\renewcommand\colorSYNTAX{\colorSYNTAXB}{{\color{\colorMATH}\ensuremath{\sS'_{1}}}}\endgroup }}\oplus ^{{\begingroup\renewcommand\colorMATH{\colorMATHB}\renewcommand\colorSYNTAX{\colorSYNTAXB}{{\color{\colorMATH}\ensuremath{\sS'_{2}}}}\endgroup }}} \tau '_{2}}}},
      and 
      {{\color{\colorMATH}\ensuremath{\tau  \sqcup  \tau ' = (\tau _{1} \sqcup  \tau '_{1}) \mathrel{^{{\begingroup\renewcommand\colorMATH{\colorMATHB}\renewcommand\colorSYNTAX{\colorSYNTAXB}{{\color{\colorMATH}\ensuremath{\sS_{1}}}}\endgroup } \sqcup  {\begingroup\renewcommand\colorMATH{\colorMATHB}\renewcommand\colorSYNTAX{\colorSYNTAXB}{{\color{\colorMATH}\ensuremath{\sS'_{1}}}}\endgroup }}\oplus ^{{\begingroup\renewcommand\colorMATH{\colorMATHB}\renewcommand\colorSYNTAX{\colorSYNTAXB}{{\color{\colorMATH}\ensuremath{\sS_{2}}}}\endgroup } \sqcup  {\begingroup\renewcommand\colorMATH{\colorMATHB}\renewcommand\colorSYNTAX{\colorSYNTAXB}{{\color{\colorMATH}\ensuremath{\sS'_{2}}}}\endgroup }}} (\tau _{2} \sqcup  \tau '_{2})}}}.
      By induction hypotheses {{\color{\colorMATH}\ensuremath{\tau _{1} <: \tau _{1} \sqcup  \tau '_{1}}}}, and {{\color{\colorMATH}\ensuremath{\tau _{2} <: \tau _{2} \sqcup  \tau '_{2}}}}, and
      by Lemma~\ref{lm:joinmeetgreen}, {{\color{\colorMATH}\ensuremath{{\begingroup\renewcommand\colorMATH{\colorMATHB}\renewcommand\colorSYNTAX{\colorSYNTAXB}{{\color{\colorMATH}\ensuremath{\sS_{1}}}}\endgroup } <: {\begingroup\renewcommand\colorMATH{\colorMATHB}\renewcommand\colorSYNTAX{\colorSYNTAXB}{{\color{\colorMATH}\ensuremath{\sS_{1}}}}\endgroup } \sqcup  {\begingroup\renewcommand\colorMATH{\colorMATHB}\renewcommand\colorSYNTAX{\colorSYNTAXB}{{\color{\colorMATH}\ensuremath{\sS'_{1}}}}\endgroup }}}}, and {{\color{\colorMATH}\ensuremath{{\begingroup\renewcommand\colorMATH{\colorMATHB}\renewcommand\colorSYNTAX{\colorSYNTAXB}{{\color{\colorMATH}\ensuremath{\sS_{2}}}}\endgroup } <: {\begingroup\renewcommand\colorMATH{\colorMATHB}\renewcommand\colorSYNTAX{\colorSYNTAXB}{{\color{\colorMATH}\ensuremath{\sS_{2}}}}\endgroup } \sqcup  {\begingroup\renewcommand\colorMATH{\colorMATHB}\renewcommand\colorSYNTAX{\colorSYNTAXB}{{\color{\colorMATH}\ensuremath{\sS'_{2}}}}\endgroup }}}}.
      Then {{\color{\colorMATH}\ensuremath{\tau '_{1} \mathrel{^{{\begingroup\renewcommand\colorMATH{\colorMATHB}\renewcommand\colorSYNTAX{\colorSYNTAXB}{{\color{\colorMATH}\ensuremath{\sS'_{1}}}}\endgroup }}\oplus ^{{\begingroup\renewcommand\colorMATH{\colorMATHB}\renewcommand\colorSYNTAX{\colorSYNTAXB}{{\color{\colorMATH}\ensuremath{\sS'_{2}}}}\endgroup }}} \tau '_{2} <: (\tau _{1} \sqcup  \tau '_{1}) \mathrel{^{{\begingroup\renewcommand\colorMATH{\colorMATHB}\renewcommand\colorSYNTAX{\colorSYNTAXB}{{\color{\colorMATH}\ensuremath{\sS_{1}}}}\endgroup } \sqcup  {\begingroup\renewcommand\colorMATH{\colorMATHB}\renewcommand\colorSYNTAX{\colorSYNTAXB}{{\color{\colorMATH}\ensuremath{\sS'_{1}}}}\endgroup }}\oplus ^{{\begingroup\renewcommand\colorMATH{\colorMATHB}\renewcommand\colorSYNTAX{\colorSYNTAXB}{{\color{\colorMATH}\ensuremath{\sS_{2}}}}\endgroup } \sqcup  {\begingroup\renewcommand\colorMATH{\colorMATHB}\renewcommand\colorSYNTAX{\colorSYNTAXB}{{\color{\colorMATH}\ensuremath{\sS'_{2}}}}\endgroup }}} (\tau _{2} \sqcup  \tau '_{2})}}} and the result holds.
    \end{subproof}
  \item  {{\color{\colorMATH}\ensuremath{\tau  = \tau _{1} \mathrel{^{{\begingroup\renewcommand\colorMATH{\colorMATHB}\renewcommand\colorSYNTAX{\colorSYNTAXB}{{\color{\colorMATH}\ensuremath{\sS_{1}}}}\endgroup }}\&^{{\begingroup\renewcommand\colorMATH{\colorMATHB}\renewcommand\colorSYNTAX{\colorSYNTAXB}{{\color{\colorMATH}\ensuremath{\sS_{2}}}}\endgroup }}} \tau _{2}}}} 
    \begin{subproof} 
      Analogous to the {{\color{\colorMATH}\ensuremath{\tau  = \tau _{1} \mathrel{^{{\begingroup\renewcommand\colorMATH{\colorMATHB}\renewcommand\colorSYNTAX{\colorSYNTAXB}{{\color{\colorMATH}\ensuremath{\sS_{1}}}}\endgroup }}\oplus ^{{\begingroup\renewcommand\colorMATH{\colorMATHB}\renewcommand\colorSYNTAX{\colorSYNTAXB}{{\color{\colorMATH}\ensuremath{\sS_{2}}}}\endgroup }}} \tau _{2}}}} case.
    \end{subproof}
  \item  {{\color{\colorMATH}\ensuremath{\tau  = \tau _{1} \mathrel{^{{\begingroup\renewcommand\colorMATH{\colorMATHB}\renewcommand\colorSYNTAX{\colorSYNTAXB}{{\color{\colorMATH}\ensuremath{\sS_{1}}}}\endgroup }}\oplus ^{{\begingroup\renewcommand\colorMATH{\colorMATHB}\renewcommand\colorSYNTAX{\colorSYNTAXB}{{\color{\colorMATH}\ensuremath{\sS_{2}}}}\endgroup }}} \tau _{2}}}} 
    \begin{subproof} 
      Analogous to the {{\color{\colorMATH}\ensuremath{\tau  = \tau _{1} \mathrel{^{{\begingroup\renewcommand\colorMATH{\colorMATHB}\renewcommand\colorSYNTAX{\colorSYNTAXB}{{\color{\colorMATH}\ensuremath{\sS_{1}}}}\endgroup }}\oplus ^{{\begingroup\renewcommand\colorMATH{\colorMATHB}\renewcommand\colorSYNTAX{\colorSYNTAXB}{{\color{\colorMATH}\ensuremath{\sS_{2}}}}\endgroup }}} \tau _{2}}}} case.
    \end{subproof}
  \end{enumerate}
\end{proof}

\begin{lemma}[Relation subsumption/weakening]\
  \label{lm:lrweakening-sensitivity}
  Consider {{\color{\colorMATH}\ensuremath{{\begingroup\renewcommand\colorMATH{\colorMATHB}\renewcommand\colorSYNTAX{\colorSYNTAXB}{{\color{\colorMATH}\ensuremath{\distance}}}\endgroup } \leq  {\begingroup\renewcommand\colorMATH{\colorMATHB}\renewcommand\colorSYNTAX{\colorSYNTAXB}{{\color{\colorMATH}\ensuremath{\distance'}}}\endgroup }}}} and {{\color{\colorMATH}\ensuremath{\sigma  <: \sigma '}}} then 
  \begin{enumerate}\item  If {{\color{\colorMATH}\ensuremath{({\begingroup\renewcommand\colorMATH{\colorMATHB}\renewcommand\colorSYNTAX{\colorSYNTAXB}{{\color{\colorMATH}\ensuremath{\sv_{1}}}}\endgroup },{\begingroup\renewcommand\colorMATH{\colorMATHB}\renewcommand\colorSYNTAX{\colorSYNTAXB}{{\color{\colorMATH}\ensuremath{\sv_{2}}}}\endgroup }) \in  {\mathcal{V}}_{{\begingroup\renewcommand\colorMATH{\colorMATHB}\renewcommand\colorSYNTAX{\colorSYNTAXB}{{\color{\colorMATH}\ensuremath{\distance}}}\endgroup }}^{k}\llbracket \sigma \rrbracket }}}, then {{\color{\colorMATH}\ensuremath{({\begingroup\renewcommand\colorMATH{\colorMATHB}\renewcommand\colorSYNTAX{\colorSYNTAXB}{{\color{\colorMATH}\ensuremath{\sv_{1}}}}\endgroup },{\begingroup\renewcommand\colorMATH{\colorMATHB}\renewcommand\colorSYNTAX{\colorSYNTAXB}{{\color{\colorMATH}\ensuremath{\sv_{2}}}}\endgroup }) \in  {\mathcal{V}}_{{\begingroup\renewcommand\colorMATH{\colorMATHB}\renewcommand\colorSYNTAX{\colorSYNTAXB}{{\color{\colorMATH}\ensuremath{\distance'}}}\endgroup }}^{k}\llbracket \sigma '\rrbracket }}}
  \item  If {{\color{\colorMATH}\ensuremath{({\begingroup\renewcommand\colorMATH{\colorMATHB}\renewcommand\colorSYNTAX{\colorSYNTAXB}{{\color{\colorMATH}\ensuremath{\se_{1}}}}\endgroup },{\begingroup\renewcommand\colorMATH{\colorMATHB}\renewcommand\colorSYNTAX{\colorSYNTAXB}{{\color{\colorMATH}\ensuremath{\se_{2}}}}\endgroup }) \in  {\mathcal{E}}_{{\begingroup\renewcommand\colorMATH{\colorMATHB}\renewcommand\colorSYNTAX{\colorSYNTAXB}{{\color{\colorMATH}\ensuremath{\distance}}}\endgroup }}^{k}\llbracket \sigma \rrbracket }}}, then {{\color{\colorMATH}\ensuremath{({\begingroup\renewcommand\colorMATH{\colorMATHB}\renewcommand\colorSYNTAX{\colorSYNTAXB}{{\color{\colorMATH}\ensuremath{\se_{1}}}}\endgroup },{\begingroup\renewcommand\colorMATH{\colorMATHB}\renewcommand\colorSYNTAX{\colorSYNTAXB}{{\color{\colorMATH}\ensuremath{\se_{2}}}}\endgroup }) \in  {\mathcal{E}}_{{\begingroup\renewcommand\colorMATH{\colorMATHB}\renewcommand\colorSYNTAX{\colorSYNTAXB}{{\color{\colorMATH}\ensuremath{\distance'}}}\endgroup }}^{k}\llbracket \sigma '\rrbracket }}}
  \item  If {{\color{\colorMATH}\ensuremath{({\begingroup\renewcommand\colorMATH{\colorMATHC}\renewcommand\colorSYNTAX{\colorSYNTAXC}{{\color{\colorMATH}\ensuremath{\pe_{1}}}}\endgroup }, {\begingroup\renewcommand\colorMATH{\colorMATHC}\renewcommand\colorSYNTAX{\colorSYNTAXC}{{\color{\colorMATH}\ensuremath{\pe_{2}}}}\endgroup }) \in  {\mathcal{E}}_{{\begingroup\renewcommand\colorMATH{\colorMATHC}\renewcommand\colorSYNTAX{\colorSYNTAXC}{{\color{\colorMATH}\ensuremath{p}}}\endgroup }}^{k}\llbracket \sigma \rrbracket }}}, then {{\color{\colorMATH}\ensuremath{({\begingroup\renewcommand\colorMATH{\colorMATHC}\renewcommand\colorSYNTAX{\colorSYNTAXC}{{\color{\colorMATH}\ensuremath{\pe_{1}}}}\endgroup },{\begingroup\renewcommand\colorMATH{\colorMATHC}\renewcommand\colorSYNTAX{\colorSYNTAXC}{{\color{\colorMATH}\ensuremath{\pe_{2}}}}\endgroup }) \in  {\mathcal{E}}_{{\begingroup\renewcommand\colorMATH{\colorMATHC}\renewcommand\colorSYNTAX{\colorSYNTAXC}{{\color{\colorMATH}\ensuremath{p'}}}\endgroup }}^{k}\llbracket \sigma '\rrbracket }}}
  \end{enumerate} 
\end{lemma}
\begin{proof}
  We only present intersting cases.
  We first prove (1) by induction on {{\color{\colorMATH}\ensuremath{\tau }}}:
  \begin{enumerate}[ncases]\item  {{\color{\colorMATH}\ensuremath{\sigma ={\begingroup\renewcommand\colorMATH{\colorMATHA}\renewcommand\colorSYNTAX{\colorSYNTAXA}{{\color{\colorSYNTAX}\texttt{{\ensuremath{{\mathbb{R}}}}}}}\endgroup }}}}
    \begin{subproof} 
      Then {{\color{\colorMATH}\ensuremath{{\begingroup\renewcommand\colorMATH{\colorMATHB}\renewcommand\colorSYNTAX{\colorSYNTAXB}{{\color{\colorMATH}\ensuremath{\sv_{1}}}}\endgroup } = {\begingroup\renewcommand\colorMATH{\colorMATHB}\renewcommand\colorSYNTAX{\colorSYNTAXB}{{\color{\colorMATH}\ensuremath{r_{1}}}}\endgroup }}}}, {{\color{\colorMATH}\ensuremath{{\begingroup\renewcommand\colorMATH{\colorMATHB}\renewcommand\colorSYNTAX{\colorSYNTAXB}{{\color{\colorMATH}\ensuremath{\sv_{2}}}}\endgroup } = {\begingroup\renewcommand\colorMATH{\colorMATHB}\renewcommand\colorSYNTAX{\colorSYNTAXB}{{\color{\colorMATH}\ensuremath{r_{2}}}}\endgroup }}}}, {{\color{\colorMATH}\ensuremath{\sigma ' = {\begingroup\renewcommand\colorMATH{\colorMATHA}\renewcommand\colorSYNTAX{\colorSYNTAXA}{{\color{\colorSYNTAX}\texttt{{\ensuremath{{\mathbb{R}}}}}}}\endgroup }}}}, and {{\color{\colorMATH}\ensuremath{({\begingroup\renewcommand\colorMATH{\colorMATHB}\renewcommand\colorSYNTAX{\colorSYNTAXB}{{\color{\colorMATH}\ensuremath{r_{1}}}}\endgroup },{\begingroup\renewcommand\colorMATH{\colorMATHB}\renewcommand\colorSYNTAX{\colorSYNTAXB}{{\color{\colorMATH}\ensuremath{r_{2}}}}\endgroup }) \in  {\mathcal{V}}_{{\begingroup\renewcommand\colorMATH{\colorMATHB}\renewcommand\colorSYNTAX{\colorSYNTAXB}{{\color{\colorMATH}\ensuremath{\distance}}}\endgroup }}^{k}\llbracket {\begingroup\renewcommand\colorMATH{\colorMATHA}\renewcommand\colorSYNTAX{\colorSYNTAXA}{{\color{\colorSYNTAX}\texttt{{\ensuremath{{\mathbb{R}}}}}}}\endgroup }\rrbracket }}}, i.e. {{\color{\colorMATH}\ensuremath{|{\begingroup\renewcommand\colorMATH{\colorMATHB}\renewcommand\colorSYNTAX{\colorSYNTAXB}{{\color{\colorMATH}\ensuremath{r_{1}}}}\endgroup } - {\begingroup\renewcommand\colorMATH{\colorMATHB}\renewcommand\colorSYNTAX{\colorSYNTAXB}{{\color{\colorMATH}\ensuremath{r_{2}}}}\endgroup }| \leq  s}}}. But if {{\color{\colorMATH}\ensuremath{s \leq  s'}}}, then it is easy to see that 
      {{\color{\colorMATH}\ensuremath{|{\begingroup\renewcommand\colorMATH{\colorMATHB}\renewcommand\colorSYNTAX{\colorSYNTAXB}{{\color{\colorMATH}\ensuremath{r_{1}}}}\endgroup } - {\begingroup\renewcommand\colorMATH{\colorMATHB}\renewcommand\colorSYNTAX{\colorSYNTAXB}{{\color{\colorMATH}\ensuremath{r_{2}}}}\endgroup }| \leq  s'}}}, therefore {{\color{\colorMATH}\ensuremath{({\begingroup\renewcommand\colorMATH{\colorMATHB}\renewcommand\colorSYNTAX{\colorSYNTAXB}{{\color{\colorMATH}\ensuremath{r_{1}}}}\endgroup },{\begingroup\renewcommand\colorMATH{\colorMATHB}\renewcommand\colorSYNTAX{\colorSYNTAXB}{{\color{\colorMATH}\ensuremath{r_{2}}}}\endgroup }) \in  {\mathcal{V}}_{{\begingroup\renewcommand\colorMATH{\colorMATHB}\renewcommand\colorSYNTAX{\colorSYNTAXB}{{\color{\colorMATH}\ensuremath{\distance'}}}\endgroup }}^{k}\llbracket {\begingroup\renewcommand\colorMATH{\colorMATHA}\renewcommand\colorSYNTAX{\colorSYNTAXA}{{\color{\colorSYNTAX}\texttt{{\ensuremath{{\mathbb{R}}}}}}}\endgroup }\rrbracket }}} and the result holds.
    \end{subproof}
  \item  {{\color{\colorMATH}\ensuremath{\sigma ={{\color{\colorSYNTAX}\texttt{unit}}}}}}
    \begin{subproof} 
      Trivial as  {{\color{\colorMATH}\ensuremath{\sigma ' = {{\color{\colorSYNTAX}\texttt{unit}}}}}} and {{\color{\colorMATH}\ensuremath{({\begingroup\renewcommand\colorMATH{\colorMATHB}\renewcommand\colorSYNTAX{\colorSYNTAXB}{{\color{\colorMATH}\ensuremath{\sv_{1}}}}\endgroup },{\begingroup\renewcommand\colorMATH{\colorMATHB}\renewcommand\colorSYNTAX{\colorSYNTAXB}{{\color{\colorMATH}\ensuremath{\sv_{2}}}}\endgroup }) \in  {\mathcal{V}}_{{\begingroup\renewcommand\colorMATH{\colorMATHB}\renewcommand\colorSYNTAX{\colorSYNTAXB}{{\color{\colorMATH}\ensuremath{\distance}}}\endgroup }}^{k}\llbracket {{\color{\colorSYNTAX}\texttt{unit}}}\rrbracket }}} does not depend on {{\color{\colorMATH}\ensuremath{s}}}, i.e. {{\color{\colorMATH}\ensuremath{\forall s', ({\begingroup\renewcommand\colorMATH{\colorMATHB}\renewcommand\colorSYNTAX{\colorSYNTAXB}{{\color{\colorMATH}\ensuremath{\sv_{1}}}}\endgroup },{\begingroup\renewcommand\colorMATH{\colorMATHB}\renewcommand\colorSYNTAX{\colorSYNTAXB}{{\color{\colorMATH}\ensuremath{\sv_{2}}}}\endgroup }) \in  {\mathcal{V}}_{{\begingroup\renewcommand\colorMATH{\colorMATHB}\renewcommand\colorSYNTAX{\colorSYNTAXB}{{\color{\colorMATH}\ensuremath{\distance'}}}\endgroup }}^{k}\llbracket {{\color{\colorSYNTAX}\texttt{unit}}}\rrbracket }}}.
    \end{subproof}
  \item  {{\color{\colorMATH}\ensuremath{\sigma =(x\mathrel{:}\sigma _{1}\mathord{\cdotp }{\begingroup\renewcommand\colorMATH{\colorMATHB}\renewcommand\colorSYNTAX{\colorSYNTAXB}{{\color{\colorMATH}\ensuremath{\distance_{2}}}}\endgroup }) \xrightarrowS { {\begingroup\renewcommand\colorMATH{\colorMATHB}\renewcommand\colorSYNTAX{\colorSYNTAXB}{{\color{\colorMATH}\ensuremath{\Distance_{1}}}}\endgroup }\mathord{\cdotp }{\begingroup\renewcommand\colorMATH{\colorMATHB}\renewcommand\colorSYNTAX{\colorSYNTAXB}{{\color{\colorMATH}\ensuremath{\sS_{2}}}}\endgroup } + {\begingroup\renewcommand\colorMATH{\colorMATHB}\renewcommand\colorSYNTAX{\colorSYNTAXB}{{\color{\colorMATH}\ensuremath{\distance_{1}}}}\endgroup }x} \sigma _{2}}}}
    \begin{subproof} 
      Then {{\color{\colorMATH}\ensuremath{\sigma ' = (x\mathrel{:}\sigma '_{1}\mathord{\cdotp }{\begingroup\renewcommand\colorMATH{\colorMATHB}\renewcommand\colorSYNTAX{\colorSYNTAXB}{{\color{\colorMATH}\ensuremath{\distance'_{2}}}}\endgroup }) \xrightarrowS { {\begingroup\renewcommand\colorMATH{\colorMATHB}\renewcommand\colorSYNTAX{\colorSYNTAXB}{{\color{\colorMATH}\ensuremath{\Distance_{1}}}}\endgroup }\mathord{\cdotp }{\begingroup\renewcommand\colorMATH{\colorMATHB}\renewcommand\colorSYNTAX{\colorSYNTAXB}{{\color{\colorMATH}\ensuremath{\sS'_{2}}}}\endgroup } + {\begingroup\renewcommand\colorMATH{\colorMATHB}\renewcommand\colorSYNTAX{\colorSYNTAXB}{{\color{\colorMATH}\ensuremath{\distance'_{1}}}}\endgroup }x} \sigma '_{2}}}} where {{\color{\colorMATH}\ensuremath{\sigma '_{1} <: \sigma _{1}, {\begingroup\renewcommand\colorMATH{\colorMATHB}\renewcommand\colorSYNTAX{\colorSYNTAXB}{{\color{\colorMATH}\ensuremath{\sS_{2}}}}\endgroup } <: {\begingroup\renewcommand\colorMATH{\colorMATHB}\renewcommand\colorSYNTAX{\colorSYNTAXB}{{\color{\colorMATH}\ensuremath{\sS'_{2}}}}\endgroup }, {\begingroup\renewcommand\colorMATH{\colorMATHB}\renewcommand\colorSYNTAX{\colorSYNTAXB}{{\color{\colorMATH}\ensuremath{\distance_{1}}}}\endgroup } \leq  {\begingroup\renewcommand\colorMATH{\colorMATHB}\renewcommand\colorSYNTAX{\colorSYNTAXB}{{\color{\colorMATH}\ensuremath{\distance'_{1}}}}\endgroup }, {\begingroup\renewcommand\colorMATH{\colorMATHB}\renewcommand\colorSYNTAX{\colorSYNTAXB}{{\color{\colorMATH}\ensuremath{\distance'_{2}}}}\endgroup } < {\begingroup\renewcommand\colorMATH{\colorMATHB}\renewcommand\colorSYNTAX{\colorSYNTAXB}{{\color{\colorMATH}\ensuremath{\distance_{2}}}}\endgroup }}}} and {{\color{\colorMATH}\ensuremath{\sigma _{2} <: \sigma '_{2}}}}.\\
      Also {{\color{\colorMATH}\ensuremath{{\begingroup\renewcommand\colorMATH{\colorMATHB}\renewcommand\colorSYNTAX{\colorSYNTAXB}{{\color{\colorMATH}\ensuremath{\sv_{1}}}}\endgroup } = \langle {\begingroup\renewcommand\colorMATH{\colorMATHB}\renewcommand\colorSYNTAX{\colorSYNTAXB}{{\color{\colorMATH}\ensuremath{\slambda}}}\endgroup } x.{\begingroup\renewcommand\colorMATH{\colorMATHB}\renewcommand\colorSYNTAX{\colorSYNTAXB}{{\color{\colorMATH}\ensuremath{\se_{1}}}}\endgroup },\gamma _{1}\rangle }}}, {{\color{\colorMATH}\ensuremath{{\begingroup\renewcommand\colorMATH{\colorMATHB}\renewcommand\colorSYNTAX{\colorSYNTAXB}{{\color{\colorMATH}\ensuremath{\sv_{2}}}}\endgroup } = \langle {\begingroup\renewcommand\colorMATH{\colorMATHB}\renewcommand\colorSYNTAX{\colorSYNTAXB}{{\color{\colorMATH}\ensuremath{\slambda}}}\endgroup } x.{\begingroup\renewcommand\colorMATH{\colorMATHB}\renewcommand\colorSYNTAX{\colorSYNTAXB}{{\color{\colorMATH}\ensuremath{\se_{2}}}}\endgroup },\gamma _{2}\rangle }}}, and {{\color{\colorMATH}\ensuremath{(\langle {\begingroup\renewcommand\colorMATH{\colorMATHB}\renewcommand\colorSYNTAX{\colorSYNTAXB}{{\color{\colorMATH}\ensuremath{\slambda}}}\endgroup } x.{\begingroup\renewcommand\colorMATH{\colorMATHB}\renewcommand\colorSYNTAX{\colorSYNTAXB}{{\color{\colorMATH}\ensuremath{\se_{1}}}}\endgroup },\gamma _{1}\rangle ,\langle {\begingroup\renewcommand\colorMATH{\colorMATHB}\renewcommand\colorSYNTAX{\colorSYNTAXB}{{\color{\colorMATH}\ensuremath{\slambda}}}\endgroup } x.{\begingroup\renewcommand\colorMATH{\colorMATHB}\renewcommand\colorSYNTAX{\colorSYNTAXB}{{\color{\colorMATH}\ensuremath{\se_{2}}}}\endgroup },\gamma _{2}\rangle ) \in  {\mathcal{V}}_{{\begingroup\renewcommand\colorMATH{\colorMATHB}\renewcommand\colorSYNTAX{\colorSYNTAXB}{{\color{\colorMATH}\ensuremath{\distance}}}\endgroup }}\llbracket (x\mathrel{:}\sigma _{1}\mathord{\cdotp }{\begingroup\renewcommand\colorMATH{\colorMATHB}\renewcommand\colorSYNTAX{\colorSYNTAXB}{{\color{\colorMATH}\ensuremath{\distance_{2}}}}\endgroup }) \xrightarrowS { {\begingroup\renewcommand\colorMATH{\colorMATHB}\renewcommand\colorSYNTAX{\colorSYNTAXB}{{\color{\colorMATH}\ensuremath{\Distance_{1}}}}\endgroup }\mathord{\cdotp }{\begingroup\renewcommand\colorMATH{\colorMATHB}\renewcommand\colorSYNTAX{\colorSYNTAXB}{{\color{\colorMATH}\ensuremath{\sS_{2}}}}\endgroup } + {\begingroup\renewcommand\colorMATH{\colorMATHB}\renewcommand\colorSYNTAX{\colorSYNTAXB}{{\color{\colorMATH}\ensuremath{\distance_{1}}}}\endgroup }x} \sigma _{2}\rrbracket }}}.\\
      
      We have to prove that {{\color{\colorMATH}\ensuremath{(\langle {\begingroup\renewcommand\colorMATH{\colorMATHB}\renewcommand\colorSYNTAX{\colorSYNTAXB}{{\color{\colorMATH}\ensuremath{\slambda}}}\endgroup } x.{\begingroup\renewcommand\colorMATH{\colorMATHB}\renewcommand\colorSYNTAX{\colorSYNTAXB}{{\color{\colorMATH}\ensuremath{\se_{1}}}}\endgroup },\gamma _{1}\rangle ,\langle {\begingroup\renewcommand\colorMATH{\colorMATHB}\renewcommand\colorSYNTAX{\colorSYNTAXB}{{\color{\colorMATH}\ensuremath{\slambda}}}\endgroup } x.{\begingroup\renewcommand\colorMATH{\colorMATHB}\renewcommand\colorSYNTAX{\colorSYNTAXB}{{\color{\colorMATH}\ensuremath{\se_{2}}}}\endgroup },\gamma _{2}\rangle ) \in  {\mathcal{V}}_{{\begingroup\renewcommand\colorMATH{\colorMATHB}\renewcommand\colorSYNTAX{\colorSYNTAXB}{{\color{\colorMATH}\ensuremath{\distance'}}}\endgroup }}^{k}\llbracket (x\mathrel{:}\sigma '_{1}\mathord{\cdotp }{\begingroup\renewcommand\colorMATH{\colorMATHB}\renewcommand\colorSYNTAX{\colorSYNTAXB}{{\color{\colorMATH}\ensuremath{\distance'_{2}}}}\endgroup }) \xrightarrowS { {\begingroup\renewcommand\colorMATH{\colorMATHB}\renewcommand\colorSYNTAX{\colorSYNTAXB}{{\color{\colorMATH}\ensuremath{\Distance_{1}}}}\endgroup }\mathord{\cdotp }{\begingroup\renewcommand\colorMATH{\colorMATHB}\renewcommand\colorSYNTAX{\colorSYNTAXB}{{\color{\colorMATH}\ensuremath{\sS'_{2}}}}\endgroup } + {\begingroup\renewcommand\colorMATH{\colorMATHB}\renewcommand\colorSYNTAX{\colorSYNTAXB}{{\color{\colorMATH}\ensuremath{\distance'_{1}}}}\endgroup }x} \sigma '_{2}\rrbracket }}}, i.e. for any {{\color{\colorMATH}\ensuremath{j < k}}}, {{\color{\colorMATH}\ensuremath{({\begingroup\renewcommand\colorMATH{\colorMATHB}\renewcommand\colorSYNTAX{\colorSYNTAXB}{{\color{\colorMATH}\ensuremath{\sv_{1}}}}\endgroup },{\begingroup\renewcommand\colorMATH{\colorMATHB}\renewcommand\colorSYNTAX{\colorSYNTAXB}{{\color{\colorMATH}\ensuremath{\sv_{2}}}}\endgroup }) \in  {\mathcal{V}}^{j}_{{\begingroup\renewcommand\colorMATH{\colorMATHB}\renewcommand\colorSYNTAX{\colorSYNTAXB}{{\color{\colorMATH}\ensuremath{\distance_{3}}}}\endgroup }}\llbracket \sigma '_{1}\rrbracket }}}, {{\color{\colorMATH}\ensuremath{{\begingroup\renewcommand\colorMATH{\colorMATHB}\renewcommand\colorSYNTAX{\colorSYNTAXB}{{\color{\colorMATH}\ensuremath{\distance_{3}}}}\endgroup } \leq  {\begingroup\renewcommand\colorMATH{\colorMATHB}\renewcommand\colorSYNTAX{\colorSYNTAXB}{{\color{\colorMATH}\ensuremath{\distance'_{2}}}}\endgroup }}}} 
      then {{\color{\colorMATH}\ensuremath{(\gamma _{1}[x\mapsto {\begingroup\renewcommand\colorMATH{\colorMATHB}\renewcommand\colorSYNTAX{\colorSYNTAXB}{{\color{\colorMATH}\ensuremath{\sv_{1}}}}\endgroup }] \vdash  {\begingroup\renewcommand\colorMATH{\colorMATHB}\renewcommand\colorSYNTAX{\colorSYNTAXB}{{\color{\colorMATH}\ensuremath{\se_{1}}}}\endgroup },\gamma _{2}[x\mapsto {\begingroup\renewcommand\colorMATH{\colorMATHB}\renewcommand\colorSYNTAX{\colorSYNTAXB}{{\color{\colorMATH}\ensuremath{\sv_{2}}}}\endgroup }] \vdash  {\begingroup\renewcommand\colorMATH{\colorMATHB}\renewcommand\colorSYNTAX{\colorSYNTAXB}{{\color{\colorMATH}\ensuremath{\se_{2}}}}\endgroup }) \in  {\mathcal{E}}^{j}_{{\begingroup\renewcommand\colorMATH{\colorMATHB}\renewcommand\colorSYNTAX{\colorSYNTAXB}{{\color{\colorMATH}\ensuremath{\distance'}}}\endgroup }+ {\begingroup\renewcommand\colorMATH{\colorMATHB}\renewcommand\colorSYNTAX{\colorSYNTAXB}{{\color{\colorMATH}\ensuremath{\Distance_{1}}}}\endgroup }\mathord{\cdotp }{\begingroup\renewcommand\colorMATH{\colorMATHB}\renewcommand\colorSYNTAX{\colorSYNTAXB}{{\color{\colorMATH}\ensuremath{\sS'_{2}}}}\endgroup }+{\begingroup\renewcommand\colorMATH{\colorMATHB}\renewcommand\colorSYNTAX{\colorSYNTAXB}{{\color{\colorMATH}\ensuremath{\distance'_{1}}}}\endgroup }{\begingroup\renewcommand\colorMATH{\colorMATHB}\renewcommand\colorSYNTAX{\colorSYNTAXB}{{\color{\colorMATH}\ensuremath{\distance_{3}}}}\endgroup }}\llbracket {\begingroup\renewcommand\colorMATH{\colorMATHB}\renewcommand\colorSYNTAX{\colorSYNTAXB}{{\color{\colorMATH}\ensuremath{\distance_{3}}}}\endgroup }x(\sigma '_{2})\rrbracket }}}.

      We know that {{\color{\colorMATH}\ensuremath{ {\begingroup\renewcommand\colorMATH{\colorMATHB}\renewcommand\colorSYNTAX{\colorSYNTAXB}{{\color{\colorMATH}\ensuremath{\distance_{3}}}}\endgroup } \leq  {\begingroup\renewcommand\colorMATH{\colorMATHB}\renewcommand\colorSYNTAX{\colorSYNTAXB}{{\color{\colorMATH}\ensuremath{\distance'_{2}}}}\endgroup } \leq  {\begingroup\renewcommand\colorMATH{\colorMATHB}\renewcommand\colorSYNTAX{\colorSYNTAXB}{{\color{\colorMATH}\ensuremath{\distance_{2}}}}\endgroup }}}}, then by induction hypothesis on {{\color{\colorMATH}\ensuremath{({\begingroup\renewcommand\colorMATH{\colorMATHB}\renewcommand\colorSYNTAX{\colorSYNTAXB}{{\color{\colorMATH}\ensuremath{\sv_{1}}}}\endgroup },{\begingroup\renewcommand\colorMATH{\colorMATHB}\renewcommand\colorSYNTAX{\colorSYNTAXB}{{\color{\colorMATH}\ensuremath{\sv_{2}}}}\endgroup }) \in  {\mathcal{V}}^{j}_{{\begingroup\renewcommand\colorMATH{\colorMATHB}\renewcommand\colorSYNTAX{\colorSYNTAXB}{{\color{\colorMATH}\ensuremath{\distance_{3}}}}\endgroup }}\llbracket \sigma '_{1}\rrbracket }}}, we know that {{\color{\colorMATH}\ensuremath{({\begingroup\renewcommand\colorMATH{\colorMATHB}\renewcommand\colorSYNTAX{\colorSYNTAXB}{{\color{\colorMATH}\ensuremath{\sv_{1}}}}\endgroup },{\begingroup\renewcommand\colorMATH{\colorMATHB}\renewcommand\colorSYNTAX{\colorSYNTAXB}{{\color{\colorMATH}\ensuremath{\sv_{2}}}}\endgroup }) \in  {\mathcal{V}}^{j}_{{\begingroup\renewcommand\colorMATH{\colorMATHB}\renewcommand\colorSYNTAX{\colorSYNTAXB}{{\color{\colorMATH}\ensuremath{\distance_{3}}}}\endgroup }}\llbracket \sigma _{1}\rrbracket }}}.
      Then we instantiate the premise with {{\color{\colorMATH}\ensuremath{{\begingroup\renewcommand\colorMATH{\colorMATHB}\renewcommand\colorSYNTAX{\colorSYNTAXB}{{\color{\colorMATH}\ensuremath{\distance''}}}\endgroup }={\begingroup\renewcommand\colorMATH{\colorMATHB}\renewcommand\colorSYNTAX{\colorSYNTAXB}{{\color{\colorMATH}\ensuremath{\distance_{3}}}}\endgroup }}}} and {{\color{\colorMATH}\ensuremath{{\begingroup\renewcommand\colorMATH{\colorMATHB}\renewcommand\colorSYNTAX{\colorSYNTAXB}{{\color{\colorMATH}\ensuremath{\sv_{1}}}}\endgroup }}}} and {{\color{\colorMATH}\ensuremath{{\begingroup\renewcommand\colorMATH{\colorMATHB}\renewcommand\colorSYNTAX{\colorSYNTAXB}{{\color{\colorMATH}\ensuremath{\sv_{2}}}}\endgroup }}}}, so we know that
      {{\color{\colorMATH}\ensuremath{(\gamma _{1}[x\mapsto {\begingroup\renewcommand\colorMATH{\colorMATHB}\renewcommand\colorSYNTAX{\colorSYNTAXB}{{\color{\colorMATH}\ensuremath{\sv_{1}}}}\endgroup }] \vdash  {\begingroup\renewcommand\colorMATH{\colorMATHB}\renewcommand\colorSYNTAX{\colorSYNTAXB}{{\color{\colorMATH}\ensuremath{\se_{1}}}}\endgroup },\gamma _{2}[x\mapsto {\begingroup\renewcommand\colorMATH{\colorMATHB}\renewcommand\colorSYNTAX{\colorSYNTAXB}{{\color{\colorMATH}\ensuremath{\sv_{2}}}}\endgroup }] \vdash  {\begingroup\renewcommand\colorMATH{\colorMATHB}\renewcommand\colorSYNTAX{\colorSYNTAXB}{{\color{\colorMATH}\ensuremath{\se_{2}}}}\endgroup }) \in  {\mathcal{E}}^{j}_{{\begingroup\renewcommand\colorMATH{\colorMATHB}\renewcommand\colorSYNTAX{\colorSYNTAXB}{{\color{\colorMATH}\ensuremath{\distance}}}\endgroup }+{\begingroup\renewcommand\colorMATH{\colorMATHB}\renewcommand\colorSYNTAX{\colorSYNTAXB}{{\color{\colorMATH}\ensuremath{\Distance_{1}}}}\endgroup }\mathord{\cdotp }{\begingroup\renewcommand\colorMATH{\colorMATHB}\renewcommand\colorSYNTAX{\colorSYNTAXB}{{\color{\colorMATH}\ensuremath{\sS_{2}}}}\endgroup }+{\begingroup\renewcommand\colorMATH{\colorMATHB}\renewcommand\colorSYNTAX{\colorSYNTAXB}{{\color{\colorMATH}\ensuremath{\distance_{1}}}}\endgroup }{\begingroup\renewcommand\colorMATH{\colorMATHB}\renewcommand\colorSYNTAX{\colorSYNTAXB}{{\color{\colorMATH}\ensuremath{\distance_{3}}}}\endgroup }}\llbracket {\begingroup\renewcommand\colorMATH{\colorMATHB}\renewcommand\colorSYNTAX{\colorSYNTAXB}{{\color{\colorMATH}\ensuremath{\distance_{3}}}}\endgroup }x(\sigma _{2})\rrbracket }}}.
      By Lemma~\ref{lm:subtypinginst}, {{\color{\colorMATH}\ensuremath{{\begingroup\renewcommand\colorMATH{\colorMATHB}\renewcommand\colorSYNTAX{\colorSYNTAXB}{{\color{\colorMATH}\ensuremath{\distance_{3}}}}\endgroup }x(\sigma _{2}) <: {\begingroup\renewcommand\colorMATH{\colorMATHB}\renewcommand\colorSYNTAX{\colorSYNTAXB}{{\color{\colorMATH}\ensuremath{\distance_{3}}}}\endgroup }x(\sigma '_{2})}}}, by Lemma~\ref{lm:dot-subt}, {{\color{\colorMATH}\ensuremath{{\begingroup\renewcommand\colorMATH{\colorMATHB}\renewcommand\colorSYNTAX{\colorSYNTAXB}{{\color{\colorMATH}\ensuremath{\Distance_{1}}}}\endgroup }\mathord{\cdotp }{\begingroup\renewcommand\colorMATH{\colorMATHB}\renewcommand\colorSYNTAX{\colorSYNTAXB}{{\color{\colorMATH}\ensuremath{\sS_{2}}}}\endgroup } \leq  {\begingroup\renewcommand\colorMATH{\colorMATHB}\renewcommand\colorSYNTAX{\colorSYNTAXB}{{\color{\colorMATH}\ensuremath{\Distance_{1}}}}\endgroup }\mathord{\cdotp }{\begingroup\renewcommand\colorMATH{\colorMATHB}\renewcommand\colorSYNTAX{\colorSYNTAXB}{{\color{\colorMATH}\ensuremath{\sS'_{2}}}}\endgroup }}}}, and as {{\color{\colorMATH}\ensuremath{{\begingroup\renewcommand\colorMATH{\colorMATHB}\renewcommand\colorSYNTAX{\colorSYNTAXB}{{\color{\colorMATH}\ensuremath{\distance}}}\endgroup }+{\begingroup\renewcommand\colorMATH{\colorMATHB}\renewcommand\colorSYNTAX{\colorSYNTAXB}{{\color{\colorMATH}\ensuremath{\Distance_{1}}}}\endgroup }\mathord{\cdotp }{\begingroup\renewcommand\colorMATH{\colorMATHB}\renewcommand\colorSYNTAX{\colorSYNTAXB}{{\color{\colorMATH}\ensuremath{\sS_{2}}}}\endgroup }+{\begingroup\renewcommand\colorMATH{\colorMATHB}\renewcommand\colorSYNTAX{\colorSYNTAXB}{{\color{\colorMATH}\ensuremath{\distance_{1}}}}\endgroup }{\begingroup\renewcommand\colorMATH{\colorMATHB}\renewcommand\colorSYNTAX{\colorSYNTAXB}{{\color{\colorMATH}\ensuremath{\distance_{3}}}}\endgroup } \leq  {\begingroup\renewcommand\colorMATH{\colorMATHB}\renewcommand\colorSYNTAX{\colorSYNTAXB}{{\color{\colorMATH}\ensuremath{\distance'}}}\endgroup }+{\begingroup\renewcommand\colorMATH{\colorMATHB}\renewcommand\colorSYNTAX{\colorSYNTAXB}{{\color{\colorMATH}\ensuremath{\Distance_{1}}}}\endgroup }\mathord{\cdotp }{\begingroup\renewcommand\colorMATH{\colorMATHB}\renewcommand\colorSYNTAX{\colorSYNTAXB}{{\color{\colorMATH}\ensuremath{\sS'_{2}}}}\endgroup }+{\begingroup\renewcommand\colorMATH{\colorMATHB}\renewcommand\colorSYNTAX{\colorSYNTAXB}{{\color{\colorMATH}\ensuremath{\distance'_{1}}}}\endgroup }{\begingroup\renewcommand\colorMATH{\colorMATHB}\renewcommand\colorSYNTAX{\colorSYNTAXB}{{\color{\colorMATH}\ensuremath{\distance_{3}}}}\endgroup }}}},
      then by induction hypothesis we know that {{\color{\colorMATH}\ensuremath{(\gamma _{1}[x\mapsto {\begingroup\renewcommand\colorMATH{\colorMATHB}\renewcommand\colorSYNTAX{\colorSYNTAXB}{{\color{\colorMATH}\ensuremath{\sv_{1}}}}\endgroup }] \vdash  {\begingroup\renewcommand\colorMATH{\colorMATHB}\renewcommand\colorSYNTAX{\colorSYNTAXB}{{\color{\colorMATH}\ensuremath{\se_{1}}}}\endgroup },\gamma _{2}[x\mapsto {\begingroup\renewcommand\colorMATH{\colorMATHB}\renewcommand\colorSYNTAX{\colorSYNTAXB}{{\color{\colorMATH}\ensuremath{\sv_{2}}}}\endgroup }] \vdash  {\begingroup\renewcommand\colorMATH{\colorMATHB}\renewcommand\colorSYNTAX{\colorSYNTAXB}{{\color{\colorMATH}\ensuremath{\se_{2}}}}\endgroup }) \in  {\mathcal{E}}^{j}_{{\begingroup\renewcommand\colorMATH{\colorMATHB}\renewcommand\colorSYNTAX{\colorSYNTAXB}{{\color{\colorMATH}\ensuremath{\distance'}}}\endgroup }+{\begingroup\renewcommand\colorMATH{\colorMATHB}\renewcommand\colorSYNTAX{\colorSYNTAXB}{{\color{\colorMATH}\ensuremath{\sS'_{1}}}}\endgroup }\mathord{\cdotp }{\begingroup\renewcommand\colorMATH{\colorMATHB}\renewcommand\colorSYNTAX{\colorSYNTAXB}{{\color{\colorMATH}\ensuremath{\sS'_{2}}}}\endgroup }+{\begingroup\renewcommand\colorMATH{\colorMATHB}\renewcommand\colorSYNTAX{\colorSYNTAXB}{{\color{\colorMATH}\ensuremath{\distance'_{1}}}}\endgroup }{\begingroup\renewcommand\colorMATH{\colorMATHB}\renewcommand\colorSYNTAX{\colorSYNTAXB}{{\color{\colorMATH}\ensuremath{\distance_{3}}}}\endgroup }}\llbracket {\begingroup\renewcommand\colorMATH{\colorMATHB}\renewcommand\colorSYNTAX{\colorSYNTAXB}{{\color{\colorMATH}\ensuremath{\distance_{3}}}}\endgroup }x(\sigma '_{2})\rrbracket }}} and the result holds.
    \end{subproof}
  \item  {{\color{\colorMATH}\ensuremath{\sigma =(x\mathrel{:}\sigma _{1}\mathord{\cdotp }{\begingroup\renewcommand\colorMATH{\colorMATHB}\renewcommand\colorSYNTAX{\colorSYNTAXB}{{\color{\colorMATH}\ensuremath{\distance_{2}}}}\endgroup }) \xrightarrowP {{\begingroup\renewcommand\colorMATH{\colorMATHB}\renewcommand\colorSYNTAX{\colorSYNTAXB}{{\color{\colorMATH}\ensuremath{\Distance_{1}}}}\endgroup }{\begingroup\renewcommand\colorMATH{\colorMATHC}\renewcommand\colorSYNTAX{\colorSYNTAXC}{{\color{\colorMATH}\ensuremath{\bigcdot}}}\endgroup }{\begingroup\renewcommand\colorMATH{\colorMATHC}\renewcommand\colorSYNTAX{\colorSYNTAXC}{{\color{\colorMATH}\ensuremath{\pS_{2}}}}\endgroup }} \sigma _{2}}}}
    \begin{subproof} 
      Then {{\color{\colorMATH}\ensuremath{\sigma ' = (x\mathrel{:}\sigma '_{1}\mathord{\cdotp }{\begingroup\renewcommand\colorMATH{\colorMATHB}\renewcommand\colorSYNTAX{\colorSYNTAXB}{{\color{\colorMATH}\ensuremath{\distance'_{2}}}}\endgroup }) \xrightarrowP {{\begingroup\renewcommand\colorMATH{\colorMATHB}\renewcommand\colorSYNTAX{\colorSYNTAXB}{{\color{\colorMATH}\ensuremath{\Distance_{1}}}}\endgroup }{\begingroup\renewcommand\colorMATH{\colorMATHC}\renewcommand\colorSYNTAX{\colorSYNTAXC}{{\color{\colorMATH}\ensuremath{\bigcdot}}}\endgroup }{\begingroup\renewcommand\colorMATH{\colorMATHC}\renewcommand\colorSYNTAX{\colorSYNTAXC}{{\color{\colorMATH}\ensuremath{\pS'_{2}}}}\endgroup }} \sigma '_{2}}}} where {{\color{\colorMATH}\ensuremath{\sigma '_{1} <: \sigma _{1}, {\begingroup\renewcommand\colorMATH{\colorMATHC}\renewcommand\colorSYNTAX{\colorSYNTAXC}{{\color{\colorMATH}\ensuremath{\pS_{2}}}}\endgroup } <: {\begingroup\renewcommand\colorMATH{\colorMATHC}\renewcommand\colorSYNTAX{\colorSYNTAXC}{{\color{\colorMATH}\ensuremath{\pS'_{2}}}}\endgroup }, {\begingroup\renewcommand\colorMATH{\colorMATHB}\renewcommand\colorSYNTAX{\colorSYNTAXB}{{\color{\colorMATH}\ensuremath{\distance'_{2}}}}\endgroup } < {\begingroup\renewcommand\colorMATH{\colorMATHB}\renewcommand\colorSYNTAX{\colorSYNTAXB}{{\color{\colorMATH}\ensuremath{\distance_{2}}}}\endgroup }}}} and {{\color{\colorMATH}\ensuremath{\sigma _{2} <: \sigma '_{2}}}}.\\
      Also {{\color{\colorMATH}\ensuremath{{\begingroup\renewcommand\colorMATH{\colorMATHB}\renewcommand\colorSYNTAX{\colorSYNTAXB}{{\color{\colorMATH}\ensuremath{\sv_{1}}}}\endgroup } = \langle {\begingroup\renewcommand\colorMATH{\colorMATHC}\renewcommand\colorSYNTAX{\colorSYNTAXC}{{\color{\colorMATH}\ensuremath{\plambda}}}\endgroup } x.{\begingroup\renewcommand\colorMATH{\colorMATHC}\renewcommand\colorSYNTAX{\colorSYNTAXC}{{\color{\colorMATH}\ensuremath{\pe_{1}}}}\endgroup },\gamma _{1}\rangle }}}, {{\color{\colorMATH}\ensuremath{{\begingroup\renewcommand\colorMATH{\colorMATHB}\renewcommand\colorSYNTAX{\colorSYNTAXB}{{\color{\colorMATH}\ensuremath{\sv_{2}}}}\endgroup } = \langle {\begingroup\renewcommand\colorMATH{\colorMATHC}\renewcommand\colorSYNTAX{\colorSYNTAXC}{{\color{\colorMATH}\ensuremath{\plambda}}}\endgroup } x.{\begingroup\renewcommand\colorMATH{\colorMATHC}\renewcommand\colorSYNTAX{\colorSYNTAXC}{{\color{\colorMATH}\ensuremath{\pe_{2}}}}\endgroup },\gamma _{2}\rangle }}}, and {{\color{\colorMATH}\ensuremath{(\langle {\begingroup\renewcommand\colorMATH{\colorMATHC}\renewcommand\colorSYNTAX{\colorSYNTAXC}{{\color{\colorMATH}\ensuremath{\plambda}}}\endgroup } x.{\begingroup\renewcommand\colorMATH{\colorMATHC}\renewcommand\colorSYNTAX{\colorSYNTAXC}{{\color{\colorMATH}\ensuremath{\pe_{1}}}}\endgroup },\gamma _{1}\rangle ,\langle {\begingroup\renewcommand\colorMATH{\colorMATHC}\renewcommand\colorSYNTAX{\colorSYNTAXC}{{\color{\colorMATH}\ensuremath{\plambda}}}\endgroup } x.{\begingroup\renewcommand\colorMATH{\colorMATHC}\renewcommand\colorSYNTAX{\colorSYNTAXC}{{\color{\colorMATH}\ensuremath{\pe_{2}}}}\endgroup },\gamma _{2}\rangle ) \in  {\mathcal{V}}_{{\begingroup\renewcommand\colorMATH{\colorMATHB}\renewcommand\colorSYNTAX{\colorSYNTAXB}{{\color{\colorMATH}\ensuremath{\distance}}}\endgroup }}\llbracket (x\mathrel{:}\sigma _{1}\mathord{\cdotp }{\begingroup\renewcommand\colorMATH{\colorMATHB}\renewcommand\colorSYNTAX{\colorSYNTAXB}{{\color{\colorMATH}\ensuremath{\distance_{2}}}}\endgroup }) \xrightarrowP {{\begingroup\renewcommand\colorMATH{\colorMATHB}\renewcommand\colorSYNTAX{\colorSYNTAXB}{{\color{\colorMATH}\ensuremath{\Distance_{1}}}}\endgroup }{\begingroup\renewcommand\colorMATH{\colorMATHC}\renewcommand\colorSYNTAX{\colorSYNTAXC}{{\color{\colorMATH}\ensuremath{\bigcdot}}}\endgroup }{\begingroup\renewcommand\colorMATH{\colorMATHC}\renewcommand\colorSYNTAX{\colorSYNTAXC}{{\color{\colorMATH}\ensuremath{\pS_{2}}}}\endgroup }} \sigma _{2}\rrbracket }}}.\\
      
      We have to prove that {{\color{\colorMATH}\ensuremath{(\langle {\begingroup\renewcommand\colorMATH{\colorMATHC}\renewcommand\colorSYNTAX{\colorSYNTAXC}{{\color{\colorMATH}\ensuremath{\plambda}}}\endgroup } x.{\begingroup\renewcommand\colorMATH{\colorMATHC}\renewcommand\colorSYNTAX{\colorSYNTAXC}{{\color{\colorMATH}\ensuremath{\pe_{1}}}}\endgroup },\gamma _{1}\rangle ,\langle {\begingroup\renewcommand\colorMATH{\colorMATHB}\renewcommand\colorSYNTAX{\colorSYNTAXB}{{\color{\colorMATH}\ensuremath{\slambda}}}\endgroup } x.{\begingroup\renewcommand\colorMATH{\colorMATHC}\renewcommand\colorSYNTAX{\colorSYNTAXC}{{\color{\colorMATH}\ensuremath{\pe_{2}}}}\endgroup },\gamma _{2}\rangle ) \in  {\mathcal{V}}_{{\begingroup\renewcommand\colorMATH{\colorMATHB}\renewcommand\colorSYNTAX{\colorSYNTAXB}{{\color{\colorMATH}\ensuremath{\distance'}}}\endgroup }}^{k}\llbracket (x\mathrel{:}\sigma '_{1}\mathord{\cdotp }{\begingroup\renewcommand\colorMATH{\colorMATHB}\renewcommand\colorSYNTAX{\colorSYNTAXB}{{\color{\colorMATH}\ensuremath{\distance'_{2}}}}\endgroup }) \xrightarrowP {{\begingroup\renewcommand\colorMATH{\colorMATHB}\renewcommand\colorSYNTAX{\colorSYNTAXB}{{\color{\colorMATH}\ensuremath{\Distance_{1}}}}\endgroup }{\begingroup\renewcommand\colorMATH{\colorMATHC}\renewcommand\colorSYNTAX{\colorSYNTAXC}{{\color{\colorMATH}\ensuremath{\bigcdot}}}\endgroup }{\begingroup\renewcommand\colorMATH{\colorMATHC}\renewcommand\colorSYNTAX{\colorSYNTAXC}{{\color{\colorMATH}\ensuremath{\pS'_{2}}}}\endgroup }} \sigma '_{2}\rrbracket }}}, i.e. for any {{\color{\colorMATH}\ensuremath{j < k}}}, {{\color{\colorMATH}\ensuremath{({\begingroup\renewcommand\colorMATH{\colorMATHB}\renewcommand\colorSYNTAX{\colorSYNTAXB}{{\color{\colorMATH}\ensuremath{\sv_{1}}}}\endgroup },{\begingroup\renewcommand\colorMATH{\colorMATHB}\renewcommand\colorSYNTAX{\colorSYNTAXB}{{\color{\colorMATH}\ensuremath{\sv_{2}}}}\endgroup }) \in  {\mathcal{V}}^{j}_{{\begingroup\renewcommand\colorMATH{\colorMATHB}\renewcommand\colorSYNTAX{\colorSYNTAXB}{{\color{\colorMATH}\ensuremath{\distance_{3}}}}\endgroup }}\llbracket \sigma '_{1}\rrbracket }}}, {{\color{\colorMATH}\ensuremath{{\begingroup\renewcommand\colorMATH{\colorMATHB}\renewcommand\colorSYNTAX{\colorSYNTAXB}{{\color{\colorMATH}\ensuremath{\distance_{3}}}}\endgroup } \leq  {\begingroup\renewcommand\colorMATH{\colorMATHB}\renewcommand\colorSYNTAX{\colorSYNTAXB}{{\color{\colorMATH}\ensuremath{\distance'_{2}}}}\endgroup }}}} 
      then {{\color{\colorMATH}\ensuremath{(\gamma _{1}[x\mapsto {\begingroup\renewcommand\colorMATH{\colorMATHB}\renewcommand\colorSYNTAX{\colorSYNTAXB}{{\color{\colorMATH}\ensuremath{\sv_{1}}}}\endgroup }] \vdash  {\begingroup\renewcommand\colorMATH{\colorMATHC}\renewcommand\colorSYNTAX{\colorSYNTAXC}{{\color{\colorMATH}\ensuremath{\pe_{1}}}}\endgroup },\gamma _{2}[x\mapsto {\begingroup\renewcommand\colorMATH{\colorMATHB}\renewcommand\colorSYNTAX{\colorSYNTAXB}{{\color{\colorMATH}\ensuremath{\sv_{2}}}}\endgroup }] \vdash  {\begingroup\renewcommand\colorMATH{\colorMATHC}\renewcommand\colorSYNTAX{\colorSYNTAXC}{{\color{\colorMATH}\ensuremath{\pe_{2}}}}\endgroup }) \in  {\mathcal{E}}^{j}_{{\begingroup\renewcommand\colorMATH{\colorMATHC}\renewcommand\colorSYNTAX{\colorSYNTAXC}{{\color{\colorMATH}\ensuremath{\rceil }}}\endgroup }{\begingroup\renewcommand\colorMATH{\colorMATHB}\renewcommand\colorSYNTAX{\colorSYNTAXB}{{\color{\colorMATH}\ensuremath{\distance'}}}\endgroup }{\begingroup\renewcommand\colorMATH{\colorMATHC}\renewcommand\colorSYNTAX{\colorSYNTAXC}{{\color{\colorMATH}\ensuremath{\lceil ^{\infty }}}}\endgroup }+{\begingroup\renewcommand\colorMATH{\colorMATHB}\renewcommand\colorSYNTAX{\colorSYNTAXB}{{\color{\colorMATH}\ensuremath{\Distance_{1}}}}\endgroup }{\begingroup\renewcommand\colorMATH{\colorMATHC}\renewcommand\colorSYNTAX{\colorSYNTAXC}{{\color{\colorMATH}\ensuremath{\bigcdot}}}\endgroup }{\begingroup\renewcommand\colorMATH{\colorMATHC}\renewcommand\colorSYNTAX{\colorSYNTAXC}{{\color{\colorMATH}\ensuremath{\pS'_{2}}}}\endgroup }}\llbracket {\begingroup\renewcommand\colorMATH{\colorMATHB}\renewcommand\colorSYNTAX{\colorSYNTAXB}{{\color{\colorMATH}\ensuremath{\distance_{3}}}}\endgroup }x(\sigma '_{2})\rrbracket }}}.

      We know that {{\color{\colorMATH}\ensuremath{ {\begingroup\renewcommand\colorMATH{\colorMATHB}\renewcommand\colorSYNTAX{\colorSYNTAXB}{{\color{\colorMATH}\ensuremath{\distance_{3}}}}\endgroup } \leq  {\begingroup\renewcommand\colorMATH{\colorMATHB}\renewcommand\colorSYNTAX{\colorSYNTAXB}{{\color{\colorMATH}\ensuremath{\distance'_{2}}}}\endgroup } \leq  {\begingroup\renewcommand\colorMATH{\colorMATHB}\renewcommand\colorSYNTAX{\colorSYNTAXB}{{\color{\colorMATH}\ensuremath{\distance_{2}}}}\endgroup }}}}, then by induction hypothesis on {{\color{\colorMATH}\ensuremath{({\begingroup\renewcommand\colorMATH{\colorMATHB}\renewcommand\colorSYNTAX{\colorSYNTAXB}{{\color{\colorMATH}\ensuremath{\sv_{1}}}}\endgroup },{\begingroup\renewcommand\colorMATH{\colorMATHB}\renewcommand\colorSYNTAX{\colorSYNTAXB}{{\color{\colorMATH}\ensuremath{\sv_{2}}}}\endgroup }) \in  {\mathcal{V}}^{j}_{{\begingroup\renewcommand\colorMATH{\colorMATHB}\renewcommand\colorSYNTAX{\colorSYNTAXB}{{\color{\colorMATH}\ensuremath{\distance_{3}}}}\endgroup }}\llbracket \sigma '_{1}\rrbracket }}}, we know that {{\color{\colorMATH}\ensuremath{({\begingroup\renewcommand\colorMATH{\colorMATHB}\renewcommand\colorSYNTAX{\colorSYNTAXB}{{\color{\colorMATH}\ensuremath{\sv_{1}}}}\endgroup },{\begingroup\renewcommand\colorMATH{\colorMATHB}\renewcommand\colorSYNTAX{\colorSYNTAXB}{{\color{\colorMATH}\ensuremath{\sv_{2}}}}\endgroup }) \in  {\mathcal{V}}^{j}_{{\begingroup\renewcommand\colorMATH{\colorMATHB}\renewcommand\colorSYNTAX{\colorSYNTAXB}{{\color{\colorMATH}\ensuremath{\distance_{3}}}}\endgroup }}\llbracket \sigma _{1}\rrbracket }}}.
      Then we instantiate the premise with {{\color{\colorMATH}\ensuremath{{\begingroup\renewcommand\colorMATH{\colorMATHB}\renewcommand\colorSYNTAX{\colorSYNTAXB}{{\color{\colorMATH}\ensuremath{\distance''}}}\endgroup }={\begingroup\renewcommand\colorMATH{\colorMATHB}\renewcommand\colorSYNTAX{\colorSYNTAXB}{{\color{\colorMATH}\ensuremath{\distance_{3}}}}\endgroup }}}} and {{\color{\colorMATH}\ensuremath{{\begingroup\renewcommand\colorMATH{\colorMATHB}\renewcommand\colorSYNTAX{\colorSYNTAXB}{{\color{\colorMATH}\ensuremath{\sv_{1}}}}\endgroup }}}} and {{\color{\colorMATH}\ensuremath{{\begingroup\renewcommand\colorMATH{\colorMATHB}\renewcommand\colorSYNTAX{\colorSYNTAXB}{{\color{\colorMATH}\ensuremath{\sv_{2}}}}\endgroup }}}}, so we know that
      {{\color{\colorMATH}\ensuremath{(\gamma _{1}[x\mapsto {\begingroup\renewcommand\colorMATH{\colorMATHB}\renewcommand\colorSYNTAX{\colorSYNTAXB}{{\color{\colorMATH}\ensuremath{\sv_{1}}}}\endgroup }] \vdash  {\begingroup\renewcommand\colorMATH{\colorMATHC}\renewcommand\colorSYNTAX{\colorSYNTAXC}{{\color{\colorMATH}\ensuremath{\pe_{1}}}}\endgroup },\gamma _{2}[x\mapsto {\begingroup\renewcommand\colorMATH{\colorMATHB}\renewcommand\colorSYNTAX{\colorSYNTAXB}{{\color{\colorMATH}\ensuremath{\sv_{2}}}}\endgroup }] \vdash  {\begingroup\renewcommand\colorMATH{\colorMATHC}\renewcommand\colorSYNTAX{\colorSYNTAXC}{{\color{\colorMATH}\ensuremath{\pe_{2}}}}\endgroup }) \in  {\mathcal{E}}^{j}_{{\begingroup\renewcommand\colorMATH{\colorMATHC}\renewcommand\colorSYNTAX{\colorSYNTAXC}{{\color{\colorMATH}\ensuremath{\rceil {\begingroup\renewcommand\colorMATH{\colorMATHA}\renewcommand\colorSYNTAX{\colorSYNTAXA}{{\color{\colorMATH}\ensuremath{{\begingroup\renewcommand\colorMATH{\colorMATHB}\renewcommand\colorSYNTAX{\colorSYNTAXB}{{\color{\colorMATH}\ensuremath{\distance}}}\endgroup }}}}\endgroup }\lceil ^{\infty }}}}\endgroup }+{\begingroup\renewcommand\colorMATH{\colorMATHB}\renewcommand\colorSYNTAX{\colorSYNTAXB}{{\color{\colorMATH}\ensuremath{\Distance_{1}}}}\endgroup }{\begingroup\renewcommand\colorMATH{\colorMATHC}\renewcommand\colorSYNTAX{\colorSYNTAXC}{{\color{\colorMATH}\ensuremath{\bigcdot}}}\endgroup }{\begingroup\renewcommand\colorMATH{\colorMATHC}\renewcommand\colorSYNTAX{\colorSYNTAXC}{{\color{\colorMATH}\ensuremath{\pS_{2}}}}\endgroup }}\llbracket {\begingroup\renewcommand\colorMATH{\colorMATHB}\renewcommand\colorSYNTAX{\colorSYNTAXB}{{\color{\colorMATH}\ensuremath{\distance_{3}}}}\endgroup }x(\sigma _{2})\rrbracket }}}.
      By Lemma~\ref{lm:subtypinginst}, {{\color{\colorMATH}\ensuremath{{\begingroup\renewcommand\colorMATH{\colorMATHB}\renewcommand\colorSYNTAX{\colorSYNTAXB}{{\color{\colorMATH}\ensuremath{\distance_{3}}}}\endgroup }x(\sigma _{2}) <: {\begingroup\renewcommand\colorMATH{\colorMATHB}\renewcommand\colorSYNTAX{\colorSYNTAXB}{{\color{\colorMATH}\ensuremath{\distance_{3}}}}\endgroup }x(\sigma '_{2})}}}, by Lemma~\ref{lm:dot-subtp}, {{\color{\colorMATH}\ensuremath{{\begingroup\renewcommand\colorMATH{\colorMATHB}\renewcommand\colorSYNTAX{\colorSYNTAXB}{{\color{\colorMATH}\ensuremath{\Distance_{1}}}}\endgroup }{\begingroup\renewcommand\colorMATH{\colorMATHC}\renewcommand\colorSYNTAX{\colorSYNTAXC}{{\color{\colorMATH}\ensuremath{\bigcdot}}}\endgroup }{\begingroup\renewcommand\colorMATH{\colorMATHC}\renewcommand\colorSYNTAX{\colorSYNTAXC}{{\color{\colorMATH}\ensuremath{\pS_{2}}}}\endgroup } \leq  {\begingroup\renewcommand\colorMATH{\colorMATHB}\renewcommand\colorSYNTAX{\colorSYNTAXB}{{\color{\colorMATH}\ensuremath{\Distance_{1}}}}\endgroup }\mathord{\cdotp }{\begingroup\renewcommand\colorMATH{\colorMATHC}\renewcommand\colorSYNTAX{\colorSYNTAXC}{{\color{\colorMATH}\ensuremath{\pS'_{2}}}}\endgroup }}}}, then {{\color{\colorMATH}\ensuremath{{\begingroup\renewcommand\colorMATH{\colorMATHC}\renewcommand\colorSYNTAX{\colorSYNTAXC}{{\color{\colorMATH}\ensuremath{\rceil {\begingroup\renewcommand\colorMATH{\colorMATHA}\renewcommand\colorSYNTAX{\colorSYNTAXA}{{\color{\colorMATH}\ensuremath{{\begingroup\renewcommand\colorMATH{\colorMATHB}\renewcommand\colorSYNTAX{\colorSYNTAXB}{{\color{\colorMATH}\ensuremath{\distance}}}\endgroup }}}}\endgroup }\lceil ^{\infty }}}}\endgroup }+{\begingroup\renewcommand\colorMATH{\colorMATHB}\renewcommand\colorSYNTAX{\colorSYNTAXB}{{\color{\colorMATH}\ensuremath{\Distance_{1}}}}\endgroup }{\begingroup\renewcommand\colorMATH{\colorMATHC}\renewcommand\colorSYNTAX{\colorSYNTAXC}{{\color{\colorMATH}\ensuremath{\bigcdot}}}\endgroup }{\begingroup\renewcommand\colorMATH{\colorMATHC}\renewcommand\colorSYNTAX{\colorSYNTAXC}{{\color{\colorMATH}\ensuremath{\pS_{2}}}}\endgroup } \leq  {\begingroup\renewcommand\colorMATH{\colorMATHC}\renewcommand\colorSYNTAX{\colorSYNTAXC}{{\color{\colorMATH}\ensuremath{\rceil {\begingroup\renewcommand\colorMATH{\colorMATHA}\renewcommand\colorSYNTAX{\colorSYNTAXA}{{\color{\colorMATH}\ensuremath{{\begingroup\renewcommand\colorMATH{\colorMATHB}\renewcommand\colorSYNTAX{\colorSYNTAXB}{{\color{\colorMATH}\ensuremath{\distance'}}}\endgroup }}}}\endgroup }\lceil ^{\infty }}}}\endgroup }+{\begingroup\renewcommand\colorMATH{\colorMATHB}\renewcommand\colorSYNTAX{\colorSYNTAXB}{{\color{\colorMATH}\ensuremath{\Distance_{1}}}}\endgroup }{\begingroup\renewcommand\colorMATH{\colorMATHC}\renewcommand\colorSYNTAX{\colorSYNTAXC}{{\color{\colorMATH}\ensuremath{\bigcdot}}}\endgroup }{\begingroup\renewcommand\colorMATH{\colorMATHC}\renewcommand\colorSYNTAX{\colorSYNTAXC}{{\color{\colorMATH}\ensuremath{\pS'_{2}}}}\endgroup }}}}.
      The result follows from induction hypothesis (3).
    \end{subproof}
  \item  {{\color{\colorMATH}\ensuremath{\sigma =\sigma _{1} \mathrel{^{{\begingroup\renewcommand\colorMATH{\colorMATHB}\renewcommand\colorSYNTAX{\colorSYNTAXB}{{\color{\colorMATH}\ensuremath{\sss_{1}}}}\endgroup }}\oplus ^{{\begingroup\renewcommand\colorMATH{\colorMATHB}\renewcommand\colorSYNTAX{\colorSYNTAXB}{{\color{\colorMATH}\ensuremath{\sss_{2}}}}\endgroup }}} \sigma _{2}}}}
    \begin{subproof} 
      Then {{\color{\colorMATH}\ensuremath{\sigma ' = \sigma '_{1} \mathrel{^{{\begingroup\renewcommand\colorMATH{\colorMATHB}\renewcommand\colorSYNTAX{\colorSYNTAXB}{{\color{\colorMATH}\ensuremath{\sss'_{1}}}}\endgroup }}\oplus ^{{\begingroup\renewcommand\colorMATH{\colorMATHB}\renewcommand\colorSYNTAX{\colorSYNTAXB}{{\color{\colorMATH}\ensuremath{\sss'_{2}}}}\endgroup }}} \sigma '_{2}}}} where {{\color{\colorMATH}\ensuremath{\sigma _{1} <: \sigma '_{1}, {\begingroup\renewcommand\colorMATH{\colorMATHB}\renewcommand\colorSYNTAX{\colorSYNTAXB}{{\color{\colorMATH}\ensuremath{\sss_{1}}}}\endgroup } \leq  {\begingroup\renewcommand\colorMATH{\colorMATHB}\renewcommand\colorSYNTAX{\colorSYNTAXB}{{\color{\colorMATH}\ensuremath{\sss'_{1}}}}\endgroup }, {\begingroup\renewcommand\colorMATH{\colorMATHB}\renewcommand\colorSYNTAX{\colorSYNTAXB}{{\color{\colorMATH}\ensuremath{\sss_{2}}}}\endgroup } \leq  {\begingroup\renewcommand\colorMATH{\colorMATHB}\renewcommand\colorSYNTAX{\colorSYNTAXB}{{\color{\colorMATH}\ensuremath{\sss'_{2}}}}\endgroup },}}} and {{\color{\colorMATH}\ensuremath{\sigma _{2} <: \sigma '_{2}}}}.\\
      We proceed by case analysis on {{\color{\colorMATH}\ensuremath{({\begingroup\renewcommand\colorMATH{\colorMATHB}\renewcommand\colorSYNTAX{\colorSYNTAXB}{{\color{\colorMATH}\ensuremath{\sv_{1}}}}\endgroup }, {\begingroup\renewcommand\colorMATH{\colorMATHB}\renewcommand\colorSYNTAX{\colorSYNTAXB}{{\color{\colorMATH}\ensuremath{\sv_{2}}}}\endgroup })}}}:
      \begin{enumerate}[ncases]\item  {{\color{\colorMATH}\ensuremath{({\begingroup\renewcommand\colorMATH{\colorMATHB}\renewcommand\colorSYNTAX{\colorSYNTAXB}{{\color{\colorMATH}\ensuremath{\sv_{1}}}}\endgroup }, {\begingroup\renewcommand\colorMATH{\colorMATHB}\renewcommand\colorSYNTAX{\colorSYNTAXB}{{\color{\colorMATH}\ensuremath{\sv_{2}}}}\endgroup }) = (\inl\hspace*{0.33em}{\begingroup\renewcommand\colorMATH{\colorMATHB}\renewcommand\colorSYNTAX{\colorSYNTAXB}{{\color{\colorMATH}\ensuremath{\sv'_{1}}}}\endgroup }, \inl\hspace*{0.33em}{\begingroup\renewcommand\colorMATH{\colorMATHB}\renewcommand\colorSYNTAX{\colorSYNTAXB}{{\color{\colorMATH}\ensuremath{\sv'_{2}}}}\endgroup })}}}
        \begin{subproof} 
          Then we know that {{\color{\colorMATH}\ensuremath{(\inl\hspace*{0.33em}{\begingroup\renewcommand\colorMATH{\colorMATHB}\renewcommand\colorSYNTAX{\colorSYNTAXB}{{\color{\colorMATH}\ensuremath{\sv'_{1}}}}\endgroup }, \inl\hspace*{0.33em}{\begingroup\renewcommand\colorMATH{\colorMATHB}\renewcommand\colorSYNTAX{\colorSYNTAXB}{{\color{\colorMATH}\ensuremath{\sv'_{2}}}}\endgroup }) \in  {\mathcal{V}}^{k}_{{\begingroup\renewcommand\colorMATH{\colorMATHB}\renewcommand\colorSYNTAX{\colorSYNTAXB}{{\color{\colorMATH}\ensuremath{\distance}}}\endgroup }+{\begingroup\renewcommand\colorMATH{\colorMATHB}\renewcommand\colorSYNTAX{\colorSYNTAXB}{{\color{\colorMATH}\ensuremath{\sss_{1}}}}\endgroup }}\llbracket \sigma _{1}\rrbracket }}}, then by induction hypothesis using {{\color{\colorMATH}\ensuremath{s+{\begingroup\renewcommand\colorMATH{\colorMATHB}\renewcommand\colorSYNTAX{\colorSYNTAXB}{{\color{\colorMATH}\ensuremath{\sss_{1}}}}\endgroup } \leq  {\begingroup\renewcommand\colorMATH{\colorMATHB}\renewcommand\colorSYNTAX{\colorSYNTAXB}{{\color{\colorMATH}\ensuremath{\sss'}}}\endgroup }+{\begingroup\renewcommand\colorMATH{\colorMATHB}\renewcommand\colorSYNTAX{\colorSYNTAXB}{{\color{\colorMATH}\ensuremath{\sss'_{1}}}}\endgroup }}}} and {{\color{\colorMATH}\ensuremath{\sigma _{1} <: \sigma _{2}}}}, then 
          {{\color{\colorMATH}\ensuremath{(\inl\hspace*{0.33em}{\begingroup\renewcommand\colorMATH{\colorMATHB}\renewcommand\colorSYNTAX{\colorSYNTAXB}{{\color{\colorMATH}\ensuremath{\sv'_{1}}}}\endgroup }, \inl\hspace*{0.33em}{\begingroup\renewcommand\colorMATH{\colorMATHB}\renewcommand\colorSYNTAX{\colorSYNTAXB}{{\color{\colorMATH}\ensuremath{\sv'_{2}}}}\endgroup }) \in  {\mathcal{V}}^{k}_{{\begingroup\renewcommand\colorMATH{\colorMATHB}\renewcommand\colorSYNTAX{\colorSYNTAXB}{{\color{\colorMATH}\ensuremath{\sss'}}}\endgroup }+{\begingroup\renewcommand\colorMATH{\colorMATHB}\renewcommand\colorSYNTAX{\colorSYNTAXB}{{\color{\colorMATH}\ensuremath{\sss'_{1}}}}\endgroup }}\llbracket \sigma '_{1}\rrbracket }}} and therefore {{\color{\colorMATH}\ensuremath{(\inl\hspace*{0.33em}{\begingroup\renewcommand\colorMATH{\colorMATHB}\renewcommand\colorSYNTAX{\colorSYNTAXB}{{\color{\colorMATH}\ensuremath{\sv'_{1}}}}\endgroup }, \inl\hspace*{0.33em}{\begingroup\renewcommand\colorMATH{\colorMATHB}\renewcommand\colorSYNTAX{\colorSYNTAXB}{{\color{\colorMATH}\ensuremath{\sv'_{2}}}}\endgroup }) \in  {\mathcal{V}}_{{\begingroup\renewcommand\colorMATH{\colorMATHB}\renewcommand\colorSYNTAX{\colorSYNTAXB}{{\color{\colorMATH}\ensuremath{\sss'}}}\endgroup }}^{k}\llbracket \sigma '_{1} \mathrel{^{{\begingroup\renewcommand\colorMATH{\colorMATHB}\renewcommand\colorSYNTAX{\colorSYNTAXB}{{\color{\colorMATH}\ensuremath{\sss'_{1}}}}\endgroup }}\oplus ^{{\begingroup\renewcommand\colorMATH{\colorMATHB}\renewcommand\colorSYNTAX{\colorSYNTAXB}{{\color{\colorMATH}\ensuremath{\sss'_{2}}}}\endgroup }}} \sigma '_{2}\rrbracket }}} and the result holds.
        \end{subproof}
      \item  {{\color{\colorMATH}\ensuremath{({\begingroup\renewcommand\colorMATH{\colorMATHB}\renewcommand\colorSYNTAX{\colorSYNTAXB}{{\color{\colorMATH}\ensuremath{\sv_{1}}}}\endgroup }, {\begingroup\renewcommand\colorMATH{\colorMATHB}\renewcommand\colorSYNTAX{\colorSYNTAXB}{{\color{\colorMATH}\ensuremath{\sv_{2}}}}\endgroup }) = (\inr\hspace*{0.33em}{\begingroup\renewcommand\colorMATH{\colorMATHB}\renewcommand\colorSYNTAX{\colorSYNTAXB}{{\color{\colorMATH}\ensuremath{\sv'_{1}}}}\endgroup }, \inr\hspace*{0.33em}{\begingroup\renewcommand\colorMATH{\colorMATHB}\renewcommand\colorSYNTAX{\colorSYNTAXB}{{\color{\colorMATH}\ensuremath{\sv'_{2}}}}\endgroup })}}}
        \begin{subproof} 
          Analogous to previous case.
        \end{subproof}
      \item  {{\color{\colorMATH}\ensuremath{({\begingroup\renewcommand\colorMATH{\colorMATHB}\renewcommand\colorSYNTAX{\colorSYNTAXB}{{\color{\colorMATH}\ensuremath{\sv_{1}}}}\endgroup }, {\begingroup\renewcommand\colorMATH{\colorMATHB}\renewcommand\colorSYNTAX{\colorSYNTAXB}{{\color{\colorMATH}\ensuremath{\sv_{2}}}}\endgroup }) = (\inl\hspace*{0.33em}{\begingroup\renewcommand\colorMATH{\colorMATHB}\renewcommand\colorSYNTAX{\colorSYNTAXB}{{\color{\colorMATH}\ensuremath{\sv'_{1}}}}\endgroup }, \inr\hspace*{0.33em}{\begingroup\renewcommand\colorMATH{\colorMATHB}\renewcommand\colorSYNTAX{\colorSYNTAXB}{{\color{\colorMATH}\ensuremath{\sv'_{2}}}}\endgroup })}}} and {{\color{\colorMATH}\ensuremath{({\begingroup\renewcommand\colorMATH{\colorMATHB}\renewcommand\colorSYNTAX{\colorSYNTAXB}{{\color{\colorMATH}\ensuremath{\sv_{1}}}}\endgroup }, {\begingroup\renewcommand\colorMATH{\colorMATHB}\renewcommand\colorSYNTAX{\colorSYNTAXB}{{\color{\colorMATH}\ensuremath{\sv_{2}}}}\endgroup }) = (\inr\hspace*{0.33em}{\begingroup\renewcommand\colorMATH{\colorMATHB}\renewcommand\colorSYNTAX{\colorSYNTAXB}{{\color{\colorMATH}\ensuremath{\sv'_{1}}}}\endgroup }, \inl\hspace*{0.33em}{\begingroup\renewcommand\colorMATH{\colorMATHB}\renewcommand\colorSYNTAX{\colorSYNTAXB}{{\color{\colorMATH}\ensuremath{\sv'_{2}}}}\endgroup })}}}
        \begin{subproof} 
          Then {{\color{\colorMATH}\ensuremath{s = \infty }}} and as {{\color{\colorMATH}\ensuremath{s \leq  {\begingroup\renewcommand\colorMATH{\colorMATHB}\renewcommand\colorSYNTAX{\colorSYNTAXB}{{\color{\colorMATH}\ensuremath{\sss'}}}\endgroup }}}}, therefore {{\color{\colorMATH}\ensuremath{{\begingroup\renewcommand\colorMATH{\colorMATHB}\renewcommand\colorSYNTAX{\colorSYNTAXB}{{\color{\colorMATH}\ensuremath{\sss'}}}\endgroup } = \infty }}}, and the result holds immediately.
        \end{subproof}
      \end{enumerate}
    \end{subproof}
  \item  {{\color{\colorMATH}\ensuremath{\sigma  = \sigma _{1} \mathrel{^{{\begingroup\renewcommand\colorMATH{\colorMATHB}\renewcommand\colorSYNTAX{\colorSYNTAXB}{{\color{\colorMATH}\ensuremath{\sss_{1}}}}\endgroup }}\&^{{\begingroup\renewcommand\colorMATH{\colorMATHB}\renewcommand\colorSYNTAX{\colorSYNTAXB}{{\color{\colorMATH}\ensuremath{\sss_{2}}}}\endgroup }}} \sigma _{2}}}}
    \begin{subproof} 
      Then {{\color{\colorMATH}\ensuremath{\sigma ' = \sigma '_{1} \mathrel{^{{\begingroup\renewcommand\colorMATH{\colorMATHB}\renewcommand\colorSYNTAX{\colorSYNTAXB}{{\color{\colorMATH}\ensuremath{\sss'_{1}}}}\endgroup }}\&^{{\begingroup\renewcommand\colorMATH{\colorMATHB}\renewcommand\colorSYNTAX{\colorSYNTAXB}{{\color{\colorMATH}\ensuremath{\sss'_{2}}}}\endgroup }}} \sigma '_{2}}}} where {{\color{\colorMATH}\ensuremath{\sigma _{1} <: \sigma '_{1}, {\begingroup\renewcommand\colorMATH{\colorMATHB}\renewcommand\colorSYNTAX{\colorSYNTAXB}{{\color{\colorMATH}\ensuremath{\sss_{1}}}}\endgroup } \leq  {\begingroup\renewcommand\colorMATH{\colorMATHB}\renewcommand\colorSYNTAX{\colorSYNTAXB}{{\color{\colorMATH}\ensuremath{\sss'_{1}}}}\endgroup }, {\begingroup\renewcommand\colorMATH{\colorMATHB}\renewcommand\colorSYNTAX{\colorSYNTAXB}{{\color{\colorMATH}\ensuremath{\sss_{2}}}}\endgroup } \leq  {\begingroup\renewcommand\colorMATH{\colorMATHB}\renewcommand\colorSYNTAX{\colorSYNTAXB}{{\color{\colorMATH}\ensuremath{\sss'_{2}}}}\endgroup },}}} and {{\color{\colorMATH}\ensuremath{\sigma _{2} <: \sigma '_{2}}}}.\\
      We know that {{\color{\colorMATH}\ensuremath{({\begingroup\renewcommand\colorMATH{\colorMATHB}\renewcommand\colorSYNTAX{\colorSYNTAXB}{{\color{\colorMATH}\ensuremath{\sv_{1}}}}\endgroup }, {\begingroup\renewcommand\colorMATH{\colorMATHB}\renewcommand\colorSYNTAX{\colorSYNTAXB}{{\color{\colorMATH}\ensuremath{\sv_{2}}}}\endgroup }) = (\langle {\begingroup\renewcommand\colorMATH{\colorMATHB}\renewcommand\colorSYNTAX{\colorSYNTAXB}{{\color{\colorMATH}\ensuremath{\sv_{1 1}}}}\endgroup }, {\begingroup\renewcommand\colorMATH{\colorMATHB}\renewcommand\colorSYNTAX{\colorSYNTAXB}{{\color{\colorMATH}\ensuremath{\sv_{1 2}}}}\endgroup }\rangle , \langle {\begingroup\renewcommand\colorMATH{\colorMATHB}\renewcommand\colorSYNTAX{\colorSYNTAXB}{{\color{\colorMATH}\ensuremath{\sv_{2 1}}}}\endgroup }, {\begingroup\renewcommand\colorMATH{\colorMATHB}\renewcommand\colorSYNTAX{\colorSYNTAXB}{{\color{\colorMATH}\ensuremath{\sv_{2 2}}}}\endgroup }\rangle )}}}, such that 
      {{\color{\colorMATH}\ensuremath{({\begingroup\renewcommand\colorMATH{\colorMATHB}\renewcommand\colorSYNTAX{\colorSYNTAXB}{{\color{\colorMATH}\ensuremath{\sv_{1 1}}}}\endgroup }, {\begingroup\renewcommand\colorMATH{\colorMATHB}\renewcommand\colorSYNTAX{\colorSYNTAXB}{{\color{\colorMATH}\ensuremath{\sv_{2 1}}}}\endgroup }) \in  {\mathcal{V}}^{k}_{{\begingroup\renewcommand\colorMATH{\colorMATHB}\renewcommand\colorSYNTAX{\colorSYNTAXB}{{\color{\colorMATH}\ensuremath{\distance}}}\endgroup }+{\begingroup\renewcommand\colorMATH{\colorMATHB}\renewcommand\colorSYNTAX{\colorSYNTAXB}{{\color{\colorMATH}\ensuremath{\sss_{1}}}}\endgroup }}\llbracket \sigma _{1}\rrbracket }}} and {{\color{\colorMATH}\ensuremath{({\begingroup\renewcommand\colorMATH{\colorMATHB}\renewcommand\colorSYNTAX{\colorSYNTAXB}{{\color{\colorMATH}\ensuremath{\sv_{1 2}}}}\endgroup }, {\begingroup\renewcommand\colorMATH{\colorMATHB}\renewcommand\colorSYNTAX{\colorSYNTAXB}{{\color{\colorMATH}\ensuremath{\sv_{2 2}}}}\endgroup }) \in  {\mathcal{V}}^{k}_{{\begingroup\renewcommand\colorMATH{\colorMATHB}\renewcommand\colorSYNTAX{\colorSYNTAXB}{{\color{\colorMATH}\ensuremath{\distance}}}\endgroup }+{\begingroup\renewcommand\colorMATH{\colorMATHB}\renewcommand\colorSYNTAX{\colorSYNTAXB}{{\color{\colorMATH}\ensuremath{\sss_{2}}}}\endgroup }}\llbracket \sigma _{2}\rrbracket }}}.
      By induction hypotheses we know that {{\color{\colorMATH}\ensuremath{({\begingroup\renewcommand\colorMATH{\colorMATHB}\renewcommand\colorSYNTAX{\colorSYNTAXB}{{\color{\colorMATH}\ensuremath{\sv_{1 1}}}}\endgroup }, {\begingroup\renewcommand\colorMATH{\colorMATHB}\renewcommand\colorSYNTAX{\colorSYNTAXB}{{\color{\colorMATH}\ensuremath{\sv_{2 1}}}}\endgroup }) \in  {\mathcal{V}}^{k}_{{\begingroup\renewcommand\colorMATH{\colorMATHB}\renewcommand\colorSYNTAX{\colorSYNTAXB}{{\color{\colorMATH}\ensuremath{\sss'}}}\endgroup }+{\begingroup\renewcommand\colorMATH{\colorMATHB}\renewcommand\colorSYNTAX{\colorSYNTAXB}{{\color{\colorMATH}\ensuremath{\sss'_{1}}}}\endgroup }}\llbracket \sigma '_{1}\rrbracket }}} and {{\color{\colorMATH}\ensuremath{({\begingroup\renewcommand\colorMATH{\colorMATHB}\renewcommand\colorSYNTAX{\colorSYNTAXB}{{\color{\colorMATH}\ensuremath{\sv_{1 2}}}}\endgroup }, {\begingroup\renewcommand\colorMATH{\colorMATHB}\renewcommand\colorSYNTAX{\colorSYNTAXB}{{\color{\colorMATH}\ensuremath{\sv_{2 2}}}}\endgroup }) \in  {\mathcal{V}}^{k}_{{\begingroup\renewcommand\colorMATH{\colorMATHB}\renewcommand\colorSYNTAX{\colorSYNTAXB}{{\color{\colorMATH}\ensuremath{\sss'}}}\endgroup }+{\begingroup\renewcommand\colorMATH{\colorMATHB}\renewcommand\colorSYNTAX{\colorSYNTAXB}{{\color{\colorMATH}\ensuremath{\sss'_{2}}}}\endgroup }}\llbracket \sigma '_{2}\rrbracket }}}, and therefore
      {{\color{\colorMATH}\ensuremath{(\langle {\begingroup\renewcommand\colorMATH{\colorMATHB}\renewcommand\colorSYNTAX{\colorSYNTAXB}{{\color{\colorMATH}\ensuremath{\sv_{1 1}}}}\endgroup }, {\begingroup\renewcommand\colorMATH{\colorMATHB}\renewcommand\colorSYNTAX{\colorSYNTAXB}{{\color{\colorMATH}\ensuremath{\sv_{1 2}}}}\endgroup }\rangle , \langle {\begingroup\renewcommand\colorMATH{\colorMATHB}\renewcommand\colorSYNTAX{\colorSYNTAXB}{{\color{\colorMATH}\ensuremath{\sv_{2 1}}}}\endgroup }, {\begingroup\renewcommand\colorMATH{\colorMATHB}\renewcommand\colorSYNTAX{\colorSYNTAXB}{{\color{\colorMATH}\ensuremath{\sv_{2 2}}}}\endgroup }\rangle ) \in  {\mathcal{V}}_{{\begingroup\renewcommand\colorMATH{\colorMATHB}\renewcommand\colorSYNTAX{\colorSYNTAXB}{{\color{\colorMATH}\ensuremath{\sss'}}}\endgroup }}^{k}\llbracket \sigma '_{1} \mathrel{^{{\begingroup\renewcommand\colorMATH{\colorMATHB}\renewcommand\colorSYNTAX{\colorSYNTAXB}{{\color{\colorMATH}\ensuremath{\sss'_{1}}}}\endgroup }}\&^{{\begingroup\renewcommand\colorMATH{\colorMATHB}\renewcommand\colorSYNTAX{\colorSYNTAXB}{{\color{\colorMATH}\ensuremath{\sss'_{2}}}}\endgroup }}} \sigma '_{2}\rrbracket }}}  and the result holds.
    \end{subproof}
  \end{enumerate}
  Now let us prove (2). We know that if {{\color{\colorMATH}\ensuremath{\gamma _{1} \vdash  {\begingroup\renewcommand\colorMATH{\colorMATHB}\renewcommand\colorSYNTAX{\colorSYNTAXB}{{\color{\colorMATH}\ensuremath{\se_{1}}}}\endgroup } \Downarrow  {\begingroup\renewcommand\colorMATH{\colorMATHB}\renewcommand\colorSYNTAX{\colorSYNTAXB}{{\color{\colorMATH}\ensuremath{\sv_{1}}}}\endgroup } \wedge  \gamma _{2} \vdash  {\begingroup\renewcommand\colorMATH{\colorMATHB}\renewcommand\colorSYNTAX{\colorSYNTAXB}{{\color{\colorMATH}\ensuremath{\se_{2}}}}\endgroup } \Downarrow  {\begingroup\renewcommand\colorMATH{\colorMATHB}\renewcommand\colorSYNTAX{\colorSYNTAXB}{{\color{\colorMATH}\ensuremath{\sv_{2}}}}\endgroup }}}} then {{\color{\colorMATH}\ensuremath{({\begingroup\renewcommand\colorMATH{\colorMATHB}\renewcommand\colorSYNTAX{\colorSYNTAXB}{{\color{\colorMATH}\ensuremath{\sv_{1}}}}\endgroup }, {\begingroup\renewcommand\colorMATH{\colorMATHB}\renewcommand\colorSYNTAX{\colorSYNTAXB}{{\color{\colorMATH}\ensuremath{\sv_{2}}}}\endgroup }) \in  {\mathcal{V}}_{{\begingroup\renewcommand\colorMATH{\colorMATHB}\renewcommand\colorSYNTAX{\colorSYNTAXB}{{\color{\colorMATH}\ensuremath{\distance}}}\endgroup }}^{k-j}\llbracket \sigma \rrbracket }}}.
  We have to prove that {{\color{\colorMATH}\ensuremath{({\begingroup\renewcommand\colorMATH{\colorMATHB}\renewcommand\colorSYNTAX{\colorSYNTAXB}{{\color{\colorMATH}\ensuremath{\sv_{1}}}}\endgroup }, {\begingroup\renewcommand\colorMATH{\colorMATHB}\renewcommand\colorSYNTAX{\colorSYNTAXB}{{\color{\colorMATH}\ensuremath{\sv_{2}}}}\endgroup }) \in  {\mathcal{V}}_{{\begingroup\renewcommand\colorMATH{\colorMATHB}\renewcommand\colorSYNTAX{\colorSYNTAXB}{{\color{\colorMATH}\ensuremath{\distance'}}}\endgroup }}^{k-j}\llbracket \sigma '\rrbracket }}}, which follows from (1).\\
  Now let us prove (3).
  We know that {{\color{\colorMATH}\ensuremath{\forall S \subseteq  val_{*}^{k} }}}, 
  if {{\color{\colorMATH}\ensuremath{{\text{Pr}}[\gamma _{1} \vdash  {\begingroup\renewcommand\colorMATH{\colorMATHC}\renewcommand\colorSYNTAX{\colorSYNTAXC}{{\color{\colorMATH}\ensuremath{\pe_{1}}}}\endgroup } \Downarrow ^{k} S] \wedge  {\text{Pr}}[\gamma _{2} \vdash  {\begingroup\renewcommand\colorMATH{\colorMATHC}\renewcommand\colorSYNTAX{\colorSYNTAXC}{{\color{\colorMATH}\ensuremath{\pe_{2}}}}\endgroup } \Downarrow ^{k} S]}}}, then
  {{\color{\colorMATH}\ensuremath{{\text{Pr}}[\gamma _{1} \vdash  {\begingroup\renewcommand\colorMATH{\colorMATHC}\renewcommand\colorSYNTAX{\colorSYNTAXC}{{\color{\colorMATH}\ensuremath{\pe_{1}}}}\endgroup } \Downarrow ^{k} S] \leq  e^{{\begingroup\renewcommand\colorMATH{\colorMATHC}\renewcommand\colorSYNTAX{\colorSYNTAXC}{{\color{\colorMATH}\ensuremath{p}}}\endgroup }.{\begingroup\renewcommand\colorMATH{\colorMATHC}\renewcommand\colorSYNTAX{\colorSYNTAXC}{{\color{\colorMATH}\ensuremath{\epsilon }}}\endgroup }} {\text{Pr}}[\gamma _{2} \vdash  {\begingroup\renewcommand\colorMATH{\colorMATHC}\renewcommand\colorSYNTAX{\colorSYNTAXC}{{\color{\colorMATH}\ensuremath{\pe_{2}}}}\endgroup } \Downarrow ^{k} S] + {\begingroup\renewcommand\colorMATH{\colorMATHC}\renewcommand\colorSYNTAX{\colorSYNTAXC}{{\color{\colorMATH}\ensuremath{p}}}\endgroup }.{\begingroup\renewcommand\colorMATH{\colorMATHC}\renewcommand\colorSYNTAX{\colorSYNTAXC}{{\color{\colorMATH}\ensuremath{\delta }}}\endgroup }}}} and
  {{\color{\colorMATH}\ensuremath{{\text{Pr}}[\gamma _{1} \vdash  {\begingroup\renewcommand\colorMATH{\colorMATHC}\renewcommand\colorSYNTAX{\colorSYNTAXC}{{\color{\colorMATH}\ensuremath{\pe_{2}}}}\endgroup } \Downarrow ^{k} S] \leq  e^{{\begingroup\renewcommand\colorMATH{\colorMATHC}\renewcommand\colorSYNTAX{\colorSYNTAXC}{{\color{\colorMATH}\ensuremath{p}}}\endgroup }.{\begingroup\renewcommand\colorMATH{\colorMATHC}\renewcommand\colorSYNTAX{\colorSYNTAXC}{{\color{\colorMATH}\ensuremath{\epsilon }}}\endgroup }} {\text{Pr}}[\gamma _{2} \vdash  {\begingroup\renewcommand\colorMATH{\colorMATHC}\renewcommand\colorSYNTAX{\colorSYNTAXC}{{\color{\colorMATH}\ensuremath{\pe_{1}}}}\endgroup } \Downarrow ^{k} S] + {\begingroup\renewcommand\colorMATH{\colorMATHC}\renewcommand\colorSYNTAX{\colorSYNTAXC}{{\color{\colorMATH}\ensuremath{p}}}\endgroup }.{\begingroup\renewcommand\colorMATH{\colorMATHC}\renewcommand\colorSYNTAX{\colorSYNTAXC}{{\color{\colorMATH}\ensuremath{\delta }}}\endgroup }}}}.
  We have to prove that,
  if {{\color{\colorMATH}\ensuremath{{\text{Pr}}[\gamma _{1} \vdash  {\begingroup\renewcommand\colorMATH{\colorMATHC}\renewcommand\colorSYNTAX{\colorSYNTAXC}{{\color{\colorMATH}\ensuremath{\pe_{1}}}}\endgroup } \Downarrow ^{k} S] \wedge  {\text{Pr}}[\gamma _{2} \vdash  {\begingroup\renewcommand\colorMATH{\colorMATHC}\renewcommand\colorSYNTAX{\colorSYNTAXC}{{\color{\colorMATH}\ensuremath{\pe_{2}}}}\endgroup } \Downarrow ^{k} S]}}}, then
  {{\color{\colorMATH}\ensuremath{{\text{Pr}}[\gamma _{1} \vdash  {\begingroup\renewcommand\colorMATH{\colorMATHC}\renewcommand\colorSYNTAX{\colorSYNTAXC}{{\color{\colorMATH}\ensuremath{\pe_{1}}}}\endgroup } \Downarrow ^{k} S] \leq  e^{{\begingroup\renewcommand\colorMATH{\colorMATHC}\renewcommand\colorSYNTAX{\colorSYNTAXC}{{\color{\colorMATH}\ensuremath{p'}}}\endgroup }.{\begingroup\renewcommand\colorMATH{\colorMATHC}\renewcommand\colorSYNTAX{\colorSYNTAXC}{{\color{\colorMATH}\ensuremath{\epsilon }}}\endgroup }} {\text{Pr}}[\gamma _{2} \vdash  {\begingroup\renewcommand\colorMATH{\colorMATHC}\renewcommand\colorSYNTAX{\colorSYNTAXC}{{\color{\colorMATH}\ensuremath{\pe_{2}}}}\endgroup } \Downarrow ^{k} S] + {\begingroup\renewcommand\colorMATH{\colorMATHC}\renewcommand\colorSYNTAX{\colorSYNTAXC}{{\color{\colorMATH}\ensuremath{p'}}}\endgroup }.{\begingroup\renewcommand\colorMATH{\colorMATHC}\renewcommand\colorSYNTAX{\colorSYNTAXC}{{\color{\colorMATH}\ensuremath{\delta }}}\endgroup }}}} and
  {{\color{\colorMATH}\ensuremath{{\text{Pr}}[\gamma _{1} \vdash  {\begingroup\renewcommand\colorMATH{\colorMATHC}\renewcommand\colorSYNTAX{\colorSYNTAXC}{{\color{\colorMATH}\ensuremath{\pe_{2}}}}\endgroup } \Downarrow ^{k} S] \leq  e^{{\begingroup\renewcommand\colorMATH{\colorMATHC}\renewcommand\colorSYNTAX{\colorSYNTAXC}{{\color{\colorMATH}\ensuremath{p'}}}\endgroup }.{\begingroup\renewcommand\colorMATH{\colorMATHC}\renewcommand\colorSYNTAX{\colorSYNTAXC}{{\color{\colorMATH}\ensuremath{\epsilon }}}\endgroup }} {\text{Pr}}[\gamma _{2} \vdash  {\begingroup\renewcommand\colorMATH{\colorMATHC}\renewcommand\colorSYNTAX{\colorSYNTAXC}{{\color{\colorMATH}\ensuremath{\pe_{1}}}}\endgroup } \Downarrow ^{k} S] + {\begingroup\renewcommand\colorMATH{\colorMATHC}\renewcommand\colorSYNTAX{\colorSYNTAXC}{{\color{\colorMATH}\ensuremath{p'}}}\endgroup }.{\begingroup\renewcommand\colorMATH{\colorMATHC}\renewcommand\colorSYNTAX{\colorSYNTAXC}{{\color{\colorMATH}\ensuremath{\delta }}}\endgroup }}}}.

  But notice that as {{\color{\colorMATH}\ensuremath{{\begingroup\renewcommand\colorMATH{\colorMATHC}\renewcommand\colorSYNTAX{\colorSYNTAXC}{{\color{\colorMATH}\ensuremath{p}}}\endgroup }.{\begingroup\renewcommand\colorMATH{\colorMATHC}\renewcommand\colorSYNTAX{\colorSYNTAXC}{{\color{\colorMATH}\ensuremath{\epsilon }}}\endgroup } \leq  {\begingroup\renewcommand\colorMATH{\colorMATHC}\renewcommand\colorSYNTAX{\colorSYNTAXC}{{\color{\colorMATH}\ensuremath{p'}}}\endgroup }.{\begingroup\renewcommand\colorMATH{\colorMATHC}\renewcommand\colorSYNTAX{\colorSYNTAXC}{{\color{\colorMATH}\ensuremath{\epsilon }}}\endgroup }}}}, then {{\color{\colorMATH}\ensuremath{e^{{\begingroup\renewcommand\colorMATH{\colorMATHC}\renewcommand\colorSYNTAX{\colorSYNTAXC}{{\color{\colorMATH}\ensuremath{p}}}\endgroup }.{\begingroup\renewcommand\colorMATH{\colorMATHC}\renewcommand\colorSYNTAX{\colorSYNTAXC}{{\color{\colorMATH}\ensuremath{\epsilon }}}\endgroup }} \leq  e^{{\begingroup\renewcommand\colorMATH{\colorMATHC}\renewcommand\colorSYNTAX{\colorSYNTAXC}{{\color{\colorMATH}\ensuremath{p'}}}\endgroup }.{\begingroup\renewcommand\colorMATH{\colorMATHC}\renewcommand\colorSYNTAX{\colorSYNTAXC}{{\color{\colorMATH}\ensuremath{\epsilon }}}\endgroup }}}}}, and
  {{\color{\colorMATH}\ensuremath{e^{{\begingroup\renewcommand\colorMATH{\colorMATHC}\renewcommand\colorSYNTAX{\colorSYNTAXC}{{\color{\colorMATH}\ensuremath{p}}}\endgroup }.{\begingroup\renewcommand\colorMATH{\colorMATHC}\renewcommand\colorSYNTAX{\colorSYNTAXC}{{\color{\colorMATH}\ensuremath{\epsilon }}}\endgroup }} {\text{Pr}}[\gamma _{2} \vdash  {\begingroup\renewcommand\colorMATH{\colorMATHC}\renewcommand\colorSYNTAX{\colorSYNTAXC}{{\color{\colorMATH}\ensuremath{\pe_{2}}}}\endgroup } \Downarrow ^{k} S] \leq  e^{{\begingroup\renewcommand\colorMATH{\colorMATHC}\renewcommand\colorSYNTAX{\colorSYNTAXC}{{\color{\colorMATH}\ensuremath{p'}}}\endgroup }.{\begingroup\renewcommand\colorMATH{\colorMATHC}\renewcommand\colorSYNTAX{\colorSYNTAXC}{{\color{\colorMATH}\ensuremath{\epsilon }}}\endgroup }} {\text{Pr}}[\gamma _{2} \vdash  {\begingroup\renewcommand\colorMATH{\colorMATHC}\renewcommand\colorSYNTAX{\colorSYNTAXC}{{\color{\colorMATH}\ensuremath{\pe_{2}}}}\endgroup } \Downarrow ^{k} S]}}}. We also know that
  {{\color{\colorMATH}\ensuremath{{\begingroup\renewcommand\colorMATH{\colorMATHC}\renewcommand\colorSYNTAX{\colorSYNTAXC}{{\color{\colorMATH}\ensuremath{p}}}\endgroup }.{\begingroup\renewcommand\colorMATH{\colorMATHC}\renewcommand\colorSYNTAX{\colorSYNTAXC}{{\color{\colorMATH}\ensuremath{\delta }}}\endgroup } \leq  {\begingroup\renewcommand\colorMATH{\colorMATHC}\renewcommand\colorSYNTAX{\colorSYNTAXC}{{\color{\colorMATH}\ensuremath{p'}}}\endgroup }.{\begingroup\renewcommand\colorMATH{\colorMATHC}\renewcommand\colorSYNTAX{\colorSYNTAXC}{{\color{\colorMATH}\ensuremath{\delta }}}\endgroup }}}}, and the result follows.
\end{proof}

\begin{lemma}
  \label{lm:weakening-index}
  Consider {{\color{\colorMATH}\ensuremath{k' \leq  k}}} then 
  \begin{enumerate}\item  If {{\color{\colorMATH}\ensuremath{({\begingroup\renewcommand\colorMATH{\colorMATHB}\renewcommand\colorSYNTAX{\colorSYNTAXB}{{\color{\colorMATH}\ensuremath{\sv_{1}}}}\endgroup },{\begingroup\renewcommand\colorMATH{\colorMATHB}\renewcommand\colorSYNTAX{\colorSYNTAXB}{{\color{\colorMATH}\ensuremath{\sv_{2}}}}\endgroup }) \in  {\mathcal{V}}_{{\begingroup\renewcommand\colorMATH{\colorMATHB}\renewcommand\colorSYNTAX{\colorSYNTAXB}{{\color{\colorMATH}\ensuremath{\sss}}}\endgroup }}^{k}\llbracket \sigma \rrbracket }}}, then {{\color{\colorMATH}\ensuremath{({\begingroup\renewcommand\colorMATH{\colorMATHB}\renewcommand\colorSYNTAX{\colorSYNTAXB}{{\color{\colorMATH}\ensuremath{\sv_{1}}}}\endgroup },{\begingroup\renewcommand\colorMATH{\colorMATHB}\renewcommand\colorSYNTAX{\colorSYNTAXB}{{\color{\colorMATH}\ensuremath{\sv_{2}}}}\endgroup }) \in  {\mathcal{V}}_{{\begingroup\renewcommand\colorMATH{\colorMATHB}\renewcommand\colorSYNTAX{\colorSYNTAXB}{{\color{\colorMATH}\ensuremath{\sss}}}\endgroup }}^{k'}\llbracket \sigma \rrbracket }}}
  \item  If {{\color{\colorMATH}\ensuremath{({\begingroup\renewcommand\colorMATH{\colorMATHB}\renewcommand\colorSYNTAX{\colorSYNTAXB}{{\color{\colorMATH}\ensuremath{\se_{1}}}}\endgroup },{\begingroup\renewcommand\colorMATH{\colorMATHB}\renewcommand\colorSYNTAX{\colorSYNTAXB}{{\color{\colorMATH}\ensuremath{\se_{2}}}}\endgroup }) \in  {\mathcal{E}}_{{\begingroup\renewcommand\colorMATH{\colorMATHB}\renewcommand\colorSYNTAX{\colorSYNTAXB}{{\color{\colorMATH}\ensuremath{\sss}}}\endgroup }}^{k}\llbracket \sigma \rrbracket }}}, then {{\color{\colorMATH}\ensuremath{({\begingroup\renewcommand\colorMATH{\colorMATHB}\renewcommand\colorSYNTAX{\colorSYNTAXB}{{\color{\colorMATH}\ensuremath{\se_{1}}}}\endgroup },{\begingroup\renewcommand\colorMATH{\colorMATHB}\renewcommand\colorSYNTAX{\colorSYNTAXB}{{\color{\colorMATH}\ensuremath{\se_{2}}}}\endgroup }) \in  {\mathcal{E}}_{{\begingroup\renewcommand\colorMATH{\colorMATHB}\renewcommand\colorSYNTAX{\colorSYNTAXB}{{\color{\colorMATH}\ensuremath{\sss}}}\endgroup }}^{k'}\llbracket \sigma \rrbracket }}}
  \item  If {{\color{\colorMATH}\ensuremath{({\begingroup\renewcommand\colorMATH{\colorMATHC}\renewcommand\colorSYNTAX{\colorSYNTAXC}{{\color{\colorMATH}\ensuremath{\pe_{1}}}}\endgroup },{\begingroup\renewcommand\colorMATH{\colorMATHC}\renewcommand\colorSYNTAX{\colorSYNTAXC}{{\color{\colorMATH}\ensuremath{\pe_{2}}}}\endgroup }) \in  {\mathcal{E}}_{{\begingroup\renewcommand\colorMATH{\colorMATHC}\renewcommand\colorSYNTAX{\colorSYNTAXC}{{\color{\colorMATH}\ensuremath{p}}}\endgroup }}^{k}\llbracket \sigma \rrbracket }}}, then {{\color{\colorMATH}\ensuremath{({\begingroup\renewcommand\colorMATH{\colorMATHC}\renewcommand\colorSYNTAX{\colorSYNTAXC}{{\color{\colorMATH}\ensuremath{\pe_{1}}}}\endgroup },{\begingroup\renewcommand\colorMATH{\colorMATHC}\renewcommand\colorSYNTAX{\colorSYNTAXC}{{\color{\colorMATH}\ensuremath{\pe_{2}}}}\endgroup }) \in  {\mathcal{E}}_{{\begingroup\renewcommand\colorMATH{\colorMATHC}\renewcommand\colorSYNTAX{\colorSYNTAXC}{{\color{\colorMATH}\ensuremath{p}}}\endgroup }}^{k'}\llbracket \sigma \rrbracket }}}
  \end{enumerate}
\end{lemma}
\begin{proof}
  By induction on {{\color{\colorMATH}\ensuremath{k}}}.
\end{proof}

\begin{lemma} 
  \label{lm:dot-subt}
  Let {{\color{\colorMATH}\ensuremath{dom({\begingroup\renewcommand\colorMATH{\colorMATHB}\renewcommand\colorSYNTAX{\colorSYNTAXB}{{\color{\colorMATH}\ensuremath{\sS}}}\endgroup }) \subseteq  dom({\begingroup\renewcommand\colorMATH{\colorMATHB}\renewcommand\colorSYNTAX{\colorSYNTAXB}{{\color{\colorMATH}\ensuremath{\Distance'}}}\endgroup })}}}, then {{\color{\colorMATH}\ensuremath{\forall  {\begingroup\renewcommand\colorMATH{\colorMATHB}\renewcommand\colorSYNTAX{\colorSYNTAXB}{{\color{\colorMATH}\ensuremath{\sS''}}}\endgroup }, {\begingroup\renewcommand\colorMATH{\colorMATHB}\renewcommand\colorSYNTAX{\colorSYNTAXB}{{\color{\colorMATH}\ensuremath{\sS}}}\endgroup } <: {\begingroup\renewcommand\colorMATH{\colorMATHB}\renewcommand\colorSYNTAX{\colorSYNTAXB}{{\color{\colorMATH}\ensuremath{\sS''}}}\endgroup }}}}, then {{\color{\colorMATH}\ensuremath{{\begingroup\renewcommand\colorMATH{\colorMATHB}\renewcommand\colorSYNTAX{\colorSYNTAXB}{{\color{\colorMATH}\ensuremath{\Distance'}}}\endgroup } \mathord{\cdotp } {\begingroup\renewcommand\colorMATH{\colorMATHB}\renewcommand\colorSYNTAX{\colorSYNTAXB}{{\color{\colorMATH}\ensuremath{\sS}}}\endgroup } \leq  {\begingroup\renewcommand\colorMATH{\colorMATHB}\renewcommand\colorSYNTAX{\colorSYNTAXB}{{\color{\colorMATH}\ensuremath{\Distance'}}}\endgroup } \mathord{\cdotp } {\begingroup\renewcommand\colorMATH{\colorMATHB}\renewcommand\colorSYNTAX{\colorSYNTAXB}{{\color{\colorMATH}\ensuremath{\sS''}}}\endgroup }}}}.
\end{lemma}
\begin{proof}
  By induction on {{\color{\colorMATH}\ensuremath{{\begingroup\renewcommand\colorMATH{\colorMATHB}\renewcommand\colorSYNTAX{\colorSYNTAXB}{{\color{\colorMATH}\ensuremath{\sS}}}\endgroup }}}}.
  \begin{enumerate}[ncases]\item  {{\color{\colorMATH}\ensuremath{{\begingroup\renewcommand\colorMATH{\colorMATHB}\renewcommand\colorSYNTAX{\colorSYNTAXB}{{\color{\colorMATH}\ensuremath{\sS}}}\endgroup } = \varnothing }}}
    \begin{subproof} 
      Trivial as {{\color{\colorMATH}\ensuremath{{\begingroup\renewcommand\colorMATH{\colorMATHB}\renewcommand\colorSYNTAX{\colorSYNTAXB}{{\color{\colorMATH}\ensuremath{\sS''}}}\endgroup } = \varnothing }}}, therefore {{\color{\colorMATH}\ensuremath{0 \leq  0}}}.
    \end{subproof}
  \item  {{\color{\colorMATH}\ensuremath{{\begingroup\renewcommand\colorMATH{\colorMATHB}\renewcommand\colorSYNTAX{\colorSYNTAXB}{{\color{\colorMATH}\ensuremath{\sS}}}\endgroup } = {\begingroup\renewcommand\colorMATH{\colorMATHB}\renewcommand\colorSYNTAX{\colorSYNTAXB}{{\color{\colorMATH}\ensuremath{\sS_{1}}}}\endgroup } + {\begingroup\renewcommand\colorMATH{\colorMATHB}\renewcommand\colorSYNTAX{\colorSYNTAXB}{{\color{\colorMATH}\ensuremath{\distance}}}\endgroup }x}}}
    \begin{subproof} 
      Let {{\color{\colorMATH}\ensuremath{{\begingroup\renewcommand\colorMATH{\colorMATHB}\renewcommand\colorSYNTAX{\colorSYNTAXB}{{\color{\colorMATH}\ensuremath{\sS''}}}\endgroup } = {\begingroup\renewcommand\colorMATH{\colorMATHB}\renewcommand\colorSYNTAX{\colorSYNTAXB}{{\color{\colorMATH}\ensuremath{\sS'_{1}}}}\endgroup } + {\begingroup\renewcommand\colorMATH{\colorMATHB}\renewcommand\colorSYNTAX{\colorSYNTAXB}{{\color{\colorMATH}\ensuremath{\distance'}}}\endgroup }x}}} such that {{\color{\colorMATH}\ensuremath{{\begingroup\renewcommand\colorMATH{\colorMATHB}\renewcommand\colorSYNTAX{\colorSYNTAXB}{{\color{\colorMATH}\ensuremath{\distance}}}\endgroup } \leq  {\begingroup\renewcommand\colorMATH{\colorMATHB}\renewcommand\colorSYNTAX{\colorSYNTAXB}{{\color{\colorMATH}\ensuremath{\distance'}}}\endgroup }}}}, then we have to prove that 
      {{\color{\colorMATH}\ensuremath{{\begingroup\renewcommand\colorMATH{\colorMATHB}\renewcommand\colorSYNTAX{\colorSYNTAXB}{{\color{\colorMATH}\ensuremath{\Distance'}}}\endgroup } \mathord{\cdotp } {\begingroup\renewcommand\colorMATH{\colorMATHB}\renewcommand\colorSYNTAX{\colorSYNTAXB}{{\color{\colorMATH}\ensuremath{\sS_{1}}}}\endgroup } + {\begingroup\renewcommand\colorMATH{\colorMATHB}\renewcommand\colorSYNTAX{\colorSYNTAXB}{{\color{\colorMATH}\ensuremath{\distance}}}\endgroup }{\begingroup\renewcommand\colorMATH{\colorMATHB}\renewcommand\colorSYNTAX{\colorSYNTAXB}{{\color{\colorMATH}\ensuremath{\Distance'}}}\endgroup }(x) \leq  {\begingroup\renewcommand\colorMATH{\colorMATHB}\renewcommand\colorSYNTAX{\colorSYNTAXB}{{\color{\colorMATH}\ensuremath{\Distance'}}}\endgroup } \mathord{\cdotp } {\begingroup\renewcommand\colorMATH{\colorMATHB}\renewcommand\colorSYNTAX{\colorSYNTAXB}{{\color{\colorMATH}\ensuremath{\sS'_{1}}}}\endgroup } + {\begingroup\renewcommand\colorMATH{\colorMATHB}\renewcommand\colorSYNTAX{\colorSYNTAXB}{{\color{\colorMATH}\ensuremath{\distance'}}}\endgroup }{\begingroup\renewcommand\colorMATH{\colorMATHB}\renewcommand\colorSYNTAX{\colorSYNTAXB}{{\color{\colorMATH}\ensuremath{\Distance'}}}\endgroup }(x)}}}, but we know by induction hypothesis that {{\color{\colorMATH}\ensuremath{{\begingroup\renewcommand\colorMATH{\colorMATHB}\renewcommand\colorSYNTAX{\colorSYNTAXB}{{\color{\colorMATH}\ensuremath{\Distance'}}}\endgroup } \mathord{\cdotp } {\begingroup\renewcommand\colorMATH{\colorMATHB}\renewcommand\colorSYNTAX{\colorSYNTAXB}{{\color{\colorMATH}\ensuremath{\sS_{1}}}}\endgroup } \leq  {\begingroup\renewcommand\colorMATH{\colorMATHB}\renewcommand\colorSYNTAX{\colorSYNTAXB}{{\color{\colorMATH}\ensuremath{\Distance'}}}\endgroup } \mathord{\cdotp } {\begingroup\renewcommand\colorMATH{\colorMATHB}\renewcommand\colorSYNTAX{\colorSYNTAXB}{{\color{\colorMATH}\ensuremath{\sS'_{1}}}}\endgroup }}}} and {{\color{\colorMATH}\ensuremath{{\begingroup\renewcommand\colorMATH{\colorMATHB}\renewcommand\colorSYNTAX{\colorSYNTAXB}{{\color{\colorMATH}\ensuremath{\distance}}}\endgroup }{\begingroup\renewcommand\colorMATH{\colorMATHB}\renewcommand\colorSYNTAX{\colorSYNTAXB}{{\color{\colorMATH}\ensuremath{\Distance'}}}\endgroup }(x) \leq  {\begingroup\renewcommand\colorMATH{\colorMATHB}\renewcommand\colorSYNTAX{\colorSYNTAXB}{{\color{\colorMATH}\ensuremath{\distance'}}}\endgroup }{\begingroup\renewcommand\colorMATH{\colorMATHB}\renewcommand\colorSYNTAX{\colorSYNTAXB}{{\color{\colorMATH}\ensuremath{\Distance'}}}\endgroup }(x)}}}, so the result holds immediately.
    \end{subproof}
  \item  {{\color{\colorMATH}\ensuremath{{\begingroup\renewcommand\colorMATH{\colorMATHB}\renewcommand\colorSYNTAX{\colorSYNTAXB}{{\color{\colorMATH}\ensuremath{\sS}}}\endgroup } = {\begingroup\renewcommand\colorMATH{\colorMATHB}\renewcommand\colorSYNTAX{\colorSYNTAXB}{{\color{\colorMATH}\ensuremath{\sS_{1}}}}\endgroup } + {\begingroup\renewcommand\colorMATH{\colorMATHB}\renewcommand\colorSYNTAX{\colorSYNTAXB}{{\color{\colorMATH}\ensuremath{\distance}}}\endgroup }}}}
    \begin{subproof} 
      Let {{\color{\colorMATH}\ensuremath{{\begingroup\renewcommand\colorMATH{\colorMATHB}\renewcommand\colorSYNTAX{\colorSYNTAXB}{{\color{\colorMATH}\ensuremath{\sS''}}}\endgroup } = {\begingroup\renewcommand\colorMATH{\colorMATHB}\renewcommand\colorSYNTAX{\colorSYNTAXB}{{\color{\colorMATH}\ensuremath{\sS'_{1}}}}\endgroup } + {\begingroup\renewcommand\colorMATH{\colorMATHB}\renewcommand\colorSYNTAX{\colorSYNTAXB}{{\color{\colorMATH}\ensuremath{\distance'}}}\endgroup }}}} such that {{\color{\colorMATH}\ensuremath{{\begingroup\renewcommand\colorMATH{\colorMATHB}\renewcommand\colorSYNTAX{\colorSYNTAXB}{{\color{\colorMATH}\ensuremath{\distance}}}\endgroup } \leq  {\begingroup\renewcommand\colorMATH{\colorMATHB}\renewcommand\colorSYNTAX{\colorSYNTAXB}{{\color{\colorMATH}\ensuremath{\distance'}}}\endgroup }}}}, then we have to prove that 
      {{\color{\colorMATH}\ensuremath{{\begingroup\renewcommand\colorMATH{\colorMATHB}\renewcommand\colorSYNTAX{\colorSYNTAXB}{{\color{\colorMATH}\ensuremath{\Distance'}}}\endgroup } \mathord{\cdotp } {\begingroup\renewcommand\colorMATH{\colorMATHB}\renewcommand\colorSYNTAX{\colorSYNTAXB}{{\color{\colorMATH}\ensuremath{\sS_{1}}}}\endgroup } + {\begingroup\renewcommand\colorMATH{\colorMATHB}\renewcommand\colorSYNTAX{\colorSYNTAXB}{{\color{\colorMATH}\ensuremath{\distance}}}\endgroup } \leq  {\begingroup\renewcommand\colorMATH{\colorMATHB}\renewcommand\colorSYNTAX{\colorSYNTAXB}{{\color{\colorMATH}\ensuremath{\Distance'}}}\endgroup } \mathord{\cdotp } {\begingroup\renewcommand\colorMATH{\colorMATHB}\renewcommand\colorSYNTAX{\colorSYNTAXB}{{\color{\colorMATH}\ensuremath{\sS'_{1}}}}\endgroup } + {\begingroup\renewcommand\colorMATH{\colorMATHB}\renewcommand\colorSYNTAX{\colorSYNTAXB}{{\color{\colorMATH}\ensuremath{\distance'}}}\endgroup }}}}, but we know by induction hypothesis that {{\color{\colorMATH}\ensuremath{{\begingroup\renewcommand\colorMATH{\colorMATHB}\renewcommand\colorSYNTAX{\colorSYNTAXB}{{\color{\colorMATH}\ensuremath{\Distance'}}}\endgroup } \mathord{\cdotp } {\begingroup\renewcommand\colorMATH{\colorMATHB}\renewcommand\colorSYNTAX{\colorSYNTAXB}{{\color{\colorMATH}\ensuremath{\sS_{1}}}}\endgroup } \leq  {\begingroup\renewcommand\colorMATH{\colorMATHB}\renewcommand\colorSYNTAX{\colorSYNTAXB}{{\color{\colorMATH}\ensuremath{\Distance'}}}\endgroup } \mathord{\cdotp } {\begingroup\renewcommand\colorMATH{\colorMATHB}\renewcommand\colorSYNTAX{\colorSYNTAXB}{{\color{\colorMATH}\ensuremath{\sS'_{1}}}}\endgroup }}}}, so the result holds immediately.
    \end{subproof}
  \end{enumerate}
\end{proof}

\begin{lemma} 
  \label{lm:dot-subtp}
  Let {{\color{\colorMATH}\ensuremath{dom({\begingroup\renewcommand\colorMATH{\colorMATHC}\renewcommand\colorSYNTAX{\colorSYNTAXC}{{\color{\colorMATH}\ensuremath{\pS}}}\endgroup }) \subseteq  dom({\begingroup\renewcommand\colorMATH{\colorMATHB}\renewcommand\colorSYNTAX{\colorSYNTAXB}{{\color{\colorMATH}\ensuremath{\Distance}}}\endgroup })}}}, then {{\color{\colorMATH}\ensuremath{\forall  {\begingroup\renewcommand\colorMATH{\colorMATHC}\renewcommand\colorSYNTAX{\colorSYNTAXC}{{\color{\colorMATH}\ensuremath{\pS'}}}\endgroup }, {\begingroup\renewcommand\colorMATH{\colorMATHC}\renewcommand\colorSYNTAX{\colorSYNTAXC}{{\color{\colorMATH}\ensuremath{\pS}}}\endgroup } <: {\begingroup\renewcommand\colorMATH{\colorMATHC}\renewcommand\colorSYNTAX{\colorSYNTAXC}{{\color{\colorMATH}\ensuremath{\pS'}}}\endgroup }}}}, then {{\color{\colorMATH}\ensuremath{{\begingroup\renewcommand\colorMATH{\colorMATHB}\renewcommand\colorSYNTAX{\colorSYNTAXB}{{\color{\colorMATH}\ensuremath{\Distance}}}\endgroup } {\begingroup\renewcommand\colorMATH{\colorMATHC}\renewcommand\colorSYNTAX{\colorSYNTAXC}{{\color{\colorMATH}\ensuremath{\bigcdot}}}\endgroup } {\begingroup\renewcommand\colorMATH{\colorMATHC}\renewcommand\colorSYNTAX{\colorSYNTAXC}{{\color{\colorMATH}\ensuremath{\pS}}}\endgroup } \leq  {\begingroup\renewcommand\colorMATH{\colorMATHB}\renewcommand\colorSYNTAX{\colorSYNTAXB}{{\color{\colorMATH}\ensuremath{\Distance}}}\endgroup } {\begingroup\renewcommand\colorMATH{\colorMATHC}\renewcommand\colorSYNTAX{\colorSYNTAXC}{{\color{\colorMATH}\ensuremath{\bigcdot}}}\endgroup } {\begingroup\renewcommand\colorMATH{\colorMATHC}\renewcommand\colorSYNTAX{\colorSYNTAXC}{{\color{\colorMATH}\ensuremath{\pS'}}}\endgroup }}}}.
\end{lemma}
\begin{proof}
  By definition of {{\color{\colorMATH}\ensuremath{{\begingroup\renewcommand\colorMATH{\colorMATHC}\renewcommand\colorSYNTAX{\colorSYNTAXC}{{\color{\colorMATH}\ensuremath{\pS}}}\endgroup } <: {\begingroup\renewcommand\colorMATH{\colorMATHC}\renewcommand\colorSYNTAX{\colorSYNTAXC}{{\color{\colorMATH}\ensuremath{\pS'}}}\endgroup }}}}.
\end{proof}

\begin{lemma}
  \label{lm:infrel}.
  \begin{enumerate}
    \item If {{\color{\colorMATH}\ensuremath{\Gamma  \mathrel{;} {\begingroup\renewcommand\colorMATH{\colorMATHB}\renewcommand\colorSYNTAX{\colorSYNTAXB}{{\color{\colorMATH}\ensuremath{\Distance}}}\endgroup } \vdash  {\begingroup\renewcommand\colorMATH{\colorMATHB}\renewcommand\colorSYNTAX{\colorSYNTAXB}{{\color{\colorMATH}\ensuremath{\se_{1}}}}\endgroup } \mathrel{:} \tau  \mathrel{;} {\begingroup\renewcommand\colorMATH{\colorMATHB}\renewcommand\colorSYNTAX{\colorSYNTAXB}{{\color{\colorMATH}\ensuremath{\sS}}}\endgroup }}}} and {{\color{\colorMATH}\ensuremath{\Gamma  \mathrel{;} {\begingroup\renewcommand\colorMATH{\colorMATHB}\renewcommand\colorSYNTAX{\colorSYNTAXB}{{\color{\colorMATH}\ensuremath{\Distance}}}\endgroup } \vdash  {\begingroup\renewcommand\colorMATH{\colorMATHB}\renewcommand\colorSYNTAX{\colorSYNTAXB}{{\color{\colorMATH}\ensuremath{\se_{2}}}}\endgroup } \mathrel{:} \tau  \mathrel{;} {\begingroup\renewcommand\colorMATH{\colorMATHB}\renewcommand\colorSYNTAX{\colorSYNTAXB}{{\color{\colorMATH}\ensuremath{\sS}}}\endgroup }}}}, then
      {{\color{\colorMATH}\ensuremath{(\gamma _{1} \vdash  {\begingroup\renewcommand\colorMATH{\colorMATHB}\renewcommand\colorSYNTAX{\colorSYNTAXB}{{\color{\colorMATH}\ensuremath{\se_{1}}}}\endgroup } \gamma _{2} \vdash  {\begingroup\renewcommand\colorMATH{\colorMATHB}\renewcommand\colorSYNTAX{\colorSYNTAXB}{{\color{\colorMATH}\ensuremath{\se_{2}}}}\endgroup }) \in  {\mathcal{E}}_{{\begingroup\renewcommand\colorMATH{\colorMATHB}\renewcommand\colorSYNTAX{\colorSYNTAXB}{{\color{\colorMATH}\ensuremath{\infty }}}\endgroup }}\llbracket {\begingroup\renewcommand\colorMATH{\colorMATHB}\renewcommand\colorSYNTAX{\colorSYNTAXB}{{\color{\colorMATH}\ensuremath{\Distance'}}}\endgroup }(\tau )\rrbracket }}}.
      \item If {{\color{\colorMATH}\ensuremath{\Gamma  \mathrel{;} {\begingroup\renewcommand\colorMATH{\colorMATHB}\renewcommand\colorSYNTAX{\colorSYNTAXB}{{\color{\colorMATH}\ensuremath{\Distance}}}\endgroup } \vdash  {\begingroup\renewcommand\colorMATH{\colorMATHC}\renewcommand\colorSYNTAX{\colorSYNTAXC}{{\color{\colorMATH}\ensuremath{\pe_{1}}}}\endgroup } \mathrel{:} \tau  \mathrel{;} {\begingroup\renewcommand\colorMATH{\colorMATHC}\renewcommand\colorSYNTAX{\colorSYNTAXC}{{\color{\colorMATH}\ensuremath{\pS}}}\endgroup }}}} and {{\color{\colorMATH}\ensuremath{\Gamma  \mathrel{;} {\begingroup\renewcommand\colorMATH{\colorMATHB}\renewcommand\colorSYNTAX{\colorSYNTAXB}{{\color{\colorMATH}\ensuremath{\Distance}}}\endgroup } \vdash  {\begingroup\renewcommand\colorMATH{\colorMATHC}\renewcommand\colorSYNTAX{\colorSYNTAXC}{{\color{\colorMATH}\ensuremath{\pe_{2}}}}\endgroup } \mathrel{:} \tau  \mathrel{;} {\begingroup\renewcommand\colorMATH{\colorMATHC}\renewcommand\colorSYNTAX{\colorSYNTAXC}{{\color{\colorMATH}\ensuremath{\pS}}}\endgroup }}}}, then
      {{\color{\colorMATH}\ensuremath{(\gamma _{1} \vdash  {\begingroup\renewcommand\colorMATH{\colorMATHC}\renewcommand\colorSYNTAX{\colorSYNTAXC}{{\color{\colorMATH}\ensuremath{\pe_{1}}}}\endgroup } \gamma _{2} \vdash  {\begingroup\renewcommand\colorMATH{\colorMATHC}\renewcommand\colorSYNTAX{\colorSYNTAXC}{{\color{\colorMATH}\ensuremath{\pe_{2}}}}\endgroup }) \in  {\mathcal{E}}_{{\begingroup\renewcommand\colorMATH{\colorMATHC}\renewcommand\colorSYNTAX{\colorSYNTAXC}{{\color{\colorMATH}\ensuremath{\infty }}}\endgroup }}\llbracket {\begingroup\renewcommand\colorMATH{\colorMATHB}\renewcommand\colorSYNTAX{\colorSYNTAXB}{{\color{\colorMATH}\ensuremath{\Distance'}}}\endgroup }(\tau )\rrbracket }}}.
  \end{enumerate}
\end{lemma}
\begin{proof}
  By induction on {{\color{\colorMATH}\ensuremath{\tau }}}.
\end{proof}

\begin{lemma}
  \label{lm:probsemanticrel}
  Let {{\color{\colorMATH}\ensuremath{j\leq k}}}, 
  {{\color{\colorMATH}\ensuremath{\gamma _{1} \vdash  {\begingroup\renewcommand\colorMATH{\colorMATHC}\renewcommand\colorSYNTAX{\colorSYNTAXC}{{\color{\colorMATH}\ensuremath{\pe_{1}}}}\endgroup } \Downarrow ^{j} \gamma '_{1} \vdash  {\begingroup\renewcommand\colorMATH{\colorMATHC}\renewcommand\colorSYNTAX{\colorSYNTAXC}{{\color{\colorMATH}\ensuremath{\pe'_{1}}}}\endgroup }, \gamma _{2} \vdash  {\begingroup\renewcommand\colorMATH{\colorMATHC}\renewcommand\colorSYNTAX{\colorSYNTAXC}{{\color{\colorMATH}\ensuremath{\pe_{2}}}}\endgroup } \Downarrow ^{*} \gamma '_{2} \vdash  {\begingroup\renewcommand\colorMATH{\colorMATHC}\renewcommand\colorSYNTAX{\colorSYNTAXC}{{\color{\colorMATH}\ensuremath{\pe'_{2}}}}\endgroup }}}}, and {{\color{\colorMATH}\ensuremath{(\gamma '_{1} \vdash  {\begingroup\renewcommand\colorMATH{\colorMATHC}\renewcommand\colorSYNTAX{\colorSYNTAXC}{{\color{\colorMATH}\ensuremath{\pe'_{1}}}}\endgroup }, \gamma '_{2} \vdash  {\begingroup\renewcommand\colorMATH{\colorMATHC}\renewcommand\colorSYNTAX{\colorSYNTAXC}{{\color{\colorMATH}\ensuremath{\pe'_{2}}}}\endgroup }) \in  {\mathcal{E}}_{{\begingroup\renewcommand\colorMATH{\colorMATHC}\renewcommand\colorSYNTAX{\colorSYNTAXC}{{\color{\colorMATH}\ensuremath{p}}}\endgroup }}^{k-j}\llbracket \tau \rrbracket }}}, then\\
  {{\color{\colorMATH}\ensuremath{(\gamma _{1} \vdash  {\begingroup\renewcommand\colorMATH{\colorMATHC}\renewcommand\colorSYNTAX{\colorSYNTAXC}{{\color{\colorMATH}\ensuremath{\pe_{1}}}}\endgroup }, \gamma _{2} \vdash  {\begingroup\renewcommand\colorMATH{\colorMATHC}\renewcommand\colorSYNTAX{\colorSYNTAXC}{{\color{\colorMATH}\ensuremath{\pe_{2}}}}\endgroup }) \in  {\mathcal{E}}_{{\begingroup\renewcommand\colorMATH{\colorMATHC}\renewcommand\colorSYNTAX{\colorSYNTAXC}{{\color{\colorMATH}\ensuremath{p}}}\endgroup }}^{k}\llbracket \tau \rrbracket }}}.
\end{lemma}
\begin{proof}
  We know that for {{\color{\colorMATH}\ensuremath{j'<k-j}}} if {{\color{\colorMATH}\ensuremath{\gamma '_{1} \vdash  {\begingroup\renewcommand\colorMATH{\colorMATHC}\renewcommand\colorSYNTAX{\colorSYNTAXC}{{\color{\colorMATH}\ensuremath{\pe'_{1}}}}\endgroup } \Downarrow ^{j'} \dist[1]}}}, then {{\color{\colorMATH}\ensuremath{\gamma '_{2} \vdash  {\begingroup\renewcommand\colorMATH{\colorMATHC}\renewcommand\colorSYNTAX{\colorSYNTAXC}{{\color{\colorMATH}\ensuremath{\pe'_{2}}}}\endgroup } \Downarrow ^{*} \dist[2]}}}, and both distributions satisfy the dp inequality. Then as {{\color{\colorMATH}\ensuremath{\gamma _{1} \vdash  {\begingroup\renewcommand\colorMATH{\colorMATHC}\renewcommand\colorSYNTAX{\colorSYNTAXC}{{\color{\colorMATH}\ensuremath{\pe_{1}}}}\endgroup } \Downarrow ^{j} \gamma '_{1} \vdash  {\begingroup\renewcommand\colorMATH{\colorMATHC}\renewcommand\colorSYNTAX{\colorSYNTAXC}{{\color{\colorMATH}\ensuremath{\pe'_{1}}}}\endgroup }}}} and {{\color{\colorMATH}\ensuremath{\gamma '_{1} \vdash  {\begingroup\renewcommand\colorMATH{\colorMATHC}\renewcommand\colorSYNTAX{\colorSYNTAXC}{{\color{\colorMATH}\ensuremath{\pe'_{1}}}}\endgroup } \Downarrow ^{j'} \dist[1]}}}, then {{\color{\colorMATH}\ensuremath{\gamma _{1} \vdash  {\begingroup\renewcommand\colorMATH{\colorMATHC}\renewcommand\colorSYNTAX{\colorSYNTAXC}{{\color{\colorMATH}\ensuremath{\pe_{1}}}}\endgroup } \Downarrow ^{j+j'} \dist[1]}}} ({{\color{\colorMATH}\ensuremath{j+j'<k}}}), and {{\color{\colorMATH}\ensuremath{\gamma _{2} \vdash  {\begingroup\renewcommand\colorMATH{\colorMATHC}\renewcommand\colorSYNTAX{\colorSYNTAXC}{{\color{\colorMATH}\ensuremath{\pe_{2}}}}\endgroup } \Downarrow ^{*} \dist[1]}}}, and the result is direct.
\end{proof}

\begin{restatable}{lemma}{novarequal}
  \label{lm:novarequal}
  If {{\color{\colorMATH}\ensuremath{\Gamma ; {\begingroup\renewcommand\colorMATH{\colorMATHB}\renewcommand\colorSYNTAX{\colorSYNTAXB}{{\color{\colorMATH}\ensuremath{\Distance}}}\endgroup } \vdash  {\begingroup\renewcommand\colorMATH{\colorMATHB}\renewcommand\colorSYNTAX{\colorSYNTAXB}{{\color{\colorMATH}\ensuremath{\se}}}\endgroup } \mathrel{:} \tau  \mathrel{;} {\begingroup\renewcommand\colorMATH{\colorMATHB}\renewcommand\colorSYNTAX{\colorSYNTAXB}{{\color{\colorMATH}\ensuremath{\sS}}}\endgroup }}}}, {{\color{\colorMATH}\ensuremath{\gamma _{1},\gamma \vdash {\begingroup\renewcommand\colorMATH{\colorMATHB}\renewcommand\colorSYNTAX{\colorSYNTAXB}{{\color{\colorMATH}\ensuremath{\se}}}\endgroup } \Downarrow ^{*} {\begingroup\renewcommand\colorMATH{\colorMATHB}\renewcommand\colorSYNTAX{\colorSYNTAXB}{{\color{\colorMATH}\ensuremath{\sv_{1}}}}\endgroup }}}}, {{\color{\colorMATH}\ensuremath{\gamma _{2},\gamma \vdash {\begingroup\renewcommand\colorMATH{\colorMATHB}\renewcommand\colorSYNTAX{\colorSYNTAXB}{{\color{\colorMATH}\ensuremath{\se}}}\endgroup } \Downarrow ^{*} {\begingroup\renewcommand\colorMATH{\colorMATHB}\renewcommand\colorSYNTAX{\colorSYNTAXB}{{\color{\colorMATH}\ensuremath{\sv_{2}}}}\endgroup }}}}, and 
        {{\color{\colorMATH}\ensuremath{ {\text{FV}}({\begingroup\renewcommand\colorMATH{\colorMATHB}\renewcommand\colorSYNTAX{\colorSYNTAXB}{{\color{\colorMATH}\ensuremath{\se}}}\endgroup }) \subseteq  dom(\gamma ) }}}, then
        {{\color{\colorMATH}\ensuremath{{\begingroup\renewcommand\colorMATH{\colorMATHB}\renewcommand\colorSYNTAX{\colorSYNTAXB}{{\color{\colorMATH}\ensuremath{\sv_{1}}}}\endgroup } = {\begingroup\renewcommand\colorMATH{\colorMATHB}\renewcommand\colorSYNTAX{\colorSYNTAXB}{{\color{\colorMATH}\ensuremath{\sv_{2}}}}\endgroup }}}}.
\end{restatable}
\begin{proof}
  By induction on {{\color{\colorMATH}\ensuremath{\Gamma ; {\begingroup\renewcommand\colorMATH{\colorMATHB}\renewcommand\colorSYNTAX{\colorSYNTAXB}{{\color{\colorMATH}\ensuremath{\Distance}}}\endgroup } \vdash  {\begingroup\renewcommand\colorMATH{\colorMATHB}\renewcommand\colorSYNTAX{\colorSYNTAXB}{{\color{\colorMATH}\ensuremath{\se}}}\endgroup } \mathrel{:} \tau  \mathrel{;} {\begingroup\renewcommand\colorMATH{\colorMATHB}\renewcommand\colorSYNTAX{\colorSYNTAXB}{{\color{\colorMATH}\ensuremath{\sS}}}\endgroup }}}}.
  \begin{enumerate}[ncases]\item  {{\color{\colorMATH}\ensuremath{\Gamma ; {\begingroup\renewcommand\colorMATH{\colorMATHB}\renewcommand\colorSYNTAX{\colorSYNTAXB}{{\color{\colorMATH}\ensuremath{\Distance}}}\endgroup } \vdash  {\begingroup\renewcommand\colorMATH{\colorMATHB}\renewcommand\colorSYNTAX{\colorSYNTAXB}{{\color{\colorMATH}\ensuremath{r}}}\endgroup } \mathrel{:} {\begingroup\renewcommand\colorMATH{\colorMATHA}\renewcommand\colorSYNTAX{\colorSYNTAXA}{{\color{\colorSYNTAX}\texttt{{\ensuremath{{\mathbb{R}}}}}}}\endgroup } \mathrel{;} \varnothing }}}  
    \begin{subproof} 
      Trivial as numbers are already values.
    \end{subproof}
  \item  {{\color{\colorMATH}\ensuremath{\Gamma ; {\begingroup\renewcommand\colorMATH{\colorMATHB}\renewcommand\colorSYNTAX{\colorSYNTAXB}{{\color{\colorMATH}\ensuremath{\Distance}}}\endgroup } \vdash  {\begingroup\renewcommand\colorMATH{\colorMATHB}\renewcommand\colorSYNTAX{\colorSYNTAXB}{{\color{\colorMATH}\ensuremath{\se_{1}}}}\endgroup } + {\begingroup\renewcommand\colorMATH{\colorMATHB}\renewcommand\colorSYNTAX{\colorSYNTAXB}{{\color{\colorMATH}\ensuremath{\se_{2}}}}\endgroup } \mathrel{:} {\begingroup\renewcommand\colorMATH{\colorMATHA}\renewcommand\colorSYNTAX{\colorSYNTAXA}{{\color{\colorSYNTAX}\texttt{{\ensuremath{{\mathbb{R}}}}}}}\endgroup } \mathrel{;} {\begingroup\renewcommand\colorMATH{\colorMATHB}\renewcommand\colorSYNTAX{\colorSYNTAXB}{{\color{\colorMATH}\ensuremath{\sS_{1}}}}\endgroup } + {\begingroup\renewcommand\colorMATH{\colorMATHB}\renewcommand\colorSYNTAX{\colorSYNTAXB}{{\color{\colorMATH}\ensuremath{\sS_{2}}}}\endgroup }}}} 
    \begin{subproof} 
      By induction hypotheses on {{\color{\colorMATH}\ensuremath{\Gamma ; {\begingroup\renewcommand\colorMATH{\colorMATHB}\renewcommand\colorSYNTAX{\colorSYNTAXB}{{\color{\colorMATH}\ensuremath{\Distance}}}\endgroup } \vdash  {\begingroup\renewcommand\colorMATH{\colorMATHB}\renewcommand\colorSYNTAX{\colorSYNTAXB}{{\color{\colorMATH}\ensuremath{\se_{1}}}}\endgroup } \mathrel{:} {\begingroup\renewcommand\colorMATH{\colorMATHA}\renewcommand\colorSYNTAX{\colorSYNTAXA}{{\color{\colorSYNTAX}\texttt{{\ensuremath{{\mathbb{R}}}}}}}\endgroup } \mathrel{;} {\begingroup\renewcommand\colorMATH{\colorMATHB}\renewcommand\colorSYNTAX{\colorSYNTAXB}{{\color{\colorMATH}\ensuremath{\sS_{1}}}}\endgroup }}}}, and {{\color{\colorMATH}\ensuremath{\Gamma ; {\begingroup\renewcommand\colorMATH{\colorMATHB}\renewcommand\colorSYNTAX{\colorSYNTAXB}{{\color{\colorMATH}\ensuremath{\Distance}}}\endgroup } \vdash  {\begingroup\renewcommand\colorMATH{\colorMATHB}\renewcommand\colorSYNTAX{\colorSYNTAXB}{{\color{\colorMATH}\ensuremath{\se_{2}}}}\endgroup } \mathrel{:} {\begingroup\renewcommand\colorMATH{\colorMATHA}\renewcommand\colorSYNTAX{\colorSYNTAXA}{{\color{\colorSYNTAX}\texttt{{\ensuremath{{\mathbb{R}}}}}}}\endgroup } \mathrel{;} {\begingroup\renewcommand\colorMATH{\colorMATHB}\renewcommand\colorSYNTAX{\colorSYNTAXB}{{\color{\colorMATH}\ensuremath{\sS_{2}}}}\endgroup }}}}.
      {{\color{\colorMATH}\ensuremath{\gamma _{1},\gamma \vdash {\begingroup\renewcommand\colorMATH{\colorMATHB}\renewcommand\colorSYNTAX{\colorSYNTAXB}{{\color{\colorMATH}\ensuremath{\se_{1}}}}\endgroup } \Downarrow ^{*} {\begingroup\renewcommand\colorMATH{\colorMATHB}\renewcommand\colorSYNTAX{\colorSYNTAXB}{{\color{\colorMATH}\ensuremath{r_{1}}}}\endgroup }}}}, {{\color{\colorMATH}\ensuremath{\gamma _{2},\gamma \vdash {\begingroup\renewcommand\colorMATH{\colorMATHB}\renewcommand\colorSYNTAX{\colorSYNTAXB}{{\color{\colorMATH}\ensuremath{\se_{1}}}}\endgroup } \Downarrow ^{*} {\begingroup\renewcommand\colorMATH{\colorMATHB}\renewcommand\colorSYNTAX{\colorSYNTAXB}{{\color{\colorMATH}\ensuremath{r_{1}}}}\endgroup }}}}, and {{\color{\colorMATH}\ensuremath{\gamma _{1},\gamma \vdash {\begingroup\renewcommand\colorMATH{\colorMATHB}\renewcommand\colorSYNTAX{\colorSYNTAXB}{{\color{\colorMATH}\ensuremath{\se_{2}}}}\endgroup } \Downarrow ^{*} {\begingroup\renewcommand\colorMATH{\colorMATHB}\renewcommand\colorSYNTAX{\colorSYNTAXB}{{\color{\colorMATH}\ensuremath{r_{2}}}}\endgroup }}}}, {{\color{\colorMATH}\ensuremath{\gamma _{2},\gamma \vdash {\begingroup\renewcommand\colorMATH{\colorMATHB}\renewcommand\colorSYNTAX{\colorSYNTAXB}{{\color{\colorMATH}\ensuremath{\se_{2}}}}\endgroup } \Downarrow ^{*} {\begingroup\renewcommand\colorMATH{\colorMATHB}\renewcommand\colorSYNTAX{\colorSYNTAXB}{{\color{\colorMATH}\ensuremath{r_{2}}}}\endgroup }}}}.
      The result holds as {{\color{\colorMATH}\ensuremath{{\begingroup\renewcommand\colorMATH{\colorMATHB}\renewcommand\colorSYNTAX{\colorSYNTAXB}{{\color{\colorMATH}\ensuremath{r_{1}}}}\endgroup } + {\begingroup\renewcommand\colorMATH{\colorMATHB}\renewcommand\colorSYNTAX{\colorSYNTAXB}{{\color{\colorMATH}\ensuremath{r_{2}}}}\endgroup } = {\begingroup\renewcommand\colorMATH{\colorMATHB}\renewcommand\colorSYNTAX{\colorSYNTAXB}{{\color{\colorMATH}\ensuremath{r_{1}}}}\endgroup } + {\begingroup\renewcommand\colorMATH{\colorMATHB}\renewcommand\colorSYNTAX{\colorSYNTAXB}{{\color{\colorMATH}\ensuremath{r_{2}}}}\endgroup }}}}.
    \end{subproof}
  \item  {{\color{\colorMATH}\ensuremath{\Gamma ; {\begingroup\renewcommand\colorMATH{\colorMATHB}\renewcommand\colorSYNTAX{\colorSYNTAXB}{{\color{\colorMATH}\ensuremath{\Distance}}}\endgroup } \vdash  {\begingroup\renewcommand\colorMATH{\colorMATHB}\renewcommand\colorSYNTAX{\colorSYNTAXB}{{\color{\colorMATH}\ensuremath{\se_{1}}}}\endgroup } * {\begingroup\renewcommand\colorMATH{\colorMATHB}\renewcommand\colorSYNTAX{\colorSYNTAXB}{{\color{\colorMATH}\ensuremath{\se_{2}}}}\endgroup } \mathrel{:} {\begingroup\renewcommand\colorMATH{\colorMATHA}\renewcommand\colorSYNTAX{\colorSYNTAXA}{{\color{\colorSYNTAX}\texttt{{\ensuremath{{\mathbb{R}}}}}}}\endgroup } \mathrel{;} {\begingroup\renewcommand\colorMATH{\colorMATHB}\renewcommand\colorSYNTAX{\colorSYNTAXB}{{\color{\colorMATH}\ensuremath{\infty }}}\endgroup }({\begingroup\renewcommand\colorMATH{\colorMATHB}\renewcommand\colorSYNTAX{\colorSYNTAXB}{{\color{\colorMATH}\ensuremath{\sS_{1}}}}\endgroup } + {\begingroup\renewcommand\colorMATH{\colorMATHB}\renewcommand\colorSYNTAX{\colorSYNTAXB}{{\color{\colorMATH}\ensuremath{\sS_{2}}}}\endgroup })}}} 
    \begin{subproof} 
      By induction hypotheses on {{\color{\colorMATH}\ensuremath{\Gamma ; {\begingroup\renewcommand\colorMATH{\colorMATHB}\renewcommand\colorSYNTAX{\colorSYNTAXB}{{\color{\colorMATH}\ensuremath{\Distance}}}\endgroup } \vdash  {\begingroup\renewcommand\colorMATH{\colorMATHB}\renewcommand\colorSYNTAX{\colorSYNTAXB}{{\color{\colorMATH}\ensuremath{\se_{1}}}}\endgroup } \mathrel{:} {\begingroup\renewcommand\colorMATH{\colorMATHA}\renewcommand\colorSYNTAX{\colorSYNTAXA}{{\color{\colorSYNTAX}\texttt{{\ensuremath{{\mathbb{R}}}}}}}\endgroup } \mathrel{;} {\begingroup\renewcommand\colorMATH{\colorMATHB}\renewcommand\colorSYNTAX{\colorSYNTAXB}{{\color{\colorMATH}\ensuremath{\sS_{1}}}}\endgroup }}}}, and {{\color{\colorMATH}\ensuremath{\Gamma ; {\begingroup\renewcommand\colorMATH{\colorMATHB}\renewcommand\colorSYNTAX{\colorSYNTAXB}{{\color{\colorMATH}\ensuremath{\Distance}}}\endgroup } \vdash  {\begingroup\renewcommand\colorMATH{\colorMATHB}\renewcommand\colorSYNTAX{\colorSYNTAXB}{{\color{\colorMATH}\ensuremath{\se_{2}}}}\endgroup } \mathrel{:} {\begingroup\renewcommand\colorMATH{\colorMATHA}\renewcommand\colorSYNTAX{\colorSYNTAXA}{{\color{\colorSYNTAX}\texttt{{\ensuremath{{\mathbb{R}}}}}}}\endgroup } \mathrel{;} {\begingroup\renewcommand\colorMATH{\colorMATHB}\renewcommand\colorSYNTAX{\colorSYNTAXB}{{\color{\colorMATH}\ensuremath{\sS_{2}}}}\endgroup }}}}.
      {{\color{\colorMATH}\ensuremath{\gamma _{1},\gamma \vdash {\begingroup\renewcommand\colorMATH{\colorMATHB}\renewcommand\colorSYNTAX{\colorSYNTAXB}{{\color{\colorMATH}\ensuremath{\se_{1}}}}\endgroup } \Downarrow ^{*} {\begingroup\renewcommand\colorMATH{\colorMATHB}\renewcommand\colorSYNTAX{\colorSYNTAXB}{{\color{\colorMATH}\ensuremath{r_{1}}}}\endgroup }}}}, {{\color{\colorMATH}\ensuremath{\gamma _{2},\gamma \vdash {\begingroup\renewcommand\colorMATH{\colorMATHB}\renewcommand\colorSYNTAX{\colorSYNTAXB}{{\color{\colorMATH}\ensuremath{\se_{1}}}}\endgroup } \Downarrow ^{*} {\begingroup\renewcommand\colorMATH{\colorMATHB}\renewcommand\colorSYNTAX{\colorSYNTAXB}{{\color{\colorMATH}\ensuremath{r_{1}}}}\endgroup }}}}, and {{\color{\colorMATH}\ensuremath{\gamma _{1},\gamma \vdash {\begingroup\renewcommand\colorMATH{\colorMATHB}\renewcommand\colorSYNTAX{\colorSYNTAXB}{{\color{\colorMATH}\ensuremath{\se_{2}}}}\endgroup } \Downarrow ^{*} {\begingroup\renewcommand\colorMATH{\colorMATHB}\renewcommand\colorSYNTAX{\colorSYNTAXB}{{\color{\colorMATH}\ensuremath{r_{2}}}}\endgroup }}}}, {{\color{\colorMATH}\ensuremath{\gamma _{2},\gamma \vdash {\begingroup\renewcommand\colorMATH{\colorMATHB}\renewcommand\colorSYNTAX{\colorSYNTAXB}{{\color{\colorMATH}\ensuremath{\se_{2}}}}\endgroup } \Downarrow ^{*} {\begingroup\renewcommand\colorMATH{\colorMATHB}\renewcommand\colorSYNTAX{\colorSYNTAXB}{{\color{\colorMATH}\ensuremath{r_{2}}}}\endgroup }}}}.
      The result holds as {{\color{\colorMATH}\ensuremath{{\begingroup\renewcommand\colorMATH{\colorMATHB}\renewcommand\colorSYNTAX{\colorSYNTAXB}{{\color{\colorMATH}\ensuremath{r_{1}}}}\endgroup } * {\begingroup\renewcommand\colorMATH{\colorMATHB}\renewcommand\colorSYNTAX{\colorSYNTAXB}{{\color{\colorMATH}\ensuremath{r_{2}}}}\endgroup } = {\begingroup\renewcommand\colorMATH{\colorMATHB}\renewcommand\colorSYNTAX{\colorSYNTAXB}{{\color{\colorMATH}\ensuremath{r_{1}}}}\endgroup } * {\begingroup\renewcommand\colorMATH{\colorMATHB}\renewcommand\colorSYNTAX{\colorSYNTAXB}{{\color{\colorMATH}\ensuremath{r_{2}}}}\endgroup }}}}.
    \end{subproof}
  \item  {{\color{\colorMATH}\ensuremath{\Gamma ; {\begingroup\renewcommand\colorMATH{\colorMATHB}\renewcommand\colorSYNTAX{\colorSYNTAXB}{{\color{\colorMATH}\ensuremath{\Distance}}}\endgroup } \vdash  {\begingroup\renewcommand\colorMATH{\colorMATHB}\renewcommand\colorSYNTAX{\colorSYNTAXB}{{\color{\colorMATH}\ensuremath{\se_{1}}}}\endgroup } \leq  {\begingroup\renewcommand\colorMATH{\colorMATHB}\renewcommand\colorSYNTAX{\colorSYNTAXB}{{\color{\colorMATH}\ensuremath{\se_{2}}}}\endgroup } \mathrel{:} {\mathbb{B}} \mathrel{;} {\begingroup\renewcommand\colorMATH{\colorMATHB}\renewcommand\colorSYNTAX{\colorSYNTAXB}{{\color{\colorMATH}\ensuremath{\infty }}}\endgroup }({\begingroup\renewcommand\colorMATH{\colorMATHB}\renewcommand\colorSYNTAX{\colorSYNTAXB}{{\color{\colorMATH}\ensuremath{\sS_{1}}}}\endgroup } + {\begingroup\renewcommand\colorMATH{\colorMATHB}\renewcommand\colorSYNTAX{\colorSYNTAXB}{{\color{\colorMATH}\ensuremath{\sS_{2}}}}\endgroup })}}} 
    \begin{subproof} 
      By induction hypotheses on {{\color{\colorMATH}\ensuremath{\Gamma ; {\begingroup\renewcommand\colorMATH{\colorMATHB}\renewcommand\colorSYNTAX{\colorSYNTAXB}{{\color{\colorMATH}\ensuremath{\Distance}}}\endgroup } \vdash  {\begingroup\renewcommand\colorMATH{\colorMATHB}\renewcommand\colorSYNTAX{\colorSYNTAXB}{{\color{\colorMATH}\ensuremath{\se_{1}}}}\endgroup } \mathrel{:} {\begingroup\renewcommand\colorMATH{\colorMATHA}\renewcommand\colorSYNTAX{\colorSYNTAXA}{{\color{\colorSYNTAX}\texttt{{\ensuremath{{\mathbb{R}}}}}}}\endgroup } \mathrel{;} {\begingroup\renewcommand\colorMATH{\colorMATHB}\renewcommand\colorSYNTAX{\colorSYNTAXB}{{\color{\colorMATH}\ensuremath{\sS_{1}}}}\endgroup }}}}, and {{\color{\colorMATH}\ensuremath{\Gamma ; {\begingroup\renewcommand\colorMATH{\colorMATHB}\renewcommand\colorSYNTAX{\colorSYNTAXB}{{\color{\colorMATH}\ensuremath{\Distance}}}\endgroup } \vdash  {\begingroup\renewcommand\colorMATH{\colorMATHB}\renewcommand\colorSYNTAX{\colorSYNTAXB}{{\color{\colorMATH}\ensuremath{\se_{2}}}}\endgroup } \mathrel{:} {\begingroup\renewcommand\colorMATH{\colorMATHA}\renewcommand\colorSYNTAX{\colorSYNTAXA}{{\color{\colorSYNTAX}\texttt{{\ensuremath{{\mathbb{R}}}}}}}\endgroup } \mathrel{;} {\begingroup\renewcommand\colorMATH{\colorMATHB}\renewcommand\colorSYNTAX{\colorSYNTAXB}{{\color{\colorMATH}\ensuremath{\sS_{2}}}}\endgroup }}}}.
      {{\color{\colorMATH}\ensuremath{\gamma _{1},\gamma \vdash {\begingroup\renewcommand\colorMATH{\colorMATHB}\renewcommand\colorSYNTAX{\colorSYNTAXB}{{\color{\colorMATH}\ensuremath{\se_{1}}}}\endgroup } \Downarrow ^{*} {\begingroup\renewcommand\colorMATH{\colorMATHB}\renewcommand\colorSYNTAX{\colorSYNTAXB}{{\color{\colorMATH}\ensuremath{r_{1}}}}\endgroup }}}}, {{\color{\colorMATH}\ensuremath{\gamma _{2},\gamma \vdash {\begingroup\renewcommand\colorMATH{\colorMATHB}\renewcommand\colorSYNTAX{\colorSYNTAXB}{{\color{\colorMATH}\ensuremath{\se_{1}}}}\endgroup } \Downarrow ^{*} {\begingroup\renewcommand\colorMATH{\colorMATHB}\renewcommand\colorSYNTAX{\colorSYNTAXB}{{\color{\colorMATH}\ensuremath{r_{1}}}}\endgroup }}}}, and {{\color{\colorMATH}\ensuremath{\gamma _{1},\gamma \vdash {\begingroup\renewcommand\colorMATH{\colorMATHB}\renewcommand\colorSYNTAX{\colorSYNTAXB}{{\color{\colorMATH}\ensuremath{\se_{2}}}}\endgroup } \Downarrow ^{*} {\begingroup\renewcommand\colorMATH{\colorMATHB}\renewcommand\colorSYNTAX{\colorSYNTAXB}{{\color{\colorMATH}\ensuremath{r_{2}}}}\endgroup }}}}, {{\color{\colorMATH}\ensuremath{\gamma _{2},\gamma \vdash {\begingroup\renewcommand\colorMATH{\colorMATHB}\renewcommand\colorSYNTAX{\colorSYNTAXB}{{\color{\colorMATH}\ensuremath{\se_{2}}}}\endgroup } \Downarrow ^{*} {\begingroup\renewcommand\colorMATH{\colorMATHB}\renewcommand\colorSYNTAX{\colorSYNTAXB}{{\color{\colorMATH}\ensuremath{r_{2}}}}\endgroup }}}}.
      The result holds as {{\color{\colorMATH}\ensuremath{{\begingroup\renewcommand\colorMATH{\colorMATHB}\renewcommand\colorSYNTAX{\colorSYNTAXB}{{\color{\colorMATH}\ensuremath{r_{1}}}}\endgroup } \leq  {\begingroup\renewcommand\colorMATH{\colorMATHB}\renewcommand\colorSYNTAX{\colorSYNTAXB}{{\color{\colorMATH}\ensuremath{r_{2}}}}\endgroup } = {\begingroup\renewcommand\colorMATH{\colorMATHB}\renewcommand\colorSYNTAX{\colorSYNTAXB}{{\color{\colorMATH}\ensuremath{r_{1}}}}\endgroup } \leq  {\begingroup\renewcommand\colorMATH{\colorMATHB}\renewcommand\colorSYNTAX{\colorSYNTAXB}{{\color{\colorMATH}\ensuremath{r_{2}}}}\endgroup }}}}.
    \end{subproof}
  \item  {{\color{\colorMATH}\ensuremath{\Gamma ; {\begingroup\renewcommand\colorMATH{\colorMATHB}\renewcommand\colorSYNTAX{\colorSYNTAXB}{{\color{\colorMATH}\ensuremath{\Distance}}}\endgroup } \vdash  x \mathrel{:} \tau  \mathrel{;} {\begingroup\renewcommand\colorMATH{\colorMATHB}\renewcommand\colorSYNTAX{\colorSYNTAXB}{{\color{\colorMATH}\ensuremath{x}}}\endgroup }}}}
    \begin{subproof} 
      Given that {{\color{\colorMATH}\ensuremath{ {\text{FV}}({\begingroup\renewcommand\colorMATH{\colorMATHB}\renewcommand\colorSYNTAX{\colorSYNTAXB}{{\color{\colorMATH}\ensuremath{\se}}}\endgroup }) = \{ x \} \subseteq  dom(\gamma ) }}}, then {{\color{\colorMATH}\ensuremath{x \in  dom(\gamma )}}}.
      Then {{\color{\colorMATH}\ensuremath{\gamma _{1},\gamma \vdash x \Downarrow ^{*} \gamma (x)}}} and {{\color{\colorMATH}\ensuremath{\gamma _{2},\gamma \vdash x \Downarrow ^{*} \gamma (x)}}}, and the result holds as {{\color{\colorMATH}\ensuremath{\gamma (x) = \gamma (x)}}}.
    \end{subproof}
  \item  {{\color{\colorMATH}\ensuremath{\Gamma ; {\begingroup\renewcommand\colorMATH{\colorMATHB}\renewcommand\colorSYNTAX{\colorSYNTAXB}{{\color{\colorMATH}\ensuremath{\Distance}}}\endgroup } \vdash  {\begingroup\renewcommand\colorMATH{\colorMATHB}\renewcommand\colorSYNTAX{\colorSYNTAXB}{{\color{\colorMATH}\ensuremath{\slambda}}}\endgroup } (x\mathrel{:}\tau _{1}\mathord{\cdotp }{\begingroup\renewcommand\colorMATH{\colorMATHB}\renewcommand\colorSYNTAX{\colorSYNTAXB}{{\color{\colorMATH}\ensuremath{\distance_{1}}}}\endgroup }).\hspace*{0.33em}{\begingroup\renewcommand\colorMATH{\colorMATHB}\renewcommand\colorSYNTAX{\colorSYNTAXB}{{\color{\colorMATH}\ensuremath{\se'}}}\endgroup } \mathrel{:} (x\mathrel{:}\tau _{1}\mathord{\cdotp }{\begingroup\renewcommand\colorMATH{\colorMATHB}\renewcommand\colorSYNTAX{\colorSYNTAXB}{{\color{\colorMATH}\ensuremath{\distance_{1}}}}\endgroup }) \xrightarrowS {{\begingroup\renewcommand\colorMATH{\colorMATHB}\renewcommand\colorSYNTAX{\colorSYNTAXB}{{\color{\colorMATH}\ensuremath{\sS''}}}\endgroup }+{\begingroup\renewcommand\colorMATH{\colorMATHB}\renewcommand\colorSYNTAX{\colorSYNTAXB}{{\color{\colorMATH}\ensuremath{\sss'}}}\endgroup }x} \tau _{2} \mathrel{;} \varnothing }}} 
    \begin{subproof} 
      We know that {{\color{\colorMATH}\ensuremath{{\text{FV}}({\begingroup\renewcommand\colorMATH{\colorMATHB}\renewcommand\colorSYNTAX{\colorSYNTAXB}{{\color{\colorMATH}\ensuremath{\se}}}\endgroup }) \subseteq  dom(\gamma )}}}, 
      then {{\color{\colorMATH}\ensuremath{\gamma _{1},\gamma \vdash {\begingroup\renewcommand\colorMATH{\colorMATHB}\renewcommand\colorSYNTAX{\colorSYNTAXB}{{\color{\colorMATH}\ensuremath{\slambda}}}\endgroup } (x\mathrel{:}\tau _{1}\mathord{\cdotp }{\begingroup\renewcommand\colorMATH{\colorMATHB}\renewcommand\colorSYNTAX{\colorSYNTAXB}{{\color{\colorMATH}\ensuremath{\distance_{1}}}}\endgroup }).\hspace*{0.33em}{\begingroup\renewcommand\colorMATH{\colorMATHB}\renewcommand\colorSYNTAX{\colorSYNTAXB}{{\color{\colorMATH}\ensuremath{\se'}}}\endgroup } \Downarrow ^{1} \langle {\begingroup\renewcommand\colorMATH{\colorMATHB}\renewcommand\colorSYNTAX{\colorSYNTAXB}{{\color{\colorMATH}\ensuremath{\slambda}}}\endgroup } (x\mathrel{:}\tau _{1}\mathord{\cdotp }{\begingroup\renewcommand\colorMATH{\colorMATHB}\renewcommand\colorSYNTAX{\colorSYNTAXB}{{\color{\colorMATH}\ensuremath{\distance_{1}}}}\endgroup }).\hspace*{0.33em}{\begingroup\renewcommand\colorMATH{\colorMATHB}\renewcommand\colorSYNTAX{\colorSYNTAXB}{{\color{\colorMATH}\ensuremath{\se'}}}\endgroup }, \gamma '\rangle }}} and 
        {{\color{\colorMATH}\ensuremath{\gamma _{2},\gamma \vdash {\begingroup\renewcommand\colorMATH{\colorMATHB}\renewcommand\colorSYNTAX{\colorSYNTAXB}{{\color{\colorMATH}\ensuremath{\slambda}}}\endgroup } (x\mathrel{:}\tau _{1}\mathord{\cdotp }{\begingroup\renewcommand\colorMATH{\colorMATHB}\renewcommand\colorSYNTAX{\colorSYNTAXB}{{\color{\colorMATH}\ensuremath{\distance_{1}}}}\endgroup }).\hspace*{0.33em}{\begingroup\renewcommand\colorMATH{\colorMATHB}\renewcommand\colorSYNTAX{\colorSYNTAXB}{{\color{\colorMATH}\ensuremath{\se'}}}\endgroup } \Downarrow ^{1} \langle {\begingroup\renewcommand\colorMATH{\colorMATHB}\renewcommand\colorSYNTAX{\colorSYNTAXB}{{\color{\colorMATH}\ensuremath{\slambda}}}\endgroup } (x\mathrel{:}\tau _{1}\mathord{\cdotp }{\begingroup\renewcommand\colorMATH{\colorMATHB}\renewcommand\colorSYNTAX{\colorSYNTAXB}{{\color{\colorMATH}\ensuremath{\distance_{1}}}}\endgroup }).\hspace*{0.33em}{\begingroup\renewcommand\colorMATH{\colorMATHB}\renewcommand\colorSYNTAX{\colorSYNTAXB}{{\color{\colorMATH}\ensuremath{\se'}}}\endgroup }, \gamma '\rangle }}}, for {{\color{\colorMATH}\ensuremath{\gamma ' \subseteq  \gamma }}} and the result holds.
    \end{subproof}
  \item  {{\color{\colorMATH}\ensuremath{\Gamma ; {\begingroup\renewcommand\colorMATH{\colorMATHB}\renewcommand\colorSYNTAX{\colorSYNTAXB}{{\color{\colorMATH}\ensuremath{\Distance}}}\endgroup } \vdash  \langle {\begingroup\renewcommand\colorMATH{\colorMATHB}\renewcommand\colorSYNTAX{\colorSYNTAXB}{{\color{\colorMATH}\ensuremath{\slambda}}}\endgroup } (x\mathrel{:}\tau _{1}\mathord{\cdotp }{\begingroup\renewcommand\colorMATH{\colorMATHB}\renewcommand\colorSYNTAX{\colorSYNTAXB}{{\color{\colorMATH}\ensuremath{\distance_{1}}}}\endgroup }).\hspace*{0.33em}{\begingroup\renewcommand\colorMATH{\colorMATHB}\renewcommand\colorSYNTAX{\colorSYNTAXB}{{\color{\colorMATH}\ensuremath{\se'}}}\endgroup }, \gamma \rangle  \mathrel{:} (x\mathrel{:}\tau _{1}\mathord{\cdotp }{\begingroup\renewcommand\colorMATH{\colorMATHB}\renewcommand\colorSYNTAX{\colorSYNTAXB}{{\color{\colorMATH}\ensuremath{\distance_{1}}}}\endgroup }) \xrightarrowS {{\begingroup\renewcommand\colorMATH{\colorMATHB}\renewcommand\colorSYNTAX{\colorSYNTAXB}{{\color{\colorMATH}\ensuremath{\sss'}}}\endgroup }x} \tau _{2} \mathrel{;} \varnothing }}} 
    \begin{subproof} 
      Trivial as closures are already values.
    \end{subproof}
  \item  {{\color{\colorMATH}\ensuremath{\Gamma ; {\begingroup\renewcommand\colorMATH{\colorMATHB}\renewcommand\colorSYNTAX{\colorSYNTAXB}{{\color{\colorMATH}\ensuremath{\Distance}}}\endgroup } \vdash  {\begingroup\renewcommand\colorMATH{\colorMATHB}\renewcommand\colorSYNTAX{\colorSYNTAXB}{{\color{\colorMATH}\ensuremath{\se_{1}}}}\endgroup }\hspace*{0.33em}{\begingroup\renewcommand\colorMATH{\colorMATHB}\renewcommand\colorSYNTAX{\colorSYNTAXB}{{\color{\colorMATH}\ensuremath{\se_{2}}}}\endgroup } \mathrel{:} [{\begingroup\renewcommand\colorMATH{\colorMATHB}\renewcommand\colorSYNTAX{\colorSYNTAXB}{{\color{\colorMATH}\ensuremath{\sS_{2}}}}\endgroup }/x]\tau _{2} \mathrel{;} {\begingroup\renewcommand\colorMATH{\colorMATHB}\renewcommand\colorSYNTAX{\colorSYNTAXB}{{\color{\colorMATH}\ensuremath{\sS_{1}}}}\endgroup } + {\begingroup\renewcommand\colorMATH{\colorMATHB}\renewcommand\colorSYNTAX{\colorSYNTAXB}{{\color{\colorMATH}\ensuremath{\sss}}}\endgroup }{\begingroup\renewcommand\colorMATH{\colorMATHB}\renewcommand\colorSYNTAX{\colorSYNTAXB}{{\color{\colorMATH}\ensuremath{\sS_{2}}}}\endgroup } + {\begingroup\renewcommand\colorMATH{\colorMATHB}\renewcommand\colorSYNTAX{\colorSYNTAXB}{{\color{\colorMATH}\ensuremath{\sS''}}}\endgroup } }}} 
    \begin{subproof} 
      By induction hypotheses on {{\color{\colorMATH}\ensuremath{\Gamma  \vdash  {\begingroup\renewcommand\colorMATH{\colorMATHB}\renewcommand\colorSYNTAX{\colorSYNTAXB}{{\color{\colorMATH}\ensuremath{\se_{1}}}}\endgroup } \mathrel{:} (x:\tau _{1}) \xrightarrowS {{\begingroup\renewcommand\colorMATH{\colorMATHB}\renewcommand\colorSYNTAX{\colorSYNTAXB}{{\color{\colorMATH}\ensuremath{\sS''}}}\endgroup }+{\begingroup\renewcommand\colorMATH{\colorMATHB}\renewcommand\colorSYNTAX{\colorSYNTAXB}{{\color{\colorMATH}\ensuremath{\sss}}}\endgroup }x} \tau _{2} \mathrel{;} {\begingroup\renewcommand\colorMATH{\colorMATHB}\renewcommand\colorSYNTAX{\colorSYNTAXB}{{\color{\colorMATH}\ensuremath{\sS_{1}}}}\endgroup }}}} and {{\color{\colorMATH}\ensuremath{\Gamma  \vdash  {\begingroup\renewcommand\colorMATH{\colorMATHB}\renewcommand\colorSYNTAX{\colorSYNTAXB}{{\color{\colorMATH}\ensuremath{\se_{2}}}}\endgroup } \mathrel{:} \tau _{1} \mathrel{;} {\begingroup\renewcommand\colorMATH{\colorMATHB}\renewcommand\colorSYNTAX{\colorSYNTAXB}{{\color{\colorMATH}\ensuremath{\sS_{2}}}}\endgroup }}}}.
      {{\color{\colorMATH}\ensuremath{\gamma _{1},\gamma \vdash {\begingroup\renewcommand\colorMATH{\colorMATHB}\renewcommand\colorSYNTAX{\colorSYNTAXB}{{\color{\colorMATH}\ensuremath{\se_{1}}}}\endgroup } \Downarrow ^{*} \langle {\begingroup\renewcommand\colorMATH{\colorMATHB}\renewcommand\colorSYNTAX{\colorSYNTAXB}{{\color{\colorMATH}\ensuremath{\slambda}}}\endgroup } (x\mathrel{:}\tau _{1}\mathord{\cdotp }{\begingroup\renewcommand\colorMATH{\colorMATHB}\renewcommand\colorSYNTAX{\colorSYNTAXB}{{\color{\colorMATH}\ensuremath{\sss_{1}}}}\endgroup }).\hspace*{0.33em}{\begingroup\renewcommand\colorMATH{\colorMATHB}\renewcommand\colorSYNTAX{\colorSYNTAXB}{{\color{\colorMATH}\ensuremath{\se'}}}\endgroup }, \gamma '\rangle }}}, {{\color{\colorMATH}\ensuremath{\gamma _{2},\gamma \vdash {\begingroup\renewcommand\colorMATH{\colorMATHB}\renewcommand\colorSYNTAX{\colorSYNTAXB}{{\color{\colorMATH}\ensuremath{\se_{1}}}}\endgroup } \Downarrow ^{*} \langle {\begingroup\renewcommand\colorMATH{\colorMATHB}\renewcommand\colorSYNTAX{\colorSYNTAXB}{{\color{\colorMATH}\ensuremath{\slambda}}}\endgroup } (x\mathrel{:}\tau _{1}\mathord{\cdotp }{\begingroup\renewcommand\colorMATH{\colorMATHB}\renewcommand\colorSYNTAX{\colorSYNTAXB}{{\color{\colorMATH}\ensuremath{\sss_{1}}}}\endgroup }).\hspace*{0.33em}{\begingroup\renewcommand\colorMATH{\colorMATHB}\renewcommand\colorSYNTAX{\colorSYNTAXB}{{\color{\colorMATH}\ensuremath{\se'}}}\endgroup }, \gamma '\rangle }}}, and {{\color{\colorMATH}\ensuremath{\gamma _{1},\gamma \vdash {\begingroup\renewcommand\colorMATH{\colorMATHB}\renewcommand\colorSYNTAX{\colorSYNTAXB}{{\color{\colorMATH}\ensuremath{\se_{2}}}}\endgroup } \Downarrow ^{*} {\begingroup\renewcommand\colorMATH{\colorMATHB}\renewcommand\colorSYNTAX{\colorSYNTAXB}{{\color{\colorMATH}\ensuremath{\sv}}}\endgroup }}}}, {{\color{\colorMATH}\ensuremath{\gamma _{2},\gamma \vdash {\begingroup\renewcommand\colorMATH{\colorMATHB}\renewcommand\colorSYNTAX{\colorSYNTAXB}{{\color{\colorMATH}\ensuremath{\se_{2}}}}\endgroup } \Downarrow ^{*} {\begingroup\renewcommand\colorMATH{\colorMATHB}\renewcommand\colorSYNTAX{\colorSYNTAXB}{{\color{\colorMATH}\ensuremath{\sv}}}\endgroup }}}}.
      The result holds as {{\color{\colorMATH}\ensuremath{\gamma '[x\mapsto {\begingroup\renewcommand\colorMATH{\colorMATHB}\renewcommand\colorSYNTAX{\colorSYNTAXB}{{\color{\colorMATH}\ensuremath{\sv}}}\endgroup }] \vdash  {\begingroup\renewcommand\colorMATH{\colorMATHB}\renewcommand\colorSYNTAX{\colorSYNTAXB}{{\color{\colorMATH}\ensuremath{\se'}}}\endgroup }\Downarrow ^{*} {\begingroup\renewcommand\colorMATH{\colorMATHB}\renewcommand\colorSYNTAX{\colorSYNTAXB}{{\color{\colorMATH}\ensuremath{\sv_{1}}}}\endgroup }}}} and {{\color{\colorMATH}\ensuremath{\gamma '[x\mapsto {\begingroup\renewcommand\colorMATH{\colorMATHB}\renewcommand\colorSYNTAX{\colorSYNTAXB}{{\color{\colorMATH}\ensuremath{\sv}}}\endgroup }] \vdash  {\begingroup\renewcommand\colorMATH{\colorMATHB}\renewcommand\colorSYNTAX{\colorSYNTAXB}{{\color{\colorMATH}\ensuremath{\se'}}}\endgroup }\Downarrow ^{*} {\begingroup\renewcommand\colorMATH{\colorMATHB}\renewcommand\colorSYNTAX{\colorSYNTAXB}{{\color{\colorMATH}\ensuremath{\sv_{2}}}}\endgroup }}}}, therefore {{\color{\colorMATH}\ensuremath{{\begingroup\renewcommand\colorMATH{\colorMATHB}\renewcommand\colorSYNTAX{\colorSYNTAXB}{{\color{\colorMATH}\ensuremath{\sv_{1}}}}\endgroup } = {\begingroup\renewcommand\colorMATH{\colorMATHB}\renewcommand\colorSYNTAX{\colorSYNTAXB}{{\color{\colorMATH}\ensuremath{\sv_{2}}}}\endgroup }}}}.
    \end{subproof}
  \item  {{\color{\colorMATH}\ensuremath{\Gamma ; {\begingroup\renewcommand\colorMATH{\colorMATHB}\renewcommand\colorSYNTAX{\colorSYNTAXB}{{\color{\colorMATH}\ensuremath{\Distance}}}\endgroup } \vdash  \ttt \mathrel{:} {{\color{\colorSYNTAX}\texttt{unit}}} \mathrel{;} \varnothing }}}  
    \begin{subproof} 
      Trivial.
    \end{subproof}
  \item  {{\color{\colorMATH}\ensuremath{\Gamma ; {\begingroup\renewcommand\colorMATH{\colorMATHB}\renewcommand\colorSYNTAX{\colorSYNTAXB}{{\color{\colorMATH}\ensuremath{\Distance}}}\endgroup } \vdash  \inl^{\tau _{2}}\hspace*{0.33em}{\begingroup\renewcommand\colorMATH{\colorMATHB}\renewcommand\colorSYNTAX{\colorSYNTAXB}{{\color{\colorMATH}\ensuremath{\se'}}}\endgroup } \mathrel{:} \tau _{1} \mathrel{^{{\begingroup\renewcommand\colorMATH{\colorMATHB}\renewcommand\colorSYNTAX{\colorSYNTAXB}{{\color{\colorMATH}\ensuremath{\sS''}}}\endgroup }}\oplus ^{\varnothing }} \tau _{2} \mathrel{;} \varnothing }}}
    \begin{subproof}
      By induction hypothesis on {{\color{\colorMATH}\ensuremath{\Gamma  \vdash  {\begingroup\renewcommand\colorMATH{\colorMATHB}\renewcommand\colorSYNTAX{\colorSYNTAXB}{{\color{\colorMATH}\ensuremath{\se'}}}\endgroup } \mathrel{:} \tau _{1} \mathrel{;} {\begingroup\renewcommand\colorMATH{\colorMATHB}\renewcommand\colorSYNTAX{\colorSYNTAXB}{{\color{\colorMATH}\ensuremath{\sS''}}}\endgroup }}}}.
      {{\color{\colorMATH}\ensuremath{\gamma _{1},\gamma \vdash {\begingroup\renewcommand\colorMATH{\colorMATHB}\renewcommand\colorSYNTAX{\colorSYNTAXB}{{\color{\colorMATH}\ensuremath{\se'}}}\endgroup } \Downarrow ^{*} {\begingroup\renewcommand\colorMATH{\colorMATHB}\renewcommand\colorSYNTAX{\colorSYNTAXB}{{\color{\colorMATH}\ensuremath{\sv}}}\endgroup }}}}, {{\color{\colorMATH}\ensuremath{\gamma _{2},\gamma \vdash {\begingroup\renewcommand\colorMATH{\colorMATHB}\renewcommand\colorSYNTAX{\colorSYNTAXB}{{\color{\colorMATH}\ensuremath{\se'}}}\endgroup } \Downarrow ^{*} {\begingroup\renewcommand\colorMATH{\colorMATHB}\renewcommand\colorSYNTAX{\colorSYNTAXB}{{\color{\colorMATH}\ensuremath{\sv}}}\endgroup }}}}, and the result holds as {{\color{\colorMATH}\ensuremath{\inl^{\tau _{2}} {\begingroup\renewcommand\colorMATH{\colorMATHB}\renewcommand\colorSYNTAX{\colorSYNTAXB}{{\color{\colorMATH}\ensuremath{\sv}}}\endgroup } = \inl^{\tau _{2}} {\begingroup\renewcommand\colorMATH{\colorMATHB}\renewcommand\colorSYNTAX{\colorSYNTAXB}{{\color{\colorMATH}\ensuremath{\sv}}}\endgroup }}}}.
    \end{subproof}
  \item  {{\color{\colorMATH}\ensuremath{\Gamma ; {\begingroup\renewcommand\colorMATH{\colorMATHB}\renewcommand\colorSYNTAX{\colorSYNTAXB}{{\color{\colorMATH}\ensuremath{\Distance}}}\endgroup } \vdash  \inr^{\tau _{1}}\hspace*{0.33em}{\begingroup\renewcommand\colorMATH{\colorMATHB}\renewcommand\colorSYNTAX{\colorSYNTAXB}{{\color{\colorMATH}\ensuremath{\se'}}}\endgroup } \mathrel{:} \tau _{1} \mathrel{^{\varnothing }\oplus ^{{\begingroup\renewcommand\colorMATH{\colorMATHB}\renewcommand\colorSYNTAX{\colorSYNTAXB}{{\color{\colorMATH}\ensuremath{\sS''}}}\endgroup }}} \tau _{2} \mathrel{;} \varnothing }}}
    \begin{subproof} 
      By induction hypothesis on {{\color{\colorMATH}\ensuremath{\Gamma  \vdash  {\begingroup\renewcommand\colorMATH{\colorMATHB}\renewcommand\colorSYNTAX{\colorSYNTAXB}{{\color{\colorMATH}\ensuremath{\se'}}}\endgroup } \mathrel{:} \tau _{1} \mathrel{;} {\begingroup\renewcommand\colorMATH{\colorMATHB}\renewcommand\colorSYNTAX{\colorSYNTAXB}{{\color{\colorMATH}\ensuremath{\sS''}}}\endgroup }}}}.
      {{\color{\colorMATH}\ensuremath{\gamma _{1},\gamma \vdash {\begingroup\renewcommand\colorMATH{\colorMATHB}\renewcommand\colorSYNTAX{\colorSYNTAXB}{{\color{\colorMATH}\ensuremath{\se'}}}\endgroup } \Downarrow ^{*} {\begingroup\renewcommand\colorMATH{\colorMATHB}\renewcommand\colorSYNTAX{\colorSYNTAXB}{{\color{\colorMATH}\ensuremath{\sv}}}\endgroup }}}}, {{\color{\colorMATH}\ensuremath{\gamma _{2},\gamma \vdash {\begingroup\renewcommand\colorMATH{\colorMATHB}\renewcommand\colorSYNTAX{\colorSYNTAXB}{{\color{\colorMATH}\ensuremath{\se'}}}\endgroup } \Downarrow ^{*} {\begingroup\renewcommand\colorMATH{\colorMATHB}\renewcommand\colorSYNTAX{\colorSYNTAXB}{{\color{\colorMATH}\ensuremath{\sv}}}\endgroup }}}}, and the result holds as {{\color{\colorMATH}\ensuremath{\inr^{\tau _{2}} {\begingroup\renewcommand\colorMATH{\colorMATHB}\renewcommand\colorSYNTAX{\colorSYNTAXB}{{\color{\colorMATH}\ensuremath{\sv}}}\endgroup } = \inr^{\tau _{2}} {\begingroup\renewcommand\colorMATH{\colorMATHB}\renewcommand\colorSYNTAX{\colorSYNTAXB}{{\color{\colorMATH}\ensuremath{\sv}}}\endgroup }}}}.
    \end{subproof}
  \item  {{\color{\colorMATH}\ensuremath{\Gamma ; {\begingroup\renewcommand\colorMATH{\colorMATHB}\renewcommand\colorSYNTAX{\colorSYNTAXB}{{\color{\colorMATH}\ensuremath{\Distance}}}\endgroup } \vdash  \ccase\hspace*{0.33em}{\begingroup\renewcommand\colorMATH{\colorMATHB}\renewcommand\colorSYNTAX{\colorSYNTAXB}{{\color{\colorMATH}\ensuremath{\se_{1}}}}\endgroup }\hspace*{0.33em}\of\hspace*{0.33em}\{ x\Rightarrow {\begingroup\renewcommand\colorMATH{\colorMATHB}\renewcommand\colorSYNTAX{\colorSYNTAXB}{{\color{\colorMATH}\ensuremath{\se_{2}}}}\endgroup }\} \hspace*{0.33em}\{ y\Rightarrow {\begingroup\renewcommand\colorMATH{\colorMATHB}\renewcommand\colorSYNTAX{\colorSYNTAXB}{{\color{\colorMATH}\ensuremath{\se_{3}}}}\endgroup }\}  \mathrel{:} [{\begingroup\renewcommand\colorMATH{\colorMATHB}\renewcommand\colorSYNTAX{\colorSYNTAXB}{{\color{\colorMATH}\ensuremath{\sS_{1}}}}\endgroup } + {\begingroup\renewcommand\colorMATH{\colorMATHB}\renewcommand\colorSYNTAX{\colorSYNTAXB}{{\color{\colorMATH}\ensuremath{\sS_{1 1}}}}\endgroup }/x]\tau _{2} \sqcup  [{\begingroup\renewcommand\colorMATH{\colorMATHB}\renewcommand\colorSYNTAX{\colorSYNTAXB}{{\color{\colorMATH}\ensuremath{\sS_{1}}}}\endgroup } + {\begingroup\renewcommand\colorMATH{\colorMATHB}\renewcommand\colorSYNTAX{\colorSYNTAXB}{{\color{\colorMATH}\ensuremath{\sS_{1 2}}}}\endgroup }/y]\tau _{3} \mathrel{;}  {\begingroup\renewcommand\colorMATH{\colorMATHB}\renewcommand\colorSYNTAX{\colorSYNTAXB}{{\color{\colorMATH}\ensuremath{\sS_{1}}}}\endgroup } \sqcup  ([{\begingroup\renewcommand\colorMATH{\colorMATHB}\renewcommand\colorSYNTAX{\colorSYNTAXB}{{\color{\colorMATH}\ensuremath{\sS_{1}}}}\endgroup } + {\begingroup\renewcommand\colorMATH{\colorMATHB}\renewcommand\colorSYNTAX{\colorSYNTAXB}{{\color{\colorMATH}\ensuremath{\sS_{1 1}}}}\endgroup }/x]({\begingroup\renewcommand\colorMATH{\colorMATHB}\renewcommand\colorSYNTAX{\colorSYNTAXB}{{\color{\colorMATH}\ensuremath{\sS_{2}}}}\endgroup }+{\begingroup\renewcommand\colorMATH{\colorMATHB}\renewcommand\colorSYNTAX{\colorSYNTAXB}{{\color{\colorMATH}\ensuremath{\sss_{2}}}}\endgroup }x) \sqcup  [{\begingroup\renewcommand\colorMATH{\colorMATHB}\renewcommand\colorSYNTAX{\colorSYNTAXB}{{\color{\colorMATH}\ensuremath{\sS_{1}}}}\endgroup } + {\begingroup\renewcommand\colorMATH{\colorMATHB}\renewcommand\colorSYNTAX{\colorSYNTAXB}{{\color{\colorMATH}\ensuremath{\sS_{1 2}}}}\endgroup }/y]({\begingroup\renewcommand\colorMATH{\colorMATHB}\renewcommand\colorSYNTAX{\colorSYNTAXB}{{\color{\colorMATH}\ensuremath{\sS_{3}}}}\endgroup }+{\begingroup\renewcommand\colorMATH{\colorMATHB}\renewcommand\colorSYNTAX{\colorSYNTAXB}{{\color{\colorMATH}\ensuremath{\sss_{3}}}}\endgroup }y))}}}
    \begin{subproof} 
        By induction hypothesis on {{\color{\colorMATH}\ensuremath{\Gamma ; {\begingroup\renewcommand\colorMATH{\colorMATHB}\renewcommand\colorSYNTAX{\colorSYNTAXB}{{\color{\colorMATH}\ensuremath{\Distance}}}\endgroup } \vdash  {\begingroup\renewcommand\colorMATH{\colorMATHB}\renewcommand\colorSYNTAX{\colorSYNTAXB}{{\color{\colorMATH}\ensuremath{\se_{1}}}}\endgroup } \mathrel{:} \tau _{1 1} \mathrel{^{{\begingroup\renewcommand\colorMATH{\colorMATHB}\renewcommand\colorSYNTAX{\colorSYNTAXB}{{\color{\colorMATH}\ensuremath{\sS_{1 1}}}}\endgroup }}\oplus ^{{\begingroup\renewcommand\colorMATH{\colorMATHB}\renewcommand\colorSYNTAX{\colorSYNTAXB}{{\color{\colorMATH}\ensuremath{\sS_{1 2}}}}\endgroup }}} \tau _{1 2} \mathrel{;} {\begingroup\renewcommand\colorMATH{\colorMATHB}\renewcommand\colorSYNTAX{\colorSYNTAXB}{{\color{\colorMATH}\ensuremath{\sS_{1}}}}\endgroup }}}}, we know that\\
        {{\color{\colorMATH}\ensuremath{\gamma _{1},\gamma \vdash {\begingroup\renewcommand\colorMATH{\colorMATHB}\renewcommand\colorSYNTAX{\colorSYNTAXB}{{\color{\colorMATH}\ensuremath{\se_{1}}}}\endgroup } \Downarrow ^{*} {\begingroup\renewcommand\colorMATH{\colorMATHB}\renewcommand\colorSYNTAX{\colorSYNTAXB}{{\color{\colorMATH}\ensuremath{\sv_{1}}}}\endgroup }}}}, {{\color{\colorMATH}\ensuremath{\gamma _{2},\gamma \vdash {\begingroup\renewcommand\colorMATH{\colorMATHB}\renewcommand\colorSYNTAX{\colorSYNTAXB}{{\color{\colorMATH}\ensuremath{\se_{1}}}}\endgroup } \Downarrow ^{*} {\begingroup\renewcommand\colorMATH{\colorMATHB}\renewcommand\colorSYNTAX{\colorSYNTAXB}{{\color{\colorMATH}\ensuremath{\sv_{1}}}}\endgroup }}}}.
        Suppose {{\color{\colorMATH}\ensuremath{{\begingroup\renewcommand\colorMATH{\colorMATHB}\renewcommand\colorSYNTAX{\colorSYNTAXB}{{\color{\colorMATH}\ensuremath{\sv_{1}}}}\endgroup } = \inl\hspace*{0.33em}{\begingroup\renewcommand\colorMATH{\colorMATHB}\renewcommand\colorSYNTAX{\colorSYNTAXB}{{\color{\colorMATH}\ensuremath{\sv'_{1}}}}\endgroup }}}} (the other case is analogous).
        Then by induction hypothesis on {{\color{\colorMATH}\ensuremath{\Gamma ,x\mathrel{:}\tau _{1 1}; {\begingroup\renewcommand\colorMATH{\colorMATHB}\renewcommand\colorSYNTAX{\colorSYNTAXB}{{\color{\colorMATH}\ensuremath{\Distance}}}\endgroup } + ({\begingroup\renewcommand\colorMATH{\colorMATHB}\renewcommand\colorSYNTAX{\colorSYNTAXB}{{\color{\colorMATH}\ensuremath{\Distance}}}\endgroup }\mathord{\cdotp }({\begingroup\renewcommand\colorMATH{\colorMATHB}\renewcommand\colorSYNTAX{\colorSYNTAXB}{{\color{\colorMATH}\ensuremath{\sS_{1}}}}\endgroup } + {\begingroup\renewcommand\colorMATH{\colorMATHB}\renewcommand\colorSYNTAX{\colorSYNTAXB}{{\color{\colorMATH}\ensuremath{\sS_{1 1}}}}\endgroup }))x \vdash  {\begingroup\renewcommand\colorMATH{\colorMATHB}\renewcommand\colorSYNTAX{\colorSYNTAXB}{{\color{\colorMATH}\ensuremath{\se_{2}}}}\endgroup } \mathrel{:} \tau _{2} \mathrel{;} {\begingroup\renewcommand\colorMATH{\colorMATHB}\renewcommand\colorSYNTAX{\colorSYNTAXB}{{\color{\colorMATH}\ensuremath{\sS_{2}}}}\endgroup }+{\begingroup\renewcommand\colorMATH{\colorMATHB}\renewcommand\colorSYNTAX{\colorSYNTAXB}{{\color{\colorMATH}\ensuremath{\sss_{2}}}}\endgroup }x}}}, we know that,
        {{\color{\colorMATH}\ensuremath{\gamma _{1},\gamma [x\mapsto {\begingroup\renewcommand\colorMATH{\colorMATHB}\renewcommand\colorSYNTAX{\colorSYNTAXB}{{\color{\colorMATH}\ensuremath{\sv'_{1}}}}\endgroup }]\vdash {\begingroup\renewcommand\colorMATH{\colorMATHB}\renewcommand\colorSYNTAX{\colorSYNTAXB}{{\color{\colorMATH}\ensuremath{\se_{2}}}}\endgroup } \Downarrow ^{*} {\begingroup\renewcommand\colorMATH{\colorMATHB}\renewcommand\colorSYNTAX{\colorSYNTAXB}{{\color{\colorMATH}\ensuremath{\sv}}}\endgroup }}}}, {{\color{\colorMATH}\ensuremath{\gamma _{2},\gamma [x\mapsto {\begingroup\renewcommand\colorMATH{\colorMATHB}\renewcommand\colorSYNTAX{\colorSYNTAXB}{{\color{\colorMATH}\ensuremath{\sv'_{1}}}}\endgroup }]\vdash  {\begingroup\renewcommand\colorMATH{\colorMATHB}\renewcommand\colorSYNTAX{\colorSYNTAXB}{{\color{\colorMATH}\ensuremath{\se_{2}}}}\endgroup } \Downarrow ^{*} {\begingroup\renewcommand\colorMATH{\colorMATHB}\renewcommand\colorSYNTAX{\colorSYNTAXB}{{\color{\colorMATH}\ensuremath{\sv}}}\endgroup }}}}, and the result holds,
    \end{subproof}
  \item  {{\color{\colorMATH}\ensuremath{\Gamma ; {\begingroup\renewcommand\colorMATH{\colorMATHB}\renewcommand\colorSYNTAX{\colorSYNTAXB}{{\color{\colorMATH}\ensuremath{\Distance}}}\endgroup } \vdash  \addProduct{{\begingroup\renewcommand\colorMATH{\colorMATHB}\renewcommand\colorSYNTAX{\colorSYNTAXB}{{\color{\colorMATH}\ensuremath{\se_{1}}}}\endgroup }}{{\begingroup\renewcommand\colorMATH{\colorMATHB}\renewcommand\colorSYNTAX{\colorSYNTAXB}{{\color{\colorMATH}\ensuremath{\se_{2}}}}\endgroup }} \mathrel{:} \tau _{1} \mathrel{^{{\begingroup\renewcommand\colorMATH{\colorMATHB}\renewcommand\colorSYNTAX{\colorSYNTAXB}{{\color{\colorMATH}\ensuremath{\sS_{1}}}}\endgroup }}\&^{{\begingroup\renewcommand\colorMATH{\colorMATHB}\renewcommand\colorSYNTAX{\colorSYNTAXB}{{\color{\colorMATH}\ensuremath{\sS_{2}}}}\endgroup }}} \tau _{2} \mathrel{;} \varnothing }}}
    \begin{subproof} 
      By induction hypotheses on {{\color{\colorMATH}\ensuremath{\Gamma  \vdash  {\begingroup\renewcommand\colorMATH{\colorMATHB}\renewcommand\colorSYNTAX{\colorSYNTAXB}{{\color{\colorMATH}\ensuremath{\se_{1}}}}\endgroup } \mathrel{:} \tau _{1} \mathrel{;} {\begingroup\renewcommand\colorMATH{\colorMATHB}\renewcommand\colorSYNTAX{\colorSYNTAXB}{{\color{\colorMATH}\ensuremath{\sS_{1}}}}\endgroup }}}} and {{\color{\colorMATH}\ensuremath{\Gamma  \vdash  {\begingroup\renewcommand\colorMATH{\colorMATHB}\renewcommand\colorSYNTAX{\colorSYNTAXB}{{\color{\colorMATH}\ensuremath{\se_{2}}}}\endgroup } \mathrel{:} \tau _{2} \mathrel{;} {\begingroup\renewcommand\colorMATH{\colorMATHB}\renewcommand\colorSYNTAX{\colorSYNTAXB}{{\color{\colorMATH}\ensuremath{\sS_{2}}}}\endgroup }}}},
      {{\color{\colorMATH}\ensuremath{\gamma _{1},\gamma \vdash {\begingroup\renewcommand\colorMATH{\colorMATHB}\renewcommand\colorSYNTAX{\colorSYNTAXB}{{\color{\colorMATH}\ensuremath{\se_{1}}}}\endgroup } \Downarrow ^{*} {\begingroup\renewcommand\colorMATH{\colorMATHB}\renewcommand\colorSYNTAX{\colorSYNTAXB}{{\color{\colorMATH}\ensuremath{\sv_{1}}}}\endgroup }}}}, {{\color{\colorMATH}\ensuremath{\gamma _{2},\gamma \vdash {\begingroup\renewcommand\colorMATH{\colorMATHB}\renewcommand\colorSYNTAX{\colorSYNTAXB}{{\color{\colorMATH}\ensuremath{\se_{1}}}}\endgroup } \Downarrow ^{*} {\begingroup\renewcommand\colorMATH{\colorMATHB}\renewcommand\colorSYNTAX{\colorSYNTAXB}{{\color{\colorMATH}\ensuremath{\sv_{1}}}}\endgroup }}}}, and {{\color{\colorMATH}\ensuremath{\gamma _{1},\gamma \vdash {\begingroup\renewcommand\colorMATH{\colorMATHB}\renewcommand\colorSYNTAX{\colorSYNTAXB}{{\color{\colorMATH}\ensuremath{\se_{2}}}}\endgroup } \Downarrow ^{*} {\begingroup\renewcommand\colorMATH{\colorMATHB}\renewcommand\colorSYNTAX{\colorSYNTAXB}{{\color{\colorMATH}\ensuremath{\sv_{2}}}}\endgroup }}}}, {{\color{\colorMATH}\ensuremath{\gamma _{2},\gamma \vdash {\begingroup\renewcommand\colorMATH{\colorMATHB}\renewcommand\colorSYNTAX{\colorSYNTAXB}{{\color{\colorMATH}\ensuremath{\se_{2}}}}\endgroup } \Downarrow ^{*} {\begingroup\renewcommand\colorMATH{\colorMATHB}\renewcommand\colorSYNTAX{\colorSYNTAXB}{{\color{\colorMATH}\ensuremath{\sv_{2}}}}\endgroup }}}}.
      The result holds as {{\color{\colorMATH}\ensuremath{\addProduct{{\begingroup\renewcommand\colorMATH{\colorMATHB}\renewcommand\colorSYNTAX{\colorSYNTAXB}{{\color{\colorMATH}\ensuremath{\sv_{1}}}}\endgroup }}{{\begingroup\renewcommand\colorMATH{\colorMATHB}\renewcommand\colorSYNTAX{\colorSYNTAXB}{{\color{\colorMATH}\ensuremath{\sv_{2}}}}\endgroup }} = \addProduct{{\begingroup\renewcommand\colorMATH{\colorMATHB}\renewcommand\colorSYNTAX{\colorSYNTAXB}{{\color{\colorMATH}\ensuremath{\sv_{1}}}}\endgroup }}{{\begingroup\renewcommand\colorMATH{\colorMATHB}\renewcommand\colorSYNTAX{\colorSYNTAXB}{{\color{\colorMATH}\ensuremath{\sv_{2}}}}\endgroup }}}}}.
    \end{subproof}
  \item  {{\color{\colorMATH}\ensuremath{\Gamma ; {\begingroup\renewcommand\colorMATH{\colorMATHB}\renewcommand\colorSYNTAX{\colorSYNTAXB}{{\color{\colorMATH}\ensuremath{\Distance}}}\endgroup } \vdash  \fst\hspace*{0.33em}{\begingroup\renewcommand\colorMATH{\colorMATHB}\renewcommand\colorSYNTAX{\colorSYNTAXB}{{\color{\colorMATH}\ensuremath{\se'}}}\endgroup } \mathrel{:} \tau _{1} \mathrel{;} {\begingroup\renewcommand\colorMATH{\colorMATHB}\renewcommand\colorSYNTAX{\colorSYNTAXB}{{\color{\colorMATH}\ensuremath{\sS''}}}\endgroup } + {\begingroup\renewcommand\colorMATH{\colorMATHB}\renewcommand\colorSYNTAX{\colorSYNTAXB}{{\color{\colorMATH}\ensuremath{\sS_{1}}}}\endgroup }}}}
    \begin{subproof} 
      By induction hypothesis on {{\color{\colorMATH}\ensuremath{\Gamma  \vdash  {\begingroup\renewcommand\colorMATH{\colorMATHB}\renewcommand\colorSYNTAX{\colorSYNTAXB}{{\color{\colorMATH}\ensuremath{\se'}}}\endgroup } \mathrel{:} \tau _{1} \mathrel{^{{\begingroup\renewcommand\colorMATH{\colorMATHB}\renewcommand\colorSYNTAX{\colorSYNTAXB}{{\color{\colorMATH}\ensuremath{\sS_{1}}}}\endgroup }}\&^{{\begingroup\renewcommand\colorMATH{\colorMATHB}\renewcommand\colorSYNTAX{\colorSYNTAXB}{{\color{\colorMATH}\ensuremath{\sS_{2}}}}\endgroup }}} \tau _{2} \mathrel{;} {\begingroup\renewcommand\colorMATH{\colorMATHB}\renewcommand\colorSYNTAX{\colorSYNTAXB}{{\color{\colorMATH}\ensuremath{\sS''}}}\endgroup }}}},
      {{\color{\colorMATH}\ensuremath{\gamma _{1},\gamma \vdash {\begingroup\renewcommand\colorMATH{\colorMATHB}\renewcommand\colorSYNTAX{\colorSYNTAXB}{{\color{\colorMATH}\ensuremath{\se'}}}\endgroup } \Downarrow ^{*} \addProduct{{\begingroup\renewcommand\colorMATH{\colorMATHB}\renewcommand\colorSYNTAX{\colorSYNTAXB}{{\color{\colorMATH}\ensuremath{\sv_{1}}}}\endgroup }}{{\begingroup\renewcommand\colorMATH{\colorMATHB}\renewcommand\colorSYNTAX{\colorSYNTAXB}{{\color{\colorMATH}\ensuremath{\sv_{2}}}}\endgroup }}}}}, {{\color{\colorMATH}\ensuremath{\gamma _{2},\gamma \vdash {\begingroup\renewcommand\colorMATH{\colorMATHB}\renewcommand\colorSYNTAX{\colorSYNTAXB}{{\color{\colorMATH}\ensuremath{\se'}}}\endgroup } \Downarrow ^{*} \addProduct{{\begingroup\renewcommand\colorMATH{\colorMATHB}\renewcommand\colorSYNTAX{\colorSYNTAXB}{{\color{\colorMATH}\ensuremath{\sv_{1}}}}\endgroup }}{{\begingroup\renewcommand\colorMATH{\colorMATHB}\renewcommand\colorSYNTAX{\colorSYNTAXB}{{\color{\colorMATH}\ensuremath{\sv_{2}}}}\endgroup }}}}}.
      The result holds as {{\color{\colorMATH}\ensuremath{\gamma _{1},\gamma \vdash {\begingroup\renewcommand\colorMATH{\colorMATHB}\renewcommand\colorSYNTAX{\colorSYNTAXB}{{\color{\colorMATH}\ensuremath{\se'}}}\endgroup } \Downarrow ^{*} {\begingroup\renewcommand\colorMATH{\colorMATHB}\renewcommand\colorSYNTAX{\colorSYNTAXB}{{\color{\colorMATH}\ensuremath{\sv_{1}}}}\endgroup }}}} and {{\color{\colorMATH}\ensuremath{\gamma _{2},\gamma \vdash {\begingroup\renewcommand\colorMATH{\colorMATHB}\renewcommand\colorSYNTAX{\colorSYNTAXB}{{\color{\colorMATH}\ensuremath{\se'}}}\endgroup } \Downarrow ^{*} {\begingroup\renewcommand\colorMATH{\colorMATHB}\renewcommand\colorSYNTAX{\colorSYNTAXB}{{\color{\colorMATH}\ensuremath{\sv_{1}}}}\endgroup }}}}.
    \end{subproof}
  \item  {{\color{\colorMATH}\ensuremath{\Gamma ; {\begingroup\renewcommand\colorMATH{\colorMATHB}\renewcommand\colorSYNTAX{\colorSYNTAXB}{{\color{\colorMATH}\ensuremath{\Distance}}}\endgroup } \vdash  \snd\hspace*{0.33em}{\begingroup\renewcommand\colorMATH{\colorMATHB}\renewcommand\colorSYNTAX{\colorSYNTAXB}{{\color{\colorMATH}\ensuremath{\se'}}}\endgroup } \mathrel{:} \tau _{1} \mathrel{;} {\begingroup\renewcommand\colorMATH{\colorMATHB}\renewcommand\colorSYNTAX{\colorSYNTAXB}{{\color{\colorMATH}\ensuremath{\sS''}}}\endgroup } + {\begingroup\renewcommand\colorMATH{\colorMATHB}\renewcommand\colorSYNTAX{\colorSYNTAXB}{{\color{\colorMATH}\ensuremath{\sS_{2}}}}\endgroup }}}}
    \begin{subproof} 
      By induction hypothesis on {{\color{\colorMATH}\ensuremath{\Gamma  \vdash  {\begingroup\renewcommand\colorMATH{\colorMATHB}\renewcommand\colorSYNTAX{\colorSYNTAXB}{{\color{\colorMATH}\ensuremath{\se'}}}\endgroup } \mathrel{:} \tau _{1} \mathrel{^{{\begingroup\renewcommand\colorMATH{\colorMATHB}\renewcommand\colorSYNTAX{\colorSYNTAXB}{{\color{\colorMATH}\ensuremath{\sS_{1}}}}\endgroup }}\&^{{\begingroup\renewcommand\colorMATH{\colorMATHB}\renewcommand\colorSYNTAX{\colorSYNTAXB}{{\color{\colorMATH}\ensuremath{\sS_{2}}}}\endgroup }}} \tau _{2} \mathrel{;} {\begingroup\renewcommand\colorMATH{\colorMATHB}\renewcommand\colorSYNTAX{\colorSYNTAXB}{{\color{\colorMATH}\ensuremath{\sS''}}}\endgroup }}}},
      {{\color{\colorMATH}\ensuremath{\gamma _{1},\gamma \vdash {\begingroup\renewcommand\colorMATH{\colorMATHB}\renewcommand\colorSYNTAX{\colorSYNTAXB}{{\color{\colorMATH}\ensuremath{\se'}}}\endgroup } \Downarrow ^{*} \addProduct{{\begingroup\renewcommand\colorMATH{\colorMATHB}\renewcommand\colorSYNTAX{\colorSYNTAXB}{{\color{\colorMATH}\ensuremath{\sv_{1}}}}\endgroup }}{{\begingroup\renewcommand\colorMATH{\colorMATHB}\renewcommand\colorSYNTAX{\colorSYNTAXB}{{\color{\colorMATH}\ensuremath{\sv_{2}}}}\endgroup }}}}}, {{\color{\colorMATH}\ensuremath{\gamma _{2},\gamma \vdash {\begingroup\renewcommand\colorMATH{\colorMATHB}\renewcommand\colorSYNTAX{\colorSYNTAXB}{{\color{\colorMATH}\ensuremath{\se'}}}\endgroup } \Downarrow ^{*} \addProduct{{\begingroup\renewcommand\colorMATH{\colorMATHB}\renewcommand\colorSYNTAX{\colorSYNTAXB}{{\color{\colorMATH}\ensuremath{\sv_{1}}}}\endgroup }}{{\begingroup\renewcommand\colorMATH{\colorMATHB}\renewcommand\colorSYNTAX{\colorSYNTAXB}{{\color{\colorMATH}\ensuremath{\sv_{2}}}}\endgroup }}}}}.
      The result holds as {{\color{\colorMATH}\ensuremath{\gamma _{1},\gamma \vdash {\begingroup\renewcommand\colorMATH{\colorMATHB}\renewcommand\colorSYNTAX{\colorSYNTAXB}{{\color{\colorMATH}\ensuremath{\se'}}}\endgroup } \Downarrow ^{*} {\begingroup\renewcommand\colorMATH{\colorMATHB}\renewcommand\colorSYNTAX{\colorSYNTAXB}{{\color{\colorMATH}\ensuremath{\sv_{2}}}}\endgroup }}}} and {{\color{\colorMATH}\ensuremath{\gamma _{2},\gamma \vdash {\begingroup\renewcommand\colorMATH{\colorMATHB}\renewcommand\colorSYNTAX{\colorSYNTAXB}{{\color{\colorMATH}\ensuremath{\se'}}}\endgroup } \Downarrow ^{*} {\begingroup\renewcommand\colorMATH{\colorMATHB}\renewcommand\colorSYNTAX{\colorSYNTAXB}{{\color{\colorMATH}\ensuremath{\sv_{2}}}}\endgroup }}}}.
    \end{subproof}
  \item  {{\color{\colorMATH}\ensuremath{\Gamma ; {\begingroup\renewcommand\colorMATH{\colorMATHB}\renewcommand\colorSYNTAX{\colorSYNTAXB}{{\color{\colorMATH}\ensuremath{\Distance}}}\endgroup } \vdash  \langle {\begingroup\renewcommand\colorMATH{\colorMATHB}\renewcommand\colorSYNTAX{\colorSYNTAXB}{{\color{\colorMATH}\ensuremath{\se_{1}}}}\endgroup },{\begingroup\renewcommand\colorMATH{\colorMATHB}\renewcommand\colorSYNTAX{\colorSYNTAXB}{{\color{\colorMATH}\ensuremath{\se_{2}}}}\endgroup }\rangle  \mathrel{:} \tau _{1} \mathrel{^{{\begingroup\renewcommand\colorMATH{\colorMATHB}\renewcommand\colorSYNTAX{\colorSYNTAXB}{{\color{\colorMATH}\ensuremath{\sS_{1}}}}\endgroup }}\otimes ^{{\begingroup\renewcommand\colorMATH{\colorMATHB}\renewcommand\colorSYNTAX{\colorSYNTAXB}{{\color{\colorMATH}\ensuremath{\sS_{2}}}}\endgroup }}} \tau _{2} \mathrel{;} \varnothing }}}
    \begin{subproof} 
      By induction hypotheses on {{\color{\colorMATH}\ensuremath{\Gamma  \vdash  {\begingroup\renewcommand\colorMATH{\colorMATHB}\renewcommand\colorSYNTAX{\colorSYNTAXB}{{\color{\colorMATH}\ensuremath{\se_{1}}}}\endgroup } \mathrel{:} \tau _{1} \mathrel{;} {\begingroup\renewcommand\colorMATH{\colorMATHB}\renewcommand\colorSYNTAX{\colorSYNTAXB}{{\color{\colorMATH}\ensuremath{\sS_{1}}}}\endgroup }}}} and {{\color{\colorMATH}\ensuremath{\Gamma  \vdash  {\begingroup\renewcommand\colorMATH{\colorMATHB}\renewcommand\colorSYNTAX{\colorSYNTAXB}{{\color{\colorMATH}\ensuremath{\se_{2}}}}\endgroup } \mathrel{:} \tau _{2} \mathrel{;} {\begingroup\renewcommand\colorMATH{\colorMATHB}\renewcommand\colorSYNTAX{\colorSYNTAXB}{{\color{\colorMATH}\ensuremath{\sS_{2}}}}\endgroup }}}},
      {{\color{\colorMATH}\ensuremath{\gamma _{1},\gamma \vdash {\begingroup\renewcommand\colorMATH{\colorMATHB}\renewcommand\colorSYNTAX{\colorSYNTAXB}{{\color{\colorMATH}\ensuremath{\se_{1}}}}\endgroup } \Downarrow ^{*} {\begingroup\renewcommand\colorMATH{\colorMATHB}\renewcommand\colorSYNTAX{\colorSYNTAXB}{{\color{\colorMATH}\ensuremath{\sv_{1}}}}\endgroup }}}}, {{\color{\colorMATH}\ensuremath{\gamma _{2},\gamma \vdash {\begingroup\renewcommand\colorMATH{\colorMATHB}\renewcommand\colorSYNTAX{\colorSYNTAXB}{{\color{\colorMATH}\ensuremath{\se_{1}}}}\endgroup } \Downarrow ^{*} {\begingroup\renewcommand\colorMATH{\colorMATHB}\renewcommand\colorSYNTAX{\colorSYNTAXB}{{\color{\colorMATH}\ensuremath{\sv_{1}}}}\endgroup }}}}, and {{\color{\colorMATH}\ensuremath{\gamma _{1},\gamma \vdash {\begingroup\renewcommand\colorMATH{\colorMATHB}\renewcommand\colorSYNTAX{\colorSYNTAXB}{{\color{\colorMATH}\ensuremath{\se_{2}}}}\endgroup } \Downarrow ^{*} {\begingroup\renewcommand\colorMATH{\colorMATHB}\renewcommand\colorSYNTAX{\colorSYNTAXB}{{\color{\colorMATH}\ensuremath{\sv_{2}}}}\endgroup }}}}, {{\color{\colorMATH}\ensuremath{\gamma _{2},\gamma \vdash {\begingroup\renewcommand\colorMATH{\colorMATHB}\renewcommand\colorSYNTAX{\colorSYNTAXB}{{\color{\colorMATH}\ensuremath{\se_{2}}}}\endgroup } \Downarrow ^{*} {\begingroup\renewcommand\colorMATH{\colorMATHB}\renewcommand\colorSYNTAX{\colorSYNTAXB}{{\color{\colorMATH}\ensuremath{\sv_{2}}}}\endgroup }}}}.
      The result holds as {{\color{\colorMATH}\ensuremath{\langle {\begingroup\renewcommand\colorMATH{\colorMATHB}\renewcommand\colorSYNTAX{\colorSYNTAXB}{{\color{\colorMATH}\ensuremath{\sv_{1}}}}\endgroup },{\begingroup\renewcommand\colorMATH{\colorMATHB}\renewcommand\colorSYNTAX{\colorSYNTAXB}{{\color{\colorMATH}\ensuremath{\sv_{2}}}}\endgroup }\rangle  = \langle {\begingroup\renewcommand\colorMATH{\colorMATHB}\renewcommand\colorSYNTAX{\colorSYNTAXB}{{\color{\colorMATH}\ensuremath{\sv_{1}}}}\endgroup },{\begingroup\renewcommand\colorMATH{\colorMATHB}\renewcommand\colorSYNTAX{\colorSYNTAXB}{{\color{\colorMATH}\ensuremath{\sv_{2}}}}\endgroup }\rangle }}}.
    \end{subproof}
  \item  {{\color{\colorMATH}\ensuremath{\Gamma ; {\begingroup\renewcommand\colorMATH{\colorMATHB}\renewcommand\colorSYNTAX{\colorSYNTAXB}{{\color{\colorMATH}\ensuremath{\Distance}}}\endgroup } \vdash  \tlet\hspace*{0.33em}x_{1},x_{2}={\begingroup\renewcommand\colorMATH{\colorMATHB}\renewcommand\colorSYNTAX{\colorSYNTAXB}{{\color{\colorMATH}\ensuremath{\se_{1}}}}\endgroup }\hspace*{0.33em}\tin\hspace*{0.33em}{\begingroup\renewcommand\colorMATH{\colorMATHB}\renewcommand\colorSYNTAX{\colorSYNTAXB}{{\color{\colorMATH}\ensuremath{\se_{2}}}}\endgroup } \mathrel{:} [{\begingroup\renewcommand\colorMATH{\colorMATHB}\renewcommand\colorSYNTAX{\colorSYNTAXB}{{\color{\colorMATH}\ensuremath{\sS_{1}}}}\endgroup }+{\begingroup\renewcommand\colorMATH{\colorMATHB}\renewcommand\colorSYNTAX{\colorSYNTAXB}{{\color{\colorMATH}\ensuremath{\sS_{1 1}}}}\endgroup }/x_{1}][{\begingroup\renewcommand\colorMATH{\colorMATHB}\renewcommand\colorSYNTAX{\colorSYNTAXB}{{\color{\colorMATH}\ensuremath{\sS_{1}}}}\endgroup }+{\begingroup\renewcommand\colorMATH{\colorMATHB}\renewcommand\colorSYNTAX{\colorSYNTAXB}{{\color{\colorMATH}\ensuremath{\sS_{1 2}}}}\endgroup }/x_{2}]\tau _{2} \mathrel{;} {\begingroup\renewcommand\colorMATH{\colorMATHB}\renewcommand\colorSYNTAX{\colorSYNTAXB}{{\color{\colorMATH}\ensuremath{\sss_{1}}}}\endgroup }({\begingroup\renewcommand\colorMATH{\colorMATHB}\renewcommand\colorSYNTAX{\colorSYNTAXB}{{\color{\colorMATH}\ensuremath{\sS_{1 1}}}}\endgroup }+{\begingroup\renewcommand\colorMATH{\colorMATHB}\renewcommand\colorSYNTAX{\colorSYNTAXB}{{\color{\colorMATH}\ensuremath{\sS_{1}}}}\endgroup })+ {\begingroup\renewcommand\colorMATH{\colorMATHB}\renewcommand\colorSYNTAX{\colorSYNTAXB}{{\color{\colorMATH}\ensuremath{\sss_{2}}}}\endgroup }({\begingroup\renewcommand\colorMATH{\colorMATHB}\renewcommand\colorSYNTAX{\colorSYNTAXB}{{\color{\colorMATH}\ensuremath{\sS_{1 2}}}}\endgroup }+{\begingroup\renewcommand\colorMATH{\colorMATHB}\renewcommand\colorSYNTAX{\colorSYNTAXB}{{\color{\colorMATH}\ensuremath{\sS_{1}}}}\endgroup }) + {\begingroup\renewcommand\colorMATH{\colorMATHB}\renewcommand\colorSYNTAX{\colorSYNTAXB}{{\color{\colorMATH}\ensuremath{\sS_{2}}}}\endgroup }}}}
    \begin{subproof} 
      By induction hypothesis on {{\color{\colorMATH}\ensuremath{\Gamma ; {\begingroup\renewcommand\colorMATH{\colorMATHB}\renewcommand\colorSYNTAX{\colorSYNTAXB}{{\color{\colorMATH}\ensuremath{\Distance}}}\endgroup } \vdash  {\begingroup\renewcommand\colorMATH{\colorMATHB}\renewcommand\colorSYNTAX{\colorSYNTAXB}{{\color{\colorMATH}\ensuremath{\se_{1}}}}\endgroup } \mathrel{:} \tau _{1 1} \mathrel{^{{\begingroup\renewcommand\colorMATH{\colorMATHB}\renewcommand\colorSYNTAX{\colorSYNTAXB}{{\color{\colorMATH}\ensuremath{\sS_{1 1}}}}\endgroup }}\otimes ^{{\begingroup\renewcommand\colorMATH{\colorMATHB}\renewcommand\colorSYNTAX{\colorSYNTAXB}{{\color{\colorMATH}\ensuremath{\sS_{1 2}}}}\endgroup }}} \tau _{1 2} \mathrel{;} {\begingroup\renewcommand\colorMATH{\colorMATHB}\renewcommand\colorSYNTAX{\colorSYNTAXB}{{\color{\colorMATH}\ensuremath{\sS_{1}}}}\endgroup }}}}, we know that\\
      {{\color{\colorMATH}\ensuremath{\gamma _{1},\gamma \vdash {\begingroup\renewcommand\colorMATH{\colorMATHB}\renewcommand\colorSYNTAX{\colorSYNTAXB}{{\color{\colorMATH}\ensuremath{\se_{1}}}}\endgroup } \Downarrow ^{*} \langle {\begingroup\renewcommand\colorMATH{\colorMATHB}\renewcommand\colorSYNTAX{\colorSYNTAXB}{{\color{\colorMATH}\ensuremath{\sv_{1}}}}\endgroup }, {\begingroup\renewcommand\colorMATH{\colorMATHB}\renewcommand\colorSYNTAX{\colorSYNTAXB}{{\color{\colorMATH}\ensuremath{\sv_{2}}}}\endgroup }\rangle }}}, {{\color{\colorMATH}\ensuremath{\gamma _{2},\gamma \vdash {\begingroup\renewcommand\colorMATH{\colorMATHB}\renewcommand\colorSYNTAX{\colorSYNTAXB}{{\color{\colorMATH}\ensuremath{\se_{1}}}}\endgroup } \Downarrow ^{*} \langle {\begingroup\renewcommand\colorMATH{\colorMATHB}\renewcommand\colorSYNTAX{\colorSYNTAXB}{{\color{\colorMATH}\ensuremath{\sv_{1}}}}\endgroup }, {\begingroup\renewcommand\colorMATH{\colorMATHB}\renewcommand\colorSYNTAX{\colorSYNTAXB}{{\color{\colorMATH}\ensuremath{\sv_{2}}}}\endgroup }\rangle }}}.
      Then by induction hypothesis on {{\color{\colorMATH}\ensuremath{\Gamma ,x:\tau _{1 1}, x:\tau _{1 2}; {\begingroup\renewcommand\colorMATH{\colorMATHB}\renewcommand\colorSYNTAX{\colorSYNTAXB}{{\color{\colorMATH}\ensuremath{\Distance}}}\endgroup } + ({\begingroup\renewcommand\colorMATH{\colorMATHB}\renewcommand\colorSYNTAX{\colorSYNTAXB}{{\color{\colorMATH}\ensuremath{\Distance}}}\endgroup }\mathord{\cdotp }({\begingroup\renewcommand\colorMATH{\colorMATHB}\renewcommand\colorSYNTAX{\colorSYNTAXB}{{\color{\colorMATH}\ensuremath{\sS_{1}}}}\endgroup }+{\begingroup\renewcommand\colorMATH{\colorMATHB}\renewcommand\colorSYNTAX{\colorSYNTAXB}{{\color{\colorMATH}\ensuremath{\sS_{1 1}}}}\endgroup }))x_{1} + ({\begingroup\renewcommand\colorMATH{\colorMATHB}\renewcommand\colorSYNTAX{\colorSYNTAXB}{{\color{\colorMATH}\ensuremath{\Distance}}}\endgroup }\mathord{\cdotp }({\begingroup\renewcommand\colorMATH{\colorMATHB}\renewcommand\colorSYNTAX{\colorSYNTAXB}{{\color{\colorMATH}\ensuremath{\sS_{1}}}}\endgroup }+{\begingroup\renewcommand\colorMATH{\colorMATHB}\renewcommand\colorSYNTAX{\colorSYNTAXB}{{\color{\colorMATH}\ensuremath{\sS_{1 2}}}}\endgroup }))x_{2} \vdash  {\begingroup\renewcommand\colorMATH{\colorMATHB}\renewcommand\colorSYNTAX{\colorSYNTAXB}{{\color{\colorMATH}\ensuremath{\se_{2}}}}\endgroup } \mathrel{:} \tau _{2} \mathrel{;} {\begingroup\renewcommand\colorMATH{\colorMATHB}\renewcommand\colorSYNTAX{\colorSYNTAXB}{{\color{\colorMATH}\ensuremath{\sS_{2}}}}\endgroup } + {\begingroup\renewcommand\colorMATH{\colorMATHB}\renewcommand\colorSYNTAX{\colorSYNTAXB}{{\color{\colorMATH}\ensuremath{\sss_{1}}}}\endgroup }x_{1} + {\begingroup\renewcommand\colorMATH{\colorMATHB}\renewcommand\colorSYNTAX{\colorSYNTAXB}{{\color{\colorMATH}\ensuremath{\sss_{2}}}}\endgroup }x_{2}}}}, we know that,
      {{\color{\colorMATH}\ensuremath{\gamma _{1},\gamma [x_{1}\mapsto {\begingroup\renewcommand\colorMATH{\colorMATHB}\renewcommand\colorSYNTAX{\colorSYNTAXB}{{\color{\colorMATH}\ensuremath{\sv_{1}}}}\endgroup },x_{2}\mapsto {\begingroup\renewcommand\colorMATH{\colorMATHB}\renewcommand\colorSYNTAX{\colorSYNTAXB}{{\color{\colorMATH}\ensuremath{\sv_{2}}}}\endgroup }]\vdash {\begingroup\renewcommand\colorMATH{\colorMATHB}\renewcommand\colorSYNTAX{\colorSYNTAXB}{{\color{\colorMATH}\ensuremath{\se_{2}}}}\endgroup } \Downarrow ^{*} {\begingroup\renewcommand\colorMATH{\colorMATHB}\renewcommand\colorSYNTAX{\colorSYNTAXB}{{\color{\colorMATH}\ensuremath{\sv}}}\endgroup }}}}, {{\color{\colorMATH}\ensuremath{\gamma _{2},\gamma [x_{1}\mapsto {\begingroup\renewcommand\colorMATH{\colorMATHB}\renewcommand\colorSYNTAX{\colorSYNTAXB}{{\color{\colorMATH}\ensuremath{\sv_{1}}}}\endgroup },x_{2}\mapsto {\begingroup\renewcommand\colorMATH{\colorMATHB}\renewcommand\colorSYNTAX{\colorSYNTAXB}{{\color{\colorMATH}\ensuremath{\sv_{2}}}}\endgroup }]\vdash  {\begingroup\renewcommand\colorMATH{\colorMATHB}\renewcommand\colorSYNTAX{\colorSYNTAXB}{{\color{\colorMATH}\ensuremath{\se_{2}}}}\endgroup } \Downarrow ^{*} {\begingroup\renewcommand\colorMATH{\colorMATHB}\renewcommand\colorSYNTAX{\colorSYNTAXB}{{\color{\colorMATH}\ensuremath{\sv}}}\endgroup }}}}, and the result holds,
    \end{subproof}
  \item  {{\color{\colorMATH}\ensuremath{\Gamma ; {\begingroup\renewcommand\colorMATH{\colorMATHB}\renewcommand\colorSYNTAX{\colorSYNTAXB}{{\color{\colorMATH}\ensuremath{\Distance}}}\endgroup } \vdash  ({\begingroup\renewcommand\colorMATH{\colorMATHB}\renewcommand\colorSYNTAX{\colorSYNTAXB}{{\color{\colorMATH}\ensuremath{\se'}}}\endgroup } \mathrel{:: } \tau ) \mathrel{:} \tau  \mathrel{;} {\begingroup\renewcommand\colorMATH{\colorMATHB}\renewcommand\colorSYNTAX{\colorSYNTAXB}{{\color{\colorMATH}\ensuremath{\sS}}}\endgroup }}}}
    \begin{subproof} 
      By induction hypothesis on {{\color{\colorMATH}\ensuremath{\Gamma  \vdash  {\begingroup\renewcommand\colorMATH{\colorMATHB}\renewcommand\colorSYNTAX{\colorSYNTAXB}{{\color{\colorMATH}\ensuremath{\se'}}}\endgroup } \mathrel{:} \tau ' \mathrel{;} {\begingroup\renewcommand\colorMATH{\colorMATHB}\renewcommand\colorSYNTAX{\colorSYNTAXB}{{\color{\colorMATH}\ensuremath{\sS}}}\endgroup }}}},
      {{\color{\colorMATH}\ensuremath{\gamma _{1},\gamma \vdash {\begingroup\renewcommand\colorMATH{\colorMATHB}\renewcommand\colorSYNTAX{\colorSYNTAXB}{{\color{\colorMATH}\ensuremath{\se'}}}\endgroup } \Downarrow ^{*} {\begingroup\renewcommand\colorMATH{\colorMATHB}\renewcommand\colorSYNTAX{\colorSYNTAXB}{{\color{\colorMATH}\ensuremath{\sv}}}\endgroup }}}}, {{\color{\colorMATH}\ensuremath{\gamma _{2},\gamma \vdash {\begingroup\renewcommand\colorMATH{\colorMATHB}\renewcommand\colorSYNTAX{\colorSYNTAXB}{{\color{\colorMATH}\ensuremath{\se'}}}\endgroup } \Downarrow ^{*} {\begingroup\renewcommand\colorMATH{\colorMATHB}\renewcommand\colorSYNTAX{\colorSYNTAXB}{{\color{\colorMATH}\ensuremath{\sv}}}\endgroup }}}}.
      The result holds as {{\color{\colorMATH}\ensuremath{\gamma _{1},\gamma \vdash ({\begingroup\renewcommand\colorMATH{\colorMATHB}\renewcommand\colorSYNTAX{\colorSYNTAXB}{{\color{\colorMATH}\ensuremath{\se'}}}\endgroup } \mathrel{:: } \tau ) \Downarrow ^{*} {\begingroup\renewcommand\colorMATH{\colorMATHB}\renewcommand\colorSYNTAX{\colorSYNTAXB}{{\color{\colorMATH}\ensuremath{\sv}}}\endgroup }}}} and {{\color{\colorMATH}\ensuremath{\gamma _{2},\gamma \vdash ({\begingroup\renewcommand\colorMATH{\colorMATHB}\renewcommand\colorSYNTAX{\colorSYNTAXB}{{\color{\colorMATH}\ensuremath{\se'}}}\endgroup } \mathrel{:: } \tau ) \Downarrow ^{*} {\begingroup\renewcommand\colorMATH{\colorMATHB}\renewcommand\colorSYNTAX{\colorSYNTAXB}{{\color{\colorMATH}\ensuremath{\sv_{1}}}}\endgroup }}}}.
    \end{subproof}
  \item  {{\color{\colorMATH}\ensuremath{\Gamma ; {\begingroup\renewcommand\colorMATH{\colorMATHB}\renewcommand\colorSYNTAX{\colorSYNTAXB}{{\color{\colorMATH}\ensuremath{\Distance}}}\endgroup } \vdash  {\begingroup\renewcommand\colorMATH{\colorMATHC}\renewcommand\colorSYNTAX{\colorSYNTAXC}{{\color{\colorMATH}\ensuremath{\plambda}}}\endgroup } (x\mathrel{:}\tau _{1}\mathord{\cdotp }{\begingroup\renewcommand\colorMATH{\colorMATHB}\renewcommand\colorSYNTAX{\colorSYNTAXB}{{\color{\colorMATH}\ensuremath{\distance'}}}\endgroup }).\hspace*{0.33em}{\begingroup\renewcommand\colorMATH{\colorMATHC}\renewcommand\colorSYNTAX{\colorSYNTAXC}{{\color{\colorMATH}\ensuremath{\pe'}}}\endgroup } \mathrel{:} (x\mathrel{:}\tau _{1}\mathord{\cdotp }{\begingroup\renewcommand\colorMATH{\colorMATHB}\renewcommand\colorSYNTAX{\colorSYNTAXB}{{\color{\colorMATH}\ensuremath{\distance'}}}\endgroup }) \xrightarrowP {{\begingroup\renewcommand\colorMATH{\colorMATHC}\renewcommand\colorSYNTAX{\colorSYNTAXC}{{\color{\colorMATH}\ensuremath{\pS''}}}\endgroup }} \tau _{2} \mathrel{;} {\begingroup\renewcommand\colorMATH{\colorMATHB}\renewcommand\colorSYNTAX{\colorSYNTAXB}{{\color{\colorMATH}\ensuremath{\varnothing }}}\endgroup }}}} 
    \begin{subproof} 
      We know that {{\color{\colorMATH}\ensuremath{{\text{FV}}({\begingroup\renewcommand\colorMATH{\colorMATHB}\renewcommand\colorSYNTAX{\colorSYNTAXB}{{\color{\colorMATH}\ensuremath{\se}}}\endgroup }) \subseteq  dom(\gamma )}}}, therefore 
      {{\color{\colorMATH}\ensuremath{\gamma _{1},\gamma \vdash {\begingroup\renewcommand\colorMATH{\colorMATHB}\renewcommand\colorSYNTAX{\colorSYNTAXB}{{\color{\colorMATH}\ensuremath{\slambda}}}\endgroup } (x\mathrel{:}\tau _{1}\mathord{\cdotp }{\begingroup\renewcommand\colorMATH{\colorMATHB}\renewcommand\colorSYNTAX{\colorSYNTAXB}{{\color{\colorMATH}\ensuremath{\distance'}}}\endgroup }).\hspace*{0.33em}{\begingroup\renewcommand\colorMATH{\colorMATHC}\renewcommand\colorSYNTAX{\colorSYNTAXC}{{\color{\colorMATH}\ensuremath{\pe'}}}\endgroup } \Downarrow ^{1} \langle {\begingroup\renewcommand\colorMATH{\colorMATHB}\renewcommand\colorSYNTAX{\colorSYNTAXB}{{\color{\colorMATH}\ensuremath{\slambda}}}\endgroup } (x\mathrel{:}\tau _{1}\mathord{\cdotp }{\begingroup\renewcommand\colorMATH{\colorMATHB}\renewcommand\colorSYNTAX{\colorSYNTAXB}{{\color{\colorMATH}\ensuremath{\distance'}}}\endgroup }).\hspace*{0.33em}{\begingroup\renewcommand\colorMATH{\colorMATHC}\renewcommand\colorSYNTAX{\colorSYNTAXC}{{\color{\colorMATH}\ensuremath{\pe'}}}\endgroup }, \gamma '\rangle }}} and {{\color{\colorMATH}\ensuremath{\gamma _{2},\gamma \vdash {\begingroup\renewcommand\colorMATH{\colorMATHB}\renewcommand\colorSYNTAX{\colorSYNTAXB}{{\color{\colorMATH}\ensuremath{\slambda}}}\endgroup } (x\mathrel{:}\tau _{1}\mathord{\cdotp }{\begingroup\renewcommand\colorMATH{\colorMATHB}\renewcommand\colorSYNTAX{\colorSYNTAXB}{{\color{\colorMATH}\ensuremath{\sss_{1}}}}\endgroup }).\hspace*{0.33em}{\begingroup\renewcommand\colorMATH{\colorMATHC}\renewcommand\colorSYNTAX{\colorSYNTAXC}{{\color{\colorMATH}\ensuremath{\pe'}}}\endgroup } \Downarrow ^{1} \langle {\begingroup\renewcommand\colorMATH{\colorMATHB}\renewcommand\colorSYNTAX{\colorSYNTAXB}{{\color{\colorMATH}\ensuremath{\slambda}}}\endgroup } (x\mathrel{:}\tau _{1}\mathord{\cdotp }{\begingroup\renewcommand\colorMATH{\colorMATHB}\renewcommand\colorSYNTAX{\colorSYNTAXB}{{\color{\colorMATH}\ensuremath{\sss_{1}}}}\endgroup }).\hspace*{0.33em}{\begingroup\renewcommand\colorMATH{\colorMATHC}\renewcommand\colorSYNTAX{\colorSYNTAXC}{{\color{\colorMATH}\ensuremath{\pe'}}}\endgroup }, \gamma '\rangle }}}, for {{\color{\colorMATH}\ensuremath{\gamma ' \subseteq  \gamma }}} and the result holds.
    \end{subproof}
  \end{enumerate}
\end{proof}

\newcommand{\kg}{k}
\newcommand{\pj}[1][]{*}
\newcommand{\pthen}{then~}
\newcommand{\pand}{and~}
%To prove the fundamental property we need to prove it along Lemma~\ref{alm:fpaux}.
\begin{restatable}[Metric Preservation]{theorem}{FundamentalProperty}\
  \label{alm:fp}
  \begin{enumerate}\item  {{\color{\colorMATH}\ensuremath{ \Gamma ,{\begingroup\renewcommand\colorMATH{\colorMATHB}\renewcommand\colorSYNTAX{\colorSYNTAXB}{{\color{\colorMATH}\ensuremath{\Distance}}}\endgroup } \vdash  {\begingroup\renewcommand\colorMATH{\colorMATHB}\renewcommand\colorSYNTAX{\colorSYNTAXB}{{\color{\colorMATH}\ensuremath{\se}}}\endgroup } \mathrel{:} \tau  \mathrel{;} {\begingroup\renewcommand\colorMATH{\colorMATHB}\renewcommand\colorSYNTAX{\colorSYNTAXB}{{\color{\colorMATH}\ensuremath{\sS}}}\endgroup } \Rightarrow  \forall  k \geq  0, \forall {\begingroup\renewcommand\colorMATH{\colorMATHB}\renewcommand\colorSYNTAX{\colorSYNTAXB}{{\color{\colorMATH}\ensuremath{\Distance'}}}\endgroup } \sqsubseteq  {\begingroup\renewcommand\colorMATH{\colorMATHB}\renewcommand\colorSYNTAX{\colorSYNTAXB}{{\color{\colorMATH}\ensuremath{\Distance}}}\endgroup }, \forall (\gamma _{1},\gamma _{2}) \in  {\mathcal{G}}_{{\begingroup\renewcommand\colorMATH{\colorMATHB}\renewcommand\colorSYNTAX{\colorSYNTAXB}{{\color{\colorMATH}\ensuremath{\Distance'}}}\endgroup }}^{\kg}\llbracket \Gamma \rrbracket . (\gamma _{1}\vdash {\begingroup\renewcommand\colorMATH{\colorMATHB}\renewcommand\colorSYNTAX{\colorSYNTAXB}{{\color{\colorMATH}\ensuremath{\se}}}\endgroup },\gamma _{2}\vdash {\begingroup\renewcommand\colorMATH{\colorMATHB}\renewcommand\colorSYNTAX{\colorSYNTAXB}{{\color{\colorMATH}\ensuremath{\se}}}\endgroup }) \in  {\mathcal{E}}_{{\begingroup\renewcommand\colorMATH{\colorMATHB}\renewcommand\colorSYNTAX{\colorSYNTAXB}{{\color{\colorMATH}\ensuremath{\Distance'}}}\endgroup }\mathord{\cdotp }{\begingroup\renewcommand\colorMATH{\colorMATHB}\renewcommand\colorSYNTAX{\colorSYNTAXB}{{\color{\colorMATH}\ensuremath{\sS}}}\endgroup }}^{k}\llbracket {\begingroup\renewcommand\colorMATH{\colorMATHB}\renewcommand\colorSYNTAX{\colorSYNTAXB}{{\color{\colorMATH}\ensuremath{\Distance'}}}\endgroup }(\tau )\rrbracket }}}
  \item  {{\color{\colorMATH}\ensuremath{ \Gamma ,{\begingroup\renewcommand\colorMATH{\colorMATHB}\renewcommand\colorSYNTAX{\colorSYNTAXB}{{\color{\colorMATH}\ensuremath{\Distance}}}\endgroup } \vdash  {\begingroup\renewcommand\colorMATH{\colorMATHC}\renewcommand\colorSYNTAX{\colorSYNTAXC}{{\color{\colorMATH}\ensuremath{\pe}}}\endgroup } \mathrel{:} \tau  \mathrel{;} {\begingroup\renewcommand\colorMATH{\colorMATHC}\renewcommand\colorSYNTAX{\colorSYNTAXC}{{\color{\colorMATH}\ensuremath{\pS}}}\endgroup } \Rightarrow  \forall  k \geq  0, \forall {\begingroup\renewcommand\colorMATH{\colorMATHB}\renewcommand\colorSYNTAX{\colorSYNTAXB}{{\color{\colorMATH}\ensuremath{\Distance'}}}\endgroup } \sqsubseteq  {\begingroup\renewcommand\colorMATH{\colorMATHB}\renewcommand\colorSYNTAX{\colorSYNTAXB}{{\color{\colorMATH}\ensuremath{\Distance}}}\endgroup }, \forall (\gamma _{1},\gamma _{2}) \in  {\mathcal{G}}_{{\begingroup\renewcommand\colorMATH{\colorMATHB}\renewcommand\colorSYNTAX{\colorSYNTAXB}{{\color{\colorMATH}\ensuremath{\Distance'}}}\endgroup }}^{\kg}\llbracket \Gamma \rrbracket . (\gamma _{1}\vdash {\begingroup\renewcommand\colorMATH{\colorMATHC}\renewcommand\colorSYNTAX{\colorSYNTAXC}{{\color{\colorMATH}\ensuremath{\pe}}}\endgroup },\gamma _{2}\vdash {\begingroup\renewcommand\colorMATH{\colorMATHC}\renewcommand\colorSYNTAX{\colorSYNTAXC}{{\color{\colorMATH}\ensuremath{\pe}}}\endgroup }) \in  {\mathcal{E}}_{{\begingroup\renewcommand\colorMATH{\colorMATHB}\renewcommand\colorSYNTAX{\colorSYNTAXB}{{\color{\colorMATH}\ensuremath{\Distance'}}}\endgroup }{\begingroup\renewcommand\colorMATH{\colorMATHC}\renewcommand\colorSYNTAX{\colorSYNTAXC}{{\color{\colorMATH}\ensuremath{\bigcdot}}}\endgroup }{\begingroup\renewcommand\colorMATH{\colorMATHC}\renewcommand\colorSYNTAX{\colorSYNTAXC}{{\color{\colorMATH}\ensuremath{\pS}}}\endgroup }}^{k}\llbracket {\begingroup\renewcommand\colorMATH{\colorMATHB}\renewcommand\colorSYNTAX{\colorSYNTAXB}{{\color{\colorMATH}\ensuremath{\Distance'}}}\endgroup }(\tau )\rrbracket }}}
  %\item  Let {{\color{\colorMATH}\ensuremath{{\begingroup\renewcommand\colorMATH{\colorMATHB}\renewcommand\colorSYNTAX{\colorSYNTAXB}{{\color{\colorMATH}\ensuremath{\Distance}}}\endgroup }_\varnothing }}}, such that {{\color{\colorMATH}\ensuremath{{\begingroup\renewcommand\colorMATH{\colorMATHB}\renewcommand\colorSYNTAX{\colorSYNTAXB}{{\color{\colorMATH}\ensuremath{\Distance}}}\endgroup }_\varnothing (x) = 0}}}, for all {{\color{\colorMATH}\ensuremath{x \in  FV(\tau )}}}. If {{\color{\colorMATH}\ensuremath{({\begingroup\renewcommand\colorMATH{\colorMATHB}\renewcommand\colorSYNTAX{\colorSYNTAXB}{{\color{\colorMATH}\ensuremath{\sv_{1}}}}\endgroup }, {\begingroup\renewcommand\colorMATH{\colorMATHB}\renewcommand\colorSYNTAX{\colorSYNTAXB}{{\color{\colorMATH}\ensuremath{\sv_{2}}}}\endgroup }) \in  {\mathcal{V}}_{0}^{k}\llbracket {\begingroup\renewcommand\colorMATH{\colorMATHB}\renewcommand\colorSYNTAX{\colorSYNTAXB}{{\color{\colorMATH}\ensuremath{\Distance}}}\endgroup }_\varnothing (\tau )\rrbracket }}} then {{\color{\colorMATH}\ensuremath{{\begingroup\renewcommand\colorMATH{\colorMATHB}\renewcommand\colorSYNTAX{\colorSYNTAXB}{{\color{\colorMATH}\ensuremath{\sv_{1}}}}\endgroup } \approx ^{k} {\begingroup\renewcommand\colorMATH{\colorMATHB}\renewcommand\colorSYNTAX{\colorSYNTAXB}{{\color{\colorMATH}\ensuremath{\sv_{2}}}}\endgroup }}}}.
  \end{enumerate}

\end{restatable}
\begin{proof}
  We prove the two parts using induction on {{\color{\colorMATH}\ensuremath{k}}} first.\\
  We start by proving Part (1)
  by induction on {{\color{\colorMATH}\ensuremath{k}}} and the typing derivation. 
  \begin{enumerate}[ncases]\item  {{\color{\colorMATH}\ensuremath{\Gamma ; {\begingroup\renewcommand\colorMATH{\colorMATHB}\renewcommand\colorSYNTAX{\colorSYNTAXB}{{\color{\colorMATH}\ensuremath{\Distance}}}\endgroup } \vdash  {\begingroup\renewcommand\colorMATH{\colorMATHB}\renewcommand\colorSYNTAX{\colorSYNTAXB}{{\color{\colorMATH}\ensuremath{r}}}\endgroup } \mathrel{:} {\begingroup\renewcommand\colorMATH{\colorMATHA}\renewcommand\colorSYNTAX{\colorSYNTAXA}{{\color{\colorSYNTAX}\texttt{{\ensuremath{{\mathbb{R}}}}}}}\endgroup } \mathrel{;} \varnothing }}}  
    \begin{subproof} 
      We have to prove that {{\color{\colorMATH}\ensuremath{\forall k, \forall (\gamma _{1},\gamma _{2}) \in  {\mathcal{G}}_{{\begingroup\renewcommand\colorMATH{\colorMATHB}\renewcommand\colorSYNTAX{\colorSYNTAXB}{{\color{\colorMATH}\ensuremath{\Distance'}}}\endgroup }}^{\kg}\llbracket \Gamma \rrbracket , (\gamma _{1}\vdash {\begingroup\renewcommand\colorMATH{\colorMATHB}\renewcommand\colorSYNTAX{\colorSYNTAXB}{{\color{\colorMATH}\ensuremath{r}}}\endgroup },\gamma _{2}\vdash {\begingroup\renewcommand\colorMATH{\colorMATHB}\renewcommand\colorSYNTAX{\colorSYNTAXB}{{\color{\colorMATH}\ensuremath{r}}}\endgroup }) \in  {\mathcal{E}}_{{\begingroup\renewcommand\colorMATH{\colorMATHB}\renewcommand\colorSYNTAX{\colorSYNTAXB}{{\color{\colorMATH}\ensuremath{\Distance'}}}\endgroup }\mathord{\cdotp }\varnothing }\llbracket {\begingroup\renewcommand\colorMATH{\colorMATHB}\renewcommand\colorSYNTAX{\colorSYNTAXB}{{\color{\colorMATH}\ensuremath{\Distance'}}}\endgroup }({\begingroup\renewcommand\colorMATH{\colorMATHA}\renewcommand\colorSYNTAX{\colorSYNTAXA}{{\color{\colorSYNTAX}\texttt{{\ensuremath{{\mathbb{R}}}}}}}\endgroup })\rrbracket }}}, for {{\color{\colorMATH}\ensuremath{{\begingroup\renewcommand\colorMATH{\colorMATHB}\renewcommand\colorSYNTAX{\colorSYNTAXB}{{\color{\colorMATH}\ensuremath{\Distance'}}}\endgroup } \sqsubseteq  {\begingroup\renewcommand\colorMATH{\colorMATHB}\renewcommand\colorSYNTAX{\colorSYNTAXB}{{\color{\colorMATH}\ensuremath{\Distance}}}\endgroup }}}}.
      Notice that {{\color{\colorMATH}\ensuremath{{\begingroup\renewcommand\colorMATH{\colorMATHB}\renewcommand\colorSYNTAX{\colorSYNTAXB}{{\color{\colorMATH}\ensuremath{\Distance'}}}\endgroup }\mathord{\cdotp }\varnothing  = 0}}} and {{\color{\colorMATH}\ensuremath{\gamma _{1}({\begingroup\renewcommand\colorMATH{\colorMATHB}\renewcommand\colorSYNTAX{\colorSYNTAXB}{{\color{\colorMATH}\ensuremath{r}}}\endgroup }) = \gamma _{2}({\begingroup\renewcommand\colorMATH{\colorMATHB}\renewcommand\colorSYNTAX{\colorSYNTAXB}{{\color{\colorMATH}\ensuremath{r}}}\endgroup }) = {\begingroup\renewcommand\colorMATH{\colorMATHB}\renewcommand\colorSYNTAX{\colorSYNTAXB}{{\color{\colorMATH}\ensuremath{r}}}\endgroup }}}}.
      Then we have to prove that 
      {{\color{\colorMATH}\ensuremath{({\begingroup\renewcommand\colorMATH{\colorMATHB}\renewcommand\colorSYNTAX{\colorSYNTAXB}{{\color{\colorMATH}\ensuremath{r}}}\endgroup },{\begingroup\renewcommand\colorMATH{\colorMATHB}\renewcommand\colorSYNTAX{\colorSYNTAXB}{{\color{\colorMATH}\ensuremath{r}}}\endgroup }) \in  {\mathcal{V}}_{0}^{k}\llbracket {\begingroup\renewcommand\colorMATH{\colorMATHA}\renewcommand\colorSYNTAX{\colorSYNTAXA}{{\color{\colorSYNTAX}\texttt{{\ensuremath{{\mathbb{R}}}}}}}\endgroup }\rrbracket }}}, i.e.  {{\color{\colorMATH}\ensuremath{|{\begingroup\renewcommand\colorMATH{\colorMATHB}\renewcommand\colorSYNTAX{\colorSYNTAXB}{{\color{\colorMATH}\ensuremath{r}}}\endgroup }-{\begingroup\renewcommand\colorMATH{\colorMATHB}\renewcommand\colorSYNTAX{\colorSYNTAXB}{{\color{\colorMATH}\ensuremath{r}}}\endgroup }| \leq  0}}} which is direct.
    \end{subproof}
  \item  {{\color{\colorMATH}\ensuremath{\Gamma ; {\begingroup\renewcommand\colorMATH{\colorMATHB}\renewcommand\colorSYNTAX{\colorSYNTAXB}{{\color{\colorMATH}\ensuremath{\Distance}}}\endgroup } \vdash  {\begingroup\renewcommand\colorMATH{\colorMATHB}\renewcommand\colorSYNTAX{\colorSYNTAXB}{{\color{\colorMATH}\ensuremath{\se_{1}}}}\endgroup } + {\begingroup\renewcommand\colorMATH{\colorMATHB}\renewcommand\colorSYNTAX{\colorSYNTAXB}{{\color{\colorMATH}\ensuremath{\se_{2}}}}\endgroup } \mathrel{:} {\begingroup\renewcommand\colorMATH{\colorMATHA}\renewcommand\colorSYNTAX{\colorSYNTAXA}{{\color{\colorSYNTAX}\texttt{{\ensuremath{{\mathbb{R}}}}}}}\endgroup } \mathrel{;} {\begingroup\renewcommand\colorMATH{\colorMATHB}\renewcommand\colorSYNTAX{\colorSYNTAXB}{{\color{\colorMATH}\ensuremath{\sS_{1}}}}\endgroup } + {\begingroup\renewcommand\colorMATH{\colorMATHB}\renewcommand\colorSYNTAX{\colorSYNTAXB}{{\color{\colorMATH}\ensuremath{\sS_{2}}}}\endgroup }}}} 
    \begin{subproof} 
      We have to prove that {{\color{\colorMATH}\ensuremath{\forall k, \forall (\gamma _{1},\gamma _{2}) \in  {\mathcal{G}}_{{\begingroup\renewcommand\colorMATH{\colorMATHB}\renewcommand\colorSYNTAX{\colorSYNTAXB}{{\color{\colorMATH}\ensuremath{\Distance'}}}\endgroup }}^{\kg}\llbracket \Gamma \rrbracket , (\gamma _{1}\vdash {\begingroup\renewcommand\colorMATH{\colorMATHB}\renewcommand\colorSYNTAX{\colorSYNTAXB}{{\color{\colorMATH}\ensuremath{\se_{1}}}}\endgroup } + {\begingroup\renewcommand\colorMATH{\colorMATHB}\renewcommand\colorSYNTAX{\colorSYNTAXB}{{\color{\colorMATH}\ensuremath{\se_{2}}}}\endgroup },\gamma _{2}\vdash {\begingroup\renewcommand\colorMATH{\colorMATHB}\renewcommand\colorSYNTAX{\colorSYNTAXB}{{\color{\colorMATH}\ensuremath{\se_{1}}}}\endgroup } + {\begingroup\renewcommand\colorMATH{\colorMATHB}\renewcommand\colorSYNTAX{\colorSYNTAXB}{{\color{\colorMATH}\ensuremath{\se_{2}}}}\endgroup }) \in  {\mathcal{E}}^{k}_{{\begingroup\renewcommand\colorMATH{\colorMATHB}\renewcommand\colorSYNTAX{\colorSYNTAXB}{{\color{\colorMATH}\ensuremath{\Distance'}}}\endgroup }\mathord{\cdotp }({\begingroup\renewcommand\colorMATH{\colorMATHB}\renewcommand\colorSYNTAX{\colorSYNTAXB}{{\color{\colorMATH}\ensuremath{\sS_{1}}}}\endgroup } + {\begingroup\renewcommand\colorMATH{\colorMATHB}\renewcommand\colorSYNTAX{\colorSYNTAXB}{{\color{\colorMATH}\ensuremath{\sS_{2}}}}\endgroup })}\llbracket {\begingroup\renewcommand\colorMATH{\colorMATHB}\renewcommand\colorSYNTAX{\colorSYNTAXB}{{\color{\colorMATH}\ensuremath{\Distance'}}}\endgroup }({\begingroup\renewcommand\colorMATH{\colorMATHA}\renewcommand\colorSYNTAX{\colorSYNTAXA}{{\color{\colorSYNTAX}\texttt{{\ensuremath{{\mathbb{R}}}}}}}\endgroup })\rrbracket }}}, for {{\color{\colorMATH}\ensuremath{{\begingroup\renewcommand\colorMATH{\colorMATHB}\renewcommand\colorSYNTAX{\colorSYNTAXB}{{\color{\colorMATH}\ensuremath{\Distance'}}}\endgroup } \sqsubseteq  {\begingroup\renewcommand\colorMATH{\colorMATHB}\renewcommand\colorSYNTAX{\colorSYNTAXB}{{\color{\colorMATH}\ensuremath{\Distance}}}\endgroup }}}}.
      By induction hypotheses \\
      {{\color{\colorMATH}\ensuremath{\Gamma ; {\begingroup\renewcommand\colorMATH{\colorMATHB}\renewcommand\colorSYNTAX{\colorSYNTAXB}{{\color{\colorMATH}\ensuremath{\Distance}}}\endgroup } \vdash  {\begingroup\renewcommand\colorMATH{\colorMATHB}\renewcommand\colorSYNTAX{\colorSYNTAXB}{{\color{\colorMATH}\ensuremath{\se_{1}}}}\endgroup } \mathrel{:} {\begingroup\renewcommand\colorMATH{\colorMATHA}\renewcommand\colorSYNTAX{\colorSYNTAXA}{{\color{\colorSYNTAX}\texttt{{\ensuremath{{\mathbb{R}}}}}}}\endgroup } \mathrel{;} {\begingroup\renewcommand\colorMATH{\colorMATHB}\renewcommand\colorSYNTAX{\colorSYNTAXB}{{\color{\colorMATH}\ensuremath{\sS_{1}}}}\endgroup } \Rightarrow  (\gamma _{1} \vdash  {\begingroup\renewcommand\colorMATH{\colorMATHB}\renewcommand\colorSYNTAX{\colorSYNTAXB}{{\color{\colorMATH}\ensuremath{\se_{1}}}}\endgroup },\gamma _{2} \vdash  {\begingroup\renewcommand\colorMATH{\colorMATHB}\renewcommand\colorSYNTAX{\colorSYNTAXB}{{\color{\colorMATH}\ensuremath{\se_{1}}}}\endgroup }) \in  {\mathcal{E}}^{k}_{{\begingroup\renewcommand\colorMATH{\colorMATHB}\renewcommand\colorSYNTAX{\colorSYNTAXB}{{\color{\colorMATH}\ensuremath{\Distance'}}}\endgroup }\mathord{\cdotp }{\begingroup\renewcommand\colorMATH{\colorMATHB}\renewcommand\colorSYNTAX{\colorSYNTAXB}{{\color{\colorMATH}\ensuremath{\sS_{1}}}}\endgroup }}\llbracket {\begingroup\renewcommand\colorMATH{\colorMATHB}\renewcommand\colorSYNTAX{\colorSYNTAXB}{{\color{\colorMATH}\ensuremath{\Distance'}}}\endgroup }({\begingroup\renewcommand\colorMATH{\colorMATHA}\renewcommand\colorSYNTAX{\colorSYNTAXA}{{\color{\colorSYNTAX}\texttt{{\ensuremath{{\mathbb{R}}}}}}}\endgroup })\rrbracket }}} and 
      {{\color{\colorMATH}\ensuremath{\Gamma ; {\begingroup\renewcommand\colorMATH{\colorMATHB}\renewcommand\colorSYNTAX{\colorSYNTAXB}{{\color{\colorMATH}\ensuremath{\Distance}}}\endgroup } \vdash  {\begingroup\renewcommand\colorMATH{\colorMATHB}\renewcommand\colorSYNTAX{\colorSYNTAXB}{{\color{\colorMATH}\ensuremath{\se_{2}}}}\endgroup } \mathrel{:} {\begingroup\renewcommand\colorMATH{\colorMATHA}\renewcommand\colorSYNTAX{\colorSYNTAXA}{{\color{\colorSYNTAX}\texttt{{\ensuremath{{\mathbb{R}}}}}}}\endgroup } \mathrel{;} {\begingroup\renewcommand\colorMATH{\colorMATHB}\renewcommand\colorSYNTAX{\colorSYNTAXB}{{\color{\colorMATH}\ensuremath{\sS_{2}}}}\endgroup } \Rightarrow  (\gamma _{1} \vdash  {\begingroup\renewcommand\colorMATH{\colorMATHB}\renewcommand\colorSYNTAX{\colorSYNTAXB}{{\color{\colorMATH}\ensuremath{\se_{2}}}}\endgroup },\gamma _{2} \vdash  {\begingroup\renewcommand\colorMATH{\colorMATHB}\renewcommand\colorSYNTAX{\colorSYNTAXB}{{\color{\colorMATH}\ensuremath{\se_{2}}}}\endgroup }) \in  {\mathcal{E}}^{k-j_{1}}_{{\begingroup\renewcommand\colorMATH{\colorMATHB}\renewcommand\colorSYNTAX{\colorSYNTAXB}{{\color{\colorMATH}\ensuremath{\Distance'}}}\endgroup }\mathord{\cdotp }{\begingroup\renewcommand\colorMATH{\colorMATHB}\renewcommand\colorSYNTAX{\colorSYNTAXB}{{\color{\colorMATH}\ensuremath{\sS_{2}}}}\endgroup }}\llbracket {\begingroup\renewcommand\colorMATH{\colorMATHB}\renewcommand\colorSYNTAX{\colorSYNTAXB}{{\color{\colorMATH}\ensuremath{\Distance'}}}\endgroup }({\begingroup\renewcommand\colorMATH{\colorMATHA}\renewcommand\colorSYNTAX{\colorSYNTAXA}{{\color{\colorSYNTAX}\texttt{{\ensuremath{{\mathbb{R}}}}}}}\endgroup })\rrbracket }}} for some {{\color{\colorMATH}\ensuremath{j_{1} \leq  k}}}.
      By unfolding {{\color{\colorMATH}\ensuremath{(\gamma _{1} \vdash  {\begingroup\renewcommand\colorMATH{\colorMATHB}\renewcommand\colorSYNTAX{\colorSYNTAXB}{{\color{\colorMATH}\ensuremath{\se_{1}}}}\endgroup },\gamma _{2} \vdash  {\begingroup\renewcommand\colorMATH{\colorMATHB}\renewcommand\colorSYNTAX{\colorSYNTAXB}{{\color{\colorMATH}\ensuremath{\se_{1}}}}\endgroup }) \in  {\mathcal{E}}^{k}_{{\begingroup\renewcommand\colorMATH{\colorMATHB}\renewcommand\colorSYNTAX{\colorSYNTAXB}{{\color{\colorMATH}\ensuremath{\Distance'}}}\endgroup }\mathord{\cdotp }{\begingroup\renewcommand\colorMATH{\colorMATHB}\renewcommand\colorSYNTAX{\colorSYNTAXB}{{\color{\colorMATH}\ensuremath{\sS_{1}}}}\endgroup }}\llbracket {\begingroup\renewcommand\colorMATH{\colorMATHB}\renewcommand\colorSYNTAX{\colorSYNTAXB}{{\color{\colorMATH}\ensuremath{\Distance'}}}\endgroup }({\begingroup\renewcommand\colorMATH{\colorMATHA}\renewcommand\colorSYNTAX{\colorSYNTAXA}{{\color{\colorSYNTAX}\texttt{{\ensuremath{{\mathbb{R}}}}}}}\endgroup })\rrbracket }}},
      we know that if {{\color{\colorMATH}\ensuremath{\gamma _{1} \vdash  {\begingroup\renewcommand\colorMATH{\colorMATHB}\renewcommand\colorSYNTAX{\colorSYNTAXB}{{\color{\colorMATH}\ensuremath{\se_{1}}}}\endgroup } \Downarrow ^{j_{1}} {\begingroup\renewcommand\colorMATH{\colorMATHB}\renewcommand\colorSYNTAX{\colorSYNTAXB}{{\color{\colorMATH}\ensuremath{r_{1 1}}}}\endgroup }}}} then {{\color{\colorMATH}\ensuremath{\gamma _{2} \vdash  {\begingroup\renewcommand\colorMATH{\colorMATHB}\renewcommand\colorSYNTAX{\colorSYNTAXB}{{\color{\colorMATH}\ensuremath{\se_{1}}}}\endgroup } \Downarrow ^{*} {\begingroup\renewcommand\colorMATH{\colorMATHB}\renewcommand\colorSYNTAX{\colorSYNTAXB}{{\color{\colorMATH}\ensuremath{r_{1 2}}}}\endgroup }}}} and {{\color{\colorMATH}\ensuremath{({\begingroup\renewcommand\colorMATH{\colorMATHB}\renewcommand\colorSYNTAX{\colorSYNTAXB}{{\color{\colorMATH}\ensuremath{r_{1 1}}}}\endgroup }, {\begingroup\renewcommand\colorMATH{\colorMATHB}\renewcommand\colorSYNTAX{\colorSYNTAXB}{{\color{\colorMATH}\ensuremath{r_{1 2}}}}\endgroup }) \in  {\mathcal{V}}^{k-j_{1}}_{{\begingroup\renewcommand\colorMATH{\colorMATHB}\renewcommand\colorSYNTAX{\colorSYNTAXB}{{\color{\colorMATH}\ensuremath{\Distance'}}}\endgroup }\mathord{\cdotp }{\begingroup\renewcommand\colorMATH{\colorMATHB}\renewcommand\colorSYNTAX{\colorSYNTAXB}{{\color{\colorMATH}\ensuremath{\sS_{1}}}}\endgroup }}\llbracket {\begingroup\renewcommand\colorMATH{\colorMATHA}\renewcommand\colorSYNTAX{\colorSYNTAXA}{{\color{\colorSYNTAX}\texttt{{\ensuremath{{\mathbb{R}}}}}}}\endgroup }\rrbracket }}}, where {{\color{\colorMATH}\ensuremath{|{\begingroup\renewcommand\colorMATH{\colorMATHB}\renewcommand\colorSYNTAX{\colorSYNTAXB}{{\color{\colorMATH}\ensuremath{r_{1 1}}}}\endgroup } - {\begingroup\renewcommand\colorMATH{\colorMATHB}\renewcommand\colorSYNTAX{\colorSYNTAXB}{{\color{\colorMATH}\ensuremath{r_{1 2}}}}\endgroup }| \leq  {\begingroup\renewcommand\colorMATH{\colorMATHB}\renewcommand\colorSYNTAX{\colorSYNTAXB}{{\color{\colorMATH}\ensuremath{\Distance'}}}\endgroup }\mathord{\cdotp }{\begingroup\renewcommand\colorMATH{\colorMATHB}\renewcommand\colorSYNTAX{\colorSYNTAXB}{{\color{\colorMATH}\ensuremath{\sS_{1}}}}\endgroup }}}}.\\
      Also, by unfolding {{\color{\colorMATH}\ensuremath{(\gamma _{1} \vdash  {\begingroup\renewcommand\colorMATH{\colorMATHB}\renewcommand\colorSYNTAX{\colorSYNTAXB}{{\color{\colorMATH}\ensuremath{\se_{2}}}}\endgroup },\gamma _{2} \vdash  {\begingroup\renewcommand\colorMATH{\colorMATHB}\renewcommand\colorSYNTAX{\colorSYNTAXB}{{\color{\colorMATH}\ensuremath{\se_{2}}}}\endgroup }) \in  {\mathcal{E}}^{k-j_{1}}_{{\begingroup\renewcommand\colorMATH{\colorMATHB}\renewcommand\colorSYNTAX{\colorSYNTAXB}{{\color{\colorMATH}\ensuremath{\Distance'}}}\endgroup }\mathord{\cdotp }{\begingroup\renewcommand\colorMATH{\colorMATHB}\renewcommand\colorSYNTAX{\colorSYNTAXB}{{\color{\colorMATH}\ensuremath{\sS_{2}}}}\endgroup }}\llbracket {\begingroup\renewcommand\colorMATH{\colorMATHB}\renewcommand\colorSYNTAX{\colorSYNTAXB}{{\color{\colorMATH}\ensuremath{\Distance'}}}\endgroup }({\begingroup\renewcommand\colorMATH{\colorMATHA}\renewcommand\colorSYNTAX{\colorSYNTAXA}{{\color{\colorSYNTAX}\texttt{{\ensuremath{{\mathbb{R}}}}}}}\endgroup })\rrbracket }}}, if {{\color{\colorMATH}\ensuremath{\gamma _{1} \vdash  {\begingroup\renewcommand\colorMATH{\colorMATHB}\renewcommand\colorSYNTAX{\colorSYNTAXB}{{\color{\colorMATH}\ensuremath{\se_{2}}}}\endgroup } \Downarrow ^{j_{2}} {\begingroup\renewcommand\colorMATH{\colorMATHB}\renewcommand\colorSYNTAX{\colorSYNTAXB}{{\color{\colorMATH}\ensuremath{r_{2 1}}}}\endgroup }}}} \pthen {{\color{\colorMATH}\ensuremath{\gamma _{2} \vdash  {\begingroup\renewcommand\colorMATH{\colorMATHB}\renewcommand\colorSYNTAX{\colorSYNTAXB}{{\color{\colorMATH}\ensuremath{\se_{2}}}}\endgroup } \Downarrow ^{\pj} {\begingroup\renewcommand\colorMATH{\colorMATHB}\renewcommand\colorSYNTAX{\colorSYNTAXB}{{\color{\colorMATH}\ensuremath{r_{2 2}}}}\endgroup }}}} \pand {{\color{\colorMATH}\ensuremath{({\begingroup\renewcommand\colorMATH{\colorMATHB}\renewcommand\colorSYNTAX{\colorSYNTAXB}{{\color{\colorMATH}\ensuremath{r_{2 1}}}}\endgroup }, {\begingroup\renewcommand\colorMATH{\colorMATHB}\renewcommand\colorSYNTAX{\colorSYNTAXB}{{\color{\colorMATH}\ensuremath{r_{2 2}}}}\endgroup }) \in  {\mathcal{V}}^{k-j_{1}-j_{2}}_{{\begingroup\renewcommand\colorMATH{\colorMATHB}\renewcommand\colorSYNTAX{\colorSYNTAXB}{{\color{\colorMATH}\ensuremath{\Distance'}}}\endgroup }\mathord{\cdotp }{\begingroup\renewcommand\colorMATH{\colorMATHB}\renewcommand\colorSYNTAX{\colorSYNTAXB}{{\color{\colorMATH}\ensuremath{\sS_{2}}}}\endgroup }}\llbracket {\begingroup\renewcommand\colorMATH{\colorMATHA}\renewcommand\colorSYNTAX{\colorSYNTAXA}{{\color{\colorSYNTAX}\texttt{{\ensuremath{{\mathbb{R}}}}}}}\endgroup }\rrbracket }}}, where {{\color{\colorMATH}\ensuremath{| {\begingroup\renewcommand\colorMATH{\colorMATHB}\renewcommand\colorSYNTAX{\colorSYNTAXB}{{\color{\colorMATH}\ensuremath{r_{2 1}}}}\endgroup } - {\begingroup\renewcommand\colorMATH{\colorMATHB}\renewcommand\colorSYNTAX{\colorSYNTAXB}{{\color{\colorMATH}\ensuremath{r_{2 2}}}}\endgroup }| \leq  {\begingroup\renewcommand\colorMATH{\colorMATHB}\renewcommand\colorSYNTAX{\colorSYNTAXB}{{\color{\colorMATH}\ensuremath{\Distance'}}}\endgroup }\mathord{\cdotp }{\begingroup\renewcommand\colorMATH{\colorMATHB}\renewcommand\colorSYNTAX{\colorSYNTAXB}{{\color{\colorMATH}\ensuremath{\sS_{2}}}}\endgroup }}}}.

      Then if {{\color{\colorMATH}\ensuremath{\gamma _{1}\vdash {\begingroup\renewcommand\colorMATH{\colorMATHB}\renewcommand\colorSYNTAX{\colorSYNTAXB}{{\color{\colorMATH}\ensuremath{\se_{1}}}}\endgroup } + {\begingroup\renewcommand\colorMATH{\colorMATHB}\renewcommand\colorSYNTAX{\colorSYNTAXB}{{\color{\colorMATH}\ensuremath{\se_{2}}}}\endgroup } \Downarrow ^{j_{1}+j_{2}+1} r'_{1}}}} and {{\color{\colorMATH}\ensuremath{\gamma _{2}\vdash {\begingroup\renewcommand\colorMATH{\colorMATHB}\renewcommand\colorSYNTAX{\colorSYNTAXB}{{\color{\colorMATH}\ensuremath{\se_{1}}}}\endgroup } + {\begingroup\renewcommand\colorMATH{\colorMATHB}\renewcommand\colorSYNTAX{\colorSYNTAXB}{{\color{\colorMATH}\ensuremath{\se_{2}}}}\endgroup } \Downarrow ^{j_{1}+j_{2}+1} r'_{2}}}}, where {{\color{\colorMATH}\ensuremath{r'_{1} = {\begingroup\renewcommand\colorMATH{\colorMATHB}\renewcommand\colorSYNTAX{\colorSYNTAXB}{{\color{\colorMATH}\ensuremath{r_{1 1}}}}\endgroup } + {\begingroup\renewcommand\colorMATH{\colorMATHB}\renewcommand\colorSYNTAX{\colorSYNTAXB}{{\color{\colorMATH}\ensuremath{r_{2 1}}}}\endgroup }}}} and {{\color{\colorMATH}\ensuremath{r'_{2} = {\begingroup\renewcommand\colorMATH{\colorMATHB}\renewcommand\colorSYNTAX{\colorSYNTAXB}{{\color{\colorMATH}\ensuremath{r_{1 2}}}}\endgroup } + {\begingroup\renewcommand\colorMATH{\colorMATHB}\renewcommand\colorSYNTAX{\colorSYNTAXB}{{\color{\colorMATH}\ensuremath{r_{2 2}}}}\endgroup }}}}, we have to prove that 
      {{\color{\colorMATH}\ensuremath{({\begingroup\renewcommand\colorMATH{\colorMATHB}\renewcommand\colorSYNTAX{\colorSYNTAXB}{{\color{\colorMATH}\ensuremath{r'_{1}}}}\endgroup }, {\begingroup\renewcommand\colorMATH{\colorMATHB}\renewcommand\colorSYNTAX{\colorSYNTAXB}{{\color{\colorMATH}\ensuremath{r'_{2}}}}\endgroup }) \in  {\mathcal{V}}^{k-j_{1}-j_{2}-1}_{{\begingroup\renewcommand\colorMATH{\colorMATHB}\renewcommand\colorSYNTAX{\colorSYNTAXB}{{\color{\colorMATH}\ensuremath{\Distance'}}}\endgroup }\mathord{\cdotp }({\begingroup\renewcommand\colorMATH{\colorMATHB}\renewcommand\colorSYNTAX{\colorSYNTAXB}{{\color{\colorMATH}\ensuremath{\sS_{1}}}}\endgroup } + {\begingroup\renewcommand\colorMATH{\colorMATHB}\renewcommand\colorSYNTAX{\colorSYNTAXB}{{\color{\colorMATH}\ensuremath{\sS_{2}}}}\endgroup })}\llbracket {\begingroup\renewcommand\colorMATH{\colorMATHA}\renewcommand\colorSYNTAX{\colorSYNTAXA}{{\color{\colorSYNTAX}\texttt{{\ensuremath{{\mathbb{R}}}}}}}\endgroup }\rrbracket }}}, i.e.   {{\color{\colorMATH}\ensuremath{|({\begingroup\renewcommand\colorMATH{\colorMATHB}\renewcommand\colorSYNTAX{\colorSYNTAXB}{{\color{\colorMATH}\ensuremath{r_{1 1}}}}\endgroup }+{\begingroup\renewcommand\colorMATH{\colorMATHB}\renewcommand\colorSYNTAX{\colorSYNTAXB}{{\color{\colorMATH}\ensuremath{r_{2 1}}}}\endgroup })-({\begingroup\renewcommand\colorMATH{\colorMATHB}\renewcommand\colorSYNTAX{\colorSYNTAXB}{{\color{\colorMATH}\ensuremath{r_{1 2}}}}\endgroup }+{\begingroup\renewcommand\colorMATH{\colorMATHB}\renewcommand\colorSYNTAX{\colorSYNTAXB}{{\color{\colorMATH}\ensuremath{r_{2 2}}}}\endgroup })| \leq  {\begingroup\renewcommand\colorMATH{\colorMATHB}\renewcommand\colorSYNTAX{\colorSYNTAXB}{{\color{\colorMATH}\ensuremath{\Distance'}}}\endgroup }\mathord{\cdotp }({\begingroup\renewcommand\colorMATH{\colorMATHB}\renewcommand\colorSYNTAX{\colorSYNTAXB}{{\color{\colorMATH}\ensuremath{\sS_{1}}}}\endgroup } + {\begingroup\renewcommand\colorMATH{\colorMATHB}\renewcommand\colorSYNTAX{\colorSYNTAXB}{{\color{\colorMATH}\ensuremath{\sS_{2}}}}\endgroup })}}}.\\
      Notice that {{\color{\colorMATH}\ensuremath{|({\begingroup\renewcommand\colorMATH{\colorMATHB}\renewcommand\colorSYNTAX{\colorSYNTAXB}{{\color{\colorMATH}\ensuremath{r_{1 1}}}}\endgroup }+{\begingroup\renewcommand\colorMATH{\colorMATHB}\renewcommand\colorSYNTAX{\colorSYNTAXB}{{\color{\colorMATH}\ensuremath{r_{2 1}}}}\endgroup })-({\begingroup\renewcommand\colorMATH{\colorMATHB}\renewcommand\colorSYNTAX{\colorSYNTAXB}{{\color{\colorMATH}\ensuremath{r_{1 2}}}}\endgroup }+{\begingroup\renewcommand\colorMATH{\colorMATHB}\renewcommand\colorSYNTAX{\colorSYNTAXB}{{\color{\colorMATH}\ensuremath{r_{2 2}}}}\endgroup })| = |({\begingroup\renewcommand\colorMATH{\colorMATHB}\renewcommand\colorSYNTAX{\colorSYNTAXB}{{\color{\colorMATH}\ensuremath{r_{1 1}}}}\endgroup }-{\begingroup\renewcommand\colorMATH{\colorMATHB}\renewcommand\colorSYNTAX{\colorSYNTAXB}{{\color{\colorMATH}\ensuremath{r_{1 2}}}}\endgroup })+({\begingroup\renewcommand\colorMATH{\colorMATHB}\renewcommand\colorSYNTAX{\colorSYNTAXB}{{\color{\colorMATH}\ensuremath{r_{2 1}}}}\endgroup }-{\begingroup\renewcommand\colorMATH{\colorMATHB}\renewcommand\colorSYNTAX{\colorSYNTAXB}{{\color{\colorMATH}\ensuremath{r_{2 2}}}}\endgroup })|}}}, and by the triangle inequality of the absolute value, 
      {{\color{\colorMATH}\ensuremath{|({\begingroup\renewcommand\colorMATH{\colorMATHB}\renewcommand\colorSYNTAX{\colorSYNTAXB}{{\color{\colorMATH}\ensuremath{r_{1 1}}}}\endgroup }-{\begingroup\renewcommand\colorMATH{\colorMATHB}\renewcommand\colorSYNTAX{\colorSYNTAXB}{{\color{\colorMATH}\ensuremath{r_{1 2}}}}\endgroup })+({\begingroup\renewcommand\colorMATH{\colorMATHB}\renewcommand\colorSYNTAX{\colorSYNTAXB}{{\color{\colorMATH}\ensuremath{r_{2 1}}}}\endgroup }-{\begingroup\renewcommand\colorMATH{\colorMATHB}\renewcommand\colorSYNTAX{\colorSYNTAXB}{{\color{\colorMATH}\ensuremath{r_{2 2}}}}\endgroup })| \leq  |({\begingroup\renewcommand\colorMATH{\colorMATHB}\renewcommand\colorSYNTAX{\colorSYNTAXB}{{\color{\colorMATH}\ensuremath{r_{1 1}}}}\endgroup }-{\begingroup\renewcommand\colorMATH{\colorMATHB}\renewcommand\colorSYNTAX{\colorSYNTAXB}{{\color{\colorMATH}\ensuremath{r_{1 2}}}}\endgroup })|+|({\begingroup\renewcommand\colorMATH{\colorMATHB}\renewcommand\colorSYNTAX{\colorSYNTAXB}{{\color{\colorMATH}\ensuremath{r_{2 1}}}}\endgroup }-{\begingroup\renewcommand\colorMATH{\colorMATHB}\renewcommand\colorSYNTAX{\colorSYNTAXB}{{\color{\colorMATH}\ensuremath{r_{2 2}}}}\endgroup })|}}}. Also as {{\color{\colorMATH}\ensuremath{| {\begingroup\renewcommand\colorMATH{\colorMATHB}\renewcommand\colorSYNTAX{\colorSYNTAXB}{{\color{\colorMATH}\ensuremath{r_{1 1}}}}\endgroup } - {\begingroup\renewcommand\colorMATH{\colorMATHB}\renewcommand\colorSYNTAX{\colorSYNTAXB}{{\color{\colorMATH}\ensuremath{r_{1 2}}}}\endgroup }| \leq  {\begingroup\renewcommand\colorMATH{\colorMATHB}\renewcommand\colorSYNTAX{\colorSYNTAXB}{{\color{\colorMATH}\ensuremath{\Distance'}}}\endgroup }\mathord{\cdotp }{\begingroup\renewcommand\colorMATH{\colorMATHB}\renewcommand\colorSYNTAX{\colorSYNTAXB}{{\color{\colorMATH}\ensuremath{\sS_{1}}}}\endgroup }}}} and {{\color{\colorMATH}\ensuremath{| {\begingroup\renewcommand\colorMATH{\colorMATHB}\renewcommand\colorSYNTAX{\colorSYNTAXB}{{\color{\colorMATH}\ensuremath{r_{2 1}}}}\endgroup } - {\begingroup\renewcommand\colorMATH{\colorMATHB}\renewcommand\colorSYNTAX{\colorSYNTAXB}{{\color{\colorMATH}\ensuremath{r_{2 2}}}}\endgroup }| \leq  {\begingroup\renewcommand\colorMATH{\colorMATHB}\renewcommand\colorSYNTAX{\colorSYNTAXB}{{\color{\colorMATH}\ensuremath{\Distance'}}}\endgroup }\mathord{\cdotp }{\begingroup\renewcommand\colorMATH{\colorMATHB}\renewcommand\colorSYNTAX{\colorSYNTAXB}{{\color{\colorMATH}\ensuremath{\sS_{2}}}}\endgroup }}}},
      then {{\color{\colorMATH}\ensuremath{|({\begingroup\renewcommand\colorMATH{\colorMATHB}\renewcommand\colorSYNTAX{\colorSYNTAXB}{{\color{\colorMATH}\ensuremath{r_{1 1}}}}\endgroup }-{\begingroup\renewcommand\colorMATH{\colorMATHB}\renewcommand\colorSYNTAX{\colorSYNTAXB}{{\color{\colorMATH}\ensuremath{r_{1 2}}}}\endgroup })|+|({\begingroup\renewcommand\colorMATH{\colorMATHB}\renewcommand\colorSYNTAX{\colorSYNTAXB}{{\color{\colorMATH}\ensuremath{r_{2 1}}}}\endgroup }-{\begingroup\renewcommand\colorMATH{\colorMATHB}\renewcommand\colorSYNTAX{\colorSYNTAXB}{{\color{\colorMATH}\ensuremath{r_{2 2}}}}\endgroup })| \leq  {\begingroup\renewcommand\colorMATH{\colorMATHB}\renewcommand\colorSYNTAX{\colorSYNTAXB}{{\color{\colorMATH}\ensuremath{\Distance'}}}\endgroup }\mathord{\cdotp }{\begingroup\renewcommand\colorMATH{\colorMATHB}\renewcommand\colorSYNTAX{\colorSYNTAXB}{{\color{\colorMATH}\ensuremath{\sS_{1}}}}\endgroup }+ {\begingroup\renewcommand\colorMATH{\colorMATHB}\renewcommand\colorSYNTAX{\colorSYNTAXB}{{\color{\colorMATH}\ensuremath{\Distance'}}}\endgroup }\mathord{\cdotp }{\begingroup\renewcommand\colorMATH{\colorMATHB}\renewcommand\colorSYNTAX{\colorSYNTAXB}{{\color{\colorMATH}\ensuremath{\sS_{2}}}}\endgroup }}}}. By Lemma~\ref{lm:associativity-inst}, {{\color{\colorMATH}\ensuremath{{\begingroup\renewcommand\colorMATH{\colorMATHB}\renewcommand\colorSYNTAX{\colorSYNTAXB}{{\color{\colorMATH}\ensuremath{\Distance'}}}\endgroup }\mathord{\cdotp }{\begingroup\renewcommand\colorMATH{\colorMATHB}\renewcommand\colorSYNTAX{\colorSYNTAXB}{{\color{\colorMATH}\ensuremath{\sS_{1}}}}\endgroup }+ {\begingroup\renewcommand\colorMATH{\colorMATHB}\renewcommand\colorSYNTAX{\colorSYNTAXB}{{\color{\colorMATH}\ensuremath{\Distance'}}}\endgroup }\mathord{\cdotp }{\begingroup\renewcommand\colorMATH{\colorMATHB}\renewcommand\colorSYNTAX{\colorSYNTAXB}{{\color{\colorMATH}\ensuremath{\sS_{2}}}}\endgroup } = {\begingroup\renewcommand\colorMATH{\colorMATHB}\renewcommand\colorSYNTAX{\colorSYNTAXB}{{\color{\colorMATH}\ensuremath{\Distance'}}}\endgroup }\mathord{\cdotp }({\begingroup\renewcommand\colorMATH{\colorMATHB}\renewcommand\colorSYNTAX{\colorSYNTAXB}{{\color{\colorMATH}\ensuremath{\sS_{1}}}}\endgroup }+ {\begingroup\renewcommand\colorMATH{\colorMATHB}\renewcommand\colorSYNTAX{\colorSYNTAXB}{{\color{\colorMATH}\ensuremath{\sS_{2}}}}\endgroup })}}}, therefore
      {{\color{\colorMATH}\ensuremath{|({\begingroup\renewcommand\colorMATH{\colorMATHB}\renewcommand\colorSYNTAX{\colorSYNTAXB}{{\color{\colorMATH}\ensuremath{r_{1 1}}}}\endgroup }+{\begingroup\renewcommand\colorMATH{\colorMATHB}\renewcommand\colorSYNTAX{\colorSYNTAXB}{{\color{\colorMATH}\ensuremath{r_{2 1}}}}\endgroup })-({\begingroup\renewcommand\colorMATH{\colorMATHB}\renewcommand\colorSYNTAX{\colorSYNTAXB}{{\color{\colorMATH}\ensuremath{r_{1 2}}}}\endgroup }+{\begingroup\renewcommand\colorMATH{\colorMATHB}\renewcommand\colorSYNTAX{\colorSYNTAXB}{{\color{\colorMATH}\ensuremath{r_{2 2}}}}\endgroup })| \leq  {\begingroup\renewcommand\colorMATH{\colorMATHB}\renewcommand\colorSYNTAX{\colorSYNTAXB}{{\color{\colorMATH}\ensuremath{\Distance'}}}\endgroup }\mathord{\cdotp }({\begingroup\renewcommand\colorMATH{\colorMATHB}\renewcommand\colorSYNTAX{\colorSYNTAXB}{{\color{\colorMATH}\ensuremath{\sS_{1}}}}\endgroup } + {\begingroup\renewcommand\colorMATH{\colorMATHB}\renewcommand\colorSYNTAX{\colorSYNTAXB}{{\color{\colorMATH}\ensuremath{\sS_{2}}}}\endgroup })}}} and the result holds.
    \end{subproof}
  \item  {{\color{\colorMATH}\ensuremath{\Gamma ; {\begingroup\renewcommand\colorMATH{\colorMATHB}\renewcommand\colorSYNTAX{\colorSYNTAXB}{{\color{\colorMATH}\ensuremath{\Distance}}}\endgroup } \vdash  {\begingroup\renewcommand\colorMATH{\colorMATHB}\renewcommand\colorSYNTAX{\colorSYNTAXB}{{\color{\colorMATH}\ensuremath{\se_{1}}}}\endgroup } * {\begingroup\renewcommand\colorMATH{\colorMATHB}\renewcommand\colorSYNTAX{\colorSYNTAXB}{{\color{\colorMATH}\ensuremath{\se_{2}}}}\endgroup } \mathrel{:} {\begingroup\renewcommand\colorMATH{\colorMATHA}\renewcommand\colorSYNTAX{\colorSYNTAXA}{{\color{\colorSYNTAX}\texttt{{\ensuremath{{\mathbb{R}}}}}}}\endgroup } \mathrel{;} {\begingroup\renewcommand\colorMATH{\colorMATHB}\renewcommand\colorSYNTAX{\colorSYNTAXB}{{\color{\colorMATH}\ensuremath{\infty }}}\endgroup }({\begingroup\renewcommand\colorMATH{\colorMATHB}\renewcommand\colorSYNTAX{\colorSYNTAXB}{{\color{\colorMATH}\ensuremath{\sS_{1}}}}\endgroup } + {\begingroup\renewcommand\colorMATH{\colorMATHB}\renewcommand\colorSYNTAX{\colorSYNTAXB}{{\color{\colorMATH}\ensuremath{\sS_{2}}}}\endgroup })}}} 
    \begin{subproof} 
      We have to prove that {{\color{\colorMATH}\ensuremath{\forall k, \forall (\gamma _{1},\gamma _{2}) \in  {\mathcal{G}}_{{\begingroup\renewcommand\colorMATH{\colorMATHB}\renewcommand\colorSYNTAX{\colorSYNTAXB}{{\color{\colorMATH}\ensuremath{\Distance'}}}\endgroup }}^{\kg}\llbracket \Gamma \rrbracket , (\gamma _{1}\vdash {\begingroup\renewcommand\colorMATH{\colorMATHB}\renewcommand\colorSYNTAX{\colorSYNTAXB}{{\color{\colorMATH}\ensuremath{\se_{1}}}}\endgroup } * {\begingroup\renewcommand\colorMATH{\colorMATHB}\renewcommand\colorSYNTAX{\colorSYNTAXB}{{\color{\colorMATH}\ensuremath{\se_{2}}}}\endgroup },\gamma _{2}\vdash {\begingroup\renewcommand\colorMATH{\colorMATHB}\renewcommand\colorSYNTAX{\colorSYNTAXB}{{\color{\colorMATH}\ensuremath{\se_{1}}}}\endgroup } * {\begingroup\renewcommand\colorMATH{\colorMATHB}\renewcommand\colorSYNTAX{\colorSYNTAXB}{{\color{\colorMATH}\ensuremath{\se_{2}}}}\endgroup }) \in  {\mathcal{E}}^{k}_{{\begingroup\renewcommand\colorMATH{\colorMATHB}\renewcommand\colorSYNTAX{\colorSYNTAXB}{{\color{\colorMATH}\ensuremath{\Distance'}}}\endgroup }\mathord{\cdotp }{\begingroup\renewcommand\colorMATH{\colorMATHB}\renewcommand\colorSYNTAX{\colorSYNTAXB}{{\color{\colorMATH}\ensuremath{\infty }}}\endgroup }({\begingroup\renewcommand\colorMATH{\colorMATHB}\renewcommand\colorSYNTAX{\colorSYNTAXB}{{\color{\colorMATH}\ensuremath{\sS_{1}}}}\endgroup } + {\begingroup\renewcommand\colorMATH{\colorMATHB}\renewcommand\colorSYNTAX{\colorSYNTAXB}{{\color{\colorMATH}\ensuremath{\sS_{2}}}}\endgroup })}\llbracket {\begingroup\renewcommand\colorMATH{\colorMATHB}\renewcommand\colorSYNTAX{\colorSYNTAXB}{{\color{\colorMATH}\ensuremath{\Distance'}}}\endgroup }({\begingroup\renewcommand\colorMATH{\colorMATHA}\renewcommand\colorSYNTAX{\colorSYNTAXA}{{\color{\colorSYNTAX}\texttt{{\ensuremath{{\mathbb{R}}}}}}}\endgroup })\rrbracket }}}, for {{\color{\colorMATH}\ensuremath{{\begingroup\renewcommand\colorMATH{\colorMATHB}\renewcommand\colorSYNTAX{\colorSYNTAXB}{{\color{\colorMATH}\ensuremath{\Distance'}}}\endgroup } \sqsubseteq  {\begingroup\renewcommand\colorMATH{\colorMATHB}\renewcommand\colorSYNTAX{\colorSYNTAXB}{{\color{\colorMATH}\ensuremath{\Distance}}}\endgroup }}}}.
      By induction hypotheses \\
      {{\color{\colorMATH}\ensuremath{\Gamma  \vdash  {\begingroup\renewcommand\colorMATH{\colorMATHB}\renewcommand\colorSYNTAX{\colorSYNTAXB}{{\color{\colorMATH}\ensuremath{\se_{1}}}}\endgroup } \mathrel{:} {\begingroup\renewcommand\colorMATH{\colorMATHA}\renewcommand\colorSYNTAX{\colorSYNTAXA}{{\color{\colorSYNTAX}\texttt{{\ensuremath{{\mathbb{R}}}}}}}\endgroup } \mathrel{;} {\begingroup\renewcommand\colorMATH{\colorMATHB}\renewcommand\colorSYNTAX{\colorSYNTAXB}{{\color{\colorMATH}\ensuremath{\sS_{1}}}}\endgroup } \Rightarrow  (\gamma _{1} \vdash  {\begingroup\renewcommand\colorMATH{\colorMATHB}\renewcommand\colorSYNTAX{\colorSYNTAXB}{{\color{\colorMATH}\ensuremath{\se_{1}}}}\endgroup },\gamma _{2} \vdash  {\begingroup\renewcommand\colorMATH{\colorMATHB}\renewcommand\colorSYNTAX{\colorSYNTAXB}{{\color{\colorMATH}\ensuremath{\se_{1}}}}\endgroup }) \in  {\mathcal{E}}^{k}_{{\begingroup\renewcommand\colorMATH{\colorMATHB}\renewcommand\colorSYNTAX{\colorSYNTAXB}{{\color{\colorMATH}\ensuremath{\Distance'}}}\endgroup }\mathord{\cdotp }{\begingroup\renewcommand\colorMATH{\colorMATHB}\renewcommand\colorSYNTAX{\colorSYNTAXB}{{\color{\colorMATH}\ensuremath{\sS_{1}}}}\endgroup }}\llbracket {\begingroup\renewcommand\colorMATH{\colorMATHB}\renewcommand\colorSYNTAX{\colorSYNTAXB}{{\color{\colorMATH}\ensuremath{\Distance'}}}\endgroup }({\begingroup\renewcommand\colorMATH{\colorMATHA}\renewcommand\colorSYNTAX{\colorSYNTAXA}{{\color{\colorSYNTAX}\texttt{{\ensuremath{{\mathbb{R}}}}}}}\endgroup })\rrbracket }}} and 
      {{\color{\colorMATH}\ensuremath{\Gamma  \vdash  {\begingroup\renewcommand\colorMATH{\colorMATHB}\renewcommand\colorSYNTAX{\colorSYNTAXB}{{\color{\colorMATH}\ensuremath{\se_{2}}}}\endgroup } \mathrel{:} {\begingroup\renewcommand\colorMATH{\colorMATHA}\renewcommand\colorSYNTAX{\colorSYNTAXA}{{\color{\colorSYNTAX}\texttt{{\ensuremath{{\mathbb{R}}}}}}}\endgroup } \mathrel{;} {\begingroup\renewcommand\colorMATH{\colorMATHB}\renewcommand\colorSYNTAX{\colorSYNTAXB}{{\color{\colorMATH}\ensuremath{\sS_{2}}}}\endgroup } \Rightarrow  (\gamma _{1} \vdash  {\begingroup\renewcommand\colorMATH{\colorMATHB}\renewcommand\colorSYNTAX{\colorSYNTAXB}{{\color{\colorMATH}\ensuremath{\se_{2}}}}\endgroup },\gamma _{2} \vdash  {\begingroup\renewcommand\colorMATH{\colorMATHB}\renewcommand\colorSYNTAX{\colorSYNTAXB}{{\color{\colorMATH}\ensuremath{\se_{2}}}}\endgroup }) \in  {\mathcal{E}}^{k-j_{1}}_{{\begingroup\renewcommand\colorMATH{\colorMATHB}\renewcommand\colorSYNTAX{\colorSYNTAXB}{{\color{\colorMATH}\ensuremath{\Distance'}}}\endgroup }\mathord{\cdotp }{\begingroup\renewcommand\colorMATH{\colorMATHB}\renewcommand\colorSYNTAX{\colorSYNTAXB}{{\color{\colorMATH}\ensuremath{\sS_{2}}}}\endgroup }}\llbracket {\begingroup\renewcommand\colorMATH{\colorMATHB}\renewcommand\colorSYNTAX{\colorSYNTAXB}{{\color{\colorMATH}\ensuremath{\Distance'}}}\endgroup }({\begingroup\renewcommand\colorMATH{\colorMATHA}\renewcommand\colorSYNTAX{\colorSYNTAXA}{{\color{\colorSYNTAX}\texttt{{\ensuremath{{\mathbb{R}}}}}}}\endgroup })\rrbracket }}} for {{\color{\colorMATH}\ensuremath{j_{1} \leq  k}}}.
      By unfolding {{\color{\colorMATH}\ensuremath{(\gamma _{1} \vdash  {\begingroup\renewcommand\colorMATH{\colorMATHB}\renewcommand\colorSYNTAX{\colorSYNTAXB}{{\color{\colorMATH}\ensuremath{\se_{1}}}}\endgroup },\gamma _{2} \vdash  {\begingroup\renewcommand\colorMATH{\colorMATHB}\renewcommand\colorSYNTAX{\colorSYNTAXB}{{\color{\colorMATH}\ensuremath{\se_{1}}}}\endgroup }) \in  {\mathcal{E}}^{k}_{{\begingroup\renewcommand\colorMATH{\colorMATHB}\renewcommand\colorSYNTAX{\colorSYNTAXB}{{\color{\colorMATH}\ensuremath{\Distance'}}}\endgroup }\mathord{\cdotp }{\begingroup\renewcommand\colorMATH{\colorMATHB}\renewcommand\colorSYNTAX{\colorSYNTAXB}{{\color{\colorMATH}\ensuremath{\sS_{1}}}}\endgroup }}\llbracket {\begingroup\renewcommand\colorMATH{\colorMATHB}\renewcommand\colorSYNTAX{\colorSYNTAXB}{{\color{\colorMATH}\ensuremath{\Distance'}}}\endgroup }({\begingroup\renewcommand\colorMATH{\colorMATHA}\renewcommand\colorSYNTAX{\colorSYNTAXA}{{\color{\colorSYNTAX}\texttt{{\ensuremath{{\mathbb{R}}}}}}}\endgroup })\rrbracket }}},
      we know that if {{\color{\colorMATH}\ensuremath{\gamma _{1} \vdash  {\begingroup\renewcommand\colorMATH{\colorMATHB}\renewcommand\colorSYNTAX{\colorSYNTAXB}{{\color{\colorMATH}\ensuremath{\se_{1}}}}\endgroup } \Downarrow ^{j_{1}} {\begingroup\renewcommand\colorMATH{\colorMATHB}\renewcommand\colorSYNTAX{\colorSYNTAXB}{{\color{\colorMATH}\ensuremath{r_{1 1}}}}\endgroup }}}} then {{\color{\colorMATH}\ensuremath{\gamma _{2} \vdash  {\begingroup\renewcommand\colorMATH{\colorMATHB}\renewcommand\colorSYNTAX{\colorSYNTAXB}{{\color{\colorMATH}\ensuremath{\se_{1}}}}\endgroup } \Downarrow ^{*} {\begingroup\renewcommand\colorMATH{\colorMATHB}\renewcommand\colorSYNTAX{\colorSYNTAXB}{{\color{\colorMATH}\ensuremath{r_{1 2}}}}\endgroup }}}} and {{\color{\colorMATH}\ensuremath{({\begingroup\renewcommand\colorMATH{\colorMATHB}\renewcommand\colorSYNTAX{\colorSYNTAXB}{{\color{\colorMATH}\ensuremath{r_{1 1}}}}\endgroup }, {\begingroup\renewcommand\colorMATH{\colorMATHB}\renewcommand\colorSYNTAX{\colorSYNTAXB}{{\color{\colorMATH}\ensuremath{r_{1 2}}}}\endgroup }) \in  {\mathcal{V}}^{k-j}_{{\begingroup\renewcommand\colorMATH{\colorMATHB}\renewcommand\colorSYNTAX{\colorSYNTAXB}{{\color{\colorMATH}\ensuremath{\Distance'}}}\endgroup }\mathord{\cdotp }{\begingroup\renewcommand\colorMATH{\colorMATHB}\renewcommand\colorSYNTAX{\colorSYNTAXB}{{\color{\colorMATH}\ensuremath{\sS_{1}}}}\endgroup }}\llbracket {\begingroup\renewcommand\colorMATH{\colorMATHA}\renewcommand\colorSYNTAX{\colorSYNTAXA}{{\color{\colorSYNTAX}\texttt{{\ensuremath{{\mathbb{R}}}}}}}\endgroup }\rrbracket }}}, where {{\color{\colorMATH}\ensuremath{|{\begingroup\renewcommand\colorMATH{\colorMATHB}\renewcommand\colorSYNTAX{\colorSYNTAXB}{{\color{\colorMATH}\ensuremath{r_{1 1}}}}\endgroup } - {\begingroup\renewcommand\colorMATH{\colorMATHB}\renewcommand\colorSYNTAX{\colorSYNTAXB}{{\color{\colorMATH}\ensuremath{r_{1 2}}}}\endgroup }| \leq  {\begingroup\renewcommand\colorMATH{\colorMATHB}\renewcommand\colorSYNTAX{\colorSYNTAXB}{{\color{\colorMATH}\ensuremath{\Distance'}}}\endgroup }\mathord{\cdotp }{\begingroup\renewcommand\colorMATH{\colorMATHB}\renewcommand\colorSYNTAX{\colorSYNTAXB}{{\color{\colorMATH}\ensuremath{\sS_{1}}}}\endgroup }}}}.\\
      Also, by unfolding {{\color{\colorMATH}\ensuremath{(\gamma _{1} \vdash  {\begingroup\renewcommand\colorMATH{\colorMATHB}\renewcommand\colorSYNTAX{\colorSYNTAXB}{{\color{\colorMATH}\ensuremath{\se_{2}}}}\endgroup },\gamma _{2} \vdash  {\begingroup\renewcommand\colorMATH{\colorMATHB}\renewcommand\colorSYNTAX{\colorSYNTAXB}{{\color{\colorMATH}\ensuremath{\se_{2}}}}\endgroup }) \in  {\mathcal{E}}^{k-j_{1}}_{{\begingroup\renewcommand\colorMATH{\colorMATHB}\renewcommand\colorSYNTAX{\colorSYNTAXB}{{\color{\colorMATH}\ensuremath{\Distance'}}}\endgroup }\mathord{\cdotp }{\begingroup\renewcommand\colorMATH{\colorMATHB}\renewcommand\colorSYNTAX{\colorSYNTAXB}{{\color{\colorMATH}\ensuremath{\sS_{2}}}}\endgroup }}\llbracket {\begingroup\renewcommand\colorMATH{\colorMATHB}\renewcommand\colorSYNTAX{\colorSYNTAXB}{{\color{\colorMATH}\ensuremath{\Distance'}}}\endgroup }({\begingroup\renewcommand\colorMATH{\colorMATHA}\renewcommand\colorSYNTAX{\colorSYNTAXA}{{\color{\colorSYNTAX}\texttt{{\ensuremath{{\mathbb{R}}}}}}}\endgroup })\rrbracket }}}, if {{\color{\colorMATH}\ensuremath{\gamma _{1} \vdash  {\begingroup\renewcommand\colorMATH{\colorMATHB}\renewcommand\colorSYNTAX{\colorSYNTAXB}{{\color{\colorMATH}\ensuremath{\se_{2}}}}\endgroup } \Downarrow ^{j_{2}} {\begingroup\renewcommand\colorMATH{\colorMATHB}\renewcommand\colorSYNTAX{\colorSYNTAXB}{{\color{\colorMATH}\ensuremath{r_{2 1}}}}\endgroup }}}} \pthen {{\color{\colorMATH}\ensuremath{\gamma _{2} \vdash  {\begingroup\renewcommand\colorMATH{\colorMATHB}\renewcommand\colorSYNTAX{\colorSYNTAXB}{{\color{\colorMATH}\ensuremath{\se_{2}}}}\endgroup } \Downarrow ^{\pj} {\begingroup\renewcommand\colorMATH{\colorMATHB}\renewcommand\colorSYNTAX{\colorSYNTAXB}{{\color{\colorMATH}\ensuremath{r_{2 2}}}}\endgroup }}}} \pand {{\color{\colorMATH}\ensuremath{({\begingroup\renewcommand\colorMATH{\colorMATHB}\renewcommand\colorSYNTAX{\colorSYNTAXB}{{\color{\colorMATH}\ensuremath{r_{2 1}}}}\endgroup }, {\begingroup\renewcommand\colorMATH{\colorMATHB}\renewcommand\colorSYNTAX{\colorSYNTAXB}{{\color{\colorMATH}\ensuremath{r_{2 2}}}}\endgroup }) \in  {\mathcal{V}}^{k-j_{1}-j_{2}}_{{\begingroup\renewcommand\colorMATH{\colorMATHB}\renewcommand\colorSYNTAX{\colorSYNTAXB}{{\color{\colorMATH}\ensuremath{\Distance'}}}\endgroup }\mathord{\cdotp }{\begingroup\renewcommand\colorMATH{\colorMATHB}\renewcommand\colorSYNTAX{\colorSYNTAXB}{{\color{\colorMATH}\ensuremath{\sS_{2}}}}\endgroup }}\llbracket {\begingroup\renewcommand\colorMATH{\colorMATHA}\renewcommand\colorSYNTAX{\colorSYNTAXA}{{\color{\colorSYNTAX}\texttt{{\ensuremath{{\mathbb{R}}}}}}}\endgroup }\rrbracket }}}, where {{\color{\colorMATH}\ensuremath{| {\begingroup\renewcommand\colorMATH{\colorMATHB}\renewcommand\colorSYNTAX{\colorSYNTAXB}{{\color{\colorMATH}\ensuremath{r_{2 1}}}}\endgroup } - {\begingroup\renewcommand\colorMATH{\colorMATHB}\renewcommand\colorSYNTAX{\colorSYNTAXB}{{\color{\colorMATH}\ensuremath{r_{2 2}}}}\endgroup }| \leq  {\begingroup\renewcommand\colorMATH{\colorMATHB}\renewcommand\colorSYNTAX{\colorSYNTAXB}{{\color{\colorMATH}\ensuremath{\Distance'}}}\endgroup }\mathord{\cdotp }{\begingroup\renewcommand\colorMATH{\colorMATHB}\renewcommand\colorSYNTAX{\colorSYNTAXB}{{\color{\colorMATH}\ensuremath{\sS_{2}}}}\endgroup }}}}.

      Then if {{\color{\colorMATH}\ensuremath{\gamma _{1}\vdash {\begingroup\renewcommand\colorMATH{\colorMATHB}\renewcommand\colorSYNTAX{\colorSYNTAXB}{{\color{\colorMATH}\ensuremath{\se_{1}}}}\endgroup } * {\begingroup\renewcommand\colorMATH{\colorMATHB}\renewcommand\colorSYNTAX{\colorSYNTAXB}{{\color{\colorMATH}\ensuremath{\se_{2}}}}\endgroup } \Downarrow ^{j_{1}+j_{2}+1} r'_{1}}}} and {{\color{\colorMATH}\ensuremath{\gamma _{2}\vdash {\begingroup\renewcommand\colorMATH{\colorMATHB}\renewcommand\colorSYNTAX{\colorSYNTAXB}{{\color{\colorMATH}\ensuremath{\se_{1}}}}\endgroup } * {\begingroup\renewcommand\colorMATH{\colorMATHB}\renewcommand\colorSYNTAX{\colorSYNTAXB}{{\color{\colorMATH}\ensuremath{\se_{2}}}}\endgroup } \Downarrow ^{j_{1}+j_{2}+1} r'_{2}}}}, where {{\color{\colorMATH}\ensuremath{r'_{1} = {\begingroup\renewcommand\colorMATH{\colorMATHB}\renewcommand\colorSYNTAX{\colorSYNTAXB}{{\color{\colorMATH}\ensuremath{r_{1 1}}}}\endgroup } * {\begingroup\renewcommand\colorMATH{\colorMATHB}\renewcommand\colorSYNTAX{\colorSYNTAXB}{{\color{\colorMATH}\ensuremath{r_{2 1}}}}\endgroup }}}} and {{\color{\colorMATH}\ensuremath{r'_{2} = {\begingroup\renewcommand\colorMATH{\colorMATHB}\renewcommand\colorSYNTAX{\colorSYNTAXB}{{\color{\colorMATH}\ensuremath{r_{1 2}}}}\endgroup } * {\begingroup\renewcommand\colorMATH{\colorMATHB}\renewcommand\colorSYNTAX{\colorSYNTAXB}{{\color{\colorMATH}\ensuremath{r_{2 2}}}}\endgroup }}}}, we have to prove that 
      {{\color{\colorMATH}\ensuremath{({\begingroup\renewcommand\colorMATH{\colorMATHB}\renewcommand\colorSYNTAX{\colorSYNTAXB}{{\color{\colorMATH}\ensuremath{r'_{1}}}}\endgroup }, {\begingroup\renewcommand\colorMATH{\colorMATHB}\renewcommand\colorSYNTAX{\colorSYNTAXB}{{\color{\colorMATH}\ensuremath{r'_{2}}}}\endgroup }) \in  {\mathcal{V}}{\mathcal{V}}^{k-j_{1}-j_{2}-1}_{{\begingroup\renewcommand\colorMATH{\colorMATHB}\renewcommand\colorSYNTAX{\colorSYNTAXB}{{\color{\colorMATH}\ensuremath{\Distance'}}}\endgroup }\mathord{\cdotp }{\begingroup\renewcommand\colorMATH{\colorMATHB}\renewcommand\colorSYNTAX{\colorSYNTAXB}{{\color{\colorMATH}\ensuremath{\infty }}}\endgroup }({\begingroup\renewcommand\colorMATH{\colorMATHB}\renewcommand\colorSYNTAX{\colorSYNTAXB}{{\color{\colorMATH}\ensuremath{\sS_{1}}}}\endgroup } + {\begingroup\renewcommand\colorMATH{\colorMATHB}\renewcommand\colorSYNTAX{\colorSYNTAXB}{{\color{\colorMATH}\ensuremath{\sS_{2}}}}\endgroup })}\llbracket {\begingroup\renewcommand\colorMATH{\colorMATHA}\renewcommand\colorSYNTAX{\colorSYNTAXA}{{\color{\colorSYNTAX}\texttt{{\ensuremath{{\mathbb{R}}}}}}}\endgroup }\rrbracket }}}, i.e.  {{\color{\colorMATH}\ensuremath{|({\begingroup\renewcommand\colorMATH{\colorMATHB}\renewcommand\colorSYNTAX{\colorSYNTAXB}{{\color{\colorMATH}\ensuremath{r_{1 1}}}}\endgroup }\mathord{\cdotp }{\begingroup\renewcommand\colorMATH{\colorMATHB}\renewcommand\colorSYNTAX{\colorSYNTAXB}{{\color{\colorMATH}\ensuremath{r_{2 1}}}}\endgroup })-({\begingroup\renewcommand\colorMATH{\colorMATHB}\renewcommand\colorSYNTAX{\colorSYNTAXB}{{\color{\colorMATH}\ensuremath{r_{1 2}}}}\endgroup } * {\begingroup\renewcommand\colorMATH{\colorMATHB}\renewcommand\colorSYNTAX{\colorSYNTAXB}{{\color{\colorMATH}\ensuremath{r_{2 2}}}}\endgroup })| \leq  {\begingroup\renewcommand\colorMATH{\colorMATHB}\renewcommand\colorSYNTAX{\colorSYNTAXB}{{\color{\colorMATH}\ensuremath{\Distance'}}}\endgroup }\mathord{\cdotp }{\begingroup\renewcommand\colorMATH{\colorMATHB}\renewcommand\colorSYNTAX{\colorSYNTAXB}{{\color{\colorMATH}\ensuremath{\infty }}}\endgroup }({\begingroup\renewcommand\colorMATH{\colorMATHB}\renewcommand\colorSYNTAX{\colorSYNTAXB}{{\color{\colorMATH}\ensuremath{\sS_{1}}}}\endgroup } + {\begingroup\renewcommand\colorMATH{\colorMATHB}\renewcommand\colorSYNTAX{\colorSYNTAXB}{{\color{\colorMATH}\ensuremath{\sS_{2}}}}\endgroup })}}}.\\
      Notice that {{\color{\colorMATH}\ensuremath{{\begingroup\renewcommand\colorMATH{\colorMATHB}\renewcommand\colorSYNTAX{\colorSYNTAXB}{{\color{\colorMATH}\ensuremath{\Distance'}}}\endgroup }\mathord{\cdotp }{\begingroup\renewcommand\colorMATH{\colorMATHB}\renewcommand\colorSYNTAX{\colorSYNTAXB}{{\color{\colorMATH}\ensuremath{\infty }}}\endgroup }({\begingroup\renewcommand\colorMATH{\colorMATHB}\renewcommand\colorSYNTAX{\colorSYNTAXB}{{\color{\colorMATH}\ensuremath{\sS_{1}}}}\endgroup } + {\begingroup\renewcommand\colorMATH{\colorMATHB}\renewcommand\colorSYNTAX{\colorSYNTAXB}{{\color{\colorMATH}\ensuremath{\sS_{2}}}}\endgroup }) = {\begingroup\renewcommand\colorMATH{\colorMATHB}\renewcommand\colorSYNTAX{\colorSYNTAXB}{{\color{\colorMATH}\ensuremath{\infty }}}\endgroup } ({\begingroup\renewcommand\colorMATH{\colorMATHB}\renewcommand\colorSYNTAX{\colorSYNTAXB}{{\color{\colorMATH}\ensuremath{\Distance'}}}\endgroup }\mathord{\cdotp }({\begingroup\renewcommand\colorMATH{\colorMATHB}\renewcommand\colorSYNTAX{\colorSYNTAXB}{{\color{\colorMATH}\ensuremath{\sS_{1}}}}\endgroup } + {\begingroup\renewcommand\colorMATH{\colorMATHB}\renewcommand\colorSYNTAX{\colorSYNTAXB}{{\color{\colorMATH}\ensuremath{\sS_{2}}}}\endgroup })) = s}}}. There are two cases to analyze, where if {{\color{\colorMATH}\ensuremath{s = {\begingroup\renewcommand\colorMATH{\colorMATHB}\renewcommand\colorSYNTAX{\colorSYNTAXB}{{\color{\colorMATH}\ensuremath{\infty }}}\endgroup }}}} then the result holds immediately. 
      Let us suppose that {{\color{\colorMATH}\ensuremath{s = 0}}}.
      Then by Lemma~\ref{lm:associativity-inst}, {{\color{\colorMATH}\ensuremath{{\begingroup\renewcommand\colorMATH{\colorMATHB}\renewcommand\colorSYNTAX{\colorSYNTAXB}{{\color{\colorMATH}\ensuremath{\Distance'}}}\endgroup }\mathord{\cdotp }({\begingroup\renewcommand\colorMATH{\colorMATHB}\renewcommand\colorSYNTAX{\colorSYNTAXB}{{\color{\colorMATH}\ensuremath{\sS_{1}}}}\endgroup } + {\begingroup\renewcommand\colorMATH{\colorMATHB}\renewcommand\colorSYNTAX{\colorSYNTAXB}{{\color{\colorMATH}\ensuremath{\sS_{2}}}}\endgroup }) = {\begingroup\renewcommand\colorMATH{\colorMATHB}\renewcommand\colorSYNTAX{\colorSYNTAXB}{{\color{\colorMATH}\ensuremath{\Distance'}}}\endgroup }\mathord{\cdotp }{\begingroup\renewcommand\colorMATH{\colorMATHB}\renewcommand\colorSYNTAX{\colorSYNTAXB}{{\color{\colorMATH}\ensuremath{\sS_{1}}}}\endgroup } + {\begingroup\renewcommand\colorMATH{\colorMATHB}\renewcommand\colorSYNTAX{\colorSYNTAXB}{{\color{\colorMATH}\ensuremath{\Distance'}}}\endgroup }\mathord{\cdotp }{\begingroup\renewcommand\colorMATH{\colorMATHB}\renewcommand\colorSYNTAX{\colorSYNTAXB}{{\color{\colorMATH}\ensuremath{\sS_{1}}}}\endgroup }}}}, also notice that {{\color{\colorMATH}\ensuremath{|{\begingroup\renewcommand\colorMATH{\colorMATHB}\renewcommand\colorSYNTAX{\colorSYNTAXB}{{\color{\colorMATH}\ensuremath{r_{1 1}}}}\endgroup } - {\begingroup\renewcommand\colorMATH{\colorMATHB}\renewcommand\colorSYNTAX{\colorSYNTAXB}{{\color{\colorMATH}\ensuremath{r_{1 2}}}}\endgroup }| \leq  {\begingroup\renewcommand\colorMATH{\colorMATHB}\renewcommand\colorSYNTAX{\colorSYNTAXB}{{\color{\colorMATH}\ensuremath{\Distance'}}}\endgroup }\mathord{\cdotp }{\begingroup\renewcommand\colorMATH{\colorMATHB}\renewcommand\colorSYNTAX{\colorSYNTAXB}{{\color{\colorMATH}\ensuremath{\sS_{1}}}}\endgroup }}}} and {{\color{\colorMATH}\ensuremath{| {\begingroup\renewcommand\colorMATH{\colorMATHB}\renewcommand\colorSYNTAX{\colorSYNTAXB}{{\color{\colorMATH}\ensuremath{r_{2 1}}}}\endgroup } - {\begingroup\renewcommand\colorMATH{\colorMATHB}\renewcommand\colorSYNTAX{\colorSYNTAXB}{{\color{\colorMATH}\ensuremath{r_{2 2}}}}\endgroup }| \leq  {\begingroup\renewcommand\colorMATH{\colorMATHB}\renewcommand\colorSYNTAX{\colorSYNTAXB}{{\color{\colorMATH}\ensuremath{\Distance'}}}\endgroup }\mathord{\cdotp }{\begingroup\renewcommand\colorMATH{\colorMATHB}\renewcommand\colorSYNTAX{\colorSYNTAXB}{{\color{\colorMATH}\ensuremath{\sS_{2}}}}\endgroup }}}}, then {{\color{\colorMATH}\ensuremath{0 \leq  {\begingroup\renewcommand\colorMATH{\colorMATHB}\renewcommand\colorSYNTAX{\colorSYNTAXB}{{\color{\colorMATH}\ensuremath{\Distance'}}}\endgroup }\mathord{\cdotp }{\begingroup\renewcommand\colorMATH{\colorMATHB}\renewcommand\colorSYNTAX{\colorSYNTAXB}{{\color{\colorMATH}\ensuremath{\sS_{1}}}}\endgroup }}}} and {{\color{\colorMATH}\ensuremath{0 \leq  {\begingroup\renewcommand\colorMATH{\colorMATHB}\renewcommand\colorSYNTAX{\colorSYNTAXB}{{\color{\colorMATH}\ensuremath{\Distance'}}}\endgroup }\mathord{\cdotp }{\begingroup\renewcommand\colorMATH{\colorMATHB}\renewcommand\colorSYNTAX{\colorSYNTAXB}{{\color{\colorMATH}\ensuremath{\sS_{2}}}}\endgroup }}}}. Therefore if {{\color{\colorMATH}\ensuremath{{\begingroup\renewcommand\colorMATH{\colorMATHB}\renewcommand\colorSYNTAX{\colorSYNTAXB}{{\color{\colorMATH}\ensuremath{\Distance'}}}\endgroup }\mathord{\cdotp }{\begingroup\renewcommand\colorMATH{\colorMATHB}\renewcommand\colorSYNTAX{\colorSYNTAXB}{{\color{\colorMATH}\ensuremath{\sS_{1}}}}\endgroup } + {\begingroup\renewcommand\colorMATH{\colorMATHB}\renewcommand\colorSYNTAX{\colorSYNTAXB}{{\color{\colorMATH}\ensuremath{\Distance'}}}\endgroup }\mathord{\cdotp }{\begingroup\renewcommand\colorMATH{\colorMATHB}\renewcommand\colorSYNTAX{\colorSYNTAXB}{{\color{\colorMATH}\ensuremath{\sS_{1}}}}\endgroup } = 0}}}, then {{\color{\colorMATH}\ensuremath{{\begingroup\renewcommand\colorMATH{\colorMATHB}\renewcommand\colorSYNTAX{\colorSYNTAXB}{{\color{\colorMATH}\ensuremath{\Distance'}}}\endgroup }\mathord{\cdotp }{\begingroup\renewcommand\colorMATH{\colorMATHB}\renewcommand\colorSYNTAX{\colorSYNTAXB}{{\color{\colorMATH}\ensuremath{\sS_{1}}}}\endgroup } = 0}}} and {{\color{\colorMATH}\ensuremath{{\begingroup\renewcommand\colorMATH{\colorMATHB}\renewcommand\colorSYNTAX{\colorSYNTAXB}{{\color{\colorMATH}\ensuremath{\Distance'}}}\endgroup }\mathord{\cdotp }{\begingroup\renewcommand\colorMATH{\colorMATHB}\renewcommand\colorSYNTAX{\colorSYNTAXB}{{\color{\colorMATH}\ensuremath{\sS_{2}}}}\endgroup } = 0}}}.
      This means that {{\color{\colorMATH}\ensuremath{{\begingroup\renewcommand\colorMATH{\colorMATHB}\renewcommand\colorSYNTAX{\colorSYNTAXB}{{\color{\colorMATH}\ensuremath{r_{1 1}}}}\endgroup } = {\begingroup\renewcommand\colorMATH{\colorMATHB}\renewcommand\colorSYNTAX{\colorSYNTAXB}{{\color{\colorMATH}\ensuremath{r_{1 2}}}}\endgroup }}}} and {{\color{\colorMATH}\ensuremath{{\begingroup\renewcommand\colorMATH{\colorMATHB}\renewcommand\colorSYNTAX{\colorSYNTAXB}{{\color{\colorMATH}\ensuremath{r_{2 1}}}}\endgroup } = {\begingroup\renewcommand\colorMATH{\colorMATHB}\renewcommand\colorSYNTAX{\colorSYNTAXB}{{\color{\colorMATH}\ensuremath{r_{2 2}}}}\endgroup }}}}, thus {{\color{\colorMATH}\ensuremath{{\begingroup\renewcommand\colorMATH{\colorMATHB}\renewcommand\colorSYNTAX{\colorSYNTAXB}{{\color{\colorMATH}\ensuremath{r_{1 1}}}}\endgroup }*{\begingroup\renewcommand\colorMATH{\colorMATHB}\renewcommand\colorSYNTAX{\colorSYNTAXB}{{\color{\colorMATH}\ensuremath{r_{2 1}}}}\endgroup } = {\begingroup\renewcommand\colorMATH{\colorMATHB}\renewcommand\colorSYNTAX{\colorSYNTAXB}{{\color{\colorMATH}\ensuremath{r_{1 2}}}}\endgroup }*{\begingroup\renewcommand\colorMATH{\colorMATHB}\renewcommand\colorSYNTAX{\colorSYNTAXB}{{\color{\colorMATH}\ensuremath{r_{2 2}}}}\endgroup }}}}, {{\color{\colorMATH}\ensuremath{|({\begingroup\renewcommand\colorMATH{\colorMATHB}\renewcommand\colorSYNTAX{\colorSYNTAXB}{{\color{\colorMATH}\ensuremath{r_{1 1}}}}\endgroup }*{\begingroup\renewcommand\colorMATH{\colorMATHB}\renewcommand\colorSYNTAX{\colorSYNTAXB}{{\color{\colorMATH}\ensuremath{r_{2 1}}}}\endgroup })-({\begingroup\renewcommand\colorMATH{\colorMATHB}\renewcommand\colorSYNTAX{\colorSYNTAXB}{{\color{\colorMATH}\ensuremath{r_{1 2}}}}\endgroup }*{\begingroup\renewcommand\colorMATH{\colorMATHB}\renewcommand\colorSYNTAX{\colorSYNTAXB}{{\color{\colorMATH}\ensuremath{r_{2 2}}}}\endgroup })| = 0}}}, and the result holds.
    \end{subproof}
  \item  {{\color{\colorMATH}\ensuremath{\Gamma ; {\begingroup\renewcommand\colorMATH{\colorMATHB}\renewcommand\colorSYNTAX{\colorSYNTAXB}{{\color{\colorMATH}\ensuremath{\Distance}}}\endgroup } \vdash  {\begingroup\renewcommand\colorMATH{\colorMATHB}\renewcommand\colorSYNTAX{\colorSYNTAXB}{{\color{\colorMATH}\ensuremath{\se_{1}}}}\endgroup } \leq  {\begingroup\renewcommand\colorMATH{\colorMATHB}\renewcommand\colorSYNTAX{\colorSYNTAXB}{{\color{\colorMATH}\ensuremath{\se_{2}}}}\endgroup } \mathrel{:} {\mathbb{B}} \mathrel{;} {\begingroup\renewcommand\colorMATH{\colorMATHB}\renewcommand\colorSYNTAX{\colorSYNTAXB}{{\color{\colorMATH}\ensuremath{\infty }}}\endgroup }({\begingroup\renewcommand\colorMATH{\colorMATHB}\renewcommand\colorSYNTAX{\colorSYNTAXB}{{\color{\colorMATH}\ensuremath{\sS_{1}}}}\endgroup } + {\begingroup\renewcommand\colorMATH{\colorMATHB}\renewcommand\colorSYNTAX{\colorSYNTAXB}{{\color{\colorMATH}\ensuremath{\sS_{2}}}}\endgroup })}}} 
    \begin{subproof} 
      We have to prove that {{\color{\colorMATH}\ensuremath{\forall k, \forall (\gamma _{1},\gamma _{2}) \in  {\mathcal{G}}_{{\begingroup\renewcommand\colorMATH{\colorMATHB}\renewcommand\colorSYNTAX{\colorSYNTAXB}{{\color{\colorMATH}\ensuremath{\Distance'}}}\endgroup }}^{\kg}\llbracket \Gamma \rrbracket , (\gamma _{1}\vdash {\begingroup\renewcommand\colorMATH{\colorMATHB}\renewcommand\colorSYNTAX{\colorSYNTAXB}{{\color{\colorMATH}\ensuremath{\se_{1}}}}\endgroup } \leq  {\begingroup\renewcommand\colorMATH{\colorMATHB}\renewcommand\colorSYNTAX{\colorSYNTAXB}{{\color{\colorMATH}\ensuremath{\se_{2}}}}\endgroup },\gamma _{2}\vdash {\begingroup\renewcommand\colorMATH{\colorMATHB}\renewcommand\colorSYNTAX{\colorSYNTAXB}{{\color{\colorMATH}\ensuremath{\se_{1}}}}\endgroup } \leq  {\begingroup\renewcommand\colorMATH{\colorMATHB}\renewcommand\colorSYNTAX{\colorSYNTAXB}{{\color{\colorMATH}\ensuremath{\se_{2}}}}\endgroup }) \in  {\mathcal{E}}^{k}_{{\begingroup\renewcommand\colorMATH{\colorMATHB}\renewcommand\colorSYNTAX{\colorSYNTAXB}{{\color{\colorMATH}\ensuremath{\Distance'}}}\endgroup }\mathord{\cdotp }\infty ({\begingroup\renewcommand\colorMATH{\colorMATHB}\renewcommand\colorSYNTAX{\colorSYNTAXB}{{\color{\colorMATH}\ensuremath{\sS_{1}}}}\endgroup } + {\begingroup\renewcommand\colorMATH{\colorMATHB}\renewcommand\colorSYNTAX{\colorSYNTAXB}{{\color{\colorMATH}\ensuremath{\sS_{2}}}}\endgroup })}\llbracket {\begingroup\renewcommand\colorMATH{\colorMATHB}\renewcommand\colorSYNTAX{\colorSYNTAXB}{{\color{\colorMATH}\ensuremath{\Distance'}}}\endgroup }({\mathbb{B}})\rrbracket }}}, for {{\color{\colorMATH}\ensuremath{{\begingroup\renewcommand\colorMATH{\colorMATHB}\renewcommand\colorSYNTAX{\colorSYNTAXB}{{\color{\colorMATH}\ensuremath{\Distance'}}}\endgroup } \sqsubseteq  {\begingroup\renewcommand\colorMATH{\colorMATHB}\renewcommand\colorSYNTAX{\colorSYNTAXB}{{\color{\colorMATH}\ensuremath{\Distance}}}\endgroup }}}}.
      By induction hypotheses \\
      {{\color{\colorMATH}\ensuremath{\Gamma  \vdash  {\begingroup\renewcommand\colorMATH{\colorMATHB}\renewcommand\colorSYNTAX{\colorSYNTAXB}{{\color{\colorMATH}\ensuremath{\se_{1}}}}\endgroup } \mathrel{:} {\begingroup\renewcommand\colorMATH{\colorMATHA}\renewcommand\colorSYNTAX{\colorSYNTAXA}{{\color{\colorSYNTAX}\texttt{{\ensuremath{{\mathbb{R}}}}}}}\endgroup } \mathrel{;} {\begingroup\renewcommand\colorMATH{\colorMATHB}\renewcommand\colorSYNTAX{\colorSYNTAXB}{{\color{\colorMATH}\ensuremath{\sS_{1}}}}\endgroup } \Rightarrow  (\gamma _{1} \vdash  {\begingroup\renewcommand\colorMATH{\colorMATHB}\renewcommand\colorSYNTAX{\colorSYNTAXB}{{\color{\colorMATH}\ensuremath{\se_{1}}}}\endgroup },\gamma _{2} \vdash  {\begingroup\renewcommand\colorMATH{\colorMATHB}\renewcommand\colorSYNTAX{\colorSYNTAXB}{{\color{\colorMATH}\ensuremath{\se_{1}}}}\endgroup }) \in  {\mathcal{E}}^{k}_{{\begingroup\renewcommand\colorMATH{\colorMATHB}\renewcommand\colorSYNTAX{\colorSYNTAXB}{{\color{\colorMATH}\ensuremath{\Distance'}}}\endgroup }\mathord{\cdotp }{\begingroup\renewcommand\colorMATH{\colorMATHB}\renewcommand\colorSYNTAX{\colorSYNTAXB}{{\color{\colorMATH}\ensuremath{\sS_{1}}}}\endgroup }}\llbracket {\begingroup\renewcommand\colorMATH{\colorMATHB}\renewcommand\colorSYNTAX{\colorSYNTAXB}{{\color{\colorMATH}\ensuremath{\Distance'}}}\endgroup }({\mathbb{B}})\rrbracket }}} and 
      {{\color{\colorMATH}\ensuremath{\Gamma  \vdash  {\begingroup\renewcommand\colorMATH{\colorMATHB}\renewcommand\colorSYNTAX{\colorSYNTAXB}{{\color{\colorMATH}\ensuremath{\se_{2}}}}\endgroup } \mathrel{:} {\begingroup\renewcommand\colorMATH{\colorMATHA}\renewcommand\colorSYNTAX{\colorSYNTAXA}{{\color{\colorSYNTAX}\texttt{{\ensuremath{{\mathbb{R}}}}}}}\endgroup } \mathrel{;} {\begingroup\renewcommand\colorMATH{\colorMATHB}\renewcommand\colorSYNTAX{\colorSYNTAXB}{{\color{\colorMATH}\ensuremath{\sS_{2}}}}\endgroup } \Rightarrow  (\gamma _{1} \vdash  {\begingroup\renewcommand\colorMATH{\colorMATHB}\renewcommand\colorSYNTAX{\colorSYNTAXB}{{\color{\colorMATH}\ensuremath{\se_{2}}}}\endgroup },\gamma _{2} \vdash  {\begingroup\renewcommand\colorMATH{\colorMATHB}\renewcommand\colorSYNTAX{\colorSYNTAXB}{{\color{\colorMATH}\ensuremath{\se_{2}}}}\endgroup }) \in  {\mathcal{E}}^{k-j_{1}}_{{\begingroup\renewcommand\colorMATH{\colorMATHB}\renewcommand\colorSYNTAX{\colorSYNTAXB}{{\color{\colorMATH}\ensuremath{\Distance'}}}\endgroup }\mathord{\cdotp }{\begingroup\renewcommand\colorMATH{\colorMATHB}\renewcommand\colorSYNTAX{\colorSYNTAXB}{{\color{\colorMATH}\ensuremath{\sS_{2}}}}\endgroup }}\llbracket {\begingroup\renewcommand\colorMATH{\colorMATHB}\renewcommand\colorSYNTAX{\colorSYNTAXB}{{\color{\colorMATH}\ensuremath{\Distance'}}}\endgroup }({\mathbb{B}})\rrbracket }}}, for some {{\color{\colorMATH}\ensuremath{j_{1} \leq  k}}}.
      By unfolding {{\color{\colorMATH}\ensuremath{(\gamma _{1} \vdash  {\begingroup\renewcommand\colorMATH{\colorMATHB}\renewcommand\colorSYNTAX{\colorSYNTAXB}{{\color{\colorMATH}\ensuremath{\se_{1}}}}\endgroup },\gamma _{2} \vdash  {\begingroup\renewcommand\colorMATH{\colorMATHB}\renewcommand\colorSYNTAX{\colorSYNTAXB}{{\color{\colorMATH}\ensuremath{\se_{1}}}}\endgroup }) \in  {\mathcal{E}}^{k}_{{\begingroup\renewcommand\colorMATH{\colorMATHB}\renewcommand\colorSYNTAX{\colorSYNTAXB}{{\color{\colorMATH}\ensuremath{\Distance'}}}\endgroup }\mathord{\cdotp }{\begingroup\renewcommand\colorMATH{\colorMATHB}\renewcommand\colorSYNTAX{\colorSYNTAXB}{{\color{\colorMATH}\ensuremath{\sS_{1}}}}\endgroup }}\llbracket {\begingroup\renewcommand\colorMATH{\colorMATHB}\renewcommand\colorSYNTAX{\colorSYNTAXB}{{\color{\colorMATH}\ensuremath{\Distance'}}}\endgroup }({\mathbb{B}})\rrbracket }}},
      we know that if {{\color{\colorMATH}\ensuremath{\gamma _{1} \vdash  {\begingroup\renewcommand\colorMATH{\colorMATHB}\renewcommand\colorSYNTAX{\colorSYNTAXB}{{\color{\colorMATH}\ensuremath{\se_{1}}}}\endgroup } \Downarrow ^{j_{1}} {\begingroup\renewcommand\colorMATH{\colorMATHB}\renewcommand\colorSYNTAX{\colorSYNTAXB}{{\color{\colorMATH}\ensuremath{r_{1 1}}}}\endgroup }}}} then {{\color{\colorMATH}\ensuremath{\gamma _{2} \vdash  {\begingroup\renewcommand\colorMATH{\colorMATHB}\renewcommand\colorSYNTAX{\colorSYNTAXB}{{\color{\colorMATH}\ensuremath{\se_{1}}}}\endgroup } \Downarrow ^{*} {\begingroup\renewcommand\colorMATH{\colorMATHB}\renewcommand\colorSYNTAX{\colorSYNTAXB}{{\color{\colorMATH}\ensuremath{r_{1 2}}}}\endgroup }}}} and {{\color{\colorMATH}\ensuremath{({\begingroup\renewcommand\colorMATH{\colorMATHB}\renewcommand\colorSYNTAX{\colorSYNTAXB}{{\color{\colorMATH}\ensuremath{r_{1 1}}}}\endgroup }, {\begingroup\renewcommand\colorMATH{\colorMATHB}\renewcommand\colorSYNTAX{\colorSYNTAXB}{{\color{\colorMATH}\ensuremath{r_{1 2}}}}\endgroup }) \in  {\mathcal{V}}^{k-j_{1}}_{{\begingroup\renewcommand\colorMATH{\colorMATHB}\renewcommand\colorSYNTAX{\colorSYNTAXB}{{\color{\colorMATH}\ensuremath{\Distance'}}}\endgroup }\mathord{\cdotp }{\begingroup\renewcommand\colorMATH{\colorMATHB}\renewcommand\colorSYNTAX{\colorSYNTAXB}{{\color{\colorMATH}\ensuremath{\sS_{1}}}}\endgroup }}\llbracket {\begingroup\renewcommand\colorMATH{\colorMATHA}\renewcommand\colorSYNTAX{\colorSYNTAXA}{{\color{\colorSYNTAX}\texttt{{\ensuremath{{\mathbb{R}}}}}}}\endgroup }\rrbracket }}}, where {{\color{\colorMATH}\ensuremath{|{\begingroup\renewcommand\colorMATH{\colorMATHB}\renewcommand\colorSYNTAX{\colorSYNTAXB}{{\color{\colorMATH}\ensuremath{r_{1 1}}}}\endgroup } - {\begingroup\renewcommand\colorMATH{\colorMATHB}\renewcommand\colorSYNTAX{\colorSYNTAXB}{{\color{\colorMATH}\ensuremath{r_{1 2}}}}\endgroup }| \leq  {\begingroup\renewcommand\colorMATH{\colorMATHB}\renewcommand\colorSYNTAX{\colorSYNTAXB}{{\color{\colorMATH}\ensuremath{\Distance'}}}\endgroup }\mathord{\cdotp }{\begingroup\renewcommand\colorMATH{\colorMATHB}\renewcommand\colorSYNTAX{\colorSYNTAXB}{{\color{\colorMATH}\ensuremath{\sS_{1}}}}\endgroup }}}}.\\
      Also, by unfolding {{\color{\colorMATH}\ensuremath{(\gamma _{1} \vdash  {\begingroup\renewcommand\colorMATH{\colorMATHB}\renewcommand\colorSYNTAX{\colorSYNTAXB}{{\color{\colorMATH}\ensuremath{\se_{2}}}}\endgroup },\gamma _{2} \vdash  {\begingroup\renewcommand\colorMATH{\colorMATHB}\renewcommand\colorSYNTAX{\colorSYNTAXB}{{\color{\colorMATH}\ensuremath{\se_{2}}}}\endgroup }) \in  {\mathcal{V}}^{k-j_{1}}_{{\begingroup\renewcommand\colorMATH{\colorMATHB}\renewcommand\colorSYNTAX{\colorSYNTAXB}{{\color{\colorMATH}\ensuremath{\Distance'}}}\endgroup }\mathord{\cdotp }{\begingroup\renewcommand\colorMATH{\colorMATHB}\renewcommand\colorSYNTAX{\colorSYNTAXB}{{\color{\colorMATH}\ensuremath{\sS_{2}}}}\endgroup }}\llbracket {\begingroup\renewcommand\colorMATH{\colorMATHB}\renewcommand\colorSYNTAX{\colorSYNTAXB}{{\color{\colorMATH}\ensuremath{\Distance'}}}\endgroup }({\mathbb{B}})\rrbracket }}}, if {{\color{\colorMATH}\ensuremath{\gamma _{1} \vdash  {\begingroup\renewcommand\colorMATH{\colorMATHB}\renewcommand\colorSYNTAX{\colorSYNTAXB}{{\color{\colorMATH}\ensuremath{\se_{2}}}}\endgroup } \Downarrow ^{j_{2}} {\begingroup\renewcommand\colorMATH{\colorMATHB}\renewcommand\colorSYNTAX{\colorSYNTAXB}{{\color{\colorMATH}\ensuremath{r_{2 1}}}}\endgroup }}}} \pthen {{\color{\colorMATH}\ensuremath{\gamma _{2} \vdash  {\begingroup\renewcommand\colorMATH{\colorMATHB}\renewcommand\colorSYNTAX{\colorSYNTAXB}{{\color{\colorMATH}\ensuremath{\se_{2}}}}\endgroup } \Downarrow ^{\pj} {\begingroup\renewcommand\colorMATH{\colorMATHB}\renewcommand\colorSYNTAX{\colorSYNTAXB}{{\color{\colorMATH}\ensuremath{r_{2 2}}}}\endgroup }}}} \pand {{\color{\colorMATH}\ensuremath{({\begingroup\renewcommand\colorMATH{\colorMATHB}\renewcommand\colorSYNTAX{\colorSYNTAXB}{{\color{\colorMATH}\ensuremath{r_{2 1}}}}\endgroup }, {\begingroup\renewcommand\colorMATH{\colorMATHB}\renewcommand\colorSYNTAX{\colorSYNTAXB}{{\color{\colorMATH}\ensuremath{r_{2 2}}}}\endgroup }) \in  {\mathcal{V}}^{k-j_{1}-j_{2}}_{{\begingroup\renewcommand\colorMATH{\colorMATHB}\renewcommand\colorSYNTAX{\colorSYNTAXB}{{\color{\colorMATH}\ensuremath{\Distance'}}}\endgroup }\mathord{\cdotp }{\begingroup\renewcommand\colorMATH{\colorMATHB}\renewcommand\colorSYNTAX{\colorSYNTAXB}{{\color{\colorMATH}\ensuremath{\sS_{2}}}}\endgroup }}\llbracket {\begingroup\renewcommand\colorMATH{\colorMATHA}\renewcommand\colorSYNTAX{\colorSYNTAXA}{{\color{\colorSYNTAX}\texttt{{\ensuremath{{\mathbb{R}}}}}}}\endgroup }\rrbracket }}}, where {{\color{\colorMATH}\ensuremath{| {\begingroup\renewcommand\colorMATH{\colorMATHB}\renewcommand\colorSYNTAX{\colorSYNTAXB}{{\color{\colorMATH}\ensuremath{r_{2 1}}}}\endgroup } - {\begingroup\renewcommand\colorMATH{\colorMATHB}\renewcommand\colorSYNTAX{\colorSYNTAXB}{{\color{\colorMATH}\ensuremath{r_{2 2}}}}\endgroup }| \leq  {\begingroup\renewcommand\colorMATH{\colorMATHB}\renewcommand\colorSYNTAX{\colorSYNTAXB}{{\color{\colorMATH}\ensuremath{\Distance'}}}\endgroup }\mathord{\cdotp }{\begingroup\renewcommand\colorMATH{\colorMATHB}\renewcommand\colorSYNTAX{\colorSYNTAXB}{{\color{\colorMATH}\ensuremath{\sS_{2}}}}\endgroup }}}}.

      Then if {{\color{\colorMATH}\ensuremath{\gamma _{1}\vdash {\begingroup\renewcommand\colorMATH{\colorMATHB}\renewcommand\colorSYNTAX{\colorSYNTAXB}{{\color{\colorMATH}\ensuremath{\se_{1}}}}\endgroup } \leq  {\begingroup\renewcommand\colorMATH{\colorMATHB}\renewcommand\colorSYNTAX{\colorSYNTAXB}{{\color{\colorMATH}\ensuremath{\se_{2}}}}\endgroup } \Downarrow ^{j_{1}+j_{2}+1} b_{1}}}} and {{\color{\colorMATH}\ensuremath{\gamma _{2}\vdash {\begingroup\renewcommand\colorMATH{\colorMATHB}\renewcommand\colorSYNTAX{\colorSYNTAXB}{{\color{\colorMATH}\ensuremath{\se_{1}}}}\endgroup } \leq  {\begingroup\renewcommand\colorMATH{\colorMATHB}\renewcommand\colorSYNTAX{\colorSYNTAXB}{{\color{\colorMATH}\ensuremath{\se_{2}}}}\endgroup } \Downarrow ^{j_{1}+j_{2}+1} b_{2}}}}, where {{\color{\colorMATH}\ensuremath{b_{i} \in  \{ \inl\hspace*{0.33em}\ttt, \inr\hspace*{0.33em}\ttt\} }}}, we have to prove that 
      {{\color{\colorMATH}\ensuremath{(b_{1}, b_{2}) \in  {\mathcal{V}}^{k-j_{1}-j_{2}-1}_{{\begingroup\renewcommand\colorMATH{\colorMATHB}\renewcommand\colorSYNTAX{\colorSYNTAXB}{{\color{\colorMATH}\ensuremath{\Distance'}}}\endgroup }\mathord{\cdotp }\infty ({\begingroup\renewcommand\colorMATH{\colorMATHB}\renewcommand\colorSYNTAX{\colorSYNTAXB}{{\color{\colorMATH}\ensuremath{\sS_{1}}}}\endgroup } + {\begingroup\renewcommand\colorMATH{\colorMATHB}\renewcommand\colorSYNTAX{\colorSYNTAXB}{{\color{\colorMATH}\ensuremath{\sS_{2}}}}\endgroup })}\llbracket {\mathbb{B}}\rrbracket }}}.\\
      Notice that {{\color{\colorMATH}\ensuremath{{\begingroup\renewcommand\colorMATH{\colorMATHB}\renewcommand\colorSYNTAX{\colorSYNTAXB}{{\color{\colorMATH}\ensuremath{\Distance'}}}\endgroup }\mathord{\cdotp }\infty ({\begingroup\renewcommand\colorMATH{\colorMATHB}\renewcommand\colorSYNTAX{\colorSYNTAXB}{{\color{\colorMATH}\ensuremath{\sS_{1}}}}\endgroup } + {\begingroup\renewcommand\colorMATH{\colorMATHB}\renewcommand\colorSYNTAX{\colorSYNTAXB}{{\color{\colorMATH}\ensuremath{\sS_{2}}}}\endgroup }) = \infty  ({\begingroup\renewcommand\colorMATH{\colorMATHB}\renewcommand\colorSYNTAX{\colorSYNTAXB}{{\color{\colorMATH}\ensuremath{\Distance'}}}\endgroup }\mathord{\cdotp }({\begingroup\renewcommand\colorMATH{\colorMATHB}\renewcommand\colorSYNTAX{\colorSYNTAXB}{{\color{\colorMATH}\ensuremath{\sS_{1}}}}\endgroup } + {\begingroup\renewcommand\colorMATH{\colorMATHB}\renewcommand\colorSYNTAX{\colorSYNTAXB}{{\color{\colorMATH}\ensuremath{\sS_{2}}}}\endgroup })) = s}}}. There are two cases to analyze, if {{\color{\colorMATH}\ensuremath{{\begingroup\renewcommand\colorMATH{\colorMATHB}\renewcommand\colorSYNTAX{\colorSYNTAXB}{{\color{\colorMATH}\ensuremath{\sss}}}\endgroup } = \infty }}} or {{\color{\colorMATH}\ensuremath{{\begingroup\renewcommand\colorMATH{\colorMATHB}\renewcommand\colorSYNTAX{\colorSYNTAXB}{{\color{\colorMATH}\ensuremath{\sss}}}\endgroup } = 0}}}.
      If {{\color{\colorMATH}\ensuremath{{\begingroup\renewcommand\colorMATH{\colorMATHB}\renewcommand\colorSYNTAX{\colorSYNTAXB}{{\color{\colorMATH}\ensuremath{\sss}}}\endgroup } = \infty }}} and {{\color{\colorMATH}\ensuremath{b_{1} \neq  b_{2}}}} then the result holds immediately. If {{\color{\colorMATH}\ensuremath{{\begingroup\renewcommand\colorMATH{\colorMATHB}\renewcommand\colorSYNTAX{\colorSYNTAXB}{{\color{\colorMATH}\ensuremath{\sss}}}\endgroup } = \infty }}} and {{\color{\colorMATH}\ensuremath{b_{1} = b_{2}}}}, let us suppose {{\color{\colorMATH}\ensuremath{b_{1} = \inl\hspace*{0.33em}\ttt}}} (the other case is analogous), then we have to prove that {{\color{\colorMATH}\ensuremath{(\inl\hspace*{0.33em}\ttt, \inl\hspace*{0.33em}\ttt) \in  {\mathcal{V}}_{\infty }^{k-j_{1}-j_{2}-1}\llbracket {{\color{\colorSYNTAX}\texttt{unit}}}\rrbracket }}}, but this is direct as {{\color{\colorMATH}\ensuremath{(\ttt, \ttt) \in  {\mathcal{V}}_{\infty }^{k-j_{1}-j_{2}-1}\llbracket {{\color{\colorSYNTAX}\texttt{unit}}}\rrbracket }}}.
      Let us suppose that {{\color{\colorMATH}\ensuremath{{\begingroup\renewcommand\colorMATH{\colorMATHB}\renewcommand\colorSYNTAX{\colorSYNTAXB}{{\color{\colorMATH}\ensuremath{\sss}}}\endgroup } = 0}}}.
      Then by Lemma~\ref{lm:associativity-inst}, {{\color{\colorMATH}\ensuremath{{\begingroup\renewcommand\colorMATH{\colorMATHB}\renewcommand\colorSYNTAX{\colorSYNTAXB}{{\color{\colorMATH}\ensuremath{\Distance'}}}\endgroup }\mathord{\cdotp }({\begingroup\renewcommand\colorMATH{\colorMATHB}\renewcommand\colorSYNTAX{\colorSYNTAXB}{{\color{\colorMATH}\ensuremath{\sS_{1}}}}\endgroup } + {\begingroup\renewcommand\colorMATH{\colorMATHB}\renewcommand\colorSYNTAX{\colorSYNTAXB}{{\color{\colorMATH}\ensuremath{\sS_{2}}}}\endgroup }) = {\begingroup\renewcommand\colorMATH{\colorMATHB}\renewcommand\colorSYNTAX{\colorSYNTAXB}{{\color{\colorMATH}\ensuremath{\Distance'}}}\endgroup }\mathord{\cdotp }{\begingroup\renewcommand\colorMATH{\colorMATHB}\renewcommand\colorSYNTAX{\colorSYNTAXB}{{\color{\colorMATH}\ensuremath{\sS_{1}}}}\endgroup } + {\begingroup\renewcommand\colorMATH{\colorMATHB}\renewcommand\colorSYNTAX{\colorSYNTAXB}{{\color{\colorMATH}\ensuremath{\Distance'}}}\endgroup }\mathord{\cdotp }{\begingroup\renewcommand\colorMATH{\colorMATHB}\renewcommand\colorSYNTAX{\colorSYNTAXB}{{\color{\colorMATH}\ensuremath{\sS_{1}}}}\endgroup }}}}, also notice that {{\color{\colorMATH}\ensuremath{|{\begingroup\renewcommand\colorMATH{\colorMATHB}\renewcommand\colorSYNTAX{\colorSYNTAXB}{{\color{\colorMATH}\ensuremath{r_{1 1}}}}\endgroup } - {\begingroup\renewcommand\colorMATH{\colorMATHB}\renewcommand\colorSYNTAX{\colorSYNTAXB}{{\color{\colorMATH}\ensuremath{r_{1 2}}}}\endgroup }| \leq  {\begingroup\renewcommand\colorMATH{\colorMATHB}\renewcommand\colorSYNTAX{\colorSYNTAXB}{{\color{\colorMATH}\ensuremath{\Distance'}}}\endgroup }\mathord{\cdotp }{\begingroup\renewcommand\colorMATH{\colorMATHB}\renewcommand\colorSYNTAX{\colorSYNTAXB}{{\color{\colorMATH}\ensuremath{\sS_{1}}}}\endgroup }}}} and {{\color{\colorMATH}\ensuremath{| {\begingroup\renewcommand\colorMATH{\colorMATHB}\renewcommand\colorSYNTAX{\colorSYNTAXB}{{\color{\colorMATH}\ensuremath{r_{2 1}}}}\endgroup } - {\begingroup\renewcommand\colorMATH{\colorMATHB}\renewcommand\colorSYNTAX{\colorSYNTAXB}{{\color{\colorMATH}\ensuremath{r_{2 2}}}}\endgroup }| \leq  {\begingroup\renewcommand\colorMATH{\colorMATHB}\renewcommand\colorSYNTAX{\colorSYNTAXB}{{\color{\colorMATH}\ensuremath{\Distance'}}}\endgroup }\mathord{\cdotp }{\begingroup\renewcommand\colorMATH{\colorMATHB}\renewcommand\colorSYNTAX{\colorSYNTAXB}{{\color{\colorMATH}\ensuremath{\sS_{2}}}}\endgroup }}}}, then {{\color{\colorMATH}\ensuremath{0 \leq  {\begingroup\renewcommand\colorMATH{\colorMATHB}\renewcommand\colorSYNTAX{\colorSYNTAXB}{{\color{\colorMATH}\ensuremath{\Distance'}}}\endgroup }\mathord{\cdotp }{\begingroup\renewcommand\colorMATH{\colorMATHB}\renewcommand\colorSYNTAX{\colorSYNTAXB}{{\color{\colorMATH}\ensuremath{\sS_{1}}}}\endgroup }}}} and {{\color{\colorMATH}\ensuremath{0 \leq  {\begingroup\renewcommand\colorMATH{\colorMATHB}\renewcommand\colorSYNTAX{\colorSYNTAXB}{{\color{\colorMATH}\ensuremath{\Distance'}}}\endgroup }\mathord{\cdotp }{\begingroup\renewcommand\colorMATH{\colorMATHB}\renewcommand\colorSYNTAX{\colorSYNTAXB}{{\color{\colorMATH}\ensuremath{\sS_{2}}}}\endgroup }}}}. Therefore if {{\color{\colorMATH}\ensuremath{{\begingroup\renewcommand\colorMATH{\colorMATHB}\renewcommand\colorSYNTAX{\colorSYNTAXB}{{\color{\colorMATH}\ensuremath{\Distance'}}}\endgroup }\mathord{\cdotp }{\begingroup\renewcommand\colorMATH{\colorMATHB}\renewcommand\colorSYNTAX{\colorSYNTAXB}{{\color{\colorMATH}\ensuremath{\sS_{1}}}}\endgroup } + {\begingroup\renewcommand\colorMATH{\colorMATHB}\renewcommand\colorSYNTAX{\colorSYNTAXB}{{\color{\colorMATH}\ensuremath{\Distance'}}}\endgroup }\mathord{\cdotp }{\begingroup\renewcommand\colorMATH{\colorMATHB}\renewcommand\colorSYNTAX{\colorSYNTAXB}{{\color{\colorMATH}\ensuremath{\sS_{1}}}}\endgroup } = 0}}}, then {{\color{\colorMATH}\ensuremath{{\begingroup\renewcommand\colorMATH{\colorMATHB}\renewcommand\colorSYNTAX{\colorSYNTAXB}{{\color{\colorMATH}\ensuremath{\Distance'}}}\endgroup }\mathord{\cdotp }{\begingroup\renewcommand\colorMATH{\colorMATHB}\renewcommand\colorSYNTAX{\colorSYNTAXB}{{\color{\colorMATH}\ensuremath{\sS_{1}}}}\endgroup } = 0}}} and {{\color{\colorMATH}\ensuremath{{\begingroup\renewcommand\colorMATH{\colorMATHB}\renewcommand\colorSYNTAX{\colorSYNTAXB}{{\color{\colorMATH}\ensuremath{\Distance'}}}\endgroup }\mathord{\cdotp }{\begingroup\renewcommand\colorMATH{\colorMATHB}\renewcommand\colorSYNTAX{\colorSYNTAXB}{{\color{\colorMATH}\ensuremath{\sS_{2}}}}\endgroup } = 0}}}.
      This means that {{\color{\colorMATH}\ensuremath{{\begingroup\renewcommand\colorMATH{\colorMATHB}\renewcommand\colorSYNTAX{\colorSYNTAXB}{{\color{\colorMATH}\ensuremath{r_{1 1}}}}\endgroup } = {\begingroup\renewcommand\colorMATH{\colorMATHB}\renewcommand\colorSYNTAX{\colorSYNTAXB}{{\color{\colorMATH}\ensuremath{r_{1 2}}}}\endgroup }}}} and {{\color{\colorMATH}\ensuremath{{\begingroup\renewcommand\colorMATH{\colorMATHB}\renewcommand\colorSYNTAX{\colorSYNTAXB}{{\color{\colorMATH}\ensuremath{r_{2 1}}}}\endgroup } = {\begingroup\renewcommand\colorMATH{\colorMATHB}\renewcommand\colorSYNTAX{\colorSYNTAXB}{{\color{\colorMATH}\ensuremath{r_{2 2}}}}\endgroup }}}}, thus {{\color{\colorMATH}\ensuremath{{\begingroup\renewcommand\colorMATH{\colorMATHB}\renewcommand\colorSYNTAX{\colorSYNTAXB}{{\color{\colorMATH}\ensuremath{r_{1 1}}}}\endgroup } \leq  {\begingroup\renewcommand\colorMATH{\colorMATHB}\renewcommand\colorSYNTAX{\colorSYNTAXB}{{\color{\colorMATH}\ensuremath{r_{2 1}}}}\endgroup } = {\begingroup\renewcommand\colorMATH{\colorMATHB}\renewcommand\colorSYNTAX{\colorSYNTAXB}{{\color{\colorMATH}\ensuremath{r_{1 2}}}}\endgroup } \leq  {\begingroup\renewcommand\colorMATH{\colorMATHB}\renewcommand\colorSYNTAX{\colorSYNTAXB}{{\color{\colorMATH}\ensuremath{r_{2 2}}}}\endgroup }}}}. Then {{\color{\colorMATH}\ensuremath{b_{1} = b_{2}}}}, let us suppose {{\color{\colorMATH}\ensuremath{b_{1} = \inl\hspace*{0.33em}\ttt}}} (the other case is analogous), then we have to prove that {{\color{\colorMATH}\ensuremath{(\inl\hspace*{0.33em}\ttt, \inl\hspace*{0.33em}\ttt) \in  {\mathcal{V}}_{0}^{k-j_{1}-j_{2}-1}\llbracket {{\color{\colorSYNTAX}\texttt{unit}}}\rrbracket }}}, but this is direct as {{\color{\colorMATH}\ensuremath{(\ttt, \ttt) \in  {\mathcal{V}}_{0}^{k-j_{1}-j_{2}-1}\llbracket {{\color{\colorSYNTAX}\texttt{unit}}}\rrbracket }}}.
    \end{subproof}
  \item  {{\color{\colorMATH}\ensuremath{\Gamma ; {\begingroup\renewcommand\colorMATH{\colorMATHB}\renewcommand\colorSYNTAX{\colorSYNTAXB}{{\color{\colorMATH}\ensuremath{\Distance}}}\endgroup } \vdash  x \mathrel{:} \tau  \mathrel{;} {\begingroup\renewcommand\colorMATH{\colorMATHB}\renewcommand\colorSYNTAX{\colorSYNTAXB}{{\color{\colorMATH}\ensuremath{x}}}\endgroup }}}}
    \begin{subproof} 
      We have to prove that {{\color{\colorMATH}\ensuremath{\forall k, \forall (\gamma _{1},\gamma _{2}) \in  {\mathcal{G}}_{{\begingroup\renewcommand\colorMATH{\colorMATHB}\renewcommand\colorSYNTAX{\colorSYNTAXB}{{\color{\colorMATH}\ensuremath{\Distance'}}}\endgroup }}^{\kg}\llbracket \Gamma \rrbracket , (\gamma _{1}\vdash x,\gamma _{2}\vdash x) \in  {\mathcal{E}}_{{\begingroup\renewcommand\colorMATH{\colorMATHB}\renewcommand\colorSYNTAX{\colorSYNTAXB}{{\color{\colorMATH}\ensuremath{\Distance'}}}\endgroup }\mathord{\cdotp }{\begingroup\renewcommand\colorMATH{\colorMATHB}\renewcommand\colorSYNTAX{\colorSYNTAXB}{{\color{\colorMATH}\ensuremath{x}}}\endgroup }}^{k}\llbracket {\begingroup\renewcommand\colorMATH{\colorMATHB}\renewcommand\colorSYNTAX{\colorSYNTAXB}{{\color{\colorMATH}\ensuremath{\Distance'}}}\endgroup }(\tau )\rrbracket }}}, where {{\color{\colorMATH}\ensuremath{\tau =\Gamma (x)}}}, for {{\color{\colorMATH}\ensuremath{{\begingroup\renewcommand\colorMATH{\colorMATHB}\renewcommand\colorSYNTAX{\colorSYNTAXB}{{\color{\colorMATH}\ensuremath{\Distance'}}}\endgroup } \sqsubseteq  {\begingroup\renewcommand\colorMATH{\colorMATHB}\renewcommand\colorSYNTAX{\colorSYNTAXB}{{\color{\colorMATH}\ensuremath{\Distance}}}\endgroup }}}}.
      Notice that {{\color{\colorMATH}\ensuremath{{\begingroup\renewcommand\colorMATH{\colorMATHB}\renewcommand\colorSYNTAX{\colorSYNTAXB}{{\color{\colorMATH}\ensuremath{\Distance'}}}\endgroup }\mathord{\cdotp }{\begingroup\renewcommand\colorMATH{\colorMATHB}\renewcommand\colorSYNTAX{\colorSYNTAXB}{{\color{\colorMATH}\ensuremath{x}}}\endgroup } = {\begingroup\renewcommand\colorMATH{\colorMATHB}\renewcommand\colorSYNTAX{\colorSYNTAXB}{{\color{\colorMATH}\ensuremath{\Distance'}}}\endgroup }\mathord{\cdotp }{\begingroup\renewcommand\colorMATH{\colorMATHB}\renewcommand\colorSYNTAX{\colorSYNTAXB}{{\color{\colorMATH}\ensuremath{1x}}}\endgroup } = {\begingroup\renewcommand\colorMATH{\colorMATHB}\renewcommand\colorSYNTAX{\colorSYNTAXB}{{\color{\colorMATH}\ensuremath{\Distance'}}}\endgroup }(x)}}}, therefore we have to prove that {{\color{\colorMATH}\ensuremath{(\gamma _{1}\vdash x,\gamma _{2}\vdash x) \in  {\mathcal{V}}_{{\begingroup\renewcommand\colorMATH{\colorMATHB}\renewcommand\colorSYNTAX{\colorSYNTAXB}{{\color{\colorMATH}\ensuremath{\Distance'}}}\endgroup }(x)}^{k-j}\llbracket {\begingroup\renewcommand\colorMATH{\colorMATHB}\renewcommand\colorSYNTAX{\colorSYNTAXB}{{\color{\colorMATH}\ensuremath{\Distance'}}}\endgroup }(\Gamma (x))\rrbracket }}}.
      But {{\color{\colorMATH}\ensuremath{\gamma _{1}\vdash x \Downarrow ^{1} \gamma _{1}(x)}}} and {{\color{\colorMATH}\ensuremath{\gamma _{2}\vdash x \Downarrow ^{1} \gamma _{2}(x)}}} therefore we have to prove that 
      {{\color{\colorMATH}\ensuremath{(\gamma _{1}(x),\gamma _{2}(x)) \in  {\mathcal{V}}_{{\begingroup\renewcommand\colorMATH{\colorMATHB}\renewcommand\colorSYNTAX{\colorSYNTAXB}{{\color{\colorMATH}\ensuremath{\Distance'}}}\endgroup }(x)}^{k-1}\llbracket {\begingroup\renewcommand\colorMATH{\colorMATHB}\renewcommand\colorSYNTAX{\colorSYNTAXB}{{\color{\colorMATH}\ensuremath{\Distance'}}}\endgroup }(\Gamma (x))\rrbracket }}}.
      The result is direct as by definition of {{\color{\colorMATH}\ensuremath{(\gamma _{1},\gamma _{2}) \in  {\mathcal{G}}_{{\begingroup\renewcommand\colorMATH{\colorMATHB}\renewcommand\colorSYNTAX{\colorSYNTAXB}{{\color{\colorMATH}\ensuremath{\Distance'}}}\endgroup }}^{\kg}\llbracket \Gamma \rrbracket }}}, we know by weakening that {{\color{\colorMATH}\ensuremath{(\gamma _{1}(x), \gamma _{2}(x)) \in  {\mathcal{V}}_{{\begingroup\renewcommand\colorMATH{\colorMATHB}\renewcommand\colorSYNTAX{\colorSYNTAXB}{{\color{\colorMATH}\ensuremath{\Distance'}}}\endgroup }(x)}^{k-1}\llbracket {\begingroup\renewcommand\colorMATH{\colorMATHB}\renewcommand\colorSYNTAX{\colorSYNTAXB}{{\color{\colorMATH}\ensuremath{\Distance'}}}\endgroup }(\Gamma (x))\rrbracket }}}.
    \end{subproof}
  \item  {{\color{\colorMATH}\ensuremath{\Gamma ; {\begingroup\renewcommand\colorMATH{\colorMATHB}\renewcommand\colorSYNTAX{\colorSYNTAXB}{{\color{\colorMATH}\ensuremath{\Distance}}}\endgroup } \vdash  {\begingroup\renewcommand\colorMATH{\colorMATHB}\renewcommand\colorSYNTAX{\colorSYNTAXB}{{\color{\colorMATH}\ensuremath{\slambda}}}\endgroup } (x\mathrel{:}\tau _{1}\mathord{\cdotp }{\begingroup\renewcommand\colorMATH{\colorMATHB}\renewcommand\colorSYNTAX{\colorSYNTAXB}{{\color{\colorMATH}\ensuremath{\sss_{1}}}}\endgroup }).\hspace*{0.33em}{\begingroup\renewcommand\colorMATH{\colorMATHB}\renewcommand\colorSYNTAX{\colorSYNTAXB}{{\color{\colorMATH}\ensuremath{\se'}}}\endgroup } \mathrel{:} (x\mathrel{:}\tau _{1}\mathord{\cdotp }{\begingroup\renewcommand\colorMATH{\colorMATHB}\renewcommand\colorSYNTAX{\colorSYNTAXB}{{\color{\colorMATH}\ensuremath{\sss_{1}}}}\endgroup }) \xrightarrowS {{\begingroup\renewcommand\colorMATH{\colorMATHB}\renewcommand\colorSYNTAX{\colorSYNTAXB}{{\color{\colorMATH}\ensuremath{\sS''}}}\endgroup }+{\begingroup\renewcommand\colorMATH{\colorMATHB}\renewcommand\colorSYNTAX{\colorSYNTAXB}{{\color{\colorMATH}\ensuremath{\sss'}}}\endgroup }x} \tau _{2} \mathrel{;} \varnothing }}} %\mt{without prepayment}
    \begin{subproof} 
      We have to prove that {{\color{\colorMATH}\ensuremath{\forall k, \forall (\gamma _{1},\gamma _{2}) \in  {\mathcal{G}}_{{\begingroup\renewcommand\colorMATH{\colorMATHB}\renewcommand\colorSYNTAX{\colorSYNTAXB}{{\color{\colorMATH}\ensuremath{\Distance'}}}\endgroup }}^{\kg}\llbracket \Gamma \rrbracket }}},\\
      {{\color{\colorMATH}\ensuremath{(\gamma _{1}\vdash {\begingroup\renewcommand\colorMATH{\colorMATHB}\renewcommand\colorSYNTAX{\colorSYNTAXB}{{\color{\colorMATH}\ensuremath{\slambda}}}\endgroup } (x\mathrel{:}\tau _{1}\mathord{\cdotp }{\begingroup\renewcommand\colorMATH{\colorMATHB}\renewcommand\colorSYNTAX{\colorSYNTAXB}{{\color{\colorMATH}\ensuremath{\sss_{1}}}}\endgroup }).\hspace*{0.33em}{\begingroup\renewcommand\colorMATH{\colorMATHB}\renewcommand\colorSYNTAX{\colorSYNTAXB}{{\color{\colorMATH}\ensuremath{\se'}}}\endgroup },\gamma _{2}\vdash {\begingroup\renewcommand\colorMATH{\colorMATHB}\renewcommand\colorSYNTAX{\colorSYNTAXB}{{\color{\colorMATH}\ensuremath{\slambda}}}\endgroup } (x\mathrel{:}\tau _{1}\mathord{\cdotp }{\begingroup\renewcommand\colorMATH{\colorMATHB}\renewcommand\colorSYNTAX{\colorSYNTAXB}{{\color{\colorMATH}\ensuremath{\sss_{1}}}}\endgroup }).\hspace*{0.33em}{\begingroup\renewcommand\colorMATH{\colorMATHB}\renewcommand\colorSYNTAX{\colorSYNTAXB}{{\color{\colorMATH}\ensuremath{\se'}}}\endgroup }) \in  {\mathcal{E}}_{{\begingroup\renewcommand\colorMATH{\colorMATHB}\renewcommand\colorSYNTAX{\colorSYNTAXB}{{\color{\colorMATH}\ensuremath{\Distance'}}}\endgroup }\mathord{\cdotp }\varnothing }^{k}\llbracket {\begingroup\renewcommand\colorMATH{\colorMATHB}\renewcommand\colorSYNTAX{\colorSYNTAXB}{{\color{\colorMATH}\ensuremath{\Distance'}}}\endgroup }(\tau _{1}\mathord{\cdotp }{\begingroup\renewcommand\colorMATH{\colorMATHB}\renewcommand\colorSYNTAX{\colorSYNTAXB}{{\color{\colorMATH}\ensuremath{\sss_{1}}}}\endgroup }) \xrightarrowS {{\begingroup\renewcommand\colorMATH{\colorMATHB}\renewcommand\colorSYNTAX{\colorSYNTAXB}{{\color{\colorMATH}\ensuremath{\Distance'}}}\endgroup }\mathord{\cdotp }{\begingroup\renewcommand\colorMATH{\colorMATHB}\renewcommand\colorSYNTAX{\colorSYNTAXB}{{\color{\colorMATH}\ensuremath{\sS''}}}\endgroup } + {\begingroup\renewcommand\colorMATH{\colorMATHB}\renewcommand\colorSYNTAX{\colorSYNTAXB}{{\color{\colorMATH}\ensuremath{\sss'}}}\endgroup }x} {\begingroup\renewcommand\colorMATH{\colorMATHB}\renewcommand\colorSYNTAX{\colorSYNTAXB}{{\color{\colorMATH}\ensuremath{\Distance'}}}\endgroup }(\tau _{2})\rrbracket }}}, for {{\color{\colorMATH}\ensuremath{{\begingroup\renewcommand\colorMATH{\colorMATHB}\renewcommand\colorSYNTAX{\colorSYNTAXB}{{\color{\colorMATH}\ensuremath{\Distance'}}}\endgroup } \sqsubseteq  {\begingroup\renewcommand\colorMATH{\colorMATHB}\renewcommand\colorSYNTAX{\colorSYNTAXB}{{\color{\colorMATH}\ensuremath{\Distance}}}\endgroup }}}}..\\
      Notice that {{\color{\colorMATH}\ensuremath{{\begingroup\renewcommand\colorMATH{\colorMATHB}\renewcommand\colorSYNTAX{\colorSYNTAXB}{{\color{\colorMATH}\ensuremath{\Distance'}}}\endgroup }\mathord{\cdotp }\varnothing  = 0}}}, that {{\color{\colorMATH}\ensuremath{{\begingroup\renewcommand\colorMATH{\colorMATHB}\renewcommand\colorSYNTAX{\colorSYNTAXB}{{\color{\colorMATH}\ensuremath{\Distance'}}}\endgroup }((x:\tau _{1}\mathord{\cdotp }{\begingroup\renewcommand\colorMATH{\colorMATHB}\renewcommand\colorSYNTAX{\colorSYNTAXB}{{\color{\colorMATH}\ensuremath{\sss_{1}}}}\endgroup }) \xrightarrowS {{\begingroup\renewcommand\colorMATH{\colorMATHB}\renewcommand\colorSYNTAX{\colorSYNTAXB}{{\color{\colorMATH}\ensuremath{\sS''}}}\endgroup } + {\begingroup\renewcommand\colorMATH{\colorMATHB}\renewcommand\colorSYNTAX{\colorSYNTAXB}{{\color{\colorMATH}\ensuremath{\sss'}}}\endgroup }x} \tau _{2}) = (x:{\begingroup\renewcommand\colorMATH{\colorMATHB}\renewcommand\colorSYNTAX{\colorSYNTAXB}{{\color{\colorMATH}\ensuremath{\Distance'}}}\endgroup }(\tau _{1})\mathord{\cdotp }{\begingroup\renewcommand\colorMATH{\colorMATHB}\renewcommand\colorSYNTAX{\colorSYNTAXB}{{\color{\colorMATH}\ensuremath{\sss_{1}}}}\endgroup }) \xrightarrowS {{\begingroup\renewcommand\colorMATH{\colorMATHB}\renewcommand\colorSYNTAX{\colorSYNTAXB}{{\color{\colorMATH}\ensuremath{\Distance'}}}\endgroup }\mathord{\cdotp }{\begingroup\renewcommand\colorMATH{\colorMATHB}\renewcommand\colorSYNTAX{\colorSYNTAXB}{{\color{\colorMATH}\ensuremath{\sS''}}}\endgroup } + {\begingroup\renewcommand\colorMATH{\colorMATHB}\renewcommand\colorSYNTAX{\colorSYNTAXB}{{\color{\colorMATH}\ensuremath{\sss'}}}\endgroup }x} {\begingroup\renewcommand\colorMATH{\colorMATHB}\renewcommand\colorSYNTAX{\colorSYNTAXB}{{\color{\colorMATH}\ensuremath{\Distance'}}}\endgroup }(\tau _{2})}}} (as {{\color{\colorMATH}\ensuremath{x \notin  dom({\begingroup\renewcommand\colorMATH{\colorMATHB}\renewcommand\colorSYNTAX{\colorSYNTAXB}{{\color{\colorMATH}\ensuremath{\Distance'}}}\endgroup })}}}), and that lambdas reduce to closures, therefore we have to prove that \\
      {{\color{\colorMATH}\ensuremath{(\langle {\begingroup\renewcommand\colorMATH{\colorMATHB}\renewcommand\colorSYNTAX{\colorSYNTAXB}{{\color{\colorMATH}\ensuremath{\slambda}}}\endgroup } (x\mathrel{:}\tau _{1}\mathord{\cdotp }{\begingroup\renewcommand\colorMATH{\colorMATHB}\renewcommand\colorSYNTAX{\colorSYNTAXB}{{\color{\colorMATH}\ensuremath{\sss_{1}}}}\endgroup }).\hspace*{0.33em}{\begingroup\renewcommand\colorMATH{\colorMATHB}\renewcommand\colorSYNTAX{\colorSYNTAXB}{{\color{\colorMATH}\ensuremath{\se'}}}\endgroup }, \gamma _{1}\rangle ,\langle {\begingroup\renewcommand\colorMATH{\colorMATHB}\renewcommand\colorSYNTAX{\colorSYNTAXB}{{\color{\colorMATH}\ensuremath{\slambda}}}\endgroup } (x\mathrel{:}\tau _{1}\mathord{\cdotp }{\begingroup\renewcommand\colorMATH{\colorMATHB}\renewcommand\colorSYNTAX{\colorSYNTAXB}{{\color{\colorMATH}\ensuremath{\sss_{1}}}}\endgroup }).\hspace*{0.33em}{\begingroup\renewcommand\colorMATH{\colorMATHB}\renewcommand\colorSYNTAX{\colorSYNTAXB}{{\color{\colorMATH}\ensuremath{\se'}}}\endgroup }, \gamma _{2}\rangle ) \in  {\mathcal{V}}_{0}^{k-1}\llbracket (x:{\begingroup\renewcommand\colorMATH{\colorMATHB}\renewcommand\colorSYNTAX{\colorSYNTAXB}{{\color{\colorMATH}\ensuremath{\Distance'}}}\endgroup }(\tau _{1})\mathord{\cdotp }{\begingroup\renewcommand\colorMATH{\colorMATHB}\renewcommand\colorSYNTAX{\colorSYNTAXB}{{\color{\colorMATH}\ensuremath{\sss_{1}}}}\endgroup }) \xrightarrowS {{\begingroup\renewcommand\colorMATH{\colorMATHB}\renewcommand\colorSYNTAX{\colorSYNTAXB}{{\color{\colorMATH}\ensuremath{\Distance'}}}\endgroup }\mathord{\cdotp }{\begingroup\renewcommand\colorMATH{\colorMATHB}\renewcommand\colorSYNTAX{\colorSYNTAXB}{{\color{\colorMATH}\ensuremath{\sS''}}}\endgroup } + {\begingroup\renewcommand\colorMATH{\colorMATHB}\renewcommand\colorSYNTAX{\colorSYNTAXB}{{\color{\colorMATH}\ensuremath{\sss'}}}\endgroup }x} {\begingroup\renewcommand\colorMATH{\colorMATHB}\renewcommand\colorSYNTAX{\colorSYNTAXB}{{\color{\colorMATH}\ensuremath{\Distance'}}}\endgroup }(\tau _{2})\rrbracket }}}. 

      Note that as {{\color{\colorMATH}\ensuremath{\Gamma , x: \tau _{1}; {\begingroup\renewcommand\colorMATH{\colorMATHB}\renewcommand\colorSYNTAX{\colorSYNTAXB}{{\color{\colorMATH}\ensuremath{\Distance}}}\endgroup } + {\begingroup\renewcommand\colorMATH{\colorMATHB}\renewcommand\colorSYNTAX{\colorSYNTAXB}{{\color{\colorMATH}\ensuremath{\sss_{1}}}}\endgroup }x \vdash  {\begingroup\renewcommand\colorMATH{\colorMATHB}\renewcommand\colorSYNTAX{\colorSYNTAXB}{{\color{\colorMATH}\ensuremath{\se'}}}\endgroup } : \tau _{2}; {\begingroup\renewcommand\colorMATH{\colorMATHB}\renewcommand\colorSYNTAX{\colorSYNTAXB}{{\color{\colorMATH}\ensuremath{\sS''}}}\endgroup } + {\begingroup\renewcommand\colorMATH{\colorMATHB}\renewcommand\colorSYNTAX{\colorSYNTAXB}{{\color{\colorMATH}\ensuremath{\sss'}}}\endgroup }x}}} and {{\color{\colorMATH}\ensuremath{(\gamma _{1},\gamma _{2}) \in  {\mathcal{G}}_{{\begingroup\renewcommand\colorMATH{\colorMATHB}\renewcommand\colorSYNTAX{\colorSYNTAXB}{{\color{\colorMATH}\ensuremath{\Distance'}}}\endgroup }}^{\kg}\llbracket \Gamma \rrbracket }}}, then {{\color{\colorMATH}\ensuremath{dom({\begingroup\renewcommand\colorMATH{\colorMATHB}\renewcommand\colorSYNTAX{\colorSYNTAXB}{{\color{\colorMATH}\ensuremath{\sS''}}}\endgroup }) \subseteq  dom({\begingroup\renewcommand\colorMATH{\colorMATHB}\renewcommand\colorSYNTAX{\colorSYNTAXB}{{\color{\colorMATH}\ensuremath{\Distance'}}}\endgroup })}}}, therefore  {{\color{\colorMATH}\ensuremath{{\begingroup\renewcommand\colorMATH{\colorMATHB}\renewcommand\colorSYNTAX{\colorSYNTAXB}{{\color{\colorMATH}\ensuremath{\Distance'}}}\endgroup }\mathord{\cdotp }{\begingroup\renewcommand\colorMATH{\colorMATHB}\renewcommand\colorSYNTAX{\colorSYNTAXB}{{\color{\colorMATH}\ensuremath{\sS''}}}\endgroup } \in  {\text{sens}}}}} (the result is a scalar).
      Consider {{\color{\colorMATH}\ensuremath{{\begingroup\renewcommand\colorMATH{\colorMATHB}\renewcommand\colorSYNTAX{\colorSYNTAXB}{{\color{\colorMATH}\ensuremath{\sss''}}}\endgroup } \leq  {\begingroup\renewcommand\colorMATH{\colorMATHB}\renewcommand\colorSYNTAX{\colorSYNTAXB}{{\color{\colorMATH}\ensuremath{\sss_{1}}}}\endgroup }, {\begingroup\renewcommand\colorMATH{\colorMATHB}\renewcommand\colorSYNTAX{\colorSYNTAXB}{{\color{\colorMATH}\ensuremath{\sv_{1}}}}\endgroup }}}} and {{\color{\colorMATH}\ensuremath{{\begingroup\renewcommand\colorMATH{\colorMATHB}\renewcommand\colorSYNTAX{\colorSYNTAXB}{{\color{\colorMATH}\ensuremath{\sv_{2}}}}\endgroup }}}} such that {{\color{\colorMATH}\ensuremath{({\begingroup\renewcommand\colorMATH{\colorMATHB}\renewcommand\colorSYNTAX{\colorSYNTAXB}{{\color{\colorMATH}\ensuremath{\sv_{1}}}}\endgroup },{\begingroup\renewcommand\colorMATH{\colorMATHB}\renewcommand\colorSYNTAX{\colorSYNTAXB}{{\color{\colorMATH}\ensuremath{\sv_{2}}}}\endgroup }) \in  {\mathcal{V}}_{{\begingroup\renewcommand\colorMATH{\colorMATHB}\renewcommand\colorSYNTAX{\colorSYNTAXB}{{\color{\colorMATH}\ensuremath{\sss''}}}\endgroup }}^{j}\llbracket {\begingroup\renewcommand\colorMATH{\colorMATHB}\renewcommand\colorSYNTAX{\colorSYNTAXB}{{\color{\colorMATH}\ensuremath{\Distance'}}}\endgroup }(\tau _{1})\rrbracket }}} for some {{\color{\colorMATH}\ensuremath{j<k}}}. We have to prove that
      {{\color{\colorMATH}\ensuremath{(\gamma _{1}, x \mapsto  {\begingroup\renewcommand\colorMATH{\colorMATHB}\renewcommand\colorSYNTAX{\colorSYNTAXB}{{\color{\colorMATH}\ensuremath{\sv_{1}}}}\endgroup } \vdash  {\begingroup\renewcommand\colorMATH{\colorMATHB}\renewcommand\colorSYNTAX{\colorSYNTAXB}{{\color{\colorMATH}\ensuremath{\se'}}}\endgroup }, \gamma _{2}, x \mapsto  {\begingroup\renewcommand\colorMATH{\colorMATHB}\renewcommand\colorSYNTAX{\colorSYNTAXB}{{\color{\colorMATH}\ensuremath{\sv_{2}}}}\endgroup } \vdash  {\begingroup\renewcommand\colorMATH{\colorMATHB}\renewcommand\colorSYNTAX{\colorSYNTAXB}{{\color{\colorMATH}\ensuremath{\se'}}}\endgroup }) \in  {\mathcal{E}}_{{\begingroup\renewcommand\colorMATH{\colorMATHB}\renewcommand\colorSYNTAX{\colorSYNTAXB}{{\color{\colorMATH}\ensuremath{\Distance'}}}\endgroup }\mathord{\cdotp }{\begingroup\renewcommand\colorMATH{\colorMATHB}\renewcommand\colorSYNTAX{\colorSYNTAXB}{{\color{\colorMATH}\ensuremath{\sS''}}}\endgroup } + {\begingroup\renewcommand\colorMATH{\colorMATHB}\renewcommand\colorSYNTAX{\colorSYNTAXB}{{\color{\colorMATH}\ensuremath{\sss'}}}\endgroup }{\begingroup\renewcommand\colorMATH{\colorMATHB}\renewcommand\colorSYNTAX{\colorSYNTAXB}{{\color{\colorMATH}\ensuremath{\sss''}}}\endgroup }}^{j-1}\llbracket {\begingroup\renewcommand\colorMATH{\colorMATHB}\renewcommand\colorSYNTAX{\colorSYNTAXB}{{\color{\colorMATH}\ensuremath{\sss''}}}\endgroup }x({\begingroup\renewcommand\colorMATH{\colorMATHB}\renewcommand\colorSYNTAX{\colorSYNTAXB}{{\color{\colorMATH}\ensuremath{\Distance'}}}\endgroup }(\tau _{2}))\rrbracket }}}. Notice that by Lemma~\ref{lm:equivsimplsubst} {{\color{\colorMATH}\ensuremath{{\begingroup\renewcommand\colorMATH{\colorMATHB}\renewcommand\colorSYNTAX{\colorSYNTAXB}{{\color{\colorMATH}\ensuremath{\sss''}}}\endgroup }x({\begingroup\renewcommand\colorMATH{\colorMATHB}\renewcommand\colorSYNTAX{\colorSYNTAXB}{{\color{\colorMATH}\ensuremath{\Distance'}}}\endgroup }(\tau _{2})) = ({\begingroup\renewcommand\colorMATH{\colorMATHB}\renewcommand\colorSYNTAX{\colorSYNTAXB}{{\color{\colorMATH}\ensuremath{\Distance'}}}\endgroup } + {\begingroup\renewcommand\colorMATH{\colorMATHB}\renewcommand\colorSYNTAX{\colorSYNTAXB}{{\color{\colorMATH}\ensuremath{\sss''}}}\endgroup }x)(\tau _{2})}}}, therefore we have to prove that 
      {{\color{\colorMATH}\ensuremath{(\gamma _{1}, x \mapsto  {\begingroup\renewcommand\colorMATH{\colorMATHB}\renewcommand\colorSYNTAX{\colorSYNTAXB}{{\color{\colorMATH}\ensuremath{\sv_{1}}}}\endgroup } \vdash  {\begingroup\renewcommand\colorMATH{\colorMATHB}\renewcommand\colorSYNTAX{\colorSYNTAXB}{{\color{\colorMATH}\ensuremath{\se'}}}\endgroup }, \gamma _{2}, x \mapsto  {\begingroup\renewcommand\colorMATH{\colorMATHB}\renewcommand\colorSYNTAX{\colorSYNTAXB}{{\color{\colorMATH}\ensuremath{\sv_{2}}}}\endgroup } \vdash  {\begingroup\renewcommand\colorMATH{\colorMATHB}\renewcommand\colorSYNTAX{\colorSYNTAXB}{{\color{\colorMATH}\ensuremath{\se'}}}\endgroup }) \in  {\mathcal{E}}_{{\begingroup\renewcommand\colorMATH{\colorMATHB}\renewcommand\colorSYNTAX{\colorSYNTAXB}{{\color{\colorMATH}\ensuremath{\Distance'}}}\endgroup }\mathord{\cdotp }{\begingroup\renewcommand\colorMATH{\colorMATHB}\renewcommand\colorSYNTAX{\colorSYNTAXB}{{\color{\colorMATH}\ensuremath{\sS''}}}\endgroup } + {\begingroup\renewcommand\colorMATH{\colorMATHB}\renewcommand\colorSYNTAX{\colorSYNTAXB}{{\color{\colorMATH}\ensuremath{\sss'}}}\endgroup }{\begingroup\renewcommand\colorMATH{\colorMATHB}\renewcommand\colorSYNTAX{\colorSYNTAXB}{{\color{\colorMATH}\ensuremath{\sss''}}}\endgroup }}^{j-1}\llbracket ({\begingroup\renewcommand\colorMATH{\colorMATHB}\renewcommand\colorSYNTAX{\colorSYNTAXB}{{\color{\colorMATH}\ensuremath{\Distance'}}}\endgroup }+{\begingroup\renewcommand\colorMATH{\colorMATHB}\renewcommand\colorSYNTAX{\colorSYNTAXB}{{\color{\colorMATH}\ensuremath{\sss''}}}\endgroup }x)(\tau _{2})\rrbracket }}}.
      
      By induction hypothesis on {{\color{\colorMATH}\ensuremath{\Gamma , x: \tau _{1}; {\begingroup\renewcommand\colorMATH{\colorMATHB}\renewcommand\colorSYNTAX{\colorSYNTAXB}{{\color{\colorMATH}\ensuremath{\Distance}}}\endgroup } + {\begingroup\renewcommand\colorMATH{\colorMATHB}\renewcommand\colorSYNTAX{\colorSYNTAXB}{{\color{\colorMATH}\ensuremath{\sss_{1}}}}\endgroup }x \vdash  {\begingroup\renewcommand\colorMATH{\colorMATHB}\renewcommand\colorSYNTAX{\colorSYNTAXB}{{\color{\colorMATH}\ensuremath{\se'}}}\endgroup } : \tau _{2}; {\begingroup\renewcommand\colorMATH{\colorMATHB}\renewcommand\colorSYNTAX{\colorSYNTAXB}{{\color{\colorMATH}\ensuremath{\sS''}}}\endgroup } + {\begingroup\renewcommand\colorMATH{\colorMATHB}\renewcommand\colorSYNTAX{\colorSYNTAXB}{{\color{\colorMATH}\ensuremath{\sss'}}}\endgroup }x}}}, and choosing {{\color{\colorMATH}\ensuremath{{\begingroup\renewcommand\colorMATH{\colorMATHB}\renewcommand\colorSYNTAX{\colorSYNTAXB}{{\color{\colorMATH}\ensuremath{{\begingroup\renewcommand\colorMATH{\colorMATHB}\renewcommand\colorSYNTAX{\colorSYNTAXB}{{\color{\colorMATH}\ensuremath{\sS}}}\endgroup }^\chi }}}\endgroup } = {\begingroup\renewcommand\colorMATH{\colorMATHB}\renewcommand\colorSYNTAX{\colorSYNTAXB}{{\color{\colorMATH}\ensuremath{\Distance'}}}\endgroup } + {\begingroup\renewcommand\colorMATH{\colorMATHB}\renewcommand\colorSYNTAX{\colorSYNTAXB}{{\color{\colorMATH}\ensuremath{\sss''}}}\endgroup }x}}}, we know that 
      {{\color{\colorMATH}\ensuremath{\forall  \gamma '_{1}, \gamma '_{2}, (\gamma '_{1}, \gamma '_{2}) \in  {\mathcal{G}}_{{\begingroup\renewcommand\colorMATH{\colorMATHB}\renewcommand\colorSYNTAX{\colorSYNTAXB}{{\color{\colorMATH}\ensuremath{{\begingroup\renewcommand\colorMATH{\colorMATHB}\renewcommand\colorSYNTAX{\colorSYNTAXB}{{\color{\colorMATH}\ensuremath{\sS}}}\endgroup }^\chi }}}\endgroup }}^{j-1}\llbracket \Gamma , x: \tau _{1}\rrbracket }}} then {{\color{\colorMATH}\ensuremath{(\gamma '_{1} \vdash  {\begingroup\renewcommand\colorMATH{\colorMATHB}\renewcommand\colorSYNTAX{\colorSYNTAXB}{{\color{\colorMATH}\ensuremath{\se'}}}\endgroup },\gamma '_{2} \vdash  {\begingroup\renewcommand\colorMATH{\colorMATHB}\renewcommand\colorSYNTAX{\colorSYNTAXB}{{\color{\colorMATH}\ensuremath{\se'}}}\endgroup }) \in  {\mathcal{E}}_{{\begingroup\renewcommand\colorMATH{\colorMATHB}\renewcommand\colorSYNTAX{\colorSYNTAXB}{{\color{\colorMATH}\ensuremath{{\begingroup\renewcommand\colorMATH{\colorMATHB}\renewcommand\colorSYNTAX{\colorSYNTAXB}{{\color{\colorMATH}\ensuremath{\sS}}}\endgroup }^\chi }}}\endgroup }\mathord{\cdotp }({\begingroup\renewcommand\colorMATH{\colorMATHB}\renewcommand\colorSYNTAX{\colorSYNTAXB}{{\color{\colorMATH}\ensuremath{\sS''}}}\endgroup } + {\begingroup\renewcommand\colorMATH{\colorMATHB}\renewcommand\colorSYNTAX{\colorSYNTAXB}{{\color{\colorMATH}\ensuremath{\sss'}}}\endgroup }x)}^{j-1}\llbracket {\begingroup\renewcommand\colorMATH{\colorMATHB}\renewcommand\colorSYNTAX{\colorSYNTAXB}{{\color{\colorMATH}\ensuremath{\sS}}}\endgroup }^\chi (\tau _{2})\rrbracket  }}}.
      Notice that\\ {{\color{\colorMATH}\ensuremath{{\begingroup\renewcommand\colorMATH{\colorMATHB}\renewcommand\colorSYNTAX{\colorSYNTAXB}{{\color{\colorMATH}\ensuremath{{\begingroup\renewcommand\colorMATH{\colorMATHB}\renewcommand\colorSYNTAX{\colorSYNTAXB}{{\color{\colorMATH}\ensuremath{\sS}}}\endgroup }^\chi }}}\endgroup }\mathord{\cdotp }({\begingroup\renewcommand\colorMATH{\colorMATHB}\renewcommand\colorSYNTAX{\colorSYNTAXB}{{\color{\colorMATH}\ensuremath{\sS''}}}\endgroup } + {\begingroup\renewcommand\colorMATH{\colorMATHB}\renewcommand\colorSYNTAX{\colorSYNTAXB}{{\color{\colorMATH}\ensuremath{\sss'}}}\endgroup }x) = ({\begingroup\renewcommand\colorMATH{\colorMATHB}\renewcommand\colorSYNTAX{\colorSYNTAXB}{{\color{\colorMATH}\ensuremath{\Distance'}}}\endgroup } + {\begingroup\renewcommand\colorMATH{\colorMATHB}\renewcommand\colorSYNTAX{\colorSYNTAXB}{{\color{\colorMATH}\ensuremath{\sss''}}}\endgroup }x)\mathord{\cdotp }({\begingroup\renewcommand\colorMATH{\colorMATHB}\renewcommand\colorSYNTAX{\colorSYNTAXB}{{\color{\colorMATH}\ensuremath{\sS''}}}\endgroup } + {\begingroup\renewcommand\colorMATH{\colorMATHB}\renewcommand\colorSYNTAX{\colorSYNTAXB}{{\color{\colorMATH}\ensuremath{\sss'}}}\endgroup }x) = {\begingroup\renewcommand\colorMATH{\colorMATHB}\renewcommand\colorSYNTAX{\colorSYNTAXB}{{\color{\colorMATH}\ensuremath{\Distance'}}}\endgroup } \mathord{\cdotp } {\begingroup\renewcommand\colorMATH{\colorMATHB}\renewcommand\colorSYNTAX{\colorSYNTAXB}{{\color{\colorMATH}\ensuremath{\sS''}}}\endgroup } + {\begingroup\renewcommand\colorMATH{\colorMATHB}\renewcommand\colorSYNTAX{\colorSYNTAXB}{{\color{\colorMATH}\ensuremath{\sss'}}}\endgroup }{\begingroup\renewcommand\colorMATH{\colorMATHB}\renewcommand\colorSYNTAX{\colorSYNTAXB}{{\color{\colorMATH}\ensuremath{\sss''}}}\endgroup }}}}.
      Therefore we know that\\
      {{\color{\colorMATH}\ensuremath{(\gamma '_{1} \vdash  {\begingroup\renewcommand\colorMATH{\colorMATHB}\renewcommand\colorSYNTAX{\colorSYNTAXB}{{\color{\colorMATH}\ensuremath{\se'}}}\endgroup },\gamma '_{2} \vdash  {\begingroup\renewcommand\colorMATH{\colorMATHB}\renewcommand\colorSYNTAX{\colorSYNTAXB}{{\color{\colorMATH}\ensuremath{\se'}}}\endgroup }) \in  {\mathcal{E}}_{{\begingroup\renewcommand\colorMATH{\colorMATHB}\renewcommand\colorSYNTAX{\colorSYNTAXB}{{\color{\colorMATH}\ensuremath{\Distance'}}}\endgroup } \mathord{\cdotp } {\begingroup\renewcommand\colorMATH{\colorMATHB}\renewcommand\colorSYNTAX{\colorSYNTAXB}{{\color{\colorMATH}\ensuremath{\sS''}}}\endgroup } + {\begingroup\renewcommand\colorMATH{\colorMATHB}\renewcommand\colorSYNTAX{\colorSYNTAXB}{{\color{\colorMATH}\ensuremath{\sss'}}}\endgroup }{\begingroup\renewcommand\colorMATH{\colorMATHB}\renewcommand\colorSYNTAX{\colorSYNTAXB}{{\color{\colorMATH}\ensuremath{\sss''}}}\endgroup }}^{j-1}\llbracket ({\begingroup\renewcommand\colorMATH{\colorMATHB}\renewcommand\colorSYNTAX{\colorSYNTAXB}{{\color{\colorMATH}\ensuremath{\Distance'}}}\endgroup } + {\begingroup\renewcommand\colorMATH{\colorMATHB}\renewcommand\colorSYNTAX{\colorSYNTAXB}{{\color{\colorMATH}\ensuremath{\sss''}}}\endgroup }x)(\tau _{2})\rrbracket  }}}.

      As {{\color{\colorMATH}\ensuremath{(\gamma _{1},\gamma _{2}) \in  {\mathcal{G}}_{{\begingroup\renewcommand\colorMATH{\colorMATHB}\renewcommand\colorSYNTAX{\colorSYNTAXB}{{\color{\colorMATH}\ensuremath{\Distance'}}}\endgroup }}^{\kg}\llbracket \Gamma \rrbracket }}}, and by Lemma~\ref{lm:lrweakening-sensitivity} {{\color{\colorMATH}\ensuremath{(\gamma _{1},\gamma _{2}) \in  {\mathcal{G}}_{{\begingroup\renewcommand\colorMATH{\colorMATHB}\renewcommand\colorSYNTAX{\colorSYNTAXB}{{\color{\colorMATH}\ensuremath{\Distance'}}}\endgroup }}^{j-1}\llbracket \Gamma \rrbracket }}}, also {{\color{\colorMATH}\ensuremath{({\begingroup\renewcommand\colorMATH{\colorMATHB}\renewcommand\colorSYNTAX{\colorSYNTAXB}{{\color{\colorMATH}\ensuremath{\sv_{1}}}}\endgroup },{\begingroup\renewcommand\colorMATH{\colorMATHB}\renewcommand\colorSYNTAX{\colorSYNTAXB}{{\color{\colorMATH}\ensuremath{\sv_{2}}}}\endgroup }) \in  {\mathcal{V}}_{{\begingroup\renewcommand\colorMATH{\colorMATHB}\renewcommand\colorSYNTAX{\colorSYNTAXB}{{\color{\colorMATH}\ensuremath{\sss''}}}\endgroup }}^{j-1}\llbracket {\begingroup\renewcommand\colorMATH{\colorMATHB}\renewcommand\colorSYNTAX{\colorSYNTAXB}{{\color{\colorMATH}\ensuremath{\Distance'}}}\endgroup }(\tau _{1})\rrbracket }}}, and {{\color{\colorMATH}\ensuremath{{\begingroup\renewcommand\colorMATH{\colorMATHB}\renewcommand\colorSYNTAX{\colorSYNTAXB}{{\color{\colorMATH}\ensuremath{\Distance'}}}\endgroup }(\tau _{1}) = ({\begingroup\renewcommand\colorMATH{\colorMATHB}\renewcommand\colorSYNTAX{\colorSYNTAXB}{{\color{\colorMATH}\ensuremath{\Distance'}}}\endgroup } + sx'')(\tau _{1})}}} (as {{\color{\colorMATH}\ensuremath{x}}} is not free in {{\color{\colorMATH}\ensuremath{\tau _{1}}}}),
      it is easy to see that {{\color{\colorMATH}\ensuremath{(\gamma _{1}, x \mapsto  {\begingroup\renewcommand\colorMATH{\colorMATHB}\renewcommand\colorSYNTAX{\colorSYNTAXB}{{\color{\colorMATH}\ensuremath{\sv_{1}}}}\endgroup }, \gamma _{2}, x \mapsto  {\begingroup\renewcommand\colorMATH{\colorMATHB}\renewcommand\colorSYNTAX{\colorSYNTAXB}{{\color{\colorMATH}\ensuremath{\sv_{2}}}}\endgroup }) \in  {\mathcal{G}}_{{\begingroup\renewcommand\colorMATH{\colorMATHB}\renewcommand\colorSYNTAX{\colorSYNTAXB}{{\color{\colorMATH}\ensuremath{\Distance'}}}\endgroup }+{\begingroup\renewcommand\colorMATH{\colorMATHB}\renewcommand\colorSYNTAX{\colorSYNTAXB}{{\color{\colorMATH}\ensuremath{\sss''}}}\endgroup }x}^{j-1}\llbracket \Gamma , x: \tau _{1}\rrbracket }}}. Finally, the result follows by choosing 
      {{\color{\colorMATH}\ensuremath{\gamma '_{1} = \gamma _{1}, x \mapsto  {\begingroup\renewcommand\colorMATH{\colorMATHB}\renewcommand\colorSYNTAX{\colorSYNTAXB}{{\color{\colorMATH}\ensuremath{\sv_{1}}}}\endgroup }}}}, and {{\color{\colorMATH}\ensuremath{ \gamma '_{2} = \gamma _{2}, x \mapsto  {\begingroup\renewcommand\colorMATH{\colorMATHB}\renewcommand\colorSYNTAX{\colorSYNTAXB}{{\color{\colorMATH}\ensuremath{\sv_{2}}}}\endgroup }}}}.
    \end{subproof}

  \item  {{\color{\colorMATH}\ensuremath{\Gamma ; {\begingroup\renewcommand\colorMATH{\colorMATHB}\renewcommand\colorSYNTAX{\colorSYNTAXB}{{\color{\colorMATH}\ensuremath{\Distance}}}\endgroup } \vdash  {\begingroup\renewcommand\colorMATH{\colorMATHB}\renewcommand\colorSYNTAX{\colorSYNTAXB}{{\color{\colorMATH}\ensuremath{\slambda}}}\endgroup } (x\mathrel{:}\tau _{1}\mathord{\cdotp }{\begingroup\renewcommand\colorMATH{\colorMATHB}\renewcommand\colorSYNTAX{\colorSYNTAXB}{{\color{\colorMATH}\ensuremath{\sss_{1}}}}\endgroup }).\hspace*{0.33em}{\begingroup\renewcommand\colorMATH{\colorMATHB}\renewcommand\colorSYNTAX{\colorSYNTAXB}{{\color{\colorMATH}\ensuremath{\se'}}}\endgroup } \mathrel{:} (x\mathrel{:}\tau _{1}\mathord{\cdotp }{\begingroup\renewcommand\colorMATH{\colorMATHB}\renewcommand\colorSYNTAX{\colorSYNTAXB}{{\color{\colorMATH}\ensuremath{\sss_{1}}}}\endgroup }) \xrightarrowS {{\begingroup\renewcommand\colorMATH{\colorMATHB}\renewcommand\colorSYNTAX{\colorSYNTAXB}{{\color{\colorMATH}\ensuremath{\sS''}}}\endgroup }+{\begingroup\renewcommand\colorMATH{\colorMATHB}\renewcommand\colorSYNTAX{\colorSYNTAXB}{{\color{\colorMATH}\ensuremath{\sss'}}}\endgroup }x} \tau _{2} \mathrel{;} {\begingroup\renewcommand\colorMATH{\colorMATHB}\renewcommand\colorSYNTAX{\colorSYNTAXB}{{\color{\colorMATH}\ensuremath{\sS'}}}\endgroup }}}} %\mt{with prepayment}
    \begin{subproof} 
      We have to prove that {{\color{\colorMATH}\ensuremath{\forall k, \forall (\gamma _{1},\gamma _{2}) \in  {\mathcal{G}}_{{\begingroup\renewcommand\colorMATH{\colorMATHB}\renewcommand\colorSYNTAX{\colorSYNTAXB}{{\color{\colorMATH}\ensuremath{\Distance'}}}\endgroup }}^{\kg}\llbracket \Gamma \rrbracket }}},\\
      {{\color{\colorMATH}\ensuremath{(\gamma _{1}\vdash {\begingroup\renewcommand\colorMATH{\colorMATHB}\renewcommand\colorSYNTAX{\colorSYNTAXB}{{\color{\colorMATH}\ensuremath{\slambda}}}\endgroup } (x\mathrel{:}\tau _{1}\mathord{\cdotp }{\begingroup\renewcommand\colorMATH{\colorMATHB}\renewcommand\colorSYNTAX{\colorSYNTAXB}{{\color{\colorMATH}\ensuremath{\sss_{1}}}}\endgroup }).\hspace*{0.33em}{\begingroup\renewcommand\colorMATH{\colorMATHB}\renewcommand\colorSYNTAX{\colorSYNTAXB}{{\color{\colorMATH}\ensuremath{\se'}}}\endgroup },\gamma _{2}\vdash {\begingroup\renewcommand\colorMATH{\colorMATHB}\renewcommand\colorSYNTAX{\colorSYNTAXB}{{\color{\colorMATH}\ensuremath{\slambda}}}\endgroup } (x\mathrel{:}\tau _{1}\mathord{\cdotp }{\begingroup\renewcommand\colorMATH{\colorMATHB}\renewcommand\colorSYNTAX{\colorSYNTAXB}{{\color{\colorMATH}\ensuremath{\sss_{1}}}}\endgroup }).\hspace*{0.33em}{\begingroup\renewcommand\colorMATH{\colorMATHB}\renewcommand\colorSYNTAX{\colorSYNTAXB}{{\color{\colorMATH}\ensuremath{\se'}}}\endgroup }) \in  {\mathcal{E}}_{{\begingroup\renewcommand\colorMATH{\colorMATHB}\renewcommand\colorSYNTAX{\colorSYNTAXB}{{\color{\colorMATH}\ensuremath{\Distance'}}}\endgroup }\mathord{\cdotp }{\begingroup\renewcommand\colorMATH{\colorMATHB}\renewcommand\colorSYNTAX{\colorSYNTAXB}{{\color{\colorMATH}\ensuremath{\sS'}}}\endgroup }}^{k}\llbracket {\begingroup\renewcommand\colorMATH{\colorMATHB}\renewcommand\colorSYNTAX{\colorSYNTAXB}{{\color{\colorMATH}\ensuremath{\Distance'}}}\endgroup }(\tau _{1}\mathord{\cdotp }{\begingroup\renewcommand\colorMATH{\colorMATHB}\renewcommand\colorSYNTAX{\colorSYNTAXB}{{\color{\colorMATH}\ensuremath{\sss_{1}}}}\endgroup }) \xrightarrowS {{\begingroup\renewcommand\colorMATH{\colorMATHB}\renewcommand\colorSYNTAX{\colorSYNTAXB}{{\color{\colorMATH}\ensuremath{\Distance'}}}\endgroup }\mathord{\cdotp }{\begingroup\renewcommand\colorMATH{\colorMATHB}\renewcommand\colorSYNTAX{\colorSYNTAXB}{{\color{\colorMATH}\ensuremath{\sS''}}}\endgroup } + {\begingroup\renewcommand\colorMATH{\colorMATHB}\renewcommand\colorSYNTAX{\colorSYNTAXB}{{\color{\colorMATH}\ensuremath{\sss'}}}\endgroup }x} {\begingroup\renewcommand\colorMATH{\colorMATHB}\renewcommand\colorSYNTAX{\colorSYNTAXB}{{\color{\colorMATH}\ensuremath{\Distance'}}}\endgroup }(\tau _{2})\rrbracket }}}, for {{\color{\colorMATH}\ensuremath{{\begingroup\renewcommand\colorMATH{\colorMATHB}\renewcommand\colorSYNTAX{\colorSYNTAXB}{{\color{\colorMATH}\ensuremath{\Distance'}}}\endgroup } \sqsubseteq  {\begingroup\renewcommand\colorMATH{\colorMATHB}\renewcommand\colorSYNTAX{\colorSYNTAXB}{{\color{\colorMATH}\ensuremath{\Distance}}}\endgroup }}}}..\\
      Notice that {{\color{\colorMATH}\ensuremath{{\begingroup\renewcommand\colorMATH{\colorMATHB}\renewcommand\colorSYNTAX{\colorSYNTAXB}{{\color{\colorMATH}\ensuremath{\Distance'}}}\endgroup }((x:\tau _{1}\mathord{\cdotp }{\begingroup\renewcommand\colorMATH{\colorMATHB}\renewcommand\colorSYNTAX{\colorSYNTAXB}{{\color{\colorMATH}\ensuremath{\sss_{1}}}}\endgroup }) \xrightarrowS {{\begingroup\renewcommand\colorMATH{\colorMATHB}\renewcommand\colorSYNTAX{\colorSYNTAXB}{{\color{\colorMATH}\ensuremath{\sS''}}}\endgroup } + {\begingroup\renewcommand\colorMATH{\colorMATHB}\renewcommand\colorSYNTAX{\colorSYNTAXB}{{\color{\colorMATH}\ensuremath{\sss'}}}\endgroup }x} \tau _{2}) = (x:{\begingroup\renewcommand\colorMATH{\colorMATHB}\renewcommand\colorSYNTAX{\colorSYNTAXB}{{\color{\colorMATH}\ensuremath{\Distance'}}}\endgroup }(\tau _{1})\mathord{\cdotp }{\begingroup\renewcommand\colorMATH{\colorMATHB}\renewcommand\colorSYNTAX{\colorSYNTAXB}{{\color{\colorMATH}\ensuremath{\sss_{1}}}}\endgroup }) \xrightarrowS {{\begingroup\renewcommand\colorMATH{\colorMATHB}\renewcommand\colorSYNTAX{\colorSYNTAXB}{{\color{\colorMATH}\ensuremath{\Distance'}}}\endgroup }\mathord{\cdotp }{\begingroup\renewcommand\colorMATH{\colorMATHB}\renewcommand\colorSYNTAX{\colorSYNTAXB}{{\color{\colorMATH}\ensuremath{\sS''}}}\endgroup } + {\begingroup\renewcommand\colorMATH{\colorMATHB}\renewcommand\colorSYNTAX{\colorSYNTAXB}{{\color{\colorMATH}\ensuremath{\sss'}}}\endgroup }x} {\begingroup\renewcommand\colorMATH{\colorMATHB}\renewcommand\colorSYNTAX{\colorSYNTAXB}{{\color{\colorMATH}\ensuremath{\Distance'}}}\endgroup }(\tau _{2})}}} (as {{\color{\colorMATH}\ensuremath{x \notin  dom({\begingroup\renewcommand\colorMATH{\colorMATHB}\renewcommand\colorSYNTAX{\colorSYNTAXB}{{\color{\colorMATH}\ensuremath{\Distance'}}}\endgroup })}}}), and that lambdas reduce to closures, therefore we have to prove that \\
      {{\color{\colorMATH}\ensuremath{(\langle {\begingroup\renewcommand\colorMATH{\colorMATHB}\renewcommand\colorSYNTAX{\colorSYNTAXB}{{\color{\colorMATH}\ensuremath{\slambda}}}\endgroup } (x\mathrel{:}\tau _{1}\mathord{\cdotp }{\begingroup\renewcommand\colorMATH{\colorMATHB}\renewcommand\colorSYNTAX{\colorSYNTAXB}{{\color{\colorMATH}\ensuremath{\sss_{1}}}}\endgroup }).\hspace*{0.33em}{\begingroup\renewcommand\colorMATH{\colorMATHB}\renewcommand\colorSYNTAX{\colorSYNTAXB}{{\color{\colorMATH}\ensuremath{\se'}}}\endgroup }, \gamma _{1}\rangle ,\langle {\begingroup\renewcommand\colorMATH{\colorMATHB}\renewcommand\colorSYNTAX{\colorSYNTAXB}{{\color{\colorMATH}\ensuremath{\slambda}}}\endgroup } (x\mathrel{:}\tau _{1}\mathord{\cdotp }{\begingroup\renewcommand\colorMATH{\colorMATHB}\renewcommand\colorSYNTAX{\colorSYNTAXB}{{\color{\colorMATH}\ensuremath{\sss_{1}}}}\endgroup }).\hspace*{0.33em}{\begingroup\renewcommand\colorMATH{\colorMATHB}\renewcommand\colorSYNTAX{\colorSYNTAXB}{{\color{\colorMATH}\ensuremath{\se'}}}\endgroup }, \gamma _{2}\rangle ) \in  {\mathcal{V}}_{{\begingroup\renewcommand\colorMATH{\colorMATHB}\renewcommand\colorSYNTAX{\colorSYNTAXB}{{\color{\colorMATH}\ensuremath{\Distance'}}}\endgroup }\mathord{\cdotp }{\begingroup\renewcommand\colorMATH{\colorMATHB}\renewcommand\colorSYNTAX{\colorSYNTAXB}{{\color{\colorMATH}\ensuremath{\sS'}}}\endgroup }}^{k-1}\llbracket (x:{\begingroup\renewcommand\colorMATH{\colorMATHB}\renewcommand\colorSYNTAX{\colorSYNTAXB}{{\color{\colorMATH}\ensuremath{\Distance'}}}\endgroup }(\tau _{1})\mathord{\cdotp }{\begingroup\renewcommand\colorMATH{\colorMATHB}\renewcommand\colorSYNTAX{\colorSYNTAXB}{{\color{\colorMATH}\ensuremath{\sss_{1}}}}\endgroup }) \xrightarrowS {{\begingroup\renewcommand\colorMATH{\colorMATHB}\renewcommand\colorSYNTAX{\colorSYNTAXB}{{\color{\colorMATH}\ensuremath{\Distance'}}}\endgroup }\mathord{\cdotp }{\begingroup\renewcommand\colorMATH{\colorMATHB}\renewcommand\colorSYNTAX{\colorSYNTAXB}{{\color{\colorMATH}\ensuremath{\sS''}}}\endgroup } + {\begingroup\renewcommand\colorMATH{\colorMATHB}\renewcommand\colorSYNTAX{\colorSYNTAXB}{{\color{\colorMATH}\ensuremath{\sss'}}}\endgroup }x} {\begingroup\renewcommand\colorMATH{\colorMATHB}\renewcommand\colorSYNTAX{\colorSYNTAXB}{{\color{\colorMATH}\ensuremath{\Distance'}}}\endgroup }(\tau _{2})\rrbracket }}}. 

      Note that as {{\color{\colorMATH}\ensuremath{\Gamma , x: \tau _{1}; {\begingroup\renewcommand\colorMATH{\colorMATHB}\renewcommand\colorSYNTAX{\colorSYNTAXB}{{\color{\colorMATH}\ensuremath{\Distance}}}\endgroup } + {\begingroup\renewcommand\colorMATH{\colorMATHB}\renewcommand\colorSYNTAX{\colorSYNTAXB}{{\color{\colorMATH}\ensuremath{\sss_{1}}}}\endgroup }x \vdash  {\begingroup\renewcommand\colorMATH{\colorMATHB}\renewcommand\colorSYNTAX{\colorSYNTAXB}{{\color{\colorMATH}\ensuremath{\se'}}}\endgroup } : \tau _{2}; {\begingroup\renewcommand\colorMATH{\colorMATHB}\renewcommand\colorSYNTAX{\colorSYNTAXB}{{\color{\colorMATH}\ensuremath{\sS''}}}\endgroup } + {\begingroup\renewcommand\colorMATH{\colorMATHB}\renewcommand\colorSYNTAX{\colorSYNTAXB}{{\color{\colorMATH}\ensuremath{\sS'}}}\endgroup } + {\begingroup\renewcommand\colorMATH{\colorMATHB}\renewcommand\colorSYNTAX{\colorSYNTAXB}{{\color{\colorMATH}\ensuremath{\sss'}}}\endgroup }x}}} and {{\color{\colorMATH}\ensuremath{(\gamma _{1},\gamma _{2}) \in  {\mathcal{G}}_{{\begingroup\renewcommand\colorMATH{\colorMATHB}\renewcommand\colorSYNTAX{\colorSYNTAXB}{{\color{\colorMATH}\ensuremath{\Distance'}}}\endgroup }}^{\kg}\llbracket \Gamma \rrbracket }}}, then {{\color{\colorMATH}\ensuremath{dom({\begingroup\renewcommand\colorMATH{\colorMATHB}\renewcommand\colorSYNTAX{\colorSYNTAXB}{{\color{\colorMATH}\ensuremath{\sS''}}}\endgroup }) \subseteq  dom({\begingroup\renewcommand\colorMATH{\colorMATHB}\renewcommand\colorSYNTAX{\colorSYNTAXB}{{\color{\colorMATH}\ensuremath{\Distance'}}}\endgroup })}}}, therefore  {{\color{\colorMATH}\ensuremath{{\begingroup\renewcommand\colorMATH{\colorMATHB}\renewcommand\colorSYNTAX{\colorSYNTAXB}{{\color{\colorMATH}\ensuremath{\Distance'}}}\endgroup }\mathord{\cdotp }{\begingroup\renewcommand\colorMATH{\colorMATHB}\renewcommand\colorSYNTAX{\colorSYNTAXB}{{\color{\colorMATH}\ensuremath{\sS''}}}\endgroup } \in  {\text{sens}}}}} (the result is a scalar).
      Consider {{\color{\colorMATH}\ensuremath{{\begingroup\renewcommand\colorMATH{\colorMATHB}\renewcommand\colorSYNTAX{\colorSYNTAXB}{{\color{\colorMATH}\ensuremath{\sss''}}}\endgroup } \leq  {\begingroup\renewcommand\colorMATH{\colorMATHB}\renewcommand\colorSYNTAX{\colorSYNTAXB}{{\color{\colorMATH}\ensuremath{\sss_{1}}}}\endgroup }, {\begingroup\renewcommand\colorMATH{\colorMATHB}\renewcommand\colorSYNTAX{\colorSYNTAXB}{{\color{\colorMATH}\ensuremath{\sv_{1}}}}\endgroup }}}} and {{\color{\colorMATH}\ensuremath{{\begingroup\renewcommand\colorMATH{\colorMATHB}\renewcommand\colorSYNTAX{\colorSYNTAXB}{{\color{\colorMATH}\ensuremath{\sv_{2}}}}\endgroup }}}} such that {{\color{\colorMATH}\ensuremath{({\begingroup\renewcommand\colorMATH{\colorMATHB}\renewcommand\colorSYNTAX{\colorSYNTAXB}{{\color{\colorMATH}\ensuremath{\sv_{1}}}}\endgroup },{\begingroup\renewcommand\colorMATH{\colorMATHB}\renewcommand\colorSYNTAX{\colorSYNTAXB}{{\color{\colorMATH}\ensuremath{\sv_{2}}}}\endgroup }) \in  {\mathcal{V}}_{{\begingroup\renewcommand\colorMATH{\colorMATHB}\renewcommand\colorSYNTAX{\colorSYNTAXB}{{\color{\colorMATH}\ensuremath{\sss''}}}\endgroup }}^{j}\llbracket {\begingroup\renewcommand\colorMATH{\colorMATHB}\renewcommand\colorSYNTAX{\colorSYNTAXB}{{\color{\colorMATH}\ensuremath{\Distance'}}}\endgroup }(\tau _{1})\rrbracket }}} for some {{\color{\colorMATH}\ensuremath{j<k}}}. We have to prove that
      {{\color{\colorMATH}\ensuremath{(\gamma _{1}, x \mapsto  {\begingroup\renewcommand\colorMATH{\colorMATHB}\renewcommand\colorSYNTAX{\colorSYNTAXB}{{\color{\colorMATH}\ensuremath{\sv_{1}}}}\endgroup } \vdash  {\begingroup\renewcommand\colorMATH{\colorMATHB}\renewcommand\colorSYNTAX{\colorSYNTAXB}{{\color{\colorMATH}\ensuremath{\se'}}}\endgroup }, \gamma _{2}, x \mapsto  {\begingroup\renewcommand\colorMATH{\colorMATHB}\renewcommand\colorSYNTAX{\colorSYNTAXB}{{\color{\colorMATH}\ensuremath{\sv_{2}}}}\endgroup } \vdash  {\begingroup\renewcommand\colorMATH{\colorMATHB}\renewcommand\colorSYNTAX{\colorSYNTAXB}{{\color{\colorMATH}\ensuremath{\se'}}}\endgroup }) \in  {\mathcal{E}}_{{\begingroup\renewcommand\colorMATH{\colorMATHB}\renewcommand\colorSYNTAX{\colorSYNTAXB}{{\color{\colorMATH}\ensuremath{\Distance'}}}\endgroup }\mathord{\cdotp }{\begingroup\renewcommand\colorMATH{\colorMATHB}\renewcommand\colorSYNTAX{\colorSYNTAXB}{{\color{\colorMATH}\ensuremath{\sS''}}}\endgroup } + {\begingroup\renewcommand\colorMATH{\colorMATHB}\renewcommand\colorSYNTAX{\colorSYNTAXB}{{\color{\colorMATH}\ensuremath{\Distance'}}}\endgroup }\mathord{\cdotp }{\begingroup\renewcommand\colorMATH{\colorMATHB}\renewcommand\colorSYNTAX{\colorSYNTAXB}{{\color{\colorMATH}\ensuremath{\sS'}}}\endgroup } + {\begingroup\renewcommand\colorMATH{\colorMATHB}\renewcommand\colorSYNTAX{\colorSYNTAXB}{{\color{\colorMATH}\ensuremath{\sss'}}}\endgroup }{\begingroup\renewcommand\colorMATH{\colorMATHB}\renewcommand\colorSYNTAX{\colorSYNTAXB}{{\color{\colorMATH}\ensuremath{\sss''}}}\endgroup }}^{j-1}\llbracket {\begingroup\renewcommand\colorMATH{\colorMATHB}\renewcommand\colorSYNTAX{\colorSYNTAXB}{{\color{\colorMATH}\ensuremath{\sss''}}}\endgroup }x({\begingroup\renewcommand\colorMATH{\colorMATHB}\renewcommand\colorSYNTAX{\colorSYNTAXB}{{\color{\colorMATH}\ensuremath{\Distance'}}}\endgroup }(\tau _{2}))\rrbracket }}}. 
      Notice that by Lemma~\ref{lm:equivsimplsubst} {{\color{\colorMATH}\ensuremath{{\begingroup\renewcommand\colorMATH{\colorMATHB}\renewcommand\colorSYNTAX{\colorSYNTAXB}{{\color{\colorMATH}\ensuremath{\sss''}}}\endgroup }x({\begingroup\renewcommand\colorMATH{\colorMATHB}\renewcommand\colorSYNTAX{\colorSYNTAXB}{{\color{\colorMATH}\ensuremath{\Distance'}}}\endgroup }(\tau _{2})) = ({\begingroup\renewcommand\colorMATH{\colorMATHB}\renewcommand\colorSYNTAX{\colorSYNTAXB}{{\color{\colorMATH}\ensuremath{\Distance'}}}\endgroup } + {\begingroup\renewcommand\colorMATH{\colorMATHB}\renewcommand\colorSYNTAX{\colorSYNTAXB}{{\color{\colorMATH}\ensuremath{\sss''}}}\endgroup }x)(\tau _{2})}}}, therefore we have to prove that 
      {{\color{\colorMATH}\ensuremath{(\gamma _{1}, x \mapsto  {\begingroup\renewcommand\colorMATH{\colorMATHB}\renewcommand\colorSYNTAX{\colorSYNTAXB}{{\color{\colorMATH}\ensuremath{\sv_{1}}}}\endgroup } \vdash  {\begingroup\renewcommand\colorMATH{\colorMATHB}\renewcommand\colorSYNTAX{\colorSYNTAXB}{{\color{\colorMATH}\ensuremath{\se'}}}\endgroup }, \gamma _{2}, x \mapsto  {\begingroup\renewcommand\colorMATH{\colorMATHB}\renewcommand\colorSYNTAX{\colorSYNTAXB}{{\color{\colorMATH}\ensuremath{\sv_{2}}}}\endgroup } \vdash  {\begingroup\renewcommand\colorMATH{\colorMATHB}\renewcommand\colorSYNTAX{\colorSYNTAXB}{{\color{\colorMATH}\ensuremath{\se'}}}\endgroup }) \in  {\mathcal{E}}_{{\begingroup\renewcommand\colorMATH{\colorMATHB}\renewcommand\colorSYNTAX{\colorSYNTAXB}{{\color{\colorMATH}\ensuremath{\Distance'}}}\endgroup }\mathord{\cdotp }{\begingroup\renewcommand\colorMATH{\colorMATHB}\renewcommand\colorSYNTAX{\colorSYNTAXB}{{\color{\colorMATH}\ensuremath{\sS''}}}\endgroup } + {\begingroup\renewcommand\colorMATH{\colorMATHB}\renewcommand\colorSYNTAX{\colorSYNTAXB}{{\color{\colorMATH}\ensuremath{\Distance'}}}\endgroup }\mathord{\cdotp }{\begingroup\renewcommand\colorMATH{\colorMATHB}\renewcommand\colorSYNTAX{\colorSYNTAXB}{{\color{\colorMATH}\ensuremath{\sS'}}}\endgroup } + {\begingroup\renewcommand\colorMATH{\colorMATHB}\renewcommand\colorSYNTAX{\colorSYNTAXB}{{\color{\colorMATH}\ensuremath{\sss'}}}\endgroup }{\begingroup\renewcommand\colorMATH{\colorMATHB}\renewcommand\colorSYNTAX{\colorSYNTAXB}{{\color{\colorMATH}\ensuremath{\sss''}}}\endgroup }}^{j-1}\llbracket ({\begingroup\renewcommand\colorMATH{\colorMATHB}\renewcommand\colorSYNTAX{\colorSYNTAXB}{{\color{\colorMATH}\ensuremath{\Distance'}}}\endgroup }+{\begingroup\renewcommand\colorMATH{\colorMATHB}\renewcommand\colorSYNTAX{\colorSYNTAXB}{{\color{\colorMATH}\ensuremath{\sss''}}}\endgroup }x)(\tau _{2})\rrbracket }}}.
      
      By induction hypothesis on {{\color{\colorMATH}\ensuremath{\Gamma , x: \tau _{1}; {\begingroup\renewcommand\colorMATH{\colorMATHB}\renewcommand\colorSYNTAX{\colorSYNTAXB}{{\color{\colorMATH}\ensuremath{\Distance}}}\endgroup } + {\begingroup\renewcommand\colorMATH{\colorMATHB}\renewcommand\colorSYNTAX{\colorSYNTAXB}{{\color{\colorMATH}\ensuremath{\sss_{1}}}}\endgroup }x \vdash  {\begingroup\renewcommand\colorMATH{\colorMATHB}\renewcommand\colorSYNTAX{\colorSYNTAXB}{{\color{\colorMATH}\ensuremath{\se'}}}\endgroup } : \tau _{2}; {\begingroup\renewcommand\colorMATH{\colorMATHB}\renewcommand\colorSYNTAX{\colorSYNTAXB}{{\color{\colorMATH}\ensuremath{\sS''}}}\endgroup } + {\begingroup\renewcommand\colorMATH{\colorMATHB}\renewcommand\colorSYNTAX{\colorSYNTAXB}{{\color{\colorMATH}\ensuremath{\sS'}}}\endgroup } + {\begingroup\renewcommand\colorMATH{\colorMATHB}\renewcommand\colorSYNTAX{\colorSYNTAXB}{{\color{\colorMATH}\ensuremath{\sss'}}}\endgroup }x}}}, and choosing {{\color{\colorMATH}\ensuremath{{\begingroup\renewcommand\colorMATH{\colorMATHB}\renewcommand\colorSYNTAX{\colorSYNTAXB}{{\color{\colorMATH}\ensuremath{{\begingroup\renewcommand\colorMATH{\colorMATHB}\renewcommand\colorSYNTAX{\colorSYNTAXB}{{\color{\colorMATH}\ensuremath{\sS}}}\endgroup }^\chi }}}\endgroup } = {\begingroup\renewcommand\colorMATH{\colorMATHB}\renewcommand\colorSYNTAX{\colorSYNTAXB}{{\color{\colorMATH}\ensuremath{\Distance'}}}\endgroup } + {\begingroup\renewcommand\colorMATH{\colorMATHB}\renewcommand\colorSYNTAX{\colorSYNTAXB}{{\color{\colorMATH}\ensuremath{\sss''}}}\endgroup }x}}}, we know that 
      {{\color{\colorMATH}\ensuremath{\forall  \gamma '_{1}, \gamma '_{2}, (\gamma '_{1}, \gamma '_{2}) \in  {\mathcal{G}}_{{\begingroup\renewcommand\colorMATH{\colorMATHB}\renewcommand\colorSYNTAX{\colorSYNTAXB}{{\color{\colorMATH}\ensuremath{{\begingroup\renewcommand\colorMATH{\colorMATHB}\renewcommand\colorSYNTAX{\colorSYNTAXB}{{\color{\colorMATH}\ensuremath{\sS}}}\endgroup }^\chi }}}\endgroup }}^{j-1}\llbracket \Gamma , x: \tau _{1}\rrbracket }}} then {{\color{\colorMATH}\ensuremath{(\gamma '_{1} \vdash  {\begingroup\renewcommand\colorMATH{\colorMATHB}\renewcommand\colorSYNTAX{\colorSYNTAXB}{{\color{\colorMATH}\ensuremath{\se'}}}\endgroup },\gamma '_{2} \vdash  {\begingroup\renewcommand\colorMATH{\colorMATHB}\renewcommand\colorSYNTAX{\colorSYNTAXB}{{\color{\colorMATH}\ensuremath{\se'}}}\endgroup }) \in  {\mathcal{E}}_{{\begingroup\renewcommand\colorMATH{\colorMATHB}\renewcommand\colorSYNTAX{\colorSYNTAXB}{{\color{\colorMATH}\ensuremath{{\begingroup\renewcommand\colorMATH{\colorMATHB}\renewcommand\colorSYNTAX{\colorSYNTAXB}{{\color{\colorMATH}\ensuremath{\sS}}}\endgroup }^\chi }}}\endgroup }\mathord{\cdotp }({\begingroup\renewcommand\colorMATH{\colorMATHB}\renewcommand\colorSYNTAX{\colorSYNTAXB}{{\color{\colorMATH}\ensuremath{\sS''}}}\endgroup } + {\begingroup\renewcommand\colorMATH{\colorMATHB}\renewcommand\colorSYNTAX{\colorSYNTAXB}{{\color{\colorMATH}\ensuremath{\sS'}}}\endgroup } + {\begingroup\renewcommand\colorMATH{\colorMATHB}\renewcommand\colorSYNTAX{\colorSYNTAXB}{{\color{\colorMATH}\ensuremath{\sss'}}}\endgroup }x)}^{j-1}\llbracket {\begingroup\renewcommand\colorMATH{\colorMATHB}\renewcommand\colorSYNTAX{\colorSYNTAXB}{{\color{\colorMATH}\ensuremath{\sS}}}\endgroup }^\chi (\tau _{2})\rrbracket  }}}.
      Notice that {{\color{\colorMATH}\ensuremath{{\begingroup\renewcommand\colorMATH{\colorMATHB}\renewcommand\colorSYNTAX{\colorSYNTAXB}{{\color{\colorMATH}\ensuremath{{\begingroup\renewcommand\colorMATH{\colorMATHB}\renewcommand\colorSYNTAX{\colorSYNTAXB}{{\color{\colorMATH}\ensuremath{\sS}}}\endgroup }^\chi }}}\endgroup }\mathord{\cdotp }({\begingroup\renewcommand\colorMATH{\colorMATHB}\renewcommand\colorSYNTAX{\colorSYNTAXB}{{\color{\colorMATH}\ensuremath{\sS''}}}\endgroup } + {\begingroup\renewcommand\colorMATH{\colorMATHB}\renewcommand\colorSYNTAX{\colorSYNTAXB}{{\color{\colorMATH}\ensuremath{\sS'}}}\endgroup } + {\begingroup\renewcommand\colorMATH{\colorMATHB}\renewcommand\colorSYNTAX{\colorSYNTAXB}{{\color{\colorMATH}\ensuremath{\sss'}}}\endgroup }x) = ({\begingroup\renewcommand\colorMATH{\colorMATHB}\renewcommand\colorSYNTAX{\colorSYNTAXB}{{\color{\colorMATH}\ensuremath{\Distance'}}}\endgroup } + {\begingroup\renewcommand\colorMATH{\colorMATHB}\renewcommand\colorSYNTAX{\colorSYNTAXB}{{\color{\colorMATH}\ensuremath{\sss''}}}\endgroup }x)\mathord{\cdotp }({\begingroup\renewcommand\colorMATH{\colorMATHB}\renewcommand\colorSYNTAX{\colorSYNTAXB}{{\color{\colorMATH}\ensuremath{\sS''}}}\endgroup } + {\begingroup\renewcommand\colorMATH{\colorMATHB}\renewcommand\colorSYNTAX{\colorSYNTAXB}{{\color{\colorMATH}\ensuremath{\sS'}}}\endgroup } + {\begingroup\renewcommand\colorMATH{\colorMATHB}\renewcommand\colorSYNTAX{\colorSYNTAXB}{{\color{\colorMATH}\ensuremath{\sss'}}}\endgroup }x) = {\begingroup\renewcommand\colorMATH{\colorMATHB}\renewcommand\colorSYNTAX{\colorSYNTAXB}{{\color{\colorMATH}\ensuremath{\Distance'}}}\endgroup } \mathord{\cdotp } {\begingroup\renewcommand\colorMATH{\colorMATHB}\renewcommand\colorSYNTAX{\colorSYNTAXB}{{\color{\colorMATH}\ensuremath{\sS''}}}\endgroup } + {\begingroup\renewcommand\colorMATH{\colorMATHB}\renewcommand\colorSYNTAX{\colorSYNTAXB}{{\color{\colorMATH}\ensuremath{\Distance'}}}\endgroup } \mathord{\cdotp } {\begingroup\renewcommand\colorMATH{\colorMATHB}\renewcommand\colorSYNTAX{\colorSYNTAXB}{{\color{\colorMATH}\ensuremath{\sS'}}}\endgroup } + {\begingroup\renewcommand\colorMATH{\colorMATHB}\renewcommand\colorSYNTAX{\colorSYNTAXB}{{\color{\colorMATH}\ensuremath{\sss'}}}\endgroup }{\begingroup\renewcommand\colorMATH{\colorMATHB}\renewcommand\colorSYNTAX{\colorSYNTAXB}{{\color{\colorMATH}\ensuremath{\sss''}}}\endgroup }}}}.
      Therefore we know that\\
      {{\color{\colorMATH}\ensuremath{(\gamma '_{1} \vdash  {\begingroup\renewcommand\colorMATH{\colorMATHB}\renewcommand\colorSYNTAX{\colorSYNTAXB}{{\color{\colorMATH}\ensuremath{\se'}}}\endgroup },\gamma '_{2} \vdash  {\begingroup\renewcommand\colorMATH{\colorMATHB}\renewcommand\colorSYNTAX{\colorSYNTAXB}{{\color{\colorMATH}\ensuremath{\se'}}}\endgroup }) \in  {\mathcal{E}}_{{\begingroup\renewcommand\colorMATH{\colorMATHB}\renewcommand\colorSYNTAX{\colorSYNTAXB}{{\color{\colorMATH}\ensuremath{\Distance'}}}\endgroup } \mathord{\cdotp } {\begingroup\renewcommand\colorMATH{\colorMATHB}\renewcommand\colorSYNTAX{\colorSYNTAXB}{{\color{\colorMATH}\ensuremath{\sS''}}}\endgroup } + {\begingroup\renewcommand\colorMATH{\colorMATHB}\renewcommand\colorSYNTAX{\colorSYNTAXB}{{\color{\colorMATH}\ensuremath{\Distance'}}}\endgroup } \mathord{\cdotp } {\begingroup\renewcommand\colorMATH{\colorMATHB}\renewcommand\colorSYNTAX{\colorSYNTAXB}{{\color{\colorMATH}\ensuremath{\sS'}}}\endgroup } + {\begingroup\renewcommand\colorMATH{\colorMATHB}\renewcommand\colorSYNTAX{\colorSYNTAXB}{{\color{\colorMATH}\ensuremath{\sss'}}}\endgroup }{\begingroup\renewcommand\colorMATH{\colorMATHB}\renewcommand\colorSYNTAX{\colorSYNTAXB}{{\color{\colorMATH}\ensuremath{\sss''}}}\endgroup }}^{j-1}\llbracket ({\begingroup\renewcommand\colorMATH{\colorMATHB}\renewcommand\colorSYNTAX{\colorSYNTAXB}{{\color{\colorMATH}\ensuremath{\Distance'}}}\endgroup } + {\begingroup\renewcommand\colorMATH{\colorMATHB}\renewcommand\colorSYNTAX{\colorSYNTAXB}{{\color{\colorMATH}\ensuremath{\sss''}}}\endgroup }x)(\tau _{2})\rrbracket  }}}.

      As {{\color{\colorMATH}\ensuremath{(\gamma _{1},\gamma _{2}) \in  {\mathcal{G}}_{{\begingroup\renewcommand\colorMATH{\colorMATHB}\renewcommand\colorSYNTAX{\colorSYNTAXB}{{\color{\colorMATH}\ensuremath{\Distance'}}}\endgroup }}^{\kg}\llbracket \Gamma \rrbracket }}}, and by Lemma~\ref{lm:lrweakening-sensitivity} {{\color{\colorMATH}\ensuremath{(\gamma _{1},\gamma _{2}) \in  {\mathcal{G}}_{{\begingroup\renewcommand\colorMATH{\colorMATHB}\renewcommand\colorSYNTAX{\colorSYNTAXB}{{\color{\colorMATH}\ensuremath{\Distance'}}}\endgroup }}^{j-1}\llbracket \Gamma \rrbracket }}}, also {{\color{\colorMATH}\ensuremath{({\begingroup\renewcommand\colorMATH{\colorMATHB}\renewcommand\colorSYNTAX{\colorSYNTAXB}{{\color{\colorMATH}\ensuremath{\sv_{1}}}}\endgroup },{\begingroup\renewcommand\colorMATH{\colorMATHB}\renewcommand\colorSYNTAX{\colorSYNTAXB}{{\color{\colorMATH}\ensuremath{\sv_{2}}}}\endgroup }) \in  {\mathcal{V}}_{{\begingroup\renewcommand\colorMATH{\colorMATHB}\renewcommand\colorSYNTAX{\colorSYNTAXB}{{\color{\colorMATH}\ensuremath{\sss''}}}\endgroup }}^{j-1}\llbracket {\begingroup\renewcommand\colorMATH{\colorMATHB}\renewcommand\colorSYNTAX{\colorSYNTAXB}{{\color{\colorMATH}\ensuremath{\Distance'}}}\endgroup }(\tau _{1})\rrbracket }}}, and {{\color{\colorMATH}\ensuremath{{\begingroup\renewcommand\colorMATH{\colorMATHB}\renewcommand\colorSYNTAX{\colorSYNTAXB}{{\color{\colorMATH}\ensuremath{\Distance'}}}\endgroup }(\tau _{1}) = ({\begingroup\renewcommand\colorMATH{\colorMATHB}\renewcommand\colorSYNTAX{\colorSYNTAXB}{{\color{\colorMATH}\ensuremath{\Distance'}}}\endgroup } + sx'')(\tau _{1})}}} (as {{\color{\colorMATH}\ensuremath{x}}} is not free in {{\color{\colorMATH}\ensuremath{\tau _{1}}}}),
      it is easy to see that {{\color{\colorMATH}\ensuremath{(\gamma _{1}, x \mapsto  {\begingroup\renewcommand\colorMATH{\colorMATHB}\renewcommand\colorSYNTAX{\colorSYNTAXB}{{\color{\colorMATH}\ensuremath{\sv_{1}}}}\endgroup }, \gamma _{2}, x \mapsto  {\begingroup\renewcommand\colorMATH{\colorMATHB}\renewcommand\colorSYNTAX{\colorSYNTAXB}{{\color{\colorMATH}\ensuremath{\sv_{2}}}}\endgroup }) \in  {\mathcal{G}}_{{\begingroup\renewcommand\colorMATH{\colorMATHB}\renewcommand\colorSYNTAX{\colorSYNTAXB}{{\color{\colorMATH}\ensuremath{\Distance'}}}\endgroup }+{\begingroup\renewcommand\colorMATH{\colorMATHB}\renewcommand\colorSYNTAX{\colorSYNTAXB}{{\color{\colorMATH}\ensuremath{\sss''}}}\endgroup }x}^{j-1}\llbracket \Gamma , x: \tau _{1}\rrbracket }}}. Finally, the result follows by choosing 
      {{\color{\colorMATH}\ensuremath{\gamma '_{1} = \gamma _{1}, x \mapsto  {\begingroup\renewcommand\colorMATH{\colorMATHB}\renewcommand\colorSYNTAX{\colorSYNTAXB}{{\color{\colorMATH}\ensuremath{\sv_{1}}}}\endgroup }}}}, and {{\color{\colorMATH}\ensuremath{ \gamma '_{2} = \gamma _{2}, x \mapsto  {\begingroup\renewcommand\colorMATH{\colorMATHB}\renewcommand\colorSYNTAX{\colorSYNTAXB}{{\color{\colorMATH}\ensuremath{\sv_{2}}}}\endgroup }}}}.
    \end{subproof}

  \item  {{\color{\colorMATH}\ensuremath{\Gamma ; {\begingroup\renewcommand\colorMATH{\colorMATHB}\renewcommand\colorSYNTAX{\colorSYNTAXB}{{\color{\colorMATH}\ensuremath{\Distance}}}\endgroup } \vdash  \langle {\begingroup\renewcommand\colorMATH{\colorMATHB}\renewcommand\colorSYNTAX{\colorSYNTAXB}{{\color{\colorMATH}\ensuremath{\slambda}}}\endgroup } (x\mathrel{:}\tau _{1}\mathord{\cdotp }{\begingroup\renewcommand\colorMATH{\colorMATHB}\renewcommand\colorSYNTAX{\colorSYNTAXB}{{\color{\colorMATH}\ensuremath{\sss_{1}}}}\endgroup }).\hspace*{0.33em}{\begingroup\renewcommand\colorMATH{\colorMATHB}\renewcommand\colorSYNTAX{\colorSYNTAXB}{{\color{\colorMATH}\ensuremath{\se'}}}\endgroup }, \gamma \rangle  \mathrel{:} (x\mathrel{:}\tau _{1}\mathord{\cdotp }{\begingroup\renewcommand\colorMATH{\colorMATHB}\renewcommand\colorSYNTAX{\colorSYNTAXB}{{\color{\colorMATH}\ensuremath{\sss_{1}}}}\endgroup }) \xrightarrowS {{\begingroup\renewcommand\colorMATH{\colorMATHB}\renewcommand\colorSYNTAX{\colorSYNTAXB}{{\color{\colorMATH}\ensuremath{\sss'}}}\endgroup }x} \tau _{2} \mathrel{;} \varnothing }}} 
    \begin{subproof} 
      \begingroup\color{\colorMATH}\begin{gather*} 
        \inferrule*[lab=
        ]{ \exists \Gamma ', {\begingroup\renewcommand\colorMATH{\colorMATHB}\renewcommand\colorSYNTAX{\colorSYNTAXB}{{\color{\colorMATH}\ensuremath{\Distance'_{0}}}}\endgroup }, dom({\begingroup\renewcommand\colorMATH{\colorMATHB}\renewcommand\colorSYNTAX{\colorSYNTAXB}{{\color{\colorMATH}\ensuremath{\sS''}}}\endgroup }) \subseteq  dom(\Gamma ') \subseteq  dom({\begingroup\renewcommand\colorMATH{\colorMATHB}\renewcommand\colorSYNTAX{\colorSYNTAXB}{{\color{\colorMATH}\ensuremath{\Distance'_{0}}}}\endgroup }) 
        \\ \forall  x_{i} \in  dom(\Gamma '), \varnothing ; \varnothing  \vdash  \gamma (x_{i}) : \tau '_{i}, \tau '_{i} <: \Gamma '(x_{i}) ; \varnothing 
        \\ \Gamma ', x: \tau _{1}; {\begingroup\renewcommand\colorMATH{\colorMATHB}\renewcommand\colorSYNTAX{\colorSYNTAXB}{{\color{\colorMATH}\ensuremath{\Distance'_{0}}}}\endgroup } + {\begingroup\renewcommand\colorMATH{\colorMATHB}\renewcommand\colorSYNTAX{\colorSYNTAXB}{{\color{\colorMATH}\ensuremath{\sss_{1}}}}\endgroup }x \vdash  {\begingroup\renewcommand\colorMATH{\colorMATHB}\renewcommand\colorSYNTAX{\colorSYNTAXB}{{\color{\colorMATH}\ensuremath{\se'}}}\endgroup } \mathrel{:} \tau _{2} \mathrel{;} {\begingroup\renewcommand\colorMATH{\colorMATHB}\renewcommand\colorSYNTAX{\colorSYNTAXB}{{\color{\colorMATH}\ensuremath{\sS''}}}\endgroup }+{\begingroup\renewcommand\colorMATH{\colorMATHB}\renewcommand\colorSYNTAX{\colorSYNTAXB}{{\color{\colorMATH}\ensuremath{\sss'}}}\endgroup }x
          }{
          \Gamma ; {\begingroup\renewcommand\colorMATH{\colorMATHB}\renewcommand\colorSYNTAX{\colorSYNTAXB}{{\color{\colorMATH}\ensuremath{\Distance}}}\endgroup } \vdash  \langle {\begingroup\renewcommand\colorMATH{\colorMATHB}\renewcommand\colorSYNTAX{\colorSYNTAXB}{{\color{\colorMATH}\ensuremath{\slambda}}}\endgroup } (x\mathrel{:}\tau _{1}\mathord{\cdotp }{\begingroup\renewcommand\colorMATH{\colorMATHB}\renewcommand\colorSYNTAX{\colorSYNTAXB}{{\color{\colorMATH}\ensuremath{\sss_{1}}}}\endgroup }).\hspace*{0.33em}{\begingroup\renewcommand\colorMATH{\colorMATHB}\renewcommand\colorSYNTAX{\colorSYNTAXB}{{\color{\colorMATH}\ensuremath{\se'}}}\endgroup }, \gamma \rangle  \mathrel{:} (x\mathrel{:} [\varnothing /x_{1},...,\varnothing /x_{n}]\tau _{1}\mathord{\cdotp }{\begingroup\renewcommand\colorMATH{\colorMATHB}\renewcommand\colorSYNTAX{\colorSYNTAXB}{{\color{\colorMATH}\ensuremath{\sss_{1}}}}\endgroup }) \xrightarrowS {{\begingroup\renewcommand\colorMATH{\colorMATHB}\renewcommand\colorSYNTAX{\colorSYNTAXB}{{\color{\colorMATH}\ensuremath{\sss'}}}\endgroup }x} [\varnothing /x_{1},...,\varnothing /x_{n}]\tau _{2} \mathrel{;} \varnothing 
        }
      \end{gather*}\endgroup
      By induction hypotheses on {{\color{\colorMATH}\ensuremath{\forall  x_{i} \in  dom(\Gamma '), \varnothing ; \varnothing  \vdash  \gamma (x_{i}) : \tau '_{i}; \varnothing }}}, we know that,
      {{\color{\colorMATH}\ensuremath{\langle \gamma (x_{i}), \gamma (x_{i})\rangle  \in  {\mathcal{V}}_{0}^{k-1}\llbracket \varnothing (\tau '_i)\rrbracket }}}, and by ~\ref{lm:lrweakening-sensitivity} 
      {{\color{\colorMATH}\ensuremath{\langle \gamma (x_{i}), \gamma (x_{i})\rangle  \in  {\mathcal{V}}_{0}^{k-1}\llbracket \varnothing (\Gamma '(x_{i}))\rrbracket }}}. %, and by Lemma~\ref{lm:lrweakening-sensitivity}, {{\color{\colorMATH}\ensuremath{\langle \gamma (x_{i}), \gamma (x_{i})\rangle  \in  {\mathcal{V}}_{{\begingroup\renewcommand\colorMATH{\colorMATHB}\renewcommand\colorSYNTAX{\colorSYNTAXB}{{\color{\colorMATH}\ensuremath{\Distance'}}}\endgroup }(x)}^{k-1}\llbracket {\begingroup\renewcommand\colorMATH{\colorMATHB}\renewcommand\colorSYNTAX{\colorSYNTAXB}{{\color{\colorMATH}\ensuremath{\Distance'}}}\endgroup }(\Gamma '(x_{i}))\rrbracket }}}.
      Therefore {{\color{\colorMATH}\ensuremath{\langle \gamma , \gamma \rangle  \in  {\mathcal{V}}_{\varnothing }^{k-1}\llbracket \Gamma '\rrbracket }}}, and by Lemma~\ref{lm:weakening-index}, {{\color{\colorMATH}\ensuremath{\langle \gamma , \gamma \rangle  \in  {\mathcal{V}}_{\varnothing }^{j}\llbracket \Gamma '\rrbracket }}} for {{\color{\colorMATH}\ensuremath{j<k}}}.\\
      Notice that {{\color{\colorMATH}\ensuremath{{\begingroup\renewcommand\colorMATH{\colorMATHB}\renewcommand\colorSYNTAX{\colorSYNTAXB}{{\color{\colorMATH}\ensuremath{\Distance'}}}\endgroup }([\varnothing /x_{1},...,\varnothing /x_{n}]\tau _{1}) = [\varnothing /x_{1},...,\varnothing /x_{n}]\tau _{1} = \varnothing ([\varnothing /x_{1},...,\varnothing /x_{n}]\tau _{1})}}}, {{\color{\colorMATH}\ensuremath{{\begingroup\renewcommand\colorMATH{\colorMATHB}\renewcommand\colorSYNTAX{\colorSYNTAXB}{{\color{\colorMATH}\ensuremath{\Distance'}}}\endgroup }([\varnothing /x_{1},...,\varnothing /x_{n}]\tau _{2}) = [\varnothing /x_{1},...,\varnothing /x_{n}]\tau _{2} = \varnothing ([\varnothing /x_{1},...,\varnothing /x_{n}]\tau _{2})}}}, as {{\color{\colorMATH}\ensuremath{\tau _{1}}}} and {{\color{\colorMATH}\ensuremath{\tau _{2}}}} have no free variables.
      We have to prove that {{\color{\colorMATH}\ensuremath{(\langle {\begingroup\renewcommand\colorMATH{\colorMATHB}\renewcommand\colorSYNTAX{\colorSYNTAXB}{{\color{\colorMATH}\ensuremath{\slambda}}}\endgroup } (x\mathrel{:}\tau _{1}\mathord{\cdotp }{\begingroup\renewcommand\colorMATH{\colorMATHB}\renewcommand\colorSYNTAX{\colorSYNTAXB}{{\color{\colorMATH}\ensuremath{\sss_{1}}}}\endgroup }).\hspace*{0.33em}{\begingroup\renewcommand\colorMATH{\colorMATHB}\renewcommand\colorSYNTAX{\colorSYNTAXB}{{\color{\colorMATH}\ensuremath{\se'}}}\endgroup }, \gamma \rangle ,\langle {\begingroup\renewcommand\colorMATH{\colorMATHB}\renewcommand\colorSYNTAX{\colorSYNTAXB}{{\color{\colorMATH}\ensuremath{\slambda}}}\endgroup } (x\mathrel{:}\tau _{1}\mathord{\cdotp }{\begingroup\renewcommand\colorMATH{\colorMATHB}\renewcommand\colorSYNTAX{\colorSYNTAXB}{{\color{\colorMATH}\ensuremath{\sss_{1}}}}\endgroup }).\hspace*{0.33em}{\begingroup\renewcommand\colorMATH{\colorMATHB}\renewcommand\colorSYNTAX{\colorSYNTAXB}{{\color{\colorMATH}\ensuremath{\se'}}}\endgroup }, \gamma \rangle ) \in  {\mathcal{V}}_{0}^{k}\llbracket (x:[\varnothing /x_{1},...,\varnothing /x_{n}]\tau _{1}\mathord{\cdotp }{\begingroup\renewcommand\colorMATH{\colorMATHB}\renewcommand\colorSYNTAX{\colorSYNTAXB}{{\color{\colorMATH}\ensuremath{\sss_{1}}}}\endgroup }) \xrightarrowS {{\begingroup\renewcommand\colorMATH{\colorMATHB}\renewcommand\colorSYNTAX{\colorSYNTAXB}{{\color{\colorMATH}\ensuremath{\sss'}}}\endgroup }x} [\varnothing /x_{1},...,\varnothing /x_{n}]\tau _{2}\rrbracket }}}. 
      Notice that if we choose {{\color{\colorMATH}\ensuremath{{\begingroup\renewcommand\colorMATH{\colorMATHB}\renewcommand\colorSYNTAX{\colorSYNTAXB}{{\color{\colorMATH}\ensuremath{\Distance'}}}\endgroup } = 0x_{1},...,0x_{n}}}}, we can use that same {{\color{\colorMATH}\ensuremath{{\begingroup\renewcommand\colorMATH{\colorMATHB}\renewcommand\colorSYNTAX{\colorSYNTAXB}{{\color{\colorMATH}\ensuremath{\Distance'}}}\endgroup }}}} in the induction hypothesis of {{\color{\colorMATH}\ensuremath{\Gamma ', x: \tau _{1}; {\begingroup\renewcommand\colorMATH{\colorMATHB}\renewcommand\colorSYNTAX{\colorSYNTAXB}{{\color{\colorMATH}\ensuremath{\Distance'_{0}}}}\endgroup } + {\begingroup\renewcommand\colorMATH{\colorMATHB}\renewcommand\colorSYNTAX{\colorSYNTAXB}{{\color{\colorMATH}\ensuremath{\sss_{1}}}}\endgroup }x \vdash  {\begingroup\renewcommand\colorMATH{\colorMATHB}\renewcommand\colorSYNTAX{\colorSYNTAXB}{{\color{\colorMATH}\ensuremath{\se'}}}\endgroup } \mathrel{:} \tau _{2} \mathrel{;} {\begingroup\renewcommand\colorMATH{\colorMATHB}\renewcommand\colorSYNTAX{\colorSYNTAXB}{{\color{\colorMATH}\ensuremath{\sS''}}}\endgroup }+{\begingroup\renewcommand\colorMATH{\colorMATHB}\renewcommand\colorSYNTAX{\colorSYNTAXB}{{\color{\colorMATH}\ensuremath{\sss'}}}\endgroup }x}}}, and then we can use the same analogous arguments of the previous case and Lemma~\ref{lm:lrweakening-sensitivity} to conclude the result.
    \end{subproof}
  \item  {{\color{\colorMATH}\ensuremath{\Gamma ; {\begingroup\renewcommand\colorMATH{\colorMATHB}\renewcommand\colorSYNTAX{\colorSYNTAXB}{{\color{\colorMATH}\ensuremath{\Distance}}}\endgroup } \vdash  {\begingroup\renewcommand\colorMATH{\colorMATHB}\renewcommand\colorSYNTAX{\colorSYNTAXB}{{\color{\colorMATH}\ensuremath{\se_{1}}}}\endgroup }\hspace*{0.33em}{\begingroup\renewcommand\colorMATH{\colorMATHB}\renewcommand\colorSYNTAX{\colorSYNTAXB}{{\color{\colorMATH}\ensuremath{\se_{2}}}}\endgroup } \mathrel{:} [{\begingroup\renewcommand\colorMATH{\colorMATHB}\renewcommand\colorSYNTAX{\colorSYNTAXB}{{\color{\colorMATH}\ensuremath{\sS_{2}}}}\endgroup }/x]\tau _{2} \mathrel{;} {\begingroup\renewcommand\colorMATH{\colorMATHB}\renewcommand\colorSYNTAX{\colorSYNTAXB}{{\color{\colorMATH}\ensuremath{\sS_{1}}}}\endgroup } + {\begingroup\renewcommand\colorMATH{\colorMATHB}\renewcommand\colorSYNTAX{\colorSYNTAXB}{{\color{\colorMATH}\ensuremath{\sss}}}\endgroup }{\begingroup\renewcommand\colorMATH{\colorMATHB}\renewcommand\colorSYNTAX{\colorSYNTAXB}{{\color{\colorMATH}\ensuremath{\sS_{2}}}}\endgroup } + {\begingroup\renewcommand\colorMATH{\colorMATHB}\renewcommand\colorSYNTAX{\colorSYNTAXB}{{\color{\colorMATH}\ensuremath{\sS''}}}\endgroup } }}} 
    \begin{subproof} 
      Notice that {{\color{\colorMATH}\ensuremath{{\begingroup\renewcommand\colorMATH{\colorMATHB}\renewcommand\colorSYNTAX{\colorSYNTAXB}{{\color{\colorMATH}\ensuremath{\Distance'}}}\endgroup }\mathord{\cdotp }({\begingroup\renewcommand\colorMATH{\colorMATHB}\renewcommand\colorSYNTAX{\colorSYNTAXB}{{\color{\colorMATH}\ensuremath{\sS_{1}}}}\endgroup } + {\begingroup\renewcommand\colorMATH{\colorMATHB}\renewcommand\colorSYNTAX{\colorSYNTAXB}{{\color{\colorMATH}\ensuremath{\sss}}}\endgroup }{\begingroup\renewcommand\colorMATH{\colorMATHB}\renewcommand\colorSYNTAX{\colorSYNTAXB}{{\color{\colorMATH}\ensuremath{\sS_{2}}}}\endgroup } + {\begingroup\renewcommand\colorMATH{\colorMATHB}\renewcommand\colorSYNTAX{\colorSYNTAXB}{{\color{\colorMATH}\ensuremath{\sS''}}}\endgroup }) = ({\begingroup\renewcommand\colorMATH{\colorMATHB}\renewcommand\colorSYNTAX{\colorSYNTAXB}{{\color{\colorMATH}\ensuremath{\Distance'}}}\endgroup }\mathord{\cdotp }{\begingroup\renewcommand\colorMATH{\colorMATHB}\renewcommand\colorSYNTAX{\colorSYNTAXB}{{\color{\colorMATH}\ensuremath{\sS_{1}}}}\endgroup } + {\begingroup\renewcommand\colorMATH{\colorMATHB}\renewcommand\colorSYNTAX{\colorSYNTAXB}{{\color{\colorMATH}\ensuremath{\sss}}}\endgroup }({\begingroup\renewcommand\colorMATH{\colorMATHB}\renewcommand\colorSYNTAX{\colorSYNTAXB}{{\color{\colorMATH}\ensuremath{\Distance'}}}\endgroup }\mathord{\cdotp }{\begingroup\renewcommand\colorMATH{\colorMATHB}\renewcommand\colorSYNTAX{\colorSYNTAXB}{{\color{\colorMATH}\ensuremath{\sS_{2}}}}\endgroup }) + {\begingroup\renewcommand\colorMATH{\colorMATHB}\renewcommand\colorSYNTAX{\colorSYNTAXB}{{\color{\colorMATH}\ensuremath{\Distance'}}}\endgroup }\mathord{\cdotp }{\begingroup\renewcommand\colorMATH{\colorMATHB}\renewcommand\colorSYNTAX{\colorSYNTAXB}{{\color{\colorMATH}\ensuremath{\sS''}}}\endgroup })}}}, for {{\color{\colorMATH}\ensuremath{{\begingroup\renewcommand\colorMATH{\colorMATHB}\renewcommand\colorSYNTAX{\colorSYNTAXB}{{\color{\colorMATH}\ensuremath{\Distance'}}}\endgroup } \sqsubseteq  {\begingroup\renewcommand\colorMATH{\colorMATHB}\renewcommand\colorSYNTAX{\colorSYNTAXB}{{\color{\colorMATH}\ensuremath{\Distance}}}\endgroup }}}}.\\
      We have to prove that {{\color{\colorMATH}\ensuremath{\forall k, \forall (\gamma _{1},\gamma _{2}) \in  {\mathcal{G}}_{{\begingroup\renewcommand\colorMATH{\colorMATHB}\renewcommand\colorSYNTAX{\colorSYNTAXB}{{\color{\colorMATH}\ensuremath{\Distance'}}}\endgroup }}^{\kg}\llbracket \Gamma \rrbracket , (\gamma _{1}\vdash {\begingroup\renewcommand\colorMATH{\colorMATHB}\renewcommand\colorSYNTAX{\colorSYNTAXB}{{\color{\colorMATH}\ensuremath{\se_{1}}}}\endgroup }\hspace*{0.33em}{\begingroup\renewcommand\colorMATH{\colorMATHB}\renewcommand\colorSYNTAX{\colorSYNTAXB}{{\color{\colorMATH}\ensuremath{\se_{2}}}}\endgroup },\gamma _{2}\vdash {\begingroup\renewcommand\colorMATH{\colorMATHB}\renewcommand\colorSYNTAX{\colorSYNTAXB}{{\color{\colorMATH}\ensuremath{\se_{1}}}}\endgroup }\hspace*{0.33em}{\begingroup\renewcommand\colorMATH{\colorMATHB}\renewcommand\colorSYNTAX{\colorSYNTAXB}{{\color{\colorMATH}\ensuremath{\se_{2}}}}\endgroup }) \in  {\mathcal{E}}^{k}_{{\begingroup\renewcommand\colorMATH{\colorMATHB}\renewcommand\colorSYNTAX{\colorSYNTAXB}{{\color{\colorMATH}\ensuremath{\Distance'}}}\endgroup }\mathord{\cdotp }{\begingroup\renewcommand\colorMATH{\colorMATHB}\renewcommand\colorSYNTAX{\colorSYNTAXB}{{\color{\colorMATH}\ensuremath{\sS_{1}}}}\endgroup } + {\begingroup\renewcommand\colorMATH{\colorMATHB}\renewcommand\colorSYNTAX{\colorSYNTAXB}{{\color{\colorMATH}\ensuremath{\sss}}}\endgroup }({\begingroup\renewcommand\colorMATH{\colorMATHB}\renewcommand\colorSYNTAX{\colorSYNTAXB}{{\color{\colorMATH}\ensuremath{\Distance'}}}\endgroup }\mathord{\cdotp }{\begingroup\renewcommand\colorMATH{\colorMATHB}\renewcommand\colorSYNTAX{\colorSYNTAXB}{{\color{\colorMATH}\ensuremath{\sS_{2}}}}\endgroup }) + {\begingroup\renewcommand\colorMATH{\colorMATHB}\renewcommand\colorSYNTAX{\colorSYNTAXB}{{\color{\colorMATH}\ensuremath{\Distance'}}}\endgroup }\mathord{\cdotp }{\begingroup\renewcommand\colorMATH{\colorMATHB}\renewcommand\colorSYNTAX{\colorSYNTAXB}{{\color{\colorMATH}\ensuremath{\sS''}}}\endgroup }}\llbracket {\begingroup\renewcommand\colorMATH{\colorMATHB}\renewcommand\colorSYNTAX{\colorSYNTAXB}{{\color{\colorMATH}\ensuremath{\Distance'}}}\endgroup }([{\begingroup\renewcommand\colorMATH{\colorMATHB}\renewcommand\colorSYNTAX{\colorSYNTAXB}{{\color{\colorMATH}\ensuremath{\sS_{2}}}}\endgroup }/x]\tau _{2})\rrbracket }}}, i.e.
      if {{\color{\colorMATH}\ensuremath{\gamma _{1}\vdash {\begingroup\renewcommand\colorMATH{\colorMATHB}\renewcommand\colorSYNTAX{\colorSYNTAXB}{{\color{\colorMATH}\ensuremath{\se_{1}}}}\endgroup }\hspace*{0.33em}{\begingroup\renewcommand\colorMATH{\colorMATHB}\renewcommand\colorSYNTAX{\colorSYNTAXB}{{\color{\colorMATH}\ensuremath{\se_{2}}}}\endgroup } \Downarrow ^{j} {\begingroup\renewcommand\colorMATH{\colorMATHB}\renewcommand\colorSYNTAX{\colorSYNTAXB}{{\color{\colorMATH}\ensuremath{\sv'_{1}}}}\endgroup }}}} \pthen {{\color{\colorMATH}\ensuremath{\gamma _{2}\vdash {\begingroup\renewcommand\colorMATH{\colorMATHB}\renewcommand\colorSYNTAX{\colorSYNTAXB}{{\color{\colorMATH}\ensuremath{\se_{1}}}}\endgroup }\hspace*{0.33em}{\begingroup\renewcommand\colorMATH{\colorMATHB}\renewcommand\colorSYNTAX{\colorSYNTAXB}{{\color{\colorMATH}\ensuremath{\se_{2}}}}\endgroup } \Downarrow ^{\pj} {\begingroup\renewcommand\colorMATH{\colorMATHB}\renewcommand\colorSYNTAX{\colorSYNTAXB}{{\color{\colorMATH}\ensuremath{\sv'_{2}}}}\endgroup }}}}, \pand
      {{\color{\colorMATH}\ensuremath{({\begingroup\renewcommand\colorMATH{\colorMATHB}\renewcommand\colorSYNTAX{\colorSYNTAXB}{{\color{\colorMATH}\ensuremath{\sv'_{1}}}}\endgroup }, {\begingroup\renewcommand\colorMATH{\colorMATHB}\renewcommand\colorSYNTAX{\colorSYNTAXB}{{\color{\colorMATH}\ensuremath{\sv'_{2}}}}\endgroup }) \in  {\mathcal{V}}^{k-j}_{{\begingroup\renewcommand\colorMATH{\colorMATHB}\renewcommand\colorSYNTAX{\colorSYNTAXB}{{\color{\colorMATH}\ensuremath{\Distance'}}}\endgroup }\mathord{\cdotp }{\begingroup\renewcommand\colorMATH{\colorMATHB}\renewcommand\colorSYNTAX{\colorSYNTAXB}{{\color{\colorMATH}\ensuremath{\sS_{1}}}}\endgroup } + s({\begingroup\renewcommand\colorMATH{\colorMATHB}\renewcommand\colorSYNTAX{\colorSYNTAXB}{{\color{\colorMATH}\ensuremath{\Distance'}}}\endgroup }\mathord{\cdotp }{\begingroup\renewcommand\colorMATH{\colorMATHB}\renewcommand\colorSYNTAX{\colorSYNTAXB}{{\color{\colorMATH}\ensuremath{\sS_{2}}}}\endgroup }) + {\begingroup\renewcommand\colorMATH{\colorMATHB}\renewcommand\colorSYNTAX{\colorSYNTAXB}{{\color{\colorMATH}\ensuremath{\Distance'}}}\endgroup }\mathord{\cdotp }{\begingroup\renewcommand\colorMATH{\colorMATHB}\renewcommand\colorSYNTAX{\colorSYNTAXB}{{\color{\colorMATH}\ensuremath{\sS''}}}\endgroup }}\llbracket {\begingroup\renewcommand\colorMATH{\colorMATHB}\renewcommand\colorSYNTAX{\colorSYNTAXB}{{\color{\colorMATH}\ensuremath{\Distance'}}}\endgroup }([{\begingroup\renewcommand\colorMATH{\colorMATHB}\renewcommand\colorSYNTAX{\colorSYNTAXB}{{\color{\colorMATH}\ensuremath{\sS_{2}}}}\endgroup }/x]\tau _{2})\rrbracket }}}.

      By induction hypotheses we know that\\
      {{\color{\colorMATH}\ensuremath{\Gamma  \vdash  {\begingroup\renewcommand\colorMATH{\colorMATHB}\renewcommand\colorSYNTAX{\colorSYNTAXB}{{\color{\colorMATH}\ensuremath{\se_{1}}}}\endgroup } \mathrel{:} (x:\tau _{1}) \xrightarrowS {{\begingroup\renewcommand\colorMATH{\colorMATHB}\renewcommand\colorSYNTAX{\colorSYNTAXB}{{\color{\colorMATH}\ensuremath{\sS''}}}\endgroup }+sx} \tau _{2} \mathrel{;} {\begingroup\renewcommand\colorMATH{\colorMATHB}\renewcommand\colorSYNTAX{\colorSYNTAXB}{{\color{\colorMATH}\ensuremath{\sS_{1}}}}\endgroup } \Rightarrow  (\gamma _{1} \vdash  {\begingroup\renewcommand\colorMATH{\colorMATHB}\renewcommand\colorSYNTAX{\colorSYNTAXB}{{\color{\colorMATH}\ensuremath{\se_{1}}}}\endgroup },\gamma _{2} \vdash  {\begingroup\renewcommand\colorMATH{\colorMATHB}\renewcommand\colorSYNTAX{\colorSYNTAXB}{{\color{\colorMATH}\ensuremath{\se_{1}}}}\endgroup }) \in  {\mathcal{E}}^{k}_{{\begingroup\renewcommand\colorMATH{\colorMATHB}\renewcommand\colorSYNTAX{\colorSYNTAXB}{{\color{\colorMATH}\ensuremath{\Distance'}}}\endgroup }\mathord{\cdotp }{\begingroup\renewcommand\colorMATH{\colorMATHB}\renewcommand\colorSYNTAX{\colorSYNTAXB}{{\color{\colorMATH}\ensuremath{\sS_{1}}}}\endgroup }}\llbracket {\begingroup\renewcommand\colorMATH{\colorMATHB}\renewcommand\colorSYNTAX{\colorSYNTAXB}{{\color{\colorMATH}\ensuremath{\Distance'}}}\endgroup }((x:\tau _{1}\mathord{\cdotp }{\begingroup\renewcommand\colorMATH{\colorMATHB}\renewcommand\colorSYNTAX{\colorSYNTAXB}{{\color{\colorMATH}\ensuremath{\sss'}}}\endgroup }) \xrightarrowS {{\begingroup\renewcommand\colorMATH{\colorMATHB}\renewcommand\colorSYNTAX{\colorSYNTAXB}{{\color{\colorMATH}\ensuremath{\sS''}}}\endgroup }+{\begingroup\renewcommand\colorMATH{\colorMATHB}\renewcommand\colorSYNTAX{\colorSYNTAXB}{{\color{\colorMATH}\ensuremath{\sss}}}\endgroup }x} \tau _{2})\rrbracket }}} and 
      {{\color{\colorMATH}\ensuremath{\Gamma  \vdash  {\begingroup\renewcommand\colorMATH{\colorMATHB}\renewcommand\colorSYNTAX{\colorSYNTAXB}{{\color{\colorMATH}\ensuremath{\se_{2}}}}\endgroup } \mathrel{:} \tau _{1} \mathrel{;} {\begingroup\renewcommand\colorMATH{\colorMATHB}\renewcommand\colorSYNTAX{\colorSYNTAXB}{{\color{\colorMATH}\ensuremath{\sS_{2}}}}\endgroup } \Rightarrow  (\gamma _{1} \vdash  {\begingroup\renewcommand\colorMATH{\colorMATHB}\renewcommand\colorSYNTAX{\colorSYNTAXB}{{\color{\colorMATH}\ensuremath{\se_{2}}}}\endgroup },\gamma _{2} \vdash  {\begingroup\renewcommand\colorMATH{\colorMATHB}\renewcommand\colorSYNTAX{\colorSYNTAXB}{{\color{\colorMATH}\ensuremath{\se_{2}}}}\endgroup }) \in  {\mathcal{E}}^{k-j_{1}}_{{\begingroup\renewcommand\colorMATH{\colorMATHB}\renewcommand\colorSYNTAX{\colorSYNTAXB}{{\color{\colorMATH}\ensuremath{\Distance'}}}\endgroup }\mathord{\cdotp }{\begingroup\renewcommand\colorMATH{\colorMATHB}\renewcommand\colorSYNTAX{\colorSYNTAXB}{{\color{\colorMATH}\ensuremath{\sS_{2}}}}\endgroup }}\llbracket {\begingroup\renewcommand\colorMATH{\colorMATHB}\renewcommand\colorSYNTAX{\colorSYNTAXB}{{\color{\colorMATH}\ensuremath{\Distance'}}}\endgroup }(\tau _{1})\rrbracket }}} for some {{\color{\colorMATH}\ensuremath{j_{1} \leq  k}}}.
      As {{\color{\colorMATH}\ensuremath{{\begingroup\renewcommand\colorMATH{\colorMATHB}\renewcommand\colorSYNTAX{\colorSYNTAXB}{{\color{\colorMATH}\ensuremath{\Distance'}}}\endgroup }((x:\tau _{1}{\begingroup\renewcommand\colorMATH{\colorMATHB}\renewcommand\colorSYNTAX{\colorSYNTAXB}{{\color{\colorMATH}\ensuremath{\sss'}}}\endgroup }) \xrightarrowS {{\begingroup\renewcommand\colorMATH{\colorMATHB}\renewcommand\colorSYNTAX{\colorSYNTAXB}{{\color{\colorMATH}\ensuremath{\sS''}}}\endgroup }+{\begingroup\renewcommand\colorMATH{\colorMATHB}\renewcommand\colorSYNTAX{\colorSYNTAXB}{{\color{\colorMATH}\ensuremath{\sss}}}\endgroup }x} \tau _{2}) = (x:{\begingroup\renewcommand\colorMATH{\colorMATHB}\renewcommand\colorSYNTAX{\colorSYNTAXB}{{\color{\colorMATH}\ensuremath{\Distance'}}}\endgroup }(\tau _{1})\mathord{\cdotp }{\begingroup\renewcommand\colorMATH{\colorMATHB}\renewcommand\colorSYNTAX{\colorSYNTAXB}{{\color{\colorMATH}\ensuremath{\sss'}}}\endgroup }) \xrightarrowS {{\begingroup\renewcommand\colorMATH{\colorMATHB}\renewcommand\colorSYNTAX{\colorSYNTAXB}{{\color{\colorMATH}\ensuremath{\Distance'}}}\endgroup }\mathord{\cdotp }{\begingroup\renewcommand\colorMATH{\colorMATHB}\renewcommand\colorSYNTAX{\colorSYNTAXB}{{\color{\colorMATH}\ensuremath{\sS''}}}\endgroup }+{\begingroup\renewcommand\colorMATH{\colorMATHB}\renewcommand\colorSYNTAX{\colorSYNTAXB}{{\color{\colorMATH}\ensuremath{\sss}}}\endgroup }x} {\begingroup\renewcommand\colorMATH{\colorMATHB}\renewcommand\colorSYNTAX{\colorSYNTAXB}{{\color{\colorMATH}\ensuremath{\Distance'}}}\endgroup }(\tau _{2})}}}, by unfolding {{\color{\colorMATH}\ensuremath{(\gamma _{1} \vdash  {\begingroup\renewcommand\colorMATH{\colorMATHB}\renewcommand\colorSYNTAX{\colorSYNTAXB}{{\color{\colorMATH}\ensuremath{\se_{1}}}}\endgroup },\gamma _{2} \vdash  {\begingroup\renewcommand\colorMATH{\colorMATHB}\renewcommand\colorSYNTAX{\colorSYNTAXB}{{\color{\colorMATH}\ensuremath{\se_{1}}}}\endgroup }) \in  {\mathcal{E}}^{k}_{{\begingroup\renewcommand\colorMATH{\colorMATHB}\renewcommand\colorSYNTAX{\colorSYNTAXB}{{\color{\colorMATH}\ensuremath{\Distance'}}}\endgroup }\mathord{\cdotp }{\begingroup\renewcommand\colorMATH{\colorMATHB}\renewcommand\colorSYNTAX{\colorSYNTAXB}{{\color{\colorMATH}\ensuremath{\sS_{1}}}}\endgroup }}\llbracket (x:{\begingroup\renewcommand\colorMATH{\colorMATHB}\renewcommand\colorSYNTAX{\colorSYNTAXB}{{\color{\colorMATH}\ensuremath{\Distance'}}}\endgroup }(\tau _{1}\mathord{\cdotp }{\begingroup\renewcommand\colorMATH{\colorMATHB}\renewcommand\colorSYNTAX{\colorSYNTAXB}{{\color{\colorMATH}\ensuremath{\sss'}}}\endgroup })) \xrightarrowS {{\begingroup\renewcommand\colorMATH{\colorMATHB}\renewcommand\colorSYNTAX{\colorSYNTAXB}{{\color{\colorMATH}\ensuremath{\Distance'}}}\endgroup }\mathord{\cdotp }{\begingroup\renewcommand\colorMATH{\colorMATHB}\renewcommand\colorSYNTAX{\colorSYNTAXB}{{\color{\colorMATH}\ensuremath{\sS''}}}\endgroup }+{\begingroup\renewcommand\colorMATH{\colorMATHB}\renewcommand\colorSYNTAX{\colorSYNTAXB}{{\color{\colorMATH}\ensuremath{\sss}}}\endgroup }x} {\begingroup\renewcommand\colorMATH{\colorMATHB}\renewcommand\colorSYNTAX{\colorSYNTAXB}{{\color{\colorMATH}\ensuremath{\Distance'}}}\endgroup }(\tau _{2})\rrbracket }}},
      we know that if {{\color{\colorMATH}\ensuremath{\gamma _{1} \vdash  {\begingroup\renewcommand\colorMATH{\colorMATHB}\renewcommand\colorSYNTAX{\colorSYNTAXB}{{\color{\colorMATH}\ensuremath{\se_{1}}}}\endgroup } \Downarrow ^{j_{1}} \langle \lambda (x\mathrel{:}\tau _{1}).\hspace*{0.33em}{\begingroup\renewcommand\colorMATH{\colorMATHB}\renewcommand\colorSYNTAX{\colorSYNTAXB}{{\color{\colorMATH}\ensuremath{\se'_{1}}}}\endgroup }, \gamma '_{1}\rangle }}} \pthen {{\color{\colorMATH}\ensuremath{\gamma _{2} \vdash  {\begingroup\renewcommand\colorMATH{\colorMATHB}\renewcommand\colorSYNTAX{\colorSYNTAXB}{{\color{\colorMATH}\ensuremath{\se_{1}}}}\endgroup } \Downarrow ^{\pj[1]} \langle \lambda (x\mathrel{:}\tau _{1}).\hspace*{0.33em}{\begingroup\renewcommand\colorMATH{\colorMATHB}\renewcommand\colorSYNTAX{\colorSYNTAXB}{{\color{\colorMATH}\ensuremath{\se'_{2}}}}\endgroup }, \gamma '_{2}\rangle }}} \pand\\ {{\color{\colorMATH}\ensuremath{(\langle {\begingroup\renewcommand\colorMATH{\colorMATHB}\renewcommand\colorSYNTAX{\colorSYNTAXB}{{\color{\colorMATH}\ensuremath{\slambda}}}\endgroup } (x\mathrel{:}\tau _{1}).\hspace*{0.33em}{\begingroup\renewcommand\colorMATH{\colorMATHB}\renewcommand\colorSYNTAX{\colorSYNTAXB}{{\color{\colorMATH}\ensuremath{\se'_{1}}}}\endgroup }, \gamma '_{1}\rangle , \langle {\begingroup\renewcommand\colorMATH{\colorMATHB}\renewcommand\colorSYNTAX{\colorSYNTAXB}{{\color{\colorMATH}\ensuremath{\slambda}}}\endgroup } (x\mathrel{:}\tau _{1}\mathord{\cdotp }{\begingroup\renewcommand\colorMATH{\colorMATHB}\renewcommand\colorSYNTAX{\colorSYNTAXB}{{\color{\colorMATH}\ensuremath{\sss'}}}\endgroup }).\hspace*{0.33em}{\begingroup\renewcommand\colorMATH{\colorMATHB}\renewcommand\colorSYNTAX{\colorSYNTAXB}{{\color{\colorMATH}\ensuremath{\se'_{2}}}}\endgroup }, \gamma '_{2}\rangle ) \in  {\mathcal{V}}^{k-j}_{{\begingroup\renewcommand\colorMATH{\colorMATHB}\renewcommand\colorSYNTAX{\colorSYNTAXB}{{\color{\colorMATH}\ensuremath{\Distance'}}}\endgroup }\mathord{\cdotp }{\begingroup\renewcommand\colorMATH{\colorMATHB}\renewcommand\colorSYNTAX{\colorSYNTAXB}{{\color{\colorMATH}\ensuremath{\sS_{1}}}}\endgroup }}\llbracket (x:{\begingroup\renewcommand\colorMATH{\colorMATHB}\renewcommand\colorSYNTAX{\colorSYNTAXB}{{\color{\colorMATH}\ensuremath{\Distance'}}}\endgroup }(\tau _{1}\mathord{\cdotp }{\begingroup\renewcommand\colorMATH{\colorMATHB}\renewcommand\colorSYNTAX{\colorSYNTAXB}{{\color{\colorMATH}\ensuremath{\sss'}}}\endgroup })) \xrightarrowS {{\begingroup\renewcommand\colorMATH{\colorMATHB}\renewcommand\colorSYNTAX{\colorSYNTAXB}{{\color{\colorMATH}\ensuremath{\Distance'}}}\endgroup }\mathord{\cdotp }{\begingroup\renewcommand\colorMATH{\colorMATHB}\renewcommand\colorSYNTAX{\colorSYNTAXB}{{\color{\colorMATH}\ensuremath{\sS''}}}\endgroup }+{\begingroup\renewcommand\colorMATH{\colorMATHB}\renewcommand\colorSYNTAX{\colorSYNTAXB}{{\color{\colorMATH}\ensuremath{\sss}}}\endgroup }x} {\begingroup\renewcommand\colorMATH{\colorMATHB}\renewcommand\colorSYNTAX{\colorSYNTAXB}{{\color{\colorMATH}\ensuremath{\Distance'}}}\endgroup }(\tau _{2})\rrbracket }}} (1).\\
      Also, by unfolding {{\color{\colorMATH}\ensuremath{(\gamma _{1} \vdash  {\begingroup\renewcommand\colorMATH{\colorMATHB}\renewcommand\colorSYNTAX{\colorSYNTAXB}{{\color{\colorMATH}\ensuremath{\se_{2}}}}\endgroup },\gamma _{2} \vdash  {\begingroup\renewcommand\colorMATH{\colorMATHB}\renewcommand\colorSYNTAX{\colorSYNTAXB}{{\color{\colorMATH}\ensuremath{\se_{2}}}}\endgroup }) \in  {\mathcal{E}}^{k-j_{1}}_{{\begingroup\renewcommand\colorMATH{\colorMATHB}\renewcommand\colorSYNTAX{\colorSYNTAXB}{{\color{\colorMATH}\ensuremath{\Distance'}}}\endgroup }\mathord{\cdotp }{\begingroup\renewcommand\colorMATH{\colorMATHB}\renewcommand\colorSYNTAX{\colorSYNTAXB}{{\color{\colorMATH}\ensuremath{\sS_{2}}}}\endgroup }}\llbracket {\begingroup\renewcommand\colorMATH{\colorMATHB}\renewcommand\colorSYNTAX{\colorSYNTAXB}{{\color{\colorMATH}\ensuremath{\Distance'}}}\endgroup }(\tau _{1}\mathord{\cdotp }{\begingroup\renewcommand\colorMATH{\colorMATHB}\renewcommand\colorSYNTAX{\colorSYNTAXB}{{\color{\colorMATH}\ensuremath{\sss'}}}\endgroup })\rrbracket }}}, if {{\color{\colorMATH}\ensuremath{\gamma _{1} \vdash  {\begingroup\renewcommand\colorMATH{\colorMATHB}\renewcommand\colorSYNTAX{\colorSYNTAXB}{{\color{\colorMATH}\ensuremath{\se_{2}}}}\endgroup } \Downarrow ^{j_{2}} v_{1}}}} \pthen {{\color{\colorMATH}\ensuremath{\gamma _{2} \vdash  {\begingroup\renewcommand\colorMATH{\colorMATHB}\renewcommand\colorSYNTAX{\colorSYNTAXB}{{\color{\colorMATH}\ensuremath{\se_{2}}}}\endgroup } \Downarrow ^{\pj[2]} v_{2}}}} \pand {{\color{\colorMATH}\ensuremath{(v_{1}, v_{2}) \in  {\mathcal{V}}^{k-j_{1}-j_{2}}_{{\begingroup\renewcommand\colorMATH{\colorMATHB}\renewcommand\colorSYNTAX{\colorSYNTAXB}{{\color{\colorMATH}\ensuremath{\Distance'}}}\endgroup }\mathord{\cdotp }{\begingroup\renewcommand\colorMATH{\colorMATHB}\renewcommand\colorSYNTAX{\colorSYNTAXB}{{\color{\colorMATH}\ensuremath{\sS_{2}}}}\endgroup }}\llbracket {\begingroup\renewcommand\colorMATH{\colorMATHB}\renewcommand\colorSYNTAX{\colorSYNTAXB}{{\color{\colorMATH}\ensuremath{\Distance'}}}\endgroup }(\tau _{1})\rrbracket }}}. %(then {{\color{\colorMATH}\ensuremath{j = j_{1}+ j_{2} + 1}}}).

      As {{\color{\colorMATH}\ensuremath{{\begingroup\renewcommand\colorMATH{\colorMATHB}\renewcommand\colorSYNTAX{\colorSYNTAXB}{{\color{\colorMATH}\ensuremath{\Distance'}}}\endgroup } \mathord{\cdotp } {\begingroup\renewcommand\colorMATH{\colorMATHB}\renewcommand\colorSYNTAX{\colorSYNTAXB}{{\color{\colorMATH}\ensuremath{\sS_{2}}}}\endgroup } \in  {\text{sens}}}}}, we instantiate (1) with {{\color{\colorMATH}\ensuremath{{\begingroup\renewcommand\colorMATH{\colorMATHB}\renewcommand\colorSYNTAX{\colorSYNTAXB}{{\color{\colorMATH}\ensuremath{\sss''}}}\endgroup } = {\begingroup\renewcommand\colorMATH{\colorMATHB}\renewcommand\colorSYNTAX{\colorSYNTAXB}{{\color{\colorMATH}\ensuremath{\Distance'}}}\endgroup }\mathord{\cdotp }{\begingroup\renewcommand\colorMATH{\colorMATHB}\renewcommand\colorSYNTAX{\colorSYNTAXB}{{\color{\colorMATH}\ensuremath{\sS_{2}}}}\endgroup }}}} (we know that {{\color{\colorMATH}\ensuremath{{\begingroup\renewcommand\colorMATH{\colorMATHB}\renewcommand\colorSYNTAX{\colorSYNTAXB}{{\color{\colorMATH}\ensuremath{\sss''}}}\endgroup } \leq  {\begingroup\renewcommand\colorMATH{\colorMATHB}\renewcommand\colorSYNTAX{\colorSYNTAXB}{{\color{\colorMATH}\ensuremath{\Distance}}}\endgroup }\mathord{\cdotp }{\begingroup\renewcommand\colorMATH{\colorMATHB}\renewcommand\colorSYNTAX{\colorSYNTAXB}{{\color{\colorMATH}\ensuremath{\sS_{2}}}}\endgroup } \leq  {\begingroup\renewcommand\colorMATH{\colorMATHB}\renewcommand\colorSYNTAX{\colorSYNTAXB}{{\color{\colorMATH}\ensuremath{\sss'}}}\endgroup }}}}), then\\
      {{\color{\colorMATH}\ensuremath{(\gamma '_{1}[x\mapsto {\begingroup\renewcommand\colorMATH{\colorMATHB}\renewcommand\colorSYNTAX{\colorSYNTAXB}{{\color{\colorMATH}\ensuremath{\sv_{1}}}}\endgroup }] \vdash  {\begingroup\renewcommand\colorMATH{\colorMATHB}\renewcommand\colorSYNTAX{\colorSYNTAXB}{{\color{\colorMATH}\ensuremath{\se'_{1}}}}\endgroup },\gamma '_{2}[x\mapsto {\begingroup\renewcommand\colorMATH{\colorMATHB}\renewcommand\colorSYNTAX{\colorSYNTAXB}{{\color{\colorMATH}\ensuremath{\sv_{2}}}}\endgroup }] \vdash  {\begingroup\renewcommand\colorMATH{\colorMATHB}\renewcommand\colorSYNTAX{\colorSYNTAXB}{{\color{\colorMATH}\ensuremath{\se'_{2}}}}\endgroup }) \in  {\mathcal{E}}^{k-j_{1}-j_{2}-1}_{{\begingroup\renewcommand\colorMATH{\colorMATHB}\renewcommand\colorSYNTAX{\colorSYNTAXB}{{\color{\colorMATH}\ensuremath{\Distance'}}}\endgroup }\mathord{\cdotp }{\begingroup\renewcommand\colorMATH{\colorMATHB}\renewcommand\colorSYNTAX{\colorSYNTAXB}{{\color{\colorMATH}\ensuremath{\sS_{1}}}}\endgroup }+{\begingroup\renewcommand\colorMATH{\colorMATHB}\renewcommand\colorSYNTAX{\colorSYNTAXB}{{\color{\colorMATH}\ensuremath{\Distance'}}}\endgroup }\mathord{\cdotp }{\begingroup\renewcommand\colorMATH{\colorMATHB}\renewcommand\colorSYNTAX{\colorSYNTAXB}{{\color{\colorMATH}\ensuremath{\sS''}}}\endgroup }+{\begingroup\renewcommand\colorMATH{\colorMATHB}\renewcommand\colorSYNTAX{\colorSYNTAXB}{{\color{\colorMATH}\ensuremath{\sss}}}\endgroup }({\begingroup\renewcommand\colorMATH{\colorMATHB}\renewcommand\colorSYNTAX{\colorSYNTAXB}{{\color{\colorMATH}\ensuremath{\Distance'}}}\endgroup }\mathord{\cdotp }{\begingroup\renewcommand\colorMATH{\colorMATHB}\renewcommand\colorSYNTAX{\colorSYNTAXB}{{\color{\colorMATH}\ensuremath{\sS_{2}}}}\endgroup })}\llbracket [{\begingroup\renewcommand\colorMATH{\colorMATHB}\renewcommand\colorSYNTAX{\colorSYNTAXB}{{\color{\colorMATH}\ensuremath{\Distance'}}}\endgroup }\mathord{\cdotp }{\begingroup\renewcommand\colorMATH{\colorMATHB}\renewcommand\colorSYNTAX{\colorSYNTAXB}{{\color{\colorMATH}\ensuremath{\sS_{2}}}}\endgroup }/x]{\begingroup\renewcommand\colorMATH{\colorMATHB}\renewcommand\colorSYNTAX{\colorSYNTAXB}{{\color{\colorMATH}\ensuremath{\Distance'}}}\endgroup }(\tau _{2})\rrbracket }}}.
      But {{\color{\colorMATH}\ensuremath{{\begingroup\renewcommand\colorMATH{\colorMATHB}\renewcommand\colorSYNTAX{\colorSYNTAXB}{{\color{\colorMATH}\ensuremath{\Distance'}}}\endgroup }\mathord{\cdotp }{\begingroup\renewcommand\colorMATH{\colorMATHB}\renewcommand\colorSYNTAX{\colorSYNTAXB}{{\color{\colorMATH}\ensuremath{\sS_{1}}}}\endgroup }+{\begingroup\renewcommand\colorMATH{\colorMATHB}\renewcommand\colorSYNTAX{\colorSYNTAXB}{{\color{\colorMATH}\ensuremath{\Distance'}}}\endgroup }\mathord{\cdotp }{\begingroup\renewcommand\colorMATH{\colorMATHB}\renewcommand\colorSYNTAX{\colorSYNTAXB}{{\color{\colorMATH}\ensuremath{\sS''}}}\endgroup }+{\begingroup\renewcommand\colorMATH{\colorMATHB}\renewcommand\colorSYNTAX{\colorSYNTAXB}{{\color{\colorMATH}\ensuremath{\sss}}}\endgroup }({\begingroup\renewcommand\colorMATH{\colorMATHB}\renewcommand\colorSYNTAX{\colorSYNTAXB}{{\color{\colorMATH}\ensuremath{\Distance'}}}\endgroup }\mathord{\cdotp }{\begingroup\renewcommand\colorMATH{\colorMATHB}\renewcommand\colorSYNTAX{\colorSYNTAXB}{{\color{\colorMATH}\ensuremath{\sS_{2}}}}\endgroup }) = {\begingroup\renewcommand\colorMATH{\colorMATHB}\renewcommand\colorSYNTAX{\colorSYNTAXB}{{\color{\colorMATH}\ensuremath{\Distance'}}}\endgroup }\mathord{\cdotp }{\begingroup\renewcommand\colorMATH{\colorMATHB}\renewcommand\colorSYNTAX{\colorSYNTAXB}{{\color{\colorMATH}\ensuremath{\sS_{1}}}}\endgroup }+{\begingroup\renewcommand\colorMATH{\colorMATHB}\renewcommand\colorSYNTAX{\colorSYNTAXB}{{\color{\colorMATH}\ensuremath{\sss}}}\endgroup }({\begingroup\renewcommand\colorMATH{\colorMATHB}\renewcommand\colorSYNTAX{\colorSYNTAXB}{{\color{\colorMATH}\ensuremath{\Distance'}}}\endgroup }\mathord{\cdotp }{\begingroup\renewcommand\colorMATH{\colorMATHB}\renewcommand\colorSYNTAX{\colorSYNTAXB}{{\color{\colorMATH}\ensuremath{\sS_{2}}}}\endgroup })+{\begingroup\renewcommand\colorMATH{\colorMATHB}\renewcommand\colorSYNTAX{\colorSYNTAXB}{{\color{\colorMATH}\ensuremath{\Distance'}}}\endgroup }\mathord{\cdotp }{\begingroup\renewcommand\colorMATH{\colorMATHB}\renewcommand\colorSYNTAX{\colorSYNTAXB}{{\color{\colorMATH}\ensuremath{\sS''}}}\endgroup }}}}, and by Lemma~\ref{lm:distrinst}, {{\color{\colorMATH}\ensuremath{[{\begingroup\renewcommand\colorMATH{\colorMATHB}\renewcommand\colorSYNTAX{\colorSYNTAXB}{{\color{\colorMATH}\ensuremath{\Distance'}}}\endgroup }\mathord{\cdotp }{\begingroup\renewcommand\colorMATH{\colorMATHB}\renewcommand\colorSYNTAX{\colorSYNTAXB}{{\color{\colorMATH}\ensuremath{\sS_{2}}}}\endgroup }/x]{\begingroup\renewcommand\colorMATH{\colorMATHB}\renewcommand\colorSYNTAX{\colorSYNTAXB}{{\color{\colorMATH}\ensuremath{\Distance'}}}\endgroup }(\tau _{2}) = {\begingroup\renewcommand\colorMATH{\colorMATHB}\renewcommand\colorSYNTAX{\colorSYNTAXB}{{\color{\colorMATH}\ensuremath{\Distance'}}}\endgroup }([{\begingroup\renewcommand\colorMATH{\colorMATHB}\renewcommand\colorSYNTAX{\colorSYNTAXB}{{\color{\colorMATH}\ensuremath{\sS_{2}}}}\endgroup }/x]\tau _{2})}}}, therefore\\ {{\color{\colorMATH}\ensuremath{(\gamma '_{1}[x\mapsto {\begingroup\renewcommand\colorMATH{\colorMATHB}\renewcommand\colorSYNTAX{\colorSYNTAXB}{{\color{\colorMATH}\ensuremath{\sv_{1}}}}\endgroup }] \vdash  {\begingroup\renewcommand\colorMATH{\colorMATHB}\renewcommand\colorSYNTAX{\colorSYNTAXB}{{\color{\colorMATH}\ensuremath{\se'_{1}}}}\endgroup },\gamma '_{2}[x\mapsto {\begingroup\renewcommand\colorMATH{\colorMATHB}\renewcommand\colorSYNTAX{\colorSYNTAXB}{{\color{\colorMATH}\ensuremath{\sv_{2}}}}\endgroup }] \vdash  {\begingroup\renewcommand\colorMATH{\colorMATHB}\renewcommand\colorSYNTAX{\colorSYNTAXB}{{\color{\colorMATH}\ensuremath{\se'_{2}}}}\endgroup }) \in  {\mathcal{E}}^{k-j_{1}-j_{2}-1}_{{\begingroup\renewcommand\colorMATH{\colorMATHB}\renewcommand\colorSYNTAX{\colorSYNTAXB}{{\color{\colorMATH}\ensuremath{\Distance'}}}\endgroup }\mathord{\cdotp }{\begingroup\renewcommand\colorMATH{\colorMATHB}\renewcommand\colorSYNTAX{\colorSYNTAXB}{{\color{\colorMATH}\ensuremath{\sS_{1}}}}\endgroup }+{\begingroup\renewcommand\colorMATH{\colorMATHB}\renewcommand\colorSYNTAX{\colorSYNTAXB}{{\color{\colorMATH}\ensuremath{\sss}}}\endgroup }({\begingroup\renewcommand\colorMATH{\colorMATHB}\renewcommand\colorSYNTAX{\colorSYNTAXB}{{\color{\colorMATH}\ensuremath{\Distance'}}}\endgroup }\mathord{\cdotp }{\begingroup\renewcommand\colorMATH{\colorMATHB}\renewcommand\colorSYNTAX{\colorSYNTAXB}{{\color{\colorMATH}\ensuremath{\sS_{2}}}}\endgroup })+{\begingroup\renewcommand\colorMATH{\colorMATHB}\renewcommand\colorSYNTAX{\colorSYNTAXB}{{\color{\colorMATH}\ensuremath{\Distance'}}}\endgroup }\mathord{\cdotp }{\begingroup\renewcommand\colorMATH{\colorMATHB}\renewcommand\colorSYNTAX{\colorSYNTAXB}{{\color{\colorMATH}\ensuremath{\sS''}}}\endgroup }}\llbracket {\begingroup\renewcommand\colorMATH{\colorMATHB}\renewcommand\colorSYNTAX{\colorSYNTAXB}{{\color{\colorMATH}\ensuremath{\Distance'}}}\endgroup }([{\begingroup\renewcommand\colorMATH{\colorMATHB}\renewcommand\colorSYNTAX{\colorSYNTAXB}{{\color{\colorMATH}\ensuremath{\sS_{2}}}}\endgroup }/x]\tau _{2})\rrbracket }}}, i.e.
      if {{\color{\colorMATH}\ensuremath{\gamma '_{1}[x\mapsto {\begingroup\renewcommand\colorMATH{\colorMATHB}\renewcommand\colorSYNTAX{\colorSYNTAXB}{{\color{\colorMATH}\ensuremath{\sv_{1}}}}\endgroup }]\vdash {\begingroup\renewcommand\colorMATH{\colorMATHB}\renewcommand\colorSYNTAX{\colorSYNTAXB}{{\color{\colorMATH}\ensuremath{\se'_{1}}}}\endgroup } \Downarrow ^{j_{3}} {\begingroup\renewcommand\colorMATH{\colorMATHB}\renewcommand\colorSYNTAX{\colorSYNTAXB}{{\color{\colorMATH}\ensuremath{\sv''_{1}}}}\endgroup }}}} \pthen {{\color{\colorMATH}\ensuremath{\gamma '_{2}[x\mapsto {\begingroup\renewcommand\colorMATH{\colorMATHB}\renewcommand\colorSYNTAX{\colorSYNTAXB}{{\color{\colorMATH}\ensuremath{\sv_{2}}}}\endgroup }]\vdash {\begingroup\renewcommand\colorMATH{\colorMATHB}\renewcommand\colorSYNTAX{\colorSYNTAXB}{{\color{\colorMATH}\ensuremath{\se'_{2}}}}\endgroup } \Downarrow ^{\pj[3]} {\begingroup\renewcommand\colorMATH{\colorMATHB}\renewcommand\colorSYNTAX{\colorSYNTAXB}{{\color{\colorMATH}\ensuremath{\sv''_{2}}}}\endgroup }}}}, \pand 
      {{\color{\colorMATH}\ensuremath{({\begingroup\renewcommand\colorMATH{\colorMATHB}\renewcommand\colorSYNTAX{\colorSYNTAXB}{{\color{\colorMATH}\ensuremath{\sv''_{1}}}}\endgroup }, {\begingroup\renewcommand\colorMATH{\colorMATHB}\renewcommand\colorSYNTAX{\colorSYNTAXB}{{\color{\colorMATH}\ensuremath{\sv''_{2}}}}\endgroup }) \in  {\mathcal{V}}^{k-j_{1}-j_{2}-j_{3}-1}_{{\begingroup\renewcommand\colorMATH{\colorMATHB}\renewcommand\colorSYNTAX{\colorSYNTAXB}{{\color{\colorMATH}\ensuremath{\Distance'}}}\endgroup }\mathord{\cdotp }{\begingroup\renewcommand\colorMATH{\colorMATHB}\renewcommand\colorSYNTAX{\colorSYNTAXB}{{\color{\colorMATH}\ensuremath{\sS_{1}}}}\endgroup } + {\begingroup\renewcommand\colorMATH{\colorMATHB}\renewcommand\colorSYNTAX{\colorSYNTAXB}{{\color{\colorMATH}\ensuremath{\sss}}}\endgroup }({\begingroup\renewcommand\colorMATH{\colorMATHB}\renewcommand\colorSYNTAX{\colorSYNTAXB}{{\color{\colorMATH}\ensuremath{\Distance'}}}\endgroup }\mathord{\cdotp }{\begingroup\renewcommand\colorMATH{\colorMATHB}\renewcommand\colorSYNTAX{\colorSYNTAXB}{{\color{\colorMATH}\ensuremath{\sS_{2}}}}\endgroup }) + {\begingroup\renewcommand\colorMATH{\colorMATHB}\renewcommand\colorSYNTAX{\colorSYNTAXB}{{\color{\colorMATH}\ensuremath{\Distance'}}}\endgroup }\mathord{\cdotp }{\begingroup\renewcommand\colorMATH{\colorMATHB}\renewcommand\colorSYNTAX{\colorSYNTAXB}{{\color{\colorMATH}\ensuremath{\sS''}}}\endgroup }}\llbracket {\begingroup\renewcommand\colorMATH{\colorMATHB}\renewcommand\colorSYNTAX{\colorSYNTAXB}{{\color{\colorMATH}\ensuremath{\Distance'}}}\endgroup }([{\begingroup\renewcommand\colorMATH{\colorMATHB}\renewcommand\colorSYNTAX{\colorSYNTAXB}{{\color{\colorMATH}\ensuremath{\sS_{2}}}}\endgroup }/x]\tau _{2})\rrbracket }}}.
      But notice that by {{\color{\colorMATH}\ensuremath{{\textsc{ app}}}}}:
      \begingroup\color{\colorMATH}\begin{gather*} 
        \inferrule*[lab=
        ]{ \gamma _{1}\vdash {\begingroup\renewcommand\colorMATH{\colorMATHB}\renewcommand\colorSYNTAX{\colorSYNTAXB}{{\color{\colorMATH}\ensuremath{\se_{1}}}}\endgroup } \Downarrow ^{j_{1}} \langle \lambda x.\hspace*{0.33em}{\begingroup\renewcommand\colorMATH{\colorMATHB}\renewcommand\colorSYNTAX{\colorSYNTAXB}{{\color{\colorMATH}\ensuremath{\se'_{1}}}}\endgroup },\gamma '_{1}\rangle 
        \\ \gamma _{1}\vdash {\begingroup\renewcommand\colorMATH{\colorMATHB}\renewcommand\colorSYNTAX{\colorSYNTAXB}{{\color{\colorMATH}\ensuremath{\se_{2}}}}\endgroup } \Downarrow ^{j_{2}} {\begingroup\renewcommand\colorMATH{\colorMATHB}\renewcommand\colorSYNTAX{\colorSYNTAXB}{{\color{\colorMATH}\ensuremath{\sv_{1}}}}\endgroup }
        \\ \gamma '_{1}[x\mapsto v_{1}]\vdash {\begingroup\renewcommand\colorMATH{\colorMATHB}\renewcommand\colorSYNTAX{\colorSYNTAXB}{{\color{\colorMATH}\ensuremath{\se'_{1}}}}\endgroup } \Downarrow ^{j_{3}} {\begingroup\renewcommand\colorMATH{\colorMATHB}\renewcommand\colorSYNTAX{\colorSYNTAXB}{{\color{\colorMATH}\ensuremath{\sv'_{1}}}}\endgroup }
          }{
          \gamma _{1}\vdash {\begingroup\renewcommand\colorMATH{\colorMATHB}\renewcommand\colorSYNTAX{\colorSYNTAXB}{{\color{\colorMATH}\ensuremath{\se_{1}}}}\endgroup }\hspace*{0.33em}{\begingroup\renewcommand\colorMATH{\colorMATHB}\renewcommand\colorSYNTAX{\colorSYNTAXB}{{\color{\colorMATH}\ensuremath{\se_{2}}}}\endgroup } \Downarrow ^{j_{1}+j_{2}+j_{3}+1} {\begingroup\renewcommand\colorMATH{\colorMATHB}\renewcommand\colorSYNTAX{\colorSYNTAXB}{{\color{\colorMATH}\ensuremath{\sv'_{1}}}}\endgroup }
        }
      \end{gather*}\endgroup
      and
      \begingroup\color{\colorMATH}\begin{gather*} 
        \inferrule*[lab=
        ]{ \gamma _{2}\vdash {\begingroup\renewcommand\colorMATH{\colorMATHB}\renewcommand\colorSYNTAX{\colorSYNTAXB}{{\color{\colorMATH}\ensuremath{\se_{1}}}}\endgroup } \Downarrow ^{*} \langle \lambda x.\hspace*{0.33em}{\begingroup\renewcommand\colorMATH{\colorMATHB}\renewcommand\colorSYNTAX{\colorSYNTAXB}{{\color{\colorMATH}\ensuremath{\se'_{2}}}}\endgroup },\gamma '_{2}\rangle 
        \\ \gamma _{2}\vdash {\begingroup\renewcommand\colorMATH{\colorMATHB}\renewcommand\colorSYNTAX{\colorSYNTAXB}{{\color{\colorMATH}\ensuremath{\se_{2}}}}\endgroup } \Downarrow ^{*} {\begingroup\renewcommand\colorMATH{\colorMATHB}\renewcommand\colorSYNTAX{\colorSYNTAXB}{{\color{\colorMATH}\ensuremath{\sv_{2}}}}\endgroup }
        \\ \gamma '_{1}[x\mapsto v_{2}]\vdash {\begingroup\renewcommand\colorMATH{\colorMATHB}\renewcommand\colorSYNTAX{\colorSYNTAXB}{{\color{\colorMATH}\ensuremath{\se'_{2}}}}\endgroup } \Downarrow ^{*} {\begingroup\renewcommand\colorMATH{\colorMATHB}\renewcommand\colorSYNTAX{\colorSYNTAXB}{{\color{\colorMATH}\ensuremath{\sv'_{2}}}}\endgroup }
          }{
          \gamma _{2}\vdash {\begingroup\renewcommand\colorMATH{\colorMATHB}\renewcommand\colorSYNTAX{\colorSYNTAXB}{{\color{\colorMATH}\ensuremath{\se_{1}}}}\endgroup }\hspace*{0.33em}{\begingroup\renewcommand\colorMATH{\colorMATHB}\renewcommand\colorSYNTAX{\colorSYNTAXB}{{\color{\colorMATH}\ensuremath{\se_{2}}}}\endgroup } \Downarrow ^{*} {\begingroup\renewcommand\colorMATH{\colorMATHB}\renewcommand\colorSYNTAX{\colorSYNTAXB}{{\color{\colorMATH}\ensuremath{\sv'_{2}}}}\endgroup }
        }
      \end{gather*}\endgroup
      for {{\color{\colorMATH}\ensuremath{_{i} \in  \{ 1,2\} }}}. Notice that {{\color{\colorMATH}\ensuremath{j = j_{1}+ j_{2} + j_{3} + 1}}}, and by premise {{\color{\colorMATH}\ensuremath{\gamma '_{1}[x\mapsto {\begingroup\renewcommand\colorMATH{\colorMATHB}\renewcommand\colorSYNTAX{\colorSYNTAXB}{{\color{\colorMATH}\ensuremath{\sv_{i}}}}\endgroup }]\vdash  {\begingroup\renewcommand\colorMATH{\colorMATHB}\renewcommand\colorSYNTAX{\colorSYNTAXB}{{\color{\colorMATH}\ensuremath{\se'_{i}}}}\endgroup } \Downarrow ^{j} {\begingroup\renewcommand\colorMATH{\colorMATHB}\renewcommand\colorSYNTAX{\colorSYNTAXB}{{\color{\colorMATH}\ensuremath{\sv'_{i}}}}\endgroup }}}}, we know that {{\color{\colorMATH}\ensuremath{{\begingroup\renewcommand\colorMATH{\colorMATHB}\renewcommand\colorSYNTAX{\colorSYNTAXB}{{\color{\colorMATH}\ensuremath{\sv''_{i}}}}\endgroup } = {\begingroup\renewcommand\colorMATH{\colorMATHB}\renewcommand\colorSYNTAX{\colorSYNTAXB}{{\color{\colorMATH}\ensuremath{\sv'_{i}}}}\endgroup }}}} and the result holds immediately.
    \end{subproof}
  \item  {{\color{\colorMATH}\ensuremath{\Gamma ; {\begingroup\renewcommand\colorMATH{\colorMATHB}\renewcommand\colorSYNTAX{\colorSYNTAXB}{{\color{\colorMATH}\ensuremath{\Distance}}}\endgroup } \vdash  \ttt \mathrel{:} {{\color{\colorSYNTAX}\texttt{unit}}} \mathrel{;} \varnothing }}}  
    \begin{subproof} 
      We have to prove that {{\color{\colorMATH}\ensuremath{\forall k, \forall (\gamma _{1},\gamma _{2}) \in  {\mathcal{G}}_{{\begingroup\renewcommand\colorMATH{\colorMATHB}\renewcommand\colorSYNTAX{\colorSYNTAXB}{{\color{\colorMATH}\ensuremath{\Distance'}}}\endgroup }}^{\kg}\llbracket \Gamma \rrbracket , (\gamma _{1}\vdash \ttt,\gamma _{2}\vdash \ttt) \in  {\mathcal{E}}_{{\begingroup\renewcommand\colorMATH{\colorMATHB}\renewcommand\colorSYNTAX{\colorSYNTAXB}{{\color{\colorMATH}\ensuremath{\Distance'}}}\endgroup }\mathord{\cdotp }\varnothing }^{k}\llbracket {\begingroup\renewcommand\colorMATH{\colorMATHB}\renewcommand\colorSYNTAX{\colorSYNTAXB}{{\color{\colorMATH}\ensuremath{\Distance'}}}\endgroup }({{\color{\colorSYNTAX}\texttt{unit}}})\rrbracket }}}, for {{\color{\colorMATH}\ensuremath{{\begingroup\renewcommand\colorMATH{\colorMATHB}\renewcommand\colorSYNTAX{\colorSYNTAXB}{{\color{\colorMATH}\ensuremath{\Distance'}}}\endgroup } \sqsubseteq  {\begingroup\renewcommand\colorMATH{\colorMATHB}\renewcommand\colorSYNTAX{\colorSYNTAXB}{{\color{\colorMATH}\ensuremath{\Distance}}}\endgroup }}}}.
      Notice that {{\color{\colorMATH}\ensuremath{{\begingroup\renewcommand\colorMATH{\colorMATHB}\renewcommand\colorSYNTAX{\colorSYNTAXB}{{\color{\colorMATH}\ensuremath{\Distance'}}}\endgroup }\mathord{\cdotp }\varnothing  = 0}}}, {{\color{\colorMATH}\ensuremath{{\begingroup\renewcommand\colorMATH{\colorMATHB}\renewcommand\colorSYNTAX{\colorSYNTAXB}{{\color{\colorMATH}\ensuremath{\sS}}}\endgroup }({{\color{\colorSYNTAX}\texttt{unit}}}) = {{\color{\colorSYNTAX}\texttt{unit}}}}}}, and {{\color{\colorMATH}\ensuremath{\gamma _{1}(\ttt) = \gamma _{2}(\ttt) = \ttt}}}.
      Then we have to prove that 
      {{\color{\colorMATH}\ensuremath{(\ttt,\ttt) \in  {\mathcal{V}}_{0}^{k}\llbracket {{\color{\colorSYNTAX}\texttt{unit}}}\rrbracket }}} which is direct.
    \end{subproof}

  \item  {{\color{\colorMATH}\ensuremath{\Gamma ; {\begingroup\renewcommand\colorMATH{\colorMATHB}\renewcommand\colorSYNTAX{\colorSYNTAXB}{{\color{\colorMATH}\ensuremath{\Distance}}}\endgroup } \vdash  \inl^{\tau _{2}}\hspace*{0.33em}{\begingroup\renewcommand\colorMATH{\colorMATHB}\renewcommand\colorSYNTAX{\colorSYNTAXB}{{\color{\colorMATH}\ensuremath{\se'}}}\endgroup } \mathrel{:} \tau _{1} \mathrel{^{{\begingroup\renewcommand\colorMATH{\colorMATHB}\renewcommand\colorSYNTAX{\colorSYNTAXB}{{\color{\colorMATH}\ensuremath{\sS''}}}\endgroup }}\oplus ^{\varnothing }} \tau _{2} \mathrel{;} \varnothing }}} %\mt{without prepayment}
    \begin{subproof} 
      We have to prove that {{\color{\colorMATH}\ensuremath{\forall k, \forall (\gamma _{1},\gamma _{2}) \in  {\mathcal{G}}_{{\begingroup\renewcommand\colorMATH{\colorMATHB}\renewcommand\colorSYNTAX{\colorSYNTAXB}{{\color{\colorMATH}\ensuremath{\Distance'}}}\endgroup }}^{\kg}\llbracket \Gamma \rrbracket , (\gamma _{1}\vdash \inl^{\tau _{2}}\hspace*{0.33em}{\begingroup\renewcommand\colorMATH{\colorMATHB}\renewcommand\colorSYNTAX{\colorSYNTAXB}{{\color{\colorMATH}\ensuremath{\se'}}}\endgroup },\gamma _{2}\vdash \inl^{\tau _{2}}\hspace*{0.33em}{\begingroup\renewcommand\colorMATH{\colorMATHB}\renewcommand\colorSYNTAX{\colorSYNTAXB}{{\color{\colorMATH}\ensuremath{\se'}}}\endgroup }) \in  {\mathcal{E}}_{{\begingroup\renewcommand\colorMATH{\colorMATHB}\renewcommand\colorSYNTAX{\colorSYNTAXB}{{\color{\colorMATH}\ensuremath{\Distance'}}}\endgroup }\mathord{\cdotp }\varnothing }^{k}\llbracket {\begingroup\renewcommand\colorMATH{\colorMATHB}\renewcommand\colorSYNTAX{\colorSYNTAXB}{{\color{\colorMATH}\ensuremath{\Distance'}}}\endgroup }(\tau _{1} \mathrel{^{{\begingroup\renewcommand\colorMATH{\colorMATHB}\renewcommand\colorSYNTAX{\colorSYNTAXB}{{\color{\colorMATH}\ensuremath{\sS''}}}\endgroup }}\oplus ^{\varnothing }} \tau _{2})\rrbracket }}}, for {{\color{\colorMATH}\ensuremath{{\begingroup\renewcommand\colorMATH{\colorMATHB}\renewcommand\colorSYNTAX{\colorSYNTAXB}{{\color{\colorMATH}\ensuremath{\Distance'}}}\endgroup } \sqsubseteq  {\begingroup\renewcommand\colorMATH{\colorMATHB}\renewcommand\colorSYNTAX{\colorSYNTAXB}{{\color{\colorMATH}\ensuremath{\Distance}}}\endgroup }}}}.
      Notice that {{\color{\colorMATH}\ensuremath{{\begingroup\renewcommand\colorMATH{\colorMATHB}\renewcommand\colorSYNTAX{\colorSYNTAXB}{{\color{\colorMATH}\ensuremath{\Distance'}}}\endgroup }\mathord{\cdotp }\varnothing  = 0}}}, and {{\color{\colorMATH}\ensuremath{{\begingroup\renewcommand\colorMATH{\colorMATHB}\renewcommand\colorSYNTAX{\colorSYNTAXB}{{\color{\colorMATH}\ensuremath{\Distance'}}}\endgroup }(\tau _{1} \mathrel{^{{\begingroup\renewcommand\colorMATH{\colorMATHB}\renewcommand\colorSYNTAX{\colorSYNTAXB}{{\color{\colorMATH}\ensuremath{\sS''}}}\endgroup }}\oplus ^{\varnothing }} \tau _{2}) = {\begingroup\renewcommand\colorMATH{\colorMATHB}\renewcommand\colorSYNTAX{\colorSYNTAXB}{{\color{\colorMATH}\ensuremath{\Distance'}}}\endgroup }(\tau _{1}) \mathrel{^{{\begingroup\renewcommand\colorMATH{\colorMATHB}\renewcommand\colorSYNTAX{\colorSYNTAXB}{{\color{\colorMATH}\ensuremath{\Distance'}}}\endgroup }\mathord{\cdotp }{\begingroup\renewcommand\colorMATH{\colorMATHB}\renewcommand\colorSYNTAX{\colorSYNTAXB}{{\color{\colorMATH}\ensuremath{\sS''}}}\endgroup }}\oplus ^{0}} {\begingroup\renewcommand\colorMATH{\colorMATHB}\renewcommand\colorSYNTAX{\colorSYNTAXB}{{\color{\colorMATH}\ensuremath{\Distance'}}}\endgroup }(\tau _{2})}}}, then we have to prove that\\
      {{\color{\colorMATH}\ensuremath{(\gamma _{1}\vdash \inl^{\tau _{2}}\hspace*{0.33em}{\begingroup\renewcommand\colorMATH{\colorMATHB}\renewcommand\colorSYNTAX{\colorSYNTAXB}{{\color{\colorMATH}\ensuremath{\se'}}}\endgroup },\gamma _{2}\vdash \inl^{\tau _{2}}\hspace*{0.33em}{\begingroup\renewcommand\colorMATH{\colorMATHB}\renewcommand\colorSYNTAX{\colorSYNTAXB}{{\color{\colorMATH}\ensuremath{\se'}}}\endgroup }) \in  {\mathcal{E}}_{0}^{k}\llbracket {\begingroup\renewcommand\colorMATH{\colorMATHB}\renewcommand\colorSYNTAX{\colorSYNTAXB}{{\color{\colorMATH}\ensuremath{\Distance'}}}\endgroup }(\tau _{1}) \mathrel{^{{\begingroup\renewcommand\colorMATH{\colorMATHB}\renewcommand\colorSYNTAX{\colorSYNTAXB}{{\color{\colorMATH}\ensuremath{\Distance'}}}\endgroup }\mathord{\cdotp }{\begingroup\renewcommand\colorMATH{\colorMATHB}\renewcommand\colorSYNTAX{\colorSYNTAXB}{{\color{\colorMATH}\ensuremath{\sS''}}}\endgroup }}\oplus ^{0}} {\begingroup\renewcommand\colorMATH{\colorMATHB}\renewcommand\colorSYNTAX{\colorSYNTAXB}{{\color{\colorMATH}\ensuremath{\Distance'}}}\endgroup }(\tau _{2})\rrbracket }}}, i.e.
      if {{\color{\colorMATH}\ensuremath{\gamma _{1}\vdash \inl^{\tau _{2}}\hspace*{0.33em}{\begingroup\renewcommand\colorMATH{\colorMATHB}\renewcommand\colorSYNTAX{\colorSYNTAXB}{{\color{\colorMATH}\ensuremath{\se'}}}\endgroup } \Downarrow ^{j} {\begingroup\renewcommand\colorMATH{\colorMATHB}\renewcommand\colorSYNTAX{\colorSYNTAXB}{{\color{\colorMATH}\ensuremath{\sv_{1}}}}\endgroup }}}} \pthen {{\color{\colorMATH}\ensuremath{\gamma _{2}\vdash \inl^{\tau _{2}}\hspace*{0.33em}{\begingroup\renewcommand\colorMATH{\colorMATHB}\renewcommand\colorSYNTAX{\colorSYNTAXB}{{\color{\colorMATH}\ensuremath{\se'}}}\endgroup } \Downarrow ^{\pj} {\begingroup\renewcommand\colorMATH{\colorMATHB}\renewcommand\colorSYNTAX{\colorSYNTAXB}{{\color{\colorMATH}\ensuremath{\sv_{2}}}}\endgroup }}}}, \pand 
      {{\color{\colorMATH}\ensuremath{({\begingroup\renewcommand\colorMATH{\colorMATHB}\renewcommand\colorSYNTAX{\colorSYNTAXB}{{\color{\colorMATH}\ensuremath{\sv_{1}}}}\endgroup }, {\begingroup\renewcommand\colorMATH{\colorMATHB}\renewcommand\colorSYNTAX{\colorSYNTAXB}{{\color{\colorMATH}\ensuremath{\sv_{2}}}}\endgroup }) \in  {\mathcal{V}}_{0}^{k-j}\llbracket {\begingroup\renewcommand\colorMATH{\colorMATHB}\renewcommand\colorSYNTAX{\colorSYNTAXB}{{\color{\colorMATH}\ensuremath{\Distance'}}}\endgroup }(\tau _{1}) \mathrel{^{{\begingroup\renewcommand\colorMATH{\colorMATHB}\renewcommand\colorSYNTAX{\colorSYNTAXB}{{\color{\colorMATH}\ensuremath{\Distance'}}}\endgroup }\mathord{\cdotp }{\begingroup\renewcommand\colorMATH{\colorMATHB}\renewcommand\colorSYNTAX{\colorSYNTAXB}{{\color{\colorMATH}\ensuremath{\sS''}}}\endgroup }}\oplus ^{0}} {\begingroup\renewcommand\colorMATH{\colorMATHB}\renewcommand\colorSYNTAX{\colorSYNTAXB}{{\color{\colorMATH}\ensuremath{\Distance'}}}\endgroup }(\tau _{2})\rrbracket }}}.

      By induction hypothesis on {{\color{\colorMATH}\ensuremath{\Gamma  \vdash  {\begingroup\renewcommand\colorMATH{\colorMATHB}\renewcommand\colorSYNTAX{\colorSYNTAXB}{{\color{\colorMATH}\ensuremath{\se'}}}\endgroup } \mathrel{:} \tau _{1} \mathrel{;} {\begingroup\renewcommand\colorMATH{\colorMATHB}\renewcommand\colorSYNTAX{\colorSYNTAXB}{{\color{\colorMATH}\ensuremath{\sS''}}}\endgroup }}}}, we know that\\ 
      {{\color{\colorMATH}\ensuremath{(\gamma _{1}\vdash {\begingroup\renewcommand\colorMATH{\colorMATHB}\renewcommand\colorSYNTAX{\colorSYNTAXB}{{\color{\colorMATH}\ensuremath{\se'}}}\endgroup },\gamma _{2}\vdash {\begingroup\renewcommand\colorMATH{\colorMATHB}\renewcommand\colorSYNTAX{\colorSYNTAXB}{{\color{\colorMATH}\ensuremath{\se'}}}\endgroup }) \in  {\mathcal{E}}_{{\begingroup\renewcommand\colorMATH{\colorMATHB}\renewcommand\colorSYNTAX{\colorSYNTAXB}{{\color{\colorMATH}\ensuremath{\Distance'}}}\endgroup }\mathord{\cdotp }{\begingroup\renewcommand\colorMATH{\colorMATHB}\renewcommand\colorSYNTAX{\colorSYNTAXB}{{\color{\colorMATH}\ensuremath{\sS''}}}\endgroup }}^{k}\llbracket {\begingroup\renewcommand\colorMATH{\colorMATHB}\renewcommand\colorSYNTAX{\colorSYNTAXB}{{\color{\colorMATH}\ensuremath{\Distance'}}}\endgroup }(\tau _{1})\rrbracket }}}, i.e. if {{\color{\colorMATH}\ensuremath{\gamma _{1}\vdash {\begingroup\renewcommand\colorMATH{\colorMATHB}\renewcommand\colorSYNTAX{\colorSYNTAXB}{{\color{\colorMATH}\ensuremath{\se'}}}\endgroup } \Downarrow ^{j} {\begingroup\renewcommand\colorMATH{\colorMATHB}\renewcommand\colorSYNTAX{\colorSYNTAXB}{{\color{\colorMATH}\ensuremath{\sv'_{1}}}}\endgroup }}}}, \pthen {{\color{\colorMATH}\ensuremath{\gamma _{2}\vdash {\begingroup\renewcommand\colorMATH{\colorMATHB}\renewcommand\colorSYNTAX{\colorSYNTAXB}{{\color{\colorMATH}\ensuremath{\se'}}}\endgroup } \Downarrow ^{\pj} {\begingroup\renewcommand\colorMATH{\colorMATHB}\renewcommand\colorSYNTAX{\colorSYNTAXB}{{\color{\colorMATH}\ensuremath{\sv'_{2}}}}\endgroup }}}} \pand 
      {{\color{\colorMATH}\ensuremath{({\begingroup\renewcommand\colorMATH{\colorMATHB}\renewcommand\colorSYNTAX{\colorSYNTAXB}{{\color{\colorMATH}\ensuremath{\sv'_{1}}}}\endgroup }, {\begingroup\renewcommand\colorMATH{\colorMATHB}\renewcommand\colorSYNTAX{\colorSYNTAXB}{{\color{\colorMATH}\ensuremath{\sv'_{2}}}}\endgroup }) \in  {\mathcal{V}}_{{\begingroup\renewcommand\colorMATH{\colorMATHB}\renewcommand\colorSYNTAX{\colorSYNTAXB}{{\color{\colorMATH}\ensuremath{\Distance'}}}\endgroup }\mathord{\cdotp }{\begingroup\renewcommand\colorMATH{\colorMATHB}\renewcommand\colorSYNTAX{\colorSYNTAXB}{{\color{\colorMATH}\ensuremath{\sS''}}}\endgroup }}^{k-j}\llbracket {\begingroup\renewcommand\colorMATH{\colorMATHB}\renewcommand\colorSYNTAX{\colorSYNTAXB}{{\color{\colorMATH}\ensuremath{\Distance'}}}\endgroup }(\tau _{1})\rrbracket }}}.

      If {{\color{\colorMATH}\ensuremath{\gamma _{i}\vdash \inl^{\tau _{2}}\hspace*{0.33em}{\begingroup\renewcommand\colorMATH{\colorMATHB}\renewcommand\colorSYNTAX{\colorSYNTAXB}{{\color{\colorMATH}\ensuremath{\se'}}}\endgroup } \Downarrow ^{j} {\begingroup\renewcommand\colorMATH{\colorMATHB}\renewcommand\colorSYNTAX{\colorSYNTAXB}{{\color{\colorMATH}\ensuremath{\sv_{i}}}}\endgroup }}}} and {{\color{\colorMATH}\ensuremath{\gamma _{i}\vdash {\begingroup\renewcommand\colorMATH{\colorMATHB}\renewcommand\colorSYNTAX{\colorSYNTAXB}{{\color{\colorMATH}\ensuremath{\se'}}}\endgroup } \Downarrow ^{\pj} {\begingroup\renewcommand\colorMATH{\colorMATHB}\renewcommand\colorSYNTAX{\colorSYNTAXB}{{\color{\colorMATH}\ensuremath{\sv'_{i}}}}\endgroup }}}}, then by {\textsc{ INL}}, {{\color{\colorMATH}\ensuremath{{\begingroup\renewcommand\colorMATH{\colorMATHB}\renewcommand\colorSYNTAX{\colorSYNTAXB}{{\color{\colorMATH}\ensuremath{\sv_{i}}}}\endgroup } = \inl^{\tau _{2}}\hspace*{0.33em}{\begingroup\renewcommand\colorMATH{\colorMATHB}\renewcommand\colorSYNTAX{\colorSYNTAXB}{{\color{\colorMATH}\ensuremath{\sv'_{i}}}}\endgroup }}}}. Then we have to prove that\\
      {{\color{\colorMATH}\ensuremath{(\inl^{\tau _{2}}\hspace*{0.33em}{\begingroup\renewcommand\colorMATH{\colorMATHB}\renewcommand\colorSYNTAX{\colorSYNTAXB}{{\color{\colorMATH}\ensuremath{\sv'_{1}}}}\endgroup }, \inl^{\tau _{2}}\hspace*{0.33em}{\begingroup\renewcommand\colorMATH{\colorMATHB}\renewcommand\colorSYNTAX{\colorSYNTAXB}{{\color{\colorMATH}\ensuremath{\sv'_{2}}}}\endgroup }) \in  {\mathcal{V}}_{0}^{k-j}\llbracket {\begingroup\renewcommand\colorMATH{\colorMATHB}\renewcommand\colorSYNTAX{\colorSYNTAXB}{{\color{\colorMATH}\ensuremath{\Distance'}}}\endgroup }(\tau _{1}) \mathrel{^{{\begingroup\renewcommand\colorMATH{\colorMATHB}\renewcommand\colorSYNTAX{\colorSYNTAXB}{{\color{\colorMATH}\ensuremath{\Distance'}}}\endgroup }\mathord{\cdotp }{\begingroup\renewcommand\colorMATH{\colorMATHB}\renewcommand\colorSYNTAX{\colorSYNTAXB}{{\color{\colorMATH}\ensuremath{\sS''}}}\endgroup }}\oplus ^{0}} {\begingroup\renewcommand\colorMATH{\colorMATHB}\renewcommand\colorSYNTAX{\colorSYNTAXB}{{\color{\colorMATH}\ensuremath{\Distance'}}}\endgroup }(\tau _{2})\rrbracket }}}, i.e. that
      {{\color{\colorMATH}\ensuremath{({\begingroup\renewcommand\colorMATH{\colorMATHB}\renewcommand\colorSYNTAX{\colorSYNTAXB}{{\color{\colorMATH}\ensuremath{\sv'_{1}}}}\endgroup }, {\begingroup\renewcommand\colorMATH{\colorMATHB}\renewcommand\colorSYNTAX{\colorSYNTAXB}{{\color{\colorMATH}\ensuremath{\sv'_{2}}}}\endgroup }) \in  {\mathcal{V}}_{{\begingroup\renewcommand\colorMATH{\colorMATHB}\renewcommand\colorSYNTAX{\colorSYNTAXB}{{\color{\colorMATH}\ensuremath{\Distance'}}}\endgroup }\mathord{\cdotp }{\begingroup\renewcommand\colorMATH{\colorMATHB}\renewcommand\colorSYNTAX{\colorSYNTAXB}{{\color{\colorMATH}\ensuremath{\sS''}}}\endgroup }+0}^{k-j}\llbracket {\begingroup\renewcommand\colorMATH{\colorMATHB}\renewcommand\colorSYNTAX{\colorSYNTAXB}{{\color{\colorMATH}\ensuremath{\Distance'}}}\endgroup }(\tau _{1})\rrbracket }}}, but as {{\color{\colorMATH}\ensuremath{{\begingroup\renewcommand\colorMATH{\colorMATHB}\renewcommand\colorSYNTAX{\colorSYNTAXB}{{\color{\colorMATH}\ensuremath{\Distance'}}}\endgroup }\mathord{\cdotp }{\begingroup\renewcommand\colorMATH{\colorMATHB}\renewcommand\colorSYNTAX{\colorSYNTAXB}{{\color{\colorMATH}\ensuremath{\sS''}}}\endgroup }+0 = {\begingroup\renewcommand\colorMATH{\colorMATHB}\renewcommand\colorSYNTAX{\colorSYNTAXB}{{\color{\colorMATH}\ensuremath{\Distance'}}}\endgroup }\mathord{\cdotp }{\begingroup\renewcommand\colorMATH{\colorMATHB}\renewcommand\colorSYNTAX{\colorSYNTAXB}{{\color{\colorMATH}\ensuremath{\sS''}}}\endgroup }}}}, the result holds immediately.
    \end{subproof}

  \item  {{\color{\colorMATH}\ensuremath{\Gamma ; {\begingroup\renewcommand\colorMATH{\colorMATHB}\renewcommand\colorSYNTAX{\colorSYNTAXB}{{\color{\colorMATH}\ensuremath{\Distance}}}\endgroup } \vdash  \inl^{\tau _{2}}\hspace*{0.33em}{\begingroup\renewcommand\colorMATH{\colorMATHB}\renewcommand\colorSYNTAX{\colorSYNTAXB}{{\color{\colorMATH}\ensuremath{\se'}}}\endgroup } \mathrel{:} \tau _{1} \mathrel{^{{\begingroup\renewcommand\colorMATH{\colorMATHB}\renewcommand\colorSYNTAX{\colorSYNTAXB}{{\color{\colorMATH}\ensuremath{\sS''}}}\endgroup }}\oplus ^{\varnothing }} \tau _{2} \mathrel{;} {\begingroup\renewcommand\colorMATH{\colorMATHB}\renewcommand\colorSYNTAX{\colorSYNTAXB}{{\color{\colorMATH}\ensuremath{\sS'}}}\endgroup }}}} %\mt{with prepayment}
    \begin{subproof} 
      We have to prove that {{\color{\colorMATH}\ensuremath{\forall k, \forall (\gamma _{1},\gamma _{2}) \in  {\mathcal{G}}_{{\begingroup\renewcommand\colorMATH{\colorMATHB}\renewcommand\colorSYNTAX{\colorSYNTAXB}{{\color{\colorMATH}\ensuremath{\Distance'}}}\endgroup }}^{\kg}\llbracket \Gamma \rrbracket , (\gamma _{1}\vdash \inl^{\tau _{2}}\hspace*{0.33em}{\begingroup\renewcommand\colorMATH{\colorMATHB}\renewcommand\colorSYNTAX{\colorSYNTAXB}{{\color{\colorMATH}\ensuremath{\se'}}}\endgroup },\gamma _{2}\vdash \inl^{\tau _{2}}\hspace*{0.33em}{\begingroup\renewcommand\colorMATH{\colorMATHB}\renewcommand\colorSYNTAX{\colorSYNTAXB}{{\color{\colorMATH}\ensuremath{\se'}}}\endgroup }) \in  {\mathcal{E}}_{{\begingroup\renewcommand\colorMATH{\colorMATHB}\renewcommand\colorSYNTAX{\colorSYNTAXB}{{\color{\colorMATH}\ensuremath{\Distance'}}}\endgroup }\mathord{\cdotp }{\begingroup\renewcommand\colorMATH{\colorMATHB}\renewcommand\colorSYNTAX{\colorSYNTAXB}{{\color{\colorMATH}\ensuremath{\sS'}}}\endgroup }}^{k}\llbracket {\begingroup\renewcommand\colorMATH{\colorMATHB}\renewcommand\colorSYNTAX{\colorSYNTAXB}{{\color{\colorMATH}\ensuremath{\Distance'}}}\endgroup }(\tau _{1} \mathrel{^{{\begingroup\renewcommand\colorMATH{\colorMATHB}\renewcommand\colorSYNTAX{\colorSYNTAXB}{{\color{\colorMATH}\ensuremath{\sS''}}}\endgroup }}\oplus ^{\varnothing }} \tau _{2})\rrbracket }}}, for {{\color{\colorMATH}\ensuremath{{\begingroup\renewcommand\colorMATH{\colorMATHB}\renewcommand\colorSYNTAX{\colorSYNTAXB}{{\color{\colorMATH}\ensuremath{\Distance'}}}\endgroup } \sqsubseteq  {\begingroup\renewcommand\colorMATH{\colorMATHB}\renewcommand\colorSYNTAX{\colorSYNTAXB}{{\color{\colorMATH}\ensuremath{\Distance}}}\endgroup }}}}.
      Notice that {{\color{\colorMATH}\ensuremath{{\begingroup\renewcommand\colorMATH{\colorMATHB}\renewcommand\colorSYNTAX{\colorSYNTAXB}{{\color{\colorMATH}\ensuremath{\Distance'}}}\endgroup }\mathord{\cdotp }\varnothing  = 0}}}, and {{\color{\colorMATH}\ensuremath{{\begingroup\renewcommand\colorMATH{\colorMATHB}\renewcommand\colorSYNTAX{\colorSYNTAXB}{{\color{\colorMATH}\ensuremath{\Distance'}}}\endgroup }(\tau _{1} \mathrel{^{{\begingroup\renewcommand\colorMATH{\colorMATHB}\renewcommand\colorSYNTAX{\colorSYNTAXB}{{\color{\colorMATH}\ensuremath{\sS''}}}\endgroup }}\oplus ^{\varnothing }} \tau _{2}) = {\begingroup\renewcommand\colorMATH{\colorMATHB}\renewcommand\colorSYNTAX{\colorSYNTAXB}{{\color{\colorMATH}\ensuremath{\Distance'}}}\endgroup }(\tau _{1}) \mathrel{^{{\begingroup\renewcommand\colorMATH{\colorMATHB}\renewcommand\colorSYNTAX{\colorSYNTAXB}{{\color{\colorMATH}\ensuremath{\Distance'}}}\endgroup }\mathord{\cdotp }{\begingroup\renewcommand\colorMATH{\colorMATHB}\renewcommand\colorSYNTAX{\colorSYNTAXB}{{\color{\colorMATH}\ensuremath{\sS''}}}\endgroup }}\oplus ^{0}} {\begingroup\renewcommand\colorMATH{\colorMATHB}\renewcommand\colorSYNTAX{\colorSYNTAXB}{{\color{\colorMATH}\ensuremath{\Distance'}}}\endgroup }(\tau _{2})}}}, then we have to prove that\\
      {{\color{\colorMATH}\ensuremath{(\gamma _{1}\vdash \inl^{\tau _{2}}\hspace*{0.33em}{\begingroup\renewcommand\colorMATH{\colorMATHB}\renewcommand\colorSYNTAX{\colorSYNTAXB}{{\color{\colorMATH}\ensuremath{\se'}}}\endgroup },\gamma _{2}\vdash \inl^{\tau _{2}}\hspace*{0.33em}{\begingroup\renewcommand\colorMATH{\colorMATHB}\renewcommand\colorSYNTAX{\colorSYNTAXB}{{\color{\colorMATH}\ensuremath{\se'}}}\endgroup }) \in  {\mathcal{E}}_{{\begingroup\renewcommand\colorMATH{\colorMATHB}\renewcommand\colorSYNTAX{\colorSYNTAXB}{{\color{\colorMATH}\ensuremath{\Distance'}}}\endgroup }\mathord{\cdotp }{\begingroup\renewcommand\colorMATH{\colorMATHB}\renewcommand\colorSYNTAX{\colorSYNTAXB}{{\color{\colorMATH}\ensuremath{\sS'}}}\endgroup }}^{k}\llbracket {\begingroup\renewcommand\colorMATH{\colorMATHB}\renewcommand\colorSYNTAX{\colorSYNTAXB}{{\color{\colorMATH}\ensuremath{\Distance'}}}\endgroup }(\tau _{1}) \mathrel{^{{\begingroup\renewcommand\colorMATH{\colorMATHB}\renewcommand\colorSYNTAX{\colorSYNTAXB}{{\color{\colorMATH}\ensuremath{\Distance'}}}\endgroup }\mathord{\cdotp }{\begingroup\renewcommand\colorMATH{\colorMATHB}\renewcommand\colorSYNTAX{\colorSYNTAXB}{{\color{\colorMATH}\ensuremath{\sS''}}}\endgroup }}\oplus ^{0}} {\begingroup\renewcommand\colorMATH{\colorMATHB}\renewcommand\colorSYNTAX{\colorSYNTAXB}{{\color{\colorMATH}\ensuremath{\Distance'}}}\endgroup }(\tau _{2})\rrbracket }}}, i.e.
      if {{\color{\colorMATH}\ensuremath{\gamma _{1}\vdash \inl^{\tau _{2}}\hspace*{0.33em}{\begingroup\renewcommand\colorMATH{\colorMATHB}\renewcommand\colorSYNTAX{\colorSYNTAXB}{{\color{\colorMATH}\ensuremath{\se'}}}\endgroup } \Downarrow ^{j} {\begingroup\renewcommand\colorMATH{\colorMATHB}\renewcommand\colorSYNTAX{\colorSYNTAXB}{{\color{\colorMATH}\ensuremath{\sv_{1}}}}\endgroup }}}} \pthen {{\color{\colorMATH}\ensuremath{\gamma _{2}\vdash \inl^{\tau _{2}}\hspace*{0.33em}{\begingroup\renewcommand\colorMATH{\colorMATHB}\renewcommand\colorSYNTAX{\colorSYNTAXB}{{\color{\colorMATH}\ensuremath{\se'}}}\endgroup } \Downarrow ^{\pj} {\begingroup\renewcommand\colorMATH{\colorMATHB}\renewcommand\colorSYNTAX{\colorSYNTAXB}{{\color{\colorMATH}\ensuremath{\sv_{2}}}}\endgroup }}}}, \pand 
      {{\color{\colorMATH}\ensuremath{({\begingroup\renewcommand\colorMATH{\colorMATHB}\renewcommand\colorSYNTAX{\colorSYNTAXB}{{\color{\colorMATH}\ensuremath{\sv_{1}}}}\endgroup }, {\begingroup\renewcommand\colorMATH{\colorMATHB}\renewcommand\colorSYNTAX{\colorSYNTAXB}{{\color{\colorMATH}\ensuremath{\sv_{2}}}}\endgroup }) \in  {\mathcal{V}}_{{\begingroup\renewcommand\colorMATH{\colorMATHB}\renewcommand\colorSYNTAX{\colorSYNTAXB}{{\color{\colorMATH}\ensuremath{\Distance'}}}\endgroup }\mathord{\cdotp }{\begingroup\renewcommand\colorMATH{\colorMATHB}\renewcommand\colorSYNTAX{\colorSYNTAXB}{{\color{\colorMATH}\ensuremath{\sS'}}}\endgroup }}^{k-j}\llbracket {\begingroup\renewcommand\colorMATH{\colorMATHB}\renewcommand\colorSYNTAX{\colorSYNTAXB}{{\color{\colorMATH}\ensuremath{\Distance'}}}\endgroup }(\tau _{1}) \mathrel{^{{\begingroup\renewcommand\colorMATH{\colorMATHB}\renewcommand\colorSYNTAX{\colorSYNTAXB}{{\color{\colorMATH}\ensuremath{\Distance'}}}\endgroup }\mathord{\cdotp }{\begingroup\renewcommand\colorMATH{\colorMATHB}\renewcommand\colorSYNTAX{\colorSYNTAXB}{{\color{\colorMATH}\ensuremath{\sS''}}}\endgroup }}\oplus ^{0}} {\begingroup\renewcommand\colorMATH{\colorMATHB}\renewcommand\colorSYNTAX{\colorSYNTAXB}{{\color{\colorMATH}\ensuremath{\Distance'}}}\endgroup }(\tau _{2})\rrbracket }}}.

      By induction hypothesis on {{\color{\colorMATH}\ensuremath{\Gamma  \vdash  {\begingroup\renewcommand\colorMATH{\colorMATHB}\renewcommand\colorSYNTAX{\colorSYNTAXB}{{\color{\colorMATH}\ensuremath{\se'}}}\endgroup } \mathrel{:} \tau _{1} \mathrel{;} {\begingroup\renewcommand\colorMATH{\colorMATHB}\renewcommand\colorSYNTAX{\colorSYNTAXB}{{\color{\colorMATH}\ensuremath{\sS''}}}\endgroup } + {\begingroup\renewcommand\colorMATH{\colorMATHB}\renewcommand\colorSYNTAX{\colorSYNTAXB}{{\color{\colorMATH}\ensuremath{\sS'}}}\endgroup }}}}, we know that\\ 
      {{\color{\colorMATH}\ensuremath{(\gamma _{1}\vdash {\begingroup\renewcommand\colorMATH{\colorMATHB}\renewcommand\colorSYNTAX{\colorSYNTAXB}{{\color{\colorMATH}\ensuremath{\se'}}}\endgroup },\gamma _{2}\vdash {\begingroup\renewcommand\colorMATH{\colorMATHB}\renewcommand\colorSYNTAX{\colorSYNTAXB}{{\color{\colorMATH}\ensuremath{\se'}}}\endgroup }) \in  {\mathcal{E}}_{{\begingroup\renewcommand\colorMATH{\colorMATHB}\renewcommand\colorSYNTAX{\colorSYNTAXB}{{\color{\colorMATH}\ensuremath{\Distance'}}}\endgroup }\mathord{\cdotp }({\begingroup\renewcommand\colorMATH{\colorMATHB}\renewcommand\colorSYNTAX{\colorSYNTAXB}{{\color{\colorMATH}\ensuremath{\sS''}}}\endgroup } + {\begingroup\renewcommand\colorMATH{\colorMATHB}\renewcommand\colorSYNTAX{\colorSYNTAXB}{{\color{\colorMATH}\ensuremath{\sS'}}}\endgroup })}^{k}\llbracket {\begingroup\renewcommand\colorMATH{\colorMATHB}\renewcommand\colorSYNTAX{\colorSYNTAXB}{{\color{\colorMATH}\ensuremath{\Distance'}}}\endgroup }(\tau _{1})\rrbracket }}}, i.e. if {{\color{\colorMATH}\ensuremath{\gamma _{1}\vdash {\begingroup\renewcommand\colorMATH{\colorMATHB}\renewcommand\colorSYNTAX{\colorSYNTAXB}{{\color{\colorMATH}\ensuremath{\se'}}}\endgroup } \Downarrow ^{j} {\begingroup\renewcommand\colorMATH{\colorMATHB}\renewcommand\colorSYNTAX{\colorSYNTAXB}{{\color{\colorMATH}\ensuremath{\sv'_{1}}}}\endgroup }}}}, \pthen {{\color{\colorMATH}\ensuremath{\gamma _{2}\vdash {\begingroup\renewcommand\colorMATH{\colorMATHB}\renewcommand\colorSYNTAX{\colorSYNTAXB}{{\color{\colorMATH}\ensuremath{\se'}}}\endgroup } \Downarrow ^{\pj} {\begingroup\renewcommand\colorMATH{\colorMATHB}\renewcommand\colorSYNTAX{\colorSYNTAXB}{{\color{\colorMATH}\ensuremath{\sv'_{2}}}}\endgroup }}}} \pand 
      {{\color{\colorMATH}\ensuremath{({\begingroup\renewcommand\colorMATH{\colorMATHB}\renewcommand\colorSYNTAX{\colorSYNTAXB}{{\color{\colorMATH}\ensuremath{\sv'_{1}}}}\endgroup }, {\begingroup\renewcommand\colorMATH{\colorMATHB}\renewcommand\colorSYNTAX{\colorSYNTAXB}{{\color{\colorMATH}\ensuremath{\sv'_{2}}}}\endgroup }) \in  {\mathcal{V}}_{{\begingroup\renewcommand\colorMATH{\colorMATHB}\renewcommand\colorSYNTAX{\colorSYNTAXB}{{\color{\colorMATH}\ensuremath{\Distance'}}}\endgroup }\mathord{\cdotp }({\begingroup\renewcommand\colorMATH{\colorMATHB}\renewcommand\colorSYNTAX{\colorSYNTAXB}{{\color{\colorMATH}\ensuremath{\sS''}}}\endgroup } + {\begingroup\renewcommand\colorMATH{\colorMATHB}\renewcommand\colorSYNTAX{\colorSYNTAXB}{{\color{\colorMATH}\ensuremath{\sS'}}}\endgroup })}^{k-j}\llbracket {\begingroup\renewcommand\colorMATH{\colorMATHB}\renewcommand\colorSYNTAX{\colorSYNTAXB}{{\color{\colorMATH}\ensuremath{\Distance'}}}\endgroup }(\tau _{1})\rrbracket }}}.

      If {{\color{\colorMATH}\ensuremath{\gamma _{i}\vdash \inl^{\tau _{2}}\hspace*{0.33em}{\begingroup\renewcommand\colorMATH{\colorMATHB}\renewcommand\colorSYNTAX{\colorSYNTAXB}{{\color{\colorMATH}\ensuremath{\se'}}}\endgroup } \Downarrow ^{j} {\begingroup\renewcommand\colorMATH{\colorMATHB}\renewcommand\colorSYNTAX{\colorSYNTAXB}{{\color{\colorMATH}\ensuremath{\sv_{i}}}}\endgroup }}}} and {{\color{\colorMATH}\ensuremath{\gamma _{i}\vdash {\begingroup\renewcommand\colorMATH{\colorMATHB}\renewcommand\colorSYNTAX{\colorSYNTAXB}{{\color{\colorMATH}\ensuremath{\se'}}}\endgroup } \Downarrow ^{\pj} {\begingroup\renewcommand\colorMATH{\colorMATHB}\renewcommand\colorSYNTAX{\colorSYNTAXB}{{\color{\colorMATH}\ensuremath{\sv'_{i}}}}\endgroup }}}}, then by {\textsc{ INL}}, {{\color{\colorMATH}\ensuremath{{\begingroup\renewcommand\colorMATH{\colorMATHB}\renewcommand\colorSYNTAX{\colorSYNTAXB}{{\color{\colorMATH}\ensuremath{\sv_{i}}}}\endgroup } = \inl^{\tau _{2}}\hspace*{0.33em}{\begingroup\renewcommand\colorMATH{\colorMATHB}\renewcommand\colorSYNTAX{\colorSYNTAXB}{{\color{\colorMATH}\ensuremath{\sv'_{i}}}}\endgroup }}}}. Then we have to prove that\\
      {{\color{\colorMATH}\ensuremath{(\inl^{\tau _{2}}\hspace*{0.33em}{\begingroup\renewcommand\colorMATH{\colorMATHB}\renewcommand\colorSYNTAX{\colorSYNTAXB}{{\color{\colorMATH}\ensuremath{\sv'_{1}}}}\endgroup }, \inl^{\tau _{2}}\hspace*{0.33em}{\begingroup\renewcommand\colorMATH{\colorMATHB}\renewcommand\colorSYNTAX{\colorSYNTAXB}{{\color{\colorMATH}\ensuremath{\sv'_{2}}}}\endgroup }) \in  {\mathcal{V}}_{{\begingroup\renewcommand\colorMATH{\colorMATHB}\renewcommand\colorSYNTAX{\colorSYNTAXB}{{\color{\colorMATH}\ensuremath{\Distance'}}}\endgroup }\mathord{\cdotp }{\begingroup\renewcommand\colorMATH{\colorMATHB}\renewcommand\colorSYNTAX{\colorSYNTAXB}{{\color{\colorMATH}\ensuremath{\sS'}}}\endgroup }}^{k-j}\llbracket {\begingroup\renewcommand\colorMATH{\colorMATHB}\renewcommand\colorSYNTAX{\colorSYNTAXB}{{\color{\colorMATH}\ensuremath{\Distance'}}}\endgroup }(\tau _{1}) \mathrel{^{{\begingroup\renewcommand\colorMATH{\colorMATHB}\renewcommand\colorSYNTAX{\colorSYNTAXB}{{\color{\colorMATH}\ensuremath{\Distance'}}}\endgroup }\mathord{\cdotp }{\begingroup\renewcommand\colorMATH{\colorMATHB}\renewcommand\colorSYNTAX{\colorSYNTAXB}{{\color{\colorMATH}\ensuremath{\sS''}}}\endgroup }}\oplus ^{0}} {\begingroup\renewcommand\colorMATH{\colorMATHB}\renewcommand\colorSYNTAX{\colorSYNTAXB}{{\color{\colorMATH}\ensuremath{\Distance'}}}\endgroup }(\tau _{2})\rrbracket }}}, i.e. that
      {{\color{\colorMATH}\ensuremath{({\begingroup\renewcommand\colorMATH{\colorMATHB}\renewcommand\colorSYNTAX{\colorSYNTAXB}{{\color{\colorMATH}\ensuremath{\sv'_{1}}}}\endgroup }, {\begingroup\renewcommand\colorMATH{\colorMATHB}\renewcommand\colorSYNTAX{\colorSYNTAXB}{{\color{\colorMATH}\ensuremath{\sv'_{2}}}}\endgroup }) \in  {\mathcal{V}}_{{\begingroup\renewcommand\colorMATH{\colorMATHB}\renewcommand\colorSYNTAX{\colorSYNTAXB}{{\color{\colorMATH}\ensuremath{\Distance'}}}\endgroup }\mathord{\cdotp }{\begingroup\renewcommand\colorMATH{\colorMATHB}\renewcommand\colorSYNTAX{\colorSYNTAXB}{{\color{\colorMATH}\ensuremath{\sS''}}}\endgroup }+{\begingroup\renewcommand\colorMATH{\colorMATHB}\renewcommand\colorSYNTAX{\colorSYNTAXB}{{\color{\colorMATH}\ensuremath{\Distance'}}}\endgroup }\mathord{\cdotp }{\begingroup\renewcommand\colorMATH{\colorMATHB}\renewcommand\colorSYNTAX{\colorSYNTAXB}{{\color{\colorMATH}\ensuremath{\sS'}}}\endgroup }}^{k-j}\llbracket {\begingroup\renewcommand\colorMATH{\colorMATHB}\renewcommand\colorSYNTAX{\colorSYNTAXB}{{\color{\colorMATH}\ensuremath{\Distance'}}}\endgroup }(\tau _{1})\rrbracket }}}, but as {{\color{\colorMATH}\ensuremath{{\begingroup\renewcommand\colorMATH{\colorMATHB}\renewcommand\colorSYNTAX{\colorSYNTAXB}{{\color{\colorMATH}\ensuremath{\Distance'}}}\endgroup }\mathord{\cdotp }({\begingroup\renewcommand\colorMATH{\colorMATHB}\renewcommand\colorSYNTAX{\colorSYNTAXB}{{\color{\colorMATH}\ensuremath{\sS''}}}\endgroup } + {\begingroup\renewcommand\colorMATH{\colorMATHB}\renewcommand\colorSYNTAX{\colorSYNTAXB}{{\color{\colorMATH}\ensuremath{\sS'}}}\endgroup }) = {\begingroup\renewcommand\colorMATH{\colorMATHB}\renewcommand\colorSYNTAX{\colorSYNTAXB}{{\color{\colorMATH}\ensuremath{\Distance'}}}\endgroup }\mathord{\cdotp }{\begingroup\renewcommand\colorMATH{\colorMATHB}\renewcommand\colorSYNTAX{\colorSYNTAXB}{{\color{\colorMATH}\ensuremath{\sS''}}}\endgroup }+{\begingroup\renewcommand\colorMATH{\colorMATHB}\renewcommand\colorSYNTAX{\colorSYNTAXB}{{\color{\colorMATH}\ensuremath{\Distance'}}}\endgroup }\mathord{\cdotp }{\begingroup\renewcommand\colorMATH{\colorMATHB}\renewcommand\colorSYNTAX{\colorSYNTAXB}{{\color{\colorMATH}\ensuremath{\sS'}}}\endgroup }}}}, the result holds immediately.
    \end{subproof}

    \item  {{\color{\colorMATH}\ensuremath{\Gamma ; {\begingroup\renewcommand\colorMATH{\colorMATHB}\renewcommand\colorSYNTAX{\colorSYNTAXB}{{\color{\colorMATH}\ensuremath{\Distance}}}\endgroup } \vdash  \inr^{\tau _{1}}\hspace*{0.33em}{\begingroup\renewcommand\colorMATH{\colorMATHB}\renewcommand\colorSYNTAX{\colorSYNTAXB}{{\color{\colorMATH}\ensuremath{\se'}}}\endgroup } \mathrel{:} \tau _{1} \mathrel{^{\varnothing }\oplus ^{{\begingroup\renewcommand\colorMATH{\colorMATHB}\renewcommand\colorSYNTAX{\colorSYNTAXB}{{\color{\colorMATH}\ensuremath{\sS''}}}\endgroup }}} \tau _{2} \mathrel{;} \varnothing }}} \mt{without prepayment}
      \begin{subproof} 
        We have to prove that {{\color{\colorMATH}\ensuremath{\forall  (\gamma _{1},\gamma _{2}) \in  {\mathcal{G}}_{{\begingroup\renewcommand\colorMATH{\colorMATHB}\renewcommand\colorSYNTAX{\colorSYNTAXB}{{\color{\colorMATH}\ensuremath{\Distance'}}}\endgroup }}^{\kg}\llbracket \Gamma \rrbracket , (\gamma _{1}\vdash \inr^{\tau _{1}}\hspace*{0.33em}{\begingroup\renewcommand\colorMATH{\colorMATHB}\renewcommand\colorSYNTAX{\colorSYNTAXB}{{\color{\colorMATH}\ensuremath{\se'}}}\endgroup },\gamma _{2}\vdash \inr^{\tau _{1}}\hspace*{0.33em}{\begingroup\renewcommand\colorMATH{\colorMATHB}\renewcommand\colorSYNTAX{\colorSYNTAXB}{{\color{\colorMATH}\ensuremath{\se'}}}\endgroup }) \in  {\mathcal{E}}_{{\begingroup\renewcommand\colorMATH{\colorMATHB}\renewcommand\colorSYNTAX{\colorSYNTAXB}{{\color{\colorMATH}\ensuremath{\Distance'}}}\endgroup }\mathord{\cdotp }\varnothing }^{k}\llbracket {\begingroup\renewcommand\colorMATH{\colorMATHB}\renewcommand\colorSYNTAX{\colorSYNTAXB}{{\color{\colorMATH}\ensuremath{\Distance'}}}\endgroup }(\tau _{1} \mathrel{^{\varnothing }\oplus ^{{\begingroup\renewcommand\colorMATH{\colorMATHB}\renewcommand\colorSYNTAX{\colorSYNTAXB}{{\color{\colorMATH}\ensuremath{\sS''}}}\endgroup }}} \tau _{2})\rrbracket }}}, for {{\color{\colorMATH}\ensuremath{{\begingroup\renewcommand\colorMATH{\colorMATHB}\renewcommand\colorSYNTAX{\colorSYNTAXB}{{\color{\colorMATH}\ensuremath{\Distance'}}}\endgroup } \sqsubseteq  {\begingroup\renewcommand\colorMATH{\colorMATHB}\renewcommand\colorSYNTAX{\colorSYNTAXB}{{\color{\colorMATH}\ensuremath{\Distance}}}\endgroup }}}}.
        Notice that {{\color{\colorMATH}\ensuremath{{\begingroup\renewcommand\colorMATH{\colorMATHB}\renewcommand\colorSYNTAX{\colorSYNTAXB}{{\color{\colorMATH}\ensuremath{\Distance'}}}\endgroup }\mathord{\cdotp }\varnothing  = 0}}}, and {{\color{\colorMATH}\ensuremath{{\begingroup\renewcommand\colorMATH{\colorMATHB}\renewcommand\colorSYNTAX{\colorSYNTAXB}{{\color{\colorMATH}\ensuremath{\Distance'}}}\endgroup }(\tau _{1} \mathrel{^{\varnothing }\oplus ^{{\begingroup\renewcommand\colorMATH{\colorMATHB}\renewcommand\colorSYNTAX{\colorSYNTAXB}{{\color{\colorMATH}\ensuremath{\sS''}}}\endgroup }}} \tau _{2}) = {\begingroup\renewcommand\colorMATH{\colorMATHB}\renewcommand\colorSYNTAX{\colorSYNTAXB}{{\color{\colorMATH}\ensuremath{\Distance'}}}\endgroup }(\tau _{1}) \mathrel{^{0}\oplus ^{{\begingroup\renewcommand\colorMATH{\colorMATHB}\renewcommand\colorSYNTAX{\colorSYNTAXB}{{\color{\colorMATH}\ensuremath{\Distance'}}}\endgroup }\mathord{\cdotp }{\begingroup\renewcommand\colorMATH{\colorMATHB}\renewcommand\colorSYNTAX{\colorSYNTAXB}{{\color{\colorMATH}\ensuremath{\sS''}}}\endgroup }}} {\begingroup\renewcommand\colorMATH{\colorMATHB}\renewcommand\colorSYNTAX{\colorSYNTAXB}{{\color{\colorMATH}\ensuremath{\Distance'}}}\endgroup }(\tau _{2})}}}, then we have to prove that\\
        {{\color{\colorMATH}\ensuremath{(\gamma _{1}\vdash \inr^{\tau _{1}}\hspace*{0.33em}{\begingroup\renewcommand\colorMATH{\colorMATHB}\renewcommand\colorSYNTAX{\colorSYNTAXB}{{\color{\colorMATH}\ensuremath{\se'}}}\endgroup },\gamma _{2}\vdash \inr^{\tau _{1}}\hspace*{0.33em}{\begingroup\renewcommand\colorMATH{\colorMATHB}\renewcommand\colorSYNTAX{\colorSYNTAXB}{{\color{\colorMATH}\ensuremath{\se'}}}\endgroup }) \in  {\mathcal{E}}_{0}^{k}\llbracket {\begingroup\renewcommand\colorMATH{\colorMATHB}\renewcommand\colorSYNTAX{\colorSYNTAXB}{{\color{\colorMATH}\ensuremath{\Distance'}}}\endgroup }(\tau _{1}) \mathrel{^{0}\oplus ^{{\begingroup\renewcommand\colorMATH{\colorMATHB}\renewcommand\colorSYNTAX{\colorSYNTAXB}{{\color{\colorMATH}\ensuremath{\Distance'}}}\endgroup }\mathord{\cdotp }{\begingroup\renewcommand\colorMATH{\colorMATHB}\renewcommand\colorSYNTAX{\colorSYNTAXB}{{\color{\colorMATH}\ensuremath{\sS''}}}\endgroup }}} {\begingroup\renewcommand\colorMATH{\colorMATHB}\renewcommand\colorSYNTAX{\colorSYNTAXB}{{\color{\colorMATH}\ensuremath{\Distance'}}}\endgroup }(\tau _{2})\rrbracket }}}, i.e.
        if {{\color{\colorMATH}\ensuremath{\gamma _{1}\vdash \inr^{\tau _{1}}\hspace*{0.33em}{\begingroup\renewcommand\colorMATH{\colorMATHB}\renewcommand\colorSYNTAX{\colorSYNTAXB}{{\color{\colorMATH}\ensuremath{\se'}}}\endgroup } \Downarrow ^{j} v_{1}}}} \pthen {{\color{\colorMATH}\ensuremath{\gamma _{2}\vdash \inr^{\tau _{1}}\hspace*{0.33em}{\begingroup\renewcommand\colorMATH{\colorMATHB}\renewcommand\colorSYNTAX{\colorSYNTAXB}{{\color{\colorMATH}\ensuremath{\se'}}}\endgroup } \Downarrow ^{\pj} v_{2}}}}, \pand 
        {{\color{\colorMATH}\ensuremath{(v_{1}, v_{2}) \in  {\mathcal{V}}_{0}^{k-j}\llbracket {\begingroup\renewcommand\colorMATH{\colorMATHB}\renewcommand\colorSYNTAX{\colorSYNTAXB}{{\color{\colorMATH}\ensuremath{\Distance'}}}\endgroup }(\tau _{1}) \mathrel{^{0}\oplus ^{{\begingroup\renewcommand\colorMATH{\colorMATHB}\renewcommand\colorSYNTAX{\colorSYNTAXB}{{\color{\colorMATH}\ensuremath{\Distance'}}}\endgroup }\mathord{\cdotp }{\begingroup\renewcommand\colorMATH{\colorMATHB}\renewcommand\colorSYNTAX{\colorSYNTAXB}{{\color{\colorMATH}\ensuremath{\sS''}}}\endgroup }}} {\begingroup\renewcommand\colorMATH{\colorMATHB}\renewcommand\colorSYNTAX{\colorSYNTAXB}{{\color{\colorMATH}\ensuremath{\Distance'}}}\endgroup }(\tau _{2})\rrbracket }}}.

        By induction hypothesis on {{\color{\colorMATH}\ensuremath{\Gamma  \vdash  {\begingroup\renewcommand\colorMATH{\colorMATHB}\renewcommand\colorSYNTAX{\colorSYNTAXB}{{\color{\colorMATH}\ensuremath{\se'}}}\endgroup } \mathrel{:} \tau _{2} \mathrel{;} {\begingroup\renewcommand\colorMATH{\colorMATHB}\renewcommand\colorSYNTAX{\colorSYNTAXB}{{\color{\colorMATH}\ensuremath{\sS''}}}\endgroup }}}}, we know that\\ 
        {{\color{\colorMATH}\ensuremath{(\gamma _{1}\vdash {\begingroup\renewcommand\colorMATH{\colorMATHB}\renewcommand\colorSYNTAX{\colorSYNTAXB}{{\color{\colorMATH}\ensuremath{\se'}}}\endgroup },\gamma _{2}\vdash {\begingroup\renewcommand\colorMATH{\colorMATHB}\renewcommand\colorSYNTAX{\colorSYNTAXB}{{\color{\colorMATH}\ensuremath{\se'}}}\endgroup }) \in  {\mathcal{E}}_{{\begingroup\renewcommand\colorMATH{\colorMATHB}\renewcommand\colorSYNTAX{\colorSYNTAXB}{{\color{\colorMATH}\ensuremath{\Distance'}}}\endgroup }\mathord{\cdotp }{\begingroup\renewcommand\colorMATH{\colorMATHB}\renewcommand\colorSYNTAX{\colorSYNTAXB}{{\color{\colorMATH}\ensuremath{\sS''}}}\endgroup }}^{k}\llbracket {\begingroup\renewcommand\colorMATH{\colorMATHB}\renewcommand\colorSYNTAX{\colorSYNTAXB}{{\color{\colorMATH}\ensuremath{\Distance'}}}\endgroup }(\tau _{2})\rrbracket }}}, i.e. if {{\color{\colorMATH}\ensuremath{\gamma _{1}\vdash {\begingroup\renewcommand\colorMATH{\colorMATHB}\renewcommand\colorSYNTAX{\colorSYNTAXB}{{\color{\colorMATH}\ensuremath{\se'}}}\endgroup } \Downarrow ^{j} {\begingroup\renewcommand\colorMATH{\colorMATHB}\renewcommand\colorSYNTAX{\colorSYNTAXB}{{\color{\colorMATH}\ensuremath{\sv'_{1}}}}\endgroup }}}}, \pthen {{\color{\colorMATH}\ensuremath{\gamma _{2}\vdash {\begingroup\renewcommand\colorMATH{\colorMATHB}\renewcommand\colorSYNTAX{\colorSYNTAXB}{{\color{\colorMATH}\ensuremath{\se'}}}\endgroup } \Downarrow ^{\pj} {\begingroup\renewcommand\colorMATH{\colorMATHB}\renewcommand\colorSYNTAX{\colorSYNTAXB}{{\color{\colorMATH}\ensuremath{\sv'_{2}}}}\endgroup }}}} \pand 
        {{\color{\colorMATH}\ensuremath{({\begingroup\renewcommand\colorMATH{\colorMATHB}\renewcommand\colorSYNTAX{\colorSYNTAXB}{{\color{\colorMATH}\ensuremath{\sv'_{1}}}}\endgroup }, {\begingroup\renewcommand\colorMATH{\colorMATHB}\renewcommand\colorSYNTAX{\colorSYNTAXB}{{\color{\colorMATH}\ensuremath{\sv'_{2}}}}\endgroup }) \in  {\mathcal{V}}_{{\begingroup\renewcommand\colorMATH{\colorMATHB}\renewcommand\colorSYNTAX{\colorSYNTAXB}{{\color{\colorMATH}\ensuremath{\Distance'}}}\endgroup }\mathord{\cdotp }{\begingroup\renewcommand\colorMATH{\colorMATHB}\renewcommand\colorSYNTAX{\colorSYNTAXB}{{\color{\colorMATH}\ensuremath{\sS''}}}\endgroup }}^{k-j}\llbracket {\begingroup\renewcommand\colorMATH{\colorMATHB}\renewcommand\colorSYNTAX{\colorSYNTAXB}{{\color{\colorMATH}\ensuremath{\Distance'}}}\endgroup }(\tau _{2})\rrbracket }}}.

        If {{\color{\colorMATH}\ensuremath{\gamma _{i}\vdash \inr^{\tau _{1}}\hspace*{0.33em}{\begingroup\renewcommand\colorMATH{\colorMATHB}\renewcommand\colorSYNTAX{\colorSYNTAXB}{{\color{\colorMATH}\ensuremath{\se'}}}\endgroup } \Downarrow ^{j} v_{i}}}} and {{\color{\colorMATH}\ensuremath{\gamma _{i}\vdash {\begingroup\renewcommand\colorMATH{\colorMATHB}\renewcommand\colorSYNTAX{\colorSYNTAXB}{{\color{\colorMATH}\ensuremath{\se'}}}\endgroup } \Downarrow ^{\pj} {\begingroup\renewcommand\colorMATH{\colorMATHB}\renewcommand\colorSYNTAX{\colorSYNTAXB}{{\color{\colorMATH}\ensuremath{\sv'_{i}}}}\endgroup }}}}, then by {\textsc{ INR}}, {{\color{\colorMATH}\ensuremath{{\begingroup\renewcommand\colorMATH{\colorMATHB}\renewcommand\colorSYNTAX{\colorSYNTAXB}{{\color{\colorMATH}\ensuremath{\sv_{i}}}}\endgroup } = \inr^{\tau _{1}}\hspace*{0.33em}{\begingroup\renewcommand\colorMATH{\colorMATHB}\renewcommand\colorSYNTAX{\colorSYNTAXB}{{\color{\colorMATH}\ensuremath{\sv'_{i}}}}\endgroup }}}}. Then we have to prove that\\
        {{\color{\colorMATH}\ensuremath{(\inr^{\tau _{1}}\hspace*{0.33em}{\begingroup\renewcommand\colorMATH{\colorMATHB}\renewcommand\colorSYNTAX{\colorSYNTAXB}{{\color{\colorMATH}\ensuremath{\sv'_{1}}}}\endgroup }, \inr^{\tau _{1}}\hspace*{0.33em}{\begingroup\renewcommand\colorMATH{\colorMATHB}\renewcommand\colorSYNTAX{\colorSYNTAXB}{{\color{\colorMATH}\ensuremath{\sv'_{2}}}}\endgroup }) \in  {\mathcal{V}}_{0}^{k-j}\llbracket {\begingroup\renewcommand\colorMATH{\colorMATHB}\renewcommand\colorSYNTAX{\colorSYNTAXB}{{\color{\colorMATH}\ensuremath{\Distance'}}}\endgroup }(\tau _{1}) \mathrel{^{0}\oplus ^{{\begingroup\renewcommand\colorMATH{\colorMATHB}\renewcommand\colorSYNTAX{\colorSYNTAXB}{{\color{\colorMATH}\ensuremath{\Distance'}}}\endgroup }\mathord{\cdotp }{\begingroup\renewcommand\colorMATH{\colorMATHB}\renewcommand\colorSYNTAX{\colorSYNTAXB}{{\color{\colorMATH}\ensuremath{\sS''}}}\endgroup }}} {\begingroup\renewcommand\colorMATH{\colorMATHB}\renewcommand\colorSYNTAX{\colorSYNTAXB}{{\color{\colorMATH}\ensuremath{\Distance'}}}\endgroup }(\tau _{2})\rrbracket }}}, i.e. that
        {{\color{\colorMATH}\ensuremath{({\begingroup\renewcommand\colorMATH{\colorMATHB}\renewcommand\colorSYNTAX{\colorSYNTAXB}{{\color{\colorMATH}\ensuremath{\sv'_{1}}}}\endgroup }, {\begingroup\renewcommand\colorMATH{\colorMATHB}\renewcommand\colorSYNTAX{\colorSYNTAXB}{{\color{\colorMATH}\ensuremath{\sv'_{2}}}}\endgroup }) \in  {\mathcal{V}}_{0+{\begingroup\renewcommand\colorMATH{\colorMATHB}\renewcommand\colorSYNTAX{\colorSYNTAXB}{{\color{\colorMATH}\ensuremath{\Distance'}}}\endgroup }\mathord{\cdotp }{\begingroup\renewcommand\colorMATH{\colorMATHB}\renewcommand\colorSYNTAX{\colorSYNTAXB}{{\color{\colorMATH}\ensuremath{\sS''}}}\endgroup }}^{k-j}\llbracket {\begingroup\renewcommand\colorMATH{\colorMATHB}\renewcommand\colorSYNTAX{\colorSYNTAXB}{{\color{\colorMATH}\ensuremath{\Distance'}}}\endgroup }(\tau _{2})\rrbracket }}}, but as {{\color{\colorMATH}\ensuremath{0+{\begingroup\renewcommand\colorMATH{\colorMATHB}\renewcommand\colorSYNTAX{\colorSYNTAXB}{{\color{\colorMATH}\ensuremath{\Distance'}}}\endgroup }\mathord{\cdotp }{\begingroup\renewcommand\colorMATH{\colorMATHB}\renewcommand\colorSYNTAX{\colorSYNTAXB}{{\color{\colorMATH}\ensuremath{\sS''}}}\endgroup } = {\begingroup\renewcommand\colorMATH{\colorMATHB}\renewcommand\colorSYNTAX{\colorSYNTAXB}{{\color{\colorMATH}\ensuremath{\Distance'}}}\endgroup }\mathord{\cdotp }{\begingroup\renewcommand\colorMATH{\colorMATHB}\renewcommand\colorSYNTAX{\colorSYNTAXB}{{\color{\colorMATH}\ensuremath{\sS''}}}\endgroup }}}}, the result holds immediately.
      \end{subproof}

    \item  {{\color{\colorMATH}\ensuremath{\Gamma ; {\begingroup\renewcommand\colorMATH{\colorMATHB}\renewcommand\colorSYNTAX{\colorSYNTAXB}{{\color{\colorMATH}\ensuremath{\Distance}}}\endgroup } \vdash  \inr^{\tau _{1}}\hspace*{0.33em}{\begingroup\renewcommand\colorMATH{\colorMATHB}\renewcommand\colorSYNTAX{\colorSYNTAXB}{{\color{\colorMATH}\ensuremath{\se'}}}\endgroup } \mathrel{:} \tau _{1} \mathrel{^{\varnothing }\oplus ^{{\begingroup\renewcommand\colorMATH{\colorMATHB}\renewcommand\colorSYNTAX{\colorSYNTAXB}{{\color{\colorMATH}\ensuremath{\sS''}}}\endgroup }}} \tau _{2} \mathrel{;} {\begingroup\renewcommand\colorMATH{\colorMATHB}\renewcommand\colorSYNTAX{\colorSYNTAXB}{{\color{\colorMATH}\ensuremath{\sS'}}}\endgroup }}}} \mt{with prepayment}
      \begin{subproof} 
        We have to prove that {{\color{\colorMATH}\ensuremath{\forall  (\gamma _{1},\gamma _{2}) \in  {\mathcal{G}}_{{\begingroup\renewcommand\colorMATH{\colorMATHB}\renewcommand\colorSYNTAX{\colorSYNTAXB}{{\color{\colorMATH}\ensuremath{\Distance'}}}\endgroup }}^{\kg}\llbracket \Gamma \rrbracket , (\gamma _{1}\vdash \inr^{\tau _{1}}\hspace*{0.33em}{\begingroup\renewcommand\colorMATH{\colorMATHB}\renewcommand\colorSYNTAX{\colorSYNTAXB}{{\color{\colorMATH}\ensuremath{\se'}}}\endgroup },\gamma _{2}\vdash \inr^{\tau _{1}}\hspace*{0.33em}{\begingroup\renewcommand\colorMATH{\colorMATHB}\renewcommand\colorSYNTAX{\colorSYNTAXB}{{\color{\colorMATH}\ensuremath{\se'}}}\endgroup }) \in  {\mathcal{E}}_{{\begingroup\renewcommand\colorMATH{\colorMATHB}\renewcommand\colorSYNTAX{\colorSYNTAXB}{{\color{\colorMATH}\ensuremath{\Distance'}}}\endgroup }\mathord{\cdotp }{\begingroup\renewcommand\colorMATH{\colorMATHB}\renewcommand\colorSYNTAX{\colorSYNTAXB}{{\color{\colorMATH}\ensuremath{\sS'}}}\endgroup }}^{k}\llbracket {\begingroup\renewcommand\colorMATH{\colorMATHB}\renewcommand\colorSYNTAX{\colorSYNTAXB}{{\color{\colorMATH}\ensuremath{\Distance'}}}\endgroup }(\tau _{1} \mathrel{^{\varnothing }\oplus ^{{\begingroup\renewcommand\colorMATH{\colorMATHB}\renewcommand\colorSYNTAX{\colorSYNTAXB}{{\color{\colorMATH}\ensuremath{\sS''}}}\endgroup }}} \tau _{2})\rrbracket }}}, for {{\color{\colorMATH}\ensuremath{{\begingroup\renewcommand\colorMATH{\colorMATHB}\renewcommand\colorSYNTAX{\colorSYNTAXB}{{\color{\colorMATH}\ensuremath{\Distance'}}}\endgroup } \sqsubseteq  {\begingroup\renewcommand\colorMATH{\colorMATHB}\renewcommand\colorSYNTAX{\colorSYNTAXB}{{\color{\colorMATH}\ensuremath{\Distance}}}\endgroup }}}}.
        Notice that {{\color{\colorMATH}\ensuremath{{\begingroup\renewcommand\colorMATH{\colorMATHB}\renewcommand\colorSYNTAX{\colorSYNTAXB}{{\color{\colorMATH}\ensuremath{\Distance'}}}\endgroup }\mathord{\cdotp }\varnothing  = 0}}}, and {{\color{\colorMATH}\ensuremath{{\begingroup\renewcommand\colorMATH{\colorMATHB}\renewcommand\colorSYNTAX{\colorSYNTAXB}{{\color{\colorMATH}\ensuremath{\Distance'}}}\endgroup }(\tau _{1} \mathrel{^{\varnothing }\oplus ^{{\begingroup\renewcommand\colorMATH{\colorMATHB}\renewcommand\colorSYNTAX{\colorSYNTAXB}{{\color{\colorMATH}\ensuremath{\sS''}}}\endgroup }}} \tau _{2}) = {\begingroup\renewcommand\colorMATH{\colorMATHB}\renewcommand\colorSYNTAX{\colorSYNTAXB}{{\color{\colorMATH}\ensuremath{\Distance'}}}\endgroup }(\tau _{1}) \mathrel{^{0}\oplus ^{{\begingroup\renewcommand\colorMATH{\colorMATHB}\renewcommand\colorSYNTAX{\colorSYNTAXB}{{\color{\colorMATH}\ensuremath{\Distance'}}}\endgroup }\mathord{\cdotp }{\begingroup\renewcommand\colorMATH{\colorMATHB}\renewcommand\colorSYNTAX{\colorSYNTAXB}{{\color{\colorMATH}\ensuremath{\sS''}}}\endgroup }}} {\begingroup\renewcommand\colorMATH{\colorMATHB}\renewcommand\colorSYNTAX{\colorSYNTAXB}{{\color{\colorMATH}\ensuremath{\Distance'}}}\endgroup }(\tau _{2})}}}, then we have to prove that\\
        {{\color{\colorMATH}\ensuremath{(\gamma _{1}\vdash \inr^{\tau _{1}}\hspace*{0.33em}{\begingroup\renewcommand\colorMATH{\colorMATHB}\renewcommand\colorSYNTAX{\colorSYNTAXB}{{\color{\colorMATH}\ensuremath{\se'}}}\endgroup },\gamma _{2}\vdash \inr^{\tau _{1}}\hspace*{0.33em}{\begingroup\renewcommand\colorMATH{\colorMATHB}\renewcommand\colorSYNTAX{\colorSYNTAXB}{{\color{\colorMATH}\ensuremath{\se'}}}\endgroup }) \in  {\mathcal{E}}_{{\begingroup\renewcommand\colorMATH{\colorMATHB}\renewcommand\colorSYNTAX{\colorSYNTAXB}{{\color{\colorMATH}\ensuremath{\Distance'}}}\endgroup }\mathord{\cdotp }{\begingroup\renewcommand\colorMATH{\colorMATHB}\renewcommand\colorSYNTAX{\colorSYNTAXB}{{\color{\colorMATH}\ensuremath{\sS'}}}\endgroup }}^{k}\llbracket {\begingroup\renewcommand\colorMATH{\colorMATHB}\renewcommand\colorSYNTAX{\colorSYNTAXB}{{\color{\colorMATH}\ensuremath{\Distance'}}}\endgroup }(\tau _{1}) \mathrel{^{0}\oplus ^{{\begingroup\renewcommand\colorMATH{\colorMATHB}\renewcommand\colorSYNTAX{\colorSYNTAXB}{{\color{\colorMATH}\ensuremath{\Distance'}}}\endgroup }\mathord{\cdotp }{\begingroup\renewcommand\colorMATH{\colorMATHB}\renewcommand\colorSYNTAX{\colorSYNTAXB}{{\color{\colorMATH}\ensuremath{\sS''}}}\endgroup }}} {\begingroup\renewcommand\colorMATH{\colorMATHB}\renewcommand\colorSYNTAX{\colorSYNTAXB}{{\color{\colorMATH}\ensuremath{\Distance'}}}\endgroup }(\tau _{2})\rrbracket }}}, i.e.
        if {{\color{\colorMATH}\ensuremath{\gamma _{1}\vdash \inr^{\tau _{1}}\hspace*{0.33em}{\begingroup\renewcommand\colorMATH{\colorMATHB}\renewcommand\colorSYNTAX{\colorSYNTAXB}{{\color{\colorMATH}\ensuremath{\se'}}}\endgroup } \Downarrow ^{j} v_{1}}}} \pthen {{\color{\colorMATH}\ensuremath{\gamma _{2}\vdash \inr^{\tau _{1}}\hspace*{0.33em}{\begingroup\renewcommand\colorMATH{\colorMATHB}\renewcommand\colorSYNTAX{\colorSYNTAXB}{{\color{\colorMATH}\ensuremath{\se'}}}\endgroup } \Downarrow ^{\pj} v_{2}}}}, \pand 
        {{\color{\colorMATH}\ensuremath{(v_{1}, v_{2}) \in  {\mathcal{V}}_{{\begingroup\renewcommand\colorMATH{\colorMATHB}\renewcommand\colorSYNTAX{\colorSYNTAXB}{{\color{\colorMATH}\ensuremath{\Distance'}}}\endgroup }\mathord{\cdotp }{\begingroup\renewcommand\colorMATH{\colorMATHB}\renewcommand\colorSYNTAX{\colorSYNTAXB}{{\color{\colorMATH}\ensuremath{\sS'}}}\endgroup }}^{k-j}\llbracket {\begingroup\renewcommand\colorMATH{\colorMATHB}\renewcommand\colorSYNTAX{\colorSYNTAXB}{{\color{\colorMATH}\ensuremath{\Distance'}}}\endgroup }(\tau _{1}) \mathrel{^{0}\oplus ^{{\begingroup\renewcommand\colorMATH{\colorMATHB}\renewcommand\colorSYNTAX{\colorSYNTAXB}{{\color{\colorMATH}\ensuremath{\Distance'}}}\endgroup }\mathord{\cdotp }{\begingroup\renewcommand\colorMATH{\colorMATHB}\renewcommand\colorSYNTAX{\colorSYNTAXB}{{\color{\colorMATH}\ensuremath{\sS''}}}\endgroup }}} {\begingroup\renewcommand\colorMATH{\colorMATHB}\renewcommand\colorSYNTAX{\colorSYNTAXB}{{\color{\colorMATH}\ensuremath{\Distance'}}}\endgroup }(\tau _{2})\rrbracket }}}.

        By induction hypothesis on {{\color{\colorMATH}\ensuremath{\Gamma  \vdash  {\begingroup\renewcommand\colorMATH{\colorMATHB}\renewcommand\colorSYNTAX{\colorSYNTAXB}{{\color{\colorMATH}\ensuremath{\se'}}}\endgroup } \mathrel{:} \tau _{2} \mathrel{;} {\begingroup\renewcommand\colorMATH{\colorMATHB}\renewcommand\colorSYNTAX{\colorSYNTAXB}{{\color{\colorMATH}\ensuremath{\sS''}}}\endgroup } + {\begingroup\renewcommand\colorMATH{\colorMATHB}\renewcommand\colorSYNTAX{\colorSYNTAXB}{{\color{\colorMATH}\ensuremath{\sS'}}}\endgroup }}}}, we know that\\ 
        {{\color{\colorMATH}\ensuremath{(\gamma _{1}\vdash {\begingroup\renewcommand\colorMATH{\colorMATHB}\renewcommand\colorSYNTAX{\colorSYNTAXB}{{\color{\colorMATH}\ensuremath{\se'}}}\endgroup },\gamma _{2}\vdash {\begingroup\renewcommand\colorMATH{\colorMATHB}\renewcommand\colorSYNTAX{\colorSYNTAXB}{{\color{\colorMATH}\ensuremath{\se'}}}\endgroup }) \in  {\mathcal{E}}_{{\begingroup\renewcommand\colorMATH{\colorMATHB}\renewcommand\colorSYNTAX{\colorSYNTAXB}{{\color{\colorMATH}\ensuremath{\Distance'}}}\endgroup }\mathord{\cdotp }({\begingroup\renewcommand\colorMATH{\colorMATHB}\renewcommand\colorSYNTAX{\colorSYNTAXB}{{\color{\colorMATH}\ensuremath{\sS''}}}\endgroup }+{\begingroup\renewcommand\colorMATH{\colorMATHB}\renewcommand\colorSYNTAX{\colorSYNTAXB}{{\color{\colorMATH}\ensuremath{\sS'}}}\endgroup })}^{k}\llbracket {\begingroup\renewcommand\colorMATH{\colorMATHB}\renewcommand\colorSYNTAX{\colorSYNTAXB}{{\color{\colorMATH}\ensuremath{\Distance'}}}\endgroup }(\tau _{2})\rrbracket }}}, i.e. if {{\color{\colorMATH}\ensuremath{\gamma _{1}\vdash {\begingroup\renewcommand\colorMATH{\colorMATHB}\renewcommand\colorSYNTAX{\colorSYNTAXB}{{\color{\colorMATH}\ensuremath{\se'}}}\endgroup } \Downarrow ^{j} {\begingroup\renewcommand\colorMATH{\colorMATHB}\renewcommand\colorSYNTAX{\colorSYNTAXB}{{\color{\colorMATH}\ensuremath{\sv'_{1}}}}\endgroup }}}}, \pthen {{\color{\colorMATH}\ensuremath{\gamma _{2}\vdash {\begingroup\renewcommand\colorMATH{\colorMATHB}\renewcommand\colorSYNTAX{\colorSYNTAXB}{{\color{\colorMATH}\ensuremath{\se'}}}\endgroup } \Downarrow ^{\pj} {\begingroup\renewcommand\colorMATH{\colorMATHB}\renewcommand\colorSYNTAX{\colorSYNTAXB}{{\color{\colorMATH}\ensuremath{\sv'_{2}}}}\endgroup }}}} \pand 
        {{\color{\colorMATH}\ensuremath{({\begingroup\renewcommand\colorMATH{\colorMATHB}\renewcommand\colorSYNTAX{\colorSYNTAXB}{{\color{\colorMATH}\ensuremath{\sv'_{1}}}}\endgroup }, {\begingroup\renewcommand\colorMATH{\colorMATHB}\renewcommand\colorSYNTAX{\colorSYNTAXB}{{\color{\colorMATH}\ensuremath{\sv'_{2}}}}\endgroup }) \in  {\mathcal{V}}_{{\begingroup\renewcommand\colorMATH{\colorMATHB}\renewcommand\colorSYNTAX{\colorSYNTAXB}{{\color{\colorMATH}\ensuremath{\Distance'}}}\endgroup }\mathord{\cdotp }({\begingroup\renewcommand\colorMATH{\colorMATHB}\renewcommand\colorSYNTAX{\colorSYNTAXB}{{\color{\colorMATH}\ensuremath{\sS''}}}\endgroup }+{\begingroup\renewcommand\colorMATH{\colorMATHB}\renewcommand\colorSYNTAX{\colorSYNTAXB}{{\color{\colorMATH}\ensuremath{\sS'}}}\endgroup })}^{k-j}\llbracket {\begingroup\renewcommand\colorMATH{\colorMATHB}\renewcommand\colorSYNTAX{\colorSYNTAXB}{{\color{\colorMATH}\ensuremath{\Distance'}}}\endgroup }(\tau _{2})\rrbracket }}}.

        If {{\color{\colorMATH}\ensuremath{\gamma _{i}\vdash \inr^{\tau _{1}}\hspace*{0.33em}{\begingroup\renewcommand\colorMATH{\colorMATHB}\renewcommand\colorSYNTAX{\colorSYNTAXB}{{\color{\colorMATH}\ensuremath{\se'}}}\endgroup } \Downarrow ^{j} v_{i}}}} and {{\color{\colorMATH}\ensuremath{\gamma _{i}\vdash {\begingroup\renewcommand\colorMATH{\colorMATHB}\renewcommand\colorSYNTAX{\colorSYNTAXB}{{\color{\colorMATH}\ensuremath{\se'}}}\endgroup } \Downarrow ^{\pj} {\begingroup\renewcommand\colorMATH{\colorMATHB}\renewcommand\colorSYNTAX{\colorSYNTAXB}{{\color{\colorMATH}\ensuremath{\sv'_{i}}}}\endgroup }}}}, then by {\textsc{ INR}}, {{\color{\colorMATH}\ensuremath{{\begingroup\renewcommand\colorMATH{\colorMATHB}\renewcommand\colorSYNTAX{\colorSYNTAXB}{{\color{\colorMATH}\ensuremath{\sv_{i}}}}\endgroup } = \inr^{\tau _{1}}\hspace*{0.33em}{\begingroup\renewcommand\colorMATH{\colorMATHB}\renewcommand\colorSYNTAX{\colorSYNTAXB}{{\color{\colorMATH}\ensuremath{\sv'_{i}}}}\endgroup }}}}. Then we have to prove that\\
        {{\color{\colorMATH}\ensuremath{(\inr^{\tau _{1}}\hspace*{0.33em}{\begingroup\renewcommand\colorMATH{\colorMATHB}\renewcommand\colorSYNTAX{\colorSYNTAXB}{{\color{\colorMATH}\ensuremath{\sv'_{1}}}}\endgroup }, \inr^{\tau _{1}}\hspace*{0.33em}{\begingroup\renewcommand\colorMATH{\colorMATHB}\renewcommand\colorSYNTAX{\colorSYNTAXB}{{\color{\colorMATH}\ensuremath{\sv'_{2}}}}\endgroup }) \in  {\mathcal{V}}_{{\begingroup\renewcommand\colorMATH{\colorMATHB}\renewcommand\colorSYNTAX{\colorSYNTAXB}{{\color{\colorMATH}\ensuremath{\Distance'}}}\endgroup }\mathord{\cdotp }{\begingroup\renewcommand\colorMATH{\colorMATHB}\renewcommand\colorSYNTAX{\colorSYNTAXB}{{\color{\colorMATH}\ensuremath{\sS'}}}\endgroup }}^{k-j}\llbracket {\begingroup\renewcommand\colorMATH{\colorMATHB}\renewcommand\colorSYNTAX{\colorSYNTAXB}{{\color{\colorMATH}\ensuremath{\Distance'}}}\endgroup }(\tau _{1}) \mathrel{^{0}\oplus ^{{\begingroup\renewcommand\colorMATH{\colorMATHB}\renewcommand\colorSYNTAX{\colorSYNTAXB}{{\color{\colorMATH}\ensuremath{\Distance'}}}\endgroup }\mathord{\cdotp }{\begingroup\renewcommand\colorMATH{\colorMATHB}\renewcommand\colorSYNTAX{\colorSYNTAXB}{{\color{\colorMATH}\ensuremath{\sS''}}}\endgroup }}} {\begingroup\renewcommand\colorMATH{\colorMATHB}\renewcommand\colorSYNTAX{\colorSYNTAXB}{{\color{\colorMATH}\ensuremath{\Distance'}}}\endgroup }(\tau _{2})\rrbracket }}}, i.e. that
        {{\color{\colorMATH}\ensuremath{({\begingroup\renewcommand\colorMATH{\colorMATHB}\renewcommand\colorSYNTAX{\colorSYNTAXB}{{\color{\colorMATH}\ensuremath{\sv'_{1}}}}\endgroup }, {\begingroup\renewcommand\colorMATH{\colorMATHB}\renewcommand\colorSYNTAX{\colorSYNTAXB}{{\color{\colorMATH}\ensuremath{\sv'_{2}}}}\endgroup }) \in  {\mathcal{V}}_{{\begingroup\renewcommand\colorMATH{\colorMATHB}\renewcommand\colorSYNTAX{\colorSYNTAXB}{{\color{\colorMATH}\ensuremath{\Distance'}}}\endgroup }\mathord{\cdotp }{\begingroup\renewcommand\colorMATH{\colorMATHB}\renewcommand\colorSYNTAX{\colorSYNTAXB}{{\color{\colorMATH}\ensuremath{\sS'}}}\endgroup }+{\begingroup\renewcommand\colorMATH{\colorMATHB}\renewcommand\colorSYNTAX{\colorSYNTAXB}{{\color{\colorMATH}\ensuremath{\Distance'}}}\endgroup }\mathord{\cdotp }{\begingroup\renewcommand\colorMATH{\colorMATHB}\renewcommand\colorSYNTAX{\colorSYNTAXB}{{\color{\colorMATH}\ensuremath{\sS''}}}\endgroup }}^{k-j}\llbracket {\begingroup\renewcommand\colorMATH{\colorMATHB}\renewcommand\colorSYNTAX{\colorSYNTAXB}{{\color{\colorMATH}\ensuremath{\Distance'}}}\endgroup }(\tau _{2})\rrbracket }}}, but as {{\color{\colorMATH}\ensuremath{{\begingroup\renewcommand\colorMATH{\colorMATHB}\renewcommand\colorSYNTAX{\colorSYNTAXB}{{\color{\colorMATH}\ensuremath{\Distance'}}}\endgroup }\mathord{\cdotp }({\begingroup\renewcommand\colorMATH{\colorMATHB}\renewcommand\colorSYNTAX{\colorSYNTAXB}{{\color{\colorMATH}\ensuremath{\sS''}}}\endgroup }+{\begingroup\renewcommand\colorMATH{\colorMATHB}\renewcommand\colorSYNTAX{\colorSYNTAXB}{{\color{\colorMATH}\ensuremath{\sS'}}}\endgroup }) = {\begingroup\renewcommand\colorMATH{\colorMATHB}\renewcommand\colorSYNTAX{\colorSYNTAXB}{{\color{\colorMATH}\ensuremath{\Distance'}}}\endgroup }\mathord{\cdotp }{\begingroup\renewcommand\colorMATH{\colorMATHB}\renewcommand\colorSYNTAX{\colorSYNTAXB}{{\color{\colorMATH}\ensuremath{\sS''}}}\endgroup } + {\begingroup\renewcommand\colorMATH{\colorMATHB}\renewcommand\colorSYNTAX{\colorSYNTAXB}{{\color{\colorMATH}\ensuremath{\Distance'}}}\endgroup }\mathord{\cdotp }{\begingroup\renewcommand\colorMATH{\colorMATHB}\renewcommand\colorSYNTAX{\colorSYNTAXB}{{\color{\colorMATH}\ensuremath{\sS'}}}\endgroup }}}}, the result holds immediately.
      \end{subproof}

    \item  {{\color{\colorMATH}\ensuremath{\Gamma ; {\begingroup\renewcommand\colorMATH{\colorMATHB}\renewcommand\colorSYNTAX{\colorSYNTAXB}{{\color{\colorMATH}\ensuremath{\Distance}}}\endgroup } \vdash  \ccase\hspace*{0.33em}{\begingroup\renewcommand\colorMATH{\colorMATHB}\renewcommand\colorSYNTAX{\colorSYNTAXB}{{\color{\colorMATH}\ensuremath{\se_{1}}}}\endgroup }\hspace*{0.33em}\of\hspace*{0.33em}\{ x\Rightarrow {\begingroup\renewcommand\colorMATH{\colorMATHB}\renewcommand\colorSYNTAX{\colorSYNTAXB}{{\color{\colorMATH}\ensuremath{\se_{2}}}}\endgroup }\} \hspace*{0.33em}\{ y\Rightarrow e_{3}\}  \mathrel{:} [{\begingroup\renewcommand\colorMATH{\colorMATHB}\renewcommand\colorSYNTAX{\colorSYNTAXB}{{\color{\colorMATH}\ensuremath{\sS_{1}}}}\endgroup } + {\begingroup\renewcommand\colorMATH{\colorMATHB}\renewcommand\colorSYNTAX{\colorSYNTAXB}{{\color{\colorMATH}\ensuremath{\sS_{1 1}}}}\endgroup }/x]\tau _{2} \sqcup  [{\begingroup\renewcommand\colorMATH{\colorMATHB}\renewcommand\colorSYNTAX{\colorSYNTAXB}{{\color{\colorMATH}\ensuremath{\sS_{1}}}}\endgroup } + {\begingroup\renewcommand\colorMATH{\colorMATHB}\renewcommand\colorSYNTAX{\colorSYNTAXB}{{\color{\colorMATH}\ensuremath{\sS_{1 2}}}}\endgroup }/y]\tau _{3} \mathrel{;}  {\begingroup\renewcommand\colorMATH{\colorMATHB}\renewcommand\colorSYNTAX{\colorSYNTAXB}{{\color{\colorMATH}\ensuremath{\sS_{1}}}}\endgroup } \sqcup  ([{\begingroup\renewcommand\colorMATH{\colorMATHB}\renewcommand\colorSYNTAX{\colorSYNTAXB}{{\color{\colorMATH}\ensuremath{\sS_{1}}}}\endgroup } + {\begingroup\renewcommand\colorMATH{\colorMATHB}\renewcommand\colorSYNTAX{\colorSYNTAXB}{{\color{\colorMATH}\ensuremath{\sS_{1 1}}}}\endgroup }/x]({\begingroup\renewcommand\colorMATH{\colorMATHB}\renewcommand\colorSYNTAX{\colorSYNTAXB}{{\color{\colorMATH}\ensuremath{\sS_{2}}}}\endgroup }+{\begingroup\renewcommand\colorMATH{\colorMATHB}\renewcommand\colorSYNTAX{\colorSYNTAXB}{{\color{\colorMATH}\ensuremath{\sss_{2}}}}\endgroup }x) \sqcup  [{\begingroup\renewcommand\colorMATH{\colorMATHB}\renewcommand\colorSYNTAX{\colorSYNTAXB}{{\color{\colorMATH}\ensuremath{\sS_{1}}}}\endgroup } + {\begingroup\renewcommand\colorMATH{\colorMATHB}\renewcommand\colorSYNTAX{\colorSYNTAXB}{{\color{\colorMATH}\ensuremath{\sS_{1 2}}}}\endgroup }/y]({\begingroup\renewcommand\colorMATH{\colorMATHB}\renewcommand\colorSYNTAX{\colorSYNTAXB}{{\color{\colorMATH}\ensuremath{\sS_{3}}}}\endgroup }+{\begingroup\renewcommand\colorMATH{\colorMATHB}\renewcommand\colorSYNTAX{\colorSYNTAXB}{{\color{\colorMATH}\ensuremath{\sss_{3}}}}\endgroup }y))}}}
      \begin{subproof} 
        We have to prove that for any {{\color{\colorMATH}\ensuremath{k}}}, {{\color{\colorMATH}\ensuremath{\forall  (\gamma _{1},\gamma _{2}) \in  {\mathcal{G}}_{{\begingroup\renewcommand\colorMATH{\colorMATHB}\renewcommand\colorSYNTAX{\colorSYNTAXB}{{\color{\colorMATH}\ensuremath{\Distance'}}}\endgroup }}^{\kg}\llbracket \Gamma \rrbracket }}},\\ 
        {{\color{\colorMATH}\ensuremath{(\gamma _{1}\vdash \ccase\hspace*{0.33em}{\begingroup\renewcommand\colorMATH{\colorMATHB}\renewcommand\colorSYNTAX{\colorSYNTAXB}{{\color{\colorMATH}\ensuremath{\se_{1}}}}\endgroup }\hspace*{0.33em}\of\hspace*{0.33em}\{ x\Rightarrow {\begingroup\renewcommand\colorMATH{\colorMATHB}\renewcommand\colorSYNTAX{\colorSYNTAXB}{{\color{\colorMATH}\ensuremath{\se_{2}}}}\endgroup }\} \hspace*{0.33em}\{ y\Rightarrow e_{3}\} ,\gamma _{2}\vdash \ccase\hspace*{0.33em}{\begingroup\renewcommand\colorMATH{\colorMATHB}\renewcommand\colorSYNTAX{\colorSYNTAXB}{{\color{\colorMATH}\ensuremath{\se_{1}}}}\endgroup }\hspace*{0.33em}\of\hspace*{0.33em}\{ x\Rightarrow {\begingroup\renewcommand\colorMATH{\colorMATHB}\renewcommand\colorSYNTAX{\colorSYNTAXB}{{\color{\colorMATH}\ensuremath{\se_{2}}}}\endgroup }\} \hspace*{0.33em}\{ y\Rightarrow {\begingroup\renewcommand\colorMATH{\colorMATHB}\renewcommand\colorSYNTAX{\colorSYNTAXB}{{\color{\colorMATH}\ensuremath{\se_{3}}}}\endgroup }\} ) \in  {\mathcal{E}}_{{\begingroup\renewcommand\colorMATH{\colorMATHB}\renewcommand\colorSYNTAX{\colorSYNTAXB}{{\color{\colorMATH}\ensuremath{\Distance'}}}\endgroup }\mathord{\cdotp }{\begingroup\renewcommand\colorMATH{\colorMATHB}\renewcommand\colorSYNTAX{\colorSYNTAXB}{{\color{\colorMATH}\ensuremath{\sS''}}}\endgroup }}^{k}\llbracket {\begingroup\renewcommand\colorMATH{\colorMATHB}\renewcommand\colorSYNTAX{\colorSYNTAXB}{{\color{\colorMATH}\ensuremath{\Distance'}}}\endgroup }(\tau ')\rrbracket }}},
        for {{\color{\colorMATH}\ensuremath{{\begingroup\renewcommand\colorMATH{\colorMATHB}\renewcommand\colorSYNTAX{\colorSYNTAXB}{{\color{\colorMATH}\ensuremath{\Distance'}}}\endgroup } \sqsubseteq  {\begingroup\renewcommand\colorMATH{\colorMATHB}\renewcommand\colorSYNTAX{\colorSYNTAXB}{{\color{\colorMATH}\ensuremath{\Distance}}}\endgroup }}}},
        where\\
        {{\color{\colorMATH}\ensuremath{{\begingroup\renewcommand\colorMATH{\colorMATHB}\renewcommand\colorSYNTAX{\colorSYNTAXB}{{\color{\colorMATH}\ensuremath{\sS''}}}\endgroup }= {\begingroup\renewcommand\colorMATH{\colorMATHB}\renewcommand\colorSYNTAX{\colorSYNTAXB}{{\color{\colorMATH}\ensuremath{\sS_{1}}}}\endgroup } \sqcup  ([{\begingroup\renewcommand\colorMATH{\colorMATHB}\renewcommand\colorSYNTAX{\colorSYNTAXB}{{\color{\colorMATH}\ensuremath{\sS_{1}}}}\endgroup } + {\begingroup\renewcommand\colorMATH{\colorMATHB}\renewcommand\colorSYNTAX{\colorSYNTAXB}{{\color{\colorMATH}\ensuremath{\sS_{1 1}}}}\endgroup }/x]({\begingroup\renewcommand\colorMATH{\colorMATHB}\renewcommand\colorSYNTAX{\colorSYNTAXB}{{\color{\colorMATH}\ensuremath{\sS_{2}}}}\endgroup }+{\begingroup\renewcommand\colorMATH{\colorMATHB}\renewcommand\colorSYNTAX{\colorSYNTAXB}{{\color{\colorMATH}\ensuremath{\sss_{2}}}}\endgroup }x) \sqcup  [{\begingroup\renewcommand\colorMATH{\colorMATHB}\renewcommand\colorSYNTAX{\colorSYNTAXB}{{\color{\colorMATH}\ensuremath{\sS_{1}}}}\endgroup } + {\begingroup\renewcommand\colorMATH{\colorMATHB}\renewcommand\colorSYNTAX{\colorSYNTAXB}{{\color{\colorMATH}\ensuremath{\sS_{1 2}}}}\endgroup }/y]({\begingroup\renewcommand\colorMATH{\colorMATHB}\renewcommand\colorSYNTAX{\colorSYNTAXB}{{\color{\colorMATH}\ensuremath{\sS_{3}}}}\endgroup }+{\begingroup\renewcommand\colorMATH{\colorMATHB}\renewcommand\colorSYNTAX{\colorSYNTAXB}{{\color{\colorMATH}\ensuremath{\sss_{3}}}}\endgroup }y)) = 
        {\begingroup\renewcommand\colorMATH{\colorMATHB}\renewcommand\colorSYNTAX{\colorSYNTAXB}{{\color{\colorMATH}\ensuremath{\sS_{1}}}}\endgroup } \sqcup  (({\begingroup\renewcommand\colorMATH{\colorMATHB}\renewcommand\colorSYNTAX{\colorSYNTAXB}{{\color{\colorMATH}\ensuremath{\sss_{2}}}}\endgroup }{\begingroup\renewcommand\colorMATH{\colorMATHB}\renewcommand\colorSYNTAX{\colorSYNTAXB}{{\color{\colorMATH}\ensuremath{\sS_{1}}}}\endgroup } + {\begingroup\renewcommand\colorMATH{\colorMATHB}\renewcommand\colorSYNTAX{\colorSYNTAXB}{{\color{\colorMATH}\ensuremath{\sss_{2}}}}\endgroup }{\begingroup\renewcommand\colorMATH{\colorMATHB}\renewcommand\colorSYNTAX{\colorSYNTAXB}{{\color{\colorMATH}\ensuremath{\sS_{1 1}}}}\endgroup } {\begingroup\renewcommand\colorMATH{\colorMATHB}\renewcommand\colorSYNTAX{\colorSYNTAXB}{{\color{\colorMATH}\ensuremath{\sS_{2}}}}\endgroup }) \sqcup  ({\begingroup\renewcommand\colorMATH{\colorMATHB}\renewcommand\colorSYNTAX{\colorSYNTAXB}{{\color{\colorMATH}\ensuremath{\sss_{3}}}}\endgroup }{\begingroup\renewcommand\colorMATH{\colorMATHB}\renewcommand\colorSYNTAX{\colorSYNTAXB}{{\color{\colorMATH}\ensuremath{\sS_{1}}}}\endgroup } + {\begingroup\renewcommand\colorMATH{\colorMATHB}\renewcommand\colorSYNTAX{\colorSYNTAXB}{{\color{\colorMATH}\ensuremath{\sss_{3}}}}\endgroup }{\begingroup\renewcommand\colorMATH{\colorMATHB}\renewcommand\colorSYNTAX{\colorSYNTAXB}{{\color{\colorMATH}\ensuremath{\sS_{1 2}}}}\endgroup } {\begingroup\renewcommand\colorMATH{\colorMATHB}\renewcommand\colorSYNTAX{\colorSYNTAXB}{{\color{\colorMATH}\ensuremath{\sS_{2}}}}\endgroup }))
        }}}, and 
        {{\color{\colorMATH}\ensuremath{\tau '=[{\begingroup\renewcommand\colorMATH{\colorMATHB}\renewcommand\colorSYNTAX{\colorSYNTAXB}{{\color{\colorMATH}\ensuremath{\sS_{1}}}}\endgroup } + {\begingroup\renewcommand\colorMATH{\colorMATHB}\renewcommand\colorSYNTAX{\colorSYNTAXB}{{\color{\colorMATH}\ensuremath{\sS_{1 1}}}}\endgroup }/x]\tau _{2} \sqcup  [{\begingroup\renewcommand\colorMATH{\colorMATHB}\renewcommand\colorSYNTAX{\colorSYNTAXB}{{\color{\colorMATH}\ensuremath{\sS_{1}}}}\endgroup } + {\begingroup\renewcommand\colorMATH{\colorMATHB}\renewcommand\colorSYNTAX{\colorSYNTAXB}{{\color{\colorMATH}\ensuremath{\sS_{1 2}}}}\endgroup }/y]\tau _{3}}}}.

        By induction hypothesis on {{\color{\colorMATH}\ensuremath{\Gamma ; {\begingroup\renewcommand\colorMATH{\colorMATHB}\renewcommand\colorSYNTAX{\colorSYNTAXB}{{\color{\colorMATH}\ensuremath{\Distance}}}\endgroup } \vdash  {\begingroup\renewcommand\colorMATH{\colorMATHB}\renewcommand\colorSYNTAX{\colorSYNTAXB}{{\color{\colorMATH}\ensuremath{\se_{1}}}}\endgroup } \mathrel{:} \tau _{1 1} \mathrel{^{{\begingroup\renewcommand\colorMATH{\colorMATHB}\renewcommand\colorSYNTAX{\colorSYNTAXB}{{\color{\colorMATH}\ensuremath{\sS_{1 1}}}}\endgroup }}\oplus ^{{\begingroup\renewcommand\colorMATH{\colorMATHB}\renewcommand\colorSYNTAX{\colorSYNTAXB}{{\color{\colorMATH}\ensuremath{\sS_{1 2}}}}\endgroup }}} \tau _{1 2} \mathrel{;} {\begingroup\renewcommand\colorMATH{\colorMATHB}\renewcommand\colorSYNTAX{\colorSYNTAXB}{{\color{\colorMATH}\ensuremath{\sS_{1}}}}\endgroup }}}}, we know that\\
        {{\color{\colorMATH}\ensuremath{(\gamma _{1}\vdash {\begingroup\renewcommand\colorMATH{\colorMATHB}\renewcommand\colorSYNTAX{\colorSYNTAXB}{{\color{\colorMATH}\ensuremath{\se_{1}}}}\endgroup },\gamma _{2}\vdash {\begingroup\renewcommand\colorMATH{\colorMATHB}\renewcommand\colorSYNTAX{\colorSYNTAXB}{{\color{\colorMATH}\ensuremath{\se_{1}}}}\endgroup }) \in  {\mathcal{E}}^{k}_{{\begingroup\renewcommand\colorMATH{\colorMATHB}\renewcommand\colorSYNTAX{\colorSYNTAXB}{{\color{\colorMATH}\ensuremath{\Distance'}}}\endgroup }\mathord{\cdotp }{\begingroup\renewcommand\colorMATH{\colorMATHB}\renewcommand\colorSYNTAX{\colorSYNTAXB}{{\color{\colorMATH}\ensuremath{\sS_{1}}}}\endgroup }}\llbracket {\begingroup\renewcommand\colorMATH{\colorMATHB}\renewcommand\colorSYNTAX{\colorSYNTAXB}{{\color{\colorMATH}\ensuremath{\Distance'}}}\endgroup }(\tau _{1 1} \mathrel{^{{\begingroup\renewcommand\colorMATH{\colorMATHB}\renewcommand\colorSYNTAX{\colorSYNTAXB}{{\color{\colorMATH}\ensuremath{\sS_{1 1}}}}\endgroup }}\oplus ^{{\begingroup\renewcommand\colorMATH{\colorMATHB}\renewcommand\colorSYNTAX{\colorSYNTAXB}{{\color{\colorMATH}\ensuremath{\sS_{1 2}}}}\endgroup }}} \tau _{1 2})\rrbracket }}}, i.e. if {{\color{\colorMATH}\ensuremath{\gamma _{1}\vdash {\begingroup\renewcommand\colorMATH{\colorMATHB}\renewcommand\colorSYNTAX{\colorSYNTAXB}{{\color{\colorMATH}\ensuremath{\se_{1}}}}\endgroup } \Downarrow ^{j_{1}} {\begingroup\renewcommand\colorMATH{\colorMATHB}\renewcommand\colorSYNTAX{\colorSYNTAXB}{{\color{\colorMATH}\ensuremath{\sv_{1 1}}}}\endgroup }}}}, \pthen {{\color{\colorMATH}\ensuremath{\gamma _{2}\vdash {\begingroup\renewcommand\colorMATH{\colorMATHB}\renewcommand\colorSYNTAX{\colorSYNTAXB}{{\color{\colorMATH}\ensuremath{\se_{1}}}}\endgroup } \Downarrow ^{\pj[1]} {\begingroup\renewcommand\colorMATH{\colorMATHB}\renewcommand\colorSYNTAX{\colorSYNTAXB}{{\color{\colorMATH}\ensuremath{\sv_{1 2}}}}\endgroup }}}} \pand 
        {{\color{\colorMATH}\ensuremath{({\begingroup\renewcommand\colorMATH{\colorMATHB}\renewcommand\colorSYNTAX{\colorSYNTAXB}{{\color{\colorMATH}\ensuremath{\sv_{1 1}}}}\endgroup }, {\begingroup\renewcommand\colorMATH{\colorMATHB}\renewcommand\colorSYNTAX{\colorSYNTAXB}{{\color{\colorMATH}\ensuremath{\sv_{1 2}}}}\endgroup }) \in  {\mathcal{V}}^{k-j_{1}}_{{\begingroup\renewcommand\colorMATH{\colorMATHB}\renewcommand\colorSYNTAX{\colorSYNTAXB}{{\color{\colorMATH}\ensuremath{\Distance'}}}\endgroup }\mathord{\cdotp }{\begingroup\renewcommand\colorMATH{\colorMATHB}\renewcommand\colorSYNTAX{\colorSYNTAXB}{{\color{\colorMATH}\ensuremath{\sS_{1}}}}\endgroup }}\llbracket {\begingroup\renewcommand\colorMATH{\colorMATHB}\renewcommand\colorSYNTAX{\colorSYNTAXB}{{\color{\colorMATH}\ensuremath{\Distance'}}}\endgroup }(\tau _{1 1}) \mathrel{^{{\begingroup\renewcommand\colorMATH{\colorMATHB}\renewcommand\colorSYNTAX{\colorSYNTAXB}{{\color{\colorMATH}\ensuremath{\Distance'}}}\endgroup }\mathord{\cdotp }{\begingroup\renewcommand\colorMATH{\colorMATHB}\renewcommand\colorSYNTAX{\colorSYNTAXB}{{\color{\colorMATH}\ensuremath{\sS_{1 1}}}}\endgroup }}\oplus ^{{\begingroup\renewcommand\colorMATH{\colorMATHB}\renewcommand\colorSYNTAX{\colorSYNTAXB}{{\color{\colorMATH}\ensuremath{\Distance'}}}\endgroup }\mathord{\cdotp }{\begingroup\renewcommand\colorMATH{\colorMATHB}\renewcommand\colorSYNTAX{\colorSYNTAXB}{{\color{\colorMATH}\ensuremath{\sS_{1 2}}}}\endgroup }}} {\begingroup\renewcommand\colorMATH{\colorMATHB}\renewcommand\colorSYNTAX{\colorSYNTAXB}{{\color{\colorMATH}\ensuremath{\Distance'}}}\endgroup }(\tau _{1 2})\rrbracket }}}.
        Either {{\color{\colorMATH}\ensuremath{{\begingroup\renewcommand\colorMATH{\colorMATHB}\renewcommand\colorSYNTAX{\colorSYNTAXB}{{\color{\colorMATH}\ensuremath{\sv_{1 1}}}}\endgroup } = \inl\hspace*{0.33em}{\begingroup\renewcommand\colorMATH{\colorMATHB}\renewcommand\colorSYNTAX{\colorSYNTAXB}{{\color{\colorMATH}\ensuremath{\sv'_{1 1}}}}\endgroup }}}} and {{\color{\colorMATH}\ensuremath{{\begingroup\renewcommand\colorMATH{\colorMATHB}\renewcommand\colorSYNTAX{\colorSYNTAXB}{{\color{\colorMATH}\ensuremath{\sv_{1 2}}}}\endgroup } = \inl\hspace*{0.33em}{\begingroup\renewcommand\colorMATH{\colorMATHB}\renewcommand\colorSYNTAX{\colorSYNTAXB}{{\color{\colorMATH}\ensuremath{\sv'_{1 2}}}}\endgroup }}}}, {{\color{\colorMATH}\ensuremath{{\begingroup\renewcommand\colorMATH{\colorMATHB}\renewcommand\colorSYNTAX{\colorSYNTAXB}{{\color{\colorMATH}\ensuremath{\sv_{1 1}}}}\endgroup } = \inr\hspace*{0.33em}{\begingroup\renewcommand\colorMATH{\colorMATHB}\renewcommand\colorSYNTAX{\colorSYNTAXB}{{\color{\colorMATH}\ensuremath{\sv'_{1 1}}}}\endgroup }}}} and {{\color{\colorMATH}\ensuremath{{\begingroup\renewcommand\colorMATH{\colorMATHB}\renewcommand\colorSYNTAX{\colorSYNTAXB}{{\color{\colorMATH}\ensuremath{\sv_{1 2}}}}\endgroup } = \inr\hspace*{0.33em}{\begingroup\renewcommand\colorMATH{\colorMATHB}\renewcommand\colorSYNTAX{\colorSYNTAXB}{{\color{\colorMATH}\ensuremath{\sv'_{1 2}}}}\endgroup }}}}, {{\color{\colorMATH}\ensuremath{{\begingroup\renewcommand\colorMATH{\colorMATHB}\renewcommand\colorSYNTAX{\colorSYNTAXB}{{\color{\colorMATH}\ensuremath{\sv_{1 1}}}}\endgroup } = \inl\hspace*{0.33em}{\begingroup\renewcommand\colorMATH{\colorMATHB}\renewcommand\colorSYNTAX{\colorSYNTAXB}{{\color{\colorMATH}\ensuremath{\sv'_{1 1}}}}\endgroup }}}} and {{\color{\colorMATH}\ensuremath{{\begingroup\renewcommand\colorMATH{\colorMATHB}\renewcommand\colorSYNTAX{\colorSYNTAXB}{{\color{\colorMATH}\ensuremath{\sv_{1 2}}}}\endgroup } = \inr\hspace*{0.33em}{\begingroup\renewcommand\colorMATH{\colorMATHB}\renewcommand\colorSYNTAX{\colorSYNTAXB}{{\color{\colorMATH}\ensuremath{\sv'_{1 2}}}}\endgroup }}}}, or {{\color{\colorMATH}\ensuremath{{\begingroup\renewcommand\colorMATH{\colorMATHB}\renewcommand\colorSYNTAX{\colorSYNTAXB}{{\color{\colorMATH}\ensuremath{\sv_{1 1}}}}\endgroup } = \inr\hspace*{0.33em}{\begingroup\renewcommand\colorMATH{\colorMATHB}\renewcommand\colorSYNTAX{\colorSYNTAXB}{{\color{\colorMATH}\ensuremath{\sv'_{1 1}}}}\endgroup }}}} and {{\color{\colorMATH}\ensuremath{{\begingroup\renewcommand\colorMATH{\colorMATHB}\renewcommand\colorSYNTAX{\colorSYNTAXB}{{\color{\colorMATH}\ensuremath{\sv_{1 2}}}}\endgroup } = \inl\hspace*{0.33em}{\begingroup\renewcommand\colorMATH{\colorMATHB}\renewcommand\colorSYNTAX{\colorSYNTAXB}{{\color{\colorMATH}\ensuremath{\sv'_{1 2}}}}\endgroup }}}}.
        We proceed by case analysis on {{\color{\colorMATH}\ensuremath{({\begingroup\renewcommand\colorMATH{\colorMATHB}\renewcommand\colorSYNTAX{\colorSYNTAXB}{{\color{\colorMATH}\ensuremath{\sv_{1 1}}}}\endgroup }, {\begingroup\renewcommand\colorMATH{\colorMATHB}\renewcommand\colorSYNTAX{\colorSYNTAXB}{{\color{\colorMATH}\ensuremath{\sv_{1 2}}}}\endgroup })}}}:

        \begin{enumerate}[ncases]\item  {{\color{\colorMATH}\ensuremath{({\begingroup\renewcommand\colorMATH{\colorMATHB}\renewcommand\colorSYNTAX{\colorSYNTAXB}{{\color{\colorMATH}\ensuremath{\sv_{1 1}}}}\endgroup }, {\begingroup\renewcommand\colorMATH{\colorMATHB}\renewcommand\colorSYNTAX{\colorSYNTAXB}{{\color{\colorMATH}\ensuremath{\sv_{1 2}}}}\endgroup }) = (\inl\hspace*{0.33em}{\begingroup\renewcommand\colorMATH{\colorMATHB}\renewcommand\colorSYNTAX{\colorSYNTAXB}{{\color{\colorMATH}\ensuremath{\sv'_{1 1}}}}\endgroup }, \inl\hspace*{0.33em}{\begingroup\renewcommand\colorMATH{\colorMATHB}\renewcommand\colorSYNTAX{\colorSYNTAXB}{{\color{\colorMATH}\ensuremath{\sv'_{1 2}}}}\endgroup })}}} (the case {{\color{\colorMATH}\ensuremath{({\begingroup\renewcommand\colorMATH{\colorMATHB}\renewcommand\colorSYNTAX{\colorSYNTAXB}{{\color{\colorMATH}\ensuremath{\sv_{1 1}}}}\endgroup }, {\begingroup\renewcommand\colorMATH{\colorMATHB}\renewcommand\colorSYNTAX{\colorSYNTAXB}{{\color{\colorMATH}\ensuremath{\sv_{1 2}}}}\endgroup }) = (\inr\hspace*{0.33em}{\begingroup\renewcommand\colorMATH{\colorMATHB}\renewcommand\colorSYNTAX{\colorSYNTAXB}{{\color{\colorMATH}\ensuremath{\sv'_{1 1}}}}\endgroup }, \inr\hspace*{0.33em}{\begingroup\renewcommand\colorMATH{\colorMATHB}\renewcommand\colorSYNTAX{\colorSYNTAXB}{{\color{\colorMATH}\ensuremath{\sv'_{1 2}}}}\endgroup })}}} is analogous)
          \begin{subproof} 
            By Lemma~\ref{lm:associativity-inst} {{\color{\colorMATH}\ensuremath{{\begingroup\renewcommand\colorMATH{\colorMATHB}\renewcommand\colorSYNTAX{\colorSYNTAXB}{{\color{\colorMATH}\ensuremath{\Distance'}}}\endgroup }\mathord{\cdotp }{\begingroup\renewcommand\colorMATH{\colorMATHB}\renewcommand\colorSYNTAX{\colorSYNTAXB}{{\color{\colorMATH}\ensuremath{\sS_{1}}}}\endgroup } + {\begingroup\renewcommand\colorMATH{\colorMATHB}\renewcommand\colorSYNTAX{\colorSYNTAXB}{{\color{\colorMATH}\ensuremath{\Distance'}}}\endgroup }\mathord{\cdotp }{\begingroup\renewcommand\colorMATH{\colorMATHB}\renewcommand\colorSYNTAX{\colorSYNTAXB}{{\color{\colorMATH}\ensuremath{\sS_{1 1}}}}\endgroup } = {\begingroup\renewcommand\colorMATH{\colorMATHB}\renewcommand\colorSYNTAX{\colorSYNTAXB}{{\color{\colorMATH}\ensuremath{\Distance'}}}\endgroup }\mathord{\cdotp }({\begingroup\renewcommand\colorMATH{\colorMATHB}\renewcommand\colorSYNTAX{\colorSYNTAXB}{{\color{\colorMATH}\ensuremath{\sS_{1}}}}\endgroup } + {\begingroup\renewcommand\colorMATH{\colorMATHB}\renewcommand\colorSYNTAX{\colorSYNTAXB}{{\color{\colorMATH}\ensuremath{\sS_{1 1}}}}\endgroup })}}}, then
            {{\color{\colorMATH}\ensuremath{({\begingroup\renewcommand\colorMATH{\colorMATHB}\renewcommand\colorSYNTAX{\colorSYNTAXB}{{\color{\colorMATH}\ensuremath{\sv'_{1 1}}}}\endgroup }, {\begingroup\renewcommand\colorMATH{\colorMATHB}\renewcommand\colorSYNTAX{\colorSYNTAXB}{{\color{\colorMATH}\ensuremath{\sv'_{1 2}}}}\endgroup }) \in  {\mathcal{V}}^{k-j_{1}}_{{\begingroup\renewcommand\colorMATH{\colorMATHB}\renewcommand\colorSYNTAX{\colorSYNTAXB}{{\color{\colorMATH}\ensuremath{\Distance'}}}\endgroup }\mathord{\cdotp }({\begingroup\renewcommand\colorMATH{\colorMATHB}\renewcommand\colorSYNTAX{\colorSYNTAXB}{{\color{\colorMATH}\ensuremath{\sS_{1}}}}\endgroup } + {\begingroup\renewcommand\colorMATH{\colorMATHB}\renewcommand\colorSYNTAX{\colorSYNTAXB}{{\color{\colorMATH}\ensuremath{\sS_{1 1}}}}\endgroup })}\llbracket {\begingroup\renewcommand\colorMATH{\colorMATHB}\renewcommand\colorSYNTAX{\colorSYNTAXB}{{\color{\colorMATH}\ensuremath{\Distance'}}}\endgroup }(\tau _{1 1})\rrbracket }}}.
            
            Also, by induction hypothesis on {{\color{\colorMATH}\ensuremath{\Gamma ,x\mathrel{:}\tau _{1 1}; {\begingroup\renewcommand\colorMATH{\colorMATHB}\renewcommand\colorSYNTAX{\colorSYNTAXB}{{\color{\colorMATH}\ensuremath{\Distance}}}\endgroup } + ({\begingroup\renewcommand\colorMATH{\colorMATHB}\renewcommand\colorSYNTAX{\colorSYNTAXB}{{\color{\colorMATH}\ensuremath{\Distance}}}\endgroup }\mathord{\cdotp }({\begingroup\renewcommand\colorMATH{\colorMATHB}\renewcommand\colorSYNTAX{\colorSYNTAXB}{{\color{\colorMATH}\ensuremath{\sS_{1}}}}\endgroup } + {\begingroup\renewcommand\colorMATH{\colorMATHB}\renewcommand\colorSYNTAX{\colorSYNTAXB}{{\color{\colorMATH}\ensuremath{\sS_{1 1}}}}\endgroup }))x \vdash  {\begingroup\renewcommand\colorMATH{\colorMATHB}\renewcommand\colorSYNTAX{\colorSYNTAXB}{{\color{\colorMATH}\ensuremath{\se_{2}}}}\endgroup } \mathrel{:} \tau _{2} \mathrel{;} {\begingroup\renewcommand\colorMATH{\colorMATHB}\renewcommand\colorSYNTAX{\colorSYNTAXB}{{\color{\colorMATH}\ensuremath{\sS_{2}}}}\endgroup }+{\begingroup\renewcommand\colorMATH{\colorMATHB}\renewcommand\colorSYNTAX{\colorSYNTAXB}{{\color{\colorMATH}\ensuremath{\sss_{2}}}}\endgroup }x}}}, by choosing {{\color{\colorMATH}\ensuremath{{\begingroup\renewcommand\colorMATH{\colorMATHB}\renewcommand\colorSYNTAX{\colorSYNTAXB}{{\color{\colorMATH}\ensuremath{\Distance'}}}\endgroup }+({\begingroup\renewcommand\colorMATH{\colorMATHB}\renewcommand\colorSYNTAX{\colorSYNTAXB}{{\color{\colorMATH}\ensuremath{\Distance'}}}\endgroup }\mathord{\cdotp }({\begingroup\renewcommand\colorMATH{\colorMATHB}\renewcommand\colorSYNTAX{\colorSYNTAXB}{{\color{\colorMATH}\ensuremath{\sS_{1}}}}\endgroup }+{\begingroup\renewcommand\colorMATH{\colorMATHB}\renewcommand\colorSYNTAX{\colorSYNTAXB}{{\color{\colorMATH}\ensuremath{\sS_{1 1}}}}\endgroup }))x \sqsubseteq  {\begingroup\renewcommand\colorMATH{\colorMATHB}\renewcommand\colorSYNTAX{\colorSYNTAXB}{{\color{\colorMATH}\ensuremath{\Distance}}}\endgroup }+({\begingroup\renewcommand\colorMATH{\colorMATHB}\renewcommand\colorSYNTAX{\colorSYNTAXB}{{\color{\colorMATH}\ensuremath{\Distance}}}\endgroup }\mathord{\cdotp }({\begingroup\renewcommand\colorMATH{\colorMATHB}\renewcommand\colorSYNTAX{\colorSYNTAXB}{{\color{\colorMATH}\ensuremath{\sS_{1}}}}\endgroup }+{\begingroup\renewcommand\colorMATH{\colorMATHB}\renewcommand\colorSYNTAX{\colorSYNTAXB}{{\color{\colorMATH}\ensuremath{\sS_{1 1}}}}\endgroup }))x}}} and {{\color{\colorMATH}\ensuremath{k = k-j_{1}}}} (and by weakening lemma ~\ref{lm:weakening-index})
            \\
            {{\color{\colorMATH}\ensuremath{(\gamma _{1}[x \mapsto  {\begingroup\renewcommand\colorMATH{\colorMATHB}\renewcommand\colorSYNTAX{\colorSYNTAXB}{{\color{\colorMATH}\ensuremath{\sv'_{1 2}}}}\endgroup }],\gamma _{2}[x \mapsto  {\begingroup\renewcommand\colorMATH{\colorMATHB}\renewcommand\colorSYNTAX{\colorSYNTAXB}{{\color{\colorMATH}\ensuremath{\sv'_{2 2}}}}\endgroup }]) \in  {\mathcal{G}}^{k-j_{1}-1}_{{\begingroup\renewcommand\colorMATH{\colorMATHB}\renewcommand\colorSYNTAX{\colorSYNTAXB}{{\color{\colorMATH}\ensuremath{\Distance'}}}\endgroup }+({\begingroup\renewcommand\colorMATH{\colorMATHB}\renewcommand\colorSYNTAX{\colorSYNTAXB}{{\color{\colorMATH}\ensuremath{\Distance'}}}\endgroup }\mathord{\cdotp }({\begingroup\renewcommand\colorMATH{\colorMATHB}\renewcommand\colorSYNTAX{\colorSYNTAXB}{{\color{\colorMATH}\ensuremath{\sS_{1}}}}\endgroup }+{\begingroup\renewcommand\colorMATH{\colorMATHB}\renewcommand\colorSYNTAX{\colorSYNTAXB}{{\color{\colorMATH}\ensuremath{\sS_{1 1}}}}\endgroup }))x}\llbracket \Gamma ,x\mathrel{:}\tau _{1 1}\rrbracket }}} (note that {{\color{\colorMATH}\ensuremath{x \notin  dom({\begingroup\renewcommand\colorMATH{\colorMATHB}\renewcommand\colorSYNTAX{\colorSYNTAXB}{{\color{\colorMATH}\ensuremath{\sS_{1}}}}\endgroup }) \cup  dom({\begingroup\renewcommand\colorMATH{\colorMATHB}\renewcommand\colorSYNTAX{\colorSYNTAXB}{{\color{\colorMATH}\ensuremath{\sS_{1 1}}}}\endgroup })}}}, therefore {{\color{\colorMATH}\ensuremath{({\begingroup\renewcommand\colorMATH{\colorMATHB}\renewcommand\colorSYNTAX{\colorSYNTAXB}{{\color{\colorMATH}\ensuremath{\Distance'}}}\endgroup }+({\begingroup\renewcommand\colorMATH{\colorMATHB}\renewcommand\colorSYNTAX{\colorSYNTAXB}{{\color{\colorMATH}\ensuremath{\Distance'}}}\endgroup }\mathord{\cdotp }({\begingroup\renewcommand\colorMATH{\colorMATHB}\renewcommand\colorSYNTAX{\colorSYNTAXB}{{\color{\colorMATH}\ensuremath{\sS_{1}}}}\endgroup }+{\begingroup\renewcommand\colorMATH{\colorMATHB}\renewcommand\colorSYNTAX{\colorSYNTAXB}{{\color{\colorMATH}\ensuremath{\sS_{1 1}}}}\endgroup }))x)(\tau _{1 1}) = {\begingroup\renewcommand\colorMATH{\colorMATHB}\renewcommand\colorSYNTAX{\colorSYNTAXB}{{\color{\colorMATH}\ensuremath{\Distance'}}}\endgroup }(\tau _{1 1})}}}) we know that\\
            {{\color{\colorMATH}\ensuremath{(\gamma _{1}[x \mapsto  {\begingroup\renewcommand\colorMATH{\colorMATHB}\renewcommand\colorSYNTAX{\colorSYNTAXB}{{\color{\colorMATH}\ensuremath{\sv'_{1 2}}}}\endgroup }]\vdash {\begingroup\renewcommand\colorMATH{\colorMATHB}\renewcommand\colorSYNTAX{\colorSYNTAXB}{{\color{\colorMATH}\ensuremath{\se_{2}}}}\endgroup },\gamma _{2}[x \mapsto  {\begingroup\renewcommand\colorMATH{\colorMATHB}\renewcommand\colorSYNTAX{\colorSYNTAXB}{{\color{\colorMATH}\ensuremath{\sv'_{2 2}}}}\endgroup }]\vdash {\begingroup\renewcommand\colorMATH{\colorMATHB}\renewcommand\colorSYNTAX{\colorSYNTAXB}{{\color{\colorMATH}\ensuremath{\se_{2}}}}\endgroup }) \in  {\mathcal{E}}^{k-j_{1}}_{({\begingroup\renewcommand\colorMATH{\colorMATHB}\renewcommand\colorSYNTAX{\colorSYNTAXB}{{\color{\colorMATH}\ensuremath{\Distance'}}}\endgroup }+({\begingroup\renewcommand\colorMATH{\colorMATHB}\renewcommand\colorSYNTAX{\colorSYNTAXB}{{\color{\colorMATH}\ensuremath{\Distance'}}}\endgroup }\mathord{\cdotp }({\begingroup\renewcommand\colorMATH{\colorMATHB}\renewcommand\colorSYNTAX{\colorSYNTAXB}{{\color{\colorMATH}\ensuremath{\sS_{1}}}}\endgroup }+{\begingroup\renewcommand\colorMATH{\colorMATHB}\renewcommand\colorSYNTAX{\colorSYNTAXB}{{\color{\colorMATH}\ensuremath{\sS_{1 1}}}}\endgroup }))x)\mathord{\cdotp }({\begingroup\renewcommand\colorMATH{\colorMATHB}\renewcommand\colorSYNTAX{\colorSYNTAXB}{{\color{\colorMATH}\ensuremath{\sS_{2}}}}\endgroup }+{\begingroup\renewcommand\colorMATH{\colorMATHB}\renewcommand\colorSYNTAX{\colorSYNTAXB}{{\color{\colorMATH}\ensuremath{\sss_{2}}}}\endgroup }x)}\llbracket ({\begingroup\renewcommand\colorMATH{\colorMATHB}\renewcommand\colorSYNTAX{\colorSYNTAXB}{{\color{\colorMATH}\ensuremath{\Distance'}}}\endgroup }+({\begingroup\renewcommand\colorMATH{\colorMATHB}\renewcommand\colorSYNTAX{\colorSYNTAXB}{{\color{\colorMATH}\ensuremath{\Distance'}}}\endgroup }\mathord{\cdotp }({\begingroup\renewcommand\colorMATH{\colorMATHB}\renewcommand\colorSYNTAX{\colorSYNTAXB}{{\color{\colorMATH}\ensuremath{\sS_{1}}}}\endgroup }+{\begingroup\renewcommand\colorMATH{\colorMATHB}\renewcommand\colorSYNTAX{\colorSYNTAXB}{{\color{\colorMATH}\ensuremath{\sS_{1 1}}}}\endgroup }))x)(\tau _{2})\rrbracket }}}.\\
            But {{\color{\colorMATH}\ensuremath{({\begingroup\renewcommand\colorMATH{\colorMATHB}\renewcommand\colorSYNTAX{\colorSYNTAXB}{{\color{\colorMATH}\ensuremath{\Distance'}}}\endgroup }+({\begingroup\renewcommand\colorMATH{\colorMATHB}\renewcommand\colorSYNTAX{\colorSYNTAXB}{{\color{\colorMATH}\ensuremath{\Distance'}}}\endgroup }\mathord{\cdotp }({\begingroup\renewcommand\colorMATH{\colorMATHB}\renewcommand\colorSYNTAX{\colorSYNTAXB}{{\color{\colorMATH}\ensuremath{\sS_{1}}}}\endgroup }+{\begingroup\renewcommand\colorMATH{\colorMATHB}\renewcommand\colorSYNTAX{\colorSYNTAXB}{{\color{\colorMATH}\ensuremath{\sS_{1 1}}}}\endgroup }))x)\mathord{\cdotp }({\begingroup\renewcommand\colorMATH{\colorMATHB}\renewcommand\colorSYNTAX{\colorSYNTAXB}{{\color{\colorMATH}\ensuremath{\sS_{2}}}}\endgroup }+{\begingroup\renewcommand\colorMATH{\colorMATHB}\renewcommand\colorSYNTAX{\colorSYNTAXB}{{\color{\colorMATH}\ensuremath{\sss_{2}}}}\endgroup }x) = {\begingroup\renewcommand\colorMATH{\colorMATHB}\renewcommand\colorSYNTAX{\colorSYNTAXB}{{\color{\colorMATH}\ensuremath{\Distance'}}}\endgroup }\mathord{\cdotp }{\begingroup\renewcommand\colorMATH{\colorMATHB}\renewcommand\colorSYNTAX{\colorSYNTAXB}{{\color{\colorMATH}\ensuremath{\sS_{2}}}}\endgroup } + {\begingroup\renewcommand\colorMATH{\colorMATHB}\renewcommand\colorSYNTAX{\colorSYNTAXB}{{\color{\colorMATH}\ensuremath{\sss_{2}}}}\endgroup }({\begingroup\renewcommand\colorMATH{\colorMATHB}\renewcommand\colorSYNTAX{\colorSYNTAXB}{{\color{\colorMATH}\ensuremath{\Distance'}}}\endgroup }\mathord{\cdotp }({\begingroup\renewcommand\colorMATH{\colorMATHB}\renewcommand\colorSYNTAX{\colorSYNTAXB}{{\color{\colorMATH}\ensuremath{\sS_{1}}}}\endgroup } + {\begingroup\renewcommand\colorMATH{\colorMATHB}\renewcommand\colorSYNTAX{\colorSYNTAXB}{{\color{\colorMATH}\ensuremath{\sS_{1 1}}}}\endgroup })) = {\begingroup\renewcommand\colorMATH{\colorMATHB}\renewcommand\colorSYNTAX{\colorSYNTAXB}{{\color{\colorMATH}\ensuremath{\Distance'}}}\endgroup }\mathord{\cdotp }({\begingroup\renewcommand\colorMATH{\colorMATHB}\renewcommand\colorSYNTAX{\colorSYNTAXB}{{\color{\colorMATH}\ensuremath{\sss_{2}}}}\endgroup }({\begingroup\renewcommand\colorMATH{\colorMATHB}\renewcommand\colorSYNTAX{\colorSYNTAXB}{{\color{\colorMATH}\ensuremath{\sS_{1}}}}\endgroup } + {\begingroup\renewcommand\colorMATH{\colorMATHB}\renewcommand\colorSYNTAX{\colorSYNTAXB}{{\color{\colorMATH}\ensuremath{\sS_{1 1}}}}\endgroup }) + {\begingroup\renewcommand\colorMATH{\colorMATHB}\renewcommand\colorSYNTAX{\colorSYNTAXB}{{\color{\colorMATH}\ensuremath{\sS_{2}}}}\endgroup })}}}, and
            by Lemma~\ref{lm:equivsimplsubst} and because {{\color{\colorMATH}\ensuremath{{\begingroup\renewcommand\colorMATH{\colorMATHB}\renewcommand\colorSYNTAX{\colorSYNTAXB}{{\color{\colorMATH}\ensuremath{\Distance'}}}\endgroup }\mathord{\cdotp }({\begingroup\renewcommand\colorMATH{\colorMATHB}\renewcommand\colorSYNTAX{\colorSYNTAXB}{{\color{\colorMATH}\ensuremath{\sS_{1}}}}\endgroup }+{\begingroup\renewcommand\colorMATH{\colorMATHB}\renewcommand\colorSYNTAX{\colorSYNTAXB}{{\color{\colorMATH}\ensuremath{\sS_{1 1}}}}\endgroup }) \in  {\text{sens}}}}}, then 
            {{\color{\colorMATH}\ensuremath{({\begingroup\renewcommand\colorMATH{\colorMATHB}\renewcommand\colorSYNTAX{\colorSYNTAXB}{{\color{\colorMATH}\ensuremath{\Distance'}}}\endgroup }+({\begingroup\renewcommand\colorMATH{\colorMATHB}\renewcommand\colorSYNTAX{\colorSYNTAXB}{{\color{\colorMATH}\ensuremath{\Distance'}}}\endgroup }\mathord{\cdotp }({\begingroup\renewcommand\colorMATH{\colorMATHB}\renewcommand\colorSYNTAX{\colorSYNTAXB}{{\color{\colorMATH}\ensuremath{\sS_{1}}}}\endgroup }+{\begingroup\renewcommand\colorMATH{\colorMATHB}\renewcommand\colorSYNTAX{\colorSYNTAXB}{{\color{\colorMATH}\ensuremath{\sS_{1 1}}}}\endgroup }))x)(\tau _{2}) = ({\begingroup\renewcommand\colorMATH{\colorMATHB}\renewcommand\colorSYNTAX{\colorSYNTAXB}{{\color{\colorMATH}\ensuremath{\Distance'}}}\endgroup }\mathord{\cdotp }({\begingroup\renewcommand\colorMATH{\colorMATHB}\renewcommand\colorSYNTAX{\colorSYNTAXB}{{\color{\colorMATH}\ensuremath{\sS_{1}}}}\endgroup }+{\begingroup\renewcommand\colorMATH{\colorMATHB}\renewcommand\colorSYNTAX{\colorSYNTAXB}{{\color{\colorMATH}\ensuremath{\sS_{1 1}}}}\endgroup }))x({\begingroup\renewcommand\colorMATH{\colorMATHB}\renewcommand\colorSYNTAX{\colorSYNTAXB}{{\color{\colorMATH}\ensuremath{\Distance'}}}\endgroup }(\tau _{2}))}}}, and by Lemma~\ref{lm:distrinst} 
            {{\color{\colorMATH}\ensuremath{({\begingroup\renewcommand\colorMATH{\colorMATHB}\renewcommand\colorSYNTAX{\colorSYNTAXB}{{\color{\colorMATH}\ensuremath{\Distance'}}}\endgroup }\mathord{\cdotp }({\begingroup\renewcommand\colorMATH{\colorMATHB}\renewcommand\colorSYNTAX{\colorSYNTAXB}{{\color{\colorMATH}\ensuremath{\sS_{1}}}}\endgroup }+{\begingroup\renewcommand\colorMATH{\colorMATHB}\renewcommand\colorSYNTAX{\colorSYNTAXB}{{\color{\colorMATH}\ensuremath{\sS_{1 1}}}}\endgroup }))x({\begingroup\renewcommand\colorMATH{\colorMATHB}\renewcommand\colorSYNTAX{\colorSYNTAXB}{{\color{\colorMATH}\ensuremath{\Distance'}}}\endgroup }(\tau _{2})) = {\begingroup\renewcommand\colorMATH{\colorMATHB}\renewcommand\colorSYNTAX{\colorSYNTAXB}{{\color{\colorMATH}\ensuremath{\Distance'}}}\endgroup }([{\begingroup\renewcommand\colorMATH{\colorMATHB}\renewcommand\colorSYNTAX{\colorSYNTAXB}{{\color{\colorMATH}\ensuremath{\sS_{1}}}}\endgroup }+{\begingroup\renewcommand\colorMATH{\colorMATHB}\renewcommand\colorSYNTAX{\colorSYNTAXB}{{\color{\colorMATH}\ensuremath{\sS_{1 1}}}}\endgroup }/x]\tau _{2})}}}.
            Therefore if {{\color{\colorMATH}\ensuremath{\gamma _{1}[x \mapsto  {\begingroup\renewcommand\colorMATH{\colorMATHB}\renewcommand\colorSYNTAX{\colorSYNTAXB}{{\color{\colorMATH}\ensuremath{\sv'_{1 2}}}}\endgroup }]\vdash {\begingroup\renewcommand\colorMATH{\colorMATHB}\renewcommand\colorSYNTAX{\colorSYNTAXB}{{\color{\colorMATH}\ensuremath{\se_{2}}}}\endgroup } \Downarrow ^{j_{2}} {\begingroup\renewcommand\colorMATH{\colorMATHB}\renewcommand\colorSYNTAX{\colorSYNTAXB}{{\color{\colorMATH}\ensuremath{\sv_{2 1}}}}\endgroup }}}}, \pthen {{\color{\colorMATH}\ensuremath{\gamma _{2}[x \mapsto  {\begingroup\renewcommand\colorMATH{\colorMATHB}\renewcommand\colorSYNTAX{\colorSYNTAXB}{{\color{\colorMATH}\ensuremath{\sv'_{2 2}}}}\endgroup }]\vdash {\begingroup\renewcommand\colorMATH{\colorMATHB}\renewcommand\colorSYNTAX{\colorSYNTAXB}{{\color{\colorMATH}\ensuremath{\se_{2}}}}\endgroup } \Downarrow ^{\pj[2]} {\begingroup\renewcommand\colorMATH{\colorMATHB}\renewcommand\colorSYNTAX{\colorSYNTAXB}{{\color{\colorMATH}\ensuremath{\sv_{2 2}}}}\endgroup }}}}, \pand\\
            {{\color{\colorMATH}\ensuremath{({\begingroup\renewcommand\colorMATH{\colorMATHB}\renewcommand\colorSYNTAX{\colorSYNTAXB}{{\color{\colorMATH}\ensuremath{\sv_{2 1}}}}\endgroup }, {\begingroup\renewcommand\colorMATH{\colorMATHB}\renewcommand\colorSYNTAX{\colorSYNTAXB}{{\color{\colorMATH}\ensuremath{\sv_{2 2}}}}\endgroup }) \in  {\mathcal{V}}^{k-j_{1}-j_{2}}_{{\begingroup\renewcommand\colorMATH{\colorMATHB}\renewcommand\colorSYNTAX{\colorSYNTAXB}{{\color{\colorMATH}\ensuremath{\Distance'}}}\endgroup }\mathord{\cdotp }({\begingroup\renewcommand\colorMATH{\colorMATHB}\renewcommand\colorSYNTAX{\colorSYNTAXB}{{\color{\colorMATH}\ensuremath{\sss_{2}}}}\endgroup }{\begingroup\renewcommand\colorMATH{\colorMATHB}\renewcommand\colorSYNTAX{\colorSYNTAXB}{{\color{\colorMATH}\ensuremath{\sS_{1}}}}\endgroup } + {\begingroup\renewcommand\colorMATH{\colorMATHB}\renewcommand\colorSYNTAX{\colorSYNTAXB}{{\color{\colorMATH}\ensuremath{\sss_{2}}}}\endgroup }{\begingroup\renewcommand\colorMATH{\colorMATHB}\renewcommand\colorSYNTAX{\colorSYNTAXB}{{\color{\colorMATH}\ensuremath{\sS_{1 1}}}}\endgroup } + {\begingroup\renewcommand\colorMATH{\colorMATHB}\renewcommand\colorSYNTAX{\colorSYNTAXB}{{\color{\colorMATH}\ensuremath{\sS_{2}}}}\endgroup })}\llbracket {\begingroup\renewcommand\colorMATH{\colorMATHB}\renewcommand\colorSYNTAX{\colorSYNTAXB}{{\color{\colorMATH}\ensuremath{\Distance'}}}\endgroup }([{\begingroup\renewcommand\colorMATH{\colorMATHB}\renewcommand\colorSYNTAX{\colorSYNTAXB}{{\color{\colorMATH}\ensuremath{\sS_{1}}}}\endgroup }+{\begingroup\renewcommand\colorMATH{\colorMATHB}\renewcommand\colorSYNTAX{\colorSYNTAXB}{{\color{\colorMATH}\ensuremath{\sS_{1 1}}}}\endgroup }/x]\tau _{2})\rrbracket }}}.

            By following the {\textsc{ case-left}} reduction rule:
            \begingroup\color{\colorMATH}\begin{gather*} 
              \inferrule*[lab=
              ]{ \gamma _{1}\vdash {\begingroup\renewcommand\colorMATH{\colorMATHB}\renewcommand\colorSYNTAX{\colorSYNTAXB}{{\color{\colorMATH}\ensuremath{\se_{1}}}}\endgroup } \Downarrow ^{j_{1}} \inl\hspace*{0.33em}{\begingroup\renewcommand\colorMATH{\colorMATHB}\renewcommand\colorSYNTAX{\colorSYNTAXB}{{\color{\colorMATH}\ensuremath{\sv'_{1 1}}}}\endgroup }
              \\ \gamma _{1}[x\mapsto {\begingroup\renewcommand\colorMATH{\colorMATHB}\renewcommand\colorSYNTAX{\colorSYNTAXB}{{\color{\colorMATH}\ensuremath{\sv'_{1 1}}}}\endgroup }]\vdash {\begingroup\renewcommand\colorMATH{\colorMATHB}\renewcommand\colorSYNTAX{\colorSYNTAXB}{{\color{\colorMATH}\ensuremath{\se_{2}}}}\endgroup } \Downarrow ^{j_{2}} {\begingroup\renewcommand\colorMATH{\colorMATHB}\renewcommand\colorSYNTAX{\colorSYNTAXB}{{\color{\colorMATH}\ensuremath{\sv_{2 1}}}}\endgroup }
                }{
                \gamma _{1}\vdash \ccase\hspace*{0.33em}{\begingroup\renewcommand\colorMATH{\colorMATHB}\renewcommand\colorSYNTAX{\colorSYNTAXB}{{\color{\colorMATH}\ensuremath{\se_{1}}}}\endgroup }\hspace*{0.33em}\of\hspace*{0.33em}\{ x\Rightarrow {\begingroup\renewcommand\colorMATH{\colorMATHB}\renewcommand\colorSYNTAX{\colorSYNTAXB}{{\color{\colorMATH}\ensuremath{\se_{2}}}}\endgroup }\} \hspace*{0.33em}\{ x\Rightarrow e_{3}\}  \Downarrow ^{j_{1}+j_{2}+1} {\begingroup\renewcommand\colorMATH{\colorMATHB}\renewcommand\colorSYNTAX{\colorSYNTAXB}{{\color{\colorMATH}\ensuremath{\sv_{2 1}}}}\endgroup }
              }
            \end{gather*}\endgroup
            and
            \begingroup\color{\colorMATH}\begin{gather*} 
              \inferrule*[lab=
              ]{ \gamma _{2}\vdash {\begingroup\renewcommand\colorMATH{\colorMATHB}\renewcommand\colorSYNTAX{\colorSYNTAXB}{{\color{\colorMATH}\ensuremath{\se_{1}}}}\endgroup } \Downarrow ^{*} \inl\hspace*{0.33em}{\begingroup\renewcommand\colorMATH{\colorMATHB}\renewcommand\colorSYNTAX{\colorSYNTAXB}{{\color{\colorMATH}\ensuremath{\sv'_{1 2}}}}\endgroup }
              \\ \gamma _{2}[x\mapsto {\begingroup\renewcommand\colorMATH{\colorMATHB}\renewcommand\colorSYNTAX{\colorSYNTAXB}{{\color{\colorMATH}\ensuremath{\sv'_{1 2}}}}\endgroup }]\vdash {\begingroup\renewcommand\colorMATH{\colorMATHB}\renewcommand\colorSYNTAX{\colorSYNTAXB}{{\color{\colorMATH}\ensuremath{\se_{2}}}}\endgroup } \Downarrow ^{*} {\begingroup\renewcommand\colorMATH{\colorMATHB}\renewcommand\colorSYNTAX{\colorSYNTAXB}{{\color{\colorMATH}\ensuremath{\sv_{2 2}}}}\endgroup }
                }{
                \gamma _{2}\vdash \ccase\hspace*{0.33em}{\begingroup\renewcommand\colorMATH{\colorMATHB}\renewcommand\colorSYNTAX{\colorSYNTAXB}{{\color{\colorMATH}\ensuremath{\se_{1}}}}\endgroup }\hspace*{0.33em}\of\hspace*{0.33em}\{ x\Rightarrow {\begingroup\renewcommand\colorMATH{\colorMATHB}\renewcommand\colorSYNTAX{\colorSYNTAXB}{{\color{\colorMATH}\ensuremath{\se_{2}}}}\endgroup }\} \hspace*{0.33em}\{ x\Rightarrow e_{3}\}  \Downarrow ^{*} {\begingroup\renewcommand\colorMATH{\colorMATHB}\renewcommand\colorSYNTAX{\colorSYNTAXB}{{\color{\colorMATH}\ensuremath{\sv_{2 2}}}}\endgroup }
              }
            \end{gather*}\endgroup
            for {{\color{\colorMATH}\ensuremath{i \in  \{ 1,2\} }}}. Then we just have to prove that\\
            {{\color{\colorMATH}\ensuremath{({\begingroup\renewcommand\colorMATH{\colorMATHB}\renewcommand\colorSYNTAX{\colorSYNTAXB}{{\color{\colorMATH}\ensuremath{\sv_{2 1}}}}\endgroup }, {\begingroup\renewcommand\colorMATH{\colorMATHB}\renewcommand\colorSYNTAX{\colorSYNTAXB}{{\color{\colorMATH}\ensuremath{\sv_{2 2}}}}\endgroup }) \in  {\mathcal{V}}^{k-j_{1}-j_{2}-1}_{{\begingroup\renewcommand\colorMATH{\colorMATHB}\renewcommand\colorSYNTAX{\colorSYNTAXB}{{\color{\colorMATH}\ensuremath{\Distance'}}}\endgroup }\mathord{\cdotp } ({\begingroup\renewcommand\colorMATH{\colorMATHB}\renewcommand\colorSYNTAX{\colorSYNTAXB}{{\color{\colorMATH}\ensuremath{\sS_{1}}}}\endgroup } \sqcup  ({\begingroup\renewcommand\colorMATH{\colorMATHB}\renewcommand\colorSYNTAX{\colorSYNTAXB}{{\color{\colorMATH}\ensuremath{\sss_{2}}}}\endgroup }{\begingroup\renewcommand\colorMATH{\colorMATHB}\renewcommand\colorSYNTAX{\colorSYNTAXB}{{\color{\colorMATH}\ensuremath{\sS_{1}}}}\endgroup } + {\begingroup\renewcommand\colorMATH{\colorMATHB}\renewcommand\colorSYNTAX{\colorSYNTAXB}{{\color{\colorMATH}\ensuremath{\sss_{2}}}}\endgroup }{\begingroup\renewcommand\colorMATH{\colorMATHB}\renewcommand\colorSYNTAX{\colorSYNTAXB}{{\color{\colorMATH}\ensuremath{\sS_{1 1}}}}\endgroup } + {\begingroup\renewcommand\colorMATH{\colorMATHB}\renewcommand\colorSYNTAX{\colorSYNTAXB}{{\color{\colorMATH}\ensuremath{\sS_{2}}}}\endgroup }) \sqcup  ({\begingroup\renewcommand\colorMATH{\colorMATHB}\renewcommand\colorSYNTAX{\colorSYNTAXB}{{\color{\colorMATH}\ensuremath{\sss_{3}}}}\endgroup }{\begingroup\renewcommand\colorMATH{\colorMATHB}\renewcommand\colorSYNTAX{\colorSYNTAXB}{{\color{\colorMATH}\ensuremath{\sS_{1}}}}\endgroup } + {\begingroup\renewcommand\colorMATH{\colorMATHB}\renewcommand\colorSYNTAX{\colorSYNTAXB}{{\color{\colorMATH}\ensuremath{\sss_{3}}}}\endgroup }{\begingroup\renewcommand\colorMATH{\colorMATHB}\renewcommand\colorSYNTAX{\colorSYNTAXB}{{\color{\colorMATH}\ensuremath{\sS_{1 2}}}}\endgroup } + {\begingroup\renewcommand\colorMATH{\colorMATHB}\renewcommand\colorSYNTAX{\colorSYNTAXB}{{\color{\colorMATH}\ensuremath{\sS_{3}}}}\endgroup }))}\llbracket {\begingroup\renewcommand\colorMATH{\colorMATHB}\renewcommand\colorSYNTAX{\colorSYNTAXB}{{\color{\colorMATH}\ensuremath{\Distance'}}}\endgroup }([{\begingroup\renewcommand\colorMATH{\colorMATHB}\renewcommand\colorSYNTAX{\colorSYNTAXB}{{\color{\colorMATH}\ensuremath{\sS_{1}}}}\endgroup } + {\begingroup\renewcommand\colorMATH{\colorMATHB}\renewcommand\colorSYNTAX{\colorSYNTAXB}{{\color{\colorMATH}\ensuremath{\sS_{1 1}}}}\endgroup }/x]\tau _{2} \sqcup  [{\begingroup\renewcommand\colorMATH{\colorMATHB}\renewcommand\colorSYNTAX{\colorSYNTAXB}{{\color{\colorMATH}\ensuremath{\sS_{1}}}}\endgroup } + {\begingroup\renewcommand\colorMATH{\colorMATHB}\renewcommand\colorSYNTAX{\colorSYNTAXB}{{\color{\colorMATH}\ensuremath{\sS_{1 2}}}}\endgroup }/y]\tau _{3})\rrbracket }}}.
            Notice that by definition of the join operator {{\color{\colorMATH}\ensuremath{({\begingroup\renewcommand\colorMATH{\colorMATHB}\renewcommand\colorSYNTAX{\colorSYNTAXB}{{\color{\colorMATH}\ensuremath{\sss_{2}}}}\endgroup }{\begingroup\renewcommand\colorMATH{\colorMATHB}\renewcommand\colorSYNTAX{\colorSYNTAXB}{{\color{\colorMATH}\ensuremath{\sS_{1}}}}\endgroup } + {\begingroup\renewcommand\colorMATH{\colorMATHB}\renewcommand\colorSYNTAX{\colorSYNTAXB}{{\color{\colorMATH}\ensuremath{\sss_{2}}}}\endgroup }{\begingroup\renewcommand\colorMATH{\colorMATHB}\renewcommand\colorSYNTAX{\colorSYNTAXB}{{\color{\colorMATH}\ensuremath{\sS_{1 1}}}}\endgroup } + {\begingroup\renewcommand\colorMATH{\colorMATHB}\renewcommand\colorSYNTAX{\colorSYNTAXB}{{\color{\colorMATH}\ensuremath{\sS_{2}}}}\endgroup }) <: ({\begingroup\renewcommand\colorMATH{\colorMATHB}\renewcommand\colorSYNTAX{\colorSYNTAXB}{{\color{\colorMATH}\ensuremath{\sss_{2}}}}\endgroup }{\begingroup\renewcommand\colorMATH{\colorMATHB}\renewcommand\colorSYNTAX{\colorSYNTAXB}{{\color{\colorMATH}\ensuremath{\sS_{1}}}}\endgroup } + {\begingroup\renewcommand\colorMATH{\colorMATHB}\renewcommand\colorSYNTAX{\colorSYNTAXB}{{\color{\colorMATH}\ensuremath{\sss_{2}}}}\endgroup }{\begingroup\renewcommand\colorMATH{\colorMATHB}\renewcommand\colorSYNTAX{\colorSYNTAXB}{{\color{\colorMATH}\ensuremath{\sS_{1 1}}}}\endgroup } + {\begingroup\renewcommand\colorMATH{\colorMATHB}\renewcommand\colorSYNTAX{\colorSYNTAXB}{{\color{\colorMATH}\ensuremath{\sS_{2}}}}\endgroup }) \sqcup  ({\begingroup\renewcommand\colorMATH{\colorMATHB}\renewcommand\colorSYNTAX{\colorSYNTAXB}{{\color{\colorMATH}\ensuremath{\sss_{3}}}}\endgroup }{\begingroup\renewcommand\colorMATH{\colorMATHB}\renewcommand\colorSYNTAX{\colorSYNTAXB}{{\color{\colorMATH}\ensuremath{\sS_{1}}}}\endgroup } + {\begingroup\renewcommand\colorMATH{\colorMATHB}\renewcommand\colorSYNTAX{\colorSYNTAXB}{{\color{\colorMATH}\ensuremath{\sss_{3}}}}\endgroup }{\begingroup\renewcommand\colorMATH{\colorMATHB}\renewcommand\colorSYNTAX{\colorSYNTAXB}{{\color{\colorMATH}\ensuremath{\sS_{1 2}}}}\endgroup } + {\begingroup\renewcommand\colorMATH{\colorMATHB}\renewcommand\colorSYNTAX{\colorSYNTAXB}{{\color{\colorMATH}\ensuremath{\sS_{3}}}}\endgroup })}}}, 
            also {{\color{\colorMATH}\ensuremath{({\begingroup\renewcommand\colorMATH{\colorMATHB}\renewcommand\colorSYNTAX{\colorSYNTAXB}{{\color{\colorMATH}\ensuremath{\sss_{2}}}}\endgroup }{\begingroup\renewcommand\colorMATH{\colorMATHB}\renewcommand\colorSYNTAX{\colorSYNTAXB}{{\color{\colorMATH}\ensuremath{\sS_{1}}}}\endgroup } + {\begingroup\renewcommand\colorMATH{\colorMATHB}\renewcommand\colorSYNTAX{\colorSYNTAXB}{{\color{\colorMATH}\ensuremath{\sss_{2}}}}\endgroup }{\begingroup\renewcommand\colorMATH{\colorMATHB}\renewcommand\colorSYNTAX{\colorSYNTAXB}{{\color{\colorMATH}\ensuremath{\sS_{1 1}}}}\endgroup } + {\begingroup\renewcommand\colorMATH{\colorMATHB}\renewcommand\colorSYNTAX{\colorSYNTAXB}{{\color{\colorMATH}\ensuremath{\sS_{2}}}}\endgroup }) \sqcup  ({\begingroup\renewcommand\colorMATH{\colorMATHB}\renewcommand\colorSYNTAX{\colorSYNTAXB}{{\color{\colorMATH}\ensuremath{\sss_{3}}}}\endgroup }{\begingroup\renewcommand\colorMATH{\colorMATHB}\renewcommand\colorSYNTAX{\colorSYNTAXB}{{\color{\colorMATH}\ensuremath{\sS_{1}}}}\endgroup } + {\begingroup\renewcommand\colorMATH{\colorMATHB}\renewcommand\colorSYNTAX{\colorSYNTAXB}{{\color{\colorMATH}\ensuremath{\sss_{3}}}}\endgroup }{\begingroup\renewcommand\colorMATH{\colorMATHB}\renewcommand\colorSYNTAX{\colorSYNTAXB}{{\color{\colorMATH}\ensuremath{\sS_{1 2}}}}\endgroup } + {\begingroup\renewcommand\colorMATH{\colorMATHB}\renewcommand\colorSYNTAX{\colorSYNTAXB}{{\color{\colorMATH}\ensuremath{\sS_{3}}}}\endgroup }) <: {\begingroup\renewcommand\colorMATH{\colorMATHB}\renewcommand\colorSYNTAX{\colorSYNTAXB}{{\color{\colorMATH}\ensuremath{\sS_{1}}}}\endgroup } \sqcup  ({\begingroup\renewcommand\colorMATH{\colorMATHB}\renewcommand\colorSYNTAX{\colorSYNTAXB}{{\color{\colorMATH}\ensuremath{\sss_{2}}}}\endgroup }{\begingroup\renewcommand\colorMATH{\colorMATHB}\renewcommand\colorSYNTAX{\colorSYNTAXB}{{\color{\colorMATH}\ensuremath{\sS_{1}}}}\endgroup } + {\begingroup\renewcommand\colorMATH{\colorMATHB}\renewcommand\colorSYNTAX{\colorSYNTAXB}{{\color{\colorMATH}\ensuremath{\sss_{2}}}}\endgroup }{\begingroup\renewcommand\colorMATH{\colorMATHB}\renewcommand\colorSYNTAX{\colorSYNTAXB}{{\color{\colorMATH}\ensuremath{\sS_{1 1}}}}\endgroup } + {\begingroup\renewcommand\colorMATH{\colorMATHB}\renewcommand\colorSYNTAX{\colorSYNTAXB}{{\color{\colorMATH}\ensuremath{\sS_{2}}}}\endgroup }) \sqcup  ({\begingroup\renewcommand\colorMATH{\colorMATHB}\renewcommand\colorSYNTAX{\colorSYNTAXB}{{\color{\colorMATH}\ensuremath{\sss_{3}}}}\endgroup }{\begingroup\renewcommand\colorMATH{\colorMATHB}\renewcommand\colorSYNTAX{\colorSYNTAXB}{{\color{\colorMATH}\ensuremath{\sS_{1}}}}\endgroup } + {\begingroup\renewcommand\colorMATH{\colorMATHB}\renewcommand\colorSYNTAX{\colorSYNTAXB}{{\color{\colorMATH}\ensuremath{\sss_{3}}}}\endgroup }{\begingroup\renewcommand\colorMATH{\colorMATHB}\renewcommand\colorSYNTAX{\colorSYNTAXB}{{\color{\colorMATH}\ensuremath{\sS_{1 2}}}}\endgroup } + {\begingroup\renewcommand\colorMATH{\colorMATHB}\renewcommand\colorSYNTAX{\colorSYNTAXB}{{\color{\colorMATH}\ensuremath{\sS_{3}}}}\endgroup })}}}
            therefore by Lemma~\ref{lm:dot-subt} {{\color{\colorMATH}\ensuremath{{\begingroup\renewcommand\colorMATH{\colorMATHB}\renewcommand\colorSYNTAX{\colorSYNTAXB}{{\color{\colorMATH}\ensuremath{\Distance'}}}\endgroup }\mathord{\cdotp }({\begingroup\renewcommand\colorMATH{\colorMATHB}\renewcommand\colorSYNTAX{\colorSYNTAXB}{{\color{\colorMATH}\ensuremath{\sss_{2}}}}\endgroup }{\begingroup\renewcommand\colorMATH{\colorMATHB}\renewcommand\colorSYNTAX{\colorSYNTAXB}{{\color{\colorMATH}\ensuremath{\sS_{1}}}}\endgroup } + {\begingroup\renewcommand\colorMATH{\colorMATHB}\renewcommand\colorSYNTAX{\colorSYNTAXB}{{\color{\colorMATH}\ensuremath{\sss_{2}}}}\endgroup }{\begingroup\renewcommand\colorMATH{\colorMATHB}\renewcommand\colorSYNTAX{\colorSYNTAXB}{{\color{\colorMATH}\ensuremath{\sS_{1 1}}}}\endgroup } + {\begingroup\renewcommand\colorMATH{\colorMATHB}\renewcommand\colorSYNTAX{\colorSYNTAXB}{{\color{\colorMATH}\ensuremath{\sS_{2}}}}\endgroup }) \leq  {\begingroup\renewcommand\colorMATH{\colorMATHB}\renewcommand\colorSYNTAX{\colorSYNTAXB}{{\color{\colorMATH}\ensuremath{\Distance'}}}\endgroup }\mathord{\cdotp }({\begingroup\renewcommand\colorMATH{\colorMATHB}\renewcommand\colorSYNTAX{\colorSYNTAXB}{{\color{\colorMATH}\ensuremath{\sS_{1}}}}\endgroup } \sqcup  ({\begingroup\renewcommand\colorMATH{\colorMATHB}\renewcommand\colorSYNTAX{\colorSYNTAXB}{{\color{\colorMATH}\ensuremath{\sss_{2}}}}\endgroup }{\begingroup\renewcommand\colorMATH{\colorMATHB}\renewcommand\colorSYNTAX{\colorSYNTAXB}{{\color{\colorMATH}\ensuremath{\sS_{1}}}}\endgroup } + {\begingroup\renewcommand\colorMATH{\colorMATHB}\renewcommand\colorSYNTAX{\colorSYNTAXB}{{\color{\colorMATH}\ensuremath{\sss_{2}}}}\endgroup }{\begingroup\renewcommand\colorMATH{\colorMATHB}\renewcommand\colorSYNTAX{\colorSYNTAXB}{{\color{\colorMATH}\ensuremath{\sS_{1 1}}}}\endgroup } + {\begingroup\renewcommand\colorMATH{\colorMATHB}\renewcommand\colorSYNTAX{\colorSYNTAXB}{{\color{\colorMATH}\ensuremath{\sS_{2}}}}\endgroup }) \sqcup  ({\begingroup\renewcommand\colorMATH{\colorMATHB}\renewcommand\colorSYNTAX{\colorSYNTAXB}{{\color{\colorMATH}\ensuremath{\sss_{3}}}}\endgroup }{\begingroup\renewcommand\colorMATH{\colorMATHB}\renewcommand\colorSYNTAX{\colorSYNTAXB}{{\color{\colorMATH}\ensuremath{\sS_{1}}}}\endgroup } + {\begingroup\renewcommand\colorMATH{\colorMATHB}\renewcommand\colorSYNTAX{\colorSYNTAXB}{{\color{\colorMATH}\ensuremath{\sss_{3}}}}\endgroup }{\begingroup\renewcommand\colorMATH{\colorMATHB}\renewcommand\colorSYNTAX{\colorSYNTAXB}{{\color{\colorMATH}\ensuremath{\sS_{1 2}}}}\endgroup } + {\begingroup\renewcommand\colorMATH{\colorMATHB}\renewcommand\colorSYNTAX{\colorSYNTAXB}{{\color{\colorMATH}\ensuremath{\sS_{3}}}}\endgroup }))}}}.\\
            Also {{\color{\colorMATH}\ensuremath{[{\begingroup\renewcommand\colorMATH{\colorMATHB}\renewcommand\colorSYNTAX{\colorSYNTAXB}{{\color{\colorMATH}\ensuremath{\sS_{1}}}}\endgroup }+{\begingroup\renewcommand\colorMATH{\colorMATHB}\renewcommand\colorSYNTAX{\colorSYNTAXB}{{\color{\colorMATH}\ensuremath{\sS_{1 1}}}}\endgroup }/x]\tau _{2} <: [{\begingroup\renewcommand\colorMATH{\colorMATHB}\renewcommand\colorSYNTAX{\colorSYNTAXB}{{\color{\colorMATH}\ensuremath{\sS_{1}}}}\endgroup }+{\begingroup\renewcommand\colorMATH{\colorMATHB}\renewcommand\colorSYNTAX{\colorSYNTAXB}{{\color{\colorMATH}\ensuremath{\sS_{1 1}}}}\endgroup }/x]\tau _{2} \sqcup  [{\begingroup\renewcommand\colorMATH{\colorMATHB}\renewcommand\colorSYNTAX{\colorSYNTAXB}{{\color{\colorMATH}\ensuremath{\sS_{1}}}}\endgroup }+{\begingroup\renewcommand\colorMATH{\colorMATHB}\renewcommand\colorSYNTAX{\colorSYNTAXB}{{\color{\colorMATH}\ensuremath{\sS_{1 2}}}}\endgroup }/y]\tau _{3}}}} therefore by Lemma~\ref{lm:subtypinginst},\\
            {{\color{\colorMATH}\ensuremath{{\begingroup\renewcommand\colorMATH{\colorMATHB}\renewcommand\colorSYNTAX{\colorSYNTAXB}{{\color{\colorMATH}\ensuremath{\Distance'}}}\endgroup }([{\begingroup\renewcommand\colorMATH{\colorMATHB}\renewcommand\colorSYNTAX{\colorSYNTAXB}{{\color{\colorMATH}\ensuremath{\sS_{1}}}}\endgroup }+{\begingroup\renewcommand\colorMATH{\colorMATHB}\renewcommand\colorSYNTAX{\colorSYNTAXB}{{\color{\colorMATH}\ensuremath{\sS_{1 1}}}}\endgroup }/x]\tau _{2}) <: {\begingroup\renewcommand\colorMATH{\colorMATHB}\renewcommand\colorSYNTAX{\colorSYNTAXB}{{\color{\colorMATH}\ensuremath{\Distance'}}}\endgroup }([{\begingroup\renewcommand\colorMATH{\colorMATHB}\renewcommand\colorSYNTAX{\colorSYNTAXB}{{\color{\colorMATH}\ensuremath{\sS_{1}}}}\endgroup } + {\begingroup\renewcommand\colorMATH{\colorMATHB}\renewcommand\colorSYNTAX{\colorSYNTAXB}{{\color{\colorMATH}\ensuremath{\sS_{1 1}}}}\endgroup }/x]\tau _{2} \sqcup  [{\begingroup\renewcommand\colorMATH{\colorMATHB}\renewcommand\colorSYNTAX{\colorSYNTAXB}{{\color{\colorMATH}\ensuremath{\sS_{1}}}}\endgroup } + {\begingroup\renewcommand\colorMATH{\colorMATHB}\renewcommand\colorSYNTAX{\colorSYNTAXB}{{\color{\colorMATH}\ensuremath{\sS_{1 2}}}}\endgroup }/y]\tau _{3})}}}.
            Finally by Lemma~\ref{lm:lrweakening-sensitivity},\\
            {{\color{\colorMATH}\ensuremath{({\begingroup\renewcommand\colorMATH{\colorMATHB}\renewcommand\colorSYNTAX{\colorSYNTAXB}{{\color{\colorMATH}\ensuremath{\sv_{2 1}}}}\endgroup }, {\begingroup\renewcommand\colorMATH{\colorMATHB}\renewcommand\colorSYNTAX{\colorSYNTAXB}{{\color{\colorMATH}\ensuremath{\sv_{2 2}}}}\endgroup }) \in  {\mathcal{V}}^{k-j_{1}-j_{2}}_{{\begingroup\renewcommand\colorMATH{\colorMATHB}\renewcommand\colorSYNTAX{\colorSYNTAXB}{{\color{\colorMATH}\ensuremath{\Distance'}}}\endgroup }\mathord{\cdotp }({\begingroup\renewcommand\colorMATH{\colorMATHB}\renewcommand\colorSYNTAX{\colorSYNTAXB}{{\color{\colorMATH}\ensuremath{\sS_{1}}}}\endgroup } \sqcup  ({\begingroup\renewcommand\colorMATH{\colorMATHB}\renewcommand\colorSYNTAX{\colorSYNTAXB}{{\color{\colorMATH}\ensuremath{\sss_{2}}}}\endgroup }{\begingroup\renewcommand\colorMATH{\colorMATHB}\renewcommand\colorSYNTAX{\colorSYNTAXB}{{\color{\colorMATH}\ensuremath{\sS_{1}}}}\endgroup } + {\begingroup\renewcommand\colorMATH{\colorMATHB}\renewcommand\colorSYNTAX{\colorSYNTAXB}{{\color{\colorMATH}\ensuremath{\sss_{2}}}}\endgroup }{\begingroup\renewcommand\colorMATH{\colorMATHB}\renewcommand\colorSYNTAX{\colorSYNTAXB}{{\color{\colorMATH}\ensuremath{\sS_{1 1}}}}\endgroup } + {\begingroup\renewcommand\colorMATH{\colorMATHB}\renewcommand\colorSYNTAX{\colorSYNTAXB}{{\color{\colorMATH}\ensuremath{\sS_{2}}}}\endgroup }) \sqcup  ({\begingroup\renewcommand\colorMATH{\colorMATHB}\renewcommand\colorSYNTAX{\colorSYNTAXB}{{\color{\colorMATH}\ensuremath{\sss_{3}}}}\endgroup }{\begingroup\renewcommand\colorMATH{\colorMATHB}\renewcommand\colorSYNTAX{\colorSYNTAXB}{{\color{\colorMATH}\ensuremath{\sS_{1}}}}\endgroup } + {\begingroup\renewcommand\colorMATH{\colorMATHB}\renewcommand\colorSYNTAX{\colorSYNTAXB}{{\color{\colorMATH}\ensuremath{\sss_{3}}}}\endgroup }{\begingroup\renewcommand\colorMATH{\colorMATHB}\renewcommand\colorSYNTAX{\colorSYNTAXB}{{\color{\colorMATH}\ensuremath{\sS_{1 2}}}}\endgroup } + {\begingroup\renewcommand\colorMATH{\colorMATHB}\renewcommand\colorSYNTAX{\colorSYNTAXB}{{\color{\colorMATH}\ensuremath{\sS_{3}}}}\endgroup }))}\llbracket {\begingroup\renewcommand\colorMATH{\colorMATHB}\renewcommand\colorSYNTAX{\colorSYNTAXB}{{\color{\colorMATH}\ensuremath{\Distance'}}}\endgroup }([{\begingroup\renewcommand\colorMATH{\colorMATHB}\renewcommand\colorSYNTAX{\colorSYNTAXB}{{\color{\colorMATH}\ensuremath{\sS_{1}}}}\endgroup } + {\begingroup\renewcommand\colorMATH{\colorMATHB}\renewcommand\colorSYNTAX{\colorSYNTAXB}{{\color{\colorMATH}\ensuremath{\sS_{1 1}}}}\endgroup }/x]\tau _{2} \sqcup  [{\begingroup\renewcommand\colorMATH{\colorMATHB}\renewcommand\colorSYNTAX{\colorSYNTAXB}{{\color{\colorMATH}\ensuremath{\sS_{1}}}}\endgroup } + {\begingroup\renewcommand\colorMATH{\colorMATHB}\renewcommand\colorSYNTAX{\colorSYNTAXB}{{\color{\colorMATH}\ensuremath{\sS_{1 2}}}}\endgroup }/y]\tau _{3})\rrbracket }}} and the result holds by Lemma~\ref{lm:weakening-index}.
          \end{subproof}
        \item  {{\color{\colorMATH}\ensuremath{({\begingroup\renewcommand\colorMATH{\colorMATHB}\renewcommand\colorSYNTAX{\colorSYNTAXB}{{\color{\colorMATH}\ensuremath{\sv_{1 1}}}}\endgroup }, {\begingroup\renewcommand\colorMATH{\colorMATHB}\renewcommand\colorSYNTAX{\colorSYNTAXB}{{\color{\colorMATH}\ensuremath{\sv_{1 2}}}}\endgroup }) = (\inl\hspace*{0.33em}{\begingroup\renewcommand\colorMATH{\colorMATHB}\renewcommand\colorSYNTAX{\colorSYNTAXB}{{\color{\colorMATH}\ensuremath{\sv'_{1 1}}}}\endgroup }, \inr\hspace*{0.33em}{\begingroup\renewcommand\colorMATH{\colorMATHB}\renewcommand\colorSYNTAX{\colorSYNTAXB}{{\color{\colorMATH}\ensuremath{\sv'_{1 2}}}}\endgroup })}}} (the case {{\color{\colorMATH}\ensuremath{({\begingroup\renewcommand\colorMATH{\colorMATHB}\renewcommand\colorSYNTAX{\colorSYNTAXB}{{\color{\colorMATH}\ensuremath{\sv_{1 1}}}}\endgroup }, {\begingroup\renewcommand\colorMATH{\colorMATHB}\renewcommand\colorSYNTAX{\colorSYNTAXB}{{\color{\colorMATH}\ensuremath{\sv_{1 2}}}}\endgroup }) = (\inr\hspace*{0.33em}{\begingroup\renewcommand\colorMATH{\colorMATHB}\renewcommand\colorSYNTAX{\colorSYNTAXB}{{\color{\colorMATH}\ensuremath{\sv'_{1 1}}}}\endgroup }, \inl\hspace*{0.33em}{\begingroup\renewcommand\colorMATH{\colorMATHB}\renewcommand\colorSYNTAX{\colorSYNTAXB}{{\color{\colorMATH}\ensuremath{\sv'_{1 2}}}}\endgroup })}}} is analogous)
          \begin{subproof} 
            Then {{\color{\colorMATH}\ensuremath{{\begingroup\renewcommand\colorMATH{\colorMATHB}\renewcommand\colorSYNTAX{\colorSYNTAXB}{{\color{\colorMATH}\ensuremath{\Distance'}}}\endgroup }\mathord{\cdotp }{\begingroup\renewcommand\colorMATH{\colorMATHB}\renewcommand\colorSYNTAX{\colorSYNTAXB}{{\color{\colorMATH}\ensuremath{\sS_{1}}}}\endgroup } = \infty }}}, and the result vacuously holds.
          \end{subproof}
        \end{enumerate}  
      \end{subproof}
    \item  {{\color{\colorMATH}\ensuremath{\Gamma ; {\begingroup\renewcommand\colorMATH{\colorMATHB}\renewcommand\colorSYNTAX{\colorSYNTAXB}{{\color{\colorMATH}\ensuremath{\Distance}}}\endgroup } \vdash  \addProduct{{\begingroup\renewcommand\colorMATH{\colorMATHB}\renewcommand\colorSYNTAX{\colorSYNTAXB}{{\color{\colorMATH}\ensuremath{\se_{1}}}}\endgroup }}{{\begingroup\renewcommand\colorMATH{\colorMATHB}\renewcommand\colorSYNTAX{\colorSYNTAXB}{{\color{\colorMATH}\ensuremath{\se_{2}}}}\endgroup }} \mathrel{:} \tau _{1} \mathrel{^{{\begingroup\renewcommand\colorMATH{\colorMATHB}\renewcommand\colorSYNTAX{\colorSYNTAXB}{{\color{\colorMATH}\ensuremath{\sS_{1}}}}\endgroup }}\&^{{\begingroup\renewcommand\colorMATH{\colorMATHB}\renewcommand\colorSYNTAX{\colorSYNTAXB}{{\color{\colorMATH}\ensuremath{\sS_{2}}}}\endgroup }}} \tau _{2} \mathrel{;} \varnothing }}}
      \begin{subproof} 
      We have to prove that\\ {{\color{\colorMATH}\ensuremath{\forall k, \forall (\gamma _{1},\gamma _{2}) \in  {\mathcal{G}}_{{\begingroup\renewcommand\colorMATH{\colorMATHB}\renewcommand\colorSYNTAX{\colorSYNTAXB}{{\color{\colorMATH}\ensuremath{\Distance'}}}\endgroup }}^{\kg}\llbracket \Gamma \rrbracket , (\gamma _{1}\vdash \addProduct{{\begingroup\renewcommand\colorMATH{\colorMATHB}\renewcommand\colorSYNTAX{\colorSYNTAXB}{{\color{\colorMATH}\ensuremath{\se_{1}}}}\endgroup }}{{\begingroup\renewcommand\colorMATH{\colorMATHB}\renewcommand\colorSYNTAX{\colorSYNTAXB}{{\color{\colorMATH}\ensuremath{\se_{2}}}}\endgroup }},\gamma _{2}\vdash \addProduct{{\begingroup\renewcommand\colorMATH{\colorMATHB}\renewcommand\colorSYNTAX{\colorSYNTAXB}{{\color{\colorMATH}\ensuremath{\se_{1}}}}\endgroup }}{{\begingroup\renewcommand\colorMATH{\colorMATHB}\renewcommand\colorSYNTAX{\colorSYNTAXB}{{\color{\colorMATH}\ensuremath{\se_{2}}}}\endgroup }}) \in  {\mathcal{E}}_{{\begingroup\renewcommand\colorMATH{\colorMATHB}\renewcommand\colorSYNTAX{\colorSYNTAXB}{{\color{\colorMATH}\ensuremath{\Distance'}}}\endgroup }\mathord{\cdotp }\varnothing }^{k}\llbracket {\begingroup\renewcommand\colorMATH{\colorMATHB}\renewcommand\colorSYNTAX{\colorSYNTAXB}{{\color{\colorMATH}\ensuremath{\Distance'}}}\endgroup }(\tau _{1} \mathrel{^{{\begingroup\renewcommand\colorMATH{\colorMATHB}\renewcommand\colorSYNTAX{\colorSYNTAXB}{{\color{\colorMATH}\ensuremath{\sS_{1}}}}\endgroup }}\&^{{\begingroup\renewcommand\colorMATH{\colorMATHB}\renewcommand\colorSYNTAX{\colorSYNTAXB}{{\color{\colorMATH}\ensuremath{\sS_{2}}}}\endgroup }}} \tau _{2})\rrbracket }}}, for {{\color{\colorMATH}\ensuremath{{\begingroup\renewcommand\colorMATH{\colorMATHB}\renewcommand\colorSYNTAX{\colorSYNTAXB}{{\color{\colorMATH}\ensuremath{\Distance'}}}\endgroup } \sqsubseteq  {\begingroup\renewcommand\colorMATH{\colorMATHB}\renewcommand\colorSYNTAX{\colorSYNTAXB}{{\color{\colorMATH}\ensuremath{\Distance}}}\endgroup }}}}.
      Notice that {{\color{\colorMATH}\ensuremath{{\begingroup\renewcommand\colorMATH{\colorMATHB}\renewcommand\colorSYNTAX{\colorSYNTAXB}{{\color{\colorMATH}\ensuremath{\Distance'}}}\endgroup }\mathord{\cdotp }\varnothing  = 0}}}, and {{\color{\colorMATH}\ensuremath{{\begingroup\renewcommand\colorMATH{\colorMATHB}\renewcommand\colorSYNTAX{\colorSYNTAXB}{{\color{\colorMATH}\ensuremath{\Distance'}}}\endgroup }(\tau _{1} \mathrel{^{{\begingroup\renewcommand\colorMATH{\colorMATHB}\renewcommand\colorSYNTAX{\colorSYNTAXB}{{\color{\colorMATH}\ensuremath{\sS_{1}}}}\endgroup }}\&^{{\begingroup\renewcommand\colorMATH{\colorMATHB}\renewcommand\colorSYNTAX{\colorSYNTAXB}{{\color{\colorMATH}\ensuremath{\sS_{2}}}}\endgroup }}} \tau _{2}) = {\begingroup\renewcommand\colorMATH{\colorMATHB}\renewcommand\colorSYNTAX{\colorSYNTAXB}{{\color{\colorMATH}\ensuremath{\Distance'}}}\endgroup }(\tau _{1}) \mathrel{^{{\begingroup\renewcommand\colorMATH{\colorMATHB}\renewcommand\colorSYNTAX{\colorSYNTAXB}{{\color{\colorMATH}\ensuremath{\Distance'}}}\endgroup }\mathord{\cdotp }{\begingroup\renewcommand\colorMATH{\colorMATHB}\renewcommand\colorSYNTAX{\colorSYNTAXB}{{\color{\colorMATH}\ensuremath{\sS_{1}}}}\endgroup }}\&^{{\begingroup\renewcommand\colorMATH{\colorMATHB}\renewcommand\colorSYNTAX{\colorSYNTAXB}{{\color{\colorMATH}\ensuremath{\Distance'}}}\endgroup }\mathord{\cdotp }{\begingroup\renewcommand\colorMATH{\colorMATHB}\renewcommand\colorSYNTAX{\colorSYNTAXB}{{\color{\colorMATH}\ensuremath{\sS_{2}}}}\endgroup }}} {\begingroup\renewcommand\colorMATH{\colorMATHB}\renewcommand\colorSYNTAX{\colorSYNTAXB}{{\color{\colorMATH}\ensuremath{\Distance'}}}\endgroup }(\tau _{2})}}}, then we have to prove that\\
      {{\color{\colorMATH}\ensuremath{(\gamma _{1}\vdash \addProduct{{\begingroup\renewcommand\colorMATH{\colorMATHB}\renewcommand\colorSYNTAX{\colorSYNTAXB}{{\color{\colorMATH}\ensuremath{\se_{1}}}}\endgroup }}{{\begingroup\renewcommand\colorMATH{\colorMATHB}\renewcommand\colorSYNTAX{\colorSYNTAXB}{{\color{\colorMATH}\ensuremath{\se_{2}}}}\endgroup }},\gamma _{2}\vdash \addProduct{{\begingroup\renewcommand\colorMATH{\colorMATHB}\renewcommand\colorSYNTAX{\colorSYNTAXB}{{\color{\colorMATH}\ensuremath{\se_{1}}}}\endgroup }}{{\begingroup\renewcommand\colorMATH{\colorMATHB}\renewcommand\colorSYNTAX{\colorSYNTAXB}{{\color{\colorMATH}\ensuremath{\se_{2}}}}\endgroup }}) \in  {\mathcal{E}}_{0}^{k}\llbracket {\begingroup\renewcommand\colorMATH{\colorMATHB}\renewcommand\colorSYNTAX{\colorSYNTAXB}{{\color{\colorMATH}\ensuremath{\Distance'}}}\endgroup }(\tau _{1}) \mathrel{^{{\begingroup\renewcommand\colorMATH{\colorMATHB}\renewcommand\colorSYNTAX{\colorSYNTAXB}{{\color{\colorMATH}\ensuremath{\Distance'}}}\endgroup }\mathord{\cdotp }{\begingroup\renewcommand\colorMATH{\colorMATHB}\renewcommand\colorSYNTAX{\colorSYNTAXB}{{\color{\colorMATH}\ensuremath{\sS_{1}}}}\endgroup }}\&^{{\begingroup\renewcommand\colorMATH{\colorMATHB}\renewcommand\colorSYNTAX{\colorSYNTAXB}{{\color{\colorMATH}\ensuremath{\Distance'}}}\endgroup }\mathord{\cdotp }{\begingroup\renewcommand\colorMATH{\colorMATHB}\renewcommand\colorSYNTAX{\colorSYNTAXB}{{\color{\colorMATH}\ensuremath{\sS_{2}}}}\endgroup }}} {\begingroup\renewcommand\colorMATH{\colorMATHB}\renewcommand\colorSYNTAX{\colorSYNTAXB}{{\color{\colorMATH}\ensuremath{\Distance'}}}\endgroup }(\tau _{2})\rrbracket }}}, i.e.
      if {{\color{\colorMATH}\ensuremath{\gamma _{1}\vdash \addProduct{{\begingroup\renewcommand\colorMATH{\colorMATHB}\renewcommand\colorSYNTAX{\colorSYNTAXB}{{\color{\colorMATH}\ensuremath{\se_{1}}}}\endgroup }}{{\begingroup\renewcommand\colorMATH{\colorMATHB}\renewcommand\colorSYNTAX{\colorSYNTAXB}{{\color{\colorMATH}\ensuremath{\se_{2}}}}\endgroup }} \Downarrow ^{j} \addProduct{{\begingroup\renewcommand\colorMATH{\colorMATHB}\renewcommand\colorSYNTAX{\colorSYNTAXB}{{\color{\colorMATH}\ensuremath{\sv_{1 1}}}}\endgroup }}{{\begingroup\renewcommand\colorMATH{\colorMATHB}\renewcommand\colorSYNTAX{\colorSYNTAXB}{{\color{\colorMATH}\ensuremath{\sv_{1 2}}}}\endgroup }}}}} and {{\color{\colorMATH}\ensuremath{\gamma _{2}\vdash \addProduct{{\begingroup\renewcommand\colorMATH{\colorMATHB}\renewcommand\colorSYNTAX{\colorSYNTAXB}{{\color{\colorMATH}\ensuremath{\se_{1}}}}\endgroup }}{{\begingroup\renewcommand\colorMATH{\colorMATHB}\renewcommand\colorSYNTAX{\colorSYNTAXB}{{\color{\colorMATH}\ensuremath{\se_{2}}}}\endgroup }} \Downarrow ^{j} \addProduct{{\begingroup\renewcommand\colorMATH{\colorMATHB}\renewcommand\colorSYNTAX{\colorSYNTAXB}{{\color{\colorMATH}\ensuremath{\sv_{2 1}}}}\endgroup }}{{\begingroup\renewcommand\colorMATH{\colorMATHB}\renewcommand\colorSYNTAX{\colorSYNTAXB}{{\color{\colorMATH}\ensuremath{\sv_{2 2}}}}\endgroup }}}}}, then 
      {{\color{\colorMATH}\ensuremath{(\addProduct{{\begingroup\renewcommand\colorMATH{\colorMATHB}\renewcommand\colorSYNTAX{\colorSYNTAXB}{{\color{\colorMATH}\ensuremath{\sv_{1 1}}}}\endgroup }}{{\begingroup\renewcommand\colorMATH{\colorMATHB}\renewcommand\colorSYNTAX{\colorSYNTAXB}{{\color{\colorMATH}\ensuremath{\sv_{1 2}}}}\endgroup }}, \addProduct{{\begingroup\renewcommand\colorMATH{\colorMATHB}\renewcommand\colorSYNTAX{\colorSYNTAXB}{{\color{\colorMATH}\ensuremath{\sv_{2 1}}}}\endgroup }}{{\begingroup\renewcommand\colorMATH{\colorMATHB}\renewcommand\colorSYNTAX{\colorSYNTAXB}{{\color{\colorMATH}\ensuremath{\sv_{2 2}}}}\endgroup }}) \in  {\mathcal{V}}_{0}^{k-j}\llbracket {\begingroup\renewcommand\colorMATH{\colorMATHB}\renewcommand\colorSYNTAX{\colorSYNTAXB}{{\color{\colorMATH}\ensuremath{\Distance'}}}\endgroup }(\tau _{1}) \mathrel{^{{\begingroup\renewcommand\colorMATH{\colorMATHB}\renewcommand\colorSYNTAX{\colorSYNTAXB}{{\color{\colorMATH}\ensuremath{\Distance'}}}\endgroup }\mathord{\cdotp }{\begingroup\renewcommand\colorMATH{\colorMATHB}\renewcommand\colorSYNTAX{\colorSYNTAXB}{{\color{\colorMATH}\ensuremath{\sS_{1}}}}\endgroup }}\&^{{\begingroup\renewcommand\colorMATH{\colorMATHB}\renewcommand\colorSYNTAX{\colorSYNTAXB}{{\color{\colorMATH}\ensuremath{\Distance'}}}\endgroup }\mathord{\cdotp }{\begingroup\renewcommand\colorMATH{\colorMATHB}\renewcommand\colorSYNTAX{\colorSYNTAXB}{{\color{\colorMATH}\ensuremath{\sS_{2}}}}\endgroup }}} {\begingroup\renewcommand\colorMATH{\colorMATHB}\renewcommand\colorSYNTAX{\colorSYNTAXB}{{\color{\colorMATH}\ensuremath{\Distance'}}}\endgroup }(\tau _{2})\rrbracket }}}, or equivalently
      {{\color{\colorMATH}\ensuremath{({\begingroup\renewcommand\colorMATH{\colorMATHB}\renewcommand\colorSYNTAX{\colorSYNTAXB}{{\color{\colorMATH}\ensuremath{\sv_{1 1}}}}\endgroup },,{\begingroup\renewcommand\colorMATH{\colorMATHB}\renewcommand\colorSYNTAX{\colorSYNTAXB}{{\color{\colorMATH}\ensuremath{\sv_{2 1}}}}\endgroup }) \in  {\mathcal{V}}^{k-j}_{0+{\begingroup\renewcommand\colorMATH{\colorMATHB}\renewcommand\colorSYNTAX{\colorSYNTAXB}{{\color{\colorMATH}\ensuremath{\Distance'}}}\endgroup }\mathord{\cdotp }{\begingroup\renewcommand\colorMATH{\colorMATHB}\renewcommand\colorSYNTAX{\colorSYNTAXB}{{\color{\colorMATH}\ensuremath{\sS_{1}}}}\endgroup }}\llbracket {\begingroup\renewcommand\colorMATH{\colorMATHB}\renewcommand\colorSYNTAX{\colorSYNTAXB}{{\color{\colorMATH}\ensuremath{\Distance'}}}\endgroup }(\tau _{1})\rrbracket }}}, and {{\color{\colorMATH}\ensuremath{({\begingroup\renewcommand\colorMATH{\colorMATHB}\renewcommand\colorSYNTAX{\colorSYNTAXB}{{\color{\colorMATH}\ensuremath{\sv_{1 2}}}}\endgroup },,{\begingroup\renewcommand\colorMATH{\colorMATHB}\renewcommand\colorSYNTAX{\colorSYNTAXB}{{\color{\colorMATH}\ensuremath{\sv_{2 2}}}}\endgroup }) \in  {\mathcal{V}}^{k-j}_{0+{\begingroup\renewcommand\colorMATH{\colorMATHB}\renewcommand\colorSYNTAX{\colorSYNTAXB}{{\color{\colorMATH}\ensuremath{\Distance'}}}\endgroup }\mathord{\cdotp }{\begingroup\renewcommand\colorMATH{\colorMATHB}\renewcommand\colorSYNTAX{\colorSYNTAXB}{{\color{\colorMATH}\ensuremath{\sS_{2}}}}\endgroup }}\llbracket {\begingroup\renewcommand\colorMATH{\colorMATHB}\renewcommand\colorSYNTAX{\colorSYNTAXB}{{\color{\colorMATH}\ensuremath{\Distance'}}}\endgroup }(\tau _{2})\rrbracket }}}.

      By induction hypothesis on {{\color{\colorMATH}\ensuremath{\Gamma  \vdash  {\begingroup\renewcommand\colorMATH{\colorMATHB}\renewcommand\colorSYNTAX{\colorSYNTAXB}{{\color{\colorMATH}\ensuremath{\se_{1}}}}\endgroup } \mathrel{:} \tau _{1} \mathrel{;} {\begingroup\renewcommand\colorMATH{\colorMATHB}\renewcommand\colorSYNTAX{\colorSYNTAXB}{{\color{\colorMATH}\ensuremath{\sS_{1}}}}\endgroup }}}} and {{\color{\colorMATH}\ensuremath{\Gamma  \vdash  {\begingroup\renewcommand\colorMATH{\colorMATHB}\renewcommand\colorSYNTAX{\colorSYNTAXB}{{\color{\colorMATH}\ensuremath{\se_{2}}}}\endgroup } \mathrel{:} \tau _{2} \mathrel{;} {\begingroup\renewcommand\colorMATH{\colorMATHB}\renewcommand\colorSYNTAX{\colorSYNTAXB}{{\color{\colorMATH}\ensuremath{\sS_{2}}}}\endgroup }}}}, we know that 
      {{\color{\colorMATH}\ensuremath{(\gamma _{1}\vdash {\begingroup\renewcommand\colorMATH{\colorMATHB}\renewcommand\colorSYNTAX{\colorSYNTAXB}{{\color{\colorMATH}\ensuremath{\se_{1}}}}\endgroup },\gamma _{2}\vdash {\begingroup\renewcommand\colorMATH{\colorMATHB}\renewcommand\colorSYNTAX{\colorSYNTAXB}{{\color{\colorMATH}\ensuremath{\se_{1}}}}\endgroup }) \in  {\mathcal{E}}^{k}_{{\begingroup\renewcommand\colorMATH{\colorMATHB}\renewcommand\colorSYNTAX{\colorSYNTAXB}{{\color{\colorMATH}\ensuremath{\Distance'}}}\endgroup }\mathord{\cdotp }{\begingroup\renewcommand\colorMATH{\colorMATHB}\renewcommand\colorSYNTAX{\colorSYNTAXB}{{\color{\colorMATH}\ensuremath{\sS_{1}}}}\endgroup }}\llbracket {\begingroup\renewcommand\colorMATH{\colorMATHB}\renewcommand\colorSYNTAX{\colorSYNTAXB}{{\color{\colorMATH}\ensuremath{\Distance'}}}\endgroup }(\tau _{1})\rrbracket }}} and {{\color{\colorMATH}\ensuremath{(\gamma _{1}\vdash {\begingroup\renewcommand\colorMATH{\colorMATHB}\renewcommand\colorSYNTAX{\colorSYNTAXB}{{\color{\colorMATH}\ensuremath{\se_{2}}}}\endgroup },\gamma _{2}\vdash {\begingroup\renewcommand\colorMATH{\colorMATHB}\renewcommand\colorSYNTAX{\colorSYNTAXB}{{\color{\colorMATH}\ensuremath{\se_{2}}}}\endgroup }) \in  {\mathcal{E}}^{k}_{{\begingroup\renewcommand\colorMATH{\colorMATHB}\renewcommand\colorSYNTAX{\colorSYNTAXB}{{\color{\colorMATH}\ensuremath{\Distance'}}}\endgroup }\mathord{\cdotp }{\begingroup\renewcommand\colorMATH{\colorMATHB}\renewcommand\colorSYNTAX{\colorSYNTAXB}{{\color{\colorMATH}\ensuremath{\sS_{2}}}}\endgroup }}\llbracket {\begingroup\renewcommand\colorMATH{\colorMATHB}\renewcommand\colorSYNTAX{\colorSYNTAXB}{{\color{\colorMATH}\ensuremath{\Distance'}}}\endgroup }(\tau _{2})\rrbracket }}} respectively. This means that
      if {{\color{\colorMATH}\ensuremath{\gamma _{1}\vdash {\begingroup\renewcommand\colorMATH{\colorMATHB}\renewcommand\colorSYNTAX{\colorSYNTAXB}{{\color{\colorMATH}\ensuremath{\se_{1}}}}\endgroup } \Downarrow ^{j_{1}} {\begingroup\renewcommand\colorMATH{\colorMATHB}\renewcommand\colorSYNTAX{\colorSYNTAXB}{{\color{\colorMATH}\ensuremath{\sv'_{1 1}}}}\endgroup }}}}, and {{\color{\colorMATH}\ensuremath{\gamma _{2}\vdash {\begingroup\renewcommand\colorMATH{\colorMATHB}\renewcommand\colorSYNTAX{\colorSYNTAXB}{{\color{\colorMATH}\ensuremath{\se_{1}}}}\endgroup } \Downarrow ^{j_{1}} {\begingroup\renewcommand\colorMATH{\colorMATHB}\renewcommand\colorSYNTAX{\colorSYNTAXB}{{\color{\colorMATH}\ensuremath{\sv'_{1 2}}}}\endgroup }}}} then 
      {{\color{\colorMATH}\ensuremath{({\begingroup\renewcommand\colorMATH{\colorMATHB}\renewcommand\colorSYNTAX{\colorSYNTAXB}{{\color{\colorMATH}\ensuremath{\sv'_{1 1}}}}\endgroup }, {\begingroup\renewcommand\colorMATH{\colorMATHB}\renewcommand\colorSYNTAX{\colorSYNTAXB}{{\color{\colorMATH}\ensuremath{\sv'_{1 2}}}}\endgroup }) \in  {\mathcal{V}}^{k-j_{1}}_{{\begingroup\renewcommand\colorMATH{\colorMATHB}\renewcommand\colorSYNTAX{\colorSYNTAXB}{{\color{\colorMATH}\ensuremath{\Distance'}}}\endgroup }\mathord{\cdotp }{\begingroup\renewcommand\colorMATH{\colorMATHB}\renewcommand\colorSYNTAX{\colorSYNTAXB}{{\color{\colorMATH}\ensuremath{\sS_{1}}}}\endgroup }}\llbracket {\begingroup\renewcommand\colorMATH{\colorMATHB}\renewcommand\colorSYNTAX{\colorSYNTAXB}{{\color{\colorMATH}\ensuremath{\Distance'}}}\endgroup }(\tau _{1})\rrbracket }}}, and that 
      if {{\color{\colorMATH}\ensuremath{\gamma _{1}\vdash {\begingroup\renewcommand\colorMATH{\colorMATHB}\renewcommand\colorSYNTAX{\colorSYNTAXB}{{\color{\colorMATH}\ensuremath{\se_{2}}}}\endgroup } \Downarrow ^{j_{2}} {\begingroup\renewcommand\colorMATH{\colorMATHB}\renewcommand\colorSYNTAX{\colorSYNTAXB}{{\color{\colorMATH}\ensuremath{\sv'_{2 1}}}}\endgroup }}}}, and {{\color{\colorMATH}\ensuremath{\gamma _{2}\vdash {\begingroup\renewcommand\colorMATH{\colorMATHB}\renewcommand\colorSYNTAX{\colorSYNTAXB}{{\color{\colorMATH}\ensuremath{\se_{2}}}}\endgroup } \Downarrow ^{j_{2}} {\begingroup\renewcommand\colorMATH{\colorMATHB}\renewcommand\colorSYNTAX{\colorSYNTAXB}{{\color{\colorMATH}\ensuremath{\sv'_{2 2}}}}\endgroup }}}} then 
      {{\color{\colorMATH}\ensuremath{({\begingroup\renewcommand\colorMATH{\colorMATHB}\renewcommand\colorSYNTAX{\colorSYNTAXB}{{\color{\colorMATH}\ensuremath{\sv'_{2 1}}}}\endgroup }, {\begingroup\renewcommand\colorMATH{\colorMATHB}\renewcommand\colorSYNTAX{\colorSYNTAXB}{{\color{\colorMATH}\ensuremath{\sv'_{2 2}}}}\endgroup }) \in  {\mathcal{V}}^{k-j_{2}}_{{\begingroup\renewcommand\colorMATH{\colorMATHB}\renewcommand\colorSYNTAX{\colorSYNTAXB}{{\color{\colorMATH}\ensuremath{\Distance'}}}\endgroup }\mathord{\cdotp }{\begingroup\renewcommand\colorMATH{\colorMATHB}\renewcommand\colorSYNTAX{\colorSYNTAXB}{{\color{\colorMATH}\ensuremath{\sS_{2}}}}\endgroup }}\llbracket {\begingroup\renewcommand\colorMATH{\colorMATHB}\renewcommand\colorSYNTAX{\colorSYNTAXB}{{\color{\colorMATH}\ensuremath{\Distance'}}}\endgroup }(\tau _{2})\rrbracket }}}. 
      As reduction is deterministic, then {{\color{\colorMATH}\ensuremath{j = j_{1}+j_{2}}}} and {{\color{\colorMATH}\ensuremath{{\begingroup\renewcommand\colorMATH{\colorMATHB}\renewcommand\colorSYNTAX{\colorSYNTAXB}{{\color{\colorMATH}\ensuremath{\sv'_{i j}}}}\endgroup } = {\begingroup\renewcommand\colorMATH{\colorMATHB}\renewcommand\colorSYNTAX{\colorSYNTAXB}{{\color{\colorMATH}\ensuremath{\sv_{i j}}}}\endgroup }}}}, therefore as {{\color{\colorMATH}\ensuremath{0+{\begingroup\renewcommand\colorMATH{\colorMATHB}\renewcommand\colorSYNTAX{\colorSYNTAXB}{{\color{\colorMATH}\ensuremath{\Distance'}}}\endgroup }\mathord{\cdotp }{\begingroup\renewcommand\colorMATH{\colorMATHB}\renewcommand\colorSYNTAX{\colorSYNTAXB}{{\color{\colorMATH}\ensuremath{\sS_{i}}}}\endgroup } = {\begingroup\renewcommand\colorMATH{\colorMATHB}\renewcommand\colorSYNTAX{\colorSYNTAXB}{{\color{\colorMATH}\ensuremath{\Distance'}}}\endgroup }\mathord{\cdotp }{\begingroup\renewcommand\colorMATH{\colorMATHB}\renewcommand\colorSYNTAX{\colorSYNTAXB}{{\color{\colorMATH}\ensuremath{\sS_{i}}}}\endgroup }}}}, the result holds immediately by Lemma~\ref{lm:weakening-index}.

      \end{subproof}
     
    %%
    %% Version with prepaid effects and linearity
    %%
    \item  {{\color{\colorMATH}\ensuremath{\Gamma ; {\begingroup\renewcommand\colorMATH{\colorMATHB}\renewcommand\colorSYNTAX{\colorSYNTAXB}{{\color{\colorMATH}\ensuremath{\Distance}}}\endgroup } \vdash  \addProduct{{\begingroup\renewcommand\colorMATH{\colorMATHB}\renewcommand\colorSYNTAX{\colorSYNTAXB}{{\color{\colorMATH}\ensuremath{\se_{1}}}}\endgroup }}{{\begingroup\renewcommand\colorMATH{\colorMATHB}\renewcommand\colorSYNTAX{\colorSYNTAXB}{{\color{\colorMATH}\ensuremath{\se_{2}}}}\endgroup }} \mathrel{:} \tau _{1} \mathrel{^{{\begingroup\renewcommand\colorMATH{\colorMATHB}\renewcommand\colorSYNTAX{\colorSYNTAXB}{{\color{\colorMATH}\ensuremath{\sS'_{1}}}}\endgroup }}\&^{{\begingroup\renewcommand\colorMATH{\colorMATHB}\renewcommand\colorSYNTAX{\colorSYNTAXB}{{\color{\colorMATH}\ensuremath{\sS_{2}'}}}\endgroup }}} \tau _{2} \mathrel{;}  \addProd{{\begingroup\renewcommand\colorMATH{\colorMATHB}\renewcommand\colorSYNTAX{\colorSYNTAXB}{{\color{\colorMATH}\ensuremath{\sS''_{1}}}}\endgroup }}{{\begingroup\renewcommand\colorMATH{\colorMATHB}\renewcommand\colorSYNTAX{\colorSYNTAXB}{{\color{\colorMATH}\ensuremath{\sS''_{2}}}}\endgroup }}}}}
      \begin{subproof}   
      \begingroup\color{\colorMATH}\begin{gather*} 
        \inferrule*[lab=
        ]{ \Gamma ; {\begingroup\renewcommand\colorMATH{\colorMATHB}\renewcommand\colorSYNTAX{\colorSYNTAXB}{{\color{\colorMATH}\ensuremath{\Distance}}}\endgroup } \vdash  {\begingroup\renewcommand\colorMATH{\colorMATHB}\renewcommand\colorSYNTAX{\colorSYNTAXB}{{\color{\colorMATH}\ensuremath{\se_{1}}}}\endgroup } \mathrel{:} \tau _{1} \mathrel{;} {\begingroup\renewcommand\colorMATH{\colorMATHB}\renewcommand\colorSYNTAX{\colorSYNTAXB}{{\color{\colorMATH}\ensuremath{\sS''_{1}}}}\endgroup } + {\begingroup\renewcommand\colorMATH{\colorMATHB}\renewcommand\colorSYNTAX{\colorSYNTAXB}{{\color{\colorMATH}\ensuremath{\sS'_{1}}}}\endgroup } 
        \\ \Gamma ; {\begingroup\renewcommand\colorMATH{\colorMATHB}\renewcommand\colorSYNTAX{\colorSYNTAXB}{{\color{\colorMATH}\ensuremath{\Distance}}}\endgroup } \vdash  {\begingroup\renewcommand\colorMATH{\colorMATHB}\renewcommand\colorSYNTAX{\colorSYNTAXB}{{\color{\colorMATH}\ensuremath{\se_{2}}}}\endgroup } \mathrel{:} \tau _{2} \mathrel{;} {\begingroup\renewcommand\colorMATH{\colorMATHB}\renewcommand\colorSYNTAX{\colorSYNTAXB}{{\color{\colorMATH}\ensuremath{\sS''_{2}}}}\endgroup } + {\begingroup\renewcommand\colorMATH{\colorMATHB}\renewcommand\colorSYNTAX{\colorSYNTAXB}{{\color{\colorMATH}\ensuremath{\sS'_{2}}}}\endgroup } 
          }{
          \Gamma ; {\begingroup\renewcommand\colorMATH{\colorMATHB}\renewcommand\colorSYNTAX{\colorSYNTAXB}{{\color{\colorMATH}\ensuremath{\Distance}}}\endgroup } \vdash  \addProduct{{\begingroup\renewcommand\colorMATH{\colorMATHB}\renewcommand\colorSYNTAX{\colorSYNTAXB}{{\color{\colorMATH}\ensuremath{\se_{1}}}}\endgroup }}{{\begingroup\renewcommand\colorMATH{\colorMATHB}\renewcommand\colorSYNTAX{\colorSYNTAXB}{{\color{\colorMATH}\ensuremath{\se_{2}}}}\endgroup }} \mathrel{:} \tau _{1} \mathrel{^{{\begingroup\renewcommand\colorMATH{\colorMATHB}\renewcommand\colorSYNTAX{\colorSYNTAXB}{{\color{\colorMATH}\ensuremath{\sS'_{1}}}}\endgroup }}\&^{{\begingroup\renewcommand\colorMATH{\colorMATHB}\renewcommand\colorSYNTAX{\colorSYNTAXB}{{\color{\colorMATH}\ensuremath{\sS_{2}'}}}\endgroup }}} \tau _{2} \mathrel{;}  \addProd{{\begingroup\renewcommand\colorMATH{\colorMATHB}\renewcommand\colorSYNTAX{\colorSYNTAXB}{{\color{\colorMATH}\ensuremath{\sS''_{1}}}}\endgroup }}{{\begingroup\renewcommand\colorMATH{\colorMATHB}\renewcommand\colorSYNTAX{\colorSYNTAXB}{{\color{\colorMATH}\ensuremath{\sS''_{2}}}}\endgroup }}
        }
      \end{gather*}\endgroup
      We have to prove that\\ {{\color{\colorMATH}\ensuremath{\forall k, \forall (\gamma _{1},\gamma _{2}) \in  {\mathcal{G}}_{{\begingroup\renewcommand\colorMATH{\colorMATHB}\renewcommand\colorSYNTAX{\colorSYNTAXB}{{\color{\colorMATH}\ensuremath{\Distance'}}}\endgroup }}^{\kg}\llbracket \Gamma \rrbracket , (\gamma _{1}\vdash \addProduct{{\begingroup\renewcommand\colorMATH{\colorMATHB}\renewcommand\colorSYNTAX{\colorSYNTAXB}{{\color{\colorMATH}\ensuremath{\se_{1}}}}\endgroup }}{{\begingroup\renewcommand\colorMATH{\colorMATHB}\renewcommand\colorSYNTAX{\colorSYNTAXB}{{\color{\colorMATH}\ensuremath{\se_{2}}}}\endgroup }},\gamma _{2}\vdash \addProduct{{\begingroup\renewcommand\colorMATH{\colorMATHB}\renewcommand\colorSYNTAX{\colorSYNTAXB}{{\color{\colorMATH}\ensuremath{\se_{1}}}}\endgroup }}{{\begingroup\renewcommand\colorMATH{\colorMATHB}\renewcommand\colorSYNTAX{\colorSYNTAXB}{{\color{\colorMATH}\ensuremath{\se_{2}}}}\endgroup }}) \in  {\mathcal{E}}^{k}_{{\begingroup\renewcommand\colorMATH{\colorMATHB}\renewcommand\colorSYNTAX{\colorSYNTAXB}{{\color{\colorMATH}\ensuremath{\Distance'}}}\endgroup }\mathord{\cdotp }(\addProd{{\begingroup\renewcommand\colorMATH{\colorMATHB}\renewcommand\colorSYNTAX{\colorSYNTAXB}{{\color{\colorMATH}\ensuremath{\sS''_{1}}}}\endgroup }}{{\begingroup\renewcommand\colorMATH{\colorMATHB}\renewcommand\colorSYNTAX{\colorSYNTAXB}{{\color{\colorMATH}\ensuremath{\sS''_{2}}}}\endgroup }})}\llbracket {\begingroup\renewcommand\colorMATH{\colorMATHB}\renewcommand\colorSYNTAX{\colorSYNTAXB}{{\color{\colorMATH}\ensuremath{\Distance'}}}\endgroup }(\tau _{1} \mathrel{^{{\begingroup\renewcommand\colorMATH{\colorMATHB}\renewcommand\colorSYNTAX{\colorSYNTAXB}{{\color{\colorMATH}\ensuremath{\sS'_{1}}}}\endgroup }}\&^{{\begingroup\renewcommand\colorMATH{\colorMATHB}\renewcommand\colorSYNTAX{\colorSYNTAXB}{{\color{\colorMATH}\ensuremath{\sS'_{2}}}}\endgroup }}} \tau _{2})\rrbracket }}}, for {{\color{\colorMATH}\ensuremath{{\begingroup\renewcommand\colorMATH{\colorMATHB}\renewcommand\colorSYNTAX{\colorSYNTAXB}{{\color{\colorMATH}\ensuremath{\Distance'}}}\endgroup } \sqsubseteq  {\begingroup\renewcommand\colorMATH{\colorMATHB}\renewcommand\colorSYNTAX{\colorSYNTAXB}{{\color{\colorMATH}\ensuremath{\Distance}}}\endgroup }}}}.
      Let {{\color{\colorMATH}\ensuremath{d'_{i} = {\begingroup\renewcommand\colorMATH{\colorMATHB}\renewcommand\colorSYNTAX{\colorSYNTAXB}{{\color{\colorMATH}\ensuremath{\Distance'}}}\endgroup }\mathord{\cdotp }{\begingroup\renewcommand\colorMATH{\colorMATHB}\renewcommand\colorSYNTAX{\colorSYNTAXB}{{\color{\colorMATH}\ensuremath{\sS'_{i}}}}\endgroup }}}} and {{\color{\colorMATH}\ensuremath{d''_{i} = {\begingroup\renewcommand\colorMATH{\colorMATHB}\renewcommand\colorSYNTAX{\colorSYNTAXB}{{\color{\colorMATH}\ensuremath{\Distance'}}}\endgroup }\mathord{\cdotp }{\begingroup\renewcommand\colorMATH{\colorMATHB}\renewcommand\colorSYNTAX{\colorSYNTAXB}{{\color{\colorMATH}\ensuremath{\sS''_{i}}}}\endgroup }}}}. 
      Notice that {{\color{\colorMATH}\ensuremath{{\begingroup\renewcommand\colorMATH{\colorMATHB}\renewcommand\colorSYNTAX{\colorSYNTAXB}{{\color{\colorMATH}\ensuremath{\Distance'}}}\endgroup }\mathord{\cdotp }(\addProd{{\begingroup\renewcommand\colorMATH{\colorMATHB}\renewcommand\colorSYNTAX{\colorSYNTAXB}{{\color{\colorMATH}\ensuremath{\sS''_{1}}}}\endgroup }}{{\begingroup\renewcommand\colorMATH{\colorMATHB}\renewcommand\colorSYNTAX{\colorSYNTAXB}{{\color{\colorMATH}\ensuremath{\sS''_{2}}}}\endgroup }}) = \addProd{d''_{1}}{d''_{2}}}}}, and {{\color{\colorMATH}\ensuremath{{\begingroup\renewcommand\colorMATH{\colorMATHB}\renewcommand\colorSYNTAX{\colorSYNTAXB}{{\color{\colorMATH}\ensuremath{\Distance'}}}\endgroup }(\tau _{1} \mathrel{^{{\begingroup\renewcommand\colorMATH{\colorMATHB}\renewcommand\colorSYNTAX{\colorSYNTAXB}{{\color{\colorMATH}\ensuremath{\sS'_{1}}}}\endgroup }}\&^{{\begingroup\renewcommand\colorMATH{\colorMATHB}\renewcommand\colorSYNTAX{\colorSYNTAXB}{{\color{\colorMATH}\ensuremath{\sS'_{2}}}}\endgroup }}} \tau _{2}) = {\begingroup\renewcommand\colorMATH{\colorMATHB}\renewcommand\colorSYNTAX{\colorSYNTAXB}{{\color{\colorMATH}\ensuremath{\Distance'}}}\endgroup }(\tau _{1}) \mathrel{^{d'_{1}}\&^{d'_{2}}} {\begingroup\renewcommand\colorMATH{\colorMATHB}\renewcommand\colorSYNTAX{\colorSYNTAXB}{{\color{\colorMATH}\ensuremath{\Distance'}}}\endgroup }(\tau _{2})}}}, then we have to prove that\\
      {{\color{\colorMATH}\ensuremath{(\gamma _{1}\vdash \addProduct{{\begingroup\renewcommand\colorMATH{\colorMATHB}\renewcommand\colorSYNTAX{\colorSYNTAXB}{{\color{\colorMATH}\ensuremath{\se_{1}}}}\endgroup }}{{\begingroup\renewcommand\colorMATH{\colorMATHB}\renewcommand\colorSYNTAX{\colorSYNTAXB}{{\color{\colorMATH}\ensuremath{\se_{2}}}}\endgroup }},\gamma _{2}\vdash \addProduct{{\begingroup\renewcommand\colorMATH{\colorMATHB}\renewcommand\colorSYNTAX{\colorSYNTAXB}{{\color{\colorMATH}\ensuremath{\se_{1}}}}\endgroup }}{{\begingroup\renewcommand\colorMATH{\colorMATHB}\renewcommand\colorSYNTAX{\colorSYNTAXB}{{\color{\colorMATH}\ensuremath{\se_{2}}}}\endgroup }}) \in  {\mathcal{E}}^{k}_{\addProd{d''_{1}}{d''_{2}}}\llbracket {\begingroup\renewcommand\colorMATH{\colorMATHB}\renewcommand\colorSYNTAX{\colorSYNTAXB}{{\color{\colorMATH}\ensuremath{\Distance'}}}\endgroup }(\tau _{1}) \mathrel{^{d'_{1}}\&^{d'_{2}}} {\begingroup\renewcommand\colorMATH{\colorMATHB}\renewcommand\colorSYNTAX{\colorSYNTAXB}{{\color{\colorMATH}\ensuremath{\Distance'}}}\endgroup }(\tau _{2})\rrbracket }}}, i.e.
      if {{\color{\colorMATH}\ensuremath{\gamma _{1}\vdash \addProduct{{\begingroup\renewcommand\colorMATH{\colorMATHB}\renewcommand\colorSYNTAX{\colorSYNTAXB}{{\color{\colorMATH}\ensuremath{\se_{1}}}}\endgroup }}{{\begingroup\renewcommand\colorMATH{\colorMATHB}\renewcommand\colorSYNTAX{\colorSYNTAXB}{{\color{\colorMATH}\ensuremath{\se_{2}}}}\endgroup }} \Downarrow ^{j} \addProduct{{\begingroup\renewcommand\colorMATH{\colorMATHB}\renewcommand\colorSYNTAX{\colorSYNTAXB}{{\color{\colorMATH}\ensuremath{\sv_{1 1}}}}\endgroup }}{{\begingroup\renewcommand\colorMATH{\colorMATHB}\renewcommand\colorSYNTAX{\colorSYNTAXB}{{\color{\colorMATH}\ensuremath{\sv_{1 2}}}}\endgroup }}}}} \pthen {{\color{\colorMATH}\ensuremath{\gamma _{2}\vdash \addProduct{{\begingroup\renewcommand\colorMATH{\colorMATHB}\renewcommand\colorSYNTAX{\colorSYNTAXB}{{\color{\colorMATH}\ensuremath{\se_{1}}}}\endgroup }}{{\begingroup\renewcommand\colorMATH{\colorMATHB}\renewcommand\colorSYNTAX{\colorSYNTAXB}{{\color{\colorMATH}\ensuremath{\se_{2}}}}\endgroup }} \Downarrow ^{\pj} \addProduct{{\begingroup\renewcommand\colorMATH{\colorMATHB}\renewcommand\colorSYNTAX{\colorSYNTAXB}{{\color{\colorMATH}\ensuremath{\sv_{2 1}}}}\endgroup }}{{\begingroup\renewcommand\colorMATH{\colorMATHB}\renewcommand\colorSYNTAX{\colorSYNTAXB}{{\color{\colorMATH}\ensuremath{\sv_{2 2}}}}\endgroup }}}}}, \pand 
      {{\color{\colorMATH}\ensuremath{(\addProduct{{\begingroup\renewcommand\colorMATH{\colorMATHB}\renewcommand\colorSYNTAX{\colorSYNTAXB}{{\color{\colorMATH}\ensuremath{\sv_{1 1}}}}\endgroup }}{{\begingroup\renewcommand\colorMATH{\colorMATHB}\renewcommand\colorSYNTAX{\colorSYNTAXB}{{\color{\colorMATH}\ensuremath{\sv_{1 2}}}}\endgroup }}, \addProduct{{\begingroup\renewcommand\colorMATH{\colorMATHB}\renewcommand\colorSYNTAX{\colorSYNTAXB}{{\color{\colorMATH}\ensuremath{\sv_{2 1}}}}\endgroup }}{{\begingroup\renewcommand\colorMATH{\colorMATHB}\renewcommand\colorSYNTAX{\colorSYNTAXB}{{\color{\colorMATH}\ensuremath{\sv_{2 2}}}}\endgroup }}) \in  {\mathcal{V}}^{k-j}_{\addProd{d''_{1}}{d''_{2}}}\llbracket {\begingroup\renewcommand\colorMATH{\colorMATHB}\renewcommand\colorSYNTAX{\colorSYNTAXB}{{\color{\colorMATH}\ensuremath{\Distance'}}}\endgroup }(\tau _{1}) \mathrel{^{d'_{1}}\&^{d'_{2}}} {\begingroup\renewcommand\colorMATH{\colorMATHB}\renewcommand\colorSYNTAX{\colorSYNTAXB}{{\color{\colorMATH}\ensuremath{\Distance'}}}\endgroup }(\tau _{2})\rrbracket }}}.

      By induction hypothesis on {{\color{\colorMATH}\ensuremath{\Gamma  \vdash  {\begingroup\renewcommand\colorMATH{\colorMATHB}\renewcommand\colorSYNTAX{\colorSYNTAXB}{{\color{\colorMATH}\ensuremath{\se_{1}}}}\endgroup } \mathrel{:} \tau _{1} \mathrel{;} {\begingroup\renewcommand\colorMATH{\colorMATHB}\renewcommand\colorSYNTAX{\colorSYNTAXB}{{\color{\colorMATH}\ensuremath{\sS''_{1}}}}\endgroup } + {\begingroup\renewcommand\colorMATH{\colorMATHB}\renewcommand\colorSYNTAX{\colorSYNTAXB}{{\color{\colorMATH}\ensuremath{\sS'_{1}}}}\endgroup }}}} and {{\color{\colorMATH}\ensuremath{\Gamma  \vdash  {\begingroup\renewcommand\colorMATH{\colorMATHB}\renewcommand\colorSYNTAX{\colorSYNTAXB}{{\color{\colorMATH}\ensuremath{\se_{2}}}}\endgroup } \mathrel{:} \tau _{2} \mathrel{;} {\begingroup\renewcommand\colorMATH{\colorMATHB}\renewcommand\colorSYNTAX{\colorSYNTAXB}{{\color{\colorMATH}\ensuremath{\sS''_{2}}}}\endgroup } + {\begingroup\renewcommand\colorMATH{\colorMATHB}\renewcommand\colorSYNTAX{\colorSYNTAXB}{{\color{\colorMATH}\ensuremath{\sS'_{2}}}}\endgroup }}}}, we know that 
      {{\color{\colorMATH}\ensuremath{(\gamma _{1}\vdash {\begingroup\renewcommand\colorMATH{\colorMATHB}\renewcommand\colorSYNTAX{\colorSYNTAXB}{{\color{\colorMATH}\ensuremath{\se_{1}}}}\endgroup },\gamma _{2}\vdash {\begingroup\renewcommand\colorMATH{\colorMATHB}\renewcommand\colorSYNTAX{\colorSYNTAXB}{{\color{\colorMATH}\ensuremath{\se_{1}}}}\endgroup }) \in  {\mathcal{E}}^{k}_{d'_{1} + d''_{1}}\llbracket {\begingroup\renewcommand\colorMATH{\colorMATHB}\renewcommand\colorSYNTAX{\colorSYNTAXB}{{\color{\colorMATH}\ensuremath{\Distance'}}}\endgroup }(\tau _{1})\rrbracket }}} and {{\color{\colorMATH}\ensuremath{(\gamma _{1}\vdash {\begingroup\renewcommand\colorMATH{\colorMATHB}\renewcommand\colorSYNTAX{\colorSYNTAXB}{{\color{\colorMATH}\ensuremath{\se_{2}}}}\endgroup },\gamma _{2}\vdash {\begingroup\renewcommand\colorMATH{\colorMATHB}\renewcommand\colorSYNTAX{\colorSYNTAXB}{{\color{\colorMATH}\ensuremath{\se_{2}}}}\endgroup }) \in  {\mathcal{E}}^{k}_{d'_{2} + d''_{2}}\llbracket {\begingroup\renewcommand\colorMATH{\colorMATHB}\renewcommand\colorSYNTAX{\colorSYNTAXB}{{\color{\colorMATH}\ensuremath{\Distance'}}}\endgroup }(\tau _{2})\rrbracket }}} respectively. This means that
      if {{\color{\colorMATH}\ensuremath{\gamma _{1}\vdash {\begingroup\renewcommand\colorMATH{\colorMATHB}\renewcommand\colorSYNTAX{\colorSYNTAXB}{{\color{\colorMATH}\ensuremath{\se_{1}}}}\endgroup } \Downarrow ^{j_{1}} {\begingroup\renewcommand\colorMATH{\colorMATHB}\renewcommand\colorSYNTAX{\colorSYNTAXB}{{\color{\colorMATH}\ensuremath{\sv'_{1 1}}}}\endgroup }}}}, \pthen {{\color{\colorMATH}\ensuremath{\gamma _{2}\vdash {\begingroup\renewcommand\colorMATH{\colorMATHB}\renewcommand\colorSYNTAX{\colorSYNTAXB}{{\color{\colorMATH}\ensuremath{\se_{1}}}}\endgroup } \Downarrow ^{\pj[1]} {\begingroup\renewcommand\colorMATH{\colorMATHB}\renewcommand\colorSYNTAX{\colorSYNTAXB}{{\color{\colorMATH}\ensuremath{\sv'_{1 2}}}}\endgroup }}}} \pand 
      {{\color{\colorMATH}\ensuremath{({\begingroup\renewcommand\colorMATH{\colorMATHB}\renewcommand\colorSYNTAX{\colorSYNTAXB}{{\color{\colorMATH}\ensuremath{\sv'_{1 1}}}}\endgroup }, {\begingroup\renewcommand\colorMATH{\colorMATHB}\renewcommand\colorSYNTAX{\colorSYNTAXB}{{\color{\colorMATH}\ensuremath{\sv'_{1 2}}}}\endgroup }) \in  {\mathcal{V}}^{k-j_{1}}_{d'_{1} + d''_{1}}\llbracket {\begingroup\renewcommand\colorMATH{\colorMATHB}\renewcommand\colorSYNTAX{\colorSYNTAXB}{{\color{\colorMATH}\ensuremath{\Distance'}}}\endgroup }(\tau _{1})\rrbracket }}}, and that 
      if {{\color{\colorMATH}\ensuremath{\gamma _{1}\vdash {\begingroup\renewcommand\colorMATH{\colorMATHB}\renewcommand\colorSYNTAX{\colorSYNTAXB}{{\color{\colorMATH}\ensuremath{\se_{2}}}}\endgroup } \Downarrow ^{j_{2}} {\begingroup\renewcommand\colorMATH{\colorMATHB}\renewcommand\colorSYNTAX{\colorSYNTAXB}{{\color{\colorMATH}\ensuremath{\sv'_{2 1}}}}\endgroup }}}}, \pthen {{\color{\colorMATH}\ensuremath{\gamma _{2}\vdash {\begingroup\renewcommand\colorMATH{\colorMATHB}\renewcommand\colorSYNTAX{\colorSYNTAXB}{{\color{\colorMATH}\ensuremath{\se_{2}}}}\endgroup } \Downarrow ^{\pj[2]} {\begingroup\renewcommand\colorMATH{\colorMATHB}\renewcommand\colorSYNTAX{\colorSYNTAXB}{{\color{\colorMATH}\ensuremath{\sv'_{2 2}}}}\endgroup }}}} \pand 
      {{\color{\colorMATH}\ensuremath{({\begingroup\renewcommand\colorMATH{\colorMATHB}\renewcommand\colorSYNTAX{\colorSYNTAXB}{{\color{\colorMATH}\ensuremath{\sv'_{2 1}}}}\endgroup }, {\begingroup\renewcommand\colorMATH{\colorMATHB}\renewcommand\colorSYNTAX{\colorSYNTAXB}{{\color{\colorMATH}\ensuremath{\sv'_{2 2}}}}\endgroup }) \in  {\mathcal{V}}^{k-j_{2}}_{d'_{2} + d''_{2}}\llbracket {\begingroup\renewcommand\colorMATH{\colorMATHB}\renewcommand\colorSYNTAX{\colorSYNTAXB}{{\color{\colorMATH}\ensuremath{\Distance'}}}\endgroup }(\tau _{2})\rrbracket }}}. 
      Notice that {{\color{\colorMATH}\ensuremath{d''_{i} \leq  \addProd{d''_{1}}{d''_{2}}}}}.
      As reduction is deterministic, then {{\color{\colorMATH}\ensuremath{j = j_{1}+j_{2}}}} and {{\color{\colorMATH}\ensuremath{{\begingroup\renewcommand\colorMATH{\colorMATHB}\renewcommand\colorSYNTAX{\colorSYNTAXB}{{\color{\colorMATH}\ensuremath{\sv'_{i j}}}}\endgroup } = {\begingroup\renewcommand\colorMATH{\colorMATHB}\renewcommand\colorSYNTAX{\colorSYNTAXB}{{\color{\colorMATH}\ensuremath{\sv_{i j}}}}\endgroup }}}}, therefore as {{\color{\colorMATH}\ensuremath{0 + d'_{i} + d''_{i} = d'_{i} + d''_{i}}}}, the result holds immediately by Lemma~\ref{lm:weakening-index}.

      \end{subproof}
    \item  {{\color{\colorMATH}\ensuremath{\Gamma ; {\begingroup\renewcommand\colorMATH{\colorMATHB}\renewcommand\colorSYNTAX{\colorSYNTAXB}{{\color{\colorMATH}\ensuremath{\Distance}}}\endgroup } \vdash  \fst\hspace*{0.33em}{\begingroup\renewcommand\colorMATH{\colorMATHB}\renewcommand\colorSYNTAX{\colorSYNTAXB}{{\color{\colorMATH}\ensuremath{\se'}}}\endgroup } \mathrel{:} \tau _{1} \mathrel{;} {\begingroup\renewcommand\colorMATH{\colorMATHB}\renewcommand\colorSYNTAX{\colorSYNTAXB}{{\color{\colorMATH}\ensuremath{\sS''}}}\endgroup } + {\begingroup\renewcommand\colorMATH{\colorMATHB}\renewcommand\colorSYNTAX{\colorSYNTAXB}{{\color{\colorMATH}\ensuremath{\sS_{1}}}}\endgroup }}}}
      \begin{subproof} 
        We have to prove that for any {{\color{\colorMATH}\ensuremath{k}}}, {{\color{\colorMATH}\ensuremath{\forall  (\gamma _{1},\gamma _{2}) \in  {\mathcal{G}}_{{\begingroup\renewcommand\colorMATH{\colorMATHB}\renewcommand\colorSYNTAX{\colorSYNTAXB}{{\color{\colorMATH}\ensuremath{\Distance'}}}\endgroup }}^{\kg}\llbracket \Gamma \rrbracket , (\gamma _{1}\vdash \fst\hspace*{0.33em}{\begingroup\renewcommand\colorMATH{\colorMATHB}\renewcommand\colorSYNTAX{\colorSYNTAXB}{{\color{\colorMATH}\ensuremath{\se'}}}\endgroup },\gamma _{2}\vdash \fst\hspace*{0.33em}{\begingroup\renewcommand\colorMATH{\colorMATHB}\renewcommand\colorSYNTAX{\colorSYNTAXB}{{\color{\colorMATH}\ensuremath{\se'}}}\endgroup }) \in  {\mathcal{E}}^{k}_{{\begingroup\renewcommand\colorMATH{\colorMATHB}\renewcommand\colorSYNTAX{\colorSYNTAXB}{{\color{\colorMATH}\ensuremath{\Distance'}}}\endgroup }\mathord{\cdotp }({\begingroup\renewcommand\colorMATH{\colorMATHB}\renewcommand\colorSYNTAX{\colorSYNTAXB}{{\color{\colorMATH}\ensuremath{\sS''}}}\endgroup } + {\begingroup\renewcommand\colorMATH{\colorMATHB}\renewcommand\colorSYNTAX{\colorSYNTAXB}{{\color{\colorMATH}\ensuremath{\sS_{1}}}}\endgroup })}\llbracket {\begingroup\renewcommand\colorMATH{\colorMATHB}\renewcommand\colorSYNTAX{\colorSYNTAXB}{{\color{\colorMATH}\ensuremath{\Distance'}}}\endgroup }(\tau _{1})\rrbracket }}}, for {{\color{\colorMATH}\ensuremath{{\begingroup\renewcommand\colorMATH{\colorMATHB}\renewcommand\colorSYNTAX{\colorSYNTAXB}{{\color{\colorMATH}\ensuremath{\Distance'}}}\endgroup } \sqsubseteq  {\begingroup\renewcommand\colorMATH{\colorMATHB}\renewcommand\colorSYNTAX{\colorSYNTAXB}{{\color{\colorMATH}\ensuremath{\Distance}}}\endgroup }}}}.
        By induction hypothesis on {{\color{\colorMATH}\ensuremath{\Gamma  \vdash  {\begingroup\renewcommand\colorMATH{\colorMATHB}\renewcommand\colorSYNTAX{\colorSYNTAXB}{{\color{\colorMATH}\ensuremath{\se'}}}\endgroup } \mathrel{:} \tau _{1} \mathrel{^{{\begingroup\renewcommand\colorMATH{\colorMATHB}\renewcommand\colorSYNTAX{\colorSYNTAXB}{{\color{\colorMATH}\ensuremath{\sS_{1}}}}\endgroup }}\&^{{\begingroup\renewcommand\colorMATH{\colorMATHB}\renewcommand\colorSYNTAX{\colorSYNTAXB}{{\color{\colorMATH}\ensuremath{\sS_{2}}}}\endgroup }}} \tau _{2} \mathrel{;} {\begingroup\renewcommand\colorMATH{\colorMATHB}\renewcommand\colorSYNTAX{\colorSYNTAXB}{{\color{\colorMATH}\ensuremath{\sS''}}}\endgroup }}}} we know that\\
        {{\color{\colorMATH}\ensuremath{(\gamma _{1}\vdash {\begingroup\renewcommand\colorMATH{\colorMATHB}\renewcommand\colorSYNTAX{\colorSYNTAXB}{{\color{\colorMATH}\ensuremath{\se'}}}\endgroup },\gamma _{2}\vdash {\begingroup\renewcommand\colorMATH{\colorMATHB}\renewcommand\colorSYNTAX{\colorSYNTAXB}{{\color{\colorMATH}\ensuremath{\se'}}}\endgroup }) \in  {\mathcal{E}}_{{\begingroup\renewcommand\colorMATH{\colorMATHB}\renewcommand\colorSYNTAX{\colorSYNTAXB}{{\color{\colorMATH}\ensuremath{\Distance'}}}\endgroup }\mathord{\cdotp }{\begingroup\renewcommand\colorMATH{\colorMATHB}\renewcommand\colorSYNTAX{\colorSYNTAXB}{{\color{\colorMATH}\ensuremath{\sS''}}}\endgroup }}^{k}\llbracket {\begingroup\renewcommand\colorMATH{\colorMATHB}\renewcommand\colorSYNTAX{\colorSYNTAXB}{{\color{\colorMATH}\ensuremath{\Distance'}}}\endgroup }(\tau _{1} \mathrel{^{{\begingroup\renewcommand\colorMATH{\colorMATHB}\renewcommand\colorSYNTAX{\colorSYNTAXB}{{\color{\colorMATH}\ensuremath{\sS_{1}}}}\endgroup }}\&^{{\begingroup\renewcommand\colorMATH{\colorMATHB}\renewcommand\colorSYNTAX{\colorSYNTAXB}{{\color{\colorMATH}\ensuremath{\sS_{2}}}}\endgroup }}} \tau _{2})\rrbracket }}}, i.e.
        if {{\color{\colorMATH}\ensuremath{\gamma _{1}\vdash {\begingroup\renewcommand\colorMATH{\colorMATHB}\renewcommand\colorSYNTAX{\colorSYNTAXB}{{\color{\colorMATH}\ensuremath{\se'}}}\endgroup } \Downarrow ^{j} \addProduct{{\begingroup\renewcommand\colorMATH{\colorMATHB}\renewcommand\colorSYNTAX{\colorSYNTAXB}{{\color{\colorMATH}\ensuremath{\sv_{1 1}}}}\endgroup }}{{\begingroup\renewcommand\colorMATH{\colorMATHB}\renewcommand\colorSYNTAX{\colorSYNTAXB}{{\color{\colorMATH}\ensuremath{\sv_{1 2}}}}\endgroup }}}}} \pthen {{\color{\colorMATH}\ensuremath{\gamma _{2}\vdash {\begingroup\renewcommand\colorMATH{\colorMATHB}\renewcommand\colorSYNTAX{\colorSYNTAXB}{{\color{\colorMATH}\ensuremath{\se'}}}\endgroup } \Downarrow ^{\pj} \addProduct{{\begingroup\renewcommand\colorMATH{\colorMATHB}\renewcommand\colorSYNTAX{\colorSYNTAXB}{{\color{\colorMATH}\ensuremath{\sv_{2 1}}}}\endgroup }}{{\begingroup\renewcommand\colorMATH{\colorMATHB}\renewcommand\colorSYNTAX{\colorSYNTAXB}{{\color{\colorMATH}\ensuremath{\sv_{2 2}}}}\endgroup }}}}}, \pand 
        {{\color{\colorMATH}\ensuremath{(\addProduct{{\begingroup\renewcommand\colorMATH{\colorMATHB}\renewcommand\colorSYNTAX{\colorSYNTAXB}{{\color{\colorMATH}\ensuremath{\sv_{1 1}}}}\endgroup }}{{\begingroup\renewcommand\colorMATH{\colorMATHB}\renewcommand\colorSYNTAX{\colorSYNTAXB}{{\color{\colorMATH}\ensuremath{\sv_{1 2}}}}\endgroup }}, \addProduct{{\begingroup\renewcommand\colorMATH{\colorMATHB}\renewcommand\colorSYNTAX{\colorSYNTAXB}{{\color{\colorMATH}\ensuremath{\sv_{2 1}}}}\endgroup }}{{\begingroup\renewcommand\colorMATH{\colorMATHB}\renewcommand\colorSYNTAX{\colorSYNTAXB}{{\color{\colorMATH}\ensuremath{\sv_{2 2}}}}\endgroup }}) \in  {\mathcal{V}}_{{\begingroup\renewcommand\colorMATH{\colorMATHB}\renewcommand\colorSYNTAX{\colorSYNTAXB}{{\color{\colorMATH}\ensuremath{\Distance'}}}\endgroup }\mathord{\cdotp }{\begingroup\renewcommand\colorMATH{\colorMATHB}\renewcommand\colorSYNTAX{\colorSYNTAXB}{{\color{\colorMATH}\ensuremath{\sS''}}}\endgroup }}^{k-j}\llbracket {\begingroup\renewcommand\colorMATH{\colorMATHB}\renewcommand\colorSYNTAX{\colorSYNTAXB}{{\color{\colorMATH}\ensuremath{\Distance'}}}\endgroup }(\tau _{1}) \mathrel{^{{\begingroup\renewcommand\colorMATH{\colorMATHB}\renewcommand\colorSYNTAX{\colorSYNTAXB}{{\color{\colorMATH}\ensuremath{\Distance'}}}\endgroup }\mathord{\cdotp }{\begingroup\renewcommand\colorMATH{\colorMATHB}\renewcommand\colorSYNTAX{\colorSYNTAXB}{{\color{\colorMATH}\ensuremath{\sS_{1}}}}\endgroup }}\&^{{\begingroup\renewcommand\colorMATH{\colorMATHB}\renewcommand\colorSYNTAX{\colorSYNTAXB}{{\color{\colorMATH}\ensuremath{\Distance'}}}\endgroup }\mathord{\cdotp }{\begingroup\renewcommand\colorMATH{\colorMATHB}\renewcommand\colorSYNTAX{\colorSYNTAXB}{{\color{\colorMATH}\ensuremath{\sS_{2}}}}\endgroup }}} {\begingroup\renewcommand\colorMATH{\colorMATHB}\renewcommand\colorSYNTAX{\colorSYNTAXB}{{\color{\colorMATH}\ensuremath{\Distance'}}}\endgroup }(\tau _{2})\rrbracket }}}, or equivalently
        {{\color{\colorMATH}\ensuremath{({\begingroup\renewcommand\colorMATH{\colorMATHB}\renewcommand\colorSYNTAX{\colorSYNTAXB}{{\color{\colorMATH}\ensuremath{\sv_{1 1}}}}\endgroup },,{\begingroup\renewcommand\colorMATH{\colorMATHB}\renewcommand\colorSYNTAX{\colorSYNTAXB}{{\color{\colorMATH}\ensuremath{\sv_{2 1}}}}\endgroup }) \in  {\mathcal{V}}^{k-j}_{{\begingroup\renewcommand\colorMATH{\colorMATHB}\renewcommand\colorSYNTAX{\colorSYNTAXB}{{\color{\colorMATH}\ensuremath{\Distance'}}}\endgroup }\mathord{\cdotp }{\begingroup\renewcommand\colorMATH{\colorMATHB}\renewcommand\colorSYNTAX{\colorSYNTAXB}{{\color{\colorMATH}\ensuremath{\sS''}}}\endgroup }+{\begingroup\renewcommand\colorMATH{\colorMATHB}\renewcommand\colorSYNTAX{\colorSYNTAXB}{{\color{\colorMATH}\ensuremath{\Distance'}}}\endgroup }\mathord{\cdotp }{\begingroup\renewcommand\colorMATH{\colorMATHB}\renewcommand\colorSYNTAX{\colorSYNTAXB}{{\color{\colorMATH}\ensuremath{\sS_{1}}}}\endgroup }}\llbracket {\begingroup\renewcommand\colorMATH{\colorMATHB}\renewcommand\colorSYNTAX{\colorSYNTAXB}{{\color{\colorMATH}\ensuremath{\Distance'}}}\endgroup }(\tau _{1})\rrbracket }}}, and {{\color{\colorMATH}\ensuremath{({\begingroup\renewcommand\colorMATH{\colorMATHB}\renewcommand\colorSYNTAX{\colorSYNTAXB}{{\color{\colorMATH}\ensuremath{\sv_{1 2}}}}\endgroup },,{\begingroup\renewcommand\colorMATH{\colorMATHB}\renewcommand\colorSYNTAX{\colorSYNTAXB}{{\color{\colorMATH}\ensuremath{\sv_{2 2}}}}\endgroup }) \in  {\mathcal{V}}^{k-j}_{{\begingroup\renewcommand\colorMATH{\colorMATHB}\renewcommand\colorSYNTAX{\colorSYNTAXB}{{\color{\colorMATH}\ensuremath{\Distance'}}}\endgroup }\mathord{\cdotp }{\begingroup\renewcommand\colorMATH{\colorMATHB}\renewcommand\colorSYNTAX{\colorSYNTAXB}{{\color{\colorMATH}\ensuremath{\sS''}}}\endgroup }+{\begingroup\renewcommand\colorMATH{\colorMATHB}\renewcommand\colorSYNTAX{\colorSYNTAXB}{{\color{\colorMATH}\ensuremath{\Distance'}}}\endgroup }\mathord{\cdotp }{\begingroup\renewcommand\colorMATH{\colorMATHB}\renewcommand\colorSYNTAX{\colorSYNTAXB}{{\color{\colorMATH}\ensuremath{\sS_{2}}}}\endgroup }}\llbracket {\begingroup\renewcommand\colorMATH{\colorMATHB}\renewcommand\colorSYNTAX{\colorSYNTAXB}{{\color{\colorMATH}\ensuremath{\Distance'}}}\endgroup }(\tau _{2})\rrbracket }}}.

        Following the {\textsc{ proj1}} reduction rule, if:
        \begingroup\color{\colorMATH}\begin{gather*} 
          \inferrule*[lab=
          ]{ \gamma _{1}\vdash {\begingroup\renewcommand\colorMATH{\colorMATHB}\renewcommand\colorSYNTAX{\colorSYNTAXB}{{\color{\colorMATH}\ensuremath{\se'}}}\endgroup } \Downarrow ^{j} \addProduct{{\begingroup\renewcommand\colorMATH{\colorMATHB}\renewcommand\colorSYNTAX{\colorSYNTAXB}{{\color{\colorMATH}\ensuremath{\sv_{1 1}}}}\endgroup }}{{\begingroup\renewcommand\colorMATH{\colorMATHB}\renewcommand\colorSYNTAX{\colorSYNTAXB}{{\color{\colorMATH}\ensuremath{\sv_{1 2}}}}\endgroup }}
            }{
            \gamma _{1}\vdash \fst\hspace*{0.33em}{\begingroup\renewcommand\colorMATH{\colorMATHB}\renewcommand\colorSYNTAX{\colorSYNTAXB}{{\color{\colorMATH}\ensuremath{\se'}}}\endgroup } \Downarrow ^{j} {\begingroup\renewcommand\colorMATH{\colorMATHB}\renewcommand\colorSYNTAX{\colorSYNTAXB}{{\color{\colorMATH}\ensuremath{\sv_{1 1}}}}\endgroup } 
          }
        \end{gather*}\endgroup
        and
        \begingroup\color{\colorMATH}\begin{gather*} 
          \inferrule*[lab=
          ]{ \gamma _{2}\vdash {\begingroup\renewcommand\colorMATH{\colorMATHB}\renewcommand\colorSYNTAX{\colorSYNTAXB}{{\color{\colorMATH}\ensuremath{\se'}}}\endgroup } \Downarrow ^{\pj} \addProduct{{\begingroup\renewcommand\colorMATH{\colorMATHB}\renewcommand\colorSYNTAX{\colorSYNTAXB}{{\color{\colorMATH}\ensuremath{\sv_{2 1}}}}\endgroup }}{{\begingroup\renewcommand\colorMATH{\colorMATHB}\renewcommand\colorSYNTAX{\colorSYNTAXB}{{\color{\colorMATH}\ensuremath{\sv_{2 2}}}}\endgroup }}
            }{
            \gamma _{2}\vdash \fst\hspace*{0.33em}{\begingroup\renewcommand\colorMATH{\colorMATHB}\renewcommand\colorSYNTAX{\colorSYNTAXB}{{\color{\colorMATH}\ensuremath{\se'}}}\endgroup } \Downarrow ^{\pj} {\begingroup\renewcommand\colorMATH{\colorMATHB}\renewcommand\colorSYNTAX{\colorSYNTAXB}{{\color{\colorMATH}\ensuremath{\sv_{2 1}}}}\endgroup } 
          }
        \end{gather*}\endgroup
        Then we have to prove that {{\color{\colorMATH}\ensuremath{({\begingroup\renewcommand\colorMATH{\colorMATHB}\renewcommand\colorSYNTAX{\colorSYNTAXB}{{\color{\colorMATH}\ensuremath{\sv_{1 1}}}}\endgroup }, {\begingroup\renewcommand\colorMATH{\colorMATHB}\renewcommand\colorSYNTAX{\colorSYNTAXB}{{\color{\colorMATH}\ensuremath{\sv_{2 1}}}}\endgroup }) \in  {\mathcal{V}}^{k-j}_{{\begingroup\renewcommand\colorMATH{\colorMATHB}\renewcommand\colorSYNTAX{\colorSYNTAXB}{{\color{\colorMATH}\ensuremath{\Distance'}}}\endgroup }\mathord{\cdotp }({\begingroup\renewcommand\colorMATH{\colorMATHB}\renewcommand\colorSYNTAX{\colorSYNTAXB}{{\color{\colorMATH}\ensuremath{\sS''}}}\endgroup } + {\begingroup\renewcommand\colorMATH{\colorMATHB}\renewcommand\colorSYNTAX{\colorSYNTAXB}{{\color{\colorMATH}\ensuremath{\sS_{1}}}}\endgroup })}\llbracket {\begingroup\renewcommand\colorMATH{\colorMATHB}\renewcommand\colorSYNTAX{\colorSYNTAXB}{{\color{\colorMATH}\ensuremath{\Distance'}}}\endgroup }(\tau _{1})\rrbracket }}}, but as by Lemma~\ref{lm:associativity-inst}, {{\color{\colorMATH}\ensuremath{{\begingroup\renewcommand\colorMATH{\colorMATHB}\renewcommand\colorSYNTAX{\colorSYNTAXB}{{\color{\colorMATH}\ensuremath{\Distance'}}}\endgroup }\mathord{\cdotp }({\begingroup\renewcommand\colorMATH{\colorMATHB}\renewcommand\colorSYNTAX{\colorSYNTAXB}{{\color{\colorMATH}\ensuremath{\sS''}}}\endgroup } + {\begingroup\renewcommand\colorMATH{\colorMATHB}\renewcommand\colorSYNTAX{\colorSYNTAXB}{{\color{\colorMATH}\ensuremath{\sS_{1}}}}\endgroup }) = {\begingroup\renewcommand\colorMATH{\colorMATHB}\renewcommand\colorSYNTAX{\colorSYNTAXB}{{\color{\colorMATH}\ensuremath{\Distance'}}}\endgroup }\mathord{\cdotp }{\begingroup\renewcommand\colorMATH{\colorMATHB}\renewcommand\colorSYNTAX{\colorSYNTAXB}{{\color{\colorMATH}\ensuremath{\sS''}}}\endgroup }+{\begingroup\renewcommand\colorMATH{\colorMATHB}\renewcommand\colorSYNTAX{\colorSYNTAXB}{{\color{\colorMATH}\ensuremath{\Distance'}}}\endgroup }\mathord{\cdotp }{\begingroup\renewcommand\colorMATH{\colorMATHB}\renewcommand\colorSYNTAX{\colorSYNTAXB}{{\color{\colorMATH}\ensuremath{\sS_{1}}}}\endgroup }}}}, the result holds immediately.
      \end{subproof}
    \item  {{\color{\colorMATH}\ensuremath{\snd\hspace*{0.33em}{\begingroup\renewcommand\colorMATH{\colorMATHB}\renewcommand\colorSYNTAX{\colorSYNTAXB}{{\color{\colorMATH}\ensuremath{\se'}}}\endgroup } \mathrel{:} \tau _{1} \mathrel{;} {\begingroup\renewcommand\colorMATH{\colorMATHB}\renewcommand\colorSYNTAX{\colorSYNTAXB}{{\color{\colorMATH}\ensuremath{\sS''}}}\endgroup } + {\begingroup\renewcommand\colorMATH{\colorMATHB}\renewcommand\colorSYNTAX{\colorSYNTAXB}{{\color{\colorMATH}\ensuremath{\sS_{2}}}}\endgroup }}}}
      \begin{subproof} 
        Analogous to previous case. 
      \end{subproof}
    \item  {{\color{\colorMATH}\ensuremath{\Gamma ; {\begingroup\renewcommand\colorMATH{\colorMATHB}\renewcommand\colorSYNTAX{\colorSYNTAXB}{{\color{\colorMATH}\ensuremath{\Distance}}}\endgroup } \vdash  \langle {\begingroup\renewcommand\colorMATH{\colorMATHB}\renewcommand\colorSYNTAX{\colorSYNTAXB}{{\color{\colorMATH}\ensuremath{\se_{1}}}}\endgroup },{\begingroup\renewcommand\colorMATH{\colorMATHB}\renewcommand\colorSYNTAX{\colorSYNTAXB}{{\color{\colorMATH}\ensuremath{\se_{2}}}}\endgroup }\rangle  \mathrel{:} \tau _{1} \mathrel{^{{\begingroup\renewcommand\colorMATH{\colorMATHB}\renewcommand\colorSYNTAX{\colorSYNTAXB}{{\color{\colorMATH}\ensuremath{\sS_{1}}}}\endgroup }}\otimes ^{{\begingroup\renewcommand\colorMATH{\colorMATHB}\renewcommand\colorSYNTAX{\colorSYNTAXB}{{\color{\colorMATH}\ensuremath{\sS_{2}}}}\endgroup }}} \tau _{2} \mathrel{;} \varnothing }}}
      \begin{subproof} 
      We have to prove that {{\color{\colorMATH}\ensuremath{\forall  (\gamma _{1},\gamma _{2}) \in  {\mathcal{G}}_{{\begingroup\renewcommand\colorMATH{\colorMATHB}\renewcommand\colorSYNTAX{\colorSYNTAXB}{{\color{\colorMATH}\ensuremath{\Distance'}}}\endgroup }}^{\kg}\llbracket \Gamma \rrbracket , (\gamma _{1}\vdash \langle {\begingroup\renewcommand\colorMATH{\colorMATHB}\renewcommand\colorSYNTAX{\colorSYNTAXB}{{\color{\colorMATH}\ensuremath{\se_{1}}}}\endgroup },{\begingroup\renewcommand\colorMATH{\colorMATHB}\renewcommand\colorSYNTAX{\colorSYNTAXB}{{\color{\colorMATH}\ensuremath{\se_{2}}}}\endgroup }\rangle ,\gamma _{2}\vdash \langle {\begingroup\renewcommand\colorMATH{\colorMATHB}\renewcommand\colorSYNTAX{\colorSYNTAXB}{{\color{\colorMATH}\ensuremath{\se_{1}}}}\endgroup },{\begingroup\renewcommand\colorMATH{\colorMATHB}\renewcommand\colorSYNTAX{\colorSYNTAXB}{{\color{\colorMATH}\ensuremath{\se_{2}}}}\endgroup }\rangle ) \in  {\mathcal{E}}_{{\begingroup\renewcommand\colorMATH{\colorMATHB}\renewcommand\colorSYNTAX{\colorSYNTAXB}{{\color{\colorMATH}\ensuremath{\Distance'}}}\endgroup }\mathord{\cdotp }\varnothing }^{k}\llbracket {\begingroup\renewcommand\colorMATH{\colorMATHB}\renewcommand\colorSYNTAX{\colorSYNTAXB}{{\color{\colorMATH}\ensuremath{\Distance'}}}\endgroup }(\tau _{1} \mathrel{^{{\begingroup\renewcommand\colorMATH{\colorMATHB}\renewcommand\colorSYNTAX{\colorSYNTAXB}{{\color{\colorMATH}\ensuremath{\sS_{1}}}}\endgroup }}\otimes ^{{\begingroup\renewcommand\colorMATH{\colorMATHB}\renewcommand\colorSYNTAX{\colorSYNTAXB}{{\color{\colorMATH}\ensuremath{\sS_{2}}}}\endgroup }}} \tau _{2})\rrbracket }}}, for {{\color{\colorMATH}\ensuremath{{\begingroup\renewcommand\colorMATH{\colorMATHB}\renewcommand\colorSYNTAX{\colorSYNTAXB}{{\color{\colorMATH}\ensuremath{\Distance'}}}\endgroup } \sqsubseteq  {\begingroup\renewcommand\colorMATH{\colorMATHB}\renewcommand\colorSYNTAX{\colorSYNTAXB}{{\color{\colorMATH}\ensuremath{\Distance}}}\endgroup }}}}.
      Notice that {{\color{\colorMATH}\ensuremath{{\begingroup\renewcommand\colorMATH{\colorMATHB}\renewcommand\colorSYNTAX{\colorSYNTAXB}{{\color{\colorMATH}\ensuremath{\Distance'}}}\endgroup }\mathord{\cdotp }\varnothing  = 0}}}, and {{\color{\colorMATH}\ensuremath{{\begingroup\renewcommand\colorMATH{\colorMATHB}\renewcommand\colorSYNTAX{\colorSYNTAXB}{{\color{\colorMATH}\ensuremath{\Distance'}}}\endgroup }(\tau _{1} \mathrel{^{{\begingroup\renewcommand\colorMATH{\colorMATHB}\renewcommand\colorSYNTAX{\colorSYNTAXB}{{\color{\colorMATH}\ensuremath{\sS_{1}}}}\endgroup }}\otimes ^{{\begingroup\renewcommand\colorMATH{\colorMATHB}\renewcommand\colorSYNTAX{\colorSYNTAXB}{{\color{\colorMATH}\ensuremath{\sS_{2}}}}\endgroup }}} \tau _{2}) = {\begingroup\renewcommand\colorMATH{\colorMATHB}\renewcommand\colorSYNTAX{\colorSYNTAXB}{{\color{\colorMATH}\ensuremath{\Distance'}}}\endgroup }(\tau _{1}) \mathrel{^{{\begingroup\renewcommand\colorMATH{\colorMATHB}\renewcommand\colorSYNTAX{\colorSYNTAXB}{{\color{\colorMATH}\ensuremath{\Distance'}}}\endgroup }\mathord{\cdotp }{\begingroup\renewcommand\colorMATH{\colorMATHB}\renewcommand\colorSYNTAX{\colorSYNTAXB}{{\color{\colorMATH}\ensuremath{\sS_{1}}}}\endgroup }}\otimes ^{{\begingroup\renewcommand\colorMATH{\colorMATHB}\renewcommand\colorSYNTAX{\colorSYNTAXB}{{\color{\colorMATH}\ensuremath{\Distance'}}}\endgroup }\mathord{\cdotp }{\begingroup\renewcommand\colorMATH{\colorMATHB}\renewcommand\colorSYNTAX{\colorSYNTAXB}{{\color{\colorMATH}\ensuremath{\sS_{2}}}}\endgroup }}} {\begingroup\renewcommand\colorMATH{\colorMATHB}\renewcommand\colorSYNTAX{\colorSYNTAXB}{{\color{\colorMATH}\ensuremath{\Distance'}}}\endgroup }(\tau _{2})}}}, then we have to prove that\\
      {{\color{\colorMATH}\ensuremath{(\gamma _{1}\vdash \langle {\begingroup\renewcommand\colorMATH{\colorMATHB}\renewcommand\colorSYNTAX{\colorSYNTAXB}{{\color{\colorMATH}\ensuremath{\se_{1}}}}\endgroup },{\begingroup\renewcommand\colorMATH{\colorMATHB}\renewcommand\colorSYNTAX{\colorSYNTAXB}{{\color{\colorMATH}\ensuremath{\se_{2}}}}\endgroup }\rangle ,\gamma _{2}\vdash \langle {\begingroup\renewcommand\colorMATH{\colorMATHB}\renewcommand\colorSYNTAX{\colorSYNTAXB}{{\color{\colorMATH}\ensuremath{\se_{1}}}}\endgroup },{\begingroup\renewcommand\colorMATH{\colorMATHB}\renewcommand\colorSYNTAX{\colorSYNTAXB}{{\color{\colorMATH}\ensuremath{\se_{2}}}}\endgroup }\rangle ) \in  {\mathcal{E}}_{0}^{k}\llbracket {\begingroup\renewcommand\colorMATH{\colorMATHB}\renewcommand\colorSYNTAX{\colorSYNTAXB}{{\color{\colorMATH}\ensuremath{\Distance'}}}\endgroup }(\tau _{1}) \mathrel{^{{\begingroup\renewcommand\colorMATH{\colorMATHB}\renewcommand\colorSYNTAX{\colorSYNTAXB}{{\color{\colorMATH}\ensuremath{\Distance'}}}\endgroup }\mathord{\cdotp }{\begingroup\renewcommand\colorMATH{\colorMATHB}\renewcommand\colorSYNTAX{\colorSYNTAXB}{{\color{\colorMATH}\ensuremath{\sS_{1}}}}\endgroup }}\otimes ^{{\begingroup\renewcommand\colorMATH{\colorMATHB}\renewcommand\colorSYNTAX{\colorSYNTAXB}{{\color{\colorMATH}\ensuremath{\Distance'}}}\endgroup }\mathord{\cdotp }{\begingroup\renewcommand\colorMATH{\colorMATHB}\renewcommand\colorSYNTAX{\colorSYNTAXB}{{\color{\colorMATH}\ensuremath{\sS_{2}}}}\endgroup }}} {\begingroup\renewcommand\colorMATH{\colorMATHB}\renewcommand\colorSYNTAX{\colorSYNTAXB}{{\color{\colorMATH}\ensuremath{\Distance'}}}\endgroup }(\tau _{2})\rrbracket }}}, i.e.
      if {{\color{\colorMATH}\ensuremath{\gamma _{1}\vdash \langle {\begingroup\renewcommand\colorMATH{\colorMATHB}\renewcommand\colorSYNTAX{\colorSYNTAXB}{{\color{\colorMATH}\ensuremath{\se_{1}}}}\endgroup },{\begingroup\renewcommand\colorMATH{\colorMATHB}\renewcommand\colorSYNTAX{\colorSYNTAXB}{{\color{\colorMATH}\ensuremath{\se_{2}}}}\endgroup }\rangle  \Downarrow ^{j_{1}} \langle {\begingroup\renewcommand\colorMATH{\colorMATHB}\renewcommand\colorSYNTAX{\colorSYNTAXB}{{\color{\colorMATH}\ensuremath{\sv_{1 1}}}}\endgroup },{\begingroup\renewcommand\colorMATH{\colorMATHB}\renewcommand\colorSYNTAX{\colorSYNTAXB}{{\color{\colorMATH}\ensuremath{\sv_{1 2}}}}\endgroup }\rangle }}} \pthen {{\color{\colorMATH}\ensuremath{\gamma _{2}\vdash \langle {\begingroup\renewcommand\colorMATH{\colorMATHB}\renewcommand\colorSYNTAX{\colorSYNTAXB}{{\color{\colorMATH}\ensuremath{\se_{1}}}}\endgroup },{\begingroup\renewcommand\colorMATH{\colorMATHB}\renewcommand\colorSYNTAX{\colorSYNTAXB}{{\color{\colorMATH}\ensuremath{\se_{2}}}}\endgroup }\rangle  \Downarrow ^{\pj[1]} \langle {\begingroup\renewcommand\colorMATH{\colorMATHB}\renewcommand\colorSYNTAX{\colorSYNTAXB}{{\color{\colorMATH}\ensuremath{\sv_{2 1}}}}\endgroup },{\begingroup\renewcommand\colorMATH{\colorMATHB}\renewcommand\colorSYNTAX{\colorSYNTAXB}{{\color{\colorMATH}\ensuremath{\sv_{2 2}}}}\endgroup }\rangle }}}, \pand 
      {{\color{\colorMATH}\ensuremath{(\langle {\begingroup\renewcommand\colorMATH{\colorMATHB}\renewcommand\colorSYNTAX{\colorSYNTAXB}{{\color{\colorMATH}\ensuremath{\sv_{1 1}}}}\endgroup },{\begingroup\renewcommand\colorMATH{\colorMATHB}\renewcommand\colorSYNTAX{\colorSYNTAXB}{{\color{\colorMATH}\ensuremath{\sv_{1 2}}}}\endgroup }\rangle , \langle {\begingroup\renewcommand\colorMATH{\colorMATHB}\renewcommand\colorSYNTAX{\colorSYNTAXB}{{\color{\colorMATH}\ensuremath{\sv_{2 1}}}}\endgroup },{\begingroup\renewcommand\colorMATH{\colorMATHB}\renewcommand\colorSYNTAX{\colorSYNTAXB}{{\color{\colorMATH}\ensuremath{\sv_{2 2}}}}\endgroup }\rangle ) \in  {\mathcal{V}}_{0}^{k-j}\llbracket {\begingroup\renewcommand\colorMATH{\colorMATHB}\renewcommand\colorSYNTAX{\colorSYNTAXB}{{\color{\colorMATH}\ensuremath{\Distance'}}}\endgroup }(\tau _{1}) \mathrel{^{{\begingroup\renewcommand\colorMATH{\colorMATHB}\renewcommand\colorSYNTAX{\colorSYNTAXB}{{\color{\colorMATH}\ensuremath{\Distance'}}}\endgroup }\mathord{\cdotp }{\begingroup\renewcommand\colorMATH{\colorMATHB}\renewcommand\colorSYNTAX{\colorSYNTAXB}{{\color{\colorMATH}\ensuremath{\sS_{1}}}}\endgroup }}\otimes ^{{\begingroup\renewcommand\colorMATH{\colorMATHB}\renewcommand\colorSYNTAX{\colorSYNTAXB}{{\color{\colorMATH}\ensuremath{\Distance'}}}\endgroup }\mathord{\cdotp }{\begingroup\renewcommand\colorMATH{\colorMATHB}\renewcommand\colorSYNTAX{\colorSYNTAXB}{{\color{\colorMATH}\ensuremath{\sS_{2}}}}\endgroup }}} {\begingroup\renewcommand\colorMATH{\colorMATHB}\renewcommand\colorSYNTAX{\colorSYNTAXB}{{\color{\colorMATH}\ensuremath{\Distance'}}}\endgroup }(\tau _{2})\rrbracket }}}, or equivalently
      {{\color{\colorMATH}\ensuremath{({\begingroup\renewcommand\colorMATH{\colorMATHB}\renewcommand\colorSYNTAX{\colorSYNTAXB}{{\color{\colorMATH}\ensuremath{\sv_{1 1}}}}\endgroup },{\begingroup\renewcommand\colorMATH{\colorMATHB}\renewcommand\colorSYNTAX{\colorSYNTAXB}{{\color{\colorMATH}\ensuremath{\sv_{2 1}}}}\endgroup }) \in  {\mathcal{V}}^{k-j}_{0+{\begingroup\renewcommand\colorMATH{\colorMATHB}\renewcommand\colorSYNTAX{\colorSYNTAXB}{{\color{\colorMATH}\ensuremath{\Distance'}}}\endgroup }\mathord{\cdotp }{\begingroup\renewcommand\colorMATH{\colorMATHB}\renewcommand\colorSYNTAX{\colorSYNTAXB}{{\color{\colorMATH}\ensuremath{\sS_{1}}}}\endgroup }}\llbracket {\begingroup\renewcommand\colorMATH{\colorMATHB}\renewcommand\colorSYNTAX{\colorSYNTAXB}{{\color{\colorMATH}\ensuremath{\Distance'}}}\endgroup }(\tau _{1})\rrbracket }}}, and {{\color{\colorMATH}\ensuremath{({\begingroup\renewcommand\colorMATH{\colorMATHB}\renewcommand\colorSYNTAX{\colorSYNTAXB}{{\color{\colorMATH}\ensuremath{\sv_{1 2}}}}\endgroup },{\begingroup\renewcommand\colorMATH{\colorMATHB}\renewcommand\colorSYNTAX{\colorSYNTAXB}{{\color{\colorMATH}\ensuremath{\sv_{2 2}}}}\endgroup }) \in  {\mathcal{V}}^{k-j}_{0+{\begingroup\renewcommand\colorMATH{\colorMATHB}\renewcommand\colorSYNTAX{\colorSYNTAXB}{{\color{\colorMATH}\ensuremath{\Distance'}}}\endgroup }\mathord{\cdotp }{\begingroup\renewcommand\colorMATH{\colorMATHB}\renewcommand\colorSYNTAX{\colorSYNTAXB}{{\color{\colorMATH}\ensuremath{\sS_{2}}}}\endgroup }}\llbracket {\begingroup\renewcommand\colorMATH{\colorMATHB}\renewcommand\colorSYNTAX{\colorSYNTAXB}{{\color{\colorMATH}\ensuremath{\Distance'}}}\endgroup }(\tau _{2})\rrbracket }}}.

      By induction hypothesis on {{\color{\colorMATH}\ensuremath{\Gamma  \vdash  {\begingroup\renewcommand\colorMATH{\colorMATHB}\renewcommand\colorSYNTAX{\colorSYNTAXB}{{\color{\colorMATH}\ensuremath{\se_{1}}}}\endgroup } \mathrel{:} \tau _{1} \mathrel{;} {\begingroup\renewcommand\colorMATH{\colorMATHB}\renewcommand\colorSYNTAX{\colorSYNTAXB}{{\color{\colorMATH}\ensuremath{\sS_{1}}}}\endgroup }}}} and {{\color{\colorMATH}\ensuremath{\Gamma  \vdash  {\begingroup\renewcommand\colorMATH{\colorMATHB}\renewcommand\colorSYNTAX{\colorSYNTAXB}{{\color{\colorMATH}\ensuremath{\se_{2}}}}\endgroup } \mathrel{:} \tau _{2} \mathrel{;} {\begingroup\renewcommand\colorMATH{\colorMATHB}\renewcommand\colorSYNTAX{\colorSYNTAXB}{{\color{\colorMATH}\ensuremath{\sS_{2}}}}\endgroup }}}}, we know that 
      {{\color{\colorMATH}\ensuremath{(\gamma _{1}\vdash {\begingroup\renewcommand\colorMATH{\colorMATHB}\renewcommand\colorSYNTAX{\colorSYNTAXB}{{\color{\colorMATH}\ensuremath{\se_{1}}}}\endgroup },\gamma _{2}\vdash {\begingroup\renewcommand\colorMATH{\colorMATHB}\renewcommand\colorSYNTAX{\colorSYNTAXB}{{\color{\colorMATH}\ensuremath{\se_{1}}}}\endgroup }) \in  {\mathcal{E}}^{k}_{{\begingroup\renewcommand\colorMATH{\colorMATHB}\renewcommand\colorSYNTAX{\colorSYNTAXB}{{\color{\colorMATH}\ensuremath{\Distance'}}}\endgroup }\mathord{\cdotp }{\begingroup\renewcommand\colorMATH{\colorMATHB}\renewcommand\colorSYNTAX{\colorSYNTAXB}{{\color{\colorMATH}\ensuremath{\sS_{1}}}}\endgroup }}\llbracket {\begingroup\renewcommand\colorMATH{\colorMATHB}\renewcommand\colorSYNTAX{\colorSYNTAXB}{{\color{\colorMATH}\ensuremath{\Distance'}}}\endgroup }(\tau _{1})\rrbracket }}} and {{\color{\colorMATH}\ensuremath{(\gamma _{1}\vdash {\begingroup\renewcommand\colorMATH{\colorMATHB}\renewcommand\colorSYNTAX{\colorSYNTAXB}{{\color{\colorMATH}\ensuremath{\se_{2}}}}\endgroup },\gamma _{2}\vdash {\begingroup\renewcommand\colorMATH{\colorMATHB}\renewcommand\colorSYNTAX{\colorSYNTAXB}{{\color{\colorMATH}\ensuremath{\se_{2}}}}\endgroup }) \in  {\mathcal{E}}^{k}_{{\begingroup\renewcommand\colorMATH{\colorMATHB}\renewcommand\colorSYNTAX{\colorSYNTAXB}{{\color{\colorMATH}\ensuremath{\Distance'}}}\endgroup }\mathord{\cdotp }{\begingroup\renewcommand\colorMATH{\colorMATHB}\renewcommand\colorSYNTAX{\colorSYNTAXB}{{\color{\colorMATH}\ensuremath{\sS_{2}}}}\endgroup }}\llbracket {\begingroup\renewcommand\colorMATH{\colorMATHB}\renewcommand\colorSYNTAX{\colorSYNTAXB}{{\color{\colorMATH}\ensuremath{\Distance'}}}\endgroup }(\tau _{2})\rrbracket }}} respectively. This means that
      if {{\color{\colorMATH}\ensuremath{\gamma _{1}\vdash {\begingroup\renewcommand\colorMATH{\colorMATHB}\renewcommand\colorSYNTAX{\colorSYNTAXB}{{\color{\colorMATH}\ensuremath{\se_{1}}}}\endgroup } \Downarrow ^{j_{1}} {\begingroup\renewcommand\colorMATH{\colorMATHB}\renewcommand\colorSYNTAX{\colorSYNTAXB}{{\color{\colorMATH}\ensuremath{\sv'_{1 1}}}}\endgroup }}}}, \pthen {{\color{\colorMATH}\ensuremath{\gamma _{2}\vdash {\begingroup\renewcommand\colorMATH{\colorMATHB}\renewcommand\colorSYNTAX{\colorSYNTAXB}{{\color{\colorMATH}\ensuremath{\se_{1}}}}\endgroup } \Downarrow ^{\pj[1]} {\begingroup\renewcommand\colorMATH{\colorMATHB}\renewcommand\colorSYNTAX{\colorSYNTAXB}{{\color{\colorMATH}\ensuremath{\sv'_{1 2}}}}\endgroup }}}} \pand 
      {{\color{\colorMATH}\ensuremath{({\begingroup\renewcommand\colorMATH{\colorMATHB}\renewcommand\colorSYNTAX{\colorSYNTAXB}{{\color{\colorMATH}\ensuremath{\sv'_{1 1}}}}\endgroup }, {\begingroup\renewcommand\colorMATH{\colorMATHB}\renewcommand\colorSYNTAX{\colorSYNTAXB}{{\color{\colorMATH}\ensuremath{\sv'_{1 2}}}}\endgroup }) \in  {\mathcal{V}}^{k-j_{1}}_{{\begingroup\renewcommand\colorMATH{\colorMATHB}\renewcommand\colorSYNTAX{\colorSYNTAXB}{{\color{\colorMATH}\ensuremath{\Distance'}}}\endgroup }\mathord{\cdotp }{\begingroup\renewcommand\colorMATH{\colorMATHB}\renewcommand\colorSYNTAX{\colorSYNTAXB}{{\color{\colorMATH}\ensuremath{\sS_{1}}}}\endgroup }}\llbracket {\begingroup\renewcommand\colorMATH{\colorMATHB}\renewcommand\colorSYNTAX{\colorSYNTAXB}{{\color{\colorMATH}\ensuremath{\Distance'}}}\endgroup }(\tau _{1})\rrbracket }}}, and that 
      if {{\color{\colorMATH}\ensuremath{\gamma _{1}\vdash {\begingroup\renewcommand\colorMATH{\colorMATHB}\renewcommand\colorSYNTAX{\colorSYNTAXB}{{\color{\colorMATH}\ensuremath{\se_{2}}}}\endgroup } \Downarrow ^{j_{2}} {\begingroup\renewcommand\colorMATH{\colorMATHB}\renewcommand\colorSYNTAX{\colorSYNTAXB}{{\color{\colorMATH}\ensuremath{\sv'_{2 1}}}}\endgroup }}}}, \pthen {{\color{\colorMATH}\ensuremath{\gamma _{2}\vdash {\begingroup\renewcommand\colorMATH{\colorMATHB}\renewcommand\colorSYNTAX{\colorSYNTAXB}{{\color{\colorMATH}\ensuremath{\se_{2}}}}\endgroup } \Downarrow ^{\pj[2]} {\begingroup\renewcommand\colorMATH{\colorMATHB}\renewcommand\colorSYNTAX{\colorSYNTAXB}{{\color{\colorMATH}\ensuremath{\sv'_{2 2}}}}\endgroup }}}} \pand 
      {{\color{\colorMATH}\ensuremath{({\begingroup\renewcommand\colorMATH{\colorMATHB}\renewcommand\colorSYNTAX{\colorSYNTAXB}{{\color{\colorMATH}\ensuremath{\sv'_{2 1}}}}\endgroup }, {\begingroup\renewcommand\colorMATH{\colorMATHB}\renewcommand\colorSYNTAX{\colorSYNTAXB}{{\color{\colorMATH}\ensuremath{\sv'_{2 2}}}}\endgroup }) \in  {\mathcal{V}}^{k-j_{2}}_{{\begingroup\renewcommand\colorMATH{\colorMATHB}\renewcommand\colorSYNTAX{\colorSYNTAXB}{{\color{\colorMATH}\ensuremath{\Distance'}}}\endgroup }\mathord{\cdotp }{\begingroup\renewcommand\colorMATH{\colorMATHB}\renewcommand\colorSYNTAX{\colorSYNTAXB}{{\color{\colorMATH}\ensuremath{\sS_{2}}}}\endgroup }}\llbracket {\begingroup\renewcommand\colorMATH{\colorMATHB}\renewcommand\colorSYNTAX{\colorSYNTAXB}{{\color{\colorMATH}\ensuremath{\Distance'}}}\endgroup }(\tau _{2})\rrbracket }}}.
      As reduction is deterministic, then {{\color{\colorMATH}\ensuremath{j = j_{1}+j_{2}}}} and {{\color{\colorMATH}\ensuremath{{\begingroup\renewcommand\colorMATH{\colorMATHB}\renewcommand\colorSYNTAX{\colorSYNTAXB}{{\color{\colorMATH}\ensuremath{\sv'_{i j}}}}\endgroup } = {\begingroup\renewcommand\colorMATH{\colorMATHB}\renewcommand\colorSYNTAX{\colorSYNTAXB}{{\color{\colorMATH}\ensuremath{\sv_{i j}}}}\endgroup }}}}, therefore as {{\color{\colorMATH}\ensuremath{0+{\begingroup\renewcommand\colorMATH{\colorMATHB}\renewcommand\colorSYNTAX{\colorSYNTAXB}{{\color{\colorMATH}\ensuremath{\Distance'}}}\endgroup }\mathord{\cdotp }{\begingroup\renewcommand\colorMATH{\colorMATHB}\renewcommand\colorSYNTAX{\colorSYNTAXB}{{\color{\colorMATH}\ensuremath{\sS_{i}}}}\endgroup } = {\begingroup\renewcommand\colorMATH{\colorMATHB}\renewcommand\colorSYNTAX{\colorSYNTAXB}{{\color{\colorMATH}\ensuremath{\Distance'}}}\endgroup }\mathord{\cdotp }{\begingroup\renewcommand\colorMATH{\colorMATHB}\renewcommand\colorSYNTAX{\colorSYNTAXB}{{\color{\colorMATH}\ensuremath{\sS_{i}}}}\endgroup }}}}, the result holds immediately by Lemma~\ref{lm:weakening-index}.

      \end{subproof}
      
    %%
    %% Version with prepaid effects and linearity
    %%
    \item  {{\color{\colorMATH}\ensuremath{\Gamma ; {\begingroup\renewcommand\colorMATH{\colorMATHB}\renewcommand\colorSYNTAX{\colorSYNTAXB}{{\color{\colorMATH}\ensuremath{\Distance}}}\endgroup } \vdash  \addProduct{{\begingroup\renewcommand\colorMATH{\colorMATHB}\renewcommand\colorSYNTAX{\colorSYNTAXB}{{\color{\colorMATH}\ensuremath{\se_{1}}}}\endgroup }}{{\begingroup\renewcommand\colorMATH{\colorMATHB}\renewcommand\colorSYNTAX{\colorSYNTAXB}{{\color{\colorMATH}\ensuremath{\se_{2}}}}\endgroup }} \mathrel{:} \tau _{1} \mathrel{^{{\begingroup\renewcommand\colorMATH{\colorMATHB}\renewcommand\colorSYNTAX{\colorSYNTAXB}{{\color{\colorMATH}\ensuremath{\sS'_{1}}}}\endgroup }}\otimes ^{{\begingroup\renewcommand\colorMATH{\colorMATHB}\renewcommand\colorSYNTAX{\colorSYNTAXB}{{\color{\colorMATH}\ensuremath{\sS_{2}'}}}\endgroup }}} \tau _{2} \mathrel{;}  \multProd{{\begingroup\renewcommand\colorMATH{\colorMATHB}\renewcommand\colorSYNTAX{\colorSYNTAXB}{{\color{\colorMATH}\ensuremath{\sS''_{1}}}}\endgroup }}{{\begingroup\renewcommand\colorMATH{\colorMATHB}\renewcommand\colorSYNTAX{\colorSYNTAXB}{{\color{\colorMATH}\ensuremath{\sS''_{2}}}}\endgroup }}}}}
      \begin{subproof}   
      \begingroup\color{\colorMATH}\begin{gather*} 
        \inferrule*[lab=
        ]{ \Gamma ; {\begingroup\renewcommand\colorMATH{\colorMATHB}\renewcommand\colorSYNTAX{\colorSYNTAXB}{{\color{\colorMATH}\ensuremath{\Distance}}}\endgroup } \vdash  {\begingroup\renewcommand\colorMATH{\colorMATHB}\renewcommand\colorSYNTAX{\colorSYNTAXB}{{\color{\colorMATH}\ensuremath{\se_{1}}}}\endgroup } \mathrel{:} \tau _{1} \mathrel{;} {\begingroup\renewcommand\colorMATH{\colorMATHB}\renewcommand\colorSYNTAX{\colorSYNTAXB}{{\color{\colorMATH}\ensuremath{\sS''_{1}}}}\endgroup } + {\begingroup\renewcommand\colorMATH{\colorMATHB}\renewcommand\colorSYNTAX{\colorSYNTAXB}{{\color{\colorMATH}\ensuremath{\sS'_{1}}}}\endgroup } 
        \\ \Gamma ; {\begingroup\renewcommand\colorMATH{\colorMATHB}\renewcommand\colorSYNTAX{\colorSYNTAXB}{{\color{\colorMATH}\ensuremath{\Distance}}}\endgroup } \vdash  {\begingroup\renewcommand\colorMATH{\colorMATHB}\renewcommand\colorSYNTAX{\colorSYNTAXB}{{\color{\colorMATH}\ensuremath{\se_{2}}}}\endgroup } \mathrel{:} \tau _{2} \mathrel{;} {\begingroup\renewcommand\colorMATH{\colorMATHB}\renewcommand\colorSYNTAX{\colorSYNTAXB}{{\color{\colorMATH}\ensuremath{\sS''_{2}}}}\endgroup } + {\begingroup\renewcommand\colorMATH{\colorMATHB}\renewcommand\colorSYNTAX{\colorSYNTAXB}{{\color{\colorMATH}\ensuremath{\sS'_{2}}}}\endgroup } 
          }{
          \Gamma ; {\begingroup\renewcommand\colorMATH{\colorMATHB}\renewcommand\colorSYNTAX{\colorSYNTAXB}{{\color{\colorMATH}\ensuremath{\Distance}}}\endgroup } \vdash  \addProduct{{\begingroup\renewcommand\colorMATH{\colorMATHB}\renewcommand\colorSYNTAX{\colorSYNTAXB}{{\color{\colorMATH}\ensuremath{\se_{1}}}}\endgroup }}{{\begingroup\renewcommand\colorMATH{\colorMATHB}\renewcommand\colorSYNTAX{\colorSYNTAXB}{{\color{\colorMATH}\ensuremath{\se_{2}}}}\endgroup }} \mathrel{:} \tau _{1} \mathrel{^{{\begingroup\renewcommand\colorMATH{\colorMATHB}\renewcommand\colorSYNTAX{\colorSYNTAXB}{{\color{\colorMATH}\ensuremath{\sS'_{1}}}}\endgroup }}\otimes ^{{\begingroup\renewcommand\colorMATH{\colorMATHB}\renewcommand\colorSYNTAX{\colorSYNTAXB}{{\color{\colorMATH}\ensuremath{\sS_{2}'}}}\endgroup }}} \tau _{2} \mathrel{;}  \multProd{{\begingroup\renewcommand\colorMATH{\colorMATHB}\renewcommand\colorSYNTAX{\colorSYNTAXB}{{\color{\colorMATH}\ensuremath{\sS''_{1}}}}\endgroup }}{{\begingroup\renewcommand\colorMATH{\colorMATHB}\renewcommand\colorSYNTAX{\colorSYNTAXB}{{\color{\colorMATH}\ensuremath{\sS''_{2}}}}\endgroup }}
        }
      \end{gather*}\endgroup
      We have to prove that\\ {{\color{\colorMATH}\ensuremath{\forall k, \forall (\gamma _{1},\gamma _{2}) \in  {\mathcal{G}}_{{\begingroup\renewcommand\colorMATH{\colorMATHB}\renewcommand\colorSYNTAX{\colorSYNTAXB}{{\color{\colorMATH}\ensuremath{\Distance'}}}\endgroup }}^{\kg}\llbracket \Gamma \rrbracket , (\gamma _{1}\vdash \addProduct{{\begingroup\renewcommand\colorMATH{\colorMATHB}\renewcommand\colorSYNTAX{\colorSYNTAXB}{{\color{\colorMATH}\ensuremath{\se_{1}}}}\endgroup }}{{\begingroup\renewcommand\colorMATH{\colorMATHB}\renewcommand\colorSYNTAX{\colorSYNTAXB}{{\color{\colorMATH}\ensuremath{\se_{2}}}}\endgroup }},\gamma _{2}\vdash \addProduct{{\begingroup\renewcommand\colorMATH{\colorMATHB}\renewcommand\colorSYNTAX{\colorSYNTAXB}{{\color{\colorMATH}\ensuremath{\se_{1}}}}\endgroup }}{{\begingroup\renewcommand\colorMATH{\colorMATHB}\renewcommand\colorSYNTAX{\colorSYNTAXB}{{\color{\colorMATH}\ensuremath{\se_{2}}}}\endgroup }}) \in  {\mathcal{E}}^{k}_{{\begingroup\renewcommand\colorMATH{\colorMATHB}\renewcommand\colorSYNTAX{\colorSYNTAXB}{{\color{\colorMATH}\ensuremath{\Distance'}}}\endgroup }\mathord{\cdotp }(\multProd{{\begingroup\renewcommand\colorMATH{\colorMATHB}\renewcommand\colorSYNTAX{\colorSYNTAXB}{{\color{\colorMATH}\ensuremath{\sS''_{1}}}}\endgroup }}{{\begingroup\renewcommand\colorMATH{\colorMATHB}\renewcommand\colorSYNTAX{\colorSYNTAXB}{{\color{\colorMATH}\ensuremath{\sS''_{2}}}}\endgroup }})}\llbracket {\begingroup\renewcommand\colorMATH{\colorMATHB}\renewcommand\colorSYNTAX{\colorSYNTAXB}{{\color{\colorMATH}\ensuremath{\Distance'}}}\endgroup }(\tau _{1} \mathrel{^{{\begingroup\renewcommand\colorMATH{\colorMATHB}\renewcommand\colorSYNTAX{\colorSYNTAXB}{{\color{\colorMATH}\ensuremath{\sS'_{1}}}}\endgroup }}\otimes ^{{\begingroup\renewcommand\colorMATH{\colorMATHB}\renewcommand\colorSYNTAX{\colorSYNTAXB}{{\color{\colorMATH}\ensuremath{\sS'_{2}}}}\endgroup }}} \tau _{2})\rrbracket }}}, for {{\color{\colorMATH}\ensuremath{{\begingroup\renewcommand\colorMATH{\colorMATHB}\renewcommand\colorSYNTAX{\colorSYNTAXB}{{\color{\colorMATH}\ensuremath{\Distance'}}}\endgroup } \sqsubseteq  {\begingroup\renewcommand\colorMATH{\colorMATHB}\renewcommand\colorSYNTAX{\colorSYNTAXB}{{\color{\colorMATH}\ensuremath{\Distance}}}\endgroup }}}}.
      Let {{\color{\colorMATH}\ensuremath{d'_{i} = {\begingroup\renewcommand\colorMATH{\colorMATHB}\renewcommand\colorSYNTAX{\colorSYNTAXB}{{\color{\colorMATH}\ensuremath{\Distance'}}}\endgroup }\mathord{\cdotp }{\begingroup\renewcommand\colorMATH{\colorMATHB}\renewcommand\colorSYNTAX{\colorSYNTAXB}{{\color{\colorMATH}\ensuremath{\sS'_{i}}}}\endgroup }}}} and {{\color{\colorMATH}\ensuremath{d''_{i} = {\begingroup\renewcommand\colorMATH{\colorMATHB}\renewcommand\colorSYNTAX{\colorSYNTAXB}{{\color{\colorMATH}\ensuremath{\Distance'}}}\endgroup }\mathord{\cdotp }{\begingroup\renewcommand\colorMATH{\colorMATHB}\renewcommand\colorSYNTAX{\colorSYNTAXB}{{\color{\colorMATH}\ensuremath{\sS''_{i}}}}\endgroup }}}}. 
      Notice that {{\color{\colorMATH}\ensuremath{{\begingroup\renewcommand\colorMATH{\colorMATHB}\renewcommand\colorSYNTAX{\colorSYNTAXB}{{\color{\colorMATH}\ensuremath{\Distance'}}}\endgroup }\mathord{\cdotp }(\multProd{{\begingroup\renewcommand\colorMATH{\colorMATHB}\renewcommand\colorSYNTAX{\colorSYNTAXB}{{\color{\colorMATH}\ensuremath{\sS''_{1}}}}\endgroup }}{{\begingroup\renewcommand\colorMATH{\colorMATHB}\renewcommand\colorSYNTAX{\colorSYNTAXB}{{\color{\colorMATH}\ensuremath{\sS''_{2}}}}\endgroup }}) = \multProd{d''_{1}}{d''_{2}}}}}, and {{\color{\colorMATH}\ensuremath{{\begingroup\renewcommand\colorMATH{\colorMATHB}\renewcommand\colorSYNTAX{\colorSYNTAXB}{{\color{\colorMATH}\ensuremath{\Distance'}}}\endgroup }(\tau _{1} \mathrel{^{{\begingroup\renewcommand\colorMATH{\colorMATHB}\renewcommand\colorSYNTAX{\colorSYNTAXB}{{\color{\colorMATH}\ensuremath{\sS'_{1}}}}\endgroup }}\otimes ^{{\begingroup\renewcommand\colorMATH{\colorMATHB}\renewcommand\colorSYNTAX{\colorSYNTAXB}{{\color{\colorMATH}\ensuremath{\sS'_{2}}}}\endgroup }}} \tau _{2}) = {\begingroup\renewcommand\colorMATH{\colorMATHB}\renewcommand\colorSYNTAX{\colorSYNTAXB}{{\color{\colorMATH}\ensuremath{\Distance'}}}\endgroup }(\tau _{1}) \mathrel{^{d'_{1}}\otimes ^{d'_{2}}} {\begingroup\renewcommand\colorMATH{\colorMATHB}\renewcommand\colorSYNTAX{\colorSYNTAXB}{{\color{\colorMATH}\ensuremath{\Distance'}}}\endgroup }(\tau _{2})}}}, then we have to prove that\\
      {{\color{\colorMATH}\ensuremath{(\gamma _{1}\vdash \addProduct{{\begingroup\renewcommand\colorMATH{\colorMATHB}\renewcommand\colorSYNTAX{\colorSYNTAXB}{{\color{\colorMATH}\ensuremath{\se_{1}}}}\endgroup }}{{\begingroup\renewcommand\colorMATH{\colorMATHB}\renewcommand\colorSYNTAX{\colorSYNTAXB}{{\color{\colorMATH}\ensuremath{\se_{2}}}}\endgroup }},\gamma _{2}\vdash \addProduct{{\begingroup\renewcommand\colorMATH{\colorMATHB}\renewcommand\colorSYNTAX{\colorSYNTAXB}{{\color{\colorMATH}\ensuremath{\se_{1}}}}\endgroup }}{{\begingroup\renewcommand\colorMATH{\colorMATHB}\renewcommand\colorSYNTAX{\colorSYNTAXB}{{\color{\colorMATH}\ensuremath{\se_{2}}}}\endgroup }}) \in  {\mathcal{E}}^{k}_{\multProd{d''_{1}}{d''_{2}}}\llbracket {\begingroup\renewcommand\colorMATH{\colorMATHB}\renewcommand\colorSYNTAX{\colorSYNTAXB}{{\color{\colorMATH}\ensuremath{\Distance'}}}\endgroup }(\tau _{1}) \mathrel{^{d'_{1}}\otimes ^{d'_{2}}} {\begingroup\renewcommand\colorMATH{\colorMATHB}\renewcommand\colorSYNTAX{\colorSYNTAXB}{{\color{\colorMATH}\ensuremath{\Distance'}}}\endgroup }(\tau _{2})\rrbracket }}}, i.e.
      if {{\color{\colorMATH}\ensuremath{\gamma _{1}\vdash \addProduct{{\begingroup\renewcommand\colorMATH{\colorMATHB}\renewcommand\colorSYNTAX{\colorSYNTAXB}{{\color{\colorMATH}\ensuremath{\se_{1}}}}\endgroup }}{{\begingroup\renewcommand\colorMATH{\colorMATHB}\renewcommand\colorSYNTAX{\colorSYNTAXB}{{\color{\colorMATH}\ensuremath{\se_{2}}}}\endgroup }} \Downarrow ^{j} \addProduct{{\begingroup\renewcommand\colorMATH{\colorMATHB}\renewcommand\colorSYNTAX{\colorSYNTAXB}{{\color{\colorMATH}\ensuremath{\sv_{1 1}}}}\endgroup }}{{\begingroup\renewcommand\colorMATH{\colorMATHB}\renewcommand\colorSYNTAX{\colorSYNTAXB}{{\color{\colorMATH}\ensuremath{\sv_{1 2}}}}\endgroup }}}}} \pthen {{\color{\colorMATH}\ensuremath{\gamma _{2}\vdash \addProduct{{\begingroup\renewcommand\colorMATH{\colorMATHB}\renewcommand\colorSYNTAX{\colorSYNTAXB}{{\color{\colorMATH}\ensuremath{\se_{1}}}}\endgroup }}{{\begingroup\renewcommand\colorMATH{\colorMATHB}\renewcommand\colorSYNTAX{\colorSYNTAXB}{{\color{\colorMATH}\ensuremath{\se_{2}}}}\endgroup }} \Downarrow ^{\pj} \addProduct{{\begingroup\renewcommand\colorMATH{\colorMATHB}\renewcommand\colorSYNTAX{\colorSYNTAXB}{{\color{\colorMATH}\ensuremath{\sv_{2 1}}}}\endgroup }}{{\begingroup\renewcommand\colorMATH{\colorMATHB}\renewcommand\colorSYNTAX{\colorSYNTAXB}{{\color{\colorMATH}\ensuremath{\sv_{2 2}}}}\endgroup }}}}}, \pand
      {{\color{\colorMATH}\ensuremath{(\addProduct{{\begingroup\renewcommand\colorMATH{\colorMATHB}\renewcommand\colorSYNTAX{\colorSYNTAXB}{{\color{\colorMATH}\ensuremath{\sv_{1 1}}}}\endgroup }}{{\begingroup\renewcommand\colorMATH{\colorMATHB}\renewcommand\colorSYNTAX{\colorSYNTAXB}{{\color{\colorMATH}\ensuremath{\sv_{1 2}}}}\endgroup }}, \addProduct{{\begingroup\renewcommand\colorMATH{\colorMATHB}\renewcommand\colorSYNTAX{\colorSYNTAXB}{{\color{\colorMATH}\ensuremath{\sv_{2 1}}}}\endgroup }}{{\begingroup\renewcommand\colorMATH{\colorMATHB}\renewcommand\colorSYNTAX{\colorSYNTAXB}{{\color{\colorMATH}\ensuremath{\sv_{2 2}}}}\endgroup }}) \in  {\mathcal{V}}^{k-j}_{\multProd{d''_{1}}{d''_{2}}}\llbracket {\begingroup\renewcommand\colorMATH{\colorMATHB}\renewcommand\colorSYNTAX{\colorSYNTAXB}{{\color{\colorMATH}\ensuremath{\Distance'}}}\endgroup }(\tau _{1}) \mathrel{^{d'_{1}}\otimes ^{d'_{2}}} {\begingroup\renewcommand\colorMATH{\colorMATHB}\renewcommand\colorSYNTAX{\colorSYNTAXB}{{\color{\colorMATH}\ensuremath{\Distance'}}}\endgroup }(\tau _{2})\rrbracket }}}.

      By induction hypothesis on {{\color{\colorMATH}\ensuremath{\Gamma  \vdash  {\begingroup\renewcommand\colorMATH{\colorMATHB}\renewcommand\colorSYNTAX{\colorSYNTAXB}{{\color{\colorMATH}\ensuremath{\se_{1}}}}\endgroup } \mathrel{:} \tau _{1} \mathrel{;} {\begingroup\renewcommand\colorMATH{\colorMATHB}\renewcommand\colorSYNTAX{\colorSYNTAXB}{{\color{\colorMATH}\ensuremath{\sS''_{1}}}}\endgroup } + {\begingroup\renewcommand\colorMATH{\colorMATHB}\renewcommand\colorSYNTAX{\colorSYNTAXB}{{\color{\colorMATH}\ensuremath{\sS'_{1}}}}\endgroup }}}} and {{\color{\colorMATH}\ensuremath{\Gamma  \vdash  {\begingroup\renewcommand\colorMATH{\colorMATHB}\renewcommand\colorSYNTAX{\colorSYNTAXB}{{\color{\colorMATH}\ensuremath{\se_{2}}}}\endgroup } \mathrel{:} \tau _{2} \mathrel{;} {\begingroup\renewcommand\colorMATH{\colorMATHB}\renewcommand\colorSYNTAX{\colorSYNTAXB}{{\color{\colorMATH}\ensuremath{\sS''_{2}}}}\endgroup } + {\begingroup\renewcommand\colorMATH{\colorMATHB}\renewcommand\colorSYNTAX{\colorSYNTAXB}{{\color{\colorMATH}\ensuremath{\sS'_{2}}}}\endgroup }}}}, we know that 
      {{\color{\colorMATH}\ensuremath{(\gamma _{1}\vdash {\begingroup\renewcommand\colorMATH{\colorMATHB}\renewcommand\colorSYNTAX{\colorSYNTAXB}{{\color{\colorMATH}\ensuremath{\se_{1}}}}\endgroup },\gamma _{2}\vdash {\begingroup\renewcommand\colorMATH{\colorMATHB}\renewcommand\colorSYNTAX{\colorSYNTAXB}{{\color{\colorMATH}\ensuremath{\se_{1}}}}\endgroup }) \in  {\mathcal{E}}^{k}_{d'_{1} + d''_{1}}\llbracket {\begingroup\renewcommand\colorMATH{\colorMATHB}\renewcommand\colorSYNTAX{\colorSYNTAXB}{{\color{\colorMATH}\ensuremath{\Distance'}}}\endgroup }(\tau _{1})\rrbracket }}} and {{\color{\colorMATH}\ensuremath{(\gamma _{1}\vdash {\begingroup\renewcommand\colorMATH{\colorMATHB}\renewcommand\colorSYNTAX{\colorSYNTAXB}{{\color{\colorMATH}\ensuremath{\se_{2}}}}\endgroup },\gamma _{2}\vdash {\begingroup\renewcommand\colorMATH{\colorMATHB}\renewcommand\colorSYNTAX{\colorSYNTAXB}{{\color{\colorMATH}\ensuremath{\se_{2}}}}\endgroup }) \in  {\mathcal{E}}^{k}_{d'_{2} + d''_{2}}\llbracket {\begingroup\renewcommand\colorMATH{\colorMATHB}\renewcommand\colorSYNTAX{\colorSYNTAXB}{{\color{\colorMATH}\ensuremath{\Distance'}}}\endgroup }(\tau _{2})\rrbracket }}} respectively. This means that
      if {{\color{\colorMATH}\ensuremath{\gamma _{1}\vdash {\begingroup\renewcommand\colorMATH{\colorMATHB}\renewcommand\colorSYNTAX{\colorSYNTAXB}{{\color{\colorMATH}\ensuremath{\se_{1}}}}\endgroup } \Downarrow ^{j_{1}} {\begingroup\renewcommand\colorMATH{\colorMATHB}\renewcommand\colorSYNTAX{\colorSYNTAXB}{{\color{\colorMATH}\ensuremath{\sv'_{1 1}}}}\endgroup }}}}, \pthen {{\color{\colorMATH}\ensuremath{\gamma _{2}\vdash {\begingroup\renewcommand\colorMATH{\colorMATHB}\renewcommand\colorSYNTAX{\colorSYNTAXB}{{\color{\colorMATH}\ensuremath{\se_{1}}}}\endgroup } \Downarrow ^{\pj[1]} {\begingroup\renewcommand\colorMATH{\colorMATHB}\renewcommand\colorSYNTAX{\colorSYNTAXB}{{\color{\colorMATH}\ensuremath{\sv'_{1 2}}}}\endgroup }}}} \pand 
      {{\color{\colorMATH}\ensuremath{({\begingroup\renewcommand\colorMATH{\colorMATHB}\renewcommand\colorSYNTAX{\colorSYNTAXB}{{\color{\colorMATH}\ensuremath{\sv'_{1 1}}}}\endgroup }, {\begingroup\renewcommand\colorMATH{\colorMATHB}\renewcommand\colorSYNTAX{\colorSYNTAXB}{{\color{\colorMATH}\ensuremath{\sv'_{1 2}}}}\endgroup }) \in  {\mathcal{V}}^{k-j_{1}}_{d'_{1} + d''_{1}}\llbracket {\begingroup\renewcommand\colorMATH{\colorMATHB}\renewcommand\colorSYNTAX{\colorSYNTAXB}{{\color{\colorMATH}\ensuremath{\Distance'}}}\endgroup }(\tau _{1})\rrbracket }}}, and that 
      if {{\color{\colorMATH}\ensuremath{\gamma _{1}\vdash {\begingroup\renewcommand\colorMATH{\colorMATHB}\renewcommand\colorSYNTAX{\colorSYNTAXB}{{\color{\colorMATH}\ensuremath{\se_{2}}}}\endgroup } \Downarrow ^{j_{2}} {\begingroup\renewcommand\colorMATH{\colorMATHB}\renewcommand\colorSYNTAX{\colorSYNTAXB}{{\color{\colorMATH}\ensuremath{\sv'_{2 1}}}}\endgroup }}}}, \pthen {{\color{\colorMATH}\ensuremath{\gamma _{2}\vdash {\begingroup\renewcommand\colorMATH{\colorMATHB}\renewcommand\colorSYNTAX{\colorSYNTAXB}{{\color{\colorMATH}\ensuremath{\se_{2}}}}\endgroup } \Downarrow ^{\pj[2]} {\begingroup\renewcommand\colorMATH{\colorMATHB}\renewcommand\colorSYNTAX{\colorSYNTAXB}{{\color{\colorMATH}\ensuremath{\sv'_{2 2}}}}\endgroup }}}} \pand 
      {{\color{\colorMATH}\ensuremath{({\begingroup\renewcommand\colorMATH{\colorMATHB}\renewcommand\colorSYNTAX{\colorSYNTAXB}{{\color{\colorMATH}\ensuremath{\sv'_{2 1}}}}\endgroup }, {\begingroup\renewcommand\colorMATH{\colorMATHB}\renewcommand\colorSYNTAX{\colorSYNTAXB}{{\color{\colorMATH}\ensuremath{\sv'_{2 2}}}}\endgroup }) \in  {\mathcal{V}}^{k-j_{2}}_{d'_{2} + d''_{2}}\llbracket {\begingroup\renewcommand\colorMATH{\colorMATHB}\renewcommand\colorSYNTAX{\colorSYNTAXB}{{\color{\colorMATH}\ensuremath{\Distance'}}}\endgroup }(\tau _{2})\rrbracket }}}. 
      Notice that {{\color{\colorMATH}\ensuremath{d''_{i} \leq  \multProd{d''_{1}}{d''_{2}}}}}.
      As reduction is deterministic, then {{\color{\colorMATH}\ensuremath{j = j_{1}+j_{2}}}} and {{\color{\colorMATH}\ensuremath{{\begingroup\renewcommand\colorMATH{\colorMATHB}\renewcommand\colorSYNTAX{\colorSYNTAXB}{{\color{\colorMATH}\ensuremath{\sv'_{i j}}}}\endgroup } = {\begingroup\renewcommand\colorMATH{\colorMATHB}\renewcommand\colorSYNTAX{\colorSYNTAXB}{{\color{\colorMATH}\ensuremath{\sv_{i j}}}}\endgroup }}}}, therefore as {{\color{\colorMATH}\ensuremath{0 + d'_{i} + d''_{i} = d'_{i} + d''_{i}}}}, the result holds immediately by Lemma~\ref{lm:weakening-index}.

      \end{subproof}
        
    \item  {{\color{\colorMATH}\ensuremath{\Gamma ; {\begingroup\renewcommand\colorMATH{\colorMATHB}\renewcommand\colorSYNTAX{\colorSYNTAXB}{{\color{\colorMATH}\ensuremath{\Distance}}}\endgroup } \vdash  \tlet\hspace*{0.33em}x_{1},x_{2}={\begingroup\renewcommand\colorMATH{\colorMATHB}\renewcommand\colorSYNTAX{\colorSYNTAXB}{{\color{\colorMATH}\ensuremath{\se_{1}}}}\endgroup }\hspace*{0.33em}\tin\hspace*{0.33em}{\begingroup\renewcommand\colorMATH{\colorMATHB}\renewcommand\colorSYNTAX{\colorSYNTAXB}{{\color{\colorMATH}\ensuremath{\se_{2}}}}\endgroup } \mathrel{:} [{\begingroup\renewcommand\colorMATH{\colorMATHB}\renewcommand\colorSYNTAX{\colorSYNTAXB}{{\color{\colorMATH}\ensuremath{\sS_{1}}}}\endgroup }+{\begingroup\renewcommand\colorMATH{\colorMATHB}\renewcommand\colorSYNTAX{\colorSYNTAXB}{{\color{\colorMATH}\ensuremath{\sS_{1 1}}}}\endgroup }/x_{1}][{\begingroup\renewcommand\colorMATH{\colorMATHB}\renewcommand\colorSYNTAX{\colorSYNTAXB}{{\color{\colorMATH}\ensuremath{\sS_{1}}}}\endgroup }+{\begingroup\renewcommand\colorMATH{\colorMATHB}\renewcommand\colorSYNTAX{\colorSYNTAXB}{{\color{\colorMATH}\ensuremath{\sS_{1 2}}}}\endgroup }/x_{2}]\tau _{2} \mathrel{;} {\begingroup\renewcommand\colorMATH{\colorMATHB}\renewcommand\colorSYNTAX{\colorSYNTAXB}{{\color{\colorMATH}\ensuremath{\sss_{1}}}}\endgroup }({\begingroup\renewcommand\colorMATH{\colorMATHB}\renewcommand\colorSYNTAX{\colorSYNTAXB}{{\color{\colorMATH}\ensuremath{\sS_{1 1}}}}\endgroup }+{\begingroup\renewcommand\colorMATH{\colorMATHB}\renewcommand\colorSYNTAX{\colorSYNTAXB}{{\color{\colorMATH}\ensuremath{\sS_{1}}}}\endgroup })+ {\begingroup\renewcommand\colorMATH{\colorMATHB}\renewcommand\colorSYNTAX{\colorSYNTAXB}{{\color{\colorMATH}\ensuremath{\sss_{2}}}}\endgroup }({\begingroup\renewcommand\colorMATH{\colorMATHB}\renewcommand\colorSYNTAX{\colorSYNTAXB}{{\color{\colorMATH}\ensuremath{\sS_{1 2}}}}\endgroup }+{\begingroup\renewcommand\colorMATH{\colorMATHB}\renewcommand\colorSYNTAX{\colorSYNTAXB}{{\color{\colorMATH}\ensuremath{\sS_{1}}}}\endgroup }) + {\begingroup\renewcommand\colorMATH{\colorMATHB}\renewcommand\colorSYNTAX{\colorSYNTAXB}{{\color{\colorMATH}\ensuremath{\sS_{2}}}}\endgroup }}}}
      \begin{subproof} 
        We have to prove that for any {{\color{\colorMATH}\ensuremath{k}}}, {{\color{\colorMATH}\ensuremath{\forall  (\gamma _{1},\gamma _{2}) \in  {\mathcal{G}}_{{\begingroup\renewcommand\colorMATH{\colorMATHB}\renewcommand\colorSYNTAX{\colorSYNTAXB}{{\color{\colorMATH}\ensuremath{\Distance'}}}\endgroup }}^{\kg}\llbracket \Gamma \rrbracket , (\gamma _{1}\vdash \tlet\hspace*{0.33em}x_{1},x_{2}={\begingroup\renewcommand\colorMATH{\colorMATHB}\renewcommand\colorSYNTAX{\colorSYNTAXB}{{\color{\colorMATH}\ensuremath{\se_{1}}}}\endgroup }\hspace*{0.33em}\tin\hspace*{0.33em}{\begingroup\renewcommand\colorMATH{\colorMATHB}\renewcommand\colorSYNTAX{\colorSYNTAXB}{{\color{\colorMATH}\ensuremath{\se_{2}}}}\endgroup },\gamma _{2}\vdash \tlet\hspace*{0.33em}x_{1},x_{2}={\begingroup\renewcommand\colorMATH{\colorMATHB}\renewcommand\colorSYNTAX{\colorSYNTAXB}{{\color{\colorMATH}\ensuremath{\se_{1}}}}\endgroup }\hspace*{0.33em}\tin\hspace*{0.33em}{\begingroup\renewcommand\colorMATH{\colorMATHB}\renewcommand\colorSYNTAX{\colorSYNTAXB}{{\color{\colorMATH}\ensuremath{\se_{2}}}}\endgroup }) \in  {\mathcal{E}}_{{\begingroup\renewcommand\colorMATH{\colorMATHB}\renewcommand\colorSYNTAX{\colorSYNTAXB}{{\color{\colorMATH}\ensuremath{\Distance'}}}\endgroup }\mathord{\cdotp }({\begingroup\renewcommand\colorMATH{\colorMATHB}\renewcommand\colorSYNTAX{\colorSYNTAXB}{{\color{\colorMATH}\ensuremath{\sS''}}}\endgroup })}^{k}\llbracket {\begingroup\renewcommand\colorMATH{\colorMATHB}\renewcommand\colorSYNTAX{\colorSYNTAXB}{{\color{\colorMATH}\ensuremath{\Distance'}}}\endgroup }(\tau )\rrbracket }}}, for {{\color{\colorMATH}\ensuremath{{\begingroup\renewcommand\colorMATH{\colorMATHB}\renewcommand\colorSYNTAX{\colorSYNTAXB}{{\color{\colorMATH}\ensuremath{\Distance'}}}\endgroup } \sqsubseteq  {\begingroup\renewcommand\colorMATH{\colorMATHB}\renewcommand\colorSYNTAX{\colorSYNTAXB}{{\color{\colorMATH}\ensuremath{\Distance}}}\endgroup }}}}.
        where {{\color{\colorMATH}\ensuremath{{\begingroup\renewcommand\colorMATH{\colorMATHB}\renewcommand\colorSYNTAX{\colorSYNTAXB}{{\color{\colorMATH}\ensuremath{\sS''}}}\endgroup } = {\begingroup\renewcommand\colorMATH{\colorMATHB}\renewcommand\colorSYNTAX{\colorSYNTAXB}{{\color{\colorMATH}\ensuremath{\sss_{1}}}}\endgroup }({\begingroup\renewcommand\colorMATH{\colorMATHB}\renewcommand\colorSYNTAX{\colorSYNTAXB}{{\color{\colorMATH}\ensuremath{\sS_{1 1}}}}\endgroup }+{\begingroup\renewcommand\colorMATH{\colorMATHB}\renewcommand\colorSYNTAX{\colorSYNTAXB}{{\color{\colorMATH}\ensuremath{\sS_{1}}}}\endgroup })+ {\begingroup\renewcommand\colorMATH{\colorMATHB}\renewcommand\colorSYNTAX{\colorSYNTAXB}{{\color{\colorMATH}\ensuremath{\sss_{2}}}}\endgroup }({\begingroup\renewcommand\colorMATH{\colorMATHB}\renewcommand\colorSYNTAX{\colorSYNTAXB}{{\color{\colorMATH}\ensuremath{\sS_{1 2}}}}\endgroup }+{\begingroup\renewcommand\colorMATH{\colorMATHB}\renewcommand\colorSYNTAX{\colorSYNTAXB}{{\color{\colorMATH}\ensuremath{\sS_{1}}}}\endgroup }) + {\begingroup\renewcommand\colorMATH{\colorMATHB}\renewcommand\colorSYNTAX{\colorSYNTAXB}{{\color{\colorMATH}\ensuremath{\sS_{2}}}}\endgroup }}}}.\\
        Suppose
        \begingroup\color{\colorMATH}\begin{gather*} 
          \inferrule*[lab=
          ]{ \gamma _{1}\vdash {\begingroup\renewcommand\colorMATH{\colorMATHB}\renewcommand\colorSYNTAX{\colorSYNTAXB}{{\color{\colorMATH}\ensuremath{\se_{1}}}}\endgroup } \Downarrow ^{j_{1}} \langle {\begingroup\renewcommand\colorMATH{\colorMATHB}\renewcommand\colorSYNTAX{\colorSYNTAXB}{{\color{\colorMATH}\ensuremath{\sv_{1 1}}}}\endgroup },{\begingroup\renewcommand\colorMATH{\colorMATHB}\renewcommand\colorSYNTAX{\colorSYNTAXB}{{\color{\colorMATH}\ensuremath{\sv_{1 2}}}}\endgroup }\rangle 
          \\ \gamma _{1}[x_{1}\mapsto {\begingroup\renewcommand\colorMATH{\colorMATHB}\renewcommand\colorSYNTAX{\colorSYNTAXB}{{\color{\colorMATH}\ensuremath{\sv_{1 1}}}}\endgroup },x_{2}\mapsto {\begingroup\renewcommand\colorMATH{\colorMATHB}\renewcommand\colorSYNTAX{\colorSYNTAXB}{{\color{\colorMATH}\ensuremath{\sv_{1 2}}}}\endgroup }] \vdash  {\begingroup\renewcommand\colorMATH{\colorMATHB}\renewcommand\colorSYNTAX{\colorSYNTAXB}{{\color{\colorMATH}\ensuremath{\se_{2}}}}\endgroup } \Downarrow ^{j_{2}} {\begingroup\renewcommand\colorMATH{\colorMATHB}\renewcommand\colorSYNTAX{\colorSYNTAXB}{{\color{\colorMATH}\ensuremath{\sv'_{1}}}}\endgroup }
            }{
            \gamma _{1}\vdash \tlet\hspace*{0.33em}x_{1},x_{2}={\begingroup\renewcommand\colorMATH{\colorMATHB}\renewcommand\colorSYNTAX{\colorSYNTAXB}{{\color{\colorMATH}\ensuremath{\se_{1}}}}\endgroup }\hspace*{0.33em}\tin\hspace*{0.33em}{\begingroup\renewcommand\colorMATH{\colorMATHB}\renewcommand\colorSYNTAX{\colorSYNTAXB}{{\color{\colorMATH}\ensuremath{\se_{2}}}}\endgroup } \Downarrow ^{j_{1}+j_{2}+1} {\begingroup\renewcommand\colorMATH{\colorMATHB}\renewcommand\colorSYNTAX{\colorSYNTAXB}{{\color{\colorMATH}\ensuremath{\sv'_{1}}}}\endgroup }
          }
        \end{gather*}\endgroup  

        By induction hypothesis on {{\color{\colorMATH}\ensuremath{\Gamma ; {\begingroup\renewcommand\colorMATH{\colorMATHB}\renewcommand\colorSYNTAX{\colorSYNTAXB}{{\color{\colorMATH}\ensuremath{\Distance}}}\endgroup } \vdash  {\begingroup\renewcommand\colorMATH{\colorMATHB}\renewcommand\colorSYNTAX{\colorSYNTAXB}{{\color{\colorMATH}\ensuremath{\se_{1}}}}\endgroup } \mathrel{:} \tau _{1 1} \mathrel{^{{\begingroup\renewcommand\colorMATH{\colorMATHB}\renewcommand\colorSYNTAX{\colorSYNTAXB}{{\color{\colorMATH}\ensuremath{\sS_{1 1}}}}\endgroup }}\otimes ^{{\begingroup\renewcommand\colorMATH{\colorMATHB}\renewcommand\colorSYNTAX{\colorSYNTAXB}{{\color{\colorMATH}\ensuremath{\sS_{1 2}}}}\endgroup }}} \tau _{1 2} \mathrel{;} {\begingroup\renewcommand\colorMATH{\colorMATHB}\renewcommand\colorSYNTAX{\colorSYNTAXB}{{\color{\colorMATH}\ensuremath{\sS_{1}}}}\endgroup }}}} we know that\\
        {{\color{\colorMATH}\ensuremath{(\gamma _{1}\vdash {\begingroup\renewcommand\colorMATH{\colorMATHB}\renewcommand\colorSYNTAX{\colorSYNTAXB}{{\color{\colorMATH}\ensuremath{\se_{1}}}}\endgroup },\gamma _{2}\vdash {\begingroup\renewcommand\colorMATH{\colorMATHB}\renewcommand\colorSYNTAX{\colorSYNTAXB}{{\color{\colorMATH}\ensuremath{\se_{1}}}}\endgroup }) \in  {\mathcal{E}}^{k}_{{\begingroup\renewcommand\colorMATH{\colorMATHB}\renewcommand\colorSYNTAX{\colorSYNTAXB}{{\color{\colorMATH}\ensuremath{\Distance'}}}\endgroup }\mathord{\cdotp }{\begingroup\renewcommand\colorMATH{\colorMATHB}\renewcommand\colorSYNTAX{\colorSYNTAXB}{{\color{\colorMATH}\ensuremath{\sS_{1}}}}\endgroup }}\llbracket {\begingroup\renewcommand\colorMATH{\colorMATHB}\renewcommand\colorSYNTAX{\colorSYNTAXB}{{\color{\colorMATH}\ensuremath{\Distance'}}}\endgroup }(\tau _{1 1} \mathrel{^{{\begingroup\renewcommand\colorMATH{\colorMATHB}\renewcommand\colorSYNTAX{\colorSYNTAXB}{{\color{\colorMATH}\ensuremath{\sS_{1 1}}}}\endgroup }}\otimes ^{{\begingroup\renewcommand\colorMATH{\colorMATHB}\renewcommand\colorSYNTAX{\colorSYNTAXB}{{\color{\colorMATH}\ensuremath{\sS_{1 2}}}}\endgroup }}} \tau _{1 2})\rrbracket }}}, i.e.
        if {{\color{\colorMATH}\ensuremath{\gamma _{1}\vdash {\begingroup\renewcommand\colorMATH{\colorMATHB}\renewcommand\colorSYNTAX{\colorSYNTAXB}{{\color{\colorMATH}\ensuremath{\se_{1}}}}\endgroup } \Downarrow ^{j_{1}} \langle {\begingroup\renewcommand\colorMATH{\colorMATHB}\renewcommand\colorSYNTAX{\colorSYNTAXB}{{\color{\colorMATH}\ensuremath{\sv_{1 1}}}}\endgroup },{\begingroup\renewcommand\colorMATH{\colorMATHB}\renewcommand\colorSYNTAX{\colorSYNTAXB}{{\color{\colorMATH}\ensuremath{\sv_{1 2}}}}\endgroup }\rangle }}} \pthen {{\color{\colorMATH}\ensuremath{\gamma _{2}\vdash {\begingroup\renewcommand\colorMATH{\colorMATHB}\renewcommand\colorSYNTAX{\colorSYNTAXB}{{\color{\colorMATH}\ensuremath{\se_{1}}}}\endgroup } \Downarrow ^{\pj[1]} \langle {\begingroup\renewcommand\colorMATH{\colorMATHB}\renewcommand\colorSYNTAX{\colorSYNTAXB}{{\color{\colorMATH}\ensuremath{\sv_{2 1}}}}\endgroup },{\begingroup\renewcommand\colorMATH{\colorMATHB}\renewcommand\colorSYNTAX{\colorSYNTAXB}{{\color{\colorMATH}\ensuremath{\sv_{2 2}}}}\endgroup }\rangle }}}, \pand
        {{\color{\colorMATH}\ensuremath{(\langle {\begingroup\renewcommand\colorMATH{\colorMATHB}\renewcommand\colorSYNTAX{\colorSYNTAXB}{{\color{\colorMATH}\ensuremath{\sv_{1 1}}}}\endgroup },{\begingroup\renewcommand\colorMATH{\colorMATHB}\renewcommand\colorSYNTAX{\colorSYNTAXB}{{\color{\colorMATH}\ensuremath{\sv_{1 2}}}}\endgroup }\rangle , \langle {\begingroup\renewcommand\colorMATH{\colorMATHB}\renewcommand\colorSYNTAX{\colorSYNTAXB}{{\color{\colorMATH}\ensuremath{\sv_{2 1}}}}\endgroup },{\begingroup\renewcommand\colorMATH{\colorMATHB}\renewcommand\colorSYNTAX{\colorSYNTAXB}{{\color{\colorMATH}\ensuremath{\sv_{2 2}}}}\endgroup }\rangle ) \in  {\mathcal{V}}^{k-j_{1}}_{{\begingroup\renewcommand\colorMATH{\colorMATHB}\renewcommand\colorSYNTAX{\colorSYNTAXB}{{\color{\colorMATH}\ensuremath{\Distance'}}}\endgroup }\mathord{\cdotp }{\begingroup\renewcommand\colorMATH{\colorMATHB}\renewcommand\colorSYNTAX{\colorSYNTAXB}{{\color{\colorMATH}\ensuremath{\sS_{1}}}}\endgroup }}\llbracket {\begingroup\renewcommand\colorMATH{\colorMATHB}\renewcommand\colorSYNTAX{\colorSYNTAXB}{{\color{\colorMATH}\ensuremath{\Distance'}}}\endgroup }(\tau _{1 1}) \mathrel{^{{\begingroup\renewcommand\colorMATH{\colorMATHB}\renewcommand\colorSYNTAX{\colorSYNTAXB}{{\color{\colorMATH}\ensuremath{\Distance'}}}\endgroup }\mathord{\cdotp }{\begingroup\renewcommand\colorMATH{\colorMATHB}\renewcommand\colorSYNTAX{\colorSYNTAXB}{{\color{\colorMATH}\ensuremath{\sS_{1 1}}}}\endgroup }}\otimes ^{{\begingroup\renewcommand\colorMATH{\colorMATHB}\renewcommand\colorSYNTAX{\colorSYNTAXB}{{\color{\colorMATH}\ensuremath{\Distance'}}}\endgroup }\mathord{\cdotp }{\begingroup\renewcommand\colorMATH{\colorMATHB}\renewcommand\colorSYNTAX{\colorSYNTAXB}{{\color{\colorMATH}\ensuremath{\sS_{1 2}}}}\endgroup }}} {\begingroup\renewcommand\colorMATH{\colorMATHB}\renewcommand\colorSYNTAX{\colorSYNTAXB}{{\color{\colorMATH}\ensuremath{\Distance'}}}\endgroup }(\tau _{1 2})\rrbracket }}}, or equivalently\\
        {{\color{\colorMATH}\ensuremath{({\begingroup\renewcommand\colorMATH{\colorMATHB}\renewcommand\colorSYNTAX{\colorSYNTAXB}{{\color{\colorMATH}\ensuremath{\sv_{1 1}}}}\endgroup },{\begingroup\renewcommand\colorMATH{\colorMATHB}\renewcommand\colorSYNTAX{\colorSYNTAXB}{{\color{\colorMATH}\ensuremath{\sv_{2 1}}}}\endgroup }) \in  {\mathcal{V}}^{k-j_{1}}_{{\begingroup\renewcommand\colorMATH{\colorMATHB}\renewcommand\colorSYNTAX{\colorSYNTAXB}{{\color{\colorMATH}\ensuremath{\Distance'}}}\endgroup }\mathord{\cdotp }{\begingroup\renewcommand\colorMATH{\colorMATHB}\renewcommand\colorSYNTAX{\colorSYNTAXB}{{\color{\colorMATH}\ensuremath{\sS_{1}}}}\endgroup }+{\begingroup\renewcommand\colorMATH{\colorMATHB}\renewcommand\colorSYNTAX{\colorSYNTAXB}{{\color{\colorMATH}\ensuremath{\Distance'}}}\endgroup }\mathord{\cdotp }{\begingroup\renewcommand\colorMATH{\colorMATHB}\renewcommand\colorSYNTAX{\colorSYNTAXB}{{\color{\colorMATH}\ensuremath{\sS_{1 1}}}}\endgroup }}\llbracket {\begingroup\renewcommand\colorMATH{\colorMATHB}\renewcommand\colorSYNTAX{\colorSYNTAXB}{{\color{\colorMATH}\ensuremath{\Distance'}}}\endgroup }(\tau _{1 1})\rrbracket }}}, and {{\color{\colorMATH}\ensuremath{({\begingroup\renewcommand\colorMATH{\colorMATHB}\renewcommand\colorSYNTAX{\colorSYNTAXB}{{\color{\colorMATH}\ensuremath{\sv_{1 2}}}}\endgroup },{\begingroup\renewcommand\colorMATH{\colorMATHB}\renewcommand\colorSYNTAX{\colorSYNTAXB}{{\color{\colorMATH}\ensuremath{\sv_{2 2}}}}\endgroup }) \in  {\mathcal{V}}^{k-j_{1}}_{{\begingroup\renewcommand\colorMATH{\colorMATHB}\renewcommand\colorSYNTAX{\colorSYNTAXB}{{\color{\colorMATH}\ensuremath{\Distance'}}}\endgroup }\mathord{\cdotp }{\begingroup\renewcommand\colorMATH{\colorMATHB}\renewcommand\colorSYNTAX{\colorSYNTAXB}{{\color{\colorMATH}\ensuremath{\sS_{1}}}}\endgroup }+{\begingroup\renewcommand\colorMATH{\colorMATHB}\renewcommand\colorSYNTAX{\colorSYNTAXB}{{\color{\colorMATH}\ensuremath{\Distance'}}}\endgroup }\mathord{\cdotp }{\begingroup\renewcommand\colorMATH{\colorMATHB}\renewcommand\colorSYNTAX{\colorSYNTAXB}{{\color{\colorMATH}\ensuremath{\sS_{1 2}}}}\endgroup }}\llbracket {\begingroup\renewcommand\colorMATH{\colorMATHB}\renewcommand\colorSYNTAX{\colorSYNTAXB}{{\color{\colorMATH}\ensuremath{\Distance'}}}\endgroup }(\tau _{1 2})\rrbracket }}}.
        By Lemma~\ref{lm:associativity-inst} 
        {{\color{\colorMATH}\ensuremath{{\begingroup\renewcommand\colorMATH{\colorMATHB}\renewcommand\colorSYNTAX{\colorSYNTAXB}{{\color{\colorMATH}\ensuremath{\Distance'}}}\endgroup }\mathord{\cdotp }{\begingroup\renewcommand\colorMATH{\colorMATHB}\renewcommand\colorSYNTAX{\colorSYNTAXB}{{\color{\colorMATH}\ensuremath{\sS_{1}}}}\endgroup } + {\begingroup\renewcommand\colorMATH{\colorMATHB}\renewcommand\colorSYNTAX{\colorSYNTAXB}{{\color{\colorMATH}\ensuremath{\Distance'}}}\endgroup }\mathord{\cdotp }{\begingroup\renewcommand\colorMATH{\colorMATHB}\renewcommand\colorSYNTAX{\colorSYNTAXB}{{\color{\colorMATH}\ensuremath{\sS_{1 1}}}}\endgroup } = {\begingroup\renewcommand\colorMATH{\colorMATHB}\renewcommand\colorSYNTAX{\colorSYNTAXB}{{\color{\colorMATH}\ensuremath{\Distance'}}}\endgroup }\mathord{\cdotp }({\begingroup\renewcommand\colorMATH{\colorMATHB}\renewcommand\colorSYNTAX{\colorSYNTAXB}{{\color{\colorMATH}\ensuremath{\sS_{1}}}}\endgroup } + {\begingroup\renewcommand\colorMATH{\colorMATHB}\renewcommand\colorSYNTAX{\colorSYNTAXB}{{\color{\colorMATH}\ensuremath{\sS_{1 1}}}}\endgroup })}}}, then
        {{\color{\colorMATH}\ensuremath{({\begingroup\renewcommand\colorMATH{\colorMATHB}\renewcommand\colorSYNTAX{\colorSYNTAXB}{{\color{\colorMATH}\ensuremath{\sv_{1 1}}}}\endgroup }, {\begingroup\renewcommand\colorMATH{\colorMATHB}\renewcommand\colorSYNTAX{\colorSYNTAXB}{{\color{\colorMATH}\ensuremath{\sv_{1 2}}}}\endgroup }) \in  {\mathcal{V}}^{k-j}_{{\begingroup\renewcommand\colorMATH{\colorMATHB}\renewcommand\colorSYNTAX{\colorSYNTAXB}{{\color{\colorMATH}\ensuremath{\Distance'}}}\endgroup }\mathord{\cdotp }({\begingroup\renewcommand\colorMATH{\colorMATHB}\renewcommand\colorSYNTAX{\colorSYNTAXB}{{\color{\colorMATH}\ensuremath{\sS_{1}}}}\endgroup } + {\begingroup\renewcommand\colorMATH{\colorMATHB}\renewcommand\colorSYNTAX{\colorSYNTAXB}{{\color{\colorMATH}\ensuremath{\sS_{1 1}}}}\endgroup })}\llbracket {\begingroup\renewcommand\colorMATH{\colorMATHB}\renewcommand\colorSYNTAX{\colorSYNTAXB}{{\color{\colorMATH}\ensuremath{\Distance'}}}\endgroup }(\tau _{1 1})\rrbracket }}}, and analogously 
        {{\color{\colorMATH}\ensuremath{{\begingroup\renewcommand\colorMATH{\colorMATHB}\renewcommand\colorSYNTAX{\colorSYNTAXB}{{\color{\colorMATH}\ensuremath{\Distance'}}}\endgroup }\mathord{\cdotp }{\begingroup\renewcommand\colorMATH{\colorMATHB}\renewcommand\colorSYNTAX{\colorSYNTAXB}{{\color{\colorMATH}\ensuremath{\sS_{1}}}}\endgroup } + {\begingroup\renewcommand\colorMATH{\colorMATHB}\renewcommand\colorSYNTAX{\colorSYNTAXB}{{\color{\colorMATH}\ensuremath{\Distance'}}}\endgroup }\mathord{\cdotp }{\begingroup\renewcommand\colorMATH{\colorMATHB}\renewcommand\colorSYNTAX{\colorSYNTAXB}{{\color{\colorMATH}\ensuremath{\sS_{1 2}}}}\endgroup } = {\begingroup\renewcommand\colorMATH{\colorMATHB}\renewcommand\colorSYNTAX{\colorSYNTAXB}{{\color{\colorMATH}\ensuremath{\Distance'}}}\endgroup }\mathord{\cdotp }({\begingroup\renewcommand\colorMATH{\colorMATHB}\renewcommand\colorSYNTAX{\colorSYNTAXB}{{\color{\colorMATH}\ensuremath{\sS_{1}}}}\endgroup } + {\begingroup\renewcommand\colorMATH{\colorMATHB}\renewcommand\colorSYNTAX{\colorSYNTAXB}{{\color{\colorMATH}\ensuremath{\sS_{1 2}}}}\endgroup })}}}, then
        {{\color{\colorMATH}\ensuremath{({\begingroup\renewcommand\colorMATH{\colorMATHB}\renewcommand\colorSYNTAX{\colorSYNTAXB}{{\color{\colorMATH}\ensuremath{\sv_{2 2}}}}\endgroup }, {\begingroup\renewcommand\colorMATH{\colorMATHB}\renewcommand\colorSYNTAX{\colorSYNTAXB}{{\color{\colorMATH}\ensuremath{\sv_{2 2}}}}\endgroup }) \in  {\mathcal{V}}^{k-j}_{{\begingroup\renewcommand\colorMATH{\colorMATHB}\renewcommand\colorSYNTAX{\colorSYNTAXB}{{\color{\colorMATH}\ensuremath{\Distance'}}}\endgroup }\mathord{\cdotp }({\begingroup\renewcommand\colorMATH{\colorMATHB}\renewcommand\colorSYNTAX{\colorSYNTAXB}{{\color{\colorMATH}\ensuremath{\sS_{1}}}}\endgroup } + {\begingroup\renewcommand\colorMATH{\colorMATHB}\renewcommand\colorSYNTAX{\colorSYNTAXB}{{\color{\colorMATH}\ensuremath{\sS_{1 2}}}}\endgroup })}\llbracket {\begingroup\renewcommand\colorMATH{\colorMATHB}\renewcommand\colorSYNTAX{\colorSYNTAXB}{{\color{\colorMATH}\ensuremath{\Distance'}}}\endgroup }(\tau _{1 2})\rrbracket }}}

        Also, by induction hypothesis on {{\color{\colorMATH}\ensuremath{\Gamma ,x:\tau _{1 1}, x:\tau _{1 2}; {\begingroup\renewcommand\colorMATH{\colorMATHB}\renewcommand\colorSYNTAX{\colorSYNTAXB}{{\color{\colorMATH}\ensuremath{\Distance}}}\endgroup } + ({\begingroup\renewcommand\colorMATH{\colorMATHB}\renewcommand\colorSYNTAX{\colorSYNTAXB}{{\color{\colorMATH}\ensuremath{\Distance}}}\endgroup }\mathord{\cdotp }({\begingroup\renewcommand\colorMATH{\colorMATHB}\renewcommand\colorSYNTAX{\colorSYNTAXB}{{\color{\colorMATH}\ensuremath{\sS_{1}}}}\endgroup }+{\begingroup\renewcommand\colorMATH{\colorMATHB}\renewcommand\colorSYNTAX{\colorSYNTAXB}{{\color{\colorMATH}\ensuremath{\sS_{1 1}}}}\endgroup }))x_{1} + ({\begingroup\renewcommand\colorMATH{\colorMATHB}\renewcommand\colorSYNTAX{\colorSYNTAXB}{{\color{\colorMATH}\ensuremath{\Distance}}}\endgroup }\mathord{\cdotp }({\begingroup\renewcommand\colorMATH{\colorMATHB}\renewcommand\colorSYNTAX{\colorSYNTAXB}{{\color{\colorMATH}\ensuremath{\sS_{1}}}}\endgroup }+{\begingroup\renewcommand\colorMATH{\colorMATHB}\renewcommand\colorSYNTAX{\colorSYNTAXB}{{\color{\colorMATH}\ensuremath{\sS_{1 2}}}}\endgroup }))x_{2} \vdash  {\begingroup\renewcommand\colorMATH{\colorMATHB}\renewcommand\colorSYNTAX{\colorSYNTAXB}{{\color{\colorMATH}\ensuremath{\se_{2}}}}\endgroup } \mathrel{:} \tau _{2} \mathrel{;} {\begingroup\renewcommand\colorMATH{\colorMATHB}\renewcommand\colorSYNTAX{\colorSYNTAXB}{{\color{\colorMATH}\ensuremath{\sS_{2}}}}\endgroup } + {\begingroup\renewcommand\colorMATH{\colorMATHB}\renewcommand\colorSYNTAX{\colorSYNTAXB}{{\color{\colorMATH}\ensuremath{\sss_{1}}}}\endgroup }x_{1} + {\begingroup\renewcommand\colorMATH{\colorMATHB}\renewcommand\colorSYNTAX{\colorSYNTAXB}{{\color{\colorMATH}\ensuremath{\sss_{2}}}}\endgroup }x_{2}}}}, by choosing {{\color{\colorMATH}\ensuremath{k = k-j_{1}}}}\\
        {{\color{\colorMATH}\ensuremath{(\gamma _{1}[x_{1} \mapsto  {\begingroup\renewcommand\colorMATH{\colorMATHB}\renewcommand\colorSYNTAX{\colorSYNTAXB}{{\color{\colorMATH}\ensuremath{\sv_{1 1}}}}\endgroup }, x_{2} \mapsto  {\begingroup\renewcommand\colorMATH{\colorMATHB}\renewcommand\colorSYNTAX{\colorSYNTAXB}{{\color{\colorMATH}\ensuremath{\sv_{1 2}}}}\endgroup }],\gamma _{2}[x_{1} \mapsto  {\begingroup\renewcommand\colorMATH{\colorMATHB}\renewcommand\colorSYNTAX{\colorSYNTAXB}{{\color{\colorMATH}\ensuremath{\sv_{2 1}}}}\endgroup }, x_{2} \mapsto  {\begingroup\renewcommand\colorMATH{\colorMATHB}\renewcommand\colorSYNTAX{\colorSYNTAXB}{{\color{\colorMATH}\ensuremath{\sv_{2 2}}}}\endgroup }]) \in  {\mathcal{G}}^{k-j_{1}-1}_{{\begingroup\renewcommand\colorMATH{\colorMATHB}\renewcommand\colorSYNTAX{\colorSYNTAXB}{{\color{\colorMATH}\ensuremath{\Distance'}}}\endgroup }+({\begingroup\renewcommand\colorMATH{\colorMATHB}\renewcommand\colorSYNTAX{\colorSYNTAXB}{{\color{\colorMATH}\ensuremath{\Distance'}}}\endgroup }\mathord{\cdotp }({\begingroup\renewcommand\colorMATH{\colorMATHB}\renewcommand\colorSYNTAX{\colorSYNTAXB}{{\color{\colorMATH}\ensuremath{\sS_{1}}}}\endgroup }+{\begingroup\renewcommand\colorMATH{\colorMATHB}\renewcommand\colorSYNTAX{\colorSYNTAXB}{{\color{\colorMATH}\ensuremath{\sS_{1 1}}}}\endgroup }))x_{1}+({\begingroup\renewcommand\colorMATH{\colorMATHB}\renewcommand\colorSYNTAX{\colorSYNTAXB}{{\color{\colorMATH}\ensuremath{\Distance'}}}\endgroup }\mathord{\cdotp }({\begingroup\renewcommand\colorMATH{\colorMATHB}\renewcommand\colorSYNTAX{\colorSYNTAXB}{{\color{\colorMATH}\ensuremath{\sS_{1}}}}\endgroup }+{\begingroup\renewcommand\colorMATH{\colorMATHB}\renewcommand\colorSYNTAX{\colorSYNTAXB}{{\color{\colorMATH}\ensuremath{\sS_{1 2}}}}\endgroup }))x_{2}}\llbracket \Gamma ,x\mathrel{:}\tau _{1 1}\rrbracket }}} (note that {{\color{\colorMATH}\ensuremath{x_{1} \notin  dom({\begingroup\renewcommand\colorMATH{\colorMATHB}\renewcommand\colorSYNTAX{\colorSYNTAXB}{{\color{\colorMATH}\ensuremath{\sS_{1}}}}\endgroup }) \cup  dom({\begingroup\renewcommand\colorMATH{\colorMATHB}\renewcommand\colorSYNTAX{\colorSYNTAXB}{{\color{\colorMATH}\ensuremath{\sS_{1 1}}}}\endgroup })}}} and {{\color{\colorMATH}\ensuremath{x_{2} \notin  dom({\begingroup\renewcommand\colorMATH{\colorMATHB}\renewcommand\colorSYNTAX{\colorSYNTAXB}{{\color{\colorMATH}\ensuremath{\sS_{1}}}}\endgroup }) \cup  dom({\begingroup\renewcommand\colorMATH{\colorMATHB}\renewcommand\colorSYNTAX{\colorSYNTAXB}{{\color{\colorMATH}\ensuremath{\sS_{1 2}}}}\endgroup })}}}, therefore {{\color{\colorMATH}\ensuremath{({\begingroup\renewcommand\colorMATH{\colorMATHB}\renewcommand\colorSYNTAX{\colorSYNTAXB}{{\color{\colorMATH}\ensuremath{\Distance'}}}\endgroup }+({\begingroup\renewcommand\colorMATH{\colorMATHB}\renewcommand\colorSYNTAX{\colorSYNTAXB}{{\color{\colorMATH}\ensuremath{\Distance'}}}\endgroup }\mathord{\cdotp }({\begingroup\renewcommand\colorMATH{\colorMATHB}\renewcommand\colorSYNTAX{\colorSYNTAXB}{{\color{\colorMATH}\ensuremath{\sS_{1}}}}\endgroup }+{\begingroup\renewcommand\colorMATH{\colorMATHB}\renewcommand\colorSYNTAX{\colorSYNTAXB}{{\color{\colorMATH}\ensuremath{\sS_{1 1}}}}\endgroup }))x_{1}+({\begingroup\renewcommand\colorMATH{\colorMATHB}\renewcommand\colorSYNTAX{\colorSYNTAXB}{{\color{\colorMATH}\ensuremath{\Distance'}}}\endgroup }\mathord{\cdotp }({\begingroup\renewcommand\colorMATH{\colorMATHB}\renewcommand\colorSYNTAX{\colorSYNTAXB}{{\color{\colorMATH}\ensuremath{\sS_{1}}}}\endgroup }+{\begingroup\renewcommand\colorMATH{\colorMATHB}\renewcommand\colorSYNTAX{\colorSYNTAXB}{{\color{\colorMATH}\ensuremath{\sS_{1 2}}}}\endgroup }))x_{2})(\tau _{1 i}) = {\begingroup\renewcommand\colorMATH{\colorMATHB}\renewcommand\colorSYNTAX{\colorSYNTAXB}{{\color{\colorMATH}\ensuremath{\Distance'}}}\endgroup }(\tau _{1 i})}}}) we know that\\
        {{\color{\colorMATH}\ensuremath{(\gamma '_{1}\vdash {\begingroup\renewcommand\colorMATH{\colorMATHB}\renewcommand\colorSYNTAX{\colorSYNTAXB}{{\color{\colorMATH}\ensuremath{\se_{2}}}}\endgroup },\gamma '_{2}\vdash {\begingroup\renewcommand\colorMATH{\colorMATHB}\renewcommand\colorSYNTAX{\colorSYNTAXB}{{\color{\colorMATH}\ensuremath{\se_{2}}}}\endgroup }) \in  {\mathcal{E}}^{k-j_{1}}_{{\begingroup\renewcommand\colorMATH{\colorMATHB}\renewcommand\colorSYNTAX{\colorSYNTAXB}{{\color{\colorMATH}\ensuremath{\sS''}}}\endgroup }\mathord{\cdotp }({\begingroup\renewcommand\colorMATH{\colorMATHB}\renewcommand\colorSYNTAX{\colorSYNTAXB}{{\color{\colorMATH}\ensuremath{\sS_{2}}}}\endgroup } + {\begingroup\renewcommand\colorMATH{\colorMATHB}\renewcommand\colorSYNTAX{\colorSYNTAXB}{{\color{\colorMATH}\ensuremath{\sss_{1}}}}\endgroup }x_{1} + {\begingroup\renewcommand\colorMATH{\colorMATHB}\renewcommand\colorSYNTAX{\colorSYNTAXB}{{\color{\colorMATH}\ensuremath{\sss_{2}}}}\endgroup }x_{2})}\llbracket {\begingroup\renewcommand\colorMATH{\colorMATHB}\renewcommand\colorSYNTAX{\colorSYNTAXB}{{\color{\colorMATH}\ensuremath{\sS''}}}\endgroup }(\tau _{2})\rrbracket }}},\\
        for {{\color{\colorMATH}\ensuremath{\gamma '_{i} = \gamma _{1}[x_{1} \mapsto  {\begingroup\renewcommand\colorMATH{\colorMATHB}\renewcommand\colorSYNTAX{\colorSYNTAXB}{{\color{\colorMATH}\ensuremath{\sv_{i 1}}}}\endgroup }, x_{2} \mapsto  {\begingroup\renewcommand\colorMATH{\colorMATHB}\renewcommand\colorSYNTAX{\colorSYNTAXB}{{\color{\colorMATH}\ensuremath{\sv_{i 2}}}}\endgroup }], {\begingroup\renewcommand\colorMATH{\colorMATHB}\renewcommand\colorSYNTAX{\colorSYNTAXB}{{\color{\colorMATH}\ensuremath{\sS''}}}\endgroup } = {\begingroup\renewcommand\colorMATH{\colorMATHB}\renewcommand\colorSYNTAX{\colorSYNTAXB}{{\color{\colorMATH}\ensuremath{\Distance'}}}\endgroup }+({\begingroup\renewcommand\colorMATH{\colorMATHB}\renewcommand\colorSYNTAX{\colorSYNTAXB}{{\color{\colorMATH}\ensuremath{\Distance'}}}\endgroup }\mathord{\cdotp }({\begingroup\renewcommand\colorMATH{\colorMATHB}\renewcommand\colorSYNTAX{\colorSYNTAXB}{{\color{\colorMATH}\ensuremath{\sS_{1}}}}\endgroup }+{\begingroup\renewcommand\colorMATH{\colorMATHB}\renewcommand\colorSYNTAX{\colorSYNTAXB}{{\color{\colorMATH}\ensuremath{\sS_{1 1}}}}\endgroup }))x_{1}+({\begingroup\renewcommand\colorMATH{\colorMATHB}\renewcommand\colorSYNTAX{\colorSYNTAXB}{{\color{\colorMATH}\ensuremath{\Distance'}}}\endgroup }\mathord{\cdotp }({\begingroup\renewcommand\colorMATH{\colorMATHB}\renewcommand\colorSYNTAX{\colorSYNTAXB}{{\color{\colorMATH}\ensuremath{\sS_{1}}}}\endgroup }+{\begingroup\renewcommand\colorMATH{\colorMATHB}\renewcommand\colorSYNTAX{\colorSYNTAXB}{{\color{\colorMATH}\ensuremath{\sS_{1 2}}}}\endgroup }))x_{2}}}}.\\
        But {{\color{\colorMATH}\ensuremath{({\begingroup\renewcommand\colorMATH{\colorMATHB}\renewcommand\colorSYNTAX{\colorSYNTAXB}{{\color{\colorMATH}\ensuremath{\Distance'}}}\endgroup }+({\begingroup\renewcommand\colorMATH{\colorMATHB}\renewcommand\colorSYNTAX{\colorSYNTAXB}{{\color{\colorMATH}\ensuremath{\Distance'}}}\endgroup }\mathord{\cdotp }({\begingroup\renewcommand\colorMATH{\colorMATHB}\renewcommand\colorSYNTAX{\colorSYNTAXB}{{\color{\colorMATH}\ensuremath{\sS_{1}}}}\endgroup }+{\begingroup\renewcommand\colorMATH{\colorMATHB}\renewcommand\colorSYNTAX{\colorSYNTAXB}{{\color{\colorMATH}\ensuremath{\sS_{1 1}}}}\endgroup }))x_{1}+({\begingroup\renewcommand\colorMATH{\colorMATHB}\renewcommand\colorSYNTAX{\colorSYNTAXB}{{\color{\colorMATH}\ensuremath{\Distance'}}}\endgroup }\mathord{\cdotp }({\begingroup\renewcommand\colorMATH{\colorMATHB}\renewcommand\colorSYNTAX{\colorSYNTAXB}{{\color{\colorMATH}\ensuremath{\sS_{1}}}}\endgroup }+{\begingroup\renewcommand\colorMATH{\colorMATHB}\renewcommand\colorSYNTAX{\colorSYNTAXB}{{\color{\colorMATH}\ensuremath{\sS_{1 2}}}}\endgroup }))x_{2})\mathord{\cdotp }({\begingroup\renewcommand\colorMATH{\colorMATHB}\renewcommand\colorSYNTAX{\colorSYNTAXB}{{\color{\colorMATH}\ensuremath{\sS_{2}}}}\endgroup } + {\begingroup\renewcommand\colorMATH{\colorMATHB}\renewcommand\colorSYNTAX{\colorSYNTAXB}{{\color{\colorMATH}\ensuremath{\sss_{1}}}}\endgroup }x_{1} + {\begingroup\renewcommand\colorMATH{\colorMATHB}\renewcommand\colorSYNTAX{\colorSYNTAXB}{{\color{\colorMATH}\ensuremath{\sss_{2}}}}\endgroup }x_{2}) = {\begingroup\renewcommand\colorMATH{\colorMATHB}\renewcommand\colorSYNTAX{\colorSYNTAXB}{{\color{\colorMATH}\ensuremath{\Distance'}}}\endgroup }\mathord{\cdotp }{\begingroup\renewcommand\colorMATH{\colorMATHB}\renewcommand\colorSYNTAX{\colorSYNTAXB}{{\color{\colorMATH}\ensuremath{\sS_{2}}}}\endgroup } + {\begingroup\renewcommand\colorMATH{\colorMATHB}\renewcommand\colorSYNTAX{\colorSYNTAXB}{{\color{\colorMATH}\ensuremath{\sss_{1}}}}\endgroup }({\begingroup\renewcommand\colorMATH{\colorMATHB}\renewcommand\colorSYNTAX{\colorSYNTAXB}{{\color{\colorMATH}\ensuremath{\Distance'}}}\endgroup }\mathord{\cdotp }({\begingroup\renewcommand\colorMATH{\colorMATHB}\renewcommand\colorSYNTAX{\colorSYNTAXB}{{\color{\colorMATH}\ensuremath{\sS_{1}}}}\endgroup } + {\begingroup\renewcommand\colorMATH{\colorMATHB}\renewcommand\colorSYNTAX{\colorSYNTAXB}{{\color{\colorMATH}\ensuremath{\sS_{1 1}}}}\endgroup })) + {\begingroup\renewcommand\colorMATH{\colorMATHB}\renewcommand\colorSYNTAX{\colorSYNTAXB}{{\color{\colorMATH}\ensuremath{\sss_{2}}}}\endgroup }({\begingroup\renewcommand\colorMATH{\colorMATHB}\renewcommand\colorSYNTAX{\colorSYNTAXB}{{\color{\colorMATH}\ensuremath{\Distance'}}}\endgroup }\mathord{\cdotp }({\begingroup\renewcommand\colorMATH{\colorMATHB}\renewcommand\colorSYNTAX{\colorSYNTAXB}{{\color{\colorMATH}\ensuremath{\sS_{1}}}}\endgroup } + {\begingroup\renewcommand\colorMATH{\colorMATHB}\renewcommand\colorSYNTAX{\colorSYNTAXB}{{\color{\colorMATH}\ensuremath{\sS_{1 2}}}}\endgroup })) = {\begingroup\renewcommand\colorMATH{\colorMATHB}\renewcommand\colorSYNTAX{\colorSYNTAXB}{{\color{\colorMATH}\ensuremath{\Distance'}}}\endgroup }\mathord{\cdotp }({\begingroup\renewcommand\colorMATH{\colorMATHB}\renewcommand\colorSYNTAX{\colorSYNTAXB}{{\color{\colorMATH}\ensuremath{\sss_{1}}}}\endgroup }({\begingroup\renewcommand\colorMATH{\colorMATHB}\renewcommand\colorSYNTAX{\colorSYNTAXB}{{\color{\colorMATH}\ensuremath{\sS_{1}}}}\endgroup } + {\begingroup\renewcommand\colorMATH{\colorMATHB}\renewcommand\colorSYNTAX{\colorSYNTAXB}{{\color{\colorMATH}\ensuremath{\sS_{1 1}}}}\endgroup })+ {\begingroup\renewcommand\colorMATH{\colorMATHB}\renewcommand\colorSYNTAX{\colorSYNTAXB}{{\color{\colorMATH}\ensuremath{\sss_{2}}}}\endgroup }({\begingroup\renewcommand\colorMATH{\colorMATHB}\renewcommand\colorSYNTAX{\colorSYNTAXB}{{\color{\colorMATH}\ensuremath{\sS_{1}}}}\endgroup } + {\begingroup\renewcommand\colorMATH{\colorMATHB}\renewcommand\colorSYNTAX{\colorSYNTAXB}{{\color{\colorMATH}\ensuremath{\sS_{1 2}}}}\endgroup }) + {\begingroup\renewcommand\colorMATH{\colorMATHB}\renewcommand\colorSYNTAX{\colorSYNTAXB}{{\color{\colorMATH}\ensuremath{\sS_{2}}}}\endgroup })}}}, and
        by Lemma~\ref{lm:equivsimplsubst} and because {{\color{\colorMATH}\ensuremath{{\begingroup\renewcommand\colorMATH{\colorMATHB}\renewcommand\colorSYNTAX{\colorSYNTAXB}{{\color{\colorMATH}\ensuremath{\Distance'}}}\endgroup }\mathord{\cdotp }({\begingroup\renewcommand\colorMATH{\colorMATHB}\renewcommand\colorSYNTAX{\colorSYNTAXB}{{\color{\colorMATH}\ensuremath{\sS_{1}}}}\endgroup }+{\begingroup\renewcommand\colorMATH{\colorMATHB}\renewcommand\colorSYNTAX{\colorSYNTAXB}{{\color{\colorMATH}\ensuremath{\sS_{1 i}}}}\endgroup }) \in  {\text{sens}}}}}, then 
        {{\color{\colorMATH}\ensuremath{({\begingroup\renewcommand\colorMATH{\colorMATHB}\renewcommand\colorSYNTAX{\colorSYNTAXB}{{\color{\colorMATH}\ensuremath{\Distance'}}}\endgroup }+({\begingroup\renewcommand\colorMATH{\colorMATHB}\renewcommand\colorSYNTAX{\colorSYNTAXB}{{\color{\colorMATH}\ensuremath{\Distance'}}}\endgroup }\mathord{\cdotp }({\begingroup\renewcommand\colorMATH{\colorMATHB}\renewcommand\colorSYNTAX{\colorSYNTAXB}{{\color{\colorMATH}\ensuremath{\sS_{1}}}}\endgroup }+{\begingroup\renewcommand\colorMATH{\colorMATHB}\renewcommand\colorSYNTAX{\colorSYNTAXB}{{\color{\colorMATH}\ensuremath{\sS_{1 1}}}}\endgroup }))x_{1}+({\begingroup\renewcommand\colorMATH{\colorMATHB}\renewcommand\colorSYNTAX{\colorSYNTAXB}{{\color{\colorMATH}\ensuremath{\Distance'}}}\endgroup }\mathord{\cdotp }({\begingroup\renewcommand\colorMATH{\colorMATHB}\renewcommand\colorSYNTAX{\colorSYNTAXB}{{\color{\colorMATH}\ensuremath{\sS_{1}}}}\endgroup }+{\begingroup\renewcommand\colorMATH{\colorMATHB}\renewcommand\colorSYNTAX{\colorSYNTAXB}{{\color{\colorMATH}\ensuremath{\sS_{1 2}}}}\endgroup }))x_{2})(\tau _{2}) = (({\begingroup\renewcommand\colorMATH{\colorMATHB}\renewcommand\colorSYNTAX{\colorSYNTAXB}{{\color{\colorMATH}\ensuremath{\Distance'}}}\endgroup }\mathord{\cdotp }({\begingroup\renewcommand\colorMATH{\colorMATHB}\renewcommand\colorSYNTAX{\colorSYNTAXB}{{\color{\colorMATH}\ensuremath{\sS_{1}}}}\endgroup }+{\begingroup\renewcommand\colorMATH{\colorMATHB}\renewcommand\colorSYNTAX{\colorSYNTAXB}{{\color{\colorMATH}\ensuremath{\sS_{1 1}}}}\endgroup }))x_{1} + ({\begingroup\renewcommand\colorMATH{\colorMATHB}\renewcommand\colorSYNTAX{\colorSYNTAXB}{{\color{\colorMATH}\ensuremath{\Distance'}}}\endgroup }\mathord{\cdotp }({\begingroup\renewcommand\colorMATH{\colorMATHB}\renewcommand\colorSYNTAX{\colorSYNTAXB}{{\color{\colorMATH}\ensuremath{\sS_{1}}}}\endgroup }+{\begingroup\renewcommand\colorMATH{\colorMATHB}\renewcommand\colorSYNTAX{\colorSYNTAXB}{{\color{\colorMATH}\ensuremath{\sS_{1 2}}}}\endgroup }))x_{2})({\begingroup\renewcommand\colorMATH{\colorMATHB}\renewcommand\colorSYNTAX{\colorSYNTAXB}{{\color{\colorMATH}\ensuremath{\Distance'}}}\endgroup }(\tau _{2}))}}}, and by Lemma~\ref{lm:distrinst} 
        {{\color{\colorMATH}\ensuremath{(({\begingroup\renewcommand\colorMATH{\colorMATHB}\renewcommand\colorSYNTAX{\colorSYNTAXB}{{\color{\colorMATH}\ensuremath{\Distance'}}}\endgroup }\mathord{\cdotp }({\begingroup\renewcommand\colorMATH{\colorMATHB}\renewcommand\colorSYNTAX{\colorSYNTAXB}{{\color{\colorMATH}\ensuremath{\sS_{1}}}}\endgroup }+{\begingroup\renewcommand\colorMATH{\colorMATHB}\renewcommand\colorSYNTAX{\colorSYNTAXB}{{\color{\colorMATH}\ensuremath{\sS_{1 1}}}}\endgroup }))x_{1} + ({\begingroup\renewcommand\colorMATH{\colorMATHB}\renewcommand\colorSYNTAX{\colorSYNTAXB}{{\color{\colorMATH}\ensuremath{\Distance'}}}\endgroup }\mathord{\cdotp }({\begingroup\renewcommand\colorMATH{\colorMATHB}\renewcommand\colorSYNTAX{\colorSYNTAXB}{{\color{\colorMATH}\ensuremath{\sS_{1}}}}\endgroup }+{\begingroup\renewcommand\colorMATH{\colorMATHB}\renewcommand\colorSYNTAX{\colorSYNTAXB}{{\color{\colorMATH}\ensuremath{\sS_{1 2}}}}\endgroup }))x_{2})({\begingroup\renewcommand\colorMATH{\colorMATHB}\renewcommand\colorSYNTAX{\colorSYNTAXB}{{\color{\colorMATH}\ensuremath{\Distance'}}}\endgroup }(\tau _{2})) = {\begingroup\renewcommand\colorMATH{\colorMATHB}\renewcommand\colorSYNTAX{\colorSYNTAXB}{{\color{\colorMATH}\ensuremath{\Distance'}}}\endgroup }([{\begingroup\renewcommand\colorMATH{\colorMATHB}\renewcommand\colorSYNTAX{\colorSYNTAXB}{{\color{\colorMATH}\ensuremath{\sS_{1}}}}\endgroup }+{\begingroup\renewcommand\colorMATH{\colorMATHB}\renewcommand\colorSYNTAX{\colorSYNTAXB}{{\color{\colorMATH}\ensuremath{\sS_{1 1}}}}\endgroup }/x_{1}][{\begingroup\renewcommand\colorMATH{\colorMATHB}\renewcommand\colorSYNTAX{\colorSYNTAXB}{{\color{\colorMATH}\ensuremath{\sS_{1}}}}\endgroup }+{\begingroup\renewcommand\colorMATH{\colorMATHB}\renewcommand\colorSYNTAX{\colorSYNTAXB}{{\color{\colorMATH}\ensuremath{\sS_{1 2}}}}\endgroup }/x_{2}]\tau _{2})}}}.\\
        Therefore if {{\color{\colorMATH}\ensuremath{\gamma '_{1}\vdash {\begingroup\renewcommand\colorMATH{\colorMATHB}\renewcommand\colorSYNTAX{\colorSYNTAXB}{{\color{\colorMATH}\ensuremath{\se_{2}}}}\endgroup } \Downarrow ^{j_{2}} {\begingroup\renewcommand\colorMATH{\colorMATHB}\renewcommand\colorSYNTAX{\colorSYNTAXB}{{\color{\colorMATH}\ensuremath{\sv'_{1}}}}\endgroup }}}}, \pthen {{\color{\colorMATH}\ensuremath{\gamma '_{2}\vdash {\begingroup\renewcommand\colorMATH{\colorMATHB}\renewcommand\colorSYNTAX{\colorSYNTAXB}{{\color{\colorMATH}\ensuremath{\se_{2}}}}\endgroup } \Downarrow ^{\pj[2]} {\begingroup\renewcommand\colorMATH{\colorMATHB}\renewcommand\colorSYNTAX{\colorSYNTAXB}{{\color{\colorMATH}\ensuremath{\sv'_{2}}}}\endgroup }}}}, \pand\\
        {{\color{\colorMATH}\ensuremath{({\begingroup\renewcommand\colorMATH{\colorMATHB}\renewcommand\colorSYNTAX{\colorSYNTAXB}{{\color{\colorMATH}\ensuremath{\sv_{2 1}}}}\endgroup }, {\begingroup\renewcommand\colorMATH{\colorMATHB}\renewcommand\colorSYNTAX{\colorSYNTAXB}{{\color{\colorMATH}\ensuremath{\sv_{2 2}}}}\endgroup }) \in  {\mathcal{V}}^{k-j_{1}-j_{2}}_{{\begingroup\renewcommand\colorMATH{\colorMATHB}\renewcommand\colorSYNTAX{\colorSYNTAXB}{{\color{\colorMATH}\ensuremath{\Distance'}}}\endgroup }\mathord{\cdotp }({\begingroup\renewcommand\colorMATH{\colorMATHB}\renewcommand\colorSYNTAX{\colorSYNTAXB}{{\color{\colorMATH}\ensuremath{\sss_{1}}}}\endgroup }({\begingroup\renewcommand\colorMATH{\colorMATHB}\renewcommand\colorSYNTAX{\colorSYNTAXB}{{\color{\colorMATH}\ensuremath{\sS_{1}}}}\endgroup } + {\begingroup\renewcommand\colorMATH{\colorMATHB}\renewcommand\colorSYNTAX{\colorSYNTAXB}{{\color{\colorMATH}\ensuremath{\sS_{1 1}}}}\endgroup })+ {\begingroup\renewcommand\colorMATH{\colorMATHB}\renewcommand\colorSYNTAX{\colorSYNTAXB}{{\color{\colorMATH}\ensuremath{\sss_{2}}}}\endgroup }({\begingroup\renewcommand\colorMATH{\colorMATHB}\renewcommand\colorSYNTAX{\colorSYNTAXB}{{\color{\colorMATH}\ensuremath{\sS_{1}}}}\endgroup } + {\begingroup\renewcommand\colorMATH{\colorMATHB}\renewcommand\colorSYNTAX{\colorSYNTAXB}{{\color{\colorMATH}\ensuremath{\sS_{1 2}}}}\endgroup }) + {\begingroup\renewcommand\colorMATH{\colorMATHB}\renewcommand\colorSYNTAX{\colorSYNTAXB}{{\color{\colorMATH}\ensuremath{\sS_{2}}}}\endgroup })}\llbracket {\begingroup\renewcommand\colorMATH{\colorMATHB}\renewcommand\colorSYNTAX{\colorSYNTAXB}{{\color{\colorMATH}\ensuremath{\Distance'}}}\endgroup }([{\begingroup\renewcommand\colorMATH{\colorMATHB}\renewcommand\colorSYNTAX{\colorSYNTAXB}{{\color{\colorMATH}\ensuremath{\sS_{1}}}}\endgroup }+{\begingroup\renewcommand\colorMATH{\colorMATHB}\renewcommand\colorSYNTAX{\colorSYNTAXB}{{\color{\colorMATH}\ensuremath{\sS_{1 1}}}}\endgroup }/x_{1}][{\begingroup\renewcommand\colorMATH{\colorMATHB}\renewcommand\colorSYNTAX{\colorSYNTAXB}{{\color{\colorMATH}\ensuremath{\sS_{1}}}}\endgroup }+{\begingroup\renewcommand\colorMATH{\colorMATHB}\renewcommand\colorSYNTAX{\colorSYNTAXB}{{\color{\colorMATH}\ensuremath{\sS_{1 2}}}}\endgroup }/x_{2}]\tau _{2})\rrbracket }}}, and the result holds by Lemma~\ref{lm:weakening-index}.
      \end{subproof}
    \item  {{\color{\colorMATH}\ensuremath{\Gamma ; {\begingroup\renewcommand\colorMATH{\colorMATHB}\renewcommand\colorSYNTAX{\colorSYNTAXB}{{\color{\colorMATH}\ensuremath{\Distance}}}\endgroup } \vdash  ({\begingroup\renewcommand\colorMATH{\colorMATHB}\renewcommand\colorSYNTAX{\colorSYNTAXB}{{\color{\colorMATH}\ensuremath{\se'}}}\endgroup } \mathrel{:: } \tau ) \mathrel{:} \tau  \mathrel{;} {\begingroup\renewcommand\colorMATH{\colorMATHB}\renewcommand\colorSYNTAX{\colorSYNTAXB}{{\color{\colorMATH}\ensuremath{\sS}}}\endgroup }}}}
      \begin{subproof} 
        We have to prove that for any {{\color{\colorMATH}\ensuremath{k}}}, {{\color{\colorMATH}\ensuremath{\forall  (\gamma _{1},\gamma _{2}) \in  {\mathcal{G}}_{{\begingroup\renewcommand\colorMATH{\colorMATHB}\renewcommand\colorSYNTAX{\colorSYNTAXB}{{\color{\colorMATH}\ensuremath{\Distance'}}}\endgroup }}^{\kg}\llbracket \Gamma \rrbracket , (\gamma _{1}\vdash ({\begingroup\renewcommand\colorMATH{\colorMATHB}\renewcommand\colorSYNTAX{\colorSYNTAXB}{{\color{\colorMATH}\ensuremath{\se'}}}\endgroup } \mathrel{:: } \tau ),\gamma _{2}\vdash ({\begingroup\renewcommand\colorMATH{\colorMATHB}\renewcommand\colorSYNTAX{\colorSYNTAXB}{{\color{\colorMATH}\ensuremath{\se'}}}\endgroup } \mathrel{:: } \tau )) \in  {\mathcal{E}}_{{\begingroup\renewcommand\colorMATH{\colorMATHB}\renewcommand\colorSYNTAX{\colorSYNTAXB}{{\color{\colorMATH}\ensuremath{\Distance'}}}\endgroup }\mathord{\cdotp }{\begingroup\renewcommand\colorMATH{\colorMATHB}\renewcommand\colorSYNTAX{\colorSYNTAXB}{{\color{\colorMATH}\ensuremath{\sS}}}\endgroup }}^{k}\llbracket {\begingroup\renewcommand\colorMATH{\colorMATHB}\renewcommand\colorSYNTAX{\colorSYNTAXB}{{\color{\colorMATH}\ensuremath{\Distance'}}}\endgroup }(\tau )\rrbracket }}}, for {{\color{\colorMATH}\ensuremath{{\begingroup\renewcommand\colorMATH{\colorMATHB}\renewcommand\colorSYNTAX{\colorSYNTAXB}{{\color{\colorMATH}\ensuremath{\Distance'}}}\endgroup } \sqsubseteq  {\begingroup\renewcommand\colorMATH{\colorMATHB}\renewcommand\colorSYNTAX{\colorSYNTAXB}{{\color{\colorMATH}\ensuremath{\Distance}}}\endgroup }}}}.
        By induction hypothesis on {{\color{\colorMATH}\ensuremath{\Gamma ; {\begingroup\renewcommand\colorMATH{\colorMATHB}\renewcommand\colorSYNTAX{\colorSYNTAXB}{{\color{\colorMATH}\ensuremath{\Distance}}}\endgroup } \vdash  {\begingroup\renewcommand\colorMATH{\colorMATHB}\renewcommand\colorSYNTAX{\colorSYNTAXB}{{\color{\colorMATH}\ensuremath{\se'}}}\endgroup } \mathrel{:} \tau ' \mathrel{;} {\begingroup\renewcommand\colorMATH{\colorMATHB}\renewcommand\colorSYNTAX{\colorSYNTAXB}{{\color{\colorMATH}\ensuremath{\sS}}}\endgroup }}}} where {{\color{\colorMATH}\ensuremath{\tau '<:\tau }}}, we know that\\
        {{\color{\colorMATH}\ensuremath{(\gamma _{1}\vdash {\begingroup\renewcommand\colorMATH{\colorMATHB}\renewcommand\colorSYNTAX{\colorSYNTAXB}{{\color{\colorMATH}\ensuremath{\se'}}}\endgroup },\gamma _{2}\vdash {\begingroup\renewcommand\colorMATH{\colorMATHB}\renewcommand\colorSYNTAX{\colorSYNTAXB}{{\color{\colorMATH}\ensuremath{\se'}}}\endgroup }) \in  {\mathcal{E}}_{{\begingroup\renewcommand\colorMATH{\colorMATHB}\renewcommand\colorSYNTAX{\colorSYNTAXB}{{\color{\colorMATH}\ensuremath{\Distance'}}}\endgroup }\mathord{\cdotp }{\begingroup\renewcommand\colorMATH{\colorMATHB}\renewcommand\colorSYNTAX{\colorSYNTAXB}{{\color{\colorMATH}\ensuremath{\sS}}}\endgroup }}^{k}\llbracket {\begingroup\renewcommand\colorMATH{\colorMATHB}\renewcommand\colorSYNTAX{\colorSYNTAXB}{{\color{\colorMATH}\ensuremath{\Distance'}}}\endgroup }(\tau ')\rrbracket }}}, i.e.
        if {{\color{\colorMATH}\ensuremath{\gamma _{1}\vdash {\begingroup\renewcommand\colorMATH{\colorMATHB}\renewcommand\colorSYNTAX{\colorSYNTAXB}{{\color{\colorMATH}\ensuremath{\se'}}}\endgroup } \Downarrow ^{j} {\begingroup\renewcommand\colorMATH{\colorMATHB}\renewcommand\colorSYNTAX{\colorSYNTAXB}{{\color{\colorMATH}\ensuremath{\sv_{1}}}}\endgroup }}}} \pthen {{\color{\colorMATH}\ensuremath{\gamma _{2}\vdash {\begingroup\renewcommand\colorMATH{\colorMATHB}\renewcommand\colorSYNTAX{\colorSYNTAXB}{{\color{\colorMATH}\ensuremath{\se'}}}\endgroup } \Downarrow ^{\pj} {\begingroup\renewcommand\colorMATH{\colorMATHB}\renewcommand\colorSYNTAX{\colorSYNTAXB}{{\color{\colorMATH}\ensuremath{\sv_{2}}}}\endgroup }}}}, \pand
        {{\color{\colorMATH}\ensuremath{({\begingroup\renewcommand\colorMATH{\colorMATHB}\renewcommand\colorSYNTAX{\colorSYNTAXB}{{\color{\colorMATH}\ensuremath{\sv_{1}}}}\endgroup }, {\begingroup\renewcommand\colorMATH{\colorMATHB}\renewcommand\colorSYNTAX{\colorSYNTAXB}{{\color{\colorMATH}\ensuremath{\sv_{2}}}}\endgroup }) \in  {\mathcal{V}}_{{\begingroup\renewcommand\colorMATH{\colorMATHB}\renewcommand\colorSYNTAX{\colorSYNTAXB}{{\color{\colorMATH}\ensuremath{\Distance'}}}\endgroup }\mathord{\cdotp }{\begingroup\renewcommand\colorMATH{\colorMATHB}\renewcommand\colorSYNTAX{\colorSYNTAXB}{{\color{\colorMATH}\ensuremath{\sS}}}\endgroup }}^{k-j}\llbracket {\begingroup\renewcommand\colorMATH{\colorMATHB}\renewcommand\colorSYNTAX{\colorSYNTAXB}{{\color{\colorMATH}\ensuremath{\Distance'}}}\endgroup }(\tau ')\rrbracket }}}.

        Following the {\textsc{ ascr}} reduction rule, if:
        \begingroup\color{\colorMATH}\begin{mathpar} 
          \inferrule*[lab=
          ]{ \gamma _{1}\vdash {\begingroup\renewcommand\colorMATH{\colorMATHB}\renewcommand\colorSYNTAX{\colorSYNTAXB}{{\color{\colorMATH}\ensuremath{\se'}}}\endgroup } \Downarrow ^{j} {\begingroup\renewcommand\colorMATH{\colorMATHB}\renewcommand\colorSYNTAX{\colorSYNTAXB}{{\color{\colorMATH}\ensuremath{\sv_{1}}}}\endgroup }
            }{
            \gamma _{1}\vdash {\begingroup\renewcommand\colorMATH{\colorMATHB}\renewcommand\colorSYNTAX{\colorSYNTAXB}{{\color{\colorMATH}\ensuremath{\se'}}}\endgroup } :: \tau  \Downarrow ^{j+1} {\begingroup\renewcommand\colorMATH{\colorMATHB}\renewcommand\colorSYNTAX{\colorSYNTAXB}{{\color{\colorMATH}\ensuremath{\sv_{1}}}}\endgroup }
          }
        \and \inferrule*[lab=
          ]{ \gamma _{2}\vdash {\begingroup\renewcommand\colorMATH{\colorMATHB}\renewcommand\colorSYNTAX{\colorSYNTAXB}{{\color{\colorMATH}\ensuremath{\se'}}}\endgroup } \Downarrow ^{*} {\begingroup\renewcommand\colorMATH{\colorMATHB}\renewcommand\colorSYNTAX{\colorSYNTAXB}{{\color{\colorMATH}\ensuremath{\sv_{2}}}}\endgroup }
            }{
            \gamma _{2}\vdash {\begingroup\renewcommand\colorMATH{\colorMATHB}\renewcommand\colorSYNTAX{\colorSYNTAXB}{{\color{\colorMATH}\ensuremath{\se'}}}\endgroup } :: \tau  \Downarrow ^{*} {\begingroup\renewcommand\colorMATH{\colorMATHB}\renewcommand\colorSYNTAX{\colorSYNTAXB}{{\color{\colorMATH}\ensuremath{\sv_{2}}}}\endgroup }
          }
        \end{mathpar}\endgroup

        Then we have to prove that {{\color{\colorMATH}\ensuremath{({\begingroup\renewcommand\colorMATH{\colorMATHB}\renewcommand\colorSYNTAX{\colorSYNTAXB}{{\color{\colorMATH}\ensuremath{\sv_{1}}}}\endgroup }, {\begingroup\renewcommand\colorMATH{\colorMATHB}\renewcommand\colorSYNTAX{\colorSYNTAXB}{{\color{\colorMATH}\ensuremath{\sv_{2}}}}\endgroup }) \in  {\mathcal{V}}_{{\begingroup\renewcommand\colorMATH{\colorMATHB}\renewcommand\colorSYNTAX{\colorSYNTAXB}{{\color{\colorMATH}\ensuremath{\Distance'}}}\endgroup }\mathord{\cdotp }{\begingroup\renewcommand\colorMATH{\colorMATHB}\renewcommand\colorSYNTAX{\colorSYNTAXB}{{\color{\colorMATH}\ensuremath{\sS}}}\endgroup }}^{k-j-1}\llbracket {\begingroup\renewcommand\colorMATH{\colorMATHB}\renewcommand\colorSYNTAX{\colorSYNTAXB}{{\color{\colorMATH}\ensuremath{\Distance'}}}\endgroup }(\tau )\rrbracket }}}. As {{\color{\colorMATH}\ensuremath{\tau '<:\tau }}}, by Lemma~\ref{lm:subtypinginst} {{\color{\colorMATH}\ensuremath{{\begingroup\renewcommand\colorMATH{\colorMATHB}\renewcommand\colorSYNTAX{\colorSYNTAXB}{{\color{\colorMATH}\ensuremath{\Distance'}}}\endgroup }(\tau ')<:{\begingroup\renewcommand\colorMATH{\colorMATHB}\renewcommand\colorSYNTAX{\colorSYNTAXB}{{\color{\colorMATH}\ensuremath{\Distance'}}}\endgroup }(\tau )}}}, and the result holds immediately by Lemma~\ref{lm:weakening-index}.
      \end{subproof}
    \item  {{\color{\colorMATH}\ensuremath{\Gamma ; {\begingroup\renewcommand\colorMATH{\colorMATHB}\renewcommand\colorSYNTAX{\colorSYNTAXB}{{\color{\colorMATH}\ensuremath{\Distance}}}\endgroup } \vdash  {\begingroup\renewcommand\colorMATH{\colorMATHC}\renewcommand\colorSYNTAX{\colorSYNTAXC}{{\color{\colorMATH}\ensuremath{\plambda}}}\endgroup } (x\mathrel{:}\tau _{1}\mathord{\cdotp }{\begingroup\renewcommand\colorMATH{\colorMATHB}\renewcommand\colorSYNTAX{\colorSYNTAXB}{{\color{\colorMATH}\ensuremath{\distance'}}}\endgroup }).\hspace*{0.33em}{\begingroup\renewcommand\colorMATH{\colorMATHC}\renewcommand\colorSYNTAX{\colorSYNTAXC}{{\color{\colorMATH}\ensuremath{\pe'}}}\endgroup } \mathrel{:} (x\mathrel{:}\tau _{1}\mathord{\cdotp }{\begingroup\renewcommand\colorMATH{\colorMATHB}\renewcommand\colorSYNTAX{\colorSYNTAXB}{{\color{\colorMATH}\ensuremath{\distance'}}}\endgroup }) \xrightarrowP {{\begingroup\renewcommand\colorMATH{\colorMATHC}\renewcommand\colorSYNTAX{\colorSYNTAXC}{{\color{\colorMATH}\ensuremath{\pS''}}}\endgroup }} \tau _{2} \mathrel{;} {\begingroup\renewcommand\colorMATH{\colorMATHB}\renewcommand\colorSYNTAX{\colorSYNTAXB}{{\color{\colorMATH}\ensuremath{\varnothing }}}\endgroup }}}} 
    \begin{subproof} 
      We have to prove that {{\color{\colorMATH}\ensuremath{\forall  (\gamma _{1},\gamma _{2}) \in  {\mathcal{G}}_{{\begingroup\renewcommand\colorMATH{\colorMATHB}\renewcommand\colorSYNTAX{\colorSYNTAXB}{{\color{\colorMATH}\ensuremath{\Distance'}}}\endgroup }}^{\kg}\llbracket \Gamma \rrbracket }}},\\
      {{\color{\colorMATH}\ensuremath{(\gamma _{1}\vdash {\begingroup\renewcommand\colorMATH{\colorMATHC}\renewcommand\colorSYNTAX{\colorSYNTAXC}{{\color{\colorMATH}\ensuremath{\plambda}}}\endgroup } (x\mathrel{:}\tau _{1}\mathord{\cdotp }{\begingroup\renewcommand\colorMATH{\colorMATHB}\renewcommand\colorSYNTAX{\colorSYNTAXB}{{\color{\colorMATH}\ensuremath{\distance'}}}\endgroup }).\hspace*{0.33em}{\begingroup\renewcommand\colorMATH{\colorMATHC}\renewcommand\colorSYNTAX{\colorSYNTAXC}{{\color{\colorMATH}\ensuremath{\pe'}}}\endgroup },\gamma _{2}\vdash {\begingroup\renewcommand\colorMATH{\colorMATHC}\renewcommand\colorSYNTAX{\colorSYNTAXC}{{\color{\colorMATH}\ensuremath{\plambda}}}\endgroup } (x\mathrel{:}\tau _{1}\mathord{\cdotp }{\begingroup\renewcommand\colorMATH{\colorMATHB}\renewcommand\colorSYNTAX{\colorSYNTAXB}{{\color{\colorMATH}\ensuremath{\distance'}}}\endgroup }).\hspace*{0.33em}{\begingroup\renewcommand\colorMATH{\colorMATHC}\renewcommand\colorSYNTAX{\colorSYNTAXC}{{\color{\colorMATH}\ensuremath{\pe'}}}\endgroup }) \in  {\mathcal{E}}_{{\begingroup\renewcommand\colorMATH{\colorMATHB}\renewcommand\colorSYNTAX{\colorSYNTAXB}{{\color{\colorMATH}\ensuremath{\Distance'}}}\endgroup }\mathord{\cdotp }{\begingroup\renewcommand\colorMATH{\colorMATHB}\renewcommand\colorSYNTAX{\colorSYNTAXB}{{\color{\colorMATH}\ensuremath{\varnothing }}}\endgroup }}^{k}\llbracket {\begingroup\renewcommand\colorMATH{\colorMATHB}\renewcommand\colorSYNTAX{\colorSYNTAXB}{{\color{\colorMATH}\ensuremath{\Distance'}}}\endgroup }(x:\tau _{1}\mathord{\cdotp }{\begingroup\renewcommand\colorMATH{\colorMATHB}\renewcommand\colorSYNTAX{\colorSYNTAXB}{{\color{\colorMATH}\ensuremath{\distance'}}}\endgroup }) \xrightarrowP {{\begingroup\renewcommand\colorMATH{\colorMATHB}\renewcommand\colorSYNTAX{\colorSYNTAXB}{{\color{\colorMATH}\ensuremath{\Distance'}}}\endgroup }\mathord{\cdotp }{\begingroup\renewcommand\colorMATH{\colorMATHC}\renewcommand\colorSYNTAX{\colorSYNTAXC}{{\color{\colorMATH}\ensuremath{\pS''}}}\endgroup }} {\begingroup\renewcommand\colorMATH{\colorMATHB}\renewcommand\colorSYNTAX{\colorSYNTAXB}{{\color{\colorMATH}\ensuremath{\Distance'}}}\endgroup }(\tau _{2})\rrbracket }}}, for {{\color{\colorMATH}\ensuremath{{\begingroup\renewcommand\colorMATH{\colorMATHB}\renewcommand\colorSYNTAX{\colorSYNTAXB}{{\color{\colorMATH}\ensuremath{\Distance'}}}\endgroup } \sqsubseteq  {\begingroup\renewcommand\colorMATH{\colorMATHB}\renewcommand\colorSYNTAX{\colorSYNTAXB}{{\color{\colorMATH}\ensuremath{\Distance}}}\endgroup }}}}.\\
      Notice that {{\color{\colorMATH}\ensuremath{{\begingroup\renewcommand\colorMATH{\colorMATHB}\renewcommand\colorSYNTAX{\colorSYNTAXB}{{\color{\colorMATH}\ensuremath{\Distance'}}}\endgroup }\mathord{\cdotp }{\begingroup\renewcommand\colorMATH{\colorMATHB}\renewcommand\colorSYNTAX{\colorSYNTAXB}{{\color{\colorMATH}\ensuremath{\varnothing }}}\endgroup } = 0}}}, and that lambdas reduce to closures, therefore we have to prove that \\
      {{\color{\colorMATH}\ensuremath{(\langle {\begingroup\renewcommand\colorMATH{\colorMATHC}\renewcommand\colorSYNTAX{\colorSYNTAXC}{{\color{\colorMATH}\ensuremath{\plambda}}}\endgroup } (x\mathrel{:}\tau _{1}\mathord{\cdotp }{\begingroup\renewcommand\colorMATH{\colorMATHB}\renewcommand\colorSYNTAX{\colorSYNTAXB}{{\color{\colorMATH}\ensuremath{\distance'}}}\endgroup }).\hspace*{0.33em}{\begingroup\renewcommand\colorMATH{\colorMATHC}\renewcommand\colorSYNTAX{\colorSYNTAXC}{{\color{\colorMATH}\ensuremath{\pe'}}}\endgroup }, \gamma _{1}\rangle ,\langle {\begingroup\renewcommand\colorMATH{\colorMATHC}\renewcommand\colorSYNTAX{\colorSYNTAXC}{{\color{\colorMATH}\ensuremath{\plambda}}}\endgroup } (x\mathrel{:}\tau _{1}\mathord{\cdotp }{\begingroup\renewcommand\colorMATH{\colorMATHB}\renewcommand\colorSYNTAX{\colorSYNTAXB}{{\color{\colorMATH}\ensuremath{\distance'}}}\endgroup }).\hspace*{0.33em}{\begingroup\renewcommand\colorMATH{\colorMATHC}\renewcommand\colorSYNTAX{\colorSYNTAXC}{{\color{\colorMATH}\ensuremath{\pe'}}}\endgroup }, \gamma _{2}\rangle ) \in  {\mathcal{V}}_{{\begingroup\renewcommand\colorMATH{\colorMATHB}\renewcommand\colorSYNTAX{\colorSYNTAXB}{{\color{\colorMATH}\ensuremath{0}}}\endgroup }}^{k}\llbracket (x:{\begingroup\renewcommand\colorMATH{\colorMATHB}\renewcommand\colorSYNTAX{\colorSYNTAXB}{{\color{\colorMATH}\ensuremath{\Distance'}}}\endgroup }(\tau _{1})\mathord{\cdotp }{\begingroup\renewcommand\colorMATH{\colorMATHB}\renewcommand\colorSYNTAX{\colorSYNTAXB}{{\color{\colorMATH}\ensuremath{\distance'}}}\endgroup }) \xrightarrowP {{\begingroup\renewcommand\colorMATH{\colorMATHB}\renewcommand\colorSYNTAX{\colorSYNTAXB}{{\color{\colorMATH}\ensuremath{\Distance'}}}\endgroup }\mathord{\cdotp }{\begingroup\renewcommand\colorMATH{\colorMATHC}\renewcommand\colorSYNTAX{\colorSYNTAXC}{{\color{\colorMATH}\ensuremath{\pS''}}}\endgroup }} {\begingroup\renewcommand\colorMATH{\colorMATHB}\renewcommand\colorSYNTAX{\colorSYNTAXB}{{\color{\colorMATH}\ensuremath{\Distance'}}}\endgroup }(\tau _{2})\rrbracket }}}. 

      Consider {{\color{\colorMATH}\ensuremath{j<k}}}, {{\color{\colorMATH}\ensuremath{{\begingroup\renewcommand\colorMATH{\colorMATHB}\renewcommand\colorSYNTAX{\colorSYNTAXB}{{\color{\colorMATH}\ensuremath{\sv_{1}}}}\endgroup }}}} and {{\color{\colorMATH}\ensuremath{{\begingroup\renewcommand\colorMATH{\colorMATHB}\renewcommand\colorSYNTAX{\colorSYNTAXB}{{\color{\colorMATH}\ensuremath{\sv_{2}}}}\endgroup }}}} such that {{\color{\colorMATH}\ensuremath{({\begingroup\renewcommand\colorMATH{\colorMATHB}\renewcommand\colorSYNTAX{\colorSYNTAXB}{{\color{\colorMATH}\ensuremath{\sv_{1}}}}\endgroup },{\begingroup\renewcommand\colorMATH{\colorMATHB}\renewcommand\colorSYNTAX{\colorSYNTAXB}{{\color{\colorMATH}\ensuremath{\sv_{2}}}}\endgroup }) \in  {\mathcal{V}}_{{\begingroup\renewcommand\colorMATH{\colorMATHB}\renewcommand\colorSYNTAX{\colorSYNTAXB}{{\color{\colorMATH}\ensuremath{\distance''}}}\endgroup }}^{j}\llbracket {\begingroup\renewcommand\colorMATH{\colorMATHB}\renewcommand\colorSYNTAX{\colorSYNTAXB}{{\color{\colorMATH}\ensuremath{\Distance'}}}\endgroup }(\tau _{1})\rrbracket }}}, for some {{\color{\colorMATH}\ensuremath{{\begingroup\renewcommand\colorMATH{\colorMATHB}\renewcommand\colorSYNTAX{\colorSYNTAXB}{{\color{\colorMATH}\ensuremath{\distance''}}}\endgroup } \leq  {\begingroup\renewcommand\colorMATH{\colorMATHB}\renewcommand\colorSYNTAX{\colorSYNTAXB}{{\color{\colorMATH}\ensuremath{\distance'}}}\endgroup }}}}. We have to prove that\\
      {{\color{\colorMATH}\ensuremath{(\gamma _{1}, x \mapsto  {\begingroup\renewcommand\colorMATH{\colorMATHB}\renewcommand\colorSYNTAX{\colorSYNTAXB}{{\color{\colorMATH}\ensuremath{\sv_{1}}}}\endgroup } \vdash  {\begingroup\renewcommand\colorMATH{\colorMATHC}\renewcommand\colorSYNTAX{\colorSYNTAXC}{{\color{\colorMATH}\ensuremath{\pe'}}}\endgroup }, \gamma _{2}, x \mapsto  {\begingroup\renewcommand\colorMATH{\colorMATHB}\renewcommand\colorSYNTAX{\colorSYNTAXB}{{\color{\colorMATH}\ensuremath{\sv_{2}}}}\endgroup } \vdash  {\begingroup\renewcommand\colorMATH{\colorMATHC}\renewcommand\colorSYNTAX{\colorSYNTAXC}{{\color{\colorMATH}\ensuremath{\pe'}}}\endgroup }) \in  {\mathcal{E}}_{({\begingroup\renewcommand\colorMATH{\colorMATHB}\renewcommand\colorSYNTAX{\colorSYNTAXB}{{\color{\colorMATH}\ensuremath{\Distance'}}}\endgroup } + {\begingroup\renewcommand\colorMATH{\colorMATHB}\renewcommand\colorSYNTAX{\colorSYNTAXB}{{\color{\colorMATH}\ensuremath{\distance''}}}\endgroup }x)\mathord{\cdotp }{\begingroup\renewcommand\colorMATH{\colorMATHC}\renewcommand\colorSYNTAX{\colorSYNTAXC}{{\color{\colorMATH}\ensuremath{\pS''}}}\endgroup }}^{j}\llbracket {\begingroup\renewcommand\colorMATH{\colorMATHB}\renewcommand\colorSYNTAX{\colorSYNTAXB}{{\color{\colorMATH}\ensuremath{\distance''}}}\endgroup }x({\begingroup\renewcommand\colorMATH{\colorMATHB}\renewcommand\colorSYNTAX{\colorSYNTAXB}{{\color{\colorMATH}\ensuremath{\Distance'}}}\endgroup }(\tau _{2}))\rrbracket }}}. Notice that by Lemma~\ref{lm:equivsimplsubst} {{\color{\colorMATH}\ensuremath{{\begingroup\renewcommand\colorMATH{\colorMATHB}\renewcommand\colorSYNTAX{\colorSYNTAXB}{{\color{\colorMATH}\ensuremath{\distance''}}}\endgroup }x({\begingroup\renewcommand\colorMATH{\colorMATHB}\renewcommand\colorSYNTAX{\colorSYNTAXB}{{\color{\colorMATH}\ensuremath{\Distance'}}}\endgroup }(\tau _{2})) = ({\begingroup\renewcommand\colorMATH{\colorMATHB}\renewcommand\colorSYNTAX{\colorSYNTAXB}{{\color{\colorMATH}\ensuremath{\Distance'}}}\endgroup } + {\begingroup\renewcommand\colorMATH{\colorMATHB}\renewcommand\colorSYNTAX{\colorSYNTAXB}{{\color{\colorMATH}\ensuremath{\distance''}}}\endgroup }x)(\tau _{2})}}}, therefore we have to prove that 
      {{\color{\colorMATH}\ensuremath{(\gamma _{1}, x \mapsto  {\begingroup\renewcommand\colorMATH{\colorMATHB}\renewcommand\colorSYNTAX{\colorSYNTAXB}{{\color{\colorMATH}\ensuremath{\sv_{1}}}}\endgroup } \vdash  {\begingroup\renewcommand\colorMATH{\colorMATHC}\renewcommand\colorSYNTAX{\colorSYNTAXC}{{\color{\colorMATH}\ensuremath{\pe'}}}\endgroup }, \gamma _{2}, x \mapsto  {\begingroup\renewcommand\colorMATH{\colorMATHB}\renewcommand\colorSYNTAX{\colorSYNTAXB}{{\color{\colorMATH}\ensuremath{\sv_{2}}}}\endgroup } \vdash  {\begingroup\renewcommand\colorMATH{\colorMATHC}\renewcommand\colorSYNTAX{\colorSYNTAXC}{{\color{\colorMATH}\ensuremath{\pe'}}}\endgroup }) \in  {\mathcal{E}}_{({\begingroup\renewcommand\colorMATH{\colorMATHB}\renewcommand\colorSYNTAX{\colorSYNTAXB}{{\color{\colorMATH}\ensuremath{\Distance'}}}\endgroup } + {\begingroup\renewcommand\colorMATH{\colorMATHB}\renewcommand\colorSYNTAX{\colorSYNTAXB}{{\color{\colorMATH}\ensuremath{\distance''}}}\endgroup }x)\mathord{\cdotp }{\begingroup\renewcommand\colorMATH{\colorMATHC}\renewcommand\colorSYNTAX{\colorSYNTAXC}{{\color{\colorMATH}\ensuremath{\pS''}}}\endgroup }}^{j}\llbracket ({\begingroup\renewcommand\colorMATH{\colorMATHB}\renewcommand\colorSYNTAX{\colorSYNTAXB}{{\color{\colorMATH}\ensuremath{\Distance'}}}\endgroup } + {\begingroup\renewcommand\colorMATH{\colorMATHB}\renewcommand\colorSYNTAX{\colorSYNTAXB}{{\color{\colorMATH}\ensuremath{\distance''}}}\endgroup }x)(\tau _{2})\rrbracket }}}.
      
      By induction hypothesis on {{\color{\colorMATH}\ensuremath{\Gamma , x: \tau _{1}; {\begingroup\renewcommand\colorMATH{\colorMATHB}\renewcommand\colorSYNTAX{\colorSYNTAXB}{{\color{\colorMATH}\ensuremath{\Distance}}}\endgroup } + {\begingroup\renewcommand\colorMATH{\colorMATHB}\renewcommand\colorSYNTAX{\colorSYNTAXB}{{\color{\colorMATH}\ensuremath{\distance'}}}\endgroup }x \vdash  {\begingroup\renewcommand\colorMATH{\colorMATHC}\renewcommand\colorSYNTAX{\colorSYNTAXC}{{\color{\colorMATH}\ensuremath{\pe'}}}\endgroup }  : \tau _{2}; {\begingroup\renewcommand\colorMATH{\colorMATHC}\renewcommand\colorSYNTAX{\colorSYNTAXC}{{\color{\colorMATH}\ensuremath{\pS''}}}\endgroup }}}}, and choosing {{\color{\colorMATH}\ensuremath{{\begingroup\renewcommand\colorMATH{\colorMATHB}\renewcommand\colorSYNTAX{\colorSYNTAXB}{{\color{\colorMATH}\ensuremath{{\begingroup\renewcommand\colorMATH{\colorMATHB}\renewcommand\colorSYNTAX{\colorSYNTAXB}{{\color{\colorMATH}\ensuremath{\sS}}}\endgroup }^\chi }}}\endgroup } = {\begingroup\renewcommand\colorMATH{\colorMATHB}\renewcommand\colorSYNTAX{\colorSYNTAXB}{{\color{\colorMATH}\ensuremath{\Distance'}}}\endgroup } + {\begingroup\renewcommand\colorMATH{\colorMATHB}\renewcommand\colorSYNTAX{\colorSYNTAXB}{{\color{\colorMATH}\ensuremath{\distance''}}}\endgroup }x \sqsubseteq  {\begingroup\renewcommand\colorMATH{\colorMATHB}\renewcommand\colorSYNTAX{\colorSYNTAXB}{{\color{\colorMATH}\ensuremath{\Distance'}}}\endgroup } + {\begingroup\renewcommand\colorMATH{\colorMATHB}\renewcommand\colorSYNTAX{\colorSYNTAXB}{{\color{\colorMATH}\ensuremath{\distance'}}}\endgroup }x}}}, we know that 
      {{\color{\colorMATH}\ensuremath{\forall  \gamma '_{1}, \gamma '_{2}, (\gamma '_{1}, \gamma '_{2}) \in  {\mathcal{G}}_{({\begingroup\renewcommand\colorMATH{\colorMATHB}\renewcommand\colorSYNTAX{\colorSYNTAXB}{{\color{\colorMATH}\ensuremath{\Distance'}}}\endgroup } + {\begingroup\renewcommand\colorMATH{\colorMATHB}\renewcommand\colorSYNTAX{\colorSYNTAXB}{{\color{\colorMATH}\ensuremath{\distance''}}}\endgroup }x)}^{j}\llbracket \Gamma , x: \tau _{1}\rrbracket }}} then {{\color{\colorMATH}\ensuremath{(\gamma '_{1} \vdash  {\begingroup\renewcommand\colorMATH{\colorMATHC}\renewcommand\colorSYNTAX{\colorSYNTAXC}{{\color{\colorMATH}\ensuremath{\pe'}}}\endgroup },\gamma '_{2} \vdash  {\begingroup\renewcommand\colorMATH{\colorMATHC}\renewcommand\colorSYNTAX{\colorSYNTAXC}{{\color{\colorMATH}\ensuremath{\pe'}}}\endgroup }) \in  {\mathcal{E}}_{({\begingroup\renewcommand\colorMATH{\colorMATHB}\renewcommand\colorSYNTAX{\colorSYNTAXB}{{\color{\colorMATH}\ensuremath{\Distance'}}}\endgroup } + {\begingroup\renewcommand\colorMATH{\colorMATHB}\renewcommand\colorSYNTAX{\colorSYNTAXB}{{\color{\colorMATH}\ensuremath{\distance''}}}\endgroup }x)\mathord{\cdotp }{\begingroup\renewcommand\colorMATH{\colorMATHC}\renewcommand\colorSYNTAX{\colorSYNTAXC}{{\color{\colorMATH}\ensuremath{\pS''}}}\endgroup }}^{j}\llbracket ({\begingroup\renewcommand\colorMATH{\colorMATHB}\renewcommand\colorSYNTAX{\colorSYNTAXB}{{\color{\colorMATH}\ensuremath{\Distance'}}}\endgroup } + {\begingroup\renewcommand\colorMATH{\colorMATHB}\renewcommand\colorSYNTAX{\colorSYNTAXB}{{\color{\colorMATH}\ensuremath{\distance''}}}\endgroup }x)(\tau _{2})\rrbracket  }}}.

      As {{\color{\colorMATH}\ensuremath{(\gamma _{1},\gamma _{2}) \in  {\mathcal{G}}_{{\begingroup\renewcommand\colorMATH{\colorMATHB}\renewcommand\colorSYNTAX{\colorSYNTAXB}{{\color{\colorMATH}\ensuremath{\Distance'}}}\endgroup }}^{j}\llbracket \Gamma \rrbracket }}} (by Lemma~\ref{lm:weakening-index}), {{\color{\colorMATH}\ensuremath{({\begingroup\renewcommand\colorMATH{\colorMATHB}\renewcommand\colorSYNTAX{\colorSYNTAXB}{{\color{\colorMATH}\ensuremath{\sv_{1}}}}\endgroup },{\begingroup\renewcommand\colorMATH{\colorMATHB}\renewcommand\colorSYNTAX{\colorSYNTAXB}{{\color{\colorMATH}\ensuremath{\sv_{2}}}}\endgroup }) \in  {\mathcal{V}}_{{\begingroup\renewcommand\colorMATH{\colorMATHB}\renewcommand\colorSYNTAX{\colorSYNTAXB}{{\color{\colorMATH}\ensuremath{\distance''}}}\endgroup }}^{j}\llbracket {\begingroup\renewcommand\colorMATH{\colorMATHB}\renewcommand\colorSYNTAX{\colorSYNTAXB}{{\color{\colorMATH}\ensuremath{\Distance'}}}\endgroup }(\tau _{1})\rrbracket }}}, and {{\color{\colorMATH}\ensuremath{ {\begingroup\renewcommand\colorMATH{\colorMATHB}\renewcommand\colorSYNTAX{\colorSYNTAXB}{{\color{\colorMATH}\ensuremath{\Distance'}}}\endgroup }(\tau _{1}) = ( {\begingroup\renewcommand\colorMATH{\colorMATHB}\renewcommand\colorSYNTAX{\colorSYNTAXB}{{\color{\colorMATH}\ensuremath{\Distance'}}}\endgroup } + {\begingroup\renewcommand\colorMATH{\colorMATHB}\renewcommand\colorSYNTAX{\colorSYNTAXB}{{\color{\colorMATH}\ensuremath{\distance''}}}\endgroup }x)(\tau _{1})}}} (as {{\color{\colorMATH}\ensuremath{x}}} is not free in {{\color{\colorMATH}\ensuremath{\tau _{1}}}}),
      it is easy to see that {{\color{\colorMATH}\ensuremath{(\gamma _{1}, x \mapsto  {\begingroup\renewcommand\colorMATH{\colorMATHB}\renewcommand\colorSYNTAX{\colorSYNTAXB}{{\color{\colorMATH}\ensuremath{\sv_{1}}}}\endgroup }, \gamma _{2}, x \mapsto  {\begingroup\renewcommand\colorMATH{\colorMATHB}\renewcommand\colorSYNTAX{\colorSYNTAXB}{{\color{\colorMATH}\ensuremath{\sv_{2}}}}\endgroup }) \in  {\mathcal{G}}_{({\begingroup\renewcommand\colorMATH{\colorMATHB}\renewcommand\colorSYNTAX{\colorSYNTAXB}{{\color{\colorMATH}\ensuremath{\Distance'}}}\endgroup } + {\begingroup\renewcommand\colorMATH{\colorMATHB}\renewcommand\colorSYNTAX{\colorSYNTAXB}{{\color{\colorMATH}\ensuremath{\distance''}}}\endgroup }x)}^{j}\llbracket \Gamma , x: \tau _{1}\rrbracket }}}. Finally, the result follows by choosing 
      {{\color{\colorMATH}\ensuremath{\gamma '_{1} = \gamma _{1}, x \mapsto  {\begingroup\renewcommand\colorMATH{\colorMATHB}\renewcommand\colorSYNTAX{\colorSYNTAXB}{{\color{\colorMATH}\ensuremath{\sv_{1}}}}\endgroup }}}}, and {{\color{\colorMATH}\ensuremath{ \gamma '_{2} = \gamma _{2}, x \mapsto  {\begingroup\renewcommand\colorMATH{\colorMATHB}\renewcommand\colorSYNTAX{\colorSYNTAXB}{{\color{\colorMATH}\ensuremath{\sv_{2}}}}\endgroup }}}}.
    \end{subproof}
  \end{enumerate}
  Let us prove now Part (2). %We always prove one direction ({{\color{\colorMATH}\ensuremath{\leq }}}) as the other is analogous.
  \begin{enumerate}[ncases]\item 
    {{\color{\colorMATH}\ensuremath{\Gamma ; {\begingroup\renewcommand\colorMATH{\colorMATHB}\renewcommand\colorSYNTAX{\colorSYNTAXB}{{\color{\colorMATH}\ensuremath{\Distance}}}\endgroup } \vdash  {\begingroup\renewcommand\colorMATH{\colorMATHC}\renewcommand\colorSYNTAX{\colorSYNTAXC}{{\color{\colorMATH}\ensuremath{\pe_{1}}}}\endgroup }\hspace*{0.33em}{\begingroup\renewcommand\colorMATH{\colorMATHC}\renewcommand\colorSYNTAX{\colorSYNTAXC}{{\color{\colorMATH}\ensuremath{\pe_{2}}}}\endgroup } \mathrel{:} [{\begingroup\renewcommand\colorMATH{\colorMATHB}\renewcommand\colorSYNTAX{\colorSYNTAXB}{{\color{\colorMATH}\ensuremath{\sS_{2}}}}\endgroup }/x]\tau _{2} \mathrel{;}   {\begingroup\renewcommand\colorMATH{\colorMATHC}\renewcommand\colorSYNTAX{\colorSYNTAXC}{{\color{\colorMATH}\ensuremath{\rceil {\begingroup\renewcommand\colorMATH{\colorMATHA}\renewcommand\colorSYNTAX{\colorSYNTAXA}{{\color{\colorMATH}\ensuremath{{\begingroup\renewcommand\colorMATH{\colorMATHB}\renewcommand\colorSYNTAX{\colorSYNTAXB}{{\color{\colorMATH}\ensuremath{\sS_{1}}}}\endgroup }}}}\endgroup }\lceil ^{\infty }}}}\endgroup } +  [{\begingroup\renewcommand\colorMATH{\colorMATHB}\renewcommand\colorSYNTAX{\colorSYNTAXB}{{\color{\colorMATH}\ensuremath{\sS_{2}}}}\endgroup }/x]{\begingroup\renewcommand\colorMATH{\colorMATHC}\renewcommand\colorSYNTAX{\colorSYNTAXC}{{\color{\colorMATH}\ensuremath{\pS''}}}\endgroup }}}} 
    \begin{subproof} 
      %Notice that {{\color{\colorMATH}\ensuremath{{\begingroup\renewcommand\colorMATH{\colorMATHB}\renewcommand\colorSYNTAX{\colorSYNTAXB}{{\color{\colorMATH}\ensuremath{\Distance'}}}\endgroup }\mathord{\cdotp }({\begingroup\renewcommand\colorMATH{\colorMATHB}\renewcommand\colorSYNTAX{\colorSYNTAXB}{{\color{\colorMATH}\ensuremath{\sS_{1}}}}\endgroup } + s{\begingroup\renewcommand\colorMATH{\colorMATHB}\renewcommand\colorSYNTAX{\colorSYNTAXB}{{\color{\colorMATH}\ensuremath{\sS_{2}}}}\endgroup } + {\begingroup\renewcommand\colorMATH{\colorMATHB}\renewcommand\colorSYNTAX{\colorSYNTAXB}{{\color{\colorMATH}\ensuremath{\sS''}}}\endgroup }) = ({\begingroup\renewcommand\colorMATH{\colorMATHB}\renewcommand\colorSYNTAX{\colorSYNTAXB}{{\color{\colorMATH}\ensuremath{\Distance'}}}\endgroup }\mathord{\cdotp }{\begingroup\renewcommand\colorMATH{\colorMATHB}\renewcommand\colorSYNTAX{\colorSYNTAXB}{{\color{\colorMATH}\ensuremath{\sS_{1}}}}\endgroup } + s({\begingroup\renewcommand\colorMATH{\colorMATHB}\renewcommand\colorSYNTAX{\colorSYNTAXB}{{\color{\colorMATH}\ensuremath{\Distance'}}}\endgroup }\mathord{\cdotp }{\begingroup\renewcommand\colorMATH{\colorMATHB}\renewcommand\colorSYNTAX{\colorSYNTAXB}{{\color{\colorMATH}\ensuremath{\sS_{2}}}}\endgroup }) + {\begingroup\renewcommand\colorMATH{\colorMATHB}\renewcommand\colorSYNTAX{\colorSYNTAXB}{{\color{\colorMATH}\ensuremath{\Distance'}}}\endgroup }\mathord{\cdotp }{\begingroup\renewcommand\colorMATH{\colorMATHB}\renewcommand\colorSYNTAX{\colorSYNTAXB}{{\color{\colorMATH}\ensuremath{\sS''}}}\endgroup })}}}.\\
      We have to prove that\\ {{\color{\colorMATH}\ensuremath{\forall k, \forall (\gamma _{1},\gamma _{2}) \in  {\mathcal{G}}_{{\begingroup\renewcommand\colorMATH{\colorMATHB}\renewcommand\colorSYNTAX{\colorSYNTAXB}{{\color{\colorMATH}\ensuremath{\Distance'}}}\endgroup }}^{\kg}\llbracket \Gamma \rrbracket }}}, for {{\color{\colorMATH}\ensuremath{{\begingroup\renewcommand\colorMATH{\colorMATHB}\renewcommand\colorSYNTAX{\colorSYNTAXB}{{\color{\colorMATH}\ensuremath{\Distance'}}}\endgroup } \sqsubseteq  {\begingroup\renewcommand\colorMATH{\colorMATHB}\renewcommand\colorSYNTAX{\colorSYNTAXB}{{\color{\colorMATH}\ensuremath{\Distance}}}\endgroup }}}}, then {{\color{\colorMATH}\ensuremath{(\gamma _{1}\vdash {\begingroup\renewcommand\colorMATH{\colorMATHC}\renewcommand\colorSYNTAX{\colorSYNTAXC}{{\color{\colorMATH}\ensuremath{\pe_{1}}}}\endgroup }\hspace*{0.33em}{\begingroup\renewcommand\colorMATH{\colorMATHC}\renewcommand\colorSYNTAX{\colorSYNTAXC}{{\color{\colorMATH}\ensuremath{\pe_{2}}}}\endgroup },\gamma _{2}\vdash {\begingroup\renewcommand\colorMATH{\colorMATHC}\renewcommand\colorSYNTAX{\colorSYNTAXC}{{\color{\colorMATH}\ensuremath{\pe_{1}}}}\endgroup }\hspace*{0.33em}{\begingroup\renewcommand\colorMATH{\colorMATHC}\renewcommand\colorSYNTAX{\colorSYNTAXC}{{\color{\colorMATH}\ensuremath{\pe_{2}}}}\endgroup }) \in  {\mathcal{E}}_{{\begingroup\renewcommand\colorMATH{\colorMATHB}\renewcommand\colorSYNTAX{\colorSYNTAXB}{{\color{\colorMATH}\ensuremath{\Distance'}}}\endgroup } {\begingroup\renewcommand\colorMATH{\colorMATHC}\renewcommand\colorSYNTAX{\colorSYNTAXC}{{\color{\colorMATH}\ensuremath{\bigcdot}}}\endgroup } {\begingroup\renewcommand\colorMATH{\colorMATHC}\renewcommand\colorSYNTAX{\colorSYNTAXC}{{\color{\colorMATH}\ensuremath{\pS}}}\endgroup }}^{k}\llbracket {\begingroup\renewcommand\colorMATH{\colorMATHB}\renewcommand\colorSYNTAX{\colorSYNTAXB}{{\color{\colorMATH}\ensuremath{\Distance'}}}\endgroup }([{\begingroup\renewcommand\colorMATH{\colorMATHB}\renewcommand\colorSYNTAX{\colorSYNTAXB}{{\color{\colorMATH}\ensuremath{\sS_{2}}}}\endgroup }/x]\tau _{2})\rrbracket }}}.

      By induction hypotheses we know that\\
      {{\color{\colorMATH}\ensuremath{\Gamma ; {\begingroup\renewcommand\colorMATH{\colorMATHB}\renewcommand\colorSYNTAX{\colorSYNTAXB}{{\color{\colorMATH}\ensuremath{\Distance}}}\endgroup } \vdash  {\begingroup\renewcommand\colorMATH{\colorMATHB}\renewcommand\colorSYNTAX{\colorSYNTAXB}{{\color{\colorMATH}\ensuremath{\se_{1}}}}\endgroup } \mathrel{:} (x:\tau _{1}\mathord{\cdotp }{\begingroup\renewcommand\colorMATH{\colorMATHB}\renewcommand\colorSYNTAX{\colorSYNTAXB}{{\color{\colorMATH}\ensuremath{\distance'}}}\endgroup }) \xrightarrowP {{\begingroup\renewcommand\colorMATH{\colorMATHC}\renewcommand\colorSYNTAX{\colorSYNTAXC}{{\color{\colorMATH}\ensuremath{\pS''}}}\endgroup }} \tau _{2} \mathrel{;} {\begingroup\renewcommand\colorMATH{\colorMATHB}\renewcommand\colorSYNTAX{\colorSYNTAXB}{{\color{\colorMATH}\ensuremath{\sS_{1}}}}\endgroup } \Rightarrow  (\gamma _{1} \vdash  {\begingroup\renewcommand\colorMATH{\colorMATHB}\renewcommand\colorSYNTAX{\colorSYNTAXB}{{\color{\colorMATH}\ensuremath{\se_{1}}}}\endgroup },\gamma _{2} \vdash  {\begingroup\renewcommand\colorMATH{\colorMATHB}\renewcommand\colorSYNTAX{\colorSYNTAXB}{{\color{\colorMATH}\ensuremath{\se_{1}}}}\endgroup }) \in  {\mathcal{E}}^{k}_{{\begingroup\renewcommand\colorMATH{\colorMATHB}\renewcommand\colorSYNTAX{\colorSYNTAXB}{{\color{\colorMATH}\ensuremath{\Distance'}}}\endgroup }\mathord{\cdotp }{\begingroup\renewcommand\colorMATH{\colorMATHB}\renewcommand\colorSYNTAX{\colorSYNTAXB}{{\color{\colorMATH}\ensuremath{\sS_{1}}}}\endgroup }}\llbracket {\begingroup\renewcommand\colorMATH{\colorMATHB}\renewcommand\colorSYNTAX{\colorSYNTAXB}{{\color{\colorMATH}\ensuremath{\Distance'}}}\endgroup }((x:\tau _{1}\mathord{\cdotp }{\begingroup\renewcommand\colorMATH{\colorMATHB}\renewcommand\colorSYNTAX{\colorSYNTAXB}{{\color{\colorMATH}\ensuremath{\distance'}}}\endgroup }) \xrightarrowP {{\begingroup\renewcommand\colorMATH{\colorMATHC}\renewcommand\colorSYNTAX{\colorSYNTAXC}{{\color{\colorMATH}\ensuremath{\pS''}}}\endgroup }} \tau _{2})\rrbracket }}} and \\
      {{\color{\colorMATH}\ensuremath{\Gamma ; {\begingroup\renewcommand\colorMATH{\colorMATHB}\renewcommand\colorSYNTAX{\colorSYNTAXB}{{\color{\colorMATH}\ensuremath{\Distance}}}\endgroup } \vdash  {\begingroup\renewcommand\colorMATH{\colorMATHB}\renewcommand\colorSYNTAX{\colorSYNTAXB}{{\color{\colorMATH}\ensuremath{\se_{2}}}}\endgroup } \mathrel{:} \tau _{1} \mathrel{;} {\begingroup\renewcommand\colorMATH{\colorMATHB}\renewcommand\colorSYNTAX{\colorSYNTAXB}{{\color{\colorMATH}\ensuremath{\sS_{2}}}}\endgroup } \Rightarrow  (\gamma _{1} \vdash  {\begingroup\renewcommand\colorMATH{\colorMATHB}\renewcommand\colorSYNTAX{\colorSYNTAXB}{{\color{\colorMATH}\ensuremath{\se_{2}}}}\endgroup },\gamma _{2} \vdash  {\begingroup\renewcommand\colorMATH{\colorMATHB}\renewcommand\colorSYNTAX{\colorSYNTAXB}{{\color{\colorMATH}\ensuremath{\se_{2}}}}\endgroup }) \in  {\mathcal{E}}^{k-j_{1}}_{{\begingroup\renewcommand\colorMATH{\colorMATHB}\renewcommand\colorSYNTAX{\colorSYNTAXB}{{\color{\colorMATH}\ensuremath{\Distance'}}}\endgroup }\mathord{\cdotp }{\begingroup\renewcommand\colorMATH{\colorMATHB}\renewcommand\colorSYNTAX{\colorSYNTAXB}{{\color{\colorMATH}\ensuremath{\sS_{2}}}}\endgroup }}\llbracket {\begingroup\renewcommand\colorMATH{\colorMATHB}\renewcommand\colorSYNTAX{\colorSYNTAXB}{{\color{\colorMATH}\ensuremath{\Distance'}}}\endgroup }(\tau _{1})\rrbracket }}}.\\
      As {{\color{\colorMATH}\ensuremath{{\begingroup\renewcommand\colorMATH{\colorMATHB}\renewcommand\colorSYNTAX{\colorSYNTAXB}{{\color{\colorMATH}\ensuremath{\Distance'}}}\endgroup }((x:\tau _{1}\mathord{\cdotp }{\begingroup\renewcommand\colorMATH{\colorMATHB}\renewcommand\colorSYNTAX{\colorSYNTAXB}{{\color{\colorMATH}\ensuremath{\distance'}}}\endgroup }) \xrightarrowP {{\begingroup\renewcommand\colorMATH{\colorMATHC}\renewcommand\colorSYNTAX{\colorSYNTAXC}{{\color{\colorMATH}\ensuremath{\pS''}}}\endgroup }} \tau _{2}) = (x:{\begingroup\renewcommand\colorMATH{\colorMATHB}\renewcommand\colorSYNTAX{\colorSYNTAXB}{{\color{\colorMATH}\ensuremath{\Distance'}}}\endgroup }(\tau _{1})\mathord{\cdotp }{\begingroup\renewcommand\colorMATH{\colorMATHB}\renewcommand\colorSYNTAX{\colorSYNTAXB}{{\color{\colorMATH}\ensuremath{\distance'}}}\endgroup }) \xrightarrowP {{\begingroup\renewcommand\colorMATH{\colorMATHB}\renewcommand\colorSYNTAX{\colorSYNTAXB}{{\color{\colorMATH}\ensuremath{\Distance'}}}\endgroup } {\begingroup\renewcommand\colorMATH{\colorMATHC}\renewcommand\colorSYNTAX{\colorSYNTAXC}{{\color{\colorMATH}\ensuremath{\bigcdot}}}\endgroup } {\begingroup\renewcommand\colorMATH{\colorMATHC}\renewcommand\colorSYNTAX{\colorSYNTAXC}{{\color{\colorMATH}\ensuremath{\pS''}}}\endgroup }} {\begingroup\renewcommand\colorMATH{\colorMATHB}\renewcommand\colorSYNTAX{\colorSYNTAXB}{{\color{\colorMATH}\ensuremath{\Distance'}}}\endgroup }(\tau _{2})}}}, by unfolding {{\color{\colorMATH}\ensuremath{(\gamma _{1} \vdash  {\begingroup\renewcommand\colorMATH{\colorMATHB}\renewcommand\colorSYNTAX{\colorSYNTAXB}{{\color{\colorMATH}\ensuremath{\se_{1}}}}\endgroup },\gamma _{2} \vdash  {\begingroup\renewcommand\colorMATH{\colorMATHB}\renewcommand\colorSYNTAX{\colorSYNTAXB}{{\color{\colorMATH}\ensuremath{\se_{1}}}}\endgroup }) \in  {\mathcal{E}}^{k}_{{\begingroup\renewcommand\colorMATH{\colorMATHB}\renewcommand\colorSYNTAX{\colorSYNTAXB}{{\color{\colorMATH}\ensuremath{\Distance'}}}\endgroup }\mathord{\cdotp }{\begingroup\renewcommand\colorMATH{\colorMATHB}\renewcommand\colorSYNTAX{\colorSYNTAXB}{{\color{\colorMATH}\ensuremath{\sS_{1}}}}\endgroup }}\llbracket (x:{\begingroup\renewcommand\colorMATH{\colorMATHB}\renewcommand\colorSYNTAX{\colorSYNTAXB}{{\color{\colorMATH}\ensuremath{\Distance'}}}\endgroup }(\tau _{1})\mathord{\cdotp }{\begingroup\renewcommand\colorMATH{\colorMATHB}\renewcommand\colorSYNTAX{\colorSYNTAXB}{{\color{\colorMATH}\ensuremath{\distance'}}}\endgroup }) \xrightarrowP {{\begingroup\renewcommand\colorMATH{\colorMATHB}\renewcommand\colorSYNTAX{\colorSYNTAXB}{{\color{\colorMATH}\ensuremath{\Distance'}}}\endgroup } {\begingroup\renewcommand\colorMATH{\colorMATHC}\renewcommand\colorSYNTAX{\colorSYNTAXC}{{\color{\colorMATH}\ensuremath{\bigcdot}}}\endgroup } {\begingroup\renewcommand\colorMATH{\colorMATHC}\renewcommand\colorSYNTAX{\colorSYNTAXC}{{\color{\colorMATH}\ensuremath{\pS''}}}\endgroup }} {\begingroup\renewcommand\colorMATH{\colorMATHB}\renewcommand\colorSYNTAX{\colorSYNTAXB}{{\color{\colorMATH}\ensuremath{\Distance'}}}\endgroup }(\tau _{2})\rrbracket }}},
      we know that if {{\color{\colorMATH}\ensuremath{\gamma _{1} \vdash  {\begingroup\renewcommand\colorMATH{\colorMATHB}\renewcommand\colorSYNTAX{\colorSYNTAXB}{{\color{\colorMATH}\ensuremath{\se_{1}}}}\endgroup } \Downarrow ^{j_{1}} \langle {\begingroup\renewcommand\colorMATH{\colorMATHC}\renewcommand\colorSYNTAX{\colorSYNTAXC}{{\color{\colorMATH}\ensuremath{\plambda}}}\endgroup } (x\mathrel{:}\tau _{1}).\hspace*{0.33em}{\begingroup\renewcommand\colorMATH{\colorMATHC}\renewcommand\colorSYNTAX{\colorSYNTAXC}{{\color{\colorMATH}\ensuremath{\pe'_{1}}}}\endgroup }, \gamma '_{1}\rangle }}} \pthen {{\color{\colorMATH}\ensuremath{\gamma _{2} \vdash  {\begingroup\renewcommand\colorMATH{\colorMATHB}\renewcommand\colorSYNTAX{\colorSYNTAXB}{{\color{\colorMATH}\ensuremath{\se_{1}}}}\endgroup } \Downarrow ^{\pj[1]} \langle {\begingroup\renewcommand\colorMATH{\colorMATHC}\renewcommand\colorSYNTAX{\colorSYNTAXC}{{\color{\colorMATH}\ensuremath{\plambda}}}\endgroup } (x\mathrel{:}\tau _{1}).\hspace*{0.33em}{\begingroup\renewcommand\colorMATH{\colorMATHC}\renewcommand\colorSYNTAX{\colorSYNTAXC}{{\color{\colorMATH}\ensuremath{\pe'_{2}}}}\endgroup }, \gamma '_{2}\rangle }}} \pand\\ {{\color{\colorMATH}\ensuremath{(\langle {\begingroup\renewcommand\colorMATH{\colorMATHC}\renewcommand\colorSYNTAX{\colorSYNTAXC}{{\color{\colorMATH}\ensuremath{\plambda}}}\endgroup } (x\mathrel{:}\tau _{1}).\hspace*{0.33em}{\begingroup\renewcommand\colorMATH{\colorMATHC}\renewcommand\colorSYNTAX{\colorSYNTAXC}{{\color{\colorMATH}\ensuremath{\pe'_{1}}}}\endgroup }, \gamma '_{1}\rangle , \langle {\begingroup\renewcommand\colorMATH{\colorMATHC}\renewcommand\colorSYNTAX{\colorSYNTAXC}{{\color{\colorMATH}\ensuremath{\plambda}}}\endgroup } (x\mathrel{:}\tau _{1}).\hspace*{0.33em}{\begingroup\renewcommand\colorMATH{\colorMATHC}\renewcommand\colorSYNTAX{\colorSYNTAXC}{{\color{\colorMATH}\ensuremath{\pe'_{2}}}}\endgroup }, \gamma '_{2}\rangle ) \in  {\mathcal{V}}^{k-j_{1}}_{{\begingroup\renewcommand\colorMATH{\colorMATHB}\renewcommand\colorSYNTAX{\colorSYNTAXB}{{\color{\colorMATH}\ensuremath{\Distance'}}}\endgroup }\mathord{\cdotp }{\begingroup\renewcommand\colorMATH{\colorMATHB}\renewcommand\colorSYNTAX{\colorSYNTAXB}{{\color{\colorMATH}\ensuremath{\sS_{1}}}}\endgroup }}\llbracket (x:{\begingroup\renewcommand\colorMATH{\colorMATHB}\renewcommand\colorSYNTAX{\colorSYNTAXB}{{\color{\colorMATH}\ensuremath{\Distance'}}}\endgroup }(\tau _{1})\mathord{\cdotp }{\begingroup\renewcommand\colorMATH{\colorMATHB}\renewcommand\colorSYNTAX{\colorSYNTAXB}{{\color{\colorMATH}\ensuremath{\distance'}}}\endgroup }) \xrightarrowP {{\begingroup\renewcommand\colorMATH{\colorMATHB}\renewcommand\colorSYNTAX{\colorSYNTAXB}{{\color{\colorMATH}\ensuremath{\Distance'}}}\endgroup } {\begingroup\renewcommand\colorMATH{\colorMATHC}\renewcommand\colorSYNTAX{\colorSYNTAXC}{{\color{\colorMATH}\ensuremath{\bigcdot}}}\endgroup } {\begingroup\renewcommand\colorMATH{\colorMATHC}\renewcommand\colorSYNTAX{\colorSYNTAXC}{{\color{\colorMATH}\ensuremath{\pS''}}}\endgroup }} {\begingroup\renewcommand\colorMATH{\colorMATHB}\renewcommand\colorSYNTAX{\colorSYNTAXB}{{\color{\colorMATH}\ensuremath{\Distance'}}}\endgroup }(\tau _{2})\rrbracket }}} (for {{\color{\colorMATH}\ensuremath{{\begingroup\renewcommand\colorMATH{\colorMATHB}\renewcommand\colorSYNTAX{\colorSYNTAXB}{{\color{\colorMATH}\ensuremath{\distance''}}}\endgroup }>={\begingroup\renewcommand\colorMATH{\colorMATHB}\renewcommand\colorSYNTAX{\colorSYNTAXB}{{\color{\colorMATH}\ensuremath{\distance'}}}\endgroup }}}}) (1).\\
      Also, by unfolding {{\color{\colorMATH}\ensuremath{(\gamma _{1} \vdash  {\begingroup\renewcommand\colorMATH{\colorMATHB}\renewcommand\colorSYNTAX{\colorSYNTAXB}{{\color{\colorMATH}\ensuremath{\se_{2}}}}\endgroup },\gamma _{2} \vdash  {\begingroup\renewcommand\colorMATH{\colorMATHB}\renewcommand\colorSYNTAX{\colorSYNTAXB}{{\color{\colorMATH}\ensuremath{\se_{2}}}}\endgroup }) \in  {\mathcal{E}}^{k-j_{1}}_{{\begingroup\renewcommand\colorMATH{\colorMATHB}\renewcommand\colorSYNTAX{\colorSYNTAXB}{{\color{\colorMATH}\ensuremath{\Distance'}}}\endgroup }\mathord{\cdotp }{\begingroup\renewcommand\colorMATH{\colorMATHB}\renewcommand\colorSYNTAX{\colorSYNTAXB}{{\color{\colorMATH}\ensuremath{\sS_{2}}}}\endgroup }}\llbracket {\begingroup\renewcommand\colorMATH{\colorMATHB}\renewcommand\colorSYNTAX{\colorSYNTAXB}{{\color{\colorMATH}\ensuremath{\Distance'}}}\endgroup }(\tau _{1})\rrbracket }}}, if {{\color{\colorMATH}\ensuremath{\gamma _{1} \vdash  {\begingroup\renewcommand\colorMATH{\colorMATHB}\renewcommand\colorSYNTAX{\colorSYNTAXB}{{\color{\colorMATH}\ensuremath{\se_{2}}}}\endgroup } \Downarrow ^{j_{2}} {\begingroup\renewcommand\colorMATH{\colorMATHB}\renewcommand\colorSYNTAX{\colorSYNTAXB}{{\color{\colorMATH}\ensuremath{\sv_{1}}}}\endgroup }}}} \pthen {{\color{\colorMATH}\ensuremath{\gamma _{2} \vdash  {\begingroup\renewcommand\colorMATH{\colorMATHB}\renewcommand\colorSYNTAX{\colorSYNTAXB}{{\color{\colorMATH}\ensuremath{\se_{2}}}}\endgroup } \Downarrow ^{\pj[2]} {\begingroup\renewcommand\colorMATH{\colorMATHB}\renewcommand\colorSYNTAX{\colorSYNTAXB}{{\color{\colorMATH}\ensuremath{\sv_{2}}}}\endgroup }}}} \pand {{\color{\colorMATH}\ensuremath{({\begingroup\renewcommand\colorMATH{\colorMATHB}\renewcommand\colorSYNTAX{\colorSYNTAXB}{{\color{\colorMATH}\ensuremath{\sv_{1}}}}\endgroup }, {\begingroup\renewcommand\colorMATH{\colorMATHB}\renewcommand\colorSYNTAX{\colorSYNTAXB}{{\color{\colorMATH}\ensuremath{\sv_{2}}}}\endgroup }) \in  {\mathcal{V}}^{k-j_{1}-j_{2}}_{{\begingroup\renewcommand\colorMATH{\colorMATHB}\renewcommand\colorSYNTAX{\colorSYNTAXB}{{\color{\colorMATH}\ensuremath{\Distance'}}}\endgroup }\mathord{\cdotp }{\begingroup\renewcommand\colorMATH{\colorMATHB}\renewcommand\colorSYNTAX{\colorSYNTAXB}{{\color{\colorMATH}\ensuremath{\sS_{2}}}}\endgroup }}\llbracket {\begingroup\renewcommand\colorMATH{\colorMATHB}\renewcommand\colorSYNTAX{\colorSYNTAXB}{{\color{\colorMATH}\ensuremath{\Distance'}}}\endgroup }(\tau _{1})\rrbracket }}}.

      By {\textsc{ p-app}} we know that {{\color{\colorMATH}\ensuremath{{\begingroup\renewcommand\colorMATH{\colorMATHB}\renewcommand\colorSYNTAX{\colorSYNTAXB}{{\color{\colorMATH}\ensuremath{\Distance}}}\endgroup } \mathord{\cdotp } {\begingroup\renewcommand\colorMATH{\colorMATHB}\renewcommand\colorSYNTAX{\colorSYNTAXB}{{\color{\colorMATH}\ensuremath{\sS_{2}}}}\endgroup } \leq  {\begingroup\renewcommand\colorMATH{\colorMATHB}\renewcommand\colorSYNTAX{\colorSYNTAXB}{{\color{\colorMATH}\ensuremath{\distance'}}}\endgroup }}}}, as {{\color{\colorMATH}\ensuremath{{\begingroup\renewcommand\colorMATH{\colorMATHB}\renewcommand\colorSYNTAX{\colorSYNTAXB}{{\color{\colorMATH}\ensuremath{\Distance'}}}\endgroup } \sqsubseteq  {\begingroup\renewcommand\colorMATH{\colorMATHB}\renewcommand\colorSYNTAX{\colorSYNTAXB}{{\color{\colorMATH}\ensuremath{\Distance}}}\endgroup }}}} then {{\color{\colorMATH}\ensuremath{{\begingroup\renewcommand\colorMATH{\colorMATHB}\renewcommand\colorSYNTAX{\colorSYNTAXB}{{\color{\colorMATH}\ensuremath{\Distance'}}}\endgroup }\mathord{\cdotp } {\begingroup\renewcommand\colorMATH{\colorMATHB}\renewcommand\colorSYNTAX{\colorSYNTAXB}{{\color{\colorMATH}\ensuremath{\sS_{2}}}}\endgroup } \leq  {\begingroup\renewcommand\colorMATH{\colorMATHB}\renewcommand\colorSYNTAX{\colorSYNTAXB}{{\color{\colorMATH}\ensuremath{\distance'}}}\endgroup }}}}.
      Then we instantiate (1) with {{\color{\colorMATH}\ensuremath{{\begingroup\renewcommand\colorMATH{\colorMATHB}\renewcommand\colorSYNTAX{\colorSYNTAXB}{{\color{\colorMATH}\ensuremath{\distance''}}}\endgroup } = {\begingroup\renewcommand\colorMATH{\colorMATHB}\renewcommand\colorSYNTAX{\colorSYNTAXB}{{\color{\colorMATH}\ensuremath{\Distance'}}}\endgroup }\mathord{\cdotp }{\begingroup\renewcommand\colorMATH{\colorMATHB}\renewcommand\colorSYNTAX{\colorSYNTAXB}{{\color{\colorMATH}\ensuremath{\sS_{2}}}}\endgroup }}}}, then for some {{\color{\colorMATH}\ensuremath{j_{3} < k-j_{1}-j_{2}}}}\\
      {{\color{\colorMATH}\ensuremath{(\gamma '_{1}[x\mapsto {\begingroup\renewcommand\colorMATH{\colorMATHB}\renewcommand\colorSYNTAX{\colorSYNTAXB}{{\color{\colorMATH}\ensuremath{\sv_{1}}}}\endgroup }] \vdash  {\begingroup\renewcommand\colorMATH{\colorMATHC}\renewcommand\colorSYNTAX{\colorSYNTAXC}{{\color{\colorMATH}\ensuremath{\pe'_{1}}}}\endgroup },\gamma '_{2}[x\mapsto {\begingroup\renewcommand\colorMATH{\colorMATHB}\renewcommand\colorSYNTAX{\colorSYNTAXB}{{\color{\colorMATH}\ensuremath{\sv_{2}}}}\endgroup }] \vdash  {\begingroup\renewcommand\colorMATH{\colorMATHC}\renewcommand\colorSYNTAX{\colorSYNTAXC}{{\color{\colorMATH}\ensuremath{\pe'_{2}}}}\endgroup }) \in  {\mathcal{E}}^{j_{3}}_{{\begingroup\renewcommand\colorMATH{\colorMATHC}\renewcommand\colorSYNTAX{\colorSYNTAXC}{{\color{\colorMATH}\ensuremath{\rceil {\begingroup\renewcommand\colorMATH{\colorMATHA}\renewcommand\colorSYNTAX{\colorSYNTAXA}{{\color{\colorMATH}\ensuremath{{\begingroup\renewcommand\colorMATH{\colorMATHB}\renewcommand\colorSYNTAX{\colorSYNTAXB}{{\color{\colorMATH}\ensuremath{\Distance'}}}\endgroup }\mathord{\cdotp }{\begingroup\renewcommand\colorMATH{\colorMATHB}\renewcommand\colorSYNTAX{\colorSYNTAXB}{{\color{\colorMATH}\ensuremath{\sS_{1}}}}\endgroup }}}}\endgroup }\lceil ^{\infty }}}}\endgroup }+({\begingroup\renewcommand\colorMATH{\colorMATHB}\renewcommand\colorSYNTAX{\colorSYNTAXB}{{\color{\colorMATH}\ensuremath{\Distance'}}}\endgroup }+ ({\begingroup\renewcommand\colorMATH{\colorMATHB}\renewcommand\colorSYNTAX{\colorSYNTAXB}{{\color{\colorMATH}\ensuremath{\Distance'}}}\endgroup }\mathord{\cdotp }{\begingroup\renewcommand\colorMATH{\colorMATHB}\renewcommand\colorSYNTAX{\colorSYNTAXB}{{\color{\colorMATH}\ensuremath{\sS_{2}}}}\endgroup })x) {\begingroup\renewcommand\colorMATH{\colorMATHC}\renewcommand\colorSYNTAX{\colorSYNTAXC}{{\color{\colorMATH}\ensuremath{\bigcdot}}}\endgroup } {\begingroup\renewcommand\colorMATH{\colorMATHC}\renewcommand\colorSYNTAX{\colorSYNTAXC}{{\color{\colorMATH}\ensuremath{\pS''}}}\endgroup }}\llbracket ({\begingroup\renewcommand\colorMATH{\colorMATHB}\renewcommand\colorSYNTAX{\colorSYNTAXB}{{\color{\colorMATH}\ensuremath{\Distance'}}}\endgroup }\mathord{\cdotp }{\begingroup\renewcommand\colorMATH{\colorMATHB}\renewcommand\colorSYNTAX{\colorSYNTAXB}{{\color{\colorMATH}\ensuremath{\sS_{2}}}}\endgroup }x)({\begingroup\renewcommand\colorMATH{\colorMATHB}\renewcommand\colorSYNTAX{\colorSYNTAXB}{{\color{\colorMATH}\ensuremath{\Distance'}}}\endgroup }(\tau _{2}))\rrbracket }}} (2).

      Suppose {{\color{\colorMATH}\ensuremath{{\begingroup\renewcommand\colorMATH{\colorMATHB}\renewcommand\colorSYNTAX{\colorSYNTAXB}{{\color{\colorMATH}\ensuremath{\Distance'}}}\endgroup } {\begingroup\renewcommand\colorMATH{\colorMATHC}\renewcommand\colorSYNTAX{\colorSYNTAXC}{{\color{\colorMATH}\ensuremath{\bigcdot}}}\endgroup }  {\begingroup\renewcommand\colorMATH{\colorMATHC}\renewcommand\colorSYNTAX{\colorSYNTAXC}{{\color{\colorMATH}\ensuremath{\rceil {\begingroup\renewcommand\colorMATH{\colorMATHA}\renewcommand\colorSYNTAX{\colorSYNTAXA}{{\color{\colorMATH}\ensuremath{{\begingroup\renewcommand\colorMATH{\colorMATHB}\renewcommand\colorSYNTAX{\colorSYNTAXB}{{\color{\colorMATH}\ensuremath{\sS_{1}}}}\endgroup }}}}\endgroup }\lceil ^{\infty }}}}\endgroup } = 0}}} (otherwise the result follows immediately).
      Then {{\color{\colorMATH}\ensuremath{{\begingroup\renewcommand\colorMATH{\colorMATHB}\renewcommand\colorSYNTAX{\colorSYNTAXB}{{\color{\colorMATH}\ensuremath{\Distance'}}}\endgroup } {\begingroup\renewcommand\colorMATH{\colorMATHC}\renewcommand\colorSYNTAX{\colorSYNTAXC}{{\color{\colorMATH}\ensuremath{\bigcdot}}}\endgroup } {\begingroup\renewcommand\colorMATH{\colorMATHC}\renewcommand\colorSYNTAX{\colorSYNTAXC}{{\color{\colorMATH}\ensuremath{\pS}}}\endgroup } = {\begingroup\renewcommand\colorMATH{\colorMATHB}\renewcommand\colorSYNTAX{\colorSYNTAXB}{{\color{\colorMATH}\ensuremath{\Distance'}}}\endgroup } {\begingroup\renewcommand\colorMATH{\colorMATHC}\renewcommand\colorSYNTAX{\colorSYNTAXC}{{\color{\colorMATH}\ensuremath{\bigcdot}}}\endgroup } ([{\begingroup\renewcommand\colorMATH{\colorMATHB}\renewcommand\colorSYNTAX{\colorSYNTAXB}{{\color{\colorMATH}\ensuremath{\sS_{2}}}}\endgroup }/x]{\begingroup\renewcommand\colorMATH{\colorMATHC}\renewcommand\colorSYNTAX{\colorSYNTAXC}{{\color{\colorMATH}\ensuremath{\pS''}}}\endgroup })}}}.
      Also as {{\color{\colorMATH}\ensuremath{{\begingroup\renewcommand\colorMATH{\colorMATHB}\renewcommand\colorSYNTAX{\colorSYNTAXB}{{\color{\colorMATH}\ensuremath{\Distance'}}}\endgroup } \sqsubseteq  {\begingroup\renewcommand\colorMATH{\colorMATHB}\renewcommand\colorSYNTAX{\colorSYNTAXB}{{\color{\colorMATH}\ensuremath{\Distance}}}\endgroup }}}}, and {{\color{\colorMATH}\ensuremath{x \notin  dom({\begingroup\renewcommand\colorMATH{\colorMATHB}\renewcommand\colorSYNTAX{\colorSYNTAXB}{{\color{\colorMATH}\ensuremath{\Distance}}}\endgroup })}}}, then by Lemma~\ref{lm:distrdotpp},
      {{\color{\colorMATH}\ensuremath{{\begingroup\renewcommand\colorMATH{\colorMATHB}\renewcommand\colorSYNTAX{\colorSYNTAXB}{{\color{\colorMATH}\ensuremath{\Distance'}}}\endgroup } {\begingroup\renewcommand\colorMATH{\colorMATHC}\renewcommand\colorSYNTAX{\colorSYNTAXC}{{\color{\colorMATH}\ensuremath{\bigcdot}}}\endgroup } ([{\begingroup\renewcommand\colorMATH{\colorMATHB}\renewcommand\colorSYNTAX{\colorSYNTAXB}{{\color{\colorMATH}\ensuremath{\sS_{2}}}}\endgroup }/x]{\begingroup\renewcommand\colorMATH{\colorMATHC}\renewcommand\colorSYNTAX{\colorSYNTAXC}{{\color{\colorMATH}\ensuremath{\pS''}}}\endgroup } = (({\begingroup\renewcommand\colorMATH{\colorMATHB}\renewcommand\colorSYNTAX{\colorSYNTAXB}{{\color{\colorMATH}\ensuremath{\Distance'}}}\endgroup }\mathord{\cdotp }{\begingroup\renewcommand\colorMATH{\colorMATHB}\renewcommand\colorSYNTAX{\colorSYNTAXB}{{\color{\colorMATH}\ensuremath{\sS_{2}}}}\endgroup })x){\begingroup\renewcommand\colorMATH{\colorMATHC}\renewcommand\colorSYNTAX{\colorSYNTAXC}{{\color{\colorMATH}\ensuremath{\bigcdot}}}\endgroup }({\begingroup\renewcommand\colorMATH{\colorMATHB}\renewcommand\colorSYNTAX{\colorSYNTAXB}{{\color{\colorMATH}\ensuremath{\Distance'}}}\endgroup } {\begingroup\renewcommand\colorMATH{\colorMATHC}\renewcommand\colorSYNTAX{\colorSYNTAXC}{{\color{\colorMATH}\ensuremath{\bigcdot}}}\endgroup } {\begingroup\renewcommand\colorMATH{\colorMATHC}\renewcommand\colorSYNTAX{\colorSYNTAXC}{{\color{\colorMATH}\ensuremath{\pS''}}}\endgroup }) = ({\begingroup\renewcommand\colorMATH{\colorMATHB}\renewcommand\colorSYNTAX{\colorSYNTAXB}{{\color{\colorMATH}\ensuremath{\Distance'}}}\endgroup }+({\begingroup\renewcommand\colorMATH{\colorMATHB}\renewcommand\colorSYNTAX{\colorSYNTAXB}{{\color{\colorMATH}\ensuremath{\Distance'}}}\endgroup }\mathord{\cdotp }{\begingroup\renewcommand\colorMATH{\colorMATHB}\renewcommand\colorSYNTAX{\colorSYNTAXB}{{\color{\colorMATH}\ensuremath{\sS_{2}}}}\endgroup })x) {\begingroup\renewcommand\colorMATH{\colorMATHC}\renewcommand\colorSYNTAX{\colorSYNTAXC}{{\color{\colorMATH}\ensuremath{\bigcdot}}}\endgroup } {\begingroup\renewcommand\colorMATH{\colorMATHC}\renewcommand\colorSYNTAX{\colorSYNTAXC}{{\color{\colorMATH}\ensuremath{\pS''}}}\endgroup }}}}.  
      By Lemma~\ref{lm:distrinst}, {{\color{\colorMATH}\ensuremath{[{\begingroup\renewcommand\colorMATH{\colorMATHB}\renewcommand\colorSYNTAX{\colorSYNTAXB}{{\color{\colorMATH}\ensuremath{\Distance'}}}\endgroup }\mathord{\cdotp }{\begingroup\renewcommand\colorMATH{\colorMATHB}\renewcommand\colorSYNTAX{\colorSYNTAXB}{{\color{\colorMATH}\ensuremath{\sS_{2}}}}\endgroup }/x]{\begingroup\renewcommand\colorMATH{\colorMATHB}\renewcommand\colorSYNTAX{\colorSYNTAXB}{{\color{\colorMATH}\ensuremath{\Distance'}}}\endgroup }(\tau _{2}) = {\begingroup\renewcommand\colorMATH{\colorMATHB}\renewcommand\colorSYNTAX{\colorSYNTAXB}{{\color{\colorMATH}\ensuremath{\Distance'}}}\endgroup }([{\begingroup\renewcommand\colorMATH{\colorMATHB}\renewcommand\colorSYNTAX{\colorSYNTAXB}{{\color{\colorMATH}\ensuremath{\sS_{2}}}}\endgroup }/x]\tau _{2})}}}, then by (1) and weakening (Lemma~\ref{lm:lrweakening-sensitivity}) {{\color{\colorMATH}\ensuremath{(\gamma '_{1}[x\mapsto {\begingroup\renewcommand\colorMATH{\colorMATHB}\renewcommand\colorSYNTAX{\colorSYNTAXB}{{\color{\colorMATH}\ensuremath{\sv_{1}}}}\endgroup }] \vdash  {\begingroup\renewcommand\colorMATH{\colorMATHC}\renewcommand\colorSYNTAX{\colorSYNTAXC}{{\color{\colorMATH}\ensuremath{\pe'_{1}}}}\endgroup },\gamma '_{2}[x\mapsto {\begingroup\renewcommand\colorMATH{\colorMATHB}\renewcommand\colorSYNTAX{\colorSYNTAXB}{{\color{\colorMATH}\ensuremath{\sv_{2}}}}\endgroup }] \vdash  {\begingroup\renewcommand\colorMATH{\colorMATHC}\renewcommand\colorSYNTAX{\colorSYNTAXC}{{\color{\colorMATH}\ensuremath{\pe'_{2}}}}\endgroup }) \in  {\mathcal{E}}_{{\begingroup\renewcommand\colorMATH{\colorMATHB}\renewcommand\colorSYNTAX{\colorSYNTAXB}{{\color{\colorMATH}\ensuremath{\Distance'}}}\endgroup } {\begingroup\renewcommand\colorMATH{\colorMATHC}\renewcommand\colorSYNTAX{\colorSYNTAXC}{{\color{\colorMATH}\ensuremath{\bigcdot}}}\endgroup } {\begingroup\renewcommand\colorMATH{\colorMATHC}\renewcommand\colorSYNTAX{\colorSYNTAXC}{{\color{\colorMATH}\ensuremath{\pS}}}\endgroup }}^{j_{3}}\llbracket {\begingroup\renewcommand\colorMATH{\colorMATHB}\renewcommand\colorSYNTAX{\colorSYNTAXB}{{\color{\colorMATH}\ensuremath{\Distance'}}}\endgroup }([{\begingroup\renewcommand\colorMATH{\colorMATHB}\renewcommand\colorSYNTAX{\colorSYNTAXB}{{\color{\colorMATH}\ensuremath{\sS_{2}}}}\endgroup }/x]\tau _{2})\rrbracket }}} (3). The result follows by Lemma~\ref{lm:probsemanticrel}.

      % Notice that {{\color{\colorMATH}\ensuremath{\llbracket {\begingroup\renewcommand\colorMATH{\colorMATHC}\renewcommand\colorSYNTAX{\colorSYNTAXC}{{\color{\colorMATH}\ensuremath{\pe_{1}}}}\endgroup }\hspace*{0.33em}{\begingroup\renewcommand\colorMATH{\colorMATHC}\renewcommand\colorSYNTAX{\colorSYNTAXC}{{\color{\colorMATH}\ensuremath{\pe_{2}}}}\endgroup }\rrbracket ^{k}_{\gamma _{i}} = \llbracket {\begingroup\renewcommand\colorMATH{\colorMATHC}\renewcommand\colorSYNTAX{\colorSYNTAXC}{{\color{\colorMATH}\ensuremath{\pe'_{i}}}}\endgroup }\rrbracket ^{k-j_{1}-j_{2}-1}_{\gamma '_{i}[x\mapsto {\begingroup\renewcommand\colorMATH{\colorMATHB}\renewcommand\colorSYNTAX{\colorSYNTAXB}{{\color{\colorMATH}\ensuremath{\sv_{i}}}}\endgroup }]}}}}, and (3) holds in particular for {{\color{\colorMATH}\ensuremath{j_{3} = k-j_{1}-j_{2}-1 < k-j_{1}-j_{2}}}}. The result follows by Lemma~\ref{lm:probsemanticrel}.
       
    \end{subproof}

    \newcommand{\tone}{{\begingroup\renewcommand\colorMATH{\colorMATHB}\renewcommand\colorSYNTAX{\colorSYNTAXB}{{\color{\colorMATH}\ensuremath{\sS_{1 1}}}}\endgroup }}
    \newcommand{\ttwo}{{\begingroup\renewcommand\colorMATH{\colorMATHB}\renewcommand\colorSYNTAX{\colorSYNTAXB}{{\color{\colorMATH}\ensuremath{\sS_{1 2}}}}\endgroup }}
    \newcommand{\sigmared}{ {\begingroup\renewcommand\colorMATH{\colorMATHC}\renewcommand\colorSYNTAX{\colorSYNTAXC}{{\color{\colorMATH}\ensuremath{\rceil {\begingroup\renewcommand\colorMATH{\colorMATHA}\renewcommand\colorSYNTAX{\colorSYNTAXA}{{\color{\colorMATH}\ensuremath{{\begingroup\renewcommand\colorMATH{\colorMATHB}\renewcommand\colorSYNTAX{\colorSYNTAXB}{{\color{\colorMATH}\ensuremath{\sS_{1}}}}\endgroup }}}}\endgroup }\lceil ^{\infty }}}}\endgroup } \sqcup  [{\begingroup\renewcommand\colorMATH{\colorMATHB}\renewcommand\colorSYNTAX{\colorSYNTAXB}{{\color{\colorMATH}\ensuremath{\sS_{1 1}}}}\endgroup }/x]{\begingroup\renewcommand\colorMATH{\colorMATHC}\renewcommand\colorSYNTAX{\colorSYNTAXC}{{\color{\colorMATH}\ensuremath{\pS_{2}}}}\endgroup } \sqcup  [{\begingroup\renewcommand\colorMATH{\colorMATHB}\renewcommand\colorSYNTAX{\colorSYNTAXB}{{\color{\colorMATH}\ensuremath{\sS_{1 2}}}}\endgroup }/x]{\begingroup\renewcommand\colorMATH{\colorMATHC}\renewcommand\colorSYNTAX{\colorSYNTAXC}{{\color{\colorMATH}\ensuremath{\pS_{3}}}}\endgroup }}
    \item  {{\color{\colorMATH}\ensuremath{\Gamma  \mathrel{;} {\begingroup\renewcommand\colorMATH{\colorMATHB}\renewcommand\colorSYNTAX{\colorSYNTAXB}{{\color{\colorMATH}\ensuremath{\Distance}}}\endgroup } \vdash  {{\color{\colorSYNTAX}\texttt{case}}}\hspace*{0.33em}{\begingroup\renewcommand\colorMATH{\colorMATHB}\renewcommand\colorSYNTAX{\colorSYNTAXB}{{\color{\colorMATH}\ensuremath{\se_{1}}}}\endgroup }\hspace*{0.33em}{{\color{\colorSYNTAX}\texttt{of}}}\hspace*{0.33em}\{ x\Rightarrow {\begingroup\renewcommand\colorMATH{\colorMATHC}\renewcommand\colorSYNTAX{\colorSYNTAXC}{{\color{\colorMATH}\ensuremath{\pe_{2}}}}\endgroup }\} \hspace*{0.33em}\{ y\Rightarrow {\begingroup\renewcommand\colorMATH{\colorMATHC}\renewcommand\colorSYNTAX{\colorSYNTAXC}{{\color{\colorMATH}\ensuremath{\pe_{3}}}}\endgroup }\}  \mathrel{:} [\tone/x]\tau _{2} \sqcup  [\ttwo/y]\tau _{3} \mathrel{;}  \sigmared}}}
      \begin{subproof} 
        We have to prove that for any {{\color{\colorMATH}\ensuremath{k}}}, {{\color{\colorMATH}\ensuremath{\forall  (\gamma _{1},\gamma _{2}) \in  {\mathcal{G}}_{{\begingroup\renewcommand\colorMATH{\colorMATHB}\renewcommand\colorSYNTAX{\colorSYNTAXB}{{\color{\colorMATH}\ensuremath{\Distance'}}}\endgroup }}^{\kg}\llbracket \Gamma \rrbracket }}}, for {{\color{\colorMATH}\ensuremath{{\begingroup\renewcommand\colorMATH{\colorMATHB}\renewcommand\colorSYNTAX{\colorSYNTAXB}{{\color{\colorMATH}\ensuremath{\Distance'}}}\endgroup } \sqsubseteq  {\begingroup\renewcommand\colorMATH{\colorMATHB}\renewcommand\colorSYNTAX{\colorSYNTAXB}{{\color{\colorMATH}\ensuremath{\Distance}}}\endgroup }}}}\\ 
        {{\color{\colorMATH}\ensuremath{(\gamma _{1}\vdash {{\color{\colorSYNTAX}\texttt{case}}}\hspace*{0.33em}{\begingroup\renewcommand\colorMATH{\colorMATHB}\renewcommand\colorSYNTAX{\colorSYNTAXB}{{\color{\colorMATH}\ensuremath{\se_{1}}}}\endgroup }\hspace*{0.33em}{{\color{\colorSYNTAX}\texttt{of}}}\hspace*{0.33em}\{ x\Rightarrow {\begingroup\renewcommand\colorMATH{\colorMATHC}\renewcommand\colorSYNTAX{\colorSYNTAXC}{{\color{\colorMATH}\ensuremath{\pe_{2}}}}\endgroup }\} \hspace*{0.33em}\{ y\Rightarrow {\begingroup\renewcommand\colorMATH{\colorMATHC}\renewcommand\colorSYNTAX{\colorSYNTAXC}{{\color{\colorMATH}\ensuremath{\pe_{3}}}}\endgroup }\} ,\gamma _{2}\vdash {{\color{\colorSYNTAX}\texttt{case}}}\hspace*{0.33em}{\begingroup\renewcommand\colorMATH{\colorMATHB}\renewcommand\colorSYNTAX{\colorSYNTAXB}{{\color{\colorMATH}\ensuremath{\se_{1}}}}\endgroup }\hspace*{0.33em}{{\color{\colorSYNTAX}\texttt{of}}}\hspace*{0.33em}\{ x\Rightarrow {\begingroup\renewcommand\colorMATH{\colorMATHC}\renewcommand\colorSYNTAX{\colorSYNTAXC}{{\color{\colorMATH}\ensuremath{\pe_{2}}}}\endgroup }\} \hspace*{0.33em}\{ y\Rightarrow {\begingroup\renewcommand\colorMATH{\colorMATHC}\renewcommand\colorSYNTAX{\colorSYNTAXC}{{\color{\colorMATH}\ensuremath{\pe_{3}}}}\endgroup }\} ) \in  {\mathcal{E}}_{{\begingroup\renewcommand\colorMATH{\colorMATHB}\renewcommand\colorSYNTAX{\colorSYNTAXB}{{\color{\colorMATH}\ensuremath{\Distance'}}}\endgroup }\mathord{\cdotp }{\begingroup\renewcommand\colorMATH{\colorMATHC}\renewcommand\colorSYNTAX{\colorSYNTAXC}{{\color{\colorMATH}\ensuremath{\pS}}}\endgroup }}^{k}\llbracket {\begingroup\renewcommand\colorMATH{\colorMATHB}\renewcommand\colorSYNTAX{\colorSYNTAXB}{{\color{\colorMATH}\ensuremath{\Distance'}}}\endgroup }(\tau )\rrbracket }}}.
                
        By induction hypothesis on {{\color{\colorMATH}\ensuremath{\Gamma \mathrel{;} {\begingroup\renewcommand\colorMATH{\colorMATHB}\renewcommand\colorSYNTAX{\colorSYNTAXB}{{\color{\colorMATH}\ensuremath{\Distance}}}\endgroup }  \vdash  {\begingroup\renewcommand\colorMATH{\colorMATHB}\renewcommand\colorSYNTAX{\colorSYNTAXB}{{\color{\colorMATH}\ensuremath{\se_{1}}}}\endgroup } \mathrel{:} \tau _{1 1} \mathrel{^{{\begingroup\renewcommand\colorMATH{\colorMATHB}\renewcommand\colorSYNTAX{\colorSYNTAXB}{{\color{\colorMATH}\ensuremath{\sS_{1 1}}}}\endgroup }}\oplus ^{{\begingroup\renewcommand\colorMATH{\colorMATHB}\renewcommand\colorSYNTAX{\colorSYNTAXB}{{\color{\colorMATH}\ensuremath{\sS_{1 2}}}}\endgroup }}} \tau _{1 2} \mathrel{;} {\begingroup\renewcommand\colorMATH{\colorMATHB}\renewcommand\colorSYNTAX{\colorSYNTAXB}{{\color{\colorMATH}\ensuremath{\sS_{1}}}}\endgroup }}}}, we know that\\
        {{\color{\colorMATH}\ensuremath{(\gamma _{1}\vdash {\begingroup\renewcommand\colorMATH{\colorMATHB}\renewcommand\colorSYNTAX{\colorSYNTAXB}{{\color{\colorMATH}\ensuremath{\se_{1}}}}\endgroup },\gamma _{2}\vdash {\begingroup\renewcommand\colorMATH{\colorMATHB}\renewcommand\colorSYNTAX{\colorSYNTAXB}{{\color{\colorMATH}\ensuremath{\se_{1}}}}\endgroup }) \in  {\mathcal{E}}^{k}_{{\begingroup\renewcommand\colorMATH{\colorMATHB}\renewcommand\colorSYNTAX{\colorSYNTAXB}{{\color{\colorMATH}\ensuremath{\Distance'}}}\endgroup }\mathord{\cdotp }{\begingroup\renewcommand\colorMATH{\colorMATHB}\renewcommand\colorSYNTAX{\colorSYNTAXB}{{\color{\colorMATH}\ensuremath{\sS_{1}}}}\endgroup }}\llbracket {\begingroup\renewcommand\colorMATH{\colorMATHB}\renewcommand\colorSYNTAX{\colorSYNTAXB}{{\color{\colorMATH}\ensuremath{\Distance'}}}\endgroup }(\tau _{1 1} \mathrel{^{{\begingroup\renewcommand\colorMATH{\colorMATHB}\renewcommand\colorSYNTAX{\colorSYNTAXB}{{\color{\colorMATH}\ensuremath{\sS_{1 1}}}}\endgroup }}\oplus ^{{\begingroup\renewcommand\colorMATH{\colorMATHB}\renewcommand\colorSYNTAX{\colorSYNTAXB}{{\color{\colorMATH}\ensuremath{\sS_{1 2}}}}\endgroup }}} \tau _{1 2})\rrbracket }}}, i.e. if {{\color{\colorMATH}\ensuremath{\gamma _{1}\vdash {\begingroup\renewcommand\colorMATH{\colorMATHB}\renewcommand\colorSYNTAX{\colorSYNTAXB}{{\color{\colorMATH}\ensuremath{\se_{1}}}}\endgroup } \Downarrow ^{j_{1}} {\begingroup\renewcommand\colorMATH{\colorMATHB}\renewcommand\colorSYNTAX{\colorSYNTAXB}{{\color{\colorMATH}\ensuremath{\sv_{1 1}}}}\endgroup }}}}, \pthen {{\color{\colorMATH}\ensuremath{\gamma _{2}\vdash {\begingroup\renewcommand\colorMATH{\colorMATHB}\renewcommand\colorSYNTAX{\colorSYNTAXB}{{\color{\colorMATH}\ensuremath{\se_{1}}}}\endgroup } \Downarrow ^{\pj[1]} {\begingroup\renewcommand\colorMATH{\colorMATHB}\renewcommand\colorSYNTAX{\colorSYNTAXB}{{\color{\colorMATH}\ensuremath{\sv_{1 2}}}}\endgroup }}}} \pand 
        {{\color{\colorMATH}\ensuremath{({\begingroup\renewcommand\colorMATH{\colorMATHB}\renewcommand\colorSYNTAX{\colorSYNTAXB}{{\color{\colorMATH}\ensuremath{\sv_{1 1}}}}\endgroup }, {\begingroup\renewcommand\colorMATH{\colorMATHB}\renewcommand\colorSYNTAX{\colorSYNTAXB}{{\color{\colorMATH}\ensuremath{\sv_{1 2}}}}\endgroup }) \in  {\mathcal{V}}^{k-j_{1}}_{{\begingroup\renewcommand\colorMATH{\colorMATHB}\renewcommand\colorSYNTAX{\colorSYNTAXB}{{\color{\colorMATH}\ensuremath{\Distance'}}}\endgroup }\mathord{\cdotp }{\begingroup\renewcommand\colorMATH{\colorMATHB}\renewcommand\colorSYNTAX{\colorSYNTAXB}{{\color{\colorMATH}\ensuremath{\sS_{1}}}}\endgroup }}\llbracket {\begingroup\renewcommand\colorMATH{\colorMATHB}\renewcommand\colorSYNTAX{\colorSYNTAXB}{{\color{\colorMATH}\ensuremath{\Distance'}}}\endgroup }(\tau _{1 1}) \mathrel{^{{\begingroup\renewcommand\colorMATH{\colorMATHB}\renewcommand\colorSYNTAX{\colorSYNTAXB}{{\color{\colorMATH}\ensuremath{\Distance'}}}\endgroup }\mathord{\cdotp }{\begingroup\renewcommand\colorMATH{\colorMATHB}\renewcommand\colorSYNTAX{\colorSYNTAXB}{{\color{\colorMATH}\ensuremath{\sS_{1 1}}}}\endgroup }}\oplus ^{{\begingroup\renewcommand\colorMATH{\colorMATHB}\renewcommand\colorSYNTAX{\colorSYNTAXB}{{\color{\colorMATH}\ensuremath{\Distance'}}}\endgroup }\mathord{\cdotp }{\begingroup\renewcommand\colorMATH{\colorMATHB}\renewcommand\colorSYNTAX{\colorSYNTAXB}{{\color{\colorMATH}\ensuremath{\sS_{1 2}}}}\endgroup }}} {\begingroup\renewcommand\colorMATH{\colorMATHB}\renewcommand\colorSYNTAX{\colorSYNTAXB}{{\color{\colorMATH}\ensuremath{\Distance'}}}\endgroup }(\tau _{1 2})\rrbracket }}}.
        Notice that {{\color{\colorMATH}\ensuremath{{\begingroup\renewcommand\colorMATH{\colorMATHB}\renewcommand\colorSYNTAX{\colorSYNTAXB}{{\color{\colorMATH}\ensuremath{\Distance'}}}\endgroup }{\begingroup\renewcommand\colorMATH{\colorMATHC}\renewcommand\colorSYNTAX{\colorSYNTAXC}{{\color{\colorMATH}\ensuremath{\bigcdot}}}\endgroup }{\begingroup\renewcommand\colorMATH{\colorMATHC}\renewcommand\colorSYNTAX{\colorSYNTAXC}{{\color{\colorMATH}\ensuremath{\pS}}}\endgroup } = {\begingroup\renewcommand\colorMATH{\colorMATHB}\renewcommand\colorSYNTAX{\colorSYNTAXB}{{\color{\colorMATH}\ensuremath{\Distance'}}}\endgroup }{\begingroup\renewcommand\colorMATH{\colorMATHC}\renewcommand\colorSYNTAX{\colorSYNTAXC}{{\color{\colorMATH}\ensuremath{\bigcdot}}}\endgroup } {\begingroup\renewcommand\colorMATH{\colorMATHC}\renewcommand\colorSYNTAX{\colorSYNTAXC}{{\color{\colorMATH}\ensuremath{\rceil {\begingroup\renewcommand\colorMATH{\colorMATHA}\renewcommand\colorSYNTAX{\colorSYNTAXA}{{\color{\colorMATH}\ensuremath{{\begingroup\renewcommand\colorMATH{\colorMATHB}\renewcommand\colorSYNTAX{\colorSYNTAXB}{{\color{\colorMATH}\ensuremath{\sS_{1}}}}\endgroup }}}}\endgroup }\lceil ^{\infty }}}}\endgroup } \sqcup  {\begingroup\renewcommand\colorMATH{\colorMATHB}\renewcommand\colorSYNTAX{\colorSYNTAXB}{{\color{\colorMATH}\ensuremath{\Distance'}}}\endgroup }{\begingroup\renewcommand\colorMATH{\colorMATHC}\renewcommand\colorSYNTAX{\colorSYNTAXC}{{\color{\colorMATH}\ensuremath{\bigcdot}}}\endgroup }([{\begingroup\renewcommand\colorMATH{\colorMATHB}\renewcommand\colorSYNTAX{\colorSYNTAXB}{{\color{\colorMATH}\ensuremath{\sS_{1 1}}}}\endgroup }/x]{\begingroup\renewcommand\colorMATH{\colorMATHC}\renewcommand\colorSYNTAX{\colorSYNTAXC}{{\color{\colorMATH}\ensuremath{\pS_{2}}}}\endgroup }) \sqcup  {\begingroup\renewcommand\colorMATH{\colorMATHB}\renewcommand\colorSYNTAX{\colorSYNTAXB}{{\color{\colorMATH}\ensuremath{\Distance'}}}\endgroup }{\begingroup\renewcommand\colorMATH{\colorMATHC}\renewcommand\colorSYNTAX{\colorSYNTAXC}{{\color{\colorMATH}\ensuremath{\bigcdot}}}\endgroup }([{\begingroup\renewcommand\colorMATH{\colorMATHB}\renewcommand\colorSYNTAX{\colorSYNTAXB}{{\color{\colorMATH}\ensuremath{\sS_{1 2}}}}\endgroup }/x]{\begingroup\renewcommand\colorMATH{\colorMATHC}\renewcommand\colorSYNTAX{\colorSYNTAXC}{{\color{\colorMATH}\ensuremath{\pS_{3}}}}\endgroup })}}}.
        If {{\color{\colorMATH}\ensuremath{{\begingroup\renewcommand\colorMATH{\colorMATHB}\renewcommand\colorSYNTAX{\colorSYNTAXB}{{\color{\colorMATH}\ensuremath{\Distance'}}}\endgroup }{\begingroup\renewcommand\colorMATH{\colorMATHC}\renewcommand\colorSYNTAX{\colorSYNTAXC}{{\color{\colorMATH}\ensuremath{\bigcdot}}}\endgroup } {\begingroup\renewcommand\colorMATH{\colorMATHC}\renewcommand\colorSYNTAX{\colorSYNTAXC}{{\color{\colorMATH}\ensuremath{\rceil {\begingroup\renewcommand\colorMATH{\colorMATHA}\renewcommand\colorSYNTAX{\colorSYNTAXA}{{\color{\colorMATH}\ensuremath{{\begingroup\renewcommand\colorMATH{\colorMATHB}\renewcommand\colorSYNTAX{\colorSYNTAXB}{{\color{\colorMATH}\ensuremath{\sS_{1}}}}\endgroup }}}}\endgroup }\lceil ^{\infty }}}}\endgroup } = \infty }}} then the result is trivial and holds immediately.
        Let us suppose that {{\color{\colorMATH}\ensuremath{{\begingroup\renewcommand\colorMATH{\colorMATHB}\renewcommand\colorSYNTAX{\colorSYNTAXB}{{\color{\colorMATH}\ensuremath{\Distance'}}}\endgroup }{\begingroup\renewcommand\colorMATH{\colorMATHC}\renewcommand\colorSYNTAX{\colorSYNTAXC}{{\color{\colorMATH}\ensuremath{\bigcdot}}}\endgroup } {\begingroup\renewcommand\colorMATH{\colorMATHC}\renewcommand\colorSYNTAX{\colorSYNTAXC}{{\color{\colorMATH}\ensuremath{\rceil {\begingroup\renewcommand\colorMATH{\colorMATHA}\renewcommand\colorSYNTAX{\colorSYNTAXA}{{\color{\colorMATH}\ensuremath{{\begingroup\renewcommand\colorMATH{\colorMATHB}\renewcommand\colorSYNTAX{\colorSYNTAXB}{{\color{\colorMATH}\ensuremath{\sS_{1}}}}\endgroup }}}}\endgroup }\lceil ^{\infty }}}}\endgroup } = 0}}}, then this means that 
        {{\color{\colorMATH}\ensuremath{{\begingroup\renewcommand\colorMATH{\colorMATHB}\renewcommand\colorSYNTAX{\colorSYNTAXB}{{\color{\colorMATH}\ensuremath{\Distance'}}}\endgroup }\mathord{\cdotp }{\begingroup\renewcommand\colorMATH{\colorMATHB}\renewcommand\colorSYNTAX{\colorSYNTAXB}{{\color{\colorMATH}\ensuremath{\sS_{1}}}}\endgroup } = 0}}}, i.e.
        either {{\color{\colorMATH}\ensuremath{{\begingroup\renewcommand\colorMATH{\colorMATHB}\renewcommand\colorSYNTAX{\colorSYNTAXB}{{\color{\colorMATH}\ensuremath{\sv_{1 1}}}}\endgroup } = \inl\hspace*{0.33em}v'_{1 1}}}} and {{\color{\colorMATH}\ensuremath{{\begingroup\renewcommand\colorMATH{\colorMATHB}\renewcommand\colorSYNTAX{\colorSYNTAXB}{{\color{\colorMATH}\ensuremath{\sv_{1 2}}}}\endgroup } = \inl\hspace*{0.33em}v'_{1 2}}}}, or {{\color{\colorMATH}\ensuremath{{\begingroup\renewcommand\colorMATH{\colorMATHB}\renewcommand\colorSYNTAX{\colorSYNTAXB}{{\color{\colorMATH}\ensuremath{\sv_{1 1}}}}\endgroup } = \inr\hspace*{0.33em}v'_{1 1}}}} and {{\color{\colorMATH}\ensuremath{{\begingroup\renewcommand\colorMATH{\colorMATHB}\renewcommand\colorSYNTAX{\colorSYNTAXB}{{\color{\colorMATH}\ensuremath{\sv_{1 2}}}}\endgroup } = \inr\hspace*{0.33em}v'_{1 2}}}}.

        %We proceed by case analysis on {{\color{\colorMATH}\ensuremath{({\begingroup\renewcommand\colorMATH{\colorMATHB}\renewcommand\colorSYNTAX{\colorSYNTAXB}{{\color{\colorMATH}\ensuremath{\sv_{1 1}}}}\endgroup }, {\begingroup\renewcommand\colorMATH{\colorMATHB}\renewcommand\colorSYNTAX{\colorSYNTAXB}{{\color{\colorMATH}\ensuremath{\sv_{1 2}}}}\endgroup })}}}:
        Let us suppose that {{\color{\colorMATH}\ensuremath{({\begingroup\renewcommand\colorMATH{\colorMATHB}\renewcommand\colorSYNTAX{\colorSYNTAXB}{{\color{\colorMATH}\ensuremath{\sv_{1 1}}}}\endgroup }, {\begingroup\renewcommand\colorMATH{\colorMATHB}\renewcommand\colorSYNTAX{\colorSYNTAXB}{{\color{\colorMATH}\ensuremath{\sv_{1 2}}}}\endgroup }) = (\inl\hspace*{0.33em}{\begingroup\renewcommand\colorMATH{\colorMATHB}\renewcommand\colorSYNTAX{\colorSYNTAXB}{{\color{\colorMATH}\ensuremath{\sv'_{1 1}}}}\endgroup }, \inl\hspace*{0.33em}{\begingroup\renewcommand\colorMATH{\colorMATHB}\renewcommand\colorSYNTAX{\colorSYNTAXB}{{\color{\colorMATH}\ensuremath{\sv'_{1 2}}}}\endgroup })}}} (the case {{\color{\colorMATH}\ensuremath{({\begingroup\renewcommand\colorMATH{\colorMATHB}\renewcommand\colorSYNTAX{\colorSYNTAXB}{{\color{\colorMATH}\ensuremath{\sv_{1 1}}}}\endgroup }, {\begingroup\renewcommand\colorMATH{\colorMATHB}\renewcommand\colorSYNTAX{\colorSYNTAXB}{{\color{\colorMATH}\ensuremath{\sv_{1 2}}}}\endgroup }) = (\inr\hspace*{0.33em}{\begingroup\renewcommand\colorMATH{\colorMATHB}\renewcommand\colorSYNTAX{\colorSYNTAXB}{{\color{\colorMATH}\ensuremath{\sv'_{1 1}}}}\endgroup }, \inr\hspace*{0.33em}{\begingroup\renewcommand\colorMATH{\colorMATHB}\renewcommand\colorSYNTAX{\colorSYNTAXB}{{\color{\colorMATH}\ensuremath{\sv'_{1 2}}}}\endgroup })}}} is analogous).
        By Lemma~\ref{lm:associativity-inst} {{\color{\colorMATH}\ensuremath{{\begingroup\renewcommand\colorMATH{\colorMATHB}\renewcommand\colorSYNTAX{\colorSYNTAXB}{{\color{\colorMATH}\ensuremath{\Distance'}}}\endgroup }\mathord{\cdotp }{\begingroup\renewcommand\colorMATH{\colorMATHB}\renewcommand\colorSYNTAX{\colorSYNTAXB}{{\color{\colorMATH}\ensuremath{\sS_{1}}}}\endgroup } + {\begingroup\renewcommand\colorMATH{\colorMATHB}\renewcommand\colorSYNTAX{\colorSYNTAXB}{{\color{\colorMATH}\ensuremath{\Distance'}}}\endgroup }\mathord{\cdotp }{\begingroup\renewcommand\colorMATH{\colorMATHB}\renewcommand\colorSYNTAX{\colorSYNTAXB}{{\color{\colorMATH}\ensuremath{\sS_{1 1}}}}\endgroup } = {\begingroup\renewcommand\colorMATH{\colorMATHB}\renewcommand\colorSYNTAX{\colorSYNTAXB}{{\color{\colorMATH}\ensuremath{\Distance'}}}\endgroup }\mathord{\cdotp }({\begingroup\renewcommand\colorMATH{\colorMATHB}\renewcommand\colorSYNTAX{\colorSYNTAXB}{{\color{\colorMATH}\ensuremath{\sS_{1}}}}\endgroup } + {\begingroup\renewcommand\colorMATH{\colorMATHB}\renewcommand\colorSYNTAX{\colorSYNTAXB}{{\color{\colorMATH}\ensuremath{\sS_{1 1}}}}\endgroup })}}}, then
        {{\color{\colorMATH}\ensuremath{({\begingroup\renewcommand\colorMATH{\colorMATHB}\renewcommand\colorSYNTAX{\colorSYNTAXB}{{\color{\colorMATH}\ensuremath{\sv'_{1 1}}}}\endgroup }, {\begingroup\renewcommand\colorMATH{\colorMATHB}\renewcommand\colorSYNTAX{\colorSYNTAXB}{{\color{\colorMATH}\ensuremath{\sv'_{1 2}}}}\endgroup }) \in  {\mathcal{V}}^{k-j_{1}}_{{\begingroup\renewcommand\colorMATH{\colorMATHB}\renewcommand\colorSYNTAX{\colorSYNTAXB}{{\color{\colorMATH}\ensuremath{\Distance'}}}\endgroup }\mathord{\cdotp }({\begingroup\renewcommand\colorMATH{\colorMATHB}\renewcommand\colorSYNTAX{\colorSYNTAXB}{{\color{\colorMATH}\ensuremath{\sS_{1}}}}\endgroup } + {\begingroup\renewcommand\colorMATH{\colorMATHB}\renewcommand\colorSYNTAX{\colorSYNTAXB}{{\color{\colorMATH}\ensuremath{\sS_{1 1}}}}\endgroup })}\llbracket {\begingroup\renewcommand\colorMATH{\colorMATHB}\renewcommand\colorSYNTAX{\colorSYNTAXB}{{\color{\colorMATH}\ensuremath{\Distance'}}}\endgroup }(\tau _{1 1})\rrbracket }}}.

        Also, by induction hypothesis on {{\color{\colorMATH}\ensuremath{\Gamma ,x\mathrel{:}\tau _{1 1}; {\begingroup\renewcommand\colorMATH{\colorMATHB}\renewcommand\colorSYNTAX{\colorSYNTAXB}{{\color{\colorMATH}\ensuremath{\Distance}}}\endgroup } + ({\begingroup\renewcommand\colorMATH{\colorMATHB}\renewcommand\colorSYNTAX{\colorSYNTAXB}{{\color{\colorMATH}\ensuremath{\Distance}}}\endgroup }\mathord{\cdotp } ({\begingroup\renewcommand\colorMATH{\colorMATHB}\renewcommand\colorSYNTAX{\colorSYNTAXB}{{\color{\colorMATH}\ensuremath{\sS_{1}}}}\endgroup } + {\begingroup\renewcommand\colorMATH{\colorMATHB}\renewcommand\colorSYNTAX{\colorSYNTAXB}{{\color{\colorMATH}\ensuremath{\sS_{1 1}}}}\endgroup }))x \vdash  {\begingroup\renewcommand\colorMATH{\colorMATHC}\renewcommand\colorSYNTAX{\colorSYNTAXC}{{\color{\colorMATH}\ensuremath{\pe_{2}}}}\endgroup } \mathrel{:} \tau _{2} \mathrel{;} {\begingroup\renewcommand\colorMATH{\colorMATHC}\renewcommand\colorSYNTAX{\colorSYNTAXC}{{\color{\colorMATH}\ensuremath{\pS_{2}}}}\endgroup }}}}, by choosing \\
        {{\color{\colorMATH}\ensuremath{{\begingroup\renewcommand\colorMATH{\colorMATHB}\renewcommand\colorSYNTAX{\colorSYNTAXB}{{\color{\colorMATH}\ensuremath{\Distance'}}}\endgroup }+({\begingroup\renewcommand\colorMATH{\colorMATHB}\renewcommand\colorSYNTAX{\colorSYNTAXB}{{\color{\colorMATH}\ensuremath{\Distance'}}}\endgroup }\mathord{\cdotp }({\begingroup\renewcommand\colorMATH{\colorMATHB}\renewcommand\colorSYNTAX{\colorSYNTAXB}{{\color{\colorMATH}\ensuremath{\sS_{1}}}}\endgroup }+{\begingroup\renewcommand\colorMATH{\colorMATHB}\renewcommand\colorSYNTAX{\colorSYNTAXB}{{\color{\colorMATH}\ensuremath{\sS_{1 1}}}}\endgroup }))x \sqsubseteq  {\begingroup\renewcommand\colorMATH{\colorMATHB}\renewcommand\colorSYNTAX{\colorSYNTAXB}{{\color{\colorMATH}\ensuremath{\Distance}}}\endgroup } + ({\begingroup\renewcommand\colorMATH{\colorMATHB}\renewcommand\colorSYNTAX{\colorSYNTAXB}{{\color{\colorMATH}\ensuremath{\Distance}}}\endgroup }\mathord{\cdotp } ({\begingroup\renewcommand\colorMATH{\colorMATHB}\renewcommand\colorSYNTAX{\colorSYNTAXB}{{\color{\colorMATH}\ensuremath{\sS_{1}}}}\endgroup } + {\begingroup\renewcommand\colorMATH{\colorMATHB}\renewcommand\colorSYNTAX{\colorSYNTAXB}{{\color{\colorMATH}\ensuremath{\sS_{1 1}}}}\endgroup }))x}}}, 
        {{\color{\colorMATH}\ensuremath{(\gamma _{1}[x \mapsto  {\begingroup\renewcommand\colorMATH{\colorMATHB}\renewcommand\colorSYNTAX{\colorSYNTAXB}{{\color{\colorMATH}\ensuremath{\sv'_{1 2}}}}\endgroup }],\gamma _{2}[x \mapsto  {\begingroup\renewcommand\colorMATH{\colorMATHB}\renewcommand\colorSYNTAX{\colorSYNTAXB}{{\color{\colorMATH}\ensuremath{\sv'_{2 2}}}}\endgroup }]) \in  {\mathcal{G}}^{k-j_{1}}_{{\begingroup\renewcommand\colorMATH{\colorMATHB}\renewcommand\colorSYNTAX{\colorSYNTAXB}{{\color{\colorMATH}\ensuremath{\Distance'}}}\endgroup }+({\begingroup\renewcommand\colorMATH{\colorMATHB}\renewcommand\colorSYNTAX{\colorSYNTAXB}{{\color{\colorMATH}\ensuremath{\Distance'}}}\endgroup }\mathord{\cdotp }({\begingroup\renewcommand\colorMATH{\colorMATHB}\renewcommand\colorSYNTAX{\colorSYNTAXB}{{\color{\colorMATH}\ensuremath{\sS_{1}}}}\endgroup }+{\begingroup\renewcommand\colorMATH{\colorMATHB}\renewcommand\colorSYNTAX{\colorSYNTAXB}{{\color{\colorMATH}\ensuremath{\sS_{1 1}}}}\endgroup }))x}\llbracket \Gamma ,x\mathrel{:}\tau _{1 1}\rrbracket }}} (note that {{\color{\colorMATH}\ensuremath{x \notin  dom({\begingroup\renewcommand\colorMATH{\colorMATHB}\renewcommand\colorSYNTAX{\colorSYNTAXB}{{\color{\colorMATH}\ensuremath{\sS_{1}}}}\endgroup }) \cup  dom({\begingroup\renewcommand\colorMATH{\colorMATHB}\renewcommand\colorSYNTAX{\colorSYNTAXB}{{\color{\colorMATH}\ensuremath{\sS_{1 1}}}}\endgroup })}}}, therefore {{\color{\colorMATH}\ensuremath{({\begingroup\renewcommand\colorMATH{\colorMATHB}\renewcommand\colorSYNTAX{\colorSYNTAXB}{{\color{\colorMATH}\ensuremath{\Distance'}}}\endgroup }+({\begingroup\renewcommand\colorMATH{\colorMATHB}\renewcommand\colorSYNTAX{\colorSYNTAXB}{{\color{\colorMATH}\ensuremath{\Distance'}}}\endgroup }\mathord{\cdotp }({\begingroup\renewcommand\colorMATH{\colorMATHB}\renewcommand\colorSYNTAX{\colorSYNTAXB}{{\color{\colorMATH}\ensuremath{\sS_{1}}}}\endgroup }+{\begingroup\renewcommand\colorMATH{\colorMATHB}\renewcommand\colorSYNTAX{\colorSYNTAXB}{{\color{\colorMATH}\ensuremath{\sS_{1 1}}}}\endgroup }))x)(\tau _{1 1}) = {\begingroup\renewcommand\colorMATH{\colorMATHB}\renewcommand\colorSYNTAX{\colorSYNTAXB}{{\color{\colorMATH}\ensuremath{\Distance'}}}\endgroup }(\tau _{1 1})}}}) we know that\\
        {{\color{\colorMATH}\ensuremath{(\gamma _{1}[x \mapsto  {\begingroup\renewcommand\colorMATH{\colorMATHB}\renewcommand\colorSYNTAX{\colorSYNTAXB}{{\color{\colorMATH}\ensuremath{\sv'_{1 2}}}}\endgroup }]\vdash {\begingroup\renewcommand\colorMATH{\colorMATHC}\renewcommand\colorSYNTAX{\colorSYNTAXC}{{\color{\colorMATH}\ensuremath{\pe_{2}}}}\endgroup },\gamma _{2}[x \mapsto  {\begingroup\renewcommand\colorMATH{\colorMATHB}\renewcommand\colorSYNTAX{\colorSYNTAXB}{{\color{\colorMATH}\ensuremath{\sv'_{2 2}}}}\endgroup }]\vdash {\begingroup\renewcommand\colorMATH{\colorMATHC}\renewcommand\colorSYNTAX{\colorSYNTAXC}{{\color{\colorMATH}\ensuremath{\pe_{2}}}}\endgroup }) \in  {\mathcal{E}}^{k-j_{1}}_{({\begingroup\renewcommand\colorMATH{\colorMATHB}\renewcommand\colorSYNTAX{\colorSYNTAXB}{{\color{\colorMATH}\ensuremath{\Distance'}}}\endgroup }+({\begingroup\renewcommand\colorMATH{\colorMATHB}\renewcommand\colorSYNTAX{\colorSYNTAXB}{{\color{\colorMATH}\ensuremath{\Distance'}}}\endgroup }\mathord{\cdotp }({\begingroup\renewcommand\colorMATH{\colorMATHB}\renewcommand\colorSYNTAX{\colorSYNTAXB}{{\color{\colorMATH}\ensuremath{\sS_{1}}}}\endgroup }+{\begingroup\renewcommand\colorMATH{\colorMATHB}\renewcommand\colorSYNTAX{\colorSYNTAXB}{{\color{\colorMATH}\ensuremath{\sS_{1 1}}}}\endgroup }))x) {\begingroup\renewcommand\colorMATH{\colorMATHC}\renewcommand\colorSYNTAX{\colorSYNTAXC}{{\color{\colorMATH}\ensuremath{\bigcdot}}}\endgroup } {\begingroup\renewcommand\colorMATH{\colorMATHC}\renewcommand\colorSYNTAX{\colorSYNTAXC}{{\color{\colorMATH}\ensuremath{\pS_{2}}}}\endgroup }}\llbracket ({\begingroup\renewcommand\colorMATH{\colorMATHB}\renewcommand\colorSYNTAX{\colorSYNTAXB}{{\color{\colorMATH}\ensuremath{\Distance'}}}\endgroup }+({\begingroup\renewcommand\colorMATH{\colorMATHB}\renewcommand\colorSYNTAX{\colorSYNTAXB}{{\color{\colorMATH}\ensuremath{\Distance'}}}\endgroup }\mathord{\cdotp }({\begingroup\renewcommand\colorMATH{\colorMATHB}\renewcommand\colorSYNTAX{\colorSYNTAXB}{{\color{\colorMATH}\ensuremath{\sS_{1}}}}\endgroup }+{\begingroup\renewcommand\colorMATH{\colorMATHB}\renewcommand\colorSYNTAX{\colorSYNTAXB}{{\color{\colorMATH}\ensuremath{\sS_{1 1}}}}\endgroup }))x)(\tau _{2})\rrbracket }}}.\\
        But {{\color{\colorMATH}\ensuremath{{\begingroup\renewcommand\colorMATH{\colorMATHB}\renewcommand\colorSYNTAX{\colorSYNTAXB}{{\color{\colorMATH}\ensuremath{\Distance'}}}\endgroup }\mathord{\cdotp }{\begingroup\renewcommand\colorMATH{\colorMATHB}\renewcommand\colorSYNTAX{\colorSYNTAXB}{{\color{\colorMATH}\ensuremath{\sS_{1}}}}\endgroup } = 0}}}, therefore {{\color{\colorMATH}\ensuremath{({\begingroup\renewcommand\colorMATH{\colorMATHB}\renewcommand\colorSYNTAX{\colorSYNTAXB}{{\color{\colorMATH}\ensuremath{\Distance'}}}\endgroup }+({\begingroup\renewcommand\colorMATH{\colorMATHB}\renewcommand\colorSYNTAX{\colorSYNTAXB}{{\color{\colorMATH}\ensuremath{\Distance'}}}\endgroup }\mathord{\cdotp }({\begingroup\renewcommand\colorMATH{\colorMATHB}\renewcommand\colorSYNTAX{\colorSYNTAXB}{{\color{\colorMATH}\ensuremath{\sS_{1}}}}\endgroup }+{\begingroup\renewcommand\colorMATH{\colorMATHB}\renewcommand\colorSYNTAX{\colorSYNTAXB}{{\color{\colorMATH}\ensuremath{\sS_{1 1}}}}\endgroup }))x){\begingroup\renewcommand\colorMATH{\colorMATHC}\renewcommand\colorSYNTAX{\colorSYNTAXC}{{\color{\colorMATH}\ensuremath{\bigcdot}}}\endgroup } {\begingroup\renewcommand\colorMATH{\colorMATHC}\renewcommand\colorSYNTAX{\colorSYNTAXC}{{\color{\colorMATH}\ensuremath{\pS_{2}}}}\endgroup } = ({\begingroup\renewcommand\colorMATH{\colorMATHB}\renewcommand\colorSYNTAX{\colorSYNTAXB}{{\color{\colorMATH}\ensuremath{\Distance'}}}\endgroup }+({\begingroup\renewcommand\colorMATH{\colorMATHB}\renewcommand\colorSYNTAX{\colorSYNTAXB}{{\color{\colorMATH}\ensuremath{\Distance'}}}\endgroup }\mathord{\cdotp }{\begingroup\renewcommand\colorMATH{\colorMATHB}\renewcommand\colorSYNTAX{\colorSYNTAXB}{{\color{\colorMATH}\ensuremath{\sS_{1 1}}}}\endgroup })x) {\begingroup\renewcommand\colorMATH{\colorMATHC}\renewcommand\colorSYNTAX{\colorSYNTAXC}{{\color{\colorMATH}\ensuremath{\bigcdot}}}\endgroup } {\begingroup\renewcommand\colorMATH{\colorMATHC}\renewcommand\colorSYNTAX{\colorSYNTAXC}{{\color{\colorMATH}\ensuremath{\pS_{2}}}}\endgroup }}}}, and
        {{\color{\colorMATH}\ensuremath{{\begingroup\renewcommand\colorMATH{\colorMATHB}\renewcommand\colorSYNTAX{\colorSYNTAXB}{{\color{\colorMATH}\ensuremath{\Distance'}}}\endgroup }\mathord{\cdotp }({\begingroup\renewcommand\colorMATH{\colorMATHB}\renewcommand\colorSYNTAX{\colorSYNTAXB}{{\color{\colorMATH}\ensuremath{\sS_{1}}}}\endgroup }+{\begingroup\renewcommand\colorMATH{\colorMATHB}\renewcommand\colorSYNTAX{\colorSYNTAXB}{{\color{\colorMATH}\ensuremath{\sS_{1 1}}}}\endgroup }) = {\begingroup\renewcommand\colorMATH{\colorMATHB}\renewcommand\colorSYNTAX{\colorSYNTAXB}{{\color{\colorMATH}\ensuremath{\Distance'}}}\endgroup }\mathord{\cdotp }{\begingroup\renewcommand\colorMATH{\colorMATHB}\renewcommand\colorSYNTAX{\colorSYNTAXB}{{\color{\colorMATH}\ensuremath{\sS_{1 1}}}}\endgroup }}}}, and
        by Lemma~\ref{lm:equivsimplsubst} and because {{\color{\colorMATH}\ensuremath{{\begingroup\renewcommand\colorMATH{\colorMATHB}\renewcommand\colorSYNTAX{\colorSYNTAXB}{{\color{\colorMATH}\ensuremath{\Distance'}}}\endgroup }\mathord{\cdotp }{\begingroup\renewcommand\colorMATH{\colorMATHB}\renewcommand\colorSYNTAX{\colorSYNTAXB}{{\color{\colorMATH}\ensuremath{\sS_{1 1}}}}\endgroup } \in  {\text{sens}}}}}, then 
        {{\color{\colorMATH}\ensuremath{({\begingroup\renewcommand\colorMATH{\colorMATHB}\renewcommand\colorSYNTAX{\colorSYNTAXB}{{\color{\colorMATH}\ensuremath{\Distance'}}}\endgroup }+({\begingroup\renewcommand\colorMATH{\colorMATHB}\renewcommand\colorSYNTAX{\colorSYNTAXB}{{\color{\colorMATH}\ensuremath{\Distance'}}}\endgroup }\mathord{\cdotp }{\begingroup\renewcommand\colorMATH{\colorMATHB}\renewcommand\colorSYNTAX{\colorSYNTAXB}{{\color{\colorMATH}\ensuremath{\sS_{1 1}}}}\endgroup })x)(\tau _{2}) = ({\begingroup\renewcommand\colorMATH{\colorMATHB}\renewcommand\colorSYNTAX{\colorSYNTAXB}{{\color{\colorMATH}\ensuremath{\Distance'}}}\endgroup }\mathord{\cdotp }{\begingroup\renewcommand\colorMATH{\colorMATHB}\renewcommand\colorSYNTAX{\colorSYNTAXB}{{\color{\colorMATH}\ensuremath{\sS_{1 1}}}}\endgroup })x({\begingroup\renewcommand\colorMATH{\colorMATHB}\renewcommand\colorSYNTAX{\colorSYNTAXB}{{\color{\colorMATH}\ensuremath{\Distance'}}}\endgroup }(\tau _{2}))}}}, and by Lemma~\ref{lm:distrinst} 
        {{\color{\colorMATH}\ensuremath{({\begingroup\renewcommand\colorMATH{\colorMATHB}\renewcommand\colorSYNTAX{\colorSYNTAXB}{{\color{\colorMATH}\ensuremath{\Distance'}}}\endgroup }\mathord{\cdotp }{\begingroup\renewcommand\colorMATH{\colorMATHB}\renewcommand\colorSYNTAX{\colorSYNTAXB}{{\color{\colorMATH}\ensuremath{\sS_{1 1}}}}\endgroup })x({\begingroup\renewcommand\colorMATH{\colorMATHB}\renewcommand\colorSYNTAX{\colorSYNTAXB}{{\color{\colorMATH}\ensuremath{\Distance'}}}\endgroup }(\tau _{2})) = {\begingroup\renewcommand\colorMATH{\colorMATHB}\renewcommand\colorSYNTAX{\colorSYNTAXB}{{\color{\colorMATH}\ensuremath{\Distance'}}}\endgroup }([{\begingroup\renewcommand\colorMATH{\colorMATHB}\renewcommand\colorSYNTAX{\colorSYNTAXB}{{\color{\colorMATH}\ensuremath{\sS_{1 1}}}}\endgroup }/x]\tau _{2})}}}.
       
        Then by weakening lemma~~\ref{lm:lrweakening-sensitivity} we know that\\
        {{\color{\colorMATH}\ensuremath{(\gamma _{1}[x \mapsto  {\begingroup\renewcommand\colorMATH{\colorMATHB}\renewcommand\colorSYNTAX{\colorSYNTAXB}{{\color{\colorMATH}\ensuremath{\sv'_{1 2}}}}\endgroup }]\vdash {\begingroup\renewcommand\colorMATH{\colorMATHC}\renewcommand\colorSYNTAX{\colorSYNTAXC}{{\color{\colorMATH}\ensuremath{\pe_{2}}}}\endgroup },\gamma _{2}[x \mapsto  {\begingroup\renewcommand\colorMATH{\colorMATHB}\renewcommand\colorSYNTAX{\colorSYNTAXB}{{\color{\colorMATH}\ensuremath{\sv'_{2 2}}}}\endgroup }]\vdash {\begingroup\renewcommand\colorMATH{\colorMATHC}\renewcommand\colorSYNTAX{\colorSYNTAXC}{{\color{\colorMATH}\ensuremath{\pe_{2}}}}\endgroup }) \in  {\mathcal{E}}^{k-j_{1}}_{({\begingroup\renewcommand\colorMATH{\colorMATHB}\renewcommand\colorSYNTAX{\colorSYNTAXB}{{\color{\colorMATH}\ensuremath{\Distance'}}}\endgroup }+({\begingroup\renewcommand\colorMATH{\colorMATHB}\renewcommand\colorSYNTAX{\colorSYNTAXB}{{\color{\colorMATH}\ensuremath{\Distance'}}}\endgroup }\mathord{\cdotp }{\begingroup\renewcommand\colorMATH{\colorMATHB}\renewcommand\colorSYNTAX{\colorSYNTAXB}{{\color{\colorMATH}\ensuremath{\sS_{1 1}}}}\endgroup })x){\begingroup\renewcommand\colorMATH{\colorMATHC}\renewcommand\colorSYNTAX{\colorSYNTAXC}{{\color{\colorMATH}\ensuremath{\bigcdot}}}\endgroup }{\begingroup\renewcommand\colorMATH{\colorMATHC}\renewcommand\colorSYNTAX{\colorSYNTAXC}{{\color{\colorMATH}\ensuremath{\pS_{2}}}}\endgroup }}\llbracket {\begingroup\renewcommand\colorMATH{\colorMATHB}\renewcommand\colorSYNTAX{\colorSYNTAXB}{{\color{\colorMATH}\ensuremath{\Distance'}}}\endgroup }([{\begingroup\renewcommand\colorMATH{\colorMATHB}\renewcommand\colorSYNTAX{\colorSYNTAXB}{{\color{\colorMATH}\ensuremath{\sS_{1 1}}}}\endgroup }/x]\tau _{2})\rrbracket }}}.
        Notice that by Lemma~\ref{lm:joinmeetred}, {{\color{\colorMATH}\ensuremath{{\begingroup\renewcommand\colorMATH{\colorMATHC}\renewcommand\colorSYNTAX{\colorSYNTAXC}{{\color{\colorMATH}\ensuremath{\pS_{2}}}}\endgroup } <: {\begingroup\renewcommand\colorMATH{\colorMATHC}\renewcommand\colorSYNTAX{\colorSYNTAXC}{{\color{\colorMATH}\ensuremath{\pS_{2}}}}\endgroup } \sqcup  {\begingroup\renewcommand\colorMATH{\colorMATHC}\renewcommand\colorSYNTAX{\colorSYNTAXC}{{\color{\colorMATH}\ensuremath{\pS_{3}}}}\endgroup }}}}, and then by Lemma~\ref{lm:dot-subtp}
        {{\color{\colorMATH}\ensuremath{({\begingroup\renewcommand\colorMATH{\colorMATHB}\renewcommand\colorSYNTAX{\colorSYNTAXB}{{\color{\colorMATH}\ensuremath{\Distance'}}}\endgroup }+({\begingroup\renewcommand\colorMATH{\colorMATHB}\renewcommand\colorSYNTAX{\colorSYNTAXB}{{\color{\colorMATH}\ensuremath{\Distance'}}}\endgroup }\mathord{\cdotp }{\begingroup\renewcommand\colorMATH{\colorMATHB}\renewcommand\colorSYNTAX{\colorSYNTAXB}{{\color{\colorMATH}\ensuremath{\sS_{1 1}}}}\endgroup })x){\begingroup\renewcommand\colorMATH{\colorMATHC}\renewcommand\colorSYNTAX{\colorSYNTAXC}{{\color{\colorMATH}\ensuremath{\bigcdot}}}\endgroup }{\begingroup\renewcommand\colorMATH{\colorMATHC}\renewcommand\colorSYNTAX{\colorSYNTAXC}{{\color{\colorMATH}\ensuremath{\pS_{2}}}}\endgroup } <: ({\begingroup\renewcommand\colorMATH{\colorMATHB}\renewcommand\colorSYNTAX{\colorSYNTAXB}{{\color{\colorMATH}\ensuremath{\Distance'}}}\endgroup }+({\begingroup\renewcommand\colorMATH{\colorMATHB}\renewcommand\colorSYNTAX{\colorSYNTAXB}{{\color{\colorMATH}\ensuremath{\Distance'}}}\endgroup }\mathord{\cdotp }{\begingroup\renewcommand\colorMATH{\colorMATHB}\renewcommand\colorSYNTAX{\colorSYNTAXB}{{\color{\colorMATH}\ensuremath{\sS_{1 1}}}}\endgroup })x){\begingroup\renewcommand\colorMATH{\colorMATHC}\renewcommand\colorSYNTAX{\colorSYNTAXC}{{\color{\colorMATH}\ensuremath{\bigcdot}}}\endgroup }{\begingroup\renewcommand\colorMATH{\colorMATHC}\renewcommand\colorSYNTAX{\colorSYNTAXC}{{\color{\colorMATH}\ensuremath{\pS_{2}}}}\endgroup } \sqcup  ({\begingroup\renewcommand\colorMATH{\colorMATHB}\renewcommand\colorSYNTAX{\colorSYNTAXB}{{\color{\colorMATH}\ensuremath{\Distance'}}}\endgroup }+({\begingroup\renewcommand\colorMATH{\colorMATHB}\renewcommand\colorSYNTAX{\colorSYNTAXB}{{\color{\colorMATH}\ensuremath{\Distance'}}}\endgroup }\mathord{\cdotp }{\begingroup\renewcommand\colorMATH{\colorMATHB}\renewcommand\colorSYNTAX{\colorSYNTAXB}{{\color{\colorMATH}\ensuremath{\sS_{1 2}}}}\endgroup })x){\begingroup\renewcommand\colorMATH{\colorMATHC}\renewcommand\colorSYNTAX{\colorSYNTAXC}{{\color{\colorMATH}\ensuremath{\bigcdot}}}\endgroup }{\begingroup\renewcommand\colorMATH{\colorMATHC}\renewcommand\colorSYNTAX{\colorSYNTAXC}{{\color{\colorMATH}\ensuremath{\pS_{3}}}}\endgroup }}}}. Also by Lemma~\ref{lm:joinmeetprops} 
        {{\color{\colorMATH}\ensuremath{[{\begingroup\renewcommand\colorMATH{\colorMATHB}\renewcommand\colorSYNTAX{\colorSYNTAXB}{{\color{\colorMATH}\ensuremath{\sS_{1 1}}}}\endgroup }/x]\tau _{2} <: [{\begingroup\renewcommand\colorMATH{\colorMATHB}\renewcommand\colorSYNTAX{\colorSYNTAXB}{{\color{\colorMATH}\ensuremath{\sS_{1 1}}}}\endgroup }/x]\tau _{2} \sqcup  [{\begingroup\renewcommand\colorMATH{\colorMATHB}\renewcommand\colorSYNTAX{\colorSYNTAXB}{{\color{\colorMATH}\ensuremath{\sS_{1 2}}}}\endgroup }/y]\tau _{3}}}}, therefore by Lemma~\ref{lm:subtypinginst},
        {{\color{\colorMATH}\ensuremath{{\begingroup\renewcommand\colorMATH{\colorMATHB}\renewcommand\colorSYNTAX{\colorSYNTAXB}{{\color{\colorMATH}\ensuremath{\Distance'}}}\endgroup }([{\begingroup\renewcommand\colorMATH{\colorMATHB}\renewcommand\colorSYNTAX{\colorSYNTAXB}{{\color{\colorMATH}\ensuremath{\sS_{1 1}}}}\endgroup }/x]\tau _{2}) <: {\begingroup\renewcommand\colorMATH{\colorMATHB}\renewcommand\colorSYNTAX{\colorSYNTAXB}{{\color{\colorMATH}\ensuremath{\Distance'}}}\endgroup }([{\begingroup\renewcommand\colorMATH{\colorMATHB}\renewcommand\colorSYNTAX{\colorSYNTAXB}{{\color{\colorMATH}\ensuremath{\sS_{1 1}}}}\endgroup }/x]\tau _{2} \sqcup  [{\begingroup\renewcommand\colorMATH{\colorMATHB}\renewcommand\colorSYNTAX{\colorSYNTAXB}{{\color{\colorMATH}\ensuremath{\sS_{1 2}}}}\endgroup }/y]\tau _{3})}}}.
        Once again by 
        Lemma~\ref{lm:lrweakening-sensitivity},
        {{\color{\colorMATH}\ensuremath{(\gamma _{1}[x \mapsto  {\begingroup\renewcommand\colorMATH{\colorMATHB}\renewcommand\colorSYNTAX{\colorSYNTAXB}{{\color{\colorMATH}\ensuremath{\sv'_{1 2}}}}\endgroup }]\vdash {\begingroup\renewcommand\colorMATH{\colorMATHC}\renewcommand\colorSYNTAX{\colorSYNTAXC}{{\color{\colorMATH}\ensuremath{\pe_{2}}}}\endgroup },\gamma _{2}[x \mapsto  {\begingroup\renewcommand\colorMATH{\colorMATHB}\renewcommand\colorSYNTAX{\colorSYNTAXB}{{\color{\colorMATH}\ensuremath{\sv'_{2 2}}}}\endgroup }]\vdash {\begingroup\renewcommand\colorMATH{\colorMATHC}\renewcommand\colorSYNTAX{\colorSYNTAXC}{{\color{\colorMATH}\ensuremath{\pe_{2}}}}\endgroup }) \in  {\mathcal{E}}_{{\begingroup\renewcommand\colorMATH{\colorMATHB}\renewcommand\colorSYNTAX{\colorSYNTAXB}{{\color{\colorMATH}\ensuremath{\Distance'}}}\endgroup }{\begingroup\renewcommand\colorMATH{\colorMATHC}\renewcommand\colorSYNTAX{\colorSYNTAXC}{{\color{\colorMATH}\ensuremath{\bigcdot}}}\endgroup }{\begingroup\renewcommand\colorMATH{\colorMATHC}\renewcommand\colorSYNTAX{\colorSYNTAXC}{{\color{\colorMATH}\ensuremath{\pS}}}\endgroup }}^{k-j_{1}}\llbracket {\begingroup\renewcommand\colorMATH{\colorMATHB}\renewcommand\colorSYNTAX{\colorSYNTAXB}{{\color{\colorMATH}\ensuremath{\Distance'}}}\endgroup }(\tau )\rrbracket }}} (3).
        The result follows by Lemma~\ref{lm:probsemanticrel}.

        %Notice that {{\color{\colorMATH}\ensuremath{\llbracket {{\color{\colorSYNTAX}\texttt{case}}}\hspace*{0.33em}{\begingroup\renewcommand\colorMATH{\colorMATHB}\renewcommand\colorSYNTAX{\colorSYNTAXB}{{\color{\colorMATH}\ensuremath{\se_{1}}}}\endgroup }\hspace*{0.33em}{{\color{\colorSYNTAX}\texttt{of}}}\hspace*{0.33em}\{ x\Rightarrow {\begingroup\renewcommand\colorMATH{\colorMATHC}\renewcommand\colorSYNTAX{\colorSYNTAXC}{{\color{\colorMATH}\ensuremath{\pe_{2}}}}\endgroup }\} \hspace*{0.33em}\{ y\Rightarrow {\begingroup\renewcommand\colorMATH{\colorMATHC}\renewcommand\colorSYNTAX{\colorSYNTAXC}{{\color{\colorMATH}\ensuremath{\pe_{3}}}}\endgroup }\} \rrbracket ^{k}_{\gamma _{i}} = \llbracket {\begingroup\renewcommand\colorMATH{\colorMATHC}\renewcommand\colorSYNTAX{\colorSYNTAXC}{{\color{\colorMATH}\ensuremath{\pe_{2}}}}\endgroup }\rrbracket ^{k-j_{1}}_{\gamma _{i}[x\mapsto {\begingroup\renewcommand\colorMATH{\colorMATHB}\renewcommand\colorSYNTAX{\colorSYNTAXB}{{\color{\colorMATH}\ensuremath{\sv'_{i 2}}}}\endgroup }]}}}}, then result follows by Lemma~\ref{lm:probsemanticrel}.
      
      \end{subproof}

    \newcommand{\thelift}{{\text{FP}}^\infty (\wideparen{x})} %({\begingroup\renewcommand\colorMATH{\colorMATHC}\renewcommand\colorSYNTAX{\colorSYNTAXC}{{\color{\colorMATH}\ensuremath{\rceil }}}\endgroup }{\begingroup\renewcommand\colorMATH{\colorMATHA}\renewcommand\colorSYNTAX{\colorSYNTAXA}{{\color{\colorMATH}\ensuremath{{\begingroup\renewcommand\colorMATH{\colorMATHB}\renewcommand\colorSYNTAX{\colorSYNTAXB}{{\color{\colorMATH}\ensuremath{\sS_{1}}}}\endgroup }}}}\endgroup }{\begingroup\renewcommand\colorMATH{\colorMATHC}\renewcommand\colorSYNTAX{\colorSYNTAXC}{{\color{\colorMATH}\ensuremath{\lceil ^{\infty }}}}\endgroup } + {\text{FP}}^\infty ({\begingroup\renewcommand\colorMATH{\colorMATHB}\renewcommand\colorSYNTAX{\colorSYNTAXB}{{\color{\colorMATH}\ensuremath{\se_{1}}}}\endgroup }))
    \item  {{\color{\colorMATH}\ensuremath{\Gamma  \mathrel{;} {\begingroup\renewcommand\colorMATH{\colorMATHB}\renewcommand\colorSYNTAX{\colorSYNTAXB}{{\color{\colorMATH}\ensuremath{\Distance}}}\endgroup } \vdash   {\begingroup\renewcommand\colorMATH{\colorMATHC}\renewcommand\colorSYNTAX{\colorSYNTAXC}{{\color{\colorSYNTAX}\texttt{return}}}\endgroup }\hspace*{0.33em}{\begingroup\renewcommand\colorMATH{\colorMATHB}\renewcommand\colorSYNTAX{\colorSYNTAXB}{{\color{\colorMATH}\ensuremath{\se_{1}}}}\endgroup } \mathrel{:} [\varnothing /\wideparen{x}]\tau  \mathrel{;}  {\text{FP}}^\infty (\wideparen{x})}}}, where {{\color{\colorMATH}\ensuremath{\wideparen{x} = {\text{FV}}({\begingroup\renewcommand\colorMATH{\colorMATHB}\renewcommand\colorSYNTAX{\colorSYNTAXB}{{\color{\colorMATH}\ensuremath{\se_{1}}}}\endgroup }) \cup  \dom({\begingroup\renewcommand\colorMATH{\colorMATHB}\renewcommand\colorSYNTAX{\colorSYNTAXB}{{\color{\colorMATH}\ensuremath{\sS_{1}}}}\endgroup })}}} %{\begingroup\renewcommand\colorMATH{\colorMATHC}\renewcommand\colorSYNTAX{\colorSYNTAXC}{{\color{\colorMATH}\ensuremath{\rceil {\begingroup\renewcommand\colorMATH{\colorMATHA}\renewcommand\colorSYNTAX{\colorSYNTAXA}{{\color{\colorMATH}\ensuremath{{\begingroup\renewcommand\colorMATH{\colorMATHB}\renewcommand\colorSYNTAX{\colorSYNTAXB}{{\color{\colorMATH}\ensuremath{\sS_{1}}}}\endgroup }}}}\endgroup }\lceil ^{\infty }}}}\endgroup } + {\text{FP}}^\infty (\wideparen{x})}}}, where {{\color{\colorMATH}\ensuremath{\wideparen{x} = {\text{FV}}({\begingroup\renewcommand\colorMATH{\colorMATHB}\renewcommand\colorSYNTAX{\colorSYNTAXB}{{\color{\colorMATH}\ensuremath{\se_{1}}}}\endgroup }) \cup  \dom({\begingroup\renewcommand\colorMATH{\colorMATHB}\renewcommand\colorSYNTAX{\colorSYNTAXB}{{\color{\colorMATH}\ensuremath{\sS_{1}}}}\endgroup })}}}.
      \begin{subproof} 
        We use notation {{\color{\colorMATH}\ensuremath{{\text{FP}}^\infty ({\begingroup\renewcommand\colorMATH{\colorMATHB}\renewcommand\colorSYNTAX{\colorSYNTAXB}{{\color{\colorMATH}\ensuremath{\se_{1}}}}\endgroup })}}} to stand for {{\color{\colorMATH}\ensuremath{{\text{FP}}^\infty (\wideparen{x})}}}.
        We have to prove that for any {{\color{\colorMATH}\ensuremath{k}}}, {{\color{\colorMATH}\ensuremath{\forall  (\gamma _{1},\gamma _{2}) \in  {\mathcal{G}}_{{\begingroup\renewcommand\colorMATH{\colorMATHB}\renewcommand\colorSYNTAX{\colorSYNTAXB}{{\color{\colorMATH}\ensuremath{\Distance'}}}\endgroup }}^{\kg}\llbracket \Gamma \rrbracket }}} and {{\color{\colorMATH}\ensuremath{{\begingroup\renewcommand\colorMATH{\colorMATHB}\renewcommand\colorSYNTAX{\colorSYNTAXB}{{\color{\colorMATH}\ensuremath{\Distance'}}}\endgroup } \sqsubseteq  {\begingroup\renewcommand\colorMATH{\colorMATHB}\renewcommand\colorSYNTAX{\colorSYNTAXB}{{\color{\colorMATH}\ensuremath{\Distance}}}\endgroup }}}},
        {{\color{\colorMATH}\ensuremath{(\gamma _{1}\vdash  {\begingroup\renewcommand\colorMATH{\colorMATHC}\renewcommand\colorSYNTAX{\colorSYNTAXC}{{\color{\colorSYNTAX}\texttt{return}}}\endgroup }\hspace*{0.33em}{\begingroup\renewcommand\colorMATH{\colorMATHB}\renewcommand\colorSYNTAX{\colorSYNTAXB}{{\color{\colorMATH}\ensuremath{\se_{1}}}}\endgroup },\gamma _{2}\vdash  {\begingroup\renewcommand\colorMATH{\colorMATHC}\renewcommand\colorSYNTAX{\colorSYNTAXC}{{\color{\colorSYNTAX}\texttt{return}}}\endgroup }\hspace*{0.33em}{\begingroup\renewcommand\colorMATH{\colorMATHB}\renewcommand\colorSYNTAX{\colorSYNTAXB}{{\color{\colorMATH}\ensuremath{\se_{1}}}}\endgroup }) \in  {\mathcal{E}}^{k}_{{\begingroup\renewcommand\colorMATH{\colorMATHB}\renewcommand\colorSYNTAX{\colorSYNTAXB}{{\color{\colorMATH}\ensuremath{\Distance'}}}\endgroup } {\begingroup\renewcommand\colorMATH{\colorMATHC}\renewcommand\colorSYNTAX{\colorSYNTAXC}{{\color{\colorMATH}\ensuremath{\bigcdot}}}\endgroup } ({\begingroup\renewcommand\colorMATH{\colorMATHC}\renewcommand\colorSYNTAX{\colorSYNTAXC}{{\color{\colorMATH}\ensuremath{\rceil }}}\endgroup }{\begingroup\renewcommand\colorMATH{\colorMATHA}\renewcommand\colorSYNTAX{\colorSYNTAXA}{{\color{\colorMATH}\ensuremath{{\begingroup\renewcommand\colorMATH{\colorMATHB}\renewcommand\colorSYNTAX{\colorSYNTAXB}{{\color{\colorMATH}\ensuremath{\sS_{1}}}}\endgroup }}}}\endgroup }{\begingroup\renewcommand\colorMATH{\colorMATHC}\renewcommand\colorSYNTAX{\colorSYNTAXC}{{\color{\colorMATH}\ensuremath{\lceil ^{\infty }}}}\endgroup }+ {\text{FP}}^\infty ({\begingroup\renewcommand\colorMATH{\colorMATHB}\renewcommand\colorSYNTAX{\colorSYNTAXB}{{\color{\colorMATH}\ensuremath{\se_{1}}}}\endgroup }))}\llbracket {\begingroup\renewcommand\colorMATH{\colorMATHB}\renewcommand\colorSYNTAX{\colorSYNTAXB}{{\color{\colorMATH}\ensuremath{\Distance'}}}\endgroup }([\varnothing /\wideparen{x}]\tau )\rrbracket }}}.

        Note that {{\color{\colorMATH}\ensuremath{{\begingroup\renewcommand\colorMATH{\colorMATHB}\renewcommand\colorSYNTAX{\colorSYNTAXB}{{\color{\colorMATH}\ensuremath{\Distance'}}}\endgroup } {\begingroup\renewcommand\colorMATH{\colorMATHC}\renewcommand\colorSYNTAX{\colorSYNTAXC}{{\color{\colorMATH}\ensuremath{\bigcdot}}}\endgroup } \thelift}}} is either {{\color{\colorMATH}\ensuremath{{\begingroup\renewcommand\colorMATH{\colorMATHB}\renewcommand\colorSYNTAX{\colorSYNTAXB}{{\color{\colorMATH}\ensuremath{\infty }}}\endgroup }}}} or {\begingroup\renewcommand\colorMATH{\colorMATHB}\renewcommand\colorSYNTAX{\colorSYNTAXB}{{\color{\colorMATH}\ensuremath{0}}}\endgroup }.
        If {{\color{\colorMATH}\ensuremath{{\begingroup\renewcommand\colorMATH{\colorMATHB}\renewcommand\colorSYNTAX{\colorSYNTAXB}{{\color{\colorMATH}\ensuremath{\Distance'}}}\endgroup } {\begingroup\renewcommand\colorMATH{\colorMATHC}\renewcommand\colorSYNTAX{\colorSYNTAXC}{{\color{\colorMATH}\ensuremath{\bigcdot}}}\endgroup } \thelift = {\begingroup\renewcommand\colorMATH{\colorMATHB}\renewcommand\colorSYNTAX{\colorSYNTAXB}{{\color{\colorMATH}\ensuremath{\infty }}}\endgroup }}}} then the result is direct by Lemma~\ref{lm:infrel}.
        Let us assume that {{\color{\colorMATH}\ensuremath{{\begingroup\renewcommand\colorMATH{\colorMATHB}\renewcommand\colorSYNTAX{\colorSYNTAXB}{{\color{\colorMATH}\ensuremath{\Distance'}}}\endgroup } {\begingroup\renewcommand\colorMATH{\colorMATHC}\renewcommand\colorSYNTAX{\colorSYNTAXC}{{\color{\colorMATH}\ensuremath{\bigcdot}}}\endgroup } \thelift =  \sum_{{\begingroup\renewcommand\colorMATH{\colorMATHC}\renewcommand\colorSYNTAX{\colorSYNTAXC}{{\color{\colorMATH}\ensuremath{p}}}\endgroup } \in  cod(\thelift)} {\begingroup\renewcommand\colorMATH{\colorMATHC}\renewcommand\colorSYNTAX{\colorSYNTAXC}{{\color{\colorMATH}\ensuremath{p}}}\endgroup } = {\begingroup\renewcommand\colorMATH{\colorMATHC}\renewcommand\colorSYNTAX{\colorSYNTAXC}{{\color{\colorMATH}\ensuremath{(0,0)}}}\endgroup }}}}.

        By induction hypothesis on {{\color{\colorMATH}\ensuremath{\Gamma ; {\begingroup\renewcommand\colorMATH{\colorMATHB}\renewcommand\colorSYNTAX{\colorSYNTAXB}{{\color{\colorMATH}\ensuremath{\Distance}}}\endgroup } \vdash  {\begingroup\renewcommand\colorMATH{\colorMATHB}\renewcommand\colorSYNTAX{\colorSYNTAXB}{{\color{\colorMATH}\ensuremath{\se_{1}}}}\endgroup } \mathrel{:} \tau  \mathrel{;} {\begingroup\renewcommand\colorMATH{\colorMATHB}\renewcommand\colorSYNTAX{\colorSYNTAXB}{{\color{\colorMATH}\ensuremath{\sS_{1}}}}\endgroup }}}}, we know that
        {{\color{\colorMATH}\ensuremath{(\gamma _{1}\vdash {\begingroup\renewcommand\colorMATH{\colorMATHB}\renewcommand\colorSYNTAX{\colorSYNTAXB}{{\color{\colorMATH}\ensuremath{\se_{1}}}}\endgroup },\gamma _{2}\vdash {\begingroup\renewcommand\colorMATH{\colorMATHB}\renewcommand\colorSYNTAX{\colorSYNTAXB}{{\color{\colorMATH}\ensuremath{\se_{1}}}}\endgroup }) \in  {\mathcal{E}}_{{\begingroup\renewcommand\colorMATH{\colorMATHB}\renewcommand\colorSYNTAX{\colorSYNTAXB}{{\color{\colorMATH}\ensuremath{\Distance'}}}\endgroup }\mathord{\cdotp }{\begingroup\renewcommand\colorMATH{\colorMATHB}\renewcommand\colorSYNTAX{\colorSYNTAXB}{{\color{\colorMATH}\ensuremath{\sS_{1}}}}\endgroup }}\llbracket {\begingroup\renewcommand\colorMATH{\colorMATHB}\renewcommand\colorSYNTAX{\colorSYNTAXB}{{\color{\colorMATH}\ensuremath{\Distance'}}}\endgroup }(\tau )\rrbracket }}}, i.e. if {{\color{\colorMATH}\ensuremath{\gamma _{1}\vdash {\begingroup\renewcommand\colorMATH{\colorMATHB}\renewcommand\colorSYNTAX{\colorSYNTAXB}{{\color{\colorMATH}\ensuremath{\se_{1}}}}\endgroup } \Downarrow ^{j} {\begingroup\renewcommand\colorMATH{\colorMATHB}\renewcommand\colorSYNTAX{\colorSYNTAXB}{{\color{\colorMATH}\ensuremath{\sv_{1}}}}\endgroup }}}}, \pthen {{\color{\colorMATH}\ensuremath{\gamma _{2}\vdash {\begingroup\renewcommand\colorMATH{\colorMATHB}\renewcommand\colorSYNTAX{\colorSYNTAXB}{{\color{\colorMATH}\ensuremath{\se_{1}}}}\endgroup } \Downarrow ^{\pj} {\begingroup\renewcommand\colorMATH{\colorMATHB}\renewcommand\colorSYNTAX{\colorSYNTAXB}{{\color{\colorMATH}\ensuremath{\sv_{2}}}}\endgroup }}}} \pand 
        {{\color{\colorMATH}\ensuremath{({\begingroup\renewcommand\colorMATH{\colorMATHB}\renewcommand\colorSYNTAX{\colorSYNTAXB}{{\color{\colorMATH}\ensuremath{\sv_{1}}}}\endgroup }, {\begingroup\renewcommand\colorMATH{\colorMATHB}\renewcommand\colorSYNTAX{\colorSYNTAXB}{{\color{\colorMATH}\ensuremath{\sv_{2}}}}\endgroup }) \in  {\mathcal{V}}^{k-j}_{{\begingroup\renewcommand\colorMATH{\colorMATHB}\renewcommand\colorSYNTAX{\colorSYNTAXB}{{\color{\colorMATH}\ensuremath{\Distance'}}}\endgroup }\mathord{\cdotp }{\begingroup\renewcommand\colorMATH{\colorMATHB}\renewcommand\colorSYNTAX{\colorSYNTAXB}{{\color{\colorMATH}\ensuremath{\sS_{1}}}}\endgroup }}\llbracket {\begingroup\renewcommand\colorMATH{\colorMATHB}\renewcommand\colorSYNTAX{\colorSYNTAXB}{{\color{\colorMATH}\ensuremath{\Distance'}}}\endgroup }(\tau )\rrbracket }}}.
        %{{\color{\colorMATH}\ensuremath{{\begingroup\renewcommand\colorMATH{\colorMATHB}\renewcommand\colorSYNTAX{\colorSYNTAXB}{{\color{\colorMATH}\ensuremath{\Distance'}}}\endgroup } \vdash  {\begingroup\renewcommand\colorMATH{\colorMATHB}\renewcommand\colorSYNTAX{\colorSYNTAXB}{{\color{\colorMATH}\ensuremath{\sv_{1}}}}\endgroup } \approx ^{k-j} {\begingroup\renewcommand\colorMATH{\colorMATHB}\renewcommand\colorSYNTAX{\colorSYNTAXB}{{\color{\colorMATH}\ensuremath{\sv_{2}}}}\endgroup } : \tau ; {\begingroup\renewcommand\colorMATH{\colorMATHB}\renewcommand\colorSYNTAX{\colorSYNTAXB}{{\color{\colorMATH}\ensuremath{\sS_{1}}}}\endgroup }}}}.
        Notice that {{\color{\colorMATH}\ensuremath{{\begingroup\renewcommand\colorMATH{\colorMATHB}\renewcommand\colorSYNTAX{\colorSYNTAXB}{{\color{\colorMATH}\ensuremath{\rceil }}}\endgroup }{\begingroup\renewcommand\colorMATH{\colorMATHB}\renewcommand\colorSYNTAX{\colorSYNTAXB}{{\color{\colorMATH}\ensuremath{\sS_{1}}}}\endgroup }{\begingroup\renewcommand\colorMATH{\colorMATHB}\renewcommand\colorSYNTAX{\colorSYNTAXB}{{\color{\colorMATH}\ensuremath{\lceil ^{\infty }}}}\endgroup } = 0}}} and {{\color{\colorMATH}\ensuremath{{\text{FP}}^\infty ({\begingroup\renewcommand\colorMATH{\colorMATHB}\renewcommand\colorSYNTAX{\colorSYNTAXB}{{\color{\colorMATH}\ensuremath{\se_{1}}}}\endgroup }) = 0}}}, then {{\color{\colorMATH}\ensuremath{ {\text{FV}}(e) = \varnothing  }}}. 
        This means that by Lemma~\ref{lm:novarequal} {{\color{\colorMATH}\ensuremath{{\begingroup\renewcommand\colorMATH{\colorMATHB}\renewcommand\colorSYNTAX{\colorSYNTAXB}{{\color{\colorMATH}\ensuremath{\sv_{1}}}}\endgroup } = {\begingroup\renewcommand\colorMATH{\colorMATHB}\renewcommand\colorSYNTAX{\colorSYNTAXB}{{\color{\colorMATH}\ensuremath{\sv_{2}}}}\endgroup }}}}.

        Let {{\color{\colorMATH}\ensuremath{\dist[1] = \begin{array}[t]{l
                                   } \lambda  x .\hspace*{0.33em}\left\{ \begin{array}{l@{\hspace*{1.00em}}c@{\hspace*{1.00em}}l
                                               } 1 &{}{\textit{when}}{}& x = {\begingroup\renewcommand\colorMATH{\colorMATHB}\renewcommand\colorSYNTAX{\colorSYNTAXB}{{\color{\colorMATH}\ensuremath{\sv_{1}}}}\endgroup }
                                               \cr  0 &{}{\textit{otherwise}}{}&
                                               \end{array}\right.
                                  \end{array}}}} and 
        {{\color{\colorMATH}\ensuremath{\dist[2] = \begin{array}[t]{l
                                   } \lambda  x .\hspace*{0.33em}\left\{ \begin{array}{l@{\hspace*{1.00em}}c@{\hspace*{1.00em}}l
                                               } 1 &{}{\textit{when}}{}& x = {\begingroup\renewcommand\colorMATH{\colorMATHB}\renewcommand\colorSYNTAX{\colorSYNTAXB}{{\color{\colorMATH}\ensuremath{\sv_{2}}}}\endgroup }
                                               \cr  0 &{}{\textit{otherwise}}{}&
                                               \end{array}\right.
                                  \end{array}}}}.
        We have to prove that {{\color{\colorMATH}\ensuremath{\dist[1](S) \leq  \dist[2](S)}}}. 
        If {{\color{\colorMATH}\ensuremath{{\begingroup\renewcommand\colorMATH{\colorMATHB}\renewcommand\colorSYNTAX{\colorSYNTAXB}{{\color{\colorMATH}\ensuremath{\sv_{1}}}}\endgroup } \notin  S}}} then the result is trivial as {{\color{\colorMATH}\ensuremath{\dist[1](S) = 0}}} and {{\color{\colorMATH}\ensuremath{0 \leq  \dist[2](S)}}} .
        Let us suppose that {{\color{\colorMATH}\ensuremath{{\begingroup\renewcommand\colorMATH{\colorMATHB}\renewcommand\colorSYNTAX{\colorSYNTAXB}{{\color{\colorMATH}\ensuremath{\sv_{1}}}}\endgroup } \in  S}}}, then as {{\color{\colorMATH}\ensuremath{{\begingroup\renewcommand\colorMATH{\colorMATHB}\renewcommand\colorSYNTAX{\colorSYNTAXB}{{\color{\colorMATH}\ensuremath{\sv_{1}}}}\endgroup }={\begingroup\renewcommand\colorMATH{\colorMATHB}\renewcommand\colorSYNTAX{\colorSYNTAXB}{{\color{\colorMATH}\ensuremath{\sv_{2}}}}\endgroup }}}}, we also know that {{\color{\colorMATH}\ensuremath{{\begingroup\renewcommand\colorMATH{\colorMATHB}\renewcommand\colorSYNTAX{\colorSYNTAXB}{{\color{\colorMATH}\ensuremath{\sv_{2}}}}\endgroup } \in  S}}}, thus
        {{\color{\colorMATH}\ensuremath{\dist[1](S) = 1}}}, and
        {{\color{\colorMATH}\ensuremath{\dist[2](S) = 1}}}, so the result holds.

      \end{subproof}

    \newcommand{\thej}{k}
    \newcommand{\thejtwo}{k-k'}
    \item  {{\color{\colorMATH}\ensuremath{\Gamma  \mathrel{;} {\begingroup\renewcommand\colorMATH{\colorMATHB}\renewcommand\colorSYNTAX{\colorSYNTAXB}{{\color{\colorMATH}\ensuremath{\Distance}}}\endgroup } \vdash  x: \tau _{1} {\begingroup\renewcommand\colorMATH{\colorMATHC}\renewcommand\colorSYNTAX{\colorSYNTAXC}{{\color{\colorMATH}\ensuremath{\leftarrow }}}\endgroup }{\begingroup\renewcommand\colorMATH{\colorMATHC}\renewcommand\colorSYNTAX{\colorSYNTAXC}{{\color{\colorMATH}\ensuremath{\pe_{1}}}}\endgroup }\mathrel{;}{\begingroup\renewcommand\colorMATH{\colorMATHC}\renewcommand\colorSYNTAX{\colorSYNTAXC}{{\color{\colorMATH}\ensuremath{\pe_{2}}}}\endgroup } \mathrel{:} [\varnothing /x]\tau _{2} \mathrel{;} {\begingroup\renewcommand\colorMATH{\colorMATHC}\renewcommand\colorSYNTAX{\colorSYNTAXC}{{\color{\colorMATH}\ensuremath{\pS_{1}}}}\endgroup } + {\begingroup\renewcommand\colorMATH{\colorMATHC}\renewcommand\colorSYNTAX{\colorSYNTAXC}{{\color{\colorMATH}\ensuremath{\pS_{2}}}}\endgroup }}}}
      \begin{subproof} 
        
        We have to prove that for any {{\color{\colorMATH}\ensuremath{k}}}, {{\color{\colorMATH}\ensuremath{\forall  (\gamma _{1},\gamma _{2}) \in  {\mathcal{G}}_{{\begingroup\renewcommand\colorMATH{\colorMATHB}\renewcommand\colorSYNTAX{\colorSYNTAXB}{{\color{\colorMATH}\ensuremath{\Distance'}}}\endgroup }}^{\kg}\llbracket \Gamma \rrbracket }}}, for {{\color{\colorMATH}\ensuremath{{\begingroup\renewcommand\colorMATH{\colorMATHB}\renewcommand\colorSYNTAX{\colorSYNTAXB}{{\color{\colorMATH}\ensuremath{\Distance'}}}\endgroup } \sqsubseteq  {\begingroup\renewcommand\colorMATH{\colorMATHB}\renewcommand\colorSYNTAX{\colorSYNTAXB}{{\color{\colorMATH}\ensuremath{\Distance}}}\endgroup }}}} it holds that\\ 
          {{\color{\colorMATH}\ensuremath{(\gamma _{1}\vdash x: \tau _{1} {\begingroup\renewcommand\colorMATH{\colorMATHC}\renewcommand\colorSYNTAX{\colorSYNTAXC}{{\color{\colorMATH}\ensuremath{\leftarrow }}}\endgroup }{\begingroup\renewcommand\colorMATH{\colorMATHC}\renewcommand\colorSYNTAX{\colorSYNTAXC}{{\color{\colorMATH}\ensuremath{\pe_{1}}}}\endgroup }\mathrel{;}{\begingroup\renewcommand\colorMATH{\colorMATHC}\renewcommand\colorSYNTAX{\colorSYNTAXC}{{\color{\colorMATH}\ensuremath{\pe_{2}}}}\endgroup },\gamma _{2}\vdash x: \tau _{1} {\begingroup\renewcommand\colorMATH{\colorMATHC}\renewcommand\colorSYNTAX{\colorSYNTAXC}{{\color{\colorMATH}\ensuremath{\leftarrow }}}\endgroup }{\begingroup\renewcommand\colorMATH{\colorMATHC}\renewcommand\colorSYNTAX{\colorSYNTAXC}{{\color{\colorMATH}\ensuremath{\pe_{1}}}}\endgroup }\mathrel{;}{\begingroup\renewcommand\colorMATH{\colorMATHC}\renewcommand\colorSYNTAX{\colorSYNTAXC}{{\color{\colorMATH}\ensuremath{\pe_{2}}}}\endgroup }) \in  {\mathcal{E}}^{k}_{{\begingroup\renewcommand\colorMATH{\colorMATHB}\renewcommand\colorSYNTAX{\colorSYNTAXB}{{\color{\colorMATH}\ensuremath{\Distance'}}}\endgroup }\mathord{\cdotp }({\begingroup\renewcommand\colorMATH{\colorMATHC}\renewcommand\colorSYNTAX{\colorSYNTAXC}{{\color{\colorMATH}\ensuremath{\pS_{1}}}}\endgroup }+[\varnothing /x]{\begingroup\renewcommand\colorMATH{\colorMATHC}\renewcommand\colorSYNTAX{\colorSYNTAXC}{{\color{\colorMATH}\ensuremath{\pS_{2}}}}\endgroup })}\llbracket {\begingroup\renewcommand\colorMATH{\colorMATHB}\renewcommand\colorSYNTAX{\colorSYNTAXB}{{\color{\colorMATH}\ensuremath{\Distance'}}}\endgroup }([\varnothing /x]\tau _{2})\rrbracket }}}.

        By induction hypothesis on {{\color{\colorMATH}\ensuremath{\Gamma  \mathrel{;} {\begingroup\renewcommand\colorMATH{\colorMATHB}\renewcommand\colorSYNTAX{\colorSYNTAXB}{{\color{\colorMATH}\ensuremath{\Distance}}}\endgroup } \vdash  {\begingroup\renewcommand\colorMATH{\colorMATHC}\renewcommand\colorSYNTAX{\colorSYNTAXC}{{\color{\colorMATH}\ensuremath{\pe_{1}}}}\endgroup } \mathrel{:} \tau _{1} \mathrel{;} {\begingroup\renewcommand\colorMATH{\colorMATHC}\renewcommand\colorSYNTAX{\colorSYNTAXC}{{\color{\colorMATH}\ensuremath{\pS_{1}}}}\endgroup }}}}, we know that\\
        {{\color{\colorMATH}\ensuremath{(\gamma _{1}\vdash {\begingroup\renewcommand\colorMATH{\colorMATHC}\renewcommand\colorSYNTAX{\colorSYNTAXC}{{\color{\colorMATH}\ensuremath{\pe_{1}}}}\endgroup },\gamma _{2}\vdash {\begingroup\renewcommand\colorMATH{\colorMATHC}\renewcommand\colorSYNTAX{\colorSYNTAXC}{{\color{\colorMATH}\ensuremath{\pe_{1}}}}\endgroup }) \in  {\mathcal{E}}^{k}_{{\begingroup\renewcommand\colorMATH{\colorMATHB}\renewcommand\colorSYNTAX{\colorSYNTAXB}{{\color{\colorMATH}\ensuremath{\Distance'}}}\endgroup }\mathord{\cdotp }{\begingroup\renewcommand\colorMATH{\colorMATHC}\renewcommand\colorSYNTAX{\colorSYNTAXC}{{\color{\colorMATH}\ensuremath{\pS_{1}}}}\endgroup }}\llbracket {\begingroup\renewcommand\colorMATH{\colorMATHB}\renewcommand\colorSYNTAX{\colorSYNTAXB}{{\color{\colorMATH}\ensuremath{\Distance'}}}\endgroup }(\tau _{1})\rrbracket }}},
        i.e. let {{\color{\colorMATH}\ensuremath{({\begingroup\renewcommand\colorMATH{\colorMATHC}\renewcommand\colorSYNTAX{\colorSYNTAXC}{{\color{\colorMATH}\ensuremath{\epsilon _{1}}}}\endgroup }, {\begingroup\renewcommand\colorMATH{\colorMATHC}\renewcommand\colorSYNTAX{\colorSYNTAXC}{{\color{\colorMATH}\ensuremath{\delta _{1}}}}\endgroup }) = {\begingroup\renewcommand\colorMATH{\colorMATHB}\renewcommand\colorSYNTAX{\colorSYNTAXB}{{\color{\colorMATH}\ensuremath{\Distance'}}}\endgroup }\mathord{\cdotp }{\begingroup\renewcommand\colorMATH{\colorMATHC}\renewcommand\colorSYNTAX{\colorSYNTAXC}{{\color{\colorMATH}\ensuremath{\pS_{1}}}}\endgroup }}}}, {{\color{\colorMATH}\ensuremath{\forall  S \subseteq  val(*)}}}, for {{\color{\colorMATH}\ensuremath{k' < k}}},
        if  {{\color{\colorMATH}\ensuremath{\gamma _{1} \vdash  {\begingroup\renewcommand\colorMATH{\colorMATHC}\renewcommand\colorSYNTAX{\colorSYNTAXC}{{\color{\colorMATH}\ensuremath{\pe_{1}}}}\endgroup } \Downarrow ^{k'} \dist[1]}}} \pthen  {{\color{\colorMATH}\ensuremath{\gamma _{2} \vdash  {\begingroup\renewcommand\colorMATH{\colorMATHC}\renewcommand\colorSYNTAX{\colorSYNTAXC}{{\color{\colorMATH}\ensuremath{\pe_{1}}}}\endgroup } \Downarrow ^{\pj[1]} \dist[2]}}}, \pand
        {{\color{\colorMATH}\ensuremath{\dist[1](S) \leq  e^{{\begingroup\renewcommand\colorMATH{\colorMATHC}\renewcommand\colorSYNTAX{\colorSYNTAXC}{{\color{\colorMATH}\ensuremath{\epsilon _{1}}}}\endgroup }}\dist[2](S) + {\begingroup\renewcommand\colorMATH{\colorMATHC}\renewcommand\colorSYNTAX{\colorSYNTAXC}{{\color{\colorMATH}\ensuremath{\delta _{1}}}}\endgroup }}}} (notice {{\color{\colorMATH}\ensuremath{k'>0}}}).

        Similarly, by induction hypothesis on {{\color{\colorMATH}\ensuremath{\Gamma , x: \tau _{1} \mathrel{;} {\begingroup\renewcommand\colorMATH{\colorMATHB}\renewcommand\colorSYNTAX{\colorSYNTAXB}{{\color{\colorMATH}\ensuremath{\Distance}}}\endgroup } + {\begingroup\renewcommand\colorMATH{\colorMATHB}\renewcommand\colorSYNTAX{\colorSYNTAXB}{{\color{\colorMATH}\ensuremath{0}}}\endgroup }x \vdash  {\begingroup\renewcommand\colorMATH{\colorMATHC}\renewcommand\colorSYNTAX{\colorSYNTAXC}{{\color{\colorMATH}\ensuremath{\pe_{2}}}}\endgroup } \mathrel{:} \tau _{2} \mathrel{;} {\begingroup\renewcommand\colorMATH{\colorMATHC}\renewcommand\colorSYNTAX{\colorSYNTAXC}{{\color{\colorMATH}\ensuremath{\pS_{2}}}}\endgroup }}}}, for all {{\color{\colorMATH}\ensuremath{(\gamma '_{1j}, \gamma '_{2j}) \in  {\mathcal{G}}_{{\begingroup\renewcommand\colorMATH{\colorMATHB}\renewcommand\colorSYNTAX{\colorSYNTAXB}{{\color{\colorMATH}\ensuremath{\Distance'}}}\endgroup } + {\begingroup\renewcommand\colorMATH{\colorMATHB}\renewcommand\colorSYNTAX{\colorSYNTAXB}{{\color{\colorMATH}\ensuremath{0}}}\endgroup }x}^{k-k_j'}\llbracket \Gamma , x: \tau _{1}\rrbracket }}}, for some {{\color{\colorMATH}\ensuremath{k_j'' < k - k'}}} and {{\color{\colorMATH}\ensuremath{{\begingroup\renewcommand\colorMATH{\colorMATHB}\renewcommand\colorSYNTAX{\colorSYNTAXB}{{\color{\colorMATH}\ensuremath{\Distance'}}}\endgroup } + {\begingroup\renewcommand\colorMATH{\colorMATHB}\renewcommand\colorSYNTAX{\colorSYNTAXB}{{\color{\colorMATH}\ensuremath{0}}}\endgroup }x \sqsubseteq  {\begingroup\renewcommand\colorMATH{\colorMATHB}\renewcommand\colorSYNTAX{\colorSYNTAXB}{{\color{\colorMATH}\ensuremath{\Distance}}}\endgroup } + {\begingroup\renewcommand\colorMATH{\colorMATHB}\renewcommand\colorSYNTAX{\colorSYNTAXB}{{\color{\colorMATH}\ensuremath{0}}}\endgroup }x}}},  we know that
        {{\color{\colorMATH}\ensuremath{(\gamma '_{1j}\vdash {\begingroup\renewcommand\colorMATH{\colorMATHC}\renewcommand\colorSYNTAX{\colorSYNTAXC}{{\color{\colorMATH}\ensuremath{\pe_{2}}}}\endgroup },\gamma '_{2j}\vdash {\begingroup\renewcommand\colorMATH{\colorMATHC}\renewcommand\colorSYNTAX{\colorSYNTAXC}{{\color{\colorMATH}\ensuremath{\pe_{2}}}}\endgroup }) \in  {\mathcal{E}}^{k-k'}_{({\begingroup\renewcommand\colorMATH{\colorMATHB}\renewcommand\colorSYNTAX{\colorSYNTAXB}{{\color{\colorMATH}\ensuremath{\Distance'}}}\endgroup } + {\begingroup\renewcommand\colorMATH{\colorMATHB}\renewcommand\colorSYNTAX{\colorSYNTAXB}{{\color{\colorMATH}\ensuremath{0}}}\endgroup }x) {\begingroup\renewcommand\colorMATH{\colorMATHC}\renewcommand\colorSYNTAX{\colorSYNTAXC}{{\color{\colorMATH}\ensuremath{\bigcdot}}}\endgroup } {\begingroup\renewcommand\colorMATH{\colorMATHC}\renewcommand\colorSYNTAX{\colorSYNTAXC}{{\color{\colorMATH}\ensuremath{\pS_{2}}}}\endgroup }}\llbracket ({\begingroup\renewcommand\colorMATH{\colorMATHB}\renewcommand\colorSYNTAX{\colorSYNTAXB}{{\color{\colorMATH}\ensuremath{\Distance'}}}\endgroup } + {\begingroup\renewcommand\colorMATH{\colorMATHB}\renewcommand\colorSYNTAX{\colorSYNTAXB}{{\color{\colorMATH}\ensuremath{0}}}\endgroup }x)(\tau _{2})\rrbracket }}}.
        Let {{\color{\colorMATH}\ensuremath{k'' = \max(k_j'')}}}.
        Notice that by Lemma~\ref{lm:distrdotpp}, {{\color{\colorMATH}\ensuremath{({\begingroup\renewcommand\colorMATH{\colorMATHB}\renewcommand\colorSYNTAX{\colorSYNTAXB}{{\color{\colorMATH}\ensuremath{\Distance'}}}\endgroup } + {\begingroup\renewcommand\colorMATH{\colorMATHB}\renewcommand\colorSYNTAX{\colorSYNTAXB}{{\color{\colorMATH}\ensuremath{0}}}\endgroup }x)\mathord{\cdotp }{\begingroup\renewcommand\colorMATH{\colorMATHC}\renewcommand\colorSYNTAX{\colorSYNTAXC}{{\color{\colorMATH}\ensuremath{\pS_{2}}}}\endgroup } = {\begingroup\renewcommand\colorMATH{\colorMATHB}\renewcommand\colorSYNTAX{\colorSYNTAXB}{{\color{\colorMATH}\ensuremath{\Distance'}}}\endgroup }\mathord{\cdotp }([\varnothing /x]{\begingroup\renewcommand\colorMATH{\colorMATHC}\renewcommand\colorSYNTAX{\colorSYNTAXC}{{\color{\colorMATH}\ensuremath{\pS_{2}}}}\endgroup })}}} as {{\color{\colorMATH}\ensuremath{x \notin  dom({\begingroup\renewcommand\colorMATH{\colorMATHB}\renewcommand\colorSYNTAX{\colorSYNTAXB}{{\color{\colorMATH}\ensuremath{\Distance'}}}\endgroup })}}}, and that by Lemma~\ref{lm:distrdot}, {{\color{\colorMATH}\ensuremath{({\begingroup\renewcommand\colorMATH{\colorMATHB}\renewcommand\colorSYNTAX{\colorSYNTAXB}{{\color{\colorMATH}\ensuremath{\Distance'}}}\endgroup } + {\begingroup\renewcommand\colorMATH{\colorMATHB}\renewcommand\colorSYNTAX{\colorSYNTAXB}{{\color{\colorMATH}\ensuremath{0}}}\endgroup }x)(\tau _{2}) = {\begingroup\renewcommand\colorMATH{\colorMATHB}\renewcommand\colorSYNTAX{\colorSYNTAXB}{{\color{\colorMATH}\ensuremath{\Distance'}}}\endgroup }([\varnothing /x]\tau _{2})}}}.
        Then let {{\color{\colorMATH}\ensuremath{({\begingroup\renewcommand\colorMATH{\colorMATHC}\renewcommand\colorSYNTAX{\colorSYNTAXC}{{\color{\colorMATH}\ensuremath{\epsilon _{2}}}}\endgroup }, {\begingroup\renewcommand\colorMATH{\colorMATHC}\renewcommand\colorSYNTAX{\colorSYNTAXC}{{\color{\colorMATH}\ensuremath{\delta _{2}}}}\endgroup }) = {\begingroup\renewcommand\colorMATH{\colorMATHB}\renewcommand\colorSYNTAX{\colorSYNTAXB}{{\color{\colorMATH}\ensuremath{\Distance'}}}\endgroup }\mathord{\cdotp }[\varnothing /x]{\begingroup\renewcommand\colorMATH{\colorMATHC}\renewcommand\colorSYNTAX{\colorSYNTAXC}{{\color{\colorMATH}\ensuremath{\pS_{2}}}}\endgroup }}}},
        {{\color{\colorMATH}\ensuremath{\forall  S' \in  val}}}, 
        if  {{\color{\colorMATH}\ensuremath{\gamma '_{1j} \vdash  {\begingroup\renewcommand\colorMATH{\colorMATHC}\renewcommand\colorSYNTAX{\colorSYNTAXC}{{\color{\colorMATH}\ensuremath{\pe_{2}}}}\endgroup } \Downarrow ^{k''} \distp[1j]}}} \pthen  {{\color{\colorMATH}\ensuremath{\gamma '_{2j} \vdash  {\begingroup\renewcommand\colorMATH{\colorMATHC}\renewcommand\colorSYNTAX{\colorSYNTAXC}{{\color{\colorMATH}\ensuremath{\pe_{2}}}}\endgroup } \Downarrow ^{\pj[2]} \distp[2]}}}, \pand
        {{\color{\colorMATH}\ensuremath{\distp[1j](S) \leq  e^{{\begingroup\renewcommand\colorMATH{\colorMATHC}\renewcommand\colorSYNTAX{\colorSYNTAXC}{{\color{\colorMATH}\ensuremath{\epsilon _{2}}}}\endgroup }}\distp[2j](S) + {\begingroup\renewcommand\colorMATH{\colorMATHC}\renewcommand\colorSYNTAX{\colorSYNTAXC}{{\color{\colorMATH}\ensuremath{\delta _{2}}}}\endgroup }}}}.

        Let {{\color{\colorMATH}\ensuremath{ 
        dval = \{{\begingroup\renewcommand\colorMATH{\colorMATHB}\renewcommand\colorSYNTAX{\colorSYNTAXB}{{\color{\colorMATH}\ensuremath{\sv}}}\endgroup } \mid \dist[1]({\begingroup\renewcommand\colorMATH{\colorMATHB}\renewcommand\colorSYNTAX{\colorSYNTAXB}{{\color{\colorMATH}\ensuremath{\sv}}}\endgroup })>0\} 
        \cup  \{{\begingroup\renewcommand\colorMATH{\colorMATHB}\renewcommand\colorSYNTAX{\colorSYNTAXB}{{\color{\colorMATH}\ensuremath{\sv}}}\endgroup } \mid \dist[2]({\begingroup\renewcommand\colorMATH{\colorMATHB}\renewcommand\colorSYNTAX{\colorSYNTAXB}{{\color{\colorMATH}\ensuremath{\sv}}}\endgroup })>0\} 
        \subseteq  val(\tau _{1}/\Gamma )
        }}}.
        Let {{\color{\colorMATH}\ensuremath{\distpp[i] = \lambda  x. \sum_{{\begingroup\renewcommand\colorMATH{\colorMATHB}\renewcommand\colorSYNTAX{\colorSYNTAXB}{{\color{\colorMATH}\ensuremath{\sv}}}\endgroup }_{j} \in  dval} \dist[i]({\begingroup\renewcommand\colorMATH{\colorMATHB}\renewcommand\colorSYNTAX{\colorSYNTAXB}{{\color{\colorMATH}\ensuremath{\sv}}}\endgroup }_{j}) \times  \distp[ij](x)}}}.
        We have to prove then that {{\color{\colorMATH}\ensuremath{\forall  S' \in  val(*)}}}, let {{\color{\colorMATH}\ensuremath{({\begingroup\renewcommand\colorMATH{\colorMATHC}\renewcommand\colorSYNTAX{\colorSYNTAXC}{{\color{\colorMATH}\ensuremath{\epsilon _{1}}}}\endgroup } + {\begingroup\renewcommand\colorMATH{\colorMATHC}\renewcommand\colorSYNTAX{\colorSYNTAXC}{{\color{\colorMATH}\ensuremath{\epsilon _{2}}}}\endgroup }, {\begingroup\renewcommand\colorMATH{\colorMATHC}\renewcommand\colorSYNTAX{\colorSYNTAXC}{{\color{\colorMATH}\ensuremath{\delta _{1}}}}\endgroup } + {\begingroup\renewcommand\colorMATH{\colorMATHC}\renewcommand\colorSYNTAX{\colorSYNTAXC}{{\color{\colorMATH}\ensuremath{\delta _{2}}}}\endgroup }) = {\begingroup\renewcommand\colorMATH{\colorMATHB}\renewcommand\colorSYNTAX{\colorSYNTAXB}{{\color{\colorMATH}\ensuremath{\Distance'}}}\endgroup }\mathord{\cdotp }{\begingroup\renewcommand\colorMATH{\colorMATHC}\renewcommand\colorSYNTAX{\colorSYNTAXC}{{\color{\colorMATH}\ensuremath{\pS_{1}}}}\endgroup } + {\begingroup\renewcommand\colorMATH{\colorMATHB}\renewcommand\colorSYNTAX{\colorSYNTAXB}{{\color{\colorMATH}\ensuremath{\Distance'}}}\endgroup }\mathord{\cdotp }[\varnothing /x]{\begingroup\renewcommand\colorMATH{\colorMATHC}\renewcommand\colorSYNTAX{\colorSYNTAXC}{{\color{\colorMATH}\ensuremath{\pS_{2}}}}\endgroup } = {\begingroup\renewcommand\colorMATH{\colorMATHB}\renewcommand\colorSYNTAX{\colorSYNTAXB}{{\color{\colorMATH}\ensuremath{\Distance'}}}\endgroup }\mathord{\cdotp }({\begingroup\renewcommand\colorMATH{\colorMATHC}\renewcommand\colorSYNTAX{\colorSYNTAXC}{{\color{\colorMATH}\ensuremath{\pS_{1}}}}\endgroup } + [\varnothing /x]{\begingroup\renewcommand\colorMATH{\colorMATHC}\renewcommand\colorSYNTAX{\colorSYNTAXC}{{\color{\colorMATH}\ensuremath{\pS_{2}}}}\endgroup })}}}
        
        \begingroup\color{\colorMATH}\begin{gather*}
\begin{array}{rcl
          } \distpp[1] S' &{}\leq {}& e^{{\begingroup\renewcommand\colorMATH{\colorMATHC}\renewcommand\colorSYNTAX{\colorSYNTAXC}{{\color{\colorMATH}\ensuremath{\epsilon _{1}}}}\endgroup } + {\begingroup\renewcommand\colorMATH{\colorMATHC}\renewcommand\colorSYNTAX{\colorSYNTAXC}{{\color{\colorMATH}\ensuremath{\epsilon _{2}}}}\endgroup }} (\distpp[2] S') + {\begingroup\renewcommand\colorMATH{\colorMATHC}\renewcommand\colorSYNTAX{\colorSYNTAXC}{{\color{\colorMATH}\ensuremath{\delta _{1}}}}\endgroup } + {\begingroup\renewcommand\colorMATH{\colorMATHC}\renewcommand\colorSYNTAX{\colorSYNTAXC}{{\color{\colorMATH}\ensuremath{\delta _{2}}}}\endgroup }
          \cr  \sum_{{\begingroup\renewcommand\colorMATH{\colorMATHB}\renewcommand\colorSYNTAX{\colorSYNTAXB}{{\color{\colorMATH}\ensuremath{\sv_{1}}}}\endgroup } \in  S'} \distpp[1]({\begingroup\renewcommand\colorMATH{\colorMATHB}\renewcommand\colorSYNTAX{\colorSYNTAXB}{{\color{\colorMATH}\ensuremath{\sv_{1}}}}\endgroup }) &{}\leq {}& e^{{\begingroup\renewcommand\colorMATH{\colorMATHC}\renewcommand\colorSYNTAX{\colorSYNTAXC}{{\color{\colorMATH}\ensuremath{\epsilon _{1}}}}\endgroup } + {\begingroup\renewcommand\colorMATH{\colorMATHC}\renewcommand\colorSYNTAX{\colorSYNTAXC}{{\color{\colorMATH}\ensuremath{\epsilon _{2}}}}\endgroup }} (\sum_{{\begingroup\renewcommand\colorMATH{\colorMATHB}\renewcommand\colorSYNTAX{\colorSYNTAXB}{{\color{\colorMATH}\ensuremath{\sv_{2}}}}\endgroup } \in  S'} \distpp[2]({\begingroup\renewcommand\colorMATH{\colorMATHB}\renewcommand\colorSYNTAX{\colorSYNTAXB}{{\color{\colorMATH}\ensuremath{\sv_{2}}}}\endgroup })) + {\begingroup\renewcommand\colorMATH{\colorMATHC}\renewcommand\colorSYNTAX{\colorSYNTAXC}{{\color{\colorMATH}\ensuremath{\delta _{1}}}}\endgroup } + {\begingroup\renewcommand\colorMATH{\colorMATHC}\renewcommand\colorSYNTAX{\colorSYNTAXC}{{\color{\colorMATH}\ensuremath{\delta _{2}}}}\endgroup }
          \cr
\end{array}
        \end{gather*}\endgroup

        As {{\color{\colorMATH}\ensuremath{k-k' < k}}}, by induction hypothesis {{\color{\colorMATH}\ensuremath{({\begingroup\renewcommand\colorMATH{\colorMATHB}\renewcommand\colorSYNTAX{\colorSYNTAXB}{{\color{\colorMATH}\ensuremath{\sv}}}\endgroup }_j,{\begingroup\renewcommand\colorMATH{\colorMATHB}\renewcommand\colorSYNTAX{\colorSYNTAXB}{{\color{\colorMATH}\ensuremath{\sv}}}\endgroup }_j) \in  {\mathcal{V}}_{0}^{k-k'}\llbracket \tau _{1}/\Gamma \rrbracket }}} (notice that {{\color{\colorMATH}\ensuremath{k'>0}}} as values are not privacy expressions), but {{\color{\colorMATH}\ensuremath{\tau _{1}/\Gamma  = {\begingroup\renewcommand\colorMATH{\colorMATHB}\renewcommand\colorSYNTAX{\colorSYNTAXB}{{\color{\colorMATH}\ensuremath{\Distance}}}\endgroup }_\varnothing (\tau _{1})}}}, and by weakening lemmas ~\ref{lm:weakening-index}, and ~\ref{lm:lrweakening-sensitivity} {{\color{\colorMATH}\ensuremath{(\gamma _{1},\gamma _{2}) \in  {\mathcal{G}}_{{\begingroup\renewcommand\colorMATH{\colorMATHB}\renewcommand\colorSYNTAX{\colorSYNTAXB}{{\color{\colorMATH}\ensuremath{\Distance'}}}\endgroup }}^{k-k'}\llbracket \Gamma \rrbracket }}}, then we can choose {{\color{\colorMATH}\ensuremath{(\gamma '_{1j}, \gamma '_{2j}) = (\gamma _{1}[x \mapsto  {\begingroup\renewcommand\colorMATH{\colorMATHB}\renewcommand\colorSYNTAX{\colorSYNTAXB}{{\color{\colorMATH}\ensuremath{\sv}}}\endgroup }_j], \gamma _{2}[x \mapsto  {\begingroup\renewcommand\colorMATH{\colorMATHB}\renewcommand\colorSYNTAX{\colorSYNTAXB}{{\color{\colorMATH}\ensuremath{\sv}}}\endgroup }_j])}}} and know that
        that, for all {{\color{\colorMATH}\ensuremath{j}}}, {{\color{\colorMATH}\ensuremath{\sum_{{\begingroup\renewcommand\colorMATH{\colorMATHB}\renewcommand\colorSYNTAX{\colorSYNTAXB}{{\color{\colorMATH}\ensuremath{\sv_{1}}}}\endgroup } \in  S'}\distp[1j]({\begingroup\renewcommand\colorMATH{\colorMATHB}\renewcommand\colorSYNTAX{\colorSYNTAXB}{{\color{\colorMATH}\ensuremath{\sv_{1}}}}\endgroup }) \leq  e^{{\begingroup\renewcommand\colorMATH{\colorMATHC}\renewcommand\colorSYNTAX{\colorSYNTAXC}{{\color{\colorMATH}\ensuremath{\epsilon _{2}}}}\endgroup }}(\sum_{{\begingroup\renewcommand\colorMATH{\colorMATHB}\renewcommand\colorSYNTAX{\colorSYNTAXB}{{\color{\colorMATH}\ensuremath{\sv_{2}}}}\endgroup } \in  S'} \distp[2j]({\begingroup\renewcommand\colorMATH{\colorMATHB}\renewcommand\colorSYNTAX{\colorSYNTAXB}{{\color{\colorMATH}\ensuremath{\sv_{2}}}}\endgroup })) + {\begingroup\renewcommand\colorMATH{\colorMATHC}\renewcommand\colorSYNTAX{\colorSYNTAXC}{{\color{\colorMATH}\ensuremath{\delta _{2}}}}\endgroup }}}}. 
        Also notice that {{\color{\colorMATH}\ensuremath{\sum_{{\begingroup\renewcommand\colorMATH{\colorMATHB}\renewcommand\colorSYNTAX{\colorSYNTAXB}{{\color{\colorMATH}\ensuremath{\sv_{1}}}}\endgroup } \in  S'}\distp[1j]({\begingroup\renewcommand\colorMATH{\colorMATHB}\renewcommand\colorSYNTAX{\colorSYNTAXB}{{\color{\colorMATH}\ensuremath{\sv_{1}}}}\endgroup })}}} is a probability therefore\\ 
        {{\color{\colorMATH}\ensuremath{\sum_{{\begingroup\renewcommand\colorMATH{\colorMATHB}\renewcommand\colorSYNTAX{\colorSYNTAXB}{{\color{\colorMATH}\ensuremath{\sv_{1}}}}\endgroup } \in  S'}\distp[1j]({\begingroup\renewcommand\colorMATH{\colorMATHB}\renewcommand\colorSYNTAX{\colorSYNTAXB}{{\color{\colorMATH}\ensuremath{\sv_{1}}}}\endgroup }) \leq  min(e^{{\begingroup\renewcommand\colorMATH{\colorMATHC}\renewcommand\colorSYNTAX{\colorSYNTAXC}{{\color{\colorMATH}\ensuremath{\epsilon _{2}}}}\endgroup }}(\sum_{{\begingroup\renewcommand\colorMATH{\colorMATHB}\renewcommand\colorSYNTAX{\colorSYNTAXB}{{\color{\colorMATH}\ensuremath{\sv_{2}}}}\endgroup } \in  S'} \distp[2j]({\begingroup\renewcommand\colorMATH{\colorMATHB}\renewcommand\colorSYNTAX{\colorSYNTAXB}{{\color{\colorMATH}\ensuremath{\sv_{2}}}}\endgroup })) + {\begingroup\renewcommand\colorMATH{\colorMATHC}\renewcommand\colorSYNTAX{\colorSYNTAXC}{{\color{\colorMATH}\ensuremath{\delta _{2}}}}\endgroup }, 1) \leq  min(e^{{\begingroup\renewcommand\colorMATH{\colorMATHC}\renewcommand\colorSYNTAX{\colorSYNTAXC}{{\color{\colorMATH}\ensuremath{\epsilon _{2}}}}\endgroup }}(\sum_{{\begingroup\renewcommand\colorMATH{\colorMATHB}\renewcommand\colorSYNTAX{\colorSYNTAXB}{{\color{\colorMATH}\ensuremath{\sv_{2}}}}\endgroup } \in  S'} \distp[2j]({\begingroup\renewcommand\colorMATH{\colorMATHB}\renewcommand\colorSYNTAX{\colorSYNTAXB}{{\color{\colorMATH}\ensuremath{\sv_{2}}}}\endgroup })), 1) + {\begingroup\renewcommand\colorMATH{\colorMATHC}\renewcommand\colorSYNTAX{\colorSYNTAXC}{{\color{\colorMATH}\ensuremath{\delta _{2}}}}\endgroup }}}}.
        Then
        \begingroup\color{\colorMATH}\begin{gather*}
\begin{array}{rcl
          }  &{}{}& \sum_{{\begingroup\renewcommand\colorMATH{\colorMATHB}\renewcommand\colorSYNTAX{\colorSYNTAXB}{{\color{\colorMATH}\ensuremath{\sv_{1}}}}\endgroup } \in  S'} \sum_{{\begingroup\renewcommand\colorMATH{\colorMATHB}\renewcommand\colorSYNTAX{\colorSYNTAXB}{{\color{\colorMATH}\ensuremath{\sv}}}\endgroup }_j \in  dval} \dist[1]({\begingroup\renewcommand\colorMATH{\colorMATHB}\renewcommand\colorSYNTAX{\colorSYNTAXB}{{\color{\colorMATH}\ensuremath{\sv}}}\endgroup }_j) \times  \distp[1j]({\begingroup\renewcommand\colorMATH{\colorMATHB}\renewcommand\colorSYNTAX{\colorSYNTAXB}{{\color{\colorMATH}\ensuremath{\sv_{1}}}}\endgroup })
          \cr   &{}={}& \sum_{{\begingroup\renewcommand\colorMATH{\colorMATHB}\renewcommand\colorSYNTAX{\colorSYNTAXB}{{\color{\colorMATH}\ensuremath{\sv}}}\endgroup }_j \in  dval} \dist[1]({\begingroup\renewcommand\colorMATH{\colorMATHB}\renewcommand\colorSYNTAX{\colorSYNTAXB}{{\color{\colorMATH}\ensuremath{\sv}}}\endgroup }_j) \times  (\sum_{{\begingroup\renewcommand\colorMATH{\colorMATHB}\renewcommand\colorSYNTAX{\colorSYNTAXB}{{\color{\colorMATH}\ensuremath{\sv_{1}}}}\endgroup } \in  S'} \distp[1j]({\begingroup\renewcommand\colorMATH{\colorMATHB}\renewcommand\colorSYNTAX{\colorSYNTAXB}{{\color{\colorMATH}\ensuremath{\sv_{1}}}}\endgroup }))
          \cr
\end{array}
        \end{gather*}\endgroup
        
        Let {{\color{\colorMATH}\ensuremath{S = dval}}}, then we know that 
        {{\color{\colorMATH}\ensuremath{\sum_{{\begingroup\renewcommand\colorMATH{\colorMATHB}\renewcommand\colorSYNTAX{\colorSYNTAXB}{{\color{\colorMATH}\ensuremath{\sv}}}\endgroup }_j \in  dval} \dist[1]({\begingroup\renewcommand\colorMATH{\colorMATHB}\renewcommand\colorSYNTAX{\colorSYNTAXB}{{\color{\colorMATH}\ensuremath{\sv}}}\endgroup }_j) \leq  e^{{\begingroup\renewcommand\colorMATH{\colorMATHC}\renewcommand\colorSYNTAX{\colorSYNTAXC}{{\color{\colorMATH}\ensuremath{\epsilon _{1}}}}\endgroup }}(\sum_{{\begingroup\renewcommand\colorMATH{\colorMATHB}\renewcommand\colorSYNTAX{\colorSYNTAXB}{{\color{\colorMATH}\ensuremath{\sv}}}\endgroup }_j \in  dval} \dist[2]({\begingroup\renewcommand\colorMATH{\colorMATHB}\renewcommand\colorSYNTAX{\colorSYNTAXB}{{\color{\colorMATH}\ensuremath{\sv}}}\endgroup }_j)) + {\begingroup\renewcommand\colorMATH{\colorMATHC}\renewcommand\colorSYNTAX{\colorSYNTAXC}{{\color{\colorMATH}\ensuremath{\delta _{1}}}}\endgroup }}}}.\\
        % Also if we choose {{\color{\colorMATH}\ensuremath{S \times  S = \{(v,v)\}}}}, then we also know that
        % {{\color{\colorMATH}\ensuremath{\llbracket {\begingroup\renewcommand\colorMATH{\colorMATHC}\renewcommand\colorSYNTAX{\colorSYNTAXC}{{\color{\colorMATH}\ensuremath{\pe_{1}}}}\endgroup }\rrbracket _{\gamma _{1}}({\begingroup\renewcommand\colorMATH{\colorMATHB}\renewcommand\colorSYNTAX{\colorSYNTAXB}{{\color{\colorMATH}\ensuremath{\sv}}}\endgroup }) \leq  e^{{\begingroup\renewcommand\colorMATH{\colorMATHC}\renewcommand\colorSYNTAX{\colorSYNTAXC}{{\color{\colorMATH}\ensuremath{\epsilon _{1}}}}\endgroup }}\llbracket {\begingroup\renewcommand\colorMATH{\colorMATHC}\renewcommand\colorSYNTAX{\colorSYNTAXC}{{\color{\colorMATH}\ensuremath{\pe_{1}}}}\endgroup }\rrbracket _{\gamma _{2}}^{*}({\begingroup\renewcommand\colorMATH{\colorMATHB}\renewcommand\colorSYNTAX{\colorSYNTAXB}{{\color{\colorMATH}\ensuremath{\sv}}}\endgroup }) + {\begingroup\renewcommand\colorMATH{\colorMATHC}\renewcommand\colorSYNTAX{\colorSYNTAXC}{{\color{\colorMATH}\ensuremath{\delta _{1}}}}\endgroup }}}}
        Let {{\color{\colorMATH}\ensuremath{\mu _{v} = (\dist[1]({\begingroup\renewcommand\colorMATH{\colorMATHB}\renewcommand\colorSYNTAX{\colorSYNTAXB}{{\color{\colorMATH}\ensuremath{\sv}}}\endgroup }) - e^{{\begingroup\renewcommand\colorMATH{\colorMATHC}\renewcommand\colorSYNTAX{\colorSYNTAXC}{{\color{\colorMATH}\ensuremath{\epsilon _{1}}}}\endgroup }}\dist[2]({\begingroup\renewcommand\colorMATH{\colorMATHB}\renewcommand\colorSYNTAX{\colorSYNTAXB}{{\color{\colorMATH}\ensuremath{\sv}}}\endgroup }))_{+}}}} (notice that {{\color{\colorMATH}\ensuremath{\mu _{v}}}} is not necessarily a measure) and {{\color{\colorMATH}\ensuremath{T_1 = \{ {\begingroup\renewcommand\colorMATH{\colorMATHB}\renewcommand\colorSYNTAX{\colorSYNTAXB}{{\color{\colorMATH}\ensuremath{\sv}}}\endgroup } \in  dval \mid \dist[1]({\begingroup\renewcommand\colorMATH{\colorMATHB}\renewcommand\colorSYNTAX{\colorSYNTAXB}{{\color{\colorMATH}\ensuremath{\sv}}}\endgroup }) - e^{{\begingroup\renewcommand\colorMATH{\colorMATHC}\renewcommand\colorSYNTAX{\colorSYNTAXC}{{\color{\colorMATH}\ensuremath{\epsilon _{1}}}}\endgroup }}\dist[2]({\begingroup\renewcommand\colorMATH{\colorMATHB}\renewcommand\colorSYNTAX{\colorSYNTAXB}{{\color{\colorMATH}\ensuremath{\sv}}}\endgroup }) > 0 \}}}}, then:

        \begingroup\color{\colorMATH}\begin{gather*}
\begin{array}{rcl
          }  &{}{}&  \sum_{{\begingroup\renewcommand\colorMATH{\colorMATHB}\renewcommand\colorSYNTAX{\colorSYNTAXB}{{\color{\colorMATH}\ensuremath{\sv}}}\endgroup } \in  dval} \mu _{v}
          \cr   &{}={}& \sum_{{\begingroup\renewcommand\colorMATH{\colorMATHB}\renewcommand\colorSYNTAX{\colorSYNTAXB}{{\color{\colorMATH}\ensuremath{\sv}}}\endgroup } \in  dval} (\dist[1]({\begingroup\renewcommand\colorMATH{\colorMATHB}\renewcommand\colorSYNTAX{\colorSYNTAXB}{{\color{\colorMATH}\ensuremath{\sv}}}\endgroup }) - e^{{\begingroup\renewcommand\colorMATH{\colorMATHC}\renewcommand\colorSYNTAX{\colorSYNTAXC}{{\color{\colorMATH}\ensuremath{\epsilon _{1}}}}\endgroup }}\dist[2]({\begingroup\renewcommand\colorMATH{\colorMATHB}\renewcommand\colorSYNTAX{\colorSYNTAXB}{{\color{\colorMATH}\ensuremath{\sv}}}\endgroup }))_{+}
          \cr   &{}={}& \sum_{{\begingroup\renewcommand\colorMATH{\colorMATHB}\renewcommand\colorSYNTAX{\colorSYNTAXB}{{\color{\colorMATH}\ensuremath{\sv}}}\endgroup } \in  T_1} \dist[1]({\begingroup\renewcommand\colorMATH{\colorMATHB}\renewcommand\colorSYNTAX{\colorSYNTAXB}{{\color{\colorMATH}\ensuremath{\sv}}}\endgroup }) - e^{{\begingroup\renewcommand\colorMATH{\colorMATHC}\renewcommand\colorSYNTAX{\colorSYNTAXC}{{\color{\colorMATH}\ensuremath{\epsilon _{1}}}}\endgroup }}\dist[2]({\begingroup\renewcommand\colorMATH{\colorMATHB}\renewcommand\colorSYNTAX{\colorSYNTAXB}{{\color{\colorMATH}\ensuremath{\sv}}}\endgroup })
          \cr  &{}\leq {}& {\begingroup\renewcommand\colorMATH{\colorMATHC}\renewcommand\colorSYNTAX{\colorSYNTAXC}{{\color{\colorMATH}\ensuremath{\delta _{1}}}}\endgroup } \quad \text{\color{black}(By def. of $({\begingroup\renewcommand\colorMATH{\colorMATHC}\renewcommand\colorSYNTAX{\colorSYNTAXC}{{\color{\colorMATH}\ensuremath{\epsilon _{1}}}}\endgroup }, {\begingroup\renewcommand\colorMATH{\colorMATHC}\renewcommand\colorSYNTAX{\colorSYNTAXC}{{\color{\colorMATH}\ensuremath{\delta _{1}}}}\endgroup })$-dp)}
          \end{array}
        \end{gather*}\endgroup
        
        Finally, 

        \begingroup\color{\colorMATH}\begin{gather*}
          \begin{array}{rcl
          }  &{}{}& \sum_{{\begingroup\renewcommand\colorMATH{\colorMATHB}\renewcommand\colorSYNTAX{\colorSYNTAXB}{{\color{\colorMATH}\ensuremath{\sv}}}\endgroup }_j \in  dval} \dist[1]({\begingroup\renewcommand\colorMATH{\colorMATHB}\renewcommand\colorSYNTAX{\colorSYNTAXB}{{\color{\colorMATH}\ensuremath{\sv}}}\endgroup }_j) \times  min(e^{{\begingroup\renewcommand\colorMATH{\colorMATHC}\renewcommand\colorSYNTAX{\colorSYNTAXC}{{\color{\colorMATH}\ensuremath{\epsilon _{2}}}}\endgroup }}(\sum_{{\begingroup\renewcommand\colorMATH{\colorMATHB}\renewcommand\colorSYNTAX{\colorSYNTAXB}{{\color{\colorMATH}\ensuremath{\sv_{1}}}}\endgroup } \in  S'} \distp[2j]({\begingroup\renewcommand\colorMATH{\colorMATHB}\renewcommand\colorSYNTAX{\colorSYNTAXB}{{\color{\colorMATH}\ensuremath{\sv_{1}}}}\endgroup })), 1) + 
          {\begingroup\renewcommand\colorMATH{\colorMATHC}\renewcommand\colorSYNTAX{\colorSYNTAXC}{{\color{\colorMATH}\ensuremath{\delta _{2}}}}\endgroup }
          \cr
\end{array}
        \end{gather*}\endgroup

        and the result holds.
      \end{subproof}
  \end{enumerate}  

\end{proof}

\begin{restatable}[Sensitivity Type Soundness at Base Types]{theorem}{SensitivityTypeSoundnessBaseTypes}
  \label{thm:SensitivityTypeSoundnessBaseTypes}
  If {{\color{\colorMATH}\ensuremath{\varnothing ; \varnothing  \vdash  {\begingroup\renewcommand\colorMATH{\colorMATHB}\renewcommand\colorSYNTAX{\colorSYNTAXB}{{\color{\colorMATH}\ensuremath{\se}}}\endgroup } \mathrel{:} (x\mathrel{:} {\begingroup\renewcommand\colorMATH{\colorMATHA}\renewcommand\colorSYNTAX{\colorSYNTAXA}{{\color{\colorSYNTAX}\texttt{{\ensuremath{{\mathbb{R}}}}}}}\endgroup } \mathord{\cdotp } \infty ) \xrightarrowS {{\begingroup\renewcommand\colorMATH{\colorMATHB}\renewcommand\colorSYNTAX{\colorSYNTAXB}{{\color{\colorMATH}\ensuremath{\sss}}}\endgroup }x} {\begingroup\renewcommand\colorMATH{\colorMATHA}\renewcommand\colorSYNTAX{\colorSYNTAXA}{{\color{\colorSYNTAX}\texttt{{\ensuremath{{\mathbb{R}}}}}}}\endgroup } \mathrel{;} \varnothing }}},\\
  {{\color{\colorMATH}\ensuremath{|{\begingroup\renewcommand\colorMATH{\colorMATHB}\renewcommand\colorSYNTAX{\colorSYNTAXB}{{\color{\colorMATH}\ensuremath{r_{1}}}}\endgroup }-{\begingroup\renewcommand\colorMATH{\colorMATHB}\renewcommand\colorSYNTAX{\colorSYNTAXB}{{\color{\colorMATH}\ensuremath{r_{2}}}}\endgroup }| \leq  {\begingroup\renewcommand\colorMATH{\colorMATHB}\renewcommand\colorSYNTAX{\colorSYNTAXB}{{\color{\colorMATH}\ensuremath{\sss^{\prime}}}}\endgroup }}}}, {{\color{\colorMATH}\ensuremath{\varnothing  \vdash  {\begingroup\renewcommand\colorMATH{\colorMATHB}\renewcommand\colorSYNTAX{\colorSYNTAXB}{{\color{\colorMATH}\ensuremath{\se}}}\endgroup }\hspace*{0.33em}{\begingroup\renewcommand\colorMATH{\colorMATHB}\renewcommand\colorSYNTAX{\colorSYNTAXB}{{\color{\colorMATH}\ensuremath{r_{1}}}}\endgroup } \Downarrow  {\begingroup\renewcommand\colorMATH{\colorMATHB}\renewcommand\colorSYNTAX{\colorSYNTAXB}{{\color{\colorMATH}\ensuremath{r_{1}^{\prime}}}}\endgroup }}}}, {{\color{\colorMATH}\ensuremath{\varnothing  \vdash  {\begingroup\renewcommand\colorMATH{\colorMATHB}\renewcommand\colorSYNTAX{\colorSYNTAXB}{{\color{\colorMATH}\ensuremath{\se}}}\endgroup }\hspace*{0.33em}{\begingroup\renewcommand\colorMATH{\colorMATHB}\renewcommand\colorSYNTAX{\colorSYNTAXB}{{\color{\colorMATH}\ensuremath{r_{2}}}}\endgroup } \Downarrow  {\begingroup\renewcommand\colorMATH{\colorMATHB}\renewcommand\colorSYNTAX{\colorSYNTAXB}{{\color{\colorMATH}\ensuremath{r_{2}^{\prime}}}}\endgroup } }}}, then {{\color{\colorMATH}\ensuremath{|{\begingroup\renewcommand\colorMATH{\colorMATHB}\renewcommand\colorSYNTAX{\colorSYNTAXB}{{\color{\colorMATH}\ensuremath{r_{1}^{\prime}}}}\endgroup }-{\begingroup\renewcommand\colorMATH{\colorMATHB}\renewcommand\colorSYNTAX{\colorSYNTAXB}{{\color{\colorMATH}\ensuremath{r_{2}^{\prime}}}}\endgroup }| \leq  {\begingroup\renewcommand\colorMATH{\colorMATHB}\renewcommand\colorSYNTAX{\colorSYNTAXB}{{\color{\colorMATH}\ensuremath{\sss}}}\endgroup }{\begingroup\renewcommand\colorMATH{\colorMATHB}\renewcommand\colorSYNTAX{\colorSYNTAXB}{{\color{\colorMATH}\ensuremath{\sss^{\prime}}}}\endgroup }}}}
\end{restatable}
%\SensitivityTypeSoundnessBaseTypes*
\begin{proof}
  We know that {{\color{\colorMATH}\ensuremath{\varnothing  \vdash  {\begingroup\renewcommand\colorMATH{\colorMATHB}\renewcommand\colorSYNTAX{\colorSYNTAXB}{{\color{\colorMATH}\ensuremath{\se}}}\endgroup } \mathrel{:} (x\mathrel{:}{\begingroup\renewcommand\colorMATH{\colorMATHA}\renewcommand\colorSYNTAX{\colorSYNTAXA}{{\color{\colorSYNTAX}\texttt{{\ensuremath{{\mathbb{R}}}}}}}\endgroup } \mathord{\cdotp } \infty ) \xrightarrowS {{\begingroup\renewcommand\colorMATH{\colorMATHB}\renewcommand\colorSYNTAX{\colorSYNTAXB}{{\color{\colorMATH}\ensuremath{\sss}}}\endgroup }x} {\begingroup\renewcommand\colorMATH{\colorMATHA}\renewcommand\colorSYNTAX{\colorSYNTAXA}{{\color{\colorSYNTAX}\texttt{{\ensuremath{{\mathbb{R}}}}}}}\endgroup } \mathrel{;} \varnothing }}} therefore by the Fundamental Property (Theorem~\ref{alm:fp}), using 
  {{\color{\colorMATH}\ensuremath{\gamma _{1} = \varnothing }}} and {{\color{\colorMATH}\ensuremath{\gamma _{2} = \varnothing }}}, then {{\color{\colorMATH}\ensuremath{(\varnothing  \vdash  {\begingroup\renewcommand\colorMATH{\colorMATHB}\renewcommand\colorSYNTAX{\colorSYNTAXB}{{\color{\colorMATH}\ensuremath{\se}}}\endgroup }, \varnothing  \vdash  {\begingroup\renewcommand\colorMATH{\colorMATHB}\renewcommand\colorSYNTAX{\colorSYNTAXB}{{\color{\colorMATH}\ensuremath{\se}}}\endgroup }) \in  {\mathcal{E}}_{0}^{k}\llbracket (x\mathrel{:}{\begingroup\renewcommand\colorMATH{\colorMATHA}\renewcommand\colorSYNTAX{\colorSYNTAXA}{{\color{\colorSYNTAX}\texttt{{\ensuremath{{\mathbb{R}}}}}}}\endgroup } \mathord{\cdotp } \infty ) \xrightarrowS {{\begingroup\renewcommand\colorMATH{\colorMATHB}\renewcommand\colorSYNTAX{\colorSYNTAXB}{{\color{\colorMATH}\ensuremath{\sss}}}\endgroup }x} {\begingroup\renewcommand\colorMATH{\colorMATHA}\renewcommand\colorSYNTAX{\colorSYNTAXA}{{\color{\colorSYNTAX}\texttt{{\ensuremath{{\mathbb{R}}}}}}}\endgroup }\rrbracket }}}, i.e.
  if {{\color{\colorMATH}\ensuremath{\varnothing  \vdash  {\begingroup\renewcommand\colorMATH{\colorMATHB}\renewcommand\colorSYNTAX{\colorSYNTAXB}{{\color{\colorMATH}\ensuremath{\se}}}\endgroup } \Downarrow ^{j_{1}} \langle \lambda x.\hspace*{0.33em}{\begingroup\renewcommand\colorMATH{\colorMATHB}\renewcommand\colorSYNTAX{\colorSYNTAXB}{{\color{\colorMATH}\ensuremath{\se'}}}\endgroup },\gamma \rangle }}} then {{\color{\colorMATH}\ensuremath{(\langle \lambda x.\hspace*{0.33em}{\begingroup\renewcommand\colorMATH{\colorMATHB}\renewcommand\colorSYNTAX{\colorSYNTAXB}{{\color{\colorMATH}\ensuremath{\se'}}}\endgroup },\gamma \rangle ,\langle \lambda x.\hspace*{0.33em}{\begingroup\renewcommand\colorMATH{\colorMATHB}\renewcommand\colorSYNTAX{\colorSYNTAXB}{{\color{\colorMATH}\ensuremath{\se'}}}\endgroup },\gamma \rangle ) \in  {\mathcal{V}}_{0}^{k-j_{1}}\llbracket (x\mathrel{:}{\begingroup\renewcommand\colorMATH{\colorMATHA}\renewcommand\colorSYNTAX{\colorSYNTAXA}{{\color{\colorSYNTAX}\texttt{{\ensuremath{{\mathbb{R}}}}}}}\endgroup }) \xrightarrowS {{\begingroup\renewcommand\colorMATH{\colorMATHB}\renewcommand\colorSYNTAX{\colorSYNTAXB}{{\color{\colorMATH}\ensuremath{\sss}}}\endgroup }x} {\begingroup\renewcommand\colorMATH{\colorMATHA}\renewcommand\colorSYNTAX{\colorSYNTAXA}{{\color{\colorSYNTAX}\texttt{{\ensuremath{{\mathbb{R}}}}}}}\endgroup }\rrbracket }}}, for some {{\color{\colorMATH}\ensuremath{k>j_{1}}}}.

  We know that {{\color{\colorMATH}\ensuremath{\varnothing  \vdash  {\begingroup\renewcommand\colorMATH{\colorMATHB}\renewcommand\colorSYNTAX{\colorSYNTAXB}{{\color{\colorMATH}\ensuremath{\se}}}\endgroup }\hspace*{0.33em}{\begingroup\renewcommand\colorMATH{\colorMATHB}\renewcommand\colorSYNTAX{\colorSYNTAXB}{{\color{\colorMATH}\ensuremath{r_{1}}}}\endgroup } \Downarrow ^{k'} {\begingroup\renewcommand\colorMATH{\colorMATHB}\renewcommand\colorSYNTAX{\colorSYNTAXB}{{\color{\colorMATH}\ensuremath{r_{1}}}}\endgroup }'}}}, and {{\color{\colorMATH}\ensuremath{\varnothing  \vdash  {\begingroup\renewcommand\colorMATH{\colorMATHB}\renewcommand\colorSYNTAX{\colorSYNTAXB}{{\color{\colorMATH}\ensuremath{\se}}}\endgroup }\hspace*{0.33em}{\begingroup\renewcommand\colorMATH{\colorMATHB}\renewcommand\colorSYNTAX{\colorSYNTAXB}{{\color{\colorMATH}\ensuremath{r_{2}}}}\endgroup } \Downarrow ^{k'} {\begingroup\renewcommand\colorMATH{\colorMATHB}\renewcommand\colorSYNTAX{\colorSYNTAXB}{{\color{\colorMATH}\ensuremath{r_{2}}}}\endgroup }'}}}, for some {{\color{\colorMATH}\ensuremath{k' = j_{1} + j_{2}}}}, and {{\color{\colorMATH}\ensuremath{k> j_{1}+ j_{2}}}} , i.e.
  \begingroup\color{\colorMATH}\begin{gather*} 
    \inferrule*[lab={\textsc{ app}}
    ]{ \varnothing \vdash e \Downarrow ^{j_{1}} \langle \lambda x.\hspace*{0.33em}{\begingroup\renewcommand\colorMATH{\colorMATHB}\renewcommand\colorSYNTAX{\colorSYNTAXB}{{\color{\colorMATH}\ensuremath{\se'}}}\endgroup },\gamma \rangle 
    \\ \gamma [x\mapsto {\begingroup\renewcommand\colorMATH{\colorMATHB}\renewcommand\colorSYNTAX{\colorSYNTAXB}{{\color{\colorMATH}\ensuremath{r_{i}}}}\endgroup }]\vdash {\begingroup\renewcommand\colorMATH{\colorMATHB}\renewcommand\colorSYNTAX{\colorSYNTAXB}{{\color{\colorMATH}\ensuremath{\se'}}}\endgroup } \Downarrow ^{j_{2}} {\begingroup\renewcommand\colorMATH{\colorMATHB}\renewcommand\colorSYNTAX{\colorSYNTAXB}{{\color{\colorMATH}\ensuremath{r'_{i}}}}\endgroup }
      }{
      \varnothing \vdash {\begingroup\renewcommand\colorMATH{\colorMATHB}\renewcommand\colorSYNTAX{\colorSYNTAXB}{{\color{\colorMATH}\ensuremath{\se}}}\endgroup }\hspace*{0.33em}{\begingroup\renewcommand\colorMATH{\colorMATHB}\renewcommand\colorSYNTAX{\colorSYNTAXB}{{\color{\colorMATH}\ensuremath{r_{i}}}}\endgroup } \Downarrow ^{k'} {\begingroup\renewcommand\colorMATH{\colorMATHB}\renewcommand\colorSYNTAX{\colorSYNTAXB}{{\color{\colorMATH}\ensuremath{r'_{i}}}}\endgroup }  
    }
  \end{gather*}\endgroup
  for some {{\color{\colorMATH}\ensuremath{\gamma }}} and {{\color{\colorMATH}\ensuremath{\Gamma }}} such that {{\color{\colorMATH}\ensuremath{\Gamma , x:{\begingroup\renewcommand\colorMATH{\colorMATHA}\renewcommand\colorSYNTAX{\colorSYNTAXA}{{\color{\colorSYNTAX}\texttt{{\ensuremath{{\mathbb{R}}}}}}}\endgroup } \vdash  e' : {\begingroup\renewcommand\colorMATH{\colorMATHA}\renewcommand\colorSYNTAX{\colorSYNTAXA}{{\color{\colorSYNTAX}\texttt{{\ensuremath{{\mathbb{R}}}}}}}\endgroup } ; {\begingroup\renewcommand\colorMATH{\colorMATHB}\renewcommand\colorSYNTAX{\colorSYNTAXB}{{\color{\colorMATH}\ensuremath{\sss}}}\endgroup }x}}}, and {{\color{\colorMATH}\ensuremath{(\gamma , \gamma ) \in  {\mathcal{G}}_{\varnothing }^{k-j_{1}-1}\llbracket \Gamma \rrbracket }}}. 
  As {{\color{\colorMATH}\ensuremath{|{\begingroup\renewcommand\colorMATH{\colorMATHB}\renewcommand\colorSYNTAX{\colorSYNTAXB}{{\color{\colorMATH}\ensuremath{r_{1}}}}\endgroup }-{\begingroup\renewcommand\colorMATH{\colorMATHB}\renewcommand\colorSYNTAX{\colorSYNTAXB}{{\color{\colorMATH}\ensuremath{r_{2}}}}\endgroup }| \leq  {\begingroup\renewcommand\colorMATH{\colorMATHB}\renewcommand\colorSYNTAX{\colorSYNTAXB}{{\color{\colorMATH}\ensuremath{\sss'}}}\endgroup }}}}, then {{\color{\colorMATH}\ensuremath{({\begingroup\renewcommand\colorMATH{\colorMATHB}\renewcommand\colorSYNTAX{\colorSYNTAXB}{{\color{\colorMATH}\ensuremath{r_{1}}}}\endgroup }, {\begingroup\renewcommand\colorMATH{\colorMATHB}\renewcommand\colorSYNTAX{\colorSYNTAXB}{{\color{\colorMATH}\ensuremath{r_{2}}}}\endgroup }) \in  {\mathcal{V}}_{{\begingroup\renewcommand\colorMATH{\colorMATHB}\renewcommand\colorSYNTAX{\colorSYNTAXB}{{\color{\colorMATH}\ensuremath{\sss'}}}\endgroup }}^{k-j_{1}-1}\llbracket {\begingroup\renewcommand\colorMATH{\colorMATHA}\renewcommand\colorSYNTAX{\colorSYNTAXA}{{\color{\colorSYNTAX}\texttt{{\ensuremath{{\mathbb{R}}}}}}}\endgroup }\rrbracket }}}.
  We instantiate {{\color{\colorMATH}\ensuremath{(\langle \lambda x.\hspace*{0.33em}{\begingroup\renewcommand\colorMATH{\colorMATHB}\renewcommand\colorSYNTAX{\colorSYNTAXB}{{\color{\colorMATH}\ensuremath{\se'}}}\endgroup },\gamma \rangle ,\langle \lambda x.\hspace*{0.33em}{\begingroup\renewcommand\colorMATH{\colorMATHB}\renewcommand\colorSYNTAX{\colorSYNTAXB}{{\color{\colorMATH}\ensuremath{\se'}}}\endgroup },\gamma \rangle ) \in  {\mathcal{V}}_{0}^{k-j_{1}}\llbracket (x\mathrel{:}{\begingroup\renewcommand\colorMATH{\colorMATHA}\renewcommand\colorSYNTAX{\colorSYNTAXA}{{\color{\colorSYNTAX}\texttt{{\ensuremath{{\mathbb{R}}}}}}}\endgroup } \mathord{\cdotp } \infty ) \xrightarrowS {{\begingroup\renewcommand\colorMATH{\colorMATHB}\renewcommand\colorSYNTAX{\colorSYNTAXB}{{\color{\colorMATH}\ensuremath{\sss}}}\endgroup }x} {\begingroup\renewcommand\colorMATH{\colorMATHA}\renewcommand\colorSYNTAX{\colorSYNTAXA}{{\color{\colorSYNTAX}\texttt{{\ensuremath{{\mathbb{R}}}}}}}\endgroup }\rrbracket }}} with {{\color{\colorMATH}\ensuremath{({\begingroup\renewcommand\colorMATH{\colorMATHB}\renewcommand\colorSYNTAX{\colorSYNTAXB}{{\color{\colorMATH}\ensuremath{r_{1}}}}\endgroup }, {\begingroup\renewcommand\colorMATH{\colorMATHB}\renewcommand\colorSYNTAX{\colorSYNTAXB}{{\color{\colorMATH}\ensuremath{r_{2}}}}\endgroup }) \in  {\mathcal{V}}_{{\begingroup\renewcommand\colorMATH{\colorMATHB}\renewcommand\colorSYNTAX{\colorSYNTAXB}{{\color{\colorMATH}\ensuremath{\sss'}}}\endgroup }}^{k-j_{1}-1}\llbracket {\begingroup\renewcommand\colorMATH{\colorMATHA}\renewcommand\colorSYNTAX{\colorSYNTAXA}{{\color{\colorSYNTAX}\texttt{{\ensuremath{{\mathbb{R}}}}}}}\endgroup }\rrbracket }}} to know that
  {{\color{\colorMATH}\ensuremath{(\gamma [x\mapsto {\begingroup\renewcommand\colorMATH{\colorMATHB}\renewcommand\colorSYNTAX{\colorSYNTAXB}{{\color{\colorMATH}\ensuremath{r_{1}}}}\endgroup }]\vdash {\begingroup\renewcommand\colorMATH{\colorMATHB}\renewcommand\colorSYNTAX{\colorSYNTAXB}{{\color{\colorMATH}\ensuremath{\se'}}}\endgroup }, \gamma [x\mapsto {\begingroup\renewcommand\colorMATH{\colorMATHB}\renewcommand\colorSYNTAX{\colorSYNTAXB}{{\color{\colorMATH}\ensuremath{r_{2}}}}\endgroup }]\vdash {\begingroup\renewcommand\colorMATH{\colorMATHB}\renewcommand\colorSYNTAX{\colorSYNTAXB}{{\color{\colorMATH}\ensuremath{\se'}}}\endgroup }) \in  {\mathcal{E}}_{0+0+{\begingroup\renewcommand\colorMATH{\colorMATHB}\renewcommand\colorSYNTAX{\colorSYNTAXB}{{\color{\colorMATH}\ensuremath{\sss}}}\endgroup }{\begingroup\renewcommand\colorMATH{\colorMATHB}\renewcommand\colorSYNTAX{\colorSYNTAXB}{{\color{\colorMATH}\ensuremath{\sss'}}}\endgroup }}^{k-j_{1}-1}\llbracket {\begingroup\renewcommand\colorMATH{\colorMATHA}\renewcommand\colorSYNTAX{\colorSYNTAXA}{{\color{\colorSYNTAX}\texttt{{\ensuremath{{\mathbb{R}}}}}}}\endgroup }\rrbracket }}}, i.e.
  if {{\color{\colorMATH}\ensuremath{\gamma [x\mapsto r_{i}]\vdash {\begingroup\renewcommand\colorMATH{\colorMATHB}\renewcommand\colorSYNTAX{\colorSYNTAXB}{{\color{\colorMATH}\ensuremath{\se'}}}\endgroup } \Downarrow ^{j_{2}} {\begingroup\renewcommand\colorMATH{\colorMATHB}\renewcommand\colorSYNTAX{\colorSYNTAXB}{{\color{\colorMATH}\ensuremath{r'_{i}}}}\endgroup }}}} then {{\color{\colorMATH}\ensuremath{({\begingroup\renewcommand\colorMATH{\colorMATHB}\renewcommand\colorSYNTAX{\colorSYNTAXB}{{\color{\colorMATH}\ensuremath{r_{1}}}}\endgroup }', {\begingroup\renewcommand\colorMATH{\colorMATHB}\renewcommand\colorSYNTAX{\colorSYNTAXB}{{\color{\colorMATH}\ensuremath{r_{2}}}}\endgroup }') \in  {\mathcal{V}}_{{\begingroup\renewcommand\colorMATH{\colorMATHB}\renewcommand\colorSYNTAX{\colorSYNTAXB}{{\color{\colorMATH}\ensuremath{\sss}}}\endgroup }{\begingroup\renewcommand\colorMATH{\colorMATHB}\renewcommand\colorSYNTAX{\colorSYNTAXB}{{\color{\colorMATH}\ensuremath{\sss'}}}\endgroup }}^{k-j_{1}-j_{2}-1}\llbracket {\begingroup\renewcommand\colorMATH{\colorMATHA}\renewcommand\colorSYNTAX{\colorSYNTAXA}{{\color{\colorSYNTAX}\texttt{{\ensuremath{{\mathbb{R}}}}}}}\endgroup }\rrbracket }}}, meaning that {{\color{\colorMATH}\ensuremath{|{\begingroup\renewcommand\colorMATH{\colorMATHB}\renewcommand\colorSYNTAX{\colorSYNTAXB}{{\color{\colorMATH}\ensuremath{r_{1}}}}\endgroup }'-{\begingroup\renewcommand\colorMATH{\colorMATHB}\renewcommand\colorSYNTAX{\colorSYNTAXB}{{\color{\colorMATH}\ensuremath{r_{2}}}}\endgroup }'| \leq  {\begingroup\renewcommand\colorMATH{\colorMATHB}\renewcommand\colorSYNTAX{\colorSYNTAXB}{{\color{\colorMATH}\ensuremath{\sss}}}\endgroup }{\begingroup\renewcommand\colorMATH{\colorMATHB}\renewcommand\colorSYNTAX{\colorSYNTAXB}{{\color{\colorMATH}\ensuremath{\sss'}}}\endgroup }}}}, which is exactly what we want to prove and the result holds immediately.
\end{proof}

 \PrivacyTypeSoundnessBaseTypes*
 \begin{proof}
  We proceed analogously to Theorem~\ref{thm:SensitivityTypeSoundnessBaseTypes}. But unfolding the definition of related computations, and using a {{\color{\colorMATH}\ensuremath{k}}} big enough so both probabilities are defined (and using the fact that two real numbers are semantically equivalent for any index).
 \end{proof}

\section{$\lang$: Type Safety}
\label{asec:type-safety}
In this section we present auxiliary definitions used in Section~\ref{sec:type-safety}, and the proof of type safety.

Non-deterministic sampling big-step semantics of privacy expressions are presented in Figure~\ref{afig:interpreter-prob-sem}. We write {{\color{\colorMATH}\ensuremath{{\begingroup\renewcommand\colorMATH{\colorMATHB}\renewcommand\colorSYNTAX{\colorSYNTAXB}{{\color{\colorMATH}\ensuremath{\se}}}\endgroup } \Downarrow  {\begingroup\renewcommand\colorMATH{\colorMATHB}\renewcommand\colorSYNTAX{\colorSYNTAXB}{{\color{\colorMATH}\ensuremath{\sv}}}\endgroup }}}}, when {{\color{\colorMATH}\ensuremath{{\begingroup\renewcommand\colorMATH{\colorMATHB}\renewcommand\colorSYNTAX{\colorSYNTAXB}{{\color{\colorMATH}\ensuremath{\se}}}\endgroup } \Downarrow ^{k} {\begingroup\renewcommand\colorMATH{\colorMATHB}\renewcommand\colorSYNTAX{\colorSYNTAXB}{{\color{\colorMATH}\ensuremath{\sv}}}\endgroup }}}} for some {{\color{\colorMATH}\ensuremath{k}}}.
This semantics is coherent with respect to the distribution semantics of Figure~\ref{afig:probabilistic-semantics}, in the following sense: 
% that {{\color{\colorMATH}\ensuremath{{\exists k. \gamma  \vdash  {\begingroup\renewcommand\colorMATH{\colorMATHC}\renewcommand\colorSYNTAX{\colorSYNTAXC}{{\color{\colorMATH}\ensuremath{\pe}}}\endgroup }\Downarrow ^{k} \dist \wedge  \dist({\begingroup\renewcommand\colorMATH{\colorMATHB}\renewcommand\colorSYNTAX{\colorSYNTAXB}{{\color{\colorMATH}\ensuremath{\sv}}}\endgroup })>0 \implies  \gamma  \vdash  {\begingroup\renewcommand\colorMATH{\colorMATHC}\renewcommand\colorSYNTAX{\colorSYNTAXC}{{\color{\colorMATH}\ensuremath{\pe}}}\endgroup } \Downarrow  {\begingroup\renewcommand\colorMATH{\colorMATHB}\renewcommand\colorSYNTAX{\colorSYNTAXB}{{\color{\colorMATH}\ensuremath{\sv}}}\endgroup }}}}}. 

\begin{lemma}[Coherence of non-deterministic semantics wrt distribution semantics]
If {{\color{\colorMATH}\ensuremath{{\exists k. \gamma  \vdash  {\begingroup\renewcommand\colorMATH{\colorMATHC}\renewcommand\colorSYNTAX{\colorSYNTAXC}{{\color{\colorMATH}\ensuremath{\pe}}}\endgroup }\Downarrow ^{k} \dist \wedge  \dist({\begingroup\renewcommand\colorMATH{\colorMATHB}\renewcommand\colorSYNTAX{\colorSYNTAXB}{{\color{\colorMATH}\ensuremath{\sv}}}\endgroup })>0 }}}} then  {{\color{\colorMATH}\ensuremath{{\gamma  \vdash  {\begingroup\renewcommand\colorMATH{\colorMATHC}\renewcommand\colorSYNTAX{\colorSYNTAXC}{{\color{\colorMATH}\ensuremath{\pe}}}\endgroup } \Downarrow  {\begingroup\renewcommand\colorMATH{\colorMATHB}\renewcommand\colorSYNTAX{\colorSYNTAXB}{{\color{\colorMATH}\ensuremath{\sv}}}\endgroup }}}}}.
\end{lemma}
\begin{proof}
We present a proof sketch for this lemma, illustrating some relevant cases. We proceed by case analysis and induction in {{\color{\colorMATH}\ensuremath{k}}} and the structure of {{\color{\colorMATH}\ensuremath{{\begingroup\renewcommand\colorMATH{\colorMATHC}\renewcommand\colorSYNTAX{\colorSYNTAXC}{{\color{\colorMATH}\ensuremath{\pe}}}\endgroup }}}}. 
\begin{enumerate}[ncases]\item  {{\color{\colorMATH}\ensuremath{{\begingroup\renewcommand\colorMATH{\colorMATHC}\renewcommand\colorSYNTAX{\colorSYNTAXC}{{\color{\colorMATH}\ensuremath{\pe}}}\endgroup } = {\begingroup\renewcommand\colorMATH{\colorMATHC}\renewcommand\colorSYNTAX{\colorSYNTAXC}{{\color{\colorSYNTAX}\texttt{return}}}\endgroup }\hspace*{0.33em}{\begingroup\renewcommand\colorMATH{\colorMATHB}\renewcommand\colorSYNTAX{\colorSYNTAXB}{{\color{\colorMATH}\ensuremath{\se}}}\endgroup }}}}
	\begin{subproof} 
		We know that 
		\begingroup\color{\colorMATH}\begin{gather*} 
		\inferrule*[lab={\textsc{ return}}
	   ]{ \gamma  \vdash  {\begingroup\renewcommand\colorMATH{\colorMATHB}\renewcommand\colorSYNTAX{\colorSYNTAXB}{{\color{\colorMATH}\ensuremath{\se}}}\endgroup } \Downarrow ^{k} {\begingroup\renewcommand\colorMATH{\colorMATHB}\renewcommand\colorSYNTAX{\colorSYNTAXB}{{\color{\colorMATH}\ensuremath{\sv'}}}\endgroup }
	    }{
	     {\begingroup\renewcommand\colorMATH{\colorMATHC}\renewcommand\colorSYNTAX{\colorSYNTAXC}{{\color{\colorSYNTAX}\texttt{return}}}\endgroup }\hspace*{0.33em}{\begingroup\renewcommand\colorMATH{\colorMATHB}\renewcommand\colorSYNTAX{\colorSYNTAXB}{{\color{\colorMATH}\ensuremath{\se}}}\endgroup } \Downarrow ^{k} \dist
	  }
		\end{gather*}\endgroup
		where 
		\begingroup\color{\colorMATH}\begin{gather*} 
			\dist = \begin{array}[t]{l
	                   } \lambda  x .\hspace*{0.33em}\left\{ \begin{array}{l@{\hspace*{1.00em}}c@{\hspace*{1.00em}}l
	                     } 1 &{}{\textit{when}}{}& x = {\begingroup\renewcommand\colorMATH{\colorMATHB}\renewcommand\colorSYNTAX{\colorSYNTAXB}{{\color{\colorMATH}\ensuremath{\sv'}}}\endgroup }
	                     \cr  0 &{}{\textit{otherwise}}{}&
	                     \end{array}\right.
	                   \end{array}	
		\end{gather*}\endgroup
		
		Let {{\color{\colorMATH}\ensuremath{{\begingroup\renewcommand\colorMATH{\colorMATHB}\renewcommand\colorSYNTAX{\colorSYNTAXB}{{\color{\colorMATH}\ensuremath{\sv}}}\endgroup }}}} such that {{\color{\colorMATH}\ensuremath{\dist({\begingroup\renewcommand\colorMATH{\colorMATHB}\renewcommand\colorSYNTAX{\colorSYNTAXB}{{\color{\colorMATH}\ensuremath{\sv}}}\endgroup })>0}}}; this means that {{\color{\colorMATH}\ensuremath{{\begingroup\renewcommand\colorMATH{\colorMATHB}\renewcommand\colorSYNTAX{\colorSYNTAXB}{{\color{\colorMATH}\ensuremath{\sv}}}\endgroup } = {\begingroup\renewcommand\colorMATH{\colorMATHB}\renewcommand\colorSYNTAX{\colorSYNTAXB}{{\color{\colorMATH}\ensuremath{\sv'}}}\endgroup }}}}, and {{\color{\colorMATH}\ensuremath{\gamma  \vdash  {\begingroup\renewcommand\colorMATH{\colorMATHB}\renewcommand\colorSYNTAX{\colorSYNTAXB}{{\color{\colorMATH}\ensuremath{\se}}}\endgroup } \Downarrow  {\begingroup\renewcommand\colorMATH{\colorMATHB}\renewcommand\colorSYNTAX{\colorSYNTAXB}{{\color{\colorMATH}\ensuremath{\sv}}}\endgroup }}}}, Then
		\begingroup\color{\colorMATH}\begin{gather*} 
		  \inferrule*[lab={\textsc{ return}}
		  ]{ \gamma  \vdash  {\begingroup\renewcommand\colorMATH{\colorMATHB}\renewcommand\colorSYNTAX{\colorSYNTAXB}{{\color{\colorMATH}\ensuremath{\se}}}\endgroup } \Downarrow  {\begingroup\renewcommand\colorMATH{\colorMATHB}\renewcommand\colorSYNTAX{\colorSYNTAXB}{{\color{\colorMATH}\ensuremath{\sv}}}\endgroup }
		    }{
		     {\begingroup\renewcommand\colorMATH{\colorMATHC}\renewcommand\colorSYNTAX{\colorSYNTAXC}{{\color{\colorSYNTAX}\texttt{return}}}\endgroup }\hspace*{0.33em}{\begingroup\renewcommand\colorMATH{\colorMATHB}\renewcommand\colorSYNTAX{\colorSYNTAXB}{{\color{\colorMATH}\ensuremath{\se}}}\endgroup } \Downarrow  {\begingroup\renewcommand\colorMATH{\colorMATHB}\renewcommand\colorSYNTAX{\colorSYNTAXB}{{\color{\colorMATH}\ensuremath{\sv}}}\endgroup }
		  }
		\end{gather*}\endgroup
		and the result holds.
	\end{subproof}
\item   {{\color{\colorMATH}\ensuremath{{\begingroup\renewcommand\colorMATH{\colorMATHC}\renewcommand\colorSYNTAX{\colorSYNTAXC}{{\color{\colorMATH}\ensuremath{\pe}}}\endgroup } = x: \tau _{1} \leftarrow  {\begingroup\renewcommand\colorMATH{\colorMATHC}\renewcommand\colorSYNTAX{\colorSYNTAXC}{{\color{\colorMATH}\ensuremath{\pe_{1}}}}\endgroup }\mathrel{;}{\begingroup\renewcommand\colorMATH{\colorMATHC}\renewcommand\colorSYNTAX{\colorSYNTAXC}{{\color{\colorMATH}\ensuremath{\pe_{2}}}}\endgroup }}}}
	\begin{subproof} 
		We know that 
		\begingroup\color{\colorMATH}\begin{gather*} 
		  \inferrule*[lab={\textsc{ bind}} 
		  ]{  \gamma  \vdash  {\begingroup\renewcommand\colorMATH{\colorMATHC}\renewcommand\colorSYNTAX{\colorSYNTAXC}{{\color{\colorMATH}\ensuremath{\pe_{1}}}}\endgroup } \Downarrow ^{k'} \dist[1]
		  \\  \forall  {\begingroup\renewcommand\colorMATH{\colorMATHB}\renewcommand\colorSYNTAX{\colorSYNTAXB}{{\color{\colorMATH}\ensuremath{\sv_{i}}}}\endgroup } \in  \Sup{\dist[1]}, \gamma [x \mapsto  {\begingroup\renewcommand\colorMATH{\colorMATHB}\renewcommand\colorSYNTAX{\colorSYNTAXB}{{\color{\colorMATH}\ensuremath{\sv_{i}}}}\endgroup }] \vdash  {\begingroup\renewcommand\colorMATH{\colorMATHC}\renewcommand\colorSYNTAX{\colorSYNTAXC}{{\color{\colorMATH}\ensuremath{\pe_{2}}}}\endgroup } \Downarrow ^{k_i} \dist[2i]
		    }{
		      \gamma  \vdash  x: \tau _{1} \leftarrow  {\begingroup\renewcommand\colorMATH{\colorMATHC}\renewcommand\colorSYNTAX{\colorSYNTAXC}{{\color{\colorMATH}\ensuremath{\pe_{1}}}}\endgroup }\mathrel{;}{\begingroup\renewcommand\colorMATH{\colorMATHC}\renewcommand\colorSYNTAX{\colorSYNTAXC}{{\color{\colorMATH}\ensuremath{\pe_{2}}}}\endgroup } \Downarrow ^{k'+\mathsf{max}_{i} k_i} \dist
		  }
		\end{gather*}\endgroup
		where {{\color{\colorMATH}\ensuremath{\dist = \lambda x. \sum_{{\begingroup\renewcommand\colorMATH{\colorMATHB}\renewcommand\colorSYNTAX{\colorSYNTAXB}{{\color{\colorMATH}\ensuremath{\sv_{i}}}}\endgroup } \in  \Sup{\dist[1]}} \dist[1]({\begingroup\renewcommand\colorMATH{\colorMATHB}\renewcommand\colorSYNTAX{\colorSYNTAXB}{{\color{\colorMATH}\ensuremath{\sv_{i}}}}\endgroup })\mathord{\cdotp }\dist[2i](x)}}}.

		Let {{\color{\colorMATH}\ensuremath{{\begingroup\renewcommand\colorMATH{\colorMATHB}\renewcommand\colorSYNTAX{\colorSYNTAXB}{{\color{\colorMATH}\ensuremath{\sv}}}\endgroup }}}} such that {{\color{\colorMATH}\ensuremath{\dist({\begingroup\renewcommand\colorMATH{\colorMATHB}\renewcommand\colorSYNTAX{\colorSYNTAXB}{{\color{\colorMATH}\ensuremath{\sv}}}\endgroup })>0}}}, then it must be the case that for some {{\color{\colorMATH}\ensuremath{{\begingroup\renewcommand\colorMATH{\colorMATHB}\renewcommand\colorSYNTAX{\colorSYNTAXB}{{\color{\colorMATH}\ensuremath{\sv_{j}}}}\endgroup } \in  \Sup{\dist[1]}}}}, {{\color{\colorMATH}\ensuremath{\dist[1]({\begingroup\renewcommand\colorMATH{\colorMATHB}\renewcommand\colorSYNTAX{\colorSYNTAXB}{{\color{\colorMATH}\ensuremath{\sv_{j}}}}\endgroup })\mathord{\cdotp }\dist[2j]({\begingroup\renewcommand\colorMATH{\colorMATHB}\renewcommand\colorSYNTAX{\colorSYNTAXB}{{\color{\colorMATH}\ensuremath{\sv}}}\endgroup }) > 0}}}, i.e. {{\color{\colorMATH}\ensuremath{\dist[1]({\begingroup\renewcommand\colorMATH{\colorMATHB}\renewcommand\colorSYNTAX{\colorSYNTAXB}{{\color{\colorMATH}\ensuremath{\sv_{j}}}}\endgroup })>0}}} and {{\color{\colorMATH}\ensuremath{\dist[2j]({\begingroup\renewcommand\colorMATH{\colorMATHB}\renewcommand\colorSYNTAX{\colorSYNTAXB}{{\color{\colorMATH}\ensuremath{\sv}}}\endgroup }) > 0}}}.
		By induction hypothesis in {{\color{\colorMATH}\ensuremath{k'<k}}} (note that every probabilistic expression takes at least one step of reduction), as {{\color{\colorMATH}\ensuremath{\dist[1]({\begingroup\renewcommand\colorMATH{\colorMATHB}\renewcommand\colorSYNTAX{\colorSYNTAXB}{{\color{\colorMATH}\ensuremath{\sv_{j}}}}\endgroup })>0}}}, we know that {{\color{\colorMATH}\ensuremath{\gamma  \vdash  {\begingroup\renewcommand\colorMATH{\colorMATHC}\renewcommand\colorSYNTAX{\colorSYNTAXC}{{\color{\colorMATH}\ensuremath{\pe_{1}}}}\endgroup } \Downarrow  {\begingroup\renewcommand\colorMATH{\colorMATHB}\renewcommand\colorSYNTAX{\colorSYNTAXB}{{\color{\colorMATH}\ensuremath{\sv_{j}}}}\endgroup }}}}. Also as {{\color{\colorMATH}\ensuremath{\gamma [x \mapsto  {\begingroup\renewcommand\colorMATH{\colorMATHB}\renewcommand\colorSYNTAX{\colorSYNTAXB}{{\color{\colorMATH}\ensuremath{\sv_{j}}}}\endgroup }] \vdash  {\begingroup\renewcommand\colorMATH{\colorMATHC}\renewcommand\colorSYNTAX{\colorSYNTAXC}{{\color{\colorMATH}\ensuremath{\pe_{2}}}}\endgroup } \Downarrow ^{k_j} \dist[2j]}}}, then by induction hypothesis in {{\color{\colorMATH}\ensuremath{k_i < k}}}, as {{\color{\colorMATH}\ensuremath{\dist[2j]({\begingroup\renewcommand\colorMATH{\colorMATHB}\renewcommand\colorSYNTAX{\colorSYNTAXB}{{\color{\colorMATH}\ensuremath{\sv}}}\endgroup }) > 0}}}, we know that {{\color{\colorMATH}\ensuremath{\gamma [x \mapsto  {\begingroup\renewcommand\colorMATH{\colorMATHB}\renewcommand\colorSYNTAX{\colorSYNTAXB}{{\color{\colorMATH}\ensuremath{\sv_{j}}}}\endgroup }] \vdash  {\begingroup\renewcommand\colorMATH{\colorMATHC}\renewcommand\colorSYNTAX{\colorSYNTAXC}{{\color{\colorMATH}\ensuremath{\pe_{2}}}}\endgroup } \Downarrow  {\begingroup\renewcommand\colorMATH{\colorMATHB}\renewcommand\colorSYNTAX{\colorSYNTAXB}{{\color{\colorMATH}\ensuremath{\sv}}}\endgroup }}}}, thus		
		\begingroup\color{\colorMATH}\begin{gather*} 
		\inferrule*[lab={\textsc{ case-left}}
		]{  \gamma  \vdash  {\begingroup\renewcommand\colorMATH{\colorMATHC}\renewcommand\colorSYNTAX{\colorSYNTAXC}{{\color{\colorMATH}\ensuremath{\pe_{1}}}}\endgroup } \Downarrow  {\begingroup\renewcommand\colorMATH{\colorMATHB}\renewcommand\colorSYNTAX{\colorSYNTAXB}{{\color{\colorMATH}\ensuremath{\sv_{1}}}}\endgroup }
		  \\  \gamma [x \mapsto  {\begingroup\renewcommand\colorMATH{\colorMATHB}\renewcommand\colorSYNTAX{\colorSYNTAXB}{{\color{\colorMATH}\ensuremath{\sv_{1}}}}\endgroup }] \vdash  {\begingroup\renewcommand\colorMATH{\colorMATHC}\renewcommand\colorSYNTAX{\colorSYNTAXC}{{\color{\colorMATH}\ensuremath{\pe_{2}}}}\endgroup } \Downarrow  {\begingroup\renewcommand\colorMATH{\colorMATHB}\renewcommand\colorSYNTAX{\colorSYNTAXB}{{\color{\colorMATH}\ensuremath{\sv}}}\endgroup }
		    }{
		      \gamma  \vdash  x: \tau _{1} \leftarrow  {\begingroup\renewcommand\colorMATH{\colorMATHC}\renewcommand\colorSYNTAX{\colorSYNTAXC}{{\color{\colorMATH}\ensuremath{\pe_{1}}}}\endgroup }\mathrel{;}{\begingroup\renewcommand\colorMATH{\colorMATHC}\renewcommand\colorSYNTAX{\colorSYNTAXC}{{\color{\colorMATH}\ensuremath{\pe_{2}}}}\endgroup } \Downarrow  {\begingroup\renewcommand\colorMATH{\colorMATHB}\renewcommand\colorSYNTAX{\colorSYNTAXB}{{\color{\colorMATH}\ensuremath{\sv}}}\endgroup }
		  }
		\end{gather*}\endgroup
		and the result holds.
	\end{subproof}
\item  {{\color{\colorMATH}\ensuremath{{\begingroup\renewcommand\colorMATH{\colorMATHC}\renewcommand\colorSYNTAX{\colorSYNTAXC}{{\color{\colorMATH}\ensuremath{\pe}}}\endgroup } = {\begingroup\renewcommand\colorMATH{\colorMATHC}\renewcommand\colorSYNTAX{\colorSYNTAXC}{{\color{\colorSYNTAX}\texttt{case}}}\endgroup }\hspace*{0.33em}{\begingroup\renewcommand\colorMATH{\colorMATHC}\renewcommand\colorSYNTAX{\colorSYNTAXC}{{\color{\colorMATH}\ensuremath{\pe_{1}}}}\endgroup }\hspace*{0.33em}{\begingroup\renewcommand\colorMATH{\colorMATHC}\renewcommand\colorSYNTAX{\colorSYNTAXC}{{\color{\colorSYNTAX}\texttt{of}}}\endgroup }\hspace*{0.33em}\{ x\Rightarrow  {\begingroup\renewcommand\colorMATH{\colorMATHC}\renewcommand\colorSYNTAX{\colorSYNTAXC}{{\color{\colorMATH}\ensuremath{\pe_{2}}}}\endgroup } \} \hspace*{0.33em}\{ x\Rightarrow  {\begingroup\renewcommand\colorMATH{\colorMATHC}\renewcommand\colorSYNTAX{\colorSYNTAXC}{{\color{\colorMATH}\ensuremath{\pe_{3}}}}\endgroup }\} }}}
	\begin{subproof} 
		We know that 
		\begingroup\color{\colorMATH}\begin{gather*} 
		   \inferrule*[lab={\textsc{ case-left}}
		   ]{ \gamma \vdash  {\begingroup\renewcommand\colorMATH{\colorMATHB}\renewcommand\colorSYNTAX{\colorSYNTAXB}{{\color{\colorMATH}\ensuremath{\se_{1}}}}\endgroup }  \Downarrow ^{k_{1}} \inl\hspace*{0.33em}{\begingroup\renewcommand\colorMATH{\colorMATHB}\renewcommand\colorSYNTAX{\colorSYNTAXB}{{\color{\colorMATH}\ensuremath{\sv_{1}}}}\endgroup }
		   \\ \gamma [x\mapsto {\begingroup\renewcommand\colorMATH{\colorMATHB}\renewcommand\colorSYNTAX{\colorSYNTAXB}{{\color{\colorMATH}\ensuremath{\sv_{1}}}}\endgroup } ]\vdash {\begingroup\renewcommand\colorMATH{\colorMATHC}\renewcommand\colorSYNTAX{\colorSYNTAXC}{{\color{\colorMATH}\ensuremath{\pe_{2}}}}\endgroup }  \Downarrow ^{k_{2}} \dist
		      }{
		      \gamma \vdash {\begingroup\renewcommand\colorMATH{\colorMATHC}\renewcommand\colorSYNTAX{\colorSYNTAXC}{{\color{\colorSYNTAX}\texttt{case}}}\endgroup }\hspace*{0.33em}{\begingroup\renewcommand\colorMATH{\colorMATHC}\renewcommand\colorSYNTAX{\colorSYNTAXC}{{\color{\colorMATH}\ensuremath{\pe}}}\endgroup }\hspace*{0.33em}{\begingroup\renewcommand\colorMATH{\colorMATHC}\renewcommand\colorSYNTAX{\colorSYNTAXC}{{\color{\colorSYNTAX}\texttt{of}}}\endgroup }\hspace*{0.33em}\{ x\Rightarrow  {\begingroup\renewcommand\colorMATH{\colorMATHC}\renewcommand\colorSYNTAX{\colorSYNTAXC}{{\color{\colorMATH}\ensuremath{\pe_{2}}}}\endgroup } \} \hspace*{0.33em}\{ x\Rightarrow  {\begingroup\renewcommand\colorMATH{\colorMATHC}\renewcommand\colorSYNTAX{\colorSYNTAXC}{{\color{\colorMATH}\ensuremath{\pe_{3}}}}\endgroup }\}  \Downarrow ^{k_{1}+k_{2}} \dist
		   }
		\end{gather*}\endgroup

		Let {{\color{\colorMATH}\ensuremath{{\begingroup\renewcommand\colorMATH{\colorMATHB}\renewcommand\colorSYNTAX{\colorSYNTAXB}{{\color{\colorMATH}\ensuremath{\sv}}}\endgroup }}}} such that {{\color{\colorMATH}\ensuremath{\dist({\begingroup\renewcommand\colorMATH{\colorMATHB}\renewcommand\colorSYNTAX{\colorSYNTAXB}{{\color{\colorMATH}\ensuremath{\sv}}}\endgroup })>0}}}.
		By induction hypothesis in {{\color{\colorMATH}\ensuremath{{\begingroup\renewcommand\colorMATH{\colorMATHC}\renewcommand\colorSYNTAX{\colorSYNTAXC}{{\color{\colorMATH}\ensuremath{\pe_{2}}}}\endgroup }}}} on {{\color{\colorMATH}\ensuremath{\gamma [x\mapsto {\begingroup\renewcommand\colorMATH{\colorMATHB}\renewcommand\colorSYNTAX{\colorSYNTAXB}{{\color{\colorMATH}\ensuremath{\sv_{1}}}}\endgroup } ]\vdash {\begingroup\renewcommand\colorMATH{\colorMATHC}\renewcommand\colorSYNTAX{\colorSYNTAXC}{{\color{\colorMATH}\ensuremath{\pe_{2}}}}\endgroup }  \Downarrow ^{k_{2}} \dist}}}, then we know that {{\color{\colorMATH}\ensuremath{\gamma [x\mapsto {\begingroup\renewcommand\colorMATH{\colorMATHB}\renewcommand\colorSYNTAX{\colorSYNTAXB}{{\color{\colorMATH}\ensuremath{\sv_{1}}}}\endgroup } ]\vdash  {\begingroup\renewcommand\colorMATH{\colorMATHC}\renewcommand\colorSYNTAX{\colorSYNTAXC}{{\color{\colorMATH}\ensuremath{\pe_{2}}}}\endgroup } \Downarrow  {\begingroup\renewcommand\colorMATH{\colorMATHB}\renewcommand\colorSYNTAX{\colorSYNTAXB}{{\color{\colorMATH}\ensuremath{\sv}}}\endgroup }}}}, thus		
		\begingroup\color{\colorMATH}\begin{gather*} 
		\inferrule*[lab={\textsc{ case-left}}
	   ]{ \gamma \vdash  {\begingroup\renewcommand\colorMATH{\colorMATHB}\renewcommand\colorSYNTAX{\colorSYNTAXB}{{\color{\colorMATH}\ensuremath{\se_{1}}}}\endgroup }  \Downarrow  \inl\hspace*{0.33em}{\begingroup\renewcommand\colorMATH{\colorMATHB}\renewcommand\colorSYNTAX{\colorSYNTAXB}{{\color{\colorMATH}\ensuremath{\sv_{1}}}}\endgroup } 
	   \\ \gamma [x\mapsto {\begingroup\renewcommand\colorMATH{\colorMATHB}\renewcommand\colorSYNTAX{\colorSYNTAXB}{{\color{\colorMATH}\ensuremath{\sv_{1}}}}\endgroup } ]\vdash {\begingroup\renewcommand\colorMATH{\colorMATHC}\renewcommand\colorSYNTAX{\colorSYNTAXC}{{\color{\colorMATH}\ensuremath{\pe_{2}}}}\endgroup }  \Downarrow  {\begingroup\renewcommand\colorMATH{\colorMATHB}\renewcommand\colorSYNTAX{\colorSYNTAXB}{{\color{\colorMATH}\ensuremath{\sv}}}\endgroup } 
	      }{
	      \gamma \vdash {{\color{\colorSYNTAX}\texttt{case}}}\hspace*{0.33em}{\begingroup\renewcommand\colorMATH{\colorMATHC}\renewcommand\colorSYNTAX{\colorSYNTAXC}{{\color{\colorMATH}\ensuremath{\pe_{1}}}}\endgroup } \hspace*{0.33em}\{ x\Rightarrow  {\begingroup\renewcommand\colorMATH{\colorMATHC}\renewcommand\colorSYNTAX{\colorSYNTAXC}{{\color{\colorMATH}\ensuremath{\pe_{2}}}}\endgroup } \} \hspace*{0.33em}\{ x\Rightarrow  {\begingroup\renewcommand\colorMATH{\colorMATHC}\renewcommand\colorSYNTAX{\colorSYNTAXC}{{\color{\colorMATH}\ensuremath{\pe_{3}}}}\endgroup }\}  \Downarrow  {\begingroup\renewcommand\colorMATH{\colorMATHB}\renewcommand\colorSYNTAX{\colorSYNTAXB}{{\color{\colorMATH}\ensuremath{\sv}}}\endgroup }
	   }
		\end{gather*}\endgroup
		and the result holds.
	\end{subproof}
\end{enumerate}

\end{proof}

Thus, type safety of the non-deterministic semantics implies type safety for the distribution semantics.\footnote{The other direction {{\color{\colorMATH}\ensuremath{{\gamma  \vdash  {\begingroup\renewcommand\colorMATH{\colorMATHC}\renewcommand\colorSYNTAX{\colorSYNTAXC}{{\color{\colorMATH}\ensuremath{\pe}}}\endgroup } \Downarrow  {\begingroup\renewcommand\colorMATH{\colorMATHB}\renewcommand\colorSYNTAX{\colorSYNTAXB}{{\color{\colorMATH}\ensuremath{\sv}}}\endgroup } \implies  \exists k. \gamma  \vdash  {\begingroup\renewcommand\colorMATH{\colorMATHC}\renewcommand\colorSYNTAX{\colorSYNTAXC}{{\color{\colorMATH}\ensuremath{\pe}}}\endgroup }\Downarrow ^{k} \dist \wedge  \dist({\begingroup\renewcommand\colorMATH{\colorMATHB}\renewcommand\colorSYNTAX{\colorSYNTAXB}{{\color{\colorMATH}\ensuremath{\sv}}}\endgroup })>0}}}} would establish soundness of the non-deterministic semantics; given that the language does not feature recursion, we believe that it holds, although this is left for future work.}

\begin{figure}[t]
\begin{small}
\begin{framed}
\begingroup\color{\colorMATH}\begin{mathpar}\inferrule*[lab={\textsc{ return}}
  ]{ \gamma  \vdash  {\begingroup\renewcommand\colorMATH{\colorMATHB}\renewcommand\colorSYNTAX{\colorSYNTAXB}{{\color{\colorMATH}\ensuremath{\se}}}\endgroup } \Downarrow  {\begingroup\renewcommand\colorMATH{\colorMATHB}\renewcommand\colorSYNTAX{\colorSYNTAXB}{{\color{\colorMATH}\ensuremath{\sv}}}\endgroup }
    }{
     {\begingroup\renewcommand\colorMATH{\colorMATHC}\renewcommand\colorSYNTAX{\colorSYNTAXC}{{\color{\colorSYNTAX}\texttt{return}}}\endgroup }\hspace*{0.33em}{\begingroup\renewcommand\colorMATH{\colorMATHB}\renewcommand\colorSYNTAX{\colorSYNTAXB}{{\color{\colorMATH}\ensuremath{\se}}}\endgroup } \Downarrow  {\begingroup\renewcommand\colorMATH{\colorMATHB}\renewcommand\colorSYNTAX{\colorSYNTAXB}{{\color{\colorMATH}\ensuremath{\sv}}}\endgroup }
  }
\and\inferrule*[lab={\textsc{ bind}}
  ]{  \gamma  \vdash  {\begingroup\renewcommand\colorMATH{\colorMATHC}\renewcommand\colorSYNTAX{\colorSYNTAXC}{{\color{\colorMATH}\ensuremath{\pe_{1}}}}\endgroup } \Downarrow  {\begingroup\renewcommand\colorMATH{\colorMATHB}\renewcommand\colorSYNTAX{\colorSYNTAXB}{{\color{\colorMATH}\ensuremath{\sv_{1}}}}\endgroup }
  \\  \gamma [x \mapsto  {\begingroup\renewcommand\colorMATH{\colorMATHB}\renewcommand\colorSYNTAX{\colorSYNTAXB}{{\color{\colorMATH}\ensuremath{\sv_{1}}}}\endgroup }] \vdash  {\begingroup\renewcommand\colorMATH{\colorMATHC}\renewcommand\colorSYNTAX{\colorSYNTAXC}{{\color{\colorMATH}\ensuremath{\pe_{2}}}}\endgroup } \Downarrow  {\begingroup\renewcommand\colorMATH{\colorMATHB}\renewcommand\colorSYNTAX{\colorSYNTAXB}{{\color{\colorMATH}\ensuremath{\sv_{2}}}}\endgroup }
    }{
      \gamma  \vdash  x: \tau _{1} \leftarrow  {\begingroup\renewcommand\colorMATH{\colorMATHC}\renewcommand\colorSYNTAX{\colorSYNTAXC}{{\color{\colorMATH}\ensuremath{\pe_{1}}}}\endgroup }\mathrel{;}{\begingroup\renewcommand\colorMATH{\colorMATHC}\renewcommand\colorSYNTAX{\colorSYNTAXC}{{\color{\colorMATH}\ensuremath{\pe_{2}}}}\endgroup } \Downarrow  {\begingroup\renewcommand\colorMATH{\colorMATHB}\renewcommand\colorSYNTAX{\colorSYNTAXB}{{\color{\colorMATH}\ensuremath{\sv_{2}}}}\endgroup }
  }
\and\inferrule*[lab={\textsc{ gauss}}
  ]{  {\begingroup\renewcommand\colorMATH{\colorMATHB}\renewcommand\colorSYNTAX{\colorSYNTAXB}{{\color{\colorMATH}\ensuremath{r}}}\endgroup } \in  {\begingroup\renewcommand\colorMATH{\colorMATHA}\renewcommand\colorSYNTAX{\colorSYNTAXA}{{\color{\colorSYNTAX}\texttt{{\ensuremath{{\mathbb{R}}}}}}}\endgroup }
    }{
      \gamma  \vdash  {{\color{\colorSYNTAX}\texttt{gauss}}} \hspace*{0.33em} \mu  \hspace*{0.33em} \sigma ^{2} \Downarrow  {\begingroup\renewcommand\colorMATH{\colorMATHB}\renewcommand\colorSYNTAX{\colorSYNTAXB}{{\color{\colorMATH}\ensuremath{r}}}\endgroup }
  }
\and\inferrule*[lab={\textsc{ if-true}}
  ]{ \gamma  \vdash  {\begingroup\renewcommand\colorMATH{\colorMATHB}\renewcommand\colorSYNTAX{\colorSYNTAXB}{{\color{\colorMATH}\ensuremath{\se_{1}}}}\endgroup } \Downarrow  {\text{true}}
  \\ \gamma  \vdash  {\begingroup\renewcommand\colorMATH{\colorMATHC}\renewcommand\colorSYNTAX{\colorSYNTAXC}{{\color{\colorMATH}\ensuremath{\pe_{2}}}}\endgroup } \Downarrow  {\begingroup\renewcommand\colorMATH{\colorMATHB}\renewcommand\colorSYNTAX{\colorSYNTAXB}{{\color{\colorMATH}\ensuremath{\sv_{2}}}}\endgroup }
     }{
      \gamma  \vdash  {{\color{\colorSYNTAX}\texttt{if}}}\hspace*{0.33em}{\begingroup\renewcommand\colorMATH{\colorMATHB}\renewcommand\colorSYNTAX{\colorSYNTAXB}{{\color{\colorMATH}\ensuremath{\se_{1}}}}\endgroup }\hspace*{0.33em}{{\color{\colorSYNTAX}\texttt{then}}}\hspace*{0.33em}\{ {\begingroup\renewcommand\colorMATH{\colorMATHC}\renewcommand\colorSYNTAX{\colorSYNTAXC}{{\color{\colorMATH}\ensuremath{\pe_{2}}}}\endgroup }\} \hspace*{0.33em}{{\color{\colorSYNTAX}\texttt{else}}}\hspace*{0.33em}\{ {\begingroup\renewcommand\colorMATH{\colorMATHC}\renewcommand\colorSYNTAX{\colorSYNTAXC}{{\color{\colorMATH}\ensuremath{\pe_{3}}}}\endgroup }\}  \Downarrow  {\begingroup\renewcommand\colorMATH{\colorMATHB}\renewcommand\colorSYNTAX{\colorSYNTAXB}{{\color{\colorMATH}\ensuremath{\sv_{2}}}}\endgroup }
  }
\and\inferrule*[lab={\textsc{ if-false}}
  ]{ \gamma  \vdash  {\begingroup\renewcommand\colorMATH{\colorMATHB}\renewcommand\colorSYNTAX{\colorSYNTAXB}{{\color{\colorMATH}\ensuremath{\se_{1}}}}\endgroup } \Downarrow  {\text{false}}
  \\ \gamma  \vdash  {\begingroup\renewcommand\colorMATH{\colorMATHC}\renewcommand\colorSYNTAX{\colorSYNTAXC}{{\color{\colorMATH}\ensuremath{\pe_{3}}}}\endgroup } \Downarrow  {\begingroup\renewcommand\colorMATH{\colorMATHB}\renewcommand\colorSYNTAX{\colorSYNTAXB}{{\color{\colorMATH}\ensuremath{\sv_{3}}}}\endgroup }
     }{
      \gamma  \vdash  {{\color{\colorSYNTAX}\texttt{if}}}\hspace*{0.33em}{\begingroup\renewcommand\colorMATH{\colorMATHB}\renewcommand\colorSYNTAX{\colorSYNTAXB}{{\color{\colorMATH}\ensuremath{\se_{1}}}}\endgroup }\hspace*{0.33em}{{\color{\colorSYNTAX}\texttt{then}}}\hspace*{0.33em}\{ {\begingroup\renewcommand\colorMATH{\colorMATHC}\renewcommand\colorSYNTAX{\colorSYNTAXC}{{\color{\colorMATH}\ensuremath{\pe_{2}}}}\endgroup }\} \hspace*{0.33em}{{\color{\colorSYNTAX}\texttt{else}}}\hspace*{0.33em}\{ {\begingroup\renewcommand\colorMATH{\colorMATHC}\renewcommand\colorSYNTAX{\colorSYNTAXC}{{\color{\colorMATH}\ensuremath{\pe_{3}}}}\endgroup }\}  \Downarrow  {\begingroup\renewcommand\colorMATH{\colorMATHB}\renewcommand\colorSYNTAX{\colorSYNTAXB}{{\color{\colorMATH}\ensuremath{\sv_{3}}}}\endgroup }
  }
\and \inferrule*[lab={\textsc{ case-left}}
   ]{ \gamma \vdash  {\begingroup\renewcommand\colorMATH{\colorMATHC}\renewcommand\colorSYNTAX{\colorSYNTAXC}{{\color{\colorMATH}\ensuremath{\pe}}}\endgroup }  \Downarrow  \inl\hspace*{0.33em}{\begingroup\renewcommand\colorMATH{\colorMATHB}\renewcommand\colorSYNTAX{\colorSYNTAXB}{{\color{\colorMATH}\ensuremath{\sv}}}\endgroup } 
   \\ \gamma [x\mapsto {\begingroup\renewcommand\colorMATH{\colorMATHB}\renewcommand\colorSYNTAX{\colorSYNTAXB}{{\color{\colorMATH}\ensuremath{\sv}}}\endgroup } ]\vdash {\begingroup\renewcommand\colorMATH{\colorMATHC}\renewcommand\colorSYNTAX{\colorSYNTAXC}{{\color{\colorMATH}\ensuremath{\pe_{2}}}}\endgroup }  \Downarrow  {\begingroup\renewcommand\colorMATH{\colorMATHB}\renewcommand\colorSYNTAX{\colorSYNTAXB}{{\color{\colorMATH}\ensuremath{\sv_{2}}}}\endgroup } 
      }{
      \gamma \vdash {{\color{\colorSYNTAX}\texttt{case}}}\hspace*{0.33em}{\begingroup\renewcommand\colorMATH{\colorMATHC}\renewcommand\colorSYNTAX{\colorSYNTAXC}{{\color{\colorMATH}\ensuremath{\pe}}}\endgroup } \hspace*{0.33em}\{ x\Rightarrow  {\begingroup\renewcommand\colorMATH{\colorMATHC}\renewcommand\colorSYNTAX{\colorSYNTAXC}{{\color{\colorMATH}\ensuremath{\pe_{2}}}}\endgroup } \} \hspace*{0.33em}\{ x\Rightarrow  {\begingroup\renewcommand\colorMATH{\colorMATHC}\renewcommand\colorSYNTAX{\colorSYNTAXC}{{\color{\colorMATH}\ensuremath{\pe_{3}}}}\endgroup }\}  \Downarrow  {\begingroup\renewcommand\colorMATH{\colorMATHB}\renewcommand\colorSYNTAX{\colorSYNTAXB}{{\color{\colorMATH}\ensuremath{\sv_{2}}}}\endgroup }
   }
\and \inferrule*[lab={\textsc{ case-right}}
   ]{ \gamma \vdash  {\begingroup\renewcommand\colorMATH{\colorMATHC}\renewcommand\colorSYNTAX{\colorSYNTAXC}{{\color{\colorMATH}\ensuremath{\pe}}}\endgroup }  \Downarrow  \inr\hspace*{0.33em}{\begingroup\renewcommand\colorMATH{\colorMATHB}\renewcommand\colorSYNTAX{\colorSYNTAXB}{{\color{\colorMATH}\ensuremath{\sv}}}\endgroup } 
   \\ \gamma [x\mapsto {\begingroup\renewcommand\colorMATH{\colorMATHB}\renewcommand\colorSYNTAX{\colorSYNTAXB}{{\color{\colorMATH}\ensuremath{\sv}}}\endgroup } ]\vdash {\begingroup\renewcommand\colorMATH{\colorMATHC}\renewcommand\colorSYNTAX{\colorSYNTAXC}{{\color{\colorMATH}\ensuremath{\pe_{3}}}}\endgroup }  \Downarrow  {\begingroup\renewcommand\colorMATH{\colorMATHB}\renewcommand\colorSYNTAX{\colorSYNTAXB}{{\color{\colorMATH}\ensuremath{\sv_{3}}}}\endgroup } 
      }{
      \gamma \vdash {{\color{\colorSYNTAX}\texttt{case}}}\hspace*{0.33em}{\begingroup\renewcommand\colorMATH{\colorMATHC}\renewcommand\colorSYNTAX{\colorSYNTAXC}{{\color{\colorMATH}\ensuremath{\pe}}}\endgroup } \hspace*{0.33em}\{ x\Rightarrow  {\begingroup\renewcommand\colorMATH{\colorMATHC}\renewcommand\colorSYNTAX{\colorSYNTAXC}{{\color{\colorMATH}\ensuremath{\pe_{2}}}}\endgroup } \} \hspace*{0.33em}\{ x\Rightarrow  {\begingroup\renewcommand\colorMATH{\colorMATHC}\renewcommand\colorSYNTAX{\colorSYNTAXC}{{\color{\colorMATH}\ensuremath{\pe_{3}}}}\endgroup }\}  \Downarrow  {\begingroup\renewcommand\colorMATH{\colorMATHB}\renewcommand\colorSYNTAX{\colorSYNTAXB}{{\color{\colorMATH}\ensuremath{\sv_{3}}}}\endgroup }
   }
\and \inferrule*[lab={\textsc{ app}}
   ]{ \gamma \vdash  {\begingroup\renewcommand\colorMATH{\colorMATHC}\renewcommand\colorSYNTAX{\colorSYNTAXC}{{\color{\colorMATH}\ensuremath{\pe_{1}}}}\endgroup }  \Downarrow  \langle \lambda x:\tau \mathord{\cdotp } {\begingroup\renewcommand\colorMATH{\colorMATHB}\renewcommand\colorSYNTAX{\colorSYNTAXB}{{\color{\colorMATH}\ensuremath{\sss}}}\endgroup }.\hspace*{0.33em}{\begingroup\renewcommand\colorMATH{\colorMATHC}\renewcommand\colorSYNTAX{\colorSYNTAXC}{{\color{\colorMATH}\ensuremath{\pe^{\prime}}}}\endgroup },\gamma ^{\prime}\rangle 
   \\ \gamma \vdash  {\begingroup\renewcommand\colorMATH{\colorMATHC}\renewcommand\colorSYNTAX{\colorSYNTAXC}{{\color{\colorMATH}\ensuremath{\pe_{2}}}}\endgroup }  \Downarrow  {\begingroup\renewcommand\colorMATH{\colorMATHB}\renewcommand\colorSYNTAX{\colorSYNTAXB}{{\color{\colorMATH}\ensuremath{\sv}}}\endgroup } 
   \\ \gamma ^{\prime}[x\mapsto {\begingroup\renewcommand\colorMATH{\colorMATHB}\renewcommand\colorSYNTAX{\colorSYNTAXB}{{\color{\colorMATH}\ensuremath{\sv}}}\endgroup } ]\vdash  {\begingroup\renewcommand\colorMATH{\colorMATHC}\renewcommand\colorSYNTAX{\colorSYNTAXC}{{\color{\colorMATH}\ensuremath{\pe^{\prime}}}}\endgroup } \Downarrow  {\begingroup\renewcommand\colorMATH{\colorMATHB}\renewcommand\colorSYNTAX{\colorSYNTAXB}{{\color{\colorMATH}\ensuremath{\sv'}}}\endgroup }
      }{
      \gamma \vdash  {\begingroup\renewcommand\colorMATH{\colorMATHC}\renewcommand\colorSYNTAX{\colorSYNTAXC}{{\color{\colorMATH}\ensuremath{\pe_{1}}}}\endgroup }\hspace*{0.33em}{\begingroup\renewcommand\colorMATH{\colorMATHC}\renewcommand\colorSYNTAX{\colorSYNTAXC}{{\color{\colorMATH}\ensuremath{\pe_{2}}}}\endgroup }  \Downarrow  {\begingroup\renewcommand\colorMATH{\colorMATHB}\renewcommand\colorSYNTAX{\colorSYNTAXB}{{\color{\colorMATH}\ensuremath{\sv'}}}\endgroup }
   }
\end{mathpar}\endgroup
\end{framed}
\end{small}
\caption{Non-deterministic sampling semantics for privacy expressions}
\label{afig:interpreter-prob-sem}
\end{figure}

\begin{figure}[t]
\begin{small}
\begin{framed}
\begingroup\color{\colorMATH}\begin{mathpar}
   \inferrule*[lab={\textsc{ }}
   ]{ \varnothing ; \varnothing  \vdash  {\begingroup\renewcommand\colorMATH{\colorMATHB}\renewcommand\colorSYNTAX{\colorSYNTAXB}{{\color{\colorMATH}\ensuremath{\sv}}}\endgroup } : \tau '; \varnothing 
   \\  \tau ' <: \tau 
      }{
      {\begingroup\renewcommand\colorMATH{\colorMATHB}\renewcommand\colorSYNTAX{\colorSYNTAXB}{{\color{\colorMATH}\ensuremath{\sv}}}\endgroup } \in  Atom\llbracket \tau \rrbracket 
   }
\and \inferrule*[lab={\textsc{ }}
   ]{ dom(\Gamma ) = dom(\gamma ) 
   \\ \forall  x \in  dom(\gamma ). 
   \\ \gamma (x) \in  {\mathcal{V}}\llbracket \Gamma (x)/\Gamma \rrbracket  
      }{
      \gamma  \in  {\mathcal{G}}\llbracket \Gamma \rrbracket 
   }
\and \inferrule*[lab={\textsc{ }}
   ]{ {\begingroup\renewcommand\colorMATH{\colorMATHB}\renewcommand\colorSYNTAX{\colorSYNTAXB}{{\color{\colorMATH}\ensuremath{r}}}\endgroup } \in  Atom\llbracket {\begingroup\renewcommand\colorMATH{\colorMATHA}\renewcommand\colorSYNTAX{\colorSYNTAXA}{{\color{\colorSYNTAX}\texttt{{\ensuremath{{\mathbb{R}}}}}}}\endgroup }\rrbracket 
      }{
      {\begingroup\renewcommand\colorMATH{\colorMATHB}\renewcommand\colorSYNTAX{\colorSYNTAXB}{{\color{\colorMATH}\ensuremath{r}}}\endgroup } \in  {\mathcal{V}}\llbracket {\begingroup\renewcommand\colorMATH{\colorMATHA}\renewcommand\colorSYNTAX{\colorSYNTAXA}{{\color{\colorSYNTAX}\texttt{{\ensuremath{{\mathbb{R}}}}}}}\endgroup }\rrbracket 
   }	
\and \inferrule*[lab={\textsc{ }}
   ]{ \ttt \in  Atom\llbracket {{\color{\colorSYNTAX}\texttt{unit}}}\rrbracket 
      }{
      \ttt \in  {\mathcal{V}}\llbracket {{\color{\colorSYNTAX}\texttt{unit}}}\rrbracket 
   }
\and \inferrule*[lab={\textsc{ }}
   ]{ \inl^{\tau '_{2}}\hspace*{0.33em}{\begingroup\renewcommand\colorMATH{\colorMATHB}\renewcommand\colorSYNTAX{\colorSYNTAXB}{{\color{\colorMATH}\ensuremath{\sv}}}\endgroup } \in  Atom\llbracket \tau _{1} \mathrel{^{{\begingroup\renewcommand\colorMATH{\colorMATHB}\renewcommand\colorSYNTAX{\colorSYNTAXB}{{\color{\colorMATH}\ensuremath{\varnothing }}}\endgroup }}\oplus ^{{\begingroup\renewcommand\colorMATH{\colorMATHB}\renewcommand\colorSYNTAX{\colorSYNTAXB}{{\color{\colorMATH}\ensuremath{\varnothing }}}\endgroup }}} \tau _{2}\rrbracket  
   \\ {\begingroup\renewcommand\colorMATH{\colorMATHB}\renewcommand\colorSYNTAX{\colorSYNTAXB}{{\color{\colorMATH}\ensuremath{\sv}}}\endgroup } \in  {\mathcal{V}}\llbracket \tau _{1}\rrbracket 
      }{
      \inl^{\tau '_{2}}\hspace*{0.33em}{\begingroup\renewcommand\colorMATH{\colorMATHB}\renewcommand\colorSYNTAX{\colorSYNTAXB}{{\color{\colorMATH}\ensuremath{\sv}}}\endgroup } \in  {\mathcal{V}}\llbracket \tau _{1} \mathrel{^{{\begingroup\renewcommand\colorMATH{\colorMATHB}\renewcommand\colorSYNTAX{\colorSYNTAXB}{{\color{\colorMATH}\ensuremath{\varnothing }}}\endgroup }}\oplus ^{{\begingroup\renewcommand\colorMATH{\colorMATHB}\renewcommand\colorSYNTAX{\colorSYNTAXB}{{\color{\colorMATH}\ensuremath{\varnothing }}}\endgroup }}} \tau _{2}\rrbracket 
   }
\and \inferrule*[lab={\textsc{ }}
   ]{ \inr^{\tau '_{1}}\hspace*{0.33em}{\begingroup\renewcommand\colorMATH{\colorMATHB}\renewcommand\colorSYNTAX{\colorSYNTAXB}{{\color{\colorMATH}\ensuremath{\sv}}}\endgroup } \in  Atom\llbracket \tau _{1} \mathrel{^{{\begingroup\renewcommand\colorMATH{\colorMATHB}\renewcommand\colorSYNTAX{\colorSYNTAXB}{{\color{\colorMATH}\ensuremath{\varnothing }}}\endgroup }}\oplus ^{{\begingroup\renewcommand\colorMATH{\colorMATHB}\renewcommand\colorSYNTAX{\colorSYNTAXB}{{\color{\colorMATH}\ensuremath{\varnothing }}}\endgroup }}} \tau _{2}\rrbracket  
   \\ {\begingroup\renewcommand\colorMATH{\colorMATHB}\renewcommand\colorSYNTAX{\colorSYNTAXB}{{\color{\colorMATH}\ensuremath{\sv}}}\endgroup } \in  {\mathcal{V}}\llbracket \tau _{2}\rrbracket 
      }{
      \inr^{\tau '_{1}}\hspace*{0.33em}{\begingroup\renewcommand\colorMATH{\colorMATHB}\renewcommand\colorSYNTAX{\colorSYNTAXB}{{\color{\colorMATH}\ensuremath{\sv}}}\endgroup } \in  {\mathcal{V}}\llbracket \tau _{1} \mathrel{^{{\begingroup\renewcommand\colorMATH{\colorMATHB}\renewcommand\colorSYNTAX{\colorSYNTAXB}{{\color{\colorMATH}\ensuremath{\varnothing }}}\endgroup }}\oplus ^{{\begingroup\renewcommand\colorMATH{\colorMATHB}\renewcommand\colorSYNTAX{\colorSYNTAXB}{{\color{\colorMATH}\ensuremath{\varnothing }}}\endgroup }}} \tau _{2}\rrbracket 
   }
\and \inferrule*[lab={\textsc{ }}
   ]{ \langle {\begingroup\renewcommand\colorMATH{\colorMATHB}\renewcommand\colorSYNTAX{\colorSYNTAXB}{{\color{\colorMATH}\ensuremath{\slambda}}}\endgroup } x:\tau \mathord{\cdotp }{\begingroup\renewcommand\colorMATH{\colorMATHB}\renewcommand\colorSYNTAX{\colorSYNTAXB}{{\color{\colorMATH}\ensuremath{\sss'}}}\endgroup }. {\begingroup\renewcommand\colorMATH{\colorMATHB}\renewcommand\colorSYNTAX{\colorSYNTAXB}{{\color{\colorMATH}\ensuremath{\se}}}\endgroup }, \gamma \rangle  \in  Atom\llbracket (x\mathrel{:}\tau _{1}\mathord{\cdotp }{\begingroup\renewcommand\colorMATH{\colorMATHB}\renewcommand\colorSYNTAX{\colorSYNTAXB}{{\color{\colorMATH}\ensuremath{\sss'}}}\endgroup }) \xrightarrowS {{\begingroup\renewcommand\colorMATH{\colorMATHB}\renewcommand\colorSYNTAX{\colorSYNTAXB}{{\color{\colorMATH}\ensuremath{\sss}}}\endgroup }x} \tau _{2}\rrbracket 
   \\ \forall   {\begingroup\renewcommand\colorMATH{\colorMATHB}\renewcommand\colorSYNTAX{\colorSYNTAXB}{{\color{\colorMATH}\ensuremath{\sv}}}\endgroup } \in  {\mathcal{V}}\llbracket \tau _{1}\rrbracket . \gamma [x \mapsto  {\begingroup\renewcommand\colorMATH{\colorMATHB}\renewcommand\colorSYNTAX{\colorSYNTAXB}{{\color{\colorMATH}\ensuremath{\sv}}}\endgroup }] \vdash  {\begingroup\renewcommand\colorMATH{\colorMATHB}\renewcommand\colorSYNTAX{\colorSYNTAXB}{{\color{\colorMATH}\ensuremath{\se}}}\endgroup } \in  {\mathcal{E}}\llbracket \tau _{2}/(x:\tau _{1})\rrbracket 
      }{
      \langle {\begingroup\renewcommand\colorMATH{\colorMATHB}\renewcommand\colorSYNTAX{\colorSYNTAXB}{{\color{\colorMATH}\ensuremath{\slambda}}}\endgroup } x:\tau _{1}\mathord{\cdotp }{\begingroup\renewcommand\colorMATH{\colorMATHB}\renewcommand\colorSYNTAX{\colorSYNTAXB}{{\color{\colorMATH}\ensuremath{\sss'}}}\endgroup }. {\begingroup\renewcommand\colorMATH{\colorMATHB}\renewcommand\colorSYNTAX{\colorSYNTAXB}{{\color{\colorMATH}\ensuremath{\se}}}\endgroup }, \gamma \rangle  \in  {\mathcal{V}}\llbracket (x\mathrel{:}\tau _{1}\mathord{\cdotp }{\begingroup\renewcommand\colorMATH{\colorMATHB}\renewcommand\colorSYNTAX{\colorSYNTAXB}{{\color{\colorMATH}\ensuremath{\sss'}}}\endgroup }) \xrightarrowS {{\begingroup\renewcommand\colorMATH{\colorMATHB}\renewcommand\colorSYNTAX{\colorSYNTAXB}{{\color{\colorMATH}\ensuremath{\sss}}}\endgroup }x} \tau _{2}\rrbracket 
   }
\and \inferrule*[lab={\textsc{ }}
   ]{ \langle {\begingroup\renewcommand\colorMATH{\colorMATHC}\renewcommand\colorSYNTAX{\colorSYNTAXC}{{\color{\colorMATH}\ensuremath{\plambda}}}\endgroup } x:\tau \mathord{\cdotp }{\begingroup\renewcommand\colorMATH{\colorMATHB}\renewcommand\colorSYNTAX{\colorSYNTAXB}{{\color{\colorMATH}\ensuremath{\sss}}}\endgroup }. {\begingroup\renewcommand\colorMATH{\colorMATHC}\renewcommand\colorSYNTAX{\colorSYNTAXC}{{\color{\colorMATH}\ensuremath{\pe}}}\endgroup }, \gamma \rangle  \in  Atom\llbracket (x\mathrel{:}\tau _{1}\mathord{\cdotp }{\begingroup\renewcommand\colorMATH{\colorMATHB}\renewcommand\colorSYNTAX{\colorSYNTAXB}{{\color{\colorMATH}\ensuremath{\sss}}}\endgroup }) \xrightarrowP {{\begingroup\renewcommand\colorMATH{\colorMATHC}\renewcommand\colorSYNTAX{\colorSYNTAXC}{{\color{\colorMATH}\ensuremath{\pS}}}\endgroup }} \tau _{2}\rrbracket 
   \\ \forall   {\begingroup\renewcommand\colorMATH{\colorMATHB}\renewcommand\colorSYNTAX{\colorSYNTAXB}{{\color{\colorMATH}\ensuremath{\sv}}}\endgroup } \in  {\mathcal{V}}\llbracket \tau _{1}\rrbracket . \gamma [x \mapsto  {\begingroup\renewcommand\colorMATH{\colorMATHB}\renewcommand\colorSYNTAX{\colorSYNTAXB}{{\color{\colorMATH}\ensuremath{\sv}}}\endgroup }] \vdash  {\begingroup\renewcommand\colorMATH{\colorMATHC}\renewcommand\colorSYNTAX{\colorSYNTAXC}{{\color{\colorMATH}\ensuremath{\pe}}}\endgroup } \in  {\mathcal{E}}\llbracket \tau _{2}/(x:\tau _{1})\rrbracket 
      }{
      \langle {\begingroup\renewcommand\colorMATH{\colorMATHC}\renewcommand\colorSYNTAX{\colorSYNTAXC}{{\color{\colorMATH}\ensuremath{\plambda}}}\endgroup } x:\tau _{1}\mathord{\cdotp }{\begingroup\renewcommand\colorMATH{\colorMATHB}\renewcommand\colorSYNTAX{\colorSYNTAXB}{{\color{\colorMATH}\ensuremath{\sss}}}\endgroup }. {\begingroup\renewcommand\colorMATH{\colorMATHC}\renewcommand\colorSYNTAX{\colorSYNTAXC}{{\color{\colorMATH}\ensuremath{\pe}}}\endgroup }, \gamma \rangle  \in  {\mathcal{V}}\llbracket (x\mathrel{:}\tau _{1}\mathord{\cdotp }{\begingroup\renewcommand\colorMATH{\colorMATHB}\renewcommand\colorSYNTAX{\colorSYNTAXB}{{\color{\colorMATH}\ensuremath{\sss}}}\endgroup }) \xrightarrowP {{\begingroup\renewcommand\colorMATH{\colorMATHC}\renewcommand\colorSYNTAX{\colorSYNTAXC}{{\color{\colorMATH}\ensuremath{\pS}}}\endgroup }} \tau _{2}\rrbracket 
   }
\and \inferrule*[lab={\textsc{ }}
   ]{ 
   \gamma  \vdash  {\begingroup\renewcommand\colorMATH{\colorMATHB}\renewcommand\colorSYNTAX{\colorSYNTAXB}{{\color{\colorMATH}\ensuremath{\se}}}\endgroup } \Downarrow  {\begingroup\renewcommand\colorMATH{\colorMATHB}\renewcommand\colorSYNTAX{\colorSYNTAXB}{{\color{\colorMATH}\ensuremath{\sv}}}\endgroup }
   \\  {\begingroup\renewcommand\colorMATH{\colorMATHB}\renewcommand\colorSYNTAX{\colorSYNTAXB}{{\color{\colorMATH}\ensuremath{\sv}}}\endgroup } \in  {\mathcal{V}}\llbracket \tau \rrbracket 
      }{
      \gamma  \vdash  {\begingroup\renewcommand\colorMATH{\colorMATHB}\renewcommand\colorSYNTAX{\colorSYNTAXB}{{\color{\colorMATH}\ensuremath{\se}}}\endgroup } \in  {\mathcal{E}}\llbracket \tau \rrbracket 
   }
\and \inferrule*[lab={\textsc{ }}
   ]{ 
		\forall {\begingroup\renewcommand\colorMATH{\colorMATHB}\renewcommand\colorSYNTAX{\colorSYNTAXB}{{\color{\colorMATH}\ensuremath{\sv}}}\endgroup }, \gamma  \vdash  {\begingroup\renewcommand\colorMATH{\colorMATHC}\renewcommand\colorSYNTAX{\colorSYNTAXC}{{\color{\colorMATH}\ensuremath{\pe}}}\endgroup } \Downarrow  {\begingroup\renewcommand\colorMATH{\colorMATHB}\renewcommand\colorSYNTAX{\colorSYNTAXB}{{\color{\colorMATH}\ensuremath{\sv}}}\endgroup } \implies {\begingroup\renewcommand\colorMATH{\colorMATHB}\renewcommand\colorSYNTAX{\colorSYNTAXB}{{\color{\colorMATH}\ensuremath{\sv}}}\endgroup } \in  {\mathcal{V}}\llbracket \tau \rrbracket 
      }{
      \gamma  \vdash  {\begingroup\renewcommand\colorMATH{\colorMATHC}\renewcommand\colorSYNTAX{\colorSYNTAXC}{{\color{\colorMATH}\ensuremath{\pe}}}\endgroup } \in  {\mathcal{E}}\llbracket \tau \rrbracket 
   }
\end{mathpar}\endgroup
\end{framed}
\end{small}
\caption{$\lang$: Type Safety Logical Relation}
\label{afig:type-safety-lr}
\end{figure}

The type safety logical relation is defined in Figure~\ref{afig:type-safety-lr}. Its definition is straightforward, split into a value relation {{\color{\colorMATH}\ensuremath{{\mathcal{V}}}}}, a computation relation {{\color{\colorMATH}\ensuremath{{\mathcal{E}}}}}, and an environment relation {{\color{\colorMATH}\ensuremath{{\mathcal{G}}}}}. As usual, the fundamental property of the type safety logical relation states that well-typed open terms are in the relation closed by an adequate environment {{\color{\colorMATH}\ensuremath{\gamma }}}:\footnote{We use the following operators to remove variables from a type:\\
$\tau /\Gamma  = [\varnothing /x_{1},...,\varnothing /x_{n}]\tau , \forall  x_{i} \in  dom(\Gamma )$ and
$\tau /\gamma  = [\varnothing /x_{1},...,\varnothing /x_{n}]\tau , \forall  x_{i} \in  dom(\gamma )$.
}

\begin{proposition}[Fundamental Property of the Type Safety Logical Relation]
  \label{lm:type-safety-FP}\;
  \begin{enumerate}[label=(\alph*)]
  \item Let {{\color{\colorMATH}\ensuremath{\Gamma ;{\begingroup\renewcommand\colorMATH{\colorMATHB}\renewcommand\colorSYNTAX{\colorSYNTAXB}{{\color{\colorMATH}\ensuremath{\sS_{0}}}}\endgroup } \vdash  {\begingroup\renewcommand\colorMATH{\colorMATHB}\renewcommand\colorSYNTAX{\colorSYNTAXB}{{\color{\colorMATH}\ensuremath{\se}}}\endgroup } : \tau  ; {\begingroup\renewcommand\colorMATH{\colorMATHB}\renewcommand\colorSYNTAX{\colorSYNTAXB}{{\color{\colorMATH}\ensuremath{\sS}}}\endgroup }}}}, and {{\color{\colorMATH}\ensuremath{\gamma  \in  {\mathcal{G}}\llbracket \Gamma \rrbracket }}}. Then
		{{\color{\colorMATH}\ensuremath{\gamma \vdash  {\begingroup\renewcommand\colorMATH{\colorMATHB}\renewcommand\colorSYNTAX{\colorSYNTAXB}{{\color{\colorMATH}\ensuremath{\se}}}\endgroup } \in  {\mathcal{E}}\llbracket \tau /\Gamma \rrbracket }}}.
  \item Let {{\color{\colorMATH}\ensuremath{\Gamma ;{\begingroup\renewcommand\colorMATH{\colorMATHB}\renewcommand\colorSYNTAX{\colorSYNTAXB}{{\color{\colorMATH}\ensuremath{\sS_{0}}}}\endgroup } \vdash  {\begingroup\renewcommand\colorMATH{\colorMATHC}\renewcommand\colorSYNTAX{\colorSYNTAXC}{{\color{\colorMATH}\ensuremath{\pe}}}\endgroup } : \tau  ; {\begingroup\renewcommand\colorMATH{\colorMATHC}\renewcommand\colorSYNTAX{\colorSYNTAXC}{{\color{\colorMATH}\ensuremath{\pS}}}\endgroup }}}}, and {{\color{\colorMATH}\ensuremath{\gamma  \in  {\mathcal{G}}\llbracket \Gamma \rrbracket }}}.  Then
		{{\color{\colorMATH}\ensuremath{\gamma \vdash  {\begingroup\renewcommand\colorMATH{\colorMATHC}\renewcommand\colorSYNTAX{\colorSYNTAXC}{{\color{\colorMATH}\ensuremath{\pe}}}\endgroup } \in  {\mathcal{E}}\llbracket \tau /\Gamma \rrbracket }}}.
  \end{enumerate}
\end{proposition}
\begin{proof}
{\bf (a) Sensitivity FP.}
We proceed by induction on {{\color{\colorMATH}\ensuremath{\Gamma ;{\begingroup\renewcommand\colorMATH{\colorMATHB}\renewcommand\colorSYNTAX{\colorSYNTAXB}{{\color{\colorMATH}\ensuremath{\sS_{0}}}}\endgroup } \vdash 	{\begingroup\renewcommand\colorMATH{\colorMATHB}\renewcommand\colorSYNTAX{\colorSYNTAXB}{{\color{\colorMATH}\ensuremath{\se}}}\endgroup } : \tau  ; {\begingroup\renewcommand\colorMATH{\colorMATHB}\renewcommand\colorSYNTAX{\colorSYNTAXB}{{\color{\colorMATH}\ensuremath{\sS}}}\endgroup }}}}.

First, to deal with the cases of sensitivity and privacy functions,  we give the typing rules for sensitivity and privacy closures below:
 \begingroup\color{\colorMATH}\begin{gather*} 
	\inferrule*[lab={\textsc{ s-closure}}
	]{ \exists \Gamma ', {\begingroup\renewcommand\colorMATH{\colorMATHB}\renewcommand\colorSYNTAX{\colorSYNTAXB}{{\color{\colorMATH}\ensuremath{\sS'_{0}}}}\endgroup }, dom({\begingroup\renewcommand\colorMATH{\colorMATHB}\renewcommand\colorSYNTAX{\colorSYNTAXB}{{\color{\colorMATH}\ensuremath{\sS''}}}\endgroup }) \subseteq  dom(\Gamma ') \subseteq  dom({\begingroup\renewcommand\colorMATH{\colorMATHB}\renewcommand\colorSYNTAX{\colorSYNTAXB}{{\color{\colorMATH}\ensuremath{\sS'_{0}}}}\endgroup }) 
	\\ \forall  x_{i} \in  dom(\Gamma '), \varnothing ; \varnothing  \vdash  \gamma (x_{i}) : \tau '_{i}, \tau '_{i} <: \Gamma '(x_{i}) ; \varnothing 
	\\ \Gamma ', x: \tau _{1}; {\begingroup\renewcommand\colorMATH{\colorMATHB}\renewcommand\colorSYNTAX{\colorSYNTAXB}{{\color{\colorMATH}\ensuremath{\sS'_{0}}}}\endgroup } + {\begingroup\renewcommand\colorMATH{\colorMATHB}\renewcommand\colorSYNTAX{\colorSYNTAXB}{{\color{\colorMATH}\ensuremath{\sss_{1}}}}\endgroup }x \vdash  {\begingroup\renewcommand\colorMATH{\colorMATHB}\renewcommand\colorSYNTAX{\colorSYNTAXB}{{\color{\colorMATH}\ensuremath{\se'}}}\endgroup } \mathrel{:} \tau _{2} \mathrel{;} {\begingroup\renewcommand\colorMATH{\colorMATHB}\renewcommand\colorSYNTAX{\colorSYNTAXB}{{\color{\colorMATH}\ensuremath{\sS''}}}\endgroup }+{\begingroup\renewcommand\colorMATH{\colorMATHB}\renewcommand\colorSYNTAX{\colorSYNTAXB}{{\color{\colorMATH}\ensuremath{\sss'}}}\endgroup }x
	  }{
	  \Gamma ; {\begingroup\renewcommand\colorMATH{\colorMATHB}\renewcommand\colorSYNTAX{\colorSYNTAXB}{{\color{\colorMATH}\ensuremath{\sS_{0}}}}\endgroup } \vdash  \langle {\begingroup\renewcommand\colorMATH{\colorMATHB}\renewcommand\colorSYNTAX{\colorSYNTAXB}{{\color{\colorMATH}\ensuremath{\slambda}}}\endgroup } (x\mathrel{:}\tau _{1}\mathord{\cdotp }{\begingroup\renewcommand\colorMATH{\colorMATHB}\renewcommand\colorSYNTAX{\colorSYNTAXB}{{\color{\colorMATH}\ensuremath{\sss_{1}}}}\endgroup }).\hspace*{0.33em}{\begingroup\renewcommand\colorMATH{\colorMATHB}\renewcommand\colorSYNTAX{\colorSYNTAXB}{{\color{\colorMATH}\ensuremath{\se'}}}\endgroup }, \gamma \rangle  \mathrel{:} (x\mathrel{:} \tau _{1}/\Gamma '\mathord{\cdotp }{\begingroup\renewcommand\colorMATH{\colorMATHB}\renewcommand\colorSYNTAX{\colorSYNTAXB}{{\color{\colorMATH}\ensuremath{\sss_{1}}}}\endgroup }) \xrightarrowS {{\begingroup\renewcommand\colorMATH{\colorMATHB}\renewcommand\colorSYNTAX{\colorSYNTAXB}{{\color{\colorMATH}\ensuremath{\sss'}}}\endgroup }x} \tau _{2}/\Gamma ' \mathrel{;} \varnothing 
	}
\end{gather*}\endgroup
\begingroup\color{\colorMATH}\begin{gather*}
	\inferrule*[lab={\textsc{ p-closure}}
	]{ \exists \Gamma ', {\begingroup\renewcommand\colorMATH{\colorMATHB}\renewcommand\colorSYNTAX{\colorSYNTAXB}{{\color{\colorMATH}\ensuremath{\sS'_{0}}}}\endgroup }, dom({\begingroup\renewcommand\colorMATH{\colorMATHB}\renewcommand\colorSYNTAX{\colorSYNTAXB}{{\color{\colorMATH}\ensuremath{\sS''}}}\endgroup }) \subseteq  dom(\Gamma ') \subseteq  dom({\begingroup\renewcommand\colorMATH{\colorMATHB}\renewcommand\colorSYNTAX{\colorSYNTAXB}{{\color{\colorMATH}\ensuremath{\sS'_{0}}}}\endgroup }) 
	\\ \forall  x_{i} \in  dom(\Gamma '), \varnothing ; \varnothing  \vdash  \gamma (x_{i}) : \tau '_{i}, \tau '_{i} <: \Gamma '(x_{i}) ; \varnothing 
	\\ \Gamma ', x: \tau _{1}; {\begingroup\renewcommand\colorMATH{\colorMATHB}\renewcommand\colorSYNTAX{\colorSYNTAXB}{{\color{\colorMATH}\ensuremath{\sS'_{0}}}}\endgroup } + {\begingroup\renewcommand\colorMATH{\colorMATHB}\renewcommand\colorSYNTAX{\colorSYNTAXB}{{\color{\colorMATH}\ensuremath{\sss_{1}}}}\endgroup }x \vdash   {\begingroup\renewcommand\colorMATH{\colorMATHC}\renewcommand\colorSYNTAX{\colorSYNTAXC}{{\color{\colorMATH}\ensuremath{\pe'}}}\endgroup } \mathrel{:} \tau _{2} \mathrel{;} {\begingroup\renewcommand\colorMATH{\colorMATHC}\renewcommand\colorSYNTAX{\colorSYNTAXC}{{\color{\colorMATH}\ensuremath{\pS''}}}\endgroup }
	  }{
	  \Gamma ; {\begingroup\renewcommand\colorMATH{\colorMATHB}\renewcommand\colorSYNTAX{\colorSYNTAXB}{{\color{\colorMATH}\ensuremath{\sS_{0}}}}\endgroup } \vdash  \langle {\begingroup\renewcommand\colorMATH{\colorMATHC}\renewcommand\colorSYNTAX{\colorSYNTAXC}{{\color{\colorMATH}\ensuremath{\plambda}}}\endgroup } (x\mathrel{:}\tau _{1}\mathord{\cdotp }{\begingroup\renewcommand\colorMATH{\colorMATHB}\renewcommand\colorSYNTAX{\colorSYNTAXB}{{\color{\colorMATH}\ensuremath{\sss_{1}}}}\endgroup }).\hspace*{0.33em} {\begingroup\renewcommand\colorMATH{\colorMATHC}\renewcommand\colorSYNTAX{\colorSYNTAXC}{{\color{\colorMATH}\ensuremath{\pe'}}}\endgroup }, \gamma \rangle  \mathrel{:} (x\mathrel{:} \tau _{1}/\Gamma '\mathord{\cdotp }{\begingroup\renewcommand\colorMATH{\colorMATHB}\renewcommand\colorSYNTAX{\colorSYNTAXB}{{\color{\colorMATH}\ensuremath{\sss_{1}}}}\endgroup }) \xrightarrowP {{\begingroup\renewcommand\colorMATH{\colorMATHC}\renewcommand\colorSYNTAX{\colorSYNTAXC}{{\color{\colorMATH}\ensuremath{\pS''}}}\endgroup }/\Gamma '} \tau _{2}/\Gamma ' \mathrel{;} \varnothing 
	}
\end{gather*}\endgroup

\begin{enumerate}[ncases]\item  {{\color{\colorMATH}\ensuremath{\Gamma ;{\begingroup\renewcommand\colorMATH{\colorMATHB}\renewcommand\colorSYNTAX{\colorSYNTAXB}{{\color{\colorMATH}\ensuremath{\sS_{0}}}}\endgroup } \vdash 	x : \Gamma (x) ; x}}}
	\begin{subproof} 
	 By {{\color{\colorMATH}\ensuremath{\gamma  \in  {\mathcal{G}}\llbracket \Gamma \rrbracket }}}, we know that {{\color{\colorMATH}\ensuremath{\gamma (x) \in  {\mathcal{V}}\llbracket \Gamma (x)/\Gamma \rrbracket }}} . By inspection of the evaluation rules, we know that
	 {{\color{\colorMATH}\ensuremath{\gamma  \vdash  x \Downarrow  \gamma (x)}}}. We have to prove that {{\color{\colorMATH}\ensuremath{\gamma (x) \in  {\mathcal{V}}\llbracket \Gamma (x)/\Gamma \rrbracket }}}, which we already know and the result holds.
	\end{subproof}
\item  {{\color{\colorMATH}\ensuremath{\Gamma ;{\begingroup\renewcommand\colorMATH{\colorMATHB}\renewcommand\colorSYNTAX{\colorSYNTAXB}{{\color{\colorMATH}\ensuremath{\sS_{0}}}}\endgroup } \vdash 	 {\begingroup\renewcommand\colorMATH{\colorMATHB}\renewcommand\colorSYNTAX{\colorSYNTAXB}{{\color{\colorMATH}\ensuremath{r}}}\endgroup } : {\begingroup\renewcommand\colorMATH{\colorMATHA}\renewcommand\colorSYNTAX{\colorSYNTAXA}{{\color{\colorSYNTAX}\texttt{{\ensuremath{{\mathbb{R}}}}}}}\endgroup } ; {\begingroup\renewcommand\colorMATH{\colorMATHB}\renewcommand\colorSYNTAX{\colorSYNTAXB}{{\color{\colorMATH}\ensuremath{\varnothing }}}\endgroup }}}}
	\begin{subproof} 
	 Trivial as {{\color{\colorMATH}\ensuremath{\varnothing ; \varnothing  \vdash  {\begingroup\renewcommand\colorMATH{\colorMATHB}\renewcommand\colorSYNTAX{\colorSYNTAXB}{{\color{\colorMATH}\ensuremath{r}}}\endgroup } : {\begingroup\renewcommand\colorMATH{\colorMATHA}\renewcommand\colorSYNTAX{\colorSYNTAXA}{{\color{\colorSYNTAX}\texttt{{\ensuremath{{\mathbb{R}}}}}}}\endgroup }; {\begingroup\renewcommand\colorMATH{\colorMATHB}\renewcommand\colorSYNTAX{\colorSYNTAXB}{{\color{\colorMATH}\ensuremath{\varnothing }}}\endgroup }}}}. 
	\end{subproof}
\item  {{\color{\colorMATH}\ensuremath{\Gamma ;{\begingroup\renewcommand\colorMATH{\colorMATHB}\renewcommand\colorSYNTAX{\colorSYNTAXB}{{\color{\colorMATH}\ensuremath{\sS_{0}}}}\endgroup } \vdash 	 \ttt : {{\color{\colorSYNTAX}\texttt{unit}}} ; {\begingroup\renewcommand\colorMATH{\colorMATHB}\renewcommand\colorSYNTAX{\colorSYNTAXB}{{\color{\colorMATH}\ensuremath{\varnothing }}}\endgroup }}}}
	\begin{subproof} 
	 Trivial as {{\color{\colorMATH}\ensuremath{\varnothing ; \varnothing  \vdash  \ttt : {{\color{\colorSYNTAX}\texttt{unit}}}; {\begingroup\renewcommand\colorMATH{\colorMATHB}\renewcommand\colorSYNTAX{\colorSYNTAXB}{{\color{\colorMATH}\ensuremath{\varnothing }}}\endgroup }}}}. 
	\end{subproof}
\item  {{\color{\colorMATH}\ensuremath{\Gamma ;{\begingroup\renewcommand\colorMATH{\colorMATHB}\renewcommand\colorSYNTAX{\colorSYNTAXB}{{\color{\colorMATH}\ensuremath{\sS_{0}}}}\endgroup } \vdash 	{\begingroup\renewcommand\colorMATH{\colorMATHB}\renewcommand\colorSYNTAX{\colorSYNTAXB}{{\color{\colorMATH}\ensuremath{\se_{1}}}}\endgroup }\hspace*{0.33em}{\begingroup\renewcommand\colorMATH{\colorMATHB}\renewcommand\colorSYNTAX{\colorSYNTAXB}{{\color{\colorMATH}\ensuremath{\se_{2}}}}\endgroup } : \tau  ; {\begingroup\renewcommand\colorMATH{\colorMATHB}\renewcommand\colorSYNTAX{\colorSYNTAXB}{{\color{\colorMATH}\ensuremath{\sS}}}\endgroup }}}}
	\begin{subproof} 
		By {{\color{\colorMATH}\ensuremath{{\textsc{ s-app}}}}} we know that
		\begingroup\color{\colorMATH}\begin{gather*} 
		   \inferrule*[lab={\textsc{ s-app}}
		   ]{ \Gamma  \mathrel{;} {\begingroup\renewcommand\colorMATH{\colorMATHB}\renewcommand\colorSYNTAX{\colorSYNTAXB}{{\color{\colorMATH}\ensuremath{\sS_{0}}}}\endgroup }\hspace*{0.33em}{\begingroup\renewcommand\colorMATH{\colorMATHB}\renewcommand\colorSYNTAX{\colorSYNTAXB}{{\color{\colorMATH}\ensuremath{\vdash }}}\endgroup }\hspace*{0.33em}{\begingroup\renewcommand\colorMATH{\colorMATHB}\renewcommand\colorSYNTAX{\colorSYNTAXB}{{\color{\colorMATH}\ensuremath{\se_{1}}}}\endgroup } \mathrel{:} (x\mathrel{:}\tau _{1}\mathord{\cdotp }{\begingroup\renewcommand\colorMATH{\colorMATHB}\renewcommand\colorSYNTAX{\colorSYNTAXB}{{\color{\colorMATH}\ensuremath{\sss_{1}}}}\endgroup }) \xrightarrowS {{\begingroup\renewcommand\colorMATH{\colorMATHB}\renewcommand\colorSYNTAX{\colorSYNTAXB}{{\color{\colorMATH}\ensuremath{\sS'}}}\endgroup }+{\begingroup\renewcommand\colorMATH{\colorMATHB}\renewcommand\colorSYNTAX{\colorSYNTAXB}{{\color{\colorMATH}\ensuremath{\sss_{2}}}}\endgroup }x } \tau _{2} \mathrel{;} {\begingroup\renewcommand\colorMATH{\colorMATHB}\renewcommand\colorSYNTAX{\colorSYNTAXB}{{\color{\colorMATH}\ensuremath{\sS_{1}}}}\endgroup }
		   \\ \Gamma  \mathrel{;} {\begingroup\renewcommand\colorMATH{\colorMATHB}\renewcommand\colorSYNTAX{\colorSYNTAXB}{{\color{\colorMATH}\ensuremath{\sS_{0}}}}\endgroup }\hspace*{0.33em}{\begingroup\renewcommand\colorMATH{\colorMATHB}\renewcommand\colorSYNTAX{\colorSYNTAXB}{{\color{\colorMATH}\ensuremath{\vdash }}}\endgroup }\hspace*{0.33em}{\begingroup\renewcommand\colorMATH{\colorMATHB}\renewcommand\colorSYNTAX{\colorSYNTAXB}{{\color{\colorMATH}\ensuremath{\se_{2}}}}\endgroup } \mathrel{:} \tau _{1} \mathrel{;} {\begingroup\renewcommand\colorMATH{\colorMATHB}\renewcommand\colorSYNTAX{\colorSYNTAXB}{{\color{\colorMATH}\ensuremath{\sS_{2}}}}\endgroup }
		   \\ {\begingroup\renewcommand\colorMATH{\colorMATHB}\renewcommand\colorSYNTAX{\colorSYNTAXB}{{\color{\colorMATH}\ensuremath{\sS_{0}}}}\endgroup }\mathord{\cdotp }{\begingroup\renewcommand\colorMATH{\colorMATHB}\renewcommand\colorSYNTAX{\colorSYNTAXB}{{\color{\colorMATH}\ensuremath{\sS_{2}}}}\endgroup } \leq  {\begingroup\renewcommand\colorMATH{\colorMATHB}\renewcommand\colorSYNTAX{\colorSYNTAXB}{{\color{\colorMATH}\ensuremath{\sss_{1}}}}\endgroup }
		      }{
		      \Gamma  \mathrel{;} {\begingroup\renewcommand\colorMATH{\colorMATHB}\renewcommand\colorSYNTAX{\colorSYNTAXB}{{\color{\colorMATH}\ensuremath{\sS_{0}}}}\endgroup }\hspace*{0.33em}{\begingroup\renewcommand\colorMATH{\colorMATHB}\renewcommand\colorSYNTAX{\colorSYNTAXB}{{\color{\colorMATH}\ensuremath{\vdash }}}\endgroup }\hspace*{0.33em}{\begingroup\renewcommand\colorMATH{\colorMATHB}\renewcommand\colorSYNTAX{\colorSYNTAXB}{{\color{\colorMATH}\ensuremath{\se_{1}}}}\endgroup }\hspace*{0.33em}{\begingroup\renewcommand\colorMATH{\colorMATHB}\renewcommand\colorSYNTAX{\colorSYNTAXB}{{\color{\colorMATH}\ensuremath{\se_{2}}}}\endgroup } \mathrel{:} [{\begingroup\renewcommand\colorMATH{\colorMATHB}\renewcommand\colorSYNTAX{\colorSYNTAXB}{{\color{\colorMATH}\ensuremath{\sS_{2}}}}\endgroup }/x]\tau _{2} \mathrel{;} {\begingroup\renewcommand\colorMATH{\colorMATHB}\renewcommand\colorSYNTAX{\colorSYNTAXB}{{\color{\colorMATH}\ensuremath{\sS_{1}}}}\endgroup } + {\begingroup\renewcommand\colorMATH{\colorMATHB}\renewcommand\colorSYNTAX{\colorSYNTAXB}{{\color{\colorMATH}\ensuremath{\sss_{2}}}}\endgroup }{\begingroup\renewcommand\colorMATH{\colorMATHB}\renewcommand\colorSYNTAX{\colorSYNTAXB}{{\color{\colorMATH}\ensuremath{\sS_{2}}}}\endgroup } + {\begingroup\renewcommand\colorMATH{\colorMATHB}\renewcommand\colorSYNTAX{\colorSYNTAXB}{{\color{\colorMATH}\ensuremath{\sS'}}}\endgroup }
		   }
		\end{gather*}\endgroup
		where {{\color{\colorMATH}\ensuremath{\tau  = [{\begingroup\renewcommand\colorMATH{\colorMATHB}\renewcommand\colorSYNTAX{\colorSYNTAXB}{{\color{\colorMATH}\ensuremath{\sS_{2}}}}\endgroup }/x]\tau _{2}}}}, and {{\color{\colorMATH}\ensuremath{{\begingroup\renewcommand\colorMATH{\colorMATHB}\renewcommand\colorSYNTAX{\colorSYNTAXB}{{\color{\colorMATH}\ensuremath{\sS}}}\endgroup } = {\begingroup\renewcommand\colorMATH{\colorMATHB}\renewcommand\colorSYNTAX{\colorSYNTAXB}{{\color{\colorMATH}\ensuremath{\sS_{1}}}}\endgroup } + {\begingroup\renewcommand\colorMATH{\colorMATHB}\renewcommand\colorSYNTAX{\colorSYNTAXB}{{\color{\colorMATH}\ensuremath{\sss_{2}}}}\endgroup }{\begingroup\renewcommand\colorMATH{\colorMATHB}\renewcommand\colorSYNTAX{\colorSYNTAXB}{{\color{\colorMATH}\ensuremath{\sS_{2}}}}\endgroup } + {\begingroup\renewcommand\colorMATH{\colorMATHB}\renewcommand\colorSYNTAX{\colorSYNTAXB}{{\color{\colorMATH}\ensuremath{\sS'}}}\endgroup }}}}.
		By induction hypotheses we know that
		{{\color{\colorMATH}\ensuremath{\gamma  \vdash  {\begingroup\renewcommand\colorMATH{\colorMATHB}\renewcommand\colorSYNTAX{\colorSYNTAXB}{{\color{\colorMATH}\ensuremath{\se_{1}}}}\endgroup } \Downarrow  {\begingroup\renewcommand\colorMATH{\colorMATHB}\renewcommand\colorSYNTAX{\colorSYNTAXB}{{\color{\colorMATH}\ensuremath{\sv_{1}}}}\endgroup }}}}, {{\color{\colorMATH}\ensuremath{\gamma  \vdash  {\begingroup\renewcommand\colorMATH{\colorMATHB}\renewcommand\colorSYNTAX{\colorSYNTAXB}{{\color{\colorMATH}\ensuremath{\se_{2}}}}\endgroup } \Downarrow  {\begingroup\renewcommand\colorMATH{\colorMATHB}\renewcommand\colorSYNTAX{\colorSYNTAXB}{{\color{\colorMATH}\ensuremath{\sv_{2}}}}\endgroup }}}},
		{{\color{\colorMATH}\ensuremath{{\begingroup\renewcommand\colorMATH{\colorMATHB}\renewcommand\colorSYNTAX{\colorSYNTAXB}{{\color{\colorMATH}\ensuremath{\sv_{1}}}}\endgroup } \in  {\mathcal{V}}\llbracket (x\mathrel{:}\tau _{1}/\Gamma \mathord{\cdotp }{\begingroup\renewcommand\colorMATH{\colorMATHB}\renewcommand\colorSYNTAX{\colorSYNTAXB}{{\color{\colorMATH}\ensuremath{\sss_{1}}}}\endgroup }) \xrightarrowS {{\begingroup\renewcommand\colorMATH{\colorMATHB}\renewcommand\colorSYNTAX{\colorSYNTAXB}{{\color{\colorMATH}\ensuremath{\sss_{2}}}}\endgroup }x } \tau _{2}/\Gamma \rrbracket }}} and 
		{{\color{\colorMATH}\ensuremath{{\begingroup\renewcommand\colorMATH{\colorMATHB}\renewcommand\colorSYNTAX{\colorSYNTAXB}{{\color{\colorMATH}\ensuremath{\sv_{2}}}}\endgroup } \in  {\mathcal{V}}\llbracket \tau _{1}/\Gamma \rrbracket }}}.

		By inspection of the function predicate, we know that 
		{{\color{\colorMATH}\ensuremath{{\begingroup\renewcommand\colorMATH{\colorMATHB}\renewcommand\colorSYNTAX{\colorSYNTAXB}{{\color{\colorMATH}\ensuremath{\sv_{1}}}}\endgroup } \in  Atom\llbracket (x\mathrel{:}\tau _{1}/\Gamma \mathord{\cdotp }{\begingroup\renewcommand\colorMATH{\colorMATHB}\renewcommand\colorSYNTAX{\colorSYNTAXB}{{\color{\colorMATH}\ensuremath{\sss_{1}}}}\endgroup }) \xrightarrowS {{\begingroup\renewcommand\colorMATH{\colorMATHB}\renewcommand\colorSYNTAX{\colorSYNTAXB}{{\color{\colorMATH}\ensuremath{\sss_{2}}}}\endgroup }x } \tau _{2}/\Gamma \rrbracket }}} and 
		{{\color{\colorMATH}\ensuremath{{\begingroup\renewcommand\colorMATH{\colorMATHB}\renewcommand\colorSYNTAX{\colorSYNTAXB}{{\color{\colorMATH}\ensuremath{\sv_{1}}}}\endgroup } = \langle {\begingroup\renewcommand\colorMATH{\colorMATHB}\renewcommand\colorSYNTAX{\colorSYNTAXB}{{\color{\colorMATH}\ensuremath{\slambda}}}\endgroup } (x\mathrel{:}\tau '_{1}\mathord{\cdotp }{\begingroup\renewcommand\colorMATH{\colorMATHB}\renewcommand\colorSYNTAX{\colorSYNTAXB}{{\color{\colorMATH}\ensuremath{\sss'_{1}}}}\endgroup }).\hspace*{0.33em}{\begingroup\renewcommand\colorMATH{\colorMATHB}\renewcommand\colorSYNTAX{\colorSYNTAXB}{{\color{\colorMATH}\ensuremath{\se'}}}\endgroup }, \gamma '\rangle }}}, for some {{\color{\colorMATH}\ensuremath{\tau '_{1}, {\begingroup\renewcommand\colorMATH{\colorMATHB}\renewcommand\colorSYNTAX{\colorSYNTAXB}{{\color{\colorMATH}\ensuremath{\sss'_{1}}}}\endgroup }, {\begingroup\renewcommand\colorMATH{\colorMATHB}\renewcommand\colorSYNTAX{\colorSYNTAXB}{{\color{\colorMATH}\ensuremath{\se'}}}\endgroup }}}} and {{\color{\colorMATH}\ensuremath{\gamma '}}}.
		We also know then that 
		{{\color{\colorMATH}\ensuremath{\gamma '[x \mapsto  {\begingroup\renewcommand\colorMATH{\colorMATHB}\renewcommand\colorSYNTAX{\colorSYNTAXB}{{\color{\colorMATH}\ensuremath{\sv_{2}}}}\endgroup }] \vdash  {\begingroup\renewcommand\colorMATH{\colorMATHB}\renewcommand\colorSYNTAX{\colorSYNTAXB}{{\color{\colorMATH}\ensuremath{\se'}}}\endgroup } \in  {\mathcal{E}}\llbracket \tau _{2}/(\Gamma ,x:\tau _{1})\rrbracket }}}, i.e. {{\color{\colorMATH}\ensuremath{\gamma '[x \mapsto  {\begingroup\renewcommand\colorMATH{\colorMATHB}\renewcommand\colorSYNTAX{\colorSYNTAXB}{{\color{\colorMATH}\ensuremath{\sv_{2}}}}\endgroup }] \vdash  {\begingroup\renewcommand\colorMATH{\colorMATHB}\renewcommand\colorSYNTAX{\colorSYNTAXB}{{\color{\colorMATH}\ensuremath{\se'}}}\endgroup } \Downarrow  {\begingroup\renewcommand\colorMATH{\colorMATHB}\renewcommand\colorSYNTAX{\colorSYNTAXB}{{\color{\colorMATH}\ensuremath{\sv'}}}\endgroup }}}} and {{\color{\colorMATH}\ensuremath{[{\begingroup\renewcommand\colorMATH{\colorMATHB}\renewcommand\colorSYNTAX{\colorSYNTAXB}{{\color{\colorMATH}\ensuremath{\sS_{2}}}}\endgroup }/x]\tau _{2}/\Gamma  = [({\begingroup\renewcommand\colorMATH{\colorMATHB}\renewcommand\colorSYNTAX{\colorSYNTAXB}{{\color{\colorMATH}\ensuremath{\sS_{2}}}}\endgroup }/\Gamma )/x](\tau _{2}/\Gamma ) = [\varnothing /x](\tau _{2}/\Gamma ) = \tau _{2}/(\Gamma ,x:\tau _{1})}}}, but
		{{\color{\colorMATH}\ensuremath{{\begingroup\renewcommand\colorMATH{\colorMATHB}\renewcommand\colorSYNTAX{\colorSYNTAXB}{{\color{\colorMATH}\ensuremath{\sv'}}}\endgroup } \in  {\mathcal{V}}\llbracket \tau _{2}/(\Gamma ,x:\tau _{1})\rrbracket }}} and the result holds.
	\end{subproof}
\item  {{\color{\colorMATH}\ensuremath{\Gamma ;{\begingroup\renewcommand\colorMATH{\colorMATHB}\renewcommand\colorSYNTAX{\colorSYNTAXB}{{\color{\colorMATH}\ensuremath{\sS_{0}}}}\endgroup } \vdash  {\begingroup\renewcommand\colorMATH{\colorMATHB}\renewcommand\colorSYNTAX{\colorSYNTAXB}{{\color{\colorMATH}\ensuremath{\slambda}}}\endgroup } (x\mathrel{:}\tau _{1}\mathord{\cdotp }{\begingroup\renewcommand\colorMATH{\colorMATHB}\renewcommand\colorSYNTAX{\colorSYNTAXB}{{\color{\colorMATH}\ensuremath{\sss}}}\endgroup }).\hspace*{0.33em}{\begingroup\renewcommand\colorMATH{\colorMATHB}\renewcommand\colorSYNTAX{\colorSYNTAXB}{{\color{\colorMATH}\ensuremath{\se'}}}\endgroup } \mathrel{:} (x:\tau _{1}\mathord{\cdotp }{\begingroup\renewcommand\colorMATH{\colorMATHB}\renewcommand\colorSYNTAX{\colorSYNTAXB}{{\color{\colorMATH}\ensuremath{\sss}}}\endgroup }) \xrightarrowS {{\begingroup\renewcommand\colorMATH{\colorMATHB}\renewcommand\colorSYNTAX{\colorSYNTAXB}{{\color{\colorMATH}\ensuremath{\sS'}}}\endgroup }} \tau _{2} ; {\begingroup\renewcommand\colorMATH{\colorMATHB}\renewcommand\colorSYNTAX{\colorSYNTAXB}{{\color{\colorMATH}\ensuremath{\varnothing }}}\endgroup }}}}
	\begin{subproof} 
		We know that 
		\begingroup\color{\colorMATH}\begin{gather*} 
			\inferrule*[lab={\textsc{ s-lam}}
		    ]{ \Gamma ,x\mathrel{:}\tau _{1} \mathrel{;} {\begingroup\renewcommand\colorMATH{\colorMATHB}\renewcommand\colorSYNTAX{\colorSYNTAXB}{{\color{\colorMATH}\ensuremath{\sS_{0}}}}\endgroup } + {\begingroup\renewcommand\colorMATH{\colorMATHB}\renewcommand\colorSYNTAX{\colorSYNTAXB}{{\color{\colorMATH}\ensuremath{\sss}}}\endgroup }x\hspace*{0.33em}{\begingroup\renewcommand\colorMATH{\colorMATHB}\renewcommand\colorSYNTAX{\colorSYNTAXB}{{\color{\colorMATH}\ensuremath{\vdash }}}\endgroup }\hspace*{0.33em}{\begingroup\renewcommand\colorMATH{\colorMATHB}\renewcommand\colorSYNTAX{\colorSYNTAXB}{{\color{\colorMATH}\ensuremath{\se'}}}\endgroup } \mathrel{:} \tau _{2} \mathrel{;} {\begingroup\renewcommand\colorMATH{\colorMATHB}\renewcommand\colorSYNTAX{\colorSYNTAXB}{{\color{\colorMATH}\ensuremath{\sS'}}}\endgroup }
		      }{
		      \Gamma  \mathrel{;} {\begingroup\renewcommand\colorMATH{\colorMATHB}\renewcommand\colorSYNTAX{\colorSYNTAXB}{{\color{\colorMATH}\ensuremath{\sS_{0}}}}\endgroup }\hspace*{0.33em}{\begingroup\renewcommand\colorMATH{\colorMATHB}\renewcommand\colorSYNTAX{\colorSYNTAXB}{{\color{\colorMATH}\ensuremath{\vdash }}}\endgroup }\hspace*{0.33em}{\begingroup\renewcommand\colorMATH{\colorMATHB}\renewcommand\colorSYNTAX{\colorSYNTAXB}{{\color{\colorMATH}\ensuremath{\slambda}}}\endgroup } (x\mathrel{:}\tau _{1}\mathord{\cdotp }{\begingroup\renewcommand\colorMATH{\colorMATHB}\renewcommand\colorSYNTAX{\colorSYNTAXB}{{\color{\colorMATH}\ensuremath{\sss}}}\endgroup }).\hspace*{0.33em}{\begingroup\renewcommand\colorMATH{\colorMATHB}\renewcommand\colorSYNTAX{\colorSYNTAXB}{{\color{\colorMATH}\ensuremath{\se'}}}\endgroup } \mathrel{:} (x\mathrel{:}\tau _{1}\mathord{\cdotp }{\begingroup\renewcommand\colorMATH{\colorMATHB}\renewcommand\colorSYNTAX{\colorSYNTAXB}{{\color{\colorMATH}\ensuremath{\sss}}}\endgroup }) \overset {\begingroup\renewcommand\colorMATH{\colorMATHB}\renewcommand\colorSYNTAX{\colorSYNTAXB}{{\color{\colorMATH}\ensuremath{\sS'}}}\endgroup }\rightarrow  \tau _{2} \mathrel{;} {\begingroup\renewcommand\colorMATH{\colorMATHB}\renewcommand\colorSYNTAX{\colorSYNTAXB}{{\color{\colorMATH}\ensuremath{\varnothing }}}\endgroup }
		    }	
		\end{gather*}\endgroup
		We know that
		{{\color{\colorMATH}\ensuremath{\gamma  \vdash  {\begingroup\renewcommand\colorMATH{\colorMATHB}\renewcommand\colorSYNTAX{\colorSYNTAXB}{{\color{\colorMATH}\ensuremath{\slambda}}}\endgroup } (x\mathrel{:}\tau _{1}\mathord{\cdotp }{\begingroup\renewcommand\colorMATH{\colorMATHB}\renewcommand\colorSYNTAX{\colorSYNTAXB}{{\color{\colorMATH}\ensuremath{\sss}}}\endgroup }).\hspace*{0.33em}{\begingroup\renewcommand\colorMATH{\colorMATHB}\renewcommand\colorSYNTAX{\colorSYNTAXB}{{\color{\colorMATH}\ensuremath{\se'}}}\endgroup } \Downarrow  \langle {\begingroup\renewcommand\colorMATH{\colorMATHB}\renewcommand\colorSYNTAX{\colorSYNTAXB}{{\color{\colorMATH}\ensuremath{\slambda}}}\endgroup } (x\mathrel{:}\tau _{1}\mathord{\cdotp }{\begingroup\renewcommand\colorMATH{\colorMATHB}\renewcommand\colorSYNTAX{\colorSYNTAXB}{{\color{\colorMATH}\ensuremath{\sss}}}\endgroup }).\hspace*{0.33em}{\begingroup\renewcommand\colorMATH{\colorMATHB}\renewcommand\colorSYNTAX{\colorSYNTAXB}{{\color{\colorMATH}\ensuremath{\se'}}}\endgroup }, \gamma \rangle }}}.
		We have to prove that 
		{{\color{\colorMATH}\ensuremath{\langle {\begingroup\renewcommand\colorMATH{\colorMATHB}\renewcommand\colorSYNTAX{\colorSYNTAXB}{{\color{\colorMATH}\ensuremath{\slambda}}}\endgroup } (x\mathrel{:}\tau _{1}\mathord{\cdotp }{\begingroup\renewcommand\colorMATH{\colorMATHB}\renewcommand\colorSYNTAX{\colorSYNTAXB}{{\color{\colorMATH}\ensuremath{\sss}}}\endgroup }).\hspace*{0.33em}{\begingroup\renewcommand\colorMATH{\colorMATHB}\renewcommand\colorSYNTAX{\colorSYNTAXB}{{\color{\colorMATH}\ensuremath{\se'}}}\endgroup }, \gamma \rangle  \in  {\mathcal{V}}\llbracket ((x:\tau _{1}\mathord{\cdotp }{\begingroup\renewcommand\colorMATH{\colorMATHB}\renewcommand\colorSYNTAX{\colorSYNTAXB}{{\color{\colorMATH}\ensuremath{\sss}}}\endgroup }) \xrightarrowS {{\begingroup\renewcommand\colorMATH{\colorMATHB}\renewcommand\colorSYNTAX{\colorSYNTAXB}{{\color{\colorMATH}\ensuremath{\sS'}}}\endgroup }} \tau _{2})/\Gamma \rrbracket }}}.
		Suppose {{\color{\colorMATH}\ensuremath{{\begingroup\renewcommand\colorMATH{\colorMATHB}\renewcommand\colorSYNTAX{\colorSYNTAXB}{{\color{\colorMATH}\ensuremath{\sS'}}}\endgroup } = {\begingroup\renewcommand\colorMATH{\colorMATHB}\renewcommand\colorSYNTAX{\colorSYNTAXB}{{\color{\colorMATH}\ensuremath{\sS''}}}\endgroup } + {\begingroup\renewcommand\colorMATH{\colorMATHB}\renewcommand\colorSYNTAX{\colorSYNTAXB}{{\color{\colorMATH}\ensuremath{\sss'}}}\endgroup }x}}}, then 
		{{\color{\colorMATH}\ensuremath{((x:\tau _{1}\mathord{\cdotp }{\begingroup\renewcommand\colorMATH{\colorMATHB}\renewcommand\colorSYNTAX{\colorSYNTAXB}{{\color{\colorMATH}\ensuremath{\sss}}}\endgroup }) \xrightarrowS {{\begingroup\renewcommand\colorMATH{\colorMATHB}\renewcommand\colorSYNTAX{\colorSYNTAXB}{{\color{\colorMATH}\ensuremath{\sS'}}}\endgroup }} \tau _{2})/\Gamma  = (x:\tau _{1}/\Gamma \mathord{\cdotp }{\begingroup\renewcommand\colorMATH{\colorMATHB}\renewcommand\colorSYNTAX{\colorSYNTAXB}{{\color{\colorMATH}\ensuremath{\sss}}}\endgroup }) \xrightarrowS {{\begingroup\renewcommand\colorMATH{\colorMATHB}\renewcommand\colorSYNTAX{\colorSYNTAXB}{{\color{\colorMATH}\ensuremath{\sss'}}}\endgroup }x} (\tau _{2}/\Gamma )}}}.

		First, we have to prove that,
		{{\color{\colorMATH}\ensuremath{\langle {\begingroup\renewcommand\colorMATH{\colorMATHB}\renewcommand\colorSYNTAX{\colorSYNTAXB}{{\color{\colorMATH}\ensuremath{\slambda}}}\endgroup } (x\mathrel{:}\tau _{1}\mathord{\cdotp }{\begingroup\renewcommand\colorMATH{\colorMATHB}\renewcommand\colorSYNTAX{\colorSYNTAXB}{{\color{\colorMATH}\ensuremath{\sss}}}\endgroup }).\hspace*{0.33em}{\begingroup\renewcommand\colorMATH{\colorMATHB}\renewcommand\colorSYNTAX{\colorSYNTAXB}{{\color{\colorMATH}\ensuremath{\se'}}}\endgroup }, \gamma \rangle  \in  Atom\llbracket (x:\tau _{1}/\Gamma \mathord{\cdotp }{\begingroup\renewcommand\colorMATH{\colorMATHB}\renewcommand\colorSYNTAX{\colorSYNTAXB}{{\color{\colorMATH}\ensuremath{\sss}}}\endgroup }) \xrightarrowS {{\begingroup\renewcommand\colorMATH{\colorMATHB}\renewcommand\colorSYNTAX{\colorSYNTAXB}{{\color{\colorMATH}\ensuremath{\sss'}}}\endgroup }x} (\tau _{2}/\Gamma )\rrbracket }}}, i.e. that
		 {{\color{\colorMATH}\ensuremath{\exists  \Gamma ', {\begingroup\renewcommand\colorMATH{\colorMATHB}\renewcommand\colorSYNTAX{\colorSYNTAXB}{{\color{\colorMATH}\ensuremath{\sS'_{0}}}}\endgroup }, dom({\begingroup\renewcommand\colorMATH{\colorMATHB}\renewcommand\colorSYNTAX{\colorSYNTAXB}{{\color{\colorMATH}\ensuremath{\sS''}}}\endgroup }) \subseteq  dom(\Gamma ') \subseteq  dom({\begingroup\renewcommand\colorMATH{\colorMATHB}\renewcommand\colorSYNTAX{\colorSYNTAXB}{{\color{\colorMATH}\ensuremath{\sS'_{0}}}}\endgroup }), \forall  x_{i} \in  dom(\Gamma '), \varnothing ; \varnothing  \vdash  \gamma (x_{i}) : \tau '_i, \tau '_i <: \Gamma '(x_{i}) ; \varnothing , \Gamma '}}}, and {{\color{\colorMATH}\ensuremath{x: \tau _{1}; {\begingroup\renewcommand\colorMATH{\colorMATHB}\renewcommand\colorSYNTAX{\colorSYNTAXB}{{\color{\colorMATH}\ensuremath{\sS'_{0}}}}\endgroup } + {\begingroup\renewcommand\colorMATH{\colorMATHB}\renewcommand\colorSYNTAX{\colorSYNTAXB}{{\color{\colorMATH}\ensuremath{\sss}}}\endgroup }x \vdash  {\begingroup\renewcommand\colorMATH{\colorMATHB}\renewcommand\colorSYNTAX{\colorSYNTAXB}{{\color{\colorMATH}\ensuremath{\se'}}}\endgroup } \mathrel{:} \tau _{2} \mathrel{;} {\begingroup\renewcommand\colorMATH{\colorMATHB}\renewcommand\colorSYNTAX{\colorSYNTAXB}{{\color{\colorMATH}\ensuremath{\sS''}}}\endgroup }+{\begingroup\renewcommand\colorMATH{\colorMATHB}\renewcommand\colorSYNTAX{\colorSYNTAXB}{{\color{\colorMATH}\ensuremath{\sss'}}}\endgroup }x}}}.
		We prove this by choosing {{\color{\colorMATH}\ensuremath{\Gamma ' = \Gamma }}}, {{\color{\colorMATH}\ensuremath{{\begingroup\renewcommand\colorMATH{\colorMATHB}\renewcommand\colorSYNTAX{\colorSYNTAXB}{{\color{\colorMATH}\ensuremath{\sS'_{0}}}}\endgroup } = {\begingroup\renewcommand\colorMATH{\colorMATHB}\renewcommand\colorSYNTAX{\colorSYNTAXB}{{\color{\colorMATH}\ensuremath{\sS_{0}}}}\endgroup }}}}, and {{\color{\colorMATH}\ensuremath{{\begingroup\renewcommand\colorMATH{\colorMATHB}\renewcommand\colorSYNTAX{\colorSYNTAXB}{{\color{\colorMATH}\ensuremath{\sS''}}}\endgroup }+{\begingroup\renewcommand\colorMATH{\colorMATHB}\renewcommand\colorSYNTAX{\colorSYNTAXB}{{\color{\colorMATH}\ensuremath{\sss'}}}\endgroup }x = {\begingroup\renewcommand\colorMATH{\colorMATHB}\renewcommand\colorSYNTAX{\colorSYNTAXB}{{\color{\colorMATH}\ensuremath{\sS'}}}\endgroup }}}}:
		 \begingroup\color{\colorMATH}\begin{gather*} 
			\inferrule*[lab=
			]{ \forall  x_{i} \in  dom(\Gamma ), \varnothing ; \varnothing  \vdash  \gamma (x_{i}) : \tau '_{i}, \tau '_{i} <: \Gamma (x_{i}) ; \varnothing 
			\\ \Gamma , x: \tau _{1}; {\begingroup\renewcommand\colorMATH{\colorMATHB}\renewcommand\colorSYNTAX{\colorSYNTAXB}{{\color{\colorMATH}\ensuremath{\sS_{0}}}}\endgroup } + {\begingroup\renewcommand\colorMATH{\colorMATHB}\renewcommand\colorSYNTAX{\colorSYNTAXB}{{\color{\colorMATH}\ensuremath{\sss_{1}}}}\endgroup }x \vdash  {\begingroup\renewcommand\colorMATH{\colorMATHB}\renewcommand\colorSYNTAX{\colorSYNTAXB}{{\color{\colorMATH}\ensuremath{\se'}}}\endgroup } \mathrel{:} \tau _{2} \mathrel{;} {\begingroup\renewcommand\colorMATH{\colorMATHB}\renewcommand\colorSYNTAX{\colorSYNTAXB}{{\color{\colorMATH}\ensuremath{\sS''}}}\endgroup }+{\begingroup\renewcommand\colorMATH{\colorMATHB}\renewcommand\colorSYNTAX{\colorSYNTAXB}{{\color{\colorMATH}\ensuremath{\sss'}}}\endgroup }x
			  }{
			  \Gamma ; {\begingroup\renewcommand\colorMATH{\colorMATHB}\renewcommand\colorSYNTAX{\colorSYNTAXB}{{\color{\colorMATH}\ensuremath{\sS_{0}}}}\endgroup } \vdash  \langle {\begingroup\renewcommand\colorMATH{\colorMATHB}\renewcommand\colorSYNTAX{\colorSYNTAXB}{{\color{\colorMATH}\ensuremath{\slambda}}}\endgroup } (x\mathrel{:}\tau _{1}\mathord{\cdotp }{\begingroup\renewcommand\colorMATH{\colorMATHB}\renewcommand\colorSYNTAX{\colorSYNTAXB}{{\color{\colorMATH}\ensuremath{\sss_{1}}}}\endgroup }).\hspace*{0.33em}{\begingroup\renewcommand\colorMATH{\colorMATHB}\renewcommand\colorSYNTAX{\colorSYNTAXB}{{\color{\colorMATH}\ensuremath{\se'}}}\endgroup }, \gamma \rangle  \mathrel{:} (x\mathrel{:} \tau _{1}/\Gamma '\mathord{\cdotp }{\begingroup\renewcommand\colorMATH{\colorMATHB}\renewcommand\colorSYNTAX{\colorSYNTAXB}{{\color{\colorMATH}\ensuremath{\sss_{1}}}}\endgroup }) \xrightarrow {{\begingroup\renewcommand\colorMATH{\colorMATHB}\renewcommand\colorSYNTAX{\colorSYNTAXB}{{\color{\colorMATH}\ensuremath{\sss'}}}\endgroup }x} \tau _{2}/\Gamma ' \mathrel{;} \varnothing 
			}
		\end{gather*}\endgroup

		Then we have prove that 
		{{\color{\colorMATH}\ensuremath{\forall  {\begingroup\renewcommand\colorMATH{\colorMATHB}\renewcommand\colorSYNTAX{\colorSYNTAXB}{{\color{\colorMATH}\ensuremath{\sv'}}}\endgroup } \in  {\mathcal{V}}\llbracket \tau _{1}/\Gamma \rrbracket , \gamma [x \mapsto  {\begingroup\renewcommand\colorMATH{\colorMATHB}\renewcommand\colorSYNTAX{\colorSYNTAXB}{{\color{\colorMATH}\ensuremath{\sv'}}}\endgroup }] \vdash  {\begingroup\renewcommand\colorMATH{\colorMATHB}\renewcommand\colorSYNTAX{\colorSYNTAXB}{{\color{\colorMATH}\ensuremath{\se'}}}\endgroup } \in  {\mathcal{E}}\llbracket \tau _{2}/(\Gamma ,x:T_{1})\rrbracket }}}.
		By induction hypothesis on 
		{{\color{\colorMATH}\ensuremath{\Gamma ,x\mathrel{:}\tau _{1} \mathrel{;} {\begingroup\renewcommand\colorMATH{\colorMATHB}\renewcommand\colorSYNTAX{\colorSYNTAXB}{{\color{\colorMATH}\ensuremath{\sS_{0}}}}\endgroup } + {\begingroup\renewcommand\colorMATH{\colorMATHB}\renewcommand\colorSYNTAX{\colorSYNTAXB}{{\color{\colorMATH}\ensuremath{\sss}}}\endgroup }x\hspace*{0.33em}{\begingroup\renewcommand\colorMATH{\colorMATHB}\renewcommand\colorSYNTAX{\colorSYNTAXB}{{\color{\colorMATH}\ensuremath{\vdash }}}\endgroup }\hspace*{0.33em}{\begingroup\renewcommand\colorMATH{\colorMATHB}\renewcommand\colorSYNTAX{\colorSYNTAXB}{{\color{\colorMATH}\ensuremath{\se'}}}\endgroup } \mathrel{:} \tau _{2} \mathrel{;} {\begingroup\renewcommand\colorMATH{\colorMATHB}\renewcommand\colorSYNTAX{\colorSYNTAXB}{{\color{\colorMATH}\ensuremath{\sS'}}}\endgroup }}}}, we know that
		for any {{\color{\colorMATH}\ensuremath{\gamma ' \in  {\mathcal{G}}\llbracket \Gamma ,x:\tau _{1}\rrbracket }}}, {{\color{\colorMATH}\ensuremath{\gamma ' \vdash  {\begingroup\renewcommand\colorMATH{\colorMATHB}\renewcommand\colorSYNTAX{\colorSYNTAXB}{{\color{\colorMATH}\ensuremath{\se'}}}\endgroup } \in  {\mathcal{E}}\llbracket \tau _{2}/(\Gamma ,x:\tau _{1})\rrbracket }}}.
		As {{\color{\colorMATH}\ensuremath{\gamma  \in  {\mathcal{G}}\llbracket \Gamma \rrbracket }}} and {{\color{\colorMATH}\ensuremath{{\begingroup\renewcommand\colorMATH{\colorMATHB}\renewcommand\colorSYNTAX{\colorSYNTAXB}{{\color{\colorMATH}\ensuremath{\sv'}}}\endgroup } \in  {\mathcal{V}}\llbracket \tau _{1}/(\Gamma , x:\tau _{1})\rrbracket }}} ({{\color{\colorMATH}\ensuremath{\tau _{1}/\Gamma  = \tau _{1}/(\Gamma , x:\tau _{1})}}}), then
		{{\color{\colorMATH}\ensuremath{\gamma [x \mapsto  {\begingroup\renewcommand\colorMATH{\colorMATHB}\renewcommand\colorSYNTAX{\colorSYNTAXB}{{\color{\colorMATH}\ensuremath{\sv'}}}\endgroup }] \in  {\mathcal{G}}\llbracket \Gamma ,x:\tau _{1}\rrbracket }}}, so we pick 
		{{\color{\colorMATH}\ensuremath{\gamma ' = \gamma [x \mapsto  {\begingroup\renewcommand\colorMATH{\colorMATHB}\renewcommand\colorSYNTAX{\colorSYNTAXB}{{\color{\colorMATH}\ensuremath{\sv'}}}\endgroup }]}}} and the result holds.
	\end{subproof}
\item  {{\color{\colorMATH}\ensuremath{\Gamma ;{\begingroup\renewcommand\colorMATH{\colorMATHB}\renewcommand\colorSYNTAX{\colorSYNTAXB}{{\color{\colorMATH}\ensuremath{\sS_{0}}}}\endgroup } \vdash  {\begingroup\renewcommand\colorMATH{\colorMATHC}\renewcommand\colorSYNTAX{\colorSYNTAXC}{{\color{\colorMATH}\ensuremath{\plambda}}}\endgroup } (x\mathrel{:}\tau _{1}\mathord{\cdotp }{\begingroup\renewcommand\colorMATH{\colorMATHB}\renewcommand\colorSYNTAX{\colorSYNTAXB}{{\color{\colorMATH}\ensuremath{\sss}}}\endgroup }).\hspace*{0.33em}{\begingroup\renewcommand\colorMATH{\colorMATHC}\renewcommand\colorSYNTAX{\colorSYNTAXC}{{\color{\colorMATH}\ensuremath{\pe'}}}\endgroup } \mathrel{:} (x:\tau _{1}\mathord{\cdotp }{\begingroup\renewcommand\colorMATH{\colorMATHB}\renewcommand\colorSYNTAX{\colorSYNTAXB}{{\color{\colorMATH}\ensuremath{\sss}}}\endgroup }) \xrightarrowP {{\begingroup\renewcommand\colorMATH{\colorMATHC}\renewcommand\colorSYNTAX{\colorSYNTAXC}{{\color{\colorMATH}\ensuremath{\pS'}}}\endgroup }} \tau _{2} ; {\begingroup\renewcommand\colorMATH{\colorMATHB}\renewcommand\colorSYNTAX{\colorSYNTAXB}{{\color{\colorMATH}\ensuremath{\varnothing }}}\endgroup }}}}
	\begin{subproof} 
		We know that 
		\begingroup\color{\colorMATH}\begin{gather*} 
			\inferrule*[lab={\textsc{ p-lam}}
		    ]{ \Gamma ,x\mathrel{:}\tau _{1} \mathrel{;} {\begingroup\renewcommand\colorMATH{\colorMATHB}\renewcommand\colorSYNTAX{\colorSYNTAXB}{{\color{\colorMATH}\ensuremath{\sS_{0}}}}\endgroup } + {\begingroup\renewcommand\colorMATH{\colorMATHB}\renewcommand\colorSYNTAX{\colorSYNTAXB}{{\color{\colorMATH}\ensuremath{\sss}}}\endgroup }x\hspace*{0.33em}{\begingroup\renewcommand\colorMATH{\colorMATHB}\renewcommand\colorSYNTAX{\colorSYNTAXB}{{\color{\colorMATH}\ensuremath{\vdash }}}\endgroup }\hspace*{0.33em} {\begingroup\renewcommand\colorMATH{\colorMATHC}\renewcommand\colorSYNTAX{\colorSYNTAXC}{{\color{\colorMATH}\ensuremath{\pe'}}}\endgroup } \mathrel{:} \tau _{2} \mathrel{;} {\begingroup\renewcommand\colorMATH{\colorMATHC}\renewcommand\colorSYNTAX{\colorSYNTAXC}{{\color{\colorMATH}\ensuremath{\pS'}}}\endgroup }
		      }{
		      \Gamma  \mathrel{;} {\begingroup\renewcommand\colorMATH{\colorMATHB}\renewcommand\colorSYNTAX{\colorSYNTAXB}{{\color{\colorMATH}\ensuremath{\sS_{0}}}}\endgroup }\hspace*{0.33em}{\begingroup\renewcommand\colorMATH{\colorMATHB}\renewcommand\colorSYNTAX{\colorSYNTAXB}{{\color{\colorMATH}\ensuremath{\vdash }}}\endgroup }\hspace*{0.33em}{\begingroup\renewcommand\colorMATH{\colorMATHC}\renewcommand\colorSYNTAX{\colorSYNTAXC}{{\color{\colorMATH}\ensuremath{\plambda}}}\endgroup } (x\mathrel{:}\tau _{1}\mathord{\cdotp }{\begingroup\renewcommand\colorMATH{\colorMATHB}\renewcommand\colorSYNTAX{\colorSYNTAXB}{{\color{\colorMATH}\ensuremath{\sss}}}\endgroup }).\hspace*{0.33em} {\begingroup\renewcommand\colorMATH{\colorMATHC}\renewcommand\colorSYNTAX{\colorSYNTAXC}{{\color{\colorMATH}\ensuremath{\pe'}}}\endgroup } \mathrel{:} (x\mathrel{:}\tau _{1}\mathord{\cdotp }{\begingroup\renewcommand\colorMATH{\colorMATHB}\renewcommand\colorSYNTAX{\colorSYNTAXB}{{\color{\colorMATH}\ensuremath{\sss}}}\endgroup }) \overset {\begingroup\renewcommand\colorMATH{\colorMATHC}\renewcommand\colorSYNTAX{\colorSYNTAXC}{{\color{\colorMATH}\ensuremath{\pS'}}}\endgroup }\rightarrow  \tau _{2} \mathrel{;} {\begingroup\renewcommand\colorMATH{\colorMATHB}\renewcommand\colorSYNTAX{\colorSYNTAXB}{{\color{\colorMATH}\ensuremath{\varnothing }}}\endgroup }
		    }	
		\end{gather*}\endgroup
		We know that
		{{\color{\colorMATH}\ensuremath{\gamma  \vdash  {\begingroup\renewcommand\colorMATH{\colorMATHC}\renewcommand\colorSYNTAX{\colorSYNTAXC}{{\color{\colorMATH}\ensuremath{\plambda}}}\endgroup } (x\mathrel{:}\tau _{1}\mathord{\cdotp }{\begingroup\renewcommand\colorMATH{\colorMATHB}\renewcommand\colorSYNTAX{\colorSYNTAXB}{{\color{\colorMATH}\ensuremath{\sss}}}\endgroup }).\hspace*{0.33em} {\begingroup\renewcommand\colorMATH{\colorMATHC}\renewcommand\colorSYNTAX{\colorSYNTAXC}{{\color{\colorMATH}\ensuremath{\pe'}}}\endgroup } \Downarrow  \langle {\begingroup\renewcommand\colorMATH{\colorMATHC}\renewcommand\colorSYNTAX{\colorSYNTAXC}{{\color{\colorMATH}\ensuremath{\plambda}}}\endgroup } (x\mathrel{:}\tau _{1}\mathord{\cdotp }{\begingroup\renewcommand\colorMATH{\colorMATHB}\renewcommand\colorSYNTAX{\colorSYNTAXB}{{\color{\colorMATH}\ensuremath{\sss}}}\endgroup }).\hspace*{0.33em} {\begingroup\renewcommand\colorMATH{\colorMATHC}\renewcommand\colorSYNTAX{\colorSYNTAXC}{{\color{\colorMATH}\ensuremath{\pe'}}}\endgroup }, \gamma \rangle }}}.
		We have to prove that 
		{{\color{\colorMATH}\ensuremath{\langle {\begingroup\renewcommand\colorMATH{\colorMATHC}\renewcommand\colorSYNTAX{\colorSYNTAXC}{{\color{\colorMATH}\ensuremath{\plambda}}}\endgroup } (x\mathrel{:}\tau _{1}\mathord{\cdotp }{\begingroup\renewcommand\colorMATH{\colorMATHB}\renewcommand\colorSYNTAX{\colorSYNTAXB}{{\color{\colorMATH}\ensuremath{\sss}}}\endgroup }).\hspace*{0.33em} {\begingroup\renewcommand\colorMATH{\colorMATHC}\renewcommand\colorSYNTAX{\colorSYNTAXC}{{\color{\colorMATH}\ensuremath{\pe'}}}\endgroup }, \gamma \rangle  \in  {\mathcal{V}}\llbracket ((x:\tau _{1}\mathord{\cdotp }{\begingroup\renewcommand\colorMATH{\colorMATHB}\renewcommand\colorSYNTAX{\colorSYNTAXB}{{\color{\colorMATH}\ensuremath{\sss}}}\endgroup }) \xrightarrowP {{\begingroup\renewcommand\colorMATH{\colorMATHC}\renewcommand\colorSYNTAX{\colorSYNTAXC}{{\color{\colorMATH}\ensuremath{\pS'}}}\endgroup }} \tau _{2})/\Gamma \rrbracket }}}.
		Then
		{{\color{\colorMATH}\ensuremath{((x:\tau _{1}\mathord{\cdotp }{\begingroup\renewcommand\colorMATH{\colorMATHB}\renewcommand\colorSYNTAX{\colorSYNTAXB}{{\color{\colorMATH}\ensuremath{\sss}}}\endgroup }) \xrightarrowP {{\begingroup\renewcommand\colorMATH{\colorMATHC}\renewcommand\colorSYNTAX{\colorSYNTAXC}{{\color{\colorMATH}\ensuremath{\pS'}}}\endgroup }} \tau _{2})/\Gamma  = (x:\tau _{1}/\Gamma \mathord{\cdotp }{\begingroup\renewcommand\colorMATH{\colorMATHB}\renewcommand\colorSYNTAX{\colorSYNTAXB}{{\color{\colorMATH}\ensuremath{\sss}}}\endgroup }) \xrightarrowP {{\begingroup\renewcommand\colorMATH{\colorMATHC}\renewcommand\colorSYNTAX{\colorSYNTAXC}{{\color{\colorMATH}\ensuremath{\pS'}}}\endgroup }/\Gamma } (\tau _{2}/\Gamma )}}}.

		First, we have to prove that,
		{{\color{\colorMATH}\ensuremath{\langle {\begingroup\renewcommand\colorMATH{\colorMATHC}\renewcommand\colorSYNTAX{\colorSYNTAXC}{{\color{\colorMATH}\ensuremath{\plambda}}}\endgroup } (x\mathrel{:}\tau _{1}\mathord{\cdotp }{\begingroup\renewcommand\colorMATH{\colorMATHB}\renewcommand\colorSYNTAX{\colorSYNTAXB}{{\color{\colorMATH}\ensuremath{\sss}}}\endgroup }).\hspace*{0.33em} {\begingroup\renewcommand\colorMATH{\colorMATHC}\renewcommand\colorSYNTAX{\colorSYNTAXC}{{\color{\colorMATH}\ensuremath{\pe'}}}\endgroup }, \gamma \rangle  \in  Atom\llbracket (x:\tau _{1}/\Gamma \mathord{\cdotp }{\begingroup\renewcommand\colorMATH{\colorMATHB}\renewcommand\colorSYNTAX{\colorSYNTAXB}{{\color{\colorMATH}\ensuremath{\sss}}}\endgroup }) \xrightarrowP {{\begingroup\renewcommand\colorMATH{\colorMATHC}\renewcommand\colorSYNTAX{\colorSYNTAXC}{{\color{\colorMATH}\ensuremath{\pS'}}}\endgroup }/\Gamma } (\tau _{2}/\Gamma )\rrbracket }}}, i.e. that
		 {{\color{\colorMATH}\ensuremath{\exists  \Gamma ', {\begingroup\renewcommand\colorMATH{\colorMATHB}\renewcommand\colorSYNTAX{\colorSYNTAXB}{{\color{\colorMATH}\ensuremath{\sS'_{0}}}}\endgroup }, dom({\begingroup\renewcommand\colorMATH{\colorMATHC}\renewcommand\colorSYNTAX{\colorSYNTAXC}{{\color{\colorMATH}\ensuremath{\pS''}}}\endgroup }) \subseteq  dom(\Gamma ') \subseteq  dom({\begingroup\renewcommand\colorMATH{\colorMATHB}\renewcommand\colorSYNTAX{\colorSYNTAXB}{{\color{\colorMATH}\ensuremath{\sS'_{0}}}}\endgroup }), \forall  x_{i} \in  dom(\Gamma '), \varnothing ; \varnothing  \vdash  \gamma (x_{i}) : \tau '_i, \tau '_i <: \Gamma '(x_{i}) ; \varnothing , \Gamma '}}}, and {{\color{\colorMATH}\ensuremath{x: \tau _{1}; {\begingroup\renewcommand\colorMATH{\colorMATHB}\renewcommand\colorSYNTAX{\colorSYNTAXB}{{\color{\colorMATH}\ensuremath{\sS'_{0}}}}\endgroup } + {\begingroup\renewcommand\colorMATH{\colorMATHB}\renewcommand\colorSYNTAX{\colorSYNTAXB}{{\color{\colorMATH}\ensuremath{\sss}}}\endgroup }x \vdash   {\begingroup\renewcommand\colorMATH{\colorMATHC}\renewcommand\colorSYNTAX{\colorSYNTAXC}{{\color{\colorMATH}\ensuremath{\pe'}}}\endgroup } \mathrel{:} \tau _{2} \mathrel{;} {\begingroup\renewcommand\colorMATH{\colorMATHC}\renewcommand\colorSYNTAX{\colorSYNTAXC}{{\color{\colorMATH}\ensuremath{\pS'}}}\endgroup }}}}.
		We prove this by choosing {{\color{\colorMATH}\ensuremath{\Gamma ' = \Gamma }}}, {{\color{\colorMATH}\ensuremath{{\begingroup\renewcommand\colorMATH{\colorMATHB}\renewcommand\colorSYNTAX{\colorSYNTAXB}{{\color{\colorMATH}\ensuremath{\sS'_{0}}}}\endgroup } = {\begingroup\renewcommand\colorMATH{\colorMATHB}\renewcommand\colorSYNTAX{\colorSYNTAXB}{{\color{\colorMATH}\ensuremath{\sS_{0}}}}\endgroup }}}}:
		 \begingroup\color{\colorMATH}\begin{gather*} 
			\inferrule*[lab=
			]{ \forall  x_{i} \in  dom(\Gamma ), \varnothing ; \varnothing  \vdash  \gamma (x_{i}) : \tau '_{i}, \tau '_{i} <: \Gamma (x_{i}) ; \varnothing 
			\\ \Gamma , x: \tau _{1}; {\begingroup\renewcommand\colorMATH{\colorMATHB}\renewcommand\colorSYNTAX{\colorSYNTAXB}{{\color{\colorMATH}\ensuremath{\sS_{0}}}}\endgroup } + {\begingroup\renewcommand\colorMATH{\colorMATHB}\renewcommand\colorSYNTAX{\colorSYNTAXB}{{\color{\colorMATH}\ensuremath{\sss_{1}}}}\endgroup }x \vdash   {\begingroup\renewcommand\colorMATH{\colorMATHC}\renewcommand\colorSYNTAX{\colorSYNTAXC}{{\color{\colorMATH}\ensuremath{\pe'}}}\endgroup } \mathrel{:} \tau _{2} \mathrel{;} {\begingroup\renewcommand\colorMATH{\colorMATHC}\renewcommand\colorSYNTAX{\colorSYNTAXC}{{\color{\colorMATH}\ensuremath{\pS'}}}\endgroup }
			  }{
			  \Gamma ; {\begingroup\renewcommand\colorMATH{\colorMATHB}\renewcommand\colorSYNTAX{\colorSYNTAXB}{{\color{\colorMATH}\ensuremath{\sS_{0}}}}\endgroup } \vdash  \langle {\begingroup\renewcommand\colorMATH{\colorMATHC}\renewcommand\colorSYNTAX{\colorSYNTAXC}{{\color{\colorMATH}\ensuremath{\plambda}}}\endgroup } (x\mathrel{:}\tau _{1}\mathord{\cdotp }{\begingroup\renewcommand\colorMATH{\colorMATHB}\renewcommand\colorSYNTAX{\colorSYNTAXB}{{\color{\colorMATH}\ensuremath{\sss_{1}}}}\endgroup }).\hspace*{0.33em} {\begingroup\renewcommand\colorMATH{\colorMATHC}\renewcommand\colorSYNTAX{\colorSYNTAXC}{{\color{\colorMATH}\ensuremath{\pe'}}}\endgroup }, \gamma \rangle  \mathrel{:} (x\mathrel{:} \tau _{1}/\Gamma \mathord{\cdotp }{\begingroup\renewcommand\colorMATH{\colorMATHB}\renewcommand\colorSYNTAX{\colorSYNTAXB}{{\color{\colorMATH}\ensuremath{\sss_{1}}}}\endgroup }) \xrightarrowP {{\begingroup\renewcommand\colorMATH{\colorMATHC}\renewcommand\colorSYNTAX{\colorSYNTAXC}{{\color{\colorMATH}\ensuremath{\pS'}}}\endgroup }/\Gamma } \tau _{2}/\Gamma  \mathrel{;} \varnothing 
			}
		\end{gather*}\endgroup

		Then we have prove that 
		{{\color{\colorMATH}\ensuremath{\forall  {\begingroup\renewcommand\colorMATH{\colorMATHB}\renewcommand\colorSYNTAX{\colorSYNTAXB}{{\color{\colorMATH}\ensuremath{\sv'}}}\endgroup } \in  {\mathcal{V}}\llbracket \tau _{1}/\Gamma \rrbracket , \gamma [x \mapsto  {\begingroup\renewcommand\colorMATH{\colorMATHB}\renewcommand\colorSYNTAX{\colorSYNTAXB}{{\color{\colorMATH}\ensuremath{\sv'}}}\endgroup }] \vdash   {\begingroup\renewcommand\colorMATH{\colorMATHC}\renewcommand\colorSYNTAX{\colorSYNTAXC}{{\color{\colorMATH}\ensuremath{\pe'}}}\endgroup } \in  {\mathcal{E}}\llbracket \tau _{2}/(\Gamma ,x:\tau _{1})\rrbracket }}}.
		By induction hypothesis on 
		{{\color{\colorMATH}\ensuremath{\Gamma ,x\mathrel{:}\tau _{1} \mathrel{;} {\begingroup\renewcommand\colorMATH{\colorMATHB}\renewcommand\colorSYNTAX{\colorSYNTAXB}{{\color{\colorMATH}\ensuremath{\sS_{0}}}}\endgroup } + {\begingroup\renewcommand\colorMATH{\colorMATHB}\renewcommand\colorSYNTAX{\colorSYNTAXB}{{\color{\colorMATH}\ensuremath{\sss}}}\endgroup }x\hspace*{0.33em}{\begingroup\renewcommand\colorMATH{\colorMATHB}\renewcommand\colorSYNTAX{\colorSYNTAXB}{{\color{\colorMATH}\ensuremath{\vdash }}}\endgroup }\hspace*{0.33em} {\begingroup\renewcommand\colorMATH{\colorMATHC}\renewcommand\colorSYNTAX{\colorSYNTAXC}{{\color{\colorMATH}\ensuremath{\pe'}}}\endgroup } \mathrel{:} \tau _{2} \mathrel{;} {\begingroup\renewcommand\colorMATH{\colorMATHB}\renewcommand\colorSYNTAX{\colorSYNTAXB}{{\color{\colorMATH}\ensuremath{\sS'}}}\endgroup }}}}, we know that
		for any {{\color{\colorMATH}\ensuremath{\gamma ' \in  {\mathcal{G}}\llbracket \Gamma ,x:\tau _{1}\rrbracket }}}, {{\color{\colorMATH}\ensuremath{\gamma ' \vdash   {\begingroup\renewcommand\colorMATH{\colorMATHC}\renewcommand\colorSYNTAX{\colorSYNTAXC}{{\color{\colorMATH}\ensuremath{\pe'}}}\endgroup } \in  {\mathcal{E}}\llbracket \tau _{2}/(\Gamma ,x:\tau _{1})\rrbracket }}}.
		As {{\color{\colorMATH}\ensuremath{\gamma  \in  {\mathcal{G}}\llbracket \Gamma \rrbracket }}} and {{\color{\colorMATH}\ensuremath{{\begingroup\renewcommand\colorMATH{\colorMATHB}\renewcommand\colorSYNTAX{\colorSYNTAXB}{{\color{\colorMATH}\ensuremath{\sv'}}}\endgroup } \in  {\mathcal{V}}\llbracket \tau _{1}/(\Gamma , x:\tau _{1})\rrbracket }}} ({{\color{\colorMATH}\ensuremath{\tau _{1}/\Gamma  = \tau _{1}/(\Gamma , x:\tau _{1})}}})
		{{\color{\colorMATH}\ensuremath{\gamma [x \mapsto  {\begingroup\renewcommand\colorMATH{\colorMATHB}\renewcommand\colorSYNTAX{\colorSYNTAXB}{{\color{\colorMATH}\ensuremath{\sv'}}}\endgroup }] \in  {\mathcal{G}}\llbracket \Gamma ,x:\tau _{1}\rrbracket }}}, so we pick 
		{{\color{\colorMATH}\ensuremath{\gamma ' = \gamma [x \mapsto  {\begingroup\renewcommand\colorMATH{\colorMATHB}\renewcommand\colorSYNTAX{\colorSYNTAXB}{{\color{\colorMATH}\ensuremath{\sv'}}}\endgroup }]}}} and the result holds.
	\end{subproof}
\item  {{\color{\colorMATH}\ensuremath{\Gamma  \mathrel{;} {\begingroup\renewcommand\colorMATH{\colorMATHB}\renewcommand\colorSYNTAX{\colorSYNTAXB}{{\color{\colorMATH}\ensuremath{\sS_{0}}}}\endgroup } \hspace*{0.33em}{\begingroup\renewcommand\colorMATH{\colorMATHB}\renewcommand\colorSYNTAX{\colorSYNTAXB}{{\color{\colorMATH}\ensuremath{\vdash }}}\endgroup }\hspace*{0.33em}\inl^{\tau _{2}}\hspace*{0.33em}{\begingroup\renewcommand\colorMATH{\colorMATHB}\renewcommand\colorSYNTAX{\colorSYNTAXB}{{\color{\colorMATH}\ensuremath{\se'}}}\endgroup } \mathrel{:} \tau _{1} \mathrel{^{{\begingroup\renewcommand\colorMATH{\colorMATHB}\renewcommand\colorSYNTAX{\colorSYNTAXB}{{\color{\colorMATH}\ensuremath{\sS}}}\endgroup }}\oplus ^{{\begingroup\renewcommand\colorMATH{\colorMATHB}\renewcommand\colorSYNTAX{\colorSYNTAXB}{{\color{\colorMATH}\ensuremath{\varnothing }}}\endgroup }}} \tau _{2} \mathrel{;} {\begingroup\renewcommand\colorMATH{\colorMATHB}\renewcommand\colorSYNTAX{\colorSYNTAXB}{{\color{\colorMATH}\ensuremath{\varnothing }}}\endgroup }}}}
	\begin{subproof} 

		We know that 
		\begingroup\color{\colorMATH}\begin{gather*} 
			\inferrule*[lab={\textsc{ inl}}
			]{ \Gamma  \mathrel{;} {\begingroup\renewcommand\colorMATH{\colorMATHB}\renewcommand\colorSYNTAX{\colorSYNTAXB}{{\color{\colorMATH}\ensuremath{\sS_{0}}}}\endgroup } \hspace*{0.33em}{\begingroup\renewcommand\colorMATH{\colorMATHB}\renewcommand\colorSYNTAX{\colorSYNTAXB}{{\color{\colorMATH}\ensuremath{\vdash }}}\endgroup }\hspace*{0.33em}{\begingroup\renewcommand\colorMATH{\colorMATHB}\renewcommand\colorSYNTAX{\colorSYNTAXB}{{\color{\colorMATH}\ensuremath{\se'}}}\endgroup } \mathrel{:} \tau _{1} \mathrel{;} {\begingroup\renewcommand\colorMATH{\colorMATHB}\renewcommand\colorSYNTAX{\colorSYNTAXB}{{\color{\colorMATH}\ensuremath{\sS}}}\endgroup }
			  }{
			  \Gamma  \mathrel{;} {\begingroup\renewcommand\colorMATH{\colorMATHB}\renewcommand\colorSYNTAX{\colorSYNTAXB}{{\color{\colorMATH}\ensuremath{\sS_{0}}}}\endgroup } \hspace*{0.33em}{\begingroup\renewcommand\colorMATH{\colorMATHB}\renewcommand\colorSYNTAX{\colorSYNTAXB}{{\color{\colorMATH}\ensuremath{\vdash }}}\endgroup }\hspace*{0.33em}\inl^{\tau _{2}}\hspace*{0.33em}{\begingroup\renewcommand\colorMATH{\colorMATHB}\renewcommand\colorSYNTAX{\colorSYNTAXB}{{\color{\colorMATH}\ensuremath{\se'}}}\endgroup } \mathrel{:} \tau _{1} \mathrel{^{{\begingroup\renewcommand\colorMATH{\colorMATHB}\renewcommand\colorSYNTAX{\colorSYNTAXB}{{\color{\colorMATH}\ensuremath{\sS}}}\endgroup }}\oplus ^{{\begingroup\renewcommand\colorMATH{\colorMATHB}\renewcommand\colorSYNTAX{\colorSYNTAXB}{{\color{\colorMATH}\ensuremath{\varnothing }}}\endgroup }}} \tau _{2} \mathrel{;} {\begingroup\renewcommand\colorMATH{\colorMATHB}\renewcommand\colorSYNTAX{\colorSYNTAXB}{{\color{\colorMATH}\ensuremath{\varnothing }}}\endgroup }
			}
		\end{gather*}\endgroup
		By induction hypothesis on {{\color{\colorMATH}\ensuremath{\Gamma  \mathrel{;} {\begingroup\renewcommand\colorMATH{\colorMATHB}\renewcommand\colorSYNTAX{\colorSYNTAXB}{{\color{\colorMATH}\ensuremath{\sS_{0}}}}\endgroup } \hspace*{0.33em}{\begingroup\renewcommand\colorMATH{\colorMATHB}\renewcommand\colorSYNTAX{\colorSYNTAXB}{{\color{\colorMATH}\ensuremath{\vdash }}}\endgroup }\hspace*{0.33em}{\begingroup\renewcommand\colorMATH{\colorMATHB}\renewcommand\colorSYNTAX{\colorSYNTAXB}{{\color{\colorMATH}\ensuremath{\se'}}}\endgroup } \mathrel{:} \tau _{1} \mathrel{;} {\begingroup\renewcommand\colorMATH{\colorMATHB}\renewcommand\colorSYNTAX{\colorSYNTAXB}{{\color{\colorMATH}\ensuremath{\sS}}}\endgroup }}}} we know that {{\color{\colorMATH}\ensuremath{\gamma  \vdash  {\begingroup\renewcommand\colorMATH{\colorMATHB}\renewcommand\colorSYNTAX{\colorSYNTAXB}{{\color{\colorMATH}\ensuremath{\se'}}}\endgroup } \Downarrow  {\begingroup\renewcommand\colorMATH{\colorMATHB}\renewcommand\colorSYNTAX{\colorSYNTAXB}{{\color{\colorMATH}\ensuremath{\sv'}}}\endgroup }}}} and {{\color{\colorMATH}\ensuremath{{\begingroup\renewcommand\colorMATH{\colorMATHB}\renewcommand\colorSYNTAX{\colorSYNTAXB}{{\color{\colorMATH}\ensuremath{\sv'}}}\endgroup } \in  {\mathcal{V}}\llbracket \tau _{1}/\Gamma \rrbracket }}}.

		As {{\color{\colorMATH}\ensuremath{\gamma  \vdash  \inl^{\tau _{2}}\hspace*{0.33em}{\begingroup\renewcommand\colorMATH{\colorMATHB}\renewcommand\colorSYNTAX{\colorSYNTAXB}{{\color{\colorMATH}\ensuremath{\se'}}}\endgroup } \Downarrow  \inl^{\tau _{2}/\gamma } {\begingroup\renewcommand\colorMATH{\colorMATHB}\renewcommand\colorSYNTAX{\colorSYNTAXB}{{\color{\colorMATH}\ensuremath{\sv'}}}\endgroup }}}}, we have to prove that 
		{{\color{\colorMATH}\ensuremath{\inl^{\tau _{2}/\gamma } {\begingroup\renewcommand\colorMATH{\colorMATHB}\renewcommand\colorSYNTAX{\colorSYNTAXB}{{\color{\colorMATH}\ensuremath{\sv'}}}\endgroup } \in  {\mathcal{V}}\llbracket (\tau _{1} \mathrel{^{{\begingroup\renewcommand\colorMATH{\colorMATHB}\renewcommand\colorSYNTAX{\colorSYNTAXB}{{\color{\colorMATH}\ensuremath{\sS}}}\endgroup }}\oplus ^{{\begingroup\renewcommand\colorMATH{\colorMATHB}\renewcommand\colorSYNTAX{\colorSYNTAXB}{{\color{\colorMATH}\ensuremath{\varnothing }}}\endgroup }}} \tau _{2})/\Gamma \rrbracket }}}.
		Notice that {{\color{\colorMATH}\ensuremath{(\tau _{1} \mathrel{^{{\begingroup\renewcommand\colorMATH{\colorMATHB}\renewcommand\colorSYNTAX{\colorSYNTAXB}{{\color{\colorMATH}\ensuremath{\sS}}}\endgroup }}\oplus ^{{\begingroup\renewcommand\colorMATH{\colorMATHB}\renewcommand\colorSYNTAX{\colorSYNTAXB}{{\color{\colorMATH}\ensuremath{\varnothing }}}\endgroup }}} \tau _{2})/\Gamma  = (\tau _{1}/\Gamma  \mathrel{^{{\begingroup\renewcommand\colorMATH{\colorMATHB}\renewcommand\colorSYNTAX{\colorSYNTAXB}{{\color{\colorMATH}\ensuremath{\varnothing }}}\endgroup }}\oplus ^{{\begingroup\renewcommand\colorMATH{\colorMATHB}\renewcommand\colorSYNTAX{\colorSYNTAXB}{{\color{\colorMATH}\ensuremath{\varnothing }}}\endgroup }}} \tau _{2}/\Gamma }}}, and that {{\color{\colorMATH}\ensuremath{\tau _{2}/\gamma  = \tau _{2}/\Gamma }}}
		Then we have to prove that {{\color{\colorMATH}\ensuremath{\inl^{\tau _{2}/\Gamma } {\begingroup\renewcommand\colorMATH{\colorMATHB}\renewcommand\colorSYNTAX{\colorSYNTAXB}{{\color{\colorMATH}\ensuremath{\sv'}}}\endgroup } \in  {\mathcal{V}}\llbracket \tau _{1}/\Gamma  \mathrel{^{{\begingroup\renewcommand\colorMATH{\colorMATHB}\renewcommand\colorSYNTAX{\colorSYNTAXB}{{\color{\colorMATH}\ensuremath{\varnothing }}}\endgroup }}\oplus ^{{\begingroup\renewcommand\colorMATH{\colorMATHB}\renewcommand\colorSYNTAX{\colorSYNTAXB}{{\color{\colorMATH}\ensuremath{\varnothing }}}\endgroup }}} \tau _{2}/\Gamma \rrbracket }}}, which is direct as we already know that {{\color{\colorMATH}\ensuremath{{\begingroup\renewcommand\colorMATH{\colorMATHB}\renewcommand\colorSYNTAX{\colorSYNTAXB}{{\color{\colorMATH}\ensuremath{\sv'}}}\endgroup } \in  {\mathcal{V}}\llbracket \tau _{1}/\Gamma \rrbracket }}}.
	\end{subproof}
\item  {{\color{\colorMATH}\ensuremath{\Gamma  \mathrel{;} {\begingroup\renewcommand\colorMATH{\colorMATHB}\renewcommand\colorSYNTAX{\colorSYNTAXB}{{\color{\colorMATH}\ensuremath{\sS_{0}}}}\endgroup } \hspace*{0.33em}{\begingroup\renewcommand\colorMATH{\colorMATHB}\renewcommand\colorSYNTAX{\colorSYNTAXB}{{\color{\colorMATH}\ensuremath{\vdash }}}\endgroup }\hspace*{0.33em}\inr^{\tau _{1}}\hspace*{0.33em}{\begingroup\renewcommand\colorMATH{\colorMATHB}\renewcommand\colorSYNTAX{\colorSYNTAXB}{{\color{\colorMATH}\ensuremath{\se'}}}\endgroup } \mathrel{:} \tau _{1} \mathrel{^{{\begingroup\renewcommand\colorMATH{\colorMATHB}\renewcommand\colorSYNTAX{\colorSYNTAXB}{{\color{\colorMATH}\ensuremath{\sS}}}\endgroup }}\oplus ^{{\begingroup\renewcommand\colorMATH{\colorMATHB}\renewcommand\colorSYNTAX{\colorSYNTAXB}{{\color{\colorMATH}\ensuremath{\varnothing }}}\endgroup }}} \tau _{2} \mathrel{;} {\begingroup\renewcommand\colorMATH{\colorMATHB}\renewcommand\colorSYNTAX{\colorSYNTAXB}{{\color{\colorMATH}\ensuremath{\varnothing }}}\endgroup }}}}
	\begin{subproof} 
		Analogous to the {{\color{\colorMATH}\ensuremath{\inl^{\tau _{2}}\hspace*{0.33em}{\begingroup\renewcommand\colorMATH{\colorMATHB}\renewcommand\colorSYNTAX{\colorSYNTAXB}{{\color{\colorMATH}\ensuremath{\se'}}}\endgroup }}}} case.
	\end{subproof}
\item  {{\color{\colorMATH}\ensuremath{\Gamma  \mathrel{;} {\begingroup\renewcommand\colorMATH{\colorMATHB}\renewcommand\colorSYNTAX{\colorSYNTAXB}{{\color{\colorMATH}\ensuremath{\sS_{0}}}}\endgroup }  \hspace*{0.33em}{\begingroup\renewcommand\colorMATH{\colorMATHB}\renewcommand\colorSYNTAX{\colorSYNTAXB}{{\color{\colorMATH}\ensuremath{\vdash }}}\endgroup }\hspace*{0.33em}\ccase\hspace*{0.33em}{\begingroup\renewcommand\colorMATH{\colorMATHB}\renewcommand\colorSYNTAX{\colorSYNTAXB}{{\color{\colorMATH}\ensuremath{\se_{1}}}}\endgroup }\hspace*{0.33em}\of\hspace*{0.33em}\{ x\Rightarrow {\begingroup\renewcommand\colorMATH{\colorMATHB}\renewcommand\colorSYNTAX{\colorSYNTAXB}{{\color{\colorMATH}\ensuremath{\se_{2}}}}\endgroup }\} \hspace*{0.33em}\{ y\Rightarrow {\begingroup\renewcommand\colorMATH{\colorMATHB}\renewcommand\colorSYNTAX{\colorSYNTAXB}{{\color{\colorMATH}\ensuremath{\se_{3}}}}\endgroup }\}  \mathrel{:} [{\begingroup\renewcommand\colorMATH{\colorMATHB}\renewcommand\colorSYNTAX{\colorSYNTAXB}{{\color{\colorMATH}\ensuremath{\sS_{1}}}}\endgroup } + {\begingroup\renewcommand\colorMATH{\colorMATHB}\renewcommand\colorSYNTAX{\colorSYNTAXB}{{\color{\colorMATH}\ensuremath{\sS_{1 1}}}}\endgroup }/x]\tau _{2} \sqcup  [{\begingroup\renewcommand\colorMATH{\colorMATHB}\renewcommand\colorSYNTAX{\colorSYNTAXB}{{\color{\colorMATH}\ensuremath{\sS_{1}}}}\endgroup } + {\begingroup\renewcommand\colorMATH{\colorMATHB}\renewcommand\colorSYNTAX{\colorSYNTAXB}{{\color{\colorMATH}\ensuremath{\sS_{1 2}}}}\endgroup }/y]\tau _{3} \mathrel{;} (({\begingroup\renewcommand\colorMATH{\colorMATHB}\renewcommand\colorSYNTAX{\colorSYNTAXB}{{\color{\colorMATH}\ensuremath{s_{2}*\Sigma _{1}}}}\endgroup } + {\begingroup\renewcommand\colorMATH{\colorMATHB}\renewcommand\colorSYNTAX{\colorSYNTAXB}{{\color{\colorMATH}\ensuremath{s_{2}\Sigma _{1 1}}}}\endgroup } + {\begingroup\renewcommand\colorMATH{\colorMATHB}\renewcommand\colorSYNTAX{\colorSYNTAXB}{{\color{\colorMATH}\ensuremath{\sS_{2}}}}\endgroup }) \sqcup  ({\begingroup\renewcommand\colorMATH{\colorMATHB}\renewcommand\colorSYNTAX{\colorSYNTAXB}{{\color{\colorMATH}\ensuremath{s_{3}*\Sigma _{1}}}}\endgroup } + {\begingroup\renewcommand\colorMATH{\colorMATHB}\renewcommand\colorSYNTAX{\colorSYNTAXB}{{\color{\colorMATH}\ensuremath{s_{3}\Sigma _{1 2}}}}\endgroup } + {\begingroup\renewcommand\colorMATH{\colorMATHB}\renewcommand\colorSYNTAX{\colorSYNTAXB}{{\color{\colorMATH}\ensuremath{\sS_{3}}}}\endgroup }))}}}
	\begin{subproof} 
		We know that 
		\begingroup\color{\colorMATH}\begin{gather*} 
		   \inferrule*[lab={\textsc{ s-case}}
		   ]{ \Gamma  \mathrel{;} {\begingroup\renewcommand\colorMATH{\colorMATHB}\renewcommand\colorSYNTAX{\colorSYNTAXB}{{\color{\colorMATH}\ensuremath{\sS_{0}}}}\endgroup }  \hspace*{0.33em}{\begingroup\renewcommand\colorMATH{\colorMATHB}\renewcommand\colorSYNTAX{\colorSYNTAXB}{{\color{\colorMATH}\ensuremath{\vdash }}}\endgroup }\hspace*{0.33em}{\begingroup\renewcommand\colorMATH{\colorMATHB}\renewcommand\colorSYNTAX{\colorSYNTAXB}{{\color{\colorMATH}\ensuremath{\se_{1}}}}\endgroup } \mathrel{:} \tau _{1 1} \mathrel{^{{\begingroup\renewcommand\colorMATH{\colorMATHB}\renewcommand\colorSYNTAX{\colorSYNTAXB}{{\color{\colorMATH}\ensuremath{\sS_{1 1}}}}\endgroup }}\oplus ^{{\begingroup\renewcommand\colorMATH{\colorMATHB}\renewcommand\colorSYNTAX{\colorSYNTAXB}{{\color{\colorMATH}\ensuremath{\sS_{1 2}}}}\endgroup }}} \tau _{1 2} \mathrel{;} {\begingroup\renewcommand\colorMATH{\colorMATHB}\renewcommand\colorSYNTAX{\colorSYNTAXB}{{\color{\colorMATH}\ensuremath{\sS_{1}}}}\endgroup }
		   \\ \Gamma ,x\mathrel{:}\tau _{1 1} \mathrel{;} {\begingroup\renewcommand\colorMATH{\colorMATHB}\renewcommand\colorSYNTAX{\colorSYNTAXB}{{\color{\colorMATH}\ensuremath{\sS_{0}}}}\endgroup } + ({\begingroup\renewcommand\colorMATH{\colorMATHB}\renewcommand\colorSYNTAX{\colorSYNTAXB}{{\color{\colorMATH}\ensuremath{\sS_{0}}}}\endgroup }\mathord{\cdotp }({\begingroup\renewcommand\colorMATH{\colorMATHB}\renewcommand\colorSYNTAX{\colorSYNTAXB}{{\color{\colorMATH}\ensuremath{\sS_{1}}}}\endgroup } + {\begingroup\renewcommand\colorMATH{\colorMATHB}\renewcommand\colorSYNTAX{\colorSYNTAXB}{{\color{\colorMATH}\ensuremath{\sS_{1 1}}}}\endgroup }))x  \hspace*{0.33em}{\begingroup\renewcommand\colorMATH{\colorMATHB}\renewcommand\colorSYNTAX{\colorSYNTAXB}{{\color{\colorMATH}\ensuremath{ \vdash  e_{2}}}}\endgroup } \mathrel{:} \tau _{2} \mathrel{;} {\begingroup\renewcommand\colorMATH{\colorMATHB}\renewcommand\colorSYNTAX{\colorSYNTAXB}{{\color{\colorMATH}\ensuremath{\sS_{2}}}}\endgroup }+{\begingroup\renewcommand\colorMATH{\colorMATHB}\renewcommand\colorSYNTAX{\colorSYNTAXB}{{\color{\colorMATH}\ensuremath{\sss_{2}}}}\endgroup }x
		   \\ \Gamma ,y\mathrel{:}\tau _{1 2} \mathrel{;} {\begingroup\renewcommand\colorMATH{\colorMATHB}\renewcommand\colorSYNTAX{\colorSYNTAXB}{{\color{\colorMATH}\ensuremath{\sS_{0}}}}\endgroup } + ({\begingroup\renewcommand\colorMATH{\colorMATHB}\renewcommand\colorSYNTAX{\colorSYNTAXB}{{\color{\colorMATH}\ensuremath{\sS_{0}}}}\endgroup }\mathord{\cdotp }({\begingroup\renewcommand\colorMATH{\colorMATHB}\renewcommand\colorSYNTAX{\colorSYNTAXB}{{\color{\colorMATH}\ensuremath{\sS_{1}}}}\endgroup } + {\begingroup\renewcommand\colorMATH{\colorMATHB}\renewcommand\colorSYNTAX{\colorSYNTAXB}{{\color{\colorMATH}\ensuremath{\sS_{1 2}}}}\endgroup }))x  \hspace*{0.33em}{\begingroup\renewcommand\colorMATH{\colorMATHB}\renewcommand\colorSYNTAX{\colorSYNTAXB}{{\color{\colorMATH}\ensuremath{ \vdash  e_{3}}}}\endgroup } \mathrel{:} \tau _{3} \mathrel{;} {\begingroup\renewcommand\colorMATH{\colorMATHB}\renewcommand\colorSYNTAX{\colorSYNTAXB}{{\color{\colorMATH}\ensuremath{\sS_{3}}}}\endgroup }+{\begingroup\renewcommand\colorMATH{\colorMATHB}\renewcommand\colorSYNTAX{\colorSYNTAXB}{{\color{\colorMATH}\ensuremath{\sss_{3}}}}\endgroup }y
		      }{
		      \Gamma  \mathrel{;} {\begingroup\renewcommand\colorMATH{\colorMATHB}\renewcommand\colorSYNTAX{\colorSYNTAXB}{{\color{\colorMATH}\ensuremath{\sS_{0}}}}\endgroup }  \hspace*{0.33em}{\begingroup\renewcommand\colorMATH{\colorMATHB}\renewcommand\colorSYNTAX{\colorSYNTAXB}{{\color{\colorMATH}\ensuremath{\vdash }}}\endgroup }\hspace*{0.33em}\ccase\hspace*{0.33em}{\begingroup\renewcommand\colorMATH{\colorMATHB}\renewcommand\colorSYNTAX{\colorSYNTAXB}{{\color{\colorMATH}\ensuremath{\se_{1}}}}\endgroup }\hspace*{0.33em}\of\hspace*{0.33em}\{ x\Rightarrow {\begingroup\renewcommand\colorMATH{\colorMATHB}\renewcommand\colorSYNTAX{\colorSYNTAXB}{{\color{\colorMATH}\ensuremath{\se_{2}}}}\endgroup }\} \hspace*{0.33em}\{ y\Rightarrow {\begingroup\renewcommand\colorMATH{\colorMATHB}\renewcommand\colorSYNTAX{\colorSYNTAXB}{{\color{\colorMATH}\ensuremath{\se_{3}}}}\endgroup }\}  \mathrel{:} \qquad\qquad\qquad\qquad\qquad\qquad
		      \\ \qquad\qquad\qquad [{\begingroup\renewcommand\colorMATH{\colorMATHB}\renewcommand\colorSYNTAX{\colorSYNTAXB}{{\color{\colorMATH}\ensuremath{\sS_{1}}}}\endgroup } + {\begingroup\renewcommand\colorMATH{\colorMATHB}\renewcommand\colorSYNTAX{\colorSYNTAXB}{{\color{\colorMATH}\ensuremath{\sS_{1 1}}}}\endgroup }/x]\tau _{2} \sqcup  [{\begingroup\renewcommand\colorMATH{\colorMATHB}\renewcommand\colorSYNTAX{\colorSYNTAXB}{{\color{\colorMATH}\ensuremath{\sS_{1}}}}\endgroup } + {\begingroup\renewcommand\colorMATH{\colorMATHB}\renewcommand\colorSYNTAX{\colorSYNTAXB}{{\color{\colorMATH}\ensuremath{\sS_{1 2}}}}\endgroup }/y]\tau _{3} \mathrel{;} (({\begingroup\renewcommand\colorMATH{\colorMATHB}\renewcommand\colorSYNTAX{\colorSYNTAXB}{{\color{\colorMATH}\ensuremath{s_{2}*\Sigma _{1}}}}\endgroup } + {\begingroup\renewcommand\colorMATH{\colorMATHB}\renewcommand\colorSYNTAX{\colorSYNTAXB}{{\color{\colorMATH}\ensuremath{s_{2}\Sigma _{1 1}}}}\endgroup } + {\begingroup\renewcommand\colorMATH{\colorMATHB}\renewcommand\colorSYNTAX{\colorSYNTAXB}{{\color{\colorMATH}\ensuremath{\sS_{2}}}}\endgroup }) \sqcup  ({\begingroup\renewcommand\colorMATH{\colorMATHB}\renewcommand\colorSYNTAX{\colorSYNTAXB}{{\color{\colorMATH}\ensuremath{s_{3}*\Sigma _{1}}}}\endgroup } + {\begingroup\renewcommand\colorMATH{\colorMATHB}\renewcommand\colorSYNTAX{\colorSYNTAXB}{{\color{\colorMATH}\ensuremath{s_{3}\Sigma _{1 2}}}}\endgroup } + {\begingroup\renewcommand\colorMATH{\colorMATHB}\renewcommand\colorSYNTAX{\colorSYNTAXB}{{\color{\colorMATH}\ensuremath{\sS_{3}}}}\endgroup }))
		   }
		\end{gather*}\endgroup
		where {{\color{\colorMATH}\ensuremath{\tau  = [{\begingroup\renewcommand\colorMATH{\colorMATHB}\renewcommand\colorSYNTAX{\colorSYNTAXB}{{\color{\colorMATH}\ensuremath{\sS_{1 1}}}}\endgroup }/x]\tau _{2} \sqcup  [{\begingroup\renewcommand\colorMATH{\colorMATHB}\renewcommand\colorSYNTAX{\colorSYNTAXB}{{\color{\colorMATH}\ensuremath{\sS_{1 2}}}}\endgroup }/y]\tau _{3}}}}.
		
		By induction hypothesis on {{\color{\colorMATH}\ensuremath{\Gamma  \mathrel{;} {\begingroup\renewcommand\colorMATH{\colorMATHB}\renewcommand\colorSYNTAX{\colorSYNTAXB}{{\color{\colorMATH}\ensuremath{\sS_{0}}}}\endgroup }\hspace*{0.33em}{\begingroup\renewcommand\colorMATH{\colorMATHB}\renewcommand\colorSYNTAX{\colorSYNTAXB}{{\color{\colorMATH}\ensuremath{\vdash }}}\endgroup }\hspace*{0.33em} {\begingroup\renewcommand\colorMATH{\colorMATHB}\renewcommand\colorSYNTAX{\colorSYNTAXB}{{\color{\colorMATH}\ensuremath{\se_{1}}}}\endgroup } \mathrel{:} \tau _{1 1} \mathrel{^{{\begingroup\renewcommand\colorMATH{\colorMATHB}\renewcommand\colorSYNTAX{\colorSYNTAXB}{{\color{\colorMATH}\ensuremath{\sS_{1 1}}}}\endgroup }}\oplus ^{{\begingroup\renewcommand\colorMATH{\colorMATHB}\renewcommand\colorSYNTAX{\colorSYNTAXB}{{\color{\colorMATH}\ensuremath{\sS_{1 2}}}}\endgroup }}} \tau _{1 2} \mathrel{;} {\begingroup\renewcommand\colorMATH{\colorMATHB}\renewcommand\colorSYNTAX{\colorSYNTAXB}{{\color{\colorMATH}\ensuremath{\sS_{1}}}}\endgroup }}}},
		we know that {{\color{\colorMATH}\ensuremath{\gamma  \vdash  {\begingroup\renewcommand\colorMATH{\colorMATHB}\renewcommand\colorSYNTAX{\colorSYNTAXB}{{\color{\colorMATH}\ensuremath{\se_{1}}}}\endgroup } \Downarrow  {\begingroup\renewcommand\colorMATH{\colorMATHB}\renewcommand\colorSYNTAX{\colorSYNTAXB}{{\color{\colorMATH}\ensuremath{\sv_{1}}}}\endgroup }}}} and {{\color{\colorMATH}\ensuremath{{\begingroup\renewcommand\colorMATH{\colorMATHB}\renewcommand\colorSYNTAX{\colorSYNTAXB}{{\color{\colorMATH}\ensuremath{\sv_{1}}}}\endgroup } \in  {\mathcal{V}}\llbracket \tau _{1 1}/\Gamma  \mathrel{^{\varnothing }\oplus ^{\varnothing }} \tau _{1 2}/\Gamma \rrbracket }}} ({{\color{\colorMATH}\ensuremath{{\begingroup\renewcommand\colorMATH{\colorMATHB}\renewcommand\colorSYNTAX{\colorSYNTAXB}{{\color{\colorMATH}\ensuremath{\sS_{1 i}}}}\endgroup }/\Gamma  = \varnothing }}}).
		Then either {{\color{\colorMATH}\ensuremath{{\begingroup\renewcommand\colorMATH{\colorMATHB}\renewcommand\colorSYNTAX{\colorSYNTAXB}{{\color{\colorMATH}\ensuremath{\sv_{1}}}}\endgroup } = \inl^{\tau _{1 2}/\Gamma }\hspace*{0.33em}{\begingroup\renewcommand\colorMATH{\colorMATHB}\renewcommand\colorSYNTAX{\colorSYNTAXB}{{\color{\colorMATH}\ensuremath{\sv_{1 1}}}}\endgroup }}}}, or {{\color{\colorMATH}\ensuremath{{\begingroup\renewcommand\colorMATH{\colorMATHB}\renewcommand\colorSYNTAX{\colorSYNTAXB}{{\color{\colorMATH}\ensuremath{\sv_{1}}}}\endgroup } = \inl^{\tau _{1 1}/\Gamma }\hspace*{0.33em}{\begingroup\renewcommand\colorMATH{\colorMATHB}\renewcommand\colorSYNTAX{\colorSYNTAXB}{{\color{\colorMATH}\ensuremath{\sv_{1 2}}}}\endgroup }}}}.
		Let us suppose {{\color{\colorMATH}\ensuremath{{\begingroup\renewcommand\colorMATH{\colorMATHB}\renewcommand\colorSYNTAX{\colorSYNTAXB}{{\color{\colorMATH}\ensuremath{\sv_{1}}}}\endgroup } = \inl^{\tau _{1 2}/\Gamma }\hspace*{0.33em}{\begingroup\renewcommand\colorMATH{\colorMATHB}\renewcommand\colorSYNTAX{\colorSYNTAXB}{{\color{\colorMATH}\ensuremath{\sv_{1 1}}}}\endgroup }}}} (the other case is similar), then
		{{\color{\colorMATH}\ensuremath{\inl^{\tau _{1 2}/\Gamma }\hspace*{0.33em}{\begingroup\renewcommand\colorMATH{\colorMATHB}\renewcommand\colorSYNTAX{\colorSYNTAXB}{{\color{\colorMATH}\ensuremath{\sv_{1 1}}}}\endgroup } \in  {\mathcal{V}}\llbracket \tau _{1 1}/\Gamma  \mathrel{^{\varnothing }\oplus ^{\varnothing }} \tau _{1 2}/\Gamma \rrbracket }}}, and thus
		{{\color{\colorMATH}\ensuremath{{\begingroup\renewcommand\colorMATH{\colorMATHB}\renewcommand\colorSYNTAX{\colorSYNTAXB}{{\color{\colorMATH}\ensuremath{\sv_{1 1}}}}\endgroup } \in  {\mathcal{V}}\llbracket \tau _{1 1}/\Gamma \rrbracket }}}.

		Then by induction hypothesis on {{\color{\colorMATH}\ensuremath{\Gamma ,x\mathrel{:}\tau _{1 1} \mathrel{;} {\begingroup\renewcommand\colorMATH{\colorMATHB}\renewcommand\colorSYNTAX{\colorSYNTAXB}{{\color{\colorMATH}\ensuremath{\sS_{0}}}}\endgroup } + ({\begingroup\renewcommand\colorMATH{\colorMATHB}\renewcommand\colorSYNTAX{\colorSYNTAXB}{{\color{\colorMATH}\ensuremath{\sS_{0}}}}\endgroup } \mathord{\cdotp } ({\begingroup\renewcommand\colorMATH{\colorMATHB}\renewcommand\colorSYNTAX{\colorSYNTAXB}{{\color{\colorMATH}\ensuremath{\sS_{1}}}}\endgroup } + {\begingroup\renewcommand\colorMATH{\colorMATHB}\renewcommand\colorSYNTAX{\colorSYNTAXB}{{\color{\colorMATH}\ensuremath{\sS_{1 1}}}}\endgroup }))x   \hspace*{0.33em}{\begingroup\renewcommand\colorMATH{\colorMATHB}\renewcommand\colorSYNTAX{\colorSYNTAXB}{{\color{\colorMATH}\ensuremath{\vdash }}}\endgroup }\hspace*{0.33em} {\begingroup\renewcommand\colorMATH{\colorMATHB}\renewcommand\colorSYNTAX{\colorSYNTAXB}{{\color{\colorMATH}\ensuremath{\se_{2}}}}\endgroup } \mathrel{:} \tau _{2} \mathrel{;} {\begingroup\renewcommand\colorMATH{\colorMATHB}\renewcommand\colorSYNTAX{\colorSYNTAXB}{{\color{\colorMATH}\ensuremath{\sS_{2}}}}\endgroup }+{\begingroup\renewcommand\colorMATH{\colorMATHB}\renewcommand\colorSYNTAX{\colorSYNTAXB}{{\color{\colorMATH}\ensuremath{\sss_{2}}}}\endgroup }x}}}, we know that
		{{\color{\colorMATH}\ensuremath{\gamma [x \mapsto  {\begingroup\renewcommand\colorMATH{\colorMATHB}\renewcommand\colorSYNTAX{\colorSYNTAXB}{{\color{\colorMATH}\ensuremath{\sv_{1 1}}}}\endgroup }] \vdash  {\begingroup\renewcommand\colorMATH{\colorMATHB}\renewcommand\colorSYNTAX{\colorSYNTAXB}{{\color{\colorMATH}\ensuremath{\se_{2}}}}\endgroup } \in  {\mathcal{E}}\llbracket \tau /\Gamma \rrbracket }}}, therefore 
		{{\color{\colorMATH}\ensuremath{\gamma  \vdash  {\begingroup\renewcommand\colorMATH{\colorMATHB}\renewcommand\colorSYNTAX{\colorSYNTAXB}{{\color{\colorMATH}\ensuremath{\se_{2}}}}\endgroup } \Downarrow  {\begingroup\renewcommand\colorMATH{\colorMATHB}\renewcommand\colorSYNTAX{\colorSYNTAXB}{{\color{\colorMATH}\ensuremath{\sv_{2}}}}\endgroup }}}} and {{\color{\colorMATH}\ensuremath{{\begingroup\renewcommand\colorMATH{\colorMATHB}\renewcommand\colorSYNTAX{\colorSYNTAXB}{{\color{\colorMATH}\ensuremath{\sv_{2}}}}\endgroup } \in  {\mathcal{V}}\llbracket \tau _{2}/(\Gamma ,x\mathrel{:}\tau _{1 1})\rrbracket }}}.

		Then by inspection of the evaluation semantics for {{\color{\colorMATH}\ensuremath{{\textsc{ p-case}}}}},
		{{\color{\colorMATH}\ensuremath{\gamma  \vdash  \ccase\hspace*{0.33em}{\begingroup\renewcommand\colorMATH{\colorMATHB}\renewcommand\colorSYNTAX{\colorSYNTAXB}{{\color{\colorMATH}\ensuremath{\se_{1}}}}\endgroup }\hspace*{0.33em}\of\hspace*{0.33em}\{ x\Rightarrow {\begingroup\renewcommand\colorMATH{\colorMATHB}\renewcommand\colorSYNTAX{\colorSYNTAXB}{{\color{\colorMATH}\ensuremath{\se_{2}}}}\endgroup }\} \hspace*{0.33em}\{ y\Rightarrow {\begingroup\renewcommand\colorMATH{\colorMATHB}\renewcommand\colorSYNTAX{\colorSYNTAXB}{{\color{\colorMATH}\ensuremath{\se_{3}}}}\endgroup }\}  \Downarrow  {\begingroup\renewcommand\colorMATH{\colorMATHB}\renewcommand\colorSYNTAX{\colorSYNTAXB}{{\color{\colorMATH}\ensuremath{\sv_{2}}}}\endgroup }}}} and we have to prove that
		{{\color{\colorMATH}\ensuremath{{\begingroup\renewcommand\colorMATH{\colorMATHB}\renewcommand\colorSYNTAX{\colorSYNTAXB}{{\color{\colorMATH}\ensuremath{\sv_{2}}}}\endgroup } \in  {\mathcal{V}}\llbracket ([{\begingroup\renewcommand\colorMATH{\colorMATHB}\renewcommand\colorSYNTAX{\colorSYNTAXB}{{\color{\colorMATH}\ensuremath{\sS_{1}}}}\endgroup } + {\begingroup\renewcommand\colorMATH{\colorMATHB}\renewcommand\colorSYNTAX{\colorSYNTAXB}{{\color{\colorMATH}\ensuremath{\sS_{1 1}}}}\endgroup }/x]\tau _{2} \sqcup  [{\begingroup\renewcommand\colorMATH{\colorMATHB}\renewcommand\colorSYNTAX{\colorSYNTAXB}{{\color{\colorMATH}\ensuremath{\sS_{1}}}}\endgroup } + {\begingroup\renewcommand\colorMATH{\colorMATHB}\renewcommand\colorSYNTAX{\colorSYNTAXB}{{\color{\colorMATH}\ensuremath{\sS_{1 2}}}}\endgroup }/y]\tau _{3}\rrbracket )/\Gamma }}}.
		But {{\color{\colorMATH}\ensuremath{([{\begingroup\renewcommand\colorMATH{\colorMATHB}\renewcommand\colorSYNTAX{\colorSYNTAXB}{{\color{\colorMATH}\ensuremath{\sS_{1}}}}\endgroup } + {\begingroup\renewcommand\colorMATH{\colorMATHB}\renewcommand\colorSYNTAX{\colorSYNTAXB}{{\color{\colorMATH}\ensuremath{\sS_{1 1}}}}\endgroup }/x]\tau _{2} \sqcup  [{\begingroup\renewcommand\colorMATH{\colorMATHB}\renewcommand\colorSYNTAX{\colorSYNTAXB}{{\color{\colorMATH}\ensuremath{\sS_{1}}}}\endgroup } + {\begingroup\renewcommand\colorMATH{\colorMATHB}\renewcommand\colorSYNTAX{\colorSYNTAXB}{{\color{\colorMATH}\ensuremath{\sS_{1 2}}}}\endgroup }/y]\tau _{3}\rrbracket )/\Gamma  = (\tau _{2} \sqcup  \tau _{3}\rrbracket )/(\Gamma ,x:\tau _{1 1})}}}. The result follows from weakening lemma~\ref{lm:weakening-type-safety}.
	\end{subproof}
\end{enumerate}
{\bf (b) Privacy FP.}
We proceed by induction on {{\color{\colorMATH}\ensuremath{\Gamma ;{\begingroup\renewcommand\colorMATH{\colorMATHB}\renewcommand\colorSYNTAX{\colorSYNTAXB}{{\color{\colorMATH}\ensuremath{\sS_{0}}}}\endgroup } \vdash 	{\begingroup\renewcommand\colorMATH{\colorMATHC}\renewcommand\colorSYNTAX{\colorSYNTAXC}{{\color{\colorMATH}\ensuremath{\pe}}}\endgroup } : \tau _{1} ; {\begingroup\renewcommand\colorMATH{\colorMATHC}\renewcommand\colorSYNTAX{\colorSYNTAXC}{{\color{\colorMATH}\ensuremath{\pS}}}\endgroup }}}}.
\begin{enumerate}[ncases]\item  {{\color{\colorMATH}\ensuremath{\Gamma  \mathrel{;} {\begingroup\renewcommand\colorMATH{\colorMATHB}\renewcommand\colorSYNTAX{\colorSYNTAXB}{{\color{\colorMATH}\ensuremath{\sS_{0}}}}\endgroup }\hspace*{0.33em}{\begingroup\renewcommand\colorMATH{\colorMATHC}\renewcommand\colorSYNTAX{\colorSYNTAXC}{{\color{\colorMATH}\ensuremath{\vdash }}}\endgroup }\hspace*{0.33em}  {\begingroup\renewcommand\colorMATH{\colorMATHC}\renewcommand\colorSYNTAX{\colorSYNTAXC}{{\color{\colorSYNTAX}\texttt{return}}}\endgroup }\hspace*{0.33em}{\begingroup\renewcommand\colorMATH{\colorMATHB}\renewcommand\colorSYNTAX{\colorSYNTAXB}{{\color{\colorMATH}\ensuremath{\se}}}\endgroup } \mathrel{:} [\varnothing /\wideparen{x}]\tau _{1} \mathrel{;} {\begingroup\renewcommand\colorMATH{\colorMATHC}\renewcommand\colorSYNTAX{\colorSYNTAXC}{{\color{\colorMATH}\ensuremath{\rceil {\begingroup\renewcommand\colorMATH{\colorMATHB}\renewcommand\colorSYNTAX{\colorSYNTAXB}{{\color{\colorMATH}\ensuremath{\sS_{1}}}}\endgroup }\lceil ^{\infty }}}}\endgroup } + {\text{FS}}^\infty (\tau _{1})}}}
	\begin{subproof} 
		By {{\color{\colorMATH}\ensuremath{{\textsc{ return}}}}} we know that
		\begingroup\color{\colorMATH}\begin{gather*} 
		  \inferrule*[lab={\textsc{ return}}
		  ]{ \Gamma  \mathrel{;} {\begingroup\renewcommand\colorMATH{\colorMATHB}\renewcommand\colorSYNTAX{\colorSYNTAXB}{{\color{\colorMATH}\ensuremath{\sS_{0}}}}\endgroup }\hspace*{0.33em}{\begingroup\renewcommand\colorMATH{\colorMATHB}\renewcommand\colorSYNTAX{\colorSYNTAXB}{{\color{\colorMATH}\ensuremath{\vdash }}}\endgroup }\hspace*{0.33em}{\begingroup\renewcommand\colorMATH{\colorMATHB}\renewcommand\colorSYNTAX{\colorSYNTAXB}{{\color{\colorMATH}\ensuremath{\se_{1}}}}\endgroup } \mathrel{:} \tau _{1} \mathrel{;} {\begingroup\renewcommand\colorMATH{\colorMATHB}\renewcommand\colorSYNTAX{\colorSYNTAXB}{{\color{\colorMATH}\ensuremath{\sS_{1}}}}\endgroup }
		  \\ \wideparen{x} = {\text{FV}}({\text{FS}}(\tau _{1}))
		    }{
		     \Gamma  \mathrel{;} {\begingroup\renewcommand\colorMATH{\colorMATHB}\renewcommand\colorSYNTAX{\colorSYNTAXB}{{\color{\colorMATH}\ensuremath{\sS_{0}}}}\endgroup }\hspace*{0.33em}{\begingroup\renewcommand\colorMATH{\colorMATHC}\renewcommand\colorSYNTAX{\colorSYNTAXC}{{\color{\colorMATH}\ensuremath{\vdash }}}\endgroup }\hspace*{0.33em}  {\begingroup\renewcommand\colorMATH{\colorMATHC}\renewcommand\colorSYNTAX{\colorSYNTAXC}{{\color{\colorSYNTAX}\texttt{return}}}\endgroup }\hspace*{0.33em}{\begingroup\renewcommand\colorMATH{\colorMATHB}\renewcommand\colorSYNTAX{\colorSYNTAXB}{{\color{\colorMATH}\ensuremath{\se_{1}}}}\endgroup } \mathrel{:} [\varnothing /\wideparen{x}]\tau _{1} \mathrel{;} {\begingroup\renewcommand\colorMATH{\colorMATHC}\renewcommand\colorSYNTAX{\colorSYNTAXC}{{\color{\colorMATH}\ensuremath{\rceil {\begingroup\renewcommand\colorMATH{\colorMATHB}\renewcommand\colorSYNTAX{\colorSYNTAXB}{{\color{\colorMATH}\ensuremath{\sS_{1}}}}\endgroup }\lceil ^{\infty }}}}\endgroup } + {\text{FS}}^\infty (\tau _{1})
		  }
		\end{gather*}\endgroup
		where {{\color{\colorMATH}\ensuremath{\tau  = [\varnothing /\wideparen{x}]\tau }}}, and {{\color{\colorMATH}\ensuremath{ {\begingroup\renewcommand\colorMATH{\colorMATHC}\renewcommand\colorSYNTAX{\colorSYNTAXC}{{\color{\colorMATH}\ensuremath{\pS}}}\endgroup } = {\begingroup\renewcommand\colorMATH{\colorMATHC}\renewcommand\colorSYNTAX{\colorSYNTAXC}{{\color{\colorMATH}\ensuremath{\rceil {\begingroup\renewcommand\colorMATH{\colorMATHB}\renewcommand\colorSYNTAX{\colorSYNTAXB}{{\color{\colorMATH}\ensuremath{\sS_{1}}}}\endgroup }\lceil ^{\infty }}}}\endgroup } + {\text{FS}}^\infty (\tau _{1})}}}.
		By induction hypotheses we know that
		{{\color{\colorMATH}\ensuremath{\gamma  \vdash  {\begingroup\renewcommand\colorMATH{\colorMATHB}\renewcommand\colorSYNTAX{\colorSYNTAXB}{{\color{\colorMATH}\ensuremath{\se_{1}}}}\endgroup } \Downarrow  {\begingroup\renewcommand\colorMATH{\colorMATHB}\renewcommand\colorSYNTAX{\colorSYNTAXB}{{\color{\colorMATH}\ensuremath{\sv_{1}}}}\endgroup }}}}, and
		{{\color{\colorMATH}\ensuremath{{\begingroup\renewcommand\colorMATH{\colorMATHB}\renewcommand\colorSYNTAX{\colorSYNTAXB}{{\color{\colorMATH}\ensuremath{\sv_{1}}}}\endgroup } \in  {\mathcal{V}}\llbracket \tau _{1}/\Gamma \rrbracket }}}.

		By inspection of the evaluation semantics {{\color{\colorMATH}\ensuremath{\gamma  \vdash  {\begingroup\renewcommand\colorMATH{\colorMATHC}\renewcommand\colorSYNTAX{\colorSYNTAXC}{{\color{\colorSYNTAX}\texttt{return}}}\endgroup }\hspace*{0.33em}{\begingroup\renewcommand\colorMATH{\colorMATHB}\renewcommand\colorSYNTAX{\colorSYNTAXB}{{\color{\colorMATH}\ensuremath{\se_{1}}}}\endgroup } \Downarrow  {\begingroup\renewcommand\colorMATH{\colorMATHB}\renewcommand\colorSYNTAX{\colorSYNTAXB}{{\color{\colorMATH}\ensuremath{\sv_{1}}}}\endgroup }}}}.
		Notice that {{\color{\colorMATH}\ensuremath{\wideparen{x} \subseteq \Gamma }}}, then
		{{\color{\colorMATH}\ensuremath{([\varnothing /\wideparen{x}]\tau _{1})/\Gamma  = \tau _{1}/\Gamma }}}, and as {{\color{\colorMATH}\ensuremath{{\begingroup\renewcommand\colorMATH{\colorMATHB}\renewcommand\colorSYNTAX{\colorSYNTAXB}{{\color{\colorMATH}\ensuremath{\sv_{1}}}}\endgroup } \in  {\mathcal{V}}\llbracket \tau _{1}/\Gamma \rrbracket }}} the result holds immediately.
	\end{subproof}
\item  {{\color{\colorMATH}\ensuremath{\Gamma  \mathrel{;} {\begingroup\renewcommand\colorMATH{\colorMATHB}\renewcommand\colorSYNTAX{\colorSYNTAXB}{{\color{\colorMATH}\ensuremath{\sS_{0}}}}\endgroup }\hspace*{0.33em}{\begingroup\renewcommand\colorMATH{\colorMATHC}\renewcommand\colorSYNTAX{\colorSYNTAXC}{{\color{\colorMATH}\ensuremath{\vdash }}}\endgroup }\hspace*{0.33em} x: \tau _{1} \leftarrow  {\begingroup\renewcommand\colorMATH{\colorMATHC}\renewcommand\colorSYNTAX{\colorSYNTAXC}{{\color{\colorMATH}\ensuremath{\pe_{1}}}}\endgroup }\mathrel{;}{\begingroup\renewcommand\colorMATH{\colorMATHC}\renewcommand\colorSYNTAX{\colorSYNTAXC}{{\color{\colorMATH}\ensuremath{\pe_{2}}}}\endgroup } \mathrel{:} [\varnothing /x]\tau _{2} \mathrel{;} {\begingroup\renewcommand\colorMATH{\colorMATHC}\renewcommand\colorSYNTAX{\colorSYNTAXC}{{\color{\colorMATH}\ensuremath{\pS_{1}}}}\endgroup } + {\begingroup\renewcommand\colorMATH{\colorMATHC}\renewcommand\colorSYNTAX{\colorSYNTAXC}{{\color{\colorMATH}\ensuremath{\pS_{2}}}}\endgroup }}}}
	\begin{subproof} 
		By {{\color{\colorMATH}\ensuremath{{\textsc{ bind}}}}} we know that
		\begingroup\color{\colorMATH}\begin{gather*} 
		  \inferrule*[lab={\textsc{ bind}}
		  ]{ \Gamma  \mathrel{;} {\begingroup\renewcommand\colorMATH{\colorMATHB}\renewcommand\colorSYNTAX{\colorSYNTAXB}{{\color{\colorMATH}\ensuremath{\sS_{0}}}}\endgroup }\hspace*{0.33em}{\begingroup\renewcommand\colorMATH{\colorMATHC}\renewcommand\colorSYNTAX{\colorSYNTAXC}{{\color{\colorMATH}\ensuremath{\vdash }}}\endgroup }\hspace*{0.33em} {\begingroup\renewcommand\colorMATH{\colorMATHC}\renewcommand\colorSYNTAX{\colorSYNTAXC}{{\color{\colorMATH}\ensuremath{\pe_{1}}}}\endgroup } \mathrel{:} \tau _{1} \mathrel{;} {\begingroup\renewcommand\colorMATH{\colorMATHC}\renewcommand\colorSYNTAX{\colorSYNTAXC}{{\color{\colorMATH}\ensuremath{\pS_{1}}}}\endgroup }
		  \\ \Gamma ,x\mathrel{:}\tau _{1} \mathrel{;} {\begingroup\renewcommand\colorMATH{\colorMATHB}\renewcommand\colorSYNTAX{\colorSYNTAXB}{{\color{\colorMATH}\ensuremath{\sS_{0}}}}\endgroup } + {\begingroup\renewcommand\colorMATH{\colorMATHB}\renewcommand\colorSYNTAX{\colorSYNTAXB}{{\color{\colorMATH}\ensuremath{0}}}\endgroup }x   \hspace*{0.33em}{\begingroup\renewcommand\colorMATH{\colorMATHC}\renewcommand\colorSYNTAX{\colorSYNTAXC}{{\color{\colorMATH}\ensuremath{\vdash }}}\endgroup }\hspace*{0.33em} {\begingroup\renewcommand\colorMATH{\colorMATHC}\renewcommand\colorSYNTAX{\colorSYNTAXC}{{\color{\colorMATH}\ensuremath{\pe_{2}}}}\endgroup } \mathrel{:} \tau _{2} \mathrel{;} {\begingroup\renewcommand\colorMATH{\colorMATHC}\renewcommand\colorSYNTAX{\colorSYNTAXC}{{\color{\colorMATH}\ensuremath{\pS_{2}}}}\endgroup }
		    }{
		     \Gamma  \mathrel{;} {\begingroup\renewcommand\colorMATH{\colorMATHB}\renewcommand\colorSYNTAX{\colorSYNTAXB}{{\color{\colorMATH}\ensuremath{\sS_{0}}}}\endgroup }\hspace*{0.33em}{\begingroup\renewcommand\colorMATH{\colorMATHC}\renewcommand\colorSYNTAX{\colorSYNTAXC}{{\color{\colorMATH}\ensuremath{\vdash }}}\endgroup }\hspace*{0.33em} x: \tau _{1} \leftarrow  {\begingroup\renewcommand\colorMATH{\colorMATHC}\renewcommand\colorSYNTAX{\colorSYNTAXC}{{\color{\colorMATH}\ensuremath{\pe_{1}}}}\endgroup }\mathrel{;}{\begingroup\renewcommand\colorMATH{\colorMATHC}\renewcommand\colorSYNTAX{\colorSYNTAXC}{{\color{\colorMATH}\ensuremath{\pe_{2}}}}\endgroup } \mathrel{:} [\varnothing /x]\tau _{2} \mathrel{;} {\begingroup\renewcommand\colorMATH{\colorMATHC}\renewcommand\colorSYNTAX{\colorSYNTAXC}{{\color{\colorMATH}\ensuremath{\pS_{1}}}}\endgroup } + [\varnothing /x]{\begingroup\renewcommand\colorMATH{\colorMATHC}\renewcommand\colorSYNTAX{\colorSYNTAXC}{{\color{\colorMATH}\ensuremath{\pS_{2}}}}\endgroup }
		  }
		\end{gather*}\endgroup
		where {{\color{\colorMATH}\ensuremath{\tau  = [\varnothing /x]\tau _{2}}}}, and {{\color{\colorMATH}\ensuremath{ {\begingroup\renewcommand\colorMATH{\colorMATHC}\renewcommand\colorSYNTAX{\colorSYNTAXC}{{\color{\colorMATH}\ensuremath{\pS}}}\endgroup } = {\begingroup\renewcommand\colorMATH{\colorMATHC}\renewcommand\colorSYNTAX{\colorSYNTAXC}{{\color{\colorMATH}\ensuremath{\pS_{1}}}}\endgroup } + [\varnothing /x]{\begingroup\renewcommand\colorMATH{\colorMATHC}\renewcommand\colorSYNTAX{\colorSYNTAXC}{{\color{\colorMATH}\ensuremath{\pS_{2}}}}\endgroup }}}}.
		By induction hypothesis on {{\color{\colorMATH}\ensuremath{\Gamma  \mathrel{;} {\begingroup\renewcommand\colorMATH{\colorMATHB}\renewcommand\colorSYNTAX{\colorSYNTAXB}{{\color{\colorMATH}\ensuremath{\sS_{0}}}}\endgroup }\hspace*{0.33em}{\begingroup\renewcommand\colorMATH{\colorMATHC}\renewcommand\colorSYNTAX{\colorSYNTAXC}{{\color{\colorMATH}\ensuremath{\vdash }}}\endgroup }\hspace*{0.33em} {\begingroup\renewcommand\colorMATH{\colorMATHC}\renewcommand\colorSYNTAX{\colorSYNTAXC}{{\color{\colorMATH}\ensuremath{\pe_{1}}}}\endgroup } \mathrel{:} \tau _{1} \mathrel{;} {\begingroup\renewcommand\colorMATH{\colorMATHC}\renewcommand\colorSYNTAX{\colorSYNTAXC}{{\color{\colorMATH}\ensuremath{\pS_{1}}}}\endgroup }}}}, we know that
		{{\color{\colorMATH}\ensuremath{\gamma  \vdash  {\begingroup\renewcommand\colorMATH{\colorMATHC}\renewcommand\colorSYNTAX{\colorSYNTAXC}{{\color{\colorMATH}\ensuremath{\pe_{1}}}}\endgroup } \in  {\mathcal{E}}\llbracket \tau _{1}/\Gamma \rrbracket }}}, therefore 
		for all {{\color{\colorMATH}\ensuremath{{\begingroup\renewcommand\colorMATH{\colorMATHB}\renewcommand\colorSYNTAX{\colorSYNTAXB}{{\color{\colorMATH}\ensuremath{\sv_{1}}}}\endgroup }}}} such that {{\color{\colorMATH}\ensuremath{\gamma  \vdash  {\begingroup\renewcommand\colorMATH{\colorMATHC}\renewcommand\colorSYNTAX{\colorSYNTAXC}{{\color{\colorMATH}\ensuremath{\pe_{1}}}}\endgroup } \Downarrow  {\begingroup\renewcommand\colorMATH{\colorMATHB}\renewcommand\colorSYNTAX{\colorSYNTAXB}{{\color{\colorMATH}\ensuremath{\sv_{1}}}}\endgroup }}}} it follows that {{\color{\colorMATH}\ensuremath{{\begingroup\renewcommand\colorMATH{\colorMATHB}\renewcommand\colorSYNTAX{\colorSYNTAXB}{{\color{\colorMATH}\ensuremath{\sv_{1}}}}\endgroup } \in  {\mathcal{V}}\llbracket \tau _{1}/\Gamma \rrbracket }}}.
		Also, by induction hypothesis on {{\color{\colorMATH}\ensuremath{\Gamma ,x\mathrel{:}\tau _{1} \mathrel{;} {\begingroup\renewcommand\colorMATH{\colorMATHB}\renewcommand\colorSYNTAX{\colorSYNTAXB}{{\color{\colorMATH}\ensuremath{\sS_{0}}}}\endgroup } + {\begingroup\renewcommand\colorMATH{\colorMATHB}\renewcommand\colorSYNTAX{\colorSYNTAXB}{{\color{\colorMATH}\ensuremath{0}}}\endgroup }x   \hspace*{0.33em}{\begingroup\renewcommand\colorMATH{\colorMATHC}\renewcommand\colorSYNTAX{\colorSYNTAXC}{{\color{\colorMATH}\ensuremath{\vdash }}}\endgroup }\hspace*{0.33em} {\begingroup\renewcommand\colorMATH{\colorMATHC}\renewcommand\colorSYNTAX{\colorSYNTAXC}{{\color{\colorMATH}\ensuremath{\pe_{2}}}}\endgroup } \mathrel{:} \tau _{2} \mathrel{;} {\begingroup\renewcommand\colorMATH{\colorMATHC}\renewcommand\colorSYNTAX{\colorSYNTAXC}{{\color{\colorMATH}\ensuremath{\pS_{2}}}}\endgroup }}}}, we know that
		{{\color{\colorMATH}\ensuremath{\gamma [x \mapsto  {\begingroup\renewcommand\colorMATH{\colorMATHB}\renewcommand\colorSYNTAX{\colorSYNTAXB}{{\color{\colorMATH}\ensuremath{\sv_{1}}}}\endgroup }] \vdash  {\begingroup\renewcommand\colorMATH{\colorMATHC}\renewcommand\colorSYNTAX{\colorSYNTAXC}{{\color{\colorMATH}\ensuremath{\pe_{2}}}}\endgroup } \in  {\mathcal{E}}\llbracket \tau _{2}/(\Gamma ,x\mathrel{:}\tau _{1})\rrbracket }}}, therefore 
		for all {{\color{\colorMATH}\ensuremath{{\begingroup\renewcommand\colorMATH{\colorMATHB}\renewcommand\colorSYNTAX{\colorSYNTAXB}{{\color{\colorMATH}\ensuremath{\sv_{2}}}}\endgroup }}}} such that {{\color{\colorMATH}\ensuremath{\gamma [x \mapsto  {\begingroup\renewcommand\colorMATH{\colorMATHB}\renewcommand\colorSYNTAX{\colorSYNTAXB}{{\color{\colorMATH}\ensuremath{\sv_{1}}}}\endgroup }] \vdash  {\begingroup\renewcommand\colorMATH{\colorMATHC}\renewcommand\colorSYNTAX{\colorSYNTAXC}{{\color{\colorMATH}\ensuremath{\pe_{2}}}}\endgroup } \Downarrow  {\begingroup\renewcommand\colorMATH{\colorMATHB}\renewcommand\colorSYNTAX{\colorSYNTAXB}{{\color{\colorMATH}\ensuremath{\sv_{2}}}}\endgroup }}}} it follows that {{\color{\colorMATH}\ensuremath{{\begingroup\renewcommand\colorMATH{\colorMATHB}\renewcommand\colorSYNTAX{\colorSYNTAXB}{{\color{\colorMATH}\ensuremath{\sv_{2}}}}\endgroup } \in  {\mathcal{V}}\llbracket \tau _{2}/(\Gamma ,x\mathrel{:}\tau _{1})\rrbracket }}}.

		Then, we have to prove that for all {{\color{\colorMATH}\ensuremath{{\begingroup\renewcommand\colorMATH{\colorMATHB}\renewcommand\colorSYNTAX{\colorSYNTAXB}{{\color{\colorMATH}\ensuremath{\sv_{2}}}}\endgroup }}}} such that {{\color{\colorMATH}\ensuremath{\gamma  \vdash  x: \tau _{1} \leftarrow  {\begingroup\renewcommand\colorMATH{\colorMATHC}\renewcommand\colorSYNTAX{\colorSYNTAXC}{{\color{\colorMATH}\ensuremath{\pe_{1}}}}\endgroup }\mathrel{;}{\begingroup\renewcommand\colorMATH{\colorMATHC}\renewcommand\colorSYNTAX{\colorSYNTAXC}{{\color{\colorMATH}\ensuremath{\pe_{2}}}}\endgroup } \Downarrow  {\begingroup\renewcommand\colorMATH{\colorMATHB}\renewcommand\colorSYNTAX{\colorSYNTAXB}{{\color{\colorMATH}\ensuremath{\sv_{2}}}}\endgroup }}}} it follows that {{\color{\colorMATH}\ensuremath{{\begingroup\renewcommand\colorMATH{\colorMATHB}\renewcommand\colorSYNTAX{\colorSYNTAXB}{{\color{\colorMATH}\ensuremath{\sv_{2}}}}\endgroup } \in  {\mathcal{V}}\llbracket ([\varnothing /x]\tau _{2})/\Gamma \rrbracket }}}.
		Let us fix {{\color{\colorMATH}\ensuremath{{\begingroup\renewcommand\colorMATH{\colorMATHB}\renewcommand\colorSYNTAX{\colorSYNTAXB}{{\color{\colorMATH}\ensuremath{\sv_{2}}}}\endgroup }}}}. 
		By inspection of the evaluation semantics for {{\color{\colorMATH}\ensuremath{{\textsc{ bind}}}}}, we know that there exist {{\color{\colorMATH}\ensuremath{{\begingroup\renewcommand\colorMATH{\colorMATHB}\renewcommand\colorSYNTAX{\colorSYNTAXB}{{\color{\colorMATH}\ensuremath{\sv_{1}}}}\endgroup }}}} such that {{\color{\colorMATH}\ensuremath{\gamma  \vdash  {\begingroup\renewcommand\colorMATH{\colorMATHC}\renewcommand\colorSYNTAX{\colorSYNTAXC}{{\color{\colorMATH}\ensuremath{\pe_{1}}}}\endgroup } \Downarrow  {\begingroup\renewcommand\colorMATH{\colorMATHB}\renewcommand\colorSYNTAX{\colorSYNTAXB}{{\color{\colorMATH}\ensuremath{\sv_{1}}}}\endgroup }}}} and {{\color{\colorMATH}\ensuremath{\gamma [x \mapsto  {\begingroup\renewcommand\colorMATH{\colorMATHB}\renewcommand\colorSYNTAX{\colorSYNTAXB}{{\color{\colorMATH}\ensuremath{\sv_{1}}}}\endgroup }] \vdash  {\begingroup\renewcommand\colorMATH{\colorMATHC}\renewcommand\colorSYNTAX{\colorSYNTAXC}{{\color{\colorMATH}\ensuremath{\pe_{2}}}}\endgroup } \Downarrow  {\begingroup\renewcommand\colorMATH{\colorMATHB}\renewcommand\colorSYNTAX{\colorSYNTAXB}{{\color{\colorMATH}\ensuremath{\sv_{2}}}}\endgroup }}}}.
		But we know that for all {{\color{\colorMATH}\ensuremath{{\begingroup\renewcommand\colorMATH{\colorMATHB}\renewcommand\colorSYNTAX{\colorSYNTAXB}{{\color{\colorMATH}\ensuremath{\sv_{1}}}}\endgroup }}}} and {{\color{\colorMATH}\ensuremath{{\begingroup\renewcommand\colorMATH{\colorMATHB}\renewcommand\colorSYNTAX{\colorSYNTAXB}{{\color{\colorMATH}\ensuremath{\sv_{2}}}}\endgroup }}}} such that {{\color{\colorMATH}\ensuremath{\gamma  \vdash  {\begingroup\renewcommand\colorMATH{\colorMATHC}\renewcommand\colorSYNTAX{\colorSYNTAXC}{{\color{\colorMATH}\ensuremath{\pe_{1}}}}\endgroup } \Downarrow  {\begingroup\renewcommand\colorMATH{\colorMATHB}\renewcommand\colorSYNTAX{\colorSYNTAXB}{{\color{\colorMATH}\ensuremath{\sv_{1}}}}\endgroup }}}} and {{\color{\colorMATH}\ensuremath{\gamma [x \mapsto  {\begingroup\renewcommand\colorMATH{\colorMATHB}\renewcommand\colorSYNTAX{\colorSYNTAXB}{{\color{\colorMATH}\ensuremath{\sv_{1}}}}\endgroup }] \vdash  {\begingroup\renewcommand\colorMATH{\colorMATHC}\renewcommand\colorSYNTAX{\colorSYNTAXC}{{\color{\colorMATH}\ensuremath{\pe_{2}}}}\endgroup } \Downarrow  {\begingroup\renewcommand\colorMATH{\colorMATHB}\renewcommand\colorSYNTAX{\colorSYNTAXB}{{\color{\colorMATH}\ensuremath{\sv_{2}}}}\endgroup }}}}, it follows that {{\color{\colorMATH}\ensuremath{{\begingroup\renewcommand\colorMATH{\colorMATHB}\renewcommand\colorSYNTAX{\colorSYNTAXB}{{\color{\colorMATH}\ensuremath{\sv_{1}}}}\endgroup } \in  {\mathcal{V}}\llbracket \tau _{1}/\Gamma \rrbracket }}} and {{\color{\colorMATH}\ensuremath{{\begingroup\renewcommand\colorMATH{\colorMATHB}\renewcommand\colorSYNTAX{\colorSYNTAXB}{{\color{\colorMATH}\ensuremath{\sv_{2}}}}\endgroup } \in  {\mathcal{V}}\llbracket \tau _{2}/(\Gamma ,x\mathrel{:}\tau _{1})\rrbracket }}}.
		Finally, notice that {{\color{\colorMATH}\ensuremath{([\varnothing /x]\tau _{2})/\Gamma  = \tau _{2}/(\Gamma ,x\mathrel{:}\tau _{1})}}} and the result holds.
		
	\end{subproof}
\item  {{\color{\colorMATH}\ensuremath{\Gamma  \mathrel{;} {\begingroup\renewcommand\colorMATH{\colorMATHB}\renewcommand\colorSYNTAX{\colorSYNTAXB}{{\color{\colorMATH}\ensuremath{\sS_{0}}}}\endgroup }\hspace*{0.33em}{\begingroup\renewcommand\colorMATH{\colorMATHC}\renewcommand\colorSYNTAX{\colorSYNTAXC}{{\color{\colorMATH}\ensuremath{\vdash }}}\endgroup }\hspace*{0.33em} {{\color{\colorSYNTAX}\texttt{gauss}}}\hspace*{0.33em}\mu \hspace*{0.33em}\sigma ^{2} \mathrel{:} {\begingroup\renewcommand\colorMATH{\colorMATHA}\renewcommand\colorSYNTAX{\colorSYNTAXA}{{\color{\colorSYNTAX}\texttt{{\ensuremath{{\mathbb{R}}}}}}}\endgroup } \mathrel{;} \varnothing }}}
	\begin{subproof} 
		By {{\color{\colorMATH}\ensuremath{{\textsc{ gauss}}}}} we know that
		\begingroup\color{\colorMATH}\begin{gather*} 
		  \inferrule*[lab={\textsc{ gauss}}
		  ]{ 
		    }{
		     \Gamma  \mathrel{;} {\begingroup\renewcommand\colorMATH{\colorMATHB}\renewcommand\colorSYNTAX{\colorSYNTAXB}{{\color{\colorMATH}\ensuremath{\sS_{0}}}}\endgroup }\hspace*{0.33em}{\begingroup\renewcommand\colorMATH{\colorMATHC}\renewcommand\colorSYNTAX{\colorSYNTAXC}{{\color{\colorMATH}\ensuremath{\vdash }}}\endgroup }\hspace*{0.33em} \gamma  \vdash  {{\color{\colorSYNTAX}\texttt{gauss}}}\hspace*{0.33em}\mu \hspace*{0.33em}\sigma ^{2} \mathrel{:} {\begingroup\renewcommand\colorMATH{\colorMATHA}\renewcommand\colorSYNTAX{\colorSYNTAXA}{{\color{\colorSYNTAX}\texttt{{\ensuremath{{\mathbb{R}}}}}}}\endgroup } \mathrel{;} \varnothing 
		  }
		\end{gather*}\endgroup
		we know that {{\color{\colorMATH}\ensuremath{\gamma  \vdash  {{\color{\colorSYNTAX}\texttt{gauss}}}\hspace*{0.33em}\mu \hspace*{0.33em}\sigma ^{2} \Downarrow  {\begingroup\renewcommand\colorMATH{\colorMATHB}\renewcommand\colorSYNTAX{\colorSYNTAXB}{{\color{\colorMATH}\ensuremath{r}}}\endgroup }}}}, for some {{\color{\colorMATH}\ensuremath{{\begingroup\renewcommand\colorMATH{\colorMATHB}\renewcommand\colorSYNTAX{\colorSYNTAXB}{{\color{\colorMATH}\ensuremath{r}}}\endgroup }}}}.
		But {{\color{\colorMATH}\ensuremath{{\begingroup\renewcommand\colorMATH{\colorMATHB}\renewcommand\colorSYNTAX{\colorSYNTAXB}{{\color{\colorMATH}\ensuremath{r}}}\endgroup } \in  {\mathcal{V}}\llbracket {\begingroup\renewcommand\colorMATH{\colorMATHA}\renewcommand\colorSYNTAX{\colorSYNTAXA}{{\color{\colorSYNTAX}\texttt{{\ensuremath{{\mathbb{R}}}}}}}\endgroup }\rrbracket }}} and the result holds.
	\end{subproof}
\item  {{\color{\colorMATH}\ensuremath{\Gamma  \mathrel{;} {\begingroup\renewcommand\colorMATH{\colorMATHB}\renewcommand\colorSYNTAX{\colorSYNTAXB}{{\color{\colorMATH}\ensuremath{\sS_{0}}}}\endgroup }\hspace*{0.33em}{\begingroup\renewcommand\colorMATH{\colorMATHC}\renewcommand\colorSYNTAX{\colorSYNTAXC}{{\color{\colorMATH}\ensuremath{\vdash }}}\endgroup }\hspace*{0.33em} if\hspace*{0.33em}{\begingroup\renewcommand\colorMATH{\colorMATHB}\renewcommand\colorSYNTAX{\colorSYNTAXB}{{\color{\colorMATH}\ensuremath{\se_{1}}}}\endgroup }\hspace*{0.33em}\{ {\begingroup\renewcommand\colorMATH{\colorMATHC}\renewcommand\colorSYNTAX{\colorSYNTAXC}{{\color{\colorMATH}\ensuremath{\pe_{2}}}}\endgroup }\} \hspace*{0.33em}\{ {\begingroup\renewcommand\colorMATH{\colorMATHC}\renewcommand\colorSYNTAX{\colorSYNTAXC}{{\color{\colorMATH}\ensuremath{\pe_{3}}}}\endgroup }\}  \mathrel{:} \tau  \mathrel{;}  {\begingroup\renewcommand\colorMATH{\colorMATHC}\renewcommand\colorSYNTAX{\colorSYNTAXC}{{\color{\colorMATH}\ensuremath{\rceil {\begingroup\renewcommand\colorMATH{\colorMATHA}\renewcommand\colorSYNTAX{\colorSYNTAXA}{{\color{\colorMATH}\ensuremath{{\begingroup\renewcommand\colorMATH{\colorMATHB}\renewcommand\colorSYNTAX{\colorSYNTAXB}{{\color{\colorMATH}\ensuremath{\sS_{1}}}}\endgroup }}}}\endgroup }\lceil ^{\infty }}}}\endgroup } \sqcup  ({\begingroup\renewcommand\colorMATH{\colorMATHC}\renewcommand\colorSYNTAX{\colorSYNTAXC}{{\color{\colorMATH}\ensuremath{\pS_{2}}}}\endgroup } \sqcup  {\begingroup\renewcommand\colorMATH{\colorMATHC}\renewcommand\colorSYNTAX{\colorSYNTAXC}{{\color{\colorMATH}\ensuremath{\pS_{3}}}}\endgroup })}}}
	\begin{subproof} 
		By {{\color{\colorMATH}\ensuremath{{\textsc{ if}}}}} we know that
		\begingroup\color{\colorMATH}\begin{gather*} 
		  \inferrule*[lab={\textsc{ if}}
		  ]{ \Gamma  \mathrel{;} {\begingroup\renewcommand\colorMATH{\colorMATHB}\renewcommand\colorSYNTAX{\colorSYNTAXB}{{\color{\colorMATH}\ensuremath{\sS_{0}}}}\endgroup } \hspace*{0.33em}{\begingroup\renewcommand\colorMATH{\colorMATHB}\renewcommand\colorSYNTAX{\colorSYNTAXB}{{\color{\colorMATH}\ensuremath{\vdash }}}\endgroup }\hspace*{0.33em}{\begingroup\renewcommand\colorMATH{\colorMATHB}\renewcommand\colorSYNTAX{\colorSYNTAXB}{{\color{\colorMATH}\ensuremath{\se_{1}}}}\endgroup } \mathrel{:} {\mathbb{B}} \mathrel{;} {\begingroup\renewcommand\colorMATH{\colorMATHB}\renewcommand\colorSYNTAX{\colorSYNTAXB}{{\color{\colorMATH}\ensuremath{\sS_{1}}}}\endgroup }
		  \\ \Gamma  \mathrel{;} {\begingroup\renewcommand\colorMATH{\colorMATHB}\renewcommand\colorSYNTAX{\colorSYNTAXB}{{\color{\colorMATH}\ensuremath{\sS_{0}}}}\endgroup } \hspace*{0.33em}{\begingroup\renewcommand\colorMATH{\colorMATHC}\renewcommand\colorSYNTAX{\colorSYNTAXC}{{\color{\colorMATH}\ensuremath{\vdash }}}\endgroup }\hspace*{0.33em} {\begingroup\renewcommand\colorMATH{\colorMATHC}\renewcommand\colorSYNTAX{\colorSYNTAXC}{{\color{\colorMATH}\ensuremath{\pe_{2}}}}\endgroup } \mathrel{:} \tau  \mathrel{;} {\begingroup\renewcommand\colorMATH{\colorMATHC}\renewcommand\colorSYNTAX{\colorSYNTAXC}{{\color{\colorMATH}\ensuremath{\pS_{2}}}}\endgroup }
		  \\ \Gamma  \mathrel{;} {\begingroup\renewcommand\colorMATH{\colorMATHB}\renewcommand\colorSYNTAX{\colorSYNTAXB}{{\color{\colorMATH}\ensuremath{\sS_{0}}}}\endgroup } \hspace*{0.33em}{\begingroup\renewcommand\colorMATH{\colorMATHC}\renewcommand\colorSYNTAX{\colorSYNTAXC}{{\color{\colorMATH}\ensuremath{\vdash }}}\endgroup }\hspace*{0.33em} {\begingroup\renewcommand\colorMATH{\colorMATHC}\renewcommand\colorSYNTAX{\colorSYNTAXC}{{\color{\colorMATH}\ensuremath{\pe_{3}}}}\endgroup } \mathrel{:} \tau  \mathrel{;} {\begingroup\renewcommand\colorMATH{\colorMATHC}\renewcommand\colorSYNTAX{\colorSYNTAXC}{{\color{\colorMATH}\ensuremath{\pS_{3}}}}\endgroup }
		     }{
		     \Gamma  \mathrel{;} {\begingroup\renewcommand\colorMATH{\colorMATHB}\renewcommand\colorSYNTAX{\colorSYNTAXB}{{\color{\colorMATH}\ensuremath{\sS_{0}}}}\endgroup }\hspace*{0.33em}{\begingroup\renewcommand\colorMATH{\colorMATHC}\renewcommand\colorSYNTAX{\colorSYNTAXC}{{\color{\colorMATH}\ensuremath{\vdash }}}\endgroup }\hspace*{0.33em} if\hspace*{0.33em}{\begingroup\renewcommand\colorMATH{\colorMATHB}\renewcommand\colorSYNTAX{\colorSYNTAXB}{{\color{\colorMATH}\ensuremath{\se_{1}}}}\endgroup }\hspace*{0.33em}\{ {\begingroup\renewcommand\colorMATH{\colorMATHC}\renewcommand\colorSYNTAX{\colorSYNTAXC}{{\color{\colorMATH}\ensuremath{\pe_{2}}}}\endgroup }\} \hspace*{0.33em}\{ {\begingroup\renewcommand\colorMATH{\colorMATHC}\renewcommand\colorSYNTAX{\colorSYNTAXC}{{\color{\colorMATH}\ensuremath{\pe_{3}}}}\endgroup }\}  \mathrel{:} \tau  \mathrel{;}  {\begingroup\renewcommand\colorMATH{\colorMATHC}\renewcommand\colorSYNTAX{\colorSYNTAXC}{{\color{\colorMATH}\ensuremath{\rceil {\begingroup\renewcommand\colorMATH{\colorMATHA}\renewcommand\colorSYNTAX{\colorSYNTAXA}{{\color{\colorMATH}\ensuremath{{\begingroup\renewcommand\colorMATH{\colorMATHB}\renewcommand\colorSYNTAX{\colorSYNTAXB}{{\color{\colorMATH}\ensuremath{\sS_{1}}}}\endgroup }}}}\endgroup }\lceil ^{\infty }}}}\endgroup } \sqcup  ({\begingroup\renewcommand\colorMATH{\colorMATHC}\renewcommand\colorSYNTAX{\colorSYNTAXC}{{\color{\colorMATH}\ensuremath{\pS_{2}}}}\endgroup } \sqcup  {\begingroup\renewcommand\colorMATH{\colorMATHC}\renewcommand\colorSYNTAX{\colorSYNTAXC}{{\color{\colorMATH}\ensuremath{\pS_{3}}}}\endgroup })
		  }
		\end{gather*}\endgroup
		where {{\color{\colorMATH}\ensuremath{ {\begingroup\renewcommand\colorMATH{\colorMATHC}\renewcommand\colorSYNTAX{\colorSYNTAXC}{{\color{\colorMATH}\ensuremath{\pS}}}\endgroup } = {\begingroup\renewcommand\colorMATH{\colorMATHC}\renewcommand\colorSYNTAX{\colorSYNTAXC}{{\color{\colorMATH}\ensuremath{\rceil {\begingroup\renewcommand\colorMATH{\colorMATHA}\renewcommand\colorSYNTAX{\colorSYNTAXA}{{\color{\colorMATH}\ensuremath{{\begingroup\renewcommand\colorMATH{\colorMATHB}\renewcommand\colorSYNTAX{\colorSYNTAXB}{{\color{\colorMATH}\ensuremath{\sS_{1}}}}\endgroup }}}}\endgroup }\lceil ^{\infty }}}}\endgroup } \sqcup  ({\begingroup\renewcommand\colorMATH{\colorMATHC}\renewcommand\colorSYNTAX{\colorSYNTAXC}{{\color{\colorMATH}\ensuremath{\pS_{2}}}}\endgroup } \sqcup  {\begingroup\renewcommand\colorMATH{\colorMATHC}\renewcommand\colorSYNTAX{\colorSYNTAXC}{{\color{\colorMATH}\ensuremath{\pS_{3}}}}\endgroup })}}}.
		By induction hypothesis on {{\color{\colorMATH}\ensuremath{\Gamma  \mathrel{;} {\begingroup\renewcommand\colorMATH{\colorMATHB}\renewcommand\colorSYNTAX{\colorSYNTAXB}{{\color{\colorMATH}\ensuremath{\sS_{0}}}}\endgroup }\hspace*{0.33em}{\begingroup\renewcommand\colorMATH{\colorMATHB}\renewcommand\colorSYNTAX{\colorSYNTAXB}{{\color{\colorMATH}\ensuremath{\vdash }}}\endgroup }\hspace*{0.33em} {\begingroup\renewcommand\colorMATH{\colorMATHB}\renewcommand\colorSYNTAX{\colorSYNTAXB}{{\color{\colorMATH}\ensuremath{\se_{1}}}}\endgroup } \mathrel{:} {\mathbb{B}} \mathrel{;} {\begingroup\renewcommand\colorMATH{\colorMATHB}\renewcommand\colorSYNTAX{\colorSYNTAXB}{{\color{\colorMATH}\ensuremath{\sS_{1}}}}\endgroup }}}},
		we know that {{\color{\colorMATH}\ensuremath{\gamma  \vdash  {\begingroup\renewcommand\colorMATH{\colorMATHB}\renewcommand\colorSYNTAX{\colorSYNTAXB}{{\color{\colorMATH}\ensuremath{\se_{1}}}}\endgroup } \Downarrow  {\begingroup\renewcommand\colorMATH{\colorMATHB}\renewcommand\colorSYNTAX{\colorSYNTAXB}{{\color{\colorMATH}\ensuremath{\sv_{1}}}}\endgroup }}}} and {{\color{\colorMATH}\ensuremath{{\begingroup\renewcommand\colorMATH{\colorMATHB}\renewcommand\colorSYNTAX{\colorSYNTAXB}{{\color{\colorMATH}\ensuremath{\sv_{1}}}}\endgroup } \in  {\mathcal{V}}\llbracket {\mathbb{B}}\rrbracket }}}.
		Unfolding booleans as sums we know that
		{{\color{\colorMATH}\ensuremath{{\begingroup\renewcommand\colorMATH{\colorMATHB}\renewcommand\colorSYNTAX{\colorSYNTAXB}{{\color{\colorMATH}\ensuremath{\sv_{1}}}}\endgroup } \in  {\mathcal{V}}\llbracket {{\color{\colorSYNTAX}\texttt{unit}}} \mathrel{^{{\begingroup\renewcommand\colorMATH{\colorMATHB}\renewcommand\colorSYNTAX{\colorSYNTAXB}{{\color{\colorMATH}\ensuremath{\varnothing }}}\endgroup }}\oplus ^{{\begingroup\renewcommand\colorMATH{\colorMATHB}\renewcommand\colorSYNTAX{\colorSYNTAXB}{{\color{\colorMATH}\ensuremath{\varnothing }}}\endgroup }}} {{\color{\colorSYNTAX}\texttt{unit}}}\rrbracket }}}.

		Let us assume that {{\color{\colorMATH}\ensuremath{{\begingroup\renewcommand\colorMATH{\colorMATHB}\renewcommand\colorSYNTAX{\colorSYNTAXB}{{\color{\colorMATH}\ensuremath{\sv_{1}}}}\endgroup } = \inl^{{{\color{\colorSYNTAX}\texttt{unit}}}}\hspace*{0.33em}\ttt}}} (the other case is analogous).
		Then by induction hypothesis on {{\color{\colorMATH}\ensuremath{\Gamma  \mathrel{;} {\begingroup\renewcommand\colorMATH{\colorMATHB}\renewcommand\colorSYNTAX{\colorSYNTAXB}{{\color{\colorMATH}\ensuremath{\sS_{0}}}}\endgroup }\hspace*{0.33em}{\begingroup\renewcommand\colorMATH{\colorMATHC}\renewcommand\colorSYNTAX{\colorSYNTAXC}{{\color{\colorMATH}\ensuremath{\vdash }}}\endgroup }\hspace*{0.33em} {\begingroup\renewcommand\colorMATH{\colorMATHC}\renewcommand\colorSYNTAX{\colorSYNTAXC}{{\color{\colorMATH}\ensuremath{\pe_{2}}}}\endgroup } \mathrel{:} \tau  \mathrel{;} {\begingroup\renewcommand\colorMATH{\colorMATHC}\renewcommand\colorSYNTAX{\colorSYNTAXC}{{\color{\colorMATH}\ensuremath{\pS_{2}}}}\endgroup }}}}, we know that
		{{\color{\colorMATH}\ensuremath{\gamma  \vdash  {\begingroup\renewcommand\colorMATH{\colorMATHC}\renewcommand\colorSYNTAX{\colorSYNTAXC}{{\color{\colorMATH}\ensuremath{\pe_{2}}}}\endgroup } \in  {\mathcal{E}}\llbracket \tau /\Gamma \rrbracket }}}, therefore 
		for all {{\color{\colorMATH}\ensuremath{{\begingroup\renewcommand\colorMATH{\colorMATHB}\renewcommand\colorSYNTAX{\colorSYNTAXB}{{\color{\colorMATH}\ensuremath{\sv_{2}}}}\endgroup }}}} such that {{\color{\colorMATH}\ensuremath{\gamma  \vdash  {\begingroup\renewcommand\colorMATH{\colorMATHC}\renewcommand\colorSYNTAX{\colorSYNTAXC}{{\color{\colorMATH}\ensuremath{\pe_{2}}}}\endgroup } \Downarrow  {\begingroup\renewcommand\colorMATH{\colorMATHB}\renewcommand\colorSYNTAX{\colorSYNTAXB}{{\color{\colorMATH}\ensuremath{\sv_{2}}}}\endgroup }}}} it follows that {{\color{\colorMATH}\ensuremath{{\begingroup\renewcommand\colorMATH{\colorMATHB}\renewcommand\colorSYNTAX{\colorSYNTAXB}{{\color{\colorMATH}\ensuremath{\sv_{2}}}}\endgroup } \in  {\mathcal{V}}\llbracket \tau /\Gamma \rrbracket }}}.

		Then we have to prove that for all {{\color{\colorMATH}\ensuremath{{\begingroup\renewcommand\colorMATH{\colorMATHB}\renewcommand\colorSYNTAX{\colorSYNTAXB}{{\color{\colorMATH}\ensuremath{\sv_{2}}}}\endgroup }}}} such that {{\color{\colorMATH}\ensuremath{\gamma  \vdash  if\hspace*{0.33em}{\begingroup\renewcommand\colorMATH{\colorMATHB}\renewcommand\colorSYNTAX{\colorSYNTAXB}{{\color{\colorMATH}\ensuremath{\se_{1}}}}\endgroup }\hspace*{0.33em}\{ {\begingroup\renewcommand\colorMATH{\colorMATHC}\renewcommand\colorSYNTAX{\colorSYNTAXC}{{\color{\colorMATH}\ensuremath{\pe_{2}}}}\endgroup }\} \hspace*{0.33em}\{ {\begingroup\renewcommand\colorMATH{\colorMATHC}\renewcommand\colorSYNTAX{\colorSYNTAXC}{{\color{\colorMATH}\ensuremath{\pe_{3}}}}\endgroup }\}  \Downarrow  {\begingroup\renewcommand\colorMATH{\colorMATHB}\renewcommand\colorSYNTAX{\colorSYNTAXB}{{\color{\colorMATH}\ensuremath{\sv_{2}}}}\endgroup }}}} it follows that {{\color{\colorMATH}\ensuremath{{\begingroup\renewcommand\colorMATH{\colorMATHB}\renewcommand\colorSYNTAX{\colorSYNTAXB}{{\color{\colorMATH}\ensuremath{\sv_{2}}}}\endgroup } \in  {\mathcal{V}}\llbracket \tau /\Gamma \rrbracket }}}.
		Then by inspection of the evaluation semantics for {{\color{\colorMATH}\ensuremath{{\textsc{ if}}}}},
		we know that if {{\color{\colorMATH}\ensuremath{\gamma  \vdash  if\hspace*{0.33em}{\begingroup\renewcommand\colorMATH{\colorMATHB}\renewcommand\colorSYNTAX{\colorSYNTAXB}{{\color{\colorMATH}\ensuremath{\se_{1}}}}\endgroup }\hspace*{0.33em}\{ {\begingroup\renewcommand\colorMATH{\colorMATHC}\renewcommand\colorSYNTAX{\colorSYNTAXC}{{\color{\colorMATH}\ensuremath{\pe_{2}}}}\endgroup }\} \hspace*{0.33em}\{ {\begingroup\renewcommand\colorMATH{\colorMATHC}\renewcommand\colorSYNTAX{\colorSYNTAXC}{{\color{\colorMATH}\ensuremath{\pe_{3}}}}\endgroup }\}  \Downarrow  {\begingroup\renewcommand\colorMATH{\colorMATHB}\renewcommand\colorSYNTAX{\colorSYNTAXB}{{\color{\colorMATH}\ensuremath{\sv_{2}}}}\endgroup }}}} then {{\color{\colorMATH}\ensuremath{\gamma  \vdash  {\begingroup\renewcommand\colorMATH{\colorMATHC}\renewcommand\colorSYNTAX{\colorSYNTAXC}{{\color{\colorMATH}\ensuremath{\pe_{2}}}}\endgroup } \Downarrow  {\begingroup\renewcommand\colorMATH{\colorMATHB}\renewcommand\colorSYNTAX{\colorSYNTAXB}{{\color{\colorMATH}\ensuremath{\sv_{2}}}}\endgroup }}}}.
		Then, the result holds immediately.
	\end{subproof}
\item  {{\color{\colorMATH}\ensuremath{\Gamma  \mathrel{;} {\begingroup\renewcommand\colorMATH{\colorMATHB}\renewcommand\colorSYNTAX{\colorSYNTAXB}{{\color{\colorMATH}\ensuremath{\sS_{0}}}}\endgroup }\hspace*{0.33em}{\begingroup\renewcommand\colorMATH{\colorMATHC}\renewcommand\colorSYNTAX{\colorSYNTAXC}{{\color{\colorMATH}\ensuremath{\vdash }}}\endgroup }\hspace*{0.33em} {{\color{\colorSYNTAX}\texttt{case}}}\hspace*{0.33em}{\begingroup\renewcommand\colorMATH{\colorMATHB}\renewcommand\colorSYNTAX{\colorSYNTAXB}{{\color{\colorMATH}\ensuremath{\se_{1}}}}\endgroup }\hspace*{0.33em}{{\color{\colorSYNTAX}\texttt{of}}}\hspace*{0.33em}\{ x\Rightarrow {\begingroup\renewcommand\colorMATH{\colorMATHC}\renewcommand\colorSYNTAX{\colorSYNTAXC}{{\color{\colorMATH}\ensuremath{\pe_{2}}}}\endgroup }\} \hspace*{0.33em}\{ y\Rightarrow {\begingroup\renewcommand\colorMATH{\colorMATHC}\renewcommand\colorSYNTAX{\colorSYNTAXC}{{\color{\colorMATH}\ensuremath{\pe_{3}}}}\endgroup }\}  \mathrel{:}[{\begingroup\renewcommand\colorMATH{\colorMATHB}\renewcommand\colorSYNTAX{\colorSYNTAXB}{{\color{\colorMATH}\ensuremath{\sS_{1 1}}}}\endgroup }/x]\tau _{2} \sqcup  [{\begingroup\renewcommand\colorMATH{\colorMATHB}\renewcommand\colorSYNTAX{\colorSYNTAXB}{{\color{\colorMATH}\ensuremath{\sS_{1 2}}}}\endgroup }/y]\tau _{3} \mathrel{;}  {\begingroup\renewcommand\colorMATH{\colorMATHC}\renewcommand\colorSYNTAX{\colorSYNTAXC}{{\color{\colorMATH}\ensuremath{\rceil {\begingroup\renewcommand\colorMATH{\colorMATHA}\renewcommand\colorSYNTAX{\colorSYNTAXA}{{\color{\colorMATH}\ensuremath{{\begingroup\renewcommand\colorMATH{\colorMATHB}\renewcommand\colorSYNTAX{\colorSYNTAXB}{{\color{\colorMATH}\ensuremath{\sS_{1}}}}\endgroup }}}}\endgroup }\lceil ^{\infty }}}}\endgroup } \sqcup  [{\begingroup\renewcommand\colorMATH{\colorMATHB}\renewcommand\colorSYNTAX{\colorSYNTAXB}{{\color{\colorMATH}\ensuremath{\sS_{1 1}}}}\endgroup }/x]{\begingroup\renewcommand\colorMATH{\colorMATHC}\renewcommand\colorSYNTAX{\colorSYNTAXC}{{\color{\colorMATH}\ensuremath{\pS_{2}}}}\endgroup } \sqcup  [{\begingroup\renewcommand\colorMATH{\colorMATHB}\renewcommand\colorSYNTAX{\colorSYNTAXB}{{\color{\colorMATH}\ensuremath{\sS_{1 2}}}}\endgroup }/y]{\begingroup\renewcommand\colorMATH{\colorMATHC}\renewcommand\colorSYNTAX{\colorSYNTAXC}{{\color{\colorMATH}\ensuremath{\pS_{3}}}}\endgroup }}}}
	\begin{subproof} 
		
		By {{\color{\colorMATH}\ensuremath{{\textsc{ p-case}}}}} we know that
		\begingroup\color{\colorMATH}\begin{gather*} 
		  \inferrule*[lab={\textsc{ p-case}}
		   ]{ \Gamma  \mathrel{;} {\begingroup\renewcommand\colorMATH{\colorMATHB}\renewcommand\colorSYNTAX{\colorSYNTAXB}{{\color{\colorMATH}\ensuremath{\sS_{0}}}}\endgroup }\hspace*{0.33em}{\begingroup\renewcommand\colorMATH{\colorMATHB}\renewcommand\colorSYNTAX{\colorSYNTAXB}{{\color{\colorMATH}\ensuremath{\vdash }}}\endgroup }\hspace*{0.33em}{\begingroup\renewcommand\colorMATH{\colorMATHB}\renewcommand\colorSYNTAX{\colorSYNTAXB}{{\color{\colorMATH}\ensuremath{\se_{1}}}}\endgroup } \mathrel{:} \tau _{1 1} \mathrel{^{{\begingroup\renewcommand\colorMATH{\colorMATHB}\renewcommand\colorSYNTAX{\colorSYNTAXB}{{\color{\colorMATH}\ensuremath{\sS_{1 1}}}}\endgroup }}\oplus ^{{\begingroup\renewcommand\colorMATH{\colorMATHB}\renewcommand\colorSYNTAX{\colorSYNTAXB}{{\color{\colorMATH}\ensuremath{\sS_{1 2}}}}\endgroup }}} \tau _{1 2} \mathrel{;} {\begingroup\renewcommand\colorMATH{\colorMATHB}\renewcommand\colorSYNTAX{\colorSYNTAXB}{{\color{\colorMATH}\ensuremath{\sS_{1}}}}\endgroup }
		   \\ \Gamma ,x\mathrel{:}\tau _{1 1} \mathrel{;} {\begingroup\renewcommand\colorMATH{\colorMATHB}\renewcommand\colorSYNTAX{\colorSYNTAXB}{{\color{\colorMATH}\ensuremath{\sS_{0}}}}\endgroup } + ({\begingroup\renewcommand\colorMATH{\colorMATHB}\renewcommand\colorSYNTAX{\colorSYNTAXB}{{\color{\colorMATH}\ensuremath{\sS_{0}}}}\endgroup } \mathord{\cdotp } ({\begingroup\renewcommand\colorMATH{\colorMATHB}\renewcommand\colorSYNTAX{\colorSYNTAXB}{{\color{\colorMATH}\ensuremath{\sS_{1}}}}\endgroup } + {\begingroup\renewcommand\colorMATH{\colorMATHB}\renewcommand\colorSYNTAX{\colorSYNTAXB}{{\color{\colorMATH}\ensuremath{\sS_{1 1}}}}\endgroup }))x   \hspace*{0.33em}{\begingroup\renewcommand\colorMATH{\colorMATHC}\renewcommand\colorSYNTAX{\colorSYNTAXC}{{\color{\colorMATH}\ensuremath{\vdash }}}\endgroup }\hspace*{0.33em} {\begingroup\renewcommand\colorMATH{\colorMATHC}\renewcommand\colorSYNTAX{\colorSYNTAXC}{{\color{\colorMATH}\ensuremath{\pe_{2}}}}\endgroup } \mathrel{:} \tau _{2} \mathrel{;} {\begingroup\renewcommand\colorMATH{\colorMATHC}\renewcommand\colorSYNTAX{\colorSYNTAXC}{{\color{\colorMATH}\ensuremath{\pS_{2}}}}\endgroup }
		   \\ \Gamma ,y\mathrel{:}\tau _{1 2} \mathrel{;} {\begingroup\renewcommand\colorMATH{\colorMATHB}\renewcommand\colorSYNTAX{\colorSYNTAXB}{{\color{\colorMATH}\ensuremath{\sS_{0}}}}\endgroup } + ({\begingroup\renewcommand\colorMATH{\colorMATHB}\renewcommand\colorSYNTAX{\colorSYNTAXB}{{\color{\colorMATH}\ensuremath{\sS_{0}}}}\endgroup } \mathord{\cdotp } ({\begingroup\renewcommand\colorMATH{\colorMATHB}\renewcommand\colorSYNTAX{\colorSYNTAXB}{{\color{\colorMATH}\ensuremath{\sS_{1}}}}\endgroup } + {\begingroup\renewcommand\colorMATH{\colorMATHB}\renewcommand\colorSYNTAX{\colorSYNTAXB}{{\color{\colorMATH}\ensuremath{\sS_{1 2}}}}\endgroup }))y   \hspace*{0.33em}{\begingroup\renewcommand\colorMATH{\colorMATHC}\renewcommand\colorSYNTAX{\colorSYNTAXC}{{\color{\colorMATH}\ensuremath{\vdash }}}\endgroup }\hspace*{0.33em} {\begingroup\renewcommand\colorMATH{\colorMATHC}\renewcommand\colorSYNTAX{\colorSYNTAXC}{{\color{\colorMATH}\ensuremath{\pe_{3}}}}\endgroup } \mathrel{:} \tau _{3} \mathrel{;} {\begingroup\renewcommand\colorMATH{\colorMATHC}\renewcommand\colorSYNTAX{\colorSYNTAXC}{{\color{\colorMATH}\ensuremath{\pS_{3}}}}\endgroup }
		      }{
		      \Gamma  \mathrel{;} {\begingroup\renewcommand\colorMATH{\colorMATHB}\renewcommand\colorSYNTAX{\colorSYNTAXB}{{\color{\colorMATH}\ensuremath{\sS_{0}}}}\endgroup }\hspace*{0.33em}{\begingroup\renewcommand\colorMATH{\colorMATHC}\renewcommand\colorSYNTAX{\colorSYNTAXC}{{\color{\colorMATH}\ensuremath{\vdash }}}\endgroup }\hspace*{0.33em} {{\color{\colorSYNTAX}\texttt{case}}}\hspace*{0.33em}{\begingroup\renewcommand\colorMATH{\colorMATHB}\renewcommand\colorSYNTAX{\colorSYNTAXB}{{\color{\colorMATH}\ensuremath{\se_{1}}}}\endgroup }\hspace*{0.33em}{{\color{\colorSYNTAX}\texttt{of}}}\hspace*{0.33em}\{ x\Rightarrow {\begingroup\renewcommand\colorMATH{\colorMATHC}\renewcommand\colorSYNTAX{\colorSYNTAXC}{{\color{\colorMATH}\ensuremath{\pe_{2}}}}\endgroup }\} \hspace*{0.33em}\{ y\Rightarrow {\begingroup\renewcommand\colorMATH{\colorMATHC}\renewcommand\colorSYNTAX{\colorSYNTAXC}{{\color{\colorMATH}\ensuremath{\pe_{3}}}}\endgroup }\}  \mathrel{:} \qquad\qquad\qquad\qquad\qquad\qquad
		    \\ \qquad\qquad\qquad [{\begingroup\renewcommand\colorMATH{\colorMATHB}\renewcommand\colorSYNTAX{\colorSYNTAXB}{{\color{\colorMATH}\ensuremath{\sS_{1 1}}}}\endgroup }/x]\tau _{2} \sqcup  [{\begingroup\renewcommand\colorMATH{\colorMATHB}\renewcommand\colorSYNTAX{\colorSYNTAXB}{{\color{\colorMATH}\ensuremath{\sS_{1 2}}}}\endgroup }/y]\tau _{3} \mathrel{;}  {\begingroup\renewcommand\colorMATH{\colorMATHC}\renewcommand\colorSYNTAX{\colorSYNTAXC}{{\color{\colorMATH}\ensuremath{\rceil {\begingroup\renewcommand\colorMATH{\colorMATHA}\renewcommand\colorSYNTAX{\colorSYNTAXA}{{\color{\colorMATH}\ensuremath{{\begingroup\renewcommand\colorMATH{\colorMATHB}\renewcommand\colorSYNTAX{\colorSYNTAXB}{{\color{\colorMATH}\ensuremath{\sS_{1}}}}\endgroup }}}}\endgroup }\lceil ^{\infty }}}}\endgroup } \sqcup  [{\begingroup\renewcommand\colorMATH{\colorMATHB}\renewcommand\colorSYNTAX{\colorSYNTAXB}{{\color{\colorMATH}\ensuremath{\sS_{1 1}}}}\endgroup }/x]{\begingroup\renewcommand\colorMATH{\colorMATHC}\renewcommand\colorSYNTAX{\colorSYNTAXC}{{\color{\colorMATH}\ensuremath{\pS_{2}}}}\endgroup } \sqcup  [{\begingroup\renewcommand\colorMATH{\colorMATHB}\renewcommand\colorSYNTAX{\colorSYNTAXB}{{\color{\colorMATH}\ensuremath{\sS_{1 2}}}}\endgroup }/y]{\begingroup\renewcommand\colorMATH{\colorMATHC}\renewcommand\colorSYNTAX{\colorSYNTAXC}{{\color{\colorMATH}\ensuremath{\pS_{3}}}}\endgroup }
		   }
		\end{gather*}\endgroup
		where {{\color{\colorMATH}\ensuremath{\tau  = [{\begingroup\renewcommand\colorMATH{\colorMATHB}\renewcommand\colorSYNTAX{\colorSYNTAXB}{{\color{\colorMATH}\ensuremath{\sS_{1 1}}}}\endgroup }/x]\tau _{2} \sqcup  [{\begingroup\renewcommand\colorMATH{\colorMATHB}\renewcommand\colorSYNTAX{\colorSYNTAXB}{{\color{\colorMATH}\ensuremath{\sS_{1 2}}}}\endgroup }/y]\tau _{3}}}}, and {{\color{\colorMATH}\ensuremath{{\begingroup\renewcommand\colorMATH{\colorMATHC}\renewcommand\colorSYNTAX{\colorSYNTAXC}{{\color{\colorMATH}\ensuremath{\pS}}}\endgroup } = {\begingroup\renewcommand\colorMATH{\colorMATHC}\renewcommand\colorSYNTAX{\colorSYNTAXC}{{\color{\colorMATH}\ensuremath{\rceil {\begingroup\renewcommand\colorMATH{\colorMATHA}\renewcommand\colorSYNTAX{\colorSYNTAXA}{{\color{\colorMATH}\ensuremath{{\begingroup\renewcommand\colorMATH{\colorMATHB}\renewcommand\colorSYNTAX{\colorSYNTAXB}{{\color{\colorMATH}\ensuremath{\sS_{1}}}}\endgroup }}}}\endgroup }\lceil ^{\infty }}}}\endgroup } \sqcup  [{\begingroup\renewcommand\colorMATH{\colorMATHB}\renewcommand\colorSYNTAX{\colorSYNTAXB}{{\color{\colorMATH}\ensuremath{\sS_{1 1}}}}\endgroup }/x]{\begingroup\renewcommand\colorMATH{\colorMATHC}\renewcommand\colorSYNTAX{\colorSYNTAXC}{{\color{\colorMATH}\ensuremath{\pS_{2}}}}\endgroup } \sqcup  [{\begingroup\renewcommand\colorMATH{\colorMATHB}\renewcommand\colorSYNTAX{\colorSYNTAXB}{{\color{\colorMATH}\ensuremath{\sS_{1 2}}}}\endgroup }/y]{\begingroup\renewcommand\colorMATH{\colorMATHC}\renewcommand\colorSYNTAX{\colorSYNTAXC}{{\color{\colorMATH}\ensuremath{\pS_{3}}}}\endgroup }}}}.
		
		By induction hypothesis on {{\color{\colorMATH}\ensuremath{\Gamma  \mathrel{;} {\begingroup\renewcommand\colorMATH{\colorMATHB}\renewcommand\colorSYNTAX{\colorSYNTAXB}{{\color{\colorMATH}\ensuremath{\sS_{0}}}}\endgroup }\hspace*{0.33em}{\begingroup\renewcommand\colorMATH{\colorMATHB}\renewcommand\colorSYNTAX{\colorSYNTAXB}{{\color{\colorMATH}\ensuremath{\vdash }}}\endgroup }\hspace*{0.33em} {\begingroup\renewcommand\colorMATH{\colorMATHB}\renewcommand\colorSYNTAX{\colorSYNTAXB}{{\color{\colorMATH}\ensuremath{\se_{1}}}}\endgroup } \mathrel{:} \tau _{1 1} \mathrel{^{{\begingroup\renewcommand\colorMATH{\colorMATHB}\renewcommand\colorSYNTAX{\colorSYNTAXB}{{\color{\colorMATH}\ensuremath{\sS_{1 1}}}}\endgroup }}\oplus ^{{\begingroup\renewcommand\colorMATH{\colorMATHB}\renewcommand\colorSYNTAX{\colorSYNTAXB}{{\color{\colorMATH}\ensuremath{\sS_{1 2}}}}\endgroup }}} \tau _{1 2} \mathrel{;} {\begingroup\renewcommand\colorMATH{\colorMATHB}\renewcommand\colorSYNTAX{\colorSYNTAXB}{{\color{\colorMATH}\ensuremath{\sS_{1}}}}\endgroup }}}},
		we know that {{\color{\colorMATH}\ensuremath{\gamma  \vdash  {\begingroup\renewcommand\colorMATH{\colorMATHB}\renewcommand\colorSYNTAX{\colorSYNTAXB}{{\color{\colorMATH}\ensuremath{\se_{1}}}}\endgroup } \Downarrow  {\begingroup\renewcommand\colorMATH{\colorMATHB}\renewcommand\colorSYNTAX{\colorSYNTAXB}{{\color{\colorMATH}\ensuremath{\sv_{1}}}}\endgroup }}}} and {{\color{\colorMATH}\ensuremath{{\begingroup\renewcommand\colorMATH{\colorMATHB}\renewcommand\colorSYNTAX{\colorSYNTAXB}{{\color{\colorMATH}\ensuremath{\sv_{1}}}}\endgroup } \in  {\mathcal{V}}\llbracket \tau _{1 1}/\Gamma  \mathrel{^{\varnothing }\oplus ^{\varnothing }} \tau _{1 2}/\Gamma \rrbracket }}} ({{\color{\colorMATH}\ensuremath{{\begingroup\renewcommand\colorMATH{\colorMATHB}\renewcommand\colorSYNTAX{\colorSYNTAXB}{{\color{\colorMATH}\ensuremath{\sS_{1 i}}}}\endgroup }/\Gamma  = \varnothing }}}).
		Then either {{\color{\colorMATH}\ensuremath{{\begingroup\renewcommand\colorMATH{\colorMATHB}\renewcommand\colorSYNTAX{\colorSYNTAXB}{{\color{\colorMATH}\ensuremath{\sv_{1}}}}\endgroup } = \inl^{\tau _{1 2}/\Gamma }\hspace*{0.33em}{\begingroup\renewcommand\colorMATH{\colorMATHB}\renewcommand\colorSYNTAX{\colorSYNTAXB}{{\color{\colorMATH}\ensuremath{\sv_{1 1}}}}\endgroup }}}}, or {{\color{\colorMATH}\ensuremath{{\begingroup\renewcommand\colorMATH{\colorMATHB}\renewcommand\colorSYNTAX{\colorSYNTAXB}{{\color{\colorMATH}\ensuremath{\sv_{1}}}}\endgroup } = \inl^{\tau _{1 1}/\Gamma }\hspace*{0.33em}{\begingroup\renewcommand\colorMATH{\colorMATHB}\renewcommand\colorSYNTAX{\colorSYNTAXB}{{\color{\colorMATH}\ensuremath{\sv_{1 2}}}}\endgroup }}}}.
		Let us suppose {{\color{\colorMATH}\ensuremath{{\begingroup\renewcommand\colorMATH{\colorMATHB}\renewcommand\colorSYNTAX{\colorSYNTAXB}{{\color{\colorMATH}\ensuremath{\sv_{1}}}}\endgroup } = \inl^{\tau _{1 2}/\Gamma }\hspace*{0.33em}{\begingroup\renewcommand\colorMATH{\colorMATHB}\renewcommand\colorSYNTAX{\colorSYNTAXB}{{\color{\colorMATH}\ensuremath{\sv_{1 1}}}}\endgroup }}}} (the other case is similar), then
		{{\color{\colorMATH}\ensuremath{\inl^{\tau _{1 2}/\Gamma }\hspace*{0.33em}{\begingroup\renewcommand\colorMATH{\colorMATHB}\renewcommand\colorSYNTAX{\colorSYNTAXB}{{\color{\colorMATH}\ensuremath{\sv_{1 1}}}}\endgroup } \in  {\mathcal{V}}\llbracket \tau _{1 1}/\Gamma  \mathrel{^{\varnothing }\oplus ^{\varnothing }} \tau _{1 2}/\Gamma \rrbracket }}}, and thus
		{{\color{\colorMATH}\ensuremath{{\begingroup\renewcommand\colorMATH{\colorMATHB}\renewcommand\colorSYNTAX{\colorSYNTAXB}{{\color{\colorMATH}\ensuremath{\sv_{1 1}}}}\endgroup } \in  {\mathcal{V}}\llbracket \tau _{1 1}/\Gamma \rrbracket }}}.

		Then by induction hypothesis on {{\color{\colorMATH}\ensuremath{\Gamma ,x\mathrel{:}\tau _{1 1} \mathrel{;} {\begingroup\renewcommand\colorMATH{\colorMATHB}\renewcommand\colorSYNTAX{\colorSYNTAXB}{{\color{\colorMATH}\ensuremath{\sS_{0}}}}\endgroup } + ({\begingroup\renewcommand\colorMATH{\colorMATHB}\renewcommand\colorSYNTAX{\colorSYNTAXB}{{\color{\colorMATH}\ensuremath{\sS_{0}}}}\endgroup } \mathord{\cdotp } ({\begingroup\renewcommand\colorMATH{\colorMATHB}\renewcommand\colorSYNTAX{\colorSYNTAXB}{{\color{\colorMATH}\ensuremath{\sS_{1}}}}\endgroup } + {\begingroup\renewcommand\colorMATH{\colorMATHB}\renewcommand\colorSYNTAX{\colorSYNTAXB}{{\color{\colorMATH}\ensuremath{\sS_{1 1}}}}\endgroup }))x\hspace*{0.33em}{\begingroup\renewcommand\colorMATH{\colorMATHC}\renewcommand\colorSYNTAX{\colorSYNTAXC}{{\color{\colorMATH}\ensuremath{\vdash }}}\endgroup }\hspace*{0.33em} {\begingroup\renewcommand\colorMATH{\colorMATHC}\renewcommand\colorSYNTAX{\colorSYNTAXC}{{\color{\colorMATH}\ensuremath{\pe_{2}}}}\endgroup } \mathrel{:} \tau _{2} \mathrel{;} {\begingroup\renewcommand\colorMATH{\colorMATHC}\renewcommand\colorSYNTAX{\colorSYNTAXC}{{\color{\colorMATH}\ensuremath{\pS_{2}}}}\endgroup }}}}, we know that
		{{\color{\colorMATH}\ensuremath{\gamma [x \mapsto  {\begingroup\renewcommand\colorMATH{\colorMATHB}\renewcommand\colorSYNTAX{\colorSYNTAXB}{{\color{\colorMATH}\ensuremath{\sv_{1 1}}}}\endgroup }] \vdash  {\begingroup\renewcommand\colorMATH{\colorMATHC}\renewcommand\colorSYNTAX{\colorSYNTAXC}{{\color{\colorMATH}\ensuremath{\pe_{2}}}}\endgroup } \in  {\mathcal{E}}\llbracket \tau /\Gamma \rrbracket }}}, therefore 
		for all {{\color{\colorMATH}\ensuremath{{\begingroup\renewcommand\colorMATH{\colorMATHB}\renewcommand\colorSYNTAX{\colorSYNTAXB}{{\color{\colorMATH}\ensuremath{\sv_{2}}}}\endgroup }}}} such that {{\color{\colorMATH}\ensuremath{\gamma  \vdash  {\begingroup\renewcommand\colorMATH{\colorMATHC}\renewcommand\colorSYNTAX{\colorSYNTAXC}{{\color{\colorMATH}\ensuremath{\pe_{2}}}}\endgroup } \Downarrow  {\begingroup\renewcommand\colorMATH{\colorMATHB}\renewcommand\colorSYNTAX{\colorSYNTAXB}{{\color{\colorMATH}\ensuremath{\sv_{2}}}}\endgroup }}}} it follows that {{\color{\colorMATH}\ensuremath{{\begingroup\renewcommand\colorMATH{\colorMATHB}\renewcommand\colorSYNTAX{\colorSYNTAXB}{{\color{\colorMATH}\ensuremath{\sv_{2}}}}\endgroup } \in  {\mathcal{V}}\llbracket \tau _{2}/(\Gamma ,x\mathrel{:}\tau _{1 1})\rrbracket }}}.

		Then we have to prove that for all {{\color{\colorMATH}\ensuremath{{\begingroup\renewcommand\colorMATH{\colorMATHB}\renewcommand\colorSYNTAX{\colorSYNTAXB}{{\color{\colorMATH}\ensuremath{\sv_{2}}}}\endgroup }}}} such that {{\color{\colorMATH}\ensuremath{\gamma  \vdash  {{\color{\colorSYNTAX}\texttt{case}}}\hspace*{0.33em}{\begingroup\renewcommand\colorMATH{\colorMATHB}\renewcommand\colorSYNTAX{\colorSYNTAXB}{{\color{\colorMATH}\ensuremath{\se_{1}}}}\endgroup }\hspace*{0.33em}{{\color{\colorSYNTAX}\texttt{of}}}\hspace*{0.33em}\{ x\Rightarrow {\begingroup\renewcommand\colorMATH{\colorMATHC}\renewcommand\colorSYNTAX{\colorSYNTAXC}{{\color{\colorMATH}\ensuremath{\pe_{2}}}}\endgroup }\} \hspace*{0.33em}\{ y\Rightarrow {\begingroup\renewcommand\colorMATH{\colorMATHC}\renewcommand\colorSYNTAX{\colorSYNTAXC}{{\color{\colorMATH}\ensuremath{\pe_{3}}}}\endgroup }\}  \Downarrow  {\begingroup\renewcommand\colorMATH{\colorMATHB}\renewcommand\colorSYNTAX{\colorSYNTAXB}{{\color{\colorMATH}\ensuremath{\sv_{2}}}}\endgroup }}}} it follows that {{\color{\colorMATH}\ensuremath{{\begingroup\renewcommand\colorMATH{\colorMATHB}\renewcommand\colorSYNTAX{\colorSYNTAXB}{{\color{\colorMATH}\ensuremath{\sv_{2}}}}\endgroup } \in  {\mathcal{V}}\llbracket ([{\begingroup\renewcommand\colorMATH{\colorMATHB}\renewcommand\colorSYNTAX{\colorSYNTAXB}{{\color{\colorMATH}\ensuremath{\sS_{1 1}}}}\endgroup }/x]\tau _{2} \sqcup  [{\begingroup\renewcommand\colorMATH{\colorMATHB}\renewcommand\colorSYNTAX{\colorSYNTAXB}{{\color{\colorMATH}\ensuremath{\sS_{1 2}}}}\endgroup }/y]\tau _{3}\rrbracket )/\Gamma }}}.
		By inspection of the evaluation semantics for {{\color{\colorMATH}\ensuremath{{\textsc{ p-case}}}}},
		if {{\color{\colorMATH}\ensuremath{\gamma  \vdash  {{\color{\colorSYNTAX}\texttt{case}}}\hspace*{0.33em}{\begingroup\renewcommand\colorMATH{\colorMATHB}\renewcommand\colorSYNTAX{\colorSYNTAXB}{{\color{\colorMATH}\ensuremath{\se_{1}}}}\endgroup }\hspace*{0.33em}{{\color{\colorSYNTAX}\texttt{of}}}\hspace*{0.33em}\{ x\Rightarrow {\begingroup\renewcommand\colorMATH{\colorMATHC}\renewcommand\colorSYNTAX{\colorSYNTAXC}{{\color{\colorMATH}\ensuremath{\pe_{2}}}}\endgroup }\} \hspace*{0.33em}\{ y\Rightarrow {\begingroup\renewcommand\colorMATH{\colorMATHC}\renewcommand\colorSYNTAX{\colorSYNTAXC}{{\color{\colorMATH}\ensuremath{\pe_{3}}}}\endgroup }\}  \Downarrow  {\begingroup\renewcommand\colorMATH{\colorMATHB}\renewcommand\colorSYNTAX{\colorSYNTAXB}{{\color{\colorMATH}\ensuremath{\sv_{2}}}}\endgroup }}}} then {{\color{\colorMATH}\ensuremath{\gamma  \vdash  {\begingroup\renewcommand\colorMATH{\colorMATHC}\renewcommand\colorSYNTAX{\colorSYNTAXC}{{\color{\colorMATH}\ensuremath{\pe_{2}}}}\endgroup } \Downarrow  {\begingroup\renewcommand\colorMATH{\colorMATHB}\renewcommand\colorSYNTAX{\colorSYNTAXB}{{\color{\colorMATH}\ensuremath{\sv_{2}}}}\endgroup }}}}.
		Then, by the induction hypothesis we know that {{\color{\colorMATH}\ensuremath{{\begingroup\renewcommand\colorMATH{\colorMATHB}\renewcommand\colorSYNTAX{\colorSYNTAXB}{{\color{\colorMATH}\ensuremath{\sv_{2}}}}\endgroup } \in  {\mathcal{V}}\llbracket \tau _{2}/(\Gamma ,x\mathrel{:}\tau _{1 1})\rrbracket }}}.
		But {{\color{\colorMATH}\ensuremath{([{\begingroup\renewcommand\colorMATH{\colorMATHB}\renewcommand\colorSYNTAX{\colorSYNTAXB}{{\color{\colorMATH}\ensuremath{\sS_{1 1}}}}\endgroup }/x]\tau _{2} \sqcup  [{\begingroup\renewcommand\colorMATH{\colorMATHB}\renewcommand\colorSYNTAX{\colorSYNTAXB}{{\color{\colorMATH}\ensuremath{\sS_{1 2}}}}\endgroup }/y]\tau _{3}\rrbracket )/\Gamma  = (\tau _{2} \sqcup  \tau _{3}\rrbracket )/(\Gamma ,x:\tau _{1 1})}}}. The result follows from weakening lemma~\ref{lm:weakening-type-safety}.
	\end{subproof}
\item  {{\color{\colorMATH}\ensuremath{\Gamma  \mathrel{;} {\begingroup\renewcommand\colorMATH{\colorMATHB}\renewcommand\colorSYNTAX{\colorSYNTAXB}{{\color{\colorMATH}\ensuremath{\sS_{0}}}}\endgroup }\hspace*{0.33em}{\begingroup\renewcommand\colorMATH{\colorMATHC}\renewcommand\colorSYNTAX{\colorSYNTAXC}{{\color{\colorMATH}\ensuremath{\vdash }}}\endgroup }\hspace*{0.33em} {\begingroup\renewcommand\colorMATH{\colorMATHC}\renewcommand\colorSYNTAX{\colorSYNTAXC}{{\color{\colorMATH}\ensuremath{\pe_{1}}}}\endgroup }\hspace*{0.33em}{\begingroup\renewcommand\colorMATH{\colorMATHC}\renewcommand\colorSYNTAX{\colorSYNTAXC}{{\color{\colorMATH}\ensuremath{\pe_{2}}}}\endgroup } \mathrel{:} [{\begingroup\renewcommand\colorMATH{\colorMATHB}\renewcommand\colorSYNTAX{\colorSYNTAXB}{{\color{\colorMATH}\ensuremath{\sS_{2}}}}\endgroup }/x]\tau _{2} \mathrel{;}  {\begingroup\renewcommand\colorMATH{\colorMATHC}\renewcommand\colorSYNTAX{\colorSYNTAXC}{{\color{\colorMATH}\ensuremath{\rceil {\begingroup\renewcommand\colorMATH{\colorMATHA}\renewcommand\colorSYNTAX{\colorSYNTAXA}{{\color{\colorMATH}\ensuremath{{\begingroup\renewcommand\colorMATH{\colorMATHB}\renewcommand\colorSYNTAX{\colorSYNTAXB}{{\color{\colorMATH}\ensuremath{\sS_{1}}}}\endgroup }}}}\endgroup }\lceil ^{\infty }}}}\endgroup } +  [{\begingroup\renewcommand\colorMATH{\colorMATHB}\renewcommand\colorSYNTAX{\colorSYNTAXB}{{\color{\colorMATH}\ensuremath{\sS_{2}}}}\endgroup }/x]{\begingroup\renewcommand\colorMATH{\colorMATHC}\renewcommand\colorSYNTAX{\colorSYNTAXC}{{\color{\colorMATH}\ensuremath{\pS}}}\endgroup }}}}
	\begin{subproof} 
		By {{\color{\colorMATH}\ensuremath{{\textsc{ p-app}}}}} we know that
		\begingroup\color{\colorMATH}\begin{gather*} 
		  \inferrule*[lab={\textsc{ p-app}}
		  ]{ \Gamma  \mathrel{;} {\begingroup\renewcommand\colorMATH{\colorMATHB}\renewcommand\colorSYNTAX{\colorSYNTAXB}{{\color{\colorMATH}\ensuremath{\sS_{0}}}}\endgroup }\hspace*{0.33em}{\begingroup\renewcommand\colorMATH{\colorMATHB}\renewcommand\colorSYNTAX{\colorSYNTAXB}{{\color{\colorMATH}\ensuremath{\vdash }}}\endgroup }\hspace*{0.33em}{\begingroup\renewcommand\colorMATH{\colorMATHB}\renewcommand\colorSYNTAX{\colorSYNTAXB}{{\color{\colorMATH}\ensuremath{\se_{1}}}}\endgroup } \mathrel{:} (x\mathrel{:}\tau _{1}\mathord{\cdotp }{\begingroup\renewcommand\colorMATH{\colorMATHB}\renewcommand\colorSYNTAX{\colorSYNTAXB}{{\color{\colorMATH}\ensuremath{\sss}}}\endgroup }) \xrightarrowP {{\begingroup\renewcommand\colorMATH{\colorMATHC}\renewcommand\colorSYNTAX{\colorSYNTAXC}{{\color{\colorMATH}\ensuremath{\pS}}}\endgroup }+{\begingroup\renewcommand\colorMATH{\colorMATHC}\renewcommand\colorSYNTAX{\colorSYNTAXC}{{\color{\colorMATH}\ensuremath{p}}}\endgroup }x} \tau _{2} \mathrel{;} {\begingroup\renewcommand\colorMATH{\colorMATHB}\renewcommand\colorSYNTAX{\colorSYNTAXB}{{\color{\colorMATH}\ensuremath{\sS_{1}}}}\endgroup }
		  \\ \Gamma  \mathrel{;} {\begingroup\renewcommand\colorMATH{\colorMATHB}\renewcommand\colorSYNTAX{\colorSYNTAXB}{{\color{\colorMATH}\ensuremath{\sS_{0}}}}\endgroup }\hspace*{0.33em}{\begingroup\renewcommand\colorMATH{\colorMATHB}\renewcommand\colorSYNTAX{\colorSYNTAXB}{{\color{\colorMATH}\ensuremath{\vdash }}}\endgroup }\hspace*{0.33em}{\begingroup\renewcommand\colorMATH{\colorMATHB}\renewcommand\colorSYNTAX{\colorSYNTAXB}{{\color{\colorMATH}\ensuremath{\se_{2}}}}\endgroup } \mathrel{:} \tau _{1} \mathrel{;} {\begingroup\renewcommand\colorMATH{\colorMATHB}\renewcommand\colorSYNTAX{\colorSYNTAXB}{{\color{\colorMATH}\ensuremath{\sS_{2}}}}\endgroup }
		  \\ {\begingroup\renewcommand\colorMATH{\colorMATHB}\renewcommand\colorSYNTAX{\colorSYNTAXB}{{\color{\colorMATH}\ensuremath{\sS_{0}}}}\endgroup } \mathord{\cdotp } {\begingroup\renewcommand\colorMATH{\colorMATHB}\renewcommand\colorSYNTAX{\colorSYNTAXB}{{\color{\colorMATH}\ensuremath{\sS_{2}}}}\endgroup } \leq  {\begingroup\renewcommand\colorMATH{\colorMATHB}\renewcommand\colorSYNTAX{\colorSYNTAXB}{{\color{\colorMATH}\ensuremath{\sss}}}\endgroup }
		     }{
		     \Gamma  \mathrel{;} {\begingroup\renewcommand\colorMATH{\colorMATHB}\renewcommand\colorSYNTAX{\colorSYNTAXB}{{\color{\colorMATH}\ensuremath{\sS_{0}}}}\endgroup }\hspace*{0.33em}{\begingroup\renewcommand\colorMATH{\colorMATHC}\renewcommand\colorSYNTAX{\colorSYNTAXC}{{\color{\colorMATH}\ensuremath{\vdash }}}\endgroup }\hspace*{0.33em} {\begingroup\renewcommand\colorMATH{\colorMATHC}\renewcommand\colorSYNTAX{\colorSYNTAXC}{{\color{\colorMATH}\ensuremath{\pe_{1}}}}\endgroup }\hspace*{0.33em}{\begingroup\renewcommand\colorMATH{\colorMATHC}\renewcommand\colorSYNTAX{\colorSYNTAXC}{{\color{\colorMATH}\ensuremath{\pe_{2}}}}\endgroup } \mathrel{:} [{\begingroup\renewcommand\colorMATH{\colorMATHB}\renewcommand\colorSYNTAX{\colorSYNTAXB}{{\color{\colorMATH}\ensuremath{\sS_{2}}}}\endgroup }/x]\tau _{2} \mathrel{;}  {\begingroup\renewcommand\colorMATH{\colorMATHC}\renewcommand\colorSYNTAX{\colorSYNTAXC}{{\color{\colorMATH}\ensuremath{\rceil {\begingroup\renewcommand\colorMATH{\colorMATHA}\renewcommand\colorSYNTAX{\colorSYNTAXA}{{\color{\colorMATH}\ensuremath{{\begingroup\renewcommand\colorMATH{\colorMATHB}\renewcommand\colorSYNTAX{\colorSYNTAXB}{{\color{\colorMATH}\ensuremath{\sS_{1}}}}\endgroup }}}}\endgroup }\lceil ^{\infty }}}}\endgroup } +  [{\begingroup\renewcommand\colorMATH{\colorMATHB}\renewcommand\colorSYNTAX{\colorSYNTAXB}{{\color{\colorMATH}\ensuremath{\sS_{2}}}}\endgroup }/x]{\begingroup\renewcommand\colorMATH{\colorMATHC}\renewcommand\colorSYNTAX{\colorSYNTAXC}{{\color{\colorMATH}\ensuremath{\pS}}}\endgroup }
		  }
		\end{gather*}\endgroup
		where {{\color{\colorMATH}\ensuremath{\tau  = [{\begingroup\renewcommand\colorMATH{\colorMATHB}\renewcommand\colorSYNTAX{\colorSYNTAXB}{{\color{\colorMATH}\ensuremath{\sS_{2}}}}\endgroup }/x]\tau _{2}}}}, and {{\color{\colorMATH}\ensuremath{ {\begingroup\renewcommand\colorMATH{\colorMATHC}\renewcommand\colorSYNTAX{\colorSYNTAXC}{{\color{\colorMATH}\ensuremath{\pS}}}\endgroup } = {\begingroup\renewcommand\colorMATH{\colorMATHC}\renewcommand\colorSYNTAX{\colorSYNTAXC}{{\color{\colorMATH}\ensuremath{\rceil {\begingroup\renewcommand\colorMATH{\colorMATHA}\renewcommand\colorSYNTAX{\colorSYNTAXA}{{\color{\colorMATH}\ensuremath{{\begingroup\renewcommand\colorMATH{\colorMATHB}\renewcommand\colorSYNTAX{\colorSYNTAXB}{{\color{\colorMATH}\ensuremath{\sS_{1}}}}\endgroup }}}}\endgroup }\lceil ^{\infty }}}}\endgroup } +  [{\begingroup\renewcommand\colorMATH{\colorMATHB}\renewcommand\colorSYNTAX{\colorSYNTAXB}{{\color{\colorMATH}\ensuremath{\sS_{2}}}}\endgroup }/x]{\begingroup\renewcommand\colorMATH{\colorMATHC}\renewcommand\colorSYNTAX{\colorSYNTAXC}{{\color{\colorMATH}\ensuremath{\pS}}}\endgroup }}}}.
		By induction hypotheses we know that
		for all {{\color{\colorMATH}\ensuremath{{\begingroup\renewcommand\colorMATH{\colorMATHB}\renewcommand\colorSYNTAX{\colorSYNTAXB}{{\color{\colorMATH}\ensuremath{\sv_{1}}}}\endgroup }}}} and {{\color{\colorMATH}\ensuremath{{\begingroup\renewcommand\colorMATH{\colorMATHB}\renewcommand\colorSYNTAX{\colorSYNTAXB}{{\color{\colorMATH}\ensuremath{\sv_{2}}}}\endgroup }}}}
		such that {{\color{\colorMATH}\ensuremath{\gamma  \vdash  {\begingroup\renewcommand\colorMATH{\colorMATHB}\renewcommand\colorSYNTAX{\colorSYNTAXB}{{\color{\colorMATH}\ensuremath{\se_{1}}}}\endgroup } \Downarrow  {\begingroup\renewcommand\colorMATH{\colorMATHB}\renewcommand\colorSYNTAX{\colorSYNTAXB}{{\color{\colorMATH}\ensuremath{\sv_{1}}}}\endgroup }}}}, {{\color{\colorMATH}\ensuremath{\gamma  \vdash  {\begingroup\renewcommand\colorMATH{\colorMATHB}\renewcommand\colorSYNTAX{\colorSYNTAXB}{{\color{\colorMATH}\ensuremath{\se_{2}}}}\endgroup } \Downarrow  {\begingroup\renewcommand\colorMATH{\colorMATHB}\renewcommand\colorSYNTAX{\colorSYNTAXB}{{\color{\colorMATH}\ensuremath{\sv_{2}}}}\endgroup }}}},
		it follows that 
		{{\color{\colorMATH}\ensuremath{{\begingroup\renewcommand\colorMATH{\colorMATHB}\renewcommand\colorSYNTAX{\colorSYNTAXB}{{\color{\colorMATH}\ensuremath{\sv_{1}}}}\endgroup } \in  {\mathcal{V}}\llbracket (x\mathrel{:}\tau _{1}/\Gamma \mathord{\cdotp }{\begingroup\renewcommand\colorMATH{\colorMATHB}\renewcommand\colorSYNTAX{\colorSYNTAXB}{{\color{\colorMATH}\ensuremath{\sss_{1}}}}\endgroup }) \xrightarrowP {{\begingroup\renewcommand\colorMATH{\colorMATHC}\renewcommand\colorSYNTAX{\colorSYNTAXC}{{\color{\colorMATH}\ensuremath{\pS}}}\endgroup }/\Gamma } \tau _{2}/\Gamma \rrbracket }}} and 
		{{\color{\colorMATH}\ensuremath{{\begingroup\renewcommand\colorMATH{\colorMATHB}\renewcommand\colorSYNTAX{\colorSYNTAXB}{{\color{\colorMATH}\ensuremath{\sv_{2}}}}\endgroup } \in  {\mathcal{V}}\llbracket \tau _{1}/\Gamma \rrbracket }}}.

		We have to prove that for all {{\color{\colorMATH}\ensuremath{{\begingroup\renewcommand\colorMATH{\colorMATHB}\renewcommand\colorSYNTAX{\colorSYNTAXB}{{\color{\colorMATH}\ensuremath{\sv'}}}\endgroup }}}} such that {{\color{\colorMATH}\ensuremath{\gamma  \vdash  {\begingroup\renewcommand\colorMATH{\colorMATHC}\renewcommand\colorSYNTAX{\colorSYNTAXC}{{\color{\colorMATH}\ensuremath{\pe_{1}}}}\endgroup }\hspace*{0.33em}{\begingroup\renewcommand\colorMATH{\colorMATHC}\renewcommand\colorSYNTAX{\colorSYNTAXC}{{\color{\colorMATH}\ensuremath{\pe_{2}}}}\endgroup } \Downarrow  {\begingroup\renewcommand\colorMATH{\colorMATHB}\renewcommand\colorSYNTAX{\colorSYNTAXB}{{\color{\colorMATH}\ensuremath{\sv'}}}\endgroup }}}} it follows that {{\color{\colorMATH}\ensuremath{{\begingroup\renewcommand\colorMATH{\colorMATHB}\renewcommand\colorSYNTAX{\colorSYNTAXB}{{\color{\colorMATH}\ensuremath{\sv'}}}\endgroup } \in  {\mathcal{V}}\llbracket [{\begingroup\renewcommand\colorMATH{\colorMATHB}\renewcommand\colorSYNTAX{\colorSYNTAXB}{{\color{\colorMATH}\ensuremath{\sS_{2}}}}\endgroup }/x]\tau _{2}/\Gamma \rrbracket }}}.
		By inspection of the evaluation semantics for {{\color{\colorMATH}\ensuremath{{\textsc{ p-app}}}}},
		we know that if {{\color{\colorMATH}\ensuremath{\gamma  \vdash  {\begingroup\renewcommand\colorMATH{\colorMATHC}\renewcommand\colorSYNTAX{\colorSYNTAXC}{{\color{\colorMATH}\ensuremath{\pe_{1}}}}\endgroup }\hspace*{0.33em}{\begingroup\renewcommand\colorMATH{\colorMATHC}\renewcommand\colorSYNTAX{\colorSYNTAXC}{{\color{\colorMATH}\ensuremath{\pe_{2}}}}\endgroup } \Downarrow  {\begingroup\renewcommand\colorMATH{\colorMATHB}\renewcommand\colorSYNTAX{\colorSYNTAXB}{{\color{\colorMATH}\ensuremath{\sv'}}}\endgroup }}}} then
		{{\color{\colorMATH}\ensuremath{{\begingroup\renewcommand\colorMATH{\colorMATHB}\renewcommand\colorSYNTAX{\colorSYNTAXB}{{\color{\colorMATH}\ensuremath{\sv_{1}}}}\endgroup } = \langle {\begingroup\renewcommand\colorMATH{\colorMATHC}\renewcommand\colorSYNTAX{\colorSYNTAXC}{{\color{\colorMATH}\ensuremath{\plambda}}}\endgroup } (x\mathrel{:}\tau '_{1}\mathord{\cdotp }{\begingroup\renewcommand\colorMATH{\colorMATHB}\renewcommand\colorSYNTAX{\colorSYNTAXB}{{\color{\colorMATH}\ensuremath{\sss'_{1}}}}\endgroup }).\hspace*{0.33em}{\begingroup\renewcommand\colorMATH{\colorMATHC}\renewcommand\colorSYNTAX{\colorSYNTAXC}{{\color{\colorMATH}\ensuremath{\pe'}}}\endgroup }, \gamma '\rangle }}} and {{\color{\colorMATH}\ensuremath{\gamma  \vdash  {\begingroup\renewcommand\colorMATH{\colorMATHB}\renewcommand\colorSYNTAX{\colorSYNTAXB}{{\color{\colorMATH}\ensuremath{\se_{2}}}}\endgroup } \Downarrow  {\begingroup\renewcommand\colorMATH{\colorMATHB}\renewcommand\colorSYNTAX{\colorSYNTAXB}{{\color{\colorMATH}\ensuremath{\sv_{2}}}}\endgroup }}}}, for some {{\color{\colorMATH}\ensuremath{\tau '_{1}, {\begingroup\renewcommand\colorMATH{\colorMATHB}\renewcommand\colorSYNTAX{\colorSYNTAXB}{{\color{\colorMATH}\ensuremath{\sss'_{1}}}}\endgroup }, {\begingroup\renewcommand\colorMATH{\colorMATHC}\renewcommand\colorSYNTAX{\colorSYNTAXC}{{\color{\colorMATH}\ensuremath{\pe'}}}\endgroup }}}} and {{\color{\colorMATH}\ensuremath{\gamma '}}}.
		By inspection of the function predicate, we know that 
		{{\color{\colorMATH}\ensuremath{\gamma '[x \mapsto  {\begingroup\renewcommand\colorMATH{\colorMATHB}\renewcommand\colorSYNTAX{\colorSYNTAXB}{{\color{\colorMATH}\ensuremath{\sv_{2}}}}\endgroup }] \vdash  {\begingroup\renewcommand\colorMATH{\colorMATHC}\renewcommand\colorSYNTAX{\colorSYNTAXC}{{\color{\colorMATH}\ensuremath{\pe'}}}\endgroup } \in  {\mathcal{E}}\llbracket \tau _{2}/(\Gamma ,x:\tau _{1})\rrbracket }}}, 
		i.e. for all {{\color{\colorMATH}\ensuremath{{\begingroup\renewcommand\colorMATH{\colorMATHB}\renewcommand\colorSYNTAX{\colorSYNTAXB}{{\color{\colorMATH}\ensuremath{\sv'}}}\endgroup }}}} such that {{\color{\colorMATH}\ensuremath{\gamma '[x \mapsto  {\begingroup\renewcommand\colorMATH{\colorMATHB}\renewcommand\colorSYNTAX{\colorSYNTAXB}{{\color{\colorMATH}\ensuremath{\sv_{2}}}}\endgroup }] \vdash  {\begingroup\renewcommand\colorMATH{\colorMATHC}\renewcommand\colorSYNTAX{\colorSYNTAXC}{{\color{\colorMATH}\ensuremath{\pe'}}}\endgroup } \Downarrow  {\begingroup\renewcommand\colorMATH{\colorMATHB}\renewcommand\colorSYNTAX{\colorSYNTAXB}{{\color{\colorMATH}\ensuremath{\sv'}}}\endgroup }}}} it follows that {{\color{\colorMATH}\ensuremath{{\begingroup\renewcommand\colorMATH{\colorMATHB}\renewcommand\colorSYNTAX{\colorSYNTAXB}{{\color{\colorMATH}\ensuremath{\sv'}}}\endgroup } \in  {\mathcal{V}}\llbracket \tau _{2}/(\Gamma ,x:\tau _{1})\rrbracket }}} 
		but {{\color{\colorMATH}\ensuremath{[{\begingroup\renewcommand\colorMATH{\colorMATHB}\renewcommand\colorSYNTAX{\colorSYNTAXB}{{\color{\colorMATH}\ensuremath{\sS_{2}}}}\endgroup }/x]\tau _{2}/\Gamma  = [({\begingroup\renewcommand\colorMATH{\colorMATHB}\renewcommand\colorSYNTAX{\colorSYNTAXB}{{\color{\colorMATH}\ensuremath{\sS_{2}}}}\endgroup }/\Gamma )/x](\tau _{2}/\Gamma ) = [\varnothing /x](\tau _{2}/\Gamma ) = \tau _{2}/(\Gamma ,x:\tau _{1})}}} and the result holds.
	\end{subproof}
\end{enumerate}
\end{proof}

\begin{lemma}[Weakening]\;
  \label{lm:weakening-type-safety}
  If {{\color{\colorMATH}\ensuremath{{\begingroup\renewcommand\colorMATH{\colorMATHB}\renewcommand\colorSYNTAX{\colorSYNTAXB}{{\color{\colorMATH}\ensuremath{\sv}}}\endgroup } \in  {\mathcal{V}}\llbracket \tau \rrbracket }}} and {{\color{\colorMATH}\ensuremath{\tau  <: \tau '}}}, {{\color{\colorMATH}\ensuremath{FV(\tau ') = \varnothing }}}, then {{\color{\colorMATH}\ensuremath{{\begingroup\renewcommand\colorMATH{\colorMATHB}\renewcommand\colorSYNTAX{\colorSYNTAXB}{{\color{\colorMATH}\ensuremath{\sv}}}\endgroup } \in  {\mathcal{V}}\llbracket \tau '\rrbracket }}}.
\end{lemma}
\begin{proof}
Straightforward induction on {{\color{\colorMATH}\ensuremath{\tau }}} such that {{\color{\colorMATH}\ensuremath{{\mathcal{V}}\llbracket \tau \rrbracket }}}.
\begin{enumerate}[ncases]\item  {{\color{\colorMATH}\ensuremath{\tau  \in  \{{\begingroup\renewcommand\colorMATH{\colorMATHA}\renewcommand\colorSYNTAX{\colorSYNTAXA}{{\color{\colorSYNTAX}\texttt{{\ensuremath{{\mathbb{R}}}}}}}\endgroup }, {{\color{\colorSYNTAX}\texttt{unit}}}\}}}}
	\begin{subproof} 
		Trivial as {{\color{\colorMATH}\ensuremath{\tau  <: \tau }}}.
	\end{subproof}
\item  {{\color{\colorMATH}\ensuremath{\tau  = (x\mathrel{:}\tau _{1}\mathord{\cdotp }{\begingroup\renewcommand\colorMATH{\colorMATHB}\renewcommand\colorSYNTAX{\colorSYNTAXB}{{\color{\colorMATH}\ensuremath{\sss}}}\endgroup }) P\xrightarrowP {{\begingroup\renewcommand\colorMATH{\colorMATHC}\renewcommand\colorSYNTAX{\colorSYNTAXC}{{\color{\colorMATH}\ensuremath{\pS_{1}}}}\endgroup }} \tau _{2}}}}
	\begin{subproof} 
		Then {{\color{\colorMATH}\ensuremath{\tau ' = (x\mathrel{:}\tau '_{1}\mathord{\cdotp }{\begingroup\renewcommand\colorMATH{\colorMATHB}\renewcommand\colorSYNTAX{\colorSYNTAXB}{{\color{\colorMATH}\ensuremath{\sss'}}}\endgroup }) \xrightarrowP {{\begingroup\renewcommand\colorMATH{\colorMATHC}\renewcommand\colorSYNTAX{\colorSYNTAXC}{{\color{\colorMATH}\ensuremath{\pS'_{1}}}}\endgroup }} \tau '_{2}}}}, for some {{\color{\colorMATH}\ensuremath{ \tau '_{1} <: \tau _{1}, {\begingroup\renewcommand\colorMATH{\colorMATHB}\renewcommand\colorSYNTAX{\colorSYNTAXB}{{\color{\colorMATH}\ensuremath{\sss'}}}\endgroup } \leq  {\begingroup\renewcommand\colorMATH{\colorMATHB}\renewcommand\colorSYNTAX{\colorSYNTAXB}{{\color{\colorMATH}\ensuremath{\sss}}}\endgroup }, {\begingroup\renewcommand\colorMATH{\colorMATHC}\renewcommand\colorSYNTAX{\colorSYNTAXC}{{\color{\colorMATH}\ensuremath{\pS_{1}}}}\endgroup } <: {\begingroup\renewcommand\colorMATH{\colorMATHC}\renewcommand\colorSYNTAX{\colorSYNTAXC}{{\color{\colorMATH}\ensuremath{\pS'_{1}}}}\endgroup },}}} and {{\color{\colorMATH}\ensuremath{\tau _{2} <: \tau '_{2}}}}.
		We know that 
		{{\color{\colorMATH}\ensuremath{{\begingroup\renewcommand\colorMATH{\colorMATHB}\renewcommand\colorSYNTAX{\colorSYNTAXB}{{\color{\colorMATH}\ensuremath{\sv}}}\endgroup } \in  {\mathcal{V}}\llbracket (x\mathrel{:}\tau _{1}\mathord{\cdotp }{\begingroup\renewcommand\colorMATH{\colorMATHB}\renewcommand\colorSYNTAX{\colorSYNTAXB}{{\color{\colorMATH}\ensuremath{\sss}}}\endgroup }) \xrightarrowP {{\begingroup\renewcommand\colorMATH{\colorMATHC}\renewcommand\colorSYNTAX{\colorSYNTAXC}{{\color{\colorMATH}\ensuremath{\pS_{1}}}}\endgroup }} \tau _{2}\rrbracket }}} and we have to prove that
		{{\color{\colorMATH}\ensuremath{{\begingroup\renewcommand\colorMATH{\colorMATHB}\renewcommand\colorSYNTAX{\colorSYNTAXB}{{\color{\colorMATH}\ensuremath{\sv}}}\endgroup } \in  {\mathcal{V}}\llbracket (x\mathrel{:}\tau '_{1}\mathord{\cdotp }{\begingroup\renewcommand\colorMATH{\colorMATHB}\renewcommand\colorSYNTAX{\colorSYNTAXB}{{\color{\colorMATH}\ensuremath{\sss'}}}\endgroup }) \xrightarrowP {{\begingroup\renewcommand\colorMATH{\colorMATHC}\renewcommand\colorSYNTAX{\colorSYNTAXC}{{\color{\colorMATH}\ensuremath{\pS'_{1}}}}\endgroup }} \tau '_{2}\rrbracket }}}.
		First we have to prove that {{\color{\colorMATH}\ensuremath{{\begingroup\renewcommand\colorMATH{\colorMATHB}\renewcommand\colorSYNTAX{\colorSYNTAXB}{{\color{\colorMATH}\ensuremath{\sv}}}\endgroup } \in  Atom\llbracket (x\mathrel{:}\tau '_{1}\mathord{\cdotp }{\begingroup\renewcommand\colorMATH{\colorMATHB}\renewcommand\colorSYNTAX{\colorSYNTAXB}{{\color{\colorMATH}\ensuremath{\sss'}}}\endgroup }) \xrightarrowP {{\begingroup\renewcommand\colorMATH{\colorMATHC}\renewcommand\colorSYNTAX{\colorSYNTAXC}{{\color{\colorMATH}\ensuremath{\pS'_{1}}}}\endgroup }} \tau '_{2}\rrbracket }}}, which is direct.

		Suppose {{\color{\colorMATH}\ensuremath{{\begingroup\renewcommand\colorMATH{\colorMATHB}\renewcommand\colorSYNTAX{\colorSYNTAXB}{{\color{\colorMATH}\ensuremath{\sv}}}\endgroup } = \langle {\begingroup\renewcommand\colorMATH{\colorMATHC}\renewcommand\colorSYNTAX{\colorSYNTAXC}{{\color{\colorMATH}\ensuremath{\plambda}}}\endgroup } x:\tau _{1}\mathord{\cdotp }{\begingroup\renewcommand\colorMATH{\colorMATHB}\renewcommand\colorSYNTAX{\colorSYNTAXB}{{\color{\colorMATH}\ensuremath{\sss}}}\endgroup }. {\begingroup\renewcommand\colorMATH{\colorMATHC}\renewcommand\colorSYNTAX{\colorSYNTAXC}{{\color{\colorMATH}\ensuremath{\pe}}}\endgroup }, \gamma \rangle }}}.
		Let {{\color{\colorMATH}\ensuremath{{\begingroup\renewcommand\colorMATH{\colorMATHB}\renewcommand\colorSYNTAX{\colorSYNTAXB}{{\color{\colorMATH}\ensuremath{\sv'}}}\endgroup } \in  {\mathcal{V}}\llbracket \tau '_{1}\rrbracket }}}, we have to prove that
		{{\color{\colorMATH}\ensuremath{\gamma [x \mapsto  {\begingroup\renewcommand\colorMATH{\colorMATHB}\renewcommand\colorSYNTAX{\colorSYNTAXB}{{\color{\colorMATH}\ensuremath{\sv'}}}\endgroup }] \vdash  {\begingroup\renewcommand\colorMATH{\colorMATHC}\renewcommand\colorSYNTAX{\colorSYNTAXC}{{\color{\colorMATH}\ensuremath{\pe}}}\endgroup } \in  {\mathcal{E}}\llbracket \tau '_{2}\rrbracket }}}, i.e. that {{\color{\colorMATH}\ensuremath{\gamma [x \mapsto  {\begingroup\renewcommand\colorMATH{\colorMATHB}\renewcommand\colorSYNTAX{\colorSYNTAXB}{{\color{\colorMATH}\ensuremath{\sv'}}}\endgroup }] \vdash  {\begingroup\renewcommand\colorMATH{\colorMATHC}\renewcommand\colorSYNTAX{\colorSYNTAXC}{{\color{\colorMATH}\ensuremath{\pe}}}\endgroup } \Downarrow  {\begingroup\renewcommand\colorMATH{\colorMATHB}\renewcommand\colorSYNTAX{\colorSYNTAXB}{{\color{\colorMATH}\ensuremath{\sv''}}}\endgroup }}}} and {{\color{\colorMATH}\ensuremath{{\begingroup\renewcommand\colorMATH{\colorMATHB}\renewcommand\colorSYNTAX{\colorSYNTAXB}{{\color{\colorMATH}\ensuremath{\sv''}}}\endgroup } \in  {\mathcal{V}}\llbracket \tau '_{2}\rrbracket }}}.
		By induction hypothesis we know that {{\color{\colorMATH}\ensuremath{{\begingroup\renewcommand\colorMATH{\colorMATHB}\renewcommand\colorSYNTAX{\colorSYNTAXB}{{\color{\colorMATH}\ensuremath{\sv'}}}\endgroup } \in  {\mathcal{V}}\llbracket \tau _{1}\rrbracket }}}, and by
		{{\color{\colorMATH}\ensuremath{{\begingroup\renewcommand\colorMATH{\colorMATHB}\renewcommand\colorSYNTAX{\colorSYNTAXB}{{\color{\colorMATH}\ensuremath{\sv}}}\endgroup } \in  {\mathcal{V}}\llbracket (x\mathrel{:}\tau _{1}\mathord{\cdotp }{\begingroup\renewcommand\colorMATH{\colorMATHB}\renewcommand\colorSYNTAX{\colorSYNTAXB}{{\color{\colorMATH}\ensuremath{\sss}}}\endgroup }) \xrightarrowP {{\begingroup\renewcommand\colorMATH{\colorMATHC}\renewcommand\colorSYNTAX{\colorSYNTAXC}{{\color{\colorMATH}\ensuremath{\pS_{1}}}}\endgroup }} \tau _{2}\rrbracket }}} we know that
		{{\color{\colorMATH}\ensuremath{\gamma [x \mapsto  {\begingroup\renewcommand\colorMATH{\colorMATHB}\renewcommand\colorSYNTAX{\colorSYNTAXB}{{\color{\colorMATH}\ensuremath{\sv'}}}\endgroup }] \vdash  {\begingroup\renewcommand\colorMATH{\colorMATHC}\renewcommand\colorSYNTAX{\colorSYNTAXC}{{\color{\colorMATH}\ensuremath{\pe}}}\endgroup } \in  {\mathcal{E}}\llbracket \tau _{2}\rrbracket }}}. 
		This means that {{\color{\colorMATH}\ensuremath{\gamma [x \mapsto  {\begingroup\renewcommand\colorMATH{\colorMATHB}\renewcommand\colorSYNTAX{\colorSYNTAXB}{{\color{\colorMATH}\ensuremath{\sv'}}}\endgroup }] \vdash  {\begingroup\renewcommand\colorMATH{\colorMATHC}\renewcommand\colorSYNTAX{\colorSYNTAXC}{{\color{\colorMATH}\ensuremath{\pe}}}\endgroup } \Downarrow  {\begingroup\renewcommand\colorMATH{\colorMATHB}\renewcommand\colorSYNTAX{\colorSYNTAXB}{{\color{\colorMATH}\ensuremath{\sv''}}}\endgroup }}}} and {{\color{\colorMATH}\ensuremath{{\begingroup\renewcommand\colorMATH{\colorMATHB}\renewcommand\colorSYNTAX{\colorSYNTAXB}{{\color{\colorMATH}\ensuremath{\sv''}}}\endgroup } \in  {\mathcal{V}}\llbracket \tau _{2}\rrbracket }}}.
		But as {{\color{\colorMATH}\ensuremath{\tau _{2} <: \tau '_{2}}}} by induction hypothesis 
		{{\color{\colorMATH}\ensuremath{{\begingroup\renewcommand\colorMATH{\colorMATHB}\renewcommand\colorSYNTAX{\colorSYNTAXB}{{\color{\colorMATH}\ensuremath{\sv''}}}\endgroup } \in  {\mathcal{V}}\llbracket \tau '_{2}\rrbracket }}} and the result holds.
	\end{subproof}
\item  {{\color{\colorMATH}\ensuremath{\tau  = (x\mathrel{:}\tau _{1}\mathord{\cdotp }{\begingroup\renewcommand\colorMATH{\colorMATHB}\renewcommand\colorSYNTAX{\colorSYNTAXB}{{\color{\colorMATH}\ensuremath{\sss}}}\endgroup }) \xrightarrowS {{\begingroup\renewcommand\colorMATH{\colorMATHB}\renewcommand\colorSYNTAX{\colorSYNTAXB}{{\color{\colorMATH}\ensuremath{\sS_{1}}}}\endgroup }} \tau _{2}}}}
	\begin{subproof} 
		Then {{\color{\colorMATH}\ensuremath{\tau ' = (x\mathrel{:}\tau '_{1}\mathord{\cdotp }{\begingroup\renewcommand\colorMATH{\colorMATHB}\renewcommand\colorSYNTAX{\colorSYNTAXB}{{\color{\colorMATH}\ensuremath{\sss'}}}\endgroup }) \xrightarrowS {{\begingroup\renewcommand\colorMATH{\colorMATHB}\renewcommand\colorSYNTAX{\colorSYNTAXB}{{\color{\colorMATH}\ensuremath{\sS'_{1}}}}\endgroup }} \tau '_{2}}}}, for some {{\color{\colorMATH}\ensuremath{ \tau '_{1} <: \tau _{1}, {\begingroup\renewcommand\colorMATH{\colorMATHB}\renewcommand\colorSYNTAX{\colorSYNTAXB}{{\color{\colorMATH}\ensuremath{\sss'}}}\endgroup } \leq  {\begingroup\renewcommand\colorMATH{\colorMATHB}\renewcommand\colorSYNTAX{\colorSYNTAXB}{{\color{\colorMATH}\ensuremath{\sss}}}\endgroup }, {\begingroup\renewcommand\colorMATH{\colorMATHB}\renewcommand\colorSYNTAX{\colorSYNTAXB}{{\color{\colorMATH}\ensuremath{\sS_{1}}}}\endgroup } <: {\begingroup\renewcommand\colorMATH{\colorMATHB}\renewcommand\colorSYNTAX{\colorSYNTAXB}{{\color{\colorMATH}\ensuremath{\sS'_{1}}}}\endgroup },}}} and {{\color{\colorMATH}\ensuremath{\tau _{2} <: \tau '_{2}}}}.
		We know that 
		{{\color{\colorMATH}\ensuremath{{\begingroup\renewcommand\colorMATH{\colorMATHB}\renewcommand\colorSYNTAX{\colorSYNTAXB}{{\color{\colorMATH}\ensuremath{\sv}}}\endgroup } \in  {\mathcal{V}}\llbracket (x\mathrel{:}\tau _{1}\mathord{\cdotp }{\begingroup\renewcommand\colorMATH{\colorMATHB}\renewcommand\colorSYNTAX{\colorSYNTAXB}{{\color{\colorMATH}\ensuremath{\sss}}}\endgroup }) \xrightarrowS {{\begingroup\renewcommand\colorMATH{\colorMATHB}\renewcommand\colorSYNTAX{\colorSYNTAXB}{{\color{\colorMATH}\ensuremath{\sS_{1}}}}\endgroup }} \tau _{2}\rrbracket }}} and we have to prove that
		{{\color{\colorMATH}\ensuremath{{\begingroup\renewcommand\colorMATH{\colorMATHB}\renewcommand\colorSYNTAX{\colorSYNTAXB}{{\color{\colorMATH}\ensuremath{\sv}}}\endgroup } \in  {\mathcal{V}}\llbracket (x\mathrel{:}\tau '_{1}\mathord{\cdotp }{\begingroup\renewcommand\colorMATH{\colorMATHB}\renewcommand\colorSYNTAX{\colorSYNTAXB}{{\color{\colorMATH}\ensuremath{\sss'}}}\endgroup }) \xrightarrowS {{\begingroup\renewcommand\colorMATH{\colorMATHB}\renewcommand\colorSYNTAX{\colorSYNTAXB}{{\color{\colorMATH}\ensuremath{\sS'_{1}}}}\endgroup }} \tau '_{2}\rrbracket }}}.
		First we have to prove that {{\color{\colorMATH}\ensuremath{{\begingroup\renewcommand\colorMATH{\colorMATHB}\renewcommand\colorSYNTAX{\colorSYNTAXB}{{\color{\colorMATH}\ensuremath{\sv}}}\endgroup } \in  Atom\llbracket (x\mathrel{:}\tau '_{1}\mathord{\cdotp }{\begingroup\renewcommand\colorMATH{\colorMATHB}\renewcommand\colorSYNTAX{\colorSYNTAXB}{{\color{\colorMATH}\ensuremath{\sss'}}}\endgroup }) \xrightarrowS {{\begingroup\renewcommand\colorMATH{\colorMATHB}\renewcommand\colorSYNTAX{\colorSYNTAXB}{{\color{\colorMATH}\ensuremath{\sS'_{1}}}}\endgroup }} \tau '_{2}\rrbracket }}}, which is direct.

		Suppose {{\color{\colorMATH}\ensuremath{{\begingroup\renewcommand\colorMATH{\colorMATHB}\renewcommand\colorSYNTAX{\colorSYNTAXB}{{\color{\colorMATH}\ensuremath{\sv}}}\endgroup } = \langle {\begingroup\renewcommand\colorMATH{\colorMATHC}\renewcommand\colorSYNTAX{\colorSYNTAXC}{{\color{\colorMATH}\ensuremath{\plambda}}}\endgroup } x:\tau _{1}\mathord{\cdotp }{\begingroup\renewcommand\colorMATH{\colorMATHB}\renewcommand\colorSYNTAX{\colorSYNTAXB}{{\color{\colorMATH}\ensuremath{\sss}}}\endgroup }. {\begingroup\renewcommand\colorMATH{\colorMATHB}\renewcommand\colorSYNTAX{\colorSYNTAXB}{{\color{\colorMATH}\ensuremath{\se}}}\endgroup }, \gamma \rangle }}}.
		Let {{\color{\colorMATH}\ensuremath{{\begingroup\renewcommand\colorMATH{\colorMATHB}\renewcommand\colorSYNTAX{\colorSYNTAXB}{{\color{\colorMATH}\ensuremath{\sv'}}}\endgroup } \in  {\mathcal{V}}\llbracket \tau '_{1}\rrbracket }}}, we have to prove that
		{{\color{\colorMATH}\ensuremath{\gamma [x \mapsto  {\begingroup\renewcommand\colorMATH{\colorMATHB}\renewcommand\colorSYNTAX{\colorSYNTAXB}{{\color{\colorMATH}\ensuremath{\sv'}}}\endgroup }] \vdash  {\begingroup\renewcommand\colorMATH{\colorMATHB}\renewcommand\colorSYNTAX{\colorSYNTAXB}{{\color{\colorMATH}\ensuremath{\se}}}\endgroup } \in  {\mathcal{E}}\llbracket \tau '_{2}\rrbracket }}}, i.e. that {{\color{\colorMATH}\ensuremath{\gamma [x \mapsto  {\begingroup\renewcommand\colorMATH{\colorMATHB}\renewcommand\colorSYNTAX{\colorSYNTAXB}{{\color{\colorMATH}\ensuremath{\sv'}}}\endgroup }] \vdash  {\begingroup\renewcommand\colorMATH{\colorMATHB}\renewcommand\colorSYNTAX{\colorSYNTAXB}{{\color{\colorMATH}\ensuremath{\se}}}\endgroup } \Downarrow  {\begingroup\renewcommand\colorMATH{\colorMATHB}\renewcommand\colorSYNTAX{\colorSYNTAXB}{{\color{\colorMATH}\ensuremath{\sv''}}}\endgroup }}}} and {{\color{\colorMATH}\ensuremath{{\begingroup\renewcommand\colorMATH{\colorMATHB}\renewcommand\colorSYNTAX{\colorSYNTAXB}{{\color{\colorMATH}\ensuremath{\sv''}}}\endgroup } \in  {\mathcal{V}}\llbracket \tau '_{2}\rrbracket }}}.
		By induction hypothesis we know that {{\color{\colorMATH}\ensuremath{{\begingroup\renewcommand\colorMATH{\colorMATHB}\renewcommand\colorSYNTAX{\colorSYNTAXB}{{\color{\colorMATH}\ensuremath{\sv'}}}\endgroup } \in  {\mathcal{V}}\llbracket \tau _{1}\rrbracket }}}, and by
		{{\color{\colorMATH}\ensuremath{{\begingroup\renewcommand\colorMATH{\colorMATHB}\renewcommand\colorSYNTAX{\colorSYNTAXB}{{\color{\colorMATH}\ensuremath{\sv}}}\endgroup } \in  {\mathcal{V}}\llbracket (x\mathrel{:}\tau _{1}\mathord{\cdotp }{\begingroup\renewcommand\colorMATH{\colorMATHB}\renewcommand\colorSYNTAX{\colorSYNTAXB}{{\color{\colorMATH}\ensuremath{\sss}}}\endgroup }) \xrightarrowS {{\begingroup\renewcommand\colorMATH{\colorMATHB}\renewcommand\colorSYNTAX{\colorSYNTAXB}{{\color{\colorMATH}\ensuremath{\sS_{1}}}}\endgroup }} \tau _{2}\rrbracket }}} we know that
		{{\color{\colorMATH}\ensuremath{\gamma [x \mapsto  {\begingroup\renewcommand\colorMATH{\colorMATHB}\renewcommand\colorSYNTAX{\colorSYNTAXB}{{\color{\colorMATH}\ensuremath{\sv'}}}\endgroup }] \vdash  {\begingroup\renewcommand\colorMATH{\colorMATHB}\renewcommand\colorSYNTAX{\colorSYNTAXB}{{\color{\colorMATH}\ensuremath{\se}}}\endgroup } \in  {\mathcal{E}}\llbracket \tau _{2}\rrbracket }}}. 
		This means that {{\color{\colorMATH}\ensuremath{\gamma [x \mapsto  {\begingroup\renewcommand\colorMATH{\colorMATHB}\renewcommand\colorSYNTAX{\colorSYNTAXB}{{\color{\colorMATH}\ensuremath{\sv'}}}\endgroup }] \vdash  {\begingroup\renewcommand\colorMATH{\colorMATHB}\renewcommand\colorSYNTAX{\colorSYNTAXB}{{\color{\colorMATH}\ensuremath{\se}}}\endgroup } \Downarrow  {\begingroup\renewcommand\colorMATH{\colorMATHB}\renewcommand\colorSYNTAX{\colorSYNTAXB}{{\color{\colorMATH}\ensuremath{\sv''}}}\endgroup }}}} and {{\color{\colorMATH}\ensuremath{{\begingroup\renewcommand\colorMATH{\colorMATHB}\renewcommand\colorSYNTAX{\colorSYNTAXB}{{\color{\colorMATH}\ensuremath{\sv''}}}\endgroup } \in  {\mathcal{V}}\llbracket \tau _{2}\rrbracket }}}.
		But as {{\color{\colorMATH}\ensuremath{\tau _{2} <: \tau '_{2}}}} by induction hypothesis 
		{{\color{\colorMATH}\ensuremath{{\begingroup\renewcommand\colorMATH{\colorMATHB}\renewcommand\colorSYNTAX{\colorSYNTAXB}{{\color{\colorMATH}\ensuremath{\sv''}}}\endgroup } \in  {\mathcal{V}}\llbracket \tau '_{2}\rrbracket }}} and the result holds.
	\end{subproof}
\item  {{\color{\colorMATH}\ensuremath{\tau  = \tau _{1} \mathrel{^{{\begingroup\renewcommand\colorMATH{\colorMATHB}\renewcommand\colorSYNTAX{\colorSYNTAXB}{{\color{\colorMATH}\ensuremath{\varnothing }}}\endgroup }}\oplus ^{{\begingroup\renewcommand\colorMATH{\colorMATHB}\renewcommand\colorSYNTAX{\colorSYNTAXB}{{\color{\colorMATH}\ensuremath{\varnothing }}}\endgroup }}} \tau _{2}}}}
	\begin{subproof} 
		Then {{\color{\colorMATH}\ensuremath{\tau ' = \tau '_{1} \mathrel{^{{\begingroup\renewcommand\colorMATH{\colorMATHB}\renewcommand\colorSYNTAX{\colorSYNTAXB}{{\color{\colorMATH}\ensuremath{\varnothing }}}\endgroup }}\oplus ^{{\begingroup\renewcommand\colorMATH{\colorMATHB}\renewcommand\colorSYNTAX{\colorSYNTAXB}{{\color{\colorMATH}\ensuremath{\varnothing }}}\endgroup }}} \tau '_{2}}}}, for some {{\color{\colorMATH}\ensuremath{\tau _{1} <: \tau '_{1}}}} and {{\color{\colorMATH}\ensuremath{\tau _{2} <: \tau '_{2}}}}.
		We know that 
		{{\color{\colorMATH}\ensuremath{{\begingroup\renewcommand\colorMATH{\colorMATHB}\renewcommand\colorSYNTAX{\colorSYNTAXB}{{\color{\colorMATH}\ensuremath{\sv}}}\endgroup } \in  {\mathcal{V}}\llbracket \tau _{1} \mathrel{^{{\begingroup\renewcommand\colorMATH{\colorMATHB}\renewcommand\colorSYNTAX{\colorSYNTAXB}{{\color{\colorMATH}\ensuremath{\varnothing }}}\endgroup }}\oplus ^{{\begingroup\renewcommand\colorMATH{\colorMATHB}\renewcommand\colorSYNTAX{\colorSYNTAXB}{{\color{\colorMATH}\ensuremath{\varnothing }}}\endgroup }}} \tau _{2}\rrbracket }}} and we have to prove that
		{{\color{\colorMATH}\ensuremath{{\begingroup\renewcommand\colorMATH{\colorMATHB}\renewcommand\colorSYNTAX{\colorSYNTAXB}{{\color{\colorMATH}\ensuremath{\sv}}}\endgroup } \in  {\mathcal{V}}\llbracket \tau '_{1} \mathrel{^{{\begingroup\renewcommand\colorMATH{\colorMATHB}\renewcommand\colorSYNTAX{\colorSYNTAXB}{{\color{\colorMATH}\ensuremath{\varnothing }}}\endgroup }}\oplus ^{{\begingroup\renewcommand\colorMATH{\colorMATHB}\renewcommand\colorSYNTAX{\colorSYNTAXB}{{\color{\colorMATH}\ensuremath{\varnothing }}}\endgroup }}} \tau '_{2}\rrbracket }}}.
		Suppose {{\color{\colorMATH}\ensuremath{{\begingroup\renewcommand\colorMATH{\colorMATHB}\renewcommand\colorSYNTAX{\colorSYNTAXB}{{\color{\colorMATH}\ensuremath{\sv}}}\endgroup } = \inl^{\tau ''_{2}}\hspace*{0.33em}{\begingroup\renewcommand\colorMATH{\colorMATHB}\renewcommand\colorSYNTAX{\colorSYNTAXB}{{\color{\colorMATH}\ensuremath{\sv'}}}\endgroup }}}} (the other case is analogous).
		We know {{\color{\colorMATH}\ensuremath{\tau ''_{2} <: \tau _{2}}}}, and as {{\color{\colorMATH}\ensuremath{\tau _{2} <: \tau '_{2}}}}, then {{\color{\colorMATH}\ensuremath{\tau ''_{2} <: \tau '_{2}}}}.
		First we have to prove that {{\color{\colorMATH}\ensuremath{{\begingroup\renewcommand\colorMATH{\colorMATHB}\renewcommand\colorSYNTAX{\colorSYNTAXB}{{\color{\colorMATH}\ensuremath{\sv}}}\endgroup } \in  Atom\llbracket \tau '_{1} \mathrel{^{{\begingroup\renewcommand\colorMATH{\colorMATHB}\renewcommand\colorSYNTAX{\colorSYNTAXB}{{\color{\colorMATH}\ensuremath{\varnothing }}}\endgroup }}\oplus ^{{\begingroup\renewcommand\colorMATH{\colorMATHB}\renewcommand\colorSYNTAX{\colorSYNTAXB}{{\color{\colorMATH}\ensuremath{\varnothing }}}\endgroup }}} \tau '_{2}\rrbracket }}}, which is direct.

		We know that {{\color{\colorMATH}\ensuremath{{\begingroup\renewcommand\colorMATH{\colorMATHB}\renewcommand\colorSYNTAX{\colorSYNTAXB}{{\color{\colorMATH}\ensuremath{\sv'}}}\endgroup } \in  {\mathcal{V}}\llbracket \tau _{1}\rrbracket }}}, then by induction hypothesis we know 
		that {{\color{\colorMATH}\ensuremath{{\begingroup\renewcommand\colorMATH{\colorMATHB}\renewcommand\colorSYNTAX{\colorSYNTAXB}{{\color{\colorMATH}\ensuremath{\sv'}}}\endgroup } \in  {\mathcal{V}}\llbracket \tau '_{1}\rrbracket }}}, therefore 
		{{\color{\colorMATH}\ensuremath{\inl^{\tau ''_{2}}\hspace*{0.33em}{\begingroup\renewcommand\colorMATH{\colorMATHB}\renewcommand\colorSYNTAX{\colorSYNTAXB}{{\color{\colorMATH}\ensuremath{\sv'}}}\endgroup } \in  {\mathcal{V}}\llbracket \tau '_{1} \mathrel{^{{\begingroup\renewcommand\colorMATH{\colorMATHB}\renewcommand\colorSYNTAX{\colorSYNTAXB}{{\color{\colorMATH}\ensuremath{\varnothing }}}\endgroup }}\oplus ^{{\begingroup\renewcommand\colorMATH{\colorMATHB}\renewcommand\colorSYNTAX{\colorSYNTAXB}{{\color{\colorMATH}\ensuremath{\varnothing }}}\endgroup }}} \tau '_{2}\rrbracket }}} and the result holds.
	\end{subproof}
\end{enumerate}
\end{proof}

Type safety for open terms follows immediately.

\begin{corollary}[Type Safety and Normalization of $\lang$]\;
  \label{lm:type-preservation-sensitivity}
  \begin{enumerate}[label=(\alph*)]
	  \item Let {{\color{\colorMATH}\ensuremath{\vdash  {\begingroup\renewcommand\colorMATH{\colorMATHB}\renewcommand\colorSYNTAX{\colorSYNTAXB}{{\color{\colorMATH}\ensuremath{\se}}}\endgroup } : \tau  ; \varnothing }}}, then 
		  {{\color{\colorMATH}\ensuremath{\vdash  {\begingroup\renewcommand\colorMATH{\colorMATHB}\renewcommand\colorSYNTAX{\colorSYNTAXB}{{\color{\colorMATH}\ensuremath{\se}}}\endgroup } \Downarrow  {\begingroup\renewcommand\colorMATH{\colorMATHB}\renewcommand\colorSYNTAX{\colorSYNTAXB}{{\color{\colorMATH}\ensuremath{\sv}}}\endgroup }}}}, and {{\color{\colorMATH}\ensuremath{\vdash  {\begingroup\renewcommand\colorMATH{\colorMATHB}\renewcommand\colorSYNTAX{\colorSYNTAXB}{{\color{\colorMATH}\ensuremath{\sv}}}\endgroup }: \tau ';\varnothing }}}, where
		  {{\color{\colorMATH}\ensuremath{\tau ' <: \tau }}}.
	  \item Let {{\color{\colorMATH}\ensuremath{\vdash  {\begingroup\renewcommand\colorMATH{\colorMATHC}\renewcommand\colorSYNTAX{\colorSYNTAXC}{{\color{\colorMATH}\ensuremath{\pe}}}\endgroup } : \tau  ; \varnothing }}}, then 
		  {{\color{\colorMATH}\ensuremath{\vdash  {\begingroup\renewcommand\colorMATH{\colorMATHC}\renewcommand\colorSYNTAX{\colorSYNTAXC}{{\color{\colorMATH}\ensuremath{\pe}}}\endgroup } \Downarrow  {\begingroup\renewcommand\colorMATH{\colorMATHB}\renewcommand\colorSYNTAX{\colorSYNTAXB}{{\color{\colorMATH}\ensuremath{\sv}}}\endgroup }}}}, and {{\color{\colorMATH}\ensuremath{\vdash  {\begingroup\renewcommand\colorMATH{\colorMATHB}\renewcommand\colorSYNTAX{\colorSYNTAXB}{{\color{\colorMATH}\ensuremath{\sv}}}\endgroup }: \tau ';\varnothing }}}, where
		  {{\color{\colorMATH}\ensuremath{\tau ' <: \tau }}}.
  \end{enumerate}
\end{corollary}
\begin{proof}
Direct consequence of Prop~~\ref{lm:type-safety-FP}.
\end{proof}

% \clearpage
% \section{$\lang$: Type Safety}
% \input{type-preservation}

\end{document}